\documentclass[11pt]{gsasthesis} 

\usepackage{etex} 

\usepackage[margin={1.2in}]{geometry}

\usepackage[titletoc]{appendix}

\usepackage{rotating}

\usepackage{longtable}

\usepackage{natbib}

\usepackage[sc]{mathpazo}
\usepackage{courier}

\usepackage[utf8]{inputenc}
\usepackage[T1]{fontenc}

\usepackage[english]{babel}
\usepackage{blindtext}

\usepackage{amsmath}

\usepackage{microtype}

\usepackage[stable,multiple]{footmisc}

\usepackage{url}   
\usepackage{hyperref}       
\usepackage{booktabs}       
\usepackage{amsfonts}       
\usepackage{nicefrac}       
\usepackage{microtype}      
\usepackage{xcolor}         
\usepackage{authblk}
\usepackage{amsthm}
\usepackage{wrapfig}
\usepackage{algpseudocode}
\usepackage{graphicx}
\usepackage{listings}
\usepackage{cancel}
\usepackage[rightcaption]{sidecap}
\usepackage{color,soul}

\newtheorem{theorem}{Theorem} 
\newtheorem{theorem*}{Theorem} 
\newtheorem{theoremalt}{Theorem}

\usepackage{colortbl}
\definecolor{Gray}{gray}{0.8}
\definecolor{LightCyan}{rgb}{0.88,1,1}
\definecolor{bananamania}{rgb}{0.98, 0.91, 0.71}
\sethlcolor{bananamania}
\usepackage{caption} 
\usepackage{array} 
\usepackage{placeins}
\hypersetup{
    colorlinks=true,
    linkcolor=black,
    citecolor=black,      
    urlcolor=blue,
    breaklinks=true
}
\usepackage[CaptionBefore]{fltpage}
\usepackage{comment}
\usepackage{nameref}
\renewcommand\_{\textunderscore\allowbreak}
\soulregister{\_}{7}
\usepackage{tcolorbox}

\RequirePackage[font=small,format=plain,labelfont=bf,textfont=it]{caption}
\expandafter\def\expandafter\UrlBreaks\expandafter{\UrlBreaks
  \do\a\do\b\do\c\do\d\do\e\do\f\do\g\do\h\do\i\do\j%
  \do\k\do\l\do\m\do\n\do\o\do\p\do\q\do\r\do\s\do\t%
  \do\u\do\v\do\w\do\x\do\y\do\z\do\A\do\B\do\C\do\D%
  \do\E\do\F\do\G\do\H\do\I\do\J\do\K\do\L\do\M\do\N%
  \do\O\do\P\do\Q\do\R\do\S\do\T\do\U\do\V\do\W\do\X%
  \do\Y\do\Z}

\title{Behavior quantification as the missing link between fields: Tools for digital psychiatry and their role in the future of neurobiology} 
\author{Michaela M. Ennis} 
\degreename{Doctor of Philosophy}
\degreefield{Neuroscience} 
\department{The Division of Medical Sciences} 
\degreemonth{April} 
\degreeyear{2023} 
\principaladvisor{Justin T. Baker}

\secondadvisor{Jean-Jacques E. Slotine}

\begin{document}
\hfuzz=100pt 


\pagenumbering{roman} 
\setcounter{tocdepth}{4}
\setcounter{secnumdepth}{3}

\thesistitlepage
\copyrightpage

\begin{abstract}
\hspace{12pt} The great behavioral heterogeneity observed between individuals with the same psychiatric disorder and even within one individual over time complicates both clinical practice and biomedical research. However, modern technologies are an exciting opportunity to improve behavioral characterization. Existing psychiatry methods that are qualitative or unscalable, such as patient surveys or clinical interviews, can now be collected at a greater capacity and analyzed to produce new quantitative measures. Furthermore, recent capabilities for continuous collection of passive sensor streams, such as phone GPS or smartwatch accelerometer, open avenues of novel questioning that were previously entirely unrealistic. Their temporally dense nature enables a cohesive study of real-time neural and behavioral signals. 

To develop comprehensive neurobiological models of psychiatric disease, it will be critical to first develop strong methods for behavioral quantification. There is huge potential in what can theoretically be captured by current technologies, but this in itself presents a large computational challenge -- one that will necessitate new data processing tools, new machine learning techniques, and ultimately a shift in how interdisciplinary work is conducted. In my thesis, I detail research projects that take different perspectives on digital psychiatry, subsequently tying ideas together with a concluding discussion on the future of the field.

\pagebreak
\newpage

\noindent Topics covered include:
\begin{itemize}
    \item Arguments for daily free-form audio journals as an underappreciated psychiatry datatype, with example scientific uses and technical validation results for a new pipeline to extract both acoustic and linguistic features from these journals. 
    \item A guide to collection of clinical interview recordings in a large, multi-site study, documenting lessons learned from the ongoing AMPSCZ project.
    \item Results on the relationship between cognitive disorganization in psychosis and linguistic disfluency use in clinical interviews.
    \item Results from a multimodal digital psychiatry dataset collected during a deep brain stimulation trial, showing detection of salient behaviors that would have otherwise gone unnoticed -- a first of its kind blueprint.
    \item Novel stability theorems for multi-area recurrent neural networks (RNNs), applied to improve RNN performance on sequential classification benchmarks; stability is important for interpretability and safety of RNNs in sensitive use cases like medicine, while multi-area networks are an obvious next step for multimodal machine learning.
\end{itemize}
\end{abstract}

\renewcommand{\contentsname}{\protect\centering\protect\Large Contents}
\renewcommand{\listfigurename}{\protect\centering\protect\Large List of Figures}
\renewcommand{\listtablename}{\protect\centering\protect\Large List of Tables}

\tableofcontents 
\listoffigures
\listoftables

\begin{acknowledgments}
I have countless people to thank from my time in the program and the many experiences that lead up to it, including:
  \begin{itemize}
    \item First and foremost my thesis advisor, Justin Baker, for all of his guidance and support throughout the PhD process. He has impressive vision for the future, and it's been a lot of fun to hear many of his ideas over the years. More importantly he has always made ample time for me and for other lab members, and has taught me a great deal about both research and non-research topics. I learned a number of qualitative skills for understanding data within the lab that I doubt I would've gotten elsewhere, and that have already served me well in other contexts. I'm also grateful for Justin's willingness to set aside some time for enriching graduate school activities, recognizing that a PhD should be more than just working in one lab for 5 straight years. I was able to TA a grad class in our field, mentor two different high school students on mini research projects, take multiple extra classes of interest, publish a paper in a different subfield, and do an internship in a new industry during my time in the program. Even within the lab, I met many other potential research collaborators because of Justin, and spent time assisting with software for clinical operations. I ultimately received a really thorough education, so I obviously have a lot to be thankful for!
    \item Josh Salvi and Einat Leibenthal, for their mentorship on the OCD and audio/language analysis projects respectively. The clinical insights, project management tools, and writing advice they provided were essential pieces for the completion of this thesis.
    \item Jean-Jacques Slotine, for his mentorship during the last 3 years of my PhD. I'm lucky to have had the chance to work on two quite different types of research during my studies, which would not have happened without Professor Slotine. While I'm thankful for his guidance throughout the project, I'm actually most thankful for him reaching out to me in the first place. He didn't need to read the final project submissions for his course, at the height of COVID when grading was pass/fail no less. But he did, and then went out of his way to see if I would want to continue work on a related project with his group. In addition to being a great opportunity, it was very reminiscent of my undergraduate experiences with Patrick Winston, and solidified the senior engineering faculty of MIT as my gold standard for what academia should be like.
    \item The members of my DAC (Sam Gershman, Randy Buckner, and Bob Datta), for their guidance on my thesis and their overall encouragement. I didn't believe it at first when we were told that the DAC is just there to help you, but I wish I did -- because I would have approached some of my earlier meetings differently, and wouldn't have put them off to the last possible second either. My committee turned out to be an invaluable resource in keeping my thesis scope tractable, helping me to pivot as COVID disrupted certain plans, and in the end making sure my writing timeline was well-planned and the different components of my thesis fit together.
    \item Leo Kozachkov, for being an amazing co-author on the nonlinear systems side project. I didn't exactly come in with the typical background, and he very easily could've brushed me off. I doubt meeting with me improved his research progress at all the first couple of months, but if he hadn't done so I probably would've fallen off the project. In the end I think we worked quite well together, and having this sort of close collaboration with a grad student peer was an important part of the PhD for me. It's not something I felt was critical when I entered, but now I'd like to see all PhD students have such an opportunity, especially any with interdisciplinary goals.  
    \item All the past and present members of the Baker Lab, who made much of the work in this thesis possible through their efforts, from data collection to infrastructure building to manual review. Truly everyone in the lab during my PhD contributed something that enabled parts of my thesis in one way or another. Bryce Gillis, Claire Foley, and Burhan Khan in particular worked directly on data review that was included in this document. Lab members also helped keep work fun, and I especially appreciate the moral support from Eric Lin, Yoonho Chung, Joanna Tao, and Lily Jeong during the ups and downs of grad school.
    \item Phil Wolff, Zarina Bilgrami, and the many people that helped make the interview component of the AMPSCZ project happen. The DPACC workers, the Pronet tech staff (Yale IT), and the folks assisting with data monitoring and site communication have all played a pivotal role in getting things off the ground, on top of being a pleasure to work with. Phil in particular has been the unsung hero of the interview project, and for me has been an effective, kind, and overall great project manager.
    \item The other Baker Lab collaborators who have contributed to my work and education, especially Darin Dougherty's group with MGH neurotherapeutics and Alik Widge's group at UMN for their critical parts in the DBS case report. I really learned something valuable from all the collaborators I interacted with, including those with OCAR at the McLean OCD Institute and members of Louis-Philippe Morency's group at CMU.
    \item The Harvard Program in Neuroscience (PiN) faculty and staff who dedicate their time to keeping the program running smoothly and ensuring that students have a good experience and remain on track. In particular the PiN-affiliated faculty members at McLean (especially Kerry Ressler and Bill Carlezon) who work to maintain a strong relationship between the program and the hospital through many means, including coursework, mentorship, and funding programs.
    \item Lacey DeLucia and Jacob Winick, the high school students whose research projects I supervised. Mentoring students was a different and really valuable perspective on the research process to gain during PhD training, and something grossly underrated by typical program requirements/suggestions in my opinion.
    \item The MIT community, for making me feel like I never left and really keeping me motivated throughout grad school, especially in the pre-pandemic years. From serving on the 2016 reunion committees and staying in touch with many of my undergrad friends pursuing parallel PhDs, to the MIT grad students I newly bonded with and the next generation of Putzen I got to meet, MIT truly remained at the center of my social circle and a critical support structure over the last 6 years. It also provided me with numerous additional opportunities to learn and to teach, so thank you to everyone involved with EC.794 (especially Rich Fletcher and Karen Hodges) as well as all the courses I took, Harvard/MIT for having the cross-registration system in the first place, and the many other professors who took time to randomly chat with me about research. I hope MIT will continue to be an open, curious, and ultimately pretty weird place in the coming years, keeping the spirit of East Campus alive.
    \item My family, especially my parents, sister, and grandparents, for the constant support and the many fun shared experiences during these years. I'm of course very thankful that I was able to move back home during the pandemic lock downs too. 
    \item Steve DeRose and the entire women's platform tennis program at Spring Brook Country Club, for not only keeping me sane in the winter of COVID, but also introducing me to an unexpected set of grad school friends and a hobby I'll stick with.
    \item The many amazing teachers and mentors I've had over the years that got me to grad school in the first place, including Patrick Winston, Boaz Barak, Guoping Feng, Brant Peterson, Michale Fee, Peter Holt, Lee-Ming Kow, Ted Scovell, Deirdre O'Mara, and Luke De. I was extremely fortunate to have numerous great research experiences throughout high school and college. I'm especially grateful for the mentors I stayed in touch with during my PhD, or whose advice regularly popped into my head; Professor Winston in particular has been a consistent source of guidance, despite passing away in summer 2019. He was such an incredible teacher at all levels, because I often find myself both pulling ideas from papers he suggested and referencing his advice on communication, meta-science, and life.
    \item Arri Landsman-Roos and the Jaguars analytics staff, for helping me kickoff the PhD right.
    \item All the cool people I met at JS, for giving me something to genuinely look forward to after grad school.
  \end{itemize}
\end{acknowledgments}

\begin{preface}
While it is not typical to have a preface within the thesis front matter, this document is substantially longer than most theses and covers a broader range of topics than usually expected. It is intended that different audiences will read different subsets of the material, and so I want to make it abundantly clear where to find various content categories. As such, I provide two lists of sections of interest -- first parts of the thesis I consider to be highlights in terms of scientific takeaways, and then parts of the thesis that have the most practical relevance for those who will be directly following up on some of my work. \\

\noindent My highlights:
\begin{itemize}
    \item In "\nameref{ch:conclusion}", I discuss a number of problems that hinder progress in neuroscience research, with a major focus on systemic flaws in modern academic science (starting at section "\nameref{sec:real-thesis}").
    \item In "\nameref{ch:intro}", I characterize specific issues in psychiatry and introduce ways that computational research could theoretically improve on many of them, to set the stage of the work within the thesis.
    \begin{itemize}
        \item While the introductory chapter focuses primarily on diagnostic labels, I provide additional background about the strengths and limitations of gold standard clinical scale ratings in section \ref{subsec:interview-today}.
    \end{itemize}
    \item Section \ref{subsec:diary-argue-recap} presents many different arguments for the use of daily app-based audio journals for psychiatric patient speech sampling, a datatype presently neglected in most acoustic and linguistic psychiatry studies. 
    \begin{itemize}
        \item Appendix \ref{cha:append-ampscz-rant} draws on some of these arguments in making a much more specific case for shifting budget priorities in the speech sampling portion of the NIMH's AMPSCZ initiative, from structured clinical interview recording analysis to audio journal analysis. Because that work helped to actually make some change to the project plan, I consider a major contribution of the thesis.
    \end{itemize}
    \item Section \ref{subsec:diary-case-study} presents pilot case reports on three Bipolar disorder patients who participated in a longitudinal digital psychiatry study including regular audio diary submissions. It provides clear proof of concept for many of the stated benefits of the datatype. 
    \begin{itemize}
        \item The preliminary modeling results of sections \ref{subsubsec:ema-diary-verb} and \ref{subsubsec:ema-diary-all} similarly provide interesting proof of concept results for the journals, as well as underscoring the importance of accounting for subject-specific factors in digital psychiatry modeling projects more generally.
    \end{itemize}
    \item Section \ref{subsec:technical-difficulties} identifies particular challenges with applying machine learning to psychiatric data, from a more technical perspective that investigates ways machine learning advances could help overcome some of those roadblocks. 
    \item Chapter \ref{ch:4} on the whole contains a number of mathematical and empirical results with relevance to both computational neuroscience and robust/interpretable machine learning applications. However, for anyone directly in those fields, the same results are presented in much more abridged form in \citep{NIPS22}. 
    \begin{itemize}
        \item I will say I am pretty pleased with the extensive set of future directions I wrote up for this thesis in section \ref{subsec:rnn-future} though. 
    \end{itemize}
\end{itemize}

\noindent Technically relevant references:
\begin{itemize}
    \item If planning an audio journal study, the information I compiled on a potential core feature set for baseline analyses in section \ref{subsec:diary-val} could be a very useful reference, as could the extensive empirical feature characterization and discussion of tricky analysis considerations throughout section \ref{subsec:diary-dists}'s results on the Baker Lab's BLS dataset. 
    \begin{itemize}
        \item If actually using the code I wrote \citep{diarygit}, supplemental section \ref{subsec:diary-code} would also be of great relevance.
        \item If you're considering using interview recordings for speech sampling instead for some reason, the comparisons made in section \ref{subsec:interview-history} might be worth reviewing. The extensive literature review on topics related to audio journals in section \ref{sec:background2} also contains background on a variety of possible speech/language sampling datatypes and their suitability for different scientific aims. 
    \end{itemize}
    \item If involved in the AMPSCZ project, please look through sections \ref{subsec:interview-methods} and \ref{sec:tool1} for basic needed context. 
    \begin{itemize}
        \item If working directly with the interview recording data collection/monitoring, please also see supplemental section \ref{sec:ampscz-pro}.
        \item If working with interview processing code \citep{interviewgit}, please also see supplemental section \ref{subsec:interview-code} and especially \ref{subsubsec:u24-next-steps-interview}.
        \begin{itemize}
            \item For setting up the code in a new interview data collection project, \ref{subsubsec:u24-dif-project} is a good resource to refer to early, in addition to looking through the rest of the mentioned code documentation.
        \end{itemize}
    \end{itemize}
    \item To use our stable multi-area recurrent neural network architecture for sequence classification tasks, as described in chapter \ref{ch:4}, see the code base at \citep{sparsegit}.
\end{itemize}

\end{preface}

\pagenumbering{arabic} 

\chapter*{Introduction}\label{ch:intro}
\addcontentsline{toc}{chapter}{Introduction}
\renewcommand\thefigure{I.\arabic{figure}}    
\setcounter{figure}{0}  
\renewcommand\thetable{I.\arabic{table}}    
\setcounter{table}{0}  
\renewcommand\thesection{I.\arabic{section}} 
\setcounter{section}{0}

Mental illness is an extremely prevalent issue, with US estimates of lifetime DSM diagnosis at nearly $50\%$ of the United States population, and over $25\%$ qualifying in any given year. Beyond recorded diagnoses, it is thought that there are a large number of sub-threshold manifestations of disorders, particularly for those categories with stringent DSM criteria. For example, there are likely more cases of sub-threshold OCD in the population than diagnosed OCD. While such cases may not cause severe impairment, they still have a negative impact on quality of life and are not yet being addressed \citep{Kessler2009}. 

Besides the great human cost, there is also an enormous economic cost of psychiatric disease. Globally, and considering both direct (e.g. healthcare) and indirect (e.g. productivity loss) costs, the annual economic impact of mental disorders was estimated at \$2.5 trillion USD in 2010. Unlike many other illness categories, such as cancer or heart disease, the indirect costs of mental illness accounted for the majority of the total costs, at $\sim 68\%$ \citep{Trautmann2016}.

Additionally, mental health problems appear to be increasing in prevalence, which is reflected in population health statistics such as life expectancy. Between 2008 and 2018, mortality rates for those ages 15-44 in the US increased, a robust effect across genders and racial/ethnic groups. This increase was in large part driven by an increase in suicides and drug overdoses, "deaths of despair" with a close link to mental illness \citep{Harper2021}. While the COVID19 pandemic has since further complicated mortality statistics, there is evidence from a number of countries that drug overdose deaths further increased during the pandemic \citep{Ardabili2022}. 

Despite the critical importance of addressing rising psychiatric illness rates and improving overall population mental health, there remain large gaps in seeking, availability, and efficacy of existing treatments. In 2020, less than half of Americans with a diagnosed mental illness received any form of treatment \citep{NIMHsurvey}. For each of Schizophrenia, Major Depressive Disorder, and OCD -- diseases of relevance to the data used in this thesis -- estimates indicate that at least $20\%$ of diagnosed patients who seek treatment are resistant to existing options \citep{Howes2022}. In order to advance treatment options, there is a deep need for improved mechanistic understanding of the underpinnings of various mental disorders.   

\section{Challenges in psychiatry}
\label{intro:psychiatry-problems}
\begin{figure}[h]
\centering
\includegraphics[width=0.75\textwidth,keepaspectratio]{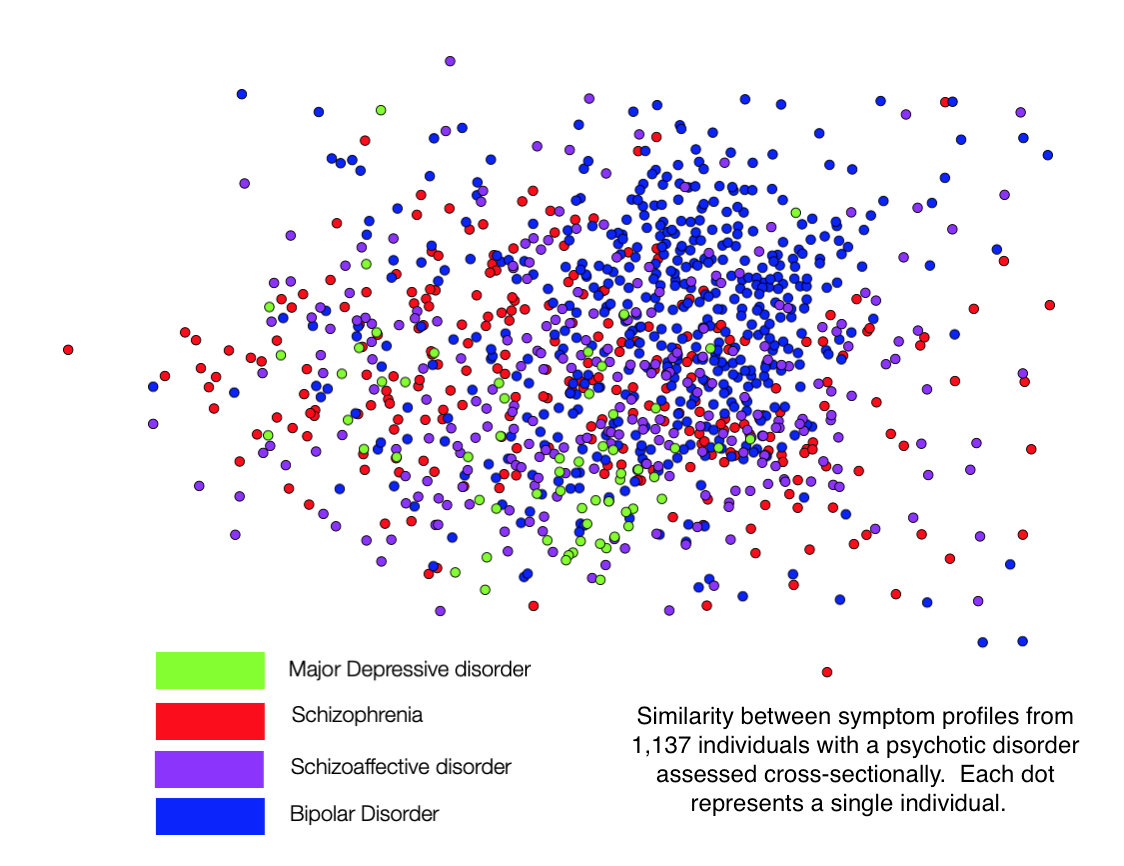}
\caption[Great heterogeneity observed in psychiatric symptoms between patients.]{\textbf{Great heterogeneity observed in psychiatric symptoms between patients.} One-time clinical scale assessments were collected from 1137 individuals diagnosed with Major Depressive Disorder (green), Schizophrenia (red), Schizoaffective Disorder (purple), or Bipolar Disorder (blue). The same questions were scored for every participant, and covered multiple different scales in common use for mood or psychotic disorders. Item-level scores across the participants were used to perform dimensionality reduction, creating the depicted 2D plot where each point represents a single patient. The point coordinates are based only on properties of the gold standard clinical scale ratings and not the diagnostic label, which is indicated by color. It is clear that there is a large amount of variability in symptom profiles between patients with the same diagnosis, especially the Schizophrenia Spectrum Disorders. Moreover, there is a large amount of overlap between all four categories, suggesting that the use of diagnosis alone as a label for research has major limitations. [Figure reproduced from Baker Lab slides, with advisor permission]}
\label{fig:intro-hetero}
\end{figure}

Indeed, to improve outcomes for current patients and develop treatments, the state of psychiatric disease diagnosis and evaluation needs to improve. Current criteria produce labels that are inconsistent between clinicians, unable to predict treatment outcomes, and difficult to track over time. This is due to a number of factors, including unavailability of objective or continuous behavioral data, and poorly defined disorder categories - which allow patients with disjoint symptoms to be classified as the same, while simultaneously having many overlapping symptoms between different diseases (Figure \ref{fig:intro-hetero}).

\begin{figure}[ht]
\centering
\includegraphics[width=0.75\textwidth,keepaspectratio]{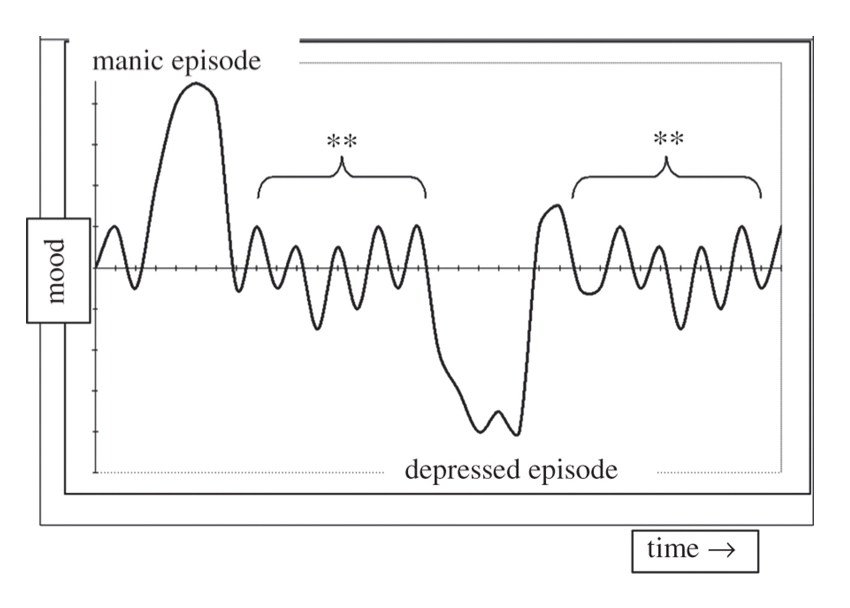}
\caption[Longitudinal fluctuations observed in psychiatric symptoms within patients.]{\textbf{Longitudinal fluctuations observed in psychiatric symptoms within patients.} This illustration depicts an example of how perceived mood (y-axis) can fluctuate in Bipolar Disorder over time (x-axis). Patients may experience sustained periods of extreme mood, with positive (manic) or negative (depressive) valence. However, while these episodes can be devastating, they are not perpetual; many people with Bipolar Disorder actually spend the majority of their life in a state with normal fluctuations of mood. Despite this level of heterogeneity that can exist over time in a single individual, longitudinal research remains in the minority of studies, and there is a real paucity of work seeking to understand the potential temporal dynamics present. [Figure reproduced from Baker Lab slides, with advisor permission]}
\label{fig:intro-long}
\end{figure}

Variation in behavior provides opportunity to better understand the brain, but without properly characterizing these behaviors, incorrect conclusions can easily be drawn. Heterogeneity between and longitudinal fluctuations within human subjects especially make the development of robust behavioral analyses critical to both research and medicine \citep{Fisher2018}. "Snapshot" studies that collect data from each individual at a single timepoint can be subject to a large amount of noise from both types of heterogeneity (Figure \ref{fig:intro-long}). Particularly because symptom severity measures are often weakly defined and temporally sparse in psychiatry, quantifying deviation in longitudinal measures of behavior could help ground otherwise difficult "gold standard" prediction tasks.

Indeed, the level of comorbidity in psychiatric diagnoses might seem shocking at first glance, but given the above limitations with existing methods it becomes entirely unsurprising. In the US, it is estimated that nearly $45\%$ of people that meet the diagnostic criteria for one psychiatric disorder in a given year will also meet the criteria for a different psychiatric disorder in the same year. Moreover, comorbid diagnosis of two different disorders is often associated with more severe symptoms in both \citep{Loo2015}. Even many of the clinical scales that are well-supported in most aspects of reliability and validity, such as the YBOCS for OCD, demonstrate poor discriminant validity \citep{Woody1995}, and further have minimal predictive power for treatment response \citep{Fontenelle2019}. 

So while it is true that technically you can build a large enough sample size to power through the noise inherent in human heterogeneity and compounded by the flaws in the present psychiatric diagnostic system, the most likely \emph{good} outcome of that is another option that can be added to the roulette wheel of trial and error treatment planning. \\

\FloatBarrier

\noindent The NIMH's Research Domain Criteria (RDoC) initiative, launched over a decade ago to reframe the study of psychiatry as a precision medicine, is an ongoing attempt to address these problems -- noting the progress made in other biomedical domains such as infectious diseases once their defining characteristics were shifted from presenting signs to underlying biology \citep{Insel2010}. However, progress in characterizing biological underpinnings or otherwise identifying psychiatric disease biomarkers has remained slow, while many of these attempts have neglected to deeply study behavior \citep{Torous2017}. The studies have also largely not been longitudinal, even though there have been calls for a greater focus on longitudinal data throughout much of the history of psychiatry \citep{Murray1985}.

Ultimately, it is probably the case that there are multiple biological pathways that can cause extremely similar behavioral presentations when disturbed. It is also probably the case that the same biological pathway abnormality can lead to quite different behavioral presentations in different people. In fact it has been noted for a long time that there is heterogeneity in behavioral responses to a particular physical disease \citep{Murray1985}. Neuropsychiatry is uniquely difficult because of the need to disentangle these factors at once, yet with a lot still unknown about each in isolation. Nevertheless, we very likely have a third complicating factor in current research, which is that development of technology for neurobiological measurements has clearly outpaced that for behavioral quantification. This has resulted in a research environment with focus that is too strongly on only the neurobiology. \\

There are almost certainly entire classes of diseases that are presently lumped together, but theoretically could be split apart by quantifiable behavior alone, such that each partition contains some set of "true" biological disorders and is disjoint from the rest. An egregious example is the Autism Spectrum, for which this is commonly acknowledged and even intended to be conveyed by the use of "spectrum". Yet in the neuroscientific literature, from animal models all the way to clinical trials, it is rare to find comprehensive behavioral information reported. Instead, the vast majority of reported results fall into one of the following categories: 

\begin{quote}
{\small A robust and interesting result, but it is specific to a rare and severe Autism subtype that is caused by a known genetic mutation}
\end{quote}

\begin{quote}
OR
\end{quote}

\begin{quote}
{\small A statistically significant and interesting result, but its effect size is small to moderate with high variance in the experimental group}
\end{quote}

\noindent AND that is forgetting about all the negative results that go unreported. Therefore there is a broader group under the Autism (ASD) umbrella that remains poorly characterized, and in my opinion will stay that way until there is a shift in the approaches primarily being employed by the neuroscience community. 

Simultaneous with the heterogeneity observed amongst individuals diagnosed with ASD, there is also concern about the discriminant validity of the ASD label. When adolescents diagnosed with ASD were compared to those diagnosed with ADHD or OCD, phenotypic overlap between labels was observed \citep{Kushki2019} quite similar to that seen with the mood and psychotic disorders discussed above (Figure \ref{fig:intro-hetero}). 

\cite{Kushki2019} used total scores on the "Social Communication Questionnaire" and the "Strengths and Weaknesses of ADHD-symptoms and Normal Behavior rating scale" along with cortical thickness measurements from 7 brain regions to cluster 226 adolescents with a primary diagnosis of ASD ($n=112$), ADHD ($n=58$), or OCD ($n=34$), or no developmental nor psychiatric diagnosis ($n=22$). Both clinical scale values differed significantly across the resulting 10 clusters, yet the diagnostic labels were not well predicted by cluster, with normalized mutual information $< 0.2$. Just one cluster, containing 13 ASD participants, was specific to a single diagnosis, while half of the clusters contained at least one participant from every category including healthy controls.   

The point of this research was not to identify new diagnostic categories, but instead to use clustering of clinically relevant variables to assess the quality of the existing labels, which did not appear promising. As primary diagnosis was confirmed by an independent clinician at the study site before enrollment, it is unlikely that these results are explainable by poor implementation of existing guidelines, though that may also be a systemic concern. Instead, evidence of unsatisfactory definitions in psychiatry continues to grow, while case/control study designs using these definitions remain prevalent. \\

Given the continuing increases in and overall large amount of funding for Autism research relative to many other DSM disorders [NIH Annual RCDC Funding Estimates], the seeming stagnation of behavioral characterization methodologies employed for these studies reflects a genuinely concerning outlook for progress in psychiatry more generally.

Despite the problems though, there is reason for optimism -- many of these shortcomings in behavior quantification were infeasible to address with the existing technologies in prior decades. Although many psychiatry researchers over the years have wanted to see more frequent clinical information covering a wider variety of symptom profiles, such a system would of course be extremely tedious with manual entry. A huge number of new avenues have thus opened up only relatively recently.

\cite{Kushki2019} argue based on their results for a research model that treats individual symptom measures as continuous outcome variables, rather than centering on categorical labels determined from a broad range of behaviors. Such an approach would benefit greatly from development of new ways to assess specific symptoms. Thus modern technological advances provide the perfect opportunity for a scientific paradigm shift in response to rising concerns about the state of psychiatry.

\section{Intersections between technology and psychiatry}
\label{sec:tie-together}
As discussed, the current inaccuracies and inconsistencies in psychiatric disease diagnosis and monitoring have a significant impact on current patients and future research; ability to reasonably classify a disorder and its current state is necessary for constructing a complete clinical model and likely for constructing a sufficiently descriptive neurobiological one as well. Up until recently, many questions about behavior were unrealistic to carefully address either longitudinally or at scale, but modern technological advances have presented the opportunity to develop a huge swath of new behavioral measures. 

The value of daily health diaries for self-report symptom tracking has been supported for a long time by both public health and psychiatry research from both the US and abroad \citep{Allen1954,Murray1985,Shepherd1991}. However, all of these studies also lamented the tedious and costly nature of collecting such diaries for the researcher, as well as the difficulty in sustaining participation levels over time with the patients. At the time, these concerns were a simple fact about daily self-reporting, but with the ubiquity of smartphones and the development of apps such as Beiwe \citep{Beiwe}, it is now possible to administer electronic daily surveys at scale and automatically curate the responses. 

Similarly, it is now not only possible to conduct clinical interviews remotely, but also to automatically generate estimates for behavioral signs that psychiatrists often look for, such as rate of speech or amount of fidgeting. Transcripts of interviews collected over time can be searched for content of interest as well as associated with simultaneous output from facial expression or acoustic analysis software. By combining clinician intuition with data science methods, this technology has great potential to both build on the existing strengths of psychiatry and discover unexpected new directions for advancing the field. \\

While technology may revolutionize (and to some extent already has) how patient self-reporting or clinical interviews are done, passive sensing technologies represent an entirely novel domain for psychiatry research. The ability to continuously track geolocation or monitor physical activity patterns makes detailed longitudinal studies that at one time would have required literally stalking someone instead among the easiest data collection protocols. 

Digital phenotyping is primed for an explosion of relevance, precisely because there are a variety of behaviorally relevant signals that it is now straightforward to collect passively and at a high temporal resolution. In addition to all the limitations in psychiatry discussed above, it is also the case that certain neuroscientific questions about behavior can only be answered using data streams that are dense in time, which hardly exists in the realm of psychiatric human behavior. As basic neuroscience research into e.g. motor circuits heavily relies on the alignment between neural recordings and simultaneous animal behavior, digital phenotyping could one day form an important component of human neuroscience research. For it to effectively fill such a role in the future, it is critical to develop a deeper understanding of various passive behavioral sensing signals now.

Beyond its important implications for the questions that future human neuroscience studies will be able to answer, digital phenotyping also has an immediate relevance to psychiatry and neuroscience research. There are a variety of low hanging fruit applications for direct use of sensor data, such as detection of sustained change in the amount of time spent at home or the frequency of phone unlocks/screen time during sleep hours. There are also a huge number of more complicated behavioral characteristics that could theoretically be extracted from these data, such as detection of specific activities of interest or correlational analyses that could lead to insights like those gained from resting state brain imaging. Through considering spontaneous fluctuations of correlated behavioral systems, we can begin to unpack how behaviors are connected to each other as well as eventual clinical outcomes. 

Although advances in hardware development of passive sensor technology will of course play a role in the future of digital phenotyping, there are already a number of relevant datatypes that can be collected with relative ease compared to more traditional outcome measures. The greatest challenge will be in establishing the quantitative methods necessary for processing and learning from these data to the fullest extent, and ultimately integrating them with existing clinical practices and biological results. 

\section{Thesis vision and aims}
\label{sec:aims}
Eventually, proposed digital psychiatry work can pave the way for a variety of advances, including:
\begin{itemize}
    \item Denser and more robust behavioral endpoints to be used in studies of biomarkers for psychiatric disease.
    \item Better methods to evaluate treatment strategies, both on a high level, and for parameter tuning in applications such as closed-loop deep brain stimulation.
    \item Tools for clinicians or patients to visualize progress and identify downward trends in behavioral markers, improving disease diagnosis and monitoring.
    \item Scalable methods for quantifying symptom serverity, to improve mental healthcare access for underserved groups.
    \item Ways for healthy individuals to improve lifestyle, such as identification of stressors.
    \item Development of novel non-invasive/non-pharmacological interventions that utilize modern tech to directly affect behavior. 
\end{itemize}
\noindent Critically, work in these domains can both occur in parallel and inform each other. \\

\begin{figure}[h]
\centering
\includegraphics[width=\textwidth,keepaspectratio]{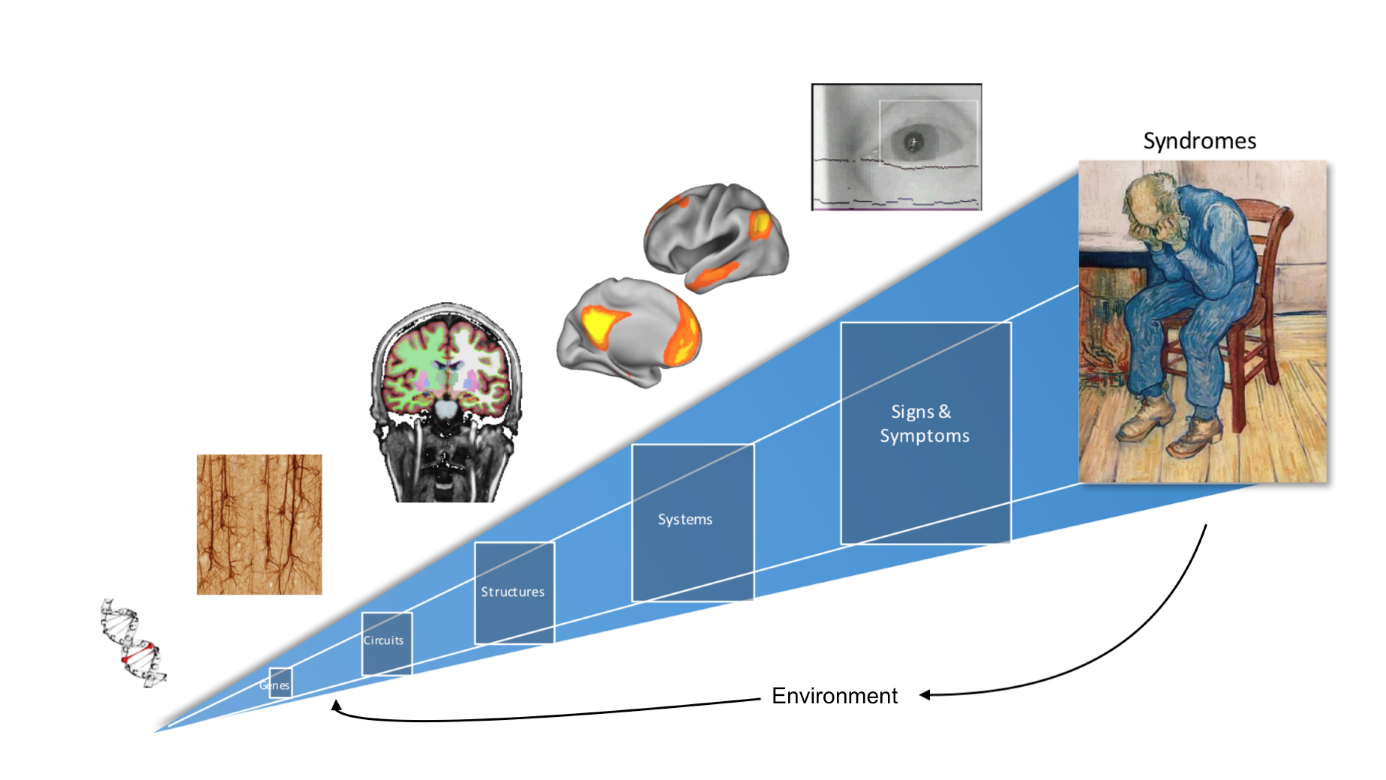}
\caption[Psychiatry research must consider many levels of abstraction.]{\textbf{Psychiatry research must consider many levels of abstraction.} Psychiatric disorder is deeply intertwined with biology, from a molecular perspective to a systems neuroscience one. The mechanisms at all these levels of abstraction not only produce disordered behavior, but disordered behavior can in turn impact these systems, and the patient's environment also interplays with both. Although there is great value in research focused on individual levels of the depicted structure, there must be research that connects them as well. For each biological level, there are some existing impressive quantification technologies as shown. For the behavioral signs and symptoms of psychiatry, there are very promising emerging methods, yet still a lack of well-validated metrics that can be aligned with the neurobiological ones. [Figure reproduced from Baker Lab slides, with advisor permission]}
\label{fig:intro-abstract}
\end{figure}

\noindent In my thesis, I will provide tools and proof of concept applications at multiple levels of abstraction for modeling psychiatric disease (Figure \ref{fig:intro-abstract}), to lay groundwork for future study in the domain of digital psychiatry. 

I will present new tools for behavioral data analysis along with the resulting applied data science research that was conducted on multiple behavioral datasets collected during psychiatry studies spanning a number of disorders -- including a case report from a patient with naturalistic neural recording data available from deep brain stimulation implants. As such, preliminary results from the literature on acoustic and linguistic properties are discussed in the introduction to chapter \ref{ch:1}, with additional context on automated clinical interview analysis provided in chapter \ref{ch:2}. Similarly, background on both digital phenotyping trials and the alignment of behavioral data with neurobiological data can be found in chapter \ref{ch:3}. 

I will also report on the development of deep learning techniques that can obtain modern performance levels while better maintaining important properties for safety-critical applications like healthcare, as well as discuss how neuroscience theory can inform such questions in machine learning. For background on relationships between advances in deep learning and advances in neuroscience, see chapter \ref{ch:4}. \\

\FloatBarrier

\noindent The specific aims of each of my chapters are as follows:
\begin{enumerate}
    \item Release of a set of tools for data monitoring and feature extraction of daily patient audio journals, including documentation and validation results for acoustics and linguistics. Couple this codebase with pilot scientific results from our Bipolar Longitudinal Study to argue for wider adoption and use of such journals.
    \item Compare and contrast research on clinical interview recordings with the preceding content on patient self-report journals, then document my code for pre-processing and quality control of audio/video recorded interviews from the large, international, multi-site AMPSCZ project that is ongoing. Present lessons from that study along with examples of impact demonstrated by the pipeline, and then report internal lab results on linguistic properties associated with disorganized thought in recorded interviews of psychotic disorders patients -- as a proof-of-concept for just one way in which the AMPSCZ dataset could later be leveraged. 
    \item Give scientific background on a novel deep brain stimulation paradigm being tested for OCD, in conjunction with case report results from a pilot patient in the stimulation trial who simultaneously contributed digital phenotyping data during the study. As this is the first report of its kind, use lessons learned to turn our work into a guide for future studies of digital phenotyping in the context of neurotherapeutics.
    \item Characterize the interplay between control theory, neuroscience theory, and machine learning, to then connect back to themes from the other chapters. Present novel mathematical and experimental results on stability of nonlinear RNNs and the recurrent combination of many nonlinear RNNs. Discuss the relevance of so called multi-area RNNs to the future of both neuroscience and machine learning, as well as the importance of stability in these models, justifying proposed future works.  
\end{enumerate} 

\noindent Given the breadth covered by these projects, note that I provide the bulk of the introductory background on each topic in the respective chapters. However, in the concluding chapter I tie the work together within a more general outlook on future directions for digital psychiatry, accompanied by commentary on the health of scientific research at large.

Ultimately, my aim is to provide foundational information on a number of lines of questioning, to contribute to future research directions in a very new subfield of science. As questions become increasingly complex and in turn increasingly interdisciplinary, it will be crucial to establish a stronger tradition of \emph{coherent} collaboration in academia. 

\section{On the connection with neurobiology}
In many ways, neurobiology research questions are akin to reverse engineering problems. Throughout engineering disciplines, an important tool for reverse engineering is the study of unintended system behaviors. This approach has been formalized as Reverse Failure Mode and Effects Analysis (RFMEA) in mechanical and electrical engineering domains \citep{Snider2008,Fyrbiak2017}, and has played a key role in many successful applications of reverse engineering to black box software systems \citep{Lorenzoli2008}. The application of similar thinking to biological problems is not a new idea; \cite{Csete2002} discussed a number of examples of failure mode analysis for elucidating biological complexity, and drew parallels to the fact that engineered systems often seem simple when working as intended but can reveal hidden implementation complexity when they break. 

The study of disease is therefore not only of practical importance for improving human well-being, but also of theoretical importance for improving our understanding of biological systems. However, study of aberrant system outputs is fundamentally limited by one's ability to measure (and interpret) those outputs. In the case of neurobiology, there is a severe lack of expressive and robust behavioral characterization tools, which in turn severely limits the "reverse engineering" that can be done. Especially because tools for probing neural activity and other biomarkers have greatly surpassed those for probing behavior, research that contributes to development of novel behavioral metrics is directly addressing a critical bottleneck currently limiting neuroscientific advancement. 

An additional complication in neuroscience that does not generally apply to reverse engineering of e.g. black box software applications, is that the internal structure of the brain can meaningfully change over time. Complex control systems are capable of demonstrating many interesting behaviors in response to their environment without ever fundamentally (and potentially irreversibly) changing their internal implementation. Of course, there are also systems that do meaningfully self-modify, and closer study of artificial neural networks is indeed one avenue for advancing this line of control theory work. Nevertheless, the fact that behavior, particularly during adolescence, can have a strong impact on underlying neurobiology, is yet another reason to increase focus on behavioral quantification in neuroscience research. \\

For my work in chapters \ref{ch:3} and \ref{ch:4}, I will directly discuss the neurobiological relevance of digital phenotyping and of theoretical recurrent network models, respectively. But a core theme throughout the thesis is that all of the topics I touch on - observable behavior, self reported internal state, simulated neural networks - are essential components to a complete understanding of the brain, and as such are components of neuroscience. It is my opinion that even without the additional neuroscience background written for chapters \ref{ch:3} and \ref{ch:4}, my thesis should unequivocally count as a neurobiology thesis.

Along these lines, the broader discussion of the state of neuroscientific research I've included in my concluding chapter will cover a variety of more specific arguments about the interdisciplinary nature of neurobiology and the need for greater diversity in what that entails. To exclude study of system-wide outputs from the set of activities that qualify as research on system $X$ would be frowned upon for most values of $X$, because it is counterproductive to scientific success. The brain has arguably the richest and most complex outputs of all, which makes it \emph{more} important for this style of work to qualify as neurobiology. To date, it has instead lead to a tendency to cast behavior aside when performing a "hard science" neurobiology study. 


\chapter{Automated analysis of daily psychiatric patient audio diaries\footnote{The foundation of this chapter is the open source code release at \citep{diarygit}, with a great deal of additional background and analyses done by me to bolster that work for my thesis. See Appendix \ref{cha:append-clarity} for detailed attributions.}}\label{ch:1}
\renewcommand\thefigure{1.\arabic{figure}}    
\setcounter{figure}{0}  
\renewcommand\thetable{1.\arabic{table}}    
\setcounter{table}{0}  
\renewcommand\thesection{1.\arabic{section}} 
\setcounter{section}{0}

New technology has introduced an abundance of new opportunities to tackle some of the fundamentally difficult problems in understanding psychiatric disease. Existing clinical scales are collected infrequently and require trained professionals to administer. Even seemingly simple tasks like collecting self-reported information on the symptoms most interfering with a patient's life are traditionally labor intensive, requiring someone who can filter out the relevant portions, record them in a clear way, and integrate any findings with the existing body of knowledge on that patient. This represents not only a major barrier in scaling care to the levels needed for our modern population, but also a limitation on the clinically relevant nuances that can be captured within a person's daily life.

With the proliferation of smartphones, it is now feasible to systematically collect daily patient self-reports via app. One particularly promising avenue for these app-based daily check-ins is the recording of audio journals. In addition to the potential for automatic cataloging of topics that are important to the patient and other personalized methods for content analysis, they can also be processed for lower level acoustic and linguistic properties that may indicate reduced functioning levels in a subconscious manner. 

While it is already straightforward to collect such journals, there is a need for tools that process them. There is a rich prior literature in acoustics and natural language processing that can be built off of, but there is currently very little work adapting and testing these tools in the context of longitudinal journal datasets, let alone with any considerations of specific interest to psychiatry. The major first step in doing so is also in line with the broader initiative recently launched by the NIH for making healthcare data "machine learning (ML) ready" - we need to understand how to monitor and preprocess incoming journals, to ensure sufficient data quality and produce validated feature sets that can be used as input for subsequent analyses. This can range from providing a small set of well-justified summary features for more traditional statistical studies to providing an organized transcript dataset for use with emerging ML techniques.

Therefore, in this chapter I will present a new tool for processing smartphone-recorded audio journals submitted daily by longitudinal study participants (note that "diary" and "journal" are used interchangeably throughout the technical sections). Our studies primarily collect recordings from psychiatric disease patients using open-ended prompts, so that will be the focus of the material here; however, the provided pipeline could just as well be applied in other study populations and/or with more targeted survey questions. I will begin in section \ref{sec:background2} with background on the use of journaling for evaluation of psychiatric disease symptoms, as well as emerging methods for automated analysis of vocal and linguistic properties in a clinical setting. I will then identify key areas of need and open questions in this space, putting my code in context in the process. \\

\noindent Because part of this code release is providing validation of and proof-of-concept results deriving from the pipeline outputs, the chapter repeatedly reports findings from our largest lab study to date, the collection of long-term and multimodal behavior data from Bipolar Disorder (BD). BD is a chronic condition characterized by mood swings with stages of both mania and depression, and general mood instability outside of episodes. It is heterogeneous in clinical presentation, response to treatment, and functional outcomes, so it is difficult to determine ideal treatment and monitoring strategies, and by extension to study. One observed trend is that the disease tends to get worse over time, with duration between episodes shortening as more episodes occur, and treatment becoming less effective after a longer duration with symptoms \citep{Salagre2018}. Therefore, early intervention is critical. 

Neurobiologically, structural abnormalities are observed in the cortex of patients, and functional hyperconnectivity in various default network modes has also been implicated. A better understanding of the different behavioral profiles and their progression over time will be critical to separate subtypes in future mechanistic research \citep{Harrison2018}. Thus the patient population for our dataset is well suited to applications of patient journal analysis. Study details are provided in section \ref{subsec:diary-methods}, to set the stage for discussion of my tool. \\

\noindent Section \ref{sec:tool2} then forms one of two cores of the chapter. Example pipeline-produced visualizations are presented first (\ref{subsec:diary-outputs}) to introduce the wide range of capabilities the tool has and further motivate this work. Next, I walk through the key metrics that the pipeline outputs and report validation results along with more general observations about using the tool (\ref{subsec:diary-val}). The most salient details about the functions of the code are interleaved within those sections, and then a detailed mechanistic write up describing both how to use (and adapt) the code and exactly how features are calculated is provided in supplemental section \ref{subsec:diary-code}. 

Once the pipeline and its outputs are well characterized, I move onto presentation of proof-of-concept scientific results in the same mood/psychotic disorders dataset, in the second core section \ref{sec:science2}. This includes a closer look at extracted feature distributions and correlation structure (\ref{subsec:diary-dists}) across the data, as well as the use of the diary features to predict ecological momentary assessment scores (\ref{subsec:diary-ema}) in a subset of participants with a primary BD diagnosis. Additionally, diaries are used along with passive sensor data streams to do a case report style investigation (\ref{subsec:diary-case-study}) into a few select patients with known periods of high symptom variation (e.g. clearly labeled depressive episodes per gold standard clinical ratings).

To tie up the chapter, section \ref{sec:discussion2} summarizes the validated pipeline use cases and pilot scientific results along with their limitations. Immediate and longer term future directions are also discussed, both for improved audio journal tool building and for the many remaining open psychiatry questions associated.

\section{Background}
\label{sec:background2}
Audio journals have a great deal of potential for improving both quantitative and qualitative tracking of psychiatric patient outcomes over time. However the apps for collecting these journals at scale, for example Beiwe \citep{Beiwe}, have existed for less than a decade, and only began to see relatively common use in research even more recently than that. Given the long duration of longitudinal study designs and the current lack of directly applicable open source tools for processing these diaries, it is not surprising that there is a paucity of existing literature to pull from. One major goal of this thesis is to argue for the utility of the audio journal datatype in a variety of research contexts, and to provide a framework to ground future work in the space. \\

\noindent To begin, I will first provide background on two topics closely related to the collection of app-based audio journals: 
\begin{enumerate}
    \item The significance of more traditional methods for patient journaling in the history of psychiatry as well as for symptom tracking diaries in public health, closing with discussion of the relation of this work to the modern audio journal (section \ref{subsec:diary-history}).
    \item The more recent major impact that both app-based data collection tools and computational tools for acoustic and linguistic processing have had on other domains, particularly when applied to other datatypes collected in psychiatry studies (section \ref{subsec:diary-lit-rev}).
\end{enumerate}
\noindent This will make clear the opportunity for said technology to capture large amounts of information in a systematic way from journals, which always had great potential but were in many ways infeasible to fully leverage historically. I will then introduce the key questions for moving the field forward and the specific aims of the present work in section \ref{subsec:diary-motivation}, to set up the rest of the chapter. 

\subsection{A historical perspective of patient journaling}
\label{subsec:diary-history}
Broadly, journaling has played a variety of roles in medicine over the years: a therapy in its own right, a tool for facilitating other interventions, and an information source for better understanding the patient. Intended prompts can vary greatly, from completely open writing to many different forms of structured worksheet, and timing can vary from infrequent long journaling sessions to daily brief ones. The intended writers can range from healthy people trying to manage daily life stress to those trying to cope with physical health issues to patients with diagnosed mental health conditions. The target readers might be a medical or social work professional, the patient at a later date, or no one at all. Thus this research has a theoretically massive scope. 

While certain methods for journaling as a therapy component or an aid to a therapy component have gained traction in modern evidence-based psychiatry, other avenues remain insufficiently explored. In particular, the use of journaling as a clinical assessment tool has been surprisingly rare in research, past or present. The utility of "health diaries" has been documented across a wide range of illnesses, both physical and mental, but these take the form of a symptom checklist much more akin to modern ecological momentary assessment (EMA) surveys than modern self-guided diary recordings.

Nevertheless, I will review literature from across the years on journaling as a therapy and on self-report surveys in medicine, to synthesize takeaways of more general relevance and motivate the new opportunities that app-collected audio journals present. 

\subsubsection{Journaling as a therapy}
Opening up about deeply personal thoughts and experiences has been an important component of psychotherapy for well over a century \citep{Freud}, and patient openness remains one of the biggest roadblocks in modern therapy \citep{Kleiven2020}. Intimate information shared by the patient is necessary for many therapy techniques, but there is reason to believe that the act of sharing this information can itself be therapeutic. From this belief came a number of applications of journaling to psychiatry.

Intensive journaling began in the 1960s with Progroff, and reached a wider audience in the 1970s via the publication of instructional textbooks on the practice. It was designed as a structured way to reflect on experiences and undergo a process of self-discovery, with multiple types of writing prompts to be worked through and build on each other \citep{Epple2007}. 

While journal therapy is sometimes still practiced as a component of therapy, this style of intensive journaling has largely fell out of fashion, and moreover there are few therapists whose primary focus is journal therapy. However, some of the most popular modern journaling techniques grew out of the intensive journaling movement as more focused offshoots. Expressive writing, for example, was formally developed by Pennebaker in the mid 1980s; it was based on the idea that opening up about past traumas could be inherently beneficial \citep{Pascoe2016}. 

\paragraph{Expressive writing.} 
Recent practices of expressive writing generally involve about 20 minutes per session of journaling on deep emotional thoughts, often centered around a particularly traumatic life event. Expressive writing sessions typically occur for only a short time period, whether as weekly journaling for a handful of months or daily journaling for a week or so. It is therefore quite different than the longitudinal open ended diaries that are the focus of this chapter, and in fact some randomized controlled trials (RCTs) of expressive writing used a prompt to write about the day as the control condition \citep{Krpan2013}. Still, the tools documented here could certainly be adapted as needed to study the content of therapeutic journals, and lessons learned from studies of therapeutic journals can apply to design of research diary prompts.

Interestingly, the effects of a few days of expressive writing have been reported to persist for weeks to months after the fact. Prior research has demonstrated a benefit to self-reported well-being and measurements of anxiety levels in healthy individuals, and somewhat more recent work on expressive writing in psychiatry has shown promised. A study of depressed individuals observed mean Beck Depression Inventory reduction of $\sim 5$ points and mean Patient Health Questionnaire-9 reduction of $\sim 2$ points in the expressive writing group compared to controls -- both $\sim 7.5\%$ reductions from respective max scale score and $\sim 15\%$ reductions from the mean baseline in the study population, where symptoms were more moderate \citep{Krpan2013}.

Other studies of expressive writing in a psychiatric population have had mixed results, and it is suggested that the benefits might be more relevant for those with moderate symptoms rather than severe mental illness \citep{Pascoe2016}. This suggests a line of questioning on the extent to which the content or style of the journals can predict clinical response, highlighting the broad utility of tools like my pipeline. \\

It is worth noting that expressive writing can have short term negative effects such as feelings of distress in some individuals. This might make it difficult to engage in the journaling practice, even if the immediate effects dissipate and in the longer term lead to benefit \citep{Baikie2005}. However as noted expressive writing may not be needed on a regular basis, but rather to reflect on specific highly emotional experiences. 

Further, the use of expressive writing in conjunction with other therapeutic methods like cognitive behavioral therapy (CBT) can help patients properly cope with negative thought processes that may arise during journaling. In practice, combining expressive writing about traumatic experiences with traditional CBT in PTSD patients has produced additional clinical improvement beyond CBT alone \citep{Pascoe2016}. \\

\paragraph{Journals in CBT.}
CBT is of course well supported by evidence to improve functioning in a number of mental illnesses, and can have positive impacts on mood and anxiety levels across many individuals, even healthy ones \citep{Beck2011}. It often includes what is arguably a form of journaling; though it is not often described in those terms it could certainly be applicable to the audio diaries framework outlined here. CBT involves working with a therapist to learn new coping mechanisms and other mental strategies for healthy processing of thoughts. As such, participants generally receive "homework" to practice these concepts in their daily life, and to track their progress so time with the therapist is most efficiently spent. CBT practitioners will give out worksheets for patients to use as a template in this process. 

Indeed, practices like the thought record are a structured form of journaling. Participants are instructed to make an entry when they notice their mood getting worse, answering questions about the situation that preceded the mood change, what their initial thoughts and associated emotional response were due to that situation, and what a healthier response to the situation using CBT principles would look like. \citep{Beck2011}. While the journaling in this instance is meant more as a tool to facilitate application of CBT concepts to the situation, particularly during the learning phase, it might additionally carry some of the discussed therapeutic effects of "journaling as an intervention". Regardless, it clearly could be leveraged in a number of ways with emerging tech. \\

There are also other styles of journaling that have independently shown clinical utility. Gratitude journaling, which is more commonly associated with CBT than expressive writing, has also been the subject of standalone RCTs \citep{Sohal2022}. The aim of gratitude journals is to focus a person's attention on the positive things in their life. These journal entries are typically shorter in duration but are to be written more often over a longer stretch of time. 

Thus gratitude journals can be directly fit into the Beiwe audio diary recording format, with a prompt like "what are you thankful for today?". Notably, gratitude prompts are also less likely to be deeply personal, a factor that may affect engagement when recording a journal out loud for a study versus writing privately. \\

\paragraph{Meta-analysis of journaling RCTs.}
A recent meta analysis of journaling interventions on diagnosed mental health conditions synthesized results from 20 peer reviewed RCTs, each of which could be broadly classified as either an expressive writing or a gratitude journal intervention. As expected there was a lot of variance in efficacy across the studies, but overall a statistically significant clinical benefit with modest effect size emerged (Cohen's $d$ estimate between 0.2 and 0.5). Curiously, there were few variables identified that explained the observed outcome variance between studies. Neither the type of journaling nor the target disorder had a significant effect, and there were also no clear patterns in demographic variables across the studies \citep{Sohal2022}. 

Interestingly, \cite{Sohal2022} found that the only statistically significant mediator in explaining some of the outcome variability amongst journaling RCTs was collection of the journals by the researchers, which occurred in about half of the considered studies. Most of these studies collected journals for basic record-keeping purposes, e.g. to reliably track participation rates, rather than for scientific analysis of the journal content. As it had a negative effect on therapeutic efficacy, the meta analysis authors suggest that patients were less open in their journaling when there was a possibility someone else would read it, thereby diminishing the therapeutic potential. 

Of course the above result is deeply relevant to app-based collection of journals, and leaves a lot to unpack about why journaling may or may not be therapeutic in different contexts. Certainly anyone implementing current journal therapies should think carefully about whether (and how) collection of the journals would enhance scientific or clinical conclusions, refraining from collection for the sake of it. However this only applies to the use of journals for an expressly therapeutic goal, and the phenomenon itself warrants closer consideration. \\

\subsubsection{Health diaries and symptom tracking}
Although journaling has well-established therapeutic potential, and journal entries are sometimes qualitatively characterized in idiographic case studies, there is very little work towards systematic analysis of diaries - whether completely open ended like the audio recordings used in this chapter or reflections on a more specific prompt like expressive writing sessions.

While filling that gap is an area of need for future work, there is both historic and current precedence for the analysis of "health diaries". These are daily journals meant to document specific medical situations. They sometimes take the form of a literal checklist for symptom tracking, as well as possibly logging relevant behaviors such as medication use, exercise routines, and dietary choices. Other times they may ask for a more detailed personal account of symptoms via free-form text. Although this latter type could fit more directly into the audio journal format described here, the former is much more common for use by a researcher or medical professional. 

Often, the goal of these health diaries is to assist the patient directly; the record of symptoms along with daily activities can be used to identify behaviors associated with worsening or improving symptoms. Having a detailed log of one's own medical history can be empowering, and reviewing progress over time with documentation can give a much more accurate picture of improvements than pure memory can.

Health diaries are at times used by clinicians, public health officials, and researchers as well. Generally health diaries intended for consumption by people besides the patient take a strict template form akin to modern EMA. This style of diary is of course more restrictive in the nuance it can feasibly capture, but it is much more quantifiable and scalable than free-form diary entries, especially in the days of pen and paper. Even then, collecting and analyzing health diaries was considered a tedious form of study in the past -- underscoring how much recent technology has enabled. 

I will now review literature on various applications of "health diary"-style journals, to synthesize takeaways on how they might be used beneficially to supplement existing therapies and how they might extend to an audio journal context for future analyses, which often goes hand in hand with administering EMA surveys. \\

\paragraph{Daily symptom logging for public health census.}
Even in the distant past, illness diaries had been tested as a public health methodology for illness/disability census, including both mental and physical ailments. A sample of the population would typically be interviewed once or a few times over a series of months about illness and symptoms experienced. Resulting statistics could vary greatly depending on whether the interviewee was asked about symptoms yesterday versus what they could recall over the last month \citep{Allen1954}. 

The public health experiment done by \cite{Allen1954} was thus to instead have a cohort keep a diary tracking illness daily, by responding to the same questions in the same order everyday in written form over the course of three months. Meanwhile an interview cohort used the standard procedure of three interviews, one per month, with those questions asked at both the "yesterday" and "in the last month" timescales. When the diaries were analyzed to determine frequency of symptoms, it was found that the census counts were between the two number produced by "yesterday" versus "in the last month". It was also found that for more debilitating illnesses with extreme symptoms, the gap between the timescales was smaller; this would be consistent with prior literature suggesting that memory decay occurs more strongly in accurately recounting less severe symptom occurrences, and it suggested that a lot of the noise in the traditional interview data is due to poor memory.

\cite{Allen1954} ultimately suggested that for pure counting purposes, the "yesterday" interview question is sufficiently accurate, but that the associated memory gaps make it much less suitable for estimating severity levels in the population than a longer time period of consideration would. The authors remark that if cost were not a factor, the diary approach would likely be preferable for accuracy of public health census. However, it was a bigger ask of participants (still true), and it was very time consuming to review (may no longer be true), making it a much more expensive procedure.

Note that because this was a survey of illness and disability more broadly, some of the conclusions may not extend to mental health. Though it is interesting to observe that what at the time is described in vague terms and considered a less severe issue might today be better understood as mental health symptoms (e.g. nervousness, headaches) -- and these are some of the symptoms that exhibited the greatest loss of clarity from infrequent interview reporting \citep{Allen1954}. Of course a lot has also changed in the intervening decades that would require a fresh look at this work. \\

\paragraph{Symptom diaries as a therapy in chronic illness.}
There are a number of more recent contexts in which health diaries have shown to be effective for clinical or personal tracking of physical health. For example, about one third of cancer patients self-monitor symptoms via diary, sometimes directly suggested by the clinician. Self-monitoring in cancer has well-documented beneficial effects on patients' ability to continue to partake in life activities on their own terms, as they gain greater knowledge of the types of symptoms they experience, their timing, and what external factors may affect them \citep{Hermansen2014}. 

The use of cancer diaries by medical professionals to assist in care planning sometimes occurs, but is less frequent than personal-only use of diaries. A survey of oncology nurses reported a common belief in the utility of self-monitoring for the patient directly, with many having heard patients directly express this sentiment. At the same time, survey results were much more mixed in whether nurses reported utilizing any of the info from patient self-monitoring themselves, or whether they thought it could be helpful in the clinical assessment process. Some even reported that the diaries led to more questions from patients that they were unequipped to answer or otherwise felt were a waste of time \citep{Hermansen2014}. 

Although tangential to the core themes of the chapter, the new research techniques described could naturally be extended to contexts like the cancer diary, allowing potentially relevant information to be uncovered without increasing burden on healthcare professionals. \\

\paragraph{Efficacy of personal sleep tracking.}
As physical health issues can have an impact on mental health issues and vice versa, there are many ways that specific physical health diaries could connect back to psychiatric health and research. For example, many disorders both physical and mental impact sleep. Sleep in turn can have predictive power for severity of future physical and mental health outcomes \citep{Nutt2008,Devnani2015,Kaskie2017}. Thus, tracking sleep is important for a variety of clinical and research problems. A common low cost and non-invasive way to do so is through sleep diaries. 

Sleep diaries are similar to symptom tracking diaries, as journalers systematically fill out a survey each morning about sleep/wake times, sleep latency, and perceived sleep quality. When kept for personal use, sleep diaries may also include free-form text prompts on topics such as perceived dreams and memory of waking events. Additionally, sleep diaries sometimes help users maintain sleep hygiene or identify activities that particularly impact their sleep quality by logging information throughout the day like caffeine intake and screen time. 

The survey questions contained in a sleep diary are regularly used by clinicians to evaluate patients with reported sleep problems before referring to more specific physiological tests, and maintaining a sleep diary during the initial phases of a new intervention is often encouraged - for example while testing a CPAP for management of sleep apnea. 

A number of research studies have employed sleep surveys to better understand population trends in perceived sleep duration and quality as they relate to demographics or other relevant health parameters. Though sleep surveys can't identify anomalies in physiological mechanisms like sleep staging, they are well-documented to align with sleep duration measurements from concurrent polysomnography or wrist actigraphy in healthy adults, often as an overestimate of $< 1$ hour \citep{Ibanez2018,Thurman2018}. However, it has been observed that the amount by which perceived sleep duration is overestimated may have variation dependent on relevant factors like demographics or total amount of sleep \citep{Patel2007,Lauderdale2008}, which also suggests care must be taken in application to psychiatric populations.

Given the relevance of sleep diaries, there is already research on app-based collection of these data through EMA prompts. In fact, the Bipolar Longitudinal Study dataset reported on in this chapter included a survey question about sleep in its EMA suite. A major benefit of the longitudinal study is that many of the research questions of interest can be asked in a manner that uses the patient as their own control. If most of the noise in sleep survey reliability can be modeled as a static function of the individual, it becomes less impactful to longitudinal EMA study results. 

Therefore the results of \cite{Lauderdale2008} require further investigation. They found that someone sleeping only around $5$ true hours would overestimate their sleep duration by $\sim 1.3$ hours on average, a difference $> 25\%$ as well as a clinically meaningful difference based on recommended sleep habits. By contrast, someone getting around $7$ true hours of sleep would overestimate the duration by just $\sim 0.3$ hours on average, a difference $< 5\%$ and within a margin of error that is entirely unsurprising for sleep self-report. Critically, \cite{Lauderdale2008} followed each participant for only 3 days, making it difficult to determine whether larger overestimates on shorter sleep sessions will generalize to most individuals when tracked over time, or if there was an external factor both limiting sleep and causing greater overestimation in the study population.

It is likely that sleep surveys will continue to be a prevalent application of EMA in the coming years, so understanding their pros and cons will be important. \\

\paragraph{ICU diaries can be a preventative measure.}
Intensive care unit (ICU) admittance is another factor related to overall health that is intertwined with psychiatry. Those admitted to the ICU have higher probability of developing a mental illness in the future; one factor associated with later psychiatric diagnosis is the extent to which memories from time in the ICU were lost. Keeping a daily diary while in the ICU is well established to decrease the severity of these memory gaps. ICU diaries typically include documentation on medical updates and daily symptom logs written from the patient's perspective, as well as information on daily happenings like family visits or watching TV \citep{Barreto2019}. They are therefore quite similar in format to the combined EMA and audio journal methodology available with Beiwe \citep{Beiwe} and similar apps. 

Because of the improvement that ICU diaries have on patients' memories of their time in the ICU, some units now recommend keeping a diary as part of clinical practice. Despite this, research on the clinical outcome of ICU diaries has been mixed. \cite{Barreto2019} thus conducted a meta-analysis of ICU diary RCTs, which suggested a statistically significant beneficial effect of ICU diaries on some psychiatric outcome measures, but not others. In particular there was lower probability of developing Major Depressive disorder after an ICU visit when a diary was kept, and an increased mean score on quality of life indicators like the SF-36 general health ratings. 

Interestingly, \cite{Barreto2019} did not find an effect of ICU diaries on risk of developing PTSD, which is one of the diagnoses with highest odds ratio for ICU admittance. As discussed above, PTSD is also one of the diagnoses most commonly targeted by expressive writing journaling interventions, which have shown moderate efficacy as a component of PTSD therapy. Of course the prompt for an ICU diary is vastly different than an expressive writing prompt, and it is done while still in the ICU rather than as a reflection on prior traumas. This once again underscores the relevance of timeline and prompt in utilizing a journal.

The information recorded in ICU diaries is not meant to be used clinically, and it is currently intended only for personal use by the patient as a form of intervention. However given the benefits of open ended journaling about daily experiences in the ICU, collecting copies of the journal entries would be a straightforward research extension. Diary content or linguistic properties might be predictive of later development of e.g. PTSD, which would enable critical early intervention strategies. \\

\paragraph{Negative side effects of health diaries.}
Although daily tracking of symptoms and life experiences can be both directly and indirectly therapeutic in various cases, this style of logging can also have a negative impact on symptom severity in other cases. The focus of the rest of this chapter is mainly on the ability to extract clinically relevant information from diaries, but unintended side effects must still must be acknowledged and properly planned for. 

One domain with potential concern around the effects of symptom tracking is chronic pain. It is not uncommon for research studies or even clinical practice to assign chronic pain patients a symptom logging template covering frequency and severity of different types of pain, to be filled out daily for some period. These diaries can assist in evaluating efficacy of new medications or identifying personal pain triggers for a patient. On the other hand, there is prior evidence that thinking more about experienced pain can increase perception of severity, so concern about the impact of closely tracking pain is warranted \citep{Ferrari2010}.

Because of this, \cite{Ferrari2010} conducted an RCT evaluating the effects of symptom tracking on clinician rating of chronic pain severity. All participants completed a pain assessment with a clinician, and then half of the participants were given a pain diary to complete daily for a 2 week period. At the end of the 2 weeks, all participants again completed a pain assessment with a clinician blinded to experimental status. Ratings of chronic pain severity were similar between the groups at the baseline assessment, but a significant increase at the 2 week mark was observed only in those that filled out the daily symptom logs.

It is unclear whether symptom tracking over a longer time period or with a concrete clinical goal (e.g. identifying pain triggers) might diminish the negative effect found by \cite{Ferrari2010}. It is also not yet well understood how sensitive perceived pain severity is to the attention effect in different individuals, yet another opportunity for improvement via personalized measures. Once individual susceptibility is better understood, a method pairing CBT-style coping techniques with chronic pain logging might be capable of improving individual perception of pain severity even as it is brought to focus more frequently. 

Overall, negative side effects of various types of journaling probably should be considered more carefully than they currently are. While it is unlikely that a completely free-form diary would have a strong impact on symptoms, more specific prompts can certainly be negative triggers for some individuals, and this is often dismissed due to journaling's safety relative to pharmacological or invasive trials. It is entirely possible to acknowledge and carefully think through side effect risks while also acknowledging the promise of clinical or research benefits often outweigh the risks.

This discussion is especially pertinent when deciding on study design, as there are many potential mitigating strategies that could be employed if properly planned. Besides severity of chronic pain, there are other types of symptoms that may be susceptible to a hyperfocus effect, especially if logging is performed by a vulnerable study population without proper clinical monitoring.

Considerations for psychiatry include reinforcing of delusions or of social anxiety. Over-analysis of prior social interactions is a major factor in worsening social anxiety for some individuals \citep{Purdon2001}, so prompting reflection on social experiences could lead to increased fear of social missteps even if extremely minor in reality. Psychosis is an even more complex and heterogeneous issue, which can be exacerbated by seemingly low risk interventions such as meditation \citep{Dyga2015}. 

It is entirely plausible that writing about delusional beliefs could help some dispel those beliefs while causing others to become more adamant about them, and this could depend on a wide variety of factors - both personal and study design. Autobiographical accounts of psychotic episodes frequently describe periods of strong delusions followed by fluctuations between belief and disbelief before finally dispelling a delusion \citep{Stanton2018}, suggesting that timing of journals may also be important. 

\cite{Stanton2018} also discuss the broader problem that autobiographical perspectives are commonly found in psychiatric records and often popular books for the general public, yet scientific use of these accounts is sparse. It is critical for the future of psychiatry that instead of scientifically neglecting autobiographical records due to their subjective nature, we find rigorous ways to use them. That could include both qualitative analysis for hypothesis generation as well as applying emerging techniques to quantify such datatypes. 

Ultimately this only increases scientific interest in diary records, but research on impact to patients must be conducted as part of this subfield if it is to gain traction, as I believe it should. Moreover, various journaling and habit tracking apps are increasingly advertised for smartphone users, including by services that are marketed to be beneficial for mental health - for example the encouragement of journaling by companies like Talkspace. For this reason alone, diaries need substantially more study, as it is likely they will continue to rise in popularity regardless of sentiment in the scientific community. \\

\paragraph{Health diaries in psychiatry research.}
Although symptom tracking diaries are more common in clinical practice for non-psychiatric disorders like cancer, epilepsy, chronic pain, sleep disturbances, and so on, there is historical work assessing the pros and cons of personal health diaries for tracking psychiatric illness, as well as much emerging work on the modern equivalent of health surveys (EMA, to be discussed further in the next section). As such, I will close this section with a review of some of the historical discourse on mental health diaries for information gathering purposes. 

Arguments for longitudinal data collection in psychiatry have existed for nearly a century, as some figures in early behaviorism called for studies to utilize a participant as their own control, to help combat the extreme heterogeneity observed in human behavior \citep{Murray1985}. At that time tools available for behavioral quantification were much more limited, making longitudinal studies either extremely tedious or full of large time gaps between data points. In many ways the former sort of longitudinal study is still strongly associated with n of 1 case reports, despite the fact that modern studies need not be so constrained.

One method proposed for relatively easy but still fairly frequent longitudinal data collection was the daily survey. As mentioned, there was already a history of successfully using such surveys for more accurate accounting of physical health issues - though in practice the methods were still tedious enough to be restricted to mostly personal use or research applications rather than practical clinical information gathering.

Early research on health diaries in psychiatry aimed to combine less frequent but detailed clinical interview assessments with simpler daily self-report symptom logs. One important example is the framework proposed by \cite{Murray1985}. There, "overt illness behavior variables" and "predisposing characteristics" were to be measured primarily by the psychiatric and biomedical records, while the interacting "acute need variables" were to be measured by the diaries. Acute need is related to self perception of symptoms, often not fully characterized by infrequent clinical visits, while overt behaviors are objectively documented negative outcomes related to illness, such as hospital admissions or unexpected absence at work. The latter may to some extent also be measured by diaries, but due to their often more extreme nature and their likelihood to be documented elsewhere, they are frequently represented already by thorough clinical ratings.

While the pilot health diary study of \cite{Murray1985} showed some promise in quantifying mental health over time, the bulk of the participant pool demonstrated minimal variation in self-perceived symptom scores from day to day across the study. The recruited subjects had more severe issues to begin with, so it is suggested that the study population was not well suited for this style of analysis. A major contribution of the work was instead a thorough analysis of potential biases in health diary data, from review of prior literature on medical surveying more generally as well as confounding factors observed that might be more specific to symptom tracking for mental health.

Health diary behavior (both participation rate and response content) will vary with symptom severity as expected, but it can also vary independently of symptoms, as a result of demographic, environmental, or other personal properties. Further, keeping symptom diaries can have an impact on the illness itself or the way the symptoms are perceived by the patient \citep{Murray1985}. 

Even the behavior of seeing a clinician can be impacted by many factors, including diary keeping, and in many historical census results it was observed that symptom severity had surprisingly low overlap with care-seeking, which was not sufficiently explainable by demographic effects and was not specific to illnesses with perceived stigma like mental ones \citep{Murray1985}. It is thus important for many experimental questions that other datatypes are collected alongside health diaries.

Interestingly many prior studies on health diaries for symptom evaluation, as well as some more recent studies on journaling as an therapeutic intervention, found women more likely to enroll and to sustain participation given enrollment - though not necessarily more likely to benefit given participation \citep{Murray1985, Krpan2013, Sohal2022}. Many open questions remain about sex-specific differences in psychiatry, so potential engagement biases ought to be minded in diary studies. 

Overall, health diary reporting can include many confounding factors that should be carefully thought through as part of future study design. One major one is the temporal effects related to the commonly observed decline in participation rate over the course of a study. These declines have been observed in studies substantially shorter than the longitudinal designs here, though easier use of modern diary platforms could be a mitigating factor. A slow decline in engagement could impact submission content and not just data availability, and furthermore the rate of participation decline will vary from subject to subject, in some cases with demographic, illness, or other experimentally relevant variables found to carry explanatory power \citep{Murray1985}. \\ 

\paragraph{Outlook on future daily survey use.}
Though health diaries have their pitfalls, they have more promise than their prevalence in psychiatry would suggest. EMA appears to be reversing this trend in the research domain, but clinical applications remain sparse despite established use in many other medical situations and the unmet demand for mental health screenings in the general population.

Indeed, health diaries have been proposed as a way to bridge general practitioners and psychiatrists multiple times in the literature, for over 30 years now. \cite{Shepherd1991} argued that survey-style diaries from a patient perspective would be much more tractable for screening by general practitioners without strong expertise in psychiatry. This would enable identification and triaging of patients in need of psychiatric evaluation, and could flag previously diagnosed patients undergoing periods of increased distress for further intervention. Yet little of the proposed framework has been scientifically codified to date and collaboration at scale between the groups in question remains limited. Now that the burden involved with this proposal has been greatly reduced by technology, there can be renewed optimism about its pursual. \\

\subsubsection{Revisiting the promise of journaling}
In physical illness great heterogeneity of behavioral responses has been observed, even when medical quantification metrics were similar \citep{Murray1985}. While it is possible that there are physiological nuances yet to be uncovered there, it is likely that biological and environmental factors unrelated to the physical illness mediate much of this effect. In psychiatry it becomes yet more complicated, as the illness is itself mental, in addition to the behavioral response to it. It is very difficult then to tease things apart, and nuanced longitudinal tracking will be a key part of understanding how behavior and biology relate. \\ 

Regular self-report surveys are one potential methodology for frequent collection of detailed health information, with historic precedence covered above. The original pen and paper format was tedious for both participants to fill out and researchers to curate. App-based EMA surveys thus open up new potential for daily symptom surveys.

\paragraph{Limitations of surveys for self-report.}
However, there is a fundamental limit to the information that can be captured by a single categorical survey question, and to the number of questions that can be realistically included before participant engagement levels diminish. 

Even if interested in only a clearly defined subset of symptoms, the number of EMA questions required to characterize their perceived severity with nuance is quite large. In a longitudinal study where a patient's responses can be normalized against their baseline, there are still many intrinsic and external factors to cover about the severity of a given symptom in a given day. At any given moment a symptom could be more or less severe, and it could persist throughout the day at a similar severity level or it could widely fluctuate, including disappearing entirely at times. 

Further, a type of symptom like intrusive thoughts can have multi-dimensional severity in a short time period, as two thoughts may have the same or differing properties of intrusiveness independent of containing content that is the same or different - and the content itself could be more or less distressing to the person. A survey asking about intrusive thoughts in general would run into this problem, where it is impossible for a single severity scale to properly distinguish quite different cases. But a survey asking about a specific recurring intrusive thought's severity only would likely miss other disease relevant episodes.  

On top of that, a truly identical underlying symptom experience could have greater impact on functioning depending on the environment the patient is in and the activities they are trying to complete that day. Experiencing moderate symptoms on a very important day might be more distressing and thus rightly assessed as more severe by the patient than severe symptoms on an inconsequential day. 

It is possible to systematically cover those and other facets of perceived severity by robustly designing targeted EMA, but as mentioned that involves inherent trade offs with ability to ask about multiple different symptoms and daily activities without losing engagement. A more realistic solution is to collect other datatypes that can complement EMA. \\

\paragraph{Passive digital phenotyping is a fundamentally different information source.}
The pairing of digital phenotyping with EMA can generate a substantially richer dataset, and can extend the definition of the "overt illness behaviors" mentioned by \cite{Murray1985} to included detection of much subtler behavioral patterns and shifts than previously feasible. EMA is regularly included in pilot digital phenotyping studies, often as an additional label source that can better align with digital phenotyping features than clinical scales alone. 

Digital phenotyping holds much promise, as will be reviewed in chapter \ref{ch:3}, but it is not a silver bullet, and to fully leverage it will require smart use of many psychiatric data sources, or otherwise a major paradigm shift. To close this section, I will thus argue for the importance of audio journals as a key piece in the future of a successful digital psychiatry movement, as they tackle important questions not currently addressable with clinical ratings, surveys, passive sensors, nor known neurobiological markers. \\

While EMA surveys have successfully complemented formal clinical ratings as labels for digital phenotyping prediction, and there are digital phenotyping features that may be inherently informative (e.g. exercise or sleep metrics), there are also fundamental limitations with the combined EMA/passive sensing approach. The sheer density of individual passive data streams and the frequent use of multiple sensor modalities creates a high dimensional and temporally fine input dataset, yet only low dimensional and temporally coarse corresponding labels. This is a difficult computational problem, and recent successes in machine learning with similar (but in some ways less complex) datasets have utilized much larger samples than could be covered by the number of subjects enrolled in a typical psychiatry study. 

The massive space of analyses that would be theoretically possible with a digital phenotyping dataset already presents a major challenge in efficiently deciding research directions, and it is further compounded by the sheer number of "unknown unknowns" that likely exist among current neurobiological models of psychiatric disease, a problem tightly linked with the great heterogeneity found between and within individuals. 

Clever machine learning approaches are one strategy for determining next steps in a data-driven fashion, and these ought to be developed in parallel with studies that use digital phenotyping to test more specific existing hypothesis. Still, there is a third category of approaches that should not be entirely ignored, and in fact can complement both the other study styles well - qualitative data exploration for idea generation and intuition building.

Furthermore, psychiatry will for the foreseeable future need to have an introspective, subjective component. Until neuroscientific models and technologies are much more advanced -- probably not in our lifetimes -- understanding a patient's experiences with disease symptoms will require the inclusion of a self-report perspective. 

Taken together, it is clear that as the field of digital psychiatry works to build better objective and quantitative metrics, there should also be parallel work on better subjective and qualitative metrics, which can in turn inform development of additional features with biological or clinical relevance. \\

\paragraph{Audio journals form a rich self-report dataset.}
As mentioned, EMA has its place as an easily quantifiable subjective instrument, but it cannot be the sole future of self-report. When well-defined "acute need variables" are to be assessed, EMA is most appropriate, but for any studies with an exploratory nature free-form self-report ought to be supported.

The beauty of the daily audio journal methodology is its dual abilities. On the one hand, audio diary prompts can be used to generate a rich dataset for qualitative exploration of patient self-report, at a timescale well-suited to longitudinal behavior tracking studies. On the other hand, acoustic and linguistic properties can be extracted from the same dataset, to obtain quantitative and objective metrics related to e.g. slowed motor functioning or subconscious cognitive impairments. Quantification of reported symptoms or other introspective experiences from the journal content is an additional possibility, as is obtaining clinician opinions of qualitative judgements they would make about the patient's external signs such as vocal properties.

Diaries not only have great potential for a variety of analysis methodologies, but they are also relatively easy to collect while maintaining some standard of quality. Although there are indeed many nuances to consider in carefully designing an audio journal study, as was discussed above along with documented historical use cases, it is very simple for a study to add an open ended diary recording option that would likely still generate some future benefit with minimal work on the part of the collecting researcher. 

From a participant perspective, the recording of audio journals is much less monotonous than daily EMA over long study periods. There exist many people who are naturally drawn to opening up about their thoughts in emotional times, hence the commonness of "tools" like the teenage diary, and thus another draw of the audio journal format.

Furthermore, collected diaries can be meaningfully utilized by groups with different levels of data science or psychiatry expertise, because of the many perspectives they invite. They are a very accessible datatype, and with modern technology even fairly qualitative methodologies are possible to perform without access to many hours of labor for careful manual review. Part of digital psychiatry tool building should involve facilitation of qualitative study of a dataset, not just strictly quantitative outputs, as the code presented in this chapter does.

By employing tools like the processing pipeline presented in this chapter, as well as taking advantage of collaboration amongst research groups, it is likewise feasible for groups with less computational knowledge to perform certain feature extraction techniques. The same dataset can be used by the same group to report on vastly different styles of research, for example a case analysis of topics that hold importance in a patient's day to day life and a correlational analysis of the relationship between speech rate and self-reported functioning levels. A demonstration of multimodal and impactful conclusions drawn from audio diaries using a dual approach can be found in chapter \ref{ch:3}. 

Overall, it is realistic for the scientific community to gain high value from diaries collected across studies at relatively minimal costs. In contrast, trying to pool resources between groups focusing on datatypes such as brain imaging and clinical interviews is much more difficult, because lack of training has a much greater impact on the quality of those datasets. Perhaps in some cases there is also a lack of effort put towards collecting modalities meant only for a data sharing agreement, which only exacerbates the issue for datatypes that require greater researcher effort or a design process tailored to a particular research question. Daily audio diary recordings are a unique method in their ability to assess patient perception while largely avoiding such issues. 

\subsubsection{Considerations for comparing journaling techniques}
Ultimately, journaling is cheap, has a long history of reported benefits both in mental illness and in the general population, and has minimal risk relative to other interventions \citep{Sohal2022}. It is already commonly integrated into evidence-based therapies like CBT, and it is advertised in many forms to many different people outside of the clinic: from bullet journal templates recommended on productivity-focused sites like LinkedIn, to the emotional benefits of keeping a diary touted on modern therapy apps like Talkspace. 

Therefore it is likely that digital records of journal entries will become increasingly commonplace in the coming years, making it important to build on what is already known about the therapeutic potential of journals and to uncover what clinically relevant information they might contain. A prerequisite for doing so is to characterize the many external factors that could affect journal content besides demographics and diagnoses, keeping in mind that other factors may mediate in a manner dependent on demographic or diagnostic factors (or each other). To aid in this research, I will now outline some of these factors to consider. \\

\paragraph{Variance by recording modality.}
One major variable that remains to be thoroughly investigated is the modality of the journal. Audio journals have the scientific benefit of enabling acoustics analysis, while still allowing for linguistic analysis on the resulting transcripts. However the vast majority of literature on journaling as a therapeutic tool focuses on writing.

Moreover, handwriting versus typing and even typing on a physical keyboard versus a smartphone keyboard could carry subtle behavioral differences - both in the content that results and in tracking its formation. Custom pens that can measure movement with high temporal resolution have already been used to detect those with early stages of dementia above chance via a drawing task, even when final drawings did not significantly differ in any detectable way \citep{6034}.

If interested in specific motor or vocal properties for a particular hypothesis, it may be clear which modality to use. However even in such a case it is important to be aware of potential effects on downstream content, if that is to be at all analyzed. Spoken diaries will often contain linguistic disfluencies like filler words or sentence restarts, something that is extremely rare to find in written text. 

Along the same lines, it is more likely that spoken diaries contain spur of the moment thoughts, often using simpler language. In contrast, written diaries afford the opportunity to reflect more carefully before putting a thought down, and most electronic diaries also make it easy to edit a previous thought after other thoughts have been added. While it is theoretically possible to track the writing process, that would add more computational complexity to the problem and ultimately increase the sample needed to draw robust conclusions. 

Besides direct effects on content, modality may also have an impact on participation rate. Participation rate as a binary variable is of course relevant to study success, but in the case of journaling, there is also the more nuanced variable of engagement. Someone with bad handwriting or arthritis may keep diaries shorter than desired if handwritten. Someone that lives with many roommates may be less comfortable discussing personal topics aloud for an audio journal. Someone who does the vast majority of their communicating on a smartphone may find it easier to type rich entries on a smartphone keyboard, but a middle-aged office worker might be more inclined to elaborate via desktop keyboard. 

There are numerous ways that the modality could alter participation rates differently for different people, so this is difficult to carefully account for. Still, it is something that needs to be kept in mind, and perhaps worth directly asking participants about in any study collecting journal data. It is also critical to consider as a broader property of the study population, as factors like age and symptom severity might suggest certain methodologies to be avoided. 

Similarly, even for a specific modality collection device could have additional effects on content and participation. Our audio diary studies always use apps like Beiwe, typically installed on a participant's own phone. In most situations, this is beneficial for engagement, as the person will usually have their phone with them and the app can send targeted notifications to record a diary each day. There may be situations though that call for a study-issued device or a simpler recording device that requires less tech savvy. \\

\paragraph{Patient access to previous records.}
Another design decision, which is sometimes indirectly tied to modality, is the ability for participants to view or listen to current or past entries. This may affect content in any journaling context, but is especially relevant for studies with a therapeutic goal. One possible explanation for therapeutic efficacy is that writing down thoughts "gets them out" in a way that repeatedly thinking them cannot. Speaking aloud in an empty room is different than creating a private audio recording and may consequently have different impact. A comparable written study could perhaps use a tablet where words quickly fade away.

Besides the potentially different subjective experiences between permanent and impermanent journals, there is also potential benefit from reflecting on thoughts recently recorded, or reviewing thoughts from over a period of journaling. It may be more natural to do this in text form, but regardless it is controllable by study design. Personal journaling apps, whether recorded or typed, of course allow access to all prior entries, but apps intended for research purposes may or may not give participants access to their historical data -- though it is typically encouraged to review a day's entry before submission, even when in audio format (with an option to rerecord). \\

\paragraph{Effects of perceived audience on journal behavior.}
In addition to participant self-review, the actual and perceived audience for the journals is deeply relevant, again to both content and, as suggested by the results of  \cite{Sohal2022}, therapeutic efficacy. If content is being tracked in real-time by a longitudinal study, there may be an impact to researcher behavior. There are ways in which this could improve a study, e.g. maintaining good engagement, but it can introduce biases as well (discussed in more depth in chapter \ref{ch:3}). 

Nevertheless, the more pervasive concern is the impact of perceived audience on patient journaling behavior. A research study on quantitative measures from audio diaries could be designed in a way that prevents any researchers from accessing the diaries themselves, and could use features that are not particularly sensitive. This would decrease the ability to clean the data and carefully check for alternative explanations for any results, and it would also eliminate the opportunity for many interesting qualitative analyses. However if there is reason for concern that content would be substantially altered because of a perception of being watched, and reason to believe that perception would be attenuated in much of the study population if they were assured researchers wouldn't listen/read, then this could be an appropriate strategy to employ. 

For studies where someone certainly will be directly reviewing diary submissions, there are still additional considerations on perceived audience. In multimodal research projects the participant may regularly interact with study staff, and may even perceive that the therapist performing recorded clinical interviews with them will also access their diaries. Both the title of the person believed to be receiving submitted journals and especially the relationship between that person and the participant could have a strong influence on what is recorded. 

Moreover, there could be temporal effects related to this relationship. A participant could have already established a good rapport with the clinician that recruits them to a study, whom they may believe has (or may actually have) access to their submissions. On the other hand, a participant could join a study with little knowledge of the staff, but over time develop a relationship and begin to view diary submissions in a similar manner as talking to that staff. In some cases this could even have dire medical implications, as although participants are told that the diaries aren't actively monitored, they may still use the recordings to report severe distress if they feel a rapport between themselves and the "listener". 

It is thus important to understand how perceived audience and the associated rapport mediate journaling participation. Much of the historical use of journaling was meant to be a deeply personal experience, and in the context of modern journaling there is little yet published. However early studies comparing therapeutic conversations with chatbots against those with a human manning the chat may inform this research in the interim. Unfortunately, these studies have largely found mixed results: some report that participants were more likely to engage with chatbots knowing they were not revealing potentially sensitive info to another human, while others report participants being more likely to engage when they knew they were speaking with a human who they could build rapport with \citep{Vaidyam2019}. 

Of course, heterogeneity in these results is not all that surprising given the many nuances already discussed. It once again underscores the importance of a personalized approach to psychiatry. It also highlights application-specific questions that could apply equally well to journal studies. Willingness to engage in therapeutic discussions with a chatbot as opposed to a human could depend vastly on the duration and frequency of conversations as well as the specific topics discussed. While a direct response is not expected to a diary entry, behavior that depends on the perceived role and personality of the chatter could mirror potential differences in diary submission already discussed. \\

\paragraph{Other demographic and study design factors.}
Indeed, other factors that could mediate behavior based on perceived audience include the journal prompt and the overall study timeline. It remains to be investigated, but it would not be unexpected if expressive writing, where deeply emotional and often negative topics are discussed, were more subject to concerns about the audience than gratitude journaling, which is about a positive outlook and often more surface level. 

Further, habituation effects are well-documented across psychology, and journaling behavior is no exception. There is a paucity of studies extending past 3 months, with many much shorter than that, but it is very plausible that concerns about diary privacy would dissipate the more habitual the recordings become -- especially if the participant never hears anything from study staff about their submissions. 

These same factors will probably have effects on other facets of collected diaries as well. Prompt design, including the question, the minimum/suggested/maximum submission length, and the frequency of the prompt, is critically important to the resulting dataset. In a vacuum, longer submissions are generally better for research, but expectations of a long submission could diminish participation rates, as could deploying prompts too frequently. 

The content itself may be affected by all these study timeline decisions, but likely more important is the prompt used (though they probably interact as well). A very open ended prompt like the ones used in this thesis can be extremely informative with just a short recording each day. It is obviously not intended to be therapeutic, but for information gathering we are able to extract what the participant decided was most salient to self-report on a daily basis. For some patients, this turns out to be immediately clinically relevant, such as discussion of treatment opinions or specific symptoms popping up. For others it is a great source for combination with multimodal digital phenotyping, as good descriptions of the person's day can be crosschecked with various predictions from passive sensors. 

There are also some participants that do not engage well with the open ended format and submit short, basic entries each day with little salient content. It is not clear how well those participants might have engaged with a different prompt style, but nonetheless it is clear that prompts need to be designed with the study aims and patient population in mind. 

\subsection{Applications of modern tech to psychiatric patient diaries}
\label{subsec:diary-lit-rev}
Daily audio journals are a promising medium for the development of digital psychiatry, fitting into both quantitative data-driven approaches and qualitative patient-driven ones. To date there have been few results on the topic, but similar to the above historical discussion, we can look to the literature on tech applications for adjacent topics in psychiatry to form expectations. 

First, it is important to note that there is good reason to believe audio journals will become a more frequent topic of near future publications, besides the many arguments for their utility. Collecting these journals is a feature included in many of the popular consumer and open source apps for ecological momentary assessment (EMA) and passive phone data collection, which are of course already being installed on numerous participant phones. Furthermore, large-scale collaborative data collection efforts like the AMPSCZ project to be described in chapter \ref{ch:2} or the AURORA project methodology by \cite{McLean2020} are including audio journals as one of the modalities of interest. While the primary intention for those diaries is to perform quantitative analyses similar to what I will describe for the project's clinical interview recordings in chapter \ref{ch:2}, it is worth mentioning that the AMPSCZ project is also encouraging the use of autobiographical patient accounts as part of its approach \citep{Larrauri2022}, an aim that aligns well with the spirit of audio diaries.

Thus, the work in this chapter is a critical step towards facilitating the best use of collected audio diary datasets. The code presented might in fact be adapted for use with the aforementioned AMPSCZ data, a possibility to be discussed further in section \ref{sec:discussion2}. In addition to addressing the need for processing tools that can provide a baseline to coordinate future research, this chapter presents a first of its kind analysis suite that takes a multi-armed approach on psychiatric patient audio journals. The longitudinal Bipolar disorder journal dataset that is characterized here is unusual not only because of the long duration for which multiple patients were followed along with the availability of multimodal signals to supplement interpretation of the journals, but also because such a dataset has not been documented to date. 

Indeed, the first result on Google Scholar (as of 1/2023) for "Bipolar audio diary" is about a filmmaker, and the second is an abstract from a 2020 Society of Biological Psychiatry poster that has nothing to do with audio journals, but happened to appear near my abstract from the same conference. That abstract was focused on the OCD case report that forms the basis for chapter \ref{ch:3} of this thesis -- which contains another novel demonstration of audio journals for psychiatry. 

\subsubsection{What does exist on audio journals?}
\label{subsubsec:journals-so-far}
As was the case with symptom tracking health diaries in the past, the use of audio journals to gather clinical or research information (rather than for personal or otherwise therapeutic use) has been rare in psychiatry, but somewhat more common in physical illnesses or to study healthy humans undergoing a time of stress. 

Many of the studies that do exist on audio journals tend to use them qualitatively, sometimes paired with quantitative self-report via surveys or with medical records. Qualitative review of audio journals has been applied to identifying primary stressors for certain occupations, evaluate perception of HIV treatment over time, and document the experiences of teenagers with muscular dystrophy \citep{Carr2019}. One advantage of the audio journal format compared to eliciting autobiographical accounts during recorded interviews is of course the greater collection frequency already discussed, but another advantage is on the willingness of participants to open up. In the HIV study reviewed by \cite{Carr2019}, it was found that some patients who reported optimism during clinical visits expressed a much more pessimistic attitude on their audio journals. 

Perhaps the first major wave of audio diary papers came out of the COVID19 pandemic. Diaries and EMA were used to document stress levels and general attitudes of healthcare workers over time \citep{Goot2021}, and diaries were also used for qualitative study of experience with COVID prevention measures in specific populations, including those acting as home caretakers for a family member \citep{Tay2021} and those entering their first year of university study during the pandemic \citep{Liebendorfer2022}. There were also a few pilot works on the use of audio recording analysis to predict presence of COVID19, both in natural speech and by recording forced coughs, though those works never reached a point where they would be usable in practice \citep{Coppock2021}. Overall, the pandemic represented a new milestone for smartphone audio recordings due to not only the relatively large amount of research output in a short timeframe, but also the media attention some of it received. \\

The extension of both qualitative and quantitative audio journal techniques to psychiatric disease is ripe for productive output, yet minimal work exists at this time. From a language perspective, early work suggested that a private and personal Wordpress blog can both generate interesting research content and benefit patients with a history of self harm \citep{Marzano2015}. However, that study followed only 5 patients for at most 3 months, and does not perform any detailed analyses on the submitted content. The biggest takeaways from \cite{Marzano2015} are the extent to which they were able to maintain daily engagement with their platform despite the lack of study compensation, and the commentary made by the participants about perceived beneficial effects. Because all entries to the platform were being monitored to detect if a patient was in imminent danger of self harm, it is difficult to separate its actual use as a medical tool from more general properties of patient self-report behavior.

A handful of other psychiatry publications have included audio journals as one of many modalities collected in mood and anxiety disorders. The work of \cite{Nickels2021} was promising due to the large number (415) of depression patients enrolled in their app-based study, and is likely the most exhaustive analysis of linguistics in voice diaries to date. Given the fully remote format of the research and the access to Verily software development resources, it is unsurprising that they were able to collect a large patient sample with good participation rates over the 12 week study period. However, they collected audio journals on only a weekly basis, and given the short duration relative to our longitudinal work, the "3779 participant-weeks of data across 384 participants" that they used in final modeling is actually smaller from a time point perspective and in some ways less rich than the journal dataset to be described in this chapter. Furthermore, their analyses covered a spectrum of modalities and did not involve a deep dive into many audio journal features. Nonetheless, it is promising that they found a statistically significant Spearman correlation between weekly journal sentiment scores and corresponding weekly PHQ-9 score \citep{Nickels2021}. 

In the acoustics domain, \cite{Place2017} collected weekly voice diaries from 73 participants diagnosed with depression or PTSD over a 12 week period. At the end of the study, participants completed a semi-structured clinical interview including clinician scored scales. This was the primary target of their prediction work, which had the advantage of greater clinical grounding but of course the disadvantage of providing only a single label per participant. They focused only on acoustic analyses of the voice recordings, and extracted features with prior depression association, as will be reviewed in an upcoming subsection (\ref{subsubsec:acoustics-linguistics-review}). In pilot modeling, the voice diary acoustic features showed predictive relevance to clinical ratings of participant depressed mood, fatigue, interest in activities, and social connectedness. Additionally, participants largely self reported feeling comfortable sharing their voice diary data with physicians, mental health providers, and medical researchers \citep{Place2017}. \\

Despite the promise of preliminary work and the evidence from many adjacent research areas that audio journals are likely to be an impactful data source, there is not much else to review at this time -- which is both unfortunate, but also a striking argument for the importance of the present work. Indeed, even large scale reviews of psychiatric patient speech and language analysis have failed to mention audio journals. A recent survey by \cite{Zhang2022} focused on applications of NLP to psychiatry nicely demonstrated this paucity. They pulled 7536 articles from across 6 academic databases, and filtered through them to obtain a list of 399 English papers containing original psychiatry research with documented NLP methods. An overwhelming $81\%$ of those works utilized social media as their data source. Recorded interviews, electronic health record (EHR) data, and screening surveys accounted for an additional $7$, $6$, and $4$ percent respectively. Just $2\% (n=8)$ of the publications included a journal-like data source, and as these were largely written the category was classified as "narrative writing". 

Furthermore, nearly $\frac{2}{3}$ of the articles reviewed focused on depression classification or suicide detection \citep{Zhang2022}, which is extremely disproportionate to diagnostic prevalence when compared with availability of NLP studies on anxiety, and which is also disproportionate to the availability of NLP research in psychotic disorders given the existing evidence base for the relevance of language in that population (to be discussed). This may be attributable in part to the overreliance on social media data sources. Luckily, audio journals can provide a strong middle ground between open social media posts and time intensive interview recording collection.

Thus due to the lack of prior work on daily audio journals for information gathering in psychiatry, especially in study of psychotic disorders, the next subsections will instead cover related topics in more detail. These include applications of tech to other historical use cases for journals: therapeutic journaling, health diary surveys, and the impact of tech on study participation rates. Additionally, results will be reviewed from previous psychiatry research that utilized acoustic or linguistic analyses with a different information source, including semi-structured clinical interview recordings and recording of speech from specific tasks (e.g. reading a passage or describing an image aloud). 

It is worth noting that the majority of recent computational work on spoken language in psychosis focuses on interview recordings, which will be further discussed in chapter \ref{ch:2}, though naturalistic recordings of phone conversations have also been considered -- including in Bipolar disorder \citep{Maxhuni2016}, so this is of clear relevance to our dataset. Because the only practical way to obtain phone recordings in many countries is to exclude the other end of the conversation however, and making/taking phone calls is a behavior that currently varies widely by demographic, there is little reason to prefer call data over audio diaries in modern study design. Therefore the comparisons made in chapter \ref{ch:2} will focus only on clinical interview recordings versus daily audio diaries. While interviews are an important datatype that has advantages over audio diaries, I believe diaries will be a much more relevant tool in the long term outlook of digital psychiatry. 

Additionally, the use of app-based open ended audio recordings in research has the potential to take a variety of forms besides the already exciting one reported on in this chapter. As part of the current section, I will also review such related concepts; for example, the app-based therapeutic journals discussed in the next subsection.  

\subsubsection{Tech extensions for therapeutic journaling}
The most obvious application of tech to journal therapies is to implement a course of therapeutic journal sessions as a web or smartphone app. The standalone journaling therapy techniques described above are already easy for an informed individual to do on their own via pen and paper, so from a therapeutic perspective it is not immediately clear what benefits a digital version would have. 

\noindent However, there are a number of indirect reasons to address this question, including:
\begin{itemize}
    \item Electronic journals are much easier to collect for research purposes, but if research on these therapies is to become digital it must first be confirmed that typing (or audio recording) of a journal entry via app retains the benefits observed for traditional handwritten journaling.
    \item Electronic journals may attract a different user base and could more generally increase awareness of journal therapy. This would improve accessibility, and from a research perspective might require new studies on whether journaling is beneficial in any newly recruited demographic groups. 
    \item Particularly for journaling techniques that benefit from sustained use such as gratitude journals, an electronic format could improve long-term engagement through mechanisms like targeted notifications.
    \begin{itemize}
        \item It is also possible that some users would benefit from the ability to more easily review and search old entries in a longitudinal setting. 
        \item Additionally, because gratitude journals focus on positivity, there would be minimal risk from testing the benefits of app generated reminders about previous moments of gratitude, a la Timehop. 
    \end{itemize}
    \item As certain components of CBT have already been successfully translated to an app setting, with evidence of efficacy in depression from multiple RCTs \citep{Denecke2022}, it could be fruitful to integrate electronic journaling techniques with these apps.
\end{itemize} 

\paragraph{Pilot results.}
Early results indeed indicate that electronic journaling has therapeutic benefits similar to expectations from traditional journaling. For example, a randomized controlled trial (RCT) of a web app implementation of positive affect journaling (PAJ) in a non-clinical sample undergoing significant life stress demonstrated the ability of electronic journaling to significantly decrease anxiety levels \citep{Smyth2018}.

PAJ is a variant on expressive writing that was developed to combine the benefits of longer journaling sessions on intimate topics with the benefits of maintaining a positive outlook on life. PAJ was previously documented to improve well-being in individuals coping with a physical illness or major environmental stressors, but \cite{Smyth2018} aimed to extend it to a digital setting. They recruited 70 participants with self-reported high stress levels, and assigned 35 to their web app intervention. The intervention group was to complete 3 journaling sessions per week for a 12 week period, with each PAJ entry taking $\sim 15$ minutes. All participants completed extended surveys on physical and psychological health at baseline and at the end of each of the 3 months. 

One relevant finding by \cite{Smyth2018} is that it was difficult to sustain participation in the trial. Average journaling sessions completed per week by the intervention group had peak $< 2.5$, was $< 1.5$ by week 6 (of 12), and was $< 1$ by the final week. The longer duration of individual PAJ sessions and the implementation as a web-based intervention rather than as a smartphone app likely made their study more onerous on participants than the short audio journal format we typically employ. Their study population was also intentionally selected to be under a high amount of stress at the time of the intervention, which could affect participation rates more than other psychological targets. 

Nevertheless, it is a positive outcome that the intervention showed efficacy even with the quickly dropping participation rates. At the end of the 12 weeks, there remained a statistically significant reduction in anxiety levels in the intervention group but not in the control group, as measured by the Hospital Anxiety and Depression Scale \citep{Smyth2018}. This is consistent with prior studies of expressive writing, that showed sustained effects many weeks after just a handful of journaling sessions \citep{Krpan2013}, so it bodes well for the extension of journaling to the digital world. \\

\paragraph{Potential concerns.}
Still, more work is needed on electronic journaling, particularly as it pertains to treatment protocols for those with psychiatric diagnoses. The recent meta analysis by \cite{Sohal2022} did not find any electronic journal RCTs in the psychiatry literature, and also noted that collection of journals by study staff was predictive of studies that found journaling to be ineffective. It is unclear then whether electronic submission of journals would demonstrate a similar pattern, and even whether digital journals implemented to be entirely private would extend to the psychiatry setting.

Given the variance in journaling response is seen not only across studies but also across patients within a study \citep{Krpan2013}, there is good reason to collect journals if possible. It remains entirely uninvestigated whether journal content might be predictive of therapeutic efficacy. If relevant features were discovered, it would open up multiple clinically promising next steps. 

For one, if content is predictive of journaling outcomes, can this be used to improve journal therapy instructions, perhaps in a personalized manner? If it is possible to change the way that ineffective journalers are writing to be more similar to the way that effective journalers are writing, will that significantly change their outcomes? If so, this is a methodology for making journal therapy a more robust and accessible clinical tool.

If it is instead not possible to use predictive features to guide ineffective journalers to efficacy, then that suggests a third factor is likely causing both the differences in journal content and the differences in journaling therapy outcomes. In that case, follow-up research on the predictive features could improve treatment planning in clinical practice, potentially not only by targeting journal therapy at the most relevant individuals, but perhaps also through identification of broader factors that are linked to outcomes of other treatment categories.

\subsubsection{EMA validation}
Just as we aim to gather information from an open ended journal format, yet similar methodologies can also be therapeutically relevant, so too can constrained daily surveys play dual roles. The use of symptom and activity logging for patient benefit is an established psychiatric practice leveraged by some therapists and clinicians as a portion of treatment. This application extends naturally to tech, where general purpose apps for logging medication or tracking habits already exist, and psychiatry-specific research has also already shown prior success. Daily symptom records as a component of app-based CBT for Bipolar disorder improved patient outcomes in a trial \citep{Matthews2015}. 

Bespoke apps for psychiatric interventions remain somewhat inaccessible by the general public, but given the large body of research on them that has been built \citep{Denecke2022}, options are likely to dramatically increase in the near future. Regardless, my primary focus here is on the use of technology to characterize behavior rather than to treat it. \\

\paragraph{Synergy between EMA and audio journals.}
Described above, previous use of "health diaries" for scientific data collection are directly analogous to modern research on EMA. Historical diaries in many cases contained space for free-form response on specific questions, to supplement the survey questions that map to EMA. Thus the pairing of audio journals and EMA is in line with established diary use cases. 

It is unsurprising then that EMA is so closely linked with audio journals in the emerging trends in digital psychiatry. As discussed, many of the qualitative studies of audio diaries that do exist consider survey responses alongside them. The majority of yet unpublished audio journal datasets are extremely likely to include EMA as well, due to the fact that the apps in common use -- for example, Beiwe, MetricWire, and MindLAMP -- are capable of collecting diaries and EMA. In some cases these apps also collect passive sensing data; the pros and cons of the different apps are reviewed in more detail as part of the background for chapter \ref{ch:3}.

The Bipolar dataset described in this chapter, as well as the OCD case report in chapter \ref{ch:3}, utilize both EMA and diary datatypes, collected via Beiwe. The proposed methods for the AMPSCZ project of chapter \ref{ch:2} also involve collecting EMA and audio journals, this time using MindLAMP. Moreover, the notification prompts for a participant to submit a daily audio recording and a daily EMA survey often go hand in hand with the research apps, and in practice we have found very good overlap between EMA and diary availability across our subjects.

Because of the relevance of EMA to current and future audio journal research, it is important to provide some context on benefits of EMA in assessing psychiatric symptom severity. \\

\paragraph{Prior results from app-based EMA.}
Health diaries have remained in use for assessing physical illness for some time, including in a purely clinical context. Medically, it is still somewhat common to come across a pen and paper diary for logging of events such as epileptic seizures or occurrences of chronic pain. However a shift to electronic methods is slowly taking place. Even when digital methods were much clunkier over a decade ago, early research demonstrated improved patient participation rates and less time spent handling data by medical staff when health diaries were moved to an electronic platform \citep{Dale2007}. Coupled with the previously established literature on the accuracy of health diaries in a number of contexts, there is good reason for optimism about the future of medical surveys.

Due to its relative simplicity, tech-enabled EMA responses have been collected by psychiatry studies for at least 15 years now, providing a strong base of prior evidence for its clinical relevance. Historical work found high compliance rates with EMA to be filled out at a particular time daily using a study-provided PDA device over 1 week \citep{Granholm2008}. For reasons to be discussed, compliance with smartphone app-based EMA should generally be even higher, particularly when expanding scope to longitudinal studies of much greater length, something made feasible only more recently.

Daily EMA designed to capture clinical scale items has been validated against the professionally-rated weekly scales in multiple major psychiatric disorders, including Schizophrenia \citep{Granholm2008} and Major Depressive Disorder \citep{Targum2021}, and patterns in mood and stress related EMA responses over the course of a longitudinal study have successfully distinguished Borderline Personality Disorder and Bipolar Disorder from each other and health controls well above chance \citep{Arribas2018}. 

Interestingly, \cite{Arribas2018} found that future self-report scores submitted by controls were predicted by past self-report scores significantly better than future self-report scores could be predicted in either clinical population. This result suggests caution in taking single day EMA responses from certain psychiatric populations at face value; though it is difficult to separate stochasticity in survey taking versus actual stochasticity in underlying mood, and it is also unclear whether more advanced sequence modelling techniques or inclusion of multimodal data sources would improve accuracy in the clinical groups. Furthermore, the effect may be specific to the EMA questions used by \cite{Arribas2018}. Ultimately, the topic warrants additional investigation.

Nevertheless, EMA responses have repeatedly shown a relationship with clinical variables of interest, and weekly mean EMA scores can successfully predict concurrent clinical ratings \citep{Targum2021}. EMA has also demonstrated treatment sensitivity in clinical trials, both for behavioral therapy in OCD \citep{Rupp2019} and for medication use in depression \citep{Targum2021}. \\

\paragraph{EMA design considerations in the app era.}
There is a breadth of literature on EMA successfully used for different purposes with questions designed to fit the study goals. Any particular EMA design will require its own validation, but there is strong reason to believe that EMA is a powerful tool for systematically collecting quantitative patient self-report data. 

There are inaccuracies and biases in how a person might self-report too. But that is largely an inherent difficulty with psychiatry. Patient perception is an important component, and inaccuracies in patient perception certainly affect the official reports that occur at clinical visits as well. Longitudinal study, recently made substantially easier by app-based EMA, can in fact better correct for patient-specific reporting biases. The frequency of EMA can also diminish the level of inaccuracies due to imperfections with recalling the events of previous days.

Memory gaps in physical illness reports have been well-documented, and are partially mediated by symptom severity. They can be improved through the use of daily health diaries \citep{Allen1954}. Mental illness might have additional complicating factors in accurate memory of prior week(s). Even in healthy controls, it is plausible that memory for mood and other properties of mental state may be worse than memory for specific events and physical phenomenon. Indeed, EMA questions assessing mood have been used to show poor retrospective recall of negative affect in Depression patients when compared to controls, though both groups exhibited some amount of inaccurate recall of both types of affect \citep{Zeev2009}.

Overall, a study planning to use EMA as a central component ought to carefully think through survey design, and base questions off of previously successful EMA studies where possible. Downstream, EMA responses can be used in many different capacities, which I discuss further in chapter \ref{ch:3}. In this chapter, I will proceed to focus on audio journals, though EMA will come up again in the context of their direct pairing with the journals.

\subsubsection{Impact of tech on journal submission rates}
The major limiting factor for many older studies of self-report was the tediousness of data collection for both participant and experimenter. There are now many tools that make data collection, organization, monitoring, processing, and analyzing much easier for study staff, as will be used throughout this chapter. Therefore the burden on experimenters has been largely lifted. However the submission of self-report still requires work on the participant's end, unlike the passive sensing tech options. 

As I've discussed, passive sensing has many promising use cases, but is not a replacement for self-report in theory and is certainly not a replacement in current practice. It is thus important to discuss how recent technology might impact study participation rates, which is mostly for the better. \\

Many prior studies of daily pen and paper health surveys have unsurprisingly found participation rates to decline over time, even with study durations often capping at a couple of months \citep{Murray1985}. It is highly unlikely that app-based health diaries would prevent participation decline entirely, but they have the potential to significantly slow the rate of decline, facilitating both more complete short-term datasets and the design of realistic longer-term studies.

One new advantage is that apps like Beiwe \citep{Beiwe} can be installed directly on a participant's own phone. Once set up and explained to a participant, self-reports can be easily submitted without any further study interaction necessary for the lifetime of the phone. By contrast, prior work would have required participants to maintain a separate notebook, or more recently a PDA, solely for the purpose of the study, and then to go out of their way to submit the diary any time the study wanted to update their database.  

Further, the apps will issue notifications to prompt for self-report submission, preventing participants from forgetting. Timing of notifications can be tailored to a participant's use patterns to minimize the probability they will ignore the notification, and additional reminders can be sent selectively if a submission hasn't been made yet for the day. \\

\paragraph{Features of current research apps.}
While we typically collect self-report submissions on a daily basis, the apps also have the capability for multiple different self-report prompts to be sent in a single day. This could be utilized to ask participants to record their thoughts on multiple distinct topics in a day, or it could be used to ask the same question multiple times in a day. This could be useful for inquiring about symptoms that are sensitive to circadian rhythm or other factors that vary naturally throughout the day -- for example, depression severity may be consistently worse in the morning or at night for some individuals \citep{Targum2021}. 

Indeed, \cite{Targum2021} distributed their app-based EMA surveys both upon waking and before bed, and still observed good participation rates. Eliciting multiple self-report submissions in a day is much more realistic using the app with notifications model than was previously possible. 

A number of other limitations to the manual self-report format that were lamented by prior researchers \citep{Allen1954,Murray1985} are now features built in to the current research apps. This includes the ability to issue custom prompts for specific subjects, and critically the ability to easily collect and parse answers to open ended questions. \\

\paragraph{Future potential for app-based incentives.}
In addition to the features that are already standard in self-report apps like Beiwe \citep{Beiwe}, there are also common properties of smartphone apps more broadly that would be straightforward to implement and might further improve participation rates. Many games and social media apps use tactics based in psychology to drive engagement, such as streak counters to encourage daily use and visual cues to increase the reward signal generated by achievements. In this case, achievements might correspond to statistics about research contribution, like total minutes submitted, or about money earned from participation where applicable. 

Depending on study population, allowing access to prior recordings and survey submissions from over the duration of participation might incentivize engagement for some subjects, as they may then feel the self-report process is valuable for their own use. Recording audio journals in particular can be a fun process that might be marketable to a wide variety of potential study participants. \\

\paragraph{Considerations unique to the audio diary format.}
For audio diaries, sustained participation is not only about submission availability, but also about submission quality. It is thus important to maintain engagement. This is discussed in more detail in chapter \ref{ch:3}, including discussion of potential biases that could be introduced by different methods for improving participation levels.

In the context of journal recordings with more specific prompts, it is also important to consider how sensitive the questions may be perceived to be. As was discussed above in section \ref{subsec:diary-history}, journal therapy seems to be less effective when patients know their journals will be collected \citep{Sohal2022}. Research on engagement levels in chatbot therapy has shown mixed results thus far \citep{Vaidyam2019}; more work is needed to understand how app-based technology might impact patient participation when journals are focused on highly personal thoughts and feelings. \\

The work detailed in this chapter centers on very free-form diary prompts that largely benefit from the app-based audio collection advantages mentioned. Stats on submission frequency and duration in our Bipolar longitudinal dataset will be reported as part of section \ref{subsec:diary-methods}.

Of course, the research smartphone apps don't only change the richness and quality of dataset that can be collected in the modern version of a self-report study; because of advances in computational analyses, they also enable novel cross-modality questions to be asked. I will now shift gears for the remainder of this background review, to focus on the use of new tech in data analysis.

\subsubsection{Acoustic and linguistic data sources}
Before proceeding to review of prior work on acoustic and linguistic features observed in mental illness, it is important to first briefly review the data sources that are used to derive these features. Because audio journals have been so scarcely researched in psychiatry to date, all work I've found on them was already covered above (subsection \ref{subsubsec:journals-so-far}). Here I will discuss more commonly used alternative data sources and their comparison with the audio diary format. Each has its own pros and cons, and for a certain style of study the choice may be obvious. A central theme of this chapter though is that audio journals are significantly underutilized, and can make a strong addition to a wide variety of study designs and analysis plans. \\

\paragraph{Social media.}
One extremely common source for patient language data is social media \citep{Zhang2022}, which has collection advantages and can be used to study interesting social dynamics. Manic episodes can be plainly visible to acquaintances because of online behavior, so certainly the datatype has relevance in this line of work. 

Still, the content of social media posts is undoubtedly affected by its visibility; the typically written, "post when you feel like it", format can also lead to a very different dataset than the type of language elicited by daily audio journal recordings. Further, many psychiatry social media studies are conducted in an environment where the authors do not have access to detailed personal health information, meaning the labels used in such studies are often less clinically grounded than other data sources. Of course, when accounting for the sheer volume of social media language studies, there very well may be a large research base in absolute terms that does combine social media with more traditional psychiatry datatypes \citep{Zhang2022}. 

Regardless, social media is much more distant from audio journals than other potential language data sources, and it is in my opinion a less fruitful source for new psychiatry researchers in most cases, precisely because it is relatively over-studied and often under-grounded. In the upcoming subsection (\ref{subsubsec:acoustics-linguistics-review}), I will draw from social media studies only where they are relevant for background on a particular linguistic feature of interest, but note it will not be an exhaustive review on social media NLP in psychiatry by any stretch. \\

\paragraph{Naturalistic sources.}
Another potential method for collecting both acoustic and linguistic information is to record everyday behavior through a smartphone app. This could involve either collection of speech and written text from the private phone calls and messages of a participant or ambient field recordings of environmental audio. Ambient recordings could be designed to occur at specific times, triggered by specific sounds, or even triggered by a multimodal signal such as recording only at certain geolocations. 

The potential for collecting natural speech and language is certainly scientifically attractive, and this promise has been repeatedly discussed in digital psychiatry circles. Pilot studies have investigated phone call analysis \citep{Maxhuni2016}, and even ambient audio has been utilized to a limited extent in some studies - for example \cite{Nickels2021} considered the volume of environmental audio as one of their features alongside those extracted from the weekly voice diaries.

However, unless implemented in a very limited form (e.g. as done by \cite{Nickels2021}), there are substantial privacy concerns surrounding these methods. Not only will it likely be more difficult to enroll subjects in a research study that involves recording of ambient audio or of private phone calls (if they even make phone calls), but it is impossible to consent all the people they interact with that may be inadvertently recorded. Handling the problem of non-participant audio is a difficult question both technically and ethically. Choosing to include such a data type should be a careful consideration based around specific research needs, rather than a general use format to widely supplement psychiatry studies like audio journals can be.

Furthermore, naturalistic voice recordings are not a sufficient replacement for the role of the audio journal in many potential digital psychiatry analyses anyway. They do not involve a self-report element, and they are less grounded in prior expectations. For ambient recording in particular, the tractability of performing exploratory data science work that still connects back to the psychiatry literature is a concern. Perhaps once an improved set of behavioral metrics has been solidified by the first wave of digital psychiatry results, those metrics could help to "bootstrap" the deeper analysis of less constrained recordings. Given the privacy headache though, I don't believe this direction is worth significant thought. \\

\paragraph{Clinical sources.}
On the other end of the spectrum, there are also audio and language sources that come directly from existing clinical practices. NLP performed on notes about the patient from EHR or on a patient's own written responses to a mental health screening form are two research directions that have already been established in the psychiatry literature \citep{Zhang2022}, though these are a fairly large departure from the audio diary methodology. 

The other major datatype in this category is the recording of patient (and interviewer) speech from semi-structured interviews. Such recordings have been used for a variety of different acoustic and linguistic analyses, and they also can often be utilized by a trained individual to manually score the gold standard clinical scales. Interview recordings as a datatype will be discussed at length as part of chapter \ref{ch:2}. Because they are a source of free patient speech that has direct grounding in existing psychiatric practice, much of the upcoming review of acoustic and linguistic features (\ref{subsubsec:acoustics-linguistics-review}) will also draw from the interview literature. \\

\paragraph{Verbal tasks.}
A final style of acoustics and linguistics data collection is to design studies around a research task. This can enable much tighter controls, for example the passage readings created to capture a phonetically balanced set of vocals. Of course the corresponding downside is the limited expressivity of the research task data.  

In the era of app-based research, regular and widespread collection of audio from verbal tasks has been made much more feasible, in turn enabling a broader variety of task prompts. Apps such as that created by Winterlight Labs have been used by multiple groups to deploy a broad range of tasks for targeted speech data collection, including object naming, phonemic fluency, semantic fluency, and image description \citep{winterlight}. In particular, the image description task format has been one of the most commonly used data sources in recent work from Winterlight. In this task, the subject is shown an image and asked to provide a brief (1-5 minute) description of the picture via audio recording. The resulting audio can be assessed for acoustic and linguistic properties, in many ways similar to the audio journal methodology here. While the image description recordings lack any self-report content and may have less qualitative potential, they do also have many of the same advantages that our diaries do. Indeed, the pipeline for objective feature extraction described in this chapter would be straightforward to apply directly to the task recordings. 

Additionally, the image description recordings have the advantage of being a response to a well-defined stimuli. Particularly as apps like Winterlight are used by many groups, a large corpus of knowledge about normal response to common prompt images could be generated and subsequently used to more easily contextualize individual patient responses and their extracted properties. This is especially important for studies that plan to collect a small number of recordings from a large number of subjects, because without longitudinal data it is not possible to use a patient's own historical trends as a baseline, but with unconstrained datatypes it is difficult to baseline otherwise. 

Of course with the dry nature of an image description task, it will be more difficult to sustain regular participant engagement over long stretches, and the individual recordings may lack a certain amount of expressiveness. As such, verbal task recordings in many ways address adjacent research questions to audio diaries, despite being collectable and processable in many similar ways. This could in fact make image description (or other verbal) task recordings a strong complement to audio diaries in a larger scale study. Synthesizing the two voice data sources would even be an inherently interesting research direction. It is plausible that certain image prompts could be tailored to clinical information about the patient, perhaps derived from relevant self-reports made via diaries. A clear example of this would be images depicting objects of anxiety in a personalized manner, although that first would require further consideration of the impact to patient well-being of such a task. \\

Currently, the image description tasks of Winterlight largely involve prompts that have already been well-studied in older laboratory-based projects, with a focus on detecting neurological deficits in older adults. One image is "The Cookie Theft Picture" from the Boston Diagnostic Aphasia Examination, which is a cartoon image of children stealing cookies from a kitchen cabinet behind their mother's back. There are a number of visual details in the drawing, and participants are instructed to describe everything they see going on in the image. Between 1983 and 1988 recordings of this task were collected from 167 Alzheimer's Disease (AD) patients and 97 healthy older adults as part of the DementiaBank project. The recordings were originally manually transcribed carefully, with disfluency and other annotations similar to the work with TranscribeMe in this chapter. Patients with dementia used significantly fewer words and demonstrated problems with word recall. More recently, the old recordings were revisited with modern analysis techniques, including acoustic feature extraction and linguistic feature extraction from a version of the transcriptions that removed disfluency elements that would be difficult to automatically extract from audio \citep{Fraser2016}. 

Although they found features relevant to AD diagnosis and could predict AD well above chance, the number of features considered makes the work of \cite{Fraser2016} purely exploratory. Nevertheless, it provided a good proof of concept for automated techniques, and there is good evidence of task clinical relevance from the original manual annotations of the dataset they revisited. More broadly, the Cookie Theft image description task has been repeatedly validated for consideration in aphasia detection \citep{Borod1980}. As such, there is a natural extension to app-based administration of the image description task to potential AD patients, and ongoing multimodal exploratory work in the AD space is presently collecting recordings via Winterlight \citep{Curcic2022}. Just as daily audio journals are likely to appear more often in future literature, short verbal task recordings are also an active area of growth in the domain of tool building and data collection work. \\

In the realm of psychiatry, image description tasks are less common than they are in neurological disorders research at this time; still, there is not much reason for skepticism that they will turn out to be relevant to psychiatric illness in some capacity. For example, it is likely that linguistic anomalies in some psychosis patients could be captured by performance on an image description task, and indeed have already been measured in other verbal tasks that could be administered by an app, such as  semantic fluency (in older lab based tasks to be reviewed next). Similarly, a number of acoustic correlates of e.g. depression severity can be measured from speech elicited in a variety of manners. Besides their Alzheimer's work, the primary push of Winterlight Labs currently is to collect a dataset from depression patients. 

A pilot publication by \cite{Tasnim2022} on that early Winterlight depression dataset showed a great deal of promise. They have collected 2,674 audio samples from across 571 participants along with self-rated mental health questionnaires for depression (PHQ-9) and anxiety (GAD-7). Five different speech tasks were recorded: sustaining a phoneme sound for as long as possible (up to 1 minute), phonemic fluency, semantic fluency, image description (using a proprietary image designed to be similar to the "Cookie Theft" cartoon), and a prompted narrative to describe the day or a hobby or event of interest. Of course this final task is essentially an audio journal, however they collected only one time point per person, and they did not ask specifically about e.g. mood, so scientifically it is quite distant from the goals of our longitudinal diary work.

Interestingly, the acoustic and linguistic model of \cite{Tasnim2022} substantially outperformed the benchmark PHQ-9 prediction accuracies reported at the most recent Audio-Visual Emotion Recognition Challenge in 2019, despite utilizing shorter recordings from their specific tasks. The distribution of PHQ-9 scores and demographics from the AVEC dataset was only modestly different, suggesting that their task recording methodology may have produced a higher quality dataset for depression detection. On the other hand, the model for PHQ-9 prediction based on acoustics and linguistics was outperformed by a model based only on subject demographic information. This highlights the importance of a personalized approach, which daily audio journals are much more well-suited for.  

Other groups have also employed similar verbal task recordings in psychiatry research using other apps like AiCure. \cite{Abbas2022} recruited 20 Schizophrenia patients to participate in scheduled assessments over a period of 2 weeks through the AiCure app. One task format used was an affective version of image description - at 3 time points during the study, patients were shown images from the Open Affective Standard Image Set and asked to both describe each image and how it made them feel. When recording participant responses during this task, video was included as well as audio, so that facial reactions to displayed images could be measured. Vocal properties during the image description task were significantly correlated with negative symptoms, and were consistent with vocal features that have been previously associated with negative symptom severity. Due to the small sample size of the study, it was not possible to do some of the richer analyses that a dataset like described could enable in a larger cohort though. \\

The strong preliminary results from the digital speech task recording datatypes, in both acoustics and linguistics, provide compelling evidence for the likelihood daily audio diaries will be able to capture features of clinical relevance. The task recordings detailed are short, app-based, and nonsocial, which covers many of the differences between interview recordings and voice journals. Of course, many of these tasks also have an established history of research in onsite collection and manual analysis, so it is expected they should translate well to the app era.

As I will now discuss a number of specific results relating to acoustic and linguistic properties in psychiatric illness, many of which are replicated in an interview recording context, it is worth keeping in mind that verbal task recordings have also seen early success in digital psychiatry. Audio diaries are thus a clear area for extension, and their ability to provide the best of both worlds in many ways makes them primed for a large uptick in research relevance - once they are noticed anyway. 

\subsubsection{Acoustic and linguistic properties}
\label{subsubsec:acoustics-linguistics-review}
Besides enabling advanced visualization and search techniques to improve qualitative analyses, and substantially decreasing the labor required to accurately compute preexisting quantitative measures, modern data science and software engineering also introduce novel quantification techniques for both audio and text datasets. These methods could be of great use in studying collected audio journals, and their development includes advances in a variety of different forms, such as:

\begin{itemize}
    \item Metrics that quantify in a consistent and defined way vocal properties that would previously be judged by a clinician qualitatively (e.g. pitch changes), and to do this at scale to obtain a larger processed dataset for less cost.
    \item The ability to compute metrics like vocal tremor with temporal resolution much finer than would be possible by even a trained human.
    \item Decreased noise in metrics that are inherently subjective, so that algorithmic estimates are closer to the average of human estimates across a dataset than any individual human is (e.g. sentiment labeling), as well as more accessible for replicability purposes.
    \item Brand new metrics arising from technical advances in other application domains that utilize audio or text, or from data-driven unsupervised approaches, possibly combined with prior psychiatry intuition (e.g. features derived from word embeddings).
    \item The emerging tractability of highly multimodal approaches that can consider many different feature types at once; perhaps eventually leading to clinically relevant features that are based solely on co-variation of base properties rather than their individual values.
\end{itemize}

\noindent Given the huge scope of acoustic and linguistic features that are now possible to consider, it is important to review the existing literature on use of these features in psychiatry. \\

\paragraph{Audio journals versus clinical interviews.}
As mentioned, most prior work on computational vocal properties or natural language processing (NLP) in a psychiatric patient population has utilized recordings or text obtained from specific study tasks, or more recently from semi-structured clinical interviews. Speech and language content produced during tasks is by definition substantially more constrained than that produced for our journals, and even in the case of clinical interviews much of the content is in response to some preset question. 

Interview data is also in the format of a dialogue, which may not only require somewhat different analysis techniques from the monologue format of the diary, but additionally may inherently elicit different behavior and ultimately a different response from a participant than the same question would if delivered as an audio journal prompt. 

Moreover, interviews are significantly longer than daily audio journals, which are intended to take only a handful of minutes each day. Journals on the other hand are submitted much more frequently and thus cover many more days -- both filling gaps in between interview dates and making longer study periods much more feasible. 

There are advantages and disadvantages to the clinical interview and to the audio journal as sources of patient speech data, as well as different nuances to consider for downstream processing of data from the two formats. Interview recordings are the subject of chapter \ref{ch:2}, and thus the details of interview analysis are discussed at much more length there. 

Nevertheless, audio diaries and interview recordings share many commonalities. They are both datasets of semi-naturalistic human speech, lending themselves to tools for extracting vocal properties from audio and for performing NLP on transcriptions of spoken word. The features that turn out to be relevant for characterizing behavior of a particular patient population during clinical interviews are indeed great candidates for features to investigate in an audio journal dataset. \\

\paragraph{Linguistics in Schizophrenia.}
Psychotic disorders, especially Schizophrenia, often present with symptoms deeply rooted in language. Schizophrenic hallucinations are overwhelmingly auditory, with more than half of diagnosed patients reporting \emph{verbal} auditory hallucinations as one of their symptoms. Research on congenitally deaf patients that develop Schizophrenia suggests that hallucinations will manifest as visual representations of language in a non-hearing population, such as imagery of disembodied hands performing signs \citep{Atkinson2006}.

Because of the relevance of language in the experience of Schizophrenia, it is not surprising that abnormalities in patient language have been well-documented by clinicians. During a psychotic episode with severe positive symptoms, delusional beliefs can be revealed by the semantic content of speech; conversely, alogia is a fundamental negative symptom of Schizophrenia defined as paucity of speech and/or poverty of content of speech, and is one of the diagnostic criteria considered in the DSM-5 \citep{DSM}.

Study of speech in Schizophrenia has thus been a common research theme in psychiatry for multiple generations now. Qualitative observation of multiple speech abnormalities in psychotic individuals was documented by 1914 with the work of Kleist \citep{Walenski2010}. More recently, research from the 1990s and early 2000s performed more systematic analysis of speech properties of interest, utilizing manual label of features like pause times \citep{Alpert1997} or grammatical structure \citep{Covington2005} to correlate with clinical scales.

Given the promise of existing literature on speech patterns in Schizophrenia, computational acoustic and linguistic analyses are a very natural extension, particularly applications of NLP to patient transcripts. It is thus unsurprising that this domain saw many of the early successes of digital psychiatry and has remained a hot topic of research interest. \\

\paragraph{Applications of NLP in Schizophrenia.}
Indeed, in the mid-2010s early NLP techniques such as latent semantic analysis were successfully used in multiple studies of psychotic disorders. \cite{Holshausen2014} performed a richer analysis of data obtained from traditional cognitive tasks in Schizophrenia patients, while \cite{Bedi2015} performed one of the first computational analyses of psychiatric patient speech from semi-structured interviews. 

Higher mean word uncommonness during semantic and phonological fluency tasks predicted greater overall impairment in Schizophrenia patients, even after controlling for other factors like cognition score at the time of the task \citep{Holshausen2014}. Word fluency tasks prompt each subject to list as many words as possible fitting into a particular category in a limited time frame, for example to name different animals or different words starting with the letter H for 90 seconds. Therefore they can assess thinking process in a more direct manner than less constrained speech can, but they also lack many forms of information such as grammatical structure. 

In the context of recorded clinical interviews, greater maximum semantic incoherence, shorter maximum phrase length, and less frequent use of determiners were identified as features most predictive of psychosis onset within 2 years of the interview date in a population of high risk youth \citep{Bedi2015}. These features are consistent with the observations of incoherent speech captured by fluency tasks as well as the observations of speech paucity in those with severe negative Schizophrenia symptoms, and they were able to outperform clinician prediction of psychosis risk in the small pilot dataset ($n=34$ with $5$ positive examples) investigated by \cite{Bedi2015}.

As NLP technology has been advanced by computer science, follow-up studies have found similar results as well as new features of potential clinical relevance. \cite{Tang2021} employed transformers to study speech from semi-structured interviews conducted with both Schizophrenia spectrum disorder patients ($n=20$) and healthy controls ($n=11$). They found features of language disturbance to distinguish patients from controls more accurately than clinical scales alone. Linguistic differences in the patient population included greater sentence to sentence incoherence, less frequent use of adjectives, adverbs, and determiners, more frequent use of pronouns, and more frequent occurrence of incomplete words \citep{Tang2021}. \\

\paragraph{Limitations of pilot work.}
One of the limitations of the existing NLP work from semi-structured interviews is that each participant is typically only recorded once, so there is little study of within-subject variability patterns in these linguistic features. A longitudinal profile of patient speech would be of interest in any population, but would be of particular note for adolescents or young adults that are at high risk of developing or already beginning to develop Schizophrenia symptoms. 

A pilot study of NLP applied to speech from over the course of development found that speech in Schizophrenia patients has a more child-like structure, which corresponds to a halt in language development rather than a regression due to the disease \citep{Mota2018} -- thereby supporting the perspective of Schizophrenia as an extended developmental disorder. Interestingly, the results of \cite{Mota2018} suggest that this halting of language development is not remediable by traditional education. More targeted early intervention strategies will require further investigation. 

Another major limitation of existing NLP work in Schizophrenia is the generally small study population sizes. It is comforting that different studies have found significant predictive power in similar linguistic features, features that also strongly overlap with expectations from historical results in psychiatry. However, datasets with so few labeled points are fundamentally incompatible with rigorous data exploration methods, while at the same time interview recordings present the opportunity to extract a tremendous number of features assessing different acoustic and linguistic properties at different timescales. 

Thus existing NLP for psychiatry research has largely been either restricted to testing hypotheses of narrow scope or incentivized to use dubious methods for claiming statistical significance of a model after a substantial feature search step. In fact the aforementioned pioneering paper from \cite{Bedi2015} is astonishingly vague about the methodology used for feature extraction and selection, with no exact count of original features nor the specific algorithm used for selection given in the main text, and only a single poorly documented Excel sheet provided as supplementary material. They then proceed to demonstrate in detail why the results of their final model on the much narrower feature set were not explainable by random chance with such a small number of features. \\

\paragraph{Can daily journals address limitations?}
Due to the current system for academic funding and publishing it is not realistic to expect truly exploratory reporting on one dataset and then rigorous follow-up on a completely separate dataset, especially given the pace of new technical developments in the NLP domain. It is therefore concerning that a good deal of recent research work in psychiatric speech analysis has focused on clinical interviews from relatively small study populations. 

Discovery of truly novel features of relevance is more difficult in this framework, and robust validation of already suspected features of relevance is more difficult to verify for the community when published work may downplay or entirely omit feature exploration steps. The scientific process then deals with more noise and ultimately takes longer to arrive at certain conclusions than it otherwise could.

Of course, collection of a large body of interview recordings is expensive and not feasible for most individual labs. The AMPSCZ project to be described in chapter \ref{ch:2} is one attempt to address this issue, by facilitating compilation of a bigger dataset from across groups. It also aims to make analysis methodologies more cohesive, to gain control over the rapidly expanding number of computer science tools available that could lead to subtly different, poorly tracked, results across studies. At the very least, it will be important to have a common language for communicating which techniques were used at different stages of a research project.

Employing audio journals to collect patient speech samples instead of clinical interviews is another approach that could help to address the issues with interview study structure. Journals certainly have their own disadvantages when compared to interviews, but overall they are substantially more realistic for a group to collect at scale. Between the reduced burden to participants, the reduced work required of study staff, and the daily prompt frequency, studies can enroll more participants for a longer duration and obtain many more distinct data points (with labels from e.g. daily EMA) per day an individual participant is enrolled, resulting in a substantially larger dataset to work with. 

Furthermore, as single diaries are much shorter than interviews and contain only a single speaker, the number of features worth exploring per recording is an order of magnitude smaller than that per interview recording. Taken together with the much larger number of labeled recordings, a big data perspective is much more tractable in an audio journal dataset. 

It is thus important to carefully consider the scientific justification for utilizing an interview dataset instead of an audio diary one; although a number of good reasons theoretically exist, most papers do not discuss why their particular work was better suited for interviews than diaries. Additionally, the current paucity of journal results in psychiatry suggests that this conversation more broadly has not been had enough. 

Nevertheless, there may be reason for optimism that upcoming work will catalyze audio diary research in psychiatry. Trends in funding and in applications available for data collection do suggest more publications on the topic are in progress. So, with limitations now in mind, I will continue the review of other promising acoustic and linguistic results from early digital psychiatry research. \\

\paragraph{Acoustic properties of clinical relevance.}
In addition to the many publications assessing linguistic features in Schizophrenia, recent tech advances have enabled clinically relevant acoustic properties to be uncovered. It has been hypothesized based on prior psychiatry literature that negative symptoms in particular will manifest changes in vocal properties \citep{Covington2012}.

Indeed, vocal analysis of speech from clinical interviews of patients with psychotic disorders has been coupled with clinical rating subscales from the "Positive and Negative Syndrome Scale" (PANSS) to identify acoustic features correlated with both positive and negative symptoms \citep{Wortwein2017, Covington2012}. 

\cite{Covington2012} found abnormalities in specific formants occurred significantly more in the speech of Schizophrenia patients experiencing severe negative symptoms than in the speech of those that received lesser negative symptom scores at the time of the recording. Formants correspond directly to physical features of speech production, so that they may have a close relationship with psychomotor signs thought to be of clinical relevance. They are easily extractable from audio data via open source tools like OpenSMILE \citep{OpenSMILE,GeMAPS}.

\cite{Wortwein2017} identified novel correlates of positive symptom severity in psychotic disorders patients, including larger vowel space and faster speaking rate. Vowel space is another property directly related to physical speech production; it correlates with perceived articulation, and is also easy to extract using open source tools, in this case Praat \citep{Greenwald2013,Son2018}. 

Speaking rate can be strongly affected by cognition, but it is yet another feature that is modulated in part by function of physical systems. Psychomotor slowing observed during episodes of depressed mood has been associated with slowed speech rate, and more recent work has found lower speech rate during clinical assessment to significantly correlate with severity of corresponding negative symptom scores \citep{Agurto2020}. Given that clinicians evaluate for various forms of speech poverty, the association of slowed speech with negative symptom severity is somewhat definitional, but still promising to see validate in practice. 

Speech rate can be extracted using Praat as well, and both \cite{Agurto2020} and \cite{Wortwein2017} have utilized Praat successfully for this purpose in their work. Though Praat speech rate scripts have been validated to be fairly accurate, and can be a good tool for assessing datasets where audio is available but high quality transcripts are not, we found syllable estimates based on the text from verbatim TranscribeMe transcripts to be more reflective of human estimates than the matching syllable counts returned by Praat as part of its speech rate calculation. This is one of many pipeline design decisions that will be discussed in section \ref{subsec:diary-val}. \\

\paragraph{Psychomotor and perceptual factors of depression.}
Though psychotic disorders may have a more fundamental basis in language abnormalities, there has also been a history of mood disorder research that would suggest a relationship between symptom severity and properties of patient speech. This is particularly true for the analysis of vocal features and of paucity of speech: descriptions of patients with severely depressed mood hardly producing any utterances date back over a century, and more recent analyses of systematic manual scoring have found slowed speech rate, slowed motor response time, overall less motor activity, and a collection of related features coined "psychomotor slowing" in patients with more severe levels of depression \citep{Sobin1997}.

Changes in physical functioning that are observed to occur during a depressive episode should therefore be captureable by the immediate extension of some of the above mentioned acoustic analyses, enabled by tools like OpenSMILE and Praat, to speech recordings from depressed individuals. Moreover, there is reason to believe other new features of relevance could be discoverable by such work. 

NLP for depression analysis is less grounded in specific prior literature, but is nevertheless of promise. A body of research on cognitive task performance by depressed patients has demonstrated that those with depression will on average perceive the magnitude of the same positive and negative cues significantly differently than the average healthy control \citep{Miller1973}. 

Though altered perception may vary across heterogeneous cases of depression, it is likely that individuals with severe depression will display a different outlook on life events than they would if healthy. In fact, one of the core therapeutic techniques that comprises cognitive behavioral therapy (CBT), an intervention with a strong base of evidence for efficacy in depression, is centered on changing how a patient perceives their life and daily environment. CBT participants are instructed to reframe negative thoughts, building up a habit to condition positive thoughts to eventually come to them automatically \citep{Beck2011}.

A negative versus positive outlook on an event is theoretically distinguishable by properties of the language used to describe it, thus opening up a number of promising directions for use of NLP in mood disorder characterization. While sentiment analysis will always have contextual and truly subjective elements, automated approaches have already demonstrated impressive (and useable) accuracy levels for some time \citep{VADER}. Because the longitudinal daily diary approach allows sentiment and other feature values over time to be normalized against the patient's own baseline, it is plausible for NLP-based techniques to detect personalized signs of worsening depressive symptoms. \\

\paragraph{Computational results in mood disorders.}
Early studies of acoustic properties typically asked participants to read specific passages aloud. These passages, such as the well-established "rainbow passage", were designed to capture a variety of vocal sounds in a phonetically balanced way \citep{Alghowinem2013}. While these tasks are acoustically well-controlled and still have relevance for certain scientific questions, they have a number of downsides for general use in psychiatry.

In modern society, it is substantially easier to collect natural speech samples than to collect controlled passage readings, so that successful use of free speech in clinical prediction tasks is a much more desirable end goal. Furthermore, free speech allows for linguistic analyses to be conducted alongside acoustic ones, and even for acoustic features to be processed in a content-dependent manner by using aligned semantic information. Additionally, there is good reason to believe that free speech is inherently a better input for detecting e.g. depression severity based on acoustics alone. It has been regularly, albeit often qualitatively, observed that participants read canned passages in a drier manner than they would typically speak in daily life. Passages are generally not written to elicit any emotion, largely intentionally as a means of tightly controlling a vocal study. 

Because of this conjecture, \cite{Alghowinem2013} conducted a study of 30 depression patients and 30 healthy controls, each of whom participated in a short semi-structured interview and read aloud a typical passage. The resulting audio was processed for acoustic features with relevance to emotion prediction, including formants, vocal loudness, jitter, and shimmer. These features exhibited substantially more variability over the course of the interview segment than during the passage reading, in both groups. Further, an SVM classification scheme with leave one out cross validation was significantly more accurate at distinguishing the depressed participants from the healthy controls using free speech-derived features than the same classification method could achieve using the passage speech-derived features.  \\

More broadly, the presence of clinically relevant signal in short snippets of vocals from semi-structured interviews has been repeatedly documented. Qualitative analysis of acoustics is not only something that psychiatrists are trained to do when evaluating symptoms, but even untrained individuals can distinguish participants diagnosed with depression from healthy controls significantly above chance by listening to brief samples of free speech. Within the same dataset, computationally extracted acoustic features explained $\sim 60\%$ of the variance in corresponding clinical severity rating \citep{Yang2012}.

Though not directly relevant for journals, there is also preliminary evidence from semi-structured interviews that acoustic properties of interviewer speech might correlate with severity of participant depression \citep{Scherer2014}. This opens up a number of questions about variance in clinician behavior based on perception of the current patient, whether intentional or unintentional. Such variance might impact patient outcomes and it might bias study results, so it certainly warrants further investigation. It also might be leverageable for better understanding of depression symptomatology, particularly in the study of social interactions. Regardless, it is yet another variable that requires careful consideration in an interview study, but would have little bearing on an audio journal study. \\

In the case of depression, there is fortunately a greater body of literature on free speech sources besides the interview. While there is not much on daily audio diaries, automated telephone answering services have been used by prior work to periodically collect a battery of depression patient speech samples at key points of a study \citep{Mundt2012}. These samples included both reading of a specific passage -- in the case of \cite{Mundt2012} the "Grandfather passage", a reading in the same spirit of the rainbow passage -- and a prompt for open-ended speech about the prior week(s).

The audio collected over the phone was analyzed for speech duration, pause count, speech rate, and vocal properties like pitch and formants, individually for the response to each prompt. Significant correlations were found between these acoustic features and time aligned clinical ratings of depression severity in one study \citep{Mundt2007}. Due to those results, an entirely new study of the acoustic processing system was conducted in a second depression population, this time in the context of a 4 week stage four clinical trial of a depression drug \citep{Mundt2012}. 

The clinical trial supplemental study performed by \cite{Mundt2012} enrolled 165 participants, and compared the change in acoustic properties between baseline and study end for each participant with the primary measured clinical outcomes. They focused on the features found to be most relevant in their prior work to increase statistical power here; narrowed down features included speech rate and pause time, but not vocal measures like formants. The select feature set demonstrated significant correlation with ratings of depression severity, whether extracted from free or automatic speech. Critically, features changed significantly more between baseline and study end in those deemed to be treatment responders than they did in the rest of the study population. 

The fact that features of speech duration, speech rate, and pauses within speech were observed to demonstrate treatment sensitivity in a clinical trial setting, in addition to replicating the prior acoustics literature associating them with scale values \citep{Mundt2012}, is a more robust computational psychiatry result than much of the above described literature on speech in psychotic disorders. Coupled with the strong reasons for prior belief in the association of these properties with depression, there is a good basis for future work on the topic at scale, and audio diaries are a closer successor to the automated phone system described than clinical interviews recordings would be. \\

Interestingly, some acoustic properties associated with depression have even generalized to detecting depressive symptoms that are secondary to another diagnosis. \cite{Tan2023} recorded speech from interviews of 23 Schizophrenia patients with comorbid depression, 19 Schizophrenia patients with no clinically significant depressive symptoms, and 22 healthy controls. They extracted 15 features across 5 categories: utterances, words, speaking rate, formulation errors, and pauses. Pause-related features were significantly different in the depression diagnosis group versus both non-depressed Schizophrenia diagnosis and healthy control groups -- including a higher number of utterances containing at least one pause and a higher number of pauses per utterance. By contrast, speech rate was not found to significantly distinguish between these groups \citep{Tan2023}.  

Because slowed speech rate has repeatedly been associated with depressive symptom severity in the past, the results of \cite{Tan2023} are worth following up on in future work. It is possible that the smaller sample size of their pilot work did not grant sufficient power to detect speech rate differences between groups, particularly as between person heterogeneity can be a large source of noise in individual sample studies. However, it is also possible that the interaction between depressive symptoms and speech rate is different in those with a primary psychotic disorder diagnosis than in those with depression alone. This could relate to how different features of speech interact with each other, how depressive symptoms are assessed in patients with comorbid disorders, or even some underlying relationship between disease mechanisms. Given the highly imperfect nature of current diagnoses, it is unlikely that a single clean answer exists; this underscores the importance of building models that can explain interactions between different speech features as well as between different symptom dimensions. \\

While analysis of acoustics in depression is an established line of research, there is less work on linguistic properties of free patient speech. The majority of NLP studies that analyze language produced in depression to date have utilized text from social media applications rather than transcribed spoken language. This research has lead to promising pilot results, particularly for the use of sentiment estimation techniques in distinguishing those with a mood disorder diagnosis from those without \citep{Babu2022}.

The literature on sentiment analysis thus supports the hypothesis that sentiment will be relevant in a journal transcript study, but it is far from confirmed. Verbatim transcripts of spoken word can differ in structure from written text, as was discussed earlier in this review. Furthermore, text written for a social media platform has a very different purpose and audience than a mostly private entry to an open-ended diary. Clinical interviews are of course a third context with other considerations and therefore additional research needed independently of any conclusions about audio journals. \\

Although most of the results I provide background on here are related to Major Depressive disorder (or previously Schizophrenia), they likely apply to other mood (or psychotic) disorders as well. Keeping in mind that results are to some extent confounded by poor labeling, there is a great deal of overlap in the broader acoustic and linguistic patterns observed across psychiatric diagnoses. For example, Bipolar disorder patients also exhibit psychomotor slowing and subsequent slowed speech rate during depressive episodes, and conversely have been documented to speak at a faster rate during manic episodes \citep{Sobin1997}. Along the same lines, NLP features that exhibit significant correlation with psychosis risk in Schizophrenia can also significantly correlate with psychosis risk assessments of Bipolar disorder patients \citep{Hitczenko2021}.

The fact that these summary level features informed by prior clinical observation appear to have a robust relationship with clinical outcomes is indeed extremely promising. At the same time, reported effect sizes are often small, and we lack computational studies that thoroughly dissect subtle differences in speech within a heterogeneous patient population, or even between patients from two distinct yet related psychiatric diagnoses. Though reviews such as the one from \cite{Hitczenko2021} call for less reliance on categorization, as have I repeatedly throughout this thesis, in current practice diagnostic labels are still relied on - and therefore research should continue to pay some mind to understanding these labels (which may in fact contain some signal of interest too), while simultaneously shifting core focus to symptom measures.

Ultimately, daily audio diaries represent the perfect opportunity for a paradigm shift in the computational acoustics and linguistics literature. Rich longitudinal datasets of psychiatric speech will enable greater application of computer science tools, to be discussed next. Critically, audio journal datasets also allow for a deeper qualitative study of open-ended patient speech over time, a perspective that has the potential to generate new insights on the dynamics of these computational features in relation to each other and external factors, as well as to inspire new forms of and uses for vocal and NLP properties. \\

\paragraph{The role of computer science research.}
Overall, there is much to be excited about in the application of computational acoustic analyses and NLP techniques to psychiatric patient speech. Many published methods already perform strongly enough on benchmarks to justify direct extension to the setting of daily audio recordings, of course with some validational review to sanity check. Other useable features will surely continue to emerge at a rapid rate with the great progress being made in machine learning currently. 

However, there are still difficult computational and engineering questions that are of greater salience for digital psychiatry than for many other applications areas of machine learning. From an algorithmic perspective, advances towards addressing challenges of learning with few data points, learning with dirty labels, and learning in an interpretable way remain extremely important limiting factors for the use of data science in psychiatry. The way these issues are addressed in an application-specific manner by interdisciplinary scientists will be one key determiner of the near future success of digital psychiatry. Another will be the progress made on these areas of research in the machine learning literature, which is a major focus for chapter \ref{ch:4}.

Aside from the fundamental theoretical challenges involved with handling dense digital psychiatry datasets, there will also be difficulties presented by the software engineering necessities as well as the social system dynamics required to implement such work in practice. Adopting good programming practices and encouraging code sharing would greatly help the field to expand, by not only making it accessible to a diversity of researchers without substantially sacrificing quality, but also by reducing the costs required for a group to conduct subsequent studies addressing different questions in different disorders. Further, good engineering would reduce the probability a stupid error is made during data analysis, and sharing that clearly written code would make any such error that did appear much easier for the community to find.

The described software improvements are largely up to the field itself to adopt, as many well-established engineering best practices do not make their way to academic research (even in computer science). Such practices are not always necessary, but there is a concerning lack of discussion about which practices ought to be used in research under what circumstances. For studies that aim to scale to many subjects or algorithms that are intended to be used for many studies, software development should be more carefully considered and probably given greater upfront budget. Digital psychiatry groups are especially likely to encounter research questions that fit this model. 

However, rapid advances in computer science and machine learning can in fact complicate the software development process, doubly so as it currently haphazardly occurs in digital psychiatry. As models for acoustic and linguistic tasks -- for example sentiment estimation -- improve, researchers might be tempted to switch to the latest methods. Researchers might even unintentionally switch models if relying on a tool that updates without paying careful attention to version control.

There can of course be very good reason to move to updated models, but that decision ought to be based on weighing of not only various accuracy metrics, but also measures of interpretability and computational complexity, as well as a deeper understanding of how observed improvements on benchmarks might differ when applied to one's own research dataset. Furthermore, the decision should take into account the degree of improvement relative to the work that switching entails for one's own group as well as the broader impact that frequent model updates can have on scientific cohesion.

The degree of negative externalities imparted by model switching will vary greatly depending on the standards developed by the digital psychiatry research community while the field is still relatively young. The more modular and readable a code base is, the easier altering specific methods or rerunning old analyses with updated methods will be. The clearer and more consistent communication about research methods is, the easier it will be to recognize differences in features used across papers and interpret accordingly. 

A major goal of this chapter's work is thus to establish a coherent set of baseline features for the analysis of free-form patient daily audio journals. Towards this end, I not only release an initial software package here along with detailed feature documentation, but discuss in depth the design decisions behind each feature included. That discussion, found throughout section \ref{subsec:diary-val}, involves not only pilot validation results from our dataset, but also a review of relevant validation results from the computer science literature, and an overview of alternative methodologies that are used by prior psychiatry literature or might have potential to surpass the methods I chose. 

The code is ultimately meant to be light weight, but can provide foundation for the development of the field because of all the reasons outline. Alternative and additional feature computation techniques can be easily compared to baselines provided by the features I use across all important domains, and further the code is written in a modular fashion so that any new methodologies of note that do become established would be straightforward to introduce into the pipeline. Ideally, this will facilitate more audio journal research by groups with a primary psychiatry focus as well as a defined process for groups at the interdisciplinary forefront to continually improve the technology without disrupting cohesion of the scientific community.

Please see section \ref{subsec:diary-val} for the background literature review on the technical aspects of the acoustic and linguistic features discussed in this introductory section. \\

\paragraph{An interdisciplinary approach.}
 Computational techniques for acoustic and linguistic processing enable many different quantitative questions to be asked of an audio diary dataset. Given the broader relevance of these modalities across many fields, feature extraction algorithms that can be easily extended to audio journals continue to improve at a rapid pace. Coupling this style of analysis with the rich qualitative perspectives that can arise from journal entries should lead to many new psychiatry insights, particularly when assisted with emerging tech tools. 
 
As such, I will close this background review with a more focused discussion of how audio diary analysis could enabled long term advances in understanding on a behavior of importance in psychiatry - sleep. This serves as just one example of the many ways that modern technology might formalize historic ideas from psychiatry in a new, more scientific fashion that was possible with previously existing tools. 

\subsubsection{The future of sleep classification tech}
As smartwatch accelerometers become increasingly commonplace to wear to bed, some of the traditional sleep questionnaire categories could become obsolete; existing actigraphy algorithms already outperform self-report on sleep measures like latency, duration, and wake events \citep{Teo2019, HabibSleep}. Yet this actually presents an opportunity for more creative EMA question design and especially freer form sleep diaries collected using methods like the audio journal recording workflow described here.

There is much that remains to be understood about sleep, and many of the aberrations that can be reliably observed in some Schizophrenia \citep{Kaskie2017} or Autism \citep{Devnani2015} patients (for example) are related to a specific stage of sleep rather than more general issues with staying or falling asleep. Wrist actigraphy studies have failed to distinguish REM from non-REM sleep, with accuracy on this binary classification problem unable to surpass $60\%$ even with a large training dataset available \citep{Boe2019}. For more nuanced staging of non-REM sleep the situation is even bleaker, and there are fundamental reasons to be skeptical of how far movement alone can go. Accelerometer data has been collected as part of polysomnography for decades now and analyses repeatedly fail to find any strong patterns linking specific stage with movement \citep{Conradt1997}.

Improved sensor technology may eventually make EEG sleep staging more practical for at home measurement, or even lead to new types of data streams with relevance to sleep stage or other difficult to detect sleep properties. Wrist watches that collect electrodermal activity have indeed already shown promise in distinguishing REM and non-REM sleep, as the still mysterious "sleep storms" that are often found in EDA signal strongly correlate with deeper stages of non-REM sleep \citep{Sano2014}. \\

\paragraph{The future of sleep diaries as an important component.}
Regardless, a new era of sleep journaling could strongly complement the established and emerging passive sensing technologies for sleep characterization. Despite the better availability of physiological monitoring paradigms and the larger existing body of work on digital phenotyping for sleep science relative to psychiatry, there are still numerous sleep issues that remain unexplained. Idiopathic hypersomnia, i.e. excessive sleepiness without explanation, is surprisingly common - diagnosed in about 1 per 10000 people, but estimated to occur at moderate severity levels in a much large percentage of the population. An analysis of the Wisconsin Sleep Cohort, a comprehensive dataset of sleep measurements from 792 individuals sampled to be representative of the population, found evidence of idiopathic hypersomnia in $2.4\%$ of participants \citep{Peppard2022}.  

Thus there is a great need for new techniques in sleep science, in addition to the scaling of older ones. Passive behavioral sensing and new biological methodologies will of course play an important role, but the utility of qualitative information in guiding hypothesis generation should not be undervalued. Individual observations from idiopathic hypersomnia patients when viewed at scale could inspire specific measurements with the potential to at least meaningfully separate those affected, if not connect back to a biological hypothesis. Just as with psychiatric disease, it is extremely likely that idiopathic hypersomnia is multiple disorders, so a good partitioning would greatly facilitate biomarker discovery and similar studies.  

Furthermore, dream interpretation was once viewed as a serious component of psychotherapy. Its foundations were primarily philosophical and anecdotal rather than rooted in scientific evidence \citep{Freud}, and consequently discussion of dreams has for the most part been taken over by pseudoscience in the popular imagination. While dream content still has legitimate relevance in a few specific contexts like traumatic dream occurrences in PTSD \citep{Holzinger2020}, there is a lack of serious recent literature on the topic that considers both what is known about dreaming neurobiologically and what information dream content could contain psychologically. 

By discarding most of what is "known" about dream interpretation, but still taking the idea of dream interpretation seriously, it is plausible that dream content could be linked with relevant psychiatric variables in some contexts. It is even plausible that down the road, information about dream content could be integrated with neurobiological models of sleep, as mechanistic nuances during REM sleep or in transitions to/from REM sleep very well might alter dreams in a predictable fashion. Using audio journals to document dreams immediately after waking would be fairly straightforward to do both longitudinally and at scale, and would produce a dataset with the potential for numerous different qualitative and quantitative analysis techniques as will be discussed. A similar framework could be used for other more restricted questions about sleep too, including descriptions of remembered waking events or thoughts and emotional state prior to falling asleep. 

 Because there is a strong intersection between sleep and psychiatry already, sleep characterization could very well be the best first focus for digital psychiatry development. It is more tractable than broader questions about characterizing e.g. psychosis, while at the same time it has potential for massive impact beyond what would be possible for a study focused on modeling a niche psychiatric issue. Further, it has strong consumer interest, thereby increasing economic incentive, ease of recruiting subjects, and potential for industry collaborations that strategically pool resources. 

\subsection{Moving forward}
\label{subsec:diary-motivation}
The above proposed use cases for audio journals in sleep analysis represent just one example of the many ways diaries could be effectively used in psychiatry research. Thus there are a large number of open questions to address. Not only exciting scientific questions, but also critical questions on how to build a strong methodological framework for this subfield of research.

 As mentioned, there are few existing works that apply the tools available for speech and language analysis more broadly to the specific application of daily audio journals. One example where app-collected journals provided great insight in a study can be found in chapter \ref{ch:3} of this thesis, where not only did a shift in language features precede an episode of highly depressed mood, but also diary content was used to identify changes in stimulant usage patterns that may have unknowingly confounded aspects of a novel deep brain stimulation trial. Those results were of course based on outputs generated by parts of the pipeline presented in this chapter, thereby further highlighting the need for our work.

Many open questions remain about best practices for both collection and analysis of longitudinal audio diaries. Most of the collection related questions are applicable to the manual style of journals too, and are therefore detailed in the above discussion on historical background.

Analysis needs to be addressed for both audio and language applications. Furthermore, because of the wide range of ways that journals can be applied, analysis techniques also need to be carefully considered at many possible levels of detail. To best facilitate future scientific questions about these journals, a number of questions about technical implementation and output validation should first be answered. \\

In this chapter, I take a concrete step in that direction, through the release of an audio journal processing pipeline that produces multimodal outputs at multiple levels of detail. The chapter demonstrates example use cases for different types of outputs and carefully documents the functionality and implementation of the current pipeline. Validation evidence for each major output is presented, and instructions for others to adapt and use the pipeline for their studies are also included. 

Furthermore, I also report proof-of-concept scientific results obtained from applying my code to a longitudinal diary dataset collected from participants with Bipolar disorder. As part of that process, I characterize the distributions of key pipeline features in that dataset, along with their correlation structure.

\subsubsection{Chapter aims}
The goal of this chapter's work is consequentially four-fold:
\begin{enumerate}
    \item To argue for more research focused on audio journals in the emerging digital psychiatry literature.
    \item To provide a blueprint for audio journal study design, from tips and monitoring tools for the data collection phase to code that implements a variety of feature extraction and visualization methods for the analysis phase - the latter including well documented pros and cons presented for each output type along with pilot validation results.
    \item To report preliminary results on properties of the diaries themselves in a Bipolar disorder cohort, to inform expectations for future psychiatry studies; as part of this, I also enumerate future directions for the pipeline that could improve upon existing limitations.
    \item To report preliminary results on the predictive power of some of the diary features towards assessing measures of more immediate clinical relevance; as part of this, I include ideas on hypotheses for future study that were generated using the proof-of-concept scientific analyses.
\end{enumerate}

\noindent In sum, these contributions will lay groundwork to facilitate research on a large set of open questions about journal analysis -- many of which were asked as part of the preceding background or will be asked in the coming discussions (section \ref{sec:discussion2}). \\

The next section (\ref{subsec:diary-methods}) will describe the data collection methodologies used here, along with basic properties of the resulting diary dataset. Information on the pipeline, which could theoretically be applied to a number of other datasets as well as the present Bipolar one, will follow in section \ref{sec:tool2}. See section \ref{sec:science2} for the meat of the analyses involving pipeline code outputs from those diaries along with some clinical information.

\section{Data collection methods}
\label{subsec:diary-methods}
While the new tools introduced in this chapter are written to generalize to other Baker Lab studies, and it would be straightforward to adapt them to many other labs' audio journal collection workflows, all of the validation and proof-of-concept applications were performed using a particular lab dataset: the Bipolar Longitudinal Study (BLS). This study was chosen as an initial focus for my pipeline due to both its size and its available budget for obtaining high-quality manual transcriptions.

BLS is an active multimodal data collection project to track behavior over long periods of time in adults with Bipolar Disorder (BD). The study aim is to collect 100 person-years of data from BD patients, including both continuously collected phone sensor and wrist actigraphy data streams, as well as monthly recorded clinical interviews and daily patient self-report surveys. The clinical interviews produce scores for the Positive and Negative Syndrome Scale (PANSS) to assess psychotic disorder symptom severity, and scores for the Montgomery-Asberg Depression Rating Scale (MADRS) to assess mood disorder symptom severity, as well as the Young Mania Rating Scale (YMRS) for Bipolar mania symptoms. The daily self-report surveys include an Ecological Momentary Assessment (EMA) prompt that asks the patient to rate their agreement with statements about various feelings and behaviors, and a prompt to record a free-form audio journal reflecting on their day. 

BLS participants are free to choose which parts of the study they want to engage with, as collection of any particular modality is considered optional, with compensation directly tied to the amount of data provided.  Patients are recruited from clinical programs of the divisions for Psychotic Disorders and Depression and Anxiety Disorders at McLean Hospital, and from the Rally with Mass General Brigham (MGB) platform. All data collection and storage procedures described in this chapter have been approved by the Institutional Review Board at McLean Hospital. While the overall study remains ongoing, it began collection in February of 2016 and has thus already collected a large amount of data to perform analyses with. As audio diary prompts for BLS participants were ceased on Beiwe in May of 2022 (and eventually transitioned to a new platform), that serves as a natural cutoff point for the dataset considered here.

The remainder of this chapter will obviously focus on the daily audio journals portion of the study. However, the clinical scales and EMA were used to ground proof-of-concept applications, and will be revisited in section \ref{sec:science2}. Note that the use of similar tools for analyzing the BLS (and other) clinical interview recordings will be the subject of chapter \ref{ch:2}. Moreover, the availability of diverse passive sensor data provides a large set of notable future study directions for BLS audio/language analysis. 

\subsection{Diary submission workflow}
\label{subsubsec:diary-submission}
Smartphone-delivered audio diary requests were sent once daily to consented BLS participants through the Beiwe mobile platform \citep{Beiwe}. Each notification asked the patient to record an audio diary for up to four minutes, using the following prompt: 

\begin{quote}
Please describe how you've felt over the last 24 hours in relation to routine or significant events that may have occurred. Remember to focus on how the events made you feel, and to avoid names of specific people or places. When you are ready, please press ‘Record’ and speak for about 2 minutes. Press ‘Stop’ when you are finished, ‘Play’ to listen to the recording, and ‘Done’ to submit the recording. The recording will stop automatically after 4 minutes. 
\end{quote}

\noindent Patients were compensated \$1 for each day an audio diary was submitted, in addition to any other compensation received for other components of the study. Tips provided to patients on creating a quality recording can be found in Appendix \ref{sec:diary-tips-supp}. \\

Submitted daily audio diaries were processed using the code described below (section \ref{subsec:diary-code}). Transcripts for language analysis were obtained from the recordings using the TranscribeMe full verbatim transcription service with sentence-level timestamps, a manually done human service. Produced transcripts are HIPAA-compliant with personally identifying information (PII) redacted. We used the "AI learning team" of transcribers to ensure a maximal level of care was taken in transcription accuracy. As will be detailed, the pipeline managed data flow to/from TranscribeMe as well as extracted acoustic and linguistic features from the diary data.

Although the present work focuses on BLS, an extremely similar design has been used in other smaller lab studies such as our Borderline Personality Disorder project and the Obsessive Compulsive Disorder case study described in Chapter \ref{ch:3}, thereby demonstrating the generality of the pipeline.

\subsection{Resulting diary dataset}
\label{subsubsec:diary-metadata}
Between the start of the study in February 2016 and the close of audio diary collection in May 2022, 74 patients were enrolled in BLS, 66 of whom consented to the audio diary protocol, with 65 submitting at least one daily diary recording. More details on BLS enrollment and the broader study protocol can be found in Appendix \ref{cha:append-chapt-refch:1}.

We received 11,324 daily diary submissions across all BLS patients, totalling 16,548 minutes of audio. A subset of diaries were filtered out by the code based on audio quality control metrics (see section \ref{subsubsec:diary-val-qc} for our QC methodology), resulting in 10,271 total diaries transcribed. In sum, there were 139,538 sentences containing 1,851,833 words in our BLS daily audio journal transcript dataset.

Of the 10,271 diaries successfully transcribed, 1,714 were less than 15 seconds, 3,193 were between 15 seconds and 1 minute, 2,167 were between 1 minute and 2 minutes, 1,524 were between 2 minutes and 3 minutes, and 1,673 were greater than 3 minutes (capped automatically at 4 minutes by recording software). While it is possible for diaries less than 15 seconds to still contain clinically meaningful content, they are less suitable for automated processing. Thus for some analyses a minimum length of 15 seconds was enforced, as noted in subsequent sections. The relationship between diary length and quality of extracted features is also discussed in more detail within section \ref{subsec:diary-val}.  

\begin{figure}[h]
\centering
\includegraphics[width=\textwidth,keepaspectratio]{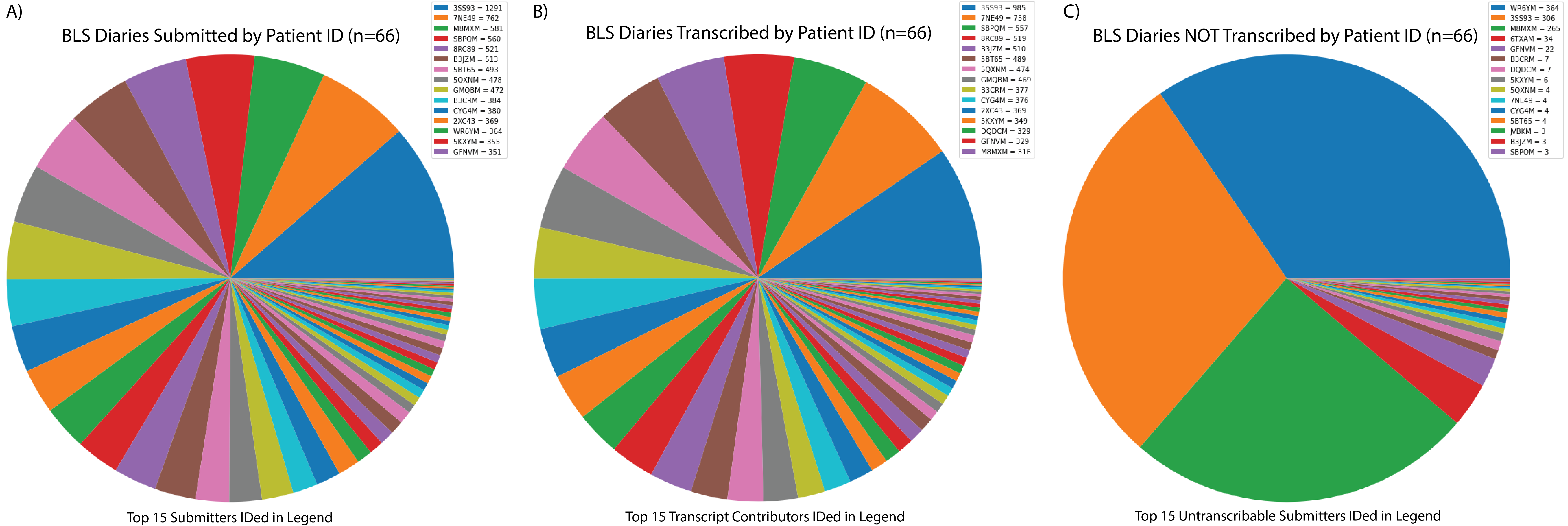}
\caption[Breakdown of daily audio diary submissions by patient in the Bipolar Longitudinal Study (BLS).]{\textbf{Breakdown of daily audio diary submissions by patient in the Bipolar Longitudinal Study (BLS).} Pie charts indicating the number of recordings submitted by each patient (A), transcribed for each patient (B), and unable to be transcribed for each patient (C). The top 15 patient IDs in each category are indicated in the legend, along with the corresponding number of diaries.}
\label{fig:diary-pt-submit-chart}
\end{figure}

One unique quality of this dataset is the extended time period over which patients were active in the study. Regular diary submission for a year or more was common, with 21 patients submitting at least 200 daily diaries (Figure \ref{fig:diary-pt-submit-chart}A), all of whom had an enrollment period of at least 365 days. The top 18 transcript contributors accounted for over $75\%$ of our BLS diary transcripts (Figure \ref{fig:diary-pt-submit-chart}B). Conversely, only a handful of patients regularly submitted poor quality diaries (Figure \ref{fig:diary-pt-submit-chart}C) - just 5 patients recorded 10 or more diaries deemed untranscribable, and 3 of those patients were also among the top 15 contributors of successful transcripts. 

\FloatBarrier

\section{A new tool for processing free form audio diaries}
\label{sec:tool2}
My code handles quality control and feature extraction/visualization from phone audio diaries collected via Beiwe, including a workflow for automated push/pull with TranscribeMe and sanity check of resulting transcripts. In this process it handles a number of file organization tasks, including logistics for preserving patient privacy such as encrypting the most sensitive outputs and removing info like real dates from those outputs marked as deidentified. It also provides additional tools to assist in data flow and QC monitoring, most notably a system for email alerts that can notify lab members whenever new audio files are processed or new transcripts are returned by TranscribeMe. 

The goal of this section is to present that code, both motivating and validating its use, as well as providing technical documentation for reference. First, I will highlight select examples of pipeline outputs and discuss specific use cases (\ref{subsec:diary-outputs}). Then I will summarize key pipeline outputs and walk through prior justification for code design decisions, in addition to our own validation on the BLS dataset for each core feature the code extracts (\ref{subsec:diary-val}). 

For those planning to directly utilize my code \citep{diarygit}, I describe all the capabilities of the pipeline along with their specific implementation details in supplemental section \ref{subsec:diary-code}.

\subsection{Final pipeline outputs}
\label{subsec:diary-outputs}
To provide initial background and motivate the upcoming sections on feature validation and pipeline architecture, a few example outputs produced by the pipeline when run on the BLS diary dataset are reproduced here. We can see from these examples the potential utility of the tool both in organizing and monitoring psychiatric patient audio journal recordings, and in facilitating scientific analysis of these data in conjunction with other clinically-relevant modalities. This chapter will therefore proceed to not only detail the extracted features in this section, but also describe a few example scientific results that have already come from the pipeline in section \ref{sec:science2}.

\begin{figure}[h]
\centering
\includegraphics[width=\textwidth,keepaspectratio]{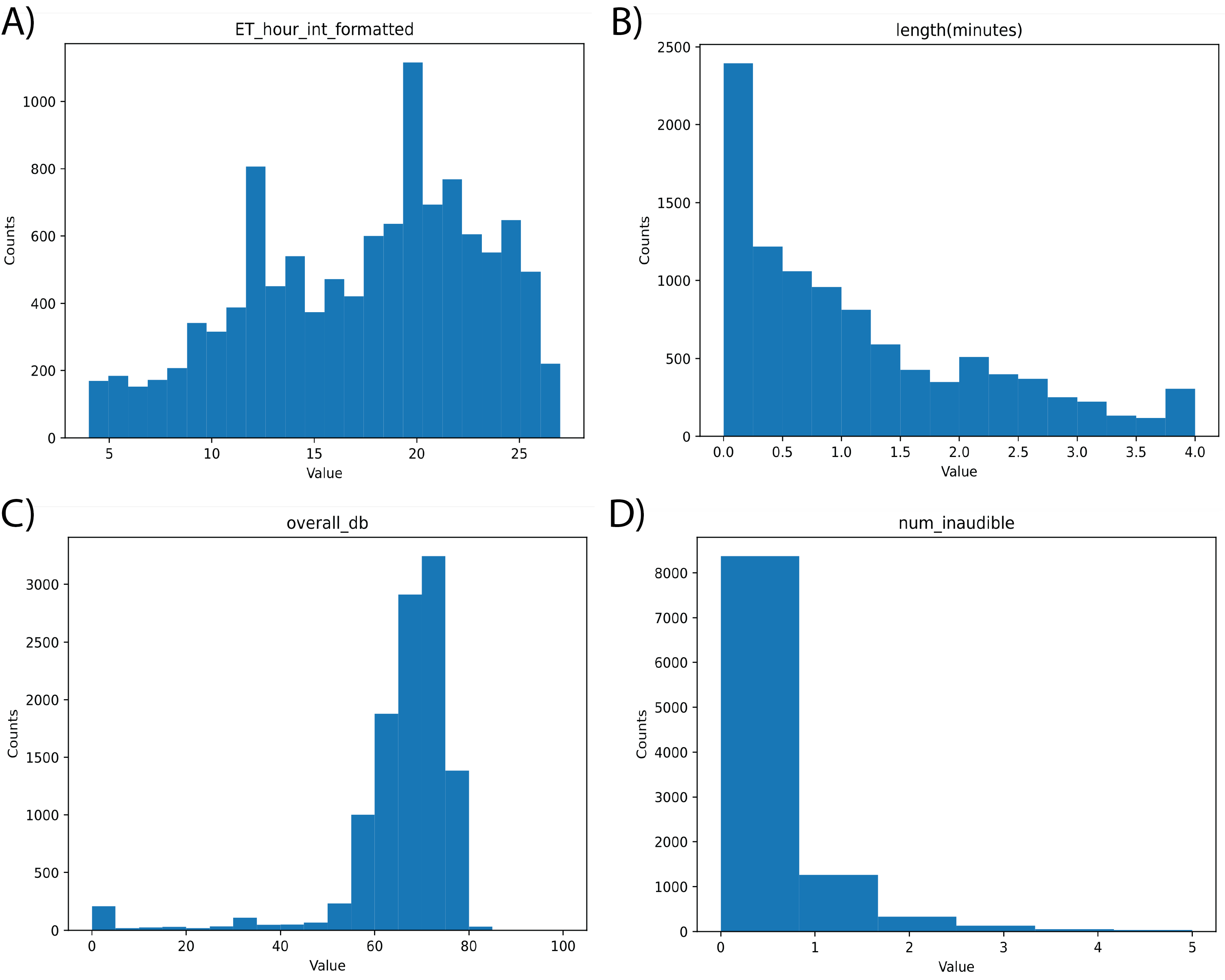}
\caption[Distributions of basic diary quality control metrics across BLS.]{\textbf{Distributions of basic diary quality control metrics across BLS.} For understanding patient participation, we track Eastern Time submission hour, encoded from 4 to 27 to better accommodate late night submissions (A). The length of each diary in minutes is considered for both gauging participation and filtering out empty submissions (B). For quality control, we also screen overall volume of each audio file, measured in decibels (C), and then track the number of words marked as inaudible by the transcription service downstream (D).}
\label{fig:diary-qc-dist-init}
\end{figure}

\begin{figure}[h]
\centering
\includegraphics[width=\textwidth,keepaspectratio]{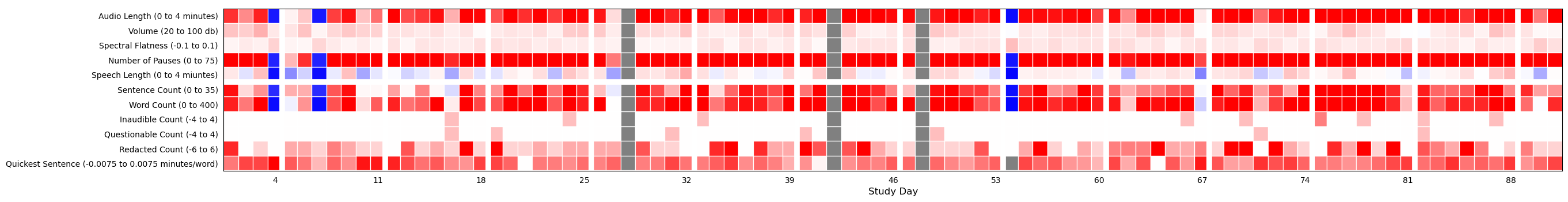}
\caption[Quality control metrics over the first 91 days of the study for an example BLS patient.]{\textbf{Quality control metrics over the first 91 days of the study for BLS patient 3S.} Each column is a day and each row a QC feature. Audio and transcript features are divided by a thicker line. The heatmap uses the bwr matplotlib colormap, with min/max bounds as specified in the row labels. Missing data is grey. Note this patient was typically consistent about submitting long recordings on a daily basis, but there were still some days in this period of both short duration (e.g. day 4) and missingness (e.g. day 28).}
\label{fig:diary-qc-heat-init}
\end{figure}

\begin{figure}[h]
\centering
\includegraphics[width=\textwidth,keepaspectratio]{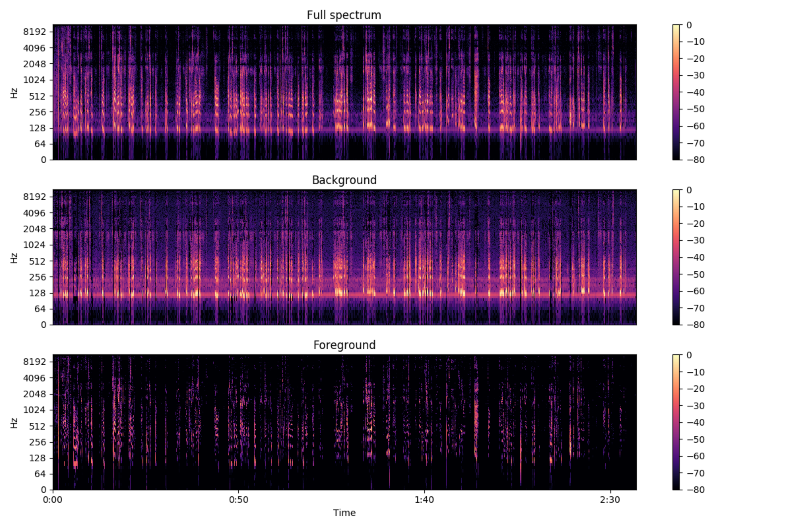}
\caption[Spectrograms from voice activity detection (VAD) on an example diary from BLS.]{\textbf{Spectrograms from voice activity detection (VAD) on an example diary by BLS patient 3S (day 16).} Comparing the raw signal (top) to the VAD-extracted foreground (bottom) and background (middle). Note that VAD may be removing aspects of human speech when filtering to isolate foreground, but is generally consistent in retaining some signal whenever speech occurs, making it a good tool for identifying pause times.}
\label{fig:diary-vad-spec}
\end{figure}

\begin{figure}[h]
\centering
\includegraphics[width=\textwidth,keepaspectratio]{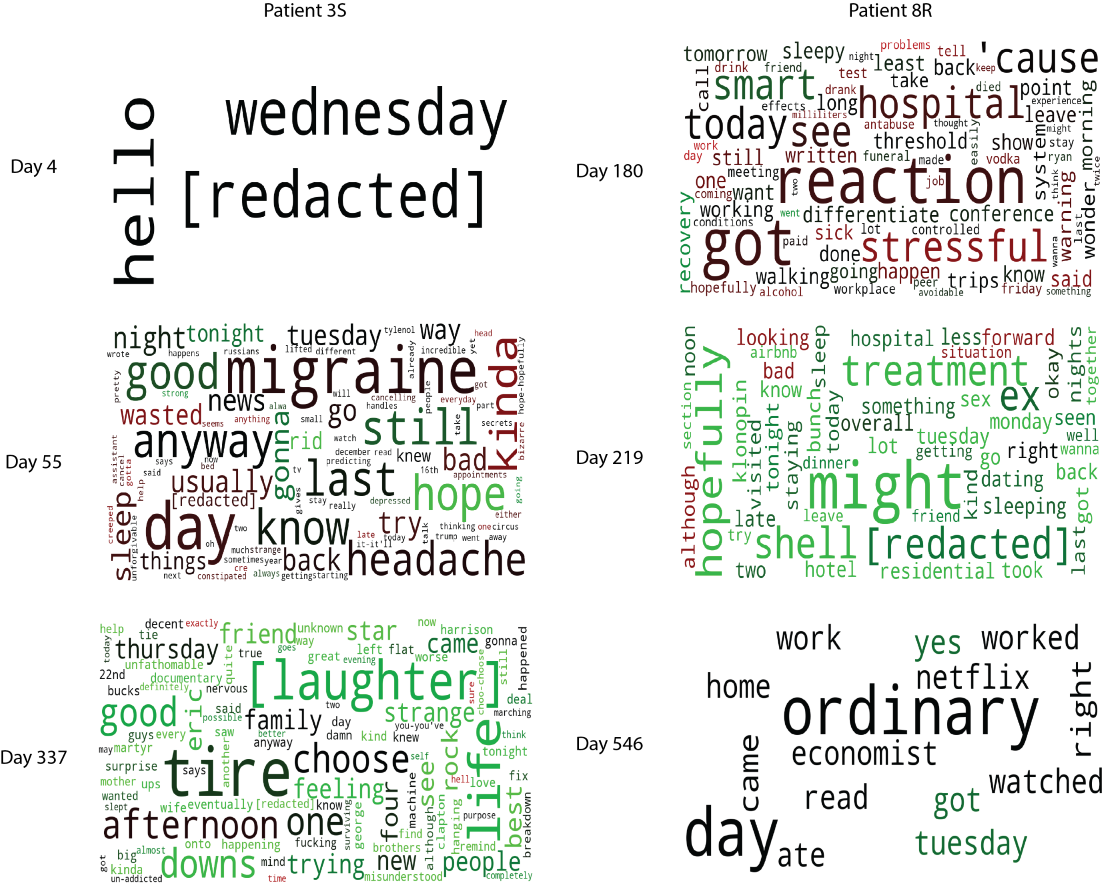}
\caption[Sentiment-colored wordclouds from a few different types of example days for two different BLS patients.]{\textbf{Sentiment-colored wordclouds from a few different types of example days for two different BLS patients.} Both 3S (left) and 8R (right) often submitted detailed diaries, but had variation in participation over the course of the study, and of course variation in the sentiment of their language. As seen in Figure \ref{fig:diary-qc-heat-init}, 3S submitted a rare very brief diary on day 4 (top). One recurring theme in their more negative diaries was migraines (example day 55, middle), but often the submitted diaries were overwhelmingly positive (example day 337, bottom). For 8R, positive diaries often discussed girlfriends (example day 180, top), and we were able to capture descriptions of hospitalization in their negative diaries (example day 219, middle). Their submissions tended to be shorter later in the study (example day 546, bottom).}
\label{fig:diary-clouds-init}
\end{figure}

\noindent Visualizations produced by the code - along with some select examples - include:
\begin{itemize}
    \item Study-wide distributions of QC-relevant metrics, such as volume and length, to provide context for data-monitoring (Figure \ref{fig:diary-qc-dist-init}). Discussion of how these distributions were utilized in determining sufficient quality cutoffs can be found in section \ref{subsubsec:diary-val-qc}. Study-wide distribution histograms are additionally generated for a number of key extracted features that may inform downstream analyses.
    \item Longitudinal heatmaps displaying the progression of audio and transcript QC measures over time, including depiction of periods with data missingness in grey (Figure \ref{fig:diary-qc-heat-init}). Looking for patches of deep blue in the top part of the heatmap can easily alert research staff to diaries with low volume, short length, or other potential quality issues. 
    \item Spectrograms of both foreground and background audio from a diary, as isolated by voice activity detection (Figure \ref{fig:diary-vad-spec}). VAD outputs are subsequently used to label pauses of $\geq 250$ ms in the audio (see section \ref{subsubsec:diary-val-aud} for validation information).
    \item Patient-specific distributions of QC metrics and other key extracted features, which can be used to inform individualized analyzes or to compare/contrast different participants. Note these histograms use the same bin settings as the study-wide ones (e.g. as shown in Figure \ref{fig:diary-qc-dist-init}).
    \item Sentiment-colored wordclouds from diaries of different days of interest in a patient's life (Figure \ref{fig:diary-clouds-init}). Conversely, by simply glancing at a set of wordclouds, interesting days can be identified that might have otherwise been missed.   
    \item Study-wide Pearson correlation matrices for acoustic and linguistic features of note, sorted by BLS-derived feature clusters. The resulting matrices can be used to quantify relationships both within and across different classes of daily diary features.
\end{itemize}

\noindent Extra examples of direct visual outputs from the use of my pipeline on the BLS dataset can be found in supplemental section \ref{subsec:sup-code-outs}. More generally, visualizations derived from pipeline outputs can be found throughout this chapter.

Indeed, the extracted feature outputs can be used in a variety of other ways beyond the built-in visualizations, by accessing the variety of CSVs produced by the code. Use cases for some of the features will be presented as part of the scientific results in section \ref{sec:science2}. To set the stage for that, in the next section (\ref{subsec:diary-val}) I will introduce key features more formally and provide pilot validation results for them. 

 Of course while distributions of and relationships between these base properties can be inherently informative, a more tailored approach involving study-specific data cleaning and normalization should be taken for final statistical analyses of any audio journal dataset (as will be demonstrated in the proof of concept scientific work).

\FloatBarrier

\subsection{Output validation}
\label{subsec:diary-val}
The audio diary pipeline is meant to be used throughout the duration of a research project, both for monitoring of data collection in active studies and for analyzing completed datasets. It also enforces file structure and naming convention requirements, and if run regularly it maintains data flow of audio files to the TranscribeMe server and resulting transcript files back to the lab server. 

As mentioned, the code outputs audio quality control metrics, extracted acoustic features, transcript quality control metrics, and extracted linguistic features. Of course, before relying on the pipeline for study facilitation or model building, thorough testing must be done. In the following subsections, quality control, audio analysis, and transcript analysis are thus each validated. Key computed features are briefly summarized, and then manual review results are reported. For mechanical details on pipeline operation and a more exhaustive list of produced outputs, see supplemental section \ref{subsec:diary-code}. 

While understanding the justification for choosing each feature extraction component included in the pipeline as well as the associated caveats is essential for planning to use the code on a diary dataset and quite useful for planning any diary dataset project, it is not essential for the report of pilot scientific results using pipeline outputs. Depending on what you are most interested in, it might be best to skip ahead to section \ref{sec:science2} and use this section only for reference as needed.

\subsubsection{Quality control metrics facilitate automated data management}
\label{subsubsec:diary-val-qc}
When conducting a longitudinal study, it is important to monitor data availability and quality in an online fashion (i.e. as it comes in). However, as studies expand the number of patients and/or the number of data modalities collected, this can become difficult for RAs to manually handle. Automated infrastructure to streamline the online data review process is thus critical, yet due to the relative newness of digital health research, such infrastructure is not well established among the scientific community. One major function of my pipeline is to fill that gap.

The data flow and quality control portions of the code can improve study management in numerous ways. Email alerts provide a quick summary for RAs of which patients are regularly submitting diaries in a given week, thereby facilitating compensation payouts as well as targeted outreach to improve engagement. Distributions of submission time (Figure \ref{fig:diary-qc-dist-init}A) can be used along with known prompt times and trends in submitted diary length (Figure \ref{fig:diary-qc-dist-init}B) to choose notifications that increase the likelihood of meaningful diary submission.

For the first few years of the BLS study, this pipeline was of course not in place. Over 5,000 diaries were collected without monitoring, and about 1,000 of the earliest ones were sent for transcription without review. Without monitoring in place, we see examples of missed opportunities and general wastefulness. Patient ID WR6YM submitted 364 total diaries, none of which were transcribable (Figure \ref{fig:diary-pt-submit-chart}C, blue). It turns out these submissions were all devoid of content, with extremely short length and low volume. Most likely the patient was intentionally submitting empty recordings to collect the participation compensation with minimal effort. 

On the other hand, patient ID 3SS93 was a prolific contributor, submitting 985 transcribable diaries (Figure \ref{fig:diary-pt-submit-chart}), the highest in the study; yet they also submitted a large number (306) of diaries that were not transcribable (Figure \ref{fig:diary-pt-submit-chart}C, orange). This participant was obviously intending full engagement, but may have encountered technical difficulties or recorded in non-ideal settings. Using the current pipeline, a minimal diary quality threshold can be easily set to qualify for payment, and patients regularly submitting low quality diaries can be given a refresher on Beiwe app operation and recording tips.

What makes for an acceptable quality diary submission is not always black and white. One example we encountered in BLS was a patient who regularly recorded diaries that only included the words "calm and peaceful". However, they did sometimes upload diaries with different language or more content. It is unclear whether the short recordings were done with the intention of gaming compensation, but the fact that variation did sometimes occur means that there could still be clinically meaningful signal captured from this patient's journals. More generally, the quality of audio and linguistic content required will depend on the goals of a study. Quality needed for a workable transcription is different than quality needed for acoustics analysis, and the amount of error acceptable in "a workable transcription" will also vary based on linguistic features to be extracted. \\

\paragraph{Audio QC introduction.}
With that in mind, I will now introduce the basic quality control features considered, and report validation results for our use case with BLS. For audio QC, we primarily considered duration, volume (dB), and spectral flatness. 

As mentioned, garbage files under 1 second in length were sometimes submitted by patients, but for the purposes of transcription (or certain low level acoustic features correlated with emotion \citep{GeMAPS}), we would like to mark anything with identifiable linguistic content as acceptable. Despite this, duration can still play a larger role in overall quality monitoring - both for gauging patient participation levels, and for determining suitability of acoustic features that require longer recordings to accurately extract (e.g. vowel space \citep{Wortwein2017}). 

Similarly, volume plays an obvious role in quality monitoring, with certain dB levels providing a clear cut threshold for files that lacked usable content. For the bulk of the BLS dataset transcription decisions, we did not impose a length cutoff, but did require a volume of at least 40 dB. In this case we prioritized specificity over sensitivity, but were still able to avoid sending 100s of files to TranscribeMe that were not in fact transcribable. The validation results will reflect on the chosen cutoffs and potential adjustments for other use cases.

Spectral flatness, the other main audio QC feature, was calculated but not evaluated during initial use of the pipeline for BLS processing. This feature is intended to capture background noise, particularly static/device noises, in the audio. It ranges from 0 to 1, where a single tone would score 0 and true white noise would score 1. As part of validation, I also discuss the inclusion of spectral flatness in future audio journal data monitoring. 

It is worth noting that the audio QC metrics here are intentionally simple. While the pipeline runs voice activity detection downstream to utilize in acoustic feature extraction, which could bolster the accuracy of quality evaluation, it also takes more computational resources with a longer runtime. Given the large dataset and compute architecture used here, it was most practical to do initial monitoring with low computational cost. However it would not be difficult to adjust the code to consider the more complicated audio features within quality decision making, if that would be more appropriate in a different circumstance. \\

\paragraph{Transcript QC introduction.}
For quality control of transcripts, it is first necessary to introduce TranscribeMe notation conventions, as transcript QC measures rely on them. Because PII must be removed from the transcripts, things like names and specific locations are replaced with "[redacted]". Transcribers are also instructed to mark words that they cannot understand within the audio as "[inaudible]". Words that they are not entirely sure about are similarly enclosed in brackets, with a "?" before the closing right bracket (referred to as "questionable" below). Other words enclosed in brackets indicate human noises such as "[laughter]", "[crying]", or "[coughing]". 

In the full verbatim setting we use, transcribers also mark disfluencies with specific punctuation. Filler words and utterances are surrounded with commas (to distinguish e.g. ", like," as a filler word from "like" as a functional word), stutters are directly transcribed using hyphens (e.g. "s-s-stutter"), and sentence restarts/interruptions are denoted with double hyphens. While this is mainly relevant to feature extraction of disfluencies, punctuation tracking also serves as a way to sanity check that TranscribeMe is correctly applying the verbatim rules to our transcripts. One issue we have encountered a few times with the quality of transcription is receiving some outputs that do not use the settings we paid for.

The other main issue we encountered on TranscribeMe's end is in accuracy of the timestamps provided at the beginning of each sentence. While reported values have ms precision (and we realistically expect accuracy to 100 ms), some returned transcripts have sentences with 0 or even negative time before the start timestamp of the following sentence. This is of course not physically possible, regardless of how short the formed sentence is. Luckily, the issue was detected by QC in 16 of the original 1000 transcripts, and upon discussing with TranscribeMe did not arise again in later transcriptions.

The word accuracy and marking of inaudibles in TranscribeMe transcripts (with the specified settings) are generally well done, as student manual review has found. Upon reviewing a random subset of transcripts with a non-zero number of inaudibles and the corresponding audio, the vast majority of inaudible markings were deemed truly inaudible. Therefore we generally consider inaudible count a good measure of potential audio quality issues rather than a screen for issues with the transcription process, although the latter could still be picked up by the feature when an uptick is noticed. 

Finally, TranscribeMe also provides speaker IDs in their manual transcriptions of our audio recordings. The vast majority of the time this is not relevant for patient audio journals, although we have occasionally detected instances like a family member interrupting the recording or a TV playing loudly in the background via monitoring of subject count.

In sum, for transcript quality control we count not only total words and total sentences, but also the number of words marked inaudible, questionable, or redacted by TranscribeMe. This forms the core set of monitored features, but transcript QC also includes measures that can be quickly glanced at for red flags (speaker count, minimum space between sentence timestamps), as well as punctuation counts that could call attention to changes in transcription detail over time (brackets, commas, hyphens). Additionally, TranscribeMe marked a fraction of our uploaded journals as not at all transcribable, providing another piece to the dataset that we can use to characterize the relationship between audio and transcript quality. \\

\paragraph{Final QC workflow.}
Over the course of BLS, we gained experience with the QC features of the pipeline and how they can best assist in the data collection process. Details of manual review of audio quality as well as correlations between audio and transcript QC features can be found in supplementary section \ref{sec:more-qc}. Here I will summarize resulting recommendations on how to use these metrics in a study and what possible issues should be looked out for.

In our hands, we have decided on QC thresholds of volume at least 40 db and duration at least 1 second as automatic filters for future audio journal studies. These thresholds are based on our results collecting recordings of at most 4 minutes through Beiwe over many years, and are based on quality of returned TranscribeMe transcriptions. Thus they should be confirmed by future users especially if a different study design might be used. 

Nevertheless, the features themselves remain of use in monitoring. Between the email alerts sent directly by the pipeline and the distribution visualizations it generates, as well as the potential for integration with DPDash and incorporating ongoing feedback from TranscribeMe, the pipeline makes it easy for humans to quickly check on quality in real-time. Using this information, participants with greater problems recording can be contacted, and those intentionally cheating the system can be excluded from study compensation. \\

\noindent Audio features we recommend watching for, especially when they are produced regularly by the same participant, include:
\begin{itemize}
    \item Volume (db) $\leq 55$
    \item Spectral flatness (file mean) $\geq 0.1$
    \item Duration (minutes) $\leq 0.25$ (i.e. 15 seconds)
\end{itemize}
\noindent While these features do not necessarily indicate low transcript quality, they do have an association with low transcript quality, and may be additionally problematic for acoustic analyses. \\

\noindent Transcript features we recommend watching for, especially when they occur more frequently in a particular TranscribeMe batch or again are associated with a particular participant, include:
\begin{itemize}
    \item A non-zero number of inaudibles, particularly if the recording is on the short side or the number is $\geq 2$.
    \item Abnormality in the number of redactions, either $\geq 5$ in one recording or a large fraction of a transcript batch ($\geq 90\%$) all containing 0 redactions.
    \item Unexpected deviations in the transcript word count based on the recording duration, or in the number of sentences based on the word count. One can also refer to the minimum and maximum sentence lengths to assist in this evaluation.
    \item More than one speaker ID appearing in the transcript.
    \item Zero dashes or particularly zero commas appearing in a transcript of reasonable length.
    \item A non-positive minimum timestamp space between sentences, or a near zero maximum timestamp space (default units in minutes, so $\leq 0.008$ is less than half of a second).
\end{itemize}
\noindent These features are again potential warning signs rather than inherent issues, and the likelihood of a problem will often depend on the context provided by the entire feature set. That said, these suggested review criteria can help to identify a small subset of transcripts to be checked more carefully. 

More often than not an issue with the number of inaudibles comes down to audio quality rather than TranscribeMe performance. The other suggested checks relate to problems, or more broadly inconsistencies, occasionally seen in returned transcripts. They are generally rare but can sometimes continue for a stretch if TranscribeMe is not made aware. Small typos like an incorrect speaker ID can simply be addressed by the lab, but larger problems like missing verbatim notation or poor sentence separation and timestamping ought to be addressed by a redo of the transcription. \\

\noindent This concludes my review of our quality control framework and the associated features. I will next discuss the larger and more computationally complex feature set that is extracted by the pipeline for downstream analysis purposes, first focusing on acoustic properties (section \ref{subsubsec:diary-val-aud}) and finally on linguistic properties (section \ref{subsubsec:diary-val-trans}). In these upcoming sections, I will discuss technical prior literature validating any tools used, as well as reviewing our own results that support the accuracy of the features - including manual review as well as some comforting observations about the distributions of and relationships between different extracted features in the BLS dataset. 

To preview, I have a strong level of trust in the relevance of our pause detection, semantic sentiment, and linguistic disfluency features, as well as the various qualitative tools to assist in interpretation of journal content - not to mention the potential clinical relevance of basic verbosity and submission metadata information too. The semantic incoherence NLP features and vocal properties extracted from OpenSMILE are also well validated on a fine timescale, but it is less clear how these features should be used to summarize patient audio diaries. These features remain relevant as returned by the pipeline, but require deeper investigation in how they ought to be used scientifically. In the case of semantic incoherence, I begin to address this as part of the research report in section \ref{sec:science2}.  

\FloatBarrier

\subsubsection{Accurately quantifying acoustic features of patient speech}
\label{subsubsec:diary-val-aud}
The quality control and data flow management described in section \ref{subsubsec:diary-val-qc} is just one component of the larger pipeline. As mentioned, it also extracts features to be used in statistical analyses addressing scientific hypotheses about the data. This section will discuss prior justification for and manual validation of acoustic features computed from BLS diaries. Section \ref{subsubsec:diary-val-trans} will do the same for linguistic features. See section \ref{sec:science2} for proof of concept scientific results utilizing many of these outputs, as well as a more detailed characterization of feature distributions in the BLS dataset (section \ref{subsec:diary-dists}). \\

\paragraph{Voice activity detection (VAD) introduction.}
The most computationally intensive and perhaps most important step of the acoustic feature extraction is voice activity detection (VAD). VAD produces separated voice (foreground) and background audio signals. For its relative computational simplicity, we chose to use Librosa's \citep{librosa} modified version of the REPET-SIM algorithm \citep{vocals, vocals2}. The foreground output is then used to label pause times (methods detailed in section \ref{subsec:diary-code}), which are not only the basis for a set of inherently interesting speech pattern features, but also are used for filtering of other downstream acoustic features to more accurately reflect properties of the patient speech.

We chose to work primarily with pause times instead of directly using foreground audio because manual review of the VAD separated files indicated that relevant human acoustics were being marked as background audio. The student reviewer reported being able to understand the majority of spoken language in the foreground audios that were selected for listening, but described it as very robotic. Conversely, they found it difficult to understand any of the voice sounds in the corresponding background audios, but described those as having more human speech-like qualities. This can be seen visually in the example spectrograms of Figure \ref{fig:diary-vad-spec}. Note that REPET-SIM was originally benchmarked against a musical dataset \citep{vocals}, so it is not too surprising that it is more conservative with what is considered foreground; this may be an advantage when it comes to accurate labeling of pause times though.

Although our extracted foreground audio was not deemed suitable for direct input to acoustic feature extraction algorithms, it was found to contain a partial signal for the vast majority of manually reviewed speech time, and critically it was very rarely found to contain any signal from non-vocal background noise. Thus the foreground signal was indeed considered a quality input from which to detect times of human speech within a given file. \\

\paragraph{Manual validation of detected pause times.}
Because pause times became the primary output of the VAD feature extraction, that was the focus of a more detailed manual validation scheme. Upon initial selection of parameters to label the foreground audio as silence or not, the resulting pause times were reviewed by a student volunteer. 24 journals from across 8 participants that were actively submitting at the time were used in this test case. The pause and speech audio files that are created using the pause times were listened to by the volunteer for those 24 submissions, along with the corresponding original audio. The student first verified that when each speech and matching pause file were combined they aligned with the expected length of the full recording. They then searched for any instances of pauses in the speech files or of speech in the pause files. 

The student found that just 2 of the 24 speech-only files contained a detectable pause (here $> \frac{1}{2}$ of a second), though 1 of those files contained multiple longer pauses. They also found that just 3 of the 24 pause-only files contained audible speech, and 2 of those 3 only contained a handful of instances of nonverbal utterances (e.g. "uhh") but no verbal speech. The times that nonverbal speech was captured, it also directly followed an actual pause in the full file. 

It is worth noting that in this initial review, 7 of the recordings were contributed by subject ID GFNVM, who consistently displayed red flags in QC monitoring and qualitatively had perhaps the poorest enunciation across the entire BLS dataset. Thus these recordings represent a particularly challenging case for the pause detection. Still, 6 of the 7 had accurate speech and pause separation, and the 1 with poor pause separation was a particularly low quality recording that accounted for the most problematic speech file that contained multiple pauses greater than 1 second.  

Based on these results, we proceeded with applying the pause detection across the dataset. The pause detection is not expected to be perfect, but it was largely very accurate in this initial review, and inaccuracies that were observed were not only mostly minor, but also mostly aligned with poorer audio QC metrics. Thus we do utilize the pause features, but recommend taking more care in interpretation when applied to low quality audio - which is of course true of any acoustic features. \\

\paragraph{Using VAD-derived features.}
Because the pause times are used in various downstream outputs, those outputs can be reviewed to further increase confidence in the utility of the algorithm. One way the pauses are used is to filter the OpenSMILE outputs that are returned in 10ms bins, which will be discussed in more detail in the voice properties section below. Broadly, the values returned by OpenSMILE in time bins marked as pauses are consistent with what one would expect for true pauses though, so that is promising.  

Another pause-derived feature set that we hoped would be useful was the volume and/or mean spectral flatness of audio taken only from pause times. The goal was to improve identification of files with more overall background noise, but student manual review did not find this to be the case. As such, these downstream features are excluded from the advertised code outputs and are not used further. This is likely a symptom of the fact that lower quality audios already tend to have poorer pause separation, and many of the quality issues we currently fail to detect (see supplementary section \ref{sec:more-qc}) would not be expected to differentiate with VAD - for example background speech from a TV or from another person in the room. 

Regardless, the diary level acoustic features calculated using the pause time estimates themselves (rather than application to other outputs) likely have use in both quality monitoring and in answering certain clinical questions. Therefore I next focus on verifying the utility of some straightforward pause-related diary summary statistics: pause count, total pause duration, and total speech duration. These features can be easily normalized to obtain \emph{percent speech} and \emph{mean pause length} metrics akin to acoustic properties commonly seen in prior literature (section \ref{subsec:diary-lit-rev}). 

For reference, the pipeline also provides summary statistics for a number of features across the dataset (see supplementary section \ref{subsec:bls-diary-summary-sup}). More than $\frac{2}{3}$ of the mean diary time was spent speaking, and there was greater relative variance in pause duration across diaries than speech duration. Note too that the computed pause time periods would be easily usable as input to other more bespoke summary feature calculations such as distributional properties of the pause lengths or the gaps between pauses across an individual diary, as needed. \\

\paragraph{OpenSMILE introduction.}
OpenSMILE outputs, the other primary set of acoustic features, are calculated in 10ms bins using the low level descriptor (lld) Geneva Minimalistic Acoustic Parameter Set (GeMAPS). The pipeline then produces a filtered copy of each OpenSMILE output by blanking out time bins that overlap with a detected pause time (which has a generally much longer timescale than 10ms). Simple summary statistics for each diary are then computed using both the raw and filtered OpenSMILE outputs, in parallel. For further technical method details, see section \ref{subsec:diary-code}. 

OpenSMILE is an open source acoustic processing tool that is very commonly used for extraction of vocal features \citep{OpenSMILE}. OpenSMILE can be utilized with a number of different configurations to implement different acoustic feature extraction algorithms; the GeMAPS configuration is perhaps the most frequently used in the emotion recognition literature, and has an abundance of validation results available on related benchmarks, including six major affective speech databases \citep{GeMAPS}.

The goal of the GeMAPS parameter set was to obtain a relatively small number of features compared to established state of the art feature extraction methods that were extremely high-dimensional. Classification performance of the GeMAPS features was compared against these high-dimensional parameter sets extracted via brute-force from the INTERSPEECH Challenges on Emotion and Paralinguistics (2009 through 2013). GeMAPS performed comparably to the benchmark sets on most tasks, and even outperformed them on a few, despite its much smaller size \citep{GeMAPS}. 

GeMAPS outputs are thus well suited to psychiatry research -- the temporally dense low level features can be utilized in more complex deep learning models when appropriate, but more interpretable summary statistics of relevance can also be computed from this feature set, unlike other high performing configuration options.

As such, OpenSMILE with the GeMAPS configuration has been the source of acoustic features for a number of existing computational psychiatry studies, and has demonstrated success in a wide variety of models characterizing symptom severity or distinguishing patients from healthy controls. This includes Schizophrenia \citep{Voppel2022}, depression \citep{Nasir2016,Syed2017}, and Bipolar disorder \citep{Syed2018}. It has also been applied to studies of vocal properties in neurological diseases such as Parkinson's \citep{Wroge2018} and Alzheimer's \citep{Haider2020}. Additionally, the GeMAPS lld outputs from OpenSMILE are also one of the primary features sets planned for the processing pipeline for the AMPSCZ clinical interview dataset to be described in chapter \ref{ch:2}, determined by a panel of emotion acoustics researchers from collaborating groups.

While many of the established results for this feature set in psychiatric applications are from a semi-structured interview context, emotion recognition benchmarks are focused on shorter duration speech from a single individual. Extension of the framework to longer dialogues carries some challenges that are not an issue with audio journals. Especially because these features have already been frequently used in interview analysis, there is little reason to believe that strong accuracy on emotion detection benchmarks would not extend to the audio journal setting.  

Thus there are a few specific details on the downstream use of OpenSMILE outputs that ought to be validated in the pipeline, which I will now discuss, but there is already good justification for the quality of the low level vocal feature returned by the code whenever input audio meets quality control standards. Of course regardless of prior validation results or any validation results presented here, each application of the pipeline to a new study should involve a sanity check of outputs before actual analyses are performed. This is true of all features, including the OpenSMILE ones.  \\

\paragraph{Computational validation of pause features.}
To assist in validation of pause detection, the properties of these features can be studied through a data science lens, to determine in what ways the dataset meets or violates expectations. As discussed above (\ref{subsubsec:diary-val-qc}), measured speech times are much more consistent with corresponding transcript word counts than the full recording durations are. 

Subject 3SS93 submitted very long diaries, but also paused a particularly large amount, thus resulting in word counts that were less high relative to the rest of BLS than one might initially expect. For example, the ratio of words contributed to the dataset by 3SS93 versus words contributed to the dataset by 8RC89 is $\sim 0.34$, while the ratio of transcribed minutes from 3SS93 to transcribed minutes from 8RC89 is only $\sim 0.18$. By restricting the duration considered in those submitted recordings to only minutes labeled as speech, the ratio comes much more in line at $\sim 0.28$. 

This theme of consistency between pause and other related features continues throughout investigation of the BLS dataset. As will be discussed (section \ref{subsec:diary-dists}), the estimation of speech rate from transcript content and timestamps can be adjusted to remove the impact of conversational pauses by using the pipeline's acoustic pause features. This is true both of the overall BLS distribution and patient-specific phenomena. For example, when speech rate is normalized by speech fraction, the abnormally low rate seen for 3SS93 becomes instead aligned with the broader BLS distribution, while an elevated speaking rate remains detectable in the 8RC89 distribution - consistent with clinician observation about the manic episodes of participant 8R. Study-wide distributions of the pause detection features and deviation within particular patients will be discussed in more detail in a scientific context in section \ref{subsec:diary-dists}, but the sanity checking results here instill more confidence in the use of these feature for those investigations. \\

To complete the characterization of the quality of pipeline pause outputs, I performed a systematic analysis in the same spirit as the preceding arguments. By looking more closely at the relationships between pause and OpenSMILE features across the journals, it was possible to not only further validate the accuracy of the pipeline overall, but also to determine specific details that were common across the submissions it did perform poorly on. As the first step, I performed a deeper dive on the alignment of VAD-identified pause fraction metric (pause minutes/recording minutes) with the words per minute metric (TranscribeMe word count/recording minutes). Then using results from this analysis, I referred back to OpenSMILE features and their pause-removed counterparts to characterize likely recording quality problems that related to pause detection performance.

\begin{figure}[h]
\centering
\includegraphics[width=0.8\textwidth,keepaspectratio]{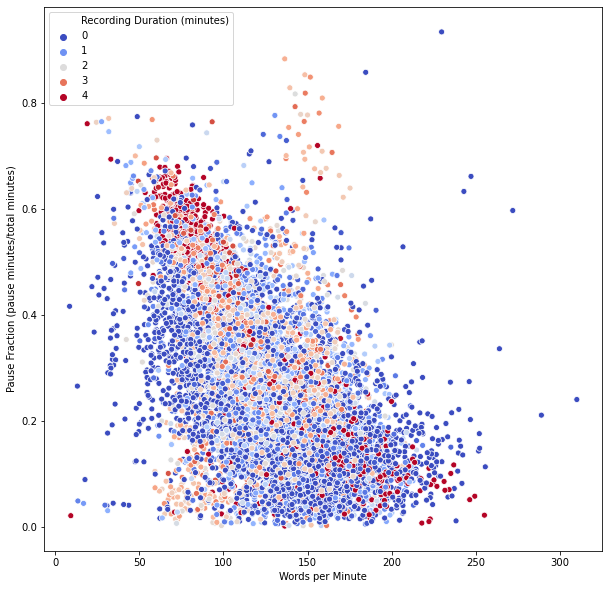}
\caption[VAD-derived pause fraction and transcript-derived words per minute features are strongly correlated, with variation modulated by recording duration.]{\textbf{VAD-derived pause fraction and transcript-derived words per minute features are strongly correlated, with variation modulated by recording duration.} Each transcribed BLS diary with at least one pause occurrence counted by VAD ($n=9872$) is plotted as a point here, with x-coordinate corresponding to the Words per Minute value for that diary and y-coordinate corresponding to the Pause Fraction value for that diary. Each diary point is colored according to its recording length in minutes, from dark blue at $0$ crossing over to dark red at $4$. Strong agreement between these features indicates good accuracy of the pause detection algorithm, though both word complexity and active speaking rate introduce additional sources of variance. Unsurprisingly, this variance is stronger in shorter diaries.}
\label{fig:pause-qc-first-scatter}
\end{figure}

Across the entire dataset of transcribable journals with at least 1 identified pause ($n=9872$), there was a moderate and highly statistically significant negative correlation (Figure \ref{fig:pause-qc-first-scatter}) between pause fraction and words per minute, both linearly (Pearson's $r=-0.562$, $p<10^{-300}$) and rank-based (Spearman's $r=-0.573$, $p<10^{-300}$). Of course with accurate pause detection we expect a high negative correlation between these features, but there are also sources of variance besides pause detection inaccuracy that would prevent perfect alignment. Pause fraction is intended to capture conversational pauses, but will not account for differences in rate of speech production outside of pauses. Additionally, word count does not consider the complexity of the words spoken, and will also not include consideration of disfluencies like stuttering or the duration of a nonverbal edit. These sources of noise are further compounded by the number of relatively short diaries, which allows for higher levels of variance in individual recordings. 

\begin{figure}[h]
\centering
\includegraphics[width=\textwidth,keepaspectratio]{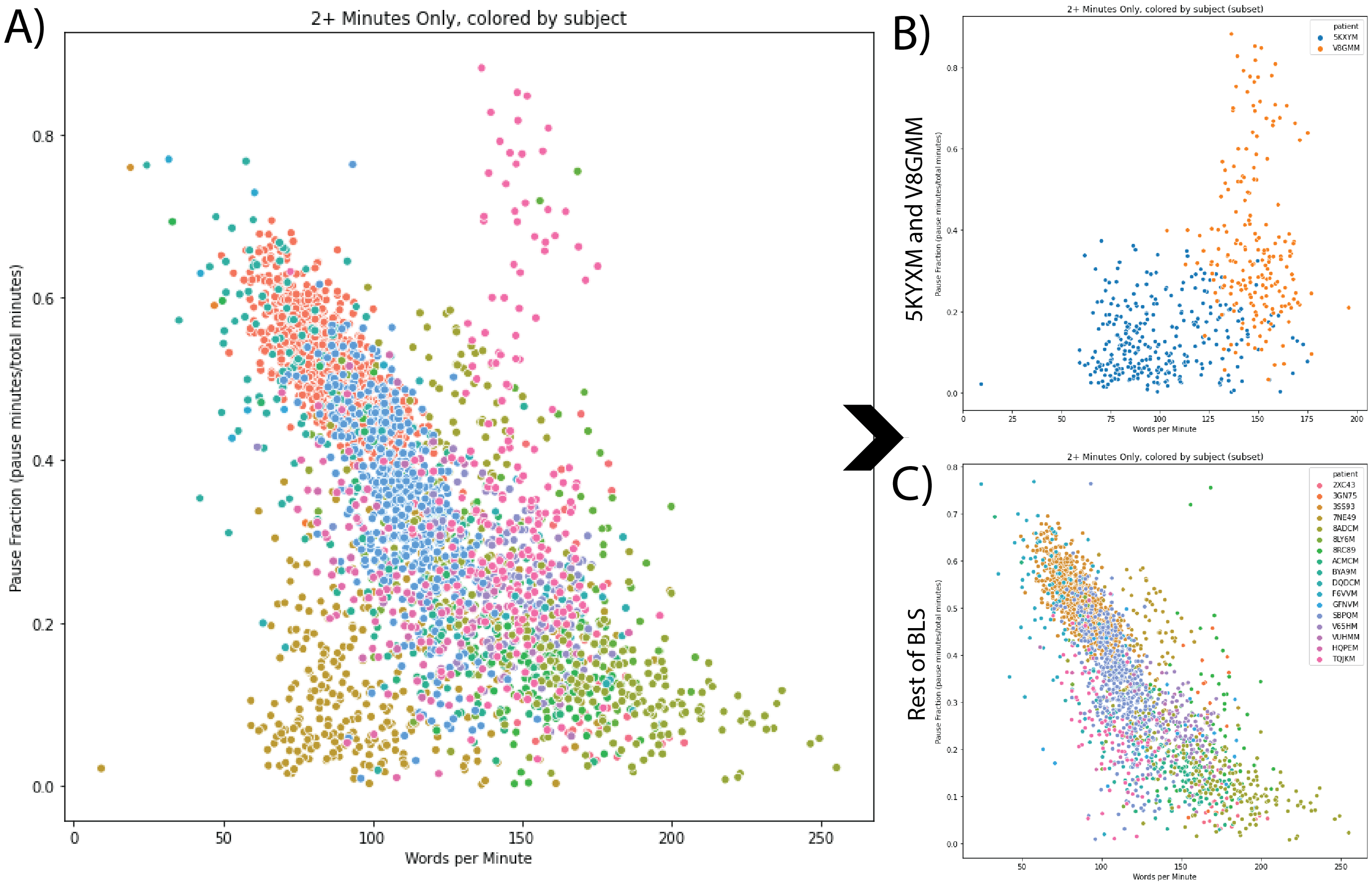}
\caption[Identifying subjects with the lowest quality pause detection results.]{\textbf{Identifying subjects with the lowest quality pause detection results.} Focusing only on the journals of recording duration at least 2 minutes (due to lesser variance), the scatter plot in Figure \ref{fig:pause-qc-first-scatter} is reproduced here, with hue now corresponding to subject ID (A). This allowed for the identification of two subjects with poor pause labeling results -- 5KXYM and V8GMM (B). When isolating the plot to these two subjects, it is clear that the pipeline consistently underestimated the amount of pausing from 5KXYM (blue) and overestimated the amount of pausing from V8GMM (orange). On the other hand, the remaining BLS subjects had consistently good pause detection results (C).}
\label{fig:pause-qc-bad-pts}
\end{figure}

The relationship depicted in Figure \ref{fig:pause-qc-first-scatter} was therefore largely validating of the pause detection accuracy. However, a couple of abnormalities outside of the cloud of natural noise raised concerns worth further investigation. To better understand the outlier clusters of Figure \ref{fig:pause-qc-first-scatter}, I next focused on diaries of 2 or minutes in duration, to decrease the level of variance in the confounding factors. By considering pause fraction against words per minute according to the corresponding subject ID in this dataset, a participant-dependent effect became apparent. Subjects 5KXYM and V8GMM were responsible for the vast majority of outlier points in the relationship, with 5K's pause rate seemingly underestimated and V8's seemingly overestimated (Figure \ref{fig:pause-qc-bad-pts}). Indeed, when removing just these two subjects from the larger dataset - with no length requirement at all - the magnitude of correlation between pause fraction and words per minute jumped meaningfully, with Pearson's $r = -0.633$ and Spearman's $r = -0.649$ (both $p < 10^{-300}$).

\begin{figure}[h]
\centering
\includegraphics[width=\textwidth,keepaspectratio]{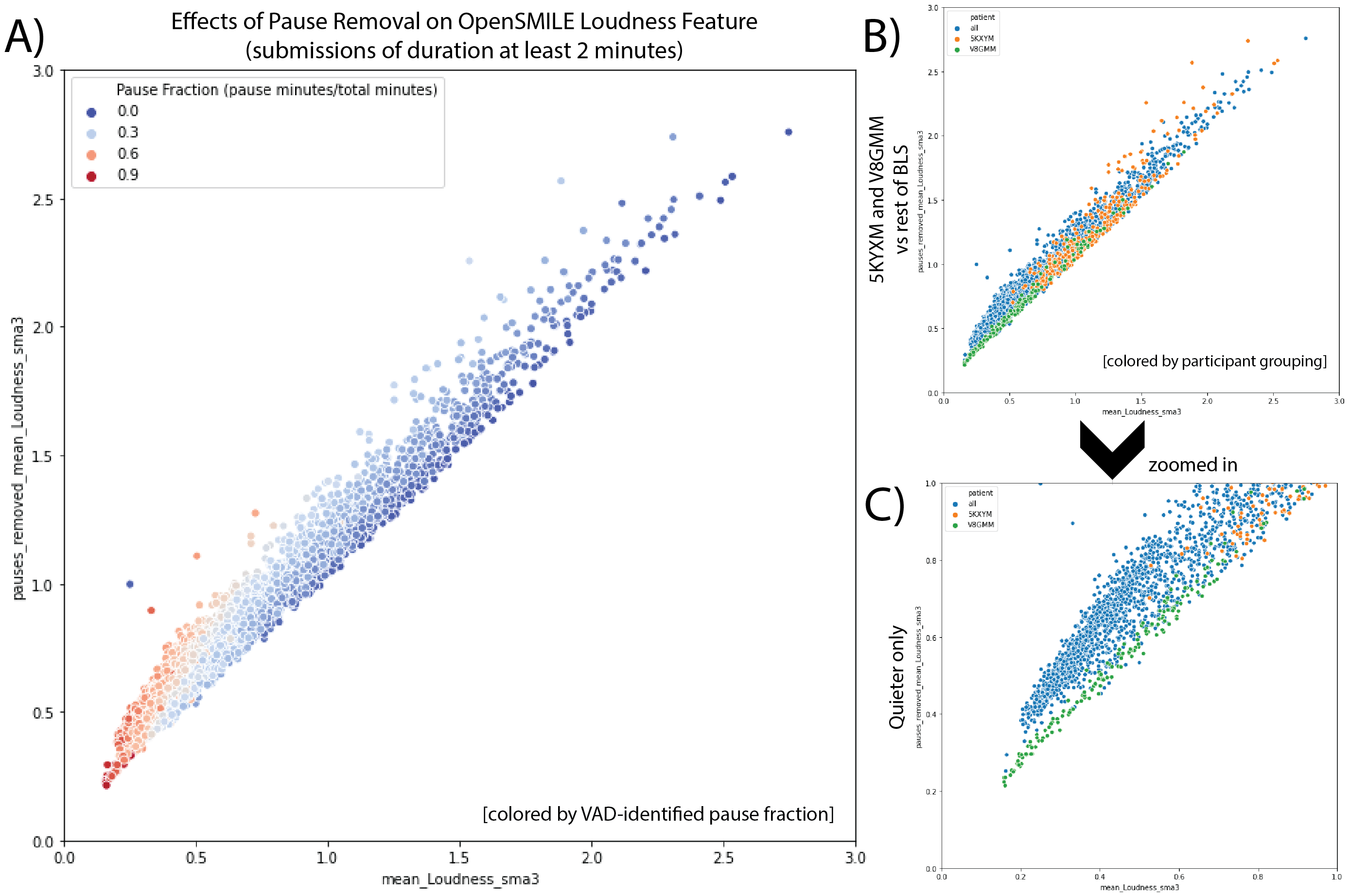}
\caption[Pervasive audio quality issues observed in OpenSMILE features from submissions of the two participants with clear pause detection inaccuracies.]{\textbf{Pervasive audio quality issues observed in OpenSMILE features from submissions of the two participants with clear pause detection inaccuracies.} Using the same diary dataset as Figure \ref{fig:pause-qc-bad-pts}, the mean of the voice loudness feature returned by OpenSMILE was plotted for each diary against the mean of the same feature after marked pause time bins were removed. As OpenSMILE should return a near-zero loudness value in bins that are true pauses, we would expect a clear shift up in mean loudness the more bins that were removed. Thus the major scatter here (A) is colored based on VAD-identified pause fraction, from dark blue at 0 to dark red at 0.8. Overall, the results were largely consistent with the stated expectations; to better understand the outlier points, I then created a version of the same scatter with hue based on subject ID instead (B). Indeed, the majority of the anomalous points belong to 5KXYM (orange) or V8GMM (green) rather than the rest of BLS (blue). Furthermore, it is clear from this plot that 5K often submitted abnormally loud diaries while V8 submitted abnormally quiet ones, such that the core OpenSMILE voice loudness feature was also impacted. By zooming into the lowest loudness diaries (C), it becomes obvious that V8 submitted a number of extremely quiet recordings relative to the rest of the dataset, so inaccuracies in pause detection are a natural consequence.}
\label{fig:pause-qc-opensmile-scatter}
\end{figure}

It was thus clear that pause detection accuracy was generally quite good, but failed specifically on two patients. In order to proceed with confidence in the pause detection results, I next investigated the reason for the subject-specific inaccuracies, and identified broader recording quality problems in the submissions of 5K and V8. As mentioned above, the results of OpenSMILE were utilized for this analysis. In particular, one feature returned by OpenSMILE for each 10 millisecond bin is the voice loudness estimate. This feature was averaged over the bins in a given diary using the raw outputs as well as using the pause-filtered versions. It is expected that pause bins should have near-zero loudness returned by OpenSMILE, while speech bins should have a reasonably high loudness value. Therefore we can evaluate the correspondence between VAD-identified pauses and OpenSMILE voice estimation by comparing the raw and pause-removed loudness averages in conjunction with the pipeline's measured pause fraction (Figure \ref{fig:pause-qc-opensmile-scatter}). 

Most notably, subject V8GMM submitted many of the quietest diaries by OpenSMILE voice loudness metric, and even after removing many of the bins labeled as pauses by the pipeline the loudness did not meaningfully increase (Figure \ref{fig:pause-qc-opensmile-scatter}). This suggests that the recordings submitted by V8GMM were exceptionally quiet compared to other participants, and would pose quality issues for acoustic analyses across the board. Similarly, 5KXYM submitted some of the loudest diaries across BLS, so it is unsurprising that they were a (less severe) outlier in the opposite direction. 

Ultimately, it is important to check for anomalies across features and to perform manual spot checks, in order to ensure quality of a given dataset and the measures extracted from it before proceeding to scientific analyses. This should always be done to some extent, even if using a previously validated open source tool. As can be seen here, a tool can be successful in certain circumstances and not others. The pipeline's pause labeling is a useful and largely accurate feature, but it does depend on a base level of recording quality in order to work correctly -- as do OpenSMILE and other commonly employed feature sets. \\

\FloatBarrier

\paragraph{Review of OpenSMILE outputs from the pipeline.}
One important part of the OpenSMILE pipeline to validate here is the use of detected pause times to generate a set of low level vocal features that correspond only to times of speech. Upon review of both the OpenSMILE summary stat distributions over diaries and the within-diary lld feature distributions for select sample diaries (all generated automatically by the code, see also \ref{subsec:sup-code-outs}), it indeed became apparent that the filtration by and large removed a subset of the near-zero values from the diary lld timeseries for key GeMAPS features, with minimal impact to the rest of the distribution. This further supports the utility of the pause detection algorithm in itself. 

As mentioned, the vocal features that are returned on the 10 millisecond level can be used directly for certain modeling applications, but summary stats computed over a broader timescale can be utilized for other research approaches. There are many ways such summary stats could be computed, and bespoke methods for a particular study would be straightforward to include in adaptations to the pipeline where needed. Currently, I focus on simple per recording stats of general utility, as is done for most features described here. 

Additionally, the OpenSMILE results generated by the pipeline include hundreds of features, per the GeMAPS lld config \citep{OpenSMILE, GeMAPS}. This may be small relative to other acoustic parameter sets, but it is still large for the purposes of a typical psychiatry study dataset. Therefore in the unique case of OpenSMILE, I also pare down the set of features for which the default summary stats are returned, to maintain interpretability of results and keep the core feature set tractable for small pilot studies. In practice, our use of the pipeline with lab datasets has focused even more narrowly on a small set of output features: \emph{Pitch (F0)}, the formants \emph{F1} and \emph{F2}, and the potential measures of vocal "shakiness" \emph{jitter} and \emph{shimmer}. These features were chosen based on both a strong base of prior literature supporting their clinical relevance (see background in section \ref{subsec:diary-lit-rev}) and a reasonable level of interpretability even for those without acoustics expertise -- properties with a good deal of overlap given the impracticality of huge feature sets up until recently. 

Complete documentation on the lld GeMAPS features that are included in the default summary stats and visualizations returned by the pipeline can be found within the code documentation in supplemental section \ref{subsec:diary-code}. For a full list of the features in the lld CSV returned by the code, one can refer directly to the OpenSMILE documentation of the GeMAPS config \citep{OpenSMILE, GeMAPS}, as the pause filtering operation does not reduce the feature set returned by OpenSMILE, it just blanks out rows estimated to lack any speech content. 

Ultimately, the summary feature set is meant for quick qualitative assessment of study datasets or pilot statistical analyses for hypothesis generation and early comparison of multimodal feature sets. Studies focused on acoustic properties should utilize the full set of low level pause filtered OpenSMILE outputs returned by the pipeline for their analyses. 

\subsubsection{Accurately quantifying linguistic features of patient speech}
\label{subsubsec:diary-val-trans}
A final result of the pipeline is the many linguistic features computed from the returned TranscribeMe transcripts. Transcription settings and the overall transcription quality level we observed are detailed in section \ref{subsubsec:diary-val-qc}. Code implementation details for natural language processing (NLP) features can be found in section \ref{subsec:diary-code}. Here, I present an overview of the key NLP pipeline outputs along with validation results. 

Broadly, the types of linguistic features the pipeline aims to capture include sentiment, disfluencies, and semantic coherence, as well as basic word usage stats (overall structure and keyword search). Features are computed at the sentence level, where sentence breakpoints are defined by the human transcriber. For counting stats, diary-level features are then just the sum of sentence ones. For other features, the diary-level summary consists of a few basic summary stats over sentences, with mean the most commonly used downstream. Validation of each core NLP feature will be discussed at both the sentence and diary scale; in some cases I will pull from prior literature that has extensively reviewed the sentence-level outputs in other datasets. \\

\paragraph{Sentiment analysis tools overview.}
Identifying how positively or negatively a patient speaks about a given topic or during a particular time period has immediate relevance to clinical contexts, so we naturally wanted to evaluate sentiment in our diary transcripts. To do so we used the Valence Aware Dictionary and sEntiment Reasoner (VADER) python package for its ease of use and prior validation results \citep{VADER}. VADER is designed for input of phrases or sentences; it handles interactions between words such as negations, but is not intended to directly capture interactions between many unique thoughts. We therefore obtain a VADER score for each sentence in a transcript, utilizing the compound score output to get a value between -1 and 1. Distance from 0 indicates the magnitude of sentiment while sign indicates valence.

When benchmarking their model, \cite{VADER} tested VADER on Tweets, movie reviews, product reviews, and NY Times editorials, and compared against the performance of sentiment scores from existing tools like Linguistic Inquiry and Word Count (LIWC). Ground truth for each benchmark dataset was based on the mean of manually derived sentiment scores from 20 human raters. VADER outperformed all established tools on all tasks, and also outperformed the majority of individual human raters on labeling the Tweets. Note that for the other three tasks, inputs were longer and thus they determined the VADER score for these by inputting individual sentences (tokenized by NLTK) and then taking the mean \citep{VADER}. We do similarly for our transcripts, but have human-quality sentence splitting outputs, and also consider other summary stats for the diary level such as maximum and minimum sentiment sentences.

The aforementioned LIWC is a very common tool to see used in psychiatry literature, and while it has capabilities besides sentiment that may make it still worth using, there seems to be little reason to continue to use its sentiment output. In addition to the consistently better benchmark performance of VADER, we had experimented with LIWC in the early parts of the BLS diary collection and found VADER to be notably better both subjectively and in correlations with same-day self-report mood score. Moreover, LIWC is not easily buildable into a pipeline such as this one, like VADER is. Thus we've decided on VADER as the sentiment classifier of choice for systematic use on our transcript datsets.  

Note that VADER can also be run very quickly across many transcripts. While there may be more complex machine learning models for sentiment estimation that outperform VADER in 2022, the lower computational cost and higher interpretability of VADER, coupled with its objectively still high quality performance, make it a good choice for a reliable pipeline piece. 

Still, there is a need to validate VADER more thoroughly in this particular context of psychiatric patient speech transcribed from daily audio diaries. There is very little existing literature on this topic, and thus the validation results from the BLS dataset reported here are of relevance to the research community. However, there is prior psychiatry work applying VADER to other sources of patient language, which lends additional credence to its use in my architecture. Much of this literature assesses language used on various social media platforms, which directly follows from its strong original validation performance on a large dataset of sentiment-labeled Tweets \citep{VADER}. 

Most relevant for the present work though are applications of VADER in the context of transcribed natural speech from psychiatric patients; past studies have indeed successfully applied it to semi-structured clinical interview data. For example, interviewees with PTSD versus MDD were distinguishable significantly above chance using a model trained on VADER sentiment scores from across the interview \citep{Sawalha2021}. While sentiment classification of interview transcripts holds promise, there will be implementation differences between less frequent longer dialogues with known questions and our aim for more frequent shorter monologues that are prompted in an open-ended way. 

Regardless, the base application of VADER on the sentence level is already quite supported across contexts, so I focus on diary-level sentiment scores derived from VADER sentence-level outputs by the pipeline in our BLS dataset validation work, as the per diary features are intended as the primary output to be used in statistical analyses at this time. The other major use for the VADER results currently, the sentiment-coloring in the word cloud visualizations (Figure \ref{fig:diary-clouds-init}), uses the sentence-level scores more directly. \\

\paragraph{Manual validation of VADER-derived transcript sentiment score.}
For two selected participants - 8RC89 and GFNVM - an RA reviewed all transcript text files to select the 5 with most positive sentiment and the 5 with most negative sentiment. These participants were chosen in order to reflect different demographics (gender, age, diagnosis), study time periods (early versus recent data collection), and general recording quality (variation in enunciation/vocal clarity). The RA looked for presence of words they considered positively-valenced (e.g. “happy”, “excited”, “enjoy”, etc.) and negatively-valenced (“sad”, “depressed”, “hate”, etc.) in each transcription, and synthesized that with the overall context to identify diaries with high magnitude sentiment. They were blinded to the automated sentiment scores, and after they composed their top and bottom 5 sentiment lists for both patients, I aligned them to the rank list for the corresponding subject when sorted by the mean sentence sentiment feature output by the pipeline.

For subject 8R, the top 5 positive days selected by manual review corresponded to rankings 2, 3, 5, 6, and 7 (in order) by descending mean sentiment score, out of 521 total journals. Similarly, the 5 most negative days selected by manual review in the 8R dataset corresponded to rankings 2, 3, 5, 8, and 9 (in order) by ascending mean sentiment score. 

For subject GF, the top 5 positive days selected by manual review corresponded to rankings 3, 4, 5, 6, and 7 (in order) by descending mean sentiment score, out of 351 total journals. Similarly, the 5 most negative days selected by manual review in the GF dataset corresponded to rankings 1, 2, 3, 4, and 6 (in order) by ascending mean sentiment score. 

Overall, the alignment results were very strong, and indeed provided even better reassurance for the automated sentiment summary statistic than I was expecting. To further confirm the results, I also checked the content of the top ranking diaries (both in mean VADER score and in manual review) after the fact. The identified transcripts were quite interesting, especially the most negative ones. For participant GF, the 4 most negative journals identified by both manual review and mean sentence sentiment perfectly coincided, and contained very negative thoughts that suggest a closer look at the larger set of low sentiment diaries submitted by GF would be worthwhile (Figure \ref{fig:negative-diaries-gf}). 

\begin{figure}[h]
\centering
\begin{tcolorbox}[top=-0.25cm,bottom=0.4cm,left=-0.25cm,right=-0.25cm]\begin{quote}\small
00:02.56  \hspace{2mm}  Having no good internet and using your phone hotspot to do all your schooling off of a crappy Chromebook is very difficult, and today was a very stressful day.

00:14.82  \hspace{2mm}  I was just very irritable and very aggravated and angry and just frustrated and upset and just felt terrible, and nothing worked out, and everything went, like, in the worst possible direction.

00:28.86  \hspace{2mm}  And I'm just in give up mode, and I just don't care anymore about anything, and I'm just done with everything, and I just don't wanna really talk right now.
\end{quote}\end{tcolorbox}

\begin{tcolorbox}[top=-0.25cm,bottom=0.4cm,left=-0.25cm,right=-0.25cm]\begin{quote}\small
00:02.47  \hspace{2mm}  I don't know why but I was very just stressed and high-strung and high anxiety, paranoia, depression, just SIB [inaudible] was just so bad today and then [inaudible] my Christmas gift in the mail and that was completely just, like, damaged and it freaked me out and so I have to reorder something, and, um, couldn't figure out an alternative and I just completely went on edge, just [started?] crying and hysterics and high anxiety and stress and just terrible [inaudible] fixated trying to find another replacement gift.

01:09.62  \hspace{2mm}  I'm just so drained.
\end{quote}\end{tcolorbox}

\begin{tcolorbox}[top=-0.25cm,bottom=0.4cm,left=-0.25cm,right=-0.25cm]\begin{quote}\small
00:01.73  \hspace{2mm}  Um, I had a bad night's sleep and was having a lot of back pain, just kept waking up.

00:08.82  \hspace{2mm}  And then, um, I had to go to work, and it was just very exhausting, and I was very stressed being by myself for half the shift on a busy day.

00:22.07  \hspace{2mm}  And, um, I got a \$15 tip and then I lost it, and it was just even more stressful and frustrating.

00:32.09  \hspace{2mm}  Uh, I was just very exhausted, and it was hard to make it through the shift.

00:36.22  \hspace{2mm}  I just felt physically ill and was in pain and just felt terrible, and I'm hoping that I can recuperate on my day off and feel better and have a better night's sleep.
\end{quote}\end{tcolorbox}

\begin{tcolorbox}[top=-0.25cm,bottom=0.4cm,left=-0.25cm,right=-0.25cm]\begin{quote}\small
00:02.22 \hspace{2mm}   I hate today.

00:04.13  \hspace{2mm}  I hate my brain.

00:06.39  \hspace{2mm}  I hate my thoughts.

00:09.11  \hspace{2mm}  I hate today.
\end{quote}\end{tcolorbox}
\caption[High agreement between mean sentence sentiment score and human manual reviewer in identifying the most negative diaries submitted by BLS participant GFNVM.]{\textbf{High agreement between mean sentence sentiment score and human manual reviewer in identifying the most negative diaries submitted by BLS participant GFNVM.} The 4 most negative GFNVM journals identified by our VADER sentiment summary feature aligned with the 4 most negative GFNVM journals picked out by a manual reviewer blinded to the automated sentiment results. Those 4 journals are reproduced here, in rank order.}
\label{fig:negative-diaries-gf}
\end{figure}

\begin{figure}[h]
\centering
\includegraphics[width=\textwidth,keepaspectratio]{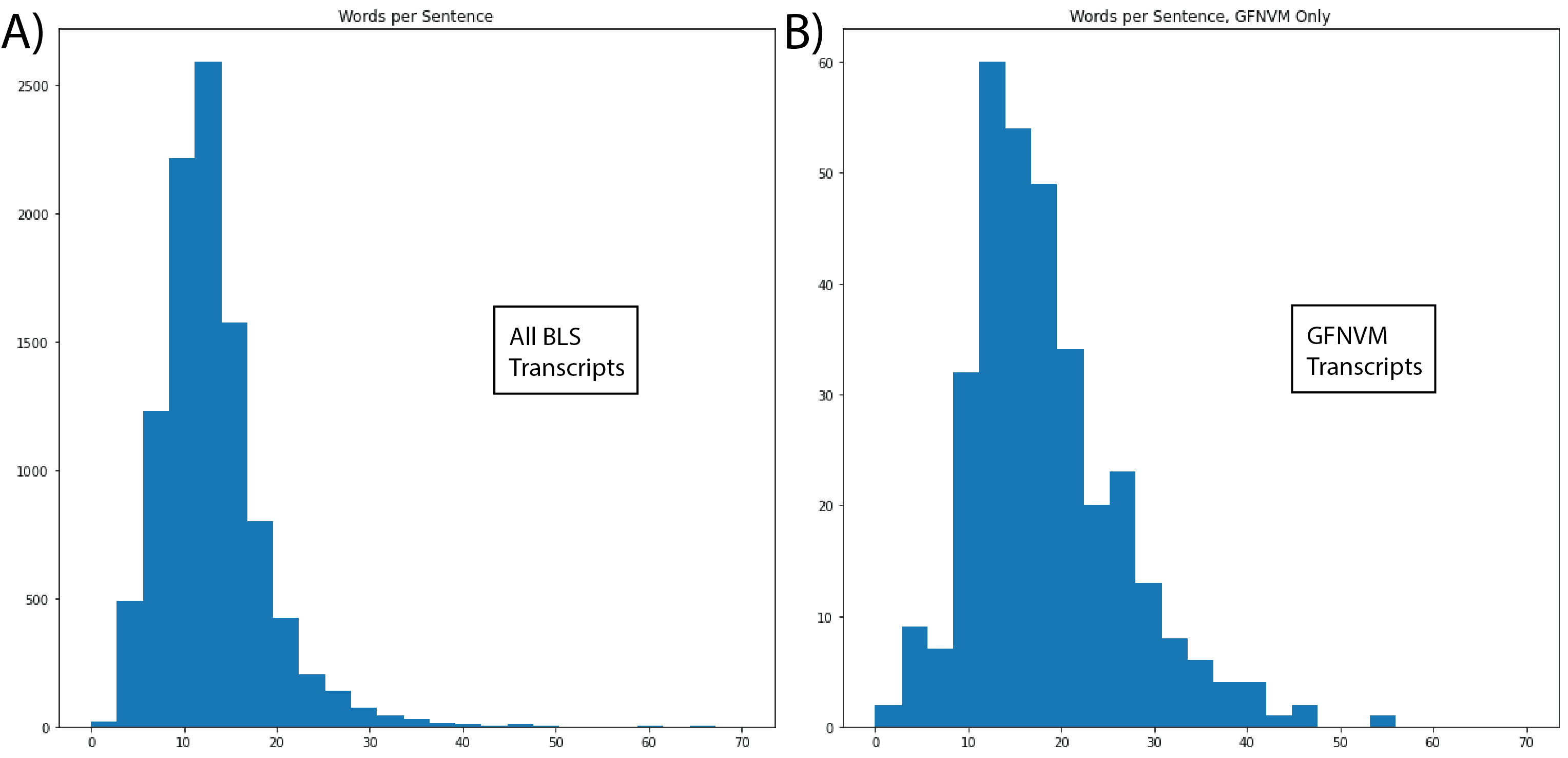}
\caption[The role of TranscribeMe sentence splitting - distribution of words per sentence across transcripts.]{\textbf{The role of TranscribeMe sentence splitting - distribution of words per sentence across transcripts.} Histograms of the words per sentence metric from journals contributed by all BLS participants (A) versus only those journals contributed by GFNVM (B) are presented here, using the same QC dataset as Figure \ref{fig:pause-qc-first-scatter} ($n=9872$ across BLS). As seen in some of the example diaries from subject GF (Figure \ref{fig:negative-diaries-gf}), the length of the individual sentences determined by TranscribeMe can vary greatly. For GF, sentences were sometimes very long, likely indicating interesting variation in speech patterns over time. The way that TranscribeMe chooses to split sentences can thus itself be a relevant feature, and it is additionally important to account for the effects of sentence splitting decisions on other pipeline outputs.}
\label{fig:diary-words-per-sentence}
\end{figure}

It was also observed via this review that GF appeared to have more very long run on sentences in their transcripts than would be expected across the broader BLS dataset, which I next confirmed by reviewing the words per sentence values that occurred across their diaries as compared to the wider BLS distribution (Figure \ref{fig:diary-words-per-sentence}). This suggests that GF's speaking pattern at times tended to cause TranscribeMe to split sentences less frequently -- another investigation that could be of note then would be to evaluate the fluctuation in words per sentence of GFNVM's journal submissions over time, as compared with both other journal features and with features of clinical relevance such as self-report survey responses or clinical scale values. \\

Although the manual review was largely in strong support of the sentiment features used by the pipeline, the examples on which the manual reviewer and the automated ranking did not exactly align are an informative source on the limitations to watch out for with our sentiment summary methods as well. The fifth most negative GF diary selected by the manual reviewer, which was ranked sixth by mean sentence sentiment, is reproduced in Figure \ref{fig:negative-diary-gf-disagree}; it contained much richer content than the fifth most negative diary by mean sentence sentiment, although what is truly most \emph{negative} is of course a matter of subjective debate (and specific definitions), as recording a diary with nothing more than "Overall, it was a bad day" carries potential negative connotations beyond the valence of the literal sentence. 

\begin{figure}[h]
\centering
\begin{tcolorbox}[top=-0.25cm,bottom=0.4cm,left=-0.25cm,right=-0.25cm]\begin{quote}\small
00:00.44    \hspace{2mm}     I had a really hard, stressful day at work and was just really irritable and on edge.

00:04.68    \hspace{2mm}     And when I got home, I was just snappy and paranoid and having flashbacks and irritable and just low in good spirits.

00:12.79    \hspace{2mm}     And then I'd start to feel better, and then I would just be triggered again-- and would just be triggered back into just irritable, angry, depressed crying.

00:21.47    \hspace{2mm}     Um, finally recovered enough where I was able to go to bed, but it was just very rocky.

00:27.51    \hspace{2mm}     And it all stemmed from just a really stressful day and, um, just lack of medicine to help me cope.
\end{quote}\end{tcolorbox}
\caption[The fifth most negative GFNVM diary selected by the manual reviewer contained more detail than that identified by the automated feature ranking.]{\textbf{The fifth most negative GFNVM diary selected by the manual reviewer contained more detail than that identified by the automated feature ranking.} The $\# 5$ most negative diary identified by the manual reviewer for GFNVM, following the first 4 presented in Figure \ref{fig:negative-diaries-gf}, is reproduced here. This diary was instead ranked $\# 6$ by ascending mean sentence sentiment score, to be contrasted with the very short diary ranked $\# 5$ by the pipeline -- which consisted of a single sentence, "Overall, it was a bad day".}
\label{fig:negative-diary-gf-disagree}
\end{figure}

Similarly, the diary with the absolute lowest mean sentence sentiment score for subject 8R, which did not make the top five selected by manual review, is reproduced in Figure \ref{fig:negative-diary-8r-disagree}. It is somewhat negative, but the relatively low number of sentences and the long word length of the most negative sentence likely inflated the negativity of its automated score.

\begin{figure}[h]
\centering
\begin{tcolorbox}[top=-0.25cm,bottom=0.4cm,left=-0.25cm,right=-0.25cm]\begin{quote}\small
00:01.35    \hspace{2mm}     Uh, Friday.

00:03.95    \hspace{2mm}     Slept most the day, tired, tired.

00:06.34    \hspace{2mm}     Um, [redacted] slept, too.

00:07.84    \hspace{2mm}     Pissed off I left my personal/everything-notebook - I just use one notebook at a time - at work 'cause I got distracted 'cause the guy was like my wife here, and I was back with [redacted] smoking, like usual.
\end{quote}\end{tcolorbox}
\caption[The most negative 8RC89 diary by mean sentence sentiment was not in the manual review top 5.]{\textbf{The most negative 8RC89 diary by mean sentence sentiment was not in the manual review top 5.} The 8R transcript with the lowest mean VADER sentiment score is reproduced here, as this diary was not amongst the 5 most negative selected by the manual reviewer. Although this diary is somewhat negative, its negativity was likely overestimated by the pipeline summary feature because it contains few sentences, and the most negative sentence is long.}
\label{fig:negative-diary-8r-disagree}
\end{figure}

The misaligned example(s) on the positive sentiment end of the rankings showed similar trends. The 8R journal with the highest mean sentence sentiment score, which was not included in the top 5 selected by the manual reviewer, contained a single sentence: "I had a pretty good day". While this is certainly a positive statement, it is unlikely any human reviewer would consider a top example of positivity in the journal transcript set. 

For subject GF, the top 2 journals by mean VADER sentiment were not in the top 5 of manual review. These 2 journals were not especially positive in a traditional sense, though it is interesting that the $\#1$ positive journal in the automated rankings mentioned mania in both sentences. Regardless, the presence of just one or two long sentences that contained in part positive phrases likely biased the score to be higher than most humans would judge.

\noindent As much of this section focuses on negative sentiment, additional detail on positive-valence sentiment transcripts can be found in supplemental section \ref{subsubsec:sentiment-diary-trans}. \\

\FloatBarrier

\paragraph{Proof of concept case report on sentiment scores.}
Because the manual sentiment scoring results were highly promising but somewhat small in scope, we took a deeper dive into sentiment scores from a participant with a particularly rich dataset (3SS93, discussed above) for additional validation. A medical student in the lab reviewed all the diary transcripts from the first $\sim 6$ months of participant 3S's dataset while blinded to the sentiment scores, and took notes on the patient profile, including recurring topics in the journals and days with journal content of particular salience. 

\begin{figure}[h]
\centering
\includegraphics[width=\textwidth,keepaspectratio]{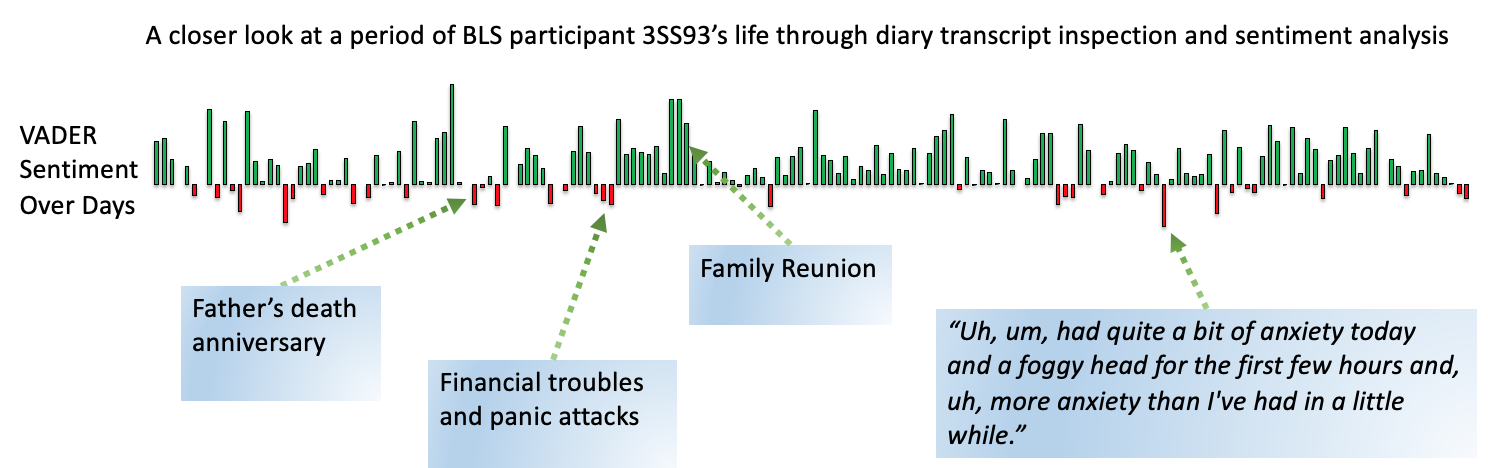}
\caption[Extreme diary sentiment scores highlight interesting content within a long-term patient dataset.]{\textbf{Extreme diary sentiment scores highlight interesting content within a long-term patient dataset.} The mean sentiment scores for diaries over the first $\sim 6$ months of BLS patient 3SS93's data are displayed as sequential bar plots, starting at day 1 of enrollment and leaving a blank space for any days missing a diary. Time points of notable content uncovered by manual review are labeled, showing concordance between human opinion on salient topics and the automated sentiment scores from those days.}
\label{fig:vader-manual-3s}
\end{figure}

Major issues reported by 3S were migraine headaches and high levels of anxiety, although the majority of diaries did not mention these topics and were more mundane accounts of daily life with neutral to positive tone, often including conversations of movies or TV shows. 3SS93 also discussed family quite a bit, in varying contexts. By overlaying timepoints marked during manual review with the sentiment scores from the same diaries, we can see that many of the transcripts with larger magnitude mean sentiment score also contain noteworthy content - particularly those with negative scores (Figure \ref{fig:vader-manual-3s}). Thus VADER sentiment could be useful not only in statistical analyses, but in quick screening of longitudinal data for diaries worth closer review as well, thereby contributing to their qualitative utility. 

To validate sentiment ratings in the context of relevant predictive power, the mean VADER sentiment score from each aforementioned 3S diary (Figure \ref{fig:vader-manual-3s}) was correlated with the raw same-day responses from self-report (EMA) survey questions assessing perceived quality of life and symptom severity using Spearman rank correlation. In particular, sentiment scores were significantly correlated with the sum score of the positively-worded (Spearman's $r=0.413$, $p<10^{-5}$) and the negatively-worded (Spearman's $r=-0.293$, $p=0.002$) EMA items, both remaining significant after Benjamini–Hochberg multiple testing correction. Thus the pilot results support the relevance of mean VADER sentiment for the pipeline's primary summary score, something that was later reinforced by the larger scale EMA and diary feature analyses performed in section \ref{sec:science2}. For more details on the review of sentiment in the first $\sim 6$ months of 3SS93's diaries, including a breakdown of item-level correlations, see supplemental section \ref{sec:3s-sentiment}. \\

\FloatBarrier

\noindent As an aside, one of the other questions we investigated with the pilot 3S transcript case study was the quality of automated transcription services, including machine learning models offered by Amazon Web Services (AWS) and by TranscribeMe. The quality of audio recorded by 3S was on the lower end of BLS (discussed within \ref{subsubsec:diary-val-qc}), but it was still good enough to produce largely high quality manual transcripts. With the automated services, the produced transcripts had many more word errors and of course lacked verbatim annotations, PII redaction, and speaker ID (though speaker ID is not really relevant for diaries, it is important for the interview datatype of chapter \ref{ch:2}). 

While the lower quality of the automated transcripts interfered with many potential analyses we consider in the linguistics piece of this pipeline, it is worth noting that the VADER sentiment scores were fairly resilient. Splitting sentence on periods produced by the automated TranscribeMe service produced a somewhat different structure than the sentence labels provided in the high quality human transcriptions. However when taking the mean of VADER sentiment across each transcript, the scores from the human and automated versions were highly correlated, and the fully automated sentiment scores retained a similar significant relationship with EMA as was reported above. Thus VADER sentiment is our most robust linguistic feature from a scalability perspective. \\ 

\FloatBarrier

\paragraph{Caveats to keep in mind with VADER sentiment.}
To summarize, one primary limitation identified by our manual review relates to the splitting of sentences. We have observed that VADER sentiment tends to give more extreme sentiment scores to particularly long sentences -- in fact, when diaries were fed in as one big paragraph to VADER sentiment in a test performed by an RA, the returned scores were almost always close to -1, 0, or 1, whereas when individual sentences were input to VADER (as the design intended), returned scores varied much more over the -1 to 1 range. This phenomenon means that when a patient speaks more quickly and is thus more likely to have run on sentences grouped together by TranscribeMe, each individual sentence is therefore more likely to receive a more extreme sentiment score (and there will also be fewer total sentences than a finer split would contain). 

On the other hand, the sentence splitting performed by TranscribeMe, when done correctly, can contain interesting information about participant speech patterns that might be missed by automated tokenization methods. As such, we do not consider this a limitation that should be directly worked against, but rather a trade-off that ought to be kept in mind, particularly in a subject like GF. One way to consider this relationship in a model of diary features would be to track the words per sentence in each diary as an additional input.

The second major limitation does not relate to VADER outputs themselves, but rather our methodology for summarizing them over a diary. As can be seen by the overall very good alignment between the mean sentence sentiment of each transcript and the manual reviewer ratings, a simple mean can capture a large amount of information. The mean has its shortcomings though, and the inclusion of additional summary metrics is a data science trade-off that needs to be considered on a case-by-case basis. Here, we aligned identification of the most positively and most negatively valenced diaries between the manual and automated methods; while in a vacuum the mean of the mean sentiment feature over a set of diaries with the same underlying sentence sentiment distribution will be the same regardless of the diary lengths, the extremes of these theoretical distributions will not be the same, because shorter diaries introduce more variance in individual diary scores. This is exemplified by the very short transcripts that are moderately positive or negative, but appear in the extremes of the present automated scoring because that single sentence is not diluted by any neutral sentences. 

To better characterize mean transcript VADER sentiment score across our BLS journal dataset, we can also look at properties of its distribution. The diary sentiment tends to be positive more often than negative and is generally fairly low in magnitude. Given that the open-ended prompts often result in neutral descriptions of the day for at least part of the recording, and the sentiment is meaned over multiple sentences, it is not surprising to see these distributional properties. 

As mentioned, the pipeline does also return minimum and maximum sentence sentiment scores for each diary by default, which can provide alternative sentiment measurements that may be worth consideration. However these features have their own drawbacks, as well as their own correlation with length -- instead biasing towards high magnitude sentiment scores in journals with more sentences. The distribution of mean scores suggests that minimum sentence sentiment in particular may be of use to isolate additional transcripts with negative thought processes that might otherwise go unnoticed. In a large enough dataset, it will be ideal to consider multiple summary statistics for the sentence sentiment distributions in conjunction with information on the verbosity and sentence structure of the transcript, thereby providing a much more complete picture of the contained sentiment. Though in a smaller scale modeling dataset the mean has performed very well in our hands at capturing a generally accurate sentiment summary via a single feature. \\

\noindent Ultimately, the scope of the current pipeline remains at a few different simple summary stats for sentiment over the whole diary transcript. This is in line with the more general principle of sharing a computationally efficient, interpretable, and well-established core feature set for future analyses. While straightforward features can go a long way, it is important as well to identify ways in which the pipeline might be extended as audio journal analysis in digital psychiatry progresses. 

One potential direction is to utilize information about content topics from the diary to inform the transcript (or sentence) level scores. Over the course of looking through BLS transcripts, we have found a number of individual instances where a more nuanced analysis method that incorporates sentence content into the score would have been valuable. For example, there was a journal that was given one of the highest mean positive sentiment ratings for a particular subject -- but turned out to contain discussion centered almost exclusively around a movie the person had seen that day, for which they had high praise. The strong positive sentiment score for this transcript is entirely in line with the intended performance of VADER, and moreover being in such high spirits about a movie probably holds some signal with regards to participant mood. 

However, it is easier to use hyperbolic adjectives when describing feelings about entertainment, and these feelings are less likely to be quite as salient as commentary the participant contributes on their own life. An improvement to the pipeline's diary sentiment rating system would thus take into account the content that the sentiment is associated with, and draw greater attention to a high magnitude sentiment score when a patient is e.g. describing their own mental state. Further examples of similar modifications would be treating sentiment of recurring topics across an individual's diaries differently than sentiment of infrequent topics, or flagging changes in sentiment about a specific topic over time. 

More broadly, as NLP tools rapidly progress, other methods can be benchmarked against VADER on diary datasets via this pipeline. New feature options for evaluating sentiment could then be integrated into the pipeline to be shared more widely, if those tools perform highly in quality control and scientific contexts. It is unclear how much of practical utility for digital psychiatry research can be gained through updates that do a one to one replacement for VADER i.e. replacing sentence-level sentiment calculations. However, there is a great deal that might be gained by the capabilities of newer models to process much longer sequences while referencing context from throughout the text \citep{transformers}. Such models could also be capable of providing a higher dimensional sentiment-style output, for example sadness versus anger, with much greater accuracy than established methods like LIWC.

Regardless, there is strong reason to believe the summary sentiment scores already extracted by the pipeline are of relevance, and this code release lays the groundwork for additional steps to be built. It is the case that care needs to be taken in ensuring modern (much less well understood) machine learning methods are doing what we think they are when applying to a digital psychiatry dataset -- both to maintain quality of results and to enable downstream scientific insights (and potential medical practice).   \\

\paragraph{Defining linguistic disfluencies.}
Another major focus area of our language feature development was counting occurrences of linguistic disfluencies, including a breakdown by disfluency category for each transcript. The clinical relevance of disfluencies is discussed at length in chapter \ref{ch:2}, so here I will report only on the technical specifics for our detection techniques and their properties in our BLS dataset. 

\noindent Each of the following disfluency types, as previously defined in section \ref{subsubsec:diary-val-qc}, are counted per sentence:

\begin{itemize}
    \item Nonverbal edits i.e. occurrences of uhs/ums
    \item Verbal edits i.e. usage of "like", "I mean", "you know" as filler words
    \item Repeats i.e. immediately reusing the same word or stuttering over part of a word
    \item Restarts i.e. changing topics or otherwise starting a new sentence before completing the current sentence
\end{itemize}

\noindent These counts are obtained using regular expressions based on the TranscribeMe verbatim notation that is designed to include linguistic disfluencies. Exact pipeline implementation details are provided within section \ref{subsec:diary-code}.

Because each counting function has a clear definition that can be checked for bugs with straightforward code testing techniques, there is not a need for close manual review of code outputs. The corresponding transcript level features were then the total counts of each type - a simple sum that also doesn't require validation beyond automated software testing. For safety across this process, some values have additionally been spot checked during analysis work.

What does require validation from our disfluency workflow is the accuracy of the TranscribeMe verbatim transcriptions in capturing each disfluency type. The code to detect the corresponding notation is straightforward, but the assumption the notation will be both sensitive and specific needs justification. Given the level of quality we observed with our transcripts more broadly, there was good reason for optimism about the quality of TranscribeMe's linguistic disfluency markings. To check that TranscribeMe's notation use appeared consistent across the dataset, a student volunteer reviewed all diaries transcribed within a one month period (April 2021). The student drafted a set of rules for detecting disfluencies based on looking at the transcripts and listening to the corresponding audio, which aligned directly with the official guidelines we obtained from TranscribeMe for verbatim notation; thus indicating that the transcribers have been correctly applying these verbatim conventions. \\

\paragraph{The future of disfluency analyses.}
For reference, the pipeline also provides summary statistics for a number of features across the dataset (see supplementary section \ref{subsec:bls-diary-summary-sup}). It is unsurprising that nonverbal edits were substantially more frequent per diary than other types of disfluencies, as use of such filler words is common in unplanned human speech. Note that for all disfluency features, the standard deviation over diaries was greater than the mean, suggesting that variation in use of disfluencies may contain interesting nonlinear signal. On the other hand, features with many 0 scores across a diary dataset will require special analysis considerations both to find relevant signal and to identifying possible confounding factors like verbosity (even when the features are officially length-normalized). Section \ref{sec:science2} will begin to address some of these considerations within the context of pilot scientific inquiries on this BLS dataset. 

Ultimately, our methodology for detecting linguistic disfluencies is quite accurate, of course with the drawback that the manual transcription service with this level of careful detail is quite expensive. However, we do not recommend automated approaches for disfluency detection at this time, and strongly suggest a protocol like ours if disfluencies are of particular interest for a research project. Most of the other feature extraction techniques that we employ here have a strong body of computer science literature and more general use cases backing them, meaning psychiatry groups can focus on the problem of how to best apply these existing tools rather than requiring the computational expertise of a traditional computer science group to develop algorithms from scratch. Disfluencies on the other hand are not of interest in various consumer applications and thus unsurprisingly are not an active area of computer science research. In fact, most work on automatic transcription tools intentionally filters out disfluencies \citep{whisper}, and natural language processing work certainly does not include them in typical training input text. Because disfluencies are also a somewhat more niche interest in the field of digital psychiatry, it is unlikely that tools for automatic scoring of different disfluency categories will be well-validated to match human performance any time soon. \\

\paragraph{Methods for estimating semantic incoherence.}
Given the relevance to psychosis as previously discussed in section \ref{sec:background2}, my code utilizes word embeddings to summarize the semantic coherence of submitted transcripts in a few different ways. 

For each word in a sentence, a vector embedding is obtained, with specific embedding methods to be described shortly. The resulting vectors are then used to summarize the semantic properties of each individual sentence via three features: the mean vector magnitude, the mean angle between vector representations of sequential words, and the mean angle between all pairings of vectors in the sentence. In an embedding, higher vector magnitude corresponds to a word estimated to be less commonly used. When comparing two different word's embedding vectors, a higher angle between them corresponds to words that are estimated to be less likely to co-occur. 

Embedding vectors for the individual words are currently obtained by the pipeline using the 300-dimensional pre-trained vectors released by \cite{Mikolov2013}. They used a dataset derived from articles on Google News -- containing about 100 billion total words -- to train a model with their skip-gram technique. The final released model provides the resulting vector representations for 3 million different words and phrases. As part of their work, \cite{Mikolov2013} evaluated the performance of their embeddings on various analogy tasks, setting a new state of the art at the time on relevant benchmarks.

Of course benchmarks measuring relationships between individual words or looking at phrase-based analogy tasks don't immediately extend to most applications with patient speech. However, there is a history of assessing patient performance on word association style tasks in the Schizophrenia literature, which can be a natural and direct application of models performing well on embedding benchmarks. Automatic analysis of speech from semantic fluency tasks in psychosis has indeed had positive results, for example with the semantic incoherence measurements used by \cite{Holshausen2014} on data from their prompt to "name as many different animals as possible in the span of 90 sec".  

In our case though, the input data is not from a constrained task but rather text from an open-ended monologue. While I would still expect an extremely nonsensical submission to be detectable by semantic incoherence from the word2vec embeddings, it is less clear if more subtle effects would be measured by the pipeline. Moreover, specific methodologies for summarizing the sentence-level semantic features per diary need to be independently validated, in addition to support needed for the embedding model. The primary summary statistic reported for each semantic feature currently is the mean over sentences, which might not be sufficiently expressive even if the embedding model performs well per sentence. The most sequentially incoherent sentence in each diary is another raw output of the pipeline that should be considered more carefully as the primary semantic summary value. Other more nuanced summary methods akin to what is described for VADER sentiment above could also be investigated if semantic coherence is of especial relevance to an intended pipeline application. 

Nevertheless, there is both prior literature and dataset-specific justification for the inclusion of the feature at this time. Even though performance may not measure up to state of the art, there are many previous demonstrations of the value of the Google News 300-dimension embeddings in related problem spaces. Even though objective validation of incoherence is an inherently more difficult task -- especially in a dataset where few patients experience severe psychotic episodes -- there is also some promise found via manual review in the BLS dataset. I will next review the applications literature, and then will report our preliminary manual review findings. \\

Broadly, the Google News 300-dimensional word2vec embeddings have reliably improved performance over baseline input representations like bag of words when used to train models on many different NLP tasks, from sentiment prediction for sentences to keyword generation for entire articles - though comparison against other word embedding methods has varied greatly depending on the task and the downstream model \citep{Khattak2019}. Further, there are tradeoffs with computational cost, ease of use, and transparency when considering various embedding models. 

To our knowledge there is no work investigating this or related models for word representation in the context of daily open-ended psychiatric patient diaries. However, we can investigate use cases from other related contexts to guide our validation process. On a high level, the application to transcripts of spoken language is outside the original scope of the model, as is use with medical data. The former is more critical for verifying good accuracy of the features extracted here, but options for well-documented openly available embeddings designed around spoken language are few and far between. The latter is the opposite, with alternative models derived from medical datasets that are available and worth discussing, but ultimately less relevance to our problem space here.

Most work applying word embeddings to biomedical questions has focused on understanding of electronic health record (EHR) notes. Prior research has successfully used the Google News vectors to improve performance on a number of EHR prediction tasks over previous baselines, but embeddings derived from medical language datasets can reliably outperform the more general embeddings on these same tasks \citep{Khattak2019}. For tasks that involve a large amount of medical jargon, specialized embedding models like the one trained on PubMed have become an obvious choice over Google News. 

Medical language analysis may have relevance to our work at times, as patients sometimes discuss specific symptoms or treatments in technical terms. However our diaries are still patient speech and largely do not involve direct discussion of their condition, so using existing models that focus on technical language from a broad body of EHR or medical literature would likely not be appropriate. Although one could imagine an ensemble approach that utilizes both generalized and specialized models for different inputs as a future direction to balance these approaches. Moreover, the results on medical language suggest that word vectors learned from a large corpus of patient audio journals would increase the quality of future analysis using this datatype, which further highlights the importance of coordinated data collection and processing techniques. 

For prior validation of embeddings on transcribed unscripted spoken word, we turn to existing results in psychiatry. This is in part because psychiatry is a rare case where verbatim transcription is desirable - the imperfections in natural human speech may contain clinically relevant information, yet these imperfections are generally unwanted when the focus is only conveying the intended content. Furthermore, much of the prior psychiatry literature on embeddings revolves around semantic incoherence in psychotic disorders, which as mentioned is the end goal of the embedding usage in my code. Thus the application of embeddings to transcripts from structured clinical interviews are the closest to our intended use case, not only because of the participant population, but also because of the way the interviews themselves are handled and subsequently processed. \\

\noindent Incoherence itself is closely related to the original benchmark embedding tasks that tested how well the angles between word vectors represented human-perceived word associations or how well word vector math could perform on human constructed analogy questions like the following:

\begin{center}\begin{tabular}{c}\begin{lstlisting}
king - man + woman = ?
\end{lstlisting}\end{tabular}\end{center}

\noindent where the answer would of course be "queen". Therefore the word vector angles that are directly used in defining semantic incoherence are in a sense already validated. The major questions about their application are how exactly the word vectors are used to estimate incoherence on the level of a sentence, paragraph, or document, and how well these various methods can differentiate degrees of incoherence in natural human speech from different environments. 

There are theoretical reasons for concern about the Google News model's ability to translate to the patient speech context. A topic switch that would sound normal in a spoken conversation may seem incoherent if instead found in a written news article. Conversely, verbatim speech can include greater use of simple words and phrases that will have small angles with each other - including filler words (e.g. "..., like, ...") and repetitions that fill time or redirect thoughts (e.g. "the, well, the ..."). Some of these issues could be addressed with data cleaning, but on the other hand their presence may be providing relevant info to the semantic incoherence calculation that should not be removed. 

Ultimately, there remains a need for a more systematic quantification of these different factors across multiple different embedding models of interest to the field, in order to develop a methodology that will truly optimize model performance and best tease apart specific properties of language being included in any incoherence metric - the latter being critical for strong interpretability. In the meantime, existing techniques have been used successfully on interview transcripts for prediction of clinically relevant variables, supporting the promise of these tools. \\

\paragraph{Semantic incoherence tools in psychiatry.}
\cite{Corona2022} in particular demonstrated clinical relevance of the exact 300-dimensional skip-gram training technique used to obtain the Google News embeddings. As their sample was Dutch, they replicated the methodology from \cite{Mikolov2013} instead on a large Dutch language corpus, to generate pre-trained vectors for use with their interviews. They then collected semi-structured interviews from 50 Schizophrenia patients and 50 control subjects, and found that incoherence metrics derived from the 300-dimension skip-gram embeddings were significantly associated with Schizophrenia diagnosis. A model utilizing only these features along with labeling of Dutch connectives was able to distinguish the patients from controls with $85\%$ accuracy. Moreover, they obtained these results without cleaning the transcriptions of the filler words or repetitions discussed, just as we have decided to do in our methodology.

There have also been multiple results demonstrating predictive power of incoherence features extracted from Schizophrenia patient speech in English. \cite{Bedi2015} utilized simple summary features such as the minimally coherent phrase across an entire semi-structured interview to identify those high-risk patients that would go on to have a psychotic episode in the next 1 to 3 years. Furthermore, they calculated incoherence using latent semantic analysis, an older technique that has been wholly outperformed by word2vec-based models like the Google News one. More recently, \cite{Tang2021} leveraged the recent major NLP advance of transformers to obtain sentence embeddings that generated sentence-to-sentence incoherence features with predictive relevance for identifying Schizophrenia spectrum disorders using speech from clinical interviews. 

Indeed, in the years since this work began transformers have taken over much of the computational NLP literature, moving machine learning focus largely away from the framework of inputting word embeddings into recurrent neural networks. The application of transformers towards analyzing patient language is an important step forward for the field, as it is the best technique for certain questions and datasets. On other hand, word2vec embeddings can be applied more flexibly and intepreted more easily, and for the time being remain a better option when it comes to accessibility, transparency, and computational cost. 

We suggest that while the field of digital psychiatry should certainly not shy away from the latest machine learning developments, it is also a field where the datasets have unique privacy, bias, and other considerations. Coupled with the fact that many groups have primary expertise in psychiatry, it is especially infeasible to constantly chase the latest technique while maintaining rigor. Thus continued work on "older" feature extraction methods is necessary to strike a proper balance, and currently the word2vec embeddings are likely the best choice for a widely-targeted code release. 

Taken together, there is good reason to believe that the incoherence measures computed by the present pipeline will be immediately applicable to study of psychotic disorders. The Google News model has strong performance on relevant word association benchmarks \citep{Mikolov2013}, and incoherence estimates that used very close \citep{Corona2022} or even worse performing \citep{Bedi2015} word similarity measurements to analyze speech from clinical interviews have found incoherence features to be significantly associated with psychotic disorders. \\

\noindent There are questions that remain to be addressed with the application of word embeddings to study incoherence in daily audio diaries. In many ways, the diaries hold greater promise, as their daily frequency creates a much larger number of labeled data points to work with, an important factor for modern machine learning success. Their shorter length can also be an advantage, as it is more tractable to compute a handful of summary features that adequately capture the semantic incoherence of the entire journal entry. Interviews contend with the opposite combination; a huge amount of speech needs to be summarized but few labels exist, which prohibits the exploration of more than a handful of potential features. 

Nevertheless, the journals have disadvantages relative to interviews as well. Participants with higher levels of functioning may not display incoherence during a short description of their day, even if they would during a longer conversation or in response to a well-designed question. Furthermore, the semi-structured interviews allow incoherence to be measured relative to known questions. It is possible that some patients are able to express coherent thoughts but do not engage in conversation in a coherent way, such that social interactions would appear bizarre but monologues wouldn't. Particularly as semantic incoherence estimates are not intended to capture coherently-expressed delusional beliefs, it is much more likely that measuring coherence within a dialogue could detect symptomatology that reflects a lack of grounding in the external environment but minimal internal inconsistency. 

Additionally, in the clinical interview literature that does exist, there is a paucity of validation on the interview level. Most work utilized models that were validated on a low level (individual words or sentences), and from there jumped to demonstrating clinical relevance of the interview level incoherence features. Therefore sanity checking the incoherence features from this pipeline on the diary level is an important contribution not only to justify the extension of these metrics to the diaries and our exact computation method, but also to provide some of the earliest evidence of validation that directly checks against a manual labeling of the clinical concept of incoherence. 

Therefore in this section I will present evidence from manual review that diary incoherence estimates align with human-perceived semantic incoherence of those diaries. I will also discuss distributional properties of these features in the BLS diary dataset, which includes Bipolar patients both with and without psychosis. That will conclude the supporting arguments for the pipeline's word2vec-based semantic coherence features, but in section \ref{sec:science2} I will present related scientific results from the BLS dataset. Further, analogous features computed from BLS interviews are discussed in chapter \ref{ch:2} (\ref{sec:disorg}). \\

\paragraph{Qualitative review of word2vec-based features produced by the pipeline.}
As argued above, the most important features to validate are the diary level semantic incoherence metrics derived from the per sentence features. Here we focus on the summary stats that correspond to taking the mean across sentences of each feature. However the pipeline also returns analogous minimum, maximum, and standard deviation summary values that could lend further insight into a dataset.

For manual review, we performed a similar procedure as that done for the transcript sentiment summary. Trying to pick a set of 5 most coherent diaries out of a mostly coherent database isn't really a sensible task however, so the RA identified only the 5 most incoherent transcripts based on a blinded manual review of all texts from each of the two participants highlighted here (8RC89 and GFNVM). I then sorted the transcripts of these two participants using their final mean sentence to sentence incoherence feature as well as their maximum within sentence incoherence (pairwise) feature, and aligned the RA's ranked list with these. The top 5 manual review rankings for both subjects were perfectly aligned with the automated between sentence incoherence rankings. The maximum within sentence incoherence produced a very different top 5 list however. 

I next reviewed the transcripts from the highest incoherence scores - of both type and from both selected subjects - to better characterize the pros and cons of these features, as was done above for sentiment analysis. The top 5 journals by \emph{sentence to sentence} incoherence (and manual review) are reproduced here for GFNVM in Figure \ref{fig:incoherent-diaries-gf} and for 8RC89 in Figure \ref{fig:incoherent-diaries-8r}. More detailed information about the qualitative review based on \emph{within sentence} incoherence measures can be found in supplemental section \ref{subsubsec:incoh-diary-trans}.

\pagebreak

\begin{FPfigure}
\centering
\begin{tcolorbox}[top=-0.5cm,bottom=0.1cm,left=-0.5cm,right=-0.5cm]\begin{quote}\scriptsize
00:01.13    \hspace{2mm}     Without having to come in on first shift to [get it?].

00:07.68    \hspace{2mm}     [inaudible].
\end{quote}\end{tcolorbox}
\vspace{-0.5cm}
\begin{tcolorbox}[top=-0.5cm,bottom=0.1cm,left=-0.5cm,right=-0.5cm]\begin{quote}\scriptsize
00:06.42    \hspace{2mm}     I don't know.

00:10.84    \hspace{2mm}    Bad mental state.
\end{quote}\end{tcolorbox}
\vspace{-0.5cm}
\begin{tcolorbox}[top=-0.5cm,bottom=0.1cm,left=-0.5cm,right=-0.5cm]\begin{quote}\tiny
(S1) 00:03.24   \hspace{2mm}    It's been a long day.

(S1) 00:04.6    \hspace{2mm}     [inaudible] [to add?].

(S2) 00:06.85   \hspace{2mm}     Do you think [inaudible] with no light [inaudible]?

(S1) 00:12.3    \hspace{2mm}     Cats have vision-- night vision.

(S2) 00:14.93   \hspace{2mm}     [inaudible].

(S1) 00:16.2    \hspace{2mm}     Yeah, they are.

(S2) 00:18.93   \hspace{2mm}     [Dont' you have something?]--

(S1) 00:18.97   \hspace{2mm}     I've seen them, yeah.

(S1) 00:20.34   \hspace{2mm}     [So I'll be?]-- yeah, I think I'm gonna be okay.

(S1) 00:25.18    \hspace{2mm}    It's gonna be okay.

(S1) 00:27.81    \hspace{2mm}     They can totally [inaudible].

(S2) 00:30.1    \hspace{2mm}     Not in the dark, dark [inaudible].

(S1) 00:31.39    \hspace{2mm}     Yes, they can.

(S2) 00:33.08   \hspace{2mm}     [inaudible].

(S1) 00:34.41     \hspace{2mm}     The green-- the green.

(S2) 00:38.12   \hspace{2mm}     What? [laughter]

(S1) 00:40.45   \hspace{2mm}     The green.

(S2) 00:40.7    \hspace{2mm}     Yeah, well, they-- the cats n-need, like a--

(S1) 00:43.345      \hspace{2mm}     The green in--

(S2) 00:43.93     \hspace{2mm}     --little light.

(S1) 00:44.67   \hspace{2mm}     It's the green in the eyes.

(S2) 00:45.57       \hspace{2mm}    That's not how night vision works.

(S2) 00:47.1        \hspace{2mm}    [inaudible]--

(S1) 00:47.41   \hspace{2mm}     Their eyes are green 'cause it's night vision.
\end{quote}\end{tcolorbox}
\vspace{-0.5cm}
\begin{tcolorbox}[top=-0.5cm,bottom=0.1cm,left=-0.5cm,right=-0.5cm]\begin{quote}\scriptsize
00:01.43    \hspace{2mm}     Over-worked with ups and downs.

00:03.65    \hspace{2mm}     And now having-- core situation that is very bad as my brakes gave out, so it's very stressful.
\end{quote}\end{tcolorbox}
\vspace{-0.5cm}
\begin{tcolorbox}[top=-0.5cm,bottom=0.1cm,left=-0.5cm,right=-0.5cm]\begin{quote}\scriptsize
00:02.93    \hspace{2mm}     Had to get up earlier-- early today for orientation.

00:05.31     \hspace{2mm}     And, um, during orientation had, like, a little PTSD trigger but I was able to do-- control of it instead of letting it take over my thoughts.

00:20.88    \hspace{2mm}     Um, I'm very irritable.

00:25.17    \hspace{2mm}     Again, [inaudible] and [inaudible] just aggravated and on edge.

00:34.92    \hspace{2mm}     It's kinda getting old.

00:37.27    \hspace{2mm}     I go back to work tomorrow, [normal work?].

00:44.97    \hspace{2mm}     Hopefully, be a good day.

00:49.1     \hspace{2mm}     Um, I just need some sleep I think 'cause I haven't slept since I woke up at 6:20, and my brain just needs to reboot.

00:56.65    \hspace{2mm}     [inaudible] helps.
\end{quote}\end{tcolorbox}
\caption[High agreement between mean sentence-to-sentence incoherence score and human manual reviewer in identifying the most incoherent diaries submitted by BLS participant GFNVM.]{\textbf{High agreement between mean sentence-to-sentence incoherence score and human manual reviewer in identifying the most incoherent diaries submitted by BLS participant GFNVM.} The 5 most incoherent GFNVM journals as ranked by our mean sentence to sentence incoherence summary feature were aligned with the 5 most incoherent GFNVM journals picked out by a manual reviewer. Those journals are reproduced here, in rank order. Most of them are on the shorter side with somewhat odd word usage or missing transition between distinct thoughts, although nothing extreme. It is unclear whether statements such as "core situation that is very bad as my brakes gave out" are intended to be metaphorical or not, but regardless the manner of expression is not entirely coherent. The $\#3$ most incoherent diary by mean sentence to sentence score contains conversation with another person, so speaker IDs are provided; this transcript appears quite incoherent at a glance, and should indeed have a high incoherence score by our automated metrics -- however given the conversational context it is difficult to interpret. Recall that checking automated QC metrics from the pipeline can help to flag issues in advance, including multiple speakers in a diary.}
\label{fig:incoherent-diaries-gf}
\end{FPfigure}

\begin{FPfigure}
\centering
\begin{tcolorbox}[top=-0.5cm,bottom=0.1cm,left=-0.5cm,right=-0.5cm]\begin{quote}\scriptsize
00:00.3      \hspace{2mm}     23rd.

00:01.26    \hspace{2mm}     Another day off.

00:02.87     \hspace{2mm}     Relaxed.

00:05.44    \hspace{2mm}     Uh. [silence]
\end{quote}\end{tcolorbox}
\vspace{-0.5cm}
\begin{tcolorbox}[top=-0.5cm,bottom=0.1cm,left=-0.5cm,right=-0.5cm]\begin{quote}\scriptsize
00:00.22    \hspace{2mm}     Tuesday the 12th.

00:01.7     \hspace{2mm}     Uh, relaxed.

00:03.43     \hspace{2mm}     [redacted] put my breathing machine together, but I didn't actually use it.

00:07.38    \hspace{2mm}     Um, had a few drinks at night.

00:14.16    \hspace{2mm}     Um, pretty uneventful day.

00:16.51    \hspace{2mm}     Went grocery shopping.

00:20.12    \hspace{2mm}     It's about it.
\end{quote}\end{tcolorbox}
\vspace{-0.5cm}
\begin{tcolorbox}[top=-0.5cm,bottom=0.1cm,left=-0.5cm,right=-0.5cm]\begin{quote}\scriptsize
(S1) 00:00.5     \hspace{2mm}     Had an uneventful day yesterday.

(S1) 00:02.39   \hspace{2mm}     Slept way too late.

(S1) 00:03.68   \hspace{2mm}     Watched Netflix's.

(S1) 00:05.1    \hspace{2mm}     Uh, did some reading like on the computer and started reading Collapse by Jared Diamond, the guy who wrote Guns, Germs, and Steel.

(S1) 00:14.2    \hspace{2mm}     Pretty good.

(S1) 00:15.79   \hspace{2mm}     Not much happened.

(S1) 00:17.42   \hspace{2mm}     Um--

(S2) 00:18.839 I think you-- I think-- I'm holding.
\end{quote}\end{tcolorbox}
\vspace{-0.5cm}
\begin{tcolorbox}[top=-0.5cm,bottom=0.1cm,left=-0.5cm,right=-0.5cm]\begin{quote}\scriptsize
00:00.26   \hspace{2mm}     Wednesday, the 27th.

00:01.86    \hspace{2mm}     Um, totally forgot to do group therapy, was tired, probably resting or something.

00:07.75    \hspace{2mm}     Just final preparations before the wedding.

00:10.57    \hspace{2mm}     Um, yeah, getting married the ni-- on Thursday the 28th.

00:16.53    \hspace{2mm}     Should be good.

00:17.47    \hspace{2mm}     Everything's prepared.

00:20.24    \hspace{2mm}     Um, yeah.

00:21.6     \hspace{2mm}     That's about it.
\end{quote}\end{tcolorbox}
\vspace{-0.5cm}
\begin{tcolorbox}[top=-0.5cm,bottom=0.1cm,left=-0.5cm,right=-0.5cm]\begin{quote}\scriptsize
00:00.83    \hspace{2mm}     Sunday.

00:01.45    \hspace{2mm}    Lazy day with [redacted].

00:03.5     \hspace{2mm}     Watched just a few things on Netflix.

00:05.07    \hspace{2mm}     Had a lot of sex.

00:07.03    \hspace{2mm}     Good relaxing day.

00:09.15    \hspace{2mm}     Use of cannabis edibles, good stuff.

00:11.7     \hspace{2mm}     Oh, we have a vape too.

00:12.64    \hspace{2mm}     The vape works pretty well.
\end{quote}\end{tcolorbox}
\caption[High agreement between mean sentence-to-sentence incoherence score and human manual reviewer in identifying the most incoherent diaries submitted by BLS participant 8RC89.]{\textbf{High agreement between mean sentence-to-sentence incoherence score and human manual reviewer in identifying the most incoherent diaries submitted by BLS participant 8RC89.} The 5 most incoherent 8RC89 journals as ranked by our mean sentence to sentence incoherence summary feature were aligned with the 5 most incoherent 8RC89 journals picked out by a manual reviewer. Those journals are reproduced here, in rank order. The properties seen are similar to the analogous top 5 diaries submitted by GFNVM (Figure \ref{fig:incoherent-diaries-gf}, with a relatively small number of sentences and also relatively shorter sentences, as well as somewhat random jumping between topics and the occasional presence of a background conversation. None of these examples stand out as particularly incoherent, but it is unlikely that we would encounter extreme word salad in the patient population of this study.}
\label{fig:incoherent-diaries-8r}
\end{FPfigure}

\FloatBarrier

A number of different insights about factors that affect the incoherence features can be gleaned from the examples identified with the highest scores in these participants. One cause for high incoherence was the presence of background conversation in the transcript, which is to be expected for an accurate incoherence measure - because the text does read as incoherent. This is instead a data collection quality problem, and a rare one at that. A similar problem that may not impact incoherence to the same extent but is much more common is the potential effect of inaudible rates on incoherence measurement. 

As mentioned, subject GF was especially likely to submit low quality audio at times: manual student review documented a number of instances of difficult to understand mumbling in GF journals, and TranscribeMe actually notified us at one point about the possibility that some GF transcripts may contain a higher number of inaudibles due to a loud background noise in some diaries that masked portions of their speech. When a large number of words are missing from a transcript, sometimes spread in an intermittent fashion, it becomes difficult for a manual review to assess \emph{semantic} incoherence, let alone for an algorithm to make an accurate judgement.

Note that at present the pipeline does not filter out the word "inaudible" before computing NLP metrics. There is no universally best way to handle the situation, as removing the word would introduce unmarked missingness in how sentences are evaluated, and removing entire sentences based on some inaudible threshold could substantially limit the dataset that is able to be considered. Different methodological decisions on data cleaning will be covered in greater detail in the discussion section \ref{sec:discussion2}, but for the bulk of this chapter I will focus on utilizing QC metrics alongside extracted features to characterize the full dataset in a generalizable but contextualized way. 

I will also include more proposed future directions in section \ref{sec:discussion2} that are related, one being a deeper investigation into the transcript inaudible rates of GFNVM over time as a promising measure of clinical value. This highlights another advantage of TranscribeMe over currently available automated transcription methods -- the ability of professional transcribers to not only handle audio with extremely poor enunciation, but to provide reliable notation of the parts that were practically impossible to understand. 

Words that are redacted for privacy reasons present an analogous problem to inaudible words, and markings of redaction are not currently being filtered out by the pipeline before feature processing either. Because there is a clear category of words that require redaction, it is easier for a manual reviewer to interpret "redacted" markings while evaluating incoherence, and it is plausible that an algorithm could be devised to reliably account for this within evaluation of incoherence. However in the scope of this pipeline, it remains a difficult decision between entirely removing the redacted words or keeping "redacted" markings in the transcript as is (and analogously for sentences with a redacted word). 

Redaction over time may be an inherently interesting feature for certain subjects, like 8RC89 here. Subject 8R spoke frequently about different girlfriends during periods of the study, and it is probably the case that the word "redacted" would result in higher incoherence and word uncommonness metrics than if a typical name were substituted. Still, less common real names would have the potential to introduce a confound even if the transcriptions were unredacted. As part of the case reports in section \ref{subsec:diary-case-study}, I will provide a pilot overview of 8R's use of redactions and how that might interact with any clinical predictive power the word2vec features may have for their dataset. 

As was discussed with the prior literature on semantic incoherence, filtering out linguistic disfluencies in preprocessing is also a judgement call with conflicting opinions in the field. This again is a decision that ought to be made based on the dataset and scientific questions at hand, but for the general pipeline release the verbatim transcriptions are input "as is" for all feature extraction. Keep in mind that this may decrease incoherence summary statistics in highly disfluent transcripts, though it can depend greatly on the disfluency category and how the disfluencies are distributed over different sentences of different lengths in the transcript. 

I will therefore take care to characterize relationships between features as a component of section \ref{sec:science2}. Sentence splitting decisions more broadly have the potential to bias incoherence metrics as well, but (as is a common theme) the human-labeled sentence breaks very well may have their own inherent clinical relevance, potentially explainable in part by acoustic factors such as speech rate and pause usage patterns being implicitly encoded in them. \\

\noindent Note I ultimately chose to exclude the mean sentence to sentence incoherence from the carefully selected subset of final modeling features because it was so highly correlated with words per sentence, while the maximum within sentence incoherence was found to provide more value independent of other studied features (see section \ref{sec:science2}). This major difference between the two incoherence summary statistics was also apparent in the examples of Figures \ref{fig:incoherent-diaries-gf} and \ref{fig:incoherent-diaries-8r}. The most incoherent journals when selected by both manual review and between sentence incoherence tended to not only contain shorter sentences, but also fewer total sentences. 

Because the vector for a given sentence is the mean of the word vectors contained in the sentence, it is not surprising that a greater number of words per sentence would have a smoothing effect on the sentence vectors leading to lower between sentence incoherence. It is also not surprising that fewer sentences would be present in the most extreme examples, because fewer sentences introduces more variance in the mean over sequential sentence pairings. It was surprising however the extent to which this relationship held in the larger dataset. Recall also the high level of variance in words per sentence found in the transcripts of participant GFNVM in particular (Figure \ref{fig:diary-words-per-sentence}).

On the other hand, the most incoherent diaries per the maximum within sentence incoherence tended to be very long, which is again expected due to the fact that more sentences present more opportunities for one with a high mean pairwise word incoherence score to occur. Qualitatively, a number of the most incoherent examples discovered by within sentence incoherence contained extreme or bizarre statements, especially for GF. For example:
\begin{quote}
    And I still-- thoughts won't leave my head.
    
    They keep attacking me and telling me just to run and just to kill-- like, I don't wanna die.
\end{quote}
\noindent Though most of these transcripts had some consistent central theme, they did bring to attention interesting signs of instability - signs that might not be well captured by even a theoretically optimal notion of between sentence incoherence. There were also potentially interesting occurrences of colorfully descriptive sentences and sentences containing less common words (e.g. "overture") in the 8R diaries that were selected based on within sentence incoherence. For example:
\begin{quote}
Got a real Nurse Ratched type up there at the psych ward.

...

Then the cats wake me up digging their razor sharp little claws into me.
\end{quote}
\noindent While these statements may not be directly clinical, it is certainly plausible that variation in such a speaking pattern over time might correlate with other features of clinical interest in a given participant. For expanded content from qualitative review of within sentence incoherence, see supplemental section \ref{subsubsec:incoh-diary-trans}.

Meanwhile, the examples discovered by the mean sentence to sentence incoherence (Figures \ref{fig:incoherent-diaries-gf} and \ref{fig:incoherent-diaries-8r}) demonstrated signs of possible incoherent thought too, and it was impressive how well they aligned with the journals selected by manual review. It was common to see rapid jumping between two (or sometimes more) topics in these diaries, or oddly specific details provided for no clear reason. Note that a lack of focus without clear transition logic can be closely linked with conceptual disorganization symptoms of psychosis, a key theme in chapter \ref{ch:2}. Although it is difficult to distinguish between unexpressed transitory thoughts and erratic internal transitions, pause time information could provide an interesting heuristic to couple with measures of topic coherence in the self-guided audio diary. 

Participant 8R also had a more general habit of ending recordings with a vague summary statement, which in some of the examples here did not match the tone of the rest of the diary -- for example complaining the whole transcript but then ending with a statement that things are good. This also underscores the benefit of qualitative review as one guiding force in determining interesting data science questions to ask in a context dependent way. \\

\paragraph{Summary of limitations with incoherence method.}
Despite some promise in the manual review results for semantic incoherence, it was not without its limitations, and it is difficult to say with much certainty yet the extent to which we can actually rely on signal contained in the incoherence metrics. Not all of the example transcripts by top incoherence score were incoherent in any expected sense, and the confounding impact of verbosity measures was high for both summary measurements. For the manual review, I had requested the most incoherent transcripts to be chosen, but that is not only highly subjective but also ill defined -- much moreso than sentiment even. True word salad may be easy to spot manually, but the patient population for BLS largely experienced only moderate psychotic symptoms at worst, if any at all. Subtler manifestations of incoherence could be defined in multiple distinct ways, and if it were more specific there is still no guarantee that there would be good agreement amongst subjective reviewers on what ought to qualify as incoherent. There is not a good notion of "least incoherent" to balance the review as there was for sentiment either, and ultimately the validation process was framed in a way that required the manual reviewer to pick 5 transcripts regardless of how incoherent they actually thought those transcripts were. 

It could be fruitful to repeat the process here with more careful instructions and a larger number of manual reviewers in order to compare notes. Ideally this would involve a trained clinical reviewer as well, but given the size of the transcript set covered in the present review that goal is somewhat unrealistic. Additionally, regardless of prior training, it is likely that reviewers will skim the transcripts in choosing a narrower set, so perhaps a more manageable subset of the journals should be chosen and reviewers should instead be asked to rank each into categorical tiers in the future. 

Another possible qualitative direction that could be interesting, most notably in a case report context, would be to look at incoherence over time in a subject at a broader timescale, and then review all the diaries in those periods that had an abnormally high e.g. monthly summary statistic. The potential of such an approach is indeed already apparent from the review of most incoherent GF journals here -- not only were there 3 extreme incoherence diaries submitted on sequential days and 4 submitted within a week's time, but these diaries spanned both metrics checked for incoherence. The GF submission on study day 114 had the second highest between sentence incoherence, while submissions on study days 115 and 116 had the fifth and second highest within sentence incoherence scores respectively. The GF submission on study day 122 then had the fourth highest between sentence incoherence. This suggests that different notions of incoherence may agree to a greater extent when compared at a different timescale, and in fact may be agreeing on a period of high clinical relevance for subject GF, despite largely producing widely different rankings of individual transcripts. \\

\noindent In sum, there is a great deal of prior literature justifying our vector embeddings, and strong reason to believe the measures of incoherence derived from them will have relevance to transcribed speech from patients with psychotic disorders. Yet there are open questions regarding application in the audio journal setting and how to fine tune the data cleaning and summarizing steps that occur before and after (respectively) the vector representations are computed. 

This work takes an initial step towards addressing those questions, by providing pilot manual validation results that support our methods, as well as reporting on our observations about the outputs when computed from a large dataset of daily audio diaries recorded by Bipolar patients. The justification is by no means rock solid, and further work is required before the semantic coherence features returned by my code can be considered anything but "preliminary". As such we plan to continue to use them when processing diary data from relevant lab studies, with the pros and cons outlined here kept in mind while planning analyses. \\

\FloatBarrier

\paragraph{Tools for content detection and metadata analysis.}
Besides the sentiment, disfluency, and incoherence NLP features described, the transcript side of the pipeline includes utilities for characterizing the structure and content of the diary text. Many of the values returned by the transcript QC functionality that was already discussed above (section \ref{subsubsec:diary-val-qc}) can additionally be used for assessing the verbosity of diaries and for normalizing the mentioned NLP outputs. Further, the diary wordclouds (Figure \ref{fig:diary-clouds-init}) and other visualizations previewed in section \ref{subsec:diary-outputs} can assist in detecting salient diary content as well as changes in submission patterns.

Beyond those capabilities of the code, there are a couple other transcript functionalities that remain to be detailed; they were similarly implemented for better understanding the structure and content of the language used in a journal. The first is syllable count computed from transcript text, which can be combined with provided sentence-level timestamps to estimate speech rate, a property well-documented to relate to psychomotor signs of depressive and manic symptom severity (discussed within section \ref{subsec:diary-lit-rev}).  
When I was initially designing the code, I tested both the current pipeline's syllable counting methodology based on TranscribeMe transcripts, and the syllable counts returned as part of the Praat speech rate script used by \citep{Wortwein2017}, among others in the field. I manually counted syllables in a handful of the diaries and found the text-based method to be substantially more accurate, which is why I proceeded with that. 

Of course, a major limitation of the transcript-based approach is that when speech rate is estimated, it uses TranscribeMe's sentence level transcripts that do not separate out longer pause periods. As such, rate of active speech is conflated with likelihood of taking long pauses in the current code's speech rate estimate methodology, and thus in a participant like 3SS93 mean speech rate is around 2 syllables per second, unexpectedly low for human speech. Nevertheless, mean speech rate across the dataset is consistent with typical human speaking rates. Moreover, by utilizing the acoustic pause detection features, speech rate for a participant with many long pauses can be largely corrected using existing pipeline outputs. Given the clear clinical relevance of speech rate, this feature will of course be investigated in more detail as part of the research results presented in section \ref{sec:science2}. \\

\noindent The other additional feature included is a keyword counting functionality. A built in setting in the code allows for counts of specific words or specific word roots to be included with the default extracted features. Multiple such counts can be processed as individual features, and at the same time multiple different words or word roots can be counted under the same single feature. Correctness was tested directly during implementation. For documentation on how to use this setting, please refer to section \ref{subsec:diary-code}.

In chapter \ref{ch:3}, this keyword setting was utilized both to track mentions of specific medications individually, and to maintain a combined count of uses of various terminology that related to the deep brain stimulation (DBS) treatment the patient was receiving. It turned out that mentions of Adderall actually held relevance to the patient's mental state and to interpretation of the outcome of the experimental DBS paradigm, providing a strong proof-of-concept for the use of simple tools like this to enable a personalized approach. Within the individualized BLS diary analyses of section \ref{sec:science2} below, additional evidence for the power of simple keyword counts as a component of diary analyses will be clear.

Not only do the keyword counts provide quantitative metrics of potential relevance, they can also help to guide more qualitative analyses. Identification of transcripts where a rarer word of interest is used could allow human review to focus on specific salient journals that might otherwise be missed, without the high overhead burden of looking through a large set of transcripts. When combined with the visual tools produced by the code, there is already great potential for technology-assisted qualitative work, and the possibility of implementing more complex tools for these purposes contains even greater promise in the long term. 

\FloatBarrier

\subsection{Code architecture overview}
\label{subsec:diary-code-intro}
A major purpose of this chapter is to provide context for a beta release of my audio diary code; in particular, an aim is to enable others to run similar studies from data collection steps down to final analysis. There is good reason to believe most of the pipeline features are technically sound and either supported by prior literature in psychiatry or useful for data collection/quality monitoring (if not both). These arguments are provided throughout the previous section \ref{subsec:diary-val}, and summarized under the "Final QC workflow" header at the end of subsection \ref{subsubsec:diary-val-qc}. Additional supporting evidence and reference material is provided by way of the pilot scientific results presented in section \ref{sec:science2}, which employ this pipeline on the dataset described above in section \ref{subsec:diary-methods}. 

\begin{figure}[h]
\centering
\includegraphics[width=\textwidth,keepaspectratio]{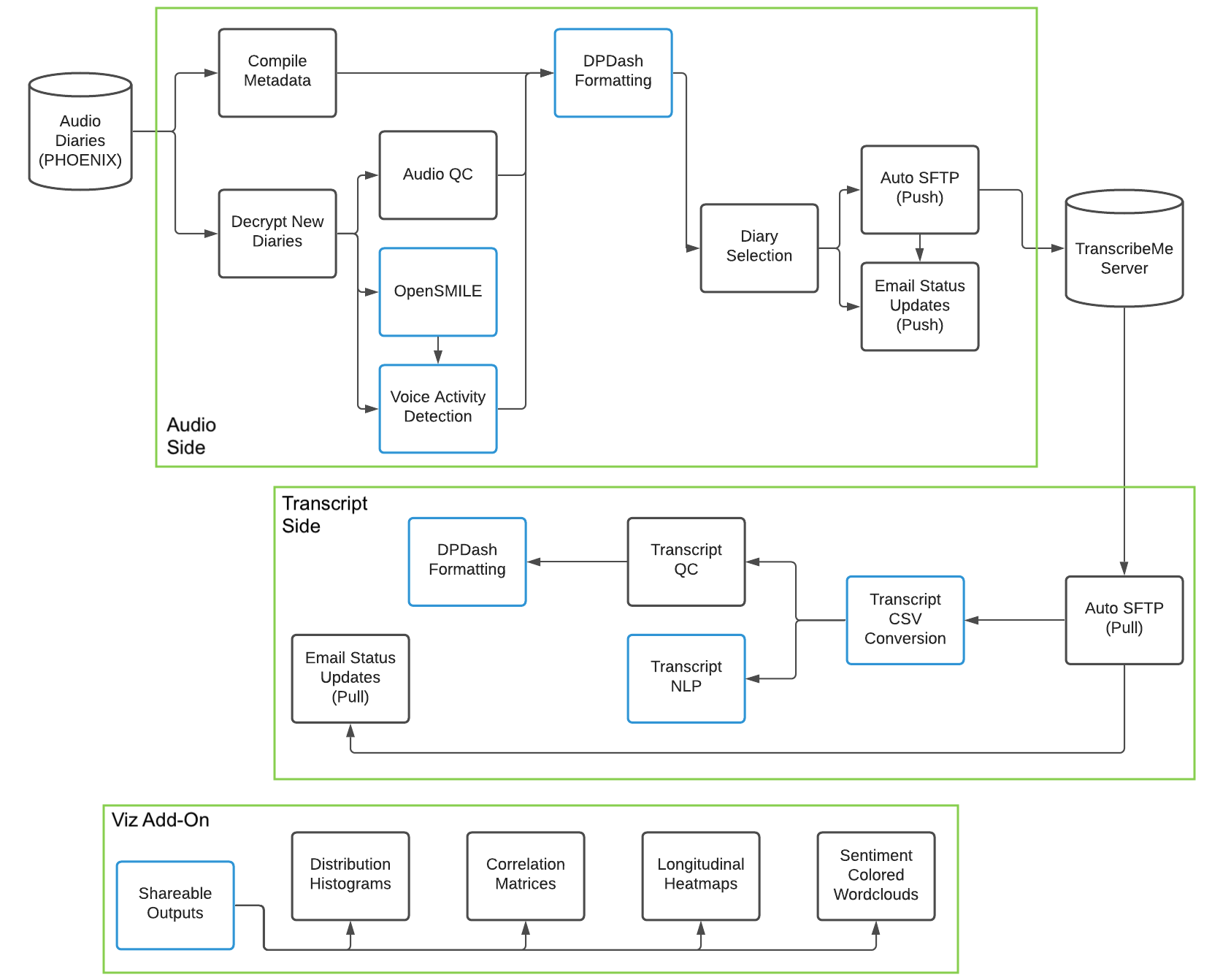}
\caption[Architecture diagram for the audio diary processing pipeline.]{\textbf{Architecture diagram for the audio diary processing pipeline.} The pipeline is broken into three major components: audio processing, transcript processing, and feature summary/visualization. Each of these components, marked by the outer green boxes, can be run in full using a provided wrapping bash script. Within each component, a number of independent functions are run. Each such function is represented by a text box here, with arrows denoting which functions rely on input from which other functions. A light blue border indicates that the function produces not only information needed by a downstream function, but also some deidentified shareable output(s) that might itself be of interest.}
\label{fig:diary-arch}
\end{figure}

It is of course also critical to document the code itself for multiple target audiences, if it is to be built on functionally or scientifically. The pipeline was written in a modular fashion to enable such future adaptations (Figure \ref{fig:diary-arch}), and supplemental section \ref{subsec:diary-code} gives extensive documentation on the architecture and its use, as well as tips on specific pieces that would most likely require adaptation for certain research projects. Although the pipeline's main purpose to date has been for ongoing use with audio diaries across Baker Lab studies, portions have already been employed to assist close collaborators in processing their audio diary data, and this sort of work should only be easier going forward in light of the now fully compiled documentation.

Within supplemental section \ref{subsec:diary-code}, I will focus on the technical details of the pipeline code, without reference to any actual outputs or broader motivating factors. Specifically, I will provide instructions for installation and use, likely troubleshooting suggestions, exact implementation details, and other relevant information that one would find in the README of a well-documented software project. For those that may run parts of this code themselves in the future, I hope it is a valuable resource. 

 \noindent Note that the current code base can be found in full at \citep{diarygit}.

\FloatBarrier

\section{Scientific application of audio diary pipeline outputs}
\label{sec:science2}
Given all the reasons for optimism about the impact that the audio journal format will have on psychiatric science, as well as the technical justifications provided for the code being released here, it is a natural step to apply this code towards scientific investigations. The long term goal of the pipeline is indeed to facilitate the answering of a number of open questions in psychiatry, like some of the questions posed within section \ref{sec:background2}. 

Thus the goal of this section is to demonstrate the use of my code in a pilot scientific project, as a proof of concept for some of its relevant capabilities and a guiding example for how downstream analyses may best utilize those capabilities. The presented results also aim to better contextualize various pipeline features, so future users can compare and contrast their study results with the results obtained here. In order for the research community to build a cohesive model of diary behavior, various acoustic and linguistic properties, and so on, it will be critical to integrate prior work in a more holistic fashion than simply citing packaged statements in a separate background section.

Furthermore, this scientific work is a first of its kind exploration into daily audio journal patterns in Bipolar Disorder (BD), and therefore the results are of independent interest for future research into modeling longitudinal fluctuations of mood or psychotic episodes -- irrespective of their great relevance for understanding future work with audio journals. The results may in particular help to inform other studies of BD that employ self report surveys or NLP on recorded clinical interviews, two tools that have been used in prior BD literature, unlike audio diaries. 

\subsection{BLS dataset recap}
The Bipolar Longitudinal Study (BLS) dataset described in section \ref{subsec:diary-methods} remains the focus in this section. The pipeline outputs reviewed within section \ref{sec:tool2} that resulted from the application of my code to BLS audio diaries will serve as the primary analysis inputs, but clinical information and daily ecological momentary assessment (EMA) self report surveys will be used in conjunction as measures of symptom severity.

Although BLS primarily enrolls participants diagnosed with BD, recall that patients with any severe psychotic or affective disorder diagnosis may be eligible. As we are more interested in particular mood and psychosis related symptoms than in disease labels (see introductory chapter), interpretation of the reported results is centered more so around relevant clinical scale and self report survey scores than diagnosis. Indeed, all BLS patients receive regular ($\sim$ monthly) clinical ratings from the same set of scales regardless of diagnosis. These scales include the MADRS and PANSS, which are gold-standard measurements for depressive and psychotic symptoms, respectively.  

However, in order to fully contextualize the results it remains important to keep in mind throughout this section the diagnoses that are contained in the considered dataset. Section \ref{subsec:diary-ema} below includes within its methodological background the available diagnostic information for each of the top 24 BLS participants whom submitted both many quality audio journals and overlapping daily EMA. These participants make up the majority of the data used in all analyses. Broadly, the set includes 14 subjects with a primary Bipolar diagnosis, 5 subjects with a primary Schizophrenia spectrum diagnosis, and 1 subject with a primary Major Depressive Disorder (MDD) diagnosis. There are also multiple patients in the set that have a secondary diagnosis of Schizophrenia or MDD, as well as multiple patients with other comorbid diagnoses such as PTSD. 

Ultimately, the dataset encompasses a variety of psychiatric issues, but a majority of the participants - including those subjects that are focused on in closer detail below - were diagnosed with BD. Bipolar patients can be further categorized into three subtypes by the DSM-5 \citep{DSM}: type I, type II, and cyclothymia. Cyclothymia is typically less severe and less commonly diagnoses, but Bipolar I ($n=8$) and Bipolar II ($n=3$) are both represented in the BLS subset used for diary modeling. Note that Bipolar I is characterized by presence of manic episodes, but does not require depressive episodes (though they may occur), while Bipolar II is characterized by presence of major depressive episodes along with periods of hypomania, but no history of manic episodes. Hypomania includes some manic behaviors, but it must not have a significant negative impact on the patient's life nor include any psychotic symptoms, otherwise it would be classified as mania \citep{DSM}. \\

BD is thus a salient patient population for longitudinal study of audio diaries, because both manic and depressive symptom fluctuations may be capturable via extracted features or qualitative factors from the diaries that vary over time. My work here represents a concrete step towards answering these questions.

In the below subsections, I will begin with closer characterization of select audio journal features in BLS (\ref{subsec:diary-dists}), and discuss ways in which journal features alone could inform scientific questions, including the information value of:
\begin{itemize}
    \item Robust differences between feature distributions of different participants (\ref{subsubsec:diary-pt-dists-comps}).
    \item Correlation structure found amongst key diary features (\ref{subsubsec:diary-corrs}).
    \item Models for the temporal dynamics of diary submissions (\ref{subsubsec:diary-time}).
\end{itemize}
\noindent I will then report results on prediction of same-day EMA responses from audio journal features, both across the BLS dataset and also utilizing patient-specific models (\ref{subsec:diary-ema}). Finally, I will integrate information from the statistical analyses of those sections with additional clinical records and the results of a qualitative review, in order to present a more comprehensive case report centered around the diaries submitted by a particular subject of interest (\ref{subsec:diary-case-study}). 

To begin each subsection, I will review the diary features used for that analysis along with the prior scientific justification for their inclusion, and will then detail the downstream steps I performed to obtain my results using those diary features. Because the features themselves were obtained from the pipeline, any methodological details that were already included above will be omitted here. Of course the results and discussion of their implications will follow, and form the bulk of each subsection. 

\subsection{Acoustic and language feature distributions in psychotic and mood disorders}
\label{subsec:diary-dists}
A major goal of this chapter is to provide important reference materials for others that might use my code. A long term aim of releasing the pipeline is to codify baseline feature definitions, in order to facilitate consistency across studies and thereby allow more meaningful comparisons to be made between the results of different research groups or from different participant populations. Towards these ends, I provide a deeper characterization of the BLS journal feature set, beginning with distributional properties. An understanding of the feature distributions will inform expectations for future users, so that existing outputs can be used most appropriately and potential code modifications of interest can be proactively identified. Further, by publishing these distributions from BLS, we lay the groundwork for compilation of a larger distribution dataset that can be applied to improve interpretation of many individual studies in the future. The distributions serve to contextualize later study results both in a technical and a scientific sense.  

Of course the distributions presented here are not without caveats. The BLS dataset does not include healthy controls, so while the longitudinal nature of the data enables patients to serve as their own controls in many analyses, there is no way to ground the reported results in "normal" journal behavior. An important future direction will thus be to collect a control dataset of audio diaries to generate a healthy reference distribution. It would be most promising if a group independent of ours were to also collect a dataset from a patient population similar to BLS along with a control dataset, so resulting distributions could be compared and replicable properties could be identified. 

Nevertheless, the BLS diary dataset is large enough and spans enough participants (Figure \ref{fig:diary-pt-submit-chart}) to draw interesting conclusions from feature distributions, and ultimately to run statistical analyses of note. All of the quantitative analyses here focus on straightforward per diary summary features, as a rich set of questions can be asked using such features alone. Another potential future direction that could be enabled by the pipeline would be to investigate temporally finer features within a diary, which could also result in new diary-level summary statistics designed to accurately capture specific phenomena of interest. Future work along these lines could involve the OpenSMILE acoustic features as well, because denser timescale analysis is especially salient for these OpenSMILE outputs.

Based on the quality control and validation results of section \ref{subsec:diary-val}, I chose to focus all of the scientific analyses in this section only on diaries that were both $> 15$ seconds in recording duration and deemed suitable for transcription, producing an initial dataset of $8557$ diaries. To further improve the quality of the considered diaries, I also filtered out any recordings that had no pauses detected, as this was associated with greater background noise or otherwise poorer recording quality in longer submissions. \\

\noindent That resulted in a final count of: 

\begin{quote}
    n=8398 journals included for scientific investigation
\end{quote}

\noindent Similarly, to retain a strong level of statistical power and ensure quality of input features, I narrowed down the per diary measures considered to the 12 listed below. These features were chosen based on both the level of confidence from the validation work reported in section \ref{subsec:diary-val} and the strength of prior evidence for clinical relevance (also discussed within \ref{subsec:diary-val}). All of the features were either directly returned by the pipeline or where noted were obtained by using one pipeline feature to normalize another. \\

\noindent The included journal features were:
\begin{itemize}
    \item Total number of words
    \item Fraction of recording duration that was spent speaking (from speech minutes/total minutes)
    \item Mean pause duration in minutes (from pause minutes/pause count)
    \item Mean sentence speech rate in syllables per second
    \item Number of restarts per word (from restart count/word count)
    \item Number of repeats per word (from repeat count/word count)
    \item Number of nonverbal edits per word (from nonverbal edit count/word count)
    \item Number of verbal edits per word (from verbal edit count/word count)
    \item Mean sentence sentiment
    \item Minimum sentence sentiment
    \item Maximum sentence pairwise incoherence
    \item Mean word uncommonness  
\end{itemize}
\noindent These features represent metrics of verbosity and rate of speech production, of linguistic disfluencies, and of semantic sentiment and coherence, which are all properties with a history of psychiatric clinical relevance and a tractability for automatic measurement and summarization. Note that over the course of exploration in the upcoming subsections, a handful of additional features closely related to these will be introduced. \\  

\noindent \textbf{On normalization.} All features with the exception of total number of words are normalized in some capacity by the length of the diary, as they each represent summary statistics over words, sentences, or minutes. However, as will be discussed at more length with regard to disfluencies in chapter \ref{ch:2}, performing a straightforward normalization technique does not mean the resulting feature will be truly independent of diary length, and the factors contributing to that underlying relationship may or may not have clinical relevance -- in some cases, e.g. restart occurrences, there are natural properties of speech production that would drive a relationship between word count and the normalized feature regardless of clinical status. 

It is also important to be mindful of exactly what each feature is normalized by, as it is possible for two diaries with similar total duration to have quite different speech duration, similar speech duration to have quite different word counts, or similar word counts to have quite different sentence counts. In all these cases, there are relationships between various features we would expect to see to account for the discrepancies; for example, if recording duration and word count are out of balance in two different journals, the pause duration, speech rate, or word length/complexity ought to be significantly different between the two. As such, by carefully studying features and their relationships, real versus artifactual differences can be identified with more confidence and observed clinical relevance of particular features can be more accurately dissected into independent components. 

For these reasons, as well as the fact that the features taking the minimum and maximum over sentences can be strongly impacted by length at the journal durations we are considering here, I include detailed analysis of the effects of diary length on the highlighted features as part of the work presented in this section. I report correlations of each feature with word count (subsection \ref{subsubsec:diary-corrs}), and run simulations to evaluate diary-level summary feature distributions when they are sampled from identical sentence-level feature distributions in diaries of differing length (within subsection \ref{subsubsec:diary-pt-dists-comps}). 

\subsubsection{Study-wide distributions}
\label{subsubsec:diary-study-wide-hists}
The distributions of each of the 12 features of focus across the final BLS journal dataset not only provide important reference for future audio journal studies, they can also help to contextualize the scientific analyses that follow. Next, I will compare these features across a few different participants of interest (\ref{subsubsec:diary-pt-dists-comps}), to better characterize between subject heterogeneity in journal behavior and connect this back to expectations from their diagnostic information. I will then analyze the correlation structure amongst these features (\ref{subsubsec:diary-corrs}), and synthesize the results so far to form even more specific hypotheses for the EMA prediction models and example case report investigations to follow. Because study-wide distributions will be included again (in less detailed form) in those subject-dependent comparisons, the BLS histograms along with a more thorough characterization of the properties of each feature distribution are found in supplemental section \ref{subsec:plain-hists}. 

\subsubsection{Between patient heterogeneity of diary features}
\label{subsubsec:diary-pt-dists-comps}

Three subject IDs were identified for potential deeper study based on data availability and observations of study staff: 3SS93, 8RC89, and 5BT65. The focus on these subjects will continue into the pilot case report analyses of section \ref{subsec:diary-case-study} below. This was done in part to preserve much of the data for future analyses whilst still enabling a rich diary set to be explored for pilot purposes here, by choosing a few participants known to really engage with the diaries. 

Indeed, all three of the selected patients participated in the study regularly for at least two years. They all have a primary diagnosis of BD; 3S and 8R with Bipolar type 1, and 5B with Bipolar type 2. I isolated the subsets of the final dataset specific to each of the three patients, which resulted in 981 time points for 3SS93, 510 time points for 8RC89, and 344 time points for 5BT65.

3SS93 was notorious for regularly submitting recordings of the maximum 4 minute duration, and also contributed the highest number of transcripts in the dataset, so their inclusion was an obvious choice for richness of content. A smaller case report focused on diary content and sentiment from the first 6 months of 3S's study participation was already reported in section \ref{subsubsec:diary-val-trans} as part of the justification for the sentiment feature. That pilot work demonstrated success in predicting EMA responses from sentiment.

8RC89 was a top 5 contributor of audio diary submissions as well, and demonstrated good variety in recording length, with $\sim \frac{3}{4}$ of their recordings exceeding 45 seconds. Because 8RC89 had documented hospitalizations for mania during their time in BLS, and was observed by other lab members to have altered geolocation activity around the times of manic episodes, I hypothesized that temporal patterns in journal features might also align with times of clinical interest for this participant.

5BT65 was a top 10 contributor of audio diary submissions. Their journal submission lengths were often relatively short -- in fact the 15 second minimum for feature quality filtered out over 100 data points here, to be revisited in subsequent sections. Note that it was reported by study staff monitoring recording lengths in DPDash that changes in diary duration for this participant tended to align with episodes of depressed mood. Thus there is already reason to believe that journal behavior in this participant has clinical relevance, though of course this needs to be analyzed more rigorously. Additionally, severe depression is the most debilitating issue for 5B, serving as good contrast with the more severe manic tendencies of 8R. \\

\noindent Ultimately, the vast majority of comparisons made were significantly different, many with large effect size, thereby indicating great between subject heterogeneity in journal properties. Note that specific methods details for all of this distributional comparison work can be found in supplemental section \ref{subsec:dist-comp-methods}, and full KS test results are reproduced in supplemental section \ref{subsec:ks-sup}.

There was a uniquely stark difference between the diary lengths across the three considered subjects, particularly when comparing 3SS93 (much longer) and 5BT65 (much shorter) to the overall distribution. This is inherently interesting, but also could be driving other observed differences. Thus I will more carefully walk through the various feature comparisons, connecting differences across features as well as back to what is known more broadly about each of the subjects. I will then more directly simulate the effects of different diary lengths on select features, to further assist in interpreting these results. 

Through this work, I will have established features with clear journal behavior differences amongst the subjects of interest, and in turn forming more concrete hypotheses for which specific features will be of most use in studying the clinical outcomes of each of the subjects. Based only on these analyses, it is not possible to definitively distinguish clinically relevant differences from other sources of behavioral heterogeneity between people; yet when coupled with existing psychiatry literature along with the basic clinical profile of each subject, these results set up nicely for the subsequent EMA prediction and case report projects.  \\

\paragraph{Core speech production features.}
Verbosity - both high and low - has been well documented to correlate with symptom ratings and other clinical outcome measures across a number of disorders, as was discussed in section \ref{sec:background2}. While there can be clinically irrelevant differences in journal verbosity between participants too, it is important regardless to establish distributional comparisons in order to understand how verbosity varies within each subject of interest. Here, the diary word count feature serves as a measure of verbosity. As mentioned, this feature is critical to understanding changes in other diary features as well. 

\begin{figure}[h]
\centering
\includegraphics[width=\textwidth,keepaspectratio]{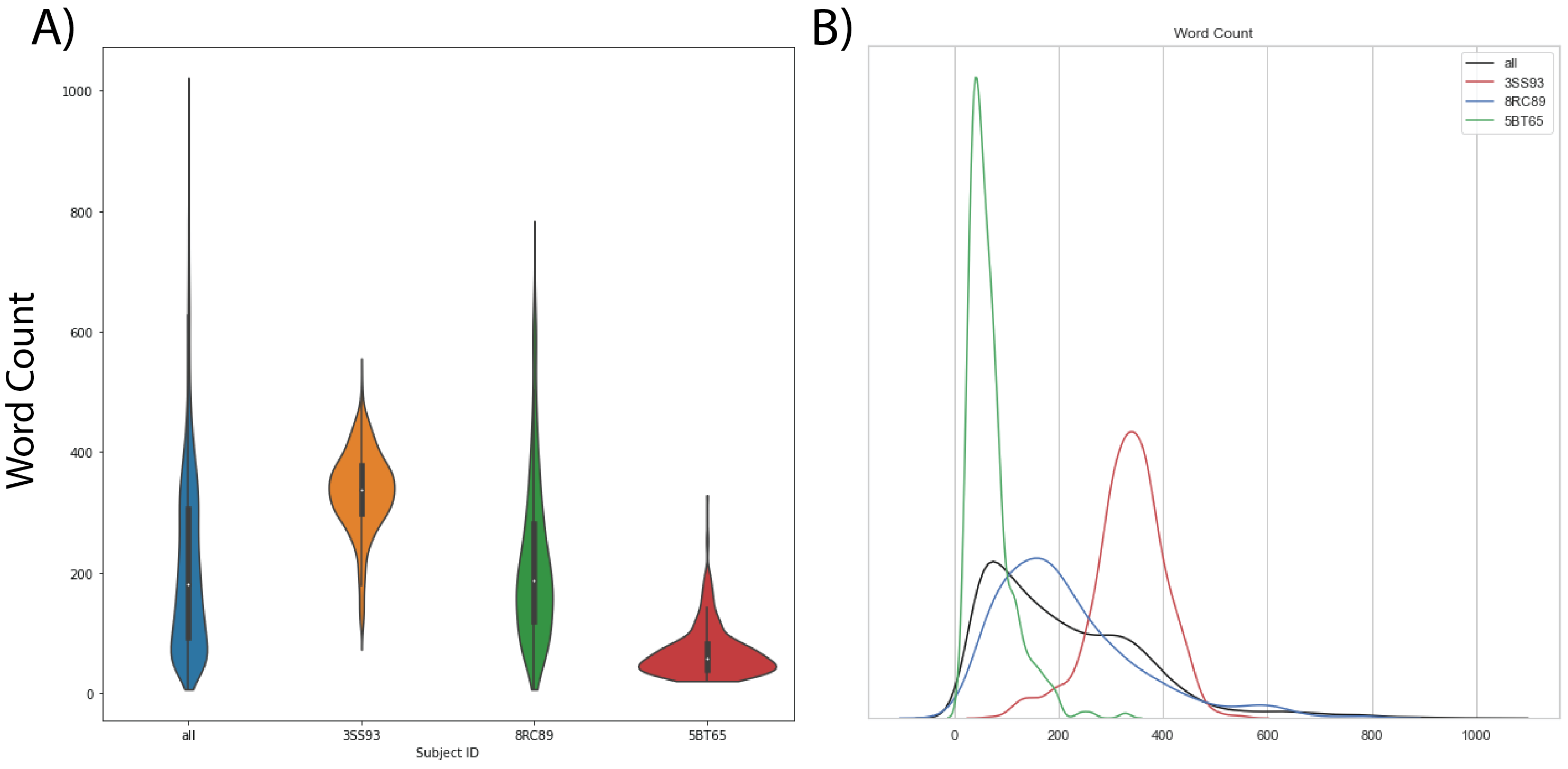}
\caption[Comparison of word count distribution between different patients of interest.]{\textbf{Comparison of word count distribution between different patients of interest.} Violin plots were created to compare the distribution of diary word counts (A) across the following quality-filtered BLS datasets, from left to right: study-wide, the journals of subject 3SS93, the journals of subject 8RC89, and the journals of subject 5BT65. Note the violins were restricted to cover only observable data using the cut=0 argument. For all violins in this section, the BLS-wide distribution contains the same range of values as the corresponding histograms in section \ref{subsubsec:diary-study-wide-hists}. To provide a different perspective on the same comparison, the distributions depicted in (A) were also input to Seaborn's kernel density estimate plotting function (B). The smoothed distributions for BLS (black), 3SS93 (red), 8RC89 (blue), and 5BT65 (green) are thus overlaid in (B).}
\label{fig:word-count-per-pt}
\end{figure}

Word count was in fact one of the journal features with the strongest subject-specific differences in the final BLS dataset (Figure \ref{fig:word-count-per-pt}). The study-wide distribution demonstrated high overall variation and was somewhat bimodal, with mean just under 200. The distribution for participant 3SS93 was concentrated strongly on the high end, with mean over 300 and only a small proportion of journals under 200 words; on the other hand, the distribution for participant 5BT65 was concentrated strongly on the low end, with mean below 100 and exceedingly few diaries above 200 words. Subjects like 3S and 5B drove a great deal of the variance seen in the study-wide distribution. However, subjects with larger verbosity variation over time can also be found in BLS. For example, participant 8RC89 had word count distribution much more closely matching that of the entirety of BLS (Figure \ref{fig:word-count-per-pt}). 8R similarly had a mean diary word count just under 200 with a broad range, though their distribution peaked around 200 rather than demonstrating hints of a bimodal shape. 

It is worth reiterating that 5B had 489 diaries transcribed (Figure \ref{fig:diary-pt-submit-chart}B) but just 344 time points included in this final dataset. Meanwhile, 3S and 8R only went from 985 to 981 and 519 to 510, respectively. This is because 5BT65 submitted a number of recordings with duration less than 15 seconds, which were filtered out above. Thus the 5B word count distribution in this analysis (Figure \ref{fig:word-count-per-pt}) is actually inflated. It is difficult to interpret many of the other journal features in such short submissions, so they remain excluded here. In later analyses, such as the EMA prediction models of section \ref{subsec:diary-ema}, the presence of short duration diaries will be treated as a feature itself for subject 5B. \\

\begin{figure}[h]
\centering
\includegraphics[width=\textwidth,keepaspectratio]{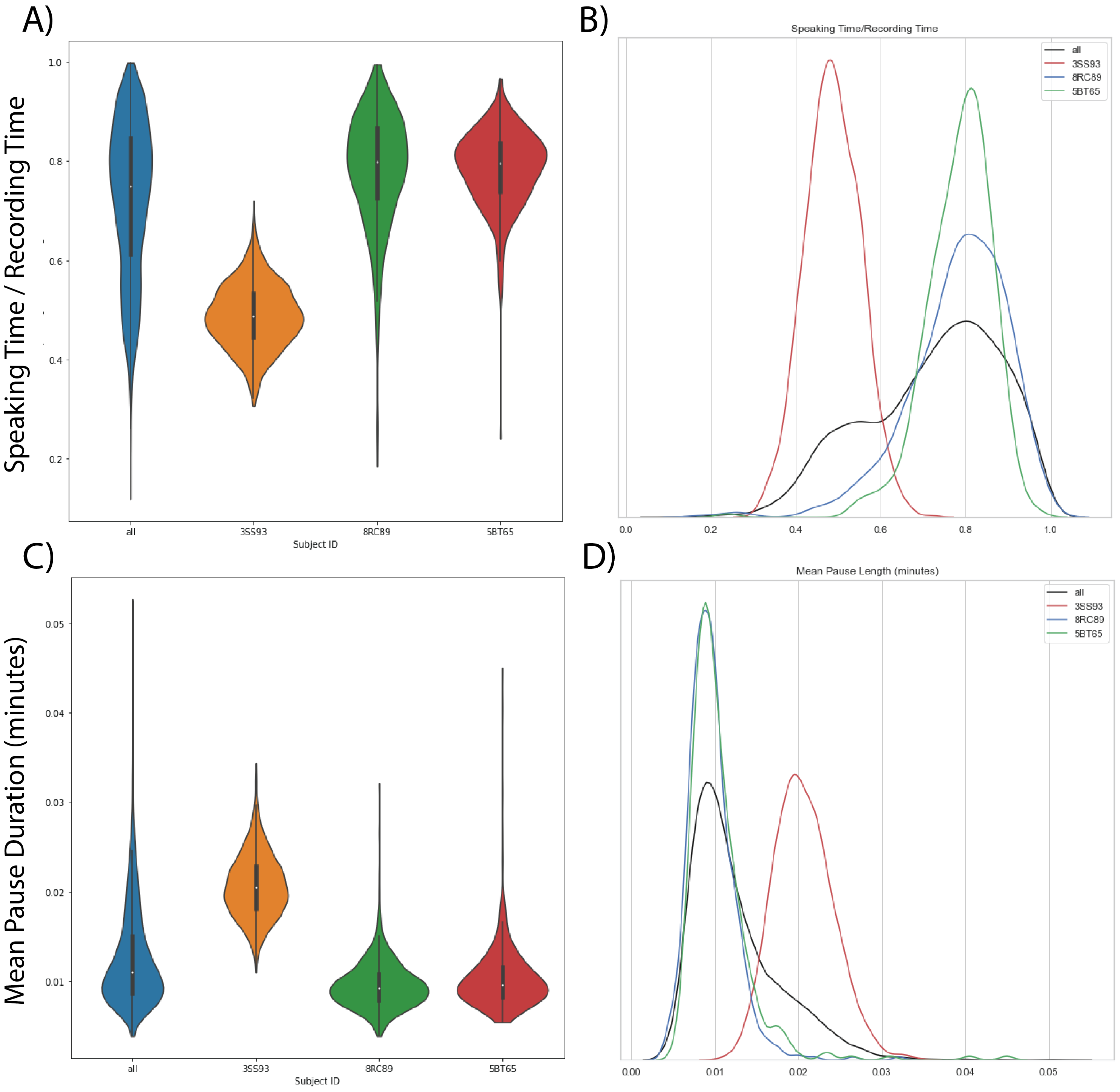}
\caption[Comparison of pause-related distributions between different patients of interest.]{\textbf{Comparison of pause-related distributions between different patients of interest.} Using the same methodology as in Figure \ref{fig:word-count-per-pt}, additional subject-specific feature distributions are compared against each other and against the distribution across the final BLS dataset, with both violin plots (left) and smoothed distribution curves (right). Here, we focus on the pause-related features: fraction of recording time spent speaking (A/B) and mean pause length (C/D).}
\label{fig:pause-props-per-pt}
\end{figure}

In addition to the established relevance of verbosity, other basic speech production features have strong ties to psychiatric symptomatology -- for example the link between pause usage and cognitive processing, or speech rate as a psychomotor sign. Interestingly, there are also stark differences between BLS patients in the distributions of other speech production related summary features across diaries. Most notably, subject 3SS93 spent substantially more time paused during journal recordings than other BLS participants, with the vast majority of their diaries including $60\%$ speech time or less, while the study-wide mean is closer to $75\%$ speech time (Figure \ref{fig:pause-props-per-pt}A/B). This is also reflected in mean pause time per diary, where 3S took average pauses exceeding 1 second in the large majority of their diaries, yet this was relatively rare across the study-wide dataset \ref{fig:pause-props-per-pt}C/D). 

There is no prior reason to suspect that these strong differences could be caused by the tendency of 3S to submit more diary content, and the observed distributions of 8R and 5B do not suggest any such trend (Figure \ref{fig:pause-props-per-pt}). Furthermore, as will be reported in section \ref{subsubsec:diary-corrs}, there was no correlation observed between word count and mean pause length over the final BLS journal set. Thus there is likely a real disparity in 3SS93's pause behavior that will warrant closer study for potential clinical relevance. 

Because participant 3SS93 contributed so many diaries with high pause duration, they had an outsized impact on the tail of the study-wide pause distributions (Figure \ref{fig:pause-props-per-pt}), which likely somewhat inflated the distribution distance metrics for the pause features of 8RC89 and 5BT65 computed by the KS test. It is unclear then whether there were any meaningful participant-specific effects on pause behavior for 8R or 5B. However, their distributions still showed promise for detection of potential within-subject variation of interest. 8R had a wider distribution of speech time percentage than the other two subjects (Figure \ref{fig:pause-props-per-pt}A/B), and 5B had a number of large outliers in mean pause duration (Figure \ref{fig:pause-props-per-pt}C/D). Of course with shorter file durations there would likely be fewer pause instances and thus more variance is expected in the mean pause length feature for 5B. Nonetheless, it would be worth checking if the extreme pause duration journals of 5BT65 reveal any patterns in the patient's life. 

\begin{figure}[h]
\centering
\includegraphics[width=0.7\textwidth,keepaspectratio]{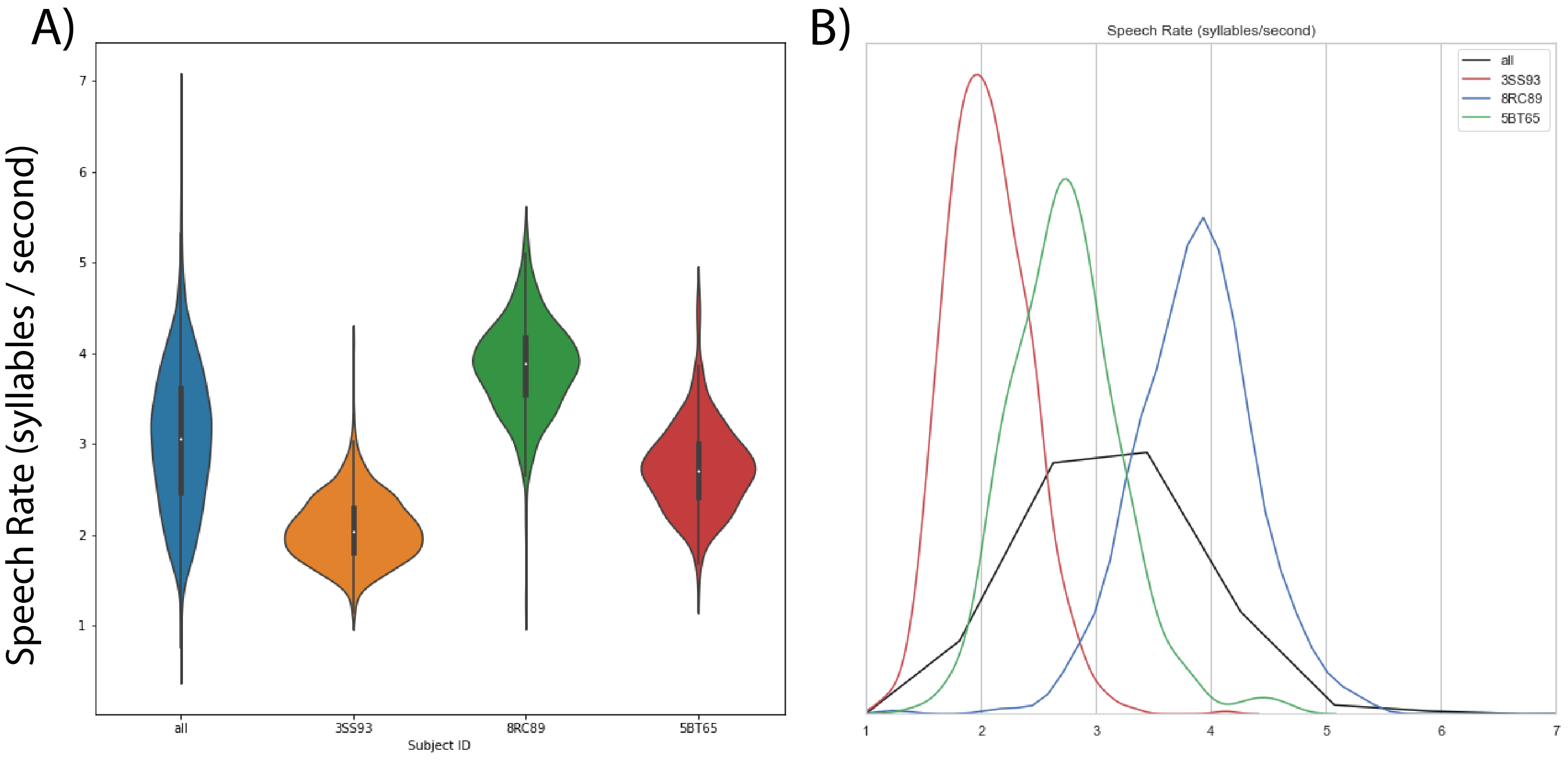}
\caption[Comparison of speech rate distribution between different patients of interest.]{\textbf{Comparison of speech rate distribution between different patients of interest.} Using the same methodology as in Figure \ref{fig:word-count-per-pt}, additional subject-specific feature distributions are compared against each other and against the distribution across the final BLS dataset, with both violin plots (A) and smoothed distribution curves (B). Here, we focus on the mean sentence speech rate, estimated in syllables/second using transcript timestamps (uncorrected for longer pauses).}
\label{fig:speech-rate-per-pt}
\end{figure}

Unsurprisingly, the speech rate distributions reflected the abnormal pause behavior of participant 3SS93 as well, with abnormally low transcript-derived speech rates observed in their dataset (Figure \ref{fig:speech-rate-per-pt}). This is expected because the speech rate metric output by the code incorporates longer pauses into its denominator, thereby representing a combination of physical speech rate and conversational pause tendencies. Interestingly, a new stark disparity also emerged in the speech rate comparison that was not attributable to pause behavior differences -- participant 8RC89 had elevated speech rate relative to the overall study distribution, with mean approaching 4 syllables/second versus the study mean of $\sim 3$ (Figure \ref{fig:speech-rate-per-pt}). Given that 8RC89 was reported to experience severe mania at times during their study involvement, an elevated speech rate distribution is consistent with signs one might expect in this patient.

It would be ideal to evaluate speech rate without the inclusion of longer pause periods, to assess motor properties of speech production more directly, while maintaining the measurement of pause behavior independently through other features. I therefore next discuss the distribution of an adjusted speech rate feature, normalized using pause data. 

\FloatBarrier

\paragraph{Adjusting the speech rate metric using pause features.}
In reality, human-perceived speech rate is likely affected both by rate of speech production and by the presence (or absence) of shorter pauses, many of which would be labeled pauses by the detection algorithm here. Still, longer pauses, especially ones that occur between complete thoughts, are plainly distinct from speech rate. Because many diaries do contain longer pause periods, it is important to have a speech rate metric that can account for that. To characterize what a pause-adjusted speech rate feature might look like, I divided the estimated speech rate (derived from the transcription syllable counts and timestamp information) by the "fraction of recording duration spent speaking" feature for each diary in the final BLS dataset. Thus a new feature was obtained that would scale the estimated speech rate to account for only the time a participant spent speaking. 

\begin{figure}[h]
\centering
\includegraphics[width=0.9\textwidth,keepaspectratio]{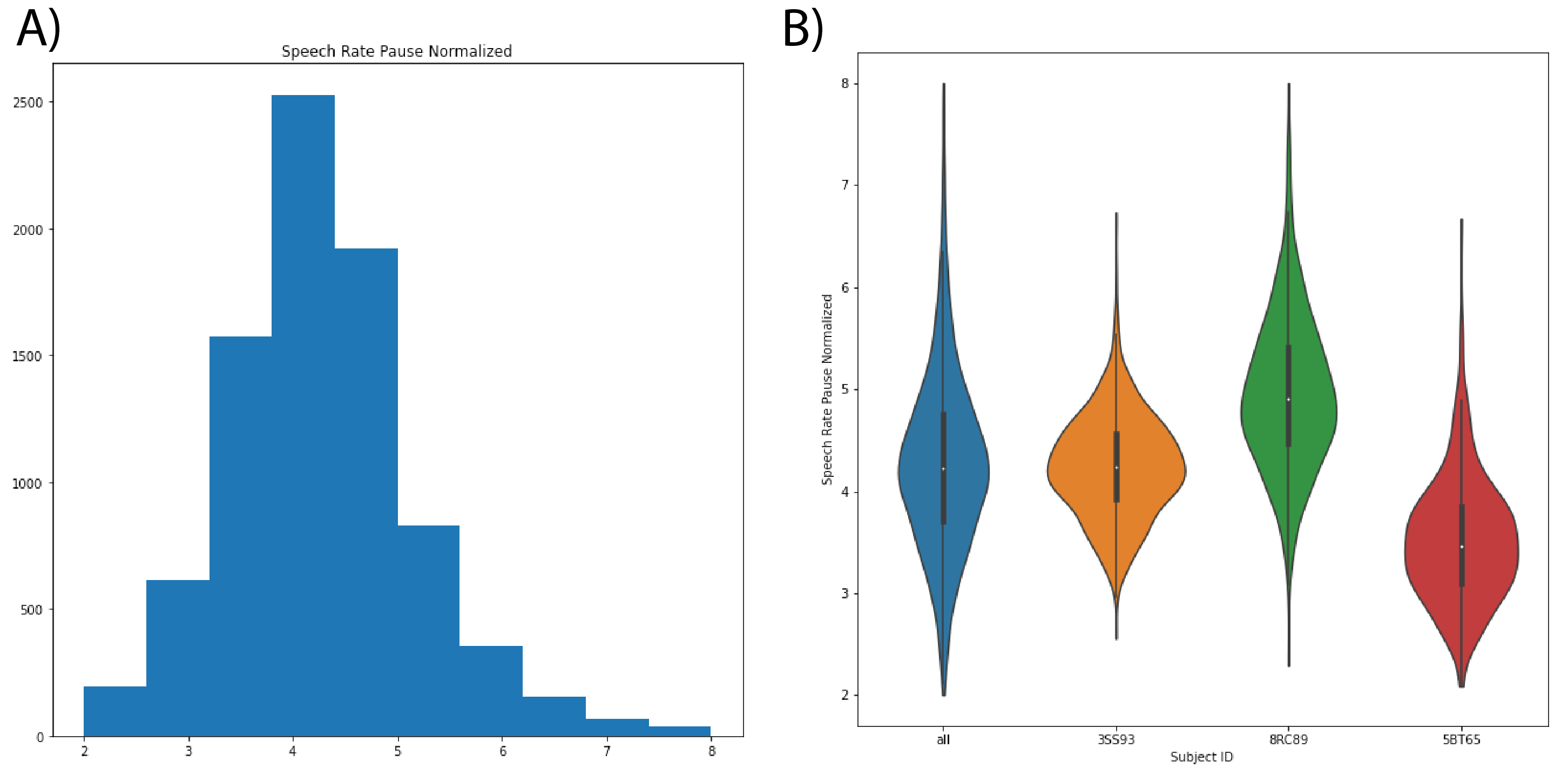}
\caption[Adjusting speech rate metric to remove longer pauses.]{\textbf{Adjusting speech rate metric to remove longer pauses.} When dividing the mean speech rate (syllables/second) feature of Figure \ref{fig:diary-verbosity-dists}B by the fraction of duration speaking feature of Figure \ref{fig:diary-verbosity-dists}C, we obtain an estimated speech rate metric that removes much of the effect of longer pauses from the original speech rate feature. A histogram of the adjusted feature is included here (A) for comparison with the original distribution in Figure \ref{fig:diary-verbosity-dists}B. Similarly, a violin plot comparing the distribution of the adjusted feature across patients (B) can be contrasted with the original feature's patient-specific distributions in Figure \ref{fig:speech-rate-per-pt}A. Note the ranges included in both (A) and (B) are limited between 2 and 8 syl/sec for clarity, analogous to the range of 1-7 used in Figure \ref{fig:diary-verbosity-dists}B.}
\label{fig:new-rate-per-pt}
\end{figure}

The adjusted speech rate feature of course shifted the entire BLS distribution up (Figure \ref{fig:new-rate-per-pt}), removing many of the unrealistically slow speaking rate values that were present in the original uncorrected feature distribution (Figure \ref{fig:speech-rate-per-pt}). In addition to the depicted shift, it is worth noting that in the original distribution there were 817 journals with speaking rate feature less than 2 syllables/second, and 13 with speaking rate greater than 8. By contrast, the adjusted speech rate distribution outliers included just 49 journals with speech rate below 2 syllables/second, but 89 with speech rate above 8. 

As mentioned, a long total pause time could occur because of many shorter pauses, pauses that might not be seen as independent of speech rate. It is therefore unsurprising to find that the adjusted distribution contains more unrealistically high speech rate values than the original one. What version to use will depend on characteristics of the dataset as well as the question at hand. Comfortingly, the pause-adjusted speech rate distribution of 3SS93 was much more in line with the study-wide adjusted speech rate distribution (Figure \ref{fig:new-rate-per-pt}B), which was one hypothesis about the adjusted feature. It is also good to see that the distribution of speech rates for subject 8RC89 remained elevated when compared against BLS with the adjusted version instead, as this strengthens the argument for consideration of speech rate in downstream analyses of 8R. 

Interestingly, 5BT65 displayed a more pronounced difference in speaking rate when using the adjusted metric (Figure \ref{fig:new-rate-per-pt}B). This is most likely because 3SS93's artificially low speaking rate (due to their abnormally long pauses) had a shifting effect on the original BLS distribution that was removed in the adjusted version. Because 5BT65's speaking rate was not as impacted by long pauses, the slowness of it subsequently became more apparent. This participant had a history of major depressive episodes, so that psychomotor slowing is indeed a potential sign to watch for. One obvious question to address about the utility of this updated speech rate feature will be whether the pause-adjusted rate has any added signal for evaluating within-subject variation of e.g. 5BT65, or if the original speech rate feature would be sufficient within the subject. \\

\FloatBarrier

\paragraph{Linguistic disfluencies.}
Disfluencies are also a part of producing everyday speech, particularly unplanned speech. Their potential clinical relevance will be a key theme of the interview linguistics analysis in chapter \ref{ch:2}. Use of disfluencies can be affected by many factors, some of which are unique to the respective interview and audio journal formats. In the case of journals, if a participant tends to say similar things in every diary or to think about what they will say prior to beginning recording, they may be less likely to produce disfluencies than they otherwise would. However, this could give disfluency prevalence greater importance when it does occur in an audio journal. It of course also could be a source of individual heterogeneity in journal behavior that may not have any connection to pathology. 

Participant variation in disfluency use is thus of especial interest. Moreover, certain categories of disfluency appeared wholly unrelated to each other - for example nonverbal and verbal edits - in both the interview dataset correlations of chapter \ref{ch:2}, and the diary correlations reported below (\ref{subsubsec:diary-corrs}). This suggests that it will be important to characterize any participant tendencies for disfluency use in terms of each individual category considered. To begin with nonverbal edits (Figure \ref{fig:disfluency-props-per-pt}A/B), the most commonly occurring type of disfluency, the only convincing difference here was a higher rate of use by subject 3SS93 than the other two subjects or the rest of BLS. The shifted-in left tail of 3S in this distribution could easily be explained by their greater verbosity: eventually a nonverbal filler will slip out and make the diary rate nonzero. Yet if the use of nonverbal fillers were purely random, one would also expect the longer diaries of 3S to shift in the right tail, which was not the case here. Further, the peak of 3SS93's distribution was right shifted as well. This effect need not have clinical significance, but it warrants further investigation of nonverbal edits in 3S.

\pagebreak

\begin{FPfigure}
\centering
\includegraphics[width=0.7\textwidth,keepaspectratio]{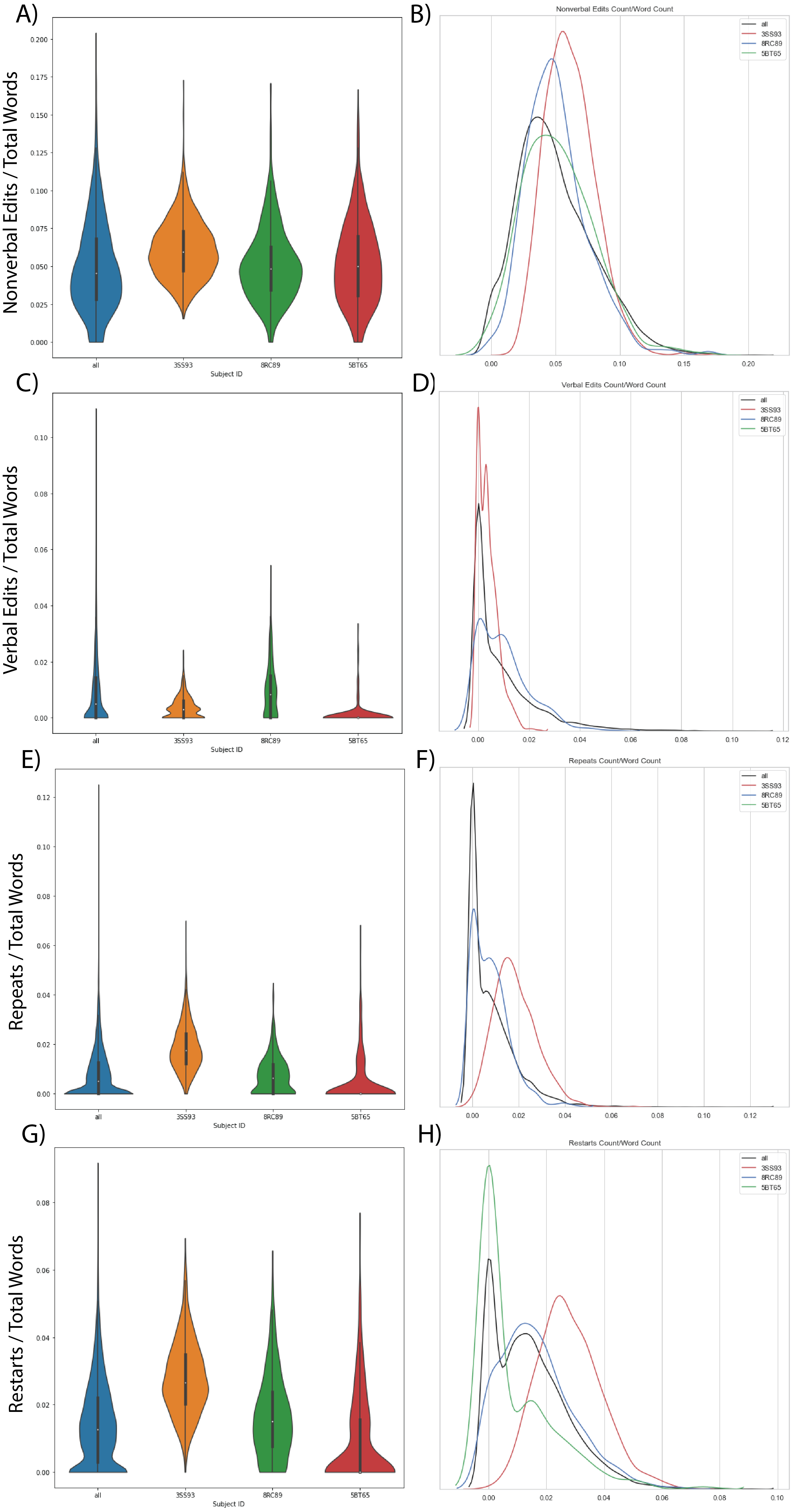}
\caption[Comparison of linguistic disfluency distributions between different patients of interest.]{\textbf{Comparison of linguistic disfluency distributions between different patients of interest.} Using the same methodology as in Figure \ref{fig:word-count-per-pt}, additional subject-specific feature distributions are compared against each other and against the distribution across the final BLS dataset, with both violin plots (left) and smoothed distribution curves (right). Here, we focus on the frequency of the various linguistic disfluencies: nonverbal edits (A/B), verbal edits (C/D), repeats (E/F), and restarts (G/H).}
\label{fig:disfluency-props-per-pt}
\end{FPfigure}

\FloatBarrier

With verbal edits (Figure \ref{fig:disfluency-props-per-pt}C/D), there were notable distributional differences in all three considered subjects. 5BT65 demonstrated the strongest distance from the BLS distribution, with substantially fewer verbal edits. Because verbal filler usage was somewhat rare and 5B tended to submit shorter diaries this is not particularly surprising though. 3SS93 also had significant distributional differences that would be reasonably explainable by verbosity alone; they had many more nonzero records than expected from the study population, but had many fewer high ($>0.01$) records too. By contrast, 8RC89 differed from the BLS distribution in a perhaps meaningful way, forming a shape much closer to uniform between 0 and $0.015$ with an otherwise similar right tail. Due to the high level of variance in verbal edit usage seen within 8R, they will be an interesting feature for potential modeling of clinically relevant feature variation in that subject (\ref{subsec:diary-ema}).

Repeats and restarts had the greatest predictive utility for the study of conceptual disorganization in chapter \ref{ch:2}, and they also demonstrated some correlation with each other. As such, they are a feature of strong interest here. 3SS93 had significant rightward shift in their distributions for both repeats (Figure \ref{fig:disfluency-props-per-pt}E/F) and especially restarts (Figure \ref{fig:disfluency-props-per-pt}G/H). The increase is unlikely to be explainable by word count alone for reasons similar to those mentioned about nonverbal edits. Furthermore, the increase was of even greater magnitude for these disfluency types, though that may be enabled by their generally lower base rate versus nonverbal edits. Nevertheless, repeats and restarts will also be of elevated interest in understanding subject 3SS93. 

Conversely, 5BT65 had a significantly greater number of diaries with 0 repeats (Figure \ref{fig:disfluency-props-per-pt}E/F) and 0 restarts (Figure \ref{fig:disfluency-props-per-pt}G/H) relative to the larger BLS population. However, this is much more likely to be explainable by word count alone, rather than an individual difference in repeat or restart use probability. Similarly, 8RC89 demonstrated minimal difference against the BLS dataset in these two disfluency categories. Regardless, there was sufficient variation in the within-subject restart distribution for both participants to consider modeling restart usage over time, which could be of interest based on its clinical significance in the results of chapter \ref{ch:2}. 

Ultimately, it is clear that 3SS93 had a propensity for disfluency usage, with the probable exception of verbal edits, but it is not yet clear whether this personal difference bears any relationship with psychiatric pathology. Along the same lines, it is interesting that 8RC89 seemed more likely to use verbal edits - but not other disfluencies - than the overall study population, but any potential clinical significance remains to be investigated. Finally, it is worth noting that the close overlap of 5BT65's nonverbal edit distribution with the BLS distribution may itself hold information. The base rate of nonverbal edits in the final BLS dataset was higher than the other disfluencies; however it was still low enough that one might expect many more journals with zero nonverbal fillers from 5BT65, given the highly limited word counts of many of the 5B diaries in this dataset (Figure \ref{fig:word-count-per-pt}).  \\

\FloatBarrier

\paragraph{Sentiment features.}
Sentence sentiment is one of the easiest features to compute automatically in this dataset, while simultaneously having good validation results and pilot evidence of a significant relationship with same day EMA responses (section \ref{subsubsec:diary-val-trans}). Thus it is also of great interest to characterize across subjects. 

\begin{figure}[h]
\centering
\includegraphics[width=0.75\textwidth,keepaspectratio]{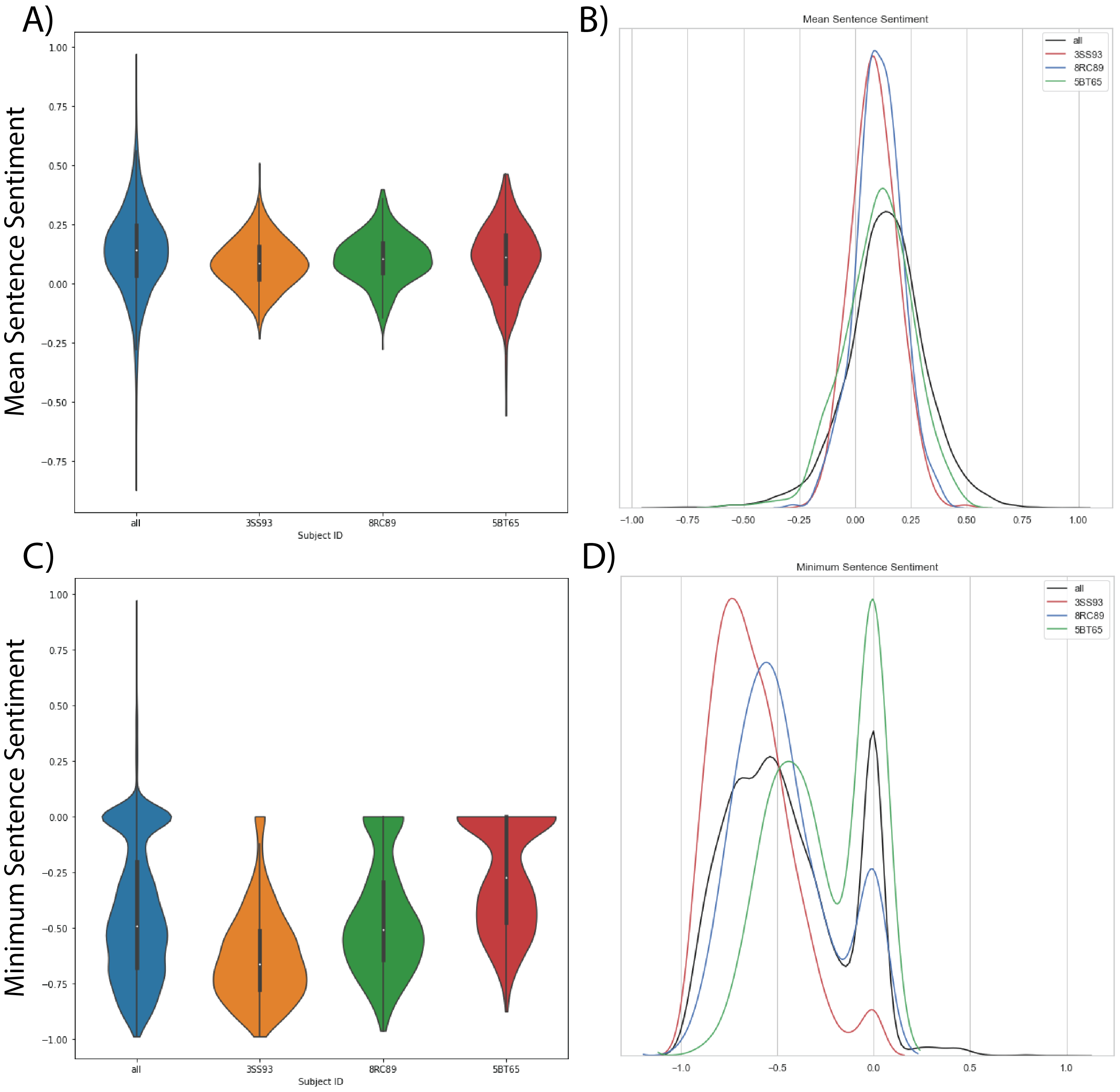}
\caption[Comparison of sentiment-related distributions between different patients of interest.]{\textbf{Comparison of sentiment-related distributions between different patients of interest.} Using the same methodology as in Figure \ref{fig:word-count-per-pt}, additional subject-specific feature distributions are compared against each other and against the distribution across the final BLS dataset, with both violin plots (left) and smoothed distribution curves (right). Here, we focus on the sentiment-related features: mean sentence sentiment (A/B) and minimum sentence sentiment (C/D). To properly interpret these results, it is important to carefully consider the effects of patient verbosity on these features - see Figure \ref{fig:random-dists-per-pt} for example distributional differences due only to verbosity.}
\label{fig:sentiment-props-per-pt}
\end{figure}

Although mean sentence sentiment did show some significant distributional differences (Figure \ref{fig:sentiment-props-per-pt}A/B), they could potentially be explained by word count differences. The longer diaries of 3SS93 would be expected to compress the mean sentence sentiment distribution, and though there was a slightly lower mean sentiment score at 3S's peak, the effect was too small to say anything definitive. The width of 8RC89's mean sentiment distribution is very similarly compressed. As 8R submitted a broader range of diary lengths, it is difficult to interpret whether the distance between their distribution and the study's has any independent meaning. 

Relatedly, it is difficult to interpret the distribution differences observed in minimum sentence sentiment (Figure \ref{fig:sentiment-props-per-pt}C/D) without more information. The distribution of 8RC89 did not differ significantly from that of the study, and while the distributions of 3SS93 and 5BT65 were substantially different, the shifts (more and less negative, respectively) would be expected to at least some degree based on diary length alone: the more sentences in a transcript, the greater opportunity for one to be highly negative, and vice versa. It is therefore necessary to do a deeper dive on expected distributional differences due to diary length, before attempting to draw any conclusions about the sentiment results here. Because the semantic incoherence features - particularly the maximum sentence incoherence - face a similar issue, I will first present the true distributional comparisons for those features. I then proceed to look at random simulation results accounting for diary length, to connect those distributions back to the empirical ones and shed new light on interpretations. \\

\FloatBarrier

\paragraph{Semantic incoherence features within-sentence.}
As discussed in section \ref{sec:background2}, semantic incoherence has been a staple of the prior literature on natural language processing (NLP) applications to psychotic disorders. Therefore, we have a clear reason for investigating these features in the BLS dataset. Given the clinical history of 8RC89, it is a good early sign that the maximum incoherence feature was more strongly differentiated from the rest of the BLS dataset in 8R than it was in 3SS93 or 5BT65. The 5B D stat when comparing their distribution with the study-wide one was in fact not significant at all, which appears consistent with their BD type II diagnosis.

\begin{figure}[h]
\centering
\includegraphics[width=0.75\textwidth,keepaspectratio]{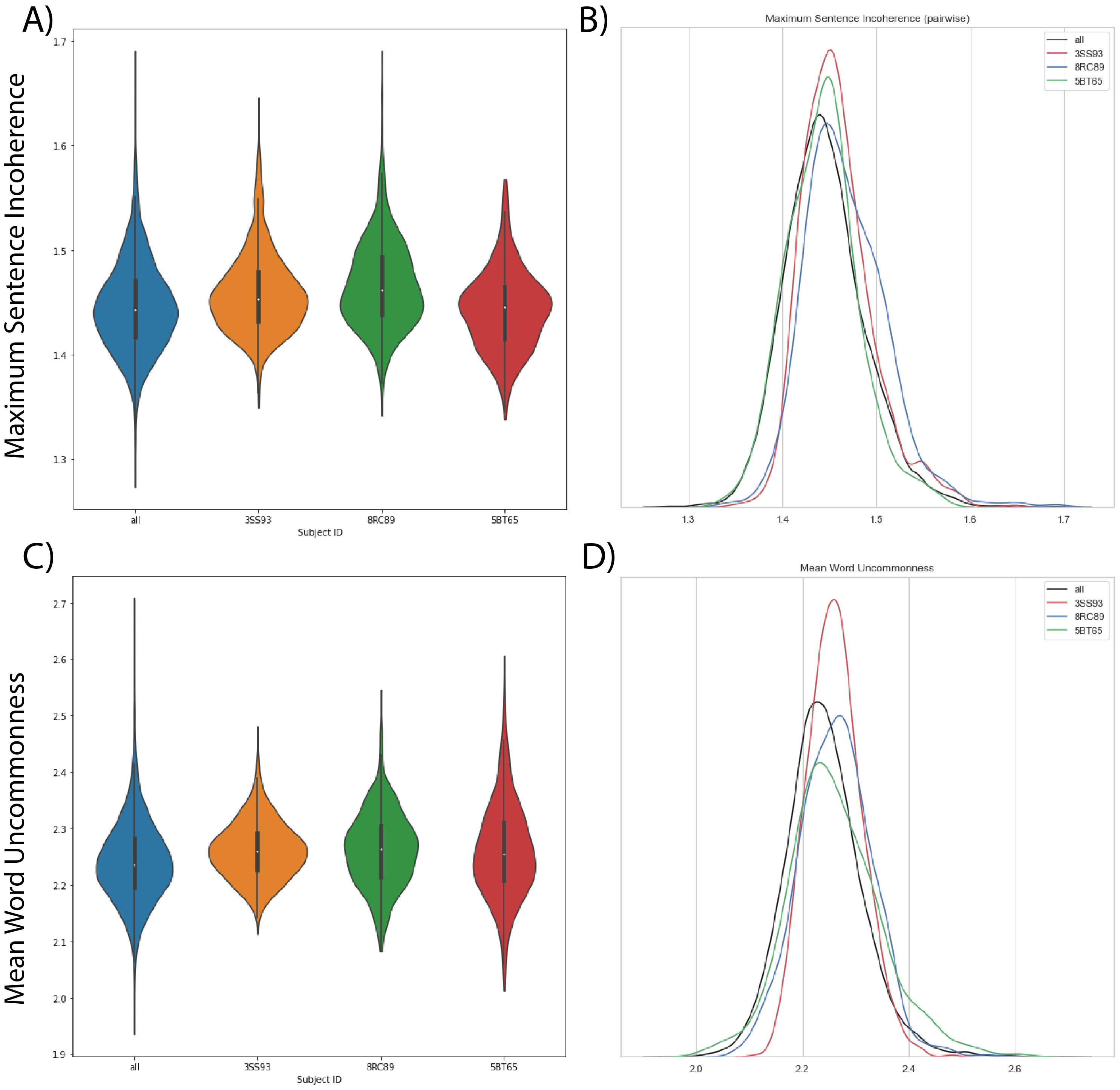}
\caption[Comparison of incoherence-related distributions between different patients of interest.]{\textbf{Comparison of incoherence-related distributions between different patients of interest.} Using the same methodology as in Figure \ref{fig:word-count-per-pt}, additional subject-specific feature distributions are compared against each other and against the distribution across the final BLS dataset, with both violin plots (left) and smoothed distribution curves (right). Here, we focus on the coherence-related features: maximum sentence incoherence (A/B) and mean word uncommonness (C/D). To properly interpret these results, it is important to carefully consider the effects of patient verbosity on these features - see Figure \ref{fig:random-dists-per-pt} for example distributional differences due only to verbosity.}
\label{fig:coherence-props-per-pt}
\end{figure}

It is of course a potential issue that greater diary length would facilitate higher maximum incoherence values, however. It's promising that the maximum incoherence distribution of 8R was clearly shifted higher than that of 3S (Figure \ref{fig:coherence-props-per-pt}A/B) - despite the latter's heavy bias towards long diaries. Nonetheless, the difference was not strong enough to make a definitive statement without a more quantitative analysis of what fraction of the distribution difference could be explained by diary length. Thus further interpretation will wait until after the upcoming random simulation results have been reviewed. 

For mean uncommonness (Figure \ref{fig:coherence-props-per-pt}C/D), the distribution widths varied as one would qualitatively expect based on diary length alone, from 3S at the narrowest to 5B at the widest. There was a small rightward shift of the peak of 8RC89's mean uncommonness distribution, which could be an interesting pairing with the potential increased max incoherence in 8R. However, these effects were again too slight to make a strong statement about without more sources of evidence. \\

\FloatBarrier

\paragraph{Revisiting length normalization.} 
Given the relatively short nature of audio journals, length variation can have a strong effect on minimum and maximum sentence features, and can also drastically change the shape of the distribution of features calculated using the mean over sentences. Because of this, it is unclear how much of the statistically significant difference in certain diary feature distributions between patients is explainable purely by the large difference in verbosity between patients. In particular, any comparisons that resulted in opposite direction differences between 3S and BLS and 5B and BLS, with 8R in the middle and fairly close to the BLS distribution, could very plausibly be explained purely by verbosity differences -- especially when the minimum is lower or the maximum is higher in the groups with longer diaries, or when the mean distribution is more spread out in the groups with shorter diaries.

To investigate the issue, I ran simulations to randomly generate diary-level feature distributions using the same underlying sentence-level distribution, to ultimately create summary stats for "diaries" of different lengths but with no other differences. The simulated distributions were therefore used to estimate the magnitude of distributional difference that could be attributed solely to underlying length difference; these estimates were subsequently compared to the magnitude of differences seen in the real dataset, so that any discrepancies much larger than found in simulation can be separated in interpretation from any discrepancies that were similar to the random results. I focused these analyses on the sentiment and incoherence related features, because those comparisons were the least clear cut between subjects in the real dataset above. The details of my methodology can again be found in supplemental section \ref{subsec:dist-comp-methods}. 

The maximum distance observed between cumulative distribution functions of the BLS-wide simulated features and the subject-specific simulated features, as measured by the KS test's $D$ statistic, was below 0.3 for all comparisons in all trials. For the mean uncommonness and mean sentiment simulations $D$ was always below 0.15, and it was also always below 0.15 for all comparisons between BLS-wide and 8RC89 simulated features. This is inline with expectations about how the minimum/maximum features would compare to the mean ones, as well as the fact that 8RC89 has a verbosity distribution much more similar to the study-wide one than 5BT65 or 3SS93 do. Additionally, comparisons between the simulated mean features were generally not statistically significant after the multiple testing corrections. Full KS test results are again reproduced in supplemental section \ref{subsec:ks-sup}.

Based on these results, statistically significant differences between mean sentiment distributions and mean uncommonness distributions found in the real data are unlikely to be a result of disparity in number of sentences per diary, particularly for those with $D$ stat that meaningfully exceeded $0.1$. This would include mean sentence sentiment and to a lesser extent mean word uncommonness, for both 3SS93 and 8RC89. Similarly for minimum sentence sentiment and maximum sentence incoherence, we should \emph{not} expect significant difference (nor $D$ stat $>> 0.1$) between 8RC89 and BLS-wide distributions purely because of sentence count differences. Thus this suggests a meaningful distinction in 8RC89's maximum incoherence distribution was found in the real dataset. 

Interpreting the minimum/maximum features for 3SS93 and 5BT65 is more challenging, but we would certainly expect a statistically significant KS test result, with $D$ likely between 0.2 and 0.3, and perhaps a bit stronger for 5BT65 - particularly when comparing incoherence. For minimum sentence sentiment, the real dataset had $D$ on the higher end of this range, so it is not conclusively a length-only effect, but also not possible to deem as an independent result. For maximum sentence incoherence on the other hand, the true results were surprisingly outside the expected range of difference \emph{on the low end}. In the case of 3SS93, it is unlikely to be meaningful, as the distance from the simulated incoherence $D$ stat was fairly small and the real KS test did reveal a statistically significant difference. However for 5BT65, the lack of difference from BLS in true incoherence distribution was stark (with $D$ well under 0.1 and no significance after testing correction), therefore warranting further investigation. \\

\noindent To assist in interpreting the simulated distributional differences against the real ones, analogous simulated distribution plots can be found in Figure \ref{fig:random-dists-per-pt}.

\pagebreak

\begin{FPfigure}
\centering
\includegraphics[width=0.7\textwidth,keepaspectratio]{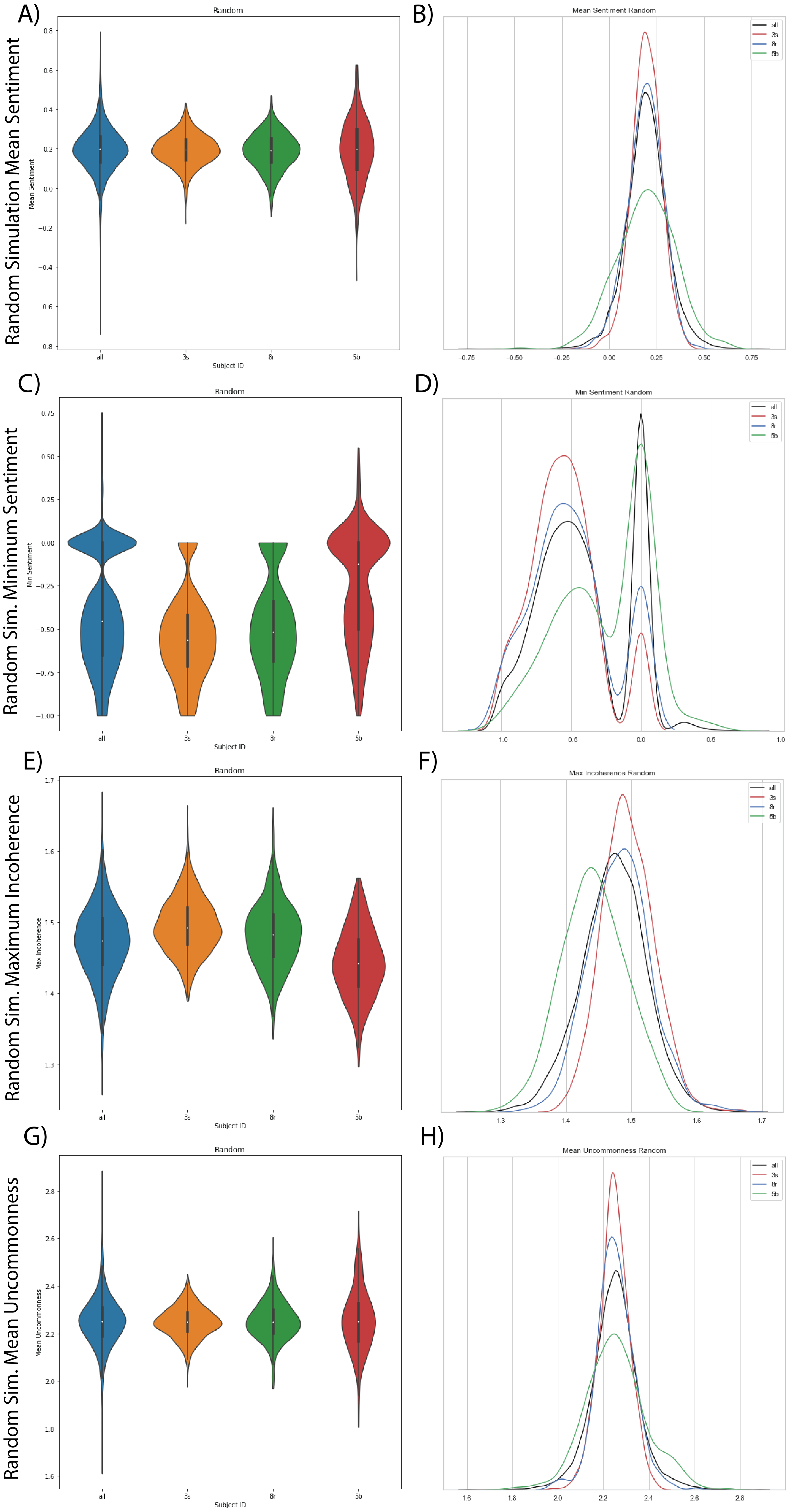}
\caption[Using random simulation to identify subject-specific distributional differences attributable to verbosity.]{\textbf{Using random simulation to identify subject-specific distributional differences attributable to verbosity.} The simulated distributions used to obtain the KS test results of Table \ref{table:random-ks-test} -- which were designed to assess the role of sentence count variation between participants in the observed distributional differences of other features -- were also visualized via the same methodology used for the real data in Figures \ref{fig:word-count-per-pt}-\ref{fig:coherence-props-per-pt}. Both violin plots (left) and smoothed distribution curves (right) are shown for the simulated distributions sampled based on the true distributions of sentence count across the final BLS dataset, and across the subsets of patients 3SS93, 8RC89, and 5BT65. The simulations focused on controlling for verbosity in the sentiment and coherence related features: mean sentence sentiment (A/B), minimum sentence sentiment (C/D), maximum sentence incoherence (E/F), and mean word uncommonness (G/H).}
\label{fig:random-dists-per-pt}
\end{FPfigure}

\FloatBarrier

 For mean and minimum sentiment, the true comparisons of Figure \ref{fig:sentiment-props-per-pt} should be reevaluated in light of the simulated distributions in Figure \ref{fig:random-dists-per-pt}A-D, which are influenced only by sentence count. It can be seen that mean sentiment curves had a leftward (i.e. negative) shift in the true distributions of both 3S and 8R, which would not be expected to appear from simulation results. For 5B, mean sentiment had a similar peak as would be expected with a random sentence sentiment distribution, but a much less stretched out positive tail.

For maximum incoherence and mean word uncommonness, the true comparisons of Figure \ref{fig:coherence-props-per-pt} should analogously be reevaluated in light of the simulated distributions in Figure \ref{fig:random-dists-per-pt}E-H. The max incoherence distribution actually observed for 8RC89 was clearly shifted right (i.e. higher) of the simulated expectation, and the true 8RC89 uncommonness distribution also had a clear rightward tilt in its peak unexplained by simulation. 3SS93's true incoherence distribution was by contrast very similar to expectations from simulation. As discussed, 5BT65 had real maximum incoherence features distributed with close overlap to the shape of the BLS-wide distribution, while simulated max incoherence indicates that we should have expected a leftward shift based on 5BT65's diary lengths. It will require further investigation to determine if this is an artifact of other properties of 5B's transcript structures or a potentially clinically relevant individual difference in incoherence feature. 

In addition to comparing relationships between distributions depicted in Figure \ref{fig:random-dists-per-pt} versus the real relationships between those depicted in Figures \ref{fig:sentiment-props-per-pt} and \ref{fig:coherence-props-per-pt}, we can also look at the overall shape of the study-wide simulated versus real distributions, in order to get a sense of how much was not captured by the simulation methodology. One thing that it did not intend to capture but that might affect the final distribution is correlation between the sentiment or incoherence score of different sentences from the same transcript. The simulation intentionally randomly sampled each sentence's scores independently, to isolate impact attributable only to differences in sentence count. Underlying properties that might differ between transcripts, such as current mood of the speaker, might cause uneven distribution of sentence features between transcripts -- for example it may be the case that most diaries contain either no negative sentences or multiple negative sentences. This potential source of discrepancy between observed and simulated study distributions is not really a limitation, though it may cause some overestimation of the effects of sentence count on the minimum/maximum features especially.

There are other potential explanations for discrepancies between real and simulated study distributions that are legitimate limitations of the current approach. One issue is that sentence sentiment and incoherence scores are very likely to be impacted by sentence length, which is another measure of verbosity that may also have subject-specific biases (probably correlated to some extent with the sentence count biases), but that was not at all modeled here. It is plausible that the lack of expected leftward shift in 5BT65's true incoherence distribution could be explained by the effects of sentence length on sentence-level incoherence, which is a major reason why that observation in particular requires closer follow-up, and will be revisited within subsequent sections (\ref{subsec:diary-ema} and \ref{subsec:diary-case-study}).

 One positive for the maximum incoherence and mean uncommonness analyses is that it would be easy in future work to break the study distribution down to the word (or word pairing) level, to systematically study the impact of sentence length in a way similar to the simulations run for diary length here. On the other hand, the lowest level output intended for sentiment score is the sentence, and furthermore the sentiment distributions observed have been more complicated to cleanly describe than many of the other features. While it would still be possible to perform correlational analyses and other related approaches to better understand the dynamics of sentence sentiment and its relationship with verbosity, it will be less exhaustive than what is possible for the incoherence-related features. For similar reasons, it is also likely that the present simulation results are a more accurate reflection of the incoherence-related features than the sentiment ones. \\

\FloatBarrier

\noindent In summary, through simulation of feature distributions amongst diaries with different sentence counts but the same underlying per sentence feature distributions, I was able to clarify the significance of subject-specific differences in distributions of sentiment and incoherence related features, and possibly even uncover meaningful differences from expectation that manifested as no difference in the true dataset. From this, confidence in the relevance of the observed lower mean sentiment scores of 3SS93 and 8RC89 compared to the full BLS diary set was increased, as was confidence in the relevance of the observed higher maximum incoherence (and to a lesser extent mean word uncommonness) scores of 8RC89 compared to the rest of the study. At the same time, no strong conclusions could be drawn about minimum sentiment, and based on those results it might be less useful to include the minimum summary statistic for sentiment alongside the mean than was originally hypothesized. Interestingly, 5BT65 should have had substantially lower maximum incoherence scores than BLS on the whole based on simulation, which was not reflected in the real data -- thereby identifying another salient question for upcoming analyses. 

Next, I will continue to build on the characterization of distributional differences between BLS participants, wrapping up with a summary of resulting observations and updated hypotheses from this section. In the subsequent section (\ref{subsubsec:diary-corrs}), I will look at correlation structure amongst diary features across BLS; one aim of doing so is to better characterize relationships between transcript word count and the rest of the final feature set, which will further elucidate to what extent interesting properties of those features are independent of the far-reaching effects of word count (and more broadly, independent of each other). \\

\paragraph{Sentence to sentence incoherence and the role of TranscribeMe's sentence splitting.}
Due to the generally short nature of the journals, it was unclear how much meaningful information the sentence to sentence incoherence feature might carry in this dataset. As such, it was originally excluded from these analyses. However due to the results of the manual review reported within section \ref{subsubsec:diary-val-trans}, where the max sentence incoherence and the mean between sentence incoherence captured quite different attributes -- with sentence to sentence incoherence better representing manual reviewer opinion on journal incoherence -- I decided to perform an initial scientific evaluation of this metric in the same spirit as the original 12 selected features. It was found that sentence to sentence incoherence correlated very strongly with transcript sentence structure though, much moreso than any of the other considered features. As we want to find features with strong modeling power beyond basic verbosity-related ones, I did not move forward with the sentence to sentence incoherence feature. The initial characterization I performed can be found in detail in supplemental section \ref{sec:sen-to-sen}.

\begin{figure}[h]
\centering
\includegraphics[width=\textwidth,keepaspectratio]{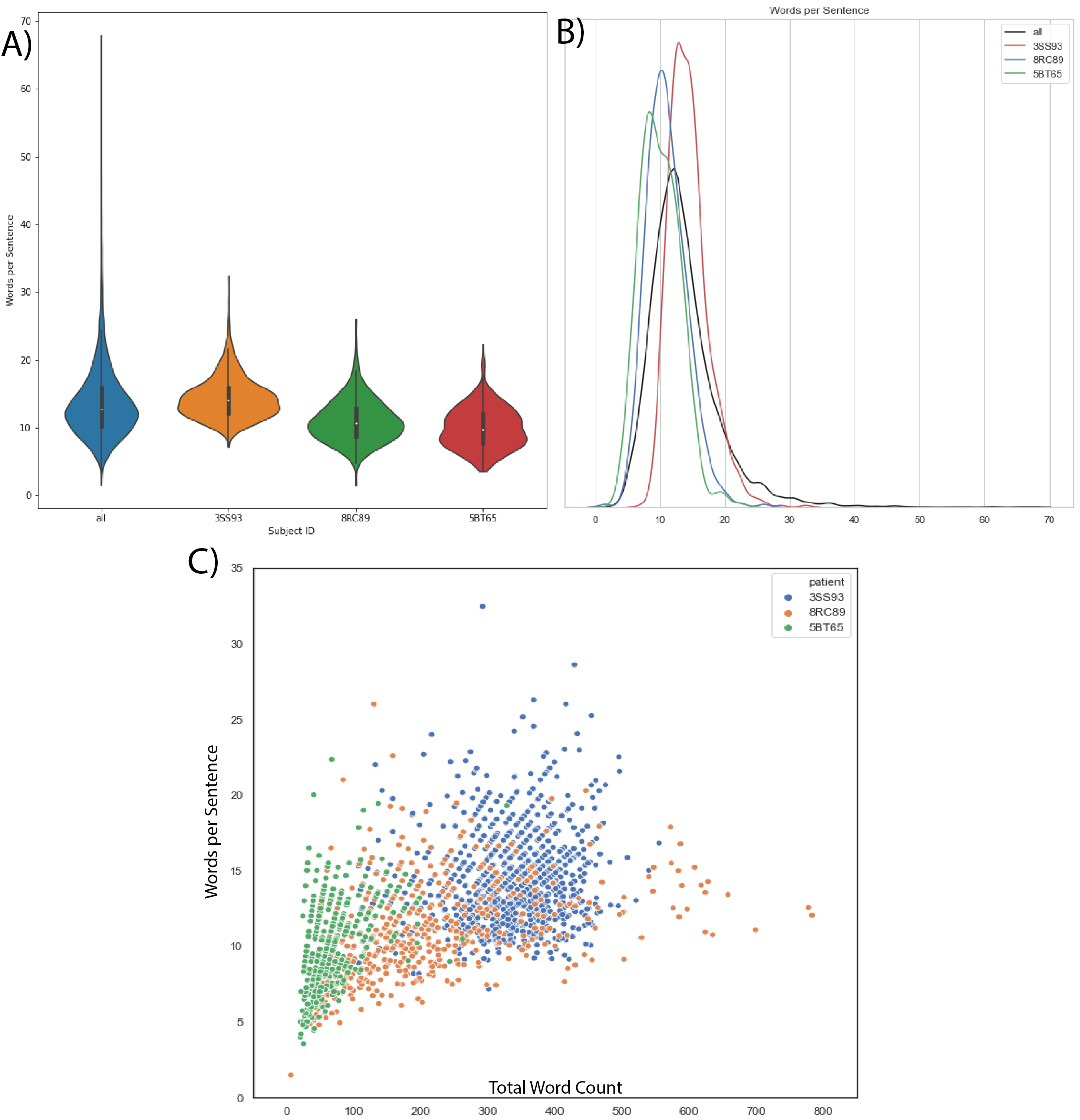}
\caption[Comparison of words per sentence distributions between different patients of interest.]{\textbf{Comparison of words per sentence distributions between different patients of interest.} Using the same methodology as in Figure \ref{fig:word-count-per-pt}, additional subject-specific feature distributions are compared against each other and against the distribution across the final BLS dataset, with both violin plots (A) and smoothed distribution curves (B). Here we consider the mean words per sentence in each transcript, as the methodology TranscribeMe uses to split sentences may reflect on clinically relevant speaking patterns, and can also directly impact downstream pipeline outputs. \newline In this figure, an additional scatter plot (C) was included, showing for each diary from 3S (blue), 8R (orange), and 5B (green) how the total word count related to the mean words per sentence. The scatter (C) was added to demonstrate that interesting subject-specific differences in mean words per sentence were not wholly explainable by differences in diary lengths.}
\label{fig:words-per-sentence-pt-dists}
\end{figure}

In light of those results, I decided to add mean words per sentence as an additional feature for consideration in the next analysis steps of sections \ref{subsec:diary-ema} and \ref{subsec:diary-case-study} instead. In terms of subject-dependent trends in mean words per sentence over diary transcripts, clear examples of individual differences were already identified within the manual review process (section \ref{subsubsec:diary-val-trans}), particularly in subject GFNVM (Figure \ref{fig:diary-words-per-sentence}) -- who was of interest for quality review due to their recording style, but was not chosen as a top subject to highlight in case report form. GFNVM was however included in the subset of patients that were used for EMA modeling in the upcoming section \ref{subsec:diary-ema}, so the potential connection between words per sentence and their same day self-reported emotional state will be discussed there. Here, I will focus on the study-wide distribution of words per sentence (Figure \ref{fig:diary-words-per-sentence}A), which displayed a great deal of variance, as well as evaluate for distributional differences in the highlighted case report subjects (Figure \ref{fig:words-per-sentence-pt-dists}).

There were indeed significant participant-dependent distributional differences in words per sentence for all three subjects: KS test results indicated a $D$-stat of $0.219$ ($p < 10^{-15}$) for 3SS93 compared to all of BLS, a $D$-stat of $0.247$ ($p < 10^{-25}$) for 8RC89 compared to all of BLS, and a $D$-stat of $0.323$ ($p < 10^{-30}$) for 5BT65 compared to all of BLS. Of course, with a higher sentence count, we would expect lower variance in mean words per sentence even though the center of the distribution should not change for this reason alone per the study-wide correlation results (to be reported in section \ref{subsubsec:diary-corrs}). Additionally, as a moderate correlation was found between word count and words per sentence, we would expect some distributional shifts for these subjects that are already explainable by word count differences alone, which were quite high (Figure \ref{fig:word-count-per-pt}). 

As such, it is unlikely that the distributional difference seen in words per sentence for 3S contains any new information (Figure \ref{fig:words-per-sentence-pt-dists}). However, for 8R there was a shift towards shorter sentences that is not explainable by previous observations, and in diaries with similar word count 8R clearly produced shorter sentences on average than 3S (Figure \ref{fig:words-per-sentence-pt-dists}C). It remains to be investigated in subsequent sections (\ref{subsec:diary-ema} and \ref{subsec:diary-case-study}) whether this is a clinically relevant individual difference, though it is worth noting that both 8R and 3S were enrolled in the study around a similar time and were demographically similar. In the case of 5B, a shift towards shorter sentences with a relatively high level of variance would be expected based on trends in their overall word count compared to the rest of BLS. Still, the variance observed even amongst similar word counts within the 5B journal dataset (Figure \ref{fig:words-per-sentence-pt-dists}C) does point to a potential variable of clinical interest when assessing the link between verbosity and symptomatology in this subject, which will also be explored in the subsequent modeling sections. \\

 Overall, the words per sentence feature likely carries clinical relevance, particularly in the way it was defined here, with individual sentences labeled by manual transcriber judgement. Mathematically, one would expect sentence splitting to affect downstream features even if it were done completely arbitrarily. A methodology that prioritizes a more even split of phrases rather than taking into account speaking pattern would produce more "pure" NLP features that would correlate less strongly with words per sentence. On the other hand, one might not want to filter out this information, as really it is a benefit of the manual transcription technique. 
 
 Thus for audio journals, it is probably best to utilize the words per sentence feature in itself, and then take care when assessing independent predictive relevance of other features that may be modulated by it -- which the results presented here can assist in. It will also be important to ensure that bias in sentence splitting does not appear in a subject or time dependent manner, which requires some coordination with TranscribeMe and periodic manual checking of resultant transcripts for blatant quality changes in sentence splitting. 
 
 If utilizing an automated transcription technique instead, it will be necessary to think more carefully about tokenization methods, both in their effect on downstream NLP features, as well as to what extent they may be able to capture speaking pattern nuances that manual transcribers can. It is likely that by default the output of automated transcription will not be able to do this, but by linking with other automated acoustic features such as finer grained speaking rate estimates, it may be possible to independently replicate. The use of automated transcription techniques is largely beyond the scope of this thesis, and the many advantages of manual transcription in present day diary analysis will be covered at greater length in the discussion (section \ref{sec:discussion2}). 
 
 As an aside, note that the consideration of sentence splitting for interview recording transcriptions is quite different due to their dialogue nature, and in fact the main pipeline of chapter \ref{ch:2} assumes turn-based transcript splitting instead of sentence-based transcript splitting, as this reduced the overall cost of the long interview transcriptions. As such, sentence splitting will largely not be covered in the interview discussion of chapter \ref{ch:2}, but it is something that ought to be included in planning of future interview studies, particularly those that aim to conduct truly open ended interviews. \\

\FloatBarrier

\noindent Now that the distributional properties of the full set of potential modeling features has been well characterized in the BLS dataset, both overall and for the subjects to be highlighted in mini case report form, I will summarize the key takeaways from this endeavour before proceeding to further analyses. To do so, I will first present an initial picture of the clinical background on 3SS93, 8RC89, and 5BT65, to link with expectations based on these observed diary distributional differences. \\

\paragraph{Cross-checking with clinical scales.}
To better contextualize all of the above feature comparisons with more information about the clinical characteristics of the participants in question, I will next report on relevant gold standard scale information collected from these patients.

The Positive and Negative Syndrome Scale (PANSS) contains positive and negative subscales each with 7 items scored from 1 to 7. Thus the PANSS positive subscale rating can range from 7 to 49, as can the PANSS negative subscale, with higher scores corresponding to more severe symptoms. The positive subscale assesses symptom severity related to delusions, conceptual disorganization, hallucinations, excitement, grandiosity, persecution, and hostility. The negative subscale assesses symptom severity related to blunted affect, emotional withdrawal, social apathy, poor rapport, lack of spontaneity, stereotyped thinking, and difficulty with abstraction \citep{PANSS}.

In practice, the original sample of 101 Schizophrenia patients measured by \cite{PANSS} found positive scores between 7 and 32 and negative scores between 8 and 38, with means (+/- standard deviations) of $18.2 (+/- 6.1)$ and $21 (+/- 6.2)$ respectively across that population. Though originally designed for Schizophrenia, PANSS is now commonly used in conjunction with other scales to assess Bipolar Disorder (BD) severity - particularly for patients that experience psychotic symptoms \citep{Fernandez2014}. 

The Montgomery-Asberg Depression Rating Scale (MADRS) is a 10 item scale, with each item scored from 0 to 6, so that the MADRS total score can range from 0 to 60. The MADRS assesses apparent sadness, reported sadness, inner tension, reduced sleep, reduced appetite, concentration difficulties, lassitude, inability to feel, pessimistic thoughts, and suicidal thoughts. It is commonly administered both to Major Depressive Disorder patients as well as those with other affective disorders, including BD. Typically used cutoff points are MADRS $\leq 6$ to indicate no symptoms and $\geq 20$ to indicate moderate or worse depression \citep{MADRS}. \\

\begin{figure}[h]
\centering
\includegraphics[width=\textwidth,keepaspectratio]{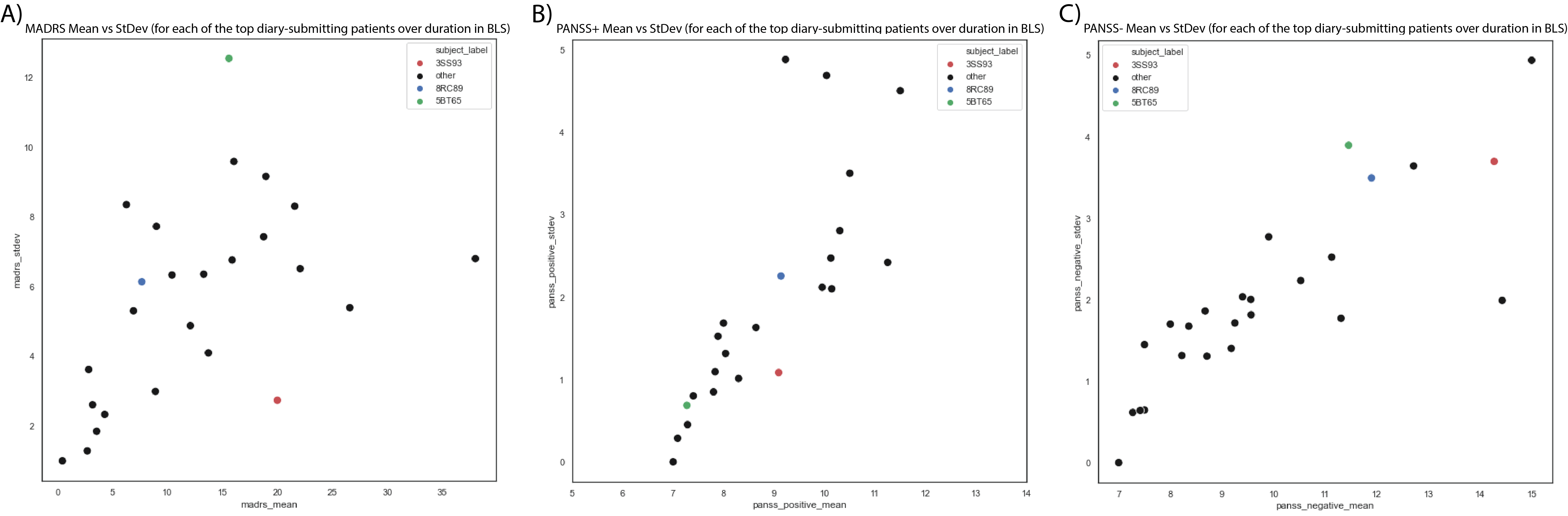}
\caption[Mean and standard deviation of key clinical scale ratings per BLS participant.]{\textbf{Mean and standard deviation of key clinical scale ratings per BLS participant.} 25 subjects submitted at least 100 journals of acceptable quality while participating in the study for at least 6 months. Monthly clinical scale ratings were also collected from the same participants in parallel. Here, the mean score over the study period is plotted against the standard deviation, for each of those patients on the following key ratings: the MADRS total (A), the PANSS positive subscale (B), and the PANSS negative subscale (C). On each scatter, the 3 patients of particular focus - 3SS93 (red), 8RC89 (blue), and 5BT65 (green) - are colored for clarity. Two subjects with mean PANSS positive score $> 15$ are excluded from the scatter (B) to prevent stretching of the x axis.}
\label{fig:bls-clinical-scatter}
\end{figure}

To provide an initial picture of the clinical profiles of the three highlighted subjects, I generated scatter plots showing the mean and standard deviation of the described clinical rating scale scores over the course of BLS for each patient with significant audio journal participation, highlighting the points corresponding to 3SS93, 8RC89, and 5BT65 (Figure \ref{fig:bls-clinical-scatter}).

5B demonstrated very high variance in MADRS total scores over the course of the study (Figure \ref{fig:bls-clinical-scatter}A, green), which in conjunction with their acoustic and linguistic feature distributions bodes well for potential detection of depressive episodes through their submitted journals. Similarly, 8R had higher variance in PANSS positive subscale scores over the course of the study (Figure \ref{fig:bls-clinical-scatter}B, blue) than most other BLS participants, and also displayed intriguing linguistic distributional properties in the context of prior literature on psychotic symptoms. Additionally, all three subjects varied highly on the PANSS negative subscale, and 3S experienced more severe negative symptoms than most (Figure \ref{fig:bls-clinical-scatter}C, red). 

Note that 3SS93 only participated in onsite interviews during their first year in the study, despite contributing data in other modalities for many years. Therefore the 3S clinical ratings summary stats include 11 total data points. During this period, their MADRS ranged between 16 and 25, their PANSS positive ranged between 7 and 11, and their PANSS negative ranged between 9 and 22.

8RC89 on the other hand continued participating in onsite interviews throughout their 5+ year enrollment in the study. Thus 8R has 59 clinical rating data points available for the calculation of summary stats. During this period, their MADRS ranged between 0 and 25, their PANSS positive ranged between 7 and 17, and their PANSS negative ranged between 7 and 22.

5BT65, whose primary diagnosis is Bipolar II rather than Bipolar I like 3S and 8R, spent $\sim 2$ years involved in BLS, and contributed both audio journals and onsite interviews through that period. 5B has 22 clinical rating data points available, with MADRS ranging between 0 and 36, PANSS positive ranging between 7 and 10, and PANSS negative ranging between 7 and 20. These ranges are consistent with expectations when compared to those of 3S and 8R, based on the type II diagnosis. \\

\FloatBarrier

\noindent The clinical features of these subjects will be revisited in greater detail in the case report work of section \ref{subsec:diary-case-study}. The consistency of general clinical and acoustic/linguistic properties with expectations that has been seen so far is promising, both for informing the immediate next steps to be reported on in wrapping this chapter and for the broader future outlook on the value of the dataset. I will now synthesize conclusions from my initial data exploration to generate more specific hypotheses that can be investigated in subsequent sections, for both EMA (\ref{subsec:diary-ema}) and the aforementioned clinical outcome relationships. \\

\paragraph{Updating (and personalizing) hypotheses.}
It is of course possible that subject-dependent distributional differences in various diary features are real but of no psychiatric relevance. Conversely, a feature might look similar between two very clinically different subjects, yet still hold predictive power when evaluated over time within each individual subject, or perhaps within one and not the other. Because of this, there is excellent promise in methodologies that can generate hypotheses with a personalized component. Still, distributional comparisons alone cannot lead to strong statements one way or the other in most cases, probably one reason this style of work is often foregone in recent digital psychiatry modeling projects.

Nevertheless, the review of distributional differences presented has produced some compelling evidence of features that have a high likelihood to carry clinical relevance in this dataset -- especially where they align with prior expectations about how features ought to interrelate and what is known about each participant's diagnosis. Thus this work has helped to establish background that can guide a principled set of next research steps, including hypotheses about which features might have the greatest predictive power in modeling self-report survey scores (section \ref{subsec:diary-ema}) or display the most compelling trends in the respective subject case reports (section \ref{subsec:diary-case-study}). 

Because word count and words per sentence can have such a strong impact on other feature distributions, they will remain a critical piece of any model. At the same time, there were highly significant distributional differences in these features, especially word count, and there is good reason to believe they have inherent psychiatric importance; it is thus natural that these verbosity metrics ought to be a major piece of the upcoming modeling regardless of subject. What is most relevant to consider here are the other diary features that showed promising variance between participants even when taking into account the expected effect of the observed verbosity differences. 

\noindent That results in the following list of features that are hypothesized to be of the highest salience for each highlighted subject:
\begin{itemize}
    \item 3SS93 
    \begin{itemize}
        \item Abnormally low speech fraction
        \item Frequent use of most disfluency types - nonverbal edits, restarts, and to a slightly lesser extent repeats
        \item Lower mean sentence sentiment
    \end{itemize}
    \item 8RC89
    \begin{itemize}
        \item Abnormally high pause-adjusted speaking rate
        \item Frequent use of verbal edits, but not necessarily other disfluency types
        \item Higher mean word uncommonness and maximum sentence incoherence (role of somewhat lower mean words per sentence in these observations requires further investigation)
    \end{itemize}
    \item 5BT65
    \begin{itemize}
        \item Abnormally low pause-adjusted speaking rate
        \item Missing an expected big positive tail in mean sentiment spread based on typical diary lengths (there is also prior reason to look for sentiment relevance in this subject)
        \item Maximum incoherence distribution closely matches BLS, against the expectations set by the many fewer sentences that it should be lower (no prior hypothesis about incoherence, but an interesting observation worth a closer look nonetheless)  
    \end{itemize}
\end{itemize}
\noindent Recall that 5BT65 had very high MADRS variance relative to the study population, while all three highlighted subjects had high PANSS negative variance relative to most of the study population, with 3SS93 having one of the higher mean PANSS negative scores. 8RC89 had notably higher variance in PANSS positive scores than the other two highlighted subjects, but only moderate variance relative to the study population.

It is worth noting again that 5BT65 had a number of diaries excluded from these analyses because they were so short as to be filtered out by the 15 second duration minimum. Of the diaries that were included from 5B, many were still quite short. This impacts the amount of information that one can reasonably expect to gain from features like mean sentence sentiment, and especially the disfluency counting features that are somewhat rare on a per sentence basis. It's therefore not surprising that 3SS93 and 8RC89 had a larger set of more robust transcript feature differences that warrant further study. On the other hand, there is good reason to believe that submission metadata and diary length alone are highly clinically relevant for 5BT65, moreso than the other subjects. \\

\noindent The hypotheses listed here will be revisited in light of the EMA modeling and case study results once those are reported on (sections \ref{subsec:diary-ema}-\ref{subsec:diary-case-study}), but first correlations between the considered features should be investigated to further inform the hypotheses and strengthen the model planning process.

\subsubsection{Correlation structure of diary features} 
\label{subsubsec:diary-corrs}
A better understanding of the empirical relationship between diary features can assist in identifying features that are largely redundant or features that may benefit from additional normalization. Using a feature set with minimal correlation is desirable for certain interpretable modeling applications; more generally, understanding existing input correlations is important for results interpretation. On the other hand, correlation between features may itself vary between patients or within a patient over time, and could contain clinically relevant information that is not apparent from any of the features in isolation. This phenomenon is leveraged in e.g. fMRI research, where concurrent activation of multiple brain regions is itself a feature of interest. While development of such features is possible through good data science work on a large dataset, it is likely that near future advances in audio journal analysis will focus on methodologies that are grounded in clinical hypotheses. 

\begin{figure}[h]
\centering
\includegraphics[width=\textwidth,keepaspectratio]{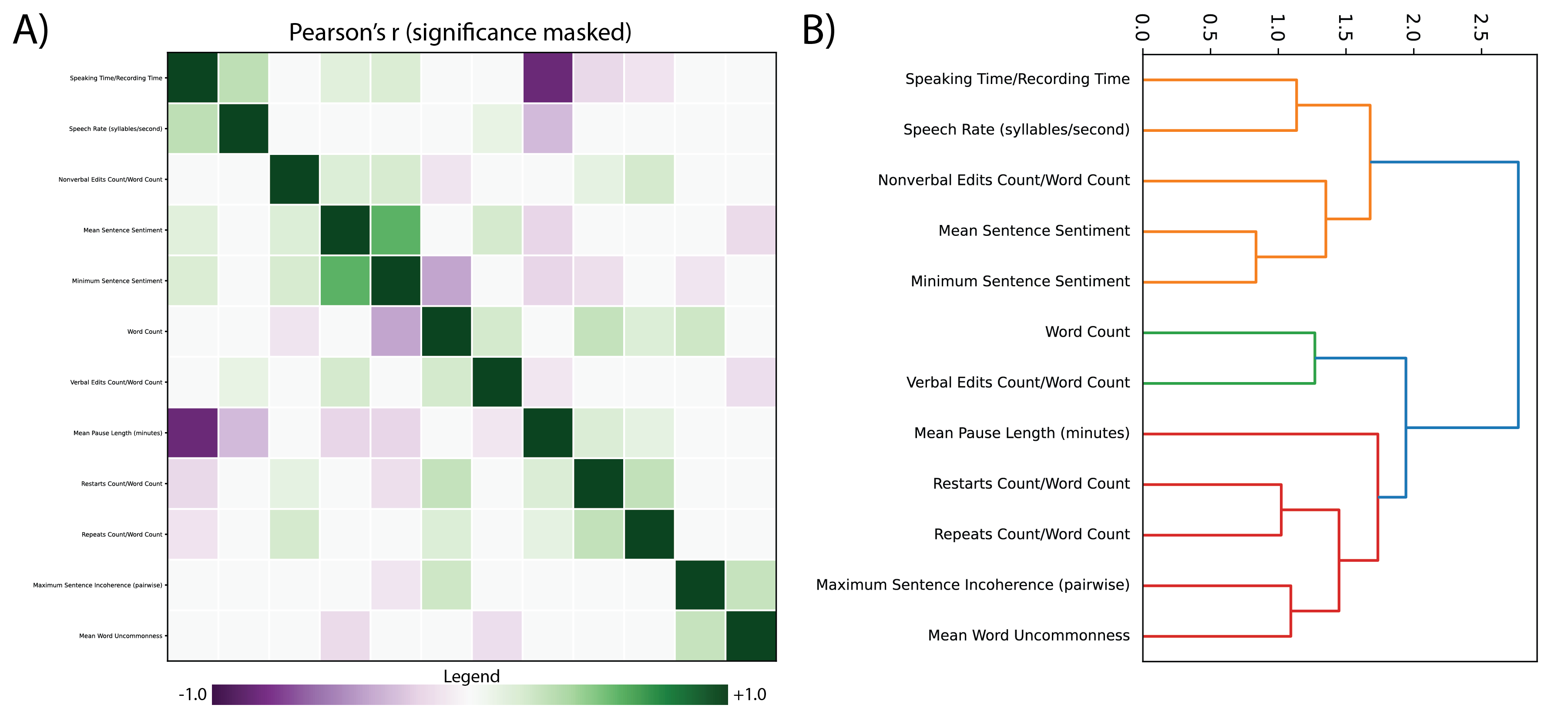}
\caption[Pearson correlation structure of curated diary features in final BLS dataset.]{\textbf{Pearson correlation structure of curated diary features in final BLS dataset.} For each pairing of the select 12 diary-level features, Pearson's $r$ was computed across the final BLS dataset ($n=8398$) using the scipy.stats python package. A significance-masked correlation matrix was constructed using the protocol detailed above, and visualized with the matplotlib PrGn colormap bounded between -1 and 1 (A). As mentioned, features were ordered based on the results of the described clustering algorithm on Pearson distances. Clustering results are presented here via dendrogram (B).}
\label{fig:pearson-diary-final}
\end{figure}

As such, I will now characterize correlation structures of diary-level features in the BLS journal dataset (Figure \ref{fig:pearson-diary-final}), and connect them back to the feature interpretation commentaries started above. For select feature pairings, I will further characterize their relationship via scatter plots in the supplemental section \ref{sec:scatters}. I will also compare correlation structures in participant-specific subsets of the dataset (as defined in section \ref{subsubsec:diary-pt-dists-comps} above) to the BLS-wide correlations. The methods I used for feature clustering and creation of correlation matrices can be found in supplemental section \ref{sec:sup-corr-meth}. As results were largely similar between Pearson and Spearman methods here, I focus the presentation in the main text on Pearson correlation.

The strongest correlation across the dataset was a negative one between "Speaking Time/Recording Time" and "Mean Pause Length" (Figure \ref{fig:pearson-diary-final}A). Coupled with the lack of unique conclusions drawn from the mean pause length distributional analysis, it is worth considering whether mean pause length should be included in further analyses when we are trying to be efficient about the number of inputs included. Similarly, amongst positive correlations the clear strongest was the relationship between mean and minimum sentiment scores (Figure \ref{fig:pearson-diary-final}A). This is not surprising given how those features are linked by definition, but in conjunction with the interpretability difficulties with minimum sentiment distribution discussed above, the "Minimum Sentence Sentiment" feature is another that would be at the top of the list for pruning if a reduction in the final features were desired. 

Considering relationships with word count, there were 4 features with a significant correlation after multiple testing correction that exceeded $4\%$ of variance explained (i.e. Pearson's $\lvert r \rvert > 0.2$): a negative correlation between word count and minimum sentence sentiment, and positive correlations between word count and verbal edits per word, restarts per word, and maximum sentence incoherence. There were an additional 2 features with statistically significant word count correlations that exceeded $1\%$ of variance explained: a negative relationship with nonverbal edits per word and a positive relationship with repeats per word (Figure \ref{fig:pearson-diary-final}A). However these correlations were so slight it is unclear whether they carry any importance in understanding the dataset, or may simply be an artifact. In the case of repeats, a stronger relationship emerged in the Spearman correlation, which does suggest there is likely some covariance of interest between repeats per word and total word count in this dataset.

The correlation of word count with the minimum/maximum features was expected based purely on the mathematical definitions, as discussed with the simulation results of the preceding subsection. The stronger correlation of word count with the "verbal edit" and "restart" subtypes of linguistic disfluency is consistent with observations made in the BLS clinical interview transcript dataset, to be reported on in chapter \ref{ch:2} (\ref{sec:disorg}). The fact that there remained a correlation between disfluencies and word count even after normalizing the disfluency count by the number of words is discussed at greater length there, but it should be kept in mind when interpreting results throughout this section as well. Note that the significant positive correlation observed between repeats and restarts, with lesser correlations (if any) found between other disfluency types and with nonverbal edits found to be essentially disjoint, is also consist with the interview transcript analysis to be presented. 

As far as other feature relationships, it is again consistent with prior expectations (see section \ref{subsec:diary-val}) that repeats per word and restarts per word were significantly positively correlated with mean pause length and significantly negatively correlated with speech fraction (Figure \ref{fig:pearson-diary-final}A). By contrast, an interesting deviation from expectations in these results was the significant positive correlation (with Pearson's $r > 0.25$) between mean word uncommonness and maximum sentence incoherence (Figure \ref{fig:pearson-diary-final}A). While it is not necessarily against original priors, this relationship was entirely absent in the interview dataset to be explored in chapter \ref{ch:2} -- not only was there no relationship on the smaller sample size interview summary level, but there was also no detectable correlation on the level of sentence scores. 

Because the interview analysis used the mean incoherence score instead of the max to summarize each interview, it is not possible to draw a direct comparison. However, as there was no correlation in these sentence-level scores for patient speech in clinical interview recordings, it remains of interest for further investigation. It is of course possible for a relationship to not exist on the sentence level but exist on the transcript level, and this might in fact be a sign of a salient underlying state variable affecting multiple parts of a recording. It is also possible that there are differences between the clinical interview and daily diary contexts that would cause the relationship between these features to vary between the two cases, even on a per sentence basis. Assessing sentence-level correlations of journal features would thus be a worthwhile future direction, though out of scope for this thesis. 

Ultimately, the alignment between these results increases confidence in interpretation of the disfluency conclusions drawn in both chapters, though of course for broader generalization it is important to emphasize that there was overlap in the subjects considered for the two datatypes. More broadly, it is comforting that correlations between word count and all other considered features across the final diary dataset were entirely negligible, thus indicating that any clinically relevant trends found downstream are unlikely to be explainable by an indirect verbosity effect. 

\FloatBarrier

\paragraph{Looping in sentence structure to correlational results.}
Since the sentence structure feature was added for consideration based on results with the original feature set, I will now characterize correlations with mean words per sentence in the final BLS dataset, along the lines of what was done for the other selected features above. Both Pearson and Spearman correlations between mean words per sentence and key related diary features are reported in Table \ref{table:words-per-sentence-corrs}. It is unsurprising that word count and words per sentence had a relatively high positive correlation, though words per sentence still provided important context outside of word count. It is comforting that sentence count had negligible correlation with words per sentence (Table \ref{table:words-per-sentence-corrs}) -- if variance in sentence length were driven primarily by a subset of randomly assigned transcribers performing splitting more lazily, we would expect longer sentences to more strongly negatively correlate with sentence count. 

\begin{table}[!htbp]
\centering
\caption[Correlation between mean words per sentence and other pipeline diary summary features of interest.]{\textbf{Correlation between mean words per sentence and other pipeline diary summary features of interest.} Across the final considered BLS journal dataset (minus the 15 transcripts with a single sentence, due to the inclusion here of Sentence to Sentence Incoherence), both Pearson and Spearman correlation values are reported between the mean words per sentence measure and a select set of other considered journal features, with each row representing one such compared feature. In each main column, the corresponding $r$-value is reported with $p$-value bound in parentheses. Based on these results, the Words per Sentence feature was utilized in downstream analyses, while Sentence to Sentence Incoherence was not further considered.}
\label{table:words-per-sentence-corrs}

\begin{tabular}{ | m{8cm} | m{2.25cm} | m{2.25cm} |  }
\hline
\textbf{Feature correlated with Words per Sentence} & \textbf{Pearson's $r$ \newline ($p$-value)} & \textbf{Spearman's $r$ \newline ($p$-value)} \\
\hline\hline
Total Word Count & $0.352$ \newline ($< 10^{-240}$) & $0.442$ \newline ($< 10^{-300}$) \\
\hline
Total Sentence Count & $-0.059$ \newline ($< 10^{-7}$) & $0.039$ \newline ($< 0.0005$) \\
\hline\hline
Mean Sentence Sentiment & $0.127$ \newline ($< 10^{-30}$) & $0.118$ \newline ($< 10^{-26}$) \\
\hline
Minimum Sentence Sentiment & $-0.111$ \newline ($< 10^{-23}$) & $-0.191$ \newline ($< 10^{-69}$) \\
\hline\hline
Mean Word Uncommonness & $-0.266$ \newline ($< 10^{-130}$) & $-0.296$ \newline ($< 10^{-165}$) \\
\hline
Maximum Sentence Incoherence & $-0.317$ \newline ($< 10^{-190}$) & $-0.309$ \newline ($< 10^{-180}$) \\
\hline
Sentence to Sentence Incoherence & $-0.733$ \newline ($< 10^{-300}$) & $-0.747$ \newline ($< 10^{-300}$) \\
\hline
\end{tabular}
\end{table}

The correlation results (Table \ref{table:words-per-sentence-corrs}) additionally supplied more context for the discussion of normalization of feature comparisons by verbosity. While the random simulation results reported above allowed for adjustment of results based on subject-specific variation in diary sentence count, and the generally small correlation of final features with total word count (see section \ref{subsubsec:diary-corrs}) also later helped to adjust expectations on the potential independent clinical relevance of participant-dependent distributional differences, these results did not sufficiently address the role of sentence length in NLP feature outcomes. Here we can see that although sentence length does have some effect on sentiment values more generally, in the present diary dataset and on the summary level considered it was a quite minimal impact. On the other hand, the relationship between words per sentence and the word2vec-derived features was somewhat stronger, and while not prohibitively so, should be kept in mind when performing and interpreting subsequent analyses. \\

\FloatBarrier

\begin{figure}[h]
\centering
\includegraphics[width=\textwidth,keepaspectratio]{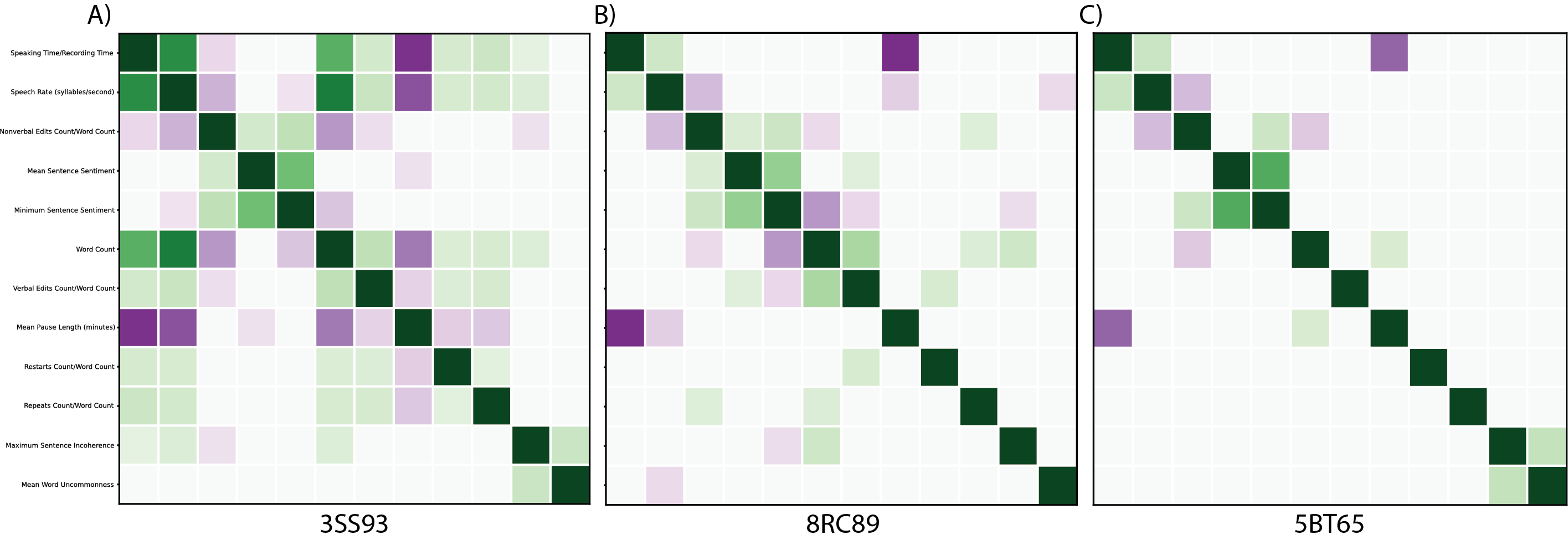}
\caption[Patient-specific Pearson correlations in select BLS subjects.]{\textbf{Patient-specific Pearson correlations in select BLS subjects.} To identify potential between participant heterogeneity in feature relationships, the same methodology as in Figure \ref{fig:pearson-diary-final}A was repeated using only the diaries from a particular subject. The chosen subjects were the same three highlighted participants as before: 3SS93 (A), 8RC89 (B), and 5BT65 (C). Note that the features are ordered the same in all three matrices, using the BLS-wide Pearson clustering results of \ref{fig:pearson-diary-final}B.}
\label{fig:pearson-diary-by-pt}
\end{figure}

\paragraph{Subject-specific correlations.}
To investigate any potential subject-dependent differences in feature correlation, I created Pearson correlation matrices using the same methodology as the BLS-wide matrix (Figure \ref{fig:pearson-diary-final}A), applied to the datasets of participants 3SS93 (Figure \ref{fig:pearson-diary-by-pt}A), 8RC89 (Figure \ref{fig:pearson-diary-by-pt}B), and 5BT65 (Figure \ref{fig:pearson-diary-by-pt}C). Note that the features in each of these matrices are ordered using the same study-wide ordering produced by the Pearson clustering described above (Figure \ref{fig:pearson-diary-final}B). 

Recall that study-wide there were 8398 data points included in the correlational analysis, while for the subject-specific analyses there were only 981 time points for 3SS93, 510 time points for 8RC89, and 344 time points for 5BT65. This makes it more difficult for correlations to remain statistically significant after multiple testing correction in the subject-specific Pearson matrices (Figure \ref{fig:pearson-diary-by-pt}), and as such the lack of a significant correlation here cannot be strongly interpreted. Nevertheless, a number of relationships of note were found by looking at correlation structure within the individual participants.

3S (Figure \ref{fig:pearson-diary-by-pt}A) had a much stronger positive correlation between "Speaking Time/Recording Time" and "Speech Rate" than was observed across the study, which was completely expected because of the abnormally long amount of time 3S spent paused in their recordings. This additionally resulted in a strong positive correlation between total word count and both of those 2 features in 3S's journals, while word count had no significant correlation with the 2 across BLS (Figure \ref{fig:pearson-diary-final}A). Because 3S often submitted recordings of maximum duration, that result is also unsurprising. With more strong word count correlations, there were thus more strong correlations within 3S's dataset in general compared to what was seen BLS-wide. 

Interestingly, a statistically significant (after Bonferroni correction) negative correlation was observed between speaking rate and nonverbal edits per word across all 3 of the considered subjects (Figure \ref{fig:pearson-diary-by-pt}), despite no such correlation appearing study-wide (Figure \ref{fig:pearson-diary-final}A). It could be the case that between-subjects variation dominated within-subjects variation in this relationship, although amongst the highlighted participants that does not appear to be the case, as 3SS93 had both an exceptionally low uncorrected speech rate stat distribution (Figure \ref{fig:speech-rate-per-pt}A) and a notably high distribution of nonverbal edits per word (Figure \ref{fig:disfluency-props-per-pt}A), which is in line with the negative correlation trend seen within each patient as well. Closer investigation into the other BLS participants would be necessary to better understand how this difference arose then. 

It is worth noting that mean word uncommonness and maximum sentence incoherence were significantly positively associated within the journals of both 3S (Figure \ref{fig:pearson-diary-by-pt}A) and 5B (Figure \ref{fig:pearson-diary-by-pt}C), as they were study-wide (Figure \ref{fig:pearson-diary-final}A), with Pearson's $r$ in all cases $> 0.25$ -- yet this relationship did not meet the Bonferroni significance threshold in 8RC89 (Figure \ref{fig:pearson-diary-by-pt}B), with $r < 0.15$ and $p=0.000775$. It is unclear how much meaning (if any) can be ascribed to this, however given that 8RC89 displayed the most compelling distributional differences against the study in both of these features (Figure \ref{fig:coherence-props-per-pt}), it is promising to find that they actually appeared less correlated in 8R. \\

\noindent For the disfluency rate features, there were a number of interesting subject-specific differences, particularly in how strongly each category of disfluency linearly correlated with total word count (Table \ref{table:disfluency-corrs-comp}). 

\begin{table}[!htbp]
\centering
\caption[Strength of linear relationship between total word count and normalized disfluency count varies by disfluency category in subject-dependent manner.]{\textbf{Strength of linear relationship between total word count and normalized disfluency count varies by disfluency category in a subject-dependent manner.} As discussed for the BLS journals above and with respect to interview transcriptions in chapter \ref{ch:2}, it is expected that some relationship between disfluency counts and verbosity will remain even after the disfluency features are normalized by a verbosity measure -- in part due to natural speech patterns, but potentially also in part due to clinically relevant underlying factors. In this table, I report the Pearson correlation values between the occurrence rate per word of each type of disfluency and the total word count of the corresponding diary transcript. Each row thus corresponds to a particular category of disfluency. The columns, from left to right, then present this $r$ for the entire final BLS journal set ($n=8398$), the subset submitted by 3SS93 only ($n=981$), the subset submitted by 8RC89 only ($n=510$), and the subset submitted by 5BT65 only ($n=344$). Correlations that were statistically significant after Bonferroni correction in the corresponding matrix are marked with an *. See Figure \ref{fig:pearson-diary-final}A for the full feature set of Pearson correlations across BLS, and Figure \ref{fig:pearson-diary-by-pt} for the subject-specific Pearson correlation matrices. \newline To best understand which disfluency categories are most salient for each subject ID, the correlational results reported here can be considered in conjunction with the comparison of disfluency distributions between participants presented in Figure \ref{fig:disfluency-props-per-pt}.}
\label{table:disfluency-corrs-comp}

\begin{tabular}{ | m{5cm} || m{1.75cm} | m{1.75cm} | m{1.75cm} |  m{1.75cm} |  }
\hline
\textbf{Disfluency category \newline (count per word): \newline Pearson w/ total words \#} & \textbf{BLS $r$} & \textbf{3SS93 $r$} & \textbf{8RC89 $r$} & \textbf{5BT65 $r$} \\
\hline\hline
Nonverbal edits & $-0.115^{*}$ & $-0.446^{*}$ & $-0.157^{*}$ & $-0.243^{*}$ \\
\hline
Verbal edits & $0.221^{*}$ & $0.299^{*}$ & $0.383^{*}$ & $0.145$ \\
\hline
Repeats & $0.168^{*}$ & $0.207^{*}$ & $0.177^{*}$ & $0.038$ \\
\hline
Restarts & $0.283^{*}$ & $0.186^{*}$ & $0.122$ & $0.179$ \\
\hline
\end{tabular}
\end{table}

Nonverbal edits per word had a notably more negative correlation with word count in participants 5BT65 and especially 3SS93 than across BLS at large. This is interesting because 3SS93 used nonverbal edits significantly more frequently than was typical in BLS (Figure \ref{fig:disfluency-props-per-pt}A), but 3SS93 also submitted journals with significantly more words than was typical for BLS (Figure \ref{fig:word-count-per-pt}A). Similarly, an abnormally high positive correlation between word count and verbal edit rate was observed specifically in participant 8RC89 (Table \ref{table:disfluency-corrs-comp}). 8R used verbal edits at a higher rate than was seen across BLS more generally (Figure \ref{fig:disfluency-props-per-pt}C) -- in contrast with 3SS93 who submitted longer journals and had higher rates in other disfluency categories, but used verbal edits much less commonly than 8R.  

The strength of linear association between repeat rate and verbosity remained within a similar range for the subject-specific and study-wide correlations, with the exception of 5B (Table \ref{table:disfluency-corrs-comp}). This, along with the smaller $r$ values for repeats more generally, could in part be explainable by the especially high rate of journals with 0 repeats compared to the other disfluency categories (Figure \ref{fig:disfluency-props-per-pt}). However it does not account for the whole story, as 5B submitted an even higher number of diaries with 0 verbal edits, yet retained a higher Pearson $r$ there. Unlike the other disfluency categories, the association between restart rate and verbosity was notably weaker within each of the subject-specific datasets considered than it was for BLS as a whole (Table \ref{table:disfluency-corrs-comp}). At the same time, 3S committed restarts at a significantly higher rate than typical for the dataset (Figure \ref{fig:disfluency-props-per-pt}G), even when accounting for some expected increase due to their generally longer submissions. 

In all of BLS, restarts per word and repeats per word were significantly positively correlated, but neither were correlated with verbal edits per word (Figure \ref{fig:pearson-diary-final}A). Meanwhile, the correlation between repeats and restarts was much weaker in each of the individual participants highlighted (Figure \ref{fig:pearson-diary-by-pt}), yet in both 3SS93 and 8RC89 there was a significant correlation between verbal edits and restarts (Figure \ref{fig:pearson-diary-by-pt}A/B). At the same time, verbal edits were easily the disfluency category that 3S showed the least distributional difference from BLS on the whole, and verbal edits were simultaneously the only disfluency category that 8R showed a strong distributional difference in. It is important to keep in mind that although these correlational differences can be of true scientific interest, they are also for the most part small correlations in absolute terms.

Consideration of the subject-specific factors in correlation structure of disfluency features has indeed lead to a number of interesting observations and new research questions. A subset of these questions will be addressed in subsequent analyses in this chapter (including both EMA modeling and case report work), as well as through the analysis of disfluency usage in clinical interviews recorded for BLS, to be covered in chapter \ref{ch:2}. Within the discussion at the end of this chapter (section \ref{sec:discussion2}), I will thus present final conclusions drawn from this work about disfluencies, along with the updated hypotheses that stem from them and other open questions that remain. \\

\FloatBarrier

\paragraph{Correlation results further inform hypotheses for modeling project.}
In summary, most features had quite small correlation with total word count across BLS, which bodes well for the potential independent information that they contain, and therefore their use in straightforward clinical modeling applications. When the correlation results reported here are taken together with the distributional results of the preceding subsection \ref{subsubsec:diary-pt-dists-comps}, we can consider a handful of the features - mean pause length, minimum sentence sentiment, and the additional test of between sentence incoherence - to have lower probability of contributing meaningful independent value to a model of the relationship between diary features and clinical outcomes. Furthermore, we can see that the pause-adjusted version of the speaking rate feature is likely more useful than the raw version returned by the pipeline, and that mean words per sentence appears to have great potential clinical utility for this dataset (and regardless should be included in order to properly account for the role of transcript verbosity structure in any features that are highly weighted by a model). As such, slight adjustments to the list of final considered features will be made for the EMA modeling application of section \ref{subsec:diary-ema}, demonstrating the upside of carefully characterizing core journal features before feeding into a model.

On the level of subject specific comparisons, a number of linear correlational differences ought to be kept in mind when further evaluating the distributional differences that were observed in the preceding subsection. Although correlation magnitudes were mostly low in absolute terms, such that further investigation is required to really interpret these findings, the effects did appear to go robustly beyond what would be plausibly attributed to meaningless statistical noise. As such, participant-dependent correlational differences will also be revisited when later interpreting final modeling results.

\noindent The most salient correlational comparison results were as follows:
\begin{itemize}
    \item The positive correlation between mean word uncommonness and maximum sentence incoherence was notably lower magnitude in 8RC89's dataset than it was for all of BLS or for either of the other two highlighted subjects.
    \begin{itemize}
        \item 8RC89 was the highlighted subject that had the strongest distributional differences in both mean word uncommonness and maximum sentence incoherence, with higher overall values observed. 
    \end{itemize}
    \item The negative correlation between word count and nonverbal utterance rate was much higher magnitude in 3SS93's dataset than it was for all of BLS or for either of the other two highlighted subjects.
    \begin{itemize}
        \item 3SS93 had substantially more total words and more frequent use of nonverbal edits than the rest of BLS or either of the other two subjects. 
    \end{itemize}
    \item The positive correlation between word count and verbal filler rate was notably higher magnitude in 8RC89's dataset than it was for all of BLS or for either of the other two highlighted subjects.
    \begin{itemize}
        \item 8RC89 used verbal edits at a notably higher rate than the rest of BLS, and was the only highlighted subject with a clear difference in verbal edit rate. The word count distribution of 8RC89's submissions was similar to the study-wide distribution. 
    \end{itemize}
    \item For all three highlighted subjects, the positive correlation between word count and frequency of restart use was notably lower than the study-wide correlation.
    \begin{itemize}
        \item 3SS93 also used restarts at a substantially higher rate than was reflected by the BLS-wide distribution.
    \end{itemize}
\end{itemize}
\noindent These correlational differences were thus different for different features and participants, independent of the strength and directionality of the corresponding distributional differences previously emphasized. An interesting scientific question raised then is whether potential differences in upcoming EMA modeling or case report results across the hypotheses of subsection \ref{subsubsec:diary-pt-dists-comps} will be explainable by the observations here.

It is also worth noting that the study-wide correlations were largely similar for Pearson and Spearman methods, but with some differences that primarily affected the disfluency features. This is not surprising given the nature of the disfluency distributions, which were largely not Gaussian-like and had a clear nonlinearity introduced due to the high number of transcripts that contained no occurrences of a particular disfluency category. Though beyond the scope of this thesis, future research on disfluencies should take care to account for possible nonlinear relationships in their modeling. \\

\noindent As a final step before proceeding to analyses of clinical outcomes, I will next characterize properties of diary data availability over time in the BLS dataset.

\subsubsection{Temporal dynamics of diary features} 
\label{subsubsec:diary-time}
Besides evaluating distributions of features and the relationships between them, another form of analysis that can be performed with an audio journal feature set alone is the analysis of temporal dynamics. The ability to accurately predict features on one day from features on previous days will be a critical building block towards eventual prediction of clinically relevant states \emph{before they happen}. Additionally, such models would have strong potential to further our understanding of the properties of audio diaries -- how they vary within an individual, how they vary between individuals, how they relate to each other, etc. This is especially true if a more interpretable modeling method is used, but it could also be true of a black box model. For example, it is possible that prediction of features from one day to the next is much harder in some individuals than in others, and this may in fact itself be related to clinically relevant underlying variables.

While modeling the dynamics of extracted journal features is beyond the scope of the present thesis, it is imperative that any such future work includes a good understanding of the role of data missingness. Effective small recurrent neural network (RNN) architectures like that described in chapter \ref{ch:4} could be directly applied to sequential prediction of audio diary features in a complete dataset, but it is less clear how to deal with various patterns of missing days that we have observed. One straightforward first pass on such a prediction problem could be to isolate all pairs of journals submitted on sequential days only, and perform a more typical supervised learning approach with the present day's features as input and the next day's features as labels; however this does not have the same expressive power that a proper sequence prediction scheme would. To lessen some of the roadblocks caused by missingness, a weekly journal summary could be considered instead of daily features -- but it would not eliminate the roadblocks, and it would also entail a sacrifice of both temporal resolution and statistical power. The exact approach to take will of course always depend on the particulars of the dataset and the scientific aims, but regardless the development of learning algorithms that can optimally handle data missingness remains an area of need. 

To contribute to the early groundwork necessary for temporal analysis of daily audio diaries, and in particular to set up future work in this domain for the BLS dataset, I characterized journal submission dynamics across the study. Beyond providing essential information on missingness for planning a temporal modeling project, the results presented in this section may also inform future study design. As such, I review not only participation over time, but also \emph{engagement} over time, estimated here by transcript word count. It is important for a successful longitudinal study to maintain quality participation over an extended period in a large enough proportion of participants. This work serves as a strong proof of concept that it is indeed very feasible to accomplish that with the audio journal format. \\

\begin{figure}[h]
\centering
\includegraphics[width=0.9\textwidth,keepaspectratio]{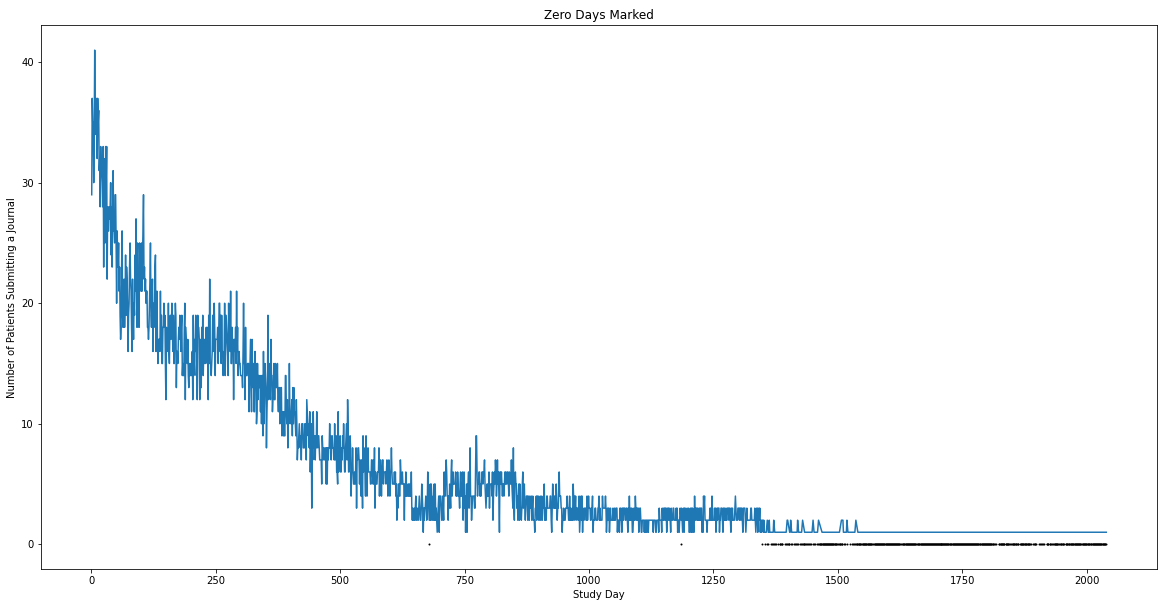}
\caption[Number of BLS participants submitting a journal by time in the study.]{\textbf{Number of BLS participants submitting a journal by time in the study.} For each participant ($n=65$ with diary submission), study day is measured relative to their day of consent, which is marked as day 1. All journal submissions (capped at one per day), regardless of quality or length, were counted across all subjects for each possible study day to generate the plot of submission numbers over the course of the study period shown here. Days with 0 submissions are excluded from the line plot but marked with black dots below. Note the x-axis here spans more than 5 years.}
\label{fig:diary-submit-day-counts}
\end{figure}

\paragraph{Submission dynamics in BLS.}
As reviewed in section \ref{subsubsec:diary-metadata}, 65 BLS participants submitted at least one usable daily audio journal. In this section, I consider the submission patterns over time from all such subjects, relative to each subject's corresponding day of consent. I begin by considering all diary submissions regardless of length or transcribability. For each study day, where 1 is the day of consent, the number of participants submitting a recording is plotted in Figure \ref{fig:diary-submit-day-counts}. Overall, participation dropped off over time in the study - as expected - but after an initial steep descent submissions stabilized for a time, with consistently $\geq 10$ subjects submitting a diary every day over the first year of the study. Furthermore, at least one participant submitted a journal on almost all study days in the first 3.5 years of enrollment, and submissions occurred from some participants well past 5 years into the study (Figure \ref{fig:diary-submit-day-counts}).

To better characterize the submission dynamics, I next considered recordings over time broken down by subject. The distribution of participation was fairly bimodal -- of the 65 patients, 30 submitted fewer than 40 audio diaries with duration at least 15 seconds, while another 25 submitted more than 100 audio diaries with duration at least 15 seconds. Recall that $n=25$ makes up more than $\frac{1}{3}$ of the recruited BLS participants, indicating strong journal participation rates from a good portion of subjects. As BLS is a highly multimodal study that recruited patients for a number of optional study components, it is likely that a project more focused on journals could obtain even better participation rates. This also provides a good proof of concept for the addition of journals to the broader digital psychiatry data collection study format. 

\begin{figure}[h]
\centering
\includegraphics[width=\textwidth,keepaspectratio]{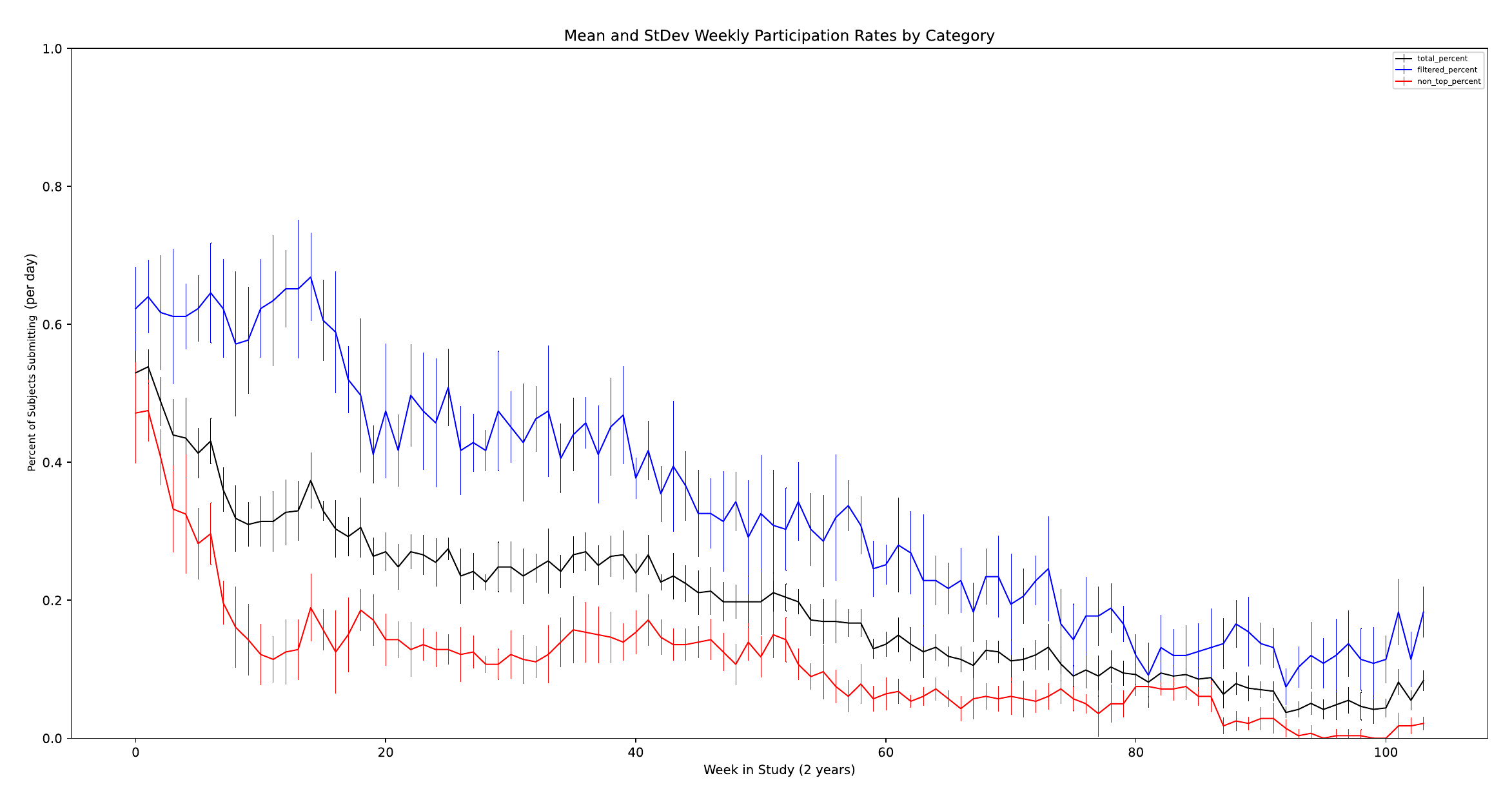}
\caption[Average fraction of participants submitting a daily journal per week in the study - characteristics of top 25 versus overall population.]{\textbf{Average fraction of participants submitting a daily journal per week in the study - characteristics of top 25 versus overall population.} The curve in Figure \ref{fig:diary-submit-day-counts} was divided by $n=65$ to obtain the fraction of diary-enrolled participants submitting a journal each study day. Here, we focus on only the first two years of collection period, thus limiting the considered days to 730. A seven day sliding window (no overlap) was applied to the two year fraction time-series, computing the mean and standard deviation to form the black line and error bars in this graph, thus depicting a smoothed weekly version of daily percent participation over time for BLS. The 25 participants submitting at least 100 diaries of duration $\geq15$ seconds (blue) were also split off from the other 40 participants (red), to compute the same time-series now restricted to only those two disjoint populations. Note that when calculating the blue curve, participation was only counted if the submitted journal exceeded 15 seconds. In the first couple of weeks the groups are only modestly different, but participation drops off rapidly in the red curve starting with week 3.}
\label{fig:diary-submit-weekly-avgs}
\end{figure}

When evaluating participation over time from the top 25 diary participants against the overall study population (Figure \ref{fig:diary-submit-weekly-avgs}), submission rates were fairly similar within the first two weeks after enrollment, but a sharp decline began in week 3 for BLS at large. Meanwhile, the top 25 subset sustained participation for a period of over 3 months before a noticeable drop around week 15, which then leveled off to a slow decline from week 20 through the end of year 2. At the end of year 1, the top 25 subset was still averaging $>2.5$ daily recording submissions \emph{per patient} each week (Figure \ref{fig:diary-submit-weekly-avgs}).    

Although maintained participation in weeks 3 to 4 was predictive of high participation rate from that subject for a much longer study period (Figure \ref{fig:diary-submit-weekly-avgs}), it is unclear whether additional measures to incentivize submissions during this period would help to lock in more longer term participants, or if it would just slightly delay drop off from the less engaged subjects. Future studies can use the pipeline monitoring systems described in this chapter to test new strategies for improving journal participation. \\

To identify potential for modeling of features from one day to the next in a straightforward supervised framework, I looked at availability stats for sequential submissions of at least 15 seconds from the top 25 diary participants. Here I focused primarily on submissions within the first year, as a year of patient data is a strong target for most studies, and one would expect sequential submissions to eventually sharply drop off for any subject if a cutoff were not set. 

Within the first year of enrollment, there were 4408 total submissions of sufficient length from across all top participants. Of those 4408 diaries, 2868 had a diary of sufficient length submitted by the corresponding participant the next day as well -- thus providing a good number of data points for future study of how features one day predict features the next day. These stats amongst the top 25 BLS subjects in their first year represented an $\sim 65\%$ probability of a participant submitting a diary $\geq 15$ seconds on a particular day, given that they submitted a diary $\geq 15$ seconds the previous day. Additionally, the base rate of participation in this subset was $\sim 48\%$, which is quite good for a daily task distributed over an entire year. 

However, there was significant behavioral heterogeneity in submission patterns from day to day, which will be a factor to consider in developing models for predicting the features of one day from those of the previous. Further context on participant-specific submission dynamics can be found in supplemental section \ref{subsec:participation-table-sup}. Other potential confounding effects on the sequential data, such as seasonality or day of week, will also require closer consideration before deciding how to proceed with a temporal dynamics modeling plan. \\

\FloatBarrier

\paragraph{Engagement over the course of a study.}
An additional concern besides maintaining study participation is maintaining \emph{quality} of participation, keeping subjects engaged in recording diaries over long timescales. Although a couple subjects were identified who routinely submitted diaries lacking real content, presumably to game the study compensation system, they were easily filtered out with a short duration minimum (section \ref{subsubsec:diary-val-qc}). Still, it remained unclear whether good faith participants might have put decreasing effort into recording diaries over time in the study, while continuing to make submissions. Fortunately, this did not appear to be the case.

\begin{figure}[h]
\centering
\includegraphics[width=0.8\textwidth,keepaspectratio]{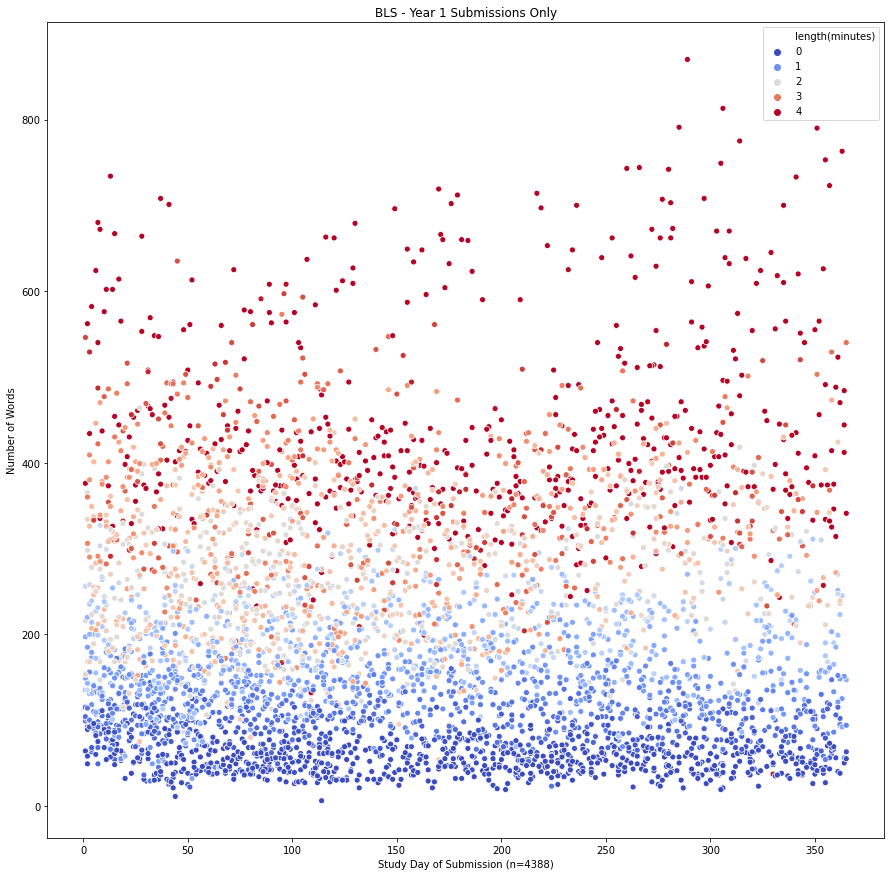}
\caption[Journal word counts, a proxy for engagement, remained steady over first year of data collection.]{\textbf{Journal word counts, a proxy for engagement, remained steady over first year of data collection.} The top 25 BLS participants described submitted 4388 diaries within their first year that were both transcribable and of duration $\geq15$ seconds ($n=4388$). Each of those diaries are a point here, with x coordinate corresponding to study day of submission and y coordinate corresponding to the word count of the transcript. Each point is also colored from blue to red to indicate recording length from lowest to highest (4 minutes). Diary content fortunately does not decrease over this time as one might have expected.}
\label{fig:engagement-time-year}
\end{figure}

To assess the question, I looked at the number of words in transcribed diaries over time in the study, as a proxy for engagement. The analysis was focused on the first year of enrollment for the above mentioned top 25 journal participants, with diaries of at least 15 seconds. Both Pearson (linear) and Spearman (rank) correlation between study day number and the number of words in the transcript were of magnitude $< 0.02$ with $p > 0.15$. Therefore, there was no relationship between time in the study and the word count of the corresponding journal, provided one was submitted (Figure \ref{fig:engagement-time-year}). 

\begin{figure}[h]
\centering
\includegraphics[width=0.8\textwidth,keepaspectratio]{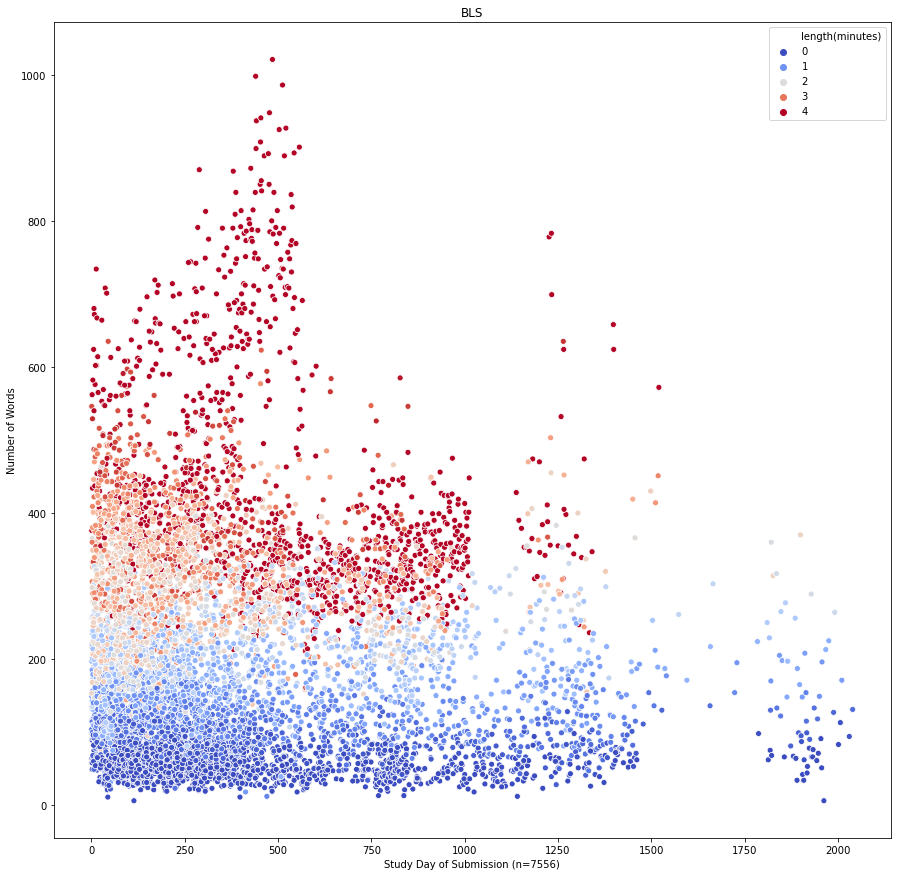}
\caption[Journal word counts across full 5+ years of possible BLS study period.]{\textbf{Journal word counts across full 5+ years of possible BLS study period.} The methodology of Figure \ref{fig:engagement-time-year} is repeated here, now considering all submitted diaries rather than restricting to the first year in the study. Word counts remain good through year 2. A dropoff is eventually observed, but practically speaking we did not find declining engagement to be an issue independent of declining participation, by this proxy.}
\label{fig:engagement-time-all}
\end{figure}

Moreover, even when considering diaries from the top 25 patient set over all time in the study, hardly any relationship was found between word count and study day (Figure \ref{fig:engagement-time-all}). Over the longer timespan, a slightly negative statistically significant Pearson correlation ($r=-0.054$, $p<0.00001$) and an even smaller negative Spearman correlation ($r=-0.027$,$p=0.018$) were observed. Note that there were only a handful of patients who remained enrolled in the study well past the 2 year mark, and any potential long term sustained dropoff in word count appeared to occur after that point.

\begin{figure}[h]
\centering
\includegraphics[width=\textwidth,keepaspectratio]{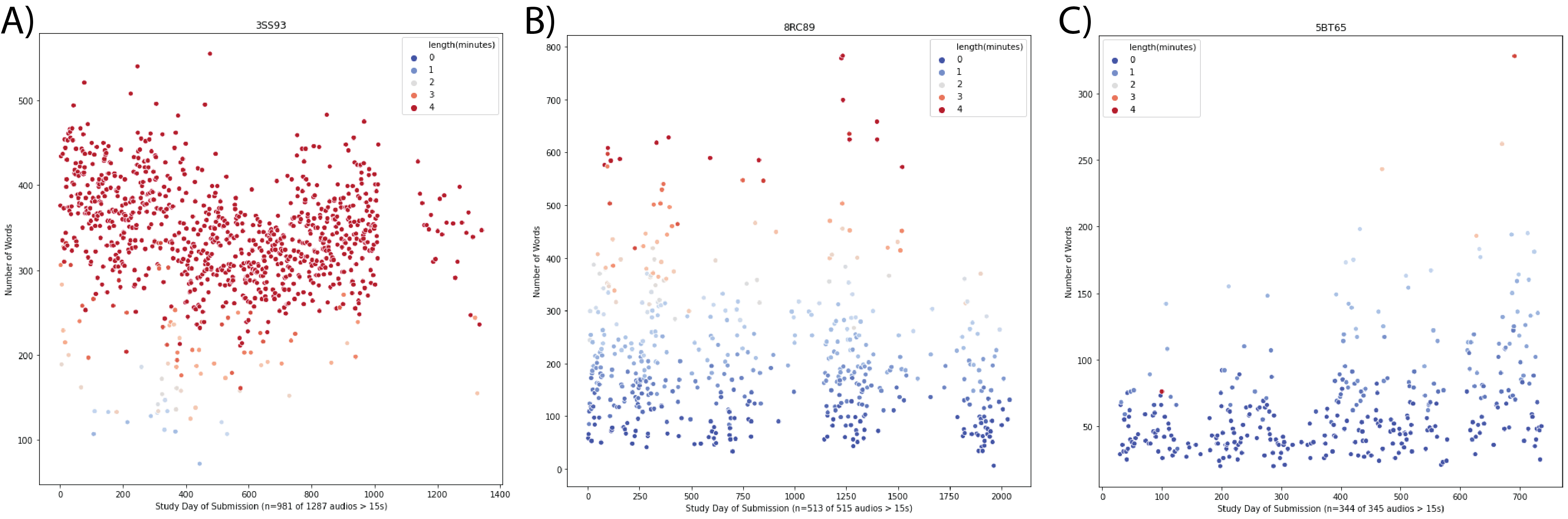}
\caption[Journal word counts over time for select BLS patients of interest.]{\textbf{Journal word counts over time for select BLS patients of interest.} The relationship shown in Figure \ref{fig:engagement-time-all} was also scattered independently for the subjects of particular focus throughout this section - 3SS93 (A), 8RC89 (B), and 5BT65 (C). Note the x and y axes have different ranges in each panel based on that participant's dataset range. Engagement trends are broadly similar to that seen across the larger group of 25 participants in Figure \ref{fig:engagement-time-all}, though 5BT65 appears to actually submit longer journals in year 2 than year 1. This may relate to changes in symptom severity.}
\label{fig:engagement-by-pt}
\end{figure}

Ultimately, it is comforting to see that not only was diary participation from a large number of BLS patients strong for an extended period, but also that engagement with the recording of diaries appeared to remain consistent over this time in most subjects, suggesting that the large dataset we have collected consists of rich content that is truly longitudinal. Of course, when considering an individual subject, transient fluctuations in diary submission patterns or lengths of submitted diaries may also carry clinical relevance. Considering word count over time in the selected participants 3S, 8R, and 5B independently (Figure \ref{fig:engagement-by-pt}), clear patient-specific dynamics in diary submission behavior can be seen. 5B in particular demonstrated periodic fluctuations in the word count of submitted diaries over their years in the study (Figure \ref{fig:engagement-by-pt}C), which was hypothesized to directly relate to disease state. 

The relationship between diary engagement fluctuation and fluctuations in 5BT65's clinical scale ratings and self reported mood will be dissected as part of subsequent sections (\ref{subsec:diary-ema} and \ref{subsec:diary-case-study}). The relevance of basic journal recording behavior - including submission time of day and submission length - to underlying clinical state will also be reviewed as it pertains to the patient that is the focus of the DBS case report in chapter \ref{ch:3}. It is thus important to consider these simple participation features as themselves of scientific interest, and not just as features of relevance for good study operation. Distinguishing between participation changes of clinical relevance and natural or study design related participation changes will be critical for optimizing collection and analyses of such data. Features like word count are noteworthy as measurement of both project progress and of patient verbosity, as well as for contextualizing many other journal features. \\

\FloatBarrier

\paragraph{Submission time of day effects.}
On the topic of diary submission behavior, I also looked at the time of day each submission was recorded, via the relevant pipeline metadata extraction described in section \ref{subsec:diary-code}. Using the $n=8398$ dataset of transcribable journals $\geq 15$ seconds from across all subjects, I considered whether diary submission time related to word count, as well as whether there were any interesting temporal dynamics or patient-specific effects in submission times over the course of the study. These analyses are important because understanding submission time trends could both facilitate better targeted notifications to improve participation rates in future studies and uncover additional behavioral patterns of clinical note. 

\begin{figure}[h]
\centering
\includegraphics[width=0.75\textwidth,keepaspectratio]{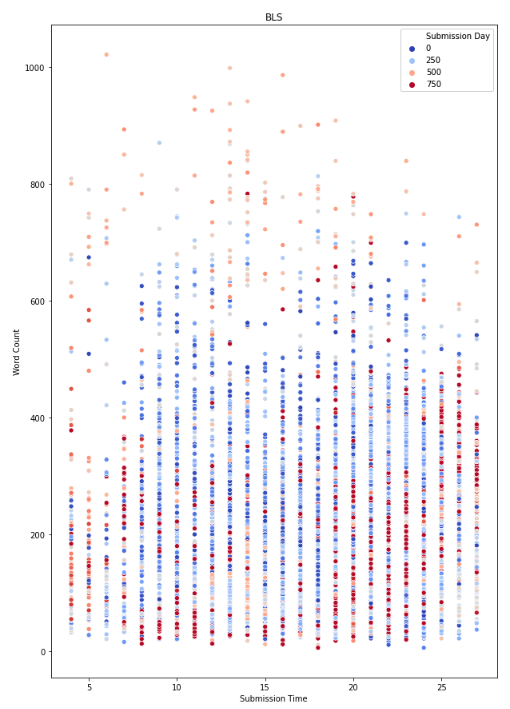}
\caption[Submission time versus word count with study day hue, across BLS journal set.]{\textbf{Submission time versus word count with study day hue, across BLS journal set.} Each diary in the final BLS dataset across all subjects (as described at the beginning of section \ref{subsec:diary-dists}) is plotted as a point here, with x-coordinate corresponding to the time of day (in Eastern Time) that the diary was submitted and y-coordinate corresponding to the word count of that diary. Each diary point is colored according to the study day it was submitted, from dark blue at the time of that patient's enrollment crossing over to dark red at the end of year 2 in the study. Diaries that were submitted beyond year 2 are included in the scatter plot, but the hue saturates past this point. Recall that the submission time integer is assigned based on the hour at time of recording, and that the integer ranges from 4 to 27 because diaries submitted after midnight but before 4am are assigned to the previous date.}
\label{fig:diary-et-vs-words}
\end{figure}

\begin{figure}[h]
\centering
\includegraphics[width=\textwidth,keepaspectratio]{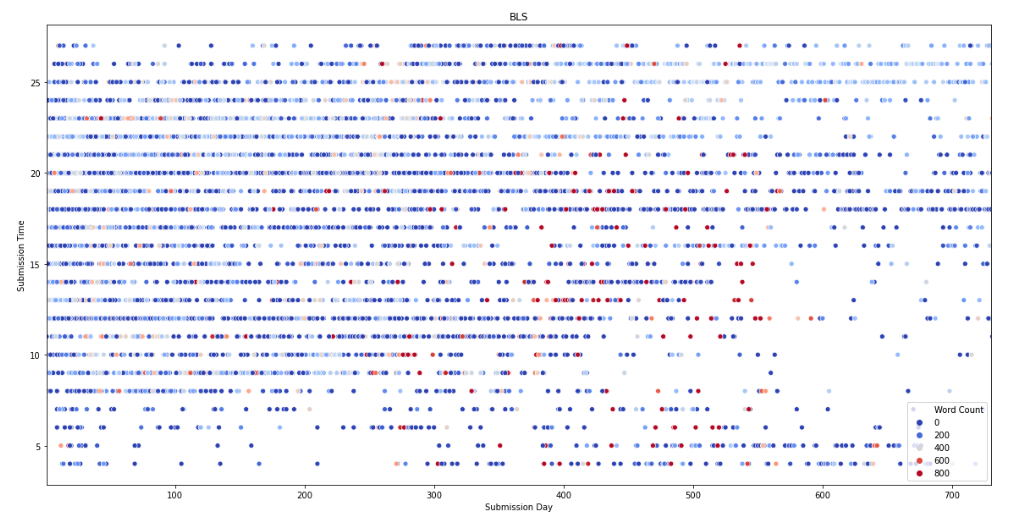}
\caption[Submission day versus submission time with word count hue, across BLS journals from first two years of enrollment.]{\textbf{Submission day versus submission time with word count hue, across BLS journals from first two years of enrollment.} The features and dataset of Figure \ref{fig:diary-et-vs-words} were also visualized as a scatter plot of study day versus submission time, with word count instead as the hue -- colored from dark blue at $leq 100$ and crossing over to dark red at $\geq 700$. Note that the x-axis here is limited to the first 2 years of enrollment in BLS, for clarity on the dynamics of greatest interest. Thus some of the points in Figure \ref{fig:diary-et-vs-words} are excluded here.}
\label{fig:diary-day-vs-time-2yr}
\end{figure}

There was a small but statistically significant correlation between study day and time of submission across the dataset, with Pearson's $r=0.111$ ($p < 10^{-23}$) and Spearman's $r=0.129$ ($p < 10^{-31}$). Similarly, the Pearson correlation between total word count and submission time was $0.093$ ($p < 10^{-16}$), and the Spearman correlation was $0.186$ ($p < 10^{-65}$). This suggests that there may be a slight shift towards later submissions as a participant stays in the study longer, and also that later submissions may contain slightly more content (Figure \ref{fig:diary-et-vs-words}). Of course, not only are these correlations slight, they are also of uncertain origin. They could be driven by a handful of participants that happened to stay in the study for a long time and also happened to submit journals later in the day, as this dataset included all participants over all time. While some small trends did still appear when restricting the view to only the first 2 years of each subject's data (Figure \ref{fig:diary-day-vs-time-2yr}), there was not enough clear evidence to make any real conclusions.

More generally, there was obvious individual heterogeneity in daily schedules, both internally and externally imposed. As such, the most fruitful methodology for looking at submission time of day patterns - whether for participation or clinical reasons - will very likely involve separate consideration for each subject. However, it is worth noting that on a typical weekday workforce schedule, a basic "describe the last 24 hours" style prompt would probably receive a more accurate detailed response in the evening than in the morning. The exact prompt of interest is therefore an additional important component to characterizing submission times. Further context on participant-specific submission dynamics in the larger BLS dataset can be found in supplemental section \ref{subsec:participation-table-sup}. 

To demonstrate the utility of a personalized approach in this domain, I reviewed the submission time data of the highlighted participants more closely. 3SS93 for example demonstrated a clear shift towards later submission times over their first year in the study. 8RC89 on the other hand displayed greater variance in diary submission times; periods of greater or lesser variability in recording time of day could be worth investigating for clinical relevance, like was done for the DBS case report subject in chapter \ref{ch:3}.

Interestingly, 5BT65 - like 3SS93 - had a clear correspondence between study day and submission time. Though 5B tended to submit journals much earlier in the day than 3SS93 did, and 5B's latest journal submissions were not necessarily at the end of the study, there was a pronounced shift to later submission times over the course of enrollment, from majority late afternoon/early evening submissions to majority late evening/early night submissions. As diary word count is of likely clinical relevance for 5BT65 in particular, it is especially salient that they submitted a number of recordings before 6pm ET, yet just 5 of those submissions had more than 100 words. On the other hand, it will be important to review more closely any possible changes in Beiwe app installation or study settings before drawing strong conclusions about submission times. \\  

\FloatBarrier

\noindent Ultimately, this pilot analysis of submission dynamics in our BLS dataset can be used as a reference point for collection procedures in future audio journal studies, as well as a starting point for future temporal prediction work on the extracted BLS journal features. In section \ref{subsec:diary-case-study} below, I will revisit fluctuations in diary features over time in a qualitative manner for the highlighted subjects (3S, 8R, and 5B). More information on their participation dynamics over the course of the study can also be found in supplemental section \ref{subsec:participation-fig-sup}. However, the primary modeling work, to be reported on next, will focus on EMA prediction from same day journal features without consideration of time, as it is beyond the scope of the thesis.

\subsection{Relationship between patient self-report survey responses and audio diary features}
\label{subsec:diary-ema}
As mentioned in section \ref{subsec:diary-methods}, patients in the Bipolar Longitudinal Study (BLS) were prompted daily by Beiwe to complete ecological momentary assessment (EMA) surveys along with the daily free-form audio journal recording prompt. The EMA questions were designed to probe patient perception of current mood and other relevant clinical signs. Given their close tie to the audio journals, EMA responses have strong potential as a target for prediction based on journal features.

I will begin by characterizing a bit more about the availability of EMA with same-day journal submissions in BLS as well as the EMA response profiles observed. Combining this information with prior knowledge about the audio journal dataset and about the clinical background of select BLS patients, I will propose a handful of modeling experiments. The results of the model(s) fitting and hold out testing will then be reported and interpreted, evaluating prediction of EMA from select journal features on the general study-wide level and when considering subject-specific factors. 

\subsubsection{Characterization of BLS EMA responses}
\label{subsubsec:bls-ema-dists}
Availability of patient daily self-report survey responses consistently overlapped well with availability of same day audio journals (Table \ref{table:bls-ema-avail}), making prediction of survey response values from diary properties a very tractable problem space. This result bodes well for the audio journal datatype more generally, as across a variety of BLS participants the likelihood of EMA response was very high given a diary recording was submitted that day. Coupled with the fact that EMA is easy to administer in the same app as audio journals, obtaining self-report survey labels to go along with recorded diaries is indeed extremely feasible. 

Many different questions were included in various EMA surveys presented to BLS subjects over the years. For the pilot modeling work in this section, I selected a subset of questions to consider more closely, based on data availability and cleanliness as well as hypothesized clinical relevance. For a full account of the list of EMA questions asked in the BLS study, along with lessons on EMA survey design in Beiwe, see supplemental section \ref{sec:bls-ema-qs}. 

\noindent The following list reports the 15 questions that I selected for further review. Each survey started with the dialogue box -
\begin{quote}
    Over the past 24 hours how much were you ...
\end{quote}
\noindent - and then presented each question in abbreviated form. For descriptive purposes I've grouped these EMA items into 4 major categories, to organize the question list:
\begin{itemize}
    \item \emph{Positively worded emotion-related EMA}
    \begin{itemize}
        \item Happy?
        \item Energetic?
        \item Inspired?
        \item Determined?
        \item Outgoing?
    \end{itemize}
    \item \emph{Negatively worded emotion-related EMA}
    \begin{itemize}
        \item Stressed?
        \item Hostile?
        \item Irritable?
        \item Ashamed?
        \item Upset?
        \item Afraid?
        \item Lonely?
    \end{itemize}
    \item \emph{Psychosis-related EMA (hallucinations)}
    \begin{itemize}
        \item Bothered by hearing voices?
        \item Bothered by seeing things others could not?
    \end{itemize}
    \item \emph{Psychosis-related EMA (delusions)}
    \begin{itemize}
        \item Bothered by feeling like other people are out to get you or cause you trouble?
    \end{itemize}
\end{itemize}
\noindent Note that each of these questions used the same set of 5 possible response options -- with the exception of "Energetic?", which had a unique wording for each corresponding option (see supplemental section \ref{sec:bls-ema-qs} for more details). The general response options were:
\begin{itemize}
    \item Very slightly or not at all
    \item A little
    \item Moderately
    \item Quite a bit
    \item Extremely
\end{itemize}
\noindent Note that by default the response to each question was encoded on the same scale, but in preprocessing the data here I flipped the valence of all positively worded emotion-related questions, so that a higher value would consistently correspond to worse symptom severity, as it is for clinical scales. 

\begin{table}[!htbp]
\centering
\caption[Good overlapping data availability for quality daily audio journals and daily self-report survey responses in the BLS dataset.]{\textbf{Good overlapping data availability for quality daily audio journals and daily self-report survey responses in the BLS dataset.} For each subject ID, I report the number of days where both an EMA response and a transcript from submitted audio diary were available, along with that patient's recorded diagnosis. IDs are ordered from lowest to highest availability. For most participants, this number closely matched the number of total transcripts available (Figure \ref{fig:diary-pt-submit-chart}B), indicating that if a diary was submitted, there was high probability of some EMA submission too. For brevity, only the top 24 patients are displayed here. All other IDs submitted both self-report types on fewer than 75 days each, with these IDs averaging a count of 29 dual submissions each. [Participants that were later chosen to be included in pilot EMA modeling are highlighted in grey. Only a subset were selected to preserve some of the dataset for future projects. I focused this subset on those with a primary diagnosis of BD and a good number of diaries above 15 seconds in duration, though not all participants meeting those criteria were included.]}
\label{table:bls-ema-avail}

\begin{tabular}{ | m{2cm} | m{3.75cm} | m{6.75cm} | }
\hline
\textbf{Patient ID} & \textbf{Days w/ diary + EMA} & \textbf{Diagnosis} \\
\hline\hline
\rowcolor{Gray}
3SS93 & 978 & Bipolar I \\
\hline
7NE49 & 705 & Schizoaffective Disorder \\
\hline
\rowcolor{Gray}
SBPQM & 534 & Bipolar I \\
\hline
B3JZM & 510 & Bipolar II \\
\hline
\rowcolor{Gray}
5BT65 & 489 & Bipolar II \\
\hline
\rowcolor{Gray}
8RC89 & 470 & Bipolar I \\
\hline
5QXNM & 465 & Bipolar (schizophrenic type) \\
\hline
\rowcolor{Gray}
GMQBM & 443 & Bipolar I (psychotic features) \\
\hline
B3CRM & 377 & Bipolar Disorder \\
\hline
CYG4M & 375 & Bipolar I \\
\hline
2XC43 & 365 & Schizoaffective (bipolar type) \\
\hline
5KXYM & 348 & Depression, Schizophrenia \\
\hline
M8MXM & 315 & Unknown \\
\hline
DQDCM & 295 & Bipolar I \\
\hline
8ADCM & 261 & Unknown \\
\hline
JVBKM & 254 & Schizophrenia \\
\hline
Z8YRM & 234 & Schizoaffective (depressive type), PTSD \\
\hline
\rowcolor{Gray}
GFNVM & 217 & Bipolar II \\
\hline
8KX53 & 194 & Bipolar I \\
\hline
\rowcolor{Gray}
ACMCM & 171 & Bipolar Disorder \\
\hline
5CR39 & 124 & Schizophrenia \\
\hline
V65HM & 105 & Borderline, Depression, PTSD, Anxiety \\
\hline
F6VVM & 99 & Unspecified affective disorder \\
\hline
9SU83 & 86 & Bipolar I \\
\hline
\end{tabular}
\end{table}

\FloatBarrier

\paragraph{Correlation structure of select EMA questions.}
For more information on how I carefully cleaned and characterized the BLS EMA dataset intended for analysis, see supplemental section \ref{sec:ema-meth}. 

\begin{figure}[h]
\centering
\includegraphics[width=\textwidth,keepaspectratio]{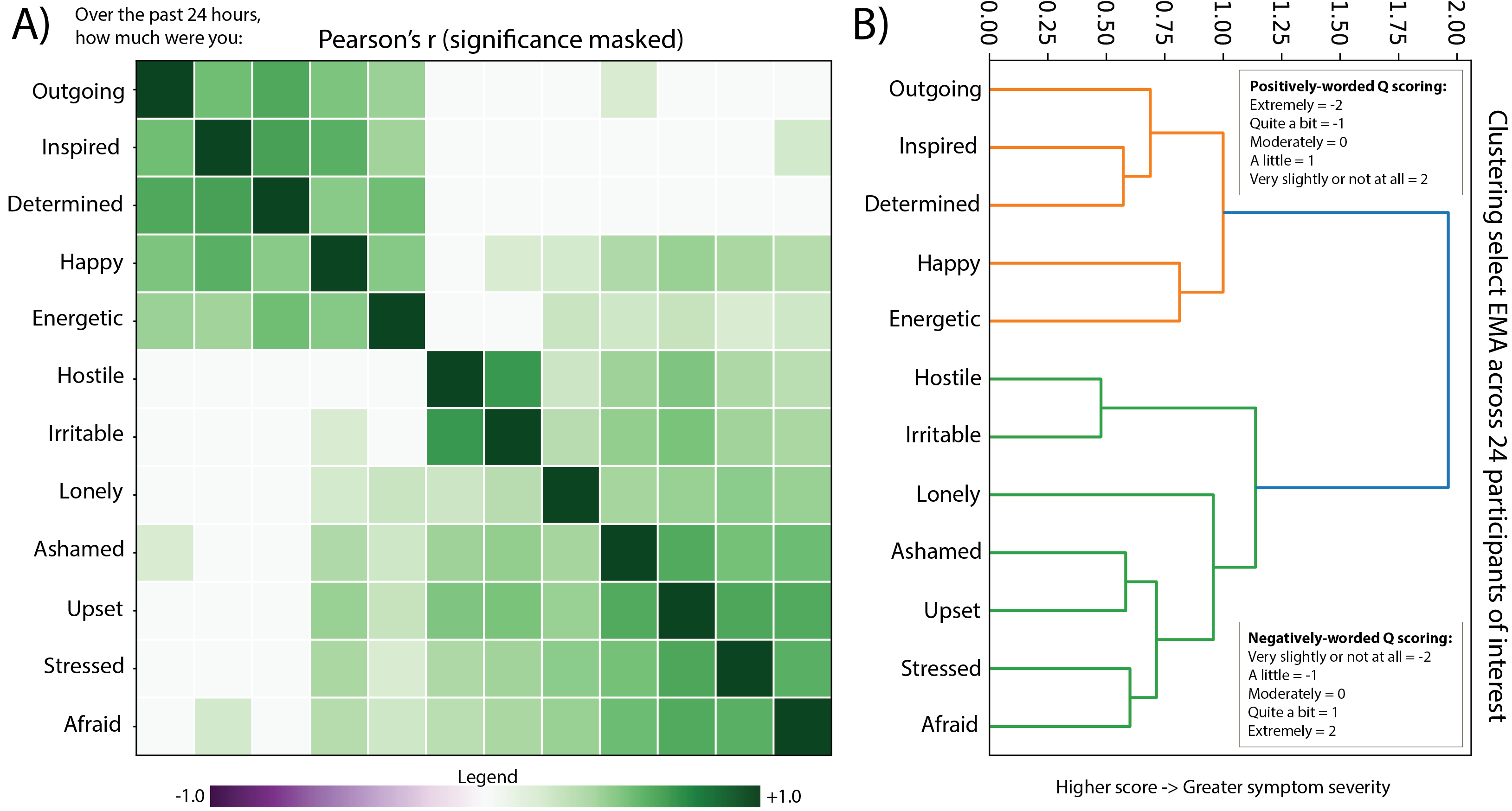}
\caption[EMA answers cluster based on question wording in BLS.]{\textbf{EMA answers cluster based on question wording in BLS.} To decide on final EMA scores for use in modeling, I investigated the Pearson correlation structure of responses to different EMA items of interest in the 24 highlighted subjects. For each pairing of the selected 12 feelings-related EMA responses, Pearson's $r$ was computed across the above described exploratory EMA dataset ($n=9628$). The Bonferroni significance-masked correlation matrix (A) was computed and visualized using an analogous method to that for the audio journal features in Figure \ref{fig:pearson-diary-final}, but with mask on $r$ magnitude thresholded at $0.2$ here instead of the previous $0.1$. As before, the items were ordered based on the results of clustering on Pearson distances. Clustering results are also presented here via dendrogram (B). This particular clustering had a cophenetic correlation score of $0.93$. Note that there is minimal linear correlation between positively-worded and negatively-worded items, thus inspiring a separate summary score with different scaling for modeling of the two different question styles.}
\label{fig:pearson-ema}
\end{figure}

From there, to determine how the selected EMA items might be best narrowed to a smaller numbers of labels for pilot modeling while retaining as much information as possible, I evaluated the correlation structure of survey responses across the cleaned EMA dataset from the 24 isolated participants. I used a similar methodology as the diary feature correlations above, first focusing on the emotion-related EMA dataset ($n=9628$ patient days). As the Pearson linear and Spearman rank correlation structures were very similar in this dataset, I focused further review on the Pearson results (Figure \ref{fig:pearson-ema}).

The primary result from the Pearson analysis was that the positively worded questions clustered together and the negatively worded questions clustered together (Figure \ref{fig:pearson-ema}) -- even with the encoding flipped so that the opposite valences would be expressed in terms of compatible symptom severity scales, related questions with opposite valence actually did not have their answers linked. This underscores the importance of careful wording of EMA questions. The separation of positively worded EMA items from negatively worded EMA items in correlational analysis (even by rank) also raises questions on exactly how different survey responses should be encoded. Although EMA typically has all items rated on a scale with the same range, it is not clear that "Very slightly or not at all \emph{happy}" should match with "extremely \emph{not happy}", let alone the complications introduced when considering actual negative emotion words like "upset". This can be connected to distributional properties of these items as well, to be discussed shortly.

Based on the results of the emotion-related EMA clustering (Figure \ref{fig:pearson-ema}), one might consider selecting a subset of items to use as labels. In conjunction with per item response distributions, I would likely choose determined, happy, energetic, irritable, lonely, and stressed as the narrowed down set here (if I were to do so). However, for the scope of the present thesis, I instead chose to consolidate to just 2 emotion-related EMA labels, in terms of smoothed summary scores: the mean positively worded EMA item score and the mean negatively worded EMA item score, for each participant day. \\

For the analogous psychosis-related EMA dataset, I performed an analogous evaluation of correlation structure. Pearson and Spearman correlations were again highly similar, and for these items every considered correlation was highly significant. Pearson's $r$ between the verbal and visual hallucination responses was $\sim 0.665$, while Pearson's $r$ for the persecutory delusions question was $\sim 0.528$ with verbal hallucination responses and $\sim 0.489$ with visual hallucination responses. Though the correlations were high enough to consider using a single summary score across all 3 of the psychosis-related questions, it was important to note that by far the most common participant response was "very slightly or not at all" to all 3 items. 

Because we are interested in how these questions might co-vary when psychotic symptoms are actually present, I performed a second correlational analysis with a filtered version of the dataset that contained only survey submissions where at least one of the psychosis-related questions was answered with "a little" or more. That dataset contained $2310$ patient days, as compared to the original dataset of size $n=8213$. The corresponding Pearson's $r$ values in the filtered dataset were $\sim 0.535$, $\sim 0.168$, and $\sim 0.157$, so that only the hallucination items remained strongly correlated with each other. As such, for the final label set I proceeded with 2 psychosis-related EMA "summary" scores as well: the mean of the hallucination responses and the response to the persecutory delusion question, for each participant day. \\

\FloatBarrier

\begin{figure}[h]
\centering
\includegraphics[width=\textwidth,keepaspectratio]{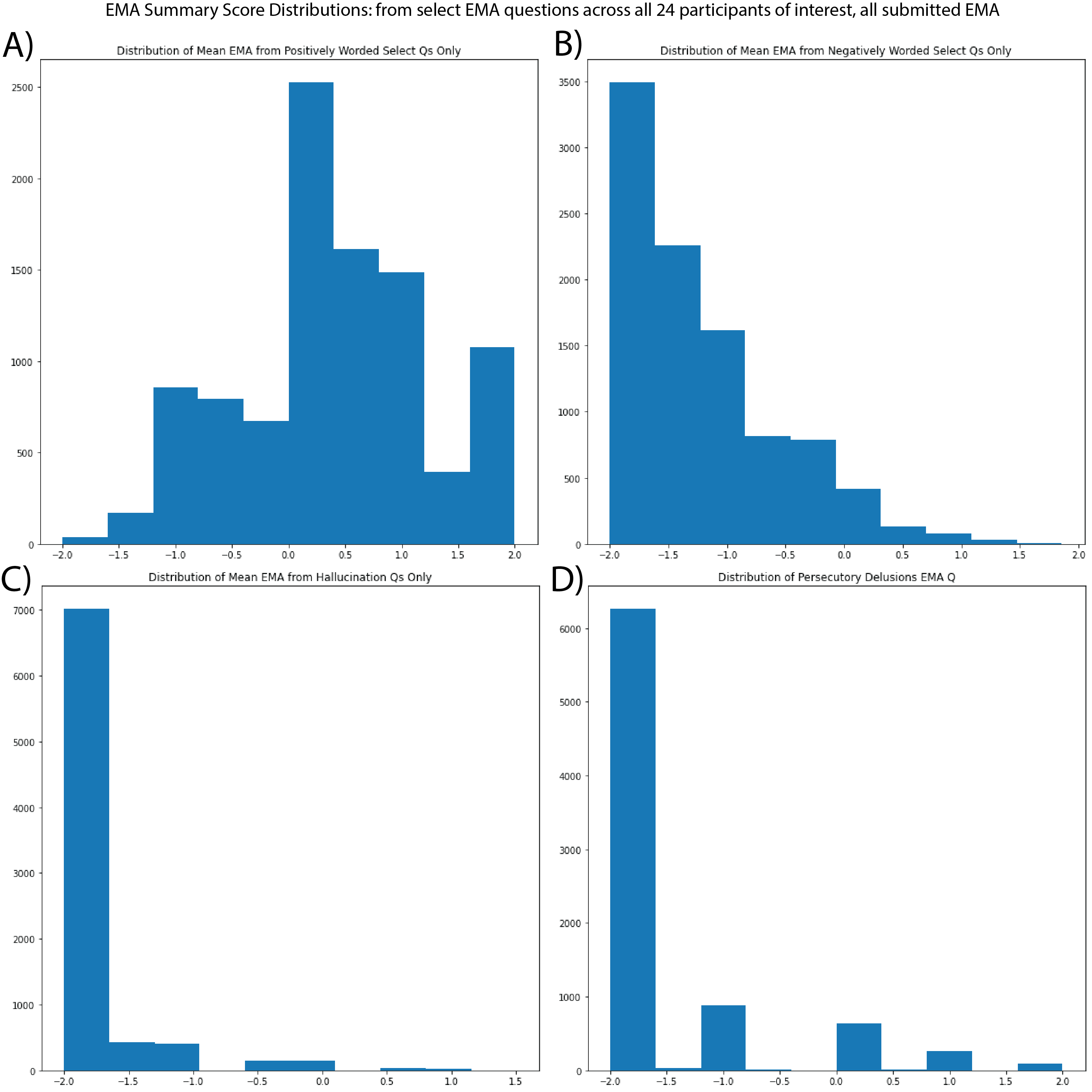}
\caption[Distribution of EMA summary scores of interest.]{\textbf{Distribution of EMA summary scores of interest.} Based on the correlational results, two feelings-related summary scores and two psychosis-related summary scores were generated from the larger set of considered EMA items across the dataset of the 24 selected BLS participants. For each time point ($n=9628$), the mean was taken independently across the positively-worded grouping (A) and the negatively-worded grouping (B) from the emotion EMA clusters of Figure \ref{fig:pearson-ema}. Similarly, for each time point with a response to the psychotic symptoms EMA items ($n=8213$), the mean of the two hallucination-related responses (C) was taken, along with the raw score for the persecutory delusions question (D). The dataset input to this process was already normalized to have higher scores correspond to more severe symptoms across all items, with the 5 potential responses assigned values from -2 to 2. Here, I present the distribution of these 4 computed summary statistics in the described exploratory EMA dataset. Note that these distributions triggered the re-indexing of negative emotion as well as psychosis-related EMAs to run from 0 to 4 for least to most symptom severity, keeping only positive emotion EMA scoring centered around 0.}
\label{fig:ema-wider-dists}
\end{figure}

\paragraph{EMA summary score distributions in processed dataset.}
The complication with interpreting answers to positively versus negatively worded questions that was discussed above can be directly connected to the summary score distributions -- the positively worded summary distribution (Figure \ref{fig:ema-wider-dists}A) was centered at the "moderately" option and was skewed towards more severe average symptom ratings (e.g. "very slightly or not at all happy"), while the negatively worded summary distribution (Figure \ref{fig:ema-wider-dists}B) had peak at the least severe response (e.g. "very slightly or not at all upset"), with only a small portion of submissions having an average response that surpassed the the midpoint of the range. Because a "moderate" response makes sense to serve as a healthy baseline for positively worded questions, but would be considered a moderate level of self-reported symptom severity (or at least a very bad day) for all other considered questions, the observed distributional patterns are not especially surprising. 

The feature distributions also raise the point that in this study population, most participants had low or no symptom severity on most days, though a good amount of variation for modeling in the emotion-related EMAs still existed fortunately. For the psychosis-related EMA summary scores (Figure \ref{fig:ema-wider-dists}C/D) though, it was rare for responses in this dataset to be anything besides "very slightly or not at all", particularly for the hallucination questions. The days with high severity responses to these questions are of great interest, but the distributional properties will require more care in model design.

One small additional caveat on the encoding of the positively worded emotion-related EMA items, was that an answer of "extremely" ($= -2$) for some of the questions may indicate pathology instead of a high level of functioning. This would be especially true for the customized wording of the "energetic" item, where the most extreme option was intentionally crafted to capture mania symptomatology, such that the severity reflected by answers was actually U-shaped (or more accurately perhaps, J-shaped). Answers of "extremely" were relatively rare in the dataset even for the positively worded questions however, and especially for the "energetic" question -- so while these answers may very well hold value the current analyses are missing, it is safely out of scope for the stated aims. 

\FloatBarrier

\subsubsection{Dataset and models used}
\label{subsubsec:bls-ema-methods}
In order to preserve much of the dataset for future predictive analyses by others, I selected a smaller subset of 7 participants to include in the pilot EMA modeling study (and more generally when evaluating clinical-related outcomes throughout this chapter). Within supplemental section \ref{sec:ema-meth}, I describe how I narrowed the dataset in more detail, as well as information on analysis methodologies and final feature selection for modeling. Here, I will next summarize the composition of the modeling datasets to be used and the distribution of the EMA summary score labels in that subset. \\

\begin{figure}[h]
\centering
\includegraphics[width=\textwidth,keepaspectratio]{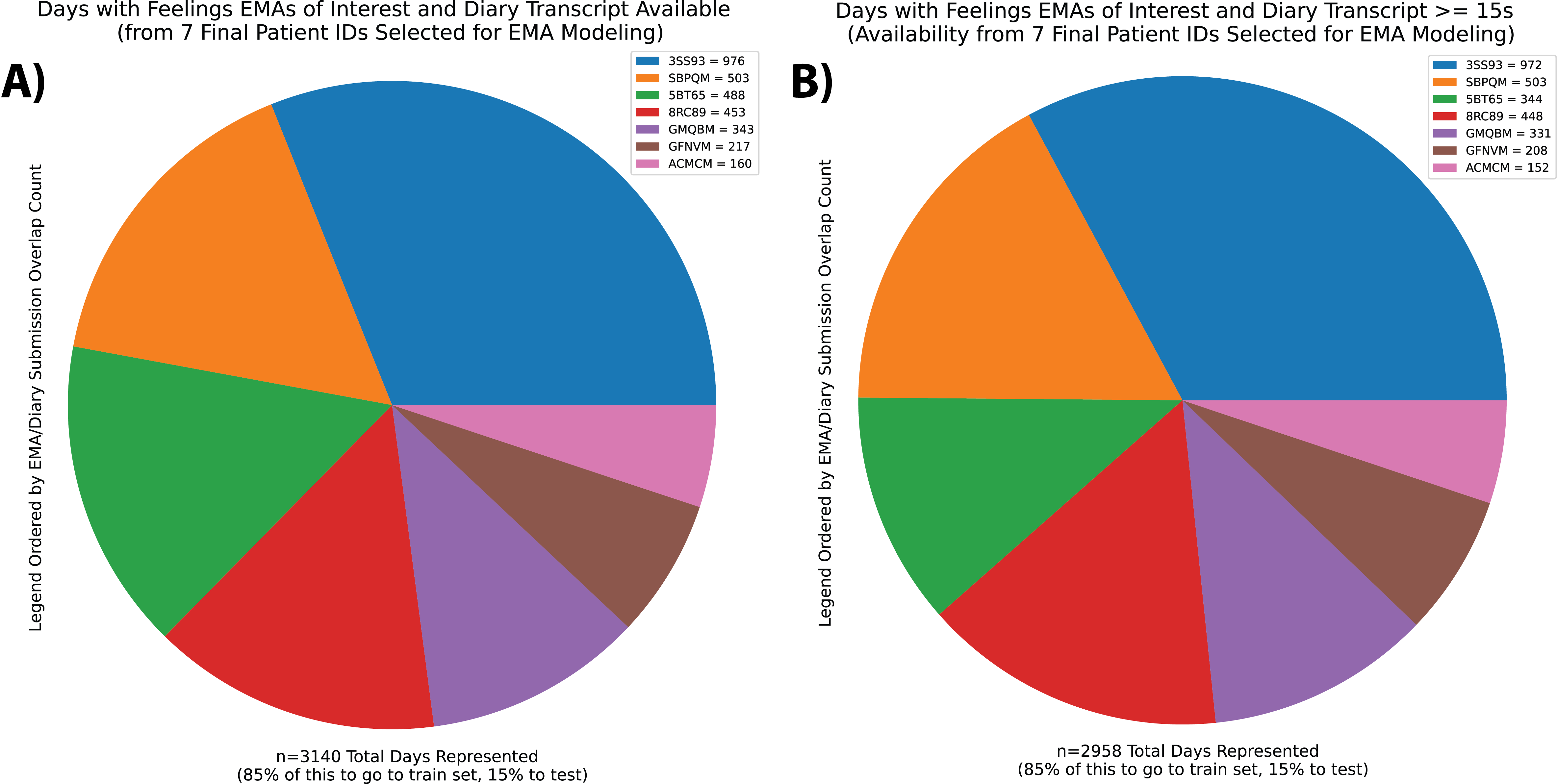}
\caption[Composition of final dataset for modeling of emotion-related EMA in 7 BLS subjects.]{\textbf{Composition of final dataset for modeling of emotion-related EMA in 7 BLS subjects.} As described, I selected the participants highlighted in Table \ref{table:bls-ema-avail} to use for the final modeling project of predicting EMA summary scores from same day audio journal features. These pie charts depict how many data points each of the 7 subjects contributed to the final modeling datasets for emotion-related EMA summary scores, after narrowing down to the set of days with both a journal transcript and relevant EMA available (A), as well as after further narrowing to those days where the journal submission was of duration $\geq 15$ seconds (B). Note that 5BT65 (green) accounted for most of the journal submissions $< 15$ seconds that were filtered out between the left and right charts. Because it is hypothesized that shorter diary length will correlate with more severe depressive symptoms, particularly for 5BT65, the dataset in (A) will be used for a model of emotion-related EMA based purely on the word count of the submission, while the duration-filtered dataset in (B) will be used for modeling with the more complete set of diary features. In both cases, the depicted dataset will be partitioned into $85\%$ training set and $15\%$ holdout test set before proceeding with model fitting.}
\label{fig:ema-mood-pie}
\end{figure}

\begin{figure}[h]
\centering
\includegraphics[width=\textwidth,keepaspectratio]{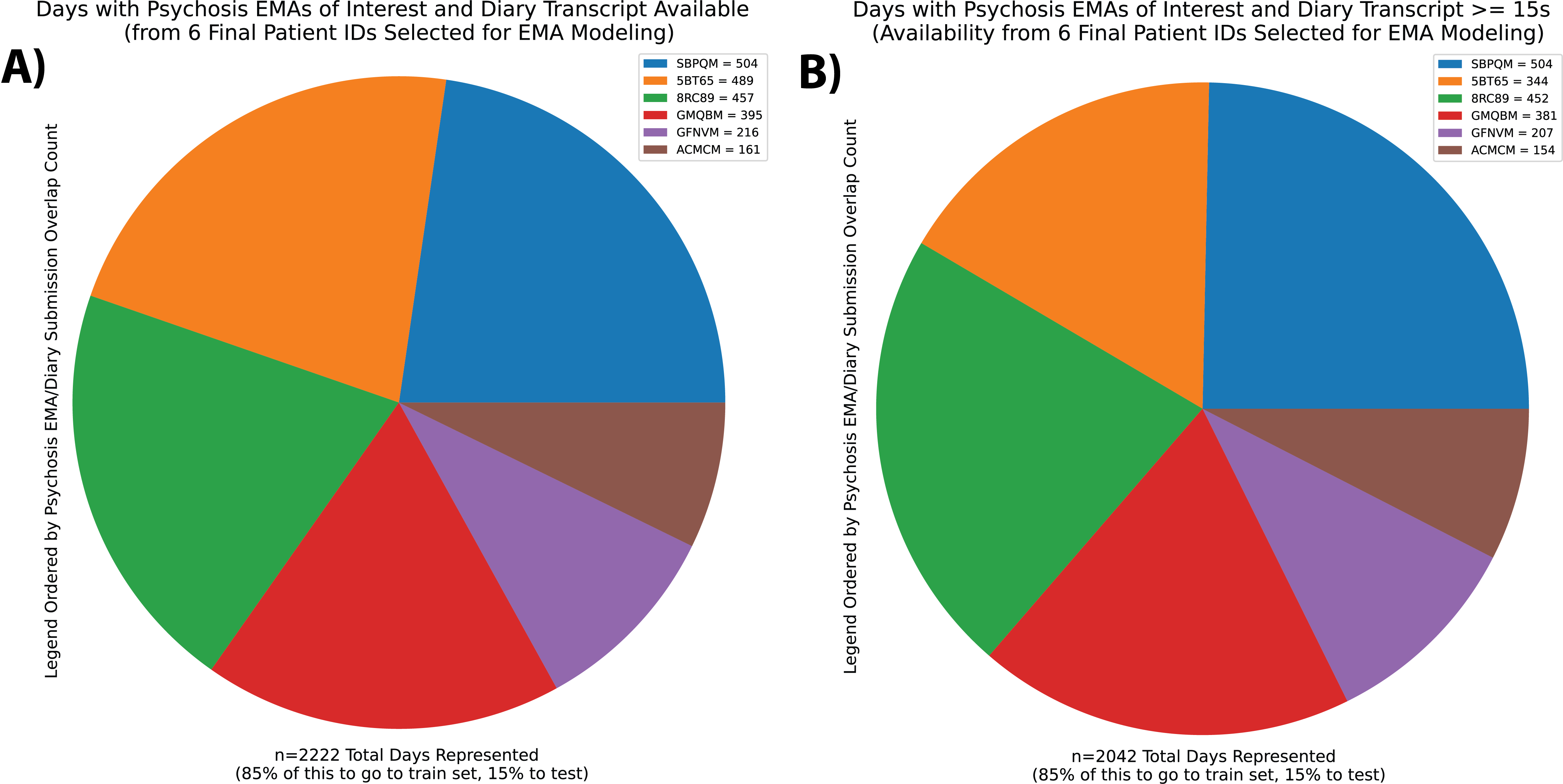}
\caption[Composition of final dataset for modeling of psychosis-related EMA in 6 BLS subjects.]{\textbf{Composition of final dataset for modeling of psychosis-related EMA in 6 BLS subjects.} Pie charts of the final modeling dataset composition for the psychosis-related EMA items were also created, analogous to those for the emotion-related EMA depicted in Figure \ref{fig:ema-mood-pie}. As participant 3SS93 submitted only a handful of responses to the psychotic symptoms EMA questions, they were excluded from this dataset. For the remaining 6 selected participants, the number of available time points was very similar between the two EMA categories. Again, 5BT65 (orange) accounted for the majority of the records that were present in the combined transcript/EMA dataset (A) but filtered out when a minimum journal duration of 15 seconds was enforced (B). Going forward, modeling of the psychotic symptoms EMA will be performed independently of the emotional symptoms EMA, but with an analogous protocol.}
\label{fig:ema-psych-pie}
\end{figure}

\paragraph{Dataset construction.}
Overall, there were $\sim 3000$ data points of overlapping EMA and diary availability in the 7 selected participants, though with some caveats to consider in formulating different modeling problems. Of the 7 subjects included, most had very similar data availability for the emotion-related and psychosis-related EMA items, but participant 3SS93 had fewer than 10 days available with both a journal submission and answers to the psychosis-related questions (mostly because they usually did not include an answer to the psychosis-related items when submitting their surveys). As such, I continued to manage the emotion-related (Figure \ref{fig:ema-mood-pie}) and psychosis-related (Figure \ref{fig:ema-psych-pie}) EMA datasets separately, with the psychosis version including only 6 subjects. Similarly, because of the hypothesized clinical relevance of diary length for participant 5BT65, along with the sizeable number of journal recordings they submitted that were $< 15$ seconds in duration, I maintained separate datasets for both minimalist modeling of diary length versus EMA summary scores across all submissions as well as for the full planned feature modeling with minimum length requirement still in place.

\noindent In sum, using the EMA data from the selected 7 subjects, I saved 4 aggregate CSVs to build the modeling datasets, as follows: 
\begin{itemize}
    \item The positively worded and negatively worded emotion-related daily summary scores, inner merged with features available from same-day journal transcript (Figure \ref{fig:ema-mood-pie}A).
    \item The above, filtered to remove all rows where the corresponding diary length was less than 15 seconds (Figure \ref{fig:ema-mood-pie}B).
    \item The hallucinations and delusions daily summary scores, inner merged with features available from same-day journal transcript, with the handful of records from subject 3SS93 removed (Figure \ref{fig:ema-psych-pie}A).
    \item The above, filtered to remove all rows where the corresponding diary length was less than 15 seconds (Figure \ref{fig:ema-psych-pie}B).
\end{itemize}
\noindent Each such CSV contained the 2 EMA summary score columns of interest to serve as labels, the core audio journal feature columns needed to compute final model input values, and any relevant metadata. For final model fitting, $85\%$ of the points in each described dataset will be used, with $15\%$ held out for testing (sampled separately for each subject ID). \\

\FloatBarrier

\begin{figure}[h]
\centering
\includegraphics[width=\textwidth,keepaspectratio]{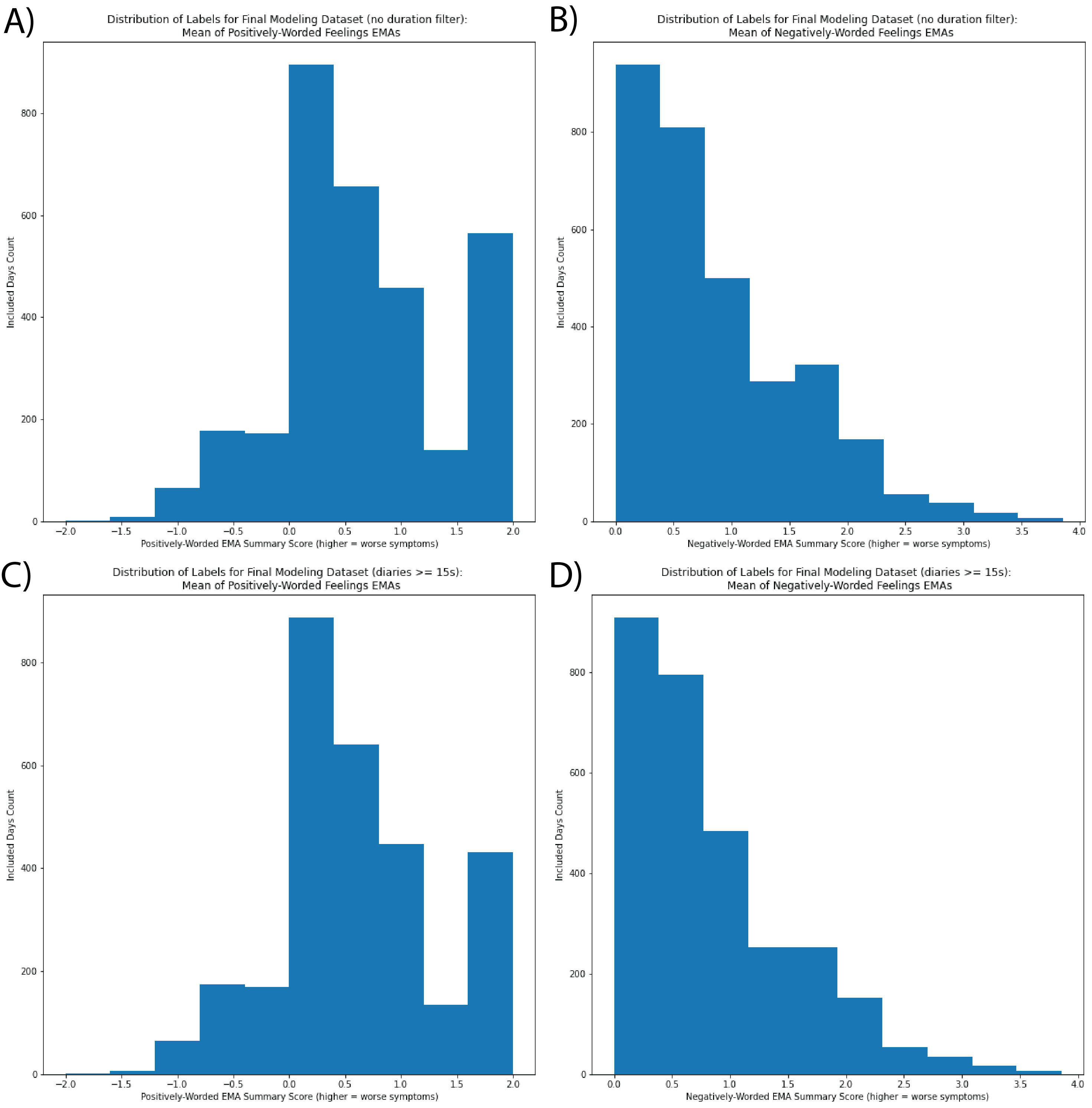}
\caption[Distribution of emotion-related EMA summary scores in final modeling dataset.]{\textbf{Distribution of emotion-related EMA summary scores in final modeling dataset.} Within the final feelings EMA datasets (Figure \ref{fig:ema-mood-pie}), I generated histograms of the positively-worded (A/C) and negatively-worded (B/D) summary scores, to be used as labels for subsequent modeling. These distributions were produced from the full datasets before train/test split, both \textbf{with} (C/D) and \textbf{without} (A/B) filtering out time points where the submitted diary was less than 15 seconds ($n = 2958$ and $3140$ respectively, across 7 participants). In all cases, higher summary scores indicates greater symptom severity. For the positively-worded summary, the scale remains $-2$ to $2$ as it was for the same feature distribution across the broader EMA dataset in Figure \ref{fig:ema-wider-dists}A. For the negatively-worded summary, the scale was shifted to be $0$ to $4$ by simply adding 2 to all values from the original scale depicted across the broader EMA dataset in Figure \ref{fig:ema-wider-dists}B. This was done to better reflect the symptom severity encoded by each option.}
\label{fig:ema-mood-dist}
\end{figure}

\begin{figure}[h]
\centering
\includegraphics[width=\textwidth,keepaspectratio]{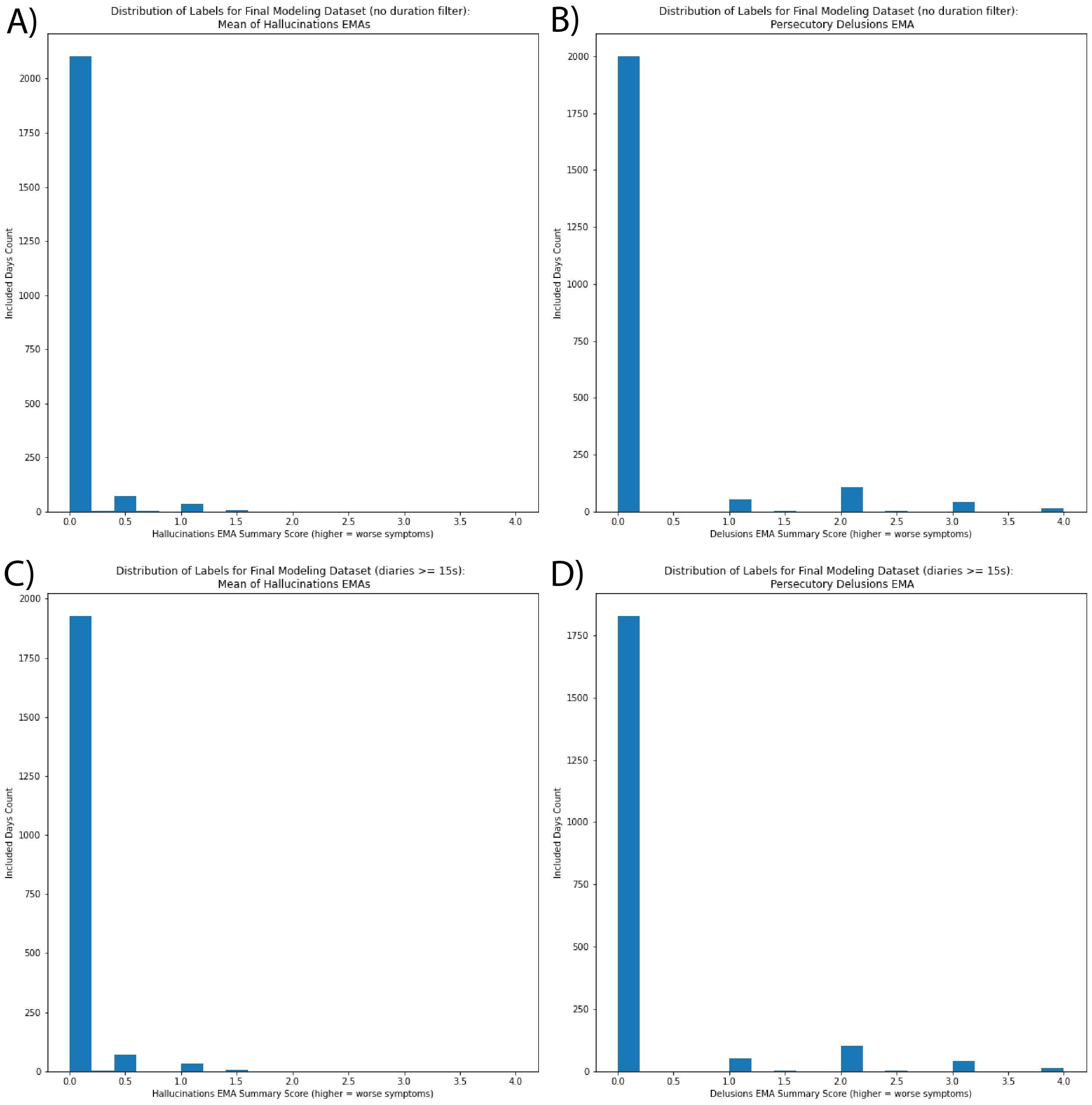}
\caption[Distribution of psychosis-related EMA summary scores in final modeling dataset.]{\textbf{Distribution of psychosis-related EMA summary scores in final modeling dataset.} As was done for the emotion-related EMA summary scores in Figure \ref{fig:ema-mood-dist}, distributions of the described psychosis-related EMA summary scores across the final modeling datasets of Figure \ref{fig:ema-psych-pie} are plotted here. This includes the hallucination (A/C) and persecutory delusion (B/D) self-report metrics, and is done both \textbf{with} (C/D) and \textbf{without} (A/B) filtering out time points where the submitted diary was less than 15 seconds ($n = 2042$ and $2222$ respectively, across 6 participants). Scores range from $0$ to $4$, representing from no symptoms to severely troubling symptoms. To compare with the distribution of the same features from the broader EMA dataset (Figure \ref{fig:ema-wider-dists}C/D), one can simply subtract 2 from all values on the x-axis here.}
\label{fig:ema-psych-dist}
\end{figure}

\paragraph{Distribution of labels in the modeling dataset.}
Once the final datasets to be utilized for EMA modeling from audio journal features were built, I looked more closely at the distributional properties of the EMA summary scores in these datasets. Histograms for the 2 emotion-related summary scores from the datasets of Figure \ref{fig:ema-mood-pie} can be found in Figure \ref{fig:ema-mood-dist}, while histograms for the 2 psychosis-related summary scores from the datasets of Figure \ref{fig:ema-psych-pie} can be found in Figure \ref{fig:ema-psych-dist}. 

The emotion-related EMA summary scores were distributed fairly similarly in the isolated 7 subjects (Figure \ref{fig:ema-mood-dist}A/B) as they were in the original 24 considered subjects (Figure \ref{fig:ema-wider-dists}A/B), though with a greater proportion of more severe symptom ratings, regardless of wording. This was not an intentional aim of the subject selection, though it is a positive development for the modeling. Conversely the psychosis-related EMA summary scores were distributed in the isolated 6 subjects (Figure \ref{fig:ema-psych-dist}A/B) with again a similar profile as they were in the original broader dataset (Figure \ref{fig:ema-wider-dists}C/D), but now proportionally fewer severe symptom ratings. This could further complicate the ability to effectively model the psychosis-related EMA summary scores, though it is not a surprising result in hindsight, as many of the biggest contributors excluded from the final dataset had a primary diagnosis of a Schizophrenia spectrum disorder, while the chosen 6 participants all had a primary Bipolar diagnosis. This context should be kept in mind when interpreting the psychosis-related EMA modeling results later. 

As can be seen in Figure \ref{fig:ema-mood-pie}, subject 5BT65 contributed the vast majority ($> 75\%$ of the filtered data points) of the records in the modeling dataset that did not meet the 15 second journal duration minimum set to maintain a reasonable quality standard for the more complex diary features. Therefore, is is unclear from these distributions alone the extent to which the differences between the modeling datasets with and without the duration minimum can be explained by a relationship with diary length versus by a participant-dependent factor related to 5B's typical EMA submissions. Nevertheless, it is interesting to observe that the overwhelming majority of the points removed from the duration filtered version of the positively worded emotion-related EMA summary score distribution had the maximum symptom severity value (rightmost bin, Figure \ref{fig:ema-mood-dist}C against Figure \ref{fig:ema-mood-dist}A). This observation already lends credence to the idea that the EMA scores can be modeled to a reasonable extent using word count alone, a question to be addressed more thoroughly below. Similarly, for the negatively worded emotion-related EMA summary score, most of the difference between the full 7 subject dataset and the duration filtered version is a decrease in the number of moderate symptom severity EMA responses found in the $\geq 15$ second distribution (Figure \ref{fig:ema-mood-dist}D compared against Figure \ref{fig:ema-mood-dist}B). 

As expected from our background on 5BT65, the length filtering had the opposite effect on the psychosis-related EMA summary distributions, reducing the number of records with zero symptoms reported for both hallucinations and delusions (Figure \ref{fig:ema-psych-dist}). Because the modeling of verbosity across all diaries was planned in large part because of 5BT65, there is thus little reason to include the psychosis-related EMA summary scores as labels in this task. I will therefore proceed with 3 modeling datasets, performing the full feature modeling for both categories of EMA but performing the minimalist verbosity modeling only for the emotion-related EMA summaries.

Note that while the effects of the duration minimum on both of the psychosis-related distributions and on the positively worded emotion-related distribution could also be consistent with participants submitting short recordings being more likely to select the top option on every EMA item so they can quickly submit (as in both cases it corresponds to "very slightly or not at all"), the fact that the negatively worded emotion-related EMA distribution shifted in another way makes it more likely that this is a true difference in self-report patterns. The consistency with clinical expectations is further comforting in this regard. 

\FloatBarrier

\paragraph{Modeling plan.}
After characterizing the EMA-derived labels across the complete modeling datasets, I next created a hold out test set for each planned model, before proceeding with any additional audio journal work. As mentioned, I chose to do an $85\%$/$15\%$ train/test split for each of the 3 final EMA modeling datasets, independently saving a separate training and test CSV for each. To ensure that distinct time periods consisting of contiguous data points were held out as part of the testing of generalization ability, I held out the first $5\%$ of data chronologically from each subject, the last $5\%$ of data chronologically from each subject, and a randomly selected $5\%$ from the remaining days. 

For input features to be used in the more extensive diary feature EMA modeling datasets, I slightly modified the list of 12 features from the distributional comparisons at the beginning of this section (\ref{sec:science2}), removing the mean pause duration and minimum sentence sentiment feature and adding the mean words per sentence feature. More information on this feature selection, along with more detailed accounting of the final datasets and specific plans for model fitting can be found in supplemental section \ref{sec:ema-meth}. Here, I will proceed with reporting of the modeling results, beginning with simple verbosity linear fits.

\subsubsection{Modeling self-reported mood with simple diary verbosity features}
\label{subsubsec:ema-diary-verb}
The first model fit utilized the training dataset with no diary duration limit, and performed ordinary least squares to find an intercept and coefficients for word count as well as words per sentence that would best fit to the same-day positively-worded feelings-related EMA mean. A similar process was then also performed for fitting the negatively-worded EMA summary score. The positive EMA summary had a highly significant fit with over $15\%$ of variance explained by the verbosity features, primarily contributed through lower word count corresponding to worse mood by this measure. The negative EMA however contained negligible linear relationship with verbosity across the training set, and no significant relationship at all with word count. \\

\paragraph{Overall model fit results.} 
The specific fit results obtained by these models were as follows:

\begin{quote}
\underline{Positive EMA model}

$R^{2} = 0.154 (p < 10^{-95})$

\emph{Model parameters} ($+/-$ standard error):

Intercept $= 1.042 +/- 0.076 (t = 26.9)$

WordCount Slope $= -0.0021 +/- 0.0001 (t = -21.75)$

WordsPerSentence Slope $= 0.012 +/- 0.006 (t = 4.32)$
\end{quote}

\begin{quote}
\underline{Negative EMA model}

$R^{2} = 0.006 (p = 0.00025)$

\emph{Model parameters} ($+/-$ standard error):

Intercept $= 0.7 +/- 0.08 (t = 17.1)$

WordCount Slope $= -10^{-5} (t = -0.135)$

WordsPerSentence Slope $= 0.011 +/- 0.006 (t = 3.9)$
\end{quote}

\noindent I then tested the positively-worded EMA verbosity model on the held out test set corresponding to this training set. Overall, the model was able to explain $\sim 12\%$ of variance in the test set EMAs, an $R^{2}$ not too far off from the training fit and still quite significant. \\

\begin{figure}[h]
\centering
\includegraphics[width=0.9\textwidth,keepaspectratio]{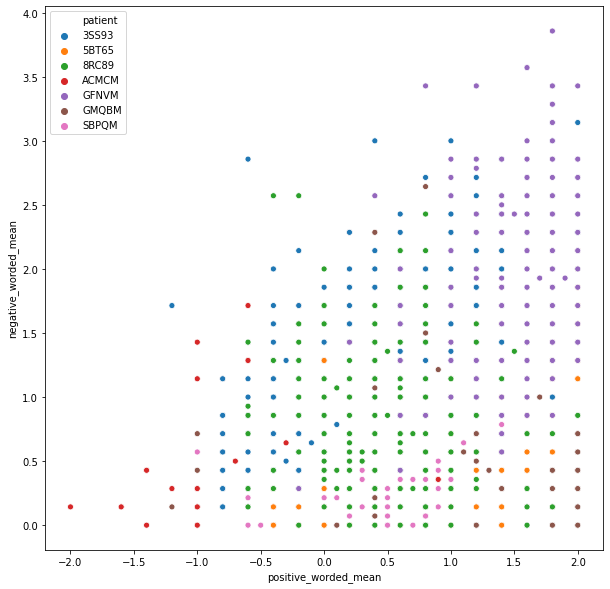}
\caption[Subject-specific patterns in EMA responses from emotions modeling dataset.]{\textbf{Subject-specific patterns in EMA responses from emotions modeling dataset.} Scatter plot showing mean positively-worded feelings-related EMA summary score versus mean negatively-worded feelings-related EMA for days across the verbosity model training dataset, colored by participant ID. Different participants demonstrated different behavior in both mean and variance across the two styles of EMA summary. For example, GF (purple) reported high symptom severity across EMAs, but showed little variance in the positively-worded ones, rarely reporting they were e.g. more than a little happy. 8R (green) had a similar shape to their scatter as GF with no real correlation between the EMA wording valences, but 8R's responses had more similar levels of variance between the two styles and overall less severe symptom scores than GF. By contrast, AC (red) covered a much wider range of the positive EMA axis than the negative one, and SB (pink) also showed much more positive than negative EMA variance. 5B (orange) and 3S (blue) on the other hand demonstrated some linear correlation between positive and negative EMA scores, with 3S showing more variance overall. Note that these seaborn scatter plots do not indicate when multiple points are overlapping. Because some points may be obfuscated here due to points at the same value from other subjects, Figure \ref{fig:ema-pt-rel-sup} subsequently shows the corresponding scatter independently for each subject, to demonstrate the described trends more clearly.}
\label{fig:ema-pt-rel}
\end{figure}

\paragraph{Subject-specific differences in EMA response patterns.}
As noted, the positive and negative EMA wordings saturate at different points of symptom severity, with positively-worded items unable to capture anything beyond e.g. "not happy" when symptoms are judged to be severe and negatively-worded items unable to capture anything beyond e.g. "not sad" when symptoms are judged to be more mild. Of course, different participants will have not only different baseline symptom severity levels, they will also have different levels of variance in symptom severity over the course of the study, which can be independent of the baseline in part. In daily self-report surveys, there is additionally an element of between-subjects heterogeneity that results from differences in self-perception. It is thus not surprising that participant-dependent differences arose in the EMA summary score relationships of the present dataset (Figure \ref{fig:ema-pt-rel}). For EMA summaries that lack much variance in a particular subject, it will be much more difficult to model relationships with diary features, as that score is providing minimal longitudinal information. It is worth noting as well that for a patient like GF who submitted positively-worded EMA responses that so rarely indicated more than a little positive emotion, filling out these surveys may be an unhealthy habit that over time serves to reinforce the lack of good feelings, like a reverse CBT. 

\begin{figure}[h]
\centering
\includegraphics[width=\textwidth,keepaspectratio]{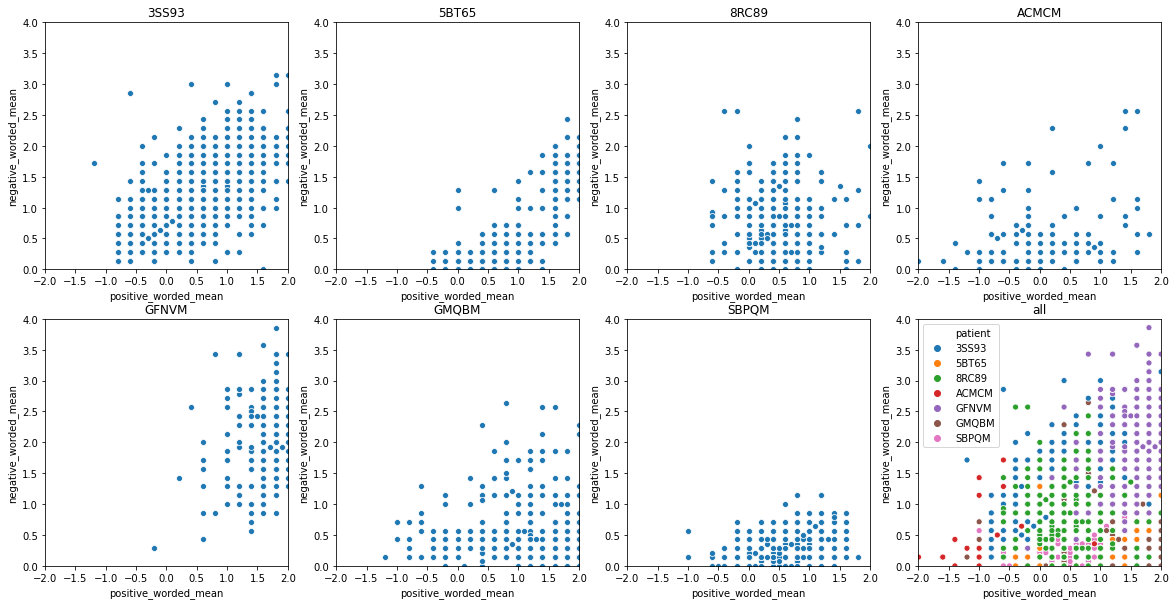}
\caption[Participant IDs cluster in scatter of positively-worded versus negatively-worded EMA summary scores.]{\textbf{Participant IDs cluster in scatter of positively-worded versus negatively-worded EMA summary scores.} Scatter plot showing mean positively-worded feelings-related EMA summary score versus mean negatively-worded feelings-related EMA for days across the verbosity model training dataset, broken down by participant ID. Figure \ref{fig:ema-pt-rel} appears in the bottom right corner here, while each other panel displays a scatter containing points from only a single subject, starting in alphabetical order from the top left. This serves to demonstrate that the subject-specific differences seen in the EMA response patterns are not an artifact of overlapping points in the combined histogram.}
\label{fig:ema-pt-rel-sup}
\end{figure}

It has already been well-documented that there were strong subject-specific differences in diary feature distributions in BLS, including in the word count and words per sentence metrics. It is therefore possible that the across subjects model is explaining variance in EMA summary scores and particularly in positively-worded EMA only because word count can be used to distinguish between participants, each of whom displayed distinct EMA response patterns (Figure \ref{fig:ema-pt-rel}). Because we are looking at just 7 subjects here, it is impossible to say if this is a clinically relevant between-subjects difference or a coincidence: while the speech differences and the EMA differences are well-powered, the alignment between them is not at all. \\

\FloatBarrier

\noindent To determine the extent to which word count and words per sentence might hold predictive relevance for EMA beyond the splitting out of individual participants, I took the z-score of each of these features relative to the corresponding participant. The resulting model fits were very small and not significant:

\begin{quote}
\underline{Positive EMA z-score model}

$R^{2} = 0.021 (p = 0.021)$

\underline{Negative EMA z-score model}

$R^{2} < 0.0001 (p = 0.958)$
\end{quote}

\noindent However, this simply means that \emph{across subjects} there was no strong/consistent \emph{linear} relationship between atypical verbosity for a given participant and their same-day EMAs. That could be the result of there actually being no longitudinal significance (linearly), or it could be the result of longitudinal significance that is restricted to particular subjects or even that takes opposite direction in different subjects. Recall also that 5B, the participant who motivated these verbosity-focused analyses in the first place, submitted a large number of very short recordings relative to the other considered subjects. As missingness can be driven by different factors for different individuals as well, it is unclear if hypothetical low word count days might just be missing for other subjects; we cannot know for sure what the verbosity relationship would look like across participants if e.g. significant extra compensation were given per week with a complete daily diary set.

\begin{figure}[h]
\centering
\includegraphics[width=\textwidth,keepaspectratio]{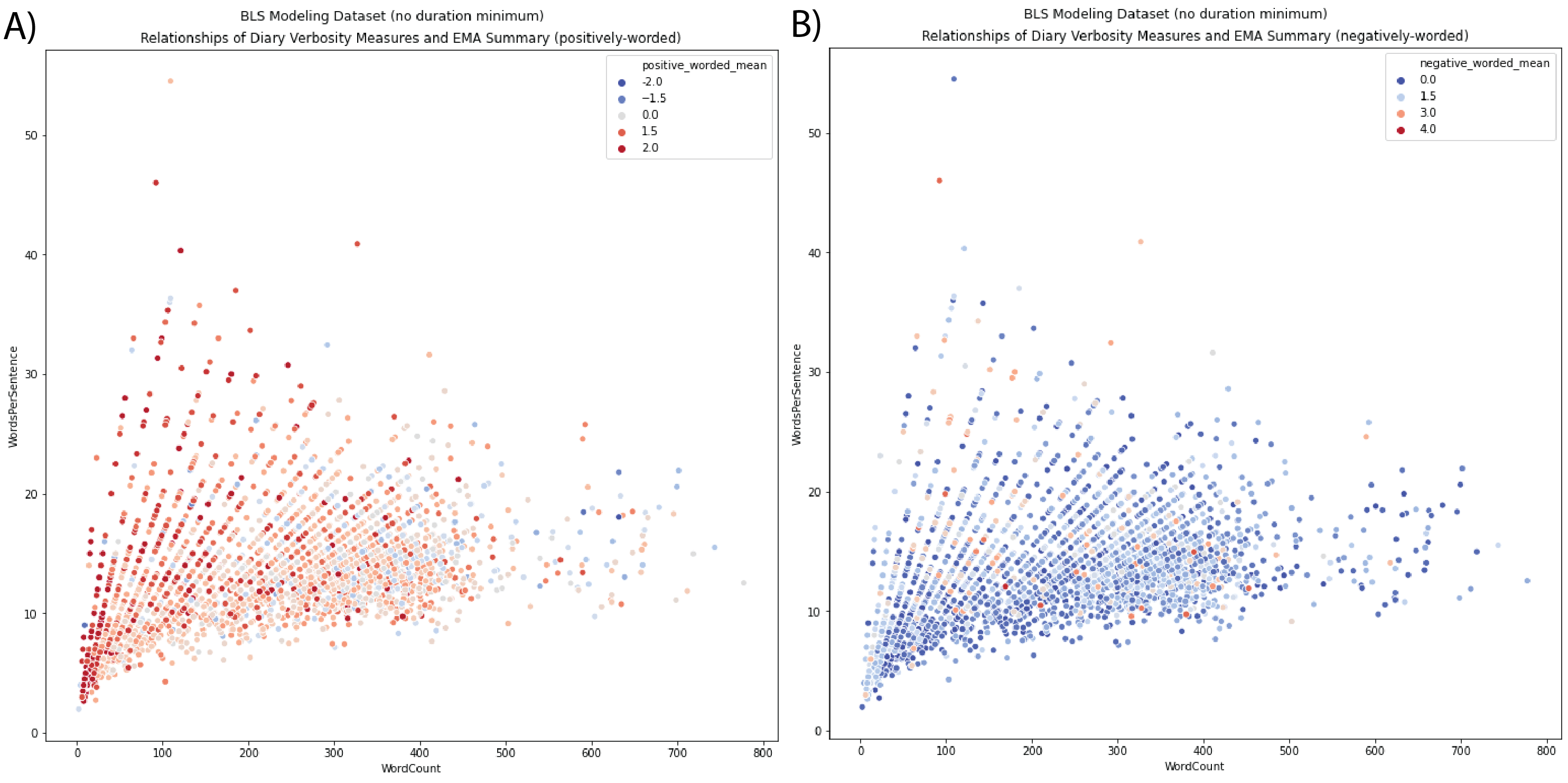}
\caption[Relationship between verbosity features and emotion-related EMA summary scores in the modeling dataset.]{\textbf{Relationship between verbosity features and emotion-related EMA summary scores in the modeling dataset.} Across the model training set with no duration limit, word count (x-axis) and words per sentence (y-axis) metrics were scattered for each diary point with seaborn. The same day mean positively-worded feelings-related EMA summary score was used to color each point with coolwarm colormap ranging between -2 and 2 (A), and then the same plot was recreated with color instead corresponding to same day mean negatively-worded feelings-related EMA ranging between 0 and 4 (B).}
\label{fig:verb-mod-scatter}
\end{figure}

Nevertheless, the relationship between verbosity and self-reported mood-related symptom severity (regardless of wording) contributed by 5B was obvious even on the dataset-wide scale, with a cluster of more severe self-reports found for word counts less than 50 (Figure \ref{fig:verb-mod-scatter}). The contribution of words per sentence outside of its correlation with total word count was unclear from the scatter plot and unlikely to be strong based on the early modeling results, but due to the large variance seen in words per sentence for participant GF specifically, I checked an individualized verbosity model for their EMA. As this yielded an insignificant fit with tiny $R^{2}$ for both EMA wordings and a quick visual did not suggest any nonlinear dynamics, it is safe to say that despite the wild subject-specific variance there was no relationship between words per sentence and self-reported mood in GF, at least not on the day level.

Thus I focus on participant-specific single variable linear models using only word count to model the EMA summaries for the remainder of this subsection. Continuing with earlier themes, I fit such models for 3S, 8R, and 5B. 3S demonstrated no significant relationship between word count and any EMA, but interesting observations were made for 8R and especially 5B. \\

\FloatBarrier

\begin{figure}[h]
\centering
\includegraphics[width=\textwidth,keepaspectratio]{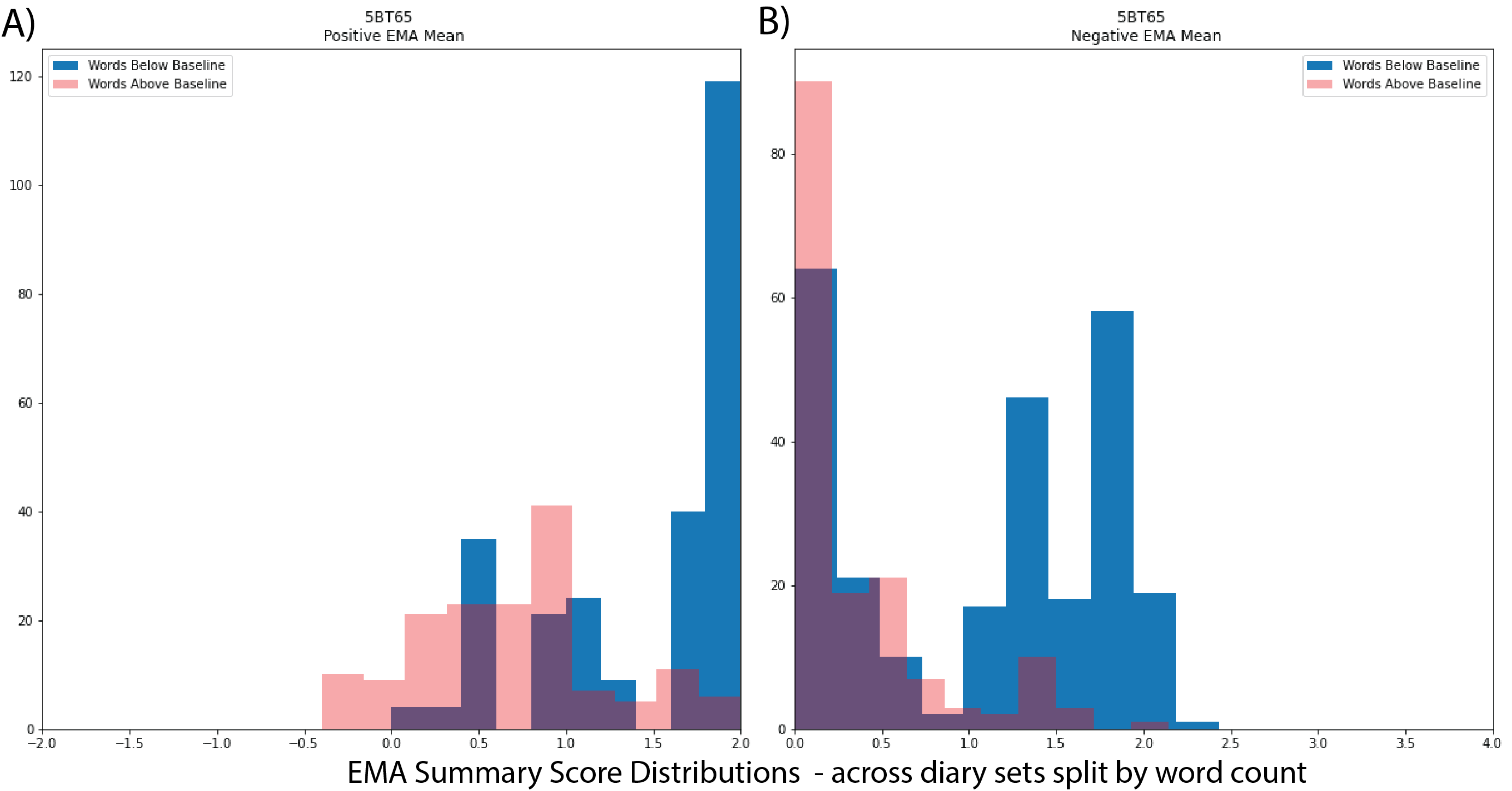}
\caption[High self-reported severity of mood symptoms by patient 5BT65 overwhelmingly corresponded with low same-day diary word count.]{\textbf{High self-reported severity of mood symptoms by patient 5BT65 overwhelmingly corresponded with low same-day diary word count.} Across the model training set with no duration limit, diaries submitted by subject ID 5BT65 were partitioned into two groups: word count $\leq 52$ (blue) and word count $> 52$ (transparent red), corresponding to verbosity below and above that participant's mean, respectively. For these two categories, the distributions of same day positively-worded (A) and negatively-worded (B) emotion-related EMA means were plotted as histograms. Days where 5B reported no positive emotions (A, far right) were primarily low verbosity diary days by a large margin. Similarly, days where 5B reported more than a little negative emotions (B, left side $> 1$) were again primarily low verbosity diary days.}
\label{fig:5b-verb-hist}
\end{figure}

\paragraph{Strong relationship between severe EMA and poverty of speech in 5BT65.}
As 5B's EMA scores tended to saturate at 0 for the negative summary score when symptoms were self-reported to be lower and at 2 for the positive summary score when symptoms were self-reported to be higher, it is immediately clear that a linear model is likely to fail to fully capture the strength of any relationship between word count and EMA. To demonstrate just how different EMA responses were when word counts were $>$ versus $\leq$ the mean 5B submission (52 words), I characterized the EMA distributions for these two groups (Figure \ref{fig:5b-verb-hist}).   

Overall, 5B had 256 diary submissions in the present training set with word count below this baseline, and 156 with word count above this baseline. The mean positive EMA in the low verbosity group was 1.44 with standard deviation 0.59, versus a mean positive EMA in the high verbosity group of 0.67 with standard deviation 0.53. Similarly, for negative EMA the low verbosity mean was $1.05 +/- 0.74$ while the high verbosity mean was $0.32 +/- 0.44$. Categorical counting stats for EMA across the two groupings of 5B's training dataset were also stark (Figure \ref{fig:5b-verb-hist}):
\begin{itemize}
\item 5B only reported having moderately or more positive emotions 4 out of the 256 days with short diaries, but 19 of the 156 days with long diaries.
\item On the other end, they reported less than a little positive emotion in 172 of the 256 days with short diaries, and only 29 of the 156 days with long diaries.
\item For negatively-worded EMA, they reported 0 negative emotions in 47 of the 256 days with short diaries and 59 of the 156 days with long diaries.
\item On the other end, they reported more than a little negative emotion in 154 of 256 days with short diaries, but only 16 of 156 days with long diaries. They also reported more than moderate negative emotion in 10 short diary days and only 1 long diary day.
\end{itemize}
\noindent Ultimately, it was very rare for 5B to submit a long diary and self-report severe symptoms relative to their norm (Figure \ref{fig:5b-verb-hist}), so the relationship between diary verbosity and same day EMA found in this subject is plainly real. 

The linear modeling results for 5B alone were unsurprisingly also strong, with more than $30\%$ of variance in positive EMA scores explainable purely by the corresponding diary word count and more than $20\%$ of variance in negative EMA scores explainable purely by the corresponding diary word count, and both relationships highly significant. Obviously lower word count was associated with more severe symptoms in these models. The results of the fit were as follows:

\begin{quote}
\underline{Positive EMA model for \textbf{5BT65}}

$R^{2} = 0.323 (p < 10^{-35})$

\emph{Model parameters} ($+/-$ standard error):

Intercept $= 1.62 +/- 0.09 (t = 37.1)$

WordCount Slope $= -0.009 +/- 0.001 (t = -14)$
\end{quote}

\begin{quote}
\underline{Negative EMA model for \textbf{5BT65}}

$R^{2} = 0.24 (p < 10^{-25})$

\emph{Model parameters} ($+/-$ standard error):

Intercept $= 1.21 +/- 0.1 (t = 24.4)$

WordCount Slope $= -0.008 +/- 0.002 (t = -11.4)$
\end{quote}

\noindent Using the fit model parameters on the 5B data points from the corresponding test set, the relationship with word count remained strong -- $31.2\%$ of test point variance in the positively-worded EMA summary and $27.7\%$ of test point variance in the negatively-worded EMA summary were explainable by the model predictions, indicating good generalization. 

Recall that the test set included hold out points from the first and last month of the 5B dataset, suggesting that a relationship between perceived symptoms and diary word count was sustained for 5B throughout their time in the study. In particular, both models had good performance on the end of study hold out set and the random hold out set, in the case of the negative model explaining a small amount of additional variance than in the training set. But for the early study hold out set, while there was still certainly some generalization, there was a small but noticeable decrease in performance. 

\begin{figure}[h]
\centering
\includegraphics[width=0.75\textwidth,keepaspectratio]{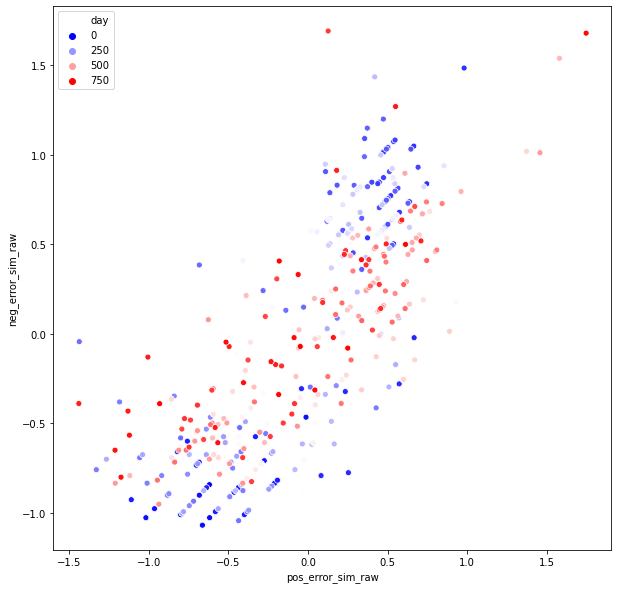}
\caption[Goodness of fit correlated with study day in verbosity model of 5BT65 mood EMAs.]{\textbf{Goodness of fit correlated with study day in verbosity model of 5BT65 mood EMAs.} This scatter plot shows the real - predicted EMA score of the linear fit model between participant ID 5BT65's diary word counts and same day EMA summary scores in the verbosity training set. The raw error for positive emotion EMA summary is found on the x-axis, and for the negative emotion EMA summary on the y-axis. Each point is colored based on the study day it was submitted, from blue to red. The model that was fit to negative EMA in particular demonstrated worse fit to the training points earlier in the study timeline, with higher magnitude error both in overestimating and underestimating severity. This suggests that a rolling slope between word count and symptom severity could have better fit the negative EMA in particular.}
\label{fig:5b-beta}
\end{figure}

It is worth noting however that the coefficient modulating this relationship (as well as the intercept) might have been better modeled with a rolling fit, as that would have not only demonstrated forward-looking prediction over a longer range of the study time period, it additionally could have been able to reflect some of the apparent changes that occurred in 5B over time (Figure \ref{fig:5b-beta}). Specifically, the negatively-worded EMA summary score produced by the model fit demonstrated worse error in general on earlier study days in the training set, and this larger error was observed in both the overestimate and underestimate directions.

Longitudinal model changes more broadly could relate to long term differences in journaling behavior (or survey behavior) as a participant acclimates to a study, long term symptom improvements due to e.g. a new treatment plan, or depending on how quickly prior fit information decays in a rolling model, it could also reflect predictions that are more directly grounded in the participant's recent disease state, thereby fluctuating parameters to capture natural behavioral variations and thus implicitly utilizing the most recent prior EMA responses in making a next prediction. The last point could make good sense for the timescale to be used for the rolling intercept, in order to improve model prediction accuracy (which as can be seen in Figure \ref{fig:5b-beta} still wasn't great, despite the impressive strength of the relationship between word count and EMA in 5B). For the slope of the relationship between word count and EMA however, a rolling slope with a much slower decay would make more sense, as this is a correlation we would only expect to significantly change over longer time periods. 

It is also worth mentioning that some of the worst error points in Figure \ref{fig:5b-beta} correspond to days with abnormally high word counts, causing impossible values to be predicted for the EMA. Because the self-report summary scores saturate, it would be necessary to introduce some non-linearity to extract all the possible value from this simple verbosity model. Regardless, the highly significant and relatively large amount of self-report variance that could be explained by the word count of the corresponding diary submission, which was relatively robust to unseen study time periods, was impressive. \\

\FloatBarrier

\paragraph{Higher verbosity correlated with more severe responses in 8RC89.}
When modeling participant 8RC89's EMA submissions against same day diary word count, there was minimal relationship for the positively-worded summary scores, but interestingly there was a moderate and significant correlation found with the negatively-worded summary scores. The linear model with word count alone explained more than $12\%$ of the variance in 8R's negative emotion EMA means. One thing that makes this result particularly notable is that the relationship was in fact in the opposite direction as 5BT65's, demonstrating the importance of modeling that is sensitive to individual differences, even with such a simple feature as word count. The model fit summary was as follows:

\begin{quote}
\underline{Negative EMA model for \textbf{8RC89}}

$R^{2} = 0.127 (p < 10^{-12})$

\emph{Model parameters} ($+/-$ standard error):

Intercept $= 0.42 +/- 0.09 (t = 9.7)$

WordCount Slope $= 0.0012 +/- 0.0005 (t = 7.4)$
\end{quote}

\noindent Using the fit model parameters to predict negative EMA summary from diary word count in the test set of 8R data points, it was possible to explain $8\%$ of variance, still a meaningful correlation.

\begin{figure}[h]
\centering
\includegraphics[width=0.9\textwidth,keepaspectratio]{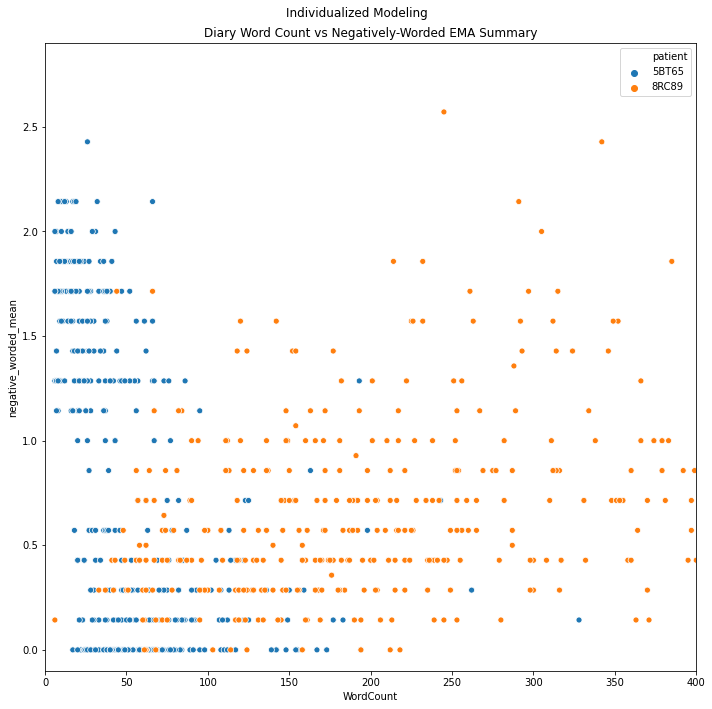}
\caption[Word count predicts negative emotions in opposite ways for 5B and 8R.]{\textbf{Word count predicts negative emotions in opposite ways for 5B and 8R.} Diary word count is scattered against same day negatively-worded emotion-related EMA mean in the verbosity modeling training set for participants 5BT65 (blue) and 8RC89 (orange). For 5B, the days with perceived greater negative emotions corresponded to lower word count diaries, whereas for 8R the relationship is reversed: the days with perceived greater negative emotions tended to correspond to higher word count diaries, albeit with a higher level of overall variance. This underscores the importance of personalized modeling in psychiatry, as participant-dependent effects occur even with such simple features as verbosity. Note the x-axis here is restricted to between 0 and 400 words for clarity, as 5B had in general very short diary submissions. For a zoomed out view, along with other less critical expository materials for this modeling, see supplemental section \ref{sec:sup-stupid}.}
\label{fig:8r-5b-comp} 
\end{figure}

The decrease in variance explained does suggest that the 8R relationship might have had more issue generalizing to unseen study time periods than the 5B model did though. Recall that 8R continued to participate in the BLS study for over 5 years and submitted more sporadically towards the end of that period, so regardless of the strength of the relationship it is not especially surprising that a simple linear model would lose some goodness of fit on hold out points from outside of the main training time period. Indeed, performance on the random hold out set was similar to the training set, but there was increased error in both the initial hold out set and the end of study hold out set. Without fitting a rolling model it is difficult to say to what extent a change in slope over time could have explained additional variance in a generalizable way. Ultimately, it seems reasonable to conclude that there was a real but weak relationship between diary word count and self-reported negative emotion severity in 8R, with opposite directionality of the strong relationship found in 5B (Figure \ref{fig:8r-5b-comp}).

The observed subject-dependence of the directionality of correlation between word count and negatively-worded emotions-related EMA (Figure \ref{fig:8r-5b-comp}) underscores the importance of personalized modeling. This is also consistent with priors from psychiatry, where symptoms of a given disorder can take quite literally opposite form (e.g. sleeping too little or sleeping too much in depression). The negative emotion items included in the considered summary score could capture both a depressed mood like what is typically associated with paucity of speech and a bad mood associated moreso with increased irritability, perhaps resulting in greater tendency for angry rants -- something that might connect with participant gender as well. Regardless of whether such differences represent fundamental mechanistic distinctions or personalized manifestations of the same underlying pathology, it is critical that studies do not average out these relationships. 

\FloatBarrier

\subsubsection{Modeling self-reported mood with core diary features}
\label{subsubsec:ema-diary-all}
To further elucidate predictors of the emotion-related EMA summary scores in the selected subjects, I next moved onto linear models fitting the full set of 11 final journal summary features. Recall that the training dataset for this task required journals to be $\geq 15$ seconds in order to be included, such that the only meaningful difference between this training set and the prior verbosity-focused modeling training set is that the current feature modeling will exclude $\sim 150$ of the shortest 5BT65 diaries. Most of these diaries had similar EMA scores on the more severe end for 5B, so keep in mind a good overall model for 5B could easily be derived from a good model for 5B on this subset. 

For initial modeling, I again began with ordinary least squares including all the (raw) features in question, fitting the training set to the positively-worded EMA summary and then the negatively-worded EMA summary. The resulting fits will in part relate to the ability of the features to distinguish between different subjects, but it remains a useful starting point. Both models were able to explain approaching $30\%$ of the variance in EMA summary across the whole training set and were highly significant. \\

\paragraph{Overall model fit results.} 
The specific fit results obtained by these models were as follows:
\begin{quote}
\underline{Positive EMA model}

$R^{2} = 0.282 (p < 10^{-165})$
\vspace{2mm}
\emph{Top feature coefficients} by t-stat (magnitude $> 3$):

Sentiment $(t = -21.1)$

WordCount $(t = -12.7)$

WordsPerSentence $(t = 6.2)$

NonverbalEdits $(t = -4.6)$

Restarts $(t = -4.3)$

Incoherence $(t = -3.4)$
\end{quote}

\begin{quote}
\underline{Negative EMA model}

$R^{2} = 0.3 (p < 10^{-182})$
\vspace{2mm}
\emph{Top feature coefficients} by t-stat (magnitude $> 3$):

Sentiment $(t = -25.8)$

WordsPerSentence $(t = 10.8)$

Uncommonness $(t = 7.1)$

VerbalEdits $(t = 5.1)$

SpeakFrac $(t = -4)$

Repeats $(t = 4)$

NonverbalEdits $(t = 4)$
\end{quote}

\noindent Although the linear correlations between the individual diary features were overall quite modest (section \ref{subsubsec:diary-corrs}), some of them were still large enough to be affecting this modeling, so to see independent contributions it will be necessary to fit some subsets of these features on the training data and see how much explained variance is actually added by an individual feature addition. The t-stats from the full feature fit can be used as a starting point for this process. It will be good as well for any potential downstream testing on the holdout set to remove features from each respective fit that do not appear to be meaningfully contributing.

Sentiment alone could explain $\sim 14\%$ of variance in the positive emotions EMA and $\sim 19\%$ of variance in the negative emotions EMA. In the duration filtered training set, the verbosity features could explain $\sim 12\%$ of the variance in the positive emotions EMA and $\sim 3\%$ of the variance in the negative emotions EMA. For the former, while words per sentence had a significant t-stat, it did not make much difference to remove it. For the latter, word count had no correlation at all with negative EMA in this training set. Thus a minimal model would be using sentiment and word count for positive EMA summary and sentiment and words per sentence for negative EMA summary. Each such linear model explained $\sim 23\%$ of variance in the EMA response scores. 

For each of nonverbal edits, restarts, and incoherence, adding the feature independently to the minimalist positive EMA model explains between 1 and 2 percent additional variance, so it is likely best to stick with the model featuring only sentiment and word count. In the negative EMA case, we can make another solid gain by including a single feature that represents the sum of all disfluency types, thus getting a quite simple linear model up to $\sim 27\%$ variance explained on the negative emotions EMA summary score. Of course these models likely have the same concern as previously discussed for verbosity-only models: a decent chunk of variance could be explained by way of distinguishing participants through speech pattern differences that may or may not be actually clinically relevant. As such, I will primarily focus on overall models of individually-normalized features, in addition to participant-specific models, in the rest of this subsection. \\  

\paragraph{Subject-specific training dataset stats.}
Before proceeding with modeling that accounts for differences between subjects, I will first summarize the extent to which both features and labels indeed do have major subject-specific differences in the training dataset. To do so, I report here on the mean and variance of each included input and label across the records contained in the main feelings-related training CSV, overall and then within each individual participant's data only (Tables \ref{table:ema-feat-stats-by-pt}-\ref{table:ema-feat-stats-by-pt2}). 

\begin{table}[!htbp]
\centering
\caption[Mean and standard deviation of core diary features per patient in the main model training dataset.]{\textbf{Mean and standard deviation of core diary features per patient in the main model training dataset.} For each of the 7 selected subject IDs as well as for the training set as a whole, I report the mean +/- the standard deviation of the 11 diary input features in this table. Each feature is a row, and each patient ID is a column (listed in alphabetical order). The depiction of the main training set in Figure \ref{fig:ema-mood-pie}B shows the relative contribution to dataset size here from across the 7 IDs. Ultimately, large differences in typical value and level of variance existed across the considered participants, so an updated general linear model was fit using input features z-scored in a participant-dependent manner instead.}
\label{table:ema-feat-stats-by-pt}

\begin{tabular}{ | m{2.75cm} || m{1.25cm} || m{1cm} | m{1cm} | m{1cm} | m{1cm} | m{1cm} | m{1cm} | m{1cm} | }
\hline
\textbf{Feature \newline Summarized} & \textbf{All \newline ($n=7$)} & \textbf{3S} & \textbf{5B} & \textbf{8R} & \textbf{AC} & \textbf{GF} & \textbf{GM} & \textbf{SBP} \\
\hline\hline
Word Count & 252 +/- 134 & 333 +/- 68 & 69 +/- 43 & 223 +/- 132 & 397 +/- 198 & 151 +/- 93 & 168 +/- 87 & 299 +/- 73 \\
\hline
Words per \newline Sentence & 14.1 +/- 4.7 & 14.4 +/- 3 & 10.2 +/- 3.1 & 11.1 +/- 3 & 14.9 +/- 4.3 & 18.1 +/- 8 & 16.2 +/- 5.6 & 15.4 +/- 4 \\
\hline
Speech \newline Fraction & 0.67 +/- 0.16 & 0.49 +/- 0.06 & 0.79 +/- 0.08 & 0.78 +/- 0.12 & 0.84 +/- 0.08 & 0.79 +/- 0.11 & 0.78 +/- 0.08 & 0.66 +/- 0.11 \\
\hline
Speech Rate & 4.3 +/- 1 & 4.2 +/- 0.5 & 3.5 +/- 0.7 & 5.1 +/- 1.1 & 4.4 +/- 2 & 4.4 +/- 1.1 & 4.5 +/- 0.6 & 3.9 +/- 0.6 \\
\hline
Nonverbal \newline Edits & 0.046 +/- 0.024 & 0.061 +/- 0.019 & 0.051 +/- 0.028 & 0.051 +/- 0.023 & 0.023 +/- 0.012 & 0.038 +/- 0.021 & 0.023 +/- 0.014 & 0.032 +/- 0.011 \\
\hline
Verbal Edits & 0.007 +/- 0.009 & 0.004 +/- 0.004 & 0.002 +/- 0.005 & 0.01 +/- 0.1 & 0.015 +/- 0.012 & 0.01 +/- 0.013 & 0.011 +/- 0.015 & 0.005 +/- 0.005 \\
\hline
Restarts & 0.018 +/- 0.013 & 0.028 +/- 0.011 & 0.01 +/- 0.014 & 0.016 +/- 0.013 & 0.013 +/- 0.008 & 0.011 +/- 0.013 & 0.01 +/- 0.009 & 0.016 +/- 0.008 \\
\hline
Repeats & 0.011 +/- 0.01 & 0.018 +/- 0.009 & 0.005 +/- 0.01 & 0.008 +/- 0.008 & 0.009 +/- 0.008 & 0.005 +/- 0.008 & 0.007 +/- 0.008 & 0.007 +/- 0.006 \\
\hline
Sentiment & 0.11 +/- 0.14 & 0.09 +/- 0.1 & 0.1 +/- 0.16 & 0.11 +/- 0.1 & 0.19 +/- 0.18 & 0.01 +/- 0.22 & 0.11 +/- 0.16 & 0.18 +/- 0.11 \\
\hline
Incoherence & 1.45 +/- 0.04 & 1.46 +/- 0.04 & 1.44 +/- 0.04 & 1.47 +/- 0.04 & 1.47 +/- 0.04 & 1.43 +/- 0.05 & 1.43 +/- 0.05 & 1.45 +/- 0.04 \\
\hline
Uncommonness & 2.24 +/- 0.07 & 2.26 +/- 0.05 & 2.26 +/- 0.09 & 2.26 +/- 0.06 & 2.25 +/- 0.08 & 2.26 +/- 0.08 & 2.2 +/- 0.07 & 2.2 +/- 0.05 \\
\hline
\end{tabular}
\end{table}

\begin{table}[!htbp]
\centering
\caption[Mean and standard deviation of summary EMA scores per patient in the main model training dataset.]{\textbf{Mean and standard deviation of summary EMA scores per patient in the main model training dataset.} For each of the 7 selected subject IDs as well as for the training set as a whole, I report the mean +/- the standard deviation of the 2 EMA labels from the primary feelings-related EMA model training set in this table, as well as the same for the 2 EMA labels from the primary psychosis-related EMA model training set, for reference. Each label is a row, and each patient ID is a column (listed in alphabetical order). Just as was the case for the diary features in Table \ref{table:ema-feat-stats-by-pt}, there were large individual differences in typical severity and variance of severity in the self-reported symptoms.}
\label{table:ema-feat-stats-by-pt2}

\begin{tabular}{ | m{2.75cm} || m{1.25cm} || m{1cm} | m{1cm} | m{1cm} | m{1cm} | m{1cm} | m{1cm} | m{1cm} | }
\hline
\textbf{Feature \newline Summarized} & \textbf{All \newline ($n=7$)} & \textbf{3S} & \textbf{5B} & \textbf{8R} & \textbf{AC} & \textbf{GF} & \textbf{GM} & \textbf{SBP} \\
\hline\hline
Positive \newline Emotions EMA & 0.65 +/- 0.69 & 0.46 +/- 0.64 & 0.87 +/- 0.61 & 0.44 +/- 0.43 & 0 +/- 0.81 & 1.56 +/- 0.34 & 1.04 +/- 0.78 & 0.63 +/- 0.48 \\
\hline
Negative \newline Emotions EMA & 0.83 +/- 0.7 & 1.15 +/- 0.56 & 0.49 +/- 0.6 & 0.69 +/- 0.46 & 0.57 +/- 0.57 & 2.08 +/- 0.69 & 0.59 +/- 0.52 & 0.27 +/- 0.24 \\
\hline\hline
Hallucinations \newline EMA & 0.04 +/- 0.18 & N/A & 0 +/- 0 & 0.01 +/- 0.09 & 0.02 +/- 0.16 & 0.38 +/- 0.42 & 0 +/- 0 & 0 +/- 0 \\
\hline
Delusions \newline EMA & 0.21 +/- 0.68 & N/A & 0 +/- 0 & 0.01 +/- 0.07 & 0.02 +/- 0.15 & 1.97 +/- 0.94 & 0.04 +/- 0.32 & 0 +/- 0 \\
\hline
\end{tabular}
\end{table}

Large differences are of course not surprising given the results of the extensive distributional comparisons for 3S, 8R, and 5B (section \ref{subsubsec:diary-pt-dists-comps}), nor given the breakdown of EMA response scores by participant in the verbosity modeling dataset above. However, these numbers are helpful to contextualize the normalization of each feature in a subject-dependent manner. As already seen, diary behavior differences between the considered participants can theoretically be utilized by a model to fit to the between patients heterogeneity observed in the EMA labels, if the input features are not normalized to the corresponding subject. 

Like for the verbosity-only modeling before, I will next take the z-score of each input feature using the calculated mean and standard deviation from each participant, thus making the values in Tables \ref{table:ema-feat-stats-by-pt} a helpful reference point. The results of the general model with personally-normalized features can then be compared to the initial general model fit in the main feelings-related EMA dataset. \\

\FloatBarrier

\paragraph{Sentiment score consistently explains EMA variance across participants.}
The z-score feature models containing all 11 diary features were able to explain $\sim 13\%$ and $\sim 9\%$ of variance in the positively-worded and negatively-worded EMA summary datasets, respectively (both $p < 10^{-41}$). 

Thus while the initial results were certainly impacted by model ability to distinguish subjects, there was also some dynamically meaningful signal across patients. Much of that signal was sentiment, which by itself in a patient-dependent z-scored manner could explain $\sim 11\%$ of positive EMA variance and $\sim 7\%$ of negative EMA variance via simple linear model (both $p < 10^{-41}$). In the positively-worded case, speech fraction and word count may each contribute an additional $\sim 1\%$; in the negatively-worded case, nonverbal edits may contribute an additional $\sim 1\%$, however these are unlikely to be worth including in a final model. The final parameters for the z-scored sentiment models across the dataset were as follows:

\begin{quote}
\underline{Positive EMA z-score model}

$R^{2} = 0.109 (p < 10^{-63})$

\emph{Model parameters} ($+/-$ standard error):

Intercept $= 0.651 +/- 0.013 (t = 50.2)$

Sentiment Slope $= -0.227 +/- 0.013 (t = -17.5)$
\end{quote}

\begin{quote}
\underline{Negative EMA z-score model}

$R^{2} = 0.071 (p < 10^{-41})$

\emph{Model parameters} ($+/-$ standard error):

Intercept $= 0.822 +/- 0.013 (t = 61.3)$

Sentiment Slope $= -0.182 +/- 0.013 (t = -13.8)$
\end{quote}

\begin{figure}[h]
\centering
\includegraphics[width=\textwidth,keepaspectratio]{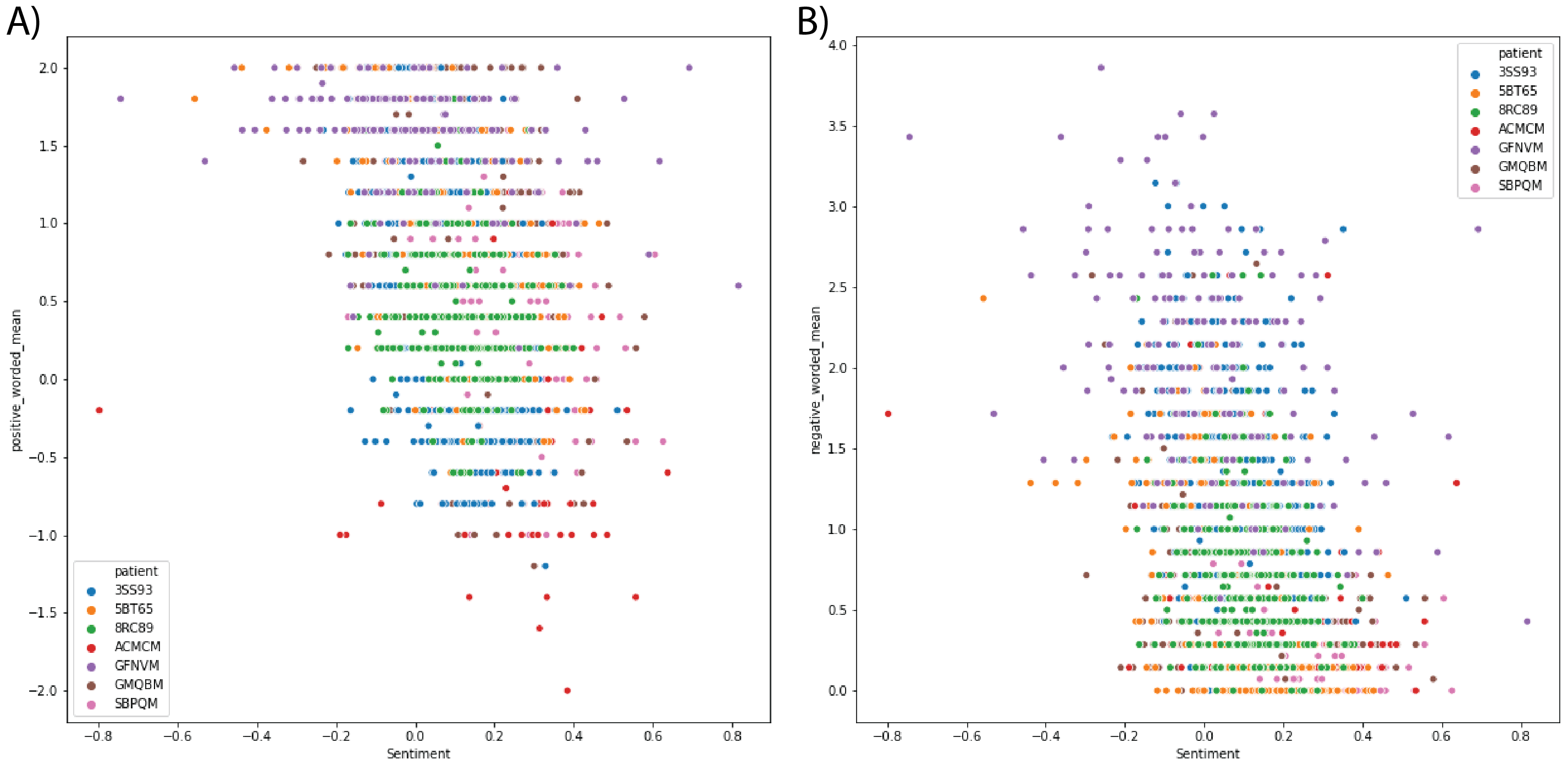}
\caption[Significant relationships between mean diary sentiment and self-reported mood symptom severity.]{\textbf{Significant relationships between mean diary sentiment and self-reported mood symptom severity.} Across the primary training dataset, mean diary sentiment (x-axis) was scattered against the same day emotion-related EMA summary scores (y-axis), for both the positively-worded (A) and negatively-worded (B) item means. Each point was colored according to the corresponding participant ID, as there were both meaningful differences in diary sentiment (and EMA) between subjects as well as meaningful differences in diary sentiment over time within a subject. Indeed, highly significant correlations between EMA and diary sentiment were found for both plotted overall relationships, with Pearson's $r=0.37$ (A) and $r=0.44$ (B) respectively. When sentiment scores were normalized to each subject's personalized baseline with a z-score, significant correlations remained but were weakened, particularly for the negative EMA, then with Pearson's $r=0.33$ (A) and $r=0.27$ (B) respectively.}
\label{fig:sentiment-ema}
\end{figure}

\noindent For the simple sentiment model of positive emotions EMA in particular, there was a relatively small decrease in variance explained when switching to the z-score based fit (Figure \ref{fig:sentiment-ema}).

On the held out test set, predictions made with this basic linear sentiment model could explain $10.4\%$ of variance in positive EMA and $6.7\%$ of variance in negative EMA, close to the training set values. Focusing only on the end of study hold out set containing $\sim 150$ data points all taken from after the individual time periods included in this modeling, the previously fit sentiment coefficient could explain $10\%$ and $6.9\%$ respectively of the variance in the positive and negative EMAs, representing good overall generalization performance. On the earliest part of the study is again where the model struggled, instead explaining only $7.5\%$ and $4.4\%$ respectively of variance in those EMA summary scores. Variance of the true scores was also higher during this period, despite it spanning a more condensed time frame for most participants than the end of study sample. It is possible that behavior during the earliest weeks of self-report collection is impacted by the novelty of the process, another argument for longer term app-based data collection.   

In light of the earlier verbosity modeling results, it is worth considering that sentiment is perhaps the feature least likely to have the "opposite directions for different participants" problem described for word count, as it is hard to imagine someone systematically speaking more positively in their diaries when they simultaneously submit EMAs indicating worse mood. On the other hand, this makes the z-score in some sense less fair for evaluation of the utility of sentiment. Individual differences in verbosity or e.g. use of filler words are intuitive to normalize against in a longitudinal study, but it is not obvious whether a person that is consistently more negative should have their journals benchmarked against a more negative expectation or not. In the end it is likely a matter that would vary depending on the study design and scientific questions being asked. Regardless, the sentiment modeling results are promising. 

It is of course also the case that everyone has some negative and positive mood days regardless of disease status, and without a control dataset it is not possible to say definitively whether journal language sentiment and self-reported mood would correlate to a similar extent in healthy individuals. While it may be questionable then to claim that the strong correlational results here indicate diary sentiment carries clinical significance, a few facts remain:
\begin{enumerate}
    \item Sustained periods of self-reported poor mood are likely to be relevant, and thus sentiment scores over a longitudinal study can be relevant in a similar way. This is arguably inherently true, but given the significant relationship found between EMA and diary sentiment even when the sentiment was normalized based on the individual participant's typical values, there is good reason to believe the pipeline's sentiment summary contains useful signal. 
    \item Extremely negative sentiment scores identified within the dataset - particularly from subject GF - during manual review (section \ref{subsubsec:diary-val-trans}) contained a number of extreme statements that would be clearly out of place in a control journal dataset, something identified by the blinded reviewer too.
\end{enumerate}
\noindent Furthermore, the combination of sentiment with other simple diary features in a nonlinear model has the potential to greatly improve on the already observed predictive power. For example, when sentiment is $< 0$ or $> 0.2$ it clearly begins to have some relevance (Figure \ref{fig:sentiment-ema}), and it becomes increasingly predictive at a nonlinear rate from there. When sentiment is between 0 and 0.2 however, we may not want to consider it much at all -- so a difference of 0.1 in sentiment can be more or less meaningful depending on the location of the comparison point. By focusing on other features when sentiment is in the "typical" range but placing great weight on sentiment when it is not, an improved overall model could be built. \\

\FloatBarrier

\paragraph{Fitting subject-specific models.}
To determine if other diary features besides word count might have unique relationships to EMA within an individual participant's dataset, I fit personalized ordinary least squares linear models on the 11 diary features to predict the 2 different EMA summary scores in each of the 7 subjects. Because this amounts to 154 potential opportunities to find an interesting relationship, we can look for Bonferroni-corrected significance cutoff of $p < 0.0003$ as an estimate of goodness of fit above chance in the training set. 

The overall modeling results for each participant are reported in Table \ref{table:ema-person-mod}. 9 of the 14 models could explain more than $20\%$ of variance in the corresponding EMA summary score with significance passing the testing-correction specified threshold. 7 of them in fact had $p < 10^{-10}$, suggesting that overfitting random noise was not an issue here. 

\begin{table}[!htbp]
\centering
\caption[Fit results for personalized linear EMA models by subject.]{\textbf{Fit results for personalized linear EMA models by subject.} For each of the 7 selected subject IDs, I fit a linear model for both the positively-worded and negatively-worded emotions-related EMA summary scores, using the 11 chosen diary features as input. The $R^{2}$ of each such model along with the F-stat derived p-value are reported in this table, with subjects as rows and EMA labels as columns.}
\label{table:ema-person-mod}

\begin{tabular}{ | m{2cm} || m{3cm} | m{3cm} | }
\hline
\textbf{Subject} & \textbf{Positive EMA \newline (all input fit)} & \textbf{Negative EMA \newline (all input fit)}  \\
\hline\hline
3SS93 & $R^{2} = 0.239 \newline (p < 10^{-40})$ & $R^{2} = 0.206 \newline (p < 10^{-33})$ \\
\hline
5BT65 & $R^{2} = 0.487 \newline (p < 10^{-33})$ & $R^{2} = 0.384 \newline (p < 10^{-23})$ \\
\hline
8RC89 & $R^{2} = 0.108 \newline (p < 10^{-4})$ & $R^{2} = 0.233 \newline (p < 10^{-15})$ \\
\hline
ACMCM & $R^{2} = 0.247 \newline (p < 0.0004)$ & $R^{2} = 0.256 \newline (p < 0.0003)$ \\
\hline
GFNVM & $R^{2} = 0.103 \newline (p < 0.1)$ & $R^{2} = 0.243 \newline (p < 10^{-5})$ \\
\hline
GMQBM & $R^{2} = 0.264 \newline (p < 10^{-12})$ & $R^{2} = 0.254 \newline (p < 10^{-11})$ \\
\hline
SBPQM & $R^{2} = 0.184 \newline (p < 10^{-12})$ & $R^{2} = 0.139 \newline (p < 10^{-8})$ \\
\hline
\end{tabular}
\end{table}

Indeed, the EMA labels used for GFNVM were originally shifted by 13 days, because the consent date for the participant in REDCap was changed between the processing of diaries and the processing of EMA. It would be quite unrealistic to predict single day EMA 13 days in the future from single day audio journal properties today for most cases. It is therefore comforting that the first iteration of GF models was completely meaningless, with $R^{2} < 0.065, p > 0.4$ for the positive EMA model, $R^{2} < 0.035, p > 0.9$ for the negative EMA model, and all coefficient slopes having $95\%$ confidence interval crossing over 0 in both models. It was only when searching for an explanation for the complete unpredictability of GF's EMAs that I found the accounting mistake. This also serves to highlight the importance of ongoing data maintenance and careful pre-analysis review. 

Simply being able to fit personalized models (Table \ref{table:ema-person-mod}) without such concern for overfitting is a testament to the value of audio diaries -- the participant with the fewest samples in the training dataset here (ACMCM) had $n=129$ points for model fitting and the participant with the second fewest (GFNVM) had $n=174$, which can be contrasted with an entire dataset size across subjects of $n=144$ for the interview disfluency modeling done in chapter \ref{ch:2}. It will be extremely difficult to thoroughly characterize speech properties in terms of between \emph{and} within patient effects using interview recordings alone. \\

Due to the upcoming case report results for subject 5B (section \ref{subsec:diary-case-study}), it is worth calling attention to their particularly strong fit in the present dataset before moving on. This is after more than 100 of their lowest word count diaries were removed for feature quality purposes, increasing the mean by 16 words and eliminating some of the clearest cut mappings between diary verbosity and EMA. I originally might have expected 5B to have lesser fit in the main training dataset for this reason, and I certainly did not expect the fit to improve so much and notably exceed all other personalized models considered. However in hindsight of the time course of their EMA responses, this becomes less surprising: the within-week variance of self-report responses from 5B was impressively low despite solid overall variance, suggesting their EMA submissions were less susceptible to random day to day noise and instead better capturing mood state changes. This bodes well for the diary features, because we aimed to use them for modeling of EMA as a proxy for clinically meaningful variance in emotions, and do not expect or even want them to explain all daily life variance in the self-report. Such variance appeared to be substantially lower in 5B than all other subjects.  

For those participants that had higher EMA variance overall as well as intra-week, it is challenging to assess whether that might be explained by truly volatile mood (i.e. itself a feature of pathology, which is potentially time-varying) or by noisy survey-taking behavior. The latter could be caused by misinterpretation of question wording or fatigue due to the fairly large number of daily survey questions presented in the full BLS EMA set. The items used here did not separate out the intensity of the emotion from the percentage of time it was felt over the last 24 hours, and they also did not provide any clear reference point for options from "a little" to "quite a bit". It would be easy to imagine day to day life "noise" swinging many of the question responses by 1 or 2 points even in healthy controls, as e.g. work is busier on some days than others. It need not be serious life matters either - it wouldn't be unreasonable to be "a little" upset because the Red Sox lost or "a little" afraid because of a horror movie. This is not necessarily a flaw in EMA, but it is a limitation of using the self-report scores as a label of interest, particularly on the daily timescale considered. 

Conversely, recall that 8R and especially SBP displayed relatively low levels of variance in their EMA summary scores, as did GF for the positively-worded EMA specifically (Table \ref{table:ema-feat-stats-by-pt2}). It is unsurprising that these models account for the majority of the weaker fit results, along with relatively lesser significance levels in the lower sample size participant AC (Table \ref{table:ema-person-mod}). Still, many of the models had results that warranted further exploration. 

The results of Table \ref{table:ema-person-mod} do not necessarily indicate that scientifically meaningful relationships have been found, as there could be a variety of statistically real but uninteresting reasons for these results. I will therefore next delve more deeply into select personalized models to determine what (if any) significant participant-specific effects are salient beyond what has already been identified. Note as well that some additional less critical expository materials for this modeling can be found in supplemental section \ref{sec:sup-stupid}. 

Models for 3S, 5B, and 8R were of particular interest due to their focus in earlier sections, as was the negatively-worded EMA summary score model for GF. GM's models were of interest as well because of their solid strength of fit and their relatively smaller reliance on sentiment in the joint model coefficients. For each of these cases, I will look more closely into features with higher significance in the respective joint model and features with potential prior relevance, to break down in greater detail the contribution of different journal features to predicting EMA in these participants. One major question to address will be the extent to which fitting personalized sentiment slopes can improve variance explained over the shared z-score based model. \\

\FloatBarrier

\paragraph{3SS93 self-reported emotion models.}
Linear fit to sentiment alone could explain $\sim 19\%$ and $\sim 15\%$ of variance in the positive and negative emotion EMAs for 3S respectively. As previously discussed in the verbosity-modeling training set, which would be near identical to the current training set for 3S specifically, neither of the transcript structure features had a meaningful relationship with either EMA summary. There was also no reason to be interested in the word2vec-derived features for 3S. We are thus down to pause-related and disfluency-related features for the last $\sim 5\%$ of variance that we might expect to explain with the present features and a linear fit for daily EMAs. 

Speech fraction is of unique interest for 3S because they had so many recordings with long pause periods. This feature in linear fit by itself could explain $\sim 5\% (p < 10^{-10})$ of variance in the positive EMA mean and $\sim 2.5\% (p < 10^{-5})$ of variance in the negative EMA mean, with a lower speech fraction (i.e. greater proportion of time spent paused) associated with worse self-reported moods in both cases. Not all of this was independent of sentiment's predictive power, though these are such distinct features that it does not necessarily take away from interpretation of either. With speech fraction added to the positive emotion EMA model, $R^{2} = 0.225$. For the negative emotion EMA model, it increased the variance explained by about 1 percent, but didn't seem meaningfully worth including.

Aside from verbal edits, 3S used disfluencies at an abnormally high rate. However, variance across the different disfluency rates over 3S diaries was relatively low (Table \ref{table:ema-feat-stats-by-pt}). It is perhaps not surprising then that neither nonverbal edits nor repeats + restarts could explain more than $1\%$ of variance in either EMA summary. While disfluencies are more likely to benefit from nonlinear modeling elements because of their atypical distribution, there was not much indication from the 3S data that disfluencies had any real relationship with same day emotion-related EMA summary score. 

Ultimately, sentiment was the primary driver of 3S EMA predictability, with the acoustic speech fraction feature making some significant contribution as well; these relationships were unrelated to word count and there was not much else worth highlighting. \\

\paragraph{5BT65 self-reported emotion models.}
For 5B, it was important to first check how strong the association between word count and EMA remained in the duration-filtered training set. The correlation was unsurprisingly significantly reduced, explaining $\sim 16\%$ of variance in positive emotion EMA summary and $\sim 7\%$ of variance in negative emotion EMA once restricted to only recordings of at least 15 seconds. In this context, the words per sentence metric had new predictive relevance, improving the variance explained in the linear verbosity fits by $\sim 2$ and $4$ percentage points respectively. A higher number of words per sentence in both cases was therefore slightly but significantly associated with worse self-reported moods. 

From there, we would of course next consider the role of sentiment. The mean diary sentiment when added to the verbosity features resulted in model variance explained of $44.5\%$ for positively-worded EMA and $33.5\%$ for negatively-worded EMA. These results already approach the joint model results. Adding a handful of other features that stood out previously such as nonverbal edits and incoherence only improved variance explained by about 1 percentage point, and when modeled with verbosity features alone, these diary metrics contributed only about $3\%$ additional variance explained. 

In sum, diary verbosity and sentiment together were powerful predictors of same-day EMA scores in 5B. A linear fit including word count, words per sentence, and mean sentence sentiment explained over $40\%$ of the variance in positively-worded EMA mean and over $30\%$ of the variance in negatively-worded EMA mean. Additionally, it is quite likely that a nonlinear model focused on only diary word count and sentiment could fit the training set even more impressively, for reasons already discussed. \\

\paragraph{8RC89 self-reported negative emotions model.}
Recall that word count could explain $\sim 14\%$ of the variance in 8RC89's same day self-report of negative emotions severity, with higher word count in this case corresponding to worse perceived symptoms. Sentiment alone explained only $\sim 6\%$ of this same variance, which was low relative to other subject-specific models. However, the explanatory power of sentiment in the training model set was independent of that from word count, so these two features in linear fit produced $R^{2} = 0.207 (p < 10^{-18})$. There were once again no further features that could be added to individually raise the variance explained by more than about 1 percentage point, so the best linear model for 8R would involve just word count and sentiment. \\

\paragraph{GFNVM self-reported negative emotions model.}
As mentioned above, verbosity-related features did not have any linear relationship with EMA for participant GFNVM. Thus we start by checking how well sentiment predicts negative emotion severity self-reported by GF. The linear model with sentiment alone was able to explain $\sim 15\%$ of the variance in GF's negative EMA summary scores. 

The other feature identified as especially salient in the coefficients and t-stats of the combined personalized linear fit was nonverbal edits. Alone, nonverbal edits could explain $\sim 6\%$ of the variance in GF's negative EMA summary scores, though with individual model $p = 0.0015$ this would not be significant after multiple testing corrections. When adding nonverbal edits to the linear fit with sentiment, $R^{2} = 0.184 (p < 10^{-7})$. In these models, a lower rate of nonverbal filler usage was consistently associated with a higher level of self-reported negative emotions. This could warrant further follow-up with reference to other information sources such as clinical ratings from GF. \\

\paragraph{GMQBM self-reported emotion models.}
Participant GM was not highlighted in the earlier verbosity modeling, but word count (though not words per sentence) indeed had a significant relationship in linear fit with both EMA summary scores. Most interesting here is that the directionality of that relationship was opposite for the two EMA wordings in this subject: lower word count was associated with lesser positive emotions reported (i.e. more severe mood symptoms by this proxy), but higher word count was associated with higher negative emotions reported. The word count only linear fits had $R^{2} = 0.065 (p = 0.000015)$ for positively-worded items and $R^{2} = 0.094 (p < 10^{-6})$ for negatively-worded items.

With sentiment added to word count in the linear model, the fit to positively-worded EMA was $R^{2} = 0.192 (p < 10^{-12})$ and the fit to negatively-worded EMA was $R^{2} = 0.219 (p < 10^{-14})$. The explanatory power of sentiment was largely separate from word count here, and the word count coefficients in this 2 feature linear model remained very similar to those in the solo models. Only the repeats per word (for the positive emotion model) and verbal edits per word (for the negative emotion model) were able to individually increase the variance explained from there by more than 1 percentage point, resulting in $21.3\%$ and $23.7\%$ respectively - so still a small improvement. 

A lower usage rate of repeats was associated with lesser positive emotions reported in the bigger model, and a higher rate of verbal edits was associated with greater negative emotions reported in the bigger model. This might be related to the same underlying dynamic as the reported word count relationship differences, though the effect is too small to read into at this time. Either way, it will be important for disfluencies - with audio journals in particular where certain rates are often 0 - to be checked using nonlinear models to make stronger conclusions. 

\subsubsection{Self-reported psychosis symptoms in GFNVM}
It turned out to be impossible to model self-reported positive symptoms of psychotic disorders for this dataset more generally, as only GFNVM demonstrated any real variance in these EMA items across the selected participants. 

Although subject selection was focused on Bipolar disorder, and a few of the patients were known to experience primarily depressive symptoms, I did not expect the divide to be so stark -- and I also did not expect GF, whose primary diagnosis was recorded as Bipolar type II, to be the subject that was submitting the vast majority of non-zero psychotic symptoms EMAs. Upon further discussion with the lab, I found this was originally correctly recorded, but GF's diagnosis is now suspected to be Bipolar I; it is apparently not uncommon to have a diagnosis move from type II to type I over time, which makes sense definitionally.

While it was sensible for the goals of this thesis to perform subject selection the way I did, this experience highlights the importance of checking participant-specific differences in the label set distribution before planning modelling in too much detail. It additionally illustrates that only so much can be assumed from diagnostic categories, though it is unclear to what extent GF interpreted the hallucination questions differently than e.g. GM, who was explicitly labeled as having psychotic features in their BD diagnosis yet never self-reported hallucinations on these surveys and only reported persecutory delusions on a handful of occasions.

Overall, participants 5B and SBP always reported 0 symptoms on the psychosis-related EMAs. Participants AC and 8R reported non-zero symptoms at some point on both hallucination and delusion items, but did so on days that were not connected in time and were ultimately very rare (6 and 3 instances in the training set respectively). As mentioned, participant 3S did not fill out these items on their EMA submissions, and GM only occasionally reported non-zero delusions in the psychosis EMA training set. For GM, this included 4 days that were all within the same one week period, and then 3 additional days at spread out points of the study. While that 4 day stretch may be of interest, it is a rare even that would be difficult to model, and furthermore it occurred near the beginning of the study (and right at the beginning of points included for training), so it is more difficult to draw even qualitative conclusions. 

\begin{figure}[h]
\centering
\includegraphics[width=0.8\textwidth,keepaspectratio]{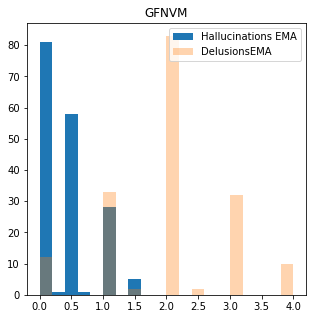}
\caption[Frequent self-reported psychotic symptoms from participant GFNVM.]{\textbf{Frequent self-reported psychotic symptoms from participant GFNVM.} While most of the participants selected for modeling rarely self-reported psychotic symptoms (if ever), subject ID GFNVM reported a non-zero amount of persecutory delusions (transparent orange) in the vast majority of their submitted EMAs, most commonly scoring the delusions item as moderate severity. They also self-reported a non-zero amount of hallucinations (blue) in about half of their submitted EMAs, though rarely with a high severity level. This histogram depicts the distributions of these two psychosis-related EMA categories for GF in the modeling training set (participant-specific $n=174$).}
\label{fig:gf-psych} 
\end{figure}

By contrast, GF very often self-reported non-zero psychotic symptoms, particularly for the delusions-related EMA item (Figure \ref{fig:gf-psych}), which was centered around moderate severity in their distribution. It was rare that GF reported hallucinations more frequently than "a little", but dividing between zero and non-zero hallucinations produced a close to 50/50 split of their EMA submissions. For GF specifically there is thus a suitable distribution to attempt to fit models of psychosis-related EMA responses using their same day diary features. \\

\FloatBarrier

\paragraph{(Attempted) personalized EMA classification model.}
Results of fitting simple models to GFNVM's psychosis-related EMA responses based on core diary features were by and large negative, though there was a potentially interesting small role for fraction of recording duration spent speaking. Because the results did not yield many insights, they are described in more detail in supplemental section \ref{sec:gf-models}. Here, I will instead provide a quick overview of the the temporal dynamics of GF's EMA submissions and their degree of alignment with clinical scale ratings, to shed some more light on the quality of the labels used in the aforementioned modeling. \\

\begin{figure}[h]
\centering
\includegraphics[width=0.9\textwidth,keepaspectratio]{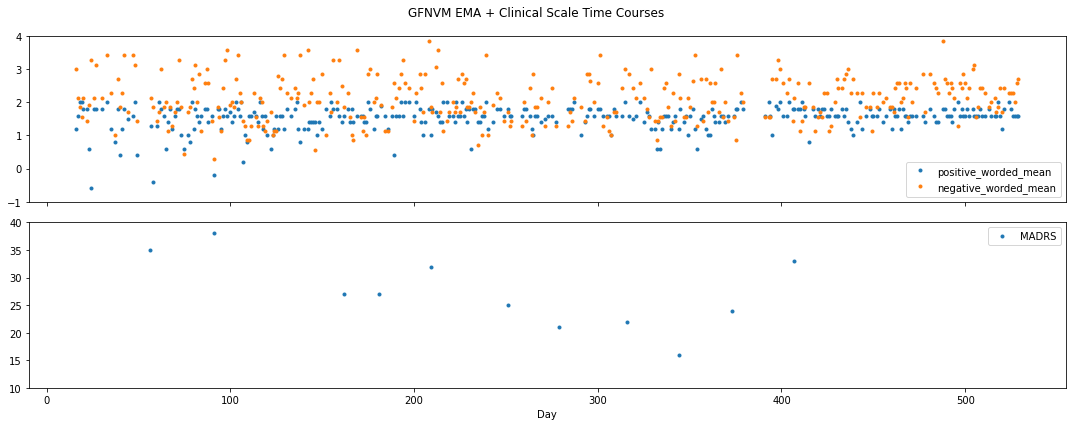}
\caption[Emotion-related EMAs and MADRS clinical ratings across the study for GFNVM.]{\textbf{Emotion-related EMAs and MADRS clinical ratings across the study for GFNVM.} This dot plot shows all positive (blue) and negative (orange) emotion summary EMAs submitted by GF, calculated as described for the modeling datasets (recall higher numbers always correspond to worse symptom severity), plotted over the course of enrollment in BLS (top). It also shows time-aligned MADRS sums where available (bottom).}
\label{fig:gf-mood-time} 
\end{figure}

\begin{figure}[h]
\centering
\includegraphics[width=0.75\textwidth,keepaspectratio]{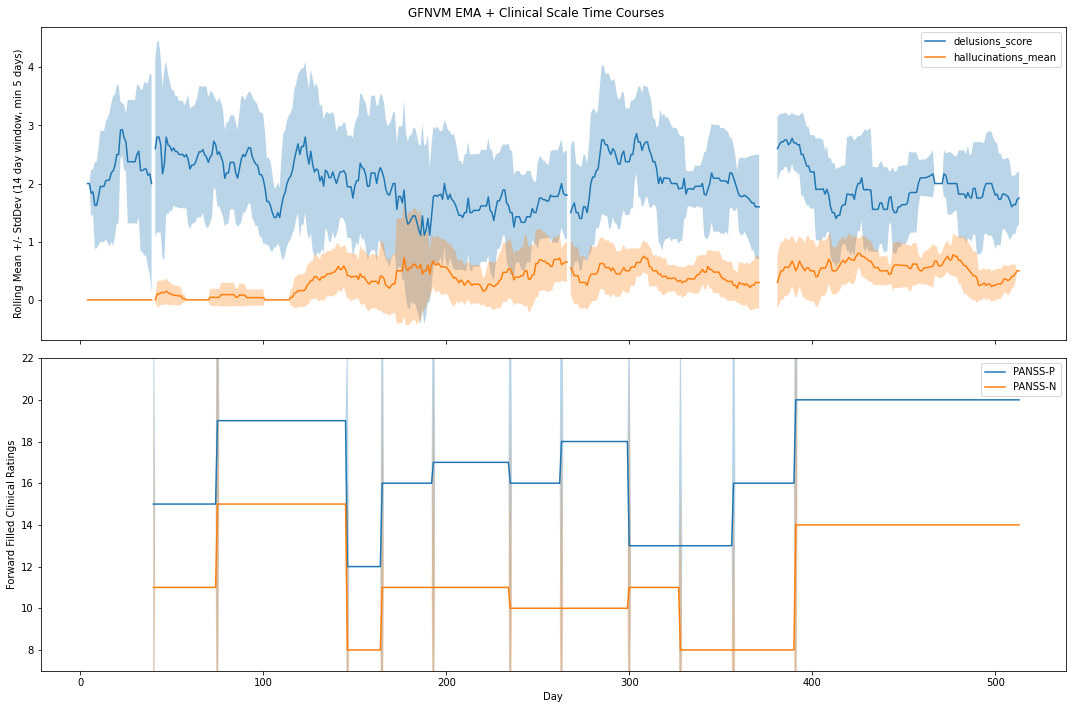}
\caption[Psychosis-related EMAs and PANSS clinical ratings across the study for GFNVM.]{\textbf{Psychosis-related EMAs and PANSS clinical ratings across the study for GFNVM.} This line plot shows a rolling mean (14 day window with right edge day number) for delusions-related EMA score (blue) and hallucinations-related EMA mean (orange) responses submitted by GF, plotted over the course of their enrollment in BLS and shaded according to the standard deviation in the same window (top). It also shows time-aligned PANSS sum scores for the positive (blue) and negative (orange) symptom subscales (bottom). For this plot, the clinical ratings were forward filled up until the next available time point, with streaks indicating the actual interview dates (in the case of an unchanged score).}
\label{fig:gf-psych-time} 
\end{figure}

\paragraph{Utilizing feature time courses to better understand GFNVM modeling results.}
It is unclear to what extent longer timescale dynamics in GF's EMA responses are explainable by varying levels of noise in survey-taking behavior versus meaningful variation. A few notable changes occurred shortly after day 100: 
\begin{enumerate}
    \item Variance in positively-worded EMA responses significantly decreased to produce mostly severe symptom ratings about lack of positive emotions, something that did not really recover even as MADRS greatly improved later in the study. Note also that this shift in variance happened many weeks after MADRS was already extremely high for GF (Figure \ref{fig:gf-mood-time}). 
    \item Hallucinations-related EMA responses went from being almost always zero to being non-zero for many days, never returning to an all-zero 7 day period. Similarly, this is despite a high PANSS positive score for GF occurring during the first 100 day stretch and relatively low PANSS positive scores occurring after this time (Figure \ref{fig:gf-psych-time}). 
\end{enumerate}
\noindent There were additional potential discrepancies in the EMA and clinical score comparisons, such as PANSS positive dropping 5 points right in the middle of one of the largest peaks (around day 300) of severe persecutory delusions self-reported through EMA (Figure \ref{fig:gf-psych-time}). Additionally, the negative emotions EMA would often span over 2 or more entire summary points in a single week, which is the difference between minimal and severe symptoms (Figure \ref{fig:gf-mood-time}). 

\begin{figure}
\centering
\includegraphics[width=\textwidth,keepaspectratio]{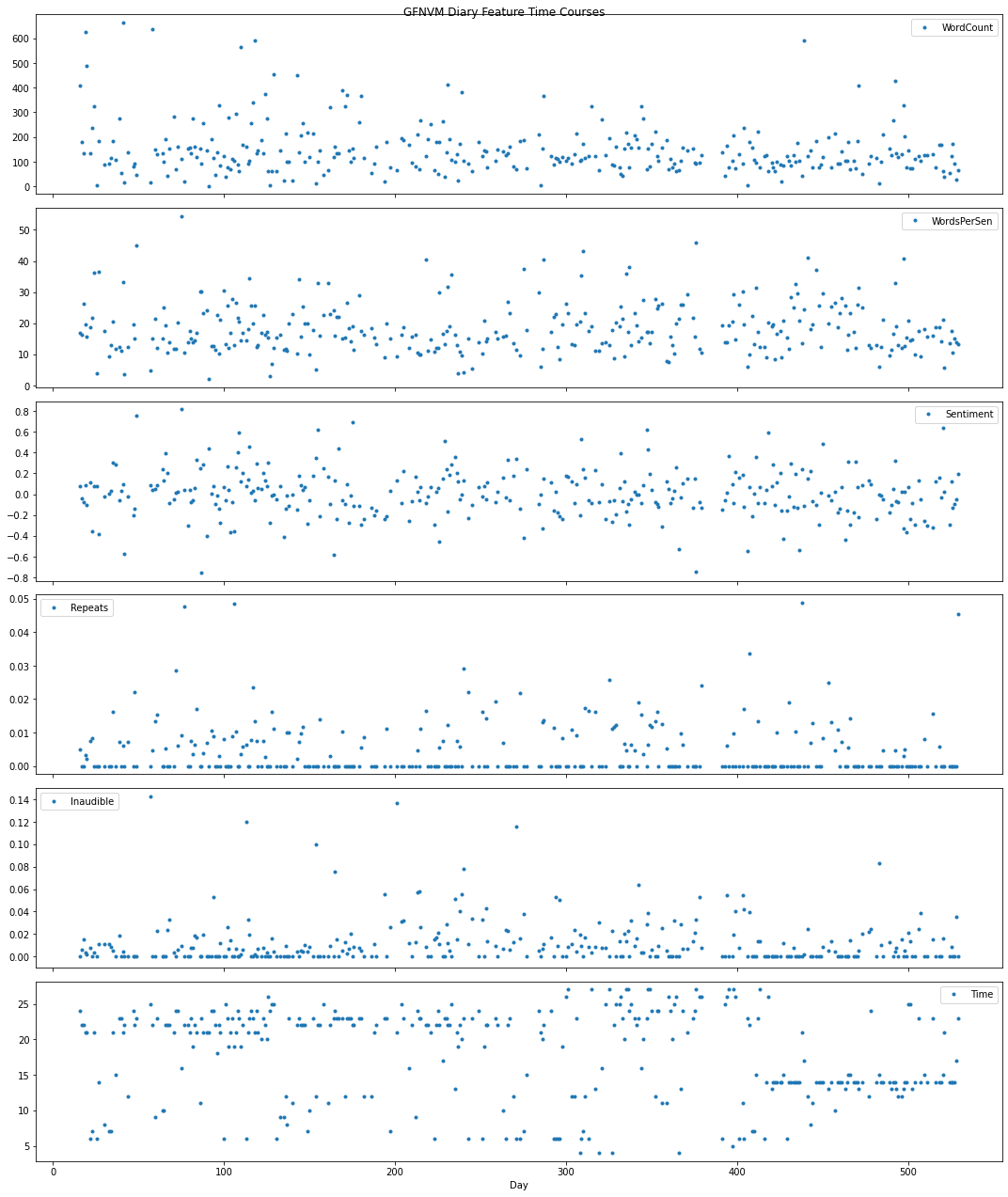}
\caption[Progression of select diary features over the course of enrollment for subject GFNVM.]{\textbf{Progression of select diary features over the course of enrollment for subject GFNVM.} With a shared study day x-axis, the following diary features were dot plotted over all GF audio journal submissions, from top to bottom: word count, mean words per sentence, mean sentence sentiment, repeats per word, inaudibles per word, and submission time (formatted as an integer per the diary pipeline described in chapter \ref{ch:1}). There were thus some patterns with some features deviating for weeks or months at a time, but the potential for clinical significance requires further investigation.}
\label{fig:gf-sup}
\end{figure}

It is possible that the EMA responses here capture some clinically-relevant information that the scales do not. Further, for the negatively-worded EMA summary there was some real alignment between changes in the clinical MADRS score and changes in self-reported symptom severity, at a broad timescale (Figure \ref{fig:gf-mood-time}). Still, these EMA appear to be overall quite noisy, so it's not clear that the above modeling failed to capture relevant signal, since it's not entirely clear how much relevant signal was there to begin with, particularly in the psychosis-related EMA patterns. Longer-term trends in the GF diary data (Figure \ref{fig:gf-sup}) did appear, but did not have any obvious correspondence with the psychosis-related EMAs, so it will require further research whether these were meaningful deviations or not. 

\FloatBarrier

\subsubsection{Takeaways from pilot linear EMA modeling}
\label{subsubsec:diary-ema-hypothesis}
Major results in modeling same-day EMA included highly significant relationships between diary verbosity and self-reported mood as well as diary sentiment and self-reported mood within multiple participants. Notably, verbosity in particular demonstrated opposite directional effects in a few different subjects, underscoring the importance of personalized modeling. 

While there were also between subjects differences captured by the overall models, with the limited number of participants chosen it was not possible to determine if these were "significant" effects because of meaningless (for our purposes) individual heterogeneity or because of individual differences that could be disease-relevant. Further, the participant-specific z-scores of features taken to fit overall models without this impact were a simple non-ideal adjustment. Z-scores will be less applicable for features with such non-normal distributions as many of the journal disfluencies, and as mentioned their justification is debatable in terms of interpretation of some other features like sentiment. 

In addition to considering more subjects to capture those factors, future work would benefit from considering nonlinear models for certain features and feature combinations, as well as from considering rolling fits to account for participant behavioral changes over very long longitudinal study enrollments (and natural early study survey/recording behavioral differences). Future work could of course also potentially benefit from introducing further new features -- though it is worth noting that EMA probably ought to be better characterized before deciding to what extent unexplained variance in these models is signal missed by the journal features versus signal that is actually irrelevant to our aims. Single day EMA is likely too variable for many subjects, especially as designed for BLS, so it might be advisable to smooth out the labels to some extent, something that could help with missingness problems that arise too. Deciding on exact parameters for formulating such an analysis will require care. \\

\paragraph{Many benefits to audio journals (again).}
It is worth noting that the larger dataset size enabled by audio journals, even after narrowing to a handful of subjects for the scope of this thesis, enabled the holdout mechanism employed. Indeed, it has been common practice in the interview recording analysis literature to perform only leave one out cross validation, so having a fully held out test set at all is a positive development attributable in part to the diary speech sampling methodology. The ability to consider contiguous sets of journal submissions as part of the generalization evaluation is yet another perk that would be difficult to achieve with interviews. 

There is of course also a large dataset of entirely held out subjects to further test generalization on, including both more participants with a primary Bipolar disorder diagnosis as well as some participants with a different psychotic or mood disorder as their primary diagnosis (Table \ref{table:bls-ema-avail}). This is something to be explored in follow up works that others will pursue using the BLS journal dataset. 

\subsection{A novel perspective on case reports}
\label{subsec:diary-case-study}
Audio journals can provide a deep and long term look into a patient's life, and thus are an invaluable resource for personalized approaches -- from both a quantitative and qualitative perspective. In the previous section (\ref{subsec:diary-ema}), models were fit both overall and with participant-dependence, which is a systematic way to consider personalized predictions. In this section, I turn to the case report format, where I use audio journal data in a variety of ways to better understand a few specific patients from the same BLS dataset. Note that review of diary recordings also plays an important role in the full-fledged OCD case report of chapter \ref{ch:3}. 

\noindent Here, I focus on the same three subjects as were investigated in the distributional analyses of section \ref{subsubsec:diary-pt-dists-comps}: 
\begin{itemize}
    \item 3SS93 (\ref{subsubsec:3s-case-study})
    \item 8RC89 (\ref{subsubsec:8r-case-study})
    \item 5BT65 (\ref{subsubsec:5b-case-study})
\end{itemize}
\noindent This process will include recurring arguments about the unique utility of audio journals in the case report context, as well as methodological lessons learned. For more details on specific methods used within these reports, see supplemental section \ref{subsubsec:case-study-methods}.

All three of these subjects contributed many transcripts to the BLS diary dataset, and displayed good availability of overlapping EMA. 3SS93 tended to submit especially long diaries thereby contributing a rich content set, 8RC89 was found in prior work to have manic episodes that coincided with abnormalities in passive digital phenotyping signals, and 5BT65 was observed by study staff to submit extremely short diaries during severe depressive episodes. As such, there were strong prior reasons to focus on these particular subjects. 

Clinically, 3S and 8R carried a Bipolar I diagnosis, while 5B's primary diagnosis was Bipolar II. 5B demonstrated very high variance in MADRS total scores over the course of the study (Figure \ref{fig:bls-clinical-scatter}A), and had distinct episodes with high depression and anxiety related clinical ratings for extended periods, as well as periods with fairly low symptom severity across the board. Similarly, 8R had higher variance in PANSS positive subscale scores over the course of the study (Figure \ref{fig:bls-clinical-scatter}B) than many other BLS participants, and had distinct periods of highly rated manic symptoms across YMRS items. 

All three subjects varied meaningfully at times on the PANSS negative subscale, and this was the only domain where 3S showed significant various over their time in BLS (Figure \ref{fig:bls-clinical-scatter}C). However, the PANSS negative subscale had worse interrater reliability than other scales on the BLS dataset, so 3S's clinical profile is more difficult to interpret. 3S stopped doing clinical interviews much earlier in the study than other datatypes precisely because not enough usable clinical variation was being captured. 

Still, the extent of 3S's participation in the audio journal recordings warrants a closer look at their data. Furthermore, the fluctuations observed in diary feature through this case report raises the question of whether occasional clinical follow-ups should have continued for this subject, and whether they in fact might have if diary monitoring were in place at the time of that decision many years ago. A different design for the BLS EMA could have helped to better clinically ground their status after they stopped official scale interviews as well.   

\subsubsection{Subject ID 3SS93}
\label{subsubsec:3s-case-study}
3S participated in BLS for an extremely long time overall, but unfortunately only received clinical scale ratings within the first 400 days of their enrollment. Thus I will consider 3 different views here -- grounding the emotions EMA summary scores in the clinical scales available, characterizing fluctuations at different timescales in diary features and corresponding EMA over the longer $1000+$ day diary dataset, and finally referencing diary content to address some questions that arise (or have already arisen) about this subject. \\

\begin{figure}[h]
\centering
\includegraphics[width=0.9\textwidth,keepaspectratio]{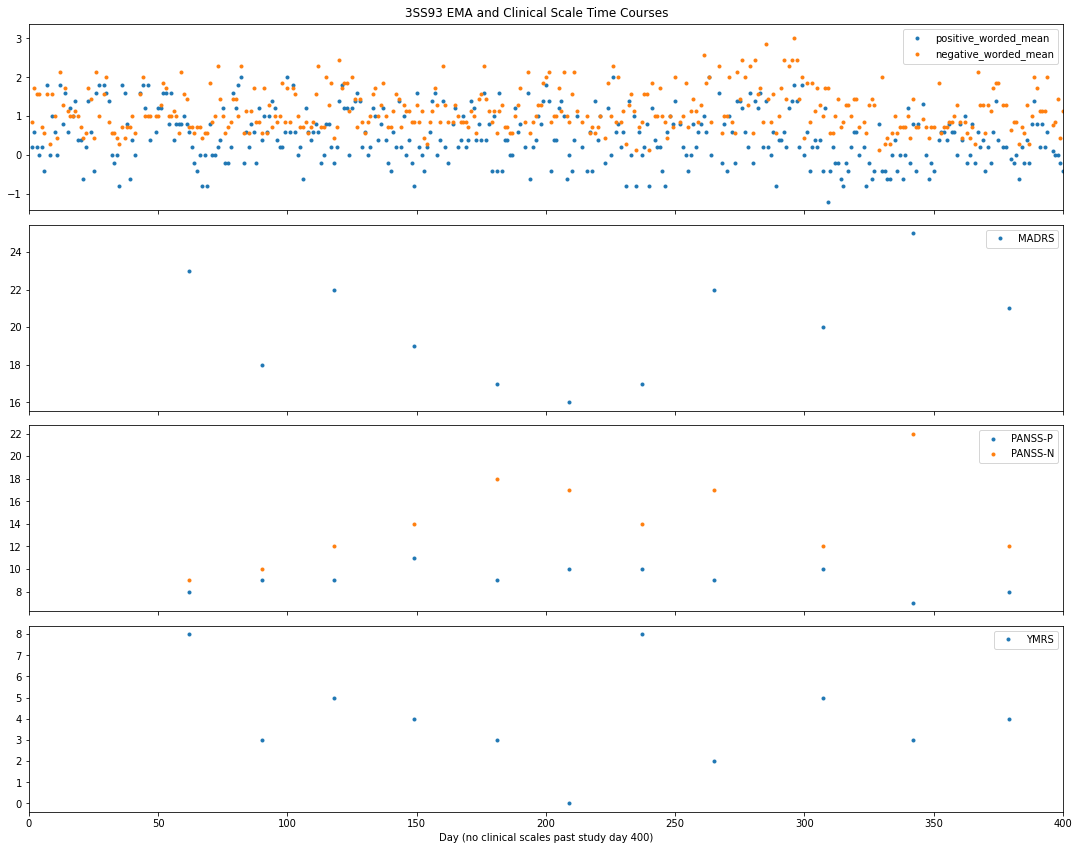}
\caption[Emotion-related EMA summary scores and clinical ratings across the study for 3SS93.]{\textbf{Emotion-related EMA summary scores and clinical ratings across the study for 3SS93.} This dot plot shows all positive (blue) and negative (orange) emotion summary EMAs submitted by 3S, plotted from their day of consent to their 400th day in the study (top). It also shows time-aligned clinical sum scores where available, in the following panel order: MADRS, PANSS positive (blue) and negative (orange), and finally YMRS. Note the y-axis limits are set to show the variance that does exist in each individual participant, not absolute severity levels.}
\label{fig:3s-scales} 
\end{figure}

\paragraph{Clinical ratings and self-report time courses.}
As mentioned, 3S did not experience many manic or positive psychotic symptoms, with minimal variation in the PANSS positive subscale and in the YMRS during the period in which they took part in clinical interviews. However, they did experience moderate depressive and negative psychotic symptoms, as measured by MADRS and PANSS negative respectively (Figure \ref{fig:3s-scales}). 

It is worth noting that the MADRS and PANSS negative symptom severities fluctuated like a mirror image in the data points collected for 3S (Figure \ref{fig:3s-scales}), up until one of the final time points at day $\sim 340$, where they both reached their highest recorded levels in 3S (25 and 22 respectively). While EMA did demonstrate some fluctuation on a broader timemscale during this period, there was minimal overlap with what one might expect based on clinical scales -- the aforementioned peak symptom time corresponded to one of the lower self-reported symptom severity periods here (as defined by the emotions-related EMA summary scores).

\begin{figure}[h]
\centering
\includegraphics[width=0.75\textwidth,keepaspectratio]{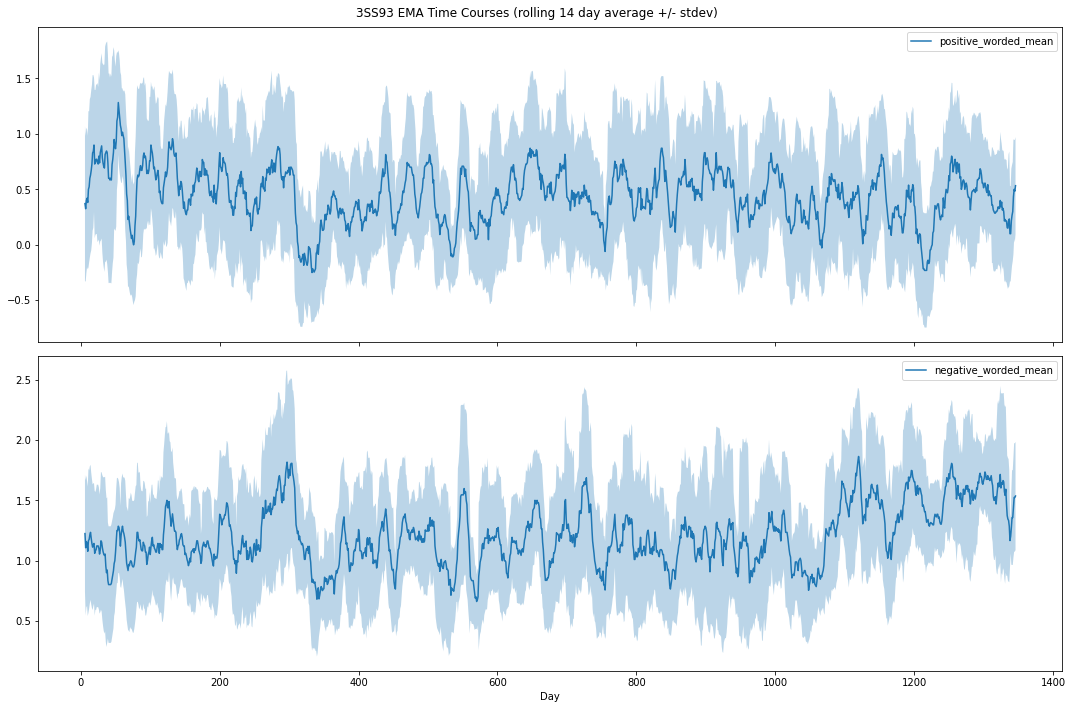}
\caption[Trends in emotion-related EMA summary scores across all 3SS93 survey submissions.]{\textbf{Trends in emotion-related EMA summary scores across all 3SS93 survey submissions.} This line plot shows a 14 day rolling average with standard deviation shading for the positively-worded EMA mean (top) and negatively-worded EMA mean (bottom) from 3S self-report submissions over their entire course of participation.}
\label{fig:3s-ema-smooth} 
\end{figure}

3SS93 submitted EMA surveys for nearly 1400 days, but stopped submitting audio journal recordings somewhat sooner. A summary of EMA fluctuations over the entire set of 3S submissions is presented in Figure \ref{fig:3s-ema-smooth}, but subsequent analyses will focus on either days 1 through 400 or days 1 through 1000 for this participant. \\

\FloatBarrier

\paragraph{Identifying longer-term trends in diary features.}
3SS93 demonstrated a number of potentially interesting longitudinal changes in some of their diary features, including both transient fluctuations on a multi-week timescale and lasting changes in diary recording behavior (Figure \ref{fig:3s-long-diary}). 

\pagebreak

\begin{FPfigure}
\centering
\includegraphics[width=0.75\textwidth,keepaspectratio]{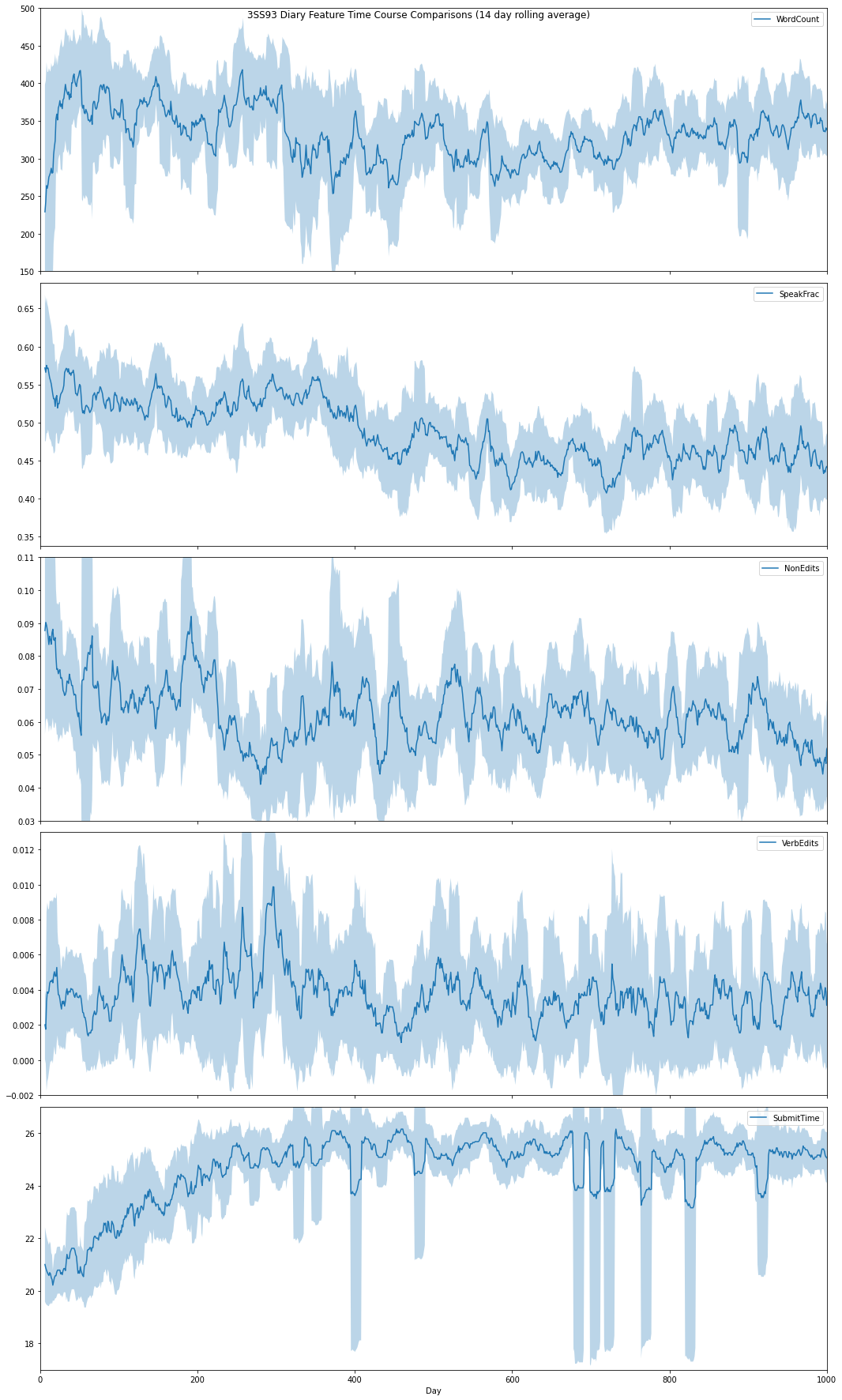}
\caption[Progression of select diary features over the course of enrollment for subject 3SS93.]{\textbf{Progression of select diary features over the course of enrollment for subject 3SS93.} With a shared study day x-axis, the following diary features were plotted as a 14 day window rolling average with standard deviation shading, over all 3S audio journal submissions between days 1 and 1000 (from top to bottom): word count, fraction of the recording sent speaking, nonverbal edits per word, verbal edits per word, and submission time (formatted as an integer per the diary pipeline).}
\label{fig:3s-long-diary}
\end{FPfigure}

\FloatBarrier

\begin{figure}[h]
\centering
\includegraphics[width=0.9\textwidth,keepaspectratio]{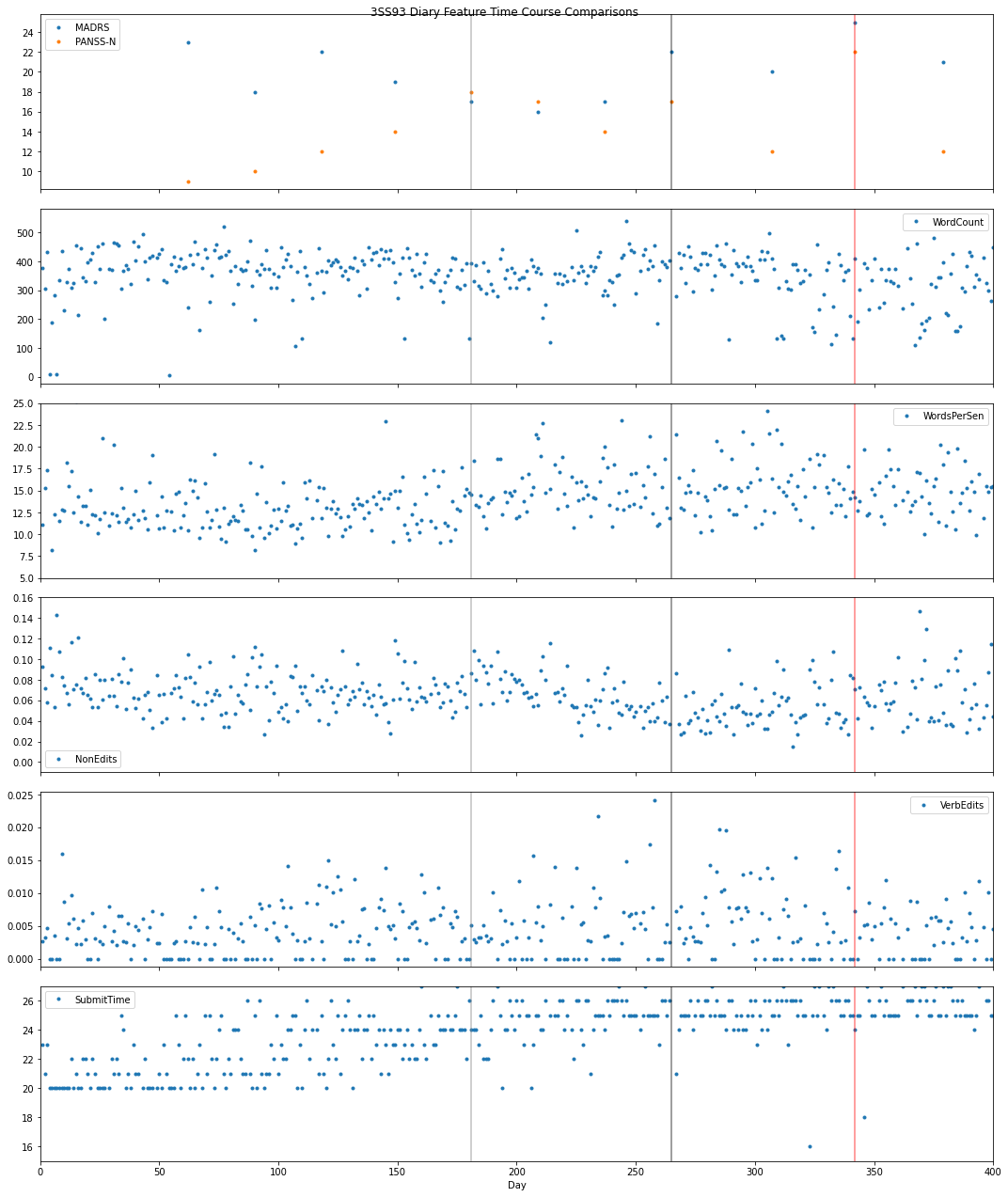}
\caption[Identifying meaningful variation in 3SS93 diary features through clinical scales.]{\textbf{Identifying meaningful variation in 3SS93 diary features through clinical scales.} Using a shared study day x-axis for days 1 through 400, the MADRS (blue) and PANSS negative (orange) clinical scales were visualized as a dot plot (top) against the stacked day-level diary feature dot plots for the first 400 days of those metrics summarized in Figure \ref{fig:3s-long-diary}, with one exception. Because the change in speech fraction occurred after day 400, that feature was replaced by mean words per sentence here (but will be revisited later). Three days were highlighted across the plots with vertical lines: day 181 (grey), day 265 (black), and day 342 (red). Day 342 was the peak severity time point for both clinical scale sums considered, day 181 was the first local peak for the PANSS negative, and day 265 was the only local MADRS peak between those points.}
\label{fig:3s-zoom-diary}
\end{figure}

For example, there was a sustained dip in fraction of recording duration spent speaking around day 400, a period of consistently lower nonverbal filler word usage rate between days 200 and 400, and a linear movement towards later submission times (something they repeatedly apologized for in the recordings, incidentally) from the start of the study to around day 250 that was subsequently sustained (Figure \ref{fig:3s-long-diary}). We can also zoom in on some of these changes in the first 400 day period that fortunately did include clinical scale ratings (Figure \ref{fig:3s-zoom-diary}).

There was a notable drop in word count surrounding the peak clinical scale severity time point at day 342 (Figure \ref{fig:3s-zoom-diary}), which is consistent with prior expectations about negative symptoms, and interesting in light of the fact that word count did not relate to same-day EMA for subject 3S. Additionally, this change may have been preceded by a change in nonverbal edit usage: shortly after the time point at day 181, nonverbal edit rate began to drop and remained lower than usual for 3S for an $\sim 100$ day period approaching the key rating at day 342 (Figure \ref{fig:3s-zoom-diary}). There was also a notable dip in verbal edit usage approaching day 181, and there is somewhat of a U-shape in verbal edits surrounding each highlighted time point in Figure \ref{fig:3s-zoom-diary}, which might warrant further investigation.  

\FloatBarrier

\paragraph{Leveraging diary content to inform feature interpretation.}
Because clinical scales lack any further context on temporal dynamics (though item-level information could be an interesting expansion) and the self-reported EMA summary scores had an opposite trend to the clinical measures of symptom severity at this time (Figure \ref{fig:3s-scales}), I next utilized the diary content itself to assist with interpretation.

\begin{figure}[h]
\centering
\includegraphics[width=0.8\textwidth,keepaspectratio]{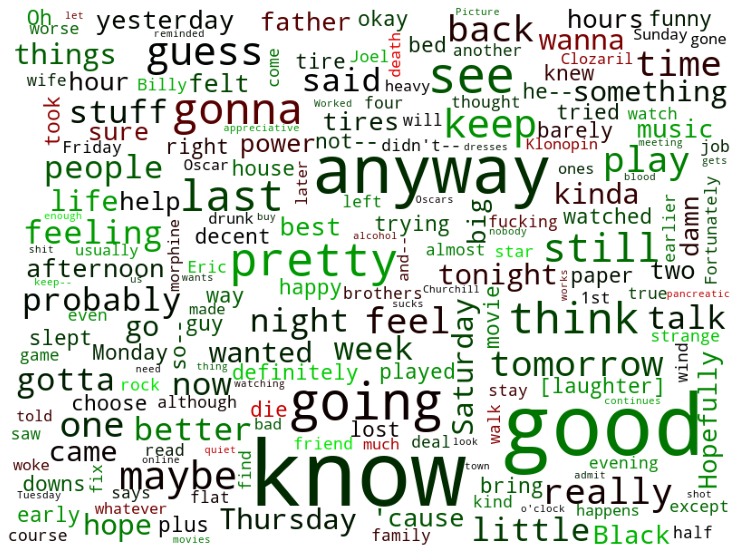}
\caption[Sentiment-colored word cloud from 3S diaries surrounding the most severe symptom period, as determined by clinical scales.]{\textbf{Sentiment-colored word cloud from 3S diaries surrounding the most severe symptom period, as determined by clinical scales.} As described for the diary pipeline, a word cloud with size corresponding to usage frequency and color corresponding to mean sentiment of sentences containing the word was generated for the concatenation of 3SS93 transcripts between days 337 and 347. Day 342 was the clinical interview that produced the highest MADRS and PANSS negative sum scores observed in this patient.}
\label{fig:main-cloud}
\end{figure}

To assist in understanding more about 3S during the time periods highlighted in Figure \ref{fig:3s-zoom-diary}, I created sentiment-colored word clouds covering $\sim 10$ days surrounding each point of interest (days 176 through 186, days 260 through 270, and days 337 through 347). Although word count was relatively low for 3S at times in this period, low for 3S was still often high on the scale of the BLS study at large. Negative sentiment words in some of the diaries surrounding the high symptom severity period stand out (Figure \ref{fig:main-cloud}), including the names of multiple medications, discussion of death, curse words, and family roles.

\begin{figure}[h]
\centering
\includegraphics[width=0.8\textwidth,keepaspectratio]{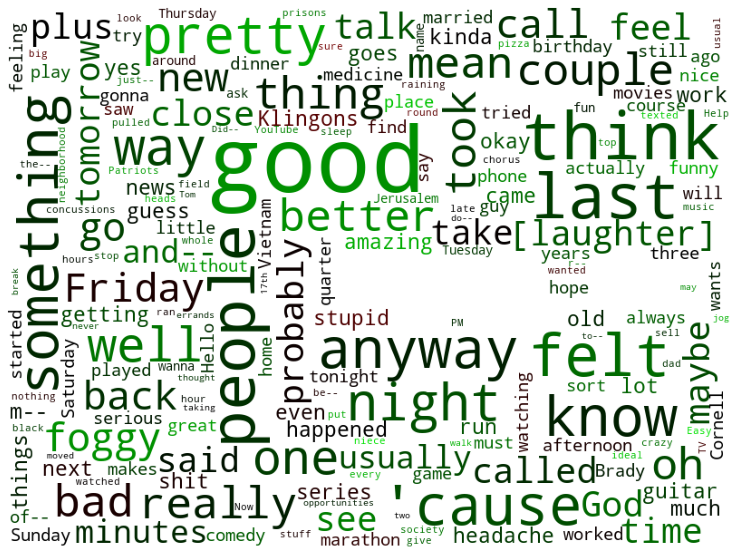}
\caption[Sentiment-colored word cloud from 3S diaries around a local PANSS negative peak.]{\textbf{Sentiment-colored word cloud from 3S diaries around a local PANSS negative peak.} As in Figure \ref{fig:main-cloud}, a sentiment-colored word cloud was also created using the 3S transcripts from days 176 through 186; this surrounds the time point highlighted with a grey bar in Figure \ref{fig:3s-zoom-diary}.}
\label{fig:early-cloud}
\end{figure}

\begin{figure}[h]
\centering
\includegraphics[width=0.8\textwidth,keepaspectratio]{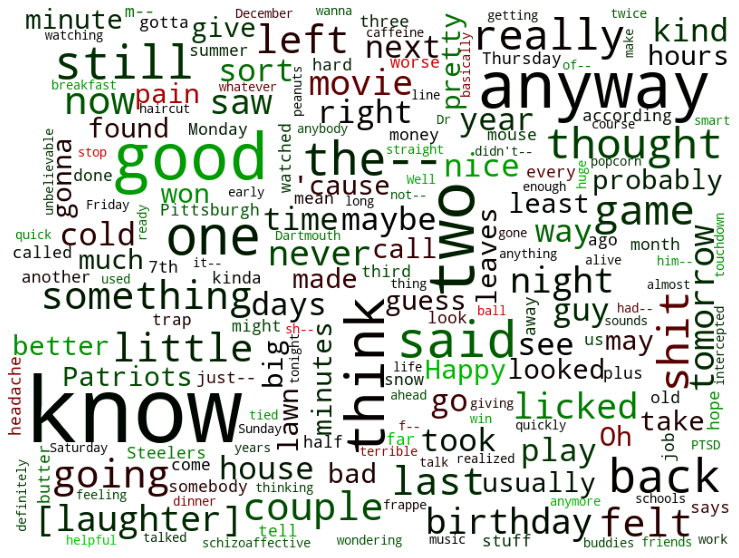}
\caption[Sentiment-colored word cloud from 3S diaries around a local MADRS peak.]{\textbf{Sentiment-colored word cloud from 3S diaries around a local MADRS peak.} As in Figures \ref{fig:main-cloud} and \ref{fig:early-cloud}, a sentiment-colored word cloud was also created using the 3S transcripts from days 260 through 270; this surrounds the time point highlighted with a black bar in Figure \ref{fig:3s-zoom-diary}.}
\label{fig:mid-cloud}
\end{figure}

Further, the content of Figure \ref{fig:main-cloud} can be contrasted with a similar word cloud surrounding day 181 (Figure \ref{fig:early-cloud}). There were fewer words strongly associated with negative sentiment in this period, and overall more generic content about basic activities or the plots of TV shows/movies. Similarly, while the word cloud for days 260 to 270 (Figure \ref{fig:mid-cloud}) had more clear red words than Figure \ref{fig:main-cloud}, the content was not salient to the extent of that in Figure \ref{fig:main-cloud}.

Looking directly at the diaries in question between days 337 and 347, there was discussion of the (past) death of the patient's father as well as the (current) behavior of an alcoholic family member. There was also discussion related directly to the patient's psychiatric condition, and in particular concerns about correctly taking suitable dosages of all of their medications. Anecdotally, they seemed both cognitively frazzled and upset about various heavy topics, so it is surprising to me that their EMA responses around the same time period did not reflect this. On the other hand, there was a high amount of variance over the attitudes expressed in the diaries even in this short period -- for example, happy discussion of "Billy Joel" that can be clearly seen on Figure \ref{fig:main-cloud}. \\

\FloatBarrier

\paragraph{Key word count measures as a basic interface between qualitative content review and quantitative feature analysis.}
Based on the results of the sentiment-colored word clouds I generated, I searched for instances of specific terms across the 3S diary dataset at large. In particular, I looked for mentions of "Klonopin" and "Clozaril" (Figure \ref{fig:main-cloud}), and through this discovery process also "Ativan". I initially thought that certain sustained diary feature changes might be tied back to medication change like what was suggested by the content about ADHD medications in the audio journals of the chapter \ref{ch:3} case report. By looking directly at the diaries that matched these key words in addition to adding occurrences as a feature plot (Figure \ref{fig:3s-meds}), I learned that 3S was taking all of them in parallel.

\pagebreak

\begin{FPfigure}
\centering
\includegraphics[width=\textwidth,keepaspectratio]{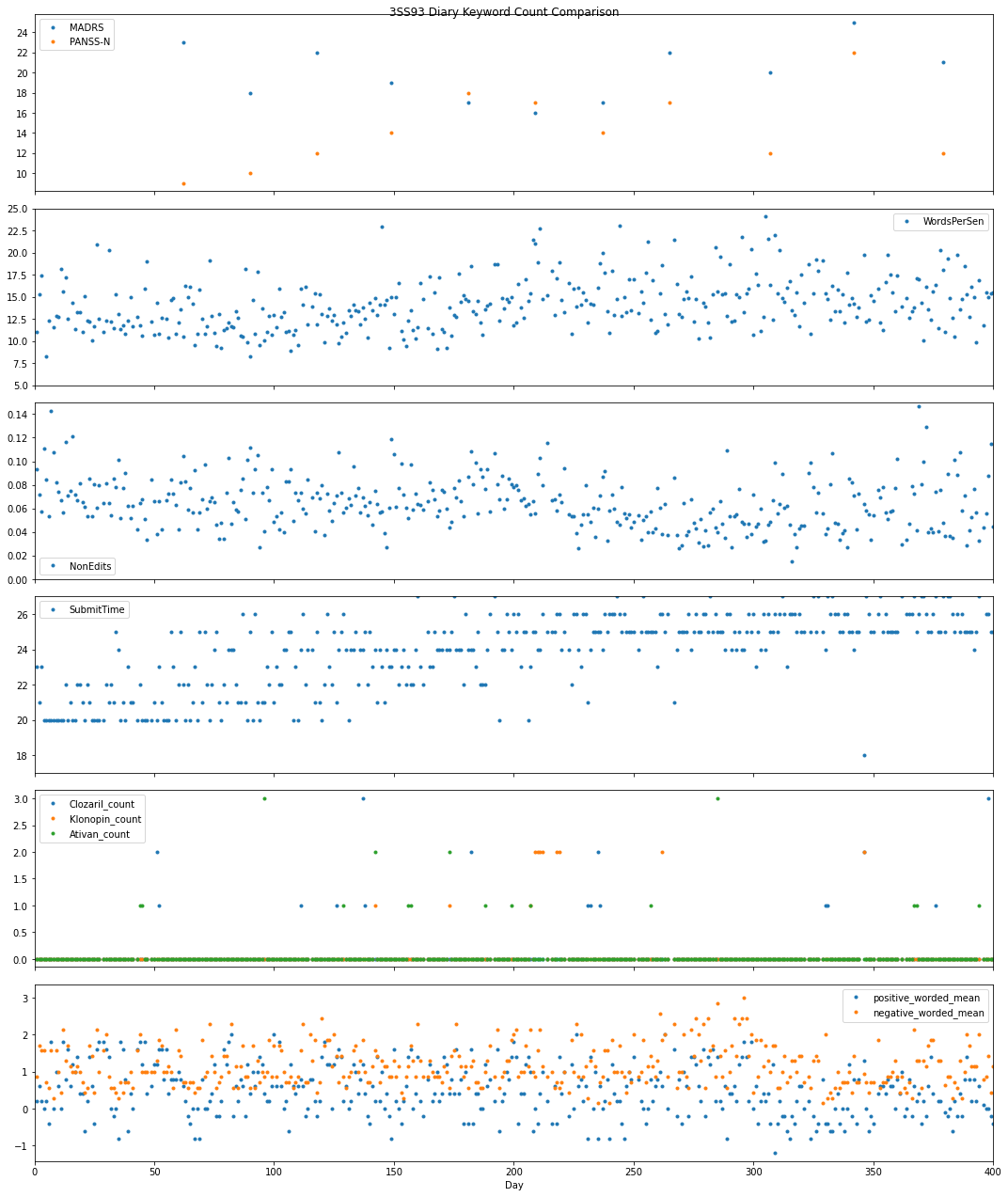}
\caption[Use of key word count diary features to identify time periods of interest - comparing 3SS93 medications.]{\textbf{Use of key word count diary features to identify time periods of interest - comparing 3SS93 medications.} Similarly to Figure \ref{fig:3s-zoom-diary}, relevant clinical scales, mean diary words per sentence, nonverbal edit usage rate, and journal submission hour are the top 4 dot plots here, again considered over the first 400 days of BLS enrollment for 3SS93. The bottom dot plot panel recreates the EMA summary score dynamics, provided for reference like the aforementioned panels. The primary new information source in this Figure then is the second panel from the bottom, which contains a dot plot for three medication-related key word count features. The number of times the word of interest occurred in each diary is on the y-axis, and each of the following key word counts are separately plotted: Clorazil (blue), Klonopin (orange), and Ativan (green).}
\label{fig:3s-meds} 
\end{FPfigure}

\FloatBarrier

\begin{figure}[h]
\centering
\includegraphics[width=\textwidth,keepaspectratio]{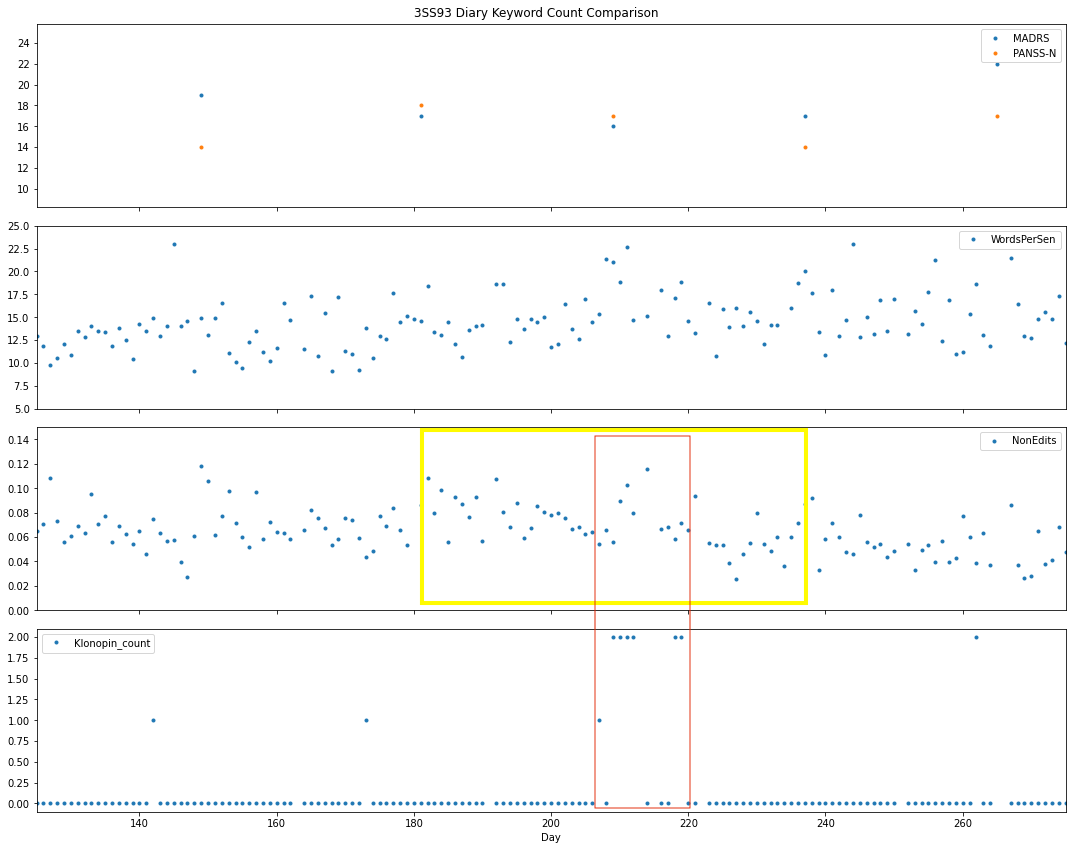}
\caption[Emotion-related EMA summary scores and clinical ratings across the study for 3SS93.]{\textbf{Emotion-related EMA summary scores and clinical ratings across the study for 3SS93.} Using a shared study day x-axis to zoom into days 125 through 275 of enrollment for 3S, I generated a dot plot of the MADRS (blue) and PANSS negative (orange) clinical scales (top) aligned against the previously described mean words per sentence (second panel) and nonverbal edits per word (third panel) diary-level feature dot plots. On the bottom panel now is a dot plot of the number of times the medication name "Klonopin" was mentioned in each journal entry. The time period of declining nonverbal filler rates noted between days 181 and 265 in Figure \ref{fig:3s-zoom-diary} is boxed here in yellow, while a period of especially frequent discussion of Klonopin is boxed in red.}
\label{fig:3s-main-meds} 
\end{figure}

Still, the feature plots proved of strong use when considered in conjunction with corresponding content. Mentions of Klonopin in particular during one of the periods of interesting journal variation focused on in Figure \ref{fig:3s-zoom-diary} was quite informative. 3S discussed Klonopin abnormally frequently in a stretch between day 200 and 220, and these mentions closely aligned with a transient change in nonverbal edit usage (Figure \ref{fig:3s-main-meds}). 

It turned out that at the start of the period in question (Figure \ref{fig:3s-main-meds}), 3SS93 wanted to cut back on their use of Klonopin. They attempted to skip doses and cut doses in half, but quickly complained about the result, and within a week was back to taking the medication as usual. The description in the diaries is vague, so it is not possible to perfectly align the days that Klonopin use was different, but it is highly plausible that medication disruption caused some transient noise in diary features, including the spike in nonverbal edits in the middle of that feature's somewhat longer term decline. If true, this would demonstrate clinically relevant fluctuations in this disfluency feature at multiple quite distinct timescales simultaneously, which is the type of aim that we hope diaries can accomplish. 

Regardless, this observation certainly opens up future directions. It was not clear what exactly triggered 3S to start trying to adjust their medications, and there is no way of knowing with the dataset here whether this was a recurring phenomenon that was not always mentioned in the journals. At later time points in the study, there are a few submissions with off-hand remarks about forgetting to take a medication and not knowing whether they should take a double dose. By more carefully poring over the diary content, utilizing more advanced models of content to detect more interesting connections, and/or accessing participant medical records that were collected for use in the study, the pilot results of Figure \ref{fig:3s-main-meds} (and Figure \ref{fig:3s-zoom-diary}) could be followed up on more rigorously. \\

\FloatBarrier

\begin{figure}[h]
\centering
\includegraphics[width=\textwidth,keepaspectratio]{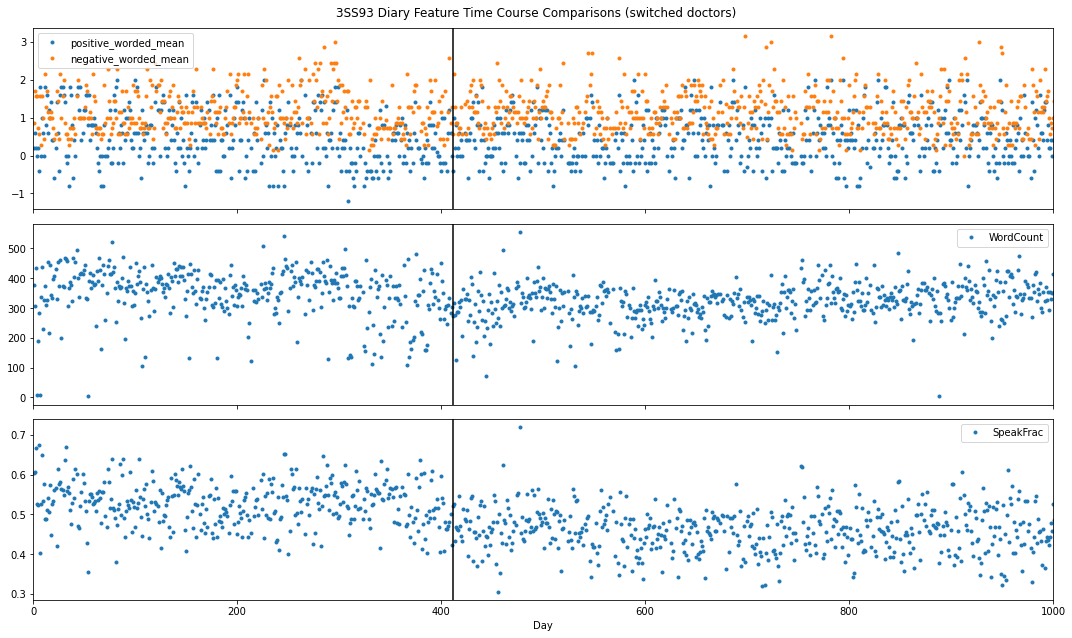}
\caption[Marking the time point that 3SS93 changed therapists on the longitudinal diary feature set.]{\textbf{Marking the time point that 3SS93 changed therapists on the longitudinal diary feature set.} Across the first 1000 days of enrollment for 3S, I generated a dot plot of positively-worded (blue) and negatively-worded (orange) EMA summary scores (top) aligned by day against diary word counts (middle) and the fraction of each recording spent speaking (bottom). A black horizontal line across all 3 panels denotes day 413, when 3S first saw a new therapist -- as recounted in their daily journal entry. On the previous day's (412) journal, they talked about seeing their old therapist for the last time.}
\label{fig:3s-provider} 
\end{figure}

\paragraph{Finding relevant content via anomalous diary features.}
Finally, to return to the observation about a long-term sustained decrease in the proportion of diary recording time 3S actually spent speaking (Figure \ref{fig:3s-long-diary}), I skimmed through diary content from days between 375 and 425 to look for mention of a specific change in treatment plan, a new phone, etc. As a result of searching for an explanation to the long-term diary feature dynamics, I found that 3S mentioned visiting his former therapist for the last time on day 412, and meeting with a new therapist on day 413 (Figure \ref{fig:3s-provider}).

It was not clear from the available information in those journals whether the new (and/or old) provider could prescribe medication, or if this represented purely a change in behavioral treatment plan. It was also not entirely clear why 3S switched (though they had complained on and off in the past about the previous therapist getting annoyed if they weren't happy enough when they showed up at the appointment), and I did not look through future records to determine if they would switch again. Furthermore, the beginnings of a decline in speech fraction can be seen shortly after their clinical symptom peak at day 374, and by day 400 the start of the sustained lower speech fraction level had been reached (Figure \ref{fig:3s-provider}). 

On the other hand, speech fraction had declined earlier in the study and then later picked back up, so it remains curious that it did not recover again after day 400 (Figure \ref{fig:3s-provider}). Although it is unlikely that switching therapists was the primary underlying factor, it does provide a good starting point for additional clinical and research questions that could be asked in a larger scope case report, in order to better unravel the story. 

\FloatBarrier

\subsubsection{Subject ID 8RC89}
\label{subsubsec:8r-case-study}
Unlike 3S, 8RC89 continued to complete clinical interview assessments alongside many years of EMA and audio journal submissions. Interestingly, 8R would sometimes go months without submitting anything at all later in enrollment, and then subsequently pick up fairly regular submissions again. Such missingness patterns over a highly longitudinal time period might be worthwhile to study as a direct information source, but that is beyond the scope of the current thesis. 

Additionally, for a number of weeks even within the first 2 years of the study, 8R participated sparsely, submitting just 1 or 2 recordings in a 7 day stretch. While collecting a large dataset less frequently but for more years is yet another possibly relevant avenue for audio journals, it is again beyond the current scope. As such, I will briefly characterize the EMA and clinical scales over a longer period of enrollment for 8R, and then I will focus this mini case report on observations from within the first year of their involvement in BLS. \\

\begin{figure}[h]
\centering
\includegraphics[width=0.95\textwidth,keepaspectratio]{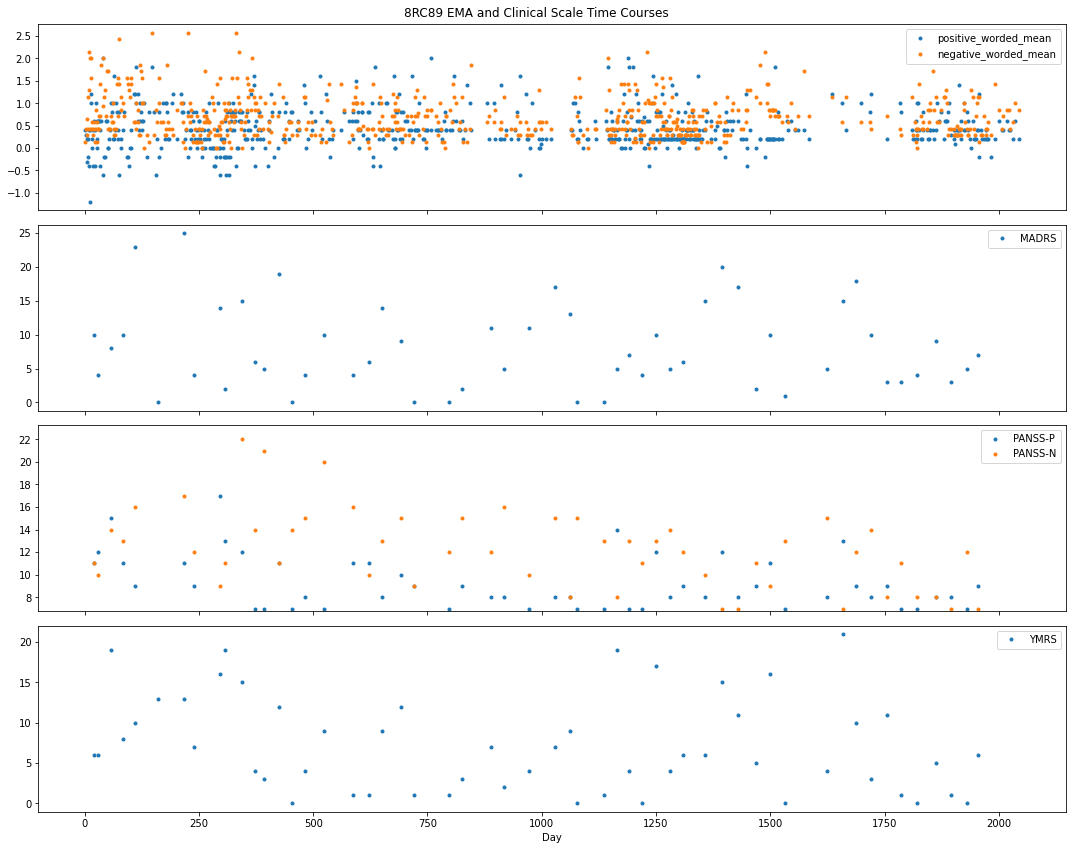}
\caption[Emotion-related EMA summary scores and clinical ratings across the study for 8RC89.]{\textbf{Emotion-related EMA summary scores and clinical ratings across the study for 8RC89.} As with Figure \ref{fig:3s-scales} for 3SS93 above, this dot plot shows all positive (blue) and negative (orange) emotion summary EMAs submitted by 8RC89, plotted from their day of consent to their final day with data available (top). It again shows time-aligned clinical scores too: MADRS, PANSS positive (blue) and negative (orange), and finally YMRS from top to bottom. In contrast to 3S, 8R had both EMA submissions and clinical scale ratings available stretching through the end of their enrollment.}
\label{fig:8r-scales} 
\end{figure}

\paragraph{Clinical ratings and self-report time courses.}
An analogous version of the EMA and scales time course plot made for 3S was repeated here for 8R (Figure \ref{fig:8r-scales}). The extreme duration of their study involvement (nearly 2000 days!), the periodic longer missingness periods, and even variance in density of available records with documented time periods can all be seen in Figure \ref{fig:8r-scales}.  

It is less immediately obvious how well clinical scale ratings aligned with trends in self-report answers. Most of the more extreme EMA values (for both bad negative emotions and good positive emotions in orange and blue respectively) occurred within the first year or so of study enrollment (Figure \ref{fig:8r-scales}). While scales also tapered off somewhat, fairly high peaks occurred later in the study and did not appear as likely to align with peaks in self-reported mood severity. On the other hand, there was a clean alignment at this distance between the first two YMRS peaks and jumps in negatively-worded EMA severity, as well as between the two particularly high MADRS ratings between those YMRS peaks and an upward bump in positively-worded EMA severity (i.e. lack of positive emotions). This bodes well for use of 8R's EMA as relevant labels in a longer duration project (Figure \ref{fig:8r-scales}). 

\FloatBarrier

\paragraph{Fluctuations of daily diary properties over year 1.}
Figure \ref{fig:8r-year-1} presents raw diary feature of interest for 8R in direct alignment with EMA and clinical scales for identification of possibly meaningful transient changes in journal properties. 

\pagebreak

\begin{FPfigure}
\centering
\includegraphics[width=\textwidth,keepaspectratio]{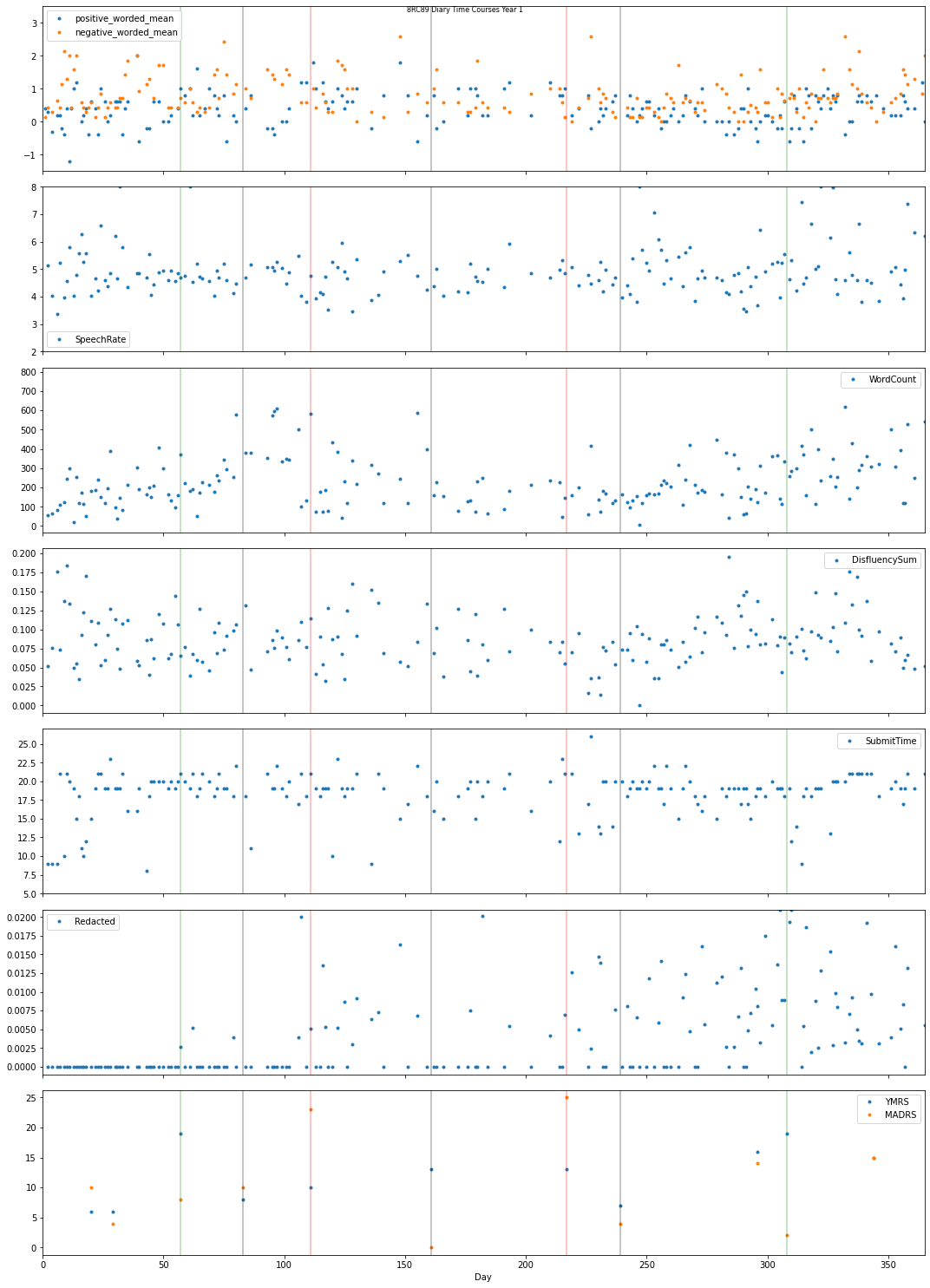}
\caption[Identifying meaningful variation in 8RC89 diary features and EMA responses over year 1 in the study - alignment with clinical scales.]{\textbf{Identifying meaningful variation in 8RC89 diary features and EMA responses over year 1 in the study - alignment with clinical scales.} Using a shared x-axis covering days 1 through 365 of 8RC89's enrollment, I generated dot plots with EMA summaries, select diary features, and then clinical scores across corresponding days. Thus the first panel of this figure depicts all of the positive (blue) and negative (orange) emotion summary EMAs submitted by 8R in their first year in the study. Then each chosen diary feature is plotted in its own panel, in the following order: the pause-corrected transcript-derived speech rate, the word count, the sum across all considered disfluency categories of the rate of disfluency use per transcript word, the submission hour as labeled by my pipeline, and the number of redactions per word. Finally, the bottom panel here shows YMRS (blue) and MADRS (orange) sum clinical scores. These scores were also used to mark some time points of particular interest with vertical lines across the Figure. The green lines map to the two days with highest YMRS sum score for 8R in the first year, and analogous for MADRS and the two red lines. Grey lines then mark all interview time points in between the selected times that have a significantly lower rating for the scale of interest. \newline See Figure \ref{fig:8r-mania} below for a different set of possible symptom time point markers that might have meaningfully supplemented the available scales here. }
\label{fig:8r-year-1}
\end{FPfigure}

\FloatBarrier

\noindent A number of diary features change on different timescales over this year for 8R. Some notes about these time courses follows, in order of the panels in Figure \ref{fig:8r-year-1}:
\begin{itemize}
    \item Surrounding both green lines corresponding to higher YMRS, there appeared to be more extreme fluctuations of EMA in both directions of self-reported severity, involving some time points with very high severity negative emotions reported and some time points with very high amounts of positive emotion, the latter of which is typical encoded as a good thing. Of course this can also be a sign of mania however, so it is consistent with expectations to find some such trends in 8R's data. This may additionally explain some of the difficult in fitting the positively-worded EMA score for 8R relative to their overall dataset size, because it has a nonlinear relationship with symptom severity in this case. Even accounting for a shifted curve, the clinical interpretation of positive emotion scores could be context dependent on the same day negative ones as well as other factors.
    \item The speech rate feature was especially high at the beginning and end of the first year, somewhat near the mentioned green YMRS marker lines. There was also an additional more transient dip in speech rate around the time of the first MADRS peak (left red line), which would be consistent with prior observations of psychomotor slowing. 
    \item Word count steadily increased in 8R diaries over the first 100 or so days, and then beginning around the first MADRS peak mentioned there was a general drop off lasting nearly 150 days with a handful of spikes in the middle. It is interesting that a more sustained increase in word count did not occur until closer to the time of the second YMRS peak. It is possible that natural ebbs and flows in word count for 8R exist on a similar timescale across the dataset, and reflect something about individual fluctuations -- it may be difficult to fully assess with the missingness in later parts of 8R's diary dataset, but this would certainly be theoretically possible for a future direction.
    \item Overall rates of disfluency usage began with relatively high values and had additional peaks surrounding both sides of the second YMRS (green) peak. There were also smaller peaks in disfluency rate over the course of the year, and some of the lowest rates could be found directly after both MADRS peaks. 
    \item Submission time was highly variable early in the study, but that is more likely to be a property of getting set up with the habit of recording regular journals on Beiwe given the timescale and other factors in this particular case.
    \item The rate of redacted words increased directly after each of the (red) MADRS peaks, and ended very high at the end of the year, while there were no redactions at all in the first 50 days of journals. It would make sense if the participant did not speak as freely at the beginning of the study in these recordings, so with the present dataset it is difficult to draw any specific conclusion. 
\end{itemize}
\noindent Regarding features at the very beginning of 8R's enrollment, they are more difficult to interpret because there is no prior baseline for the participant and little time for diary recording behavior to "acclimate", which could potentially explain some of the early discrepancies. 

Hold outs from the start of data collection period were also the hardest for the basic linear models of EMA from these diary features to generalize to, across multiple subjects. Because recruitment for BLS often stems from admission to the psychiatric facilities at McLean, it is unclear if the earliest study periods may contain a clinical bias too. Referring to medical records could help to address this, or for the present aims, referring to diary content could also elucidate participant situation around the time of enrollment. \\

\paragraph{Benchmarking with mentions of mania.}
By looking for discussion of mania in 8R's diaries as described in Figure \ref{fig:8r-mania}, I was able to mark two additional time points of possible clinical relevance on the year 1 time courses of Figure \ref{fig:8r-year-1}. 

\pagebreak

\begin{FPfigure}
\centering
\includegraphics[width=\textwidth,keepaspectratio]{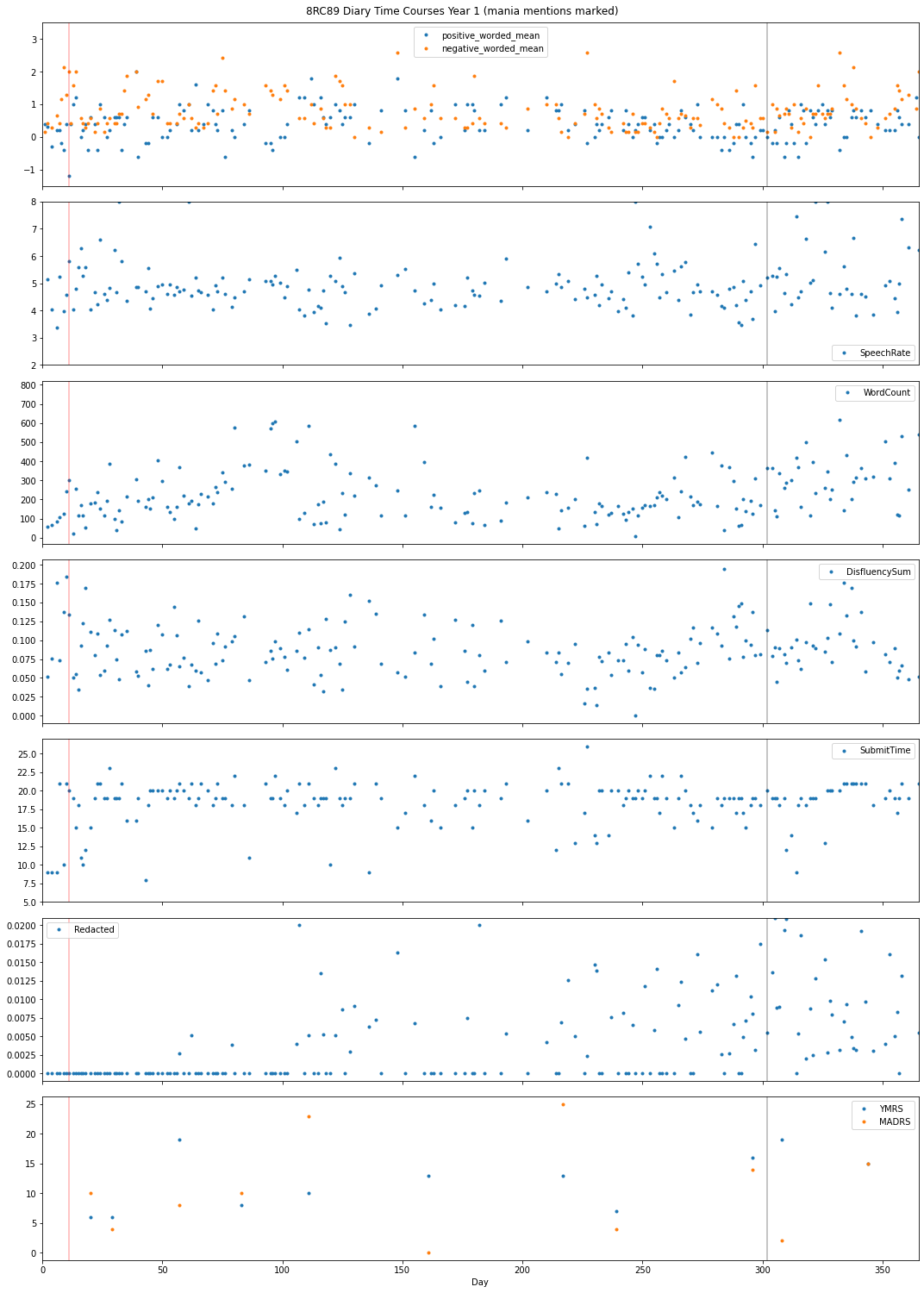}
\caption[Correspondence between diary mentions of mania and other diary features at those times in 8RC89's first year dataset.]{\textbf{Correspondence between diary mentions of mania and other diary features at those times in 8RC89's first year dataset.} Figure \ref{fig:8r-year-1} is reproduced here, but with highlighted time points now related to discussion of mania in the corresponding diary. Of the four diaries from the first year where 8R mentioned "mania" or "manic" (or these words embedded in other words, e.g. "hypomania"), 2 were describing other people and 2 were self-referential. The latter 2 time points, day 11 and day 302, are marked with red and grey vertical lines here respectively, to reference feature variation against akin to the clinical scale score peak markings of Figure \ref{fig:8r-year-1}. Day 11 is marked in red because in that diary, 8R agreed with their mother when she said they were becoming hypomanic. To contrast this day 302 is marked with grey, because 8R mentioned their mom was worried about them having a manic episode then, but in that entry they dispute the notion.}
\label{fig:8r-mania}
\end{FPfigure}

\FloatBarrier

It was very interesting to find that many of the early study abnormalities aligned well with the day 11 marker of potential hypomania (Figure \ref{fig:8r-mania}), a statement 8R themselves agreed with at that time. Thus combining the objective diary features like disfluencies and speech rate with the content of diaries, we can label likely symptom time points that were missed by the official clinical interview timeline. This is important not only for increasing our understanding of the patient, but also for properly evaluating features of interest -- it is now much less attractive to assume the feature peaks in the early days of 8R's enrollment were initial data collection noise or otherwise a "false positive". 

The sentiment is especially true because similar patterns in a number of the diary features occurred again near the end of the first year, a time period we now know was also identified by 8R's mother as potential oncoming hypomania (Figure \ref{fig:8r-mania}), in addition to being near multiple relatively high YMRS ratings for this participant. The discussion in that diary from day 302 is additionally of some inherent interest, as it identified small personalized signs that 8R's mother believed correlated with the onset of their manic episodes (Figures \ref{fig:8r-diary-mania-content} and \ref{fig:8r-diary-mania-content2}). In sum, it is highly recommended to use diary content and extracted features to each inform the other. \\

\begin{figure}[h]
\centering
\begin{tcolorbox}[top=-0.25cm,bottom=0.4cm,left=-0.25cm,right=-0.25cm]\begin{quote}\small
00:02.07 \hspace{2mm} Um, my mother thinks I might be getting hypomanic and she might be right.

00:08.38 \hspace{2mm} The other night, I-I'm not sure which night that was, I slept only six hours or so and then last night I didn't sleep at all.

00:14.82 \hspace{2mm} I was, uh, up all night, um, you know, watching shows on streaming and, uh, um, I, um, had been having cravings so I'd actually bought a half-pint of vodka.

00:31.21 \hspace{2mm} So I consumed [inaudible] and a half shots in, uh, orange juice and I-- it was somewhat pleasant but also somewhat emotional.

00:40.37 \hspace{2mm} I'm still dealing with the fact that my girlfriend's considering moving to, um, [redacted].

00:46.1 \hspace{2mm} Um, I decided-- it was partially to help me sleep and I didn't feel sleepy but I decided it was too risky to stay awake-- I mean stay-- to go to sleep so I stayed awake.

00:55.08 \hspace{2mm} Didn't sleep at all and, um, you know, I feel pretty good.

00:59.61 \hspace{2mm} I was having more ideas today, um, but you know, I'm tired so hopefully I'll sleep tonight.

01:05.13 \hspace{2mm} If I don't, I will take like more Seroquel gradually, 100 milligrams per hour when, uh, I have trouble sleeping and, uh, you know, I might call, um the on-call person at, uh, [redacted] because my psychiatrist is, um, um, um, on vacation.

01:24.34 \hspace{2mm} I'm not going to drink again.

01:25.47 \hspace{2mm} It was not worth it.

01:26.43 \hspace{2mm} Um, um, that's about all.

01:29.45 \hspace{2mm} Doing all right.

01:29.78 \hspace{2mm} Saw my girlfriend today.

01:31.01 \hspace{2mm} Um, went to a m-- a mus-- a festival and it was free in-- at [redacted].

01:37.57 \hspace{2mm} It was fun.

01:38.44 \hspace{2mm} Um, I'm just gonna go with it for now.

01:40.62 \hspace{2mm} Better to have loved and lost than never at all and, uh, you know, I have more time than she does because, uh, she's five years older than I am.
\end{quote}\end{tcolorbox}
\caption[BLS participant 8RC89 identifies hypomanic episode time points directly through diary content (day 11).]{\textbf{BLS participant 8RC89 identifies hypomanic episode time points directly through diary content (day 11).} Transcripts of the diaries recorded by subject 8RC89 on day 11 and then day 302 are reproduced, with day 11 in this Figure and day 302 next in Figure \ref{fig:8r-diary-mania-content2}. These were the two diaries in the first year of 8R's study enrollment that directly discussed their own manic symptoms -- both in the context of their mother's thoughts. The two days are marked on the journal features time course of Figure \ref{fig:8r-mania}, aligning with some compelling trends in measures like speech rate. This was an especially compelling use for journals because day 11 preceded available clinical scale ratings, while on day 302 8R was actually arguing that they were not getting manic, despite a number of signals to the contrary (including a relatively high YMRS for this subject). The behaviors they identified that their mother associates with their mania (e.g. grocery shopping) also demonstrates how possible hypotheses to be tested in passive digital phenotyping data streams can be obtained from journal content.}
\label{fig:8r-diary-mania-content}
\end{figure}

\begin{figure}[h]
\centering
\begin{tcolorbox}[top=-0.25cm,bottom=0.4cm,left=-0.25cm,right=-0.25cm]\begin{quote}\small
00:01.51 \hspace{2mm} Uh, so Tuesday, Wednesday, Thursday.

00:03.5 \hspace{2mm} Tuesday, I, uh, probably did some programming, and, um, I went to, um, um, AA at 8 o'clock at [redacted], walked there and back.

00:15.54 \hspace{2mm} And, um, that was good.

00:17.49 \hspace{2mm} Um, Wednesday, um, group therapy was canceled, so I went to DBSA, but, um, again, I wasn't really enthusiastic about-- it was only an hour group.

00:28.73 \hspace{2mm} And I figured I could talk to anybody I wanted to talk to, which I did, um, you know, before the group started.

00:34.26 \hspace{2mm} So I left early before group actually started, and, um, went to [redacted]'s house, and we, uh, lay in her bed, and cuddled, and watched Law \& Order. It was nice.

00:45.14 \hspace{2mm} Um, and I think she-- yeah, she cooked me some eggs. Yeah, that was good.

00:50.69 \hspace{2mm} And, um, I've been eating a lot, but that's okay. I know I can lose weight if I want to.

00:56.06 \hspace{2mm} Um, and, uh, Thursday, I, uh, I went to the grocery store and bought some, um--

01:04.15 \hspace{2mm} I had a therapist appointment, and we discussed my mom's concerns.

01:07.22 \hspace{2mm} She thinks my mood's elevated, I might get manic.

01:09.46 \hspace{2mm} But, um, I'm not really hypomanic because I still get tired. I don't have a decreased need for sleep.

01:16.91 \hspace{2mm} And, um, I went to the grocery store and bought some jelly because I like that on my bread, and some pickled herring - and whenever she sees me buy herring, she gets worried that I'm manic - and, um, some cereal that's better than, um, what I-- we have at the house, and a can of frosting to replace a can of frosting that I ate most of when I was hungry at night.

01:38.78 \hspace{2mm} Um, and, um, so she thought I might be-- have an elevated mood because I rarely shop for groceries.

01:45.92 \hspace{2mm} And, um, I went over to [redacted]'s in the evening, around 6:30, and, um, she cooked, um, um, steak and lettuce.

01:57.69 \hspace{2mm} It was like a Thai dish. It was good.

02:00.31 \hspace{2mm} And, um, we, uh, fooled around a little bit.

02:02.42 \hspace{2mm} We didn't have sex, but we did do the other things, which was nice because we rarely do that.

02:05.35 \hspace{2mm} And, um, yeah, that's about it.
\end{quote}\end{tcolorbox}
\caption[BLS participant 8RC89 identifies hypomanic episode time points directly through diary content (day 302).]{\textbf{BLS participant 8RC89 identifies hypomanic episode time points directly through diary content (day 302).} As described for the day 11 transcript reproduced in Figure \ref{fig:8r-diary-mania-content}, this Figure presents the transcript of 8R's diary submission from the other time point during their first year where they discussed potentially being hypomanic: day 302.}
\label{fig:8r-diary-mania-content2}
\end{figure}

 \FloatBarrier

 \paragraph{Contextualizing features with counts of important discussion topics.}
Both of 8RC89's self-referential mentions of mania in the first year dataset were presented in terms of the thoughts of their mother. At times they have commented similarly on e.g. their mother worrying about them drinking too much. Separately, they talk about various girlfriends quite a lot, though with PII redaction it would require good use of context clues (perhaps sourced multimodally) to track these identities across the journals. Time spent with girlfriend(s) is repeatedly expressed to be important to the participant via the journals. 

\begin{figure}[h]
\centering
\includegraphics[width=\textwidth,keepaspectratio]{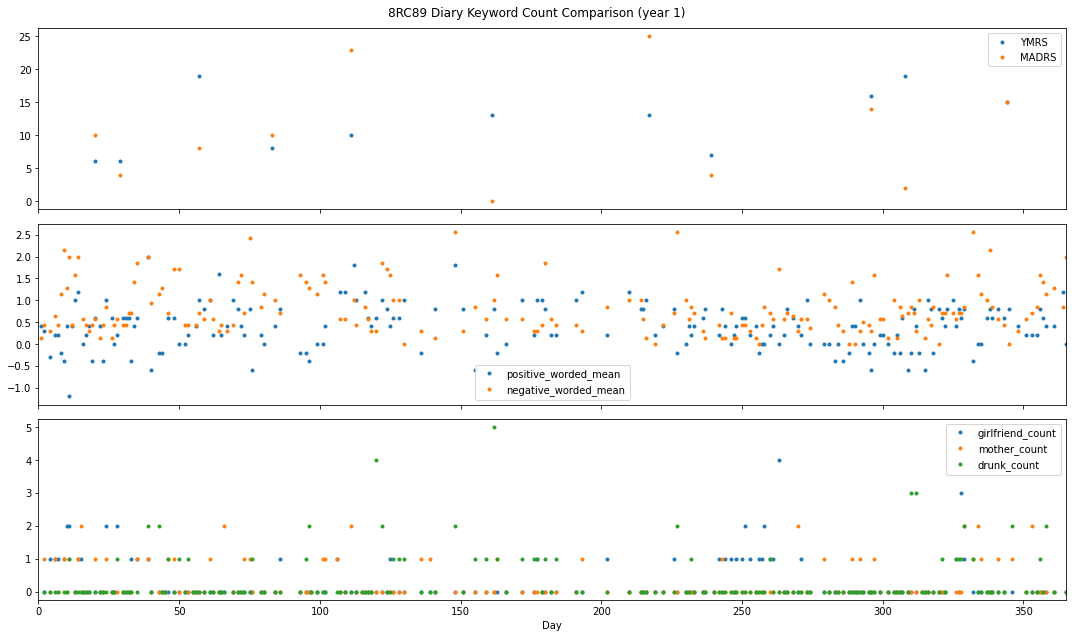}
\caption[Tallying mentions of important topics for 8RC89 in diaries submitted over the first year.]{\textbf{Tallying mentions of important topics for 8RC89 in diaries submitted over the first year.} This dot plot reproduces YMRS (blue) and MADRS (orange) clinical scores from 8R's first year in the study, as well as all their submitted positive (blue) and negative (orange) emotion summary EMAs during that time. It then plots in the bottom panel counts of keywords for 8R contained in each diary. The selected keywords here were: "girlfriend" (blue), "mother" (orange), and "drunk" (green). Note that counts for drunk also included uses of the words "drink" or "drinking".}
\label{fig:8r-keywords} 
\end{figure}

In fact all of these mentioned topics are easily spottable just by skimming some of the journal transcripts from 8R, unlike the less common topics focused on in the analysis of 3S above and 5B below. However, it is not at all clear to what extent mentioning these topics is systematically relevant to 8R's status. As such I plotted clinical scale and EMA values over the first year of data along with daily diary keyword occurrence counts for these 3 topics of particular interest (Figure \ref{fig:8r-keywords}).

Notably, a period with low clinical scale ratings and low self-report symptom severity around study day 250 contained a cluster of many diaries with discussion of a girlfriend (Figure \ref{fig:8r-keywords}). Meanwhile, short periods containing both journals with non-zero mentions of the participant's mother and journals with non-zero mentions of drinking seemed to frequently coincide with EMA peaks, some of which were probably reflecting transient annoyance (potentially directly related to thinking about the topics during self-report). It would again require further investigation to determine if specific word count correspondences are significant, and especially if they may be in a combination-only way. Still, the early results here show promise for the role of diary content in modeling, even if that begins with very tractable and understandable, yet ultimately simplistic, methods. 

 \FloatBarrier

\subsubsection{Subject ID 5BT65}
\label{subsubsec:5b-case-study}
5BT65 had by far the best model fits in the EMA modeling section, so it is especially promising to consider fluctuations in diary and EMA time courses for this subject. Further, strong connection between clinical scales and EMA in 5B would in turn provide good statistical evidence for patient-specific clinical predictive power of diary features. As it turns out, the EMA, clinical scale, and diary dynamics over 5B's dataset are even more convincing than could have been expected. I will present some early results that are withing scope here, but there is likely much additional room for case report work with 5B in particular. 

\begin{figure}[h]
\centering
\includegraphics[width=\textwidth,keepaspectratio]{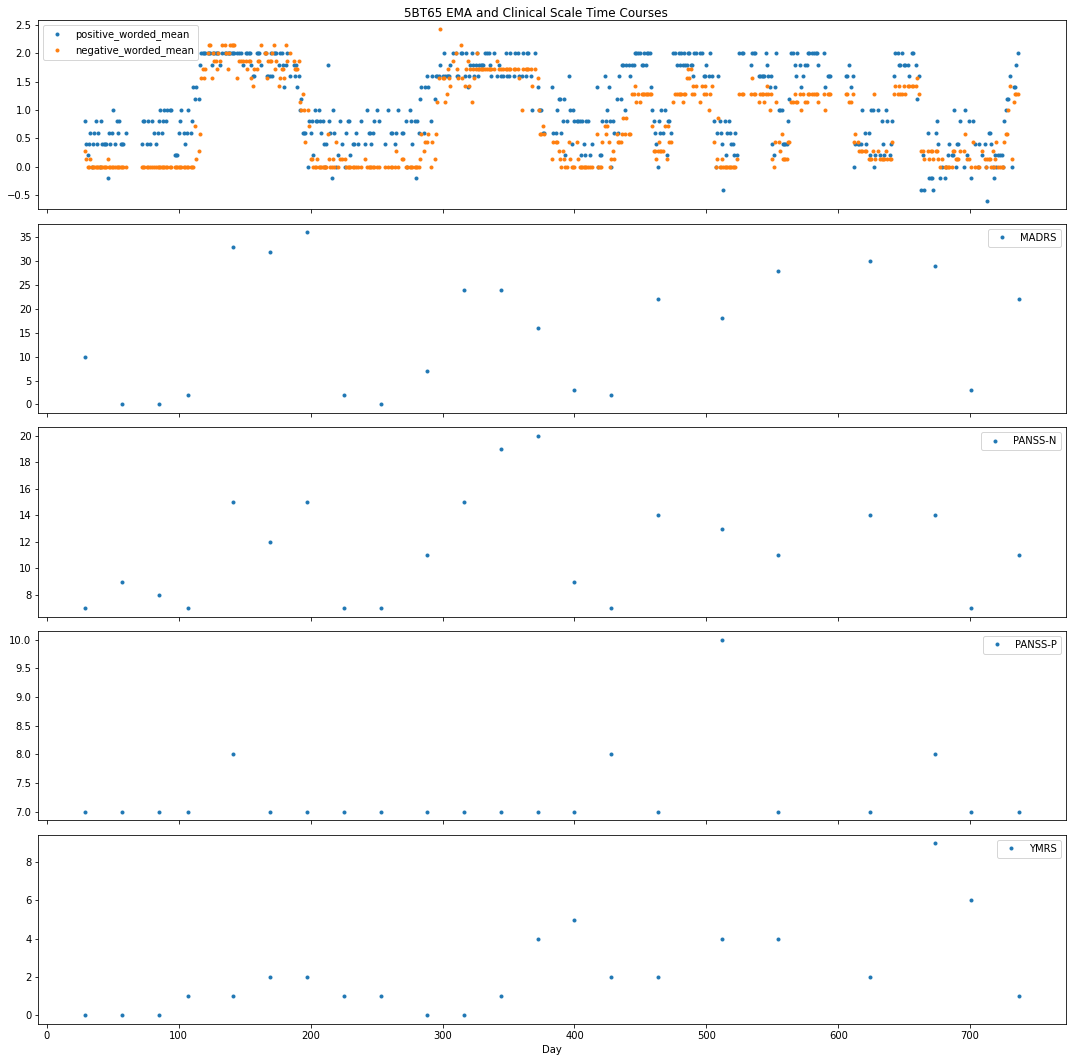}
\caption[Emotion-related EMA summary scores and clinical ratings across the study for 5BT65.]{\textbf{Emotion-related EMA summary scores and clinical ratings across the study for 5BT65.} As with Figures \ref{fig:3s-scales} and \ref{fig:8r-scales} above, this dot plot shows all positive (blue) and negative (orange) emotion summary EMA scores for 5BT65 across study days (top). It then shows time-aligned clinical scores where available, in the following order: MADRS, PANSS negative, PANSS positive, and YMRS.}
\label{fig:5b-scales} 
\end{figure}

\paragraph{Clinical ratings and self-report time courses.}
To begin, similar dot plot to the EMA and scales time course Figures made for 3S and 8R was repeated here for 5B (Figure \ref{fig:5b-scales}). Longer timescale fluctuations in EMA "state" were immediately obvious in 5B's summary score plots, and these fluctuations also aligned impressively well with changes over time in their clinical scale ratings \ref{fig:5b-scales}.  

Participant 5BT65 primarily experienced depressive symptoms, with large fluctuations observed from very mild to extremely severe MADRS scores over the course of their enrollment (Figure \ref{fig:5b-scales}). 5B also experienced moderate negative psychotic symptoms at times, but positive psychotic symptoms and mania symptoms were minimal through their enrollment in the study. Remarkably, over the course of their first $\sim 400$ days in the study, 5B had 2 very clear episodes with moderate to severe depressive symptoms and 2 very clear periods with mild to no depressive symptoms, each lasting months at a time. What is perhaps most surprising about this is just how neatly EMA response severity fluctuated in line with these clinical scale fluctuations. It is worth noting as well that the first such episode was primarily characterized by extreme MADRS score, with a modest uptick in PANSS negative at the same time, while the second such episode was characterized by a larger increase in PANSS negative score and a (relatively) more moderate increase in MADRS sum then (Figure \ref{fig:5b-scales}). 

\begin{figure}[h]
\centering
\includegraphics[width=\textwidth,keepaspectratio]{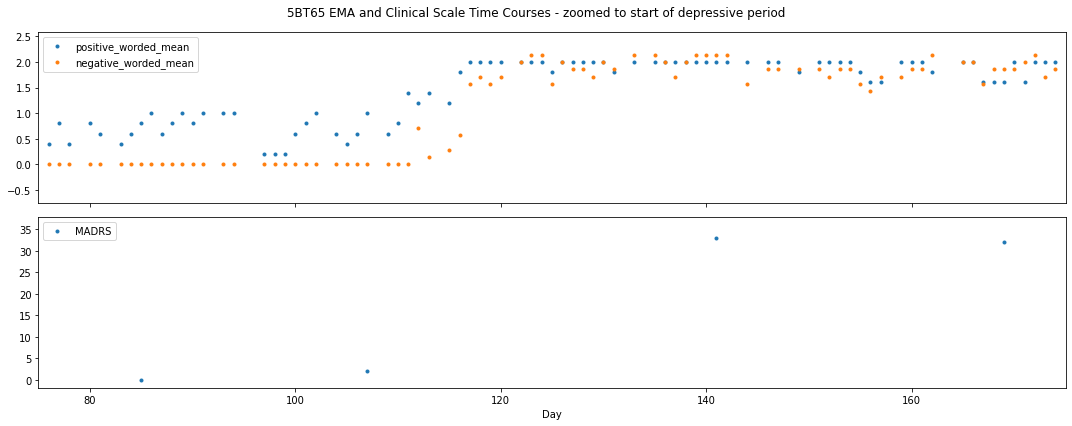}
\caption[EMA summary scores successfully filled gaps in clinical scale temporal resolution, for detection of depressive episode in 5BT65.]{\textbf{EMA summary scores successfully filled gaps in clinical scale temporal resolution, for detection of depressive episode in 5BT65.} The plot in Figure \ref{fig:5b-scales} was zoomed in to study days 75 through 175, to focus on the dynamics of the beginning of 5B's first observed depressive episode in the dataset. As such, the plot was also limited to focus on only EMA summary scores (top) and MADRS (bottom). EMA very clearly could have alerted to the start of the episode weeks before MADRS in this instance, simply because there were large temporal gaps in ratings of the clinical scales.}
\label{fig:5b-madrs} 
\end{figure}

One question we can ask about such strong EMA trends is to what extent their temporal resolution might assist in catching episodes earlier, potentially without much requirement at all for modeling in the traditional sense. For 5BT65 specifically at least it would have been possible, and in fact could have proved clinically usefully purely as a gap filler rather than a data point with future predictive power. Zooming in on the first and most severe depressive episode observed for 5B, we can easily see that the near-zero MADRS score directly preceding the episode was measured at a time where there was not yet a clear indication of symptoms from the EMA summary; but within 2 weeks 5B's EMA responses would already be at the saturated peak, and MADRS on the other hand was not rated again until more than 30 days after (Figure \ref{fig:5b-madrs}). 

Another highly interesting property of the self-report dynamics of 5B was the seemingly abrupt change in the periodicity and magnitude of symptom fluctuations following the end of the second major episode around study day 400 (Figure \ref{fig:5b-scales}). For an $\sim 250$ day stretch, MADRS and PANSS negative scores both remained around moderate levels relative to the early study scores demonstrated by 5B, no longer appearing to oscillate. While we cannot say for sure - as certainly something meaningful did change around day 400 - it seems apparent that this is primarily a sample rate issue. If it were feasible to rate so frequently, the gold standard scales would have likely continued to match EMA much more closely than they did in practice here, and it is even more likely that the scales should want to capture this. The properties of the EMA fluctuations after day 400 suggest that symptoms have become more moderate, as the peak of the negative emotion EMA mean diminished at the same time. Conversely, the peak of the positive emotion EMA mean stayed near the saturated value of "no positive emotions" in episodes during this period (Figure \ref{fig:5b-scales}). 

Near the end of year 2 (and the available data for 5B), a longer period with mild symptom severity EMA scores emerged, as did a significant drop in the clinical scales. At this time around day 700 was additionally the only time in the study period that the positive emotion EMA score dropped below 0 for a cluster of survey responses, indicating an above average level of positive emotions (Figure \ref{fig:5b-scales}). In the last section of available data, EMA scores rise again, and so it is unfortunately unclear how 5B progressed. It did very much seem that over the multi-year study timescale they were consistently getting a little better, based on the described changes to the EMA dynamics. 

Alternatively, it is of course plausible that the described 250 day period following day 400 in Figure \ref{fig:5b-scales} was itself an episode with a symptomatology involving greater emotional volatility. Still, it is likely functionally preferable for many to have moderately severe symptom magnitudes that rapidly come and go than to have extremely severe symptom magnitudes that linger for months. Regardless of whether the change was natural or an intervention, and regardless of whether it represented a step towards improvement or itself part of a cycle, it is critical to collectively document and work towards understanding such fluctuation patterns at multiple timescales, in order to truly understand psychiatric disease or even neurobiology more broadly. The EMA in 5B has provided one (glaring) example of a case where app-based measures can already facilitate that research but clinical scales cannot (in their current form) alone. 

Overall, 5BT65 would be an ideal participant for an extended digital phenotyping case report, as their EMA surveys were not only submitted daily fairly reliably, but the answers themselves were highly trustworthy at face value relative to most app-based self-reports, as was exemplified by the characterization in section \ref{subsec:diary-ema}. Even just restricted to EMA and corresponding diaries there is a great deal that could be done beyond the current scope, including a deeper item-level account of EMA over time and a deeper dive into diary content to identify autobiographical information that might be relevant to the long-term observed changes. For the remainder of this section though, I will investigate how the pipeline's computed journal features might connect with the document EMA dynamics. \\

\FloatBarrier

\paragraph{How did 5B's most significant diary features vary prior to a major depressive episode?}
The first obvious question is to revisit Figure \ref{fig:5b-madrs} with the most relevant diary features from 5BT65's EMA modeling results (word count, sentiment), to determine how those properties moved at the onset of the most severe and extended (and first) depressive episode experienced by 5B during the study. Here we consider the mood episode start and end days to be defined by the EMA signal, as far too many days passed between MADRS scores (Figure \ref{fig:5b-madrs}) to use those as ground truth for this question. 

\begin{figure}[h]
\centering
\includegraphics[width=\textwidth,keepaspectratio]{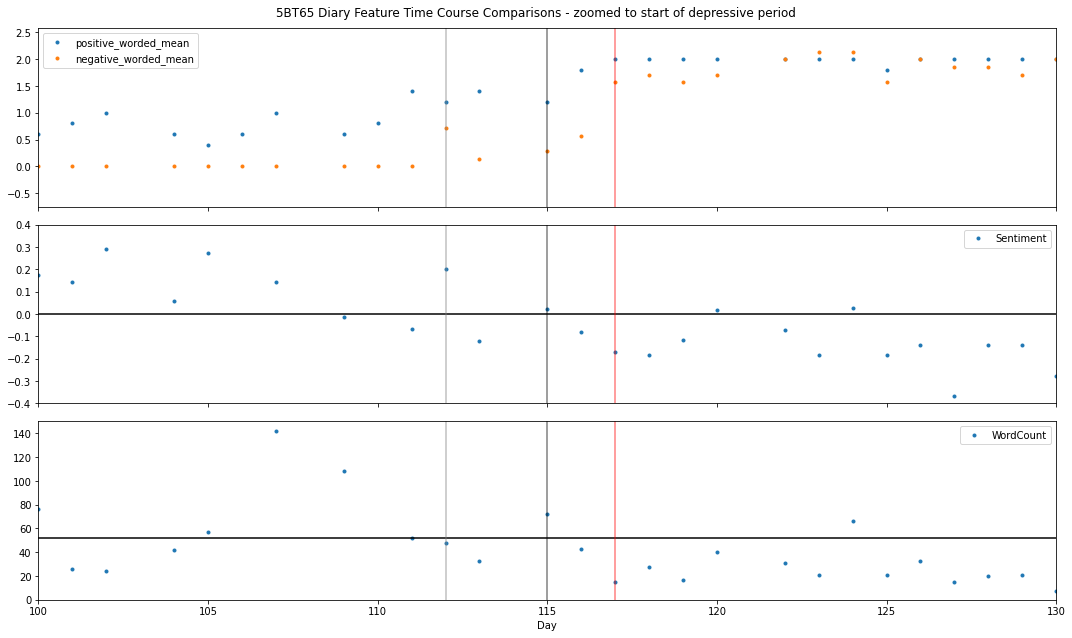}
\caption[Basic diary features changed in sync with EMA at the start of 5BT65's first depressive episode.]{\textbf{Basic diary features changed in sync with EMA at the start of 5BT65's first depressive episode.} A follow-up to the plots of Figure \ref{fig:5b-madrs}, this Figure zooms into study days 100 through 130 of 5B's enrollment in BLS, aiming to compare key dairy features against same day EMA to determine if those features might change ahead of or perhaps behind EMA. Positive (blue) and negative (orange) self-report emotion summary scores were thus plotted over time in the top panel, followed by two additional dot plots showing diary mean sentiment and total word count on the same days. Markers on day 115 (black) and 117 (red) denote the base and the top respectively of the first major depressive episode seen in the zoomed out view of Figure \ref{fig:5b-scales}. A marker was also placed on day 112, as this was the last day with non-zero negative symptoms that recovered before the episode. Another advantage of the zoomed in view is that short periods of data missingness are clearly visible. \newline Note that the MADRS clinical interview point depicted in Figure \ref{fig:5b-madrs} occurred on day 107. Ultimately it is not surprising that there were not clear signs of an episode at that time, nor is it all that surprising that the diary features that fit well with same-day EMA adjusted to the episode in sync with EMA.}
\label{fig:5b-diary-madrs} 
\end{figure}

I therefore plotted an even further zoomed in version of the EMA summary scores along with select diary features, representing about one month of real time from between days 100 and 130 (Figure \ref{fig:5b-diary-madrs}). From days 100 to 107 (the clinical interview day for the directly preceding MADRS rating of minimal symptoms), 5B self-reported no negative emotions, at least a little positive emotions, and had positive diary sentiment scores. That was followed by 2 days of Beiwe missingness, not something abnormal. On days 109 and 111 recorded diaries had negative mean sentiment, and a diary (but not an EMA) was missing from day 110. Day 112 represented the last day for a "local" jump in negatively-worded EMA summary, as well as the last day in the period with a diary sentiment score meaningfully above 0. Beiwe self-report data was missing on day 114 and then day 115 started the monotonic ascent, i.e. the base of the episode as viewed from the EMA fluctuations. Sentiment was $\sim 0$ on this day and diary word count was a bit above average, and the EMA summary scores were not notably changed either. The increase between days 115 and 117 in self-reported severity was well-reflected by a decrease in sentiment and word count of diaries from the same period, but there was nothing from within these data alone to suggest the diaries could have preemptively detected the beginning of an episode (Figure \ref{fig:5b-diary-madrs}). \\

\FloatBarrier

\paragraph{What did 5B say prior to a major depressive episode?}
Fortunately, diaries can be used in many ways beyond core summary features. While they certainly have great potential for eventual more complex models, and while the summary features certainly have their own value (as repeatedly demonstrated), it is worth reiterating that simple qualitative reviews can yield interesting results when it comes to the audio journal format. Upon seeing some negative sentiment scores preceding the major changes to EMA scores in Figure \ref{fig:5b-diary-madrs}, I checked how frequently negative sentiment scores of that magnitude were found outside of the core episode time periods, and it was frequent enough that it would be a lousy predictor. Still, there is content behind each score, and I was curious whether the negative sentiment journals directly preceding the start of the episode might differ from other slightly negative sentiment journals recorded by 5B. 

I found that on days 112, 113, the "official" start of the episode on day 115, and the "transitory" point of moderate EMA on day 116, 5BT65 mentioned in their recordings that they were 
\begin{quote}
    kinda down
\end{quote}
\noindent sometimes multiple times in a recording and sometimes as an entirely standalone sentence without much of a transition. 

Because of this, I counted occurrences of "kinda down" across the entire diary set, to see if it was unique to right before that episode. Which in itself would be mildly interesting, since it is also possible they use the phrase a lot and finding it here was $100\%$ a coincidence. What I did not expect to find though was for it to generalize to the other, somewhat clinically different, prolonged episode observed during 5B's time in the study. Indeed, the phrase "kinda down" was used in only 7 total diaries across the entire set of $>500$ transcribed from this participant:
\begin{quote}
    112, 113, 115, 116, 289, 292, and 294
\end{quote}
\noindent so the usage truly came out of nowhere to appear in 4 days of a contiguous 5 day stretch (the middle of which did not have any submission), then did not appear again for nearly 6 months, and then popped back up in 3 days of a 6 day stretch containing only 4 active Beiwe submissions, after which it was never used again. 

\begin{figure}[h]
\centering
\includegraphics[width=0.8\textwidth,keepaspectratio]{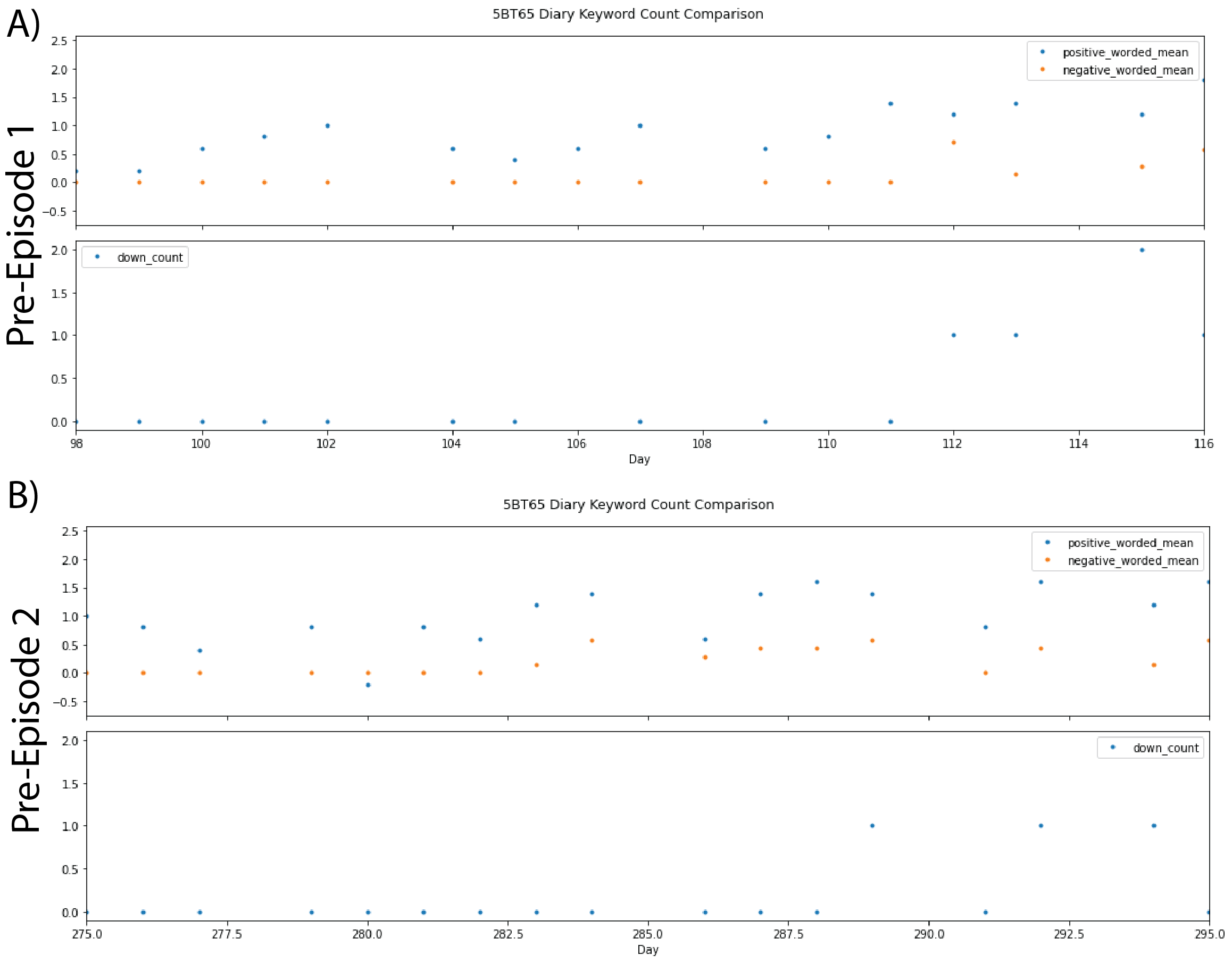}
\caption[Signal of mood episode from 5BT65's choice of language preceded any reliable signal from self-report score changes.]{\textbf{Signal of mood episode from 5BT65's choice of language preceded any reliable signal from self-report score changes.} For the 20 days (96 through 116) leading up to the beginning of the first mood episode (A) as well as an analogous range (days 275 through 295) leading up to the beginning of the second major 5B episode (B), I plotted positively-worded (blue) and negatively-worded (orange) EMA summary scores across days in one panel (top) and then plotted the number of occurrences of the phrase "kinda down" in the journal from each day within the other panel (bottom). Note that the final peak of each mood episode was intentionally excluded here to demonstrate that without forward looking, it would be very difficult to make a confident statement about the upcoming progression of self-report emotions based on the values of those self-reports up to the end of these time periods. However the repeated and concentrated appearance of "kinda down" could alert to the impending mood episodes in each case. See Figure \ref{fig:kinda-down2} for the wider view.}
\label{fig:kinda-down} 
\end{figure}

\begin{figure}[h]
\centering
\includegraphics[width=\textwidth,keepaspectratio]{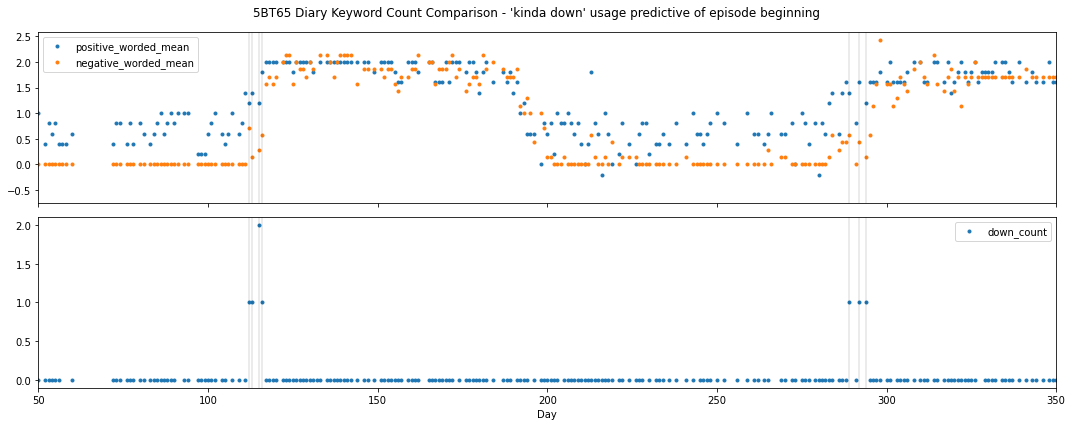}
\caption[Signal of mood episode from 5BT65's choice of language reliably coincided with impending self-report score changes.]{\textbf{Signal of mood episode from 5BT65's choice of language reliably coincided with impending self-report score changes.} For a period of 300 days surrounding the beginning of 5BT65's first major mood episode and reaching well into the peak of their second such episode, The progression of the emotion-related EMA summary scores is plotted (top), aligned with a plot of occurrences of the phrase "kinda down" across 5B's diaries (bottom). This is a zoomed out version of the data presented in Figure \ref{fig:kinda-down}, demonstrating the actual onset of each episode and how the EMA and diary dynamics fit together more broadly. A light grey vertical line spans both plots at each of the 7 days with a diary containing the phrase of interest. Note that "kinda down" did not appear in any of 5B's diaries outside of the depicted range here.}
\label{fig:kinda-down2} 
\end{figure}

Aligned with the EMA data, it is a strong signal at both times, prior to a strong signal being available from the EMA or more fundamental extracted diary properties (Figures \ref{fig:kinda-down}-\ref{fig:kinda-down2}). This was the case even for the second episode, which had a less immediate jump to the high severity negatively-worded EMA scores than the first episode did (Figure \ref{fig:kinda-down2}). 

5B did not have another prolonged mood episode like their first two during BLS, so it is not clear whether they would have used such phrasing again if they were in that state again. Regardless, this result would obviously not directly generalize to other individuals. It serves primarily as a powerful proof of concept for simple techniques to take advantage of unique benefits of audio journals. 

It is likely that many individuals have similar types of quirks within their language use that could be uncovered through a targeted content discovery process using time-aligned features extracted from diaries, EMA, clinical scales, digital phenotyping, etc. The journal format permits an iterative process for data analysis to inform content search and content to in turn inform next data analysis steps. Visualization tools like the sentiment-colored word cloud can additionally help specific content of note stand out in a large transcript set. I genuinely spent a very small percentage of my time on these case sections looking directly through diaries. \\

\begin{figure}[h]
\centering
\includegraphics[width=\textwidth,keepaspectratio]{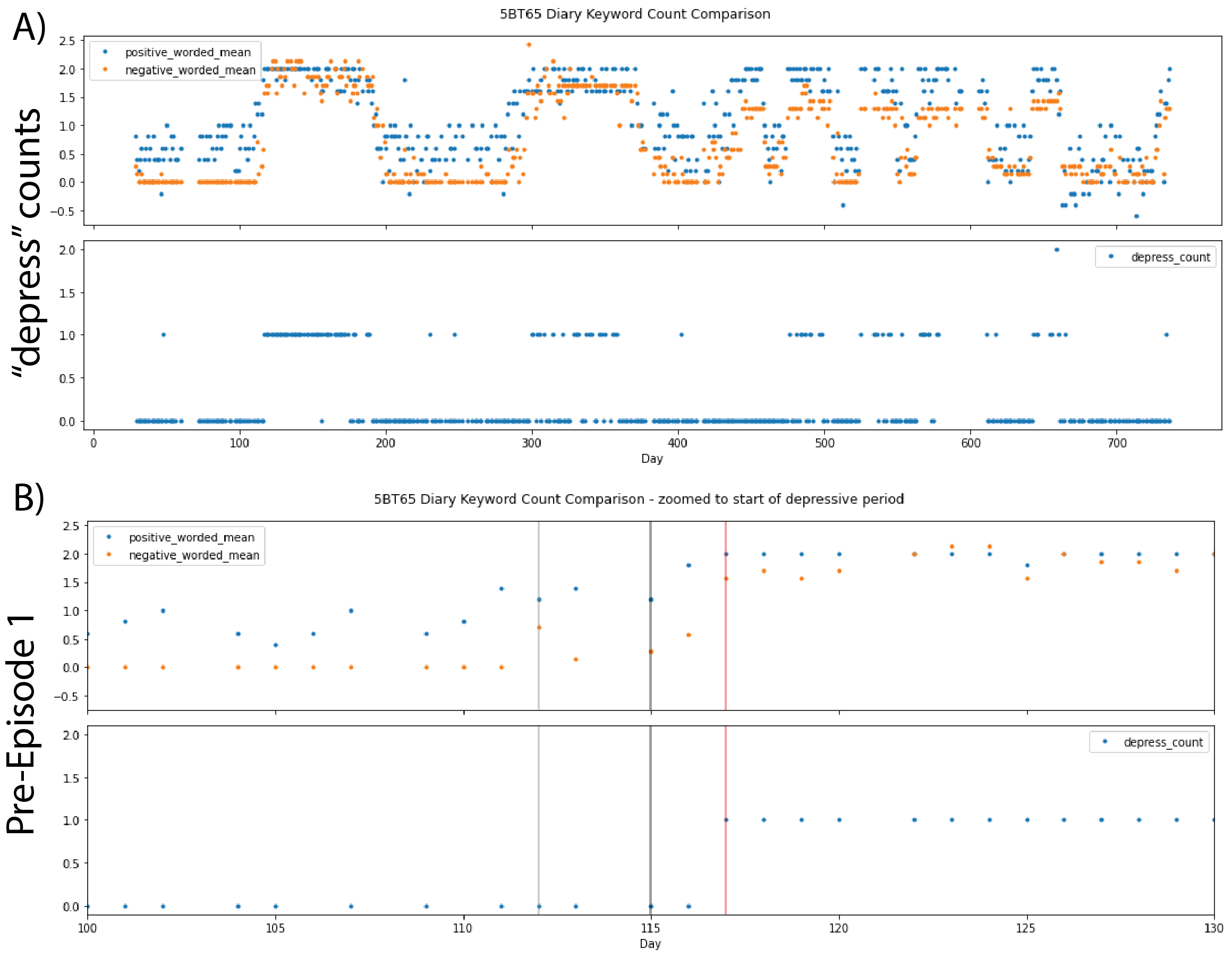}
\caption[Use of "depress"-prefixed words in audio journals strongly overlapped with mood episodes in 5BT65.]{\textbf{Use of "depress"-prefixed words in audio journals strongly overlapped with mood episodes in 5BT65.} Across the entire 5B self-report dataset (A), positive (blue) and negative (orange) emotion EMA summary scores were plotted over days (top), with time-aligned word counts from the corresponding diary submissions (bottom). Here words containing "depress" were tallied, and the vast majority of the cases where this count was non-zero was also a time of severe self-report mood results (A). To determine whether discussion of depression might precede an episode, I also zoomed in to the above plot to center around the lead up to 5B's first major episode (B). This plot is analogous to those for word count and sentiment in Figure \ref{fig:5b-diary-madrs}, but now instead using "depress" occurrences as the comparison variable. The count remained at 0 for weeks leading up to the mood episode, and then jumped up on the same day (117) labeled as the start of the EMA peak and thus already clearly into the episode from a data perspective (B). Still, it is interesting that the topic came up so immediately for the participant here, and doubly so because it continued to be mentioned in the 12 straight diaries after that point that can be seen in the present Figure.}
\label{fig:5b-depress} 
\end{figure}

\noindent Moreover, there are obvious search terms based on prior knowledge of a given participant's diagnosis, for example marking mentions of mania in 8RC89 above. For 5BT65, counts across diaries of uses of "depress" (which includes depression, depressed, depressive, etc. per the pipeline's built-in key word counting) showed a very clear overlap with identified mood episodes in 5B (Figure \ref{fig:5b-depress}). 

This was not especially useful here because 5B had such a rich structure already present in their EMA responses, and mentions of depression coincided so closely with already identified mood state -- to the point that "depressed" popped up immediately on day 117, the time point previously flagged as the first severe symptom point of the initial mood episode (Figure \ref{fig:5b-depress}). Such terms were used highly frequently within episodes and especially the first episode too, and overall many diaries in a row were submitted containing related language.

Conversely it is obviously not a $100\%$ accurate mapping in the more general participant case. Nevertheless, this is yet another strong proof of concept for meaningful relationship between simple diary content patterns and clinical state. Future works should of course confirm information about content in flagged diaries, and should ultimately build improved models for not only labeling content, but also integrating information from content accounting with information from e.g. time-aligned automated feature value or the temporal dynamics over the dataset for a given content pattern. \\

\FloatBarrier

\begin{figure}[h]
\centering
\includegraphics[width=\textwidth,keepaspectratio]{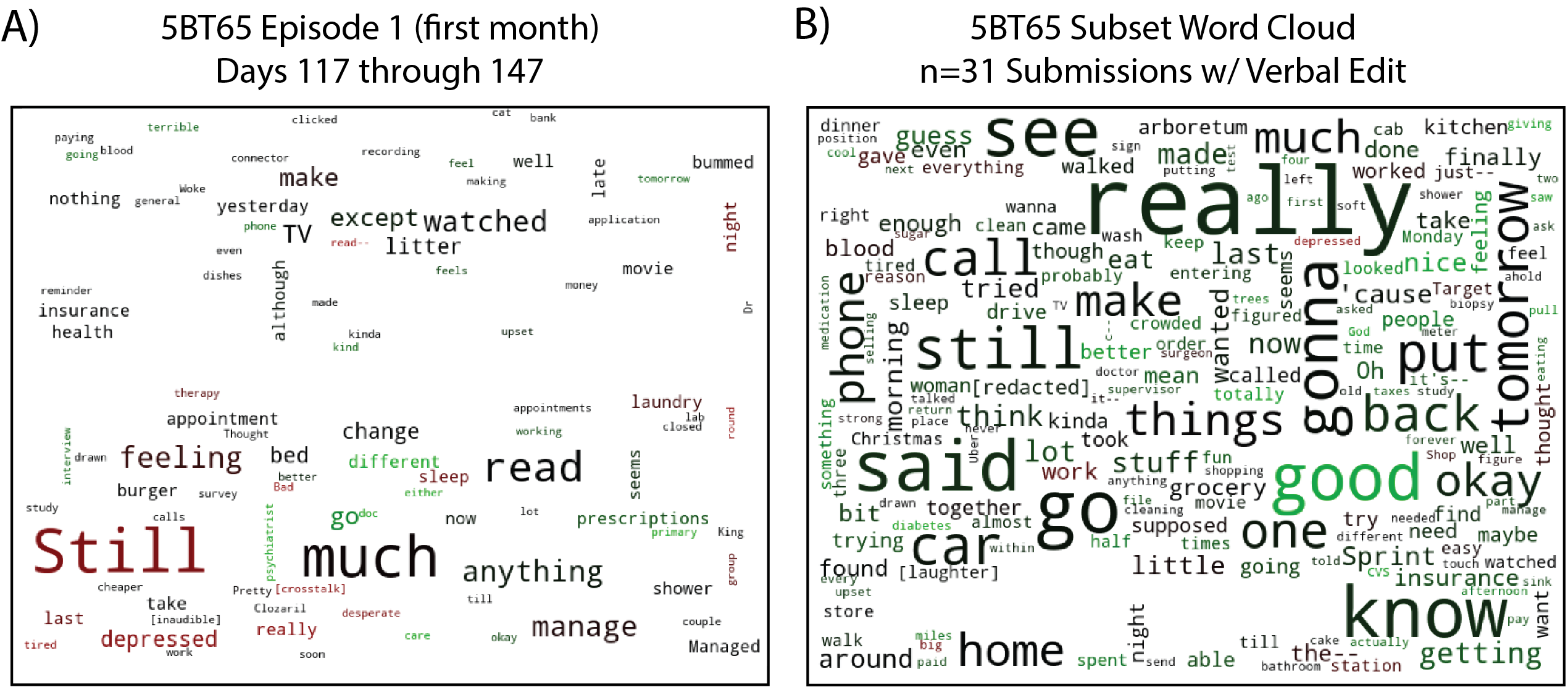}
\caption[Sentiment-colored word clouds well-represent 5B's variance in journal behavior.]{\textbf{Sentiment-colored word clouds well-represent 5B's variance in journal behavior.} Using the functionality built into the pipeline, I generated sentiment-colored word clouds for some transcript subsets of interest. The first 30 days of the early depressive episode experienced by 5B are represented by (A), while the 31 days across the 5B dataset containing non-zero use of verbal edits are represented by (B). Large differences in verbosity and sentiment in addition to content style are immediately apparent, demonstrating the relevance of this visualization tool.}
\label{fig:5b-cloud} 
\end{figure}

\paragraph{Sentiment-colored word clouds.}
Because verbosity and sentiment were so strongly associated with EMA in 5B, and we observed sustained periods with highly similar self-reporting behavior, we might expect sentiment-colored word clouds across different study periods (as well as skimming day-level thumbnails) to be of particular value in getting a sense for 5B's dataset. While a deep dive on word clouds would be beyond the scope here, differences were already apparent in word clouds generated between different straightforwardly-defined study periods (Figure \ref{fig:5b-cloud}). 

\begin{figure}[h]
\centering
\includegraphics[width=\textwidth,keepaspectratio]{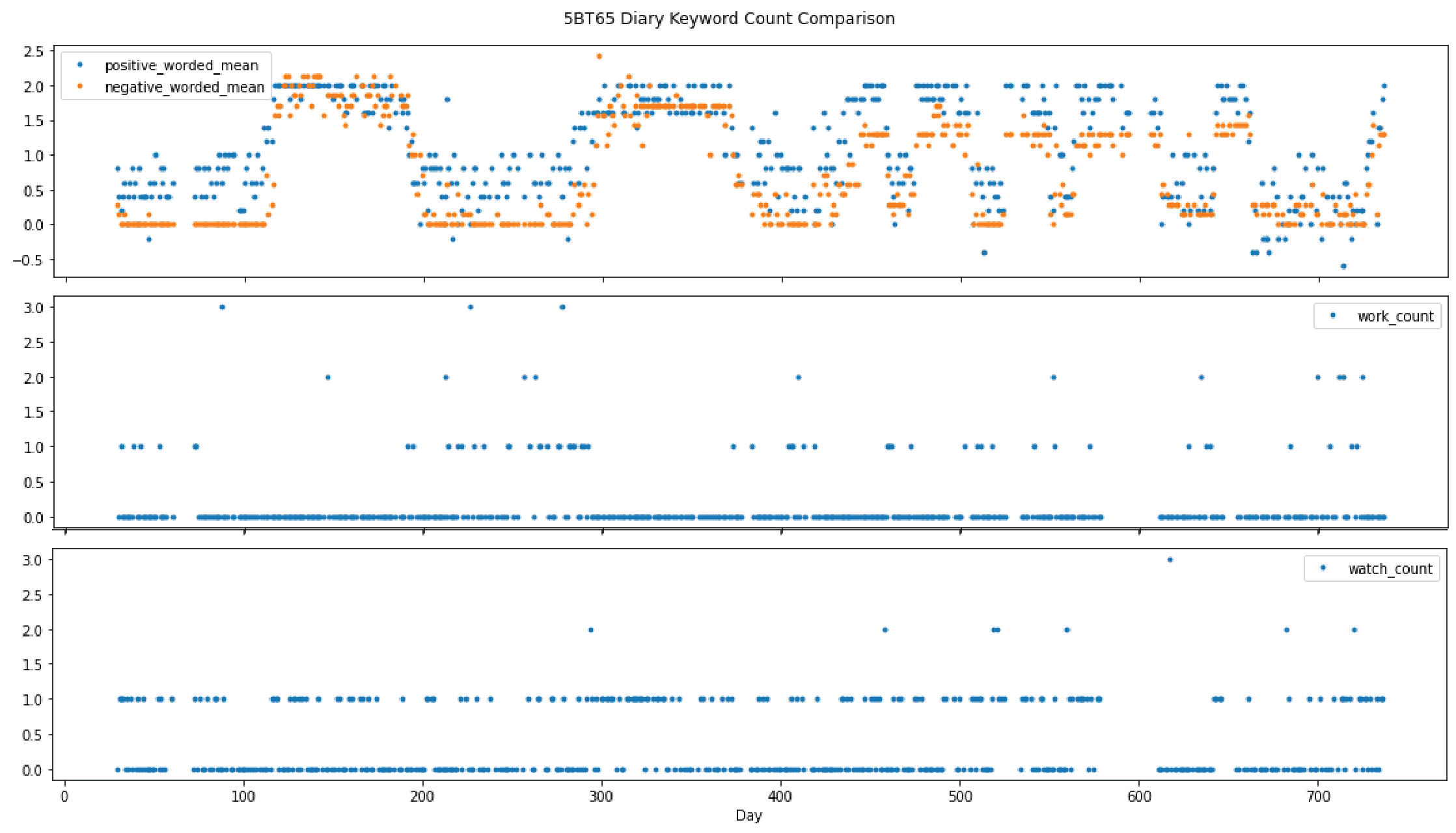}
\caption[Relationships between activity-related word counts and EMA.]{\textbf{Relationships between activity-related word counts and EMA.} Across the entire 5B self-report dataset, positive (blue) and negative (orange) emotion EMA summary scores were plotted over days (top), with time-aligned word counts from the corresponding diary submissions (bottom). Here "work" (middle) and "watch" (bottom) were tallied. Discussion of work was largely restricted to periods without severe mood symptoms, but discussion of watching things was more uniformly distributed.}
\label{fig:5b-count-sup} 
\end{figure}

The frequent use of "depressed" in diaries recorded during 5B's episodes that was discussed above (Figure \ref{fig:5b-depress}) can be easily spotted in Figure \ref{fig:5b-cloud}. Further, identification of potential new key words of interest for quantification was facilitated by the word clouds in Figure \ref{fig:5b-cloud}, for example checking words like "computer", "watch", and "work". I expected work to coincide squarely with healthier psychological states (as estimated by EMA), which it did (Figure \ref{fig:5b-count-sup}). However I was not sure whether talk of watching e.g. TV might have some state-dependent component; which if it did, that was not obvious from the visualization of EMA dynamics versus word counts (Figure \ref{fig:5b-count-sup}). \\

\FloatBarrier

\paragraph{Fluctuations in diary-extracted features.}
Looking at variations over time in diary properties already proved quite salient in both 3S and 8R. Because EMA variation contained so much relevant information at different timescales in 5B, and a number of diary features had significant (but certainly not exhaustive) relationship with EMA in 5B, considering their temporal fluctuations is likely to prove especially fruitful. Indeed, expected underlying state-related variation that coincides with the EMA-identified mood episodes was found across a number of EMA features in 5B (Figure \ref{fig:5b-diary-time}).

\pagebreak

\begin{FPfigure}
\centering
\includegraphics[width=\textwidth,keepaspectratio]{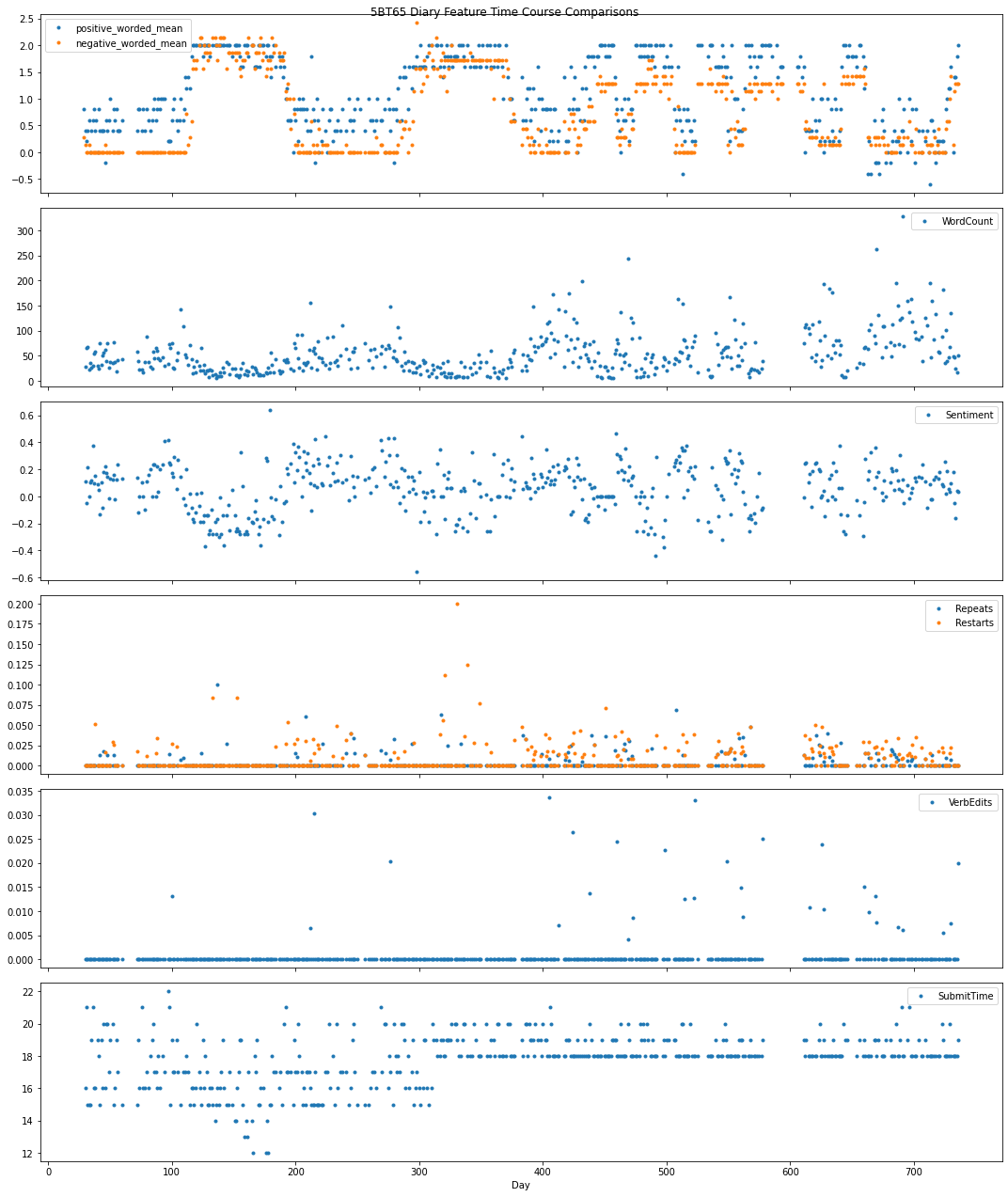}
\caption[Meaningful fluctuations at multiple timescales in 5BT65 diary features.]{\textbf{Meaningful fluctuations at multiple timescales in 5BT65 diary features.} Using a shared x-axis to align the data across 5BT65's enrollment in BLS, I generated dot plots with EMA summaries and select diary features. Thus the first panel of this figure depicts all of the positive (blue) and negative (orange) emotion summary EMAs, as already discussed extensively with Figure \ref{fig:5b-scales}. It then serves as a reference for each chosen diary feature plotted in its own panel, in the following order: the word count, the mean sentence sentiment, the number of repeats (blue) and restarts (red) used per word, the number of verbal edits used per word, and the submission hour as labeled by my pipeline. Fluctuations corresponding to the EMA-defined mood episodes can also be easily found within this diary data, as can longer term changes over the course of the study that might help with explaining the temporal dynamics of the temporal dynamics, along with shorter term changes in e.g. sentiment that can capture typical fluctuations of mood in daily life and thus some of the variance within-state of the EMA summaries.}
\label{fig:5b-diary-time}
\end{FPfigure}

\FloatBarrier

In addition to the obvious fluctuations in word count and sentiment that align with major mood periods, there were a number of changes in diary features over the course of the study that might be of salience in explaining the dynamics of 5B on the whole. For example, shortly after day 300 their diary submission time patterns changed abruptly and did not return to their original late afternoon habits (Figure \ref{fig:5b-diary-time}, bottom). This could be due to a life change like a new job, a technical change with e.g. Beiwe prompting, or it may even carry more surprising significant than that. It would require further investigation, but the shift in submission time was suspiciously close to the onset of the second mood episode, promptly after which a number of other dynamics changed in the 5B feature set. 

\begin{figure}[h]
\centering
\includegraphics[width=\textwidth,keepaspectratio]{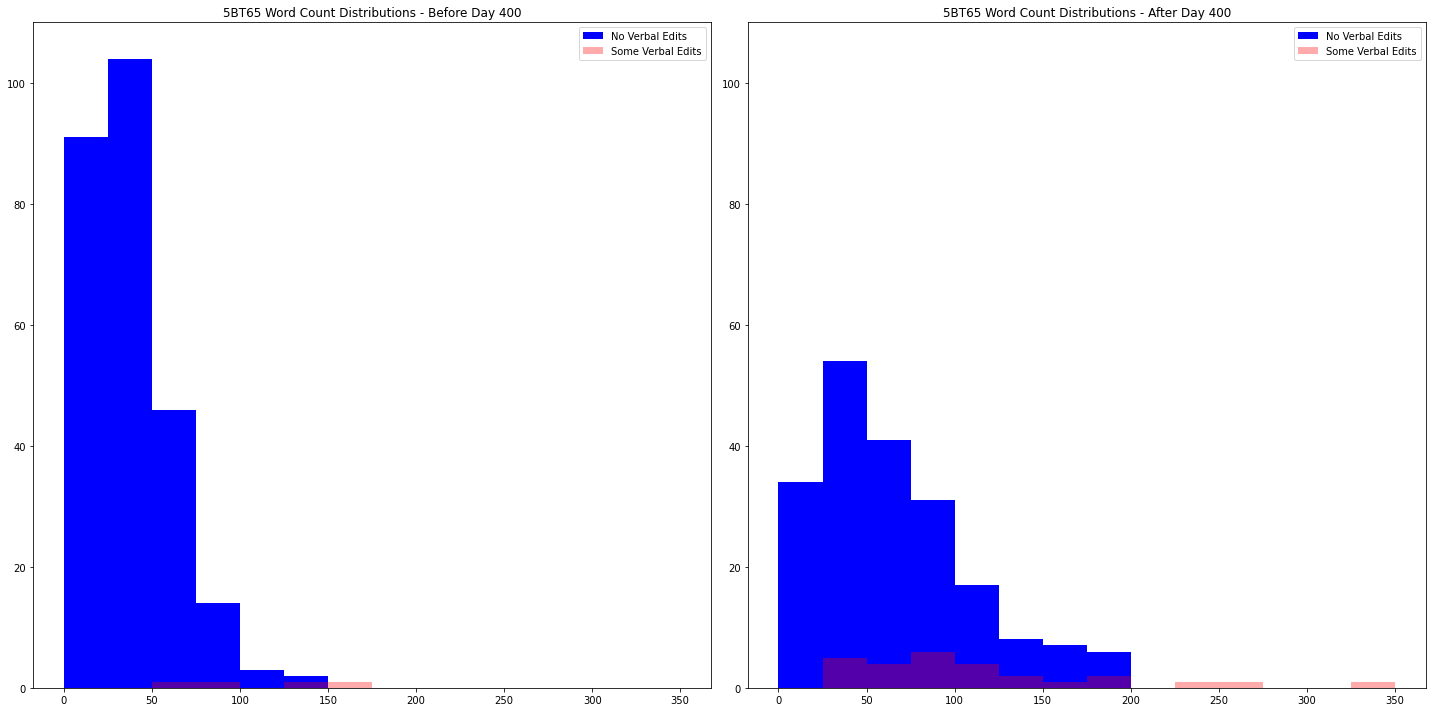}
\caption[Increased verbal edit usage in the second year of 5B's enrollment does not appear to be explainable by increased word count alone.]{\textbf{Increased verbal edit usage in the second year of 5B's enrollment does not appear to be explainable by increased word count alone.} Histograms of word counts from 5BT65's diary transcripts before day 400 (left) and from day 400 on (right) are plotted here, using the same x-axis and y-axis limits and the same bins. On each histogram, the days with no verbal edits (blue) and days with verbal edits (transparent red) are plotted separately. Overall, word count was higher for 5B in the latter half of the study and more verbose diaries are more likely to contain verbal edits all else being equal - however there were also more days with verbal filler usage at lower word counts during the latter half, as can be seen here, and the verbal edits that did appear in the first half were closer in time to transitory periods than one might expect from random chance (Figure \ref{fig:smooth-verb}).}
\label{fig:5b-verb-edits} 
\end{figure}

Along those lines, it is worth highlighting that the vast majority of journal entries containing verbal edits (Figure \ref{fig:5b-diary-time}, second from bottom) occurred shortly after day 400, when the shift in dynamics emerged. Of course a higher baseline word count, something that also arose in 5B right after day 400 (Figure \ref{fig:5b-diary-time}, second from top), would naturally facilitate a higher probability of a verbal edit occurring within the same of a recording. The effect does not appear to be fully explainable by word count alone though, because the "low word count" points in the latter half of the data had quite similar verbosity distributions to the "high word count points" in the earlier half of the data, yet did not seem as likely to contain verbal edits (Figure \ref{fig:5b-verb-edits}). 

Related potential transients that ought to be investigated are the \emph{relative} jumps in word count that appeared to closely surround transitions into and out of EMA-defined mood states and the extent to which those diaries in particular might be correlated with verbal edit use. Additionally, restarts and repeats (Figure \ref{fig:5b-diary-time}, third from bottom) both had large transient increases starting in the early phases of a peak severity mood episode, returning to middling levels outside of episodes. That in particular would call for nonlinear modeling consideration, as the absolute highest values but actually not at all the middle values may be associated with symptom severity. Better separating out the nonlinear effects that shorter diary length have on disfluency tracking will be an important future step in general, so that it can be better determined which feature changes (here and elsewhere) are truly meaningful. 

Interestingly, while baseline word counts shifted up for 5B later in BLS (calling back to the discussion of rolling models as well), there was not such a clear magnitude shift in the sentiment (Figure \ref{fig:5b-diary-time}, third from top), which still covered largely the same range just (unsurprisingly) varied much faster. Furthermore, as word count remained relatively high during the later, briefer negative self-report periods, there was eventually an extended stretch around day 600 where 5B continued to submit surveys but did not record any journals. It is possible that the observed shift to talking more is a good sign in general for improving symptoms of 5B, but it also may put pressure to keep up with recording diaries of reasonable duration, ultimately leading to a period with missing data entirely. Although that can still be informative, it was extremely helpful that 5B continued to record \emph{something} regardless of length even in the midst of severe symptoms in the earlier parts of the study.   

\begin{figure}[h]
\centering
\includegraphics[width=\textwidth,keepaspectratio]{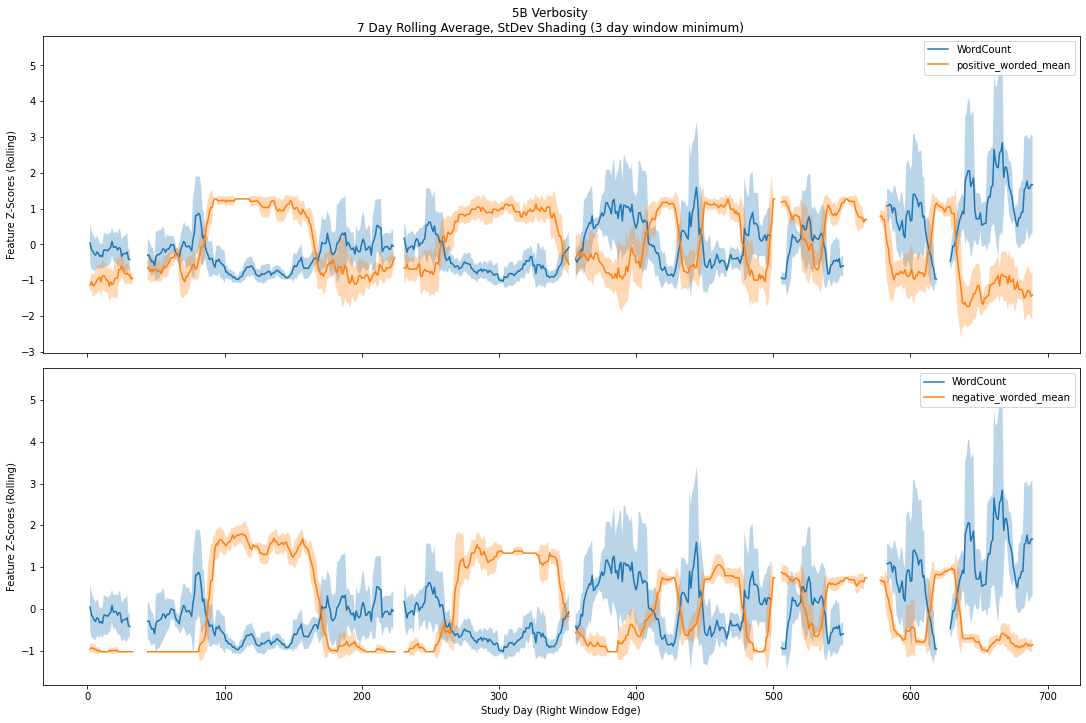}
\caption[Trends in word count directly reflect trends in emotion-related EMA summary scores over the 5BT65 dataset.]{\textbf{Trends in word count directly reflect trends in emotion-related EMA summary scores over the 5BT65 dataset.} This line plot shows a 7 day rolling average with standard deviation shading for the positively-worded EMA mean (top, orange) and negatively-worded EMA mean (bottom, orange) from 5B self-report submissions over their entire course of participation in BLS. On the same panels a time-aligned 7 day rolling average also with standard deviation shading is plotted for the total word count of the 5B diaries over time (blue). The blue curves will thus be identical between top and bottom plots. Note that the base feature being used for the rolling average and ultimately plotted here is the z-score of the raw feature relative to 5B's overall mean and standard deviation for that feature. This was done to enable a variety of different feature types to have their dynamics plotted on the same graph, so overlap in time could be leveraged for qualitative insights.}
\label{fig:smooth-word} 
\end{figure}

\begin{figure}[h]
\centering
\includegraphics[width=\textwidth,keepaspectratio]{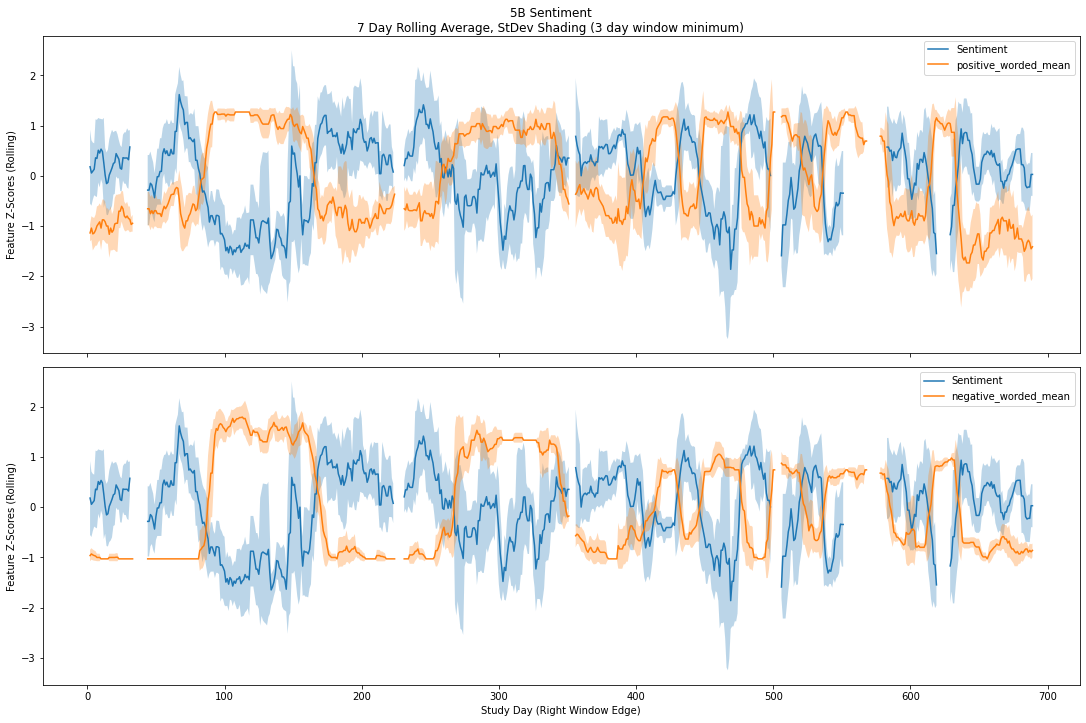}
\caption[Trends in sentiment capture smaller timescale variations in the 5BT65 EMA summary dataset.]{\textbf{Trends in sentiment capture smaller timescale variations in the 5BT65 EMA summary dataset.} The same function as done for word count in Figure \ref{fig:smooth-word} was repeated here for mean diary sentiment.}
\label{fig:smooth-sentiment} 
\end{figure}

\begin{figure}[h]
\centering
\includegraphics[width=\textwidth,keepaspectratio]{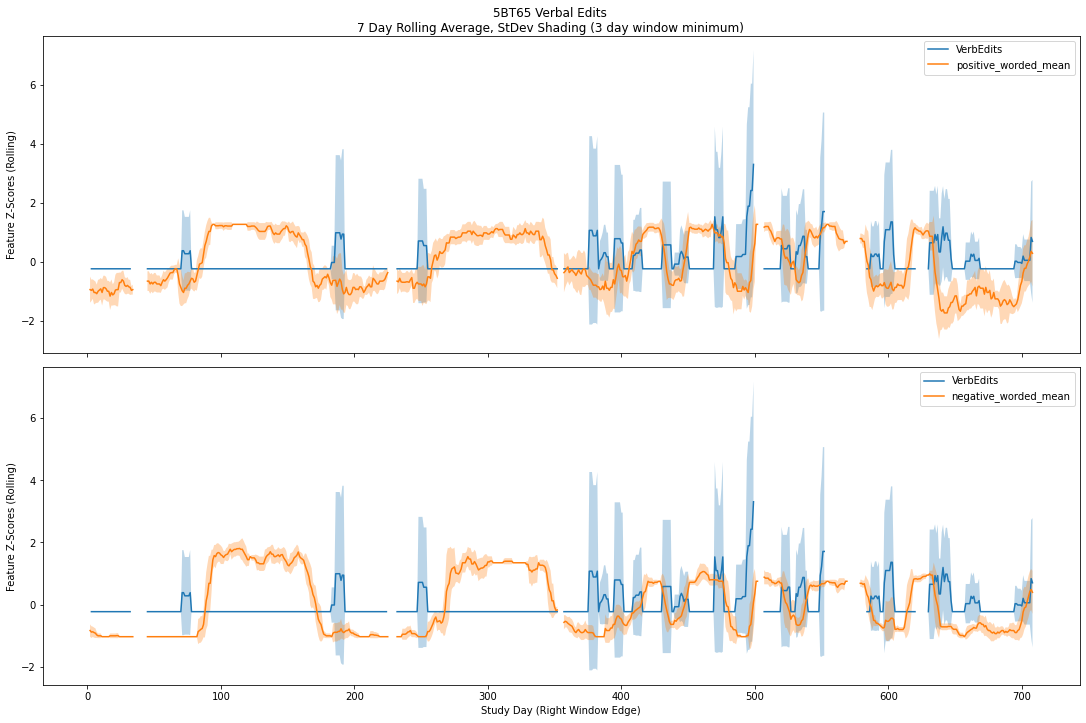}
\caption[Trends in sentiment capture smaller timescale variations in the 5BT65 EMA summary dataset.]{\textbf{Trends in sentiment capture smaller timescale variations in the 5BT65 EMA summary dataset.} The same function as done for word count in Figure \ref{fig:smooth-word} was repeated here for verbal edits per word. As verbal edit counts were most often 0, the z-score is not ideal for this situation, but it suffices to get the high level picture across. Directly comparing this plot to Figure \ref{fig:smooth-word} demonstrates that the more frequent verbal edit spikes occurring in the latter part of the study are unlikely to be fully explained by larger word counts alone.}
\label{fig:smooth-verb} 
\end{figure}

Smoothed comparisons over time of the word count, sentiment, and verbal edits against summary EMA scores can further impart some of the above points (Figures \ref{fig:smooth-word}-\ref{fig:smooth-verb}) and perhaps suggest additional follow-ups. Explicit modeling of the temporal dynamics of psychiatric state in terms of observable behavior - along the lines of the tools to be introduced in chapter \ref{ch:4} - would be most tractable to approach by starting out with subjects like 5B. Taking into account additional known clinical information beyond the scope of the thesis and performing a deeper dive into the diary content over time with an eye for insights on the observed fluctuations would make for a powerful "best of both worlds" approach that is both clinically and mathematically explainable. 

\FloatBarrier

\section{Discussion}
\label{sec:discussion2}
The core aims of this chapter were to both motivate the use of audio journals in more psychiatry studies going forward and to provide foundational material to enable such studies. This section will thus summarize the main introductory arguments for audio journals (\ref{subsec:diary-argue-recap}), the abilities and validation of the supporting code base released with this chapter (\ref{subsec:diary-code-recap}), and the pilot scientific results characterizing extracted features in a Bipolar Disorder dataset (\ref{subsec:diary-science-recap}). I will then discuss current limitations of the work to set up an extensive discussion of the many future directions it enables (\ref{subsec:diary-future}). Finally, the chapter will close with a recap of the key contributions herein (\ref{subsec:diary-contrib}).

\subsection{The important role of audio journals}
\label{subsec:diary-argue-recap}
While various forms of patient journaling have been longstanding therapeutic components of some psychiatric interventions (section \ref{subsec:diary-history}), and there is an abundance of evidence for associations between various patient acoustic/linguistic properties and severity of some psychiatric symptoms (section \ref{subsec:diary-lit-rev}), there is a surprising paucity of work analyzing psychiatric patient journal entries. With the recent rise of multiple digital psychiatry apps equipped to record daily audio journals along with other data modalities, collection of audio diaries both at scale and as an add-on to existing studies has become relatively cheap and overall very feasible. 

Despite this, daily app-based audio recordings have yet to be a focus of much speech analysis or digital psychiatry literature. One might worry this is due to uninspiring pilot results in the large scale exploratory projects where they have been collected, but in my experience audio diaries have mostly been left to collect dust while other datatypes, such as interview recordings for speech sampling or EMA for daily symptom self-reporting, receive a great deal of analysis attention; the AMPSCZ project, to be detailed in chapter \ref{ch:2}, is a major recent example, though time remains for the collected journals to receive the attention they deserve.

Throughout the chapter, I have provided theoretical arguments as well as practical pilot results demonstrating the many ways that audio journals are uniquely well suited to present scientific opportunities for many different psychiatry studies. Compared to interview recordings, diaries have better temporal resolution, are easier to collect from a larger number of participants, are easier to collect from a given participant over a longer study period, enable capture of open ended self-reporting in a truly patient driven manner, and produce individual time points that are more realistic to adequately summarize without requiring a large feature set or complicated context-aware algorithms. 

Although there are advantages to the interview recording format for a few particular types of scientific question (see chapter \ref{ch:2}), in the general case audio journals are substantially more tractable for exploratory data science work as well as for hypothesis-driven work that addresses longitudinal questions or requires a more heterogeneous study population. Furthermore, they have great promise for novel insights both qualitative and quantitative, as they remain much less studied than clinical interviews; to truly realize the potential of emerging digital psychiatry tools, it will be necessary to obtain data that is more naturalistic and more temporally dense.

By contrast, digital phenotyping datatypes like phone GPS or wrist accelerometry have much greater temporal coverage than audio journals and are even easier to collect from many subjects due to their passive nature. However, it is difficult to interpret these datatypes without clinical or self-report context, and it is difficult for primarily psychiatry-focused groups to carefully analyze such large timeseries with relatively infrequent labels. There are certainly scientific questions of interest to be addressed with passive sensing datatypes alone, but there are also many relevant questions that would benefit from the availability of audio journal data. 

Analysis of patient speech is an important ongoing direction for psychiatry due to its established clinical evidence base and its relationship with well characterized biological principles. The connection between psychomotor signs and impacted speech patterns alone warrants more work linking audio journals with passive sensing data. Additionally, the open ended patient self-reporting enabled by daily diaries can be used to contextualize passive data in unique ways that would be difficult to achieve with surveys alone (see chapter \ref{ch:3} for proof of concept examples in a multimodal case report). While it may not be appropriate for every group, the digital psychiatry field at large should not underrate the potential value in autobiographical accounts, as it has thus far. Quantitative and qualitative work need not be mutually exclusive, nor do objective and subjective metrics. The daily audio diary format can be well utilized in all 4 ways, and it can be well utilized in conjunction with many different datatypes or even on its own. \\

Now that I have summarized many of the major arguments for increased use of audio journals, both for groups primarily focused on speech sampling and groups primarily focused on digital phenotyping, I will spend the rest of the section expanding on the dynamics of these fields at present and how audio journals ought to fit into the future. This will also include a discussion of the importance of software tools like my pipeline in that future, as well as a more detailed set of arguments on the value of autobiographical patient perspectives. The latter point is not necessary to support the high level claim of the chapter on the severely underrated status of daily audio diaries, but it is an important point for psychiatry more broadly in my opinion, and it remains very relevant to the journal format. Ultimately, by carefully characterizing the great value of audio diaries, I will have further motivated the contributions of this chapter, along with the upcoming discussion section on future directions for this datatype at multiple scientific timescales (\ref{subsec:diary-future}).

\subsubsection{Clinical and computational advantages make for the best of both worlds}
Audio journals provide scientific advantages across a number of domains, yet this ironically might be the reason they've remained relatively ignored. As will be reviewed below in section \ref{subsubsec:diary-u24}, the issue has apparently impacted audio journals on an organizational level in the early stages of AMPSCZ. With such a large collaborative project it is expected that responsibilities be partitioned and certain subgroups interact only rarely, but because audio journals don't fit neatly in a single box, they have not fit neatly under a single team's responsibilities either. Journals are simultaneously a speech sampling datatype and a phone app datatype, and as is consistent with prior psychology research, people are less likely to do something that they think might be someone else's responsibility. 

This is probably not the only factor limiting progress with audio diaries in AMPSCZ, however it is clean representation of a more abstract problem that I feel currently pervades relevant research projects. Many traditional speech sampling groups have transitioned to the digital age through the use of interview or task recordings for analysis -- formats that have a more immediate connection to prior psychiatry literature, but lack a number of the advantages that digital psychiatry can provide. It is especially problematic when this involves application of recent data science/machine learning techniques that are not appropriate for the sample sizes they are able to obtain with their chosen datatypes. 

On the other hand, many emerging digital psychiatry groups have put the majority of their focus on the datatypes that can generate the largest amount of data, and with a highly quantitative lens. Despite many such studies already collecting EMA surveys to use as one label source for symptom severity, the audio diaries that can often be collected by the same app don't seem to receive much consideration -- meanwhile these diaries are a potential multidirectional threat, from objective and quantitative vocal feature extraction all the way to subjective and qualitative report of clinically relevant phenomena. They could serve as inputs or labels, and they may be seen in terms of their potential for advanced machine learning applications like transformers or in terms of their potential for very human driven psychiatric analysis. If much better smart watches are available in 5 years or if industry data processing tools soundly beat anything developed by academic science at e.g. labeling a particular movement pattern of note, the digital phenotyping datasets and papers in progress right now might not age well. An audio journal dataset or a truly multifaceted audio journal study is much less likely to age poorly. Intermediate results can be extremely important, but again when thinking for the field as a whole, it is important for generational processes that some contributions are more timeless.

Proper collection and particularly analysis of daily audio diaries can indeed be more resource intensive than is warranted for some digital phenotyping studies, and it can also be well beyond the scope of rigorously designed interview or task recording studies that focus on addressing scientific questions directly related to the respective formats. I see little reason to believe such decisions are being made in a principled manner at present though, which is what has motivated much of this section: I aim to present detailed arguments for the use of audio journals that address both perspectives, such that future research plans may hopefully address some of these considerations outright. In doing so, I suspect that a number of hypothetical studies would legitimately increase the role of app-based audio diaries in their investigations. In the upcoming snippets I will address clinical and big data agendas, and provide specific examples of how both inputs and labels of interest could be obtained from diaries in a wide variety of study contexts, with some drawn directly from demonstrated properties of the BLS journal set. \\

\paragraph{Psychiatrists must not underrate data science tractability nor overrely on existing clinical measures.}
In practice, I have witnessed interview recordings prioritized over audio diaries at greater cost on multiple occasions. As such, the following list presents concerns that ought to be considered more often before making such a judgement:
\begin{itemize}
    \item Interview recordings have been prioritized as a speech sampling method over audio journals to date, likely because they seem less risky from a perspective focused on previous psychiatric literature. However it is very difficult to collect a large number of interview recordings, and those recordings are typically long -- which introduces a huge number of potential summary metrics over a small number of data points. Therefore almost all existing literature on the topic has either tested a small handful of already well known features or (more common recently) leaned entirely in to being exploratory with no attempt to make statements on relevant features with any statistical power. It is not sustainable for a whole field to behave this way for an extended time period, in particular as the features tested in this exploratory way completely change with nearly every paper (under the justification that technology advances rapidly and we must keep up). 
    \item Audio journals meanwhile can be collected at lesser cost from a greater number of subjects, and participating subjects will record many more journals providing much better temporal resolution than interview recordings can during the same study period. Overall, audio journal recordings can generate a substantially larger number of data points than interview recordings can, even when the journals are treated as an afterthought. Because each individual journal is much shorter, it is also much more tractable to compare journals over time and to summarize them with a reasonably-sized feature set. 
    \item Self report EMA surveys collected via the same app as audio journal recordings facilitate the collection of patient-provided labels nearly simultaneous with journal submission. While self report surveys have been fairly popular in recent literature, audio journals have not been. Additionally, self-administered scales with extensive prior literature, for example the Beck Depression Inventory (BDI), can be collected using these apps. The timescale and data availability alignment that EMA can be easily designed to have with audio journals is very conducive to quality study design, without requiring high costs or researcher labor burdens. It is odd then that EMA has become quite common to collect, often using an app that could be enabled to administer audio journal prompts too, yet the resulting EMA is used with other datatypes instead. For any proposal with a scientific question involving speech sampling and a study plan involving EMA, prioritization of audio journals should be highly encouraged. 
    \item When interview recordings correspond more closely with clinical scale ratings, they tend to contain a somewhat stunted conversation style that provides less rich content for analysis. These interviews also ask already well established questions, and while such questions obviously do have value in assessing psychiatric illness, it will be very difficult to discover new psychological factors of relevance without considering any datatype that allows more freedom. In my opinion, unconstrained autobiographical reporting from patients should be considered of greater interest as one portion of the academic literature.
    \item More broadly, if every study uses only established clinical scale ratings (or even worse diagnostic categories) as their labels, often looking at scale totals instead of symptom items at that, it will be extremely difficult for any breakthroughs to occur. As may be a trend throughout modern science, progress will likely be incremental, and incremental in a direction that perhaps cannot ever advance past a certain point due to fundamental limitations on that path. 
    \item Clinical scale ratings can of course still be used to contextualize long term diary collection studies, without introducing the additional cost and complexity of recording, transcribing, and processing the clinical interviews. Collecting interview recordings can still have its uses as well, and indeed there should be some studies that do prioritize these sources. At the same time, study design decisions should follow from scientific questions, not from supposed universal "best practices". Doubly so because the field of digital psychiatry is so new that it is difficult to understand why there even should be such strong trends yet in what datatypes are favored across a wide variety of studies. 
\end{itemize}
\noindent Note that interview recordings and audio diaries are compared at length within chapter \ref{ch:2}, which focuses on interview recording analysis. Examples of good possible reasons for recording interviews instead of or in addition to journal collection are provided there, but ideally future interview studies will address many of the questions in this list to some extent regardless. \\

\paragraph{Technical formalization must not eschew the value of subjective experiences and qualitative insights.}
It is rare to see the aforementioned fishing expedition studies include a careful characterization of the features they have deemed to be useful, despite the fact such a characterization could increase confidence in the stated conclusions by demonstrating a cohesive account of how different features contributed to observed clinical correlations as well as how consistent the features that did not correlate would be with expectations if one were to accept that the positive results were real. This need not be done in a highly algorithmic way to have value, and at times even a few well designed scatter plots would help immensely with assessing how confident one should be in an under-powered reported result. There is an obvious incentive on the authors' part to leave these plots out if they look too noisy, but there is also a cultural phenomenon at play here from what I can tell: rigor has been confounded with quantification, especially quantification that sounds technically complex.

Obviously quantitative methods can be used to test and ultimately make rigorous statements that would otherwise not be possible. However quantitative methods are not inherently rigorous. It is easy for them to be applied incorrectly in a way that reviewers do not notice when working in interdisciplinary fields, and even when applied correctly there are many caveats to keep in mind in interpreting the results. I have seen extremely mathematically savvy people make incorrect starting assumptions due to lack of domain specific knowledge, which led to bad models in spite of theoretically good quantitative methods. One preventative measure against this is to be willing to sanity check yourself in what might seem a "dumb" way. Qualitative review is indeed a fantastic avenue for humans to come up with ideas, not only for future studies, but also for questions that ought to be asked in order to adequately challenge your present conclusions. Furthermore, there is nothing wrong with including qualitative evidence to supplement the strength of your claims, as this enables other scientists to utilize the perspective in forming an opinion on said claims, and to potentially draw new insights from the study results that would not be possible with a paper that reports only traditional statistics. 

As mentioned, it is uncommon to find such qualitative characterization even in the supplement of a paper, and thus there is really no way to know if the authors attempted to carefully investigate their data beyond standard "plug and play" modeling. There is not enough collaboration with true machine learning experts in digital psychiatry at present, which exacerbates the issue. But even if there were, I highly doubt it would fully resolve this concern. On the contrary, I have seen incorrect Theorem proofs from highly technical groups make it into top machine learning conferences. A few examples are presented in chapter \ref{ch:4}, one of which has a very simple counterexample in the form of a 2 unit network; really a counterexample that anyone thinking about the problem from the perspective of the network as opposed to an abstracted-away math problem could have seen. 

Additionally, within the context of applying ML to psychiatry, it is not uncommon for technically interesting problems to diverge from clinically practical problems given the current state of the respective fields. Machine learning tools like the facial feature labeling models introduced in chapter \ref{ch:2} were interesting ML problems when released 5+ years ago, but their use in study of psychiatric illness has been underwhelming to date, with psychiatry groups left in large part alone to figure out how to effectively wield frame by frame facial action unit features from just a double digit number of labeled clinical interviews. The latter is in some sense impossible to cast in terms of a machine learning problem, and regardless most modern ML experts are not focused on learning from small amounts of data, which in practice here would likely involve domain knowledge and careful application of older data science tools. 

This brings us to yet another important role for audio journals: they are uniquely well suited for in depth domain specific qualitative analyses that can be paired with feature modeling in a manner that permits widespread interest. Diaries have great potential to better unite the different disciplines making up digital psychiatry, facilitating collaboration that could actually meld some methodological frameworks instead of just passing data back and forth. \\

\paragraph{The power of combining audio journals with other datatypes.}
It should be clear at this point that audio journals are a powerful datatype for future research into psychiatric patient speech, whether that research is very clinically grounded, very data driven, or somewhere in between. The versatility of audio journals does not stop there though, because there are many potential upsides for the use of daily diaries in a wide range of digital psychiatry studies, even those with minimal prior interest in speech.

One major perk of modern digital phenotyping data is the ability to focus analyses at many different timescales. Audio journals have this advantage over interviews to some extent, as journal features can be considered daily, but they can also be summarized weekly, biweekly, or - to align well with many clinical interview protocols - monthly. What makes this even more valuable in the context of multimodal work is the ability to align insights from audio journals with highly detailed measures of the same day from passive sensors. 

I will discuss the role of audio journal content in informing clinical ground truth next, but that is only one specific way in which they can provide value. More broadly, they can provide insightful context about a participant's day that can strongly synergize with trends detected in passive sensing. This can be utilized both for validation and for generation of multimodal feature sets that enable greater clinical predictive power than either could alone. Such benefits are not unique to the combination of audio journal and passive sensing, as different passive data streams can supplement each other similarly. However the type of information audio journals supplement with is in large part impossible to obtain with passive objective measures or highly constrained quantifications like EMA. 

A few proof of concept examples of the use of audio journals to supplement digital phenotyping data in different ways and at different time scales can be found in chapter \ref{ch:3}'s case report on an OCD patient undergoing a novel deep brain stimulation treatment. The patient submitted audio diaries very regularly, and on good days would often mention going out for a walk or for a bike ride, while bad days often mentioned not leaving the upstairs. Thus for one method for validation of the collected phone accelerometry data, we checked movement patterns against the specific physical activities (or lackthereof) reported on the days a clear statement was made. 

The journal dataset of chapter \ref{ch:3} was also used to assist in interpretation of observed GPS patterns, as the subject was quite religious and would often mention specific religious commitments or periods of relevant holiday. In the broader dataset, briefer visits to their religious institution were sometimes observed, and they often coincided with a related discussion in the submitted journal. However when the subject spent very long days at a religious institution for a number of days straight immediately after they were unblinded to the stimulation status of the device, there was no religious activity mentioned in any of the diaries that could explain the behavior. Given that the participant would regularly describe their day in detail, this was an interesting omission. 

Similarly, the participant in chapter \ref{ch:3} mentioned taking ADHD medications in their diaries on a regular basis with some periodicity, but started mentioning a different ADHD medication instead for a period in the middle of the study. A medication switch was not clear from clinical records we had available, but this sort of context is extremely relevant for a study involving a neurological intervention along with neural data collection. 

These examples all arose from just one case report, a case report that, as it turns out, focused on a dataset containing much less interesting variation than has been observed in a number of BLS subjects to date. This is true not only from the perspective of the richness of audio journal content, as can be seen from the results reported in the thesis, but also from the perspective of passive sensing data analyzed as part of other lab publications drawing from the study \citep{HabibGPS,HabibSleep}. Furthermore, the ability to estimate objective properties like vocal speech rate on a daily basis via diaries opens up numerous opportunities for deeper study of phenomena like psychomotor slowing, by creating a dataset of these objective speech features that achieves a temporal resolution actually well-suited for combination with the common passive sensing features of early digital psychiatry. 

Ultimately, there is practically an endless number of ways that a diary dataset could end up supplementing a passive sensing one, as journals introduce a flexibility into the subjective self-report data collected for each subject. It is not practical (and at some point scientifically unsound) to collect massive EMA survey responses that cover a wide range of potential labels of interest. Through audio diaries, and alternative filtering approach is taken, where the subject limits what is covered. Unexpected discoveries can still pop up in a way not truly possible for highly constrained study designs, while at the same time the scale of the diary data collected remains very tractable. \\

\paragraph{Audio journals can supply labels, not just inputs.}
As detailed, audio journals can provide inputs for modeling of EMA responses, clinical scores, diagnostic categories, and other common labels used in digital psychiatry. They can be an interesting sole source for input information, or they can synergize with inputs extracted from passive sensing signals. They can also provide interesting context for a case report utilizing such multimodal data sources, and can open up a unique additional avenue for feature validation. 

Beyond all that, I believe for certain scientific questions they can stand entirely alone, supplying both inputs and labels -- and for other questions they can serve as a primary source of assumed ground truth clinical labels to use with passive sensing derived inputs. As such, they are one of the most versatile digital psychiatry datatypes to date. However, this requires them to be used that way, which again falls into a similar dual threat trap. Most machine learning specialists will not have the clinical expertise necessary to frame audio journal content as a label in a psychiatrically motivated way and will prefer to avoid relatively labor intensive data processing approaches irrespective of scientific potential. Most psychiatrists will be hesitant to depart from existing scales and will not have an informed data science perspective on what audio journal labels might look like.  

While such labels would be far from perfect, I feel they have clear scientific uses and at times fairly unique ones. In reviewing examples from throughout this chapter, and in looking at journal content more broadly during the process of executing the research, I found numerous statements that were inarguably pathological, and many more that raised an eyebrow. In my opinion, a trained clinician (or perhaps even an RA) would be very capable of looking at a set of these journals and identifying those days that display a particular symptom of note with high probability. While a generic journal with no signs is not interpretable as a lack of symptoms, the positive signal inherent in those free form recordings where a patient chose to describe symptoms in detail would be highly valuable if it could be predicted, in my opinion. 

I would argue that the choice of what to discuss in a given journal is itself a relevant factor of current disease state for many patients. This is particularly true when that choice can be benchmarked against the patient's own baseline recording decisions from a long period of journal submissions. In some ways it might serve as a useful middle ground between the very challenging problem of predicting objectively severe rare events - like hospitalization from a manic episode - and the philosophically incremental problem of predicting biweekly clinical scale ratings or daily EMA symptom checklists. Regardless, this style of self-report enables symptoms to be captured that weren't even being looked for, whether that means nuance in distinguishing symptoms of a targeted disease beyond what EMA can realistically provide or discovering symptoms that were not previously on the radar for that patient. 

Given the many issues with current diagnostic labels in psychiatry, having some subset of research take this style of approach seems quite important. It allows for the potential upside of a fishing expedition, but introduces more principle in the sense that real patient experiences can drive the exploration. We need ways to try some things that are not just incremental, but we also can't exactly afford to aimlessly throw shit at the wall forever, with the resources required to build digital psychiatry datasets of even modest size. Outside of psychiatry Einstein magically appearing to cause an intellectual paradigm shift, it is not likely we can avoid this problem through only modern formal scientific mechanisms. Yet listening more to subjective patient experiences in a research context is a plausible way to bridge some of this gap.

One researcher expressed to me that psychiatric science is under a much greater pressure than other fields to avoid appearing unscientific, and so there is a modern fear surrounding certain styles of thinking and discovery. While this may have understandable roots, it is telling that the fear is centered around "appearing" rather than around "being", as is the extent to which the baby seems to have been thrown out with the bathwater. A lot is going to be incorrect when a field is new, so it is infeasible to attempt to enforce universal scientific standards; in order to get to a more established spot, it is critical that early failures are allowed to happen. Additionally, while theoretical philosophizing and instinctual reasoning are certainly not a sufficient methodology for advancing a field of research, I do suspect it is a necessary component that would synergize well with formal scientific processes today if it were no longer defacto stigmatized. \\

\noindent Nevertheless, there are extremely concrete examples of inherently clinical information that was contained in some BLS audio diaries, potential use cases which do not require a leap of faith in psychiatric definitions or an especially qualitative perspective. Take for example the following sentences from a journal submission by 8RC89:
\begin{quote}
    Um, my mother thinks I might be getting hypomanic and she might be right.
\end{quote}
\noindent If a patient chooses to say they think they are becoming manic in a diary recording, why would that be less trustworthy than a patient submitting a self-report survey response rating their level of mania symptoms as moderate? Why is that less of a clinical ground truth than the patient saying as much in a structured interview? The fact that it was not an explicit response to a question is a double-edged sword for the reasons discussed above, but it does not make sense to so heavily discount the evidence provided by someone choosing to report something like this largely unprompted. It is in many ways less likely to be a false positive, because they are speaking organically in what is experienced in the moment as a private personal expression. I don't see why such a statement should be considered unusable in a direct clinical sense for psychiatry research.

This sort of content is not too difficult to obtain or to track either. With a simple search for "manic" across the BLS journal transcript dataset, I identified 120 submissions across 16 subjects that explicitly discussed feeling manic. With more advanced computational methods, it would be possible to more exhaustively characterize mentions of mania over the study and in a participant-specific context while remaining strictly quantitative. The crucial point here is that this wasn't especially rare. Not all BLS participants experience mania at all, and not all BLS participants regularly submitted quality audio journals, so to find 16 distinct subjects is promising. Moreover, the average number of records using "manic" per participant worked out to $\sim 2.5\%$ of the average number of total records submitted per participant that engaged well with the audio journal collection longitudinally -- rare enough to be potentially salient but more common than manic episodes officially documented in the hospital's records.

Notably, the above quote gives further background about the participant and the role of their mother in both their life and in specifically managing their Bipolar disorder. The full diary contains a number of additional thoughts the patient chose to share that day, which can be extra interesting when taken together with their statement about hypomania. This exemplifies the many ways that diaries can be relevant. Fascinating clinical notes could probably be taken on the submissions of 8R without external information sources about the subject, and there are also very explicit mentions of psychopathology that can be directly counted. \\

\noindent The use of extracted features like sentence sentiment has the potential to identify diaries that can be evaluated in a clinically-established way as well. While considering automated sentence sentiment outputs as a form of ground truth may not be appropriate for many studies at present, a reading of the associated journals could certainly identify established signs of psychiatric pathology. The following sentences are just one example of the extreme statements found within many of GFNVM's lowest sentiment journal submissions, per the pipeline:
\begin{quote}
I hate my brain.

I hate my thoughts.
\end{quote}
\noindent If someone chooses to make these statements when asked to recount their last 24 hours, how is that not inherently clinically relevant? 

Related arguments can be made for a number of foundational passive sensing features; however, there is much room for nuance in the extent to which e.g. decreased physical activity or increased time at home is directly usable as a marker of psychiatric illness severity. Certainly these phenomena could be explained by external factors, and do occur at times in perfectly healthy individuals. While some such features have been found to correlate with e.g. depressive symptom ratings in psychiatric illness patients, and they do have direct links to overall health outcomes, it is not yet clear how they fit into models of psychiatric pathology specifically. This is quite distinct from the sort of statements made in the above quoted GF journal. It would be extremely difficult to find a presently mentally healthy individual, by any clinically accepted standards, who would record a daily free-form audio journal that amounts to a handful of such sentences. If submitted in good faith it is hard to imagine that making these statements would even be definitionally compatible with being psychiatric symptom free. \\

These examples of course do not address the big inverse limitation -- that even subjects who participate a lot may intentionally try to hide symptom severity in the content of what they say, or may have symptoms that cause them to submit journals with less information (or not at all) when more severe. In these cases, the journals (or lackthereof) can still be a powerful input information source, but cannot serve as a label in the way I am proposing. Still, trends over time and relationships between features in the patient daily diaries can be quite interesting without any clear clinical label available. 

To be clear, it is absolutely the case that for the field at large to build complete models of journal features in the context of psychiatry, some studies must obtain clinical ground truth labels in a more typical way that is indeed more robust presently (though also not perfect). Likewise, some studies must obtain e.g. neurobiological features in order for mechanistic models to improve. These facts do not mean that all studies must do so however. \\

\noindent In summary, audio journals are highly versatile in the value they can provide to different studies that address different psychiatric questions in different ways. They can simultaneously improve tractability of speech sampling work and enable passive sensing work to generate psychiatric insights that are outside the scope of the most established clinical structures. They can simultaneously be an interesting supplement to nearly any digital psychiatry data collection project and serve as a promising primary data source for hypothesis-driven work. They can even be fascinating as an entirely standalone dataset, to an interdisciplinary audience no less. They are severely underutilized currently in my opinion, not helped in this matter by systemic structures. As such, I will next discuss the role of software infrastructure development in enabling audio diaries, and really all digital psychiatry datatypes, to reach their full potential. 

\subsubsection{A need for quality software infrastructure}
A major theme of chapter \ref{ch:2} is the importance of good software infrastructure for organizing data, maintaining quality standards, and monitoring progress across groups in a large collaborative data collection project. This discussion occurs in the context of interview recordings and more specifically the infrastructure I developed for interview data collected in the AMPSCZ \citep{AMPSCZ} project, but the same points apply to audio journals. 

Given the lack of existing audio diary analysis literature, as well as the daily data flow expected of the format, good data management infrastructure is even more critical to develop for journal research, if the datatype is to leveraged as it should. While the code released with this chapter was written for processing of studies across the Baker Lab, it can serve as a foundational starting point for a more general open source audio diary processing pipeline that could serve the broader scientific community. Potential future software for the journals being collected but not yet processed by AMPSCZ is in fact covered in this chapter, in section \ref{subsubsec:diary-u24} below -- including a list of steps necessary to transform the diary code of this chapter into a central tool like the interview code of chapter \ref{ch:2}. 

Purely in the realm of data management and quality assurance, there are indeed a number of benefits conferred by using software like my pipeline, many of which were demonstrated on the single study scale in the process of collecting the BLS dataset (see section \ref{subsec:diary-val}). These benefits can directly lead to larger datasets of better quality, while simultaneously cutting down on wasteful expenses. Example functionalities of the pipeline are as follows:
\begin{itemize}
    \item Identifying participants that are intentionally submitting garbage recordings to game the compensation system.
    \item Automatically filtering out recordings of unacceptable quality prior to spending money on transcriptions or including truly unreliable features in analyses.
    \item Intervening quickly if subjects have a recurring technical issue with Beiwe (or other recording app).
    \item Noting common types of quality issues and updating participant instructions accordingly.
    \item Monitoring and comparing progress across multiple studies in parallel - of particular interest when managing data collection across multiple different locations like in AMPSCZ.
    \item Flagging suspect transcripts that may arrive to ensure TranscribeMe is maintaining their good quality standards over a long time period. 
\end{itemize}
\noindent Note also that due to the particularly sensitive nature of personal audio recordings, it is important that privacy and security standards are upheld when managing these data, another reason that well written core infrastructure for speech sampling research should be a major priority. 

Especially as more big data collection projects like AMPSCZ are funded across different disease areas, good centralized software for managing audio journal data would pay rich dividends. Such projects are great for building large datasets that open up additional analysis options not feasible for individual studies; imagine the interesting directions that could be pursued with a dataset of similar longitudinal depth as the 25 or so consistently participating BLS subjects, but constructed from 1000 similar subjects across 40 different international locations rather than 25 at 1. This would lead to a dataset with strong power for many more complex features to be utilized in within \emph{and} between subjects analyses. It would also enable interesting questions to be asked about generalization across cities, countries, and even languages, while as many data collection related factors as possible were held constant. 

However, the practical success of this style of project - and especially the efficiency of repeating similar processes for different disease areas - relies on having good infrastructure in place, for two distinct reasons. The first is that such broad exploratory work has amorphous aims and a variety of researchers that will hopefully one day take advantage of the dataset. Without a clear goal to carefully evaluate against, it becomes necessary to make the clear goal for the collected data to be of the highest quality in all regards possible, thereby allowing for a large number of analysis methods to be employed downstream without issue. The second reason is that in order to recruit many collaborating sites for multimodal psychiatry data collection work at scale, there will need to be many researchers involved in building the dataset who have no prior expertise in speech sampling. Similarly, there will likely be data collection occurring in places with different cultural and legal considerations surrounding speech sampling, not to mention the very obvious relevance of different languages to speech data in particular. Therefore in order to maximize participation as desired without forgoing quality, it is essential to have robust and user friendly software for data management and monitoring. 

Because audio diaries have the advantage of being a technically very similar datatype regardless of the disease area considered (more like wrist actigraphy than clinical interview recordings), well architected software would be extremely straightforward to extend to new projects. The question then becomes whether an appropriate initial investment is being made into ensuring such software is written, sooner rather than later. The fact that a single grad student working part time on the project has been responsible for the bulk of the interview recording software infrastructure for AMPSCZ to date, and a plan for the journals has not been fully settled yet almost a year into production data collection, certainly does not bode well. 

Unfortunately, this is only the tip of the iceberg. At this time, feature extraction software has not been implemented in any usable form for the AMPSCZ interviews. The arguments on software leading up to this point have focused on dataset management, which is a crucial component but is not sufficient for the project to run smoothly, especially because sharing of raw participant audio recordings is subject to understandably strict guidelines that exclude even most of the researchers involved in data collection and study design from accessing the larger set of them. There are additionally a number of methodological arguments for the availability of open source digital psychiatry processing pipelines that output core sets of validated features from raw inputs. It is unclear why the larger group has not yet come to a conclusion on some sort of base feature set, but it has made clear to me that the need for this sort of software design is much greater than I originally anticipated.  \\

\paragraph{Establishing common foundations for the next era of psychiatric speech sampling research.}
An indirect drawback of the rapid rate of progress in machine learning - and most notably here natural language processing - is the difficulty in building a cohesive knowledge base of rigorous applied research results while simultaneously not allowing applied research proposals to fall too far behind present technological capabilities without good reason. There are many factors in play contributing to this issue, including:
\begin{itemize}
    \item It is challenging for many interdisciplinary groups to properly assess when a new tool is well-suited for their dataset, because that often requires keeping up to date with bleeding edge computational research. At the same time, there is career pressure for many in this capacity to quickly move to the latest tools.
    \item Exploratory works are much less likely to be followed up on when they utilize methods that will rapidly be replaced and subsequently require their own exploration -- thereby destroying much of the possible value of these underpowered characterizations. 
    \item On the opposite end of the spectrum, many established psychiatry groups will not have the technical background from even a software perspective to run more recent methodologies. Therefore much traditional hypothesis-driven work still relies on only older methods with more user friendly software interfaces (e.g. LIWC). 
    \item With minimal established benchmarks for digital psychiatry, there is little consistency in what methods exploratory papers compare against, and there is not a good common language shared across the field that would allow works to build on each other in a sensible manner. This problem is further exacerbated when it comes to the critical task of communication that can bridge groups with clinical versus computational expertise; presently, their aims seem to fly past each other. It might relate as well to the relative paucity of pure machine learning experts attempting to wade into the digital psychiatry space, as opposed to other difficult but currently more tractable real world applied problems.
    \item Maintaining good practices for collaboration is not only an issue with analysis methods, but also with clear specification of data collection protocols. For example, structured clinical interviews and open ended interviews (discussed more in chapter \ref{ch:2}) certainly have shared properties of interest, but they still diverge in ways highly relevant to analysis too -- yet there are not many cohesive standards in the field for planning nor communicating data collection protocols around relevant data science considerations. This is again a major pain point when it comes to interdisciplinary communication, because psychiatrists do not often have the necessary technical background to identify seemingly small details that are important for software design or machine learning outcomes, while computer scientists do not often have the necessary clinical background to identify when they are abstracting away details of high psychiatric importance or focusing on features that are relatively unimportant. Without a solid framework for communicating on such matters, both groups will remain frustrated.
    \item Many of the most impressively accurate modern machine learning techniques are at this time difficult to interpret, make very different types of mistakes than classical data science techniques, and have greatly benefited from access to massive datasets for training and evaluation. Medical applications on the other hand have unique considerations in all of these domains, and for psychiatry in particular there is limited availability of clean labels. This is something I will discuss at greater length in the background for chapter \ref{ch:4}.
    \item A lot of the above points are compounded by the challenges with data sharing in psychiatry. Due to the sensitive nature of patient data and the lack of agreed upon benchmarks or even foundational feature extraction methods for deidentifying data, published results are generally in broad terms that do not allow future works to make comparisons such as identifying distributional differences in core input properties or included clinical labels. It is therefore especially challenging to build on exploratory results outside of the group that performed them, and groups that publish exploratory works are not always interested in pursuing this style of follow up work. 
\end{itemize}
\noindent As noted, many of these concerns apply to digital psychiatry at large. However, most of them are even worse for speech sampling than for other datatypes. Patient speech datasets are generally much smaller than passive sensing datasets, they are much more sensitive when it comes to data sharing considerations, and they involve multiple quite different modalities that warrant different expertises -- acoustics, linguistics, and clinically relevant patient subjective reporting. 

Of course shifting speech sampling data collection more towards the audio journal format would be a big step towards closing some of the interdisciplinary gap described, but we also very much need more established software tools to fully address concerns. We need these tools across digital psychiatry, but I am most concerned with audio journals in particular, because of their high potential upside and their surprising lack of usage to date. \\

The pipeline I documented in this chapter already lays some initial groundwork towards addressing such audio diary analysis concerns. This includes pilot feature validation work, careful characterization of output distributions in a relevant dataset, and an exhaustive detailing of ways the pipeline can presently be used and ways it might be built on in the future. Perhaps the most key component of the feature extraction code itself though is simply the choice (and justification) of specific core metrics that others can now more easily run on any audio journals they collect. My hope is that this will serve as a foundational light weight yet relatively modern feature set to bridge results from various future studies with distinct aims.

More importantly for the future of the field, the work in this chapter ought to be followed up on in spirit, whether intentionally or not. There is great need for accessible and well validated software that can provide a common baseline methodology across psychiatric speech sampling research in this increasingly digital era. If lead and maintained by a strong and committed software group, such a code base could become a place where the research community is able to contribute newer features they think have particular promise, and users in turn can help to further validate such features, thereby regularly expanding the common language of the core feature set. 

Though it is far from perfect, websites like PapersWithCode have made it easier for groups with different academic backgrounds from around the world to find papers that tackle a particular benchmark machine learning task, and to easily find information on the set of existing popular tasks that attempt to measure a particular type of learning challenge. Not all research does or should fit into such a box, but when a field attempting to utilize machine learning lacks any sort of organization like this it is natural, especially with modern academic incentives, that published research progress resembles a treadmill.  

Of course, many machine learning benchmarks can be run by following a simple Python package tutorial, while processing digital psychiatry datatypes in different ways often requires work from scratch. This is exactly the reason that a good open source code base is needed to not only establish baselines, but facilitate them for a variety of researchers. A properly maintained pipeline would greatly reduce the burden of integrating features that showed promise in prior exploratory works into new tests, to combat the problem of perpetual exploration. 

Along the same lines of the preliminary visualization functions provided in my pipeline, tools to assist in clinical evaluation and qualitative hypothesis generation can be part of such an open source project as well, helping computationally-inclined researchers to learn more about what clinicians are looking at. Furthermore, when integrating such a pipeline with data collection apps like Beiwe or MindLAMP, collection procedures could be more standardized by default, so that groups will gain a common understanding on important data considerations both clinical and technical. If one does not have a good reason to reinvent the wheel, they shouldn't.

Were such a pipeline to come to fruition, it would also open up the possibility of sharing basic distributional properties across the research community without sharing anything of a sensitive nature or anything that might be considered proprietary by the research group. As state of the art features are improved upon, the core feature set can still remain of strong scientific relevance, as demonstrated throughout this chapter. All research could indeed be improved by sharing a more detailed understanding of such core features, in parallel with development of increasingly complex algorithms on customized datasets. \\

\noindent Developing such software well in the modern academic system will likely carry a high upfront financial cost. However it would pay great dividends to do correctly now, both in terms of true scientific progress and in terms of the cost of data collection of future related studies. These costs are discussed at more length in the context of the large collaborative AMPSCZ data collection project within chapter \ref{ch:2}. I also review the state of academic software in broader terms in the concluding chapter, along with potential social factors that might be addressable independently of funding available for programming contributions.

Ultimately, if psychiatry wants to really take advantage of technology, it cannot refuse the cost of software development, whether that is paid in terms of a restructuring of the academic credit model or in terms of cold hard cash. You get what you pay for, and bad software requires more maintenance and more monitoring, and it is more difficult to adapt it to new situations or add new features. It is also more difficult for others to use it even if they don't attempt to change anything. A lack of good software now will cause research undertakings to bleed money later, if there is any true belief in a digital future of psychiatry. Then again, a lack of commitment might turn into a self-fulfilling prophecy in the upcoming decade, because it will be challenging to overcome all of the systemic problems mentioned without one.

\subsubsection{Leveraging the broad relevance of human story telling}
Regardless of how the current brand of interdisciplinary digital psychiatry plays out, it is extremely probable that at some point in the future, some form of digital psychiatry produces truly important results. I expect a frequently collected autobiographical format like the daily audio diary to play a role in such successes, and in particular to help actually bridge the interdisciplinary gap. Journals can be very useful for both "sides" for all the reasons detailed above. There are also interesting philosophical links between journaling and fundamental properties of human cognition - some of which were instrumental in the early days of AI research. \\

\paragraph{Story understanding in models of human intelligence.}
Many believe that story understanding is a key component to human intelligence and communication. A key hypothesis of \cite{Winston} was that:
\begin{quote}
The mechanisms that enable humans to tell, understand, and recombine stories separate human intelligence from that of other primates
\end{quote}
\noindent and that these mechanisms ought to be a source of inspiration for making advances in AI. 

As \cite{Winston} points out, education is full of stories throughout development, from moralizing children's tales to the case reports studied in law and medical school. Indeed, the salient things that people remember and later utilize are often story based. You will get most people angrier with a well-told story of one injustice than statistics about thousands of them. Symbolic reasoning, a pervasive theme in early AI results, can be closely linked with story composition and with the human propensity for creating art \citep{Winston}.

Based on evidence that a variety of types of arts emerged rather suddenly and subsequently spread rapidly $\sim 60,000$ years ago, paleoanthropologists have hypothesized that a large nonlinear effect on intellectual capabilities arose when humans became symbolic, setting them apart from prior humans with close structural similarities as well as other primates like Neanderthals \citep{Tattersall}. The ability for humans to efficiently share knowledge through symbols and to generalize via their composition, sharing increasingly complex information, enabled a social evolution of sorts perhaps -- spread through many forms of story telling.  \\

\paragraph{Story telling perspectives within scientific practice.}
In some ways, the human story telling method of learning and teaching can lead to irrational "normal" behavior, and subsequent conclusions that are wholly unscientific. At the same time, scientific results can be reported effectively in a very dry way that would be a stretch to call a story. Still, that does not mean that the story telling perspective should be avoided in science, as it sometimes seems. Human cognitive biases are not universally bad, and can in fact be a source of great insight at times; a source that might become increasingly important to take advantage of as "by the book" thinking can be increasingly automated. They can also be leveraged to improve communication of science, a requirement for good ideas and strong results to actually become used. 

This is especially true for interdisciplinary communication, because as much as scientists may want to believe they are substantially more rational than the general public (and maybe they are), it remains difficult to fight human instincts in an alien field. In 1963, Edward Lorenz published "Deterministic non-periodic flow" to a modest response restricted to his own field of meteorology. With minimal addition to the underlying mathematical principles, his work took off to become an entire interdisciplinary field of study - chaos theory - only after he told a silly story about it at the 1972 meeting of the American Association for the Advancement of Science, at the urging of a colleague. The title of that talk of course was "Does the flap of a butterfly's wings in Brazil set off a tornado in Texas?" \citep{Lorenz1963,Lorenz1972}. 

The story of Edward Lorenz was originally told to me as an example of the irrational memetic behavior of scientists that apparently transcends generations. It is silly in many respects, and it can lead to some legitimately bad outcomes when bad science is spread in a similar fashion. Yet at the same time, it is not unreasonable to me that the work didn't spread until the story came along. The story made the central theme of the work clear -- that there exist deterministic systems for which very small perturbations we cannot reliably measure are capable of producing largely different and thus seemingly random outcomes. It made it clear in a way that would be memorable for the rest of someone's career, in case something relevant ever came up. It also made it clear in a way that could be immediately paralleled (in abstract terms) to many other possible scenarios across a range of fields, regardless of one's prior exposure to nonlinear stability theory, which is a great way to inform hypothesis generation.

The audio journal datatype can tap into this style of field-transcending communication, because it provides source material both for detailed quantitative analyses and for compelling story telling about the results of such analyses in subject(s) of interest. Journals can in fact generate a huge variety of stories because they are a series of stories themselves, which in my opinion has important implications for the future of hypothesis generation in psychiatry. Moreover, the relevance of story telling and story understanding in human intelligence unsurprisingly links back to some psychiatric disease states, making the study of patient driven open ended self-reporting of direct (and not just meta) relevance on this topic. \\

\paragraph{Story recall abnormalities in psychotic disorders.}
Given the hypothesized uniqueness of story telling to human intelligence, Schizophrenia may be of particular note for its potential uniquely human angle. To our knowledge, Schizophrenia is not found in other animals, and in fact many of the genes implicated in the disorder are found in human accelerated regions, the parts of the genome likely associated with human evolution \citep{Insel2010SCZ,Xu2015}. If stories are an important part of human nature, and Schizophrenia manifests uniquely in humans, then exploring story telling in Schizophrenia would be a research direction of broad interest. As can be seen throughout this chapter, many audio journals recorded by psychotic disorders patients do take the form of a story, and thus may capture interesting properties of patient story telling over time in an exploratory fashion.  

Delusions found in Schizophrenia patients are very often self referential, and commonly contain themes of persecution and thought reading and control \citep{Gold2000}. As such, autobiographical story telling like what is captured by audio diaries could be especially interesting. Moreover, general story understanding deficits have also been found in Schizophrenia, as patients will rate unlikely explanations as more likely than controls do, and will have a harder time dismissing contradicted hypotheses than controls do, when presented with basic stories in research tasks \citep{Moritz2006}. 

Memory of stories also appears to be impaired in Schizophrenia patients, as demonstrated by performance on story recall tasks where details were poorly remembered and concepts were incorrectly mixed across tasks \citep{Matsui2007}. \cite{Hoffman2011} hypothesizes hyperlearning as a key mechanism underlying the story understanding and recall deficits and ultimately the psychotic delusions observed in Schizophrenia patients, based in part on biological evidence of hippocampal over-activation and over-release of dopamine in Schizophrenia, both of which are crucial to learning and reinforcement. Even though patients may not remember many details, the details they do learn may be extended across different stories that should be kept distinct. Hyperlearning in the context of an artificial neural network architecture indeed resulted in the over application of learned concepts in their computational experiments to test this hypothesis \citep{Hoffman2011}. 

As diaries enable extended longitudinal data collection, one might consider an analysis of events that are repeatedly mentioned or of how recurring concepts are used by the participant over the duration of the study. A participant-directed autobiographical datatype that can be feasibly obtained relatively frequently for multiple years lends itself to untold potential subjective discoveries about a person, while at the same time presenting opportunities to be connected with a number of potential mechanistic studies in cognitive science and neurobiology, especially when collected in conjunction with other data modalities. 

\subsection{Summary of pipeline release}
\label{subsec:diary-code-recap}
Given the salience of the audio journal format and its lack of attention to date, coupled with the general need for common foundational infrastructure in digital psychiatry, there is great potential impact for the development of comprehensive audio journal processing software. In this chapter, I took a large first step towards that end, by both documenting my diary pipeline (section \ref{sec:tool2}) and reporting on early scientific results found through the use of the pipeline's outputs (section \ref{sec:science2}). In this discussion section, I will review key takeaways on the capabilities of the pipeline for assisting an audio diary study, from its role in the data collection process through to final statistical analyses. At a high level, the primary functionalities of the code are represented in the following list.
\begin{itemize}
    \item Data management:
    \begin{itemize}
        \item Recognize newly available recordings as they come in, handling decryption of files where needed and ensuring sensitive files do not remain on the server unencrypted for an extended period.
        \item Parse Beiwe and subject metadata to assign audio files to the correct study day (real dates removed) with submission timestamp in correct timezone.
        \item Transfer audio files that satisfy transcription settings to TranscribeMe.
        \item Track pending transcriptions and pull back completed transcripts when done, converting to CSV format.
        \item Maintain standardized folder structure and file naming conventions for all outputs.
        \item Send monitoring emails to study staff with information on newly processed (audio and transcript) files per subject ID upon each run.
    \end{itemize}
    \item Quality control:
    \begin{itemize}
        \item Compute lightweight quality measures for all recognized audio diaries, with the option to make transcription decisions based on these values.
        \item Analogously compute basic transcript quality measures on all returned transcripts - introductions to both audio and transcript QC can be found in section \ref{subsubsec:diary-val-qc} above.
        \item Send monitoring emails to study staff to notify on relevant QC metrics for newly processed diaries, also optionally integrating with DPDash to easily monitor QC values over time via web app as desired. 
    \end{itemize}
    \item Acoustics feature extraction:
    \begin{itemize}
        \item Run OpenSMILE to obtain low level descriptor acoustic features from the GeMAPS config.
        \item Run voice activity detection algorithm with Librosa, to subsequently label pause times in the recording and derive diary-level features from that.
        \item Use pause times to create filtered copies of the low level OpenSMILE features, also computing diary-level summary stats from OpenSMILE outputs.
    \end{itemize}
    \item Linguistics feature extraction:
    \begin{itemize}
        \item Leverage TranscribeMe notation to count occurrences of disfluency by category in the transcript via simple regular expressions.
        \item Run VADER sentiment on each transcript sentence to calculate per diary sentiment summary statistics.
        \item Employ the Google News word2vec embedding model to compute measures of incoherence (within sentence and between) as well as word uncommonness.
        \item Retrieve basic QC data of potential relevance such as word and sentence counts, optionally add counting of specific keywords.
        \item Count syllables in each sentence using NLTK, combine with TranscribeMe sentence timestamps to generate an estimate of speaking rate from transcription alone. 
    \end{itemize}
    \item Visualizations to summarize features:
    \begin{itemize}
        \item Create up to date histograms and correlation matrices showing the latest properties of select diary level features across the dataset.
        \item Create heatmaps showing progression of select features over time in each subject along with clear picture of data missingness.
        \item Create spectrogram images and sentiment colored word clouds for each individual diary - examples of these and all other visualization functions built into the pipeline by default can be found in section \ref{subsec:diary-outputs} above.
    \end{itemize}
\end{itemize}
\noindent Note that a major advantage of using the pipeline for feature extraction is not only that it packages together a number of tools of interest, but also that the selection of these tools can serve as a common core of features to be benchmarked against as new methodologies are developed. Within section \ref{subsec:diary-val}, I provided a review of prior validation literature for each applicable tool, and justified its inclusion in the pipeline over other choices. The features included are meant to be feasible for any lab to run and approachable for interpretation, while still maintaining a fairly high quality standard. Thus the pipeline can simultaneously be used by psychiatry groups with minimal technical background as the primary resource for evaluating a number of scientific questions and be used by hybrid groups with a machine learning focus as a comparison point for new techniques -- thereby helping to reduce the large amount of unorganized and at times unscientific churn that currently happens with tech exploration in the digital psychiatry literature.

Perhaps more importantly, I have provided preliminary validation results for a number of the included features on our large Bipolar Longitudinal Study (BLS) journal recording dataset, along with a nuanced characterization of the distributions of and relationships between the most key features derived from the pipeline, across BLS and in terms of participant-dependent variability. Because the entire BLS data collection process would be replicable by using the Beiwe app for diary recording and TranscribeMe for high quality professional transcriptions, these results are directly relevant to a number of researchers that may be interested in utilizing audio journals in the future. Additionally, data monitoring and quality assurance pipeline capabilities can improve the final dataset collected over what would be obtained from using Beiwe alone. 

To facilitate future pipeline applications, I will next review the results of our work validating the pipeline features (\ref{subsubsec:diary-val-recap}), and then I will summarize and discuss the feature characterizations and other scientific results obtained thus far from the BLS pipeline outputs (\ref{subsec:diary-science-recap}). Software implementation details will not be included in these discussions, so please see supplemental section \ref{subsec:diary-code} for exhaustive code documentation if needed.

\subsubsection{Validation overview}
\label{subsubsec:diary-val-recap}
In section \ref{subsec:diary-val}, I explained how the major code design decisions were made and reported on our confidence level in various pipeline outputs based on experiences with our lab data to date. \ref{subsubsec:diary-val-qc} covered the quality control metrics, \ref{subsubsec:diary-val-aud} covered the acoustic extracted features, and \ref{subsubsec:diary-val-trans} covered the linguistic extracted features. Where applicable, I reviewed prior literature on the accuracy and relevance of each tool chosen to be integrated into the pipeline, including comparison to potential alternatives. 

I also reported on how well different pipeline outputs aligned with the results of manual review performed by lab members, as well as sanity checking of expected relationships between certain features. In summary, the QC tools provided by the pipeline are not perfect, but are a large improvement over the previously non-existent options for active monitoring of audio journal study submissions. They repeatedly demonstrated the ability to improve our final BLS dataset and save project money in practice, by filtering out bad uploads prior to issuing to TranscribeMe and detecting participants regularly submitting low quality diaries (whether accidental or gaming the compensation). 

For statistical analyses, there was strong evidence for the continued use of linguistic disfluency, semantic sentiment, and pause detection features, as well as the various qualitative functions provided by the pipeline to assist in interpretation of journal content. These features were internally consistent with expectations across the sizeable diary feature set, in addition to passing pilot manual review checks for agreement with human-provided ratings. Furthermore, the potential clinical relevance of basic verbosity and diary submission metadata features should not be underrated due to their technical simplicity. Ultimately, the scientific results to be recapped shortly are another very strong testament to the quality of the reviewed pipeline features. There is certainly room for expansion and improvement, but the aim to establish a solid basline has been accomplished. 

On the other hand, while the semantic incoherence NLP features and vocal properties extracted from OpenSMILE have been well validated on technical benchmarks in the past, they were more difficult to assess in our hands. Semantic incoherence outputs certainly showed promise, but were also particularly susceptible to some sources of noise, and were difficult to even define expectations for given the general lack of severe psychosis symptoms seen in the BLS dataset. OpenSMILE outputs were not closely investigated in this thesis, as technical details of acoustic signal processing were beyond the scope and the high number of features coupled with their high temporal resolution would make a data-based evaluation a more involved project. Given that GeMAPS OpenSMILE features have been used by many other studies, there should be no problem in using the low level pipeline outputs in a future study, but it is less clear how much value the summary features carry. Regardless, the primary interest with speech sampling here is in linguistic analyses, which will again be a major theme with the interview recordings in chapter \ref{ch:2}. \\

\noindent Limitations of the pipeline at present are covered in more depth in the documentation of supplemental section \ref{subsec:diary-code}, but for the most part the pilot validation evidence for the included features and other functionalities was strong, providing further support for the use of my code in future works. Still, any study using any open source code should of course perform their own confirmation that measures they intend to use are working as expected. Outside of technical setup discrepancies between research groups and potential undetected bugs in the code, there very well could be certain study populations or specific psychiatry research questions that the features as is would be especially poorly suited for. Given the simplicity of most included features, their utility should for the most part generalize -- which is an important additional advantage to this code. Still, some underlying messages remain the same through much of this chapter, including how crucial it is to check for unexpected patterns (whether good or bad) within the details of any research dataset

\subsection{Pilot scientific results recap}
\label{subsec:diary-science-recap}
One major contribution of my work was simply characterizing the primary feature set in a large dataset of journals from patients with psychotic and mood disorders, with a focus on Bipolar disorder. The dataset size in itself was a powerful argument for audio journal speech sampling, with over 10,000 high quality transcripts obtained from across 66 consenting BLS participants. 25 of the subjects submitted over 100 journals with a length above 15 seconds during their time in the study, and some subjects submitted hundreds of journals that were well over 2 minutes and/or continued to submit journals well beyond 2 years of enrollment. As reported along with other participation dynamics in section \ref{subsubsec:diary-submission}, within the first year of enrollment for those 25 participants, there were 2868 data points that had both same day and next day quality transcribed audio of duration $\geq 15$ seconds -- so even for tight timescale temporal prediction modeling we obtained a healthy dataset size. Engagement (as measure by word count) did not drop off over the course of the first 2 years of submitted diaries in the main participant set either. This bodes very well for the feasibility of collecting large and rich journal datasets across people and time, and it was clear from recorded content that some subjects truly enjoyed recording the journals.

Scientifically, the distributions (\ref{subsubsec:diary-pt-dists-comps}) and correlation structure (\ref{subsubsec:diary-corrs}) of features produced can inform future works on feature selection and model design process, and can ultimately contextualize features that are obtained by other groups and from different study populations. Importantly, there were highly significant distributional differences in select features across all subjects highlighted -- including large differences in features like diary word count and mean words per sentence that can impact distributions of many other extracted metrics, and thereby downstream modeling. At the same time, there were also significant distributional differences in downstream features like disfluency usage or sentence sentiment that were not clearly attributable to differences in the considered verbosity measures. These results highlight just how important it is to carefully consider within versus between subject effects. While a large scale study like AMPSCZ may produce a diary dataset of sufficient size to investigate between subject effects clinically, it is difficult to say much about this with the BLS diary set. Not only was the total number of engaged patients relatively small, but also there was a great deal of diversity in diagnosis and other demographics amongst them. Many feature differences between subjects were themselves quite stark and convincingly real, in the sense that the participants are truly different in how their journaling speech affects even relatively simple features over time. 

However, there is no way at all to confidently comment on clinical relevance of any between subjects differences in the present work. This was further exacerbated by the lack of control subjects, though there was good heterogeneity of clinical presentation amongst the considered BLS participants. Differences in pause patterns or use of uncommon words or of linguistic disfluency rates (and so on) can occur between individuals for reasons unrelated to pathology, and a large number of subjects would be needed to distinguish. Still, the distributional differences found are critical for a number of reasons. Subject-dependent feature variation might help to identify those metrics that are most salient for each participant's within-subject modeling, before invoking any ground truth labels -- though this remains an open question and is probably highly dependent on the feature and the difference. Additionally, relationships between features and pathology within a patient need not have the same direction of effect, something else it may be possible to glean information about from patients' individual distributions. Perhaps most importantly though are the benefits for planning and interpreting models. The shape of the input feature distributions will inform parameter choices, as will the total amount of variance. Correlation between different input features is important to characterize as well to facilitate explainability or possibly reduce the dimensions of input features in an unsupervised manner. This too demonstrated subject-dependent differences in our pilot results, at times in opposite directions. \\

\noindent A second major contribution of my work was to model EMA scores from same day diary features in a smaller subset of participants (to preserve the dataset for others' future analyses), thereby identifying features with grounding in self-reported mood on the diary timescale, both overall and for different individuals (section \ref{subsec:diary-ema}). While ultimately simple linear modeling work, this again highlighted a number of important takeaways for future diary work. Additionally, the strong overlap found between diary and EMA response rates in patients engaging with diaries emphasized just how tractable app-based journal modeling can be, even in such a small subset of patients.

The many highly participant-specific effects mentioned above, which extended to patterns in EMA labels of wildly different periodicities and typical self-report severities, demonstrated how important it is that digital psychiatry works properly account for many distinct and interacting explanations in their modeling, or at least show the data carefully enough such that the community can properly judge the origins of significant correlations. For a simple example from this chapter, a linear fit on diary word count alone was able to explain more than $15\%$ of the variance in same-day positive EMA summary score in the selected subset of BLS participants, with off the charts significance. Yet when word count was first normalized to the patient's own baseline via z-score, the relationship went entirely away, dropping to barely $1\%$ variance explained and no significance. It turned out that word count was very different between subjects, as was the distribution of EMA summary scores. The fact that this moderately correlated in such a small total population ($n=7$) does not say much about a real \emph{in the clinical sense} relationship, and it does not really matter how many diaries each of those individuals submitted. 

Conversely, diary word count did have an impressively real clinical relationship for one particular subject (5BT65), with a linear fit to word count alone explaining $24-33\%$ of the variance in their longitudinal EMA summary features and persisting in hold out test set as well checking out with the eye test. Further, it was likely from visualization that a nonlinear and/or rolling beta model would have even better captured their EMA variance. Looking at longer timescale diary feature fluctuations, the correspondence of word count with depressive episode periods in this participant was remarkably strong. Within the same dataset however, most subjects had minimal relationship between EMA summary and same-day journal word count. 

Interestingly, one subject (8RC89) had a modest and highly significant linear relationship between EMA and word count, but with higher word count corresponding to more severe self-reported negative EMA symptoms, thereby contrasting with the participant having a highly significant connection between fewer words and more severe self-reported symptoms. This may be explainable by differences in their pathology, as 5B experienced primarily severe depressive episodes, while 8R experienced clear manic episodes at times, and may have been inclined to record longer diaries when e.g. irritable as opposed to a depressed person that may record shorter diaries when they are lacking energy. Indeed looking at feature timecourses in 8R, strong self-reported emotions of both positive and negative valence coincided well with clinical markers for hypomania symptoms. Regardless, it is telling that even in this very pilot exploration examples were easily found that would benefit from hierarchical modeling that includes personalized components for different participants. 

Other diary features also contributed significantly to explaining variance in EMA summary scores, with sentiment in particular sharing significance (and unsurprisingly directionality) in relationships with EMA across subjects, even when normalizing features independently within each subject via z-score. This provides strong validation for the use of the VADER sentiment tool in the pipeline. Pause and disfluency-related features demonstrated relevance within some particular subjects as well, and may be even better utilized by a nonlinear model based on properties of their underlying distributions. The features that the scientific analyses showed the least overall promise for were the word2vec-derived ones, though it is worth noting that I focused almost exclusively on participants with heavily mood-related symptoms, so the pipeline incoherence features remain to be properly tested on cases with clear positive psychotic symptoms. \\

The final major scientific impact of this chapter is through the proof-of-concept case reports of section \ref{subsec:diary-case-study}, not because of the results directly, but rather because of the simple ideas they consistently demonstrated about the relevance of multi-timescale dynamics and the power of even the simplest qualitative tools to utilize diary content for the improvement of quantitative understanding. Examples include:
\begin{itemize}
    \item Using word clouds generated over periods of unexpected diary feature changes in 3S to identify topics of relevance - it turned out that a transient disruption to a potential relationship with linguistic disfluencies could be matched with a series of diaries that discussed a temporary attempt to cut back on medication as well as its subsequent failure over an $\sim 1$ week period. 
    \item Automatically accounting the usage rates over time for clinically-relevant words like mentions of mania in the 8R diaries. By narrowing them down in this way, I found 2 entries with compelling statements related to active mania, and when adding those points to the markers of high clinical symptoms, trends in diary and EMA features potentially related to mania became much more in line with expectations. 
    \item Characterizing variance directly at the different timescales and in the different (relevant) datatypes in which it occurred in the 5BT65 dataset, not only identifying striking examples for the utility of EMA in future psychiatry research, but also finding new compelling relationships between diary content and onset of the major mood episodes in 5B. Specific phrasing was found to be highly specific to only the time periods directly before long-term mood episodes identified in 5B, and could flag these days before reliably detectable changes in EMA (in a highly patient-specific manner of course). 
\end{itemize}
\noindent The overarching theme across these results is to think carefully about different modeling timescales that could capture different levels of disease-relevant variation, and to take advantage of the iterative process that diaries can enable between content search and statistical analyses.

\subsection{Future directions}
\label{subsec:diary-future}
Throughout the background, validation work, results, and now discussion, I've proposed a variety of broad ideas on ways in which diaries could be used for future digital psychiatry research, as well as suggested a number of specific improvements that could be made directly to the current core pipeline workflow I've described. In this section, I will go into more depth about future applications and new features of particular interest. Broadly, the audio diary pipeline could be directly applied to a wide variety of clinical (or other behavioral) contexts, as free-form journaling prompts can be of interest across a range of disorders and short daily recordings on specific prompts have a wide set of options and use cases that would fit with the current code too. This might include the more traditional clinical use of journaling as a therapeutic intervention. \\

One major direction suggested by these pilot results is to focus more on fully individualized modeling using these longitudinal data, but considering a larger sample of participants in which the goal is to have a procedure that \emph{meta-generalizes}. In other words, it is not a prerequisite that any particular relationship is true for a larger set of people, as long as the models generalize well enough over time in the respective subjects and the method for finding those models generalizes to many others. If it indeed would be possible to identify relevant features in an individualized fashion from a larger base feature set using rolling models, one could imagine making patient-specific predictions in real time at relevant points over the course of a study and seeing how accurate they turn out to be. In practice, clinicians would (ideally) update their ideas about a patient over time and may make predictions about e.g. what subtle signs indicate a particular patient might be trending in a particular direction. The fact that research tends to have a single stop point with analysis at the end is a side effect of the very specific model for scientific results publication that is used system-wide. In that model, it is indeed poor practice to check many hypotheses over time, because targeted stopping is an easy way to p-hack. But there are plenty of other ways to quantify accuracy of continual predictions -- one might look to literature from the online learning subfield of ML or to financial market model testing practices. Of course, these are largely new problems in the psychiatry research space, because up until recently such longitudinal work was hardly feasible. Unfortunately, analyses of digital methods to date have concentrated too much in the psychiatry research style of evaluation; with the paucity of serious audio journal work, there is hardly anything like this at all in the speech sampling space especially. A greater diversity of methodologies to cover the limitations of each other will be critical to future success. 

Of course, if it were possible to learn the behaviors of many participants with a similar overarching method, that could then go a long way in early intervention studies, clinical trial evaluations, and so on. Less obviously (but still critical), it would also be very useful for neurobiology research. A very high level major open question is the extent to which patients with clearly different symptom manifestations in the same disorder are classified incorrectly versus a mechanistically similar underlying pathology causing different behavioral results for non-pathological personal reasons. While it is highly likely that there is quite a bit of bad classification with current disease labels, it is also highly plausible that the same biological mechanism could cause vastly different responses when other non-clinical biology or the outside environment are drastically different. Using individualized longitudinal models, we could have behavioral signal that is both much more appropriate for neural data timescales than traditional scales and much more clinically grounded than a generalized model based on current disease labels could hope to be. This would enable comparing neurobiological signals at times of different model output, regardless of what behavioral features most contributed, to look for shared mechanisms across individuals. 

For comparisons across individuals, it will be important for some future diary work to include control subjects, and to overall scale up the number of subjects included. Because BLS included regular ratings of a battery of clinical scales, MRI scans, wrist watch data collection swaps, etc. there was a feasibility limit imposed on the study population size that was much smaller than what could occur with diaries alone. While clinical grounding and multimodal data sources are also important directions, another style of study could simply collect diaries and self-report surveys (that draw from a validated scale rather than a broader swath of bespoke questions) in as large of a population as possible with some diagnosis of interest available, along with a set of controls. This would allow comparisons of temporal dynamics by group to occur in much greater detail, and could provide a bridge between currently popular large population snapshot studies and the hopefully emerging personalized longitudinal modeling work along the lines of what could be achieved with BLS. \\

A clear more immediate next step of this work would be to continue with the present BLS dataset. It would be ideal to do a deeper dive on variations of ways to clean and summarize the core sentence level feature set produced by the code (and work in additional features like OpenSMILE outputs), including a better understanding of how the features relate to each other within the sentence structure of a diary. It would also be very beneficial to develop methods so that features might have temporal trends quantified in different patients at different timescales, including testing more simple summary techniques like peak finding algorithms or auto-correlations. Given the large amount of participants held out entirely from the EMA modeling and other closer comparisons I did, there are of course many ways the existing work could be extended to the rest of the set, which would enable more formal modeling paradigms that consider both individual and overall fits, in addition to capturing nonlinear relationships that weren't fully fit by the linear modeling in this chapter.

Once initial hypotheses are tested as desired, the BLS dataset could then be used for further exploratory works on entirely new features of potential interest, most notably content-related ones. By developing personalized features derived from topic model style algorithms, as well as performing more qualitative exploration to come up with longer term ideas for content-based feature generation methods, early testing on a variety of ideas could occur, setting up hypotheses for the next lab diary dataset in psychotic and/or mood disorders. Diary content could even be directly compared across different days via modern methods like transformers, something uniquely enabled by this format for patient speech sampling. 

Similarly, predictions of future diary features from prior ones and characterization of how feature relationships themselves can change over time in an individual will be quite important topics to eventually explore that BLS could provide a foundation for. Alongside that, there are the many other modalities of data that were collected during BLS and might prove interesting to cross-reference with diary-related discoveries. Ultimately, digital psychiatry is still such an early field and audio diaries are still such a new format that curious exploration by some will be instrumental to formalizing some research processes. Because of the strong size and duration of the BLS journal set and the already well-documented relevant trends observed when selecting only a handful of individuals to look at, it is likely that this could be a very good source for idea generation and method fine tuning as new projects collect data in parallel. \\

Finally, though there are a huge number of possible benefits of audio journals for psychiatric speech sampling, they are of course not a silver bullet. Studies focusing on other data formats like the clinical interview should of course continue as well, and we should aim to understand how extracted features from these different forms might compare/contrast. The most crucial part is that studies are designed with clear justifications for the data collection and analysis choices they will make - all of which have limitation - in terms of the specific parts of the overall psychiatry search space they wish to tackle. As technology becomes ever more pervasive, digital methods will probably be difficult to entirely separate from psychiatry research eventually, but the specific digital methods employed and their varying design details can have a very wide range to consider many possibilities as appropriate. Thus this highlights the value of early works that lay out many of these details, like what I've done here.

Related characterization work could obviously go much further: not just in considering different disease populations and control datasets, but also to cover many other data collection nuances. It is an open question what sort of participation rates would be achievable in a more severe psychiatric study pool (though note the case report participant of chapter \ref{ch:3} submitted journals regularly despite severe treatment resistant symptoms), or how demographic and cultural factors might impact diary behaviors. Besides having a large impact on possible scientific value of a given journal dataset, these issues can make it difficult to budget for a study in advance. However, an upside is that the cost of the audio journal study is largely dependent on the participation rates, so as long as poor quality submissions are rejected as the pipeline can do, the cost will be proportional to the completeness of the final dataset. Regardless, it would be very beneficial for the research community to build up a shared knowledge base about diary engagement tendencies in different patient populations, as well as to address questions about how schemes to increase participation mid-study may or may not bias results. The latter question connects to broader technical considerations on handling missingness that remain to be worked out, both because it can be a difficult machine learning problem in general and because in this case missingness may itself be inherently meaningful.

By coupling such works with extensions to feature characterization like described above, future digital psychiatry research would be very well positioned to employ audio journals across those settings where they are likely to be salient, and to do so in a way most likely to address their specific aims. A key piece of building up this knowledge (and creating mew tools for study efficiency like domain-adapted and validated automated transcription tools) will be larger scale collaborative data collection projects, which can facilitate data sharing and subsequently knowledge sharing much better than individual lab projects can. One initiative of immediate relevance then is the AMPSCZ project, as it includes collection of an audio journal datatype and fortunately now has plans to make real use of that datatype.

\subsubsection{A note on adapting the pipeline to the AMPSCZ project}
\label{subsubsec:diary-u24}
The NIMH's Accelerating Medicines Partnership Program - Schizophrenia (AMPSCZ) project is a multimodal big data collection endeavour spanning many research locations internationally, with a focus on young people at clinical high risk for developing Schizophrenia \citep{AMPSCZ}. This project is a major focus of chapter \ref{ch:2}, where I will document the data flow and quality assurance pipeline I wrote as infrastructure for the interview recording portion of AMPSCZ. Meanwhile, audio journals are also being collected across many participants, and are discussed at greater length in a scientific planning sense in Appendix \ref{cha:append-ampscz-rant}.

As of mid-February 2023, just over 1000 diaries had been successfully submitted by AMPSCZ subjects from across 19 locations. The project is currently in early stages, with only $\sim 2\%$ of planned interviews for the lifetime of AMPSCZ recorded yet (chapter \ref{ch:2}) -- suggesting that the journal dataset is on pace to reach an impressive size of 50,000 recordings. Despite the large amount of data collected to date, there remains no pipeline for data flow or quality monitoring of the diaries, let alone a plan for feature extraction for the project. Only much more recently was a decision made about getting the diaries transcribed, something that resulted from arguments similar to those I've made throughout this chapter along with arguments built on early interview dataset observations (to be covered in chapter \ref{ch:2}). 

This situation only further goes to show that audio journals are a datatype with great scientific value, and moreover a datatype that has been largely neglected and certainly underrated by recent research. We do not yet have any information about the quality of the $> 1000$ recordings collected thus far; not even their typical length nor much about submission trends over subjects and time. There is minimal clarity on leadership for the journal datatype in AMPSCZ, with it awkwardly falling between the self report survey collection and the audio recording collection. 

However, given the recent budget shift (resulting from the arguments of Appendix \ref{cha:append-ampscz-rant}) to enable gold standard transcripts to be obtained for the AMPSCZ diaries, there is certainly hope for the future of the format, as well as this initiative. The presented arguments could also assist in determining to what extent budgeting for high quality professional transcription for a particular speech sampling research project makes sense. \\

One future direction for the audio diary pipeline presented in this chapter is therefore adaptation to the AMPSCZ project. While tips for adapting the code to other probable study contexts were provided in supplemental section \ref{subsubsec:diary-adapt}, those were focused on usage of the pipeline by individual lab groups. Infrastructure for a large scale international collaborative project can be a much more complicated undertaking. Yet because many of the adaptations needed are analogous to work already implemented for the interview recordings to be described in chapter \ref{ch:2}, a blending of these two code bases may be more feasible to complete even given that the person to do so will likely have limited bandwidth and/or limited software experience. 

To facilitate that process, I have provided a detailed list of the major differences that exist between the current audio journal pipeline and what will be needed for the AMPSCZ project. That can be found in supplemental section \ref{sec:ampscz-diary-todo}.

\subsection{Contributions}
\label{subsec:diary-contrib}
In sum, I believe that audio diaries have been severely underutilized in early digital psychiatry research, and especially so in studies that wish to focus on analysis of patient speech. As such, this chapter aims to serve as a thorough account of not only why audio journals should be a priority for the field in the future, but also how audio journals might be best leveraged in practice -- including potential considerations from perspectives of study design all the way to final dataset analysis, and supported by a variety of pilot results in a longitudinal Bipolar disorder journal dataset containing over 10,000 submissions from across more than 50 subjects. \\

\noindent My primary contributions from this chapter are as follows:
\begin{itemize}
    \item Constructed a detailed series of arguments for the scientific utility of the audio journal format in digital psychiatry, synthesizing points from both psychiatry/neuroscience and data science/machine learning, along with highlighting a variety of practical considerations and making direct comparisons to other datatypes that might be considered alternatives.
    \begin{itemize}
        \item Despite the paucity of directly relevant prior research, I also included an extensive literature review on indirectly related topics, as well as a discussion of why digital psychiatry has perhaps underrated audio journals as a datatype to date. 
        \item To further these arguments, I proposed a number of follow up questions of potential scientific interest at the end of the chapter -- both those with immediate connection to the audio journal results reported here, and longer term questions that might leverage the unique advantages of diary data. 
        \item Note that a context-specific version of these arguments was later used to make a large impact on the direction of the patient language analysis portion of the NIMH's AMPSCZ project, shifting over \$100,000 of transcription budget from the interview protocol to instead build a full diary transcript set. AMPSCZ will be a primary topic in chapter \ref{ch:2}, and the document I wrote arguing for the change can be found in Appendix \ref{cha:append-ampscz-rant}. 
    \end{itemize}
    \item Wrote, tested, documented, and publicly released a pipeline for data management, quality control, feature extraction, and summary visualization of daily audio journal recordings.
    \begin{itemize}
        \item Performed pilot validation of the key included features, and wrote an extended discussion of the pros and cons of these features, justifying their inclusion in the pipeline.
        \item In particular, the code can not only facilitate future studies in building a quality audio journal research dataset, but it also can serve as a lightweight feature extraction code base that provides a set of core benchmark metrics, to hopefully improve cohesion between future studies in this space.
        \item I also identified future directions for the pipeline to continue making progress towards these constructive aims. 
    \end{itemize}
    \item Utilized my pipeline to obtain features for scientific analyses from the audio journals that were collected in our Bipolar Longitudinal Study (BLS). One major aim of this endeavour was simply to more deeply characterize properties of the diary features empirically, to serve as an important reference point for future analyses and a blueprint to influence design of future data collection. 
    \begin{itemize}
        \item Characterized key feature distributions and correlation structures, identifying a variety of highly significant participant-dependent distributional differences in our dataset. Some differences in factors like word count, speaking rate, duration of pauses, and disfluency use were in line with expectations from the considered participants' clinical profiles. 
        \item More generally the observed feature correlations and distributional differences raised a number of considerations about how to handle modeling in a longitudinal dataset like constructed with audio journals.
        \item As part of this, I also identified differences in submission patterns across the larger dataset, showing very impressive participation and engagement with the diary format in over $\frac{1}{3}$ of enrolled patients in the multimodal BLS project, as well as isolating when participation dropped off for those who were less engaged. This could facilitate future predictive modeling work, particularly the characterization of missingness distribution over different participant timelines. 
    \end{itemize}
    \item Fit linear models of self-report survey scores based on extracted features from same day diary recordings in a subset of participants, and tested best model fits on hold out data from those same subjects. Through this process I found a number of participant-dependent significant relationships -- including with opposite directionalities -- in addition to a relevance for sentiment score across considered patients.
    \begin{itemize}
        \item To do so, I characterized responses to emotion-related EMAs in a participant-independent manner across a larger set of subjects, finding that positively-worded and negatively-worded questions had minimal correlations with each other, but strong correlations within group. The distributions were also differently shaped, likely because of the symptom severity that the two types of questions will saturate at.
        \item When looking at the positive and negative emotion self-report summary scores within the 7 chosen patients, there were obvious participant-specific differences in the mean and variance of each type and the correspondence between the two types of EMA prompts, such that models correlating diary features with EMA directly could perform well based only on telling apart patients, which with $n=7$ is probably not a meaningful distinction.
        \item For that reason, I fit overall slope models using z-scored versions of the input diary features based on each participants' own mean and variance. When doing so, there was no meaningful relationship between verbosity and positive nor negative EMA across the dataset. At the same time, subject 5B had highly significant relationships with lower word count corresponding to worse symptoms across self-report, explaining $\sim 32\%$ and $\sim 24\%$ of variance with a simple linear fit in the positive and negative summary scores respectively. Meanwhile, subject 8R had a significant relationship between word count and negative EMA mean that explained $\sim 12\%$ of variance, but with higher word count corresponding to worse reported negative emotions.
        \item Both verbosity models generalized well to hold out sets from random points within the main period and in large part to the end of study hold out set for each participant. There was some degradation of generalization in the beginning of study hold out set however. Looking more closely at 5B, it is apparent that even the training set fit was worse earlier in the study timeline, and a rolling beta model would like be more appropriate to capture the slower changes that occurred in their disease profile over the course of BLS.
        \item Diary sentiment on the other hand was able to linearly explain $\sim 11\%$ and $\sim 7\%$ of variance in positive and negative emotion EMA responses with a single slope across all of BLS when the sentiment was corrected to the participant's baseline using z-score, which was also highly significant and generalizable to the hold out set. This suggests a consistent role for sentiment across the considered subjects and providing confidence in the VADER sentiment summary output. Sentiment is particularly well suited to this because it is much less likely to have opposite direction effects and much more likely to vary with self-reported emotion on the day level timescale even in typical adults.
        \item Ultimately, it was clearly important to consider factors specific to within a patient and to take care with temporal dynamics when fitting such longitudinal models, and there is a paucity of psychiatry research to date that use modeling techniques across a number of participants that would avoid smoothing over these interesting signals. Calling attention to the examples uncovered here is important for future work, and further it underscores the importance of diaries in the broader speech sampling research space, because interview recordings cannot feasibly get the longitudinal sample size necessary for a hierarchical style of modeling.  
    \end{itemize}
    \item Explored early n of 1 studies in three participants of especial interest, highlighting interesting properties about how their diary features varied over time and at different scales, both in isolation and as compared to available self-report and clinical rating time courses. By referring to diary content at study points of interest and quantifying occurrences of topics of interest across the diary, I additionally provided proof of concept for some of the powerful ways diaries can be used in a more qualitative sense, to synergize with produced features.
    \begin{itemize}
        \item Participant 5B, diagnosed with Bipolar II, experienced clear depressive episodes clinically during their time in BLS. They also demonstrated some overall improvements in the latter part of their 2 year BLS enrollment. Their EMA scores fluctuated very clearly with clinical episodes, and in the second year the EMA-defined episodes began to fluctuate more rapidly than the $\sim$ once per month clinical ratings could capture, which already demonstrates the power of digital psychiatry.
        \begin{itemize}
        \item Notably, 5B had analogous fluctuations in features like word count and sentiment over the course of their time in BLS, such that those features could have been used to mark the mood episode onsets just as well. Furthermore, there was an uptick in their baseline word count in the second half of the study, and an uptick in some other features that could not be entirely explained by higher word count journals such as verbal filler word usage. Some of these features appeared to spike around the times of transition into and out of an episode.
        \item 5B's diary content was also highly informative even at a cursory look. They used the phrase "kinda down" only 7 times across all recorded journals, but 4 of these occurrences were right before the first major mood episode and 3 were right before the second one, highly concentrated and preceding a clear signal in the EMA by a couple days. 
         \end{itemize}
        \item Participant 8R, diagnosed with Bipolar I, experienced some depressive periods and some hypomanic episodes during their time in BLS. While their EMA was not as clean cut as 5B, a pattern did emerge with reporting high levels of negative emotions and positive emotions simultaneously at times in the study corresponding to higher mania-related symptoms. In the first year of the study, they discussed a possible hypomanic episode themselves at two distinct times in their diaries, which aligned well with peaks in diary features like speaking rate and rate of disfluency usage, suggesting diary content could help fill in gaps in the YMRS along with EMA trends -- thereby increasing confidence in hypotheses about the relevance of some extracted diary features too.
        \item Participant 3S was also diagnosed with Bipolar I, and submitted many long journals over the course of the study, but did not have a whole lot of clinical variation, which is why the scales stopped being collected much earlier in their enrollment period than other features. Their EMA was also quite noisy, but there were interesting fluctuations as well as sustained over time changes in their extracted diary features. Additionally, there was a noticeable drop in word counts around the time they experienced the highest negative symptom clinical ratings, even though their self-reports indicated lesser symptoms at that same time. Interestingly, by checking journal content around a shorter period of abnormal diary features, a time where 3S attempted to cut back on medications and then quickly changed their mind was uncovered.
    \end{itemize}
\end{itemize}
\noindent These contributions are all important steps towards establishing a future of audio diary analysis in psychiatry, including theoretical arguments for the datatype and pilot demonstration of many of the proposed advantages, as well as tools to assist in replicating our work and identification of a broad range of important considerations for future projects to improve upon them. In the next chapter (\ref{ch:2}), I will focus on interview recordings as a speech sampling datatype, and introduce the large scale data collection initiative of AMPSCZ. Fortunately, that project has recently reconsidered the value of the audio diary format, and so there will be many exciting opportunities in the future for data at such a scale to be analyzed in a manner that can capture clinically relevant differences both between and within individuals -- and perhaps even use the longitudinal fluctuation patterns within each individual as between group predictors. As various interview contexts have their own scientific value too, it will be especially fruitful to see how audio journal and interview recording features from the same person may differ within and across study periods; thus further motivating the relevance of the datatype characterizations I provided across the two chapters.

\chapter{Automated analysis of clinical interview recordings\footnote{This chapter is centered on publicly releasing the software infrastructure I wrote for the NIMH's AMPSCZ project, now open sourced at \citep{interviewgit}. Portions of the chapter also discuss relevant work from a Schizophrenia Research publication that I contributed to, lead by \cite{disorg22}. See Appendix \ref{cha:append-clarity} for detailed attributions.}}\label{ch:2}
\renewcommand\thefigure{2.\arabic{figure}}    
\setcounter{figure}{0}  
\renewcommand\thetable{2.\arabic{table}}    
\setcounter{table}{0}  
\renewcommand\thesection{2.\arabic{section}} 
\setcounter{section}{0}

Mirroring chapter \ref{ch:1} in many ways, this chapter focuses on analysis of clinical interview recordings. While audio diaries have a number of advantages over clinical interviews, clinical interviews also have a number of advantages themselves. Interview datasets are often directly connected to the gold standard clinical scale ratings, as these scores are assigned based on observation of the very interview that was recorded. As such, much of the literature review on acoustics and linguistics already provided in chapter \ref{ch:1} drew on results from interview recordings, not audio journals, because there is a substantially larger body of existing research on interviews. Further, clinical interviews provide an opportunity to assess social functioning through the interview dialogue, and are much more likely to include recorded video for other modalities of analysis like facial expression dynamics. Given the recent rise of telehealth precipitated by the COVID19 pandemic, collection and standardization of interview recording datasets for psychiatry research has become significantly more feasible. 

There is thus a need for open source software tools in this domain, similar to the previously discussed needs for coherent collaboration in the audio journal space. Although there is a much stronger groundwork already available for the use of packages like OpenSMILE in the clinical interview context, there remains a paucity of software options for managing the many multimodal processing tools across an interview dataset, and there is minimal progress on a unifying framework for baseline feature definitions. Because clinical interviews are additionally a more difficult datatype to collect in large samples, success in answering bigger scientific questions with interview recordings will likely require a higher level of data sharing amongst the community. A major aim of this chapter is to release and document my interview processing code base, and ultimately to contribute towards a system for monitoring collection of structured interview recordings across collaborative psychiatry studies. \\

\noindent The chapter will have two major components: the description of and example use cases for my interview data management pipeline (section \ref{sec:tool1}), and proof-of-concept scientific results produced using features I extracted from a lab dataset (section \ref{sec:disorg}). The former is currently running to facilitate a large, international multi-site study that includes interview recording data (the AMPSCZ project), and the latter is already published \citep{disorg22}.

Before diving into the details of my contributions and results, I will provide needed context in section \ref{sec:background1} on the many differences to consider between the audio diaries already discussed at length in chapter \ref{ch:1} and the interview recording format of interest here, as well as the relevant similarities between these datatypes. For key differences, I will also review related prior literature to give background that is not already available in \ref{sec:background2}. To finalize background section \ref{sec:background1}, I will discuss previous psychiatry results connected to the primary hypothesis of \cite{disorg22} -- about conceptual disorganization symptoms in psychotic illness. 

Once background on existing work has been covered, I will introduce the aforementioned AMPSCZ project along with the associated interview data collection methods in section \ref{subsec:interview-methods}. At that time, I will also recap the Bipolar Longitudinal Study (BLS) dataset described in chapter \ref{ch:1}, and provide details on the interview recording methodologies used for that study, as it is the source for the work of \cite{disorg22}. That will set the stage for the core sections of the chapter, \ref{sec:tool1} and \ref{sec:disorg}. Finally, I will close the chapter with a summary of the takeaways, limitations, and overall impact of this work, as well as related future directions of promise, in section \ref{sec:discussion1}.

\section{Background}
\label{sec:background1}
Because of the strong grounding the interview format already has in prior psychiatric practice, it is the area with perhaps the shortest leap between current psychiatry and emerging technologies, and thus has been a popular application of early research. There is a well established evidence base for differences in some acoustic and linguistic properties across many psychiatric illnesses, dating back to older manual scoring studies and further supported by the pilot computational results in this new era of research -- background that was covered at length in chapter \ref{ch:1} (particularly \ref{subsubsec:acoustics-linguistics-review}). With the recent rise of telehealth, expedited by the COVID19 pandemic, collecting a large dataset of interview recordings has become even more feasible, as will be demonstrated with the processing of Zoom interviews in this chapter. Advances generated from such research have strong potential to greatly increase accessibility of present psychiatric resources, in addition to the hopes of uncovering new scientific information. 

To further contextualize the interview recording datatype, I will next (\ref{subsec:interview-history}) compare and contrast these recordings with the audio journals previously described in chapter \ref{ch:1}. Gaps in background material highlighted by the comparison will then be covered, including a discussion of different interview formats and their role in modern psychiatry (\ref{subsec:interview-today}) and a brief review of video processing tools for and results from interview recordings (\ref{subsec:interview-lit-rev}). Throughout these sections I will provide a broader overview along with deeper background on subtopics of most relevance to the projects that will be reported on in this chapter. I will close the background section with a discussion of the key aims of my work and how they relate to the need areas identified for future psychiatric interview research (\ref{subsec:interview-motivation}).

\subsection{Comparing structured interview recordings with audio journals}
\label{subsec:interview-history}
Clinical interview recordings and daily audio diaries share many commonalities. They both lend themselves to a variety of different analysis techniques, whether qualitative or quantitative, and they both contribute objectively measurable information like speech rate as well as subjective introspective information from the patient. In many respects, the features extracted by the pipeline described in chapter \ref{ch:1} could be immediately applied to the interview context and be used fruitfully by a research investigation. Comparisons of similar features in these two different contexts is in fact an interesting open research question in itself. 

Still, there are a number of major differences that affect both many details of how processing of the respective datatypes should be approached  and more broadly what types of questions each datatype can best answer. Key differences include:
\begin{itemize}
    \item Audio journals are typically an order of magnitude shorter in duration - the maximum length allowed by the BLS diary protocol in chapter \ref{ch:1} was 4 minutes, while the majority of interview recordings collected in BLS easily exceeded 30 minutes.
    \begin{itemize}
        \item To maintain good participation rates in a journal study, it is likely necessary to keep suggested durations in line with these prior numbers. 
        \item On the other hand, it is possible interview recording durations could decrease on average as remote interviews become an increasingly common occurrence. This will likely depend on study-specific nuances. 
    \end{itemize}
    \item Audio journals are much easier to collect, both for researchers and participants. Thus clinical interviews tend to cover many fewer days in a given period and also may be more difficult to retain participants over a longer study design. 
    \begin{itemize}
        \item Short virtual interviews conducted by an automated system (like is already done at some companies for initial fully-automated steps of job screening) could be a middle ground datatype between audio diaries and structured interview recordings in the future.
        \item However for many structured interview formats of prior interest in psychiatry, there remains a more labor intensive component requiring a longer period of human interaction - making it near impossible to build up a dense day to day dataset like was done with the diaries in chapter \ref{ch:1}. 
    \end{itemize}
    \item Structured interview time points can be directly tied to well-established clinical outcome measures, as clinical scales are generally scored based on patient interviews. There is no good analog for this in audio journals, with the closest thing being self-report survey responses that are often submitted around the same time as the daily diary submission.
    \begin{itemize}
        \item It is of course possible to fit a model that attempts to predict clinical scale ratings based on diary properties from that and/or preceding days, but such a model would be missing a link to connect patient behavior with how clinicians score symptoms.
        \item Because of this clear grounding for interview recordings, it is perhaps also considered a less risky approach to the symptom score prediction problem; if clinicians are rating symptoms based on interview content, it would be shocking if extractable features to predict scores \emph{didn't} exist in these interview recordings.
        \item I think interview recordings are actually much more risky from an operational/data science perspective, even if they are less risky from a psychiatry one. Regardless, they are a much more common datatype to find in the early computational psychiatry literature, and as such there are more established research protocols and results for interviews. 
    \end{itemize}
    \item Recorded interviews often include video of the participant and sometimes the interviewer along with the audio recording. While it is theoretically possible to include video collection in recording of diaries on the smartphone app, it is more of an added burden to the participant than it is in the interview setting, and it is not included in the typical data collection workflow described in chapter \ref{ch:1}.
    \begin{itemize}
        \item The video modality of course opens up a host of new potential metrics related to physical movement, and in particular facial affect. Considerations for analysis of interview videos along with relevant prior literature are thus reviewed below in section \ref{subsec:interview-lit-rev}.
    \end{itemize}
    \item Structured interviews are able to assess properties of social functioning that audio journals have no information on.
    \begin{itemize}
        \item Any acoustic and linguistic properties discussed on the journal level in chapter \ref{ch:1} can be considered in the interview in this additional context - evaluating the connection of these features between interviewer prompts and participant responses over the course of the interview.
        \item Additional features specific to the interview setting may also be designed to capture properties of human dialogue that are undefined in a monologue setting. A basic example would be measuring the frequency of interruptions, which can be done easily using TranscribeMe's verbatim notation for dialogues.
        \item Features can be extracted from the video modality to assess social functioning as well. One common prior use case is to quantify gaze direction as an estimation of eye contact behavior, which will be discussed as part of the review in section \ref{subsec:interview-lit-rev}. Particularly when video of the interviewer is available aligned with that of the subject, a number of interesting social cues can be derived from the interview video modality.
        \item The social context may itself change participant behavior in a way that alters the base features that can be compared directly between interview and diary speech. The specifics of the clinical or research setting may also have an impact that is different from a more general social setting. These properties of the clinical interview are neither better nor worse -- as was discussed in chapter \ref{ch:1} there is good reason to believe some participants will be more honest in a journal recording than in an interview conversation, but on the other hand behavior specific to the conversational setting can be of inherent interest.
    \end{itemize}
    \item App-based daily audio journal prompts are often very open ended, as was the case for the submissions analyzed in chapter \ref{ch:1}. Interviews on the other hand are usually at least partially structured with a pre-specified list of questions to address, so that one can guarantee certain topics will be covered in each interview time point, and further can cover a fairly large number of them. 
    \begin{itemize}
        \item Although it is possible to prompt for a diary recording about a more specific topic (e.g. experienced hallucinations, drug use, etc.), it would be difficult to maintain study engagement while asking for multiple different types of daily journal submissions. It is also unlikely that a participant will have much to say every single day about many of the potential individual topic choices.
        \item In addition to the questions of clinical interest that are included by design in each interview recording for a particular study, the interview format also allows for follow-up questions to be asked based on what the participant just said. This allows for clarifications to be made as well as for a trained interviewer to uncover information that otherwise wouldn't have been revealed by a patient. 
        \item Ultimately, the information prompted for within an interview can vary greatly based on design, and has quite different considerations than design of journal prompts. As such, various interview structures are reviewed in the next section (\ref{subsec:interview-today}), both to discuss their downstream affects on analysis decisions and contextualize their relationship within existing clinical and research practices.
        \item It is worth noting that what exactly participants choose to discuss in a free form audio journal format carries interesting information that would be infeasible to obtain in most interview settings, so again these differences are not clear cut advantages or disadvantages for one format over the other. 
    \end{itemize}
\end{itemize}
\noindent Given all of that, there is strong argument for collection of both datatypes in a study where possible. Indeed, both the BLS and AMPSCZ projects described in this chapter do include interview recordings and daily audio diaries, as did the OCD case report that is the subject of chapter \ref{ch:3}. 

While I demonstrate the utility of diaries and interviews independently throughout the thesis, at times in capacities that would clearly not be feasible with the other format, ways that their analysis can truly synergize is an important open question that is largely beyond the scope of this thesis. \\

\noindent As the background relevant for audio journals has already been established in chapter \ref{ch:1} (\ref{sec:background2}), I will now proceed with review of the new topics brought to light here: different ways of structuring an interview for different clinical and research purposes (\ref{subsec:interview-today}), and then analysis methods for video recording data (\ref{subsec:interview-lit-rev}).

\subsection{Clinical interviews in modern psychiatry}
\label{subsec:interview-today}
Clinical scales, to be discussed in greater depth in subsection \ref{subsubsec:clin-scales-intro} next, are often used in psychiatry research and especially clinical trials \citep{Wood2017}. The goal standard scales rated by trained observers are now largely collected through the use of a semi-standardized structured interview conducted by a trained interviewer, though this was not always the case \citep{Williams2008}. Consistent guidelines for a particular scale will suggest questions to ask in order to assess each item and will also suggest follow-up questions that might be asked depending on patient response and potentially other behaviors. For rating scales, there are often guidelines related directly to the patient response, and for some of these instruments there will also be guidelines for rating observed behaviors outside of the response content. For example, the MADRS is designed to exclude consideration of psychomotor signs, but other scales for rating depression severity do ask trained raters to consider e.g. the extent to which speech was slow \citep{Jain2007}.

Besides the structured clinical interviews associated with particular symptom severity scales, there are a number of other use cases for interviews in psychiatry. Perhaps the most relevant informational interview in actual practice is the Structured Clinical Interview for DSM-5 Disorders (SCID-5), which is a structured interview format somewhat similar to the clinical scales, but intended for clinicians to determine a potential DSM-5 diagnosis at the end of the interview \citep{SCID}. Some studies use the SCID-5 to confirm participant diagnoses, but it is not a large data source for speech sampling work. 

In many cases one might consider psychotherapy sessions as interviews too. Though the primary purpose of "talk therapy" is to be therapeutic, there is still information gathering involved and more generally content that could be quite interesting. In 2019 $\sim 10\%$ of Americans took part in some form of psychotherapy (including both specific evidence-based techniques like cognitive behavioral therapy and other forms of therapy such as modern psychoanalysis) \citep{therapy-stats}; the rise of remote therapy options during COVID19 has likely only increased these numbers. At the same time, "psychotherapy" is a very broad term, and most forms of psychotherapy in real practice involve only highly clinician-dependent qualitative information gathering \citep{Lewis2019}. It is also likely that certain forms of talk therapy contain more sensitive personal conversation than typical clinical interviews, and require a longer term relationship building between therapist and patient. These factors make it difficult for academic science to consider analysis of many types of recorded psychotherapy sessions in research, though in the longer term psychiatry research ought to utilize technology to systematically study techniques that were previously more of an art -- and to take care in doing so, as it is easier to throw the baby out with the bathwater than it is to capture nuance in what works and what doesn't. 

For research at present, besides recording official clinical interviews, some studies will record free-form interviews specifically for the purposes of analysis, to capture patient speech in dialogue. These interviews could take the form of discussion on a particular prompt, though they are often more like the interview equivalent of our audio journal prompt in chapter \ref{ch:1}: meant to be participant-driven, eliciting conversation on whatever topic they wish to discuss. It is important for open-ended interviewers to encourage engagement and keep the conversation going as much as possible, without interrupting the participant's line of thinking or unnecessarily directing the conversation topics. So conducting these interviews does require some skill, but it does not require the same sort of formal training and labor burden as clinical interviews do, and while clinical interviews often take an hour or more, it is uncommon for a free-form interview to extend past 30 minutes. 

There are of course pros and cons to these two interview recording sources, and the AMPSCZ project discussed in this chapter collects both clinical and open ended recordings due to its wide set of aims as a large collaborative project. There are also technical differences in how one might process and extract features from these different types of interview recordings, something that will be covered in the context of AMPSCZ in the results of section \ref{sec:tool1}. One hope for the content in this chapter is that it could serve as a valuable resource for future study design considerations, leading to more informed choices about speech sampling data sources based on both scientific aims and practical resource limitations. 

\subsubsection{More on clinical scales}
\label{subsubsec:clin-scales-intro}
Broadly, clinical scales are the most common and often the only method available for quantifying psychiatric disease symptom severity. Some scales, like the PHQ-9, are scored purely through self-report; these are unsurprisingly much cheaper to administer and less time consuming for the patient, but they are more susceptible to false negatives and are likely more noisy in both the margin of error for the total score and the accuracy of symptom-specific break downs. Self-report scales are used relatively often in clinical practice, particularly in early care as screening tools and to assist with treatment planning \citep{Wood2017}. 

However, a number of independent surveys of different populations of mental health care providers (e.g. psychiatrists, psychologists) from different locations (e.g. US, UK) have found that a minority of practitioners utilize \emph{any} scales or other validated measurements, with estimates consistently coming in under $20\%$, and just $\sim 5\%$ estimated to administer measurements at the suggested frequency \citep{Lewis2019}. Despite the shortcomings of current scales, they are still a major improvement over no quantification at all, and there is a large enough body of established evidence at this point that the lag between practice and research is perhaps even more alarming than the gaps in research. At the same time, there is potential for digital psychiatry tools to directly address the problem. \cite{Lewis2019} proposed a series of actions that might improve underutilization of measurement-based care, one of which was the improvement of electronic medical record systems to facilitate psychiatric scales. Support for administration of self-report scales via a surveys app like those reviewed in chapter \ref{ch:1} could greatly increase adoption of longitudinal measurements by mental health care providers. \\

The other category of clinical scale is the type most relevant here: the practitioner-scored scale. These scales involve the aforementioned clinical interview, where a trained interviewer asks about specific symptom dimensions and probes for details and exposition as needed. A trained rater (who may or may not be the same person) scores the corresponding scale items by synthesizing information from the patient's responses as well as observing other behavioral properties. These interviews can be long and carefully rating them adds further labor burden. Still, they are the current gold standard for rigorous measurement of symptom severity, and as such their use is widespread in research, especially so for demonstrating treatment efficacy in clinical trials \citep{Wood2017}. 

Note that throughout the rest of the thesis, "clinical scales" is used to refer specifically to scales rated by a trained observer based on a formal interview. We do not collect any self-report scales, as our self-report EMA surveys are bespoke and thus not wholly validated nor necessarily meant to be generalized. In a similar vein, note that when referring to psychiatric/clinical norms through most of the thesis, I am referring primarily to clinical research practices, unless otherwise stated. \\

Gold standard clinical scales are generally very good measurements in the sense that they have high interrater reliability, strong consistency of scores between items measuring similar symptoms, high validity in terms of identifying diagnosis versus control as well as agreement between different scales assessing the same disorder, and some correlation with broader features of functioning level. For many of the commonly used observer-rated scales, there is an abundance of evidence that they are capturing real information about psychiatric disease state in a robust way, and they have been successful in detecting treatment response in a number of clinical trails \citep{Wood2017}. For scales of particular interest to the thesis, I have provided additional background on this sort of validation. In chapter \ref{ch:1}, the MADRS for depression symptoms, the YMRS for mania symptoms, and the PANSS for psychotic disorders symptoms were of focus. In this chapter, the PANSS is again used, and the CAARMS and SIPS scales for those at clinical high risk of developing psychosis are also discussed. In chapter \ref{ch:3}, the MADRS is again of relevance, and the YBOCS for assessing OCD symptoms will be of major focus. 

For all of these scales, I will also discuss downsides however. For example, the YBOCS has shown poor discriminant validity with measures of depression and anxiety, to the point that samples have found greater correlation between the YBOCS and Hamilton depression scale than between the YBOCS and other OCD scales \citep{Woody1995}. This is an extension of the fuzzy label problem discussed for diagnostic labels in the introductory chapter, something especially challenging for mechanistic research. It is highly likely that depressive symptoms can arise both as a secondary consequence of unrelated debilitating symptoms and as a primary neurobiological pathology, and this is the sort of behavioral distinction that is missing from much present research.

In this chapter, the AMPSCZ project is a large focus, and one major goal of that project is to evaluate a new clinical scale for assessing high psychosis risk that is based on the existing CAARMS and SIPS scales. CAARMS and SIPS both score well on traditional evaluations for clinical scales like those described, but they are not very accurate at predicting who will actually develop psychosis in the future \citep{Malda2019}, which is the real (100+) million dollar question \citep{AMPSCZ}. Similarly, when evaluating emerging treatment options like deep brain stimulation for severe depression, clinical scales such as the MADRS appear to be missing information, making them insufficient alone for important downstream decision making. There have been a number of case reports where individuals with severe treatment resistant depression had failed to maintain a job (or other major functioning benchmarks) through a variety of treatment attempts, but then DBS (or other experimental therapy) rectified this -- yet there was no clinically significant improvement in scale ratings, so the person was labeled a non-responder \citep{Rabin2022}. 

It is clear that the scales are not capturing all clinically relevant information that exists, and I doubt anyone would dispute that. However, these case stories suggest a potential bias in what the scales are capable of reliably detecting, such that real improvements in more severe cases might not register due to saturation of item scores. This would disproportionately affect last line treatment research, including DBS. The case report of chapter \ref{ch:3} is a situation where we did not find any major behavioral difference in response to treatment but did find some anomalies directly after unblinding, suggesting that in our case the patient's self-reported improvement was indeed just placebo. At the same time, it is clear that certain functional improvements could have easily been quantified in our data if they had occurred, and with the duration of the study we could have determined if they persisted. If the patient had started leaving the house regularly to go to a job or a volunteer position or even take part in some hobby, this would have been objectively observable through their GPS. We would not have to rely on self-report and it would not need to be such a stark functional improvement as going from staying home all day to having a 9 to 5 career. \\

Regardless, the above are all concerns about clinical scales that we would hope digital psychiatry can improve on in the context of joint collection. The fundamental problem that limits such research is the difficulty in obtaining clinical scale ratings, in terms of participant and study staff burden and in terms of fundamental limitations in the timescale these ratings are designed to capture. The MADRS for example was designed to capture treatment response in clinical trials \citep{MADRS}, which is a relatively slow process that need not be sensitive to meaningful natural variation e.g. an emerging potential depressive episode or a good day for the patient. Modern structured interviews for the MADRS recommend rating the scale based on the patient's experiences over the course of the prior week \citep{Williams2008}, and in most longitudinal studies scales are rated biweekly or monthly. Not only does this limit the number of timepoints that can be realistically collected per participant, but it seriously limits the number of participants that can realistically be enrolled for a given budget. The ability to conduct interviews virtually can help to some extent with participant burden impacting recruiting, but it remains a large undertaking for trained individuals to conduct, record/document, and score these interviews. 

The small sample size of labeled scale data presents a challenging machine learning problem in how to reconcile label timescales with dense, multimodal digital psychiatry input data. Even restricted to the context of modern patient speech sampling, most of the studies reviewed in chapter \ref{ch:1} tested far more acoustic and/or linguistic features per recording (usually in interview format) than the total number of labeled ratings available, making their results an extremely preliminary exploration. Other studies that retained statistical power were quite limited in scope, covering only a handful of well-documented speech features. It is imperative to develop a hybrid approach utilizing tools like self-report surveys or functionally interpretable features, in order to fill the label availability gap and leverage obtained gold standard clinical scale ratings most efficiently. This tractability problem is in fact a major theme throughout the thesis.

Furthermore - not unlike the gap between research and real life clinical practice - gold standard scale ratings are themselves used in a subpar way in a great deal of psychiatry research, exacerbating the fundamental issues they do have. As reviewed by \cite{Hitczenko2021} within the most relevant domain of computational speech sampling research in psychotic disorders, it is most common for studies to collect data from each participant at a single time, which removes one of the primary advantages that scales have over the (still too often used for prediction research) diagnostic label. On top of that, even studies that do use scales and use them at multiple time points tend to focus on just the total symptom severity score from a handful of scales (e.g. PANSS positive and PANSS negative), which continues the trend of lumping quite distinct behavioral profiles together. Of course as I just pointed out, the small dataset sizes do introduce practical challenges towards this end, but there are many ways to approach the issue and yet published research lacks much needed diversity \citep{Hitczenko2021}. \\

The limitations of utilizing gold standard scales as the primary labels in psychiatry are one thing impacting current speech sampling research. The other major issue is the format of the clinical interview itself. As mentioned, structured interviews for rating clinical scales involve a number of suggested questions, many of which are fairly matter of fact prompts about e.g. how often a symptom was experienced in the last week. This results in a conversational structure that is quite stunted at times relative to a natural dialogue, and it hinders the extent to which the patient is able to uniquely express themselves. Therefore the conversations themselves are often less interesting content-wise, with lesser opportunity for linguistic analyses designed around complex (or even full sentences) speech, and less likelihood for emotional responses to develop in e.g. vocal acoustics. Additionally, it is not clear how much information the analyses would be able to extract beyond what is already being rated for the scale, and any attempt with a pure automation goal would need to demonstrate that it can predict gold standard scales based on the recording better than a self-rated version of the scale could, to be remotely practically useful. To limit the research burden in a paradigm shifting way, it would need to be possible to replace the trained interviewer as well, and either still beat self-reports or drastically reduce the amount of time required from the participants -- in which case the analyzed recording is hardly taking the form of a structured interview anyway.

It is the case that structured interviews often involve freer-form discussion components, but the implementation of this can vary widely by interviewer, and extracting these portions from the structured interview is inefficient for many research purposes, if one could record a dedicated open ended interview instead. Dedicated open interviews can also be more participant-driven and do not need to involve any direct discussion of symptoms. When entirely general open ended interviews are separately conducted, they are likely to elicit a different (and I'd guess often richer) response in some participants than a freer form portion of a symptom rating interview would. Participants with negative symptoms affecting verbosity may not engage much with open ended interviews, but they are unlikely to engage much in the structured interviews either, and in the open case it is a far clearer distinguishing sign. For example, someone experiencing minimal symptoms at a given time will probably not have much to say about disease experiences, but if prompted to talk about something of meaning to them would have no trouble going into detail. 

Ultimately, open dialogue could center around some topic of study interest or of relevance to the patient's symptom profile, it could be designed around some creative task, or it could be entirely free form. There are many options for interview protocol, and these decisions could be made independently of decisions about clinical scale ratings to use. While there isn't much cost to roughly recording and approximately transcribing a structured clinical interview that is already occurring, there is a real cost to obtaining a high quality dataset of such recordings, and there is also a real cost associated with putting significant analysis efforts into these recordings. As such, one goal of this chapter is to call for more careful consideration of speech sampling study aims and when exactly analysis of clinical interviews is the right option. This connects to the overarching question of chapter \ref{ch:1}: why have audio journals been rarely used in early digital psychiatry work, and can we be more intentional about choosing speech sampling methodologies based on study aim, using interview recordings only where actually warranted? Fortunately, some initial progress has already resulted on this front, as AMPSCZ recently moved some of the \emph{speech sampling} focus (and budget) from structured clinical interviews to the previously neglected audio journals. The arguments that resulted in this change can be found in Appendix \ref{cha:append-ampscz-rant}. \\

An additional goal of this chapter is in pushing for greater consideration of specific scale items when using gold standard ratings as a source of ground truth information. Accurately linking automated methods to the fuzziest of labels can only advance the state of psychiatry so far. Of course such research can be built on, but to do so efficiently there should be more digital psychiatry work focusing on symptom dissection in parallel. Diversity of aims and approaches is perhaps the largest overarching theme of the entire thesis, as it is all too common to hear groups working on fundamentally similar speech sampling goals that explore a wide array of "cutting edge" features with only a very coarse clinical outcome set and far too few timepoints to make real conclusions. A small step in this direction here is the study of different types of linguistic disfluencies and their link with psychotic disorders clinical ratings focused specifically on symptoms related to conceptual disorganization. I will therefore next provide a brief overview of prior literature on conceptual disorganization in psychosis.

\subsubsection{Measuring specific symptoms: a focus on conceptual disorganization}
\label{subsubsec:disorg-intro-rev}
Because of the great heterogeneity in symptom profiles that has been observed between individuals with psychiatric illness, is important that digital psychiatry research take a more symptom driven approach, rather than relying on prediction of diagnostic labels, total clinical scale ratings, or even particular subscale ratings. For example, factor analysis of the Positive and Negative Syndrome Scale (PANSS) has suggested that not all items currently under the positive (or negative) subscales belong grouped together \citep{Wallwork2012}. One symptom domain that this analysis suggested should be split out was disorganization, a focus of the work presented in section \ref{sec:disorg}.

Disorganization, which may present as an impairment in thought (conceptual disorganization) and/or as non-goal directed behavior (behavioral disorganization), is one of the core syndromes of psychotic disorders, including Schizophrenia and Bipolar Disorder \citep{Morgan2017,Yalincetin2017}. Conceptual disorganization comprises difficulties in the coherent sequencing of thoughts, which can manifest as increases in typical features of spoken language such as verbosity, and atypical features such as illogical, derailed or tangential speech, distractible speech, and peculiar use of words and sentence constructions \citep{Andreasen1986,Kuperberg2010,Liddle2002}.

 In addition to disorganized speech, psychotic illness may involve impoverished speech, presenting as a reduction in the rate and quantity of language production. In Schizophrenia, it has been suggested that the disorganization and impoverished dimensions of spoken language co-occur in early stages of illness, but disorganization diminishes while impoverishment persists with progression of the disease \citep{Palaniyappan2022,Roche2016}.

 As such, there is a clear link between language use and psychotic illness, and in particular between language use and disorganized thought. While there are a number of prior works performing computational analysis of patient language during semi-structured interviews, as was discussed in chapter \ref{ch:1}, there is minimal such work focusing on specific symptom domains \citep{Hitczenko2021} -- including conceptual disorganization. It is therefore an important next step in this line of research to characterize patient speech features that may predict disorganized thought. 

 Beyond verbosity and incoherence related features, there is good reason to believe that use of linguistic disfluencies could be linked to conceptual disorganization. Sentence restarts, one category of disfluencies, are directly connected to disorganization in some sense, as they are by definition a derailed thought; though they of course may appear in normal speech to some extent as well. Additionally, it is plausible that use of nonverbal (e.g. "umm") and verbal (e.g. "like") filler words and of word repetitions could relate to disorganized thought. More broadly, both fillers \citep{Tang2021} and repetitions \citep{Hong2015} have been shown to have relevance to Schizoprenia diagnosis. \\

\noindent To elucidate which specific disfluency types correlate with conceptual disorganization, as well as to quantify the relationship between verbosity and conceptual disorganization in the setting of a recorded semi-structured interview, we performed language processing techniques on clinical interview transcripts and used the resulting features to model the specific PANSS items of interest. Those results are presented in section \ref{sec:disorg}, but more generally it is important for additional studies in this space to take an approach driven primarily by specific symptom domains rather than by widely defined sum severity scores or diagnostic labels. 

\subsection{Analysis of interview videos}
\label{subsec:interview-lit-rev}
As was discussed in chapter \ref{ch:1}, there is a growing body of literature on acoustic and linguistic analyses of psychiatric patient speech from semi-structured interview recordings. Much moreso than exists at this time for audio journals, hence the detailed review of acoustics and linguistics processing of interview recordings within section \ref{sec:background2}. Unsurprisingly, there is also strong prior evidence for the clinical relevance of a number of features that could be extracted from interview videos, including early successful applications of open source tools for automatically labeling e.g. facial expression. Because video was beyond the scope of chapter \ref{ch:1}, I will provide a brief review of some results of interest in this space here instead, along with some of the primary feature extraction tools available. \\

In Bipolar Disorder, psychomotor slowing is common in depressive states, and degree of slowness correlates with depression severity, as does deficit in effortful attention. Mood, movement, and motivation are therefore tightly linked in patients during depressive states, to a stronger degree than that seen in unipolar depression \citep{Burdick2009}. Mood is also predictable via psychomotor properties in healthy subjects. Both self-reported mood and emotional valence of stimuli were predictable via analysis of posture. Downward gaze was a key feature in negative situations, and hand tapping in positive ones \citep{Thrasher2011}. 

Coding of movement speed, fidgeting, posture, facial expressions, and other such features was done in a number of earlier studies by hand. The facial action coding system \citep{Ekman1978} for systematic labeling of facial expressions, perhaps the most famous example, has been used for over 40 years now. Modern machine learning techniques now allow this kind of labeling - including automatic labeling of the facial action units (FAUs) defined by \cite{Ekman1978} - to occur at scale and at a much finer temporal resolution. In the context of an onsite interview recording, more advanced hardware like eye tracking devices or high resolution cameras could also be used to enable additional and more accurate features to be extracted. Physiological properties like heart rate have even been successfully estimated via video with an appropriate camera setup \citep{Haque2016}.

Thus there is good reason to believe, both clinically and technically, that video analysis could yield a number of interesting new insights in research of clinical interviews. Automatic video processing has indeed already been employed to detect relationships between mood and facial expressions. Depression severity was predicted well above chance using a combination of facial features and vocal properties, all recorded from accessible phone cameras \citep{Haque2018}. Facial recognition software was also used to evaluate the effects of deep brain stimulation in OCD patients, where an automated approach for assessing negative affect was able to accurately predict treatment condition \citep{Goodman}.

In the context of clinical interviews, analysis of video recordings from psychotic disorders inpatients has been previously used to successfully predict clinical scores. PANSS general, negative, and positive ratings were fit with linear support vector regression using the facial expression features active during each interview question as inputs. $\sim 50\%$ of the variance in the clinical scales was explainable by the extracted facial expression information. Particular features were isolated that best predicted overall and subitem scores, such as lack of smile when asked about self-confidence correlating with negative symptoms, and brow raise when alone correlating with unusual thought content \citep{Vijay2016}. \\

In addition to the promising pilot research results that have used video recordings to predict psychiatrically salient features, there is also a strong existing ecosystem of open source software for feature extraction from video data. Most notably, there are a number of options for automatic processing of facial features, including labeling of FAUs, head pose, and gaze direction. One such package, the one employed by \cite{Vijay2016}, is OpenFace \citep{OpenFace} -- which is also the software used for the interview analysis performed in chapter \ref{ch:3}. OpenFace had strong early benchmarking results, runs in a very reasonable timeframe given its reliance on CPU, and was written to be user friendly on Windows machines. 

While OpenFace remains a good option for some use cases, it can be extremely difficult to install on some Linux-based compute clusters. We were unable to install it on the primary compute cluster for Partners Healthcare, ERIS, even after contacting the IT department to attempt the install for us. Because of this, we considered alternative options, and decided on PyFeat \citep{pyfeat} for use in future pipelines. As a python package, PyFeat is easy to install, and it is actively maintained. Not only are updates regularly pushed to the core infrastructure, but there is a mechanism for model customization and open source model contribution, such that many different model options for the main video processing tasks are available to the user, some of which are also capable of GPU utilization where available. Different models can be specified (or altogether excluded) for face detection, facial landmark recognition, head pose and gaze detection, facial action unit labeling, and emotion recognition. 

Ultimately, there are already good options available for many video feature extraction tasks, including tools that remain in active development with more updates coming in the future - like PyFeat \citep{pyfeat}. Many of the models included in PyFeat, particularly the ones performing concretely defined labeling tasks like head pose and facial action unit coding have obtained strong benchmarking results. Still, benchmarking is often performed on higher quality data than one might obtain in practical recordings of clinical interviews, so results should of course continue to be sanity checked for a particular study before proceeding with downstream analyses. 

Given the great promise for psychiatric video analysis but the relatively early stage of the field, there are a number of reasons to call for a more unified framework that could better contextualize results across studies. Especially as models may rapidly change and research may become increasingly multimodal, effective collaboration - both intellectually and in data sharing - will be critical. In turn, continuing to build on the open source software available to facilitate such digital psychiatry studies is an important research goal.

\subsection{Moving forward}
\label{subsec:interview-motivation}
In sum, recordings of psychiatric patient interviews are a highly salient datatype for digital psychiatry. Many results with an existing evidence base from prior psychiatry work that utilized manual scoring and qualitative assessments of observed patient speech, language, facial affect, and overall movement can now be directly translated to study in the automated interview recording analysis setting. Recordings of clinical interviews further have directly associated ratings from gold standard clinical scales, to serve as well-grounded labels for statistical analyses. Pilot work in this domain has shown great promise for the translation of certain relevant behavioral features to the context of automatic metric extraction. With the rise of telehealth and remote research administration during the COVID19 pandemic, it is also now significantly easier to collect a dataset of recorded interviews, particularly open-ended or other research interview formats that do not require extensive training to conduct and a long time commitment to score. Moreover, given recent big advances in the capabilities of machine learning (ML) techniques, it is plausible that subtle behavioral signs not yet discovered could be found to be of clinical relevance via exploratory machine learning applications to interview recordings.

At the same time, there are still significant challenges to overcome in psychiatric interview recording research, in addition to the many open scientific questions expected from such a young research topic. A major focus of this chapter is the study of speech in psychotic disorders, which has a number of interesting historical results and promising pilot work based on those results (presented within chapter \ref{ch:1}). However, as reviewed by \cite{Hitczenko2021}, the subfield also has large strides it still needs to make. The majority of prior interview recording studies have focused on prediction of overall clinical scale scores or even categorical classification of disease diagnosis, thereby including a large amount of potential variation into the label set that is unlikely to be possible to explain with the available inputs, as well as preventing any statistically significant relationship that might be found from being explained in terms of mechanistic details related to specific symptom profiles. Further, the majority of existing work has collected only a single, or at best a handful of, interview recording(s) per participant, so that it is not possible to carefully study variation over time within an individual. The work I present in section \ref{sec:disorg} of this chapter - along with the broader study dataset that it uses - begins to improve upon these limitations, by analyzing a truly longitudinal set of interview recordings and focusing this analysis only on a particular symptom domain of psychosis, disorganization.

An even more concerning limitation of pilot work (that was more carefully avoided by \cite{disorg22}) is that many of the landmark studies on automated analysis of speech in psychotic disorders used a small dataset of $\sim 100$ interviews, if that, and performed initial feature searches spanning well over 100 different acoustic and linguistic features. This is in part due to the many ways of summarizing finer timescale features on the level of the interview, but to my knowledge these works never deeply characterize the uncovered relationships to assuage even some of the concerns about potential statistical noise contributing to the final results -- often instead obfuscating the amount of feature searching they did via technical jargon. Because supplementary materials are regularly not exhaustive nor clearly written, papers from different groups tend not to closely build on each other in a computational sense, and models in pure data science fields change at a rapid pace relative to the timescale of collecting a psychiatry dataset, this practice has resulted in many pilot studies publishing potentially useful features from their explorations that do not subsequently receive a proper follow-up study, while publication after publication constitutes another pilot study exploring slightly different features in a slightly different context. 

Surprisingly, despite the large number of features that have been tested in relatively small interview recording datasets, genuinely novel feature discovery has largely not been attempted. The vast majority of automatically extracted features tested in the described exploratory works are based directly on speech phenomena already well-established in psychiatric disease, with a focus on testing specific methodologies for automatically and carefully quantifying such features. There is certainly value to research dissecting the finer points of particular features that already have a general basis in psychiatry, as will be argued with the study of linguistic disfluencies in section \ref{sec:disorg}. The core concern is not that interview research studies are too conservative with their grounding in the psychiatry literature, but rather that even those studies leveraging their exploratory nature in order to avoid careful statistical or clinical analysis (whether intentional or not) are still remaining extremely close to concepts already utilized in modern psychiatric practice, thus occupying a space that is in some sense the worst of both worlds. \\

\noindent In the concluding chapter, I will discuss at greater length my opinions on the future of digital psychiatry and the broader implications of various study design and analysis practices. To close the background of this chapter, I will now focus on more specific considerations and open questions that ought to be addressed for the study of interview recordings, tying this together with the aims of the work I present here.

One major distinction that is important to make before embarking on an interview research project is to determine what sort of primary outcome is desired -- not just for the study itself, but for the downstream work that will draw on the study. Broadly, the methodologies necessary for optimal predictive power in machine learning are (at present) often different from those necessary to elucidate underlying principles that might offer eventual mechanistic insight. The work I will present in chapter \ref{ch:4} has a close connection with interpretability of deep learning systems, which is still a new and poorly understood area of research; ML interpretability likely will remain separated from practical state of the art models in many domains for a long time. As such, it is detrimental to long term research success in digital psychiatry to confound these two aims. While they can still contribute to each other in certain ways, most studies will be better served by committing to a primary focus and using that to drive all design decisions. 

Note that by \emph{prediction}, I mean any prediction of a label from the input data, with labels often representing clinical "ground truth" that corresponds to the same time point as the associated input. An area that is so far largely unexplored (and may be difficult to properly explore with current knowledge) is the \emph{prediction} of future clinical state or other behavioral properties from information available at the present time point. Throughout this chapter, I will use the former definition for the word \emph{prediction} unless otherwise specified.

Although the prediction/understanding distinction is relevant to many modalities in digital psychiatry, it is uniquely important to clarify this point in the interview literature, for two primary reasons. The first is that interview datasets are - as already highlighted - small relative to many other potential modalities, thereby limiting available analysis tools and making statistical power a precious resource in any study that aims to tackle a specific question. Not "picking a lane" prior to working with a typical interview dataset can severely limit what can be convincingly accomplished with that dataset. The second key factor is that because of the intimate link between recorded interviews and clinical scales, it is one of the most tempting datasets to fit into a prediction framework. Indeed, clinical ratings are expensive to obtain and as such high accuracy automation of scales from recorded interviews would have great value in both research and clinical practice. Of course, this makes it difficult to obtain a large labeled dataset, which makes the prediction problem harder. 

In my opinion, many existing works on automatic interview processing have not made optimal decisions for predictive power nor for understanding, instead (again) occupying a less effective middle ground. This issue is perpetuated not only by computational decisions, but also by the decision in some groups to focus on recordings of structured clinical interviews without a clear commitment to the research aims that are best suited for such interviews. Because of their more structured format, clinical interviews provide the closely matched scale labels but consequently sacrifice a degree of expressivity compared to recordings of open-ended conversations. 

Along these lines, it will be important for the advancement of interview recording research to better characterize differences between dataset collection protocols: Which differences will be impactful to which scientific questions? How might differences affect the outputs of established open source feature extraction tools and how might potential biases introduced by this be controlled for? If such unknowns are not understood and research directions remain siloed into individual lab groups, it will remain difficult to assess generalization of a paper's results, whether those results are prediction models or data-derived mechanistic hypotheses. Interview datasets can vary extremely widely even if the population of participants were held static; numerous differences can arise from the collection hardware used to the prompts used for conducting the interviews. Seemingly small changes like allowing an outside observer to sit in the interview can have impacts on both subject behavior and technical details to consider when processing the recording. 

It is obviously infeasible to completely evaluate all potential interview differences, but there is not enough work that attempts to address these concerns at all, some of which could be better understood without necessitating a complicated dataset e.g. with fake interview recordings taken between lab members. Future research should not only include some studies that have a direct focus on analysis of interviews as a datatype, but also should involve a greater frequency of publications that provide thoughtful discussion on how they considered such factors in both study design and analysis (at least as part of the supplement). As the field of digital psychiatry progresses, it will eventually become critical to elucidate connections between features of different research modalities, which will introduce even more problems if a strong foundation for automatic interview analysis is not built. \\

Amongst the recurring themes that underlie the various limitations in early interview recording work is the inherent challenge in identifying tractable yet interesting research questions to study with these recordings. Further complicating this issue however is the lack of cohesion between groups that should be building on each others' works, an effect that is especially strong when looking across interdisciplinary lines, i.e. between groups that are primarily computational and those that focus primarily on psychiatry. Problems like variable dataset quality and model churn naturally arise when interdisciplinary communication is hampered by inconsistent definitions and unacknowledged philosophical disagreements. 

In the future directions of section \ref{sec:discussion1}, I will discuss a number of salient open interview research questions and connect them back to these abstract concerns. Here, I will next focus on a scientific study format - the large scale collaborative data collection project - that might be able to address a subset of the concerns. To set the stage for the work of section \ref{sec:tool1}, which involves tool building for quality interview data collection in one such project, I will review the potential advantages of the format along with open questions that remain about it. After that, I will close the background section by presenting the chapter aims that connect back to identified gaps, in subsection \ref{subsubsec:interview-aims-intro}.

\subsubsection{Large scale collaborative data collection}
\label{subsubsec:ampscz-intro}
An important way to address some of the limitations of pilot interview recording analyses is to build much larger datasets through collaboration between research groups. While large scale collaborative data collection is an important component for advancing digital psychiatry more generally, it is uniquely relevant for datatypes that are costly to build even a modest dataset of - and interview recordings (particularly clinical interviews) are a prime example. To efficiently run such a project, good software infrastructure is critical. Due to the cost per interview, the many potential causes of poor recording quality, the deeply sensitive nature of personal information that can be found in audio/video recordings, and the potential confounds introduced by language and cultural differences in an international study, good infrastructure at the initial data collection stage is even more critical for success of the interview recording component of a multimodal large scale collaborative data collection project.

Because computational analysis of interview recordings is a relatively new area of research within psychiatry, managing a multi-site study to collect and process these recordings is fairly uncharted territory. While good open source tools exist for extraction of particular acoustic, linguistic, and visual features, there remains a need for software to manage the data collection and quality assurance process in this domain, and to integrate the different individual tools of interest into a cohesive framework. This is important for standardizing quality of collected interview recording data, developing a common language (and benchmark) set for interview-derived features, and otherwise monitoring study progress to improve efficiency of spending and accountability of contributors when many different groups are receiving funding. It is thus a key piece to advancing the NIH's initiative for improving data readiness for machine learning processes. In this chapter (section \ref{sec:tool1}), I will contribute to filling the described infrastructure gap, through the software I wrote for the NIMH's Accelerating Medicines Partnership Program - Schizophrenia (AMPSCZ) project; note the code to be documented is already available as an open source repository \citep{interviewgit}.

AMPSCZ is a highly multimodal data collection project spanning many research sites, with a focus on improving understanding and treatment of psychotic illness through the study of young people at clinical high risk (CHR) for developing Schizophrenia. Clinical interviews are a key component of this project, and analysis aims include recordings of these clinical interviews in addition to recordings of more expressive open-ended interviews with participants. As stated in the project aims \citep{AMPSCZ}:

\begin{quote}
    People with Schizophrenia often experience a delay between diagnosis and the start of treatment. This delay, which can range from 1 to 3 years, is often associated with poorer response to treatment and significantly worse long-term outcomes.
\end{quote}

\begin{quote}
    Young people may start to show signs of risk for psychosis months or even years before they receive a diagnosis. A primary goal of the AMPSCZ program is to establish an international research network focused on recruiting young people who are clinical high risk for schizophrenia.
\end{quote} \leavevmode\newline

The Accelerating Medicines Partnership (AMP) program has been an NIH initiative since 2014, aiming to fund large scale collaborative data collection projects through both public funds and private contributions from partner pharmaceutical and biotech companies as well as non-profit donations \citep{AMP}. AMPSCZ is a more recent project launched under the AMP umbrella, the first to target psychiatric illness and consequently the first to be funded by the NIMH. AMPSCZ is planned to run for 5 years, with \$117.7 million in funding over that period -- \$99.4 million of which is pledged by the NIMH \citep{AMPSCZ}. Note the NIH has consistently spent $\sim 250$ million USD annually on Schizophrenia research over the last decade, and a similar budget is planned for the coming few years [NIH Annual RCDC Funding Estimates]. As AMPSCZ funding accounts for $\sim 20$ million USD a year, the project is about $8\%$ of the NIH funding that will go towards Schizophrenia research throughout its duration. It is therefore especially crucial that this project is well run. 
\nocite{categorical-spending}

\noindent Other disease areas targeted by AMP projects are \citep{AMP}:
\begin{itemize}
    \item Alzheimer's Disease (AD), started with project 1.0 in 2014 and continuing as project 2.0 today.
    \item Type 2 Diabetes, started in 2014 and extended into today's Common Metabolic Diseases AMP project.
    \item Rheumatoid Arthritis and Lupus, started in 2014 and extended into today's Autoimmune and Immune-Mediated Diseases AMP project.
    \item Parkinson's Disease, started in 2018 and ongoing.
    \item Bespoke Gene Therapy Consortium, a recent AMP project started after AMPSCZ.
    \item Heart Failure, a recent AMP project started after AMPSCZ.
\end{itemize}

The renewal of previous AMP projects suggests that the original projects were a success. Indeed, AMP AD 1.0, the original AMP project with by far the largest funding base, has generated a large dataset of multi-omic human data for AD research, which has been accessed by over 3000 researchers to date and resulted in over 200 publications thus far, in addition to generating a set of 542 unique potential drug targets \citep{AMPAD}. On the other hand, AMP AD 1.0 was launched under the umbrella of the 2012 National Plan to Address Alzheimer's Disease initiative, when the US government set a goal of treating and preventing AD by 2025 \citep{ad-blog}. Since then, AD research has not made anywhere near the gains that the broader initiative hoped for. In itself, this is not a criticism of AMP AD specifically, and it is likely that more time is needed to properly judge the project, as scientific results stemming from it require further follow-up. Certainly the large collaborative dataset generated has great potential to advance understanding of AD, and is likely a needed parallel direction to typical single lab studies, a number of which have been the subject of recent controversy in the AD field.

Still, questions remain about the fundamental frameworks currently being used for AD research, which could very well benefit from better characterization of behavior along the lines of the proposed directions for digital psychiatry. More broadly, questions also remain about how the AMP funding model could be improved to better facilitate collaboration, encourage robustness, and foster creativity in the large collaborative data collection project style. These issues are for the most part beyond the scope of this thesis, although I try to touch on a few related topics throughout. \\

Ultimately, the AMPSCZ project is unique amongst AMP initiatives not only in the target disease area, but also in the types of data collected. The infrastructure built for managing interview recordings within AMPSCZ, a specific focus of this chapter, could be easily extended to future AMP projects in other domains of psychiatry (or more general behavioral quantification applications) -- even easier than utilizing the code in a distinct project structure, due to the interacting pieces of infrastructure that have been built by others to support the core operations of AMPSCZ. Given the extensions of AMP that have occurred to date, this pipeline could thus be a long term contribution that reaches many areas of psychiatry, particularly if it helps to facilitate the success of the interview recording component of AMPSCZ. For similar reasons, the successful use of the audio journals being collected in AMPSCZ could substantially bolster the argument for their wider use in psychiatry research; while it may not change much philosophically or scientifically about the points made in chapter \ref{ch:1}, the availability of large scale data collection infrastructure and the existence of a well-regarded proof of concept would go a long way in addressing any practical concerns about journals. 

For all of these reasons, it must again be emphasized that efficiently collecting (and eventually analyzing) a quality dataset within the context of AMPSCZ will be highly relevant for the near future prospects of digital psychiatry. In documenting my pipeline, I will demonstrate its utility in a number of real situations already encountered during the early stages of AMPSCZ data collection. Through both the reported initial positive results and the detailing of issues to keep in mind as the project progresses, the present chapter will contribute a good deal to the development of quality interview collection frameworks for large scale collaborative psychiatry projects that incorporate recording of clinical or research interviews.

In fact, as of late January 2023 - early in the overall project's data collection timeline - 398 interview recordings from 149 subjects across 17 global locations had been successfully processed by the pipeline. This already far exceeds the number of interviews included in prior works studying automatic processing of speech sampled from interviews with psychotic disorders patients (e.g. \citep{Bedi2015,Tang2021,disorg22}), thereby demonstrating the power of the AMPSCZ-style approach to combat many of the identified limitations of prior digital psychiatry interview literature. Over the lifetime of the project, more than 20,000 interview recordings are planned to be collected from a CHR population of nearly 2,000, with an additional 1,500+ interview recordings to be collected from 600+ control subjects \citep{Brady2023}.

\subsubsection{Chapter aims}
\label{subsubsec:interview-aims-intro}
Taken together, the goal of this chapter's work is as follows:
\begin{enumerate}
    \item Provide a detailed review of the pros and cons of the interview recording format as compared to the audio journals of chapter \ref{ch:1}, and also of different approaches for recording and conducting interviews. In the discussion, I will synthesize this background with the results of the chapter to argue for my own opinions on the use of interview recordings in future digital psychiatry work.
    \item Document the interview recording data flow and quality monitoring pipeline I wrote for the NIMH's AMPSCZ project, with the following subaims:
    \begin{itemize}
        \item Release the code for potential use by other research groups in similar projects, with added context for adaptations that might be necessary for different research settings.
        \item Demonstrate the utility of the tool in the AMPSCZ project to date, and reflect on lessons learned in the process of writing it - to hopefully inform design considerations of similar future studies.
        \item Provide detailed information to support the continued use of the pipeline as the AMPSCZ project progresses, including instructions for staff performing quality monitoring and tips for others that may need to debug parts of the code or handle edge cases in the future.  
    \end{itemize}
    \item Report on our results from \citep{disorg22} linking specific categories of linguistic disfluency to conceptual disorganization in psychotic illness, and expand on those results with additional context from the extracted interview transcript features.
\end{enumerate}

\noindent In sum, these contributions will facilitate future research on interview recordings, and in fact already have greatly facilitated the data collection stage of the large AMPSCZ collaboration -- including the early identification of a number of quality issues, which will in turn result in a higher quality final dataset for the NIH's data sharing repository.

\section{Interview data collection methods}
\label{subsec:interview-methods}
As broader context for the interview recording datatype has now been provided in section \ref{sec:background1}, I will next describe details of the hardware setups used to record the interview datasets that are considered in this chapter, as well as the protocols for conducting those interviews. The material here serves as an overview, while more extensive procedural information can be found for AMPSCZ interview conduct and recording in supplemental section \ref{sec:ampscz-pro}.

When discussing hardware, I will also review some advantages and disadvantages we've encountered with the different equipment options. This discussion of recording setup will be divided into two parts - in person (\ref{subsec:onsite-hardware}) and remote (\ref{subsec:offsite-hardware}) recordings.

When reviewing protocols for conducting interviews (\ref{subsec:interview-protocols-used}), I will include some general study-wide information along with interview-specific details for both BLS (\ref{subsubsec:bls-interview-protocol}) and AMPSCZ (\ref{subsubsec:ampscz-interview-protocol}). These study methodologies will be referred to throughout the remainder of the chapter, with AMPSCZ being the ongoing project at the center of the pipeline reported on in section \ref{sec:tool1}, and BLS being the source dataset for the scientific results of section \ref{sec:disorg}. \\

\noindent \textbf{Important!}
\begin{quote}
    Anyone reading this chapter for the purposes of work on AMPSCZ or with the intention of running an AMPSCZ-like project is highly encouraged to review the materials in that supplemental procedures section (\ref{sec:ampscz-pro}). It has a great deal of important protocol information for understanding the details of various recording quality and code problems that we have encountered thus far. It is vital that all provided instructions for data collection are correctly followed!
\end{quote}

\subsection{Onsite interview recording set up}
\label{subsec:onsite-hardware}
In the Baker lab, we have historically employed two systems for acquisition of audio/visual (AV) data from in-person clinical interviews, both of which were used at times during BLS. In both setups, participant speech was recorded using a high-quality cardioid headset microphone (Sennheiser HSP4), with a separate headset microphone worn by the interviewer to enable speaker separation necessary for extraction of participant voice features. Secondary audio was acquired either from a stereo shotgun microphone (ZOOM SSH6) attached to a free standing camcorder (ZOOM Q8 3.0MP) or via a webcam (Panasonic) attached to a high performance Windows Desktop (Dell Alienware X51), where audio and video signals were synchronized using software developed by a collaborator team at Carnegie Mellon University. Video data from interviews were acquired at 1080p from both setups, at a framerate of 30 fps.

Our microphone recording setup can result in very high quality data when set up properly, capturing each individual's speech alone and with high fidelity. On the other hand, it is much easier to accidentally end up with very low quality data due to improper use; this equipment is meant for professional use by e.g. filmmakers and requires some training to properly tune. With incorrect settings, for example poorly adjusted gain, we have frequently encountered audio files that had a large amount of loud background noise present - making it impossible to even transcribe. If the primary study goal is linguistics analysis from the resulting audio, it is likely preferable to use a simpler and more user friendly recording setup, especially if the transcriptions will be done by the high quality TranscribeMe service that can perform speaker separation manually. Similarly, certain acoustic feature estimations like speech rate and pause times can be done sufficiently accurately for many purposes with recordings of standard consumer quality (chapter \ref{ch:1}). However if detailed acoustics analyses are an aim, recording with the described equipment would likely prove worthwhile. In that case, it is important that RAs are well trained in using the hardware, including completing practice interview recordings and knowing how to sanity check for quality before beginning each interview.  \\

For the AMPSCZ project, in-person clinical (i.e. psychs) interviews are captured with a single hardware audio-only recorder, the EVISTR recorder. When first using the EVISTR recorder, it is important to set the date and time accurately so that recording metadata can be automatically processed by the data management code for the project. There are also recommended settings to use for ensuring acceptable recording quality. These settings should only need to be changed once. Instructions to do so can be found in supplemental section \ref{sec:evistr}. \\

Note this onsite procedure for AMPSCZ psychs interviews will not produced diarized (i.e. speaker specific) audio streams, although automatic separation to a moderate degree of quality is possible with some existing software. By supplementing these outputs with TranscribeMe-provided speaker IDs and turn timestamps, it is plausible that high quality diarized audio might be obtained from the mono recording. Nevertheless, the acoustic analyses planned for clinical interviews in this project are currently simpler and not of highest priority. Video for the psychs interviews are even less of a priority and will not be available at all for those recordings collected onsite. Some linguistics analyses are planned, and the psychs interview recordings are by far the biggest transcription expense for AMPSCZ due to their length and count. However, the bigger focus for research analyses in the AMPSCZ project is the open ended interview format that will also be recorded with participants. 

For open interviews in AMPSCZ, it is required to obtain diarized audio, as well as video, of both the participant and the interviewer. When open interviews occur in-person, they are therefore conducted using Zoom, with the same general setup procedure as will be described shortly for remote interviews. The advantage of in-person Zoom interviews is that the participant will be using a lab-provided computer with attached headset, and the interviewer can personally ensure everything is set up properly before beginning -- so that quality will be higher and more standardized. Particularly for a multimodal study where participants will need to be onsite for other data collection such as fMRI, conducting an in-person Zoom interview can be worthwhile. Of course, for the interview itself, the participant and interviewer need to be in separate rooms onsite, communicating only over Zoom.

Using Zoom for in-person interviewing will very likely result in behavior more similar to that seen in remote interviews rather than the behavior seen in traditional onsite ones. In many respects, the remote and onsite formats are closely related, but certain features - especially video features like gaze direction - will require different treatment in the two cases. It is also possible that for some individuals the social pressures felt in the two formats are different, which could subsequently impact participant responses in less predictable ways. Depending on the study aims, it may be worth considering more closely if both formats should be collected or if only one should be the focus. With AMPSCZ, using Zoom for the onsite open interviews keeps the dataset much more standardized while still allowing for both in-person and remote options, which in turn enables greater scale. \\

\subsection{Offsite interview recordings via Zoom}
\label{subsec:offsite-hardware}
As mentioned, recording remote clinical and other research interviews allows for significant expansion of study scale. During the height of the COVID19 pandemic, it was also the only way to collect such data, which lead to the rapid implementation of a Zoom-based protocol in our lab, and a mix of onsite and offsite recordings within the BLS dataset. In the Baker lab, onsite and offsite interviews have been treated as separate interview types, in a similar manner to the psychs and open interview distinction drawn for AMPSCZ. Depending on study aims and how onsite interviews are conducted, this may or may not make sense for a different project. Certain BLS analyses, like the results of \cite{disorg22}, ultimately pooled together onsite and offsite interview transcripts, but other analyses may treat these types separately.

For AMPSCZ, remote interviews are also conducted over Zoom, with the same general recording protocol for open or remote clinical interviews. The protocol for the Baker lab Zoom recordings is very similar to the AMPSCZ protocol, so I will focus on our instructions and overall recommendations from the Zoom recordings of the AMPSCZ project here. Note that Zoom was chosen as a recording platform in part because it is the highest quality meeting software that offers diarized audio files in the built-in recording capabilities. Some sites of the AMPSCZ project wanted to use Microsoft Teams, but Teams does not have any option for saving speaker specific audio recordings. Diarization after the fact using machine learning techniques will be subject to much greater error than the recordings provided by the Zoom software application, which can leverage information from each individual's device separately during the recording process. Quality diarization is critical for a number of acoustic analyses, as well as for future use of automated transcription services (that often do not perform speaker identification). \\

To use Zoom for collecting interview recordings, all interviewers should use their work machines with a HIPPA compliant Zoom account provided by their institution. It is recommended that participants are reminded of their scheduled interview in advance via email, to ensure they will be able to join the call in a private place at that time, ideally on a computer. If the subject would be better able to participate in the interview at a different time, it is preferable to reschedule rather than have them join the interview on their phone or from a location with background activity. 

At each session, quality checks should be performed with the participants and any logistical discussions should be handled before beginning the interview recording. It is important when clicking the record button to select "record to computer" rather than "record to cloud", because speaker specific audio is not available with cloud recording in Zoom. The interviewer should be the first person to speak after beginning recording and the participant the second person to speak, to facilitate easy labeling of transcript speaker IDs. Detailed information on Zoom account settings and recording protocol can be found in supplemental section \ref{sec:zoom-settings}.

When the interview is completed, recording should be stopped within Zoom, ideally before the participant leaves the meeting to keep diarized audio time alignment correct without any adjustments needed. When the meeting ends, Zoom will automatically convert the recording to the standard expected file formats and save them in a meeting subfolder under the folder path pre-specified in your settings. This process can take some time, especially for long meetings, and it is important that it is not interrupted or the data from that interview could be permanently lost. Do not quit Zoom, close your laptop, or power down your computer until the conversion is complete. A progress bar will appear when the meeting ends, so you can estimate and keep track of the time required. When the files are correctly saved, the folder containing them will pop up in your file browser. It is only at this point that the folder should be copied to the data upload location. It is especially important that while Zoom is still in the process of saving the recording, you do not move or modify the folder that the files are being saved into (or any of its contents). Per the AMPSCZ SOP, for this pipeline the folder should never be modified and should be moved or copied as is into the appropriate Box/Mediaflux location for server upload, once it is fully saved.

\subsection{Protocols for conducting research interviews}
\label{subsec:interview-protocols-used}
As introduced in section \ref{subsec:interview-today}, there are a number of different ways that interviews may be conducted in both psychiatric research and clinical practice. Different styles of interview are better suited to different research questions and different analysis techniques, and can also introduce numerous technical details that need to be considered in implementing a software pipeline for handling interview recordings. Because of this, it is important to review how interviews are conducted for the present studies.

The (semi-)structured clinical interview is perhaps the most common format of interview for recording analysis to date, likely due to its immediate link with the gold standard clinical scales that are rated during such interviews. Protocol details for conducting a clinical interview can vary based on the scales involved and the study aims, but generally these interviews generate longer recordings with periods of brief back and forth on questions of subjective symptom severity and frequency. Clinicians or otherwise trained interviewers will then rate the scales based on both direct patient responses as well as qualitative observations of patient behavior during the interview, e.g. speaking rate or eye contact.

The clinical interviews thus provide an attractive opportunity to automatically measure symptom-related behaviors throughout the interview like a clinician might, and to ultimately attempt to link these features with automatic processing of patient responses in order to predict the scale ratings that will be given for that interview. As clinical scales can be time consuming to rate and require trained observers, such an automated system could be extremely valuable in practice. To date however, it has been difficult to train an automated system on clinical interviews due to the generally low sample sizes of individual studies. One hope of a large scale project like AMPSCZ is to overcome the hurdles that limit smaller datasets -- although even for such an expensive and expansive project the number of interviews collected will be on the small end for modern machine learning applications. Nevertheless, clinical interviews are a good starting ground for testing automatic feature extraction of behavioral properties previously associated with psychiatric illness via manual scoring.

Another way to address the limits of clinical interview recordings is to reconsider the goal of collecting them, and only focus on their analysis when it is most scientifically relevant. There exist recording formats that are cheaper to collect, recording formats that can elicit more expressive patient responses, and recording formats that can more deeply explore particular scientific questions beyond the current scope of psychiatric practice. One potential format, the daily audio journal, was discussed at length in chapter \ref{ch:1}, and contrasted with recorded interviews above in section \ref{subsec:interview-history}. Of course, alternative interview styles also exist that can check some of these boxes while still providing many of the benefits that interviews do have over audio journals. Accordingly, the other major category of interview protocol that will be described at length in this chapter is the open-ended interview.

The procedure for conducting an open interview can vary in details, but at a high level the goal of an open interview is to get the participant to share a lot of information about their thoughts and feelings on topic(s) of personal interest. The interview is intended to be driven by the participant, similar to the basic principles in conducting a Rogerian therapy session. Of course the goal is not to be therapeutic, but simply to collect rich information about the participant. It therefore does not require much training to conduct -- the ELIZA chatbot created in the mid-1960s could do something similar in fact. It also does not require a long time commitment after the session, and the interviews themselves tend to be shorter and likely less tedious for the participant. At the same time, the stories told by the interviewee in the allotted time can be much longer and more expressive than what is contained in a typical clinical interview. \\

For BLS, to be detailed in subsection \ref{subsubsec:bls-interview-protocol}, the study format involved semi-structured clinical interviews intended to rate a battery of scales. For AMPSCZ, to be detailed in subsection \ref{subsubsec:ampscz-interview-protocol}, both open-ended and clinical (i.e. psychs) interview procedures are included for all subjects. In that project, open and psychs interviews are both recorded, transcribed, and planned to be analyzed, with initial processing following a largely similar workflow for both types; though it is possible for future projects to consider treating these interview recording datatypes more differently while still collecting the clinical interviews for the grounding their associated scale ratings provide. 

Note that it is critical to all interview procedures that detailed information on each session is recorded in REDCap/RPMS (or your database of choice) in a timely fashion. Here, the Baker lab and the Pronet sites of the AMPSCZ project use REDCap, while the Prescient sites of the AMPSCZ project use RPMS (supplemental section \ref{subsec:redcap}).  

\subsubsection{Bipolar longitudinal study (BLS)}
\label{subsubsec:bls-interview-protocol}
The scientific investigation in this chapter (section \ref{sec:disorg}) utilized the interview recording dataset from BLS. For an introduction to the broader BLS study aims and protocol, see chapter \ref{ch:1} (\ref{subsec:diary-methods}). 

Semi-structured interview recordings were conducted monthly as one of the collected BLS datatypes, and though enrolled subjects had flexibility in which portions of the study they participated in, a large interview dataset was successfully collected from 59 patients covering about 1 year (i.e. 12 interviews) per patient on average. Note that transcribed interviews for BLS were produced via the same TranscribeMe protocol as was used for audio journals in chapter \ref{ch:1}. \\

\noindent Information on the scales included in BLS research interviews and the instructions given to study staff for conducting them can be found in supplemental section \ref{sec:sup-bls-prot}.

\subsubsection{The AMPSCZ project}
\label{subsubsec:ampscz-interview-protocol}
The code architecture presented in this chapter (section \ref{sec:tool1}) was primarily written for the NIMH's Accelerating Medicines Partnership Program - Schizophrenia (AMPSCZ) project, which aims to collect dense multimodal behavioral data from both patients at high risk of developing Schizophrenia and healthy controls, in a systematic fashion across many different locations. The large, international multi-site study will run for a period of 5 years, and aim to collect data from over 1000 experimental subjects, with an analysis focus on characterizing youth at clinical high risk for psychosis \citep{Cotter2022}. The project is currently in the early stages of data collection, with a substantial amount of prep work already completed; a handful of sites are $\sim 6$ months into the collection protocol and others are being on-boarded as of late 2022. The major aim for the data analysis team at this time is thus to demonstrate working data flow and monitoring tools to facilitate quality data collection across modalities by the sites. 

One major piece of the study is the recording of structured interviews with participants. Interview recording and other raw data collection are to be performed by individual study sites, who then submit their data to one of two central servers for organization and processing. The resulting outputs will later be shared for analysis across sites. The central servers here are Pronet, in charge of 26 total sites (20 English-speaking), and Prescient, in charge of 11 total sites (4 English-speaking). The data management and quality control code, to be described below, is installed on each central server to run on all submitted interviews. In sum we are supporting 37 different groups and 8 possible languages: English (24 sites), French (2 sites), Spanish (2 sites), German (3 sites), Italian (1 site), Danish (1 site), Mandarin (2 sites), and Korean (2 sites). These also span a wider range of cultural differences, including Spanish sites in South American and Spain, French sites in Canada and France, and English sites from across the US, UK, and Australia. \\

Given the great diversity of participating sites, it is especially important that the protocol for conducting study interviews is clearly defined. Towards this goal, two different types of interview recordings are collected as part of the study - open interviews and psychs interviews. Each enrolled AMPSCZ participant should complete 2 open interviews and 9 psychs interviews. The open interviews will be about 15 minutes, while the psychs interviews will be about 40 minutes. For open interviews, stricter controls on the recording procedure are enforced, for example having exactly two interview participants (interviewer and interviewee) with video on through the duration of the recording. For psychs interviews on the other hand, it is permitted for video to be left off or for there to be multiple interviewers present. The hardware audio-only recorder for onsite interviews is also only permitted for use with psychs interviews, as open interviews are expected to be conducted on Zoom in a consistent fashion. 

\noindent The two interview types have different scientific purposes as well, and include a very different style of content. As stated by the AMPSCZ project aims in the SOP:
\begin{quote}
Open-ended language samples provide insight into naturally occurring thoughts. Language samples from the PSYCHS allow us to investigate changes in language over time and provide insight into the measurement properties of the PSYCHS itself.
\end{quote}
 
 \noindent Because of this, I have reviewed information on how these two interview types should actually be conducted and what we expect to get out of them within supplement section \ref{sec:ampscz-pro}: a primer on the intended flow of open interviews along with example questions can be found in \ref{subsec:open-sup-ra}, and analogously for psychs interviews in \ref{subsec:psychs-sup-ra}. 
 
 As part of the early project monitoring that will be presented later in section \ref{subsec:interview-outputs}, comparisons between the data collected so far for these two interview types were made, and the results of these comparisons were largely in line with expectations based on the aims of the two interview types - so it is helpful to be aware of details like those covered in the supplement \ref{sec:ampscz-pro}. If one is to be closely involved with AMPSCZ interview collection or processing, it will be especially critical to know the basic protocols. It will be very clear over the course of the chapter that we have seen repeatedly what happens when one doesn't attend to procedural details, from frequent mistakes in uploaded recordings to surprising lack of knowledge on what to expect with the transcripts of different interview types from a data science perspective. Interview structure is ultimately highly relevant to how extracted features can best be used.
 
 Along these lines, I will use both the scientific background and the preliminary monitoring results to argue that the substantially larger amount of money being spent on psychs transcriptions versus open is highly disproportionate to their relative scientific potentials, and furthermore the neglect of the collected audio journals while psychs transcriptions receive a large budget is a serious oversight in the speech sampling portion of the AMPSCZ project plan. The arguments are presented in full in Appendix \ref{cha:append-ampscz-rant} -- fortunately, the transcription budget was just recently updated in part because of these points.
 
 Nevertheless, it is critical to establish the basic instructions on conducting both types of AMPSCZ interviews, in addition to recording them to the highest quality possible. Open interviews in particular can be more of an art than a science, and detecting possible low quality interviewer performance in open interviews is a topic that will be touched on in the results on my pipeline's monitoring functionality.   \\

\paragraph{A note on transcription conventions in AMPSCZ.} 
For transcription of the collected interviews, TranscribeMe is again used, with a similar but slightly different protocol as that used for the BLS transcriptions here and in chapter \ref{ch:1}. As some of these differences impact pipeline design decisions, the details for transcribing the AMPSCZ interviews are similarly important for working with the code presented here, or even the resulting transcripts downstream. They are helpful also for contextualizing some of the points made in the review of pipeline operations upcoming (section \ref{sec:tool1}). 

Once again, these difference are characterized within the supplemental section \ref{sec:ampscz-pro} (this time in \ref{subsec:u24-tm-ap}). 

\section{A new tool for clinical interview processing}
\label{sec:tool1}
Because of the large scale of the AMPSCZ project and the diverse set of concerns that sites may have or of issues that may arise, quality software infrastructure for managing data flow and monitoring in a cohesive manner is critical. The goal of this section is to document my publicly available code \citep{interviewgit} that serves this purpose for the interview recording datatype. It fills two separate separate needs: providing important information for those that will continue to interface with my pipeline in various ways through participation in the AMPSCZ project and detailing the software implementation for those that might consider using it to manage data collection for another study. From a broader perspective, the work of this section also provides insight into key considerations for maintaining quality of interview recordings and for facilitating large scale collaborations in this domain. 

To begin, I will provide a basic overview of the technical infrastructure supporting the AMPSCZ project at large (\ref{subsec:u24-server-overview}), thereby giving important background for understanding the example pipeline outputs and description of common issues to date for AMPSCZ interview recording data organization/quality control (\ref{subsec:interview-outputs}). This will enable demonstration of the logistical and scientific utility of the code, which has not just facilitated collection of a higher quality recording set for an important initiative, but has helped guide updates to the speech sampling research design as well, ultimately resulting in a shifting of transcription budget priorities. Such points will be made clear through the process of presenting the pipeline's outputs (and a related broader update on the AMPSCZ interview data collection progress) within the current section - but a standalone report more specifically on that topic can also be found in Appendix \ref{cha:append-ampscz-rant}.  

Finally, documentation on my code's implementation, along with instructions for using it in and adapting it to new projects can be found in supplemental section \ref{subsec:interview-code} as needed. This is a critical contribution done to facilitate a smooth transition for the future of interview recording data management with AMPSCZ. The project is set to continue for $\sim 3$ more years, with an increasing ramp up of participants and subsequently more monitoring to do and more edge cases for the code to tackle. 

\subsection{Background on technical infrastructure for the AMPSCZ project}
\label{subsec:u24-server-overview}
The NIMH's Accelerating Medicines Partnership Program - Schizophrenia (AMPSCZ) project is a recently launched collaborative digital psychiatry study, spanning 37 sites internationally. It aims to collect a number of datatypes longitudinally in both healthy controls and those with clinical high risk for Schizophrenia diagnosis. One such datatype is of course recorded semi-structured interviews, both for eliciting free speech (open interviews) and for scoring clinical scales (psychs interviews). For introductory information on the makeup of the AMPSCZ project and the protocol to be used across sites for collecting these interview recordings, please see section \ref{subsubsec:ampscz-interview-protocol} above (as well as supplemental section \ref{sec:ampscz-pro}). 

\begin{figure}[h]
\centering
\includegraphics[width=0.9\textwidth,keepaspectratio]{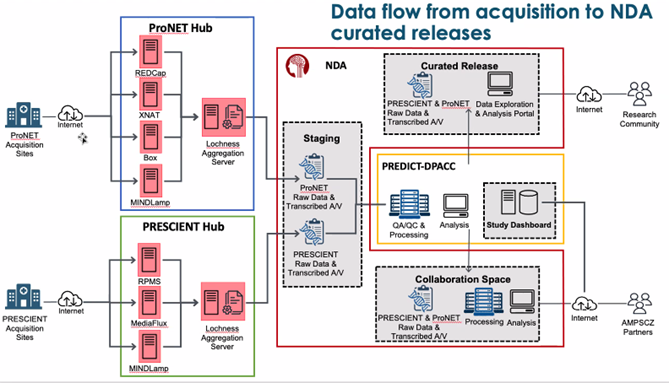}
\caption[Configuration of servers for data management and processing in the international multi-site AMPSCZ project.]{\textbf{Configuration of servers for data management and processing in the international multi-site AMPSCZ project.} As mentioned, the Lochness software handles the upload of raw data from acquisition sites to the corresponding central server, for all datatypes. In this chapter, we focus on recordings of clinical interviews. The code described here runs on both "Lochness Aggregation Server" machines depicted; there, it organizes the uploaded data, handles back and forth with TranscribeMe, performs lightweight quality assurance for audio, video, and transcript datatypes, and assists human monitors in assessing each site's status over the course of the study. Select deidentified datatypes (including redacted transcriptions) are then set aside by the pipeline to be pushed by Lochness to a staging server under control of the DPACC group, which among other things manages data sharing to different entities. Eventually a more powerful computer will be set up for each central aggregation server, to mount its hard disk space and allow for computationally complex low level feature extraction to run on the raw audio and video data, to be shared with the staging server as well. However this chapter focuses on the interview management/QC software running on the Lochness Aggregation Servers currently, which will continue to handle the interview portion of study management for the duration of the AMPSCZ project. See Figure \ref{fig:zoom-arch} for an architecture diagram of that code.}
\label{fig:u24-arch}
\end{figure}

To run such a large operation involving such sensitive datatypes, a fairly complex data flow plan is required, with servers at multiple different institutions involved (Figure \ref{fig:u24-arch}). Raw data from across sites needs to be aggregated and eventually processed to deidentified outputs that can be shared with the network of sites and eventually the larger NIH data repository (NDA). Quality control monitoring must occur throughout this process, both for the sake of future research quality as well as for the assurance that participant privacy is maintained. Communication with those collecting data at sites and with those at each critical point in the data flow path is required to keep operations running smoothly, especially at the early stages of the study -- something else that automated processes can (and do) assist with.

To contextualize the rest of this section, recall that there are two different central servers managing data flow from two different sets of sites, as depicted in Figure \ref{fig:u24-arch}: Pronet and Prescient. The Lochness pipeline implemented by collaborators handles the pull of raw interview data from Box (for Pronet) and Mediaflux (for Prescient) to the corresponding data aggregation server. Each raw interview at this stage is associated with a site, type (open versus psychs), and subject ID, but is otherwise an untouched file as output by Zoom (or in some cases the permitted hardware audio recorder). 

My code, which will be described in detail next, runs on the Pronet and Prescient data aggregation servers. Outputs it produces that are deidentified are placed in a location such that Lochness can transfer them to the next stage of the larger data flow path (Figure \ref{fig:u24-arch}). The version of Lochness utilized for the AMPSCZ project is an adaptation of the Lochness software described for internal lab use in chapter \ref{ch:1}. Similarly, the DPDash software to be discussed for study monitoring was adapted by collaborators for this project from the version used by the Baker lab.

In order to set up the production data management servers and processing pipelines that are now running for true data collection by many project sites, a development testing scheme was initially utilized. Development versions of each server in the project architecture (Figure \ref{fig:u24-arch}) were created for both the Pronet and Prescient sides, and early versions of all necessary code were installed. Sites were asked to submit mock interview recordings containing fake interviews between study staff, to be processed as if they were real -- including transcription service by TranscribeMe. Many bugs in both code and workflow were addressed through this process, before moving on to set up of the production infrastructure to be used in practice for the AMPSCZ project. 

There were also a number of issues not detected or not sufficiently resolved during the development testing phase; I will primarily focus on present day considerations in the upcoming sections (and supplement \ref{subsec:interview-code}), but at times I will discuss design decisions made because of dev testing results. In the discussion section (\ref{sec:discussion1}), I will include a review of benefits gained from this project structure and ways it might be improved for future projects. This speaks to an overarching goal of the chapter, which is to provide helpful reference information not only for the continuation of AMPSCZ speech sampling, but additionally for the establishment of other similar initiatives.

It is worth noting before moving forward that the data aggregation servers have modest compute resources and many processes for data management across the many modalities running simultaneously. Thus my pipeline focuses on necessary data flow operations and calculation of light weight quality control (QC) measures only. Development of feature extraction code will take place on provisioned audio/video processing servers that have their respective data aggregation server's briefcase mounted. This development process remains in progress and is not a focus of this chapter, though it is discussed in some more detail as part of the future directions within section \ref{sec:discussion1}.

\FloatBarrier

\subsection{Pipeline monitoring progress report}
\label{subsec:interview-outputs}
To facilitate interview recording collection, the pipeline has a few major functionalities, performed daily across sites and interview types on the described data aggregation server(s):
\begin{enumerate}
    \item Tracking interview recordings submitted so far, and in this process flagging any major upload issues such as missing modalities, incorrect file formats, or missing metadata. As part of this, newly submitted interviews are logged daily, and proceed to the next steps of the pipeline if the upload does not have fatal flaws.
    \item Mapping interview date metadata to a study day number based on the days since the subject in question consented - this is done to remove potentially identifiable real date information from any outputs that are considered deidentified, while still maintaining an invertible mapping to the stored raw data.
    \item Extracting light weight quality control (QC) metrics from the primary audio and video files of each interview, to monitor for major quality issues in submitted recordings. Poor quality can result from hardware issues or from failure to follow the standard operating procedure (SOP) for how the interviews are supposed to be conducted.
    \item Managing data flow to TranscribeMe's server of the mono interview audio recordings that pass basic file accounting and QC checks. This includes managing the routing of recordings in different languages to the correct set of transcribers.
    \item Managing data flow from TranscribeMe's server of any returned transcripts. This includes active monitoring of current pending transcripts, and management of the manual site redaction review process that a subset of transcripts must undergo before being considered finalized.
    \item Generating redacted copies of finalized transcripts using TranscribeMe's redaction notation (curly braces around PII), and ensuring the redacted versions of the transcripts are in place for transfer to downstream AMPSCZ servers that handle the raw deidentified datatypes meant for use in analysis.
    \item Computing QC metrics on the final redacted transcripts, to further inform the quality monitoring process.
    \item Utilizing information from the entire pipeline run to communicate updates and new potential issues for each site - by both generating email alerts directly, as well as ensuring all QC metrics are organized and in place to be transferred to downstream AMPSCZ servers that handle the QC monitoring dashboards.
    \item Summarizing the file accounting information, warnings, and QC metrics across sites on the server, for central monitoring purposes of each site's progress to date on each interview type. This involves a daily summary email with all information for a select group of recipients and a weekly summary email with QC-related information for a broader group of people involved with the project. For the weekly email, distributional visualizations and a heavily pruned html-formatted stats table are generated from the denser set of QC outputs.
\end{enumerate}
\noindent This code is essentially identical between the Pronet and Prescient central servers, but it is run independently on each. Central monitoring is performed by myself and others in the tool building arm of the project, in conjunction with the relevant researchers overseeing Pronet and Prescient, respectively. \\

I will next review how some of the outputs that are produced by my code have actually been used in active monitoring of the AMPSCZ data collection process so far (\ref{subsubsec:u24-monitor}) -- information that would assist any future researchers tasked with a central monitoring role in the AMPSCZ project as well as anyone considering using this or similar code for a different study. Based on the results of monitoring, I will then discuss the frequent issues that have been encountered during interview data collection in this project so far (\ref{subsubsec:u24-issues}). That section is largely written as a guide for collecting sites or those communicating with sites; for a focus on software issues, please see the documentation within section \ref{subsec:interview-code}. Finally, the progress so far has lead to better estimates of storage requirements, as well as identified potential issues that could eat up more storage than actually necessary, so these lessons are reported in subsection \ref{subsubsec:u24-storage}.  

\subsubsection{Example output uses}
\label{subsubsec:u24-monitor}
There are a few distinct types of outputs generated by my code to be exemplified, as follows. Note the details of such outputs are covered at greater length from a software implementation perspective in supplemental section \ref{subsec:interview-code}.
\begin{itemize}
    \item Intermediate outputs that may be used for additional future processing steps, but which cannot be transferred to downstream servers in their current form due to potential personally identifiable information (PII).
    \item Final outputs that are deidentified and can be transferred to downstream servers for use in future analyses. This primarily refers to the finalized redacted transcripts, but some of the computed QC metrics may also have use in later analysis work.
    \item Communication, which includes email alerts for monitoring code status and interfaces on DPDash (an online visualization tool written by collaborators) for checking data quality.
\end{itemize}
\noindent To demonstrate how different outputs can be used to monitor data collection and address potential quality issues, examples of the most user-relevant outputs are provided from some of our mock interview test data in this subsection -- thus focusing primarily on the "communication" category. As part of this, I will discuss a number of scenarios that can be detected by the code. In the following subsection (\ref{subsubsec:u24-issues}) I will then detail issues we have frequently encountered in production use and how they can be resolved. \\ 

\begin{figure}[h]
\centering
\includegraphics[width=0.8\textwidth,keepaspectratio]{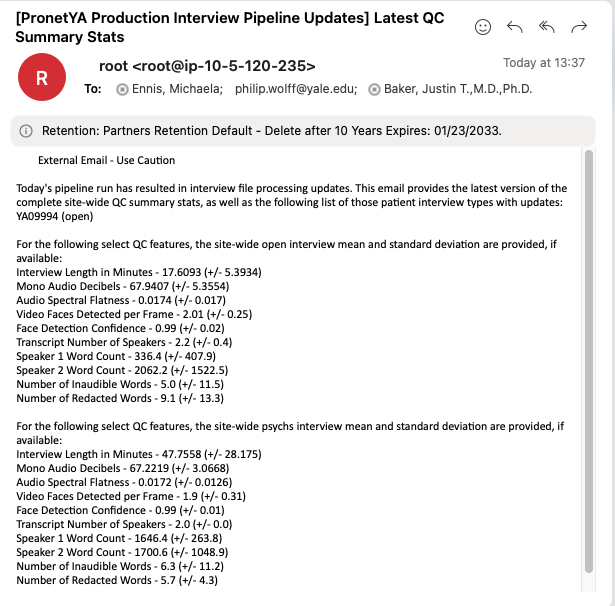}
\caption[Example of a site-specific daily update email with the latest interview QC stats.]{\textbf{Example of a site-specific daily update email with the latest interview QC stats.} One of the five daily update emails designed to go to sites is sent whenever there are any newly extracted QC metrics for an interview contributed by that site. In this email a list of subject IDs and corresponding interview types that had any new QC that day is first provided, and then for both open and psychs interviews the mean and standard deviation of key QC values across the site are reported. The example here was screen captured from an update about the Pronet YA site on 1/23/2023. Note that the depicted QC update email does not yet get directed to site-specific contacts, as only the separate email about transcripts awaiting manual redaction review is sent to sites currently. However the other five site-specific update emails, including this one, are an important piece of central monitoring; they are intended to be redirected to management by individual sites later in the study.}
\label{fig:example-qc-email}
\end{figure}

\paragraph{Site-specific emails.}
To assist sites in monitoring their interview recording uploads, a series of daily emails meant for individual sites was built into the pipeline. Each site, if they have an applicable update for any open or psychs interview that day, will receive the following 5 distinct email logs:
\begin{itemize}
    \item An audio update, which gives the final file names (deidentified versions) of all newly recognized mono audio files that were processed and successfully uploaded to TranscribeMe from across that site. These file names include metadata information like subject ID and interview type. The email alert also gives the total number of minutes of audio newly uploaded for cost estimation, and lists any files that either had SFTP upload fail or were rejected by the QC script (whether due to missing the quality threshold or the QC itself crashing), so they can be further troubleshot.
    \item A transcript update, which lists the final file names of any transcripts for that site that are still pending a TranscribeMe output, were newly returned by TranscribeMe that day, or were newly returned by site review that day. In the case of a return by TranscribeMe, the alert will denote if the transcript was selected for manual redaction review.
    \item A video update, which lists the final file names of all newly recognized video files that were successfully processed for QC for that site. If this email sends but the list is empty, it indicates that a new interview was detected but QC failed to complete, which will need to be troubleshot.
    \item A QC stats update, which will send whenever there is an update to QC metrics for any modality and interview type for that site. This email lists the subject IDs and corresponding interview type that had a new update that day, and then it provides site-wide mean and standard deviation of select QC features across open interviews and across psychs interviews. A demonstration of this email from a recent update for Pronet site YA is provided in Figure \ref{fig:example-qc-email}.
    \item A list of new warnings, sent only if specific SOP or other major quality related concerns arise in an interview that was processed for that site on that day. Note that this email can contain text that is PII, commonly dates, but even possibly names - as the speaker specific audio files provided by Zoom are automatically labeled using display name, which is often a person's full name. It is thus important to ensure that only those who would have access to these rawer data receive the warnings emails, whereas by contrast QC metrics even when reported per interview (as in DPDash, to be described below) are already deidentified. On the other hand, these warnings are vital to monitoring for issues that would be impossible to detect on DPDash, because many of them involve pre-QC problems that prevent QC output from being generated in the first place. A common set of encountered issues along these lines is provided in section \ref{subsubsec:u24-issues}. The specific problems that the warnings email will cover, which includes some early SOP violations and some key QC red flags, are as follows:
    \begin{itemize}
        \item Any SOP violation that prevents a newly detected interview upload from having video and/or mono audio processed will cause that upload's name to be listed in the email as an SOP violation.
        \item If an uploaded Zoom interview is missing the diarized speaker-specific files in the Audio Record subfolder, the interview name will be listed as missing files. Though these are non-essential to this QC/dataflow pipeline, they will be important to downstream acoustic processing, and must be uploaded as expected.
        \item If a newly processed interview creates some inconsistency with the day and session numbers used in the deidentified names, the final name will be listed under an appropriate heading to denote the issue. Problems include:
        \begin{itemize}
            \item Video and audio files from the same interview folder receiving different day or session numbers.
            \item The consent date registered when processing older files for a given participant no longer matches the current consent date in the metadata for that participant.
            \item There is a repeat in session number assignment for a given participant and interview type, or day and session numbers have incongruent orderings - indicating interviews were successfully uploaded out of order.
        \end{itemize}
        \item If a transcript returned by TranscribeMe is not UTF-8 encoded, or (though less of a concern) an English transcript is not ASCII encoded, the transcript filename will be listed.
        \item Processed interviews with the following QC issues of note are listed (using their final deidentified name): 
        \begin{itemize}
            \item Rejected by audio QC script.
            \item Recording length $< 4$ minutes.
            \item Extracted video frames had no faces detected.
            \item Redacted version of transcript had no redactions within.
        \end{itemize}
    \end{itemize}
\end{itemize}
\noindent Note that the warnings email is far from exhaustive, and also includes problems of varying severity, so a warning email does not always indicate a major issue. Nevertheless, this email is a critical part of the monitoring workflow, and is designed to be used by sites in conjunction with their DPDash QC dashboard view (to be discussed) in order to catch and troubleshoot any problems as they arise. The other 4 site-specific emails serve to primarily update sites on processes that are running correctly, though they may also contextualize detected issues. \\

Recall that sites are required to review a subset of transcripts for any potential issues with PII redaction (supplemental section \ref{subsec:u24-tm-ap}). On the data aggregation server(s), this data management process is handled by my pipeline, though an outside script performs the push of set aside transcripts to the appropriate site's Box/Mediaflux, and Lochness handles the initial pull of approved transcripts back to the raw side of the data aggregation server. Because a major issue encountered during mock interview testing was sites failing to return transcripts sent for redaction review or doing so incorrectly within Box/Mediaflux, an email feature was added into my pipeline to send a daily alert to the relevant site representatives when any transcriptions are still awaiting site redaction review, with a list of their (already deidentified) file names. 

In the real data collection process thus far, there have been fewer problems with transcriptions going missing for long periods during the site manual review phase, perhaps in part because of this email system -- though for a few specific sites response time problems remain, ongoing for many months. The alerts about manual site redaction review are currently the only alerts that are actually sent to contacts at each individual site, though all the other described site-specific email alerts have the capability to be sent to site-specific contacts. There are currently two email address list setting for each site config, one for the addresses to send this transcript review alert email, and one for the addresses to send all the rest of the above-described alerts; the latter is only being sent to central monitoring rather than actual sites at this time. For more information, please see the code use instructions in supplemental section \ref{subsec:interview-code}. \\

Although sites are not currently receiving many of these site-specific emails, the information being directed to central monitoring has been invaluable in detecting problems that need to be communicated to sites, particularly the warnings email. The SOP information that the warnings alert contains includes violations that are not possible to detect via the more broadly shareable QC metrics that are the larger focus in this subsection. Ideally, sites will begin receiving these emails on days when new warnings are detected, to ease the burden on central monitoring. In the meantime, I scan through new site-specific update emails, using email filters to manage the process across all sites. Future monitors can review the code details provided within section \ref{subsec:interview-code} to design filters that will capture only particular site-specific emails as needed. When I find problems with newly uploaded interviews, I manually compile them into a contextualized list to send weekly to a collaborator who then communicates with the sites about recurring issues and necessary action items for interview upload mistakes or bad quality standards.

For a list of common mistakes I've encountered (largely detected via site-specific warning emails) and how they should be avoided per the SOP, please see the upcoming subsection \ref{subsubsec:u24-issues}. The SOP itself is also described in more detail above (section \ref{subsubsec:ampscz-interview-protocol}). Note that a concatenated CSV with information on all pre-QC SOP violations detected across each site for Pronet and for Prescient is sent as a separate weekly update email to central monitoring, alongside the QC weekly server-wide summary emails that will be discussed at length shortly. The pre-QC warnings log is sent separately because as mentioned it can contain PII in some instances, and thus needs to be restricted to a limited recipient list. As the project continues, the frequency of these violations should decrease, and sites should be better equipped to address problems themselves. 

However, because of the many types of issues and the potential involvement of PII, as well as the necessity of reaching a wide diversity of sites to actually correct their issues and limit future mistakes, much of this violation review process is intended to remain manual throughout the project lifetime, assisted by the described email alerts. Furthermore, due again to the breadth of issues encountered, as well as the implementation details of Lochness (to be discussed in \ref{subsubsec:u24-storage}), it is not feasible to automatically determine when a previously detected issue in a given interview has been fixed. The updates made by the pipeline focus on natural steps in the processing of correct files and on newly detected warnings. Information from these updates can be used by a human to assess whether problems with prior interviews have since been addressed, as part of the monitoring workflow. 

Now that the daily email functionalities of my code and their use in the AMPSCZ project has been detailed, I will move onto closer discussion of QC metrics and their use -- both through values visualized per interview instance on the project's DPDash tool and through a weekly distributional characterization update built into my pipeline. These resources assist in quality monitoring, and together with the pre-QC warnings described, allow for a great variety of problems to be quickly and centrally noticed. \\

\begin{figure}[h]
\centering
\includegraphics[width=\textwidth,keepaspectratio]{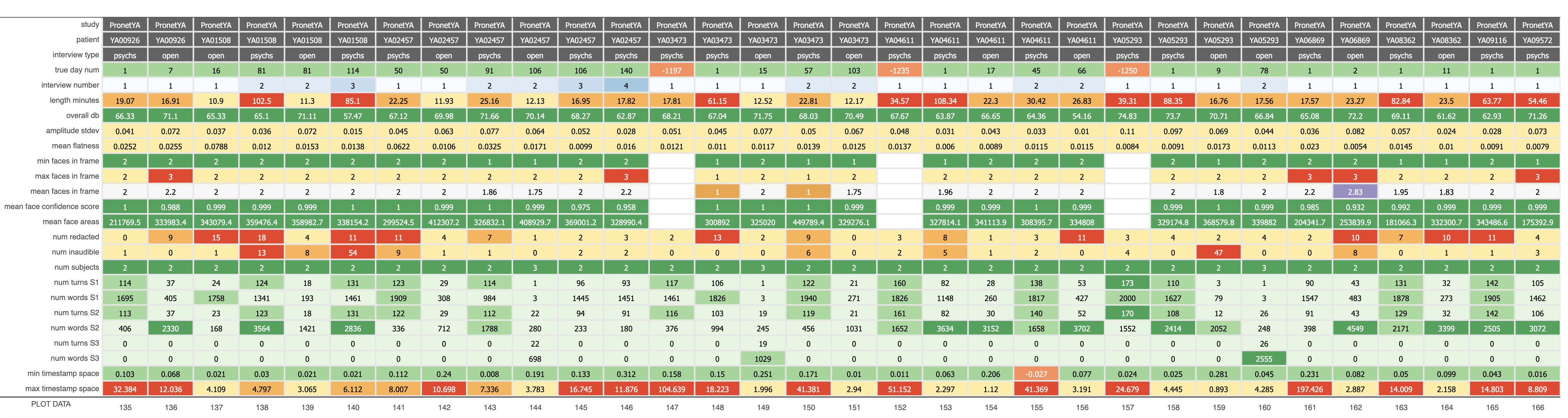}
\caption[Example use case of DPDash for monitoring interview recordings submitted by the YA site of the AMPSCZ project.]{\textbf{Example use case of DPDash for monitoring interview recordings submitted by the YA site of the AMPSCZ project.} In this DPDash view, all interviews submitted by site YA - regardless of subject ID or interview type - can be viewed side by side. Each colored row corresponds to a QC feature for audio, video, or transcript data from each interview, with a green to red color scheme designed to draw attention to outlier or potentially problematic features in red. White indicates data missingness (not all psychs interviews include video per the project protocol). The dashboard is accessible to study staff with accounts via web browser, and it updates automatically as my code process new interviews and Lochness subsequently pushes the updated QC outputs to the DPACC server (Figure \ref{fig:u24-arch}) where DPDash can import them. Note there is also a per subject DPDash view that displays interview QC stats on the study day (days since consent) it was conducted. This is well-suited for more frequent digital phenotyping datatypes, but it leaves large gaps in the interview timeline and is therefore less useful here.}
\label{fig:dpdash}
\end{figure}

\paragraph{DPDash CSVs.}
Another major output of the code is a larger set of QC features for the audio, video, and transcript modalities, formatted as a CSV for direct import into DPDash (Figure \ref{fig:dpdash}). The online study dashboard is thus automatically updated with quality control information about all modalities of an interview as the data comes through, making it much easier for staff to keep tabs on interview data collection progress and spot potential issues quickly.  

The QC metrics computed by the code include duration (minutes), volume (db), and spectral flatness (related to presence of background static) from the mono interview audio recording; basic face-detection related information via PyFeat \citep{pyfeat} outputs extracted from a subset of sample frames of the interview video; and transcript summary information derived from TranscribeMe notation, including information about the number of speakers IDed, the verbosity, the per turn timestamps, and the number of words that TranscribeMe redacted or marked as inaudible. All QC features extracted by the pipeline are detailed in the code documentation of supplemental section \ref{subsec:interview-code}. Many of them are included in the DPDash heatmap view, as can be seen in Figure \ref{fig:dpdash}. 

A subset of the QC metrics have been of particular relevance in identifying problematic files. For those features, I will do a deeper dive next. Before proceeding though, it is important to reiterate that only interviews with QC successfully processed for at least one modality will appear in the DPDash view or in the QC summary statistics to be described imminently. Interviews with major SOP violations that prevent processing are instead caught only by the pipeline's site-specific monitoring system detailed above, in part because logging information about these interviews may contain PII. 

Ideally, sites will monitor the interview QC DPDash dashboards of their own interviews, to function as their own quality assurers. However, just as was mentioned about the site-specific pre-QC warnings, sites are not independently managing monitoring at this time. Central monitoring staff presently review DPDash for each site on a periodic basis, and alert sites with issues accordingly. I supplement this process through monitoring for outliers on the server-wide QC metric distributions to be detailed shortly. Many of the observations I make serve as a good proof of concept for the use of these QC metrics in easily detecting a variety of concerns, and should lay the foundation for an even quicker monitoring process for future central project staff. \\

\FloatBarrier

\paragraph{DPChart summaries.}
Before diving into more detail on the QC features extracted and related observations to date, it is worth mentioning that for the AMPSCZ project, DPDash has been adapted to include a chart view for bird's eye project monitoring. The aim is to define an acceptable interview or an on-track subject ID using simple criteria, so that bar charts for each site can be quickly scanned together to determine if particular sites are failing basic requirements or otherwise falling behind. This can be helpful in quickly spotting issues that may require closer follow-up, and is additionally useful in evaluating overall project status for the interview modality (and other modalities which are all performing a similar process). Such a bird's eye view is particularly important for those that are overseeing the broader AMPSCZ project and want to track basic updates across the different datatypes and aims. Summaries displayed in DPChart are included in progress review meetings with the NIMH, and as such they can add accountability for proper execution of individual portions of the project as well as more broadly influence future funding decisions in this long term data collection initiative.

\begin{figure}[h]
\centering
\includegraphics[width=\textwidth,keepaspectratio]{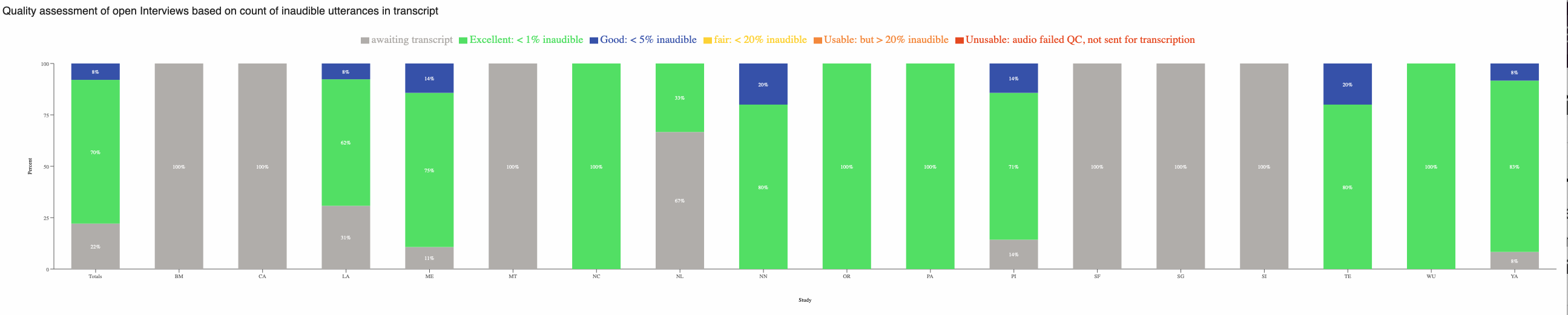}
\caption[Example use case of DPChart for monitoring data collection across sites.]{\textbf{Example use case of DPChart for monitoring data collection across sites.} For a high level view of project data collection progress, CSVs are also generated for import into DPChart, a new functionality added to DPDash. Depicted here is a snapshot of open interview quality monitoring across all 17 active interviewing sites (14 from Pronet and 3 from Prescient) as of 1/31/2023. Each successfully uploaded open interview is given a numeric category based on the quality of the returned transcription, which are mapped by DPChart to the following labels: no transcript available yet (grey), fewer than 1 inaudible occurrence per 100 transcribed words (green), between 1 and 5 inaudibles per 100 words (blue), between 5 and 20 inaudibles per 100 words (yellow), greater than 20 inaudibles per 100 words (orange), or audio failed QC and was intentionally not sent for transcription (red). This produces bars for each site with the percentage of open interviews falling into each quality category, along with a total bar on the far left representing project-wide percentages. The corresponding absolute numbers can be found by hovering over each bar, or referring to the table that appears below the bars. It is clear that while some sites have greater issues with availability, quality of returned transcripts has been very good thus far. An analogous chart view also exists on DPDash for the psychs interviews, as well as a chart view that tallies open transcript availability per subject ID and expected study timepoint - the latter is described in more detail in the text. \newline As an aside, note that there is an odd bug in configuring DPChart that prevents the plaintext labels chosen to represent each category in the UI from containing a '-' character. If a dash is included in any chart label, counts of that label will be excluded from the figure without warning.}
\label{fig:dpchart}
\end{figure}

My interview pipeline integrates with DPChart by way of an additional script I wrote within the repository for AMPSCZ's DPDash utility functions. The script takes the DPDash QC CSVs described above that are output by the pipeline, and generates from them CSVs that are importable as a DPChart view. It does this for three different views of key interest to project organizers -- overall quality level of open interviews, overall quality level of psychs interviews, and ratings for subject IDs based on the availability of open interview transcripts at expected time points. These CSVs are regenerated on a daily basis, so that DPChart can be used to monitor high level project progress in real time. Figure \ref{fig:dpchart} shows an example of the open interview quality DPChart and details its implementation; an analogous chart is also generated for psychs interviews.

The DPCharts that focus on quality category are a simplified and visualized version of the interview quality categories used for the tables in the weekly summary emails, which are to be described shortly; more details on current project status and identified issues with respect to recording quality assurance will be covered at that time. First though, I will focus on the chart view of open transcript availability per subject ID, as this contributes information distinct from the monitoring of recording quality. \\

\FloatBarrier

\noindent For determining time points of open transcript availability based only on the interview QC CSVs output by my pipeline, I use the following (now automated) process:
\begin{enumerate}
    \item Identify all subject IDs with some QC metrics available for at least one interview recording.
    \item For each such subject ID, determine how many interviews there have been of open type with transcript QC metrics available.
    \item If none, the subject ID is assigned to the 0 open interviews category and the process continues to the next subject.
    \item If one, the study day number corresponding to that interview is checked. If the interview occurred less than 40 days since study enrollment, the ID is assigned to the baseline only category, otherwise it is assigned to the 2 month follow-up only category. The process then continues to the next subject.
    \item Otherwise, two conditions are checked on the set of open transcriptions (which really should have size 2, but that is not strictly enforced) to determine the subject ID's category before moving to the next one:
    \begin{itemize}
        \item The interview with the smallest study day number should have day $< 40$ - if so the script moves on to the next check, otherwise the subject is assigned to the 2 month follow-up only category.
        \item The gap between days of the interview with the largest study day number and the interview with the smallest study day number should be $\geq 40$ - if so the subject is assigned to the ideal category of completed open interview records, otherwise the subject is considered to have baseline transcript only.
    \end{itemize}
\end{enumerate}
\noindent DPChart then uses the category for each subject ID to create a view very similar to the quality chart of Figure \ref{fig:dpchart}, but expressing the percentage of subject IDs that fall into each of the four categories described. To supplement the DPChart view, I also automatically record more detailed information per subject - for example information on whether an interview time point is available but missing a transcript - into a CSV that is included along with the other attachments in the weekly server progress update emails (to be described).

The subject ID open availability accounting has already helped to identify certain site-specific issues. One such problem affected many of our other metrics as well: site LA has repeatedly been uploading psychs interviews incorrectly under the open category, with 4 such instances already verified as of 1/31/2023 and another suspected. This manifests in multiple ways in the subject ID transcript accounting; for one participant LA has uploaded 4 supposed open interview folders all from separate dates (only 2 open interviews should occur per subject), and for 3 other participants LA has uploaded 2 open interviews each with an abnormally short gap between them. In the latter case, the gaps are 1, 7, and 21 days for each of the 3 affected participants -- while a gap of 21 days is arguably acceptable for the second open interview even if not ideal, the others are a clear sign of incorrect procedure.

Another problem detected in the availability accounting is 2 separate PA subjects that had multiple open interview uploads from a single day, resulting in 2 transcripts for each ID corresponding to the exact same time point. This is likely explained by accidental splitting of the recording, which would explain the short length observed in some of PA's open interviews - because 4 of the data points are actually covering just 2 real time points. More generally, this accounting has provided a detailed picture of study collection progress. For details on the status observations made from across the 14 active Pronet sites as of 1/31/2023, please see supplemental section \ref{sec:open-timepoint-counts}.

Based on these results, I am now in the process of verifying guidelines for open interview time points, including what range of gaps between first and second interview would be considered ideal and how closely after consent the first interview should be occurring. In parallel, other monitoring staff are looking more closely into specific subject IDs identified as possibly having baseline missingness issues. With additional clarity, the exact procedure for defining availability categories can be updated and guidance can be issued to sites as needed.

In the future, closer integration with metadata and other interview notes being entered into REDCap/RPMS could further improve the accuracy of availability accounting and elucidate a site's intentions with a seemingly incorrect upload. Presently, some steps of the manual monitoring workflow do involve referring back to the REDCap/RPMS interface, but there is not any code that filters through raw data pulled from REDCap/RPMS about the interviews, meaning it would be difficult to merge that information with interview pipeline outputs at this time. 

Nevertheless, with only the info it extracts from raw interview upload data, the pipeline is still able to perform a wide variety of organizational and quality assurance accounting. To demonstrate, I will now discuss properties of QC metrics from across the AMPSCZ interview dataset thus far, along with examples of problems these metrics have been able to identify in practice. \\

\paragraph{Detailed project-wide monitoring.}
Utilizing the QC outputs, distributions of key QC features across all sites and within each site for a given central server (here Pronet vs Prescient) are generated weekly and saved in a PDF, which is then emailed to those in charge of monitoring interviews and communicating with sites for that central server. Some example distributions from Pronet data collection thus far will be presented shortly.

Based on the SOP and the observed QC distributions, summary counting statistics were designed so that updated counts of the number of quality (based on modality-specific QC thresholds) interviews processed to date per site and interview type are included in that same email. Tables \ref{table:pronet-counts} and \ref{table:prescient-counts} (for Pronet and Prescient respectively) provide this information from AMPSCZ collection progress as of mid-January 2023. Because Prescient has just once site actively collecting production interview data at this time, and the quality of their collection process has been good, the discussion of monitoring results going forward will focus on Pronet.

\begin{table}[!htbp]
\centering
\caption[Example server-wide summary table from Pronet weekly interview status report email.]{\textbf{Server-wide summary table from Pronet weekly interview status report email.} An HTML-formatted table is generated weekly tallying quality interview counts across Pronet sites and open versus psychs interviews. The values from the monitoring report of 1/18/2023 are reproduced here. As described above, the total count of successfully processed interviews so far (overall and for each modality) is given, along with counts of interviews that meet the good or partial quality thresholds listed. Finally, mean interview length for each site and interview type is reported in minutes.}
\label{table:pronet-counts}

\begin{tabular}{ | m{0.6cm} | m{1.1cm} || m{0.8cm} || m{0.8cm} | m{0.9cm} | m{0.8cm} || m{0.8cm} | m{0.9cm} | m{0.9cm} | m{0.9cm} || m{1.1cm} | }
\hline
\textbf{Site ID} & \textbf{Type} & \textbf{Total \#} & \textbf{Vid. \#} & \textbf{Good Vid. \#} & \textbf{Ok Vid. \#} & \textbf{Aud. \#} & \textbf{Trans. \#} & \textbf{Good Trans. \#} & \textbf{Ok Trans. \#} & \textbf{Mean Aud. \newline Length} \\
\hline\hline
LA & open & 17 & 15 & 12 & 3 & 15 & 11 & 11 & 0 & 36.1 \\
\hline
LA & psychs & 22 & 20 & 14 & 3 & 20 & 13 & 13 & 0 & 76 \\
\hline\hline
NC & open & 13 & 13 & 12 & 0 & 13 & 13 & 13 & 0 & 20 \\
\hline
NC & psychs & 28 & 28 & 7 & 21 & 28 & 28 & 28 & 0 & 39.6 \\
\hline\hline
NL & open & 4 & 4 & 4 & 0 & 4 & 0 & 0 & 0 & 13.5 \\
\hline
NL & psychs & 4 & 4 & 3 & 0 & 4 & 2 & 2 & 0 & 60.8 \\
\hline\hline
NN & open & 8 & 8 & 7 & 1 & 8 & 8 & 7 & 1 & 15.9 \\
\hline
NN & psychs & 13 & 13 & 8 & 1 & 13 & 9 & 9 & 0 & 78.9 \\
\hline\hline
OR & open & 6 & 6 & 3 & 3 & 6 & 6 & 6 & 0 & 16.1 \\
\hline
OR & psychs & 11 & 3 & 2 & 0 & 11 & 11 & 11 & 0 & 39.9 \\
\hline\hline
PA & open & 8 & 8 & 6 & 1 & 7 & 7 & 7 & 0 & 14.1 \\
\hline
PA & psychs & 11 & 11 & 1 & 7 & 8 & 8 & 7 & 1 & 77.1 \\
\hline\hline
PI & open & 5 & 5 & 0 & 0 & 5 & 1 & 0 & 1 & 10 \\
\hline
PI & psychs & 9 & 9 & 1 & 1 & 9 & 7 & 7 & 0 & 53 \\
\hline\hline
TE & open & 5 & 4 & 4 & 0 & 5 & 5 & 4 & 1 & 15.6 \\
\hline
TE & psychs & 8 & 8 & 5 & 3 & 8 & 6 & 6 & 0 & 71.6 \\
\hline\hline
WU & open & 8 & 8 & 7 & 1 & 8 & 8 & 8 & 0 & 15.6 \\
\hline
WU & psychs & 27 & 9 & 6 & 3 & 25 & 25 & 23 & 2 & 45.5 \\
\hline\hline
YA & open & 14 & 14 & 10 & 4 & 14 & 14 & 13 & 1 & 17.2 \\
\hline
YA & psychs & 27 & 20 & 18 & 0 & 26 & 26 & 25 & 1 & 46.3 \\
\hline
\end{tabular}
\end{table}

\begin{table}[!htbp]
\centering
\caption[Example server-wide summary table from Prescient weekly interview status report email.]{\textbf{Server-wide summary table from Prescient weekly interview status report email.} As in Table \ref{table:pronet-counts}, an analogous workflow occurs for the Prescient central server. The values from the monitoring report of 1/18/2023 are reproduced here. Because there is only one Prescient site actively collecting interview data at this time, there is minimal additional information to report from that weekly summary email. However, it is worth noting that site ME is doing a good job with quality data collection thus far.}
\label{table:prescient-counts}

\begin{tabular}{ | m{0.6cm} | m{1.1cm} || m{0.8cm} || m{0.8cm} | m{0.9cm} | m{0.8cm} || m{0.8cm} | m{0.9cm} | m{0.9cm} | m{0.9cm} || m{1.1cm} | }
\hline
\textbf{Site ID} & \textbf{Type} & \textbf{Total \#} & \textbf{Vid. \#} & \textbf{Good Vid. \#} & \textbf{Ok Vid. \#} & \textbf{Aud. \#} & \textbf{Trans. \#} & \textbf{Good Trans. \#} & \textbf{Ok Trans. \#} & \textbf{Mean Aud. \newline Length} \\
\hline\hline
ME & open & 12 & 12 & 11 & 1 & 12 & 12 & 11 & 1 & 24.5 \\
\hline
ME & psychs & 25 & 23 & 19 & 2 & 25 & 23 & 22 & 1 & 46.2 \\
\hline
\end{tabular}
\end{table}

Ultimately, the goal of the high level summary portion of monitoring is to identify sites that are repeatedly making certain types of mistakes or otherwise encountering recurring quality problems of a particular kind, so that future interviews can hopefully be improved upon. It also helps give a broader view of study progress and general properties of the collected interviews. Detailed information on how the weekly accounting tables are compiled can be found in supplemental section \ref{sec:dumb-table}. 

The rest of the current subsection will focus on feature histograms produced for the same weekly summary email, as these allow for a larger number of features to be quickly checked across sites and interview type with more potential nuance. Indeed, they were used to decide on thresholds for the table in the first place! Of course the goal of the counting stats are quite different, meant moreso for project monitors that want a quick sense of absolute progress each site is making towards interview recording collection. \\

\noindent The server-wide distribution PDFs generated as part of the weekly monitoring report email are another great source for a more nuanced understanding of the current dataset composition. The (original) full set of features that histograms are generated for, to inform and expand upon the summary counts, are as follows:
\begin{itemize}
    \item Audio length in minutes
    \item Mean number of faces detected per extracted video frame
    \item Number of words marked inaudible by TranscribeMe per total words in the transcript
    \item Number of words redacted by TranscribeMe per total words in the transcript
    \item Number of unique speakers identified by TranscribeMe
    \item Total words in the transcript
    \item Number of words marked inaudible by TranscribeMe
    \item Number of words redacted by TranscribeMe
    \item Overall audio volume in decibels
    \item Mean spectral flatness of the audio
    \item Total number of speaker turns in the transcript
    \item Mean area of the faces detected across extracted video frames
\end{itemize}

\noindent Note that each of these features has a stacked histogram created both for all interviews processed from that central server, colored by interview type (open or psychs), as well as stacked histograms created separately for only open and for only psychs interviews, with each of those colored by interview site. For select features with clearly interpretable units, specific bin ranges are pre-specified to highlight relevant QC categories. 

\begin{figure}[h]
\centering
\includegraphics[width=\textwidth,keepaspectratio]{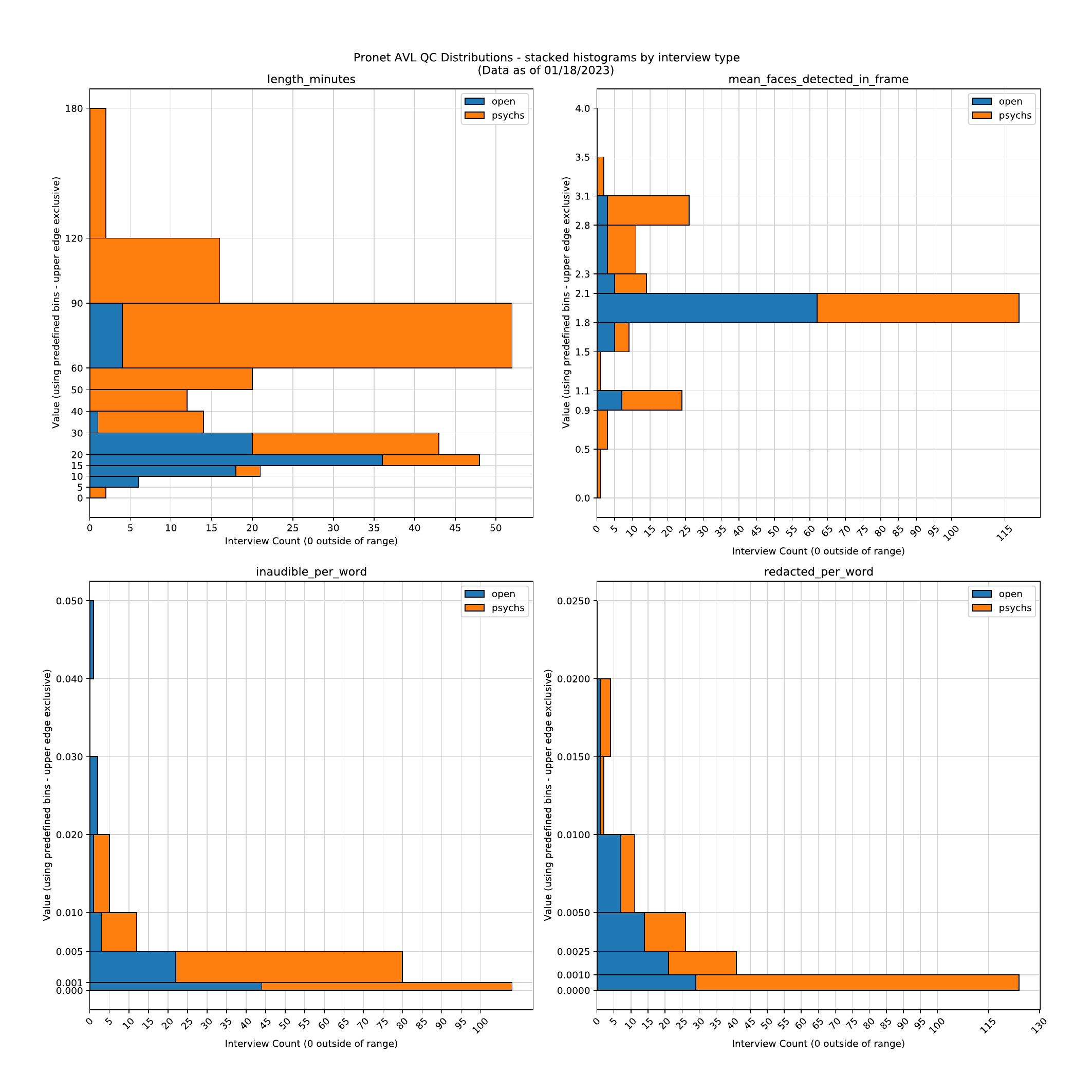}
\caption[Snapshot of key interview QC feature distributions from the Pronet server, colored by interview type.]{\textbf{Snapshot of key interview QC feature distributions from the Pronet server, colored by interview type.} Stacked histograms depicting the distributions of relevant quality control metrics across interviews are generated weekly for each central server. Within each distribution, interviews are marked by their type - for the AMPSCZ project, open (blue) versus psychs (orange). An example of the first page of this weekly distributions summary is presented here, taken from the Pronet project on 1/18/2023. It represents the four most important features for high level monitoring: interview duration in minutes (top left), mean number of faces detected from extracted video frames (top right), number of words marked inaudible per total transcript words (bottom left), and number of words redacted per total transcript words (bottom right).}
\label{fig:pronet-key-dists}
\end{figure}

\begin{figure}[h]
\centering
\includegraphics[width=\textwidth,keepaspectratio]{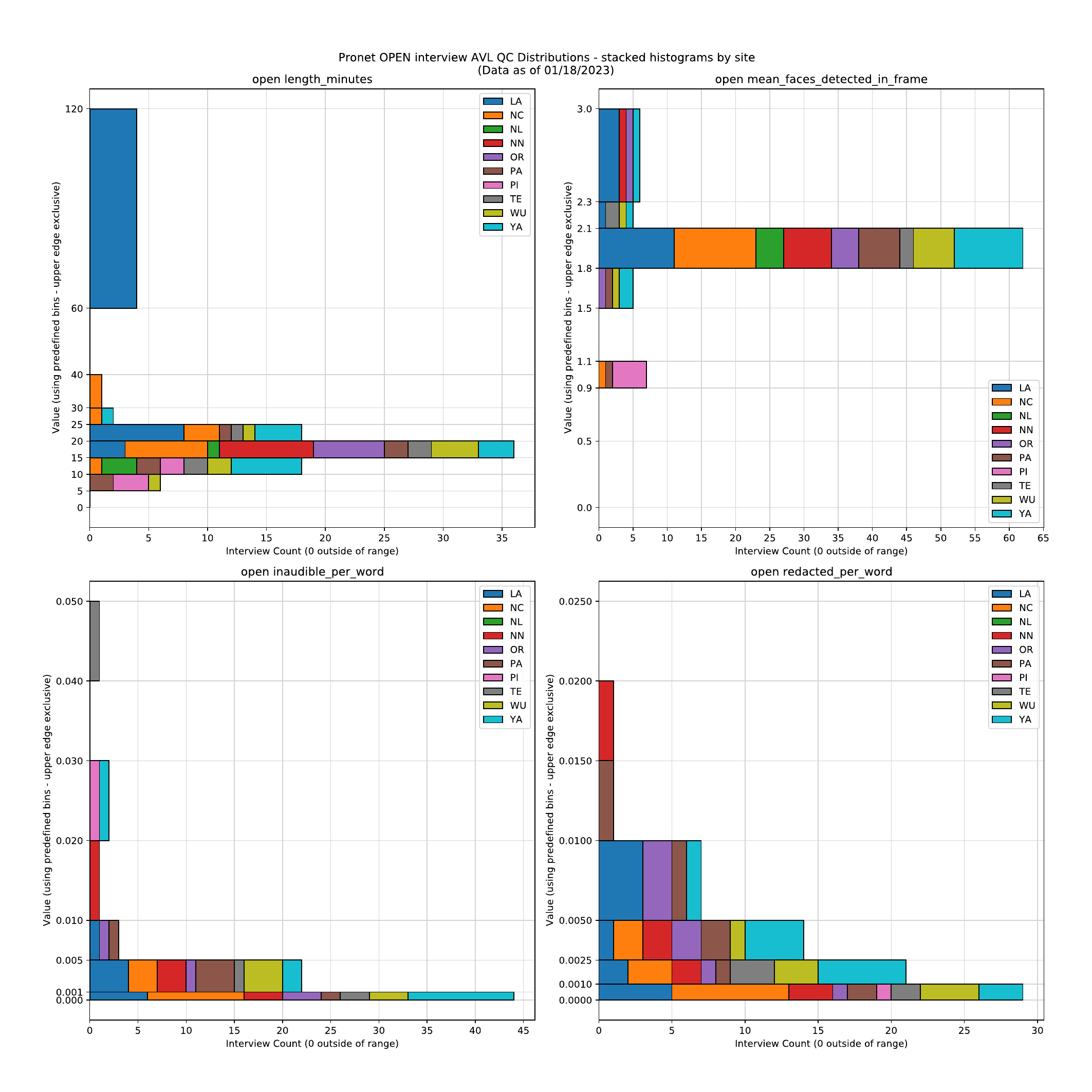}
\caption[Snapshot of key interview QC feature distributions from open interviews on the Pronet server, colored by source site.]{\textbf{Snapshot of key interview QC feature distributions from open interviews on the Pronet server, colored by source site.} Distributions for the same quality control metrics found in Figure \ref{fig:pronet-key-dists} were also generated restricted to open interviews only, with stacked histograms now marking interview site.}
\label{fig:pronet-open-key-dists}
\end{figure}

\begin{figure}[h]
\centering
\includegraphics[width=\textwidth,keepaspectratio]{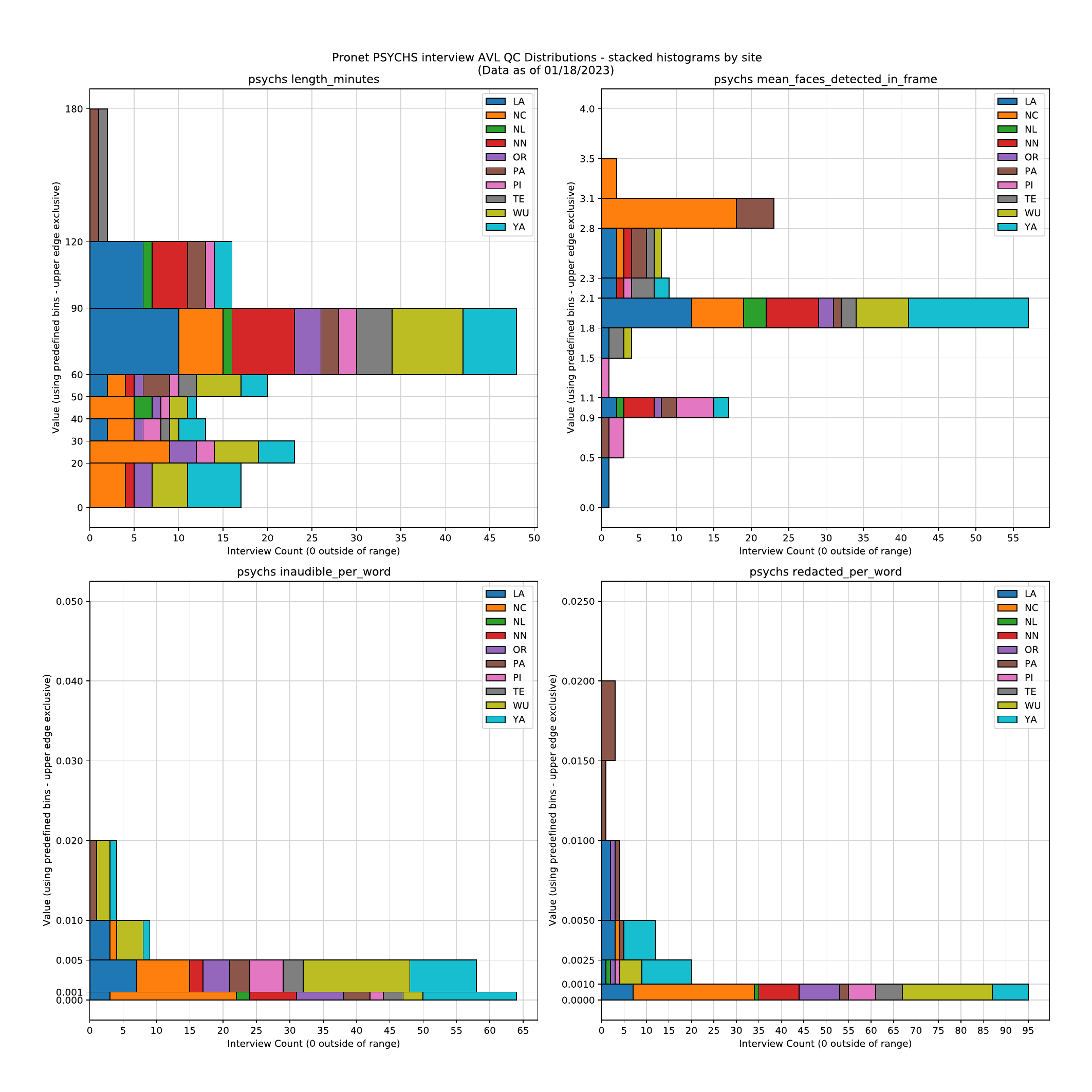}
\caption[Snapshot of key interview QC feature distributions from psychs interviews on the Pronet server, colored by source site.]{\textbf{Snapshot of key interview QC feature distributions from psychs interviews on the Pronet server, colored by source site.} As was shown for open interviews in Figure \ref{fig:pronet-open-key-dists}, the same site-labeled QC feature distributions were also generated for psychs interviews in the same way.}
\label{fig:pronet-psychs-key-dists}
\end{figure}

\begin{figure}[h]
\centering
\includegraphics[width=\textwidth,keepaspectratio]{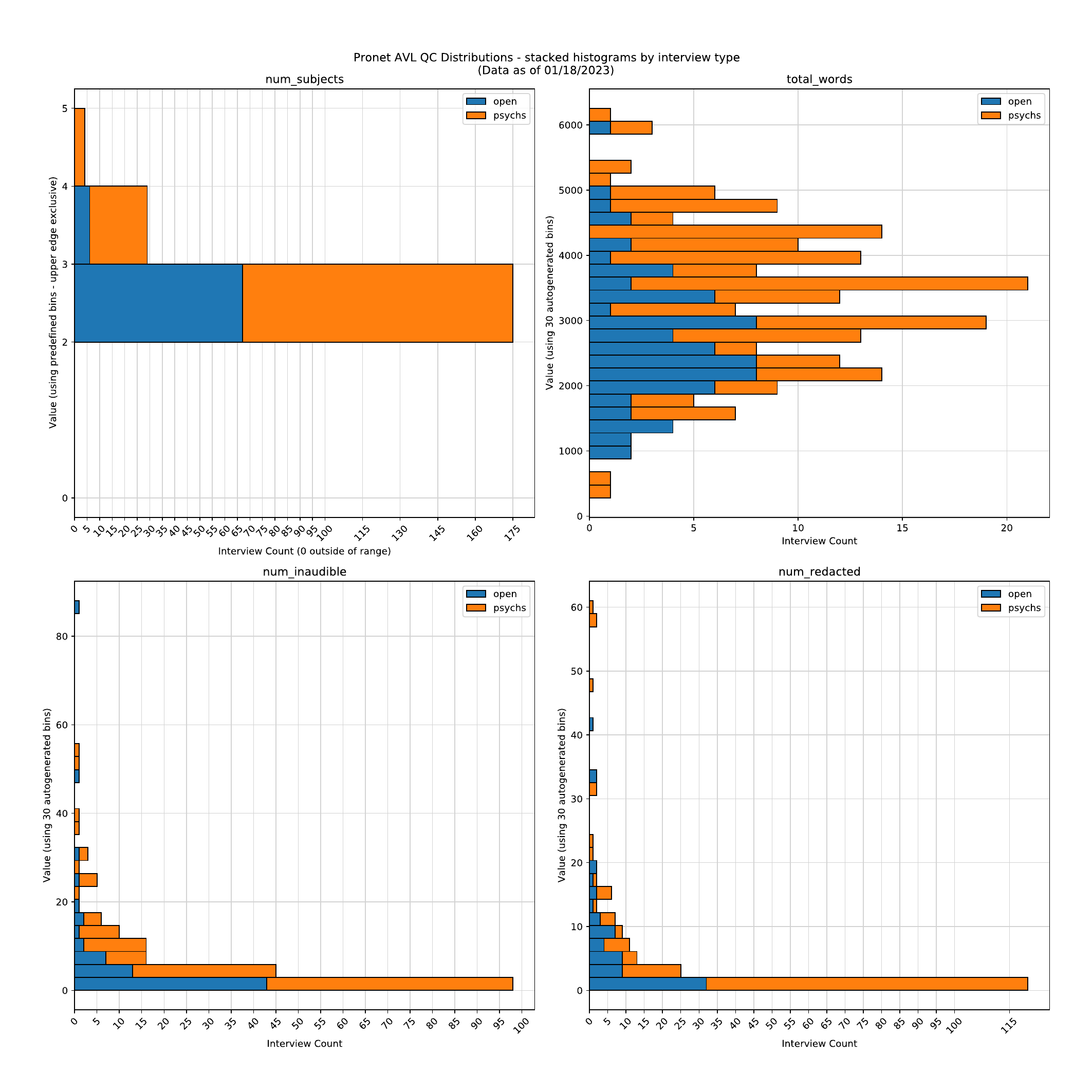}
\caption[Snapshot of additional interview transcript feature distributions from the Pronet server, colored by interview type.]{\textbf{Snapshot of additional interview transcript feature distributions from the Pronet server, colored by interview type.} Alongside the distributions depicted in Figure \ref{fig:pronet-key-dists}, distributions of other potentially relevant QC metrics are also generated weekly. Sourced from the same PDF report as Figure \ref{fig:pronet-key-dists}, stacked histograms for the following features are shown here: number of unique speakers in the transcript (top left), total word count of the transcript (top right), unnormalized inaudible word count (bottom left), and unnormalized redacted word count (bottom right).}
\label{fig:pronet-supp-dists-trans}
\end{figure}

\begin{figure}[h]
\centering
\includegraphics[width=\textwidth,keepaspectratio]{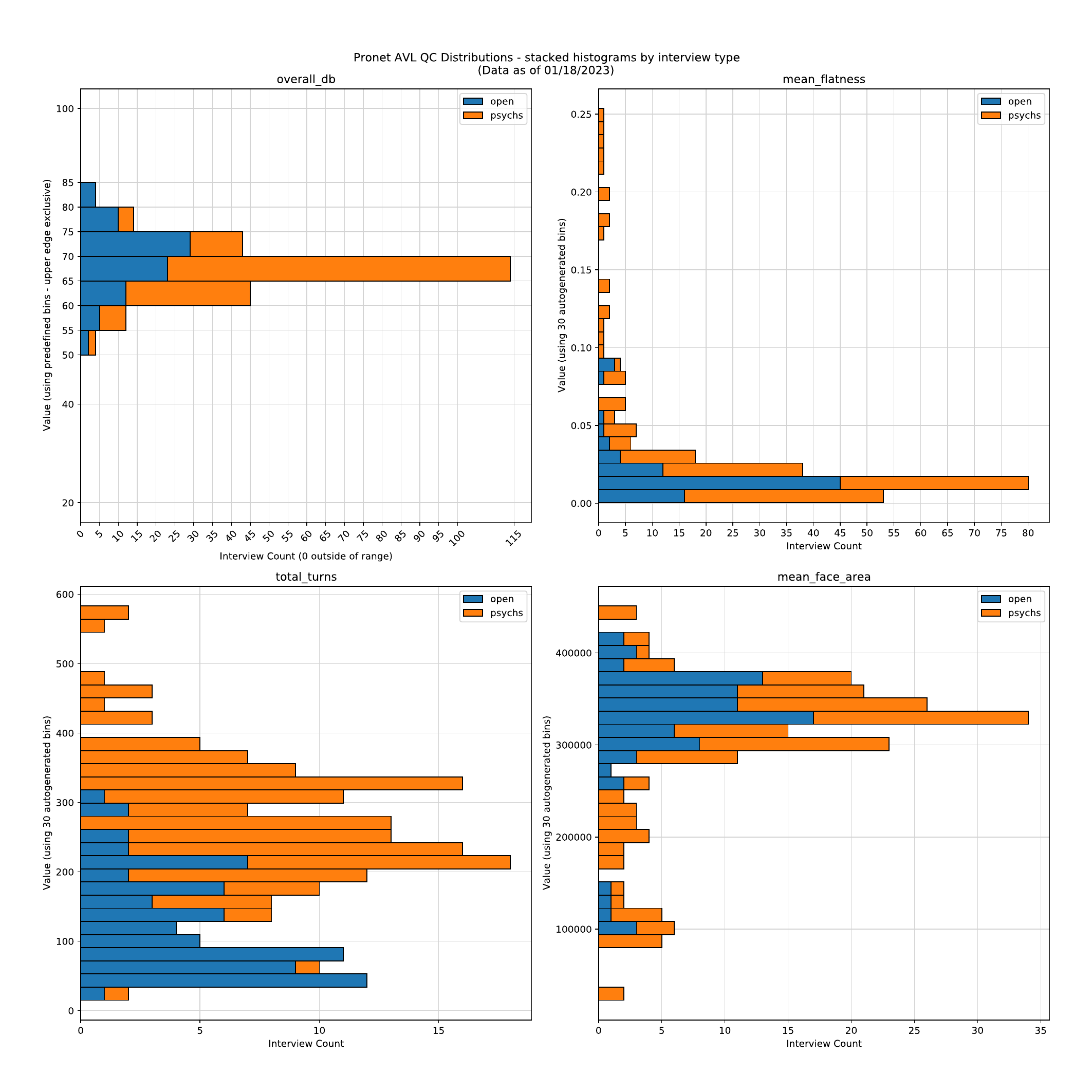}
\caption[Snapshot of additional multimodal interview feature distributions of note from the Pronet server, colored by interview type.]{\textbf{Snapshot of additional multimodal interview feature distributions of note from the Pronet server, colored by interview type.} Like in Figure \ref{fig:pronet-supp-dists-trans}, distributions of other potentially relevant features, to supplement Figure \ref{fig:pronet-key-dists}, are presented. The final set of QC metrics are volume in dB (top left) and mean spectral flatness (top right) of the audio, number of speaker turns in the transcript (bottom left), and mean area of detected faces (bottom right).}
\label{fig:pronet-supp-dists-avl}
\end{figure}

For the emailed status update, histograms are grouped 4 to a page in the order of the above list, and a separate 3-page PDF is generated for each of the "all by type", "open by site", and "psychs by site" distribution views. Figures \ref{fig:pronet-key-dists}, \ref{fig:pronet-open-key-dists}, and \ref{fig:pronet-psychs-key-dists} depict an example of the first page of each of these respective PDFs. Figures \ref{fig:pronet-supp-dists-trans} and \ref{fig:pronet-supp-dists-avl} depict pages two and three, respectively, of the "all by type" PDF. \\

\noindent These distributions can provide valuable information on what issues different sites are struggling with, and give a good reference point for expected values of various QC features. In fact, a number of relevant observations have already come from a review of the server-wide distribution plots and table generated by the code for weekly summary, including:
\begin{itemize}
    \item There were a few very long open interviews submitted by LA, which are probably mistaken uploads that were supposed to go under the psychs interview type. LA has made a large number of mistakes with the interview upload process in general.  
    \item Certain sites have submitted open interviews under 10 minutes, which is not an acceptable length per the SOP. The original aim was for open interviews to be $\sim 20$ minutes, and even while this goal has been relaxed somewhat, open interviews should really be close to 15 minutes at least. Site PI is particularly problematic in this regard so far.
    \item PI has also submitted many interviews in speaker view, and as a result the number of faces per frame feature has hovered around 1 for this site in both distributions. Other sites have also had this issue pop up in one or two interviews. For all Zoom interviews the view should be in gallery mode, and for open interviews there should be consistently 2 faces per frame. 
    \item There were 9 interviews (5 open) in the early dataset with between 1.5 and 1.8 mean faces per frame, which indicates a correct Zoom view but potential quality issues with faces staying on camera. Indeed, manual review indicated in these open interviews that the participant drifted partially out of view on multiple occasions, demonstrating the usefulness of the automatic QC.
    \item The single open transcript that PI had returned was particularly bad quality, with more than 1 inaudible every 50 words. This is not necessarily unusable, but it is non-ideal, and 3 of their 5 open audio recordings submitted have had volume below 60 db, on the low end of the distribution. As more transcripts are returned, there may turn out to be a trend with well above average inaudible frequency for PI transcripts. 
    \begin{itemize}
        \item Furthermore, there has been an extended delay in their manual redaction review process, which is why they still have so few transcripts reflected -- thus emphasizing the importance of following the interview upload protocol so that regular monitoring can catch issues before quality concerns pile up.  
    \end{itemize}
    \item The rate of inaudibles in psychs interviews appears to be lower overall, but within this distribution sites WU and LA have the worst statistics, particularly WU. Interestingly, LA audio tends to have higher mean flatness than normal, but good volume, while WU has a few audio files with db < 60 and a sizeable portion with db < 65, but is low on the flatness distribution. WU does not any lower volume submissions on the open interviews by contrast, and it can be seen from the number of psychs videos submitted by WU that they are particularly likely to use the hardware recorder for psychs interviews instead of Zoom. OR is the other site most likely to use the hardware recorder and does not have as much of a pronounced trend with volume or inaudibles, but it would be consistent with our experience with lab hardware for the onsite methodology to be more finicky based on how it was set up.
    \item There might be something odd about the way PA conducts interviews or the way those interviews are being transcribed, though it is still early in the study and could be a participant-specific factor. Their transcripts consistently have higher inaudible frequency, though not to an extent that would be considered problematic for analysis. Moreover, they are one of the highest sites on the number of turns in open interview transcripts, but one of the lowest on the number of words in those transcripts, which could indicate more interjection on the part of the interviewer. 
    \begin{itemize}
        \item This observation about the discrepancy in PA's word count and turn count distributions relative to other sites it what triggered a closer look at relationships between difference QC features across interviews from different sites, perhaps the main result of this chapter due to its (site non-specific) impactful downstream results on AMPSCZ speech sampling. 
    \end{itemize}
    \item Note also that in psychs interviews, PA has had one of the highest rates of redaction, which might relate to actual differences in interview content or to differences in assigned transcriber. It is important to double check that systematic bias cannot be introduced by TranscribeMe's method for assigning transcribers (across English-language sites), which should be random and not influenced by site, subject ID, study day, or file size.
    \begin{itemize}
        \item Given the multiple distinct anomalies found in PA's early distributions across the interview types, it would be a good idea to check in on their interview content more closely with a trained human reviewer.
    \end{itemize}
    \item Especially early on in a project, there will also be false alarms triggered by QC metrics. An early potential issue flagged here that was later assuaged was the high number of transcripts coming back with 0 redactions. It is of course much better to flag this and realize upon manual review that it is not a problem than vice versa. 
    \begin{itemize}
        \item When I discuss common problems encountered across stages of the pipeline in section \ref{subsubsec:u24-issues}, I will provide more information on our observations with redactions. In sum, a manual review of transcripts that contained no redactions indicated that there was indeed no redaction-worthy PII found, and TranscribeMe is generally doing a good job with PII. In hindsight, it is not too surprising that many of the long psychs interviews would come back with no redactions, because of the structure of those interviews (to be dissected shortly) and the current transcription plan. 
        \item Occasionally a transcriber will be extremely aggressive with redactions, removing words like "design" in the sentence "Um. Ever since high school, I've always just been into [redacted]". Therefore we now keep a closer eye out for extreme high redaction counts in the QC, to be prepared for intervention if redactions begin needlessly and regularly impacting the usability of transcripts for analyses - though this is ultimately unlikely. 
        \item The described manual site review of a random sample of transcripts for redaction accuracy will of course continue, and if anything arises there or the frequency of 0 redaction transcripts further increases we can revisit whether these transcripts are a potential problem. 
    \end{itemize} 
\end{itemize}
\noindent The above notes were generated from reviewing the Pronet summaries as of 1/18/2023, using stats and visuals generated by the pipeline, as exemplified in Tables \ref{table:pronet-counts}-\ref{table:prescient-counts} and Figures \ref{fig:pronet-key-dists}-\ref{fig:pronet-supp-dists-avl}. They are exemplary of the process that goes into AMPSCZ interview quality monitoring and the actionable suggests that can results for sites and their coordinators. 

Recall though that this is just one piece of overall data monitoring for the project, because a relatively large number of interviews do not make it through the QC process at all due to issues with e.g. file organization. Those issues are also tracked (in part) by my pipeline, but they are not especially interesting, whereas a variety of insights can be gained from reviewing the interview recording QC features from the early phases of the project. One important point is that results of quality metric monitoring can in turn lead to added features for future QC, to more easily capture issues initially noticed from careful visual review. In this way, building and maintaining the interview infrastructure can be a cyclical, iterative process that involves in some capacity all the necessary components of a collaborative interview speech sampling initiative. 

I will next detail a major example where further exploration of QC metrics lead to new important data collection and monitoring considerations. Two new pages of histograms were recently added to the weekly distribution PDFs with additional features of use for monitoring, in large part prompted by the upcoming observations. This review is also what truly began my push for changes to the transcription plan, which resulted in the report (and positive project outcome) of Appendix \ref{cha:append-ampscz-rant}. 

\FloatBarrier

\paragraph{Relationships between QC metrics in the interview dataset.}
To further characterize some of the QC feature observations made above, I also looked more closely into relationships between different metrics across the early interview dataset, through generation of a set of scatter plots. These plots depict QC features from all interviews successfully processed on the Pronet central server as of 1/23/2023. Each was generated using the scatterplot function from the Seaborn python package. Recall that some portion of interviews will not have QC data available across modalities (e.g. still awaiting transcript, no video because EVISTR was used, etc.) and therefore will not appear on scatters that compare metrics multimodally. Additionally, in cases with overlapping interview points, only one will be visible per the function defaults. 

Still, the majority of the interviews successfully processed by pipeline were able to be wholly represented on these scatters, and they have indeed assisted in forming a better understanding of the interview QC features, as well as of trends specific to site or interview type -- which ultimately has had an impact on how we will monitor the code's QC outputs and the interviews themselves going forward. Furthermore, the scatter plots can be connected to relationships previously uncovered between different audio journal QC metrics in chapter \ref{ch:1} (\ref{subsubsec:diary-val-qc})

While the scatter plots presented here are not built into the main pipeline and are thus not automatically generated each week nor included in any email summaries, a Jupyter notebook found in the supplementary materials of the repository will create a variety of inline scatter plots from the per interview combined QC CSV, which is both emailed out weekly and imported into DPDash daily. As such, the notebook can be easily used to review the latest versions of these relationships as the project continues. Moreover, additional feature distributions were added to the weekly histogram PDFs based on relationships uncovered during the review of these relationships, as already mentioned above.

\noindent The 8 new interview-level features included as 2 additional pages on all weekly interview histogram PDFs generated across AMPSCZ by \cite{interviewgit} are: 
\begin{itemize}
    \item The mean words per turn of the transcript.
    \item The fraction of total words spoken by the individual with the most words attributed in the transcript.
    \item The TranscribeMe speaker ID corresponding to the individual who spoke the most words (represents the order that each unique speaker spoke first in the recording).
    \item The study day (days since participant consent) that the interview was recorded on.
    \item The mean PyFeat face confidence score across the faces detected in sampled frames from the video.
    \item The lowest number of faces that PyFeat detected in a frame sampled from the video.
    \item The highest numbers of faces that PyFeat detected in a frame sampled from the video.
    \item The number of unique speaker IDs in the transcript divided by the highest number of faces detected in a frame from the corresponding video. 
\end{itemize}
\noindent This list in part serves as a preview for the feature relationships to be evaluated next, in order to wrap up the current subsection on example uses of AMPSCZ QC outputs. \\ 

\begin{figure}[h]
\centering
\includegraphics[width=\textwidth,keepaspectratio]{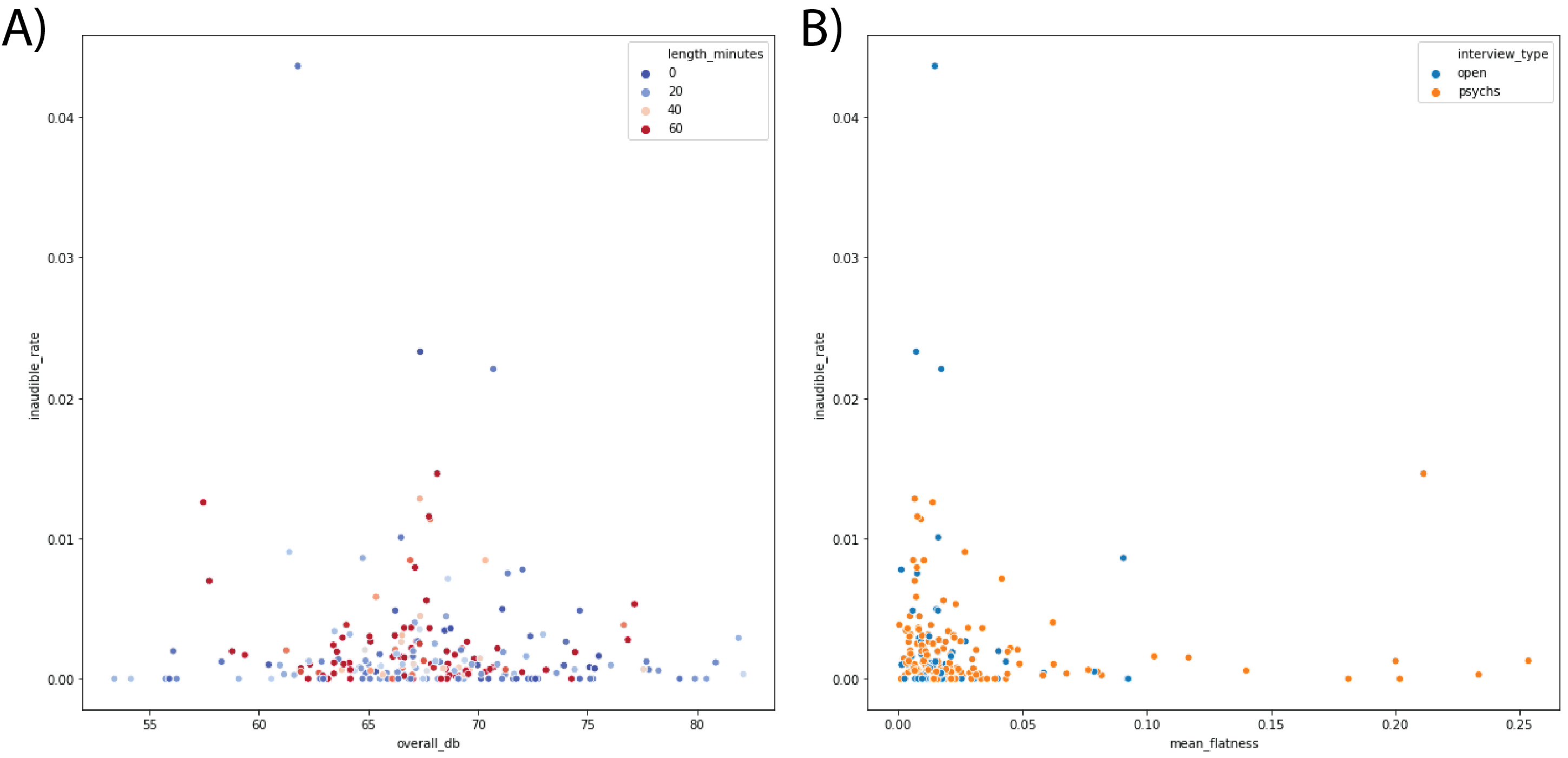}
\caption[Relationships between audio QC and transcript QC metrics across interviews processed on Pronet.]{\textbf{Relationships between audio QC and transcript QC metrics across interviews processed on Pronet.} As described above, scatter plots were generated to compare QC features of interest across all Pronet interviews to date. Here, key relationships amongst metrics derived from the mono interview audio files (and resulting transcripts) are depicted. On both plots, each point represents a single interview, with y-coordinate corresponding to the rate at which words were marked inaudible by TranscribeMe (units$=\frac{inaudibles}{words}$). On the left (A), the x-coordinate corresponds to the interview's overall volume (units$=db$) and the point's hue corresponds to its duration in minutes, colored from dark blue at 0 minutes to dark red at 60 or more minutes. On the right (B), the x-coordinate corresponds to the interview's mean spectral flatness feature and the point's hue corresponds to interview type: open (blue) or psychs (orange).}
\label{fig:pronet-aud-scatter}
\end{figure}

\noindent One important comparison to make was the relationship between our audio QC features and the quality of the resulting transcriptions, measured by the frequency of inaudible markings made by TranscribeMe (Figure \ref{fig:pronet-aud-scatter}). It is worth noting that an inaudible occurrence may correspond to a single inaudible word or it may correspond to an entire inaudible phrase, so that the units of the inaudible rate feature (number of inaudibles/total words) should not be taken entirely literally when interpreting sufficient quality thresholds. On the other hand, a high rate of inaudibles may be due to participant speech patterns that are not "correctable" or perhaps are even inherently clinically relevant, such as severe mumbling. Thus the visualized relationships are only meant to guide interpretation rather than to devise strict rules. In fact, the nuance in these features is exactly why it is critical to consider how different metrics may contextualize each other, rather than only considering each QC metric in isolation.

Although there was no definitive quality cutoff identified, lower volume ($< 68 db$) was associated with the higher inaudible rate transcriptions, particularly for those recordings of longer duration (Figure \ref{fig:pronet-aud-scatter}A). Similarly, audio files with a high spectral flatness value were more likely to come back with a high rate of inaudibles, but the relationship was far from deterministic (Figure \ref{fig:pronet-aud-scatter}B). Interestingly, high mean flatness was more common in psychs interviews, even when accounting for their longer duration, suggesting something about the recording protocol might impact the psychs audio files. This may have to do with the additional recording format allowed for psychs interviews, or it may relate to the number of speakers allowed on the call, or perhaps even the type of response expected from participants. Further investigation into the QC metrics and metadata produced by the pipeline would clarify how this information should (or should not) be used in future monitoring workflows.

\begin{figure}[h]
\centering
\includegraphics[width=\textwidth,keepaspectratio]{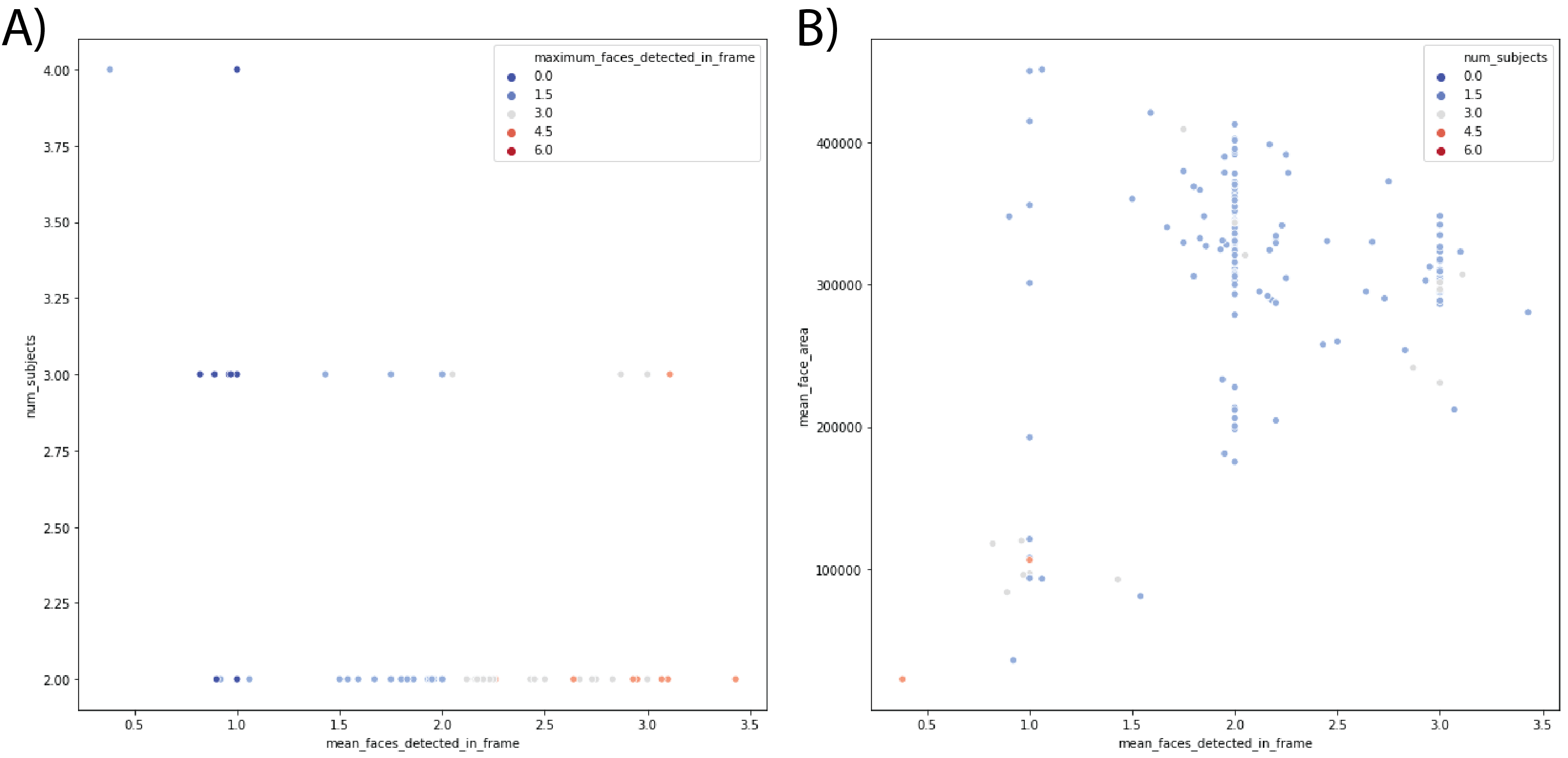}
\caption[Relationships between video QC metrics across interviews processed on Pronet.]{\textbf{Relationships between video QC metrics across interviews processed on Pronet.} Analogous to the audio scatters of Figure \ref{fig:pronet-aud-scatter}, similar plots were created to elucidate the properties of important video QC features across all Pronet interviews. The mean number of faces in extracted frames was plotted against both the number of speaker IDs found in the corresponding transcript (A) and the mean area of the boxes enclosing all such detected faces (B). For the comparison with speaker counts, each point was colored by the maximum number of faces detected across extracted frames (A), with 3 faces colored grey and less than/greater than 3 faces colored increasingly blue/red respectively. For the comparison with face area, each point was instead colored by the number of transcript speaker IDs (B), using the same color bounds as in (A).}
\label{fig:pronet-vid-scatter}
\end{figure}

Besides audio, video is also captured during the majority of interviews recorded for this project. Relationships between different video QC metrics could also be of interest in understanding the properties of uploaded videos thus far. In particular, the number of faces detected, the estimated size of the detected faces, and the number of speakers identified in the corresponding transcript should all relate to each other in a manner dependent on the Zoom view mode, the number of people on the call, and the number of people participating in video and/or audio recording (Figure \ref{fig:pronet-vid-scatter}). Of course, TranscribeMe also sometimes misidentifies a speaker ID - they transcribe only based on the mono audio file - so in cases where the speaker count is unexpected manual review may be warranted.

However, sites are allowed to have more than 2 Zoom participants during psychs interviews if they desire, and we have indeed observed that this happens fairly regularly, for at least part of the Zoom call. Further, interviews are not required to have their cameras on during psychs interviews. Thus the many transcripts with 3 or more speaker IDs are probably largely correct even if the mean number of faces does not match, and conversely the transcripts with 2 speaker IDs but more than 2 faces detected are likely related to the $\sim 30$ minute transcription cutoff, which excludes sizeable portions of many of the psychs recordings (Figure \ref{fig:pronet-vid-scatter}A). Interestingly, of the interviews with $\leq 1$ face detected on average per extracted frame, the transcriptions with $> 2$ speaker IDs are all concentrated towards the cluster of smaller mean face area (Figure \ref{fig:pronet-vid-scatter}B). Face area may therefore be usable to reconcile differences between video and transcription attendee counts, highlighting only a subset of discrepancies that are worth a closer manual review.

\begin{figure}[h]
\centering
\includegraphics[width=\textwidth,keepaspectratio]{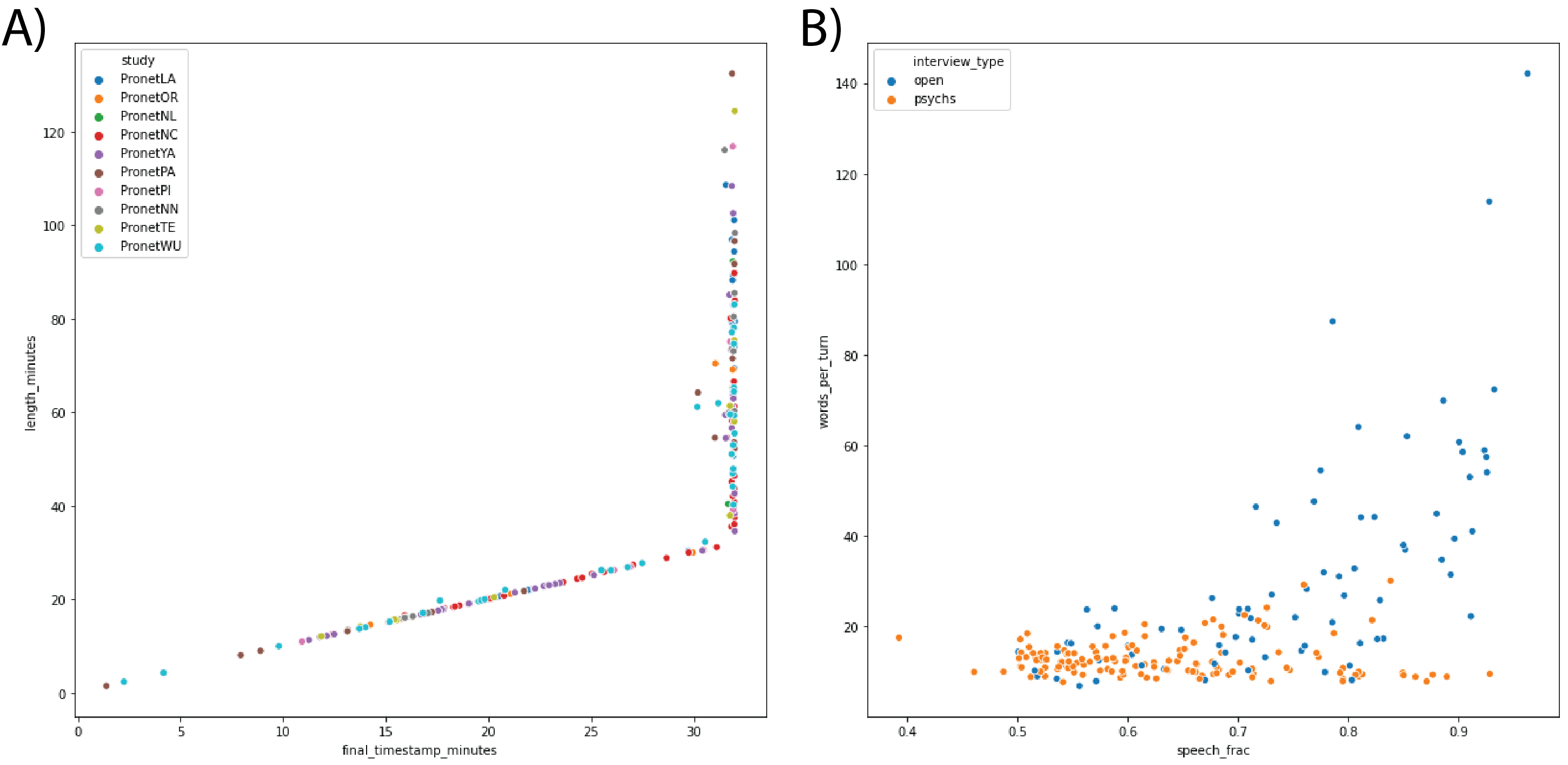}
\caption[Using transcript summary features to better understand interview structure.]{\textbf{Using transcript summary features to better understand interview structure.} As in Figures \ref{fig:pronet-aud-scatter} and \ref{fig:pronet-vid-scatter}, scatter plots were generated to visualize transcript summary feature relationships across all Pronet interviews. First, the start timestamp of the final turn in the transcript (in minutes) was plotted against the corresponding audio length (in minutes) and colored by contributing site (A). This was done to verify TranscribeMe's implementation of the length cutoff described earlier. Then, additional features about the structure of the transcription were derived from the existing QC metrics: the word count from the most verbose speaker ID was divided by the total word count to obtain "speech\_frac", and the total number of words was divided by the total number of turns to obtain "words\_per\_turn". Note that the former feature does not contain any information about participant versus interviewer, it simply relays how balanced the verbosity was amongst speakers. These two features were plotted against each other, with points colored by interview type (B).}
\label{fig:pronet-trans-scatter}
\end{figure}

In addition to typical quality control, it is important to understand the structure of the returned transcriptions -- which may in turn affect our interpretation of e.g. how severe a given inaudible rate is. Both the way an interview is conducted and the consistency of TranscribeMe's methodology will impact how many distinct conversational turns are identified, which consequently alters the number of timestamps received in the transcript. Moreover, the fraction of speech attributed to each speaker ID can vary due to interview conduct, but also due to features of clinical relevance such as participant verbosity. Because of these factors, it is highly relevant to review properties of transcription structure from each interview (Figure \ref{fig:pronet-trans-scatter}). 

One purpose of these plots was simply to verify that operational protocols are being properly followed. Indeed, the variable length cutoff of $30-32$ minutes has been enforced by TranscribeMe every time, and it is comforting to see that most of these transcripts are stopped at very close to the full $32$ minutes charged (Figure \ref{fig:pronet-trans-scatter}A). It is worth noting that although the final turn timestamps tend to saturate at $\sim 32$ minutes, psychs interviews routinely have well over 100 total turns, suggesting generally short conversational turns - as will be investigated in more detail next. 

The other major purpose was to better understand the turn structure of the conducted interviews, and how evenly distributed linguistic content has been between speakers. Note that the interviewer/participant identities of the speakers do not directly map to transcript speaker IDs, so an imbalance detected here could relate to low or high participant verbosity. Critically, verbosity has clinical relevance, but it also has a direct relationship with the interviewer's style. Psychs interviews by design include a number of questions that are meant to be answered with a survey-like categorical response (e.g. "agree"), which makes it unsurprising that we have observed many fewer words per turn and a much more common near-even balance of word usage in psychs interviews (Figure \ref{fig:pronet-trans-scatter}B). Still, the high level of variance in words per turn and in the fraction of words contributed by the most verbose speaker ID amongst the open interviews is of great interest. As the study progresses, any potential discrepancies in interviewer conduct should be teased out so that the clinical salience of such features can be rigorously explored. \\

\begin{figure}[h]
\centering
\includegraphics[width=\textwidth,keepaspectratio]{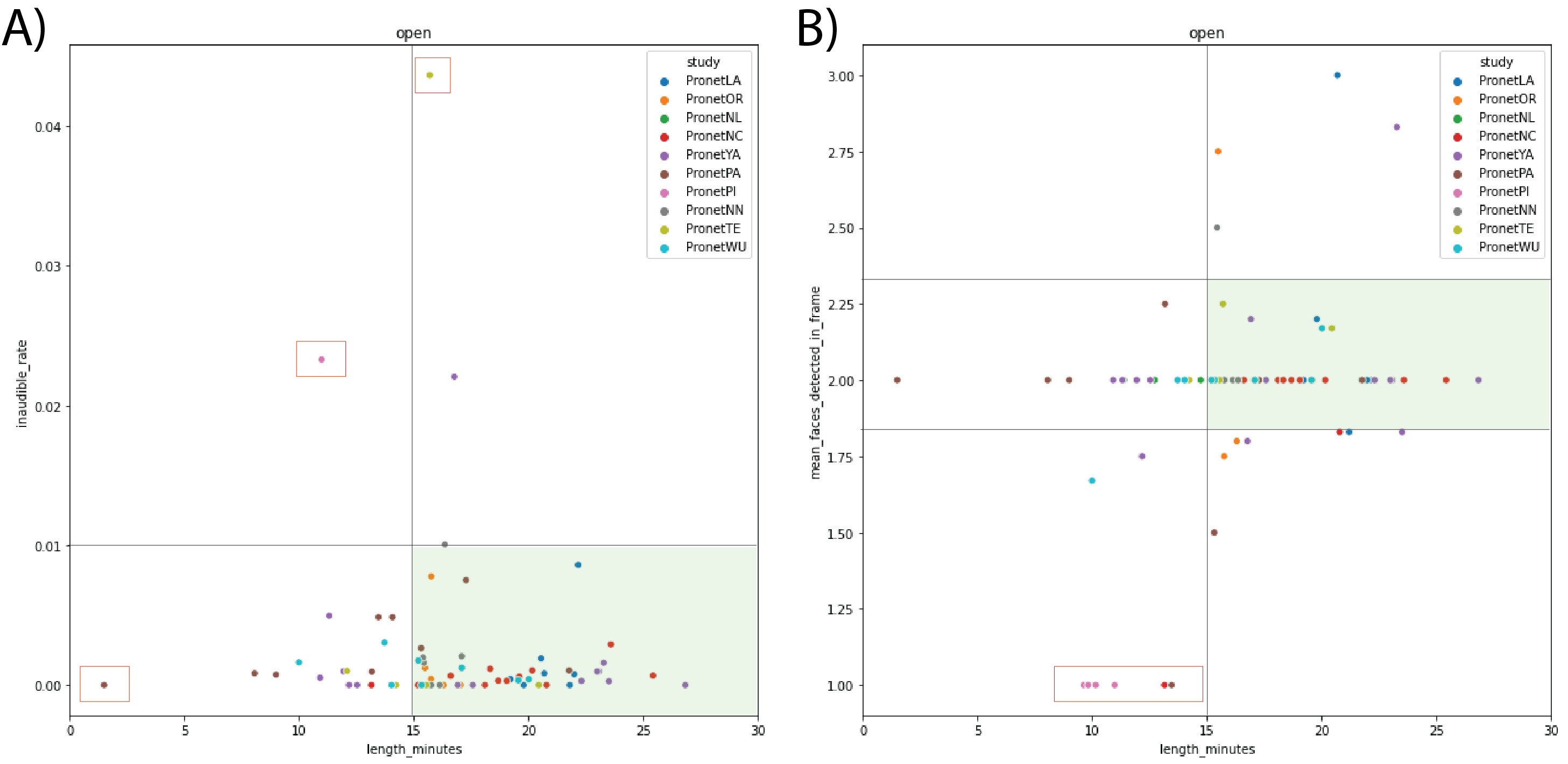}
\caption[Relationships between key quality thresholds across Pronet's open interviews.]{\textbf{Relationships between key quality thresholds across Pronet's open interviews.} Because guidelines for collecting open interviews are more strictly defined, it is particularly important to understand the properties of quality measures amongst open interviews. Using the same methodology as in Figures \ref{fig:pronet-aud-scatter}-\ref{fig:pronet-trans-scatter}, scatter plots were also generated representing only the open interviews processed on Pronet to date. Here, the scatters focus on the key quality thresholds used for Table \ref{table:pronet-counts} and their relationship with interview length, since the latter has been lacking in a number of submitted open interviews as well. Thus these visualizations help to highlight those interviews that are of good quality both by length and by the main audio (A) and video (B) QC features. In both plots, the quadrant with a light green background contains the interview points that are of the highest quality category (including $\geq 15$ minutes), while a few particularly problematic points are boxed in light red. The left scatter (A) plots recording duration against inaudible rate (inaudible count/total words). The right scatter (B) plots recording duration against the mean number of detected faces across extracted frames. Both scatters have the interview points colored by contributing site. Note the x-axis in these plots is artificially limited to between 0 and 30 minutes, to exclude the outlier interviews submitted by site YA that were intended to be uploaded as psychs interviews instead.}
\label{fig:pronet-open-cutoffs-scatter}
\end{figure}

Because open interviews are subject to more specific protocol expectations, the central monitoring team has more power to intervene on open interview quality issues. Because open interviews are conducted only twice per participant, and are of greater experimental interest for speech processing due to the longer form responses elicited, it is uniquely important that open interviews are indeed carefully monitored for quality and other research standards. Categorizing audio and video quality using the available lightweight QC metrics was discussed above (Table \ref{table:pronet-counts}), but for open interviews, satisfaction of a modest length requirement is an additional potential consideration -- as many interviews to date have failed to reach the 15 minute target (Figure \ref{fig:pronet-open-cutoffs-scatter}).

Just as was discussed about inaudible rate (revisited in Figure \ref{fig:pronet-open-cutoffs-scatter}A), open interview duration can be shorter than desired directly because of either procedural (interviewer failing to follow protocol) or clinical (participant demonstrating paucity of speech) reasons. Open interviews are particularly susceptible to both of these duration factors, because the questioning is much less scripted than in psychs interviews. A harsh length cutoff is consequently not ideal; however, it is still important to monitor for abnormally short open interviews and especially a recurring pattern of specific sites submitting short open interviews. In the current dataset, one strong red flag was the failure of site PI to follow the SOP for video recording of open interviews, in conjunction with the same interviews tending to be well under 15 minutes (Figure \ref{fig:pronet-open-cutoffs-scatter}B).

\begin{figure}[h]
\centering
\includegraphics[width=\textwidth,keepaspectratio]{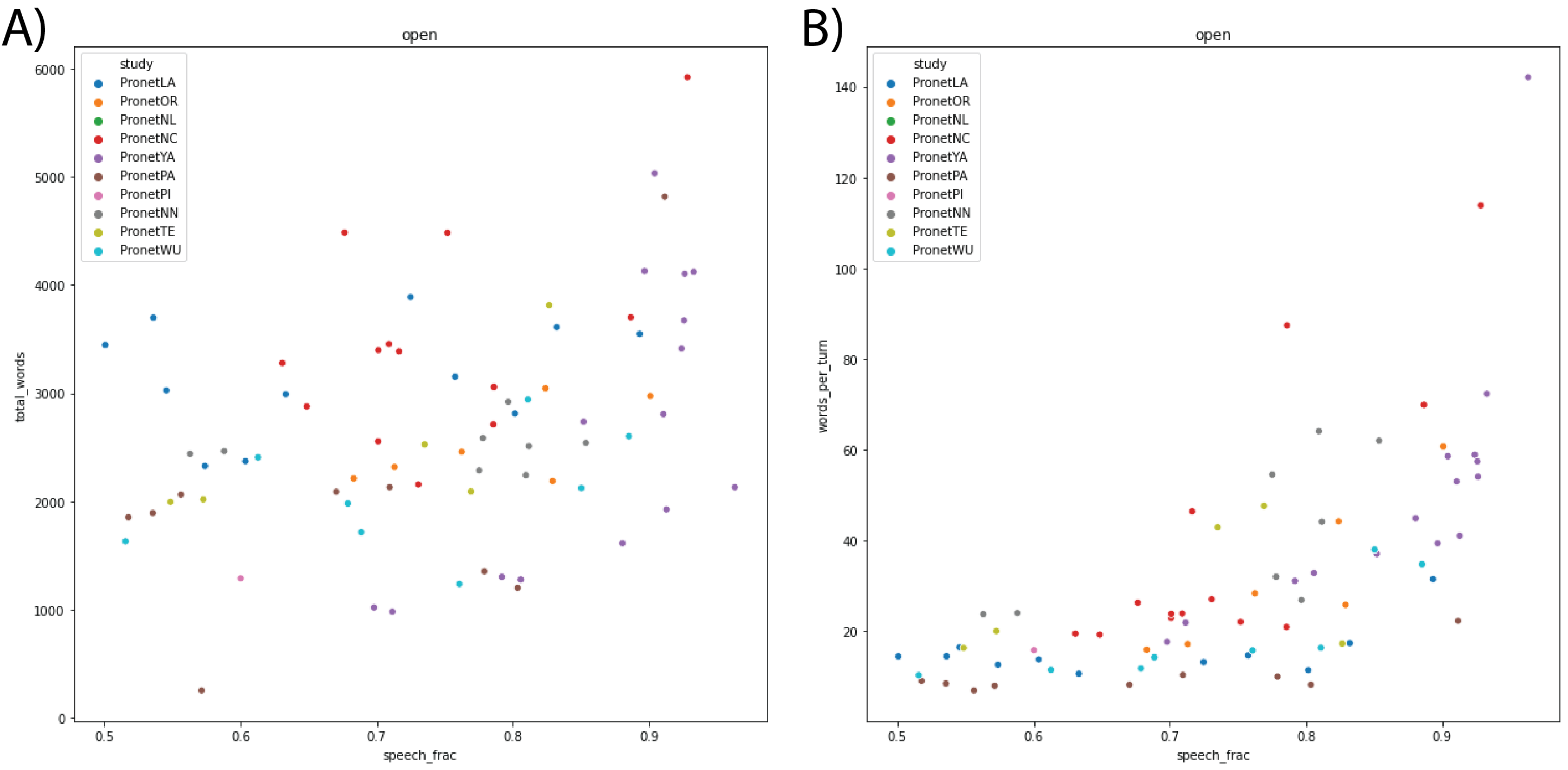}
\caption[Features of open interview structure by Pronet site.]{\textbf{Features of open interview structure by Pronet site.} As was done across all interviews in Figure \ref{fig:pronet-trans-scatter}, transcription structure related metrics were also scattered for only open interviews, the type of particular interest. Here, the plots focus on site-specific trends, with both using site ID as the hue for interview points. The right scatter (B) again plots "speech\_frac" versus "words\_per\_turn", features derived from the QC metrics as described in Figure \ref{fig:pronet-trans-scatter}. The left scatter (A) instead plots "speech\_frac" against the total number of words in the transcript.}
\label{fig:pronet-open-struct-scatter}
\end{figure}

A site-specific pattern of further interest in open interviews is the turn-based structure of the transcriptions and its relationship with verbosity of the primary speaker, as was investigated by interview type in Figure \ref{fig:pronet-trans-scatter}B. Due to those results as well as priors about the structure of open interviews, I considered per site the relationship between the fraction of words contributed by the most wordy speaker ID and both the total words in the transcript as well as the words per turn (Figure \ref{fig:pronet-open-struct-scatter}).

Something mentioned in the distributional analysis was the oddly high number of turns in the generally short open interviews submitted by Pronet site PA. Unsurprisingly, this was apparent in the words per turn scatter plot (Figure \ref{fig:pronet-open-struct-scatter}B, brown). However, it is interesting that a lower words per turn value was consistently seen for PA regardless of the speech fraction feature, indicating there may indeed be something off about their interview conduct (or assigned transcriber), rather than this being a natural result of the participants they happen to have enrolled early in the study. The perfect lower bound that can be created in Figure \ref{fig:pronet-open-struct-scatter}B by connecting the 9 open interview points from PA - which span $5+$ participants - is a surprisingly clean result that should be closely monitored as data collection progresses.

A similar pattern to PA was seen to a lesser extent for sites LA and WU, while sites such as NC, YA, OR, and NN demonstrated a much higher number of words per turn in interviews with higher speech fraction value (Figure \ref{fig:pronet-open-struct-scatter}B). This phenomenon is not simply modulated by interview length; LA has not submitted any short duration recordings whereas YA has submitted a number of them, and sites could not be divided in the same way on the scatter of speech fraction versus total transcript word count (Figure \ref{fig:pronet-open-struct-scatter}A). A qualitative review of the open interviews conducting by these different sites so far could better elucidate contributing factors towards the systematically differing transcript structures. \\

To further investigate the results of Figures \ref{fig:pronet-trans-scatter}B and \ref{fig:pronet-open-struct-scatter}B, I generated additional scatter plots derived from the transcript QC metrics, breaking down the turn-based structure by specific speaker IDs. One way to do this is by assigned TranscribeMe speaker ID: S1 is the ID for the first speaker, S2 is the ID for the second unique speaker, and so on. Per the protocol, the primary interviewer should speak first and the participant second in all interview recordings, though given the many other SOP violations that have occurred (see \ref{subsubsec:u24-issues}) it is not advisable to rely solely on this convention for identifying interviewer and participant IDs. 

An alternative analysis method is to identify the speaker ID that has spoken the most words in the interview and the speaker ID that has spoken the second most words in the interview, and derive features for them in terms of these IDs instead. Indeed, the maximum speaker ID was already used in the above mentioned "speech\_frac" feature. Of course this is not intended to be a direct ID of the participant or interviewer, as participant verbosity can vary greatly; however, it provides important context for other features of interview structure.

For these reasons, I first characterized the TranscribeMe provided speaker IDs versus the IDs obtained from the verbosity-derived mapping in the present dataset. The vast majority of interview transcripts had $\leq 3$ unique speakers identified, and for the open interview type it was rare to encounter more than 2 speakers - as it should be per the SOP (Table \ref{table:u24-speaker-counts}). Notably, a clear majority of open interviews had TranscribeMe ID S2 as the speaker contributing the most words, while psychs interviews were more balanced and had a slight majority towards ID S1 (Table \ref{table:u24-speaker-max-ids}). This is consistent with what one would expect if S1 were typically the interviewer, because the freer form open interviews should be more participant driven.

\begin{table}[!htbp]
\centering
\caption[Transcription counts by number of speakers present, for open versus psychs interviews.]{\textbf{Transcription counts by number of speakers present, for open versus psychs interviews.} Using TranscribeMe's speaker IDs, the number of unique speakers present in each Pronet transcript (as of 1/23/2023) was tallied. Every transcript had at least 2 speaker IDs and at most 5. This table reports the number of open and the number of psychs interviews for each possible speaker count. Note that site WU was responsible for 3 of the 5 psychs interviews with $>3$ speakers, including the one with 5 IDs. LA and PI were the other two sites contributing to that count. Recall also that site LA has submitted a handful of psychs interviews as open interviews accidentally, which did have more than 2 faces detected (Figure \ref{fig:pronet-open-key-dists}).}
\label{table:u24-speaker-counts}

\begin{tabular}{ | m{3cm} || m{2.2cm} | m{2.2cm} | m{2.2cm} | m{2.2cm} | }
\hline
\textbf{Interview Type} & \textbf{2 Speakers \#} & \textbf{3 Speakers \#} & \textbf{4 Speakers \#} & \textbf{5 Speakers \#} \\
\hline\hline
open & 70 & 6 & 0 & 0 \\
\hline
psychs & 113 & 23 & 4 & 1 \\
\hline
\end{tabular}
\end{table}

\begin{table}[!htbp]
\centering
\caption[Transcription counts by ID of maximum speaker, for open versus psychs interviews.]{\textbf{Transcription counts by ID of maximum speaker, for open versus psychs interviews.} The TranscribeMe speaker ID of the most verbose speaker in each Pronet transcript (as of 1/23/2023) was identified, defined using total words contributed per unique ID. Recall that S1 is the ID assigned to the first person to speak, S2 the second, and so on. Every transcript had S1, S2, or S3 as the maximum speaker. This table reports the number of open and the number of psychs interviews for each max speaker ID. Note that site WU was responsible for 2 of the 4 psychs interviews with max ID S3, and YA was responsible for 3 of the 4 open interviews with max ID S3. PI contributed one interview of each type to this count, and NC was the final site responsible for a psychs interview with max ID S3.}
\label{table:u24-speaker-max-ids}

\begin{tabular}{ | m{3cm} || m{3.2cm} | m{3.2cm} | m{3.2cm} | }
\hline
\textbf{Interview Type} & \textbf{S1 Most Verbose \#} & \textbf{S2 Most Verbose \#} & \textbf{S3 Most Verbose \#} \\
\hline\hline
open & 11 & 61 & 4  \\
\hline
psychs & 80 & 57 & 4  \\
\hline
\end{tabular}
\end{table}

\noindent Based on these results, I moved forward with the following transcript structural features derived from QC metrics:
\begin{itemize}
    \item "S1\_speech\_frac": The number of words in the transcript attributed to TranscribeMe speaker ID S1 (first speaker) divided by the total word count of the transcript.
    \item "S2\_speech\_frac": The number of words in the transcript attributed to TranscribeMe speaker ID S2 divided by the total word count of the transcript.
    \item "S1\_words\_per\_turn": The number of words in the transcript attributed to ID S1 divided by the number of turns attributed to S1.
    \item "S2\_words\_per\_turn": The number of words in the transcript attributed to ID S2 divided by the number of turns attributed to S2.
    \item "max\_speaker\_id\_string": The TranscribeMe speaker ID that contributed the most words in the transcript.
    \item "max\_speaker\_words\_per\_turn": The number of words attributed to the max speaker ID divided by the number of turns attributed to the max speaker ID.
    \item "second\_speaker\_words\_per\_turn": The number of words attributed to the second most verbose speaker ID divided by the number of turns attributed to the second most verbose speaker ID.
\end{itemize}
\noindent I then used these features to create the additional scatter plots.

\begin{figure}[h]
\centering
\includegraphics[width=\textwidth,keepaspectratio]{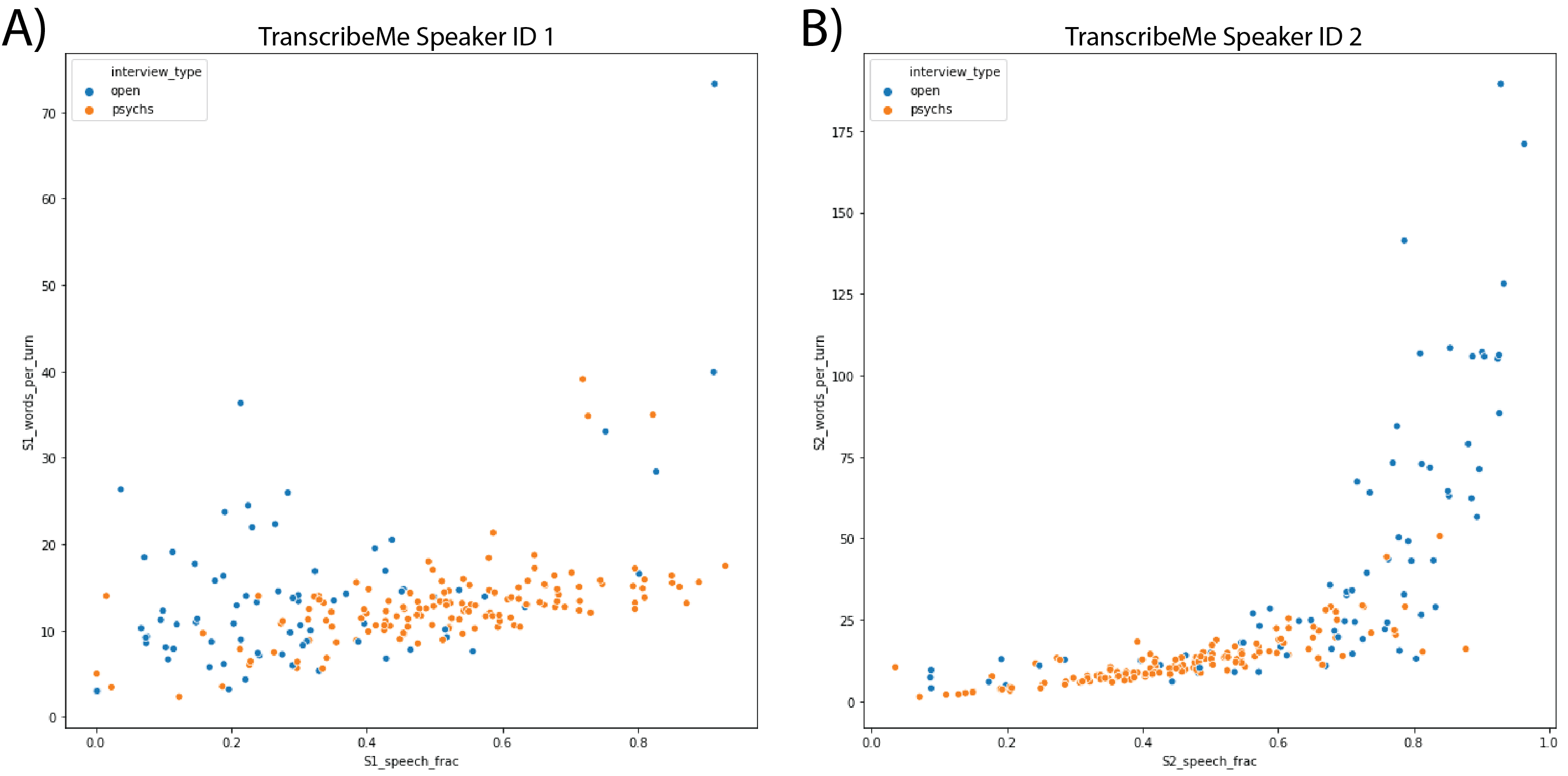}
\caption[Interview transcript structure by TranscribeMe speaker ID across Pronet.]{\textbf{Interview transcript structure by TranscribeMe speaker ID across Pronet.} Using the speaker ID specific versions of the features depicted in Figure \ref{fig:pronet-trans-scatter}B, which are detailed above, I broke down the turn structure trends seen in the open (blue) and psychs (orange) interviews by the first (A) versus second (B) unique speakers chronologically. By protocol, the first speaker should be the primary interviewer and the second the participant. Though this has not been perfectly followed, it has been consistent enough to identify potential high level trends attributable to interviewer versus participant. Note that the y-axis of (A) extends to just under 80 words per turn, while the y-axis of (B) extends as far as nearly 200 words per turn.}
\label{fig:pronet-all-ids-scatter}
\end{figure}

The first set of comparisons I made here was to better characterize the systematic differences in turn-based structure between the two interview types (Figure \ref{fig:pronet-all-ids-scatter}), which also served to more closely evaluate how often sites are following the correct convention of having the primary interviewer speak first.

The distinction between open and psychs interviews reported in Figure \ref{fig:pronet-trans-scatter}B was even more stark when split by speaker ID. The first speaker, which should largely be the interviewer, was concentrated at less than $40\%$ of the contributed words for open interviews, and this was inverted in psychs interviews (Figure \ref{fig:pronet-all-ids-scatter}A). Interestingly, the words per turn for S1 was often higher in open interviews than psychs interviews despite this fact. Furthermore, this view can be used to flag interviews that are likely to have the participant incorrectly speaking first -- the handful of blue dots in the upper right quadrant of Figure \ref{fig:pronet-all-ids-scatter}A are particularly suspect. 

The second speaker, which should largely be the participant, demonstrated the opposite trend on the x-axis (Figure \ref{fig:pronet-all-ids-scatter}B). Obviously, the "S1\_speech\_frac" and "S2\_speech\_frac" features will sum to 1 in the majority of cases because of the relative rarity of more than 2 speakers in a transcript, so this is unsurprising. Still, it is interesting to see the high level of variation in the words per turn of TranscribeMe ID S2 during open interviews (Figure \ref{fig:pronet-all-ids-scatter}B); much of the site-specific relationships in Figure \ref{fig:pronet-open-struct-scatter}B appear to be driven by S2.

\begin{figure}[h]
\centering
\includegraphics[width=\textwidth,keepaspectratio]{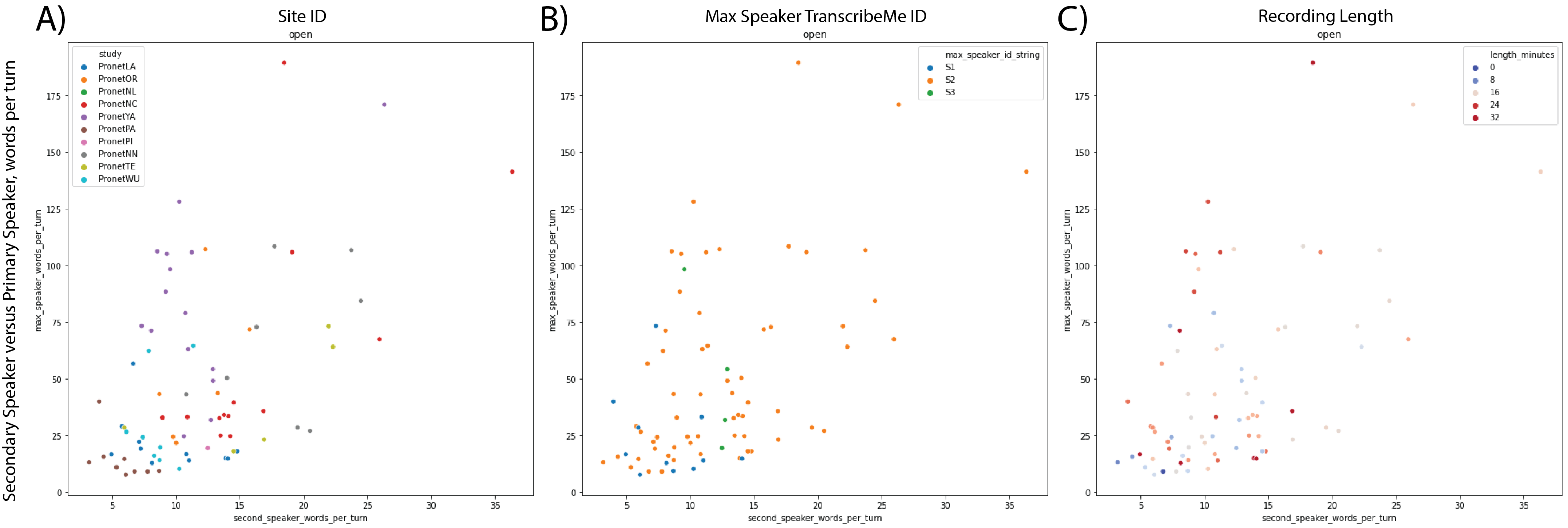}
\caption[Speaker-specific factors in open interview structure.]{\textbf{Speaker-specific factors in open interview structure.} As was done for overall transcript structure in Figure \ref{fig:pronet-open-struct-scatter}, I looked more closely at the speaker-specific versions of transcript structure metrics (defined above) in only the open interviews of the dataset. All 3 plots here scatter "max\_speaker\_words\_per\_turn" (y-coordinate) against "second\_speaker\_words\_per\_turn" (x-coordinate). Each uses hue to highlight a different potential third factor in this relationship: contributing site (A), "max\_speaker\_id\_string" (B), and recording duration (C). Sites are color coded as in Figure \ref{fig:pronet-open-struct-scatter}, "max\_speaker\_id\_string" is a categorical value that can equal S1 (blue), S2 (orange) or S3 (green), and duration is colored from darkest blue ($\leq 5$ minutes) to darkest red ($\geq 30$ minutes).}
\label{fig:pronet-open-ids-scatter}
\end{figure}

Because of the promising results of the speaker-specific analysis and the interesting site-specific factors seen in the overall transcript structure analysis from open interviews, it is a natural next step to investigate the speaker ID features more closely in the open format. To clean up some of the outliers by TranscribeMe speaker ID (Figure \ref{fig:pronet-all-ids-scatter}), I focused on scatter plots of features from the verbosity defined speaker IDs for the open interview case (Figure \ref{fig:pronet-open-ids-scatter}).

When considering words per turn separately for the primary and secondary speakers, clear site-specific patterns again emerged (Figure \ref{fig:pronet-open-ids-scatter}A). Pronet site PA was extremely low on words per turn from the primary speaker in 8 of 9 open interviews, but they were also extremely low on words per turn from the secondary speaker in the majority of their interviews, with all of the latter values falling in the bottom half (well below $10$ words per turn). Site WU similarly tended to have low words per turn from both the primary and secondary speakers, to a lesser extent. On the other hand, YA had very high words per turn from the primary speaker despite middling words per turn from the secondary speaker, while NC had relatively high secondary speaker words per turn yet the primary speaker's words per turn were generally moderate in their transcripts (Figure \ref{fig:pronet-open-ids-scatter}A).   

It is not clear how to interpret these patterns, as interviewer behavior and participant behavior can both impact each other. Conversely, even with extra effort to elicit long responses some participants may remain very brief, which is an inherently interesting phenomenon. There is thus nothing implicitly wrong with the wide spread of transcripts in Figure \ref{fig:pronet-open-ids-scatter}. However, it is surprising and perhaps concerning that individual site's transcripts tended to cluster together, across multiple subjects (recall that each participant contributes at most 2 open interviews). The sites that have consistently low words per turn from both speakers likely warrant a manual review follow-up on the site's interview conduct as well as TranscribeMe's protocol adherence and transcript assignment procedure. 

It is worth noting that the relationship with recording length here (Figure \ref{fig:pronet-open-ids-scatter}C) was less clean than that with contributing site, though there was unsurprisingly some correlation. It may be the case that certain interviewers are more likely to interject or to ask questions piecewise, which is not objectively better or worse than the longer individual response format; though it is important to keep this in mind for certain analyses, as it obviously affects transcript structure and can in fact impact elicited content. Depending on study goals it may even be desirable to step in and instruct better standardization across sites. Regardless, interviews that are both short and have short turns from both speakers are suggestive of a lower quality open interview protocol. 

\FloatBarrier

In summary, a number of QC features displayed interesting relationships with each other, often dependent on interview type, submitting site, or another QC factor. It is thus important to keep these relationships in mind (and revisit them) when monitoring interviews over the course of the long timescale AMPSCZ project. They can guide decisions for appropriately accounting for recording quality in downstream modeling, and more generally help to detect possible confounds. They can even inform high level project decisions, because the obvious structural differences between the interview types would very likely relate to the value obtained from the corresponding transcripts. While certain transcription plans have since been updated in light of my report utilizing these plots (see also Appendix \ref{cha:append-ampscz-rant}), there still remain many potential concerns that have not yet been closely followed-up on, including the patterns of site-specific differences in open interview transcript structure. \\

\noindent With background now clear on the pipeline's functionality and the data collection monitoring process being employed by AMPSCZ for interview recordings, I will next present an exhaustive list of issues encountered thus far, along with ways to hopefully prevent them going forward. Some issues are discussed at all stages of the data flow, but by and large the present problems require adjustments to be made by the sites collecting data. As such, the bulk of the upcoming subsection (\ref{subsubsec:u24-issues}) is targeted at the sites, focusing on improvements that they could make to their collection workflow and to how they submit interviews to the central server. Naturally, it is targeted at whoever might replace my role in monitoring the interview infrastructure for AMPSCZ in the future as well, along with (to some extent) those involved with central monitoring at a higher level who may be able to succeed with a different style of approach for adjusting problematic site behavior than what has been tried to date. 

\subsubsection{Common issues encountered in the AMPSCZ project}
\label{subsubsec:u24-issues}
Most times that an interview is not processed when expected there is an issue with it violating expected upload conventions, either because the interview recording output was incorrectly placed so Lochness could not pull it to the server, or the interview recording output was somehow modified from the default so that my interview code could not extract the necessary metadata for processing from what Lochness uploaded. It was also common during the testing phase for this project to find subject IDs that were missing metadata - if there is a new subject ID it needs to be entered in REDCap/RPMS along with a valid consent date before Lochness can pull any of its interviews.

For more information on problems that prevent Lochness from pulling interviews from Box/Mediaflux to the data aggregation server, see that code's documentation. In this section, I will detail the numerous potential problems that can prevent the interview pipeline from processing data that have been uploaded to the central server. Many of these are issues with sites not following the standard operating procedure (SOP) that have been observed at a high frequency in this multi-site study to date. Others are issues that can impact recording quality, which sites should keep in mind. Still others are edge cases that might cause problems with the code, which should be monitored by the central personnel overseeing interview collection across the sites. The purpose of this documentation is to serve as a guide for the present AMPSCZ study, as well as a blueprint for future studies taking a similar approach, particularly any planning to use my code in some capacity. \\

\paragraph{Site interview uploads.}
First, the list of file organization related SOP violations frequently committed by sites. Most of these prevent processing of at least one interview modality, if not the entire interview:
\begin{itemize}
    \item Using the wrong software for virtual interviews, e.g. Microsoft Teams instead of Zoom.
    \item Putting interview sessions under the wrong type, e.g. uploading an open interview as if it were a psychs interview.
    \item Unnecessarily modifying the recording folders produced by Zoom:
    \begin{itemize}
        \item Uploading a compressed version of the Zoom folder.
        \item Renaming the Zoom folder, disrupting the date/time metadata provided there.
        \item Uploading the contents individually instead of inside the produced folder, again removing necessary metadata as well as disrupting expected folder structure.
        \item Removing or otherwise modifying files from the Zoom folder before uploading, so that some data types are not found - note the primary audio M4A and primary video MP4 should not be renamed!
        \begin{itemize}
        \item The speaker specific M4A files should also be included in the upload and kept in the 'Audio Record' subfolder, named as is. Although this pipeline does not utilize the diarized audios, they will be critical to future feature extraction, and my code does check for their presence.
        \end{itemize}
        \item Not waiting until Zoom has finished saving the interview folder locally before moving it into Box/Mediaflux for upload can also cause one or more of the relevant interview contents to be missing from the interview on the server. This may also lead to unusable partial versions of interview files taking up storage space (likely related to the large .zoom files discussed in section \ref{subsubsec:u24-storage}).
        \begin{itemize}
            \item If the folder is copied to Box/Mediaflux too soon the pipeline processing will fail, but the interview should at least be accessible to the site for re-upload at a later time. 
            \item If the process of saving the local version is interrupted by e.g. modifying the in progress interview folder, quitting Zoom, or restarting the computer, the recording could be entirely lost. 
        \end{itemize}
        \item Adding files into a folder that were not generated by Zoom, or combining the files from multiple Zoom interviews into one interview folder.
    \end{itemize}
    \item Stopping and restarting recording within a single Zoom interview session, which is not permitted in the SOP. For reasons that will be detailed, the code does not currently support split single interview sessions. For eventual alignment of diarized audios, it is also important that recording is not started until after the participant has joined the call.
    \begin{itemize}
        \item Pausing a recording and later resuming does not cause this problem, however it will result in "dead air" in the resulting files for the duration of the pause. It is thus not advisable to pause for an extended period.
        \item Conversely, sites are not supposed to upload multiple interview folders for a single subject day, which can occur if the meeting itself is ended and a new meeting is started for a second recording. While the pipeline does technically support this and there may be cases where it is appropriate (accidentally stopping a recording), it could cause budgetary issues with long interviews if it is done systematically.
    \end{itemize}
    \item Having incorrect Zoom account settings, which could cause the untouched automatically generated files to not follow the SOP. 
    \begin{itemize}
        \item See \ref{sec:zoom-settings} for initial Zoom setup instructions.
    \end{itemize}
    \item Unnecessarily modifying the recording WAV file produced by the approved hardware recorder, or otherwise configuring the recorder incorrectly:
    \begin{itemize}
        \item Renaming the file to break the default convention produced by the recorder with specifically formatted date and time metadata.
        \item Not setting up the recorder with the correct date and time, and proceeding to upload a file with the wrong (possibly pre-consent) date.
    \end{itemize}
    \item With transcripts sent for manual site review, either renaming the file after review or not putting it into the correct folder structure for returning approved transcripts.
\end{itemize}

\noindent All of these listed violations were committed multiple times by multiple different sites in production data collection, and violations spanned all sites. Most of the above issues will be flagged by the pipeline's error logging (detailed in section \ref{subsec:interview-code}), so that the central monitoring group can easily notice them and communicate with the sites as needed to fix. However this process can substantially delay the time it takes to receive interview transcriptions, and to catch other issues with recording quality at a new site. It also requires otherwise unnecessary manual work on the part of both the site and the monitors, and can eat up disk space as will be discussed in the next section (\ref{subsubsec:u24-storage}). 

Overall, more care should be taken to avoid these mistakes on such a regular basis. Further, a few of them -- incorrect interview type identification and incorrect interview date metadata -- are not feasible to systematically detect and may cause wrong labels to propagate to later analyses for affected interviews. Similar issues that could theoretically occur include uploading an interview for the incorrect subject ID or entering an incorrect consent date for a subject ID in REDCap. Even when detected manually, whether by a site noticing a mistake or monitoring protocols noticing an egregious set of anomalies from one site, these particular issues currently require tedious manual intervention. It is not sufficient for the site to just move the interview to the correct upload location after a mistaken upload took effect, as many changes will need to be made on the data aggregation server; it becomes necessary to fix the labels for that interview across pipeline intermediates and final outputs, as well as to ensure the raw data repository is fixed. 

Along these lines, note that if interviews are uploaded out of order or an earlier interview has major naming issues that are only fixed after a later interview has been processed, session numbers assigned to the interviews in pipeline outputs will be incorrect. Session number is based on the set of interviews in raw that meet the top level naming requirements per interview type and subject ID, when placed in alphanumeric order. If a more recent interview is processed first and then an old one is subsequently introduced, both will be labeled as session 1 within the internal workings of the pipeline. There is currently no way to automatically change the assigned session number used in the processed outputs, including the transcript filename, so for the purposes of this code the session number will remain incorrect. However, correct session number is not essential to downstream uses, as the day number assigned based on subject consent date is much more relevant to aligning the data and generally can be used without any reference to assigned session number. In rare cases there may be multiple interviews uploaded from the same day for the same subject and interview type, in which case it will be necessary to pay closer attention to potential anomalies in session number. Of course if the interview upload protocol is followed the session number shifting problem will never occur. \\

\paragraph{Recording quality.} 
After an interview has been correctly uploaded, a major component of the code is to process each interview for basic quality control. Additionally, manual reviewers with access to the raw recordings were tasked with checking a handful of random interviews from each site towards the beginning of the study. Their observations were used to help identify the QC features of most relevance, and also to identify more broadly the recurring quality concerns that sites should be instructed to avoid. Abnormal QC features in turn identified additional interviews to be reviewed. Through this process, we devised the following list of considerations for collecting good interview recordings and verifying quality of the corresponding transcripts:

\begin{itemize}
    \item Per the SOP, when using Zoom sites should ensure they are recording on gallery view rather than speaker view.
    \item Interviewers should relay instructions to new participants on maintaining recording quality, and be on the lookout for recurring problems to correct (mid-interview if needed):
    \begin{itemize}
        \item Face moving fully or partially out of frame for portions of the interview.
        \item Shaky camera due to unstable device placement.
        \item Echoing audio.
        \item Speaking too closely into the mic, creating breathy/popping plosives.
        \item Speaking too far away from the mic or otherwise not enunciating.
        \item Recording in an area with a lot of background noise.
        \item Poor internet connectivity.
        \item Using low quality hardware or having poorly optimized input settings on Zoom - the latter should be confirmed with participants at the start of every session (see \ref{sec:zoom-settings}).
    \end{itemize}
    \item Many of the above problems can be made much worse if the participant is using a phone instead of a computer to join the call. Participants should be encouraged to use a computer if possible, and if not they should be encouraged to use headphones with a microphone instead of the built-in phone hardware.
    \item Submitted open interviews should ideally be at least 15 minutes in length, and interviews less than 10 minutes should be rare. There may be specific participants that do not speak much, which could of course be clinically interesting. However we have also noticed early issues with some sites inadequately conducting the open interviews, resulting in repeated shorter recordings.
    \begin{itemize}
        \item Note also that open interviews should contain exactly one interviewer and one participant, and Zoom video should be available with detectable and unobstructed (i.e. no mask) faces for both "attendees". These interviews should have a more standardized level of quality than psychs and will be monitored accordingly.
    \end{itemize}
    \item Conversely, some very long psychs interviews have been uploaded, and it is unclear if sites are planning content appropriately around time limits - for budgetary reasons, TranscribeMe will only transcribe $\sim 30$ minutes of these recordings (using an intelligent cutoff point of at most $32$ minutes). Additionally, with especially long interviews ($> 120$ minutes) it is possible for AV processing functions to crash on the low RAM aggregation server, causing the interview to be skipped by the data flow and instead just log an error.
    \begin{itemize}
        \item One thing that should certainly not be included in the interview recording is the run through of the initial settings checklist for quality assurance. Settings should be finalized and interview quality should be manually determined to be acceptable by the interviewer before beginning the actual recording. 
        \item The recording should not capture any description of the interview plan to the participant either - it should begin with actual interview content.
        \item Another potential source of unnecessary file length is excessively long recording pauses. TranscribeMe uses timestamps to determine transcript cutoff, and also to determine billing. Including long pauses within the first 30 minutes of an uploaded interview should be avoided.
        \item The clinical interviews may also naturally take a long time, especially at the baseline timepoint; it is then important to consider which parts should be prioritized for transcription and other analyses when collecting recordings that are expected to be very long.
    \end{itemize}
    \item To facilitate automatic analyses from TranscribeMe outputs, it is important that the primary interviewer speaks first in the recording and the subject speaks second. In psychs interviews, additional interviewers (or participating parents) may enter the conversation later.
    \item TranscribeMe is not always consistent in how they mark personally identifiable information (PII), with some interviews heavily redacted, and other long interviews coming back with 0 redactions. While the transcriber will sometimes redact more harshly, we luckily have found very few instances where something necessary to redact was missed. 
    \begin{itemize}
        \item A close manual review of a handful of 0 redaction transcripts as well as a review of all capitalized words across all 0 redaction transcripts was completed by a senior member of the Pronet team, yet only 1 instance of missed PII was discovered, and it was a borderline case that was only potential PII if triangulated with other information.
        \item Regardless, it remains critical that sites perform careful and timely manual redaction review on those transcripts that are randomly returned to them; reviewers should primarily be on the lookout for missed PII, but should also make a note if instances of overzealous redaction appear. 
    \end{itemize} 
\end{itemize}

\noindent Again, many of these problems were observed multiple times across multiple sites. While the video, audio, and transcript QC measures can capture some of them (particularly when severe), the QC computation was designed to be lightweight, and is thus not a perfect indicator for quality of downstream feature extraction. Additionally, catching these issues with QC does not necessarily mean they will be resolvable in the affected interviews, so sites need to get ahead of them as much as possible. The QC workflow is meant to assist sites in identifying problems for future reference, as well as to filter out particularly problematic interviews before they reach the stage of further processing. 

Another potential issue tracked by the pipeline is interviews (with an experimental focus on the open interview type) being conducted at the wrong part of the study timeline for a particular subject. Open interviews should occur at baseline and two month follow-up points for each subject. It is important that interview timing is in a reasonable range of the target points, and that for each interview the recording information is appropriately entered in REDCap/RPMS, so it can be crosschecked with the raw interview data processed by this pipeline. For more details on the interview protocol instructions provided to sites, see the information within section \ref{subsec:interview-methods} above. \\

\paragraph{Software maintenance.} 
Outside of users not following the SOP and quality issues caused by challenges with interview environment or recording hardware, problems can also arise because of bugs in the interview code. Regular monitoring of study data flow has already addressed many such bugs, often requiring only small tweaks to the code. However that monitoring process should continue as more edge cases will likely arise over time, particularly as the study expands to more sites that use non-English languages or might otherwise conduct interviews in subtly different ways that will provide new test cases for the code. 

Most problems with the code will be initially detectable via the automatic email alerts. If a code issue needs to be further investigated, detailed log files are automatically saved under a subfolder of the repository install, organized by site and labeled with pipeline name and Unix run time. As will be explained in section \ref{subsec:interview-code}, each portion of the pipeline generates its own logs with both basic status/runtime updates as well as any error messages encountered. 

There have also been a few cases where troubleshooting required tweaks to server configuration rather than to the code itself. It required some coordination with Lochness and server IT to ensure that the account running the interview code had sufficient folder permissions to read all needed inputs and save all needed processed outputs (themselves with correct permissions for further transfer where applicable), while preventing the account from modifying raw interview data or outputs from other datatypes. Any future project that plans to run the code on a specialized account, as was required for security here, should be aware of various permissions issues or otherwise missing top-level study folders that might prevent the pipeline from running at all. 

Similarly, low storage availability on the data aggregation server halted all processing for a time, until disk space was expanded by IT. Going forward, storage should be more closely monitored for both this and future studies, as will be discussed further in the next section. \\

Finally, there are some situations that might arise at times but that the code intentionally does not currently handle, nor are there specific plans to do so in the future. The most common one that affects sites' recording protocols is the splitting of Zoom interview recordings; this is an issue that could also impact future users of the code. If recording of a given Zoom interview is stopped and later restarted, whether intentional or due to host disconnect, the resultant interview folder will contain multiple audio and video files. It is not clear how this should be handled systematically, as concatenation of the files may or may not be appropriate depending on the exact circumstances, and exact time alignment can be difficult to handle across the split diarized audio files generated by Zoom. Further, sites have so often renamed files in our hands that it is difficult to rely on the presence of the extra metadata needed from the default filenames when the interview is split. 

In general, the issue can be avoided during planned periods of non-recording by using the pause functionality on Zoom instead of stopping. However long pauses within the recording are non-ideal for transcription cost and analysis reasons, and should be used sparingly. Ideally, interviews will always be recorded as one continuous session with minimal (if any) pauses, but of course there may occasionally be a need for a longer break or some sort of accident or technical difficulty that causes recording to completely stop. When there is a stoppage in recording, the interview will only be processed by the code if the parts are uploaded as independent sessions. This can be automatically handled by Zoom if the interviewer ends the meeting and starts a new one before beginning the second recording. That will create two separate folders, which can both be uploaded if they both contain content of relevance. Alternatively, the interviewer could manually move the second set of recording files into a different, appropriately named, folder after the interview is completed. Though that method is not recommended due to the high risk of moving a mismatched set of files or making a mistake with the manual folder name entry. 

The other case sites might encounter that the code is intentionally not guaranteed to handle at this time is failure on excessively long interviews. The QC code is meant to be lightweight, and if used on a more powerful machine for a different study this should not be a problem. However on the machines used for data flow for the AMPSCZ project (particularly Pronet), compute power is quite limited intentionally. Because of this, it has been observed that a few uniquely long submissions were unable to complete QC without killing the kernel, and thus those interviews ended up skipped over for not only quality measures but also downstream steps like transcription upload. There is not a hard cutoff on what length interview causes this, but all interviews that did encounter the problem exceeded 120 minutes. In these rare cases, the issue could be resolved by manually truncating the file and reuploading, the current suggested workaround. 

However, it is much more strongly suggested that such long interview recordings should be avoided entirely. The SOP guidelines state that psychs interview recordings should be about $40$ minutes, while the maximum length that is guaranteed to be included in transcription by TranscribeMe is 30 minutes due to budgetary constraints. This is significantly shorter than the interviews of $2+$ hours crashing the code. The high end of expected duration ranges mentioned by the CAARMS and SIPS manuals, on which PSYCHS is based, do not even exceed $90$ minutes.

While there is no upper bound on length strictly required, there is little reason to be submitting such long interview recordings on a regular basis for this study's protocol. Additionally, sites that regularly submit dragged out interviews may be difficult to compare with sites that more closely followed the SOP, as the automatic truncation protocol TranscribeMe is following might be unintentionally removing responses to some key questions that would have been fully included in an interview of normal duration. As the project continues, AMPSCZ sites should really be mindful of when they start and end the recording of an interview if they anticipate the session itself will be abnormally long. Note that sites should absolutely not use splitting the interview into separate session folders as a way to get around restrictions on duration. Split sessions from a single date should be rare, and if every site did this the project would risk going over the transcription budget. \\

\paragraph{Transcript review.}
Beyond recording protocol situations that are not supported by the code, there are also potential (albeit unlikely) problems with returned transcription quality from TranscribeMe that are not currently possible to address or in many cases even accurately detect automatically. While development testing and later spot checks of TranscribeMe output quality have mostly been good, with redactions as the biggest pain point mentioned above, a number of other problems have been found in rare cases across projects. Of course TranscribeMe typos or other inconsistencies in the provided transcripts can impact transcript processing, so it is important to be aware of failure modes. The following list presents considerations to be aware of and how they might best be addressed with the current pipeline. 

\begin{itemize}
    \item Uploaded transcript may use incorrect file format or include characters in the text that are not UTF-8 compatible: the latter can be flagged by the code at least, the former will cause the transcript to be labeled as missing until fixed by TranscribeMe on their server. In either case TranscribeMe should be contacted.
    \item Transcriber may deviate from expected notation for verbatim transcription with speaker ID and timestamps: if egregious it might crash the transcript QC code, but would be straightforward to replace with a redone transcription, whereas if subtle it could easily escape detection and introduce inaccuracy in downstream analyses such as disfluency counts. Produced transcript QC counts of related punctuation like dashes and commas can be tracked over time to assist with detection where feasible. 
    \begin{itemize}
        \item In the past we have occasionally detected typos with speaker identification as well, which requires manual update to ensure each turn is assigned to the correct speaker ID.
    \end{itemize}
    \item Transcript word error rate and timestamp accuracy could vary with assigned transcriber or over time: while some mistakes are expected and there will be noise in where they occur, it is important to ensure that a large dip in quality of a given transcript is detectable ASAP, as well as that a smaller dip in quality systematically affecting a particular set of transcripts (e.g. a site that submits most interviews at the beginning of the project versus the end) cannot introduce unwanted bias into downstream analyses. 
    \begin{itemize}
        \item A less careful transcriber might include more inaudible markings or an egregious case of paraphrasing might return a smaller word count than one might expect for the interview's length. Although these red flags could also pop up for natural reasons, it would be ideal for the central quality monitor to check on the accuracy of transcripts with aberrations in any of our present QC measures.
        \item Hopefully major quality issues will be detectable by the current pipeline, especially as it is better integrated into the overall collection workflow. Regardless, it will be difficult to systematically catch smaller problems at the initial screening stage this code is designed for, so future analyses should keep that in mind. 
        \item It is worth noting that details of working with TranscribeMe and how transcribers are assigned to audio files may require different consideration for a large multi-site project versus a single group's study. It will be important to maintain a relationship with TranscribeMe for the duration of this project to keep on top of any issues.
    \end{itemize}
    \item There could also be unanticipated problems with foreign languages as mentioned, especially those with non-Latin characters: this needs to be watched for when relevant sites are added, and code can be adapted where reasonable. Of course without knowledge of the relevant foreign languages, it will be difficult to guarantee quality assurance to the same degree as English submissions, hence the inclusion here.
\end{itemize} 

\noindent Ultimately, TranscribeMe has largely returned high accuracy transcriptions, so as of now there is not reason for outright concern about the above list. These are simply factors to be aware of, because as stated the base QC provided by this pipeline is not intended as a catch-all, and shouldn't be mistaken for such.

\subsubsection{Managing storage}
\label{subsubsec:u24-storage}
As mentioned, during initial testing the central server ran out of storage space for a time, which obviously impacted all data processing. Interviews are larger than most of the other datatypes being collected for the AMPSCZ and similar digital psychiatry studies. It is therefore important to have a good idea about the amount of space that will be required for a planned interview dataset. Following the initial storage space expansion for this study, I documented file size properties of the collected Zoom interviews for future reference. In this section, I present that information, both to help with anticipation of additional storage issues that could arise as this study continues, as well as to assist the planning of future studies. \\

\noindent The following is an estimate of the required hard disk storage space for the Zoom interview protocol:

\begin{itemize}
    \item 1 MB per minute of audio for the main audio file, but there are also the speaker specific audios, which generally take up 2x that space, so 3 MB per minute of audio total. This part is quite consistent.
    \item For video it varies much more, anywhere from 4 MB to 12 MB per minute.
    \item A safe estimate then would be 15 MB total per minute of interview for an entire Zoom raw recording, although on average I’d say it is closer to 10 MB per minute.
    \item For reference, open interviews average about 15 minutes in length and are consistently between 10 and 25, while psychs interviews average 50 minutes but are routinely shorter than 20 minutes or longer than 1 hour.
\end{itemize}

\noindent So there is a lot of variation, but pure averages I estimate needing $\sim 150$ MB for an open interview folder and $\sim 500$ MB for a psychs one. 

Most of the pipeline outputs are negligible in size, however if the WAV-converted version of the main audio file is to be retained under processed outputs once transcription is complete, more storage should be set aside. WAV is uncompressed and for quality audio takes $\sim 10$ MB per minute. It might be somewhat smaller in practice here, but reserving 300 MB of disk space per open interview planned and 1 GB of disk space per psychs interview planned is recommended when using the pipeline as is. However, retaining the WAVs is purely for convenience, as they can easily be reconstructed from the input and the other pipeline outputs corresponding to it. WAV format will be needed for downstream feature extraction on the compute server, e.g. application of OpenSMILE, but it could be reconverted each time as needed if storage is a concern. \\ 

\noindent Regardless, we encountered many, many GB of completely wasted space to be cleaned up on the data aggregation server when running the multi-site study in practice, which likely more than doubled the amount of space being used per true interview. This is in large part a function of human error in interview uploads, so as protocols are solidified the magnitude of the problem should decrease. For example, as of 12/12/2022, the Pronet LA site had successfully submitted 7 psychs interviews and 6 open interviews, and had over 20 GB of space taken by raw interview files that were \emph{not} used at all in processing the 13 correct interviews. This is obviously the most egregious case, but it illustrates the relevance of these problems in attempting to manage storage space. A few specific things to watch out for are unneeded files in the Zoom folder, and Zoom folders that are not processable for reasons mentioned in the preceding section \ref{subsubsec:u24-issues}. 

As far as files created automatically by Zoom, the double\_click\_to\_convert*.zoom files within an interview folder are of unclear utility and not used by the pipeline at all, yet sometimes take up a huge amount of space -- even exceeding the rest of the interview. We're not sure if there was an element of human error involved in the .zoom file size ballooning, but either way Lochness was updated to no longer pull these files when it pulls the rest of an interview, and we would recommend future studies do similarly.

For handling other hard disk cleanup issues, it is important to note the design of Lochness; it is the reason that storage is long-term impacted by interviews uploaded with the file organization violations discussed in the preceding section. Lochness moves interviews from Box to the data aggregation server, but it does not modify or delete anything. If something is renamed, it is treated as an entirely new file by Lochness. Thus interviews with incorrect folder structure, naming, or file format will persist on the data aggregation server even after fixed in Box by the site. The corrected interview will be pulled by Lochness and subsequently processed by the code, but the prior version of the interview will remain a duplicate on the server, ignored. While the interview pipeline code does log many types of SOP violations with the date they were first detected, it is not intended to delete raw data files and should not be run on an account that would even make that possible. Further, the pipeline cannot see the state of Box, and if the offending interview had not yet been fixed there, it would just be re-pulled to the server the next day. 

Therefore, we recommend that as part of the manual monitoring workflow, someone with access to both Box and the data aggregation server should follow-up on critical SOP violations logged by the interview code. Besides contacting sites to fix their mistakes and hopefully prevent the same issues from becoming regular occurrences, the person monitoring raw data flow should manually (or semi-manually) delete the incorrect versions of interview sessions duplicated on the server. Alternatively, harsher upload requirements could have been incorporated into the Lochness code, thereby preventing upload of any unprocessable interview folders to the server until fixed in Box. Taking this route instead would remove much of the server disk maintenance burden, but it would also prevent the interview pipeline from flagging these issues. The Lochness team would then need to handle or at least relay communication about incorrect interviews placed in Box, or the interview team would need to do more careful manual checking of files added to Box to ensure they are pulled by Lochness or fixed when needed. The latter would be more work than the current setup, and while the former workflow could be considered in the future, it should be noted that there is greater risk of accidental interview deletion when misformatted interviews are withheld from the server entirely. \\

\noindent Because of the highly personal nature of recorded clinical interviews, the central server's disk space is the final destination for raw audio and video files; this is unlike most of the other study datatypes, which are less sensitive and as such can have raw data funnelled to downstream servers for analysis and sharing. Thus it is critical to provision enough central server disk space for long-term storage of the raw interview files, and to keep on top of the necessary tasks for maintaining organization of that disk. As of now, only QC features and the redacted versions of returned transcripts are sent to the downstream locations.

Similarly, it is especially important for interviews to have good data flow and quality control monitoring infrastructure, so that issues can be flagged and progress can be accurately reported even while raw data access permissions are kept highly limited. My code presented in this chapter is designed to fit those criteria. Furthermore, the central data aggregation servers here were set up to be computationally light weight, while simultaneously handling data flow for many modalities. Therefore the interview organization and QC code is written to use as minimal compute resources and take as minimal time per interview as feasible. Keeping this in mind, I will detail my code architecture and provide instructions for its further use in the next section (\ref{subsec:interview-code}). 

My interview monitoring/management pipeline is indeed the focus of the tool building portion of the present chapter. However, a second big part of handling the interview recording data collected during this project will be a comprehensive feature extraction protocol that runs from a processing server administered by the central server personnel and with the aforementioned disk mounted. Any extracted features, for example frame-by-frame facial expression measures or low level audio descriptors, can subsequently be shared with downstream servers as is already done for raw files of other datatypes. This part of the code wasn't required for data collection to start, and will require input from contributing research groups to determine the full feature set. Once it is completed, required disk space at the central server might further increase. Discussion of upcoming steps for that part of the project more broadly can be found in section \ref{subsec:interview-future}.

\subsection{Code architecture overview}
\label{subsec:interview-code-intro}
\begin{figure}
\centering
\includegraphics[width=\textwidth,keepaspectratio]{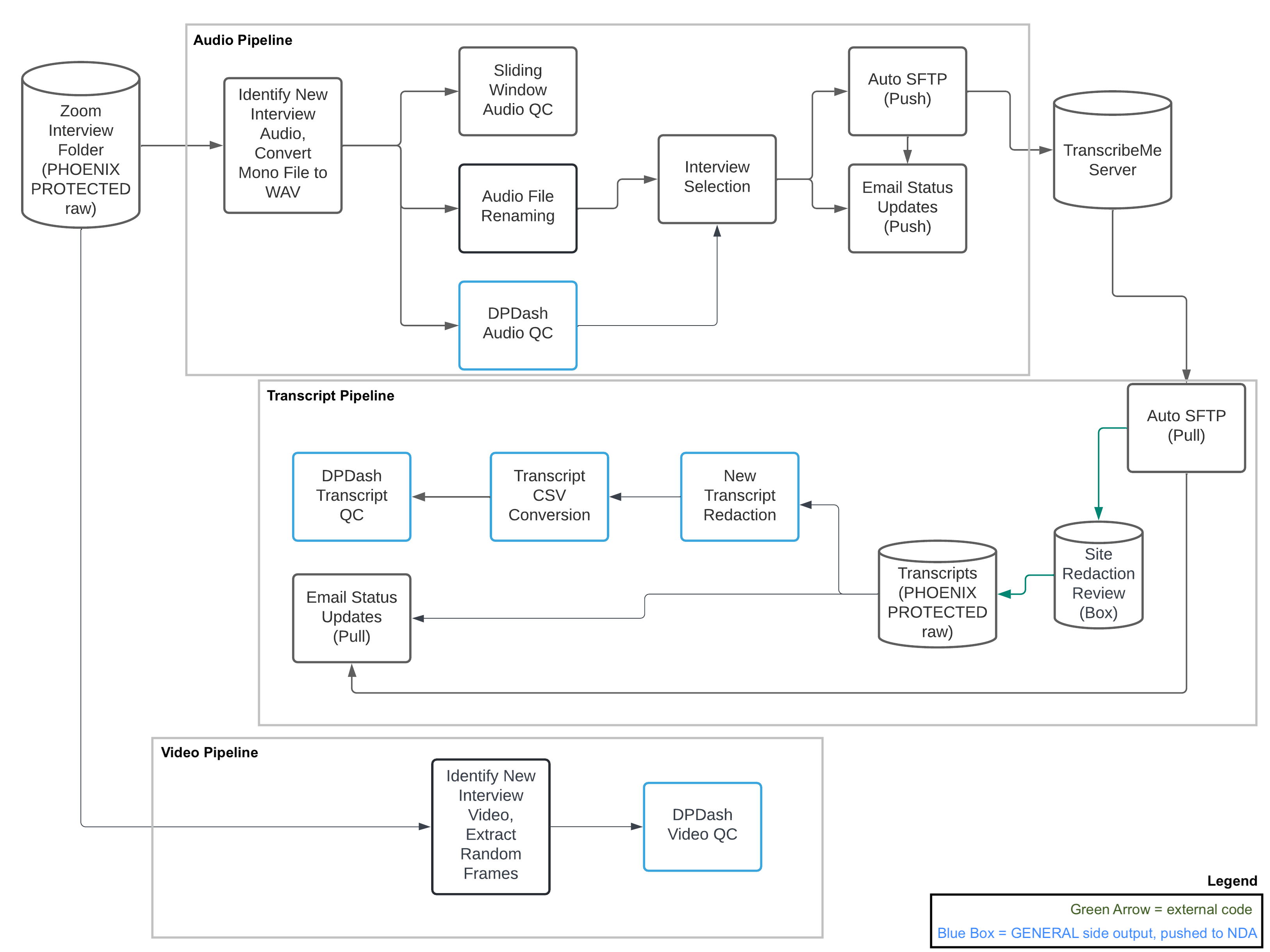}
\caption[Architecture diagram for clinical interview data aggregation and quality control software.]{\textbf{Architecture diagram for clinical interview data aggregation and quality control software.} The pipeline is broken into three major components: audio processing, transcript processing, and video processing. Each of these components, marked by the outer grey boxes, can be run in full for a specific site using a provided wrapping bash script and settings file template. On the AMPSCZ central data aggregation servers, all components are run for all sites as part of a daily cron job. Within each component, a number of independent functions are run. Each such function is represented by a text box here, with arrows denoting which functions rely on input from which other functions. Black arrows indicate that the corresponding data flow is handled by the pipeline code, while green arrows indicate that external code (provided by other groups working on the AMPSCZ project) is required to completely bridge the modules. A light blue border indicates that the function produces not only information needed by a downstream function or otherwise saved for future internal reference on the PROTECTED side, but also some deidentified shareable output(s) that will be stored on the GENERAL side of processed and thereby get sent to the later servers depicted in Figure \ref{fig:u24-arch} - which can be used for data sharing with AMPSCZ partners and eventually the broader research community. As will be described, additional utility functions for monitoring the AMPSCZ inputs/outputs are also run daily and shared in the pipeline repository, though they are not part of the core data management and quality control framework depicted here.}
\label{fig:zoom-arch}
\end{figure}

Like the code release of chapter \ref{ch:1} (\ref{subsec:diary-code}), this pipeline was implemented using a modular structure (Figure \ref{fig:zoom-arch}), to facilitate future updates and adaptations. Also similar to the audio diary pipeline, I provide extensive documentation in supplemental section \ref{subsec:interview-code}. The goal here is two fold:
\begin{enumerate}
    \item Ensure that others working on AMPSCZ can continue to efficiently monitor and improve on the current code, as well as debug where needed. While the pipeline is complete in the sense that it accomplishes core data flow and quality control aims, there are still a number of overarching speech sampling infrastructure TODOs remaining for the project, so it is important that I am effectively replaced. Thus an aim of this chapter is to get my hypothetical replacement up to speed. 
    \item Enable different projects, such as a Wellcome Leap initiative that has expressed interest or perhaps even a distinct future AMP initiative, to be able to use this code. Further, it is a crucial aim that the pipeline be reasonably approachable to adapt to unique situations with e.g. different interview formats. Handling all of the AMPSCZ needs made it difficult to find time to focus on further generalizing the code itself, but this thorough documentation is I hope the second best thing.
\end{enumerate}
\noindent Note also that within these two major code release priorities, there may be multiple distinct target audiences, hence the many details in this chapter. The current code being used for the ongoing AMPSCZ interview collection process can be found at \citep{interviewgit}. 

\FloatBarrier

\section{Relationships between clinical interview language features and disorganized thought}
\label{sec:disorg}
Now that important software infrastructure has been well established, I will provide an example analysis demonstrating some of the ways that high quality interview transcript datasets might be leveraged for scientific study, from the clinical interview recordings collected for the lab's BLS study. 

A major limitation of prior work analyzing the linguistic features of patient interview speech in psychotic disorders has been the focus on predicting broadly defined measures of overall symptom severity, if not a binary classification of diagnostic category, using a small sample of interviews with a single time point per patient \citep{Hitczenko2021}. We take a first step towards addressing many of those problems by focusing on a specific symptom domain of psychosis, disorganized thought, and by assessing its relationship with interview speech in a longitudinal dataset. Furthermore, we take a principled approach towards feature selection, using clearly defined and clearly justified predictors in our experimental models. A large focus of this effort is the use of linguistic disfluencies, which we break down into different categories, and subsequently investigate their relationships not only with clinically rated measures of disorganized thought, but also with each other. 

In the upcoming sections, I will introduce the interview dataset (\ref{subsec:disorg-dataset}) and the feature extraction and modeling methodologies used (\ref{subsec:disorg-methods}) before diving into a detailed presentation of our results (\ref{subsec:interview-results}). Note that this work has been in part already published by \cite{disorg22}, but I provide a deeper look at properties of the linguistic features in question here, particularly in subsection \ref{subsubsec:disorg-nlp-feats}, as well as my own interpretations of the main results. 

\subsection{Reviewing BLS}
\label{subsec:disorg-dataset}
The same Bipolar Longitudinal (BLS) study described in chapter \ref{ch:1} (section \ref{subsec:diary-methods}) was also utilized in this work. Of the 74 BLS participants, 59 consented to recorded monthly clinical interviews. Duration in the study varied greatly between participants, but on average one year of data was collected per subject. There were 737 total data points that included complete clinical scale ratings and same-day recorded interview across the entire dataset. 

Because of the heterogeneity of BLS patients in diagnosis and symptom profiles, not all subjects were a good match for this investigation. For example, the full dataset included a number of participants with severe affective symptoms but no history of psychotic symptoms. While availability of controls with other psychiatric symptoms is valuable, it is not desirable to have those with no relevant symptoms greatly outnumber those that do show variability in the symptoms of interest. Furthermore, it is preferable to focus on subjects that stayed in the study for a number of months, in order to implement more effective mixed models.  

Given the relatively small number of interviews available -- two different subjects contributed more than 737 daily audio journal transcript data points \emph{individually} -- it might have been ideal to use all available data in some capacity. However, a major part of our hypothesis was the relevance of linguistic disfluencies and a major aim was to investigate potential differences by disfluency category. To accurately count disfluencies with high confidence, we utilized a service for careful verbatim transcription done manually. As discussed in chapter \ref{ch:1} (section \ref{subsubsec:diary-val-trans}), disfluencies are more difficult to capture with current automated techniques than many other linguistic features. 

Thus it was necessary for the quality of our analysis to pay an additional \$1.75 per minute of interview included in the project. Due to the length of individual interviews, the cost quickly adds up; as such, the interviews were filtered to prioritize time points with greater negative symptom severity and participants with a longer period of study participation, and also to remove recordings of questionable quality using a workflow similar to that described for the AMPSCZ project in section \ref{sec:tool1}. 

The described interviews were conducted onsite and later via Zoom using the recording protocols detailed in section \ref{subsec:interview-methods}. Video was not used here, nor were acoustic properties, and transcriptions were done by a human. The recording methods are therefore less critical to this analysis than they might be to other research involving interview recordings. 

\subsubsection{Considered dataset}
The final dataset included 144 interviews spanning 18 subjects. The number of interviews contributed per subject ranged from 1 to 20, with a mean of 8.06 and a standard deviation of 6.92. Each of the 144 raw data points to be analyzed represented an interview transcript plus a complete set of Positive and Negative Syndrome Scale (PANSS) ratings, along with any associated metadata. These raw samples were then processed to obtain the input for statistical analyses. 

From the PANSS, the positive subscale item assessing conceptual disorganization (P2), the negative subscale item assessing difficulty in abstract thinking (N5), and the general item assessing poor attention (G11) were retained, to serve as the labels for each point. These three items were chosen because of our primary interest in disorganized thought and prior factor analysis results associating them \citep{Wallwork2012}. The subject ID of each data point was also retained from the metadata, for use in mixed effects modeling.

Note that there were a handful of additional BLS recorded interviews that did not have a complete set of clinical scales associated. In the final dataset, there was 1 such additional interview included, which had a corresponding P2 score but not a G11 or N5 score. Therefore the models for P2 prediction use 145 input time points, while the other models use the 144 already described. \\

\noindent Ultimately, 144 labeled time points is actually a large dataset for a study of linguistic properties from semi-structured interviews in psychotic disorders patients. Much of the prior work, from the paper that kicked off the scientific excitement in this domain \citep{Bedi2015} to the more recent results that employed large deep learning models for feature extraction \citep{Tang2021}, has utilized only smaller datasets of well under 100 interviews. 

Moreover, similar prior works have tended to include only a single time point per participant and to focus only on diagnosis-based category labels \citep{Hitczenko2021}. Despite the limitations of our final dataset, it still represents an advance towards the framework called for by \cite{Hitczenko2021} in multiple capacities. The dataset size is a relative strength and thereby a positive scientific contribution, even as it is also a weakness in absolute terms.

\subsection{Analysis methods}
\label{subsec:disorg-methods}
For the input features, 11 summary statistics were derived from the associated interview transcript, broadly related to verbosity, disfluencies, and incoherence. The number of low level language properties included was kept limited and the summary methods used to obtain per interview properties were kept simple in order to preserve statistical power as well as interpretability.

The eleven primary language features calculated per interview were - patient words divided by total words, patient words per second, patient words per sentence, patient disfluencies per sentence, patient nonverbal edits per sentence, patient verbal edits per sentence, patient repeats per sentence, patient restarts per sentence, mean patient sequential word incoherence per sentence, mean patient pairwise word incoherence per sentence, and mean patient word uncommonness per sentence. 

The modeling features were calculated using a similar protocol to the audio diary features of chapter \ref{ch:1}. Specific details of the transcript feature extraction techniques used for extracting these features from the selected BLS clinical interview transcripts can be found in supplemental section \ref{sec:disorg-sup}. The results were then used in statistical modeling of the clinical scale items of interest, as will be described next. 

\subsubsection{Statistics}
As part of my work, I evaluated the distributions and correlation structure of the obtained per sentence and per interview features within this dataset. The methodology for those investigations is the same as that for the default histograms and correlation matrices generated by the audio diary pipeline detailed in chapter \ref{ch:1} (within section \ref{subsec:diary-code}). For modeling the clinical scores from the interview language features, the following protocol was used. This description was reproduced from the relevant portions of the methods in \citep{disorg22}:

\begin{quote}
The analyses were designed to estimate the associations between each language feature and each disorganization clinical measure. Each clinical measure was regressed on each feature in a separate mixed effects linear regression model, taking participant as a random variable (allowing for different intercepts and slopes for each participant). To minimize the risk of type I errors, a Bonferroni correction was applied to account for the number of language features (n = 11) that were tested with each clinical measure. Thus, the statistical significance threshold was set at p < 0.0045. 
\end{quote}

\subsection{Results}
\label{subsec:interview-results}
In summary, we found a significant relationship between patient use of repeats and severity of conceptual disorganization (P2), which was independent of the significant relationship we found between the overall verbosity features and conceptual disorganization. We also found significant relationships between P2 and both restart and verbal edit use, though it is yet unclear to what extent that relationship is independent of verbosity (and in the case of restarts, independent of repeats). Nevertheless, there were promising signs of potential discriminatory utility in the distributions of these features, particularly that of verbal edits. Furthermore, while these features had dataset-wide significance to P2 score, there was also significant participant-specific variation found via the mixed effects model. 

To contextualize these results, I will report on properties of the clinical scale score distributions found in our dataset (\ref{subsubsec:disorg-scales}) and characterize both distributions and correlation structure within the linguistics features of the dataset (\ref{subsubsec:disorg-nlp-feats}) in the following subsections. I will then present the modeling results in more detail and discuss their interpretation in light of the feature and label characterizations I presented (\ref{subsubsec:disorg-modeling}), to close the results section.

\subsubsection{Clinical scales}
\label{subsubsec:disorg-scales}
I will first report on the distribution of P2 (conceptual disorganization), N5 (difficulty in abstract thinking), and G11 (poor attention) PANSS scores in the BLS dataset, both overall and in the final dataset isolated for this linguistics investigation. \\

\noindent Recall that each PANSS item is scored from 1 to 7, with the following designations:
\begin{enumerate}
    \item Absent
    \item Minimal
    \item Mild
    \item Moderate
    \item Moderate severe
    \item Severe
    \item Extreme
\end{enumerate}
\noindent A score of 3 or more indicates a symptom that is clearly established to be present, while a score of 4 or more indicates a symptom that is interfering with that patient's daily life. Increasing score represents a symptom with a higher frequency of occurrence or a stronger impact on the patient's life \citep{PANSS}.

In the original survey of 101 Schizophrenia patients by \cite{PANSS}, the mean PANSS item score was $\sim 2.5$. Of course not all items are expected to be high for an individual patient, but it is important to note that very high item scores are quite rare, and would be expected to be especially so in the outpatient BLS study population. \\ 

\noindent With that context in mind, I will summarize the clinical scale distribution results reported by \cite{disorg22}. 

Across the 59 BLS patients with interview data, P2 scores ranged between 1 and 5, with mean 1.43 and standard deviation 0.89. Similarly, N5 scores ranged between 1 and 6, with mean 1.41 and standard deviation 0.8. With somewhat more severity observed, G11 scores ranged between 1 and 5 with mean 1.7 and standard deviation 0.99. Indeed, 39 of the 59 patients had G11 $\geq 3$ at at least one point in the study, while 26 and 25 had N5 and P2 scores respectively $\geq 3$.

\begin{figure}[h]
\centering
\includegraphics[width=0.8\textwidth,keepaspectratio]{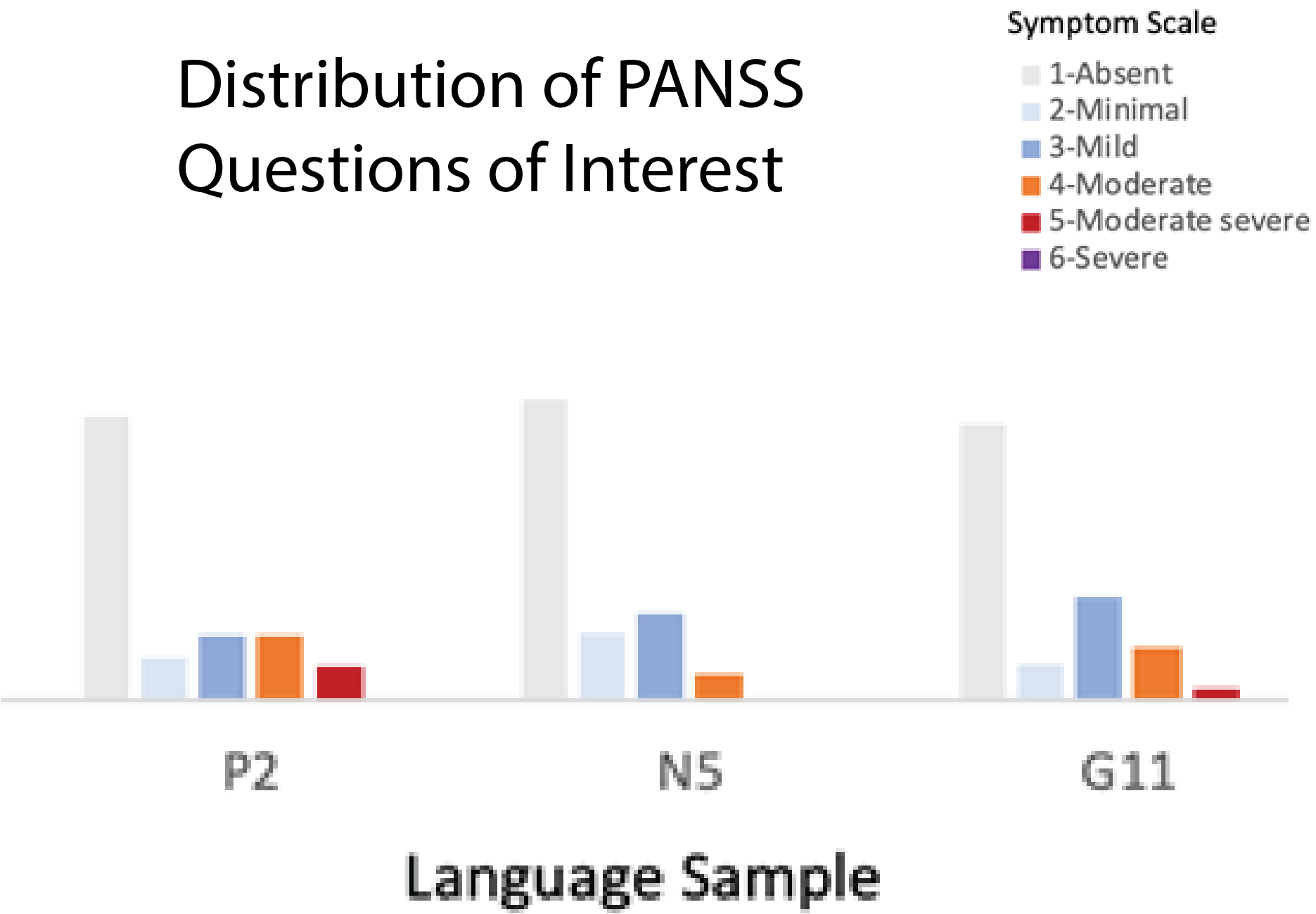}
\caption[Distribution of disorganization-related PANSS scores in the BLS interview language dataset.]{\textbf{Distribution of disorganization-related PANSS scores in the BLS interview language dataset.} Distribution of the PANSS P2 (conceptual disorganization), N5 (difficulty in abstract thinking), and G11 (poor attention) clinical scores in the sample of interviews used for analysis of language. Figure and caption reproduced from a panel of Figure 1 in \citep{disorg22}, on which I am a contributing author. There are 145 total data points making up the P2 histogram, and 144 total data points making up each of the N5 and G11 histograms.}
\label{fig:disorg-scales-dist}
\end{figure}

With a focus on variation in these scale items, the final dataset of 144 interviews from 18 BLS patients contained a greater proportion of symptomatic time points in the distributions (Figure \ref{fig:disorg-scales-dist}). The rating ranges remained the same, but in the final distribution P2 had mean 2.01 and standard deviation 1.36, N5 had mean 1.7 and standard deviation 1.03, and G11 had mean 1.96 and standard deviation 1.22. 11 of the 18 subjects had at least one time point with P2 $\geq 3$, 11 of the 18 subjects had at least one time point with N5 $\geq 3$, and 15 of the 18 subjects had at least one time point with G11 $\geq 3$. \\

For each item of interest, about half of the interviews in the final dataset have that symptom scored as absent (Figure \ref{fig:disorg-scales-dist}). There are a few different factors potentially contributing to this final composition, even after a number of symptom-free data points were filtered from the full set of BLS clinical interviews. 

One is that some BLS patients exhibited strong positive symptoms but not strong negative symptoms, or vice versa, such that an interview included because of moderate P2 score might have N5 absent, or vice versa. Though factor analysis has grouped these items together - and separately from other positive and other negative items - in the past \citep{Wallwork2012}, this correlation is of course not perfect and may in fact only exist in a subset of patients. As will be discussed in chapter \ref{ch:3}, there is evidence that different clinical scale factoring models fit different populations; in that chapter, the focus is on splitting the YBOCS for OCD along versus against the obsessive and compulsive subscales.

Another consideration in the high number of absent ratings is the longitudinal nature of the study. Single time point studies often recruit patients that are actively experiencing symptoms. By contrast, BLS follows participants for on average a year after recruitment, and during that time some participants do experience periods free of many symptoms. Because the final dataset intended to capture a number of interviews per included participant (with mean about 8), there may be extra samples with all symptoms of interest absent that were included because of subject ID. This decision was made not only for the mixed effects modeling done here, but also to build a transcript dataset for subjects of salience that will be usable for future investigations into different psychiatry topics. \\

Overall, most of the interviews with P2, N5, or G11 scored $\geq 4$ across BLS were used in the language modeling analysis. Of those excluded, many had been flagged for exceptionally poor audio quality - as discussed in section \ref{subsec:interview-methods}, there were recurring issues with some of the equipment used for onsite interview recordings. 

For scores of 2 or 3, inclusion decisions also considered the balance of subject IDs, with the majority of these data points across BLS not included in the final dataset. Nevertheless, for P2 and N5 the ratio of mild symptom scores to absent symptom scores more than doubled in the filtered set, and the ratio of minimal symptom scores to absent symptom scores also increased. For G11 there was a greater incidence of symptoms to begin with, but again the ratio of mild symptom scores to absent symptom scores increased in the filtered set. \\

Ultimately, amongst the roughly 50\% of time points for each item in the final dataset with non-absent symptom scores, there were somewhat different distributions observed. For P2, there was a good variety of scores between 2 and 5, with the greatest availability centered at scores of 3 and 4 (Figure \ref{fig:disorg-scales-dist}, left). For N2, there was a larger bias towards scores of 2 and 3, with a small set of 4's available (Figure \ref{fig:disorg-scales-dist}, middle). For G11, scores of 3 were substantially more available, and scores of 4 also had good availability relative to the other items; but scores of 5 were rare, and scores of 2 were more often filtered out (Figure \ref{fig:disorg-scales-dist}, right).

The final distribution properties are a reflection both on selection criteria and on the original distribution. In the case of mild or stronger ($\geq 3$) symptom scores on any of the items, a lack of availability in the final dataset indicates a lack of availability across the entirety of BLS however. Given the study population, it is not surprising that scores $\geq 5$ were generally rare. This of course may affect what we are able to detect. Differences in the distributions across items may hold some clinical interest, but they also may affect how well each label is able to be modeled. 

The fact that scores of 2, 3, and 4 were more common for G11 than for P2 or N5 was fairly robust across BLS and may be a reflection of either how that item was rated or of the composition of the study population. It would not be surprising for those with primarily affective symptoms to sometimes demonstrate poor attention without demonstrating cognitive symptoms more directly related to psychotic disorders. Difficulty concentrating is in fact an item on the MADRS rating scale for depression, while P2 and N5 don't have a clear analog \citep{MADRS}. 

The fact that 5's were relatively more common for conceptual disorganization than for the other two items across BLS is likely a random effect, in that it is driven by only a few patients that happened to demonstrate more severity in disorganized thought than in the other domains. It is unlikely that a scaled up replica of BLS would continue to display this disparity between items, though it is impossible to draw any conclusions without more research. Yet in the present dataset, it is important to keep this disparity in mind, as it may contribute to the modeling results that follow.  

\FloatBarrier

\subsubsection{Characterizing linguistic features}
\label{subsubsec:disorg-nlp-feats}
Now that the labels in the analysis dataset have been characterized, I will better characterize the inputs -- both through evaluating the distributions and internal correlation structure amongst the per interview linguistic features, as well as through a closer look at properties of the per sentence features that were used to generate the per interview ones. Once that is complete, there will be sufficient context to dive into the modeling results using these inputs and labels, in section \ref{subsubsec:disorg-modeling}. \\

\paragraph{Interview feature distributions.}
To evaluate the distributions of the linguistics features we used in the models of Figure \ref{fig:disorg-nlp-regression}, I generated and inspected histograms of those interview-level features over the considered transcript dataset ($n=145$).

\begin{figure}[h]
\centering
\includegraphics[width=\textwidth,keepaspectratio]{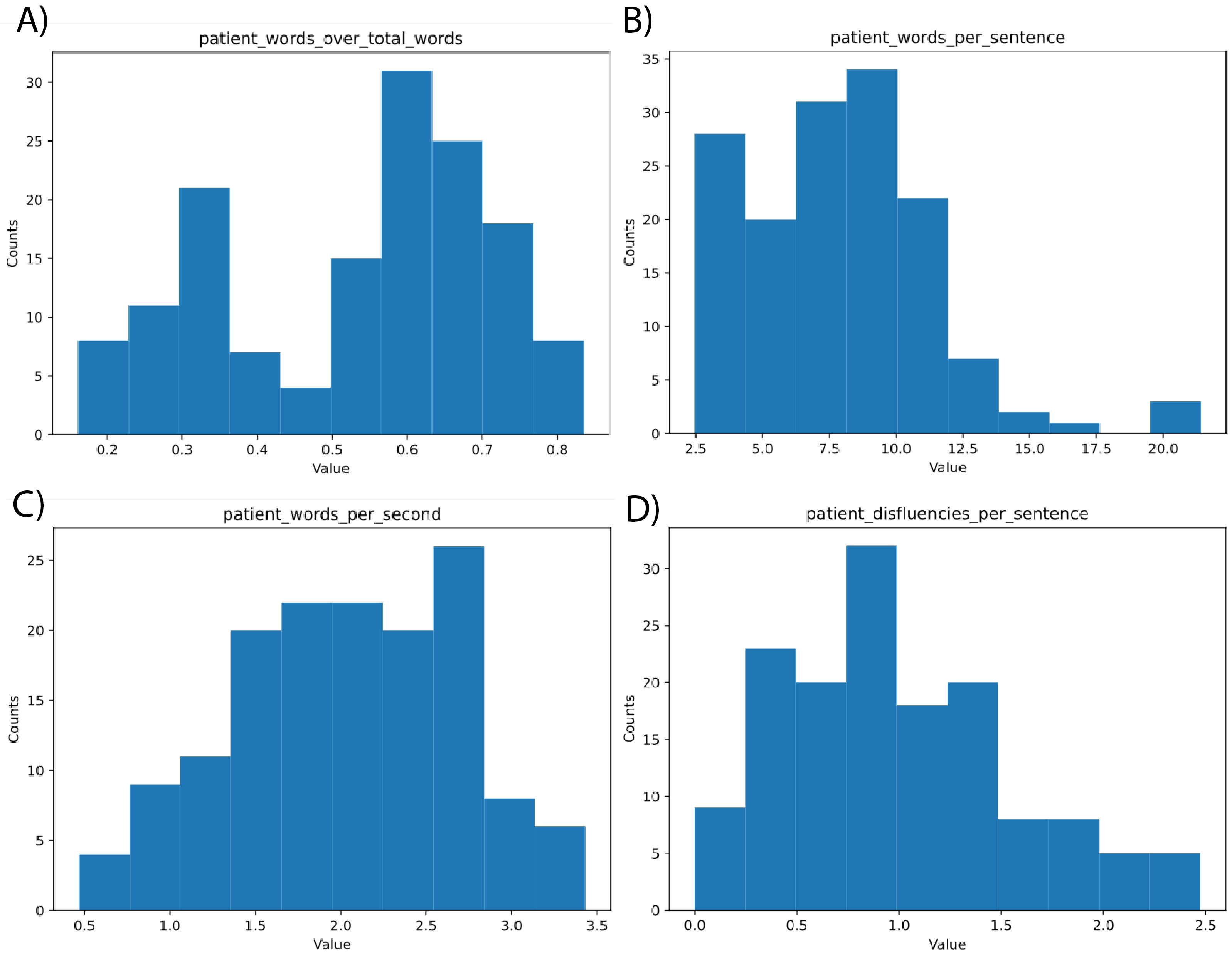}
\caption[Distribution of interview-level speech production features across BLS transcripts.]{\textbf{Distribution of interview-level speech production features across BLS transcripts.} For each linguistic feature tested by \cite{disorg22}, a histogram over the dataset of 145 BLS interview transcripts was generated using matplotlib (Figures \ref{fig:disorg-verbosity-dists}-\ref{fig:disorg-nlp-dists}). Here, we focus on basic speech production related features: the fraction of total words in the interview spoken by the patient (A), the average number of words per patient sentence (B), the average number of words per second of patient speech (C), and the average number of total disfluencies per patient sentence (D). The distribution broken down by disfluency category can be found in the next Figure \ref{fig:disorg-disfluency-dists}. Note that the features (and data) in these histograms directly correspond to features that were used to model conceptual disorganization in Figure \ref{fig:disorg-nlp-regression}. The first 4 relationships starting from the bottom of that figure match (A)-(D) here, in order.}
\label{fig:disorg-verbosity-dists}
\end{figure}

First, I considered speech production related features (Figure \ref{fig:disorg-verbosity-dists}). The fraction of words in the interview transcript attributable to the patient varied widely in this dataset, with a bimodal shape (Figure \ref{fig:disorg-verbosity-dists}A). The larger of the two peaks in the distribution was centered around $\sim 0.6$ and the smaller around $\sim 0.3$. About $\frac{1}{3}$ of the dataset was transcripts with patient accounting for less than half of total speech, which is somewhat unexpected in the semi-structured interview setting. Although some of the questions call for one word responses (e.g. rating specific symptom severities), there are a number of open ended questions as well that could prompt much longer patient responses. 

Unsurprisingly, the variance in speech fraction was also reflected in high variance over the number of words per patient sentence distribution (Figure \ref{fig:disorg-verbosity-dists}B). Interestingly, there were over 25 transcripts in the dataset with only $2.5$-$4.4$ mean words per sentence, potentially indicating extremely abbreviated responses that could be themselves of clinical interest. Conversely, $\sim \frac{1}{4}$ of the transcripts had mean words per sentence exceeding $10$. It may be worth further investigating the transcripts on both ends of this distribution to better understand how they arose. As the interview protocol did not have an entirely rigid structure and different interviewers were involved with BLS (often in a time or subject dependent manner), it would be useful to rule out any verbosity variance effects that could be attributable to interviewer behavior.

Variation in verbosity, especially at the extreme ends, could impact disfluency behavior in a way that cannot be corrected for with standard normalization techniques. An interviewee that repeatedly gives very short answers in a particular interview would hardly even have the chance to utter a verbal edit or restart, and in many cases would be less likely (per word) to repeat themselves too - though we do count stuttering over an individual word as a repeat instance. Conversely, extremely long sentences, which would have been classified that way by a human transcriber at TranscribeMe, likely indicate either more disfluency use (why it was grouped together in the first place) or a rambling style of "run on" thoughts (may or may not be accompanied by disfluencies). Because we normalized disfluency counts over sentences instead of words, such long sentences are in a sense penalized; yet in another sense they are weighted more fairly, as a restart by definition leads to additional sentence length that perhaps should not be normalized out by a word count divisor.   

Ultimately, it is difficult to perfectly normalize certain features, as what they should or should not express has dataset and question dependent factors. As I proceed in this section, I will discuss other potential adjustments to disfluency counts that we could have considered. However these are study-specific decisions; for inclusion in a general use software release like work I describe in this thesis, there is clear reason to keep the base returned outputs as straightforwardly defined as possible, so others can use as they see fit. \\

\begin{figure}[h]
\centering
\includegraphics[width=\textwidth,keepaspectratio]{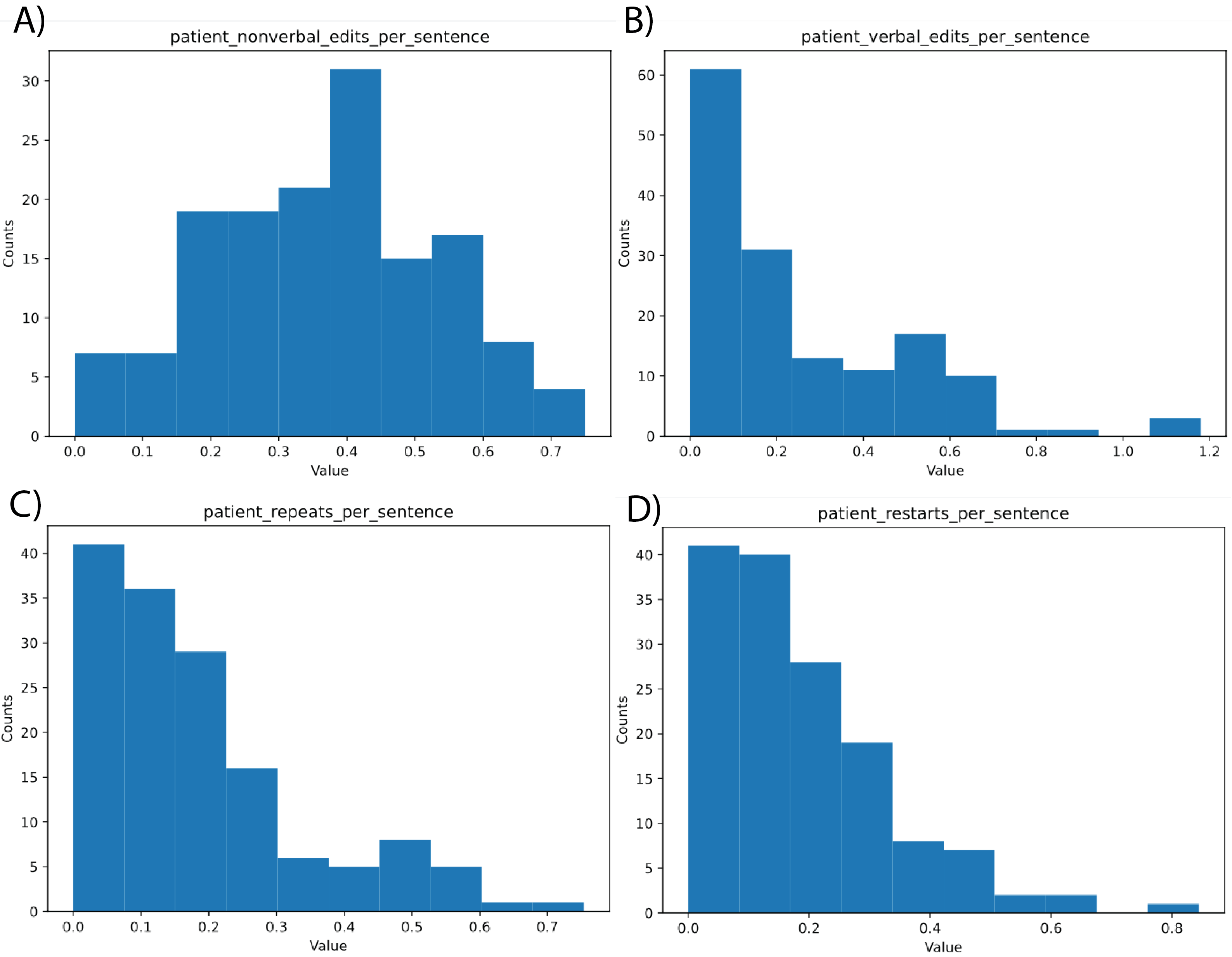}
\caption[Distribution of interview-level disfluency features across BLS transcripts.]{\textbf{Distribution of interview-level disfluency features across BLS transcripts.} As was introduced in Figure \ref{fig:disorg-verbosity-dists}, we move onto distributions of the individual disfluency categories in this dataset. Histograms of the mean number of nonverbal edits (A), verbal edits (B), repeats (C), and restarts (D), computed over patient sentences for each interview transcript, are depicted here.}
\label{fig:disorg-disfluency-dists}
\end{figure}

Disfluencies, the core feature of interest here, were quite common across the BLS interview transcript dataset ($n=145$), with the distribution over patient speech centered just under 1 disfluency (of any type) per sentence on average (Figure \ref{fig:disorg-verbosity-dists}D). To characterize these disfluencies per category, I next reviewed the distributions of the interview-level nonverbal edit, verbal edit, repeat, and restart features (Figure \ref{fig:disorg-disfluency-dists}).

Nonverbal edits were the most commonly used disfluency by patients in this interview dataset (Figure \ref{fig:disorg-disfluency-dists}A), which is in line with prior expectations about conversational nonverbal filler use. The distribution here was centered around $\sim 0.4$ mean nonverbal edits, with $\sim \frac{2}{3}$ of the transcripts averaging $0.3$ or more nonverbal edits per patient sentence. 

All other disfluency categories had peak at the smallest bin ($< 0.1$ or smaller). Verbal edits (Figure \ref{fig:disorg-disfluency-dists}B) and repeats (Figure \ref{fig:disorg-disfluency-dists}C) shared some distributional properties, with an initially sharp dropoff followed by a fat tail. More than half of the interview transcripts had mean verbal edits per patient sentence $< 0.2$, but there were also over 30 transcripts with mean verbal edits $> 0.4$ and a handful $> 1$. As there were no other disfluency types with any transcript exceeding 1 per patient sentence, it may be worth further investigating these verbal edit outlier interviews.

Mean repeats per patient sentence displayed a similar pattern to a lesser extent, with just over half the distribution $< 0.15$ (Figure \ref{fig:disorg-disfluency-dists}C), yet more than 20 transcripts $> 0.3$. Restarts on the other hand were moderately common, with a slighter initial dropoff but a small tail (Figure \ref{fig:disorg-disfluency-dists}D). There were substantially more transcripts with mean restarts per patient sentence between $0.2$ and $0.4$ than for verbal edits or especially for repeats, but less than $10\%$ of the interviews had mean restarts above $0.4$. 

The observed distributions of course have an impact on interpreting the eventual modeling results. Distributions with fat tails, like those for verbal edits and repeats, often warrant different regression models than predictors with a more typical Gaussian-like shape would. Because we considered $\sim 8$ interviews on average from each of 18 participants, there is also a potential subject-specific effect to these distributions. While we did use mixed effects modeling in the prediction of clinical scores from these features, it would be ideal in the future to better understand the dataset by investigating where different subjects fall within these distributions of interest.

It is also worth noting that given the generally low symptom severity in this population, the distribution of nonverbal edits would itself suggest a minimal role -- unless subtle increases in nonverbal filler use are significant, which would be surprising to find in a small and heterogeneous population, there is no way for the distribution depicted in Figure \ref{fig:disorg-disfluency-dists}A to predict the distribution of Figure \ref{fig:disorg-scales-dist}. However, our research was very focused on a specific category of symptoms, such that it remains highly plausible that features like nonverbal edits could hold predictive power for other important symptom types. \\

\begin{figure}[h]
\centering
\includegraphics[width=\textwidth,keepaspectratio]{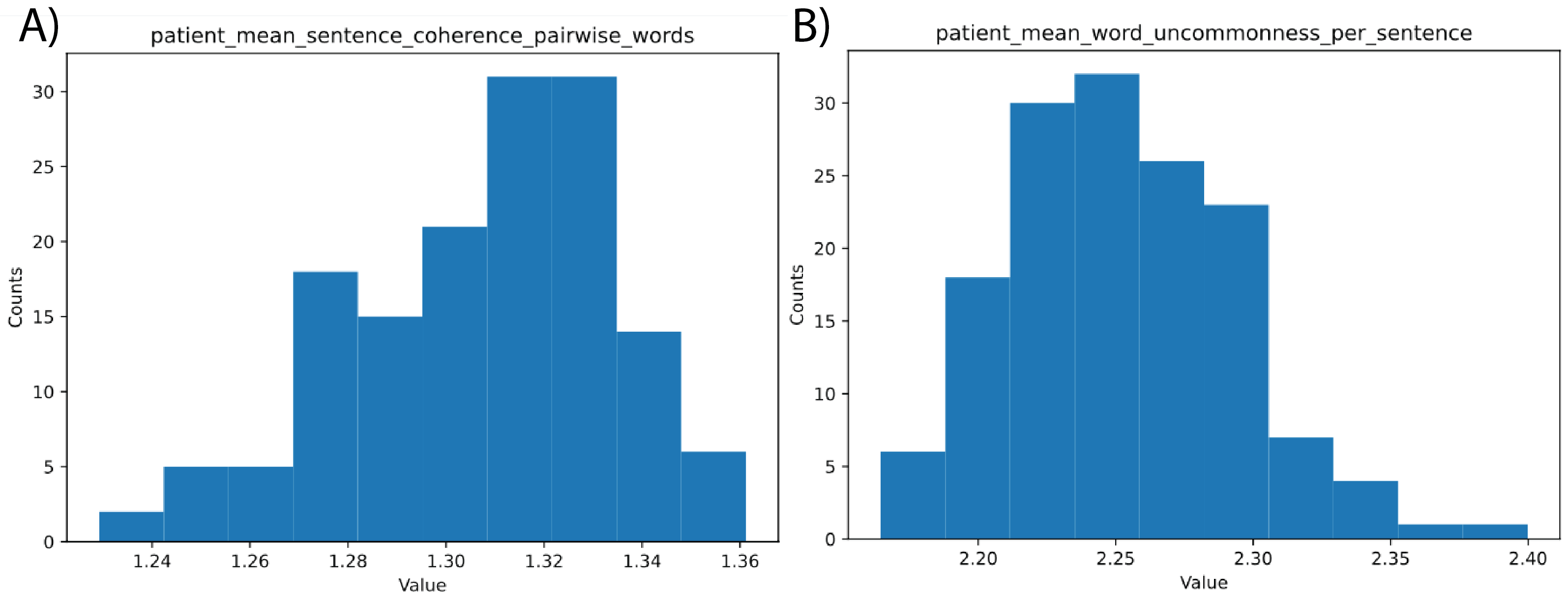}
\caption[Distribution of interview-level semantic incoherence features across BLS transcripts.]{\textbf{Distribution of interview-level semantic incoherence features across BLS transcripts.} Histograms were also generated for the word2vec-derived semantic coherence interview features, as was done for the other key features in Figures \ref{fig:disorg-verbosity-dists} and \ref{fig:disorg-disfluency-dists}. The mean patient sentence pairwise incoherence (A) and mean patient sentence word uncommonness (B) are depicted here. The mean incoherence metric generated using angles between sequential words only was extremely similar to the pairwise distribution in (A), so it is excluded from this figure.}
\label{fig:disorg-nlp-dists}
\end{figure}

The final category of feature included in the clinical modeling (Figure \ref{fig:disorg-nlp-regression}) was the word2vec-derived semantic incoherence features. As the mean sequential sentence incoherence and mean pairwise sentence incoherence had extremely similar distributions (and as will be discussed turned out to be highly correlated), only the latter is investigated more closely here, along with the mean word uncommonness feature (Figure \ref{fig:disorg-nlp-dists}). 

The units of these features are not as immediately interpretable, so it is difficult to make many strong statements about these distributions without additional context. As each is the mean (over sentences) of a mean (over words/word pairings), it is not too surprising that we observed a Gaussian-like shape. The word uncommonness feature (Figure \ref{fig:disorg-nlp-dists}B) was centered around $\sim 2.25$ with a standard deviation of roughly 0.05, which represented a fairly low coefficient of variation. 

The sentence incoherence feature (Figure \ref{fig:disorg-nlp-dists}A) was similarly narrow, with range 1.23 to 1.36 and more than half of the dataset falling between 1.3 and 1.34. The outliers found on the low end of the incoherence feature are likely a sign of low verbosity, so it is difficult to imagine this distribution containing much independent information. In the next section, I will review sentence-level feature distributions, as this could help clear up whether a different summary statistic on sentence-level incoherence might yield more relevant interview features in the future. \\

\FloatBarrier

\begin{figure}[h]
\centering
\includegraphics[width=\textwidth,keepaspectratio]{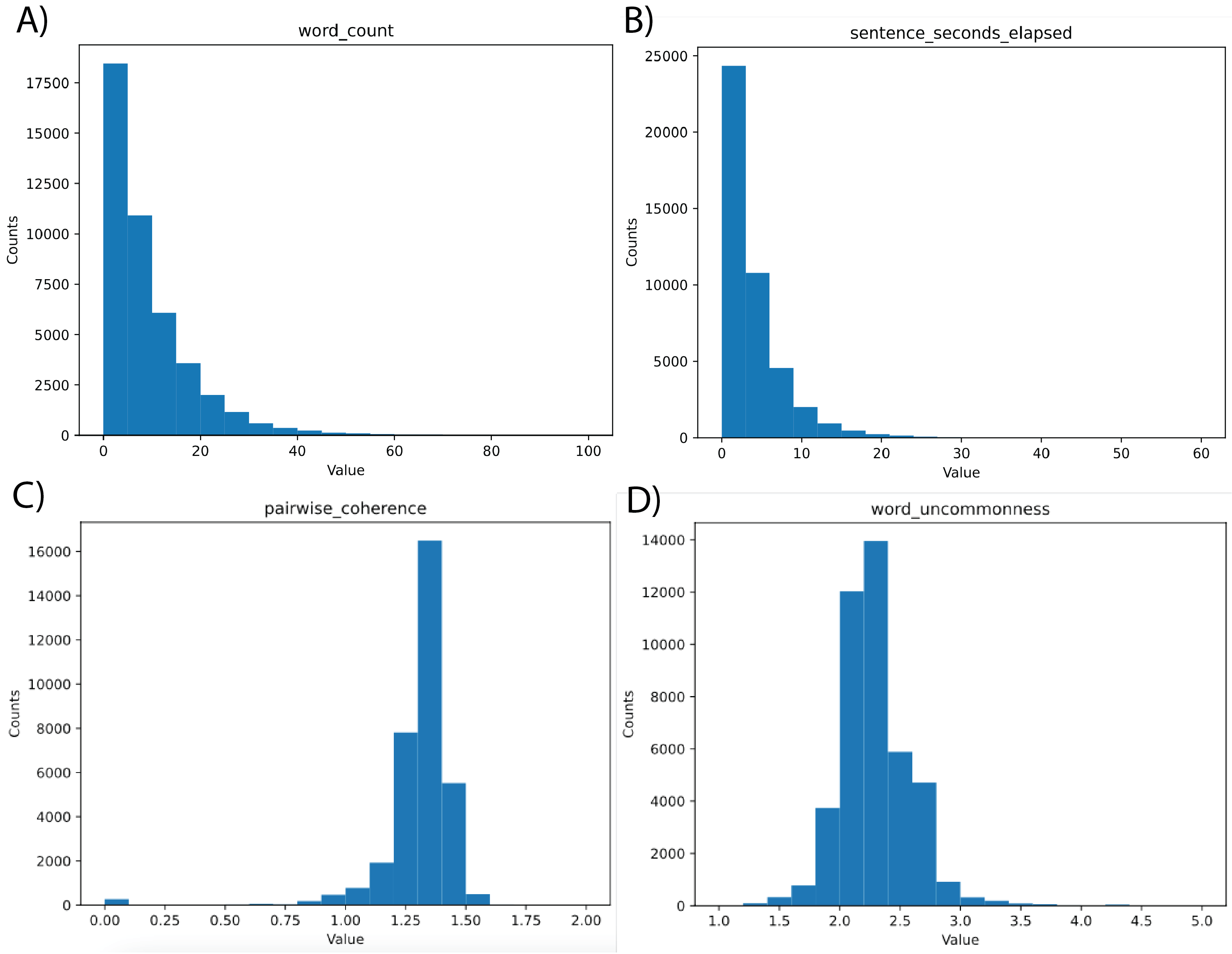}
\caption[Distribution of select sentence-level features across patient sentences in the BLS interview transcript dataset.]{\textbf{Distribution of select sentence-level features across patient sentences in the BLS interview transcript dataset.} To better understand the interview features used by \cite{disorg22}, we can characterize the distributions of the underlying sentence-level features in the same dataset. Across all patient sentences in the transcripts of the filtered BLS interview set, histograms were generated for the following metrics: sentence word count (A), sentence duration in seconds (B), sentence incoherence measured via mean angle between all word embedding pairings (C), and sentence word uncommonness measured via mean vector magnitude of all word embeddings (D).}
\label{fig:disorg-sentence-dists}
\end{figure}

\paragraph{Sentence-level components.}
There were approximately 42500 total patient sentences across the BLS interview transcript dataset that we considered. To get a better understanding of where the above interview-level metrics were coming from, I thus evaluated the distributions of underlying sentence-level features, beginning with speech production and semantic coherence related sentence properties (Figure \ref{fig:disorg-sentence-dists}).

Unsurprisingly, the mean word uncommonness per sentence distribution remained centered around $\sim 2.25$, but with greatly expanded range (Figure \ref{fig:disorg-sentence-dists}D). The pairwise incoherence per sentence distribution was expanded as well, though to a lesser degree (Figure \ref{fig:disorg-sentence-dists}C). The peak was between 1.3 and 1.4, but with a left-skew around this peak. There was a small bump at 0 incoherence attributable to one word responses. 

Interestingly, there was also a small bump at sentence incoherence $> 1.5$, corresponding to nearly 500 sentences with higher incoherence (Figure \ref{fig:disorg-sentence-dists}C). Understanding how these distribute across transcripts would be an important next step to determining if a summary stat like maximum or perhaps 90th percentile sentence incoherence might better capture coherence variability in the dataset. Because there are few examples of symptom severity above moderate amongst these transcripts, a stark difference in incoherence would not be expected anyway. It therefore remains a future direction to assess whether a feature designed to capture more subtle changes in coherence could be relevant to symptoms of disorganized thought. \\

\begin{figure}[h]
\centering
\includegraphics[width=\textwidth,keepaspectratio]{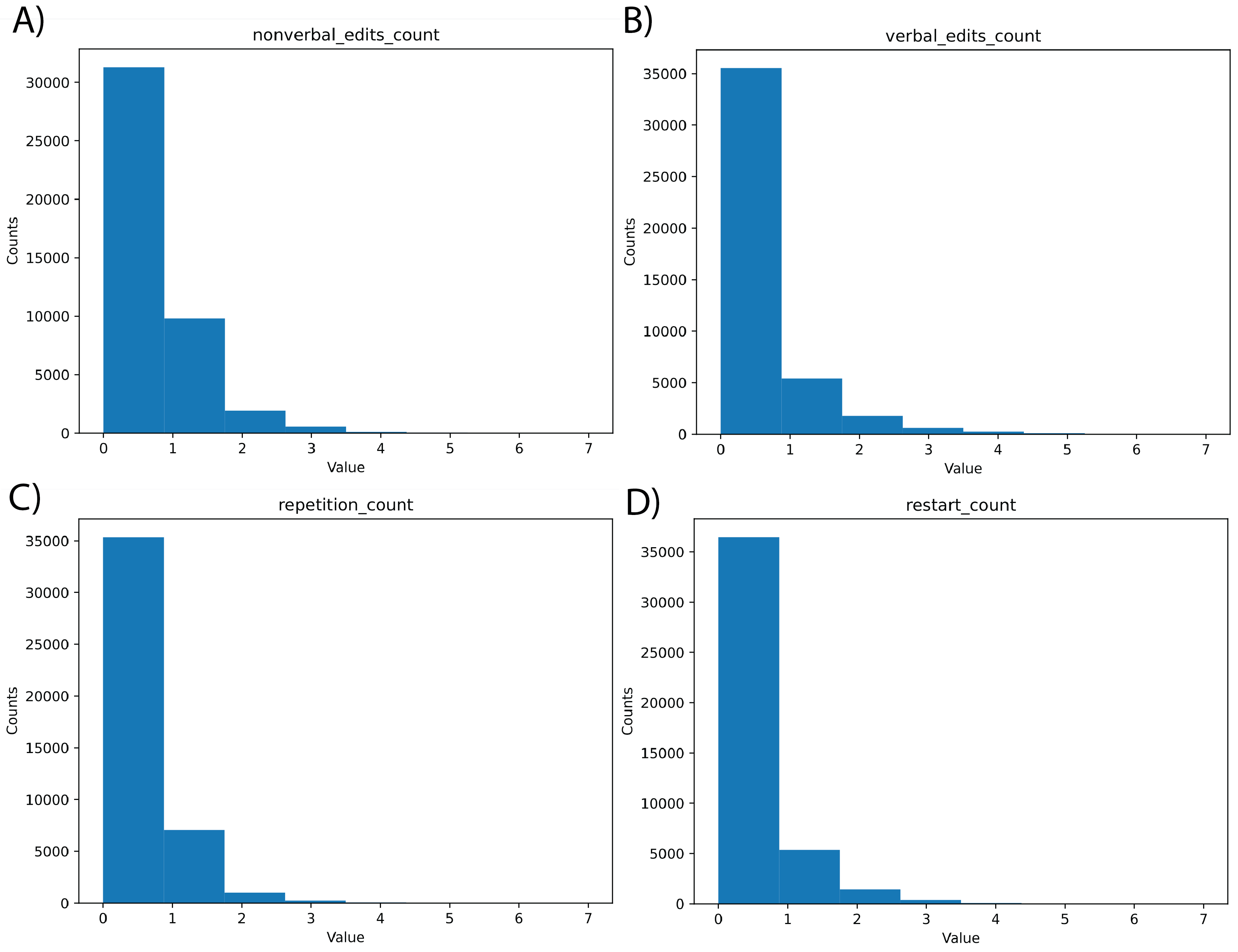}
\caption[Distribution of disfluency counts across patient \emph{sentences} in the BLS interview transcript dataset.]{\textbf{Distribution of disfluency counts across patient \emph{sentences} in the BLS interview transcript dataset.} Similarly to Figure \ref{fig:disorg-sentence-dists}, histograms were also generated for the sentence-level counts of each disfluency type: nonverbal edits (A), verbal edits (B), repeats (C), and restarts (D).}
\label{fig:disorg-sentence-disfl-dists}
\end{figure}

Occurrences of different categories of linguistic disfluency across patient sentences were similarly investigated (Figure \ref{fig:disorg-sentence-disfl-dists}), due to their importance to our aims for this study. As expected based on the interview distributions, nonverbal edits (Figure \ref{fig:disorg-sentence-disfl-dists}A) appeared in notably more sentences than the other three types. Of sentences with at least one nonverbal edit, $> \frac{3}{4}$ of them contained exactly one.

Comparing the other three types of disfluencies yielded much more interesting results. All had a similar number of sentences with nonzero value. Verbal edits (Figure \ref{fig:disorg-sentence-disfl-dists}B) were the most likely to reoccur in a single sentence, as $\sim \frac{1}{3}$ of all sentences with at least one verbal edit contained more than one. By contrast, repeats (Figure \ref{fig:disorg-sentence-disfl-dists}C) were the least likely to occur more than once in a single sentence, which is somewhat surprising because of the similarity between the repeat and verbal edit distribution shapes on the interview level (Figure \ref{fig:disorg-disfluency-dists}). This suggests that the larger tail observed in mean repeats per sentence was due to stronger between sentence correlation in repeats, while the tail in verbal edits would be at least partially explained by a correlation within sentences.

Restarts (Figure \ref{fig:disorg-sentence-disfl-dists}D) had only a moderate likelihood of occurring more than once in a single sentence, although higher than one might expect based on the definition of a restart. Surprisingly, restarts also had a slightly greater number of zero count sentences in the dataset than verbal edits or repeats did, despite having fewer very low data points on the transcript level (Figure \ref{fig:disorg-disfluency-dists}). This suggests that restarts were less related across sentences within a transcript in our dataset. Whether that could be explained by participant-specific behavioral patterns, by some underlying pathology, or by something else entirely would require a deeper dive into the dataset. Nevertheless, these distributional characterizations help set up for interpreting our modeling results. \\

As part of our transcript preprocessing workflow, we extracted a number of additional sentence-level features within the interview transcripts that were never utilized in modeling disorganized thought nor even summarized per interview. These extra features were obtained from existing infrastructure code, mainly based on the audio diary pipeline described in chapter \ref{ch:1}. The final interview features to use in this research were decided upon because of prior justification in the literature related specifically to the symptoms of interest to us. However, for understanding properties of linguistic features more broadly, the larger sentence feature set can be worth exploring. 

\begin{figure}[h]
\centering
\includegraphics[width=0.8\textwidth,keepaspectratio]{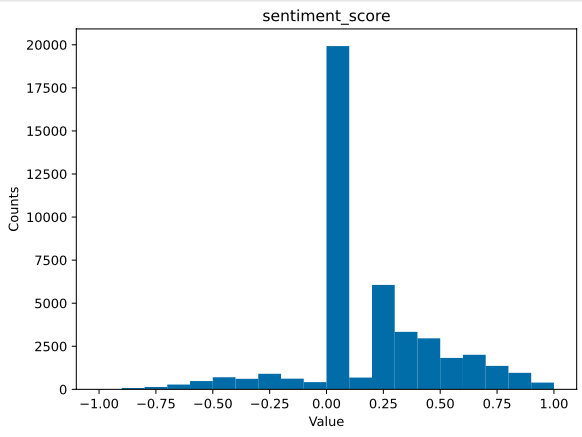}
\caption[Distribution of VADER sentiment output over patient interview sentences.]{\textbf{Distribution of VADER sentiment output over patient interview sentences.} Although it was not included in the final feature set or even summarized at the interview level, VADER sentiment (among other features that were described in chapter \ref{ch:1}) was computed for each sentence of each interview transcript for future use. Due to its relevance with audio journals in chapter \ref{ch:1}, the distribution of sentence sentiment scores over all patient sentences across interview transcripts is plotted here for reference.}
\label{fig:sentence-sentiment-pt-dist}
\end{figure}

Because VADER sentiment was a key part of the transcript analysis work of chapter \ref{ch:1}, its sentence-level distribution over patient interview speech is of especial interest (Figure \ref{fig:sentence-sentiment-pt-dist}). Nearly half of the sentences in this dataset had a sentiment score of 0. Amongst sentences with meaningful sentence magnitude, the large majority were of positive valence. This has clear implications for the discussion of summary statistics taken across sentences in audio journals, though of course there are also potential context differences that might affect per sentence sentiment distributions between journals and interviews. \\

\FloatBarrier

\begin{figure}[h]
\centering
\includegraphics[width=0.8\textwidth,keepaspectratio]{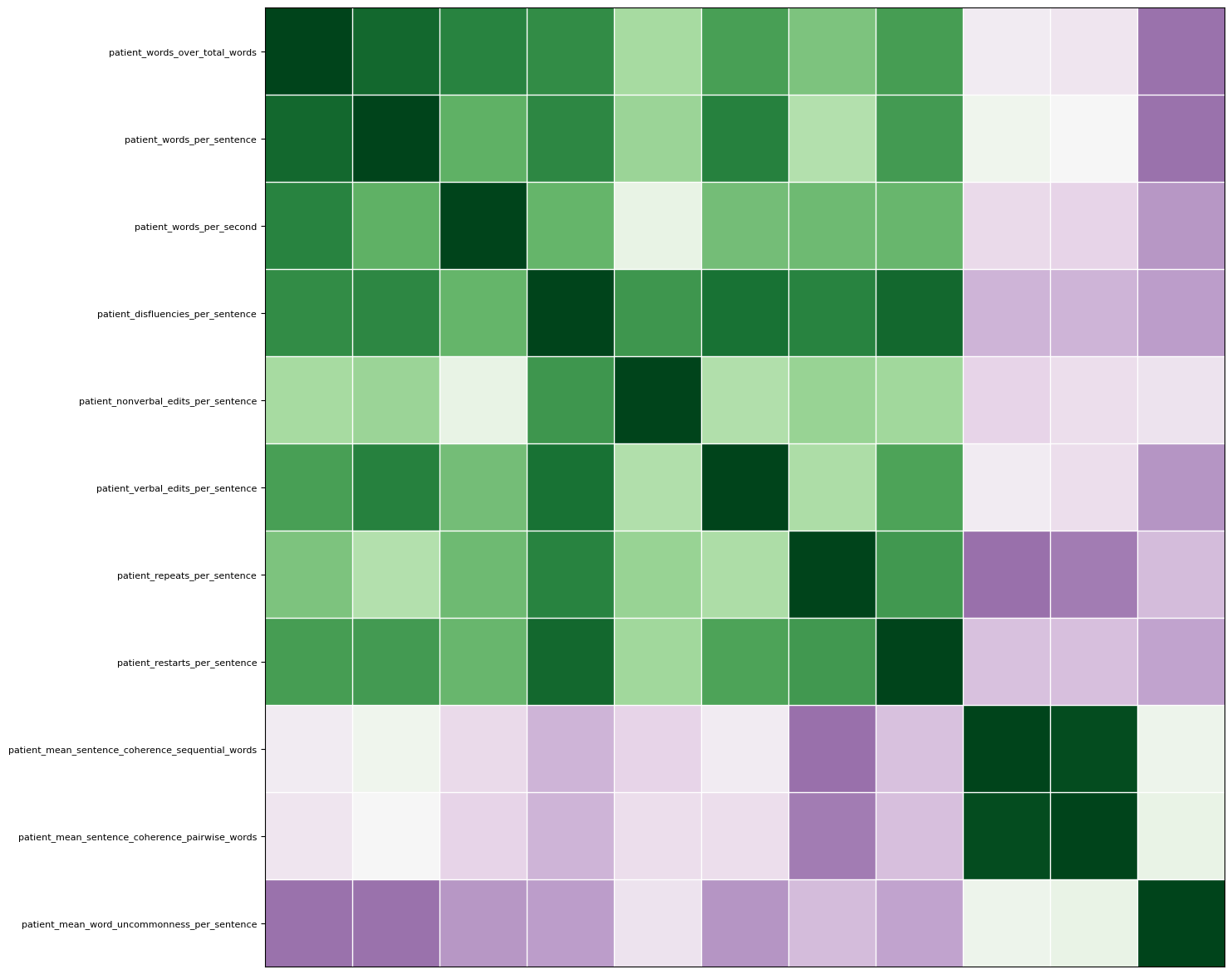}
\caption[Correlation structure of chosen interview-level linguistic features in the BLS transcript dataset.]{\textbf{Correlation structure of chosen interview-level linguistic features in the BLS transcript dataset.} Pearson's $r$ was computed with the scipy.stats python package for each pairing of the 11 final interview transcript features used by \cite{disorg22}. The resulting correlation matrix is visualized here using the matplotlib PrGn colormap with bounds -1 to 1. The features here are presented in the same order as the corresponding distributions were presented -- from bottom to top of the modeling results listed in Figure \ref{fig:disorg-nlp-regression}.}
\label{fig:disorg-interview-corr}
\end{figure}

\paragraph{Correlation structures.}
Finally, to gain some basic understanding on the relationships between the 11 features used to model conceptual disorganization in Figure \ref{fig:disorg-nlp-regression}, a Pearson correlation matrix was constructed to present the linear $r$ between each input linguistic feature pairing across the 145 interview transcripts considered (Figure \ref{fig:disorg-interview-corr}). 

One major question of interest for this correlation analysis is how related each disfluency feature was to the metrics of overall verbosity, as it is critical to understand how much of any observed predictive power for disfluencies was independent of the well-documented clinical relevance of verbosity. We found that nonverbal edits and repeats per patient sentence had minimal linear correlation with patient words per sentence and fraction patient speech features, while verbal edits and restarts demonstrated moderate positive correlation with these verbosity metrics (Figure \ref{fig:disorg-interview-corr}). Notably, restarts had a moderate positive correlation with both repeats and verbal edits, but there was minimal correlation between disfluency types observed otherwise. It is possible that much of the correlation between restarts and verbal edits is explainable by their shared correlation with verbosity features. It is less clear what could be driving the association between repeats and restarts.  

Other correlations found in Figure \ref{fig:disorg-interview-corr} include strong positive relationships between patient words/total words and both words per sentence and words per second, though words per sentence and words per second had a more minimal relationship. The strongest linear correlation observed was between the pairwise and sequential incoherence per sentence metrics, which were so tightly linked as to be essentially redundant in the context of these interview features. Repeats per sentence demonstrated a negative correlation with both features and especially the sequential calculation method, which is unsurprising given the definition of both features: the angle between a word and itself is obviously 0. Besides this, there were minimal correlations between the two per sentence incoherence metrics and all the other features. That included no relationship between incoherence and uncommonness, which was a little surprising. Mean word uncommonness on the other hand was negatively correlated with a number of features, most strongly with the patient words/total words and the patient words per sentence features. Those negative correlations likely explain the negative correlations that were seen between uncommonness and some disfluency types -- though it does make sense by definition that the insertion of simple verbal filler words might negative correlate with a word uncommonness feature, regardless of any other verbosity relationship. 

Overall, it is difficult to fully evaluate these results without a more nuanced breakdown. Disfluencies by definition add words to a sentence, so regardless of any effects related to disfluency likelihood in extended or more complex sentences, one would still expect to observe some sort of correlation between verbosity and disfluencies in any dataset. This sort of complication is true of many of the features being compared here, for example uncommonness and disfluencies as well. Disfluencies especially may require nonlinear correlation methods to best capture relationships with verbosity. Furthermore, the number of patients included in this dataset was relatively small, so an apparent moderate study-wide correlation could be driven only by a strong subject-specific relationship. Nonetheless, the correlations we observed provide an important first step towards guiding that future research, and remain quite relevant in interpreting the upcoming modeling results. \\

Indeed, in the future - particularly with a larger dataset - a deeper analysis could much better characterize the relationship of these features in a clinical interview context. Considering potential nonlinear relationships both quantitatively and through qualitative inspection of scatter plots could elucidate the nature of different relationships suggested by the correlations of Figure \ref{fig:disorg-interview-corr}. 

\FloatBarrier

Moreover, scatter plots with a third feature used for hue could indicate when one feature has a moderating effect on the relationship between the other two. Such a methodology could also be used to spot patient-specific variation. More systematically, dimensionality reduction techniques could be employed to better probe the features that were extracted here. One could also learn more by investigating outlier transcripts more closely, whether an outlier in individual features of salience or in the alignment of values for a relationship of salience.

With a better understanding of the dataset and/or with a very clear question in mind, one could also start from the feature extraction stage to precisely define independent features. For example, mean word uncommonness and semantic incoherence features could be restricted to consider embeddings only of certain types of words, excluding disfluencies and common words like articles. Verbosity features could also be calculated with words attributable to a disfluency excluded, though it is not entirely clear how to define this in the case of restarts. One might consider splitting up repeats of entire words and repeats of syllables into two separate variables as well, calling the latter a stuttering metric. On the interview level, an alternative analysis could involve normalizing each feature based on what was typical for the participant contributing it. 

There are a large number of ways that features assessing a more specific construct could be extracted, whether from first principles or in a data driven matter. How to proceed in future studies could thus vary greatly; however, it would be ideal for such work to still consider baseline feature definitions in line with prior work for comparison purposes. This should not be framed as extra multiple testing to be corrected for, as the features themselves are highly related, and a careful review and presentation of their properties would likely reveal if an observed correlation was spurious. Random simulations like the ones performed in chapter \ref{ch:1} could additionally help clear up concerns on the robustness of results. \\

\noindent For some pilot discussion about correlation of interview features on the level of the individual participant sentence - another line of investigation that could help elucidate our mental model for underlying causes of interview feature relationships - please see supplemental section \ref{sec:sentence-disorg-feats}

\subsubsection{Disfluencies predict conceptual disorganization}
\label{subsubsec:disorg-modeling}
The input interview linguistic features just characterized in section \ref{subsubsec:disorg-nlp-feats} were used to predict the same-day clinical scale labels that were characterized in section \ref{subsubsec:disorg-scales}. 

Recall that the modeling dataset was comprised of 144 data points (145 for P2), each with 11 input features and 3 clinical labels, and coming from one of 18 selected BLS participants (section \ref{subsec:disorg-dataset}). Each feature/label pairing was tested in a separate linear mixed effects model with participant as a random variable, and significance was assessed using a Bonferroni-corrected cutoff (section \ref{subsec:disorg-methods}). I will now report on those results and discuss their implications. \\

\begin{figure}[h]
\centering
\includegraphics[width=\textwidth,keepaspectratio]{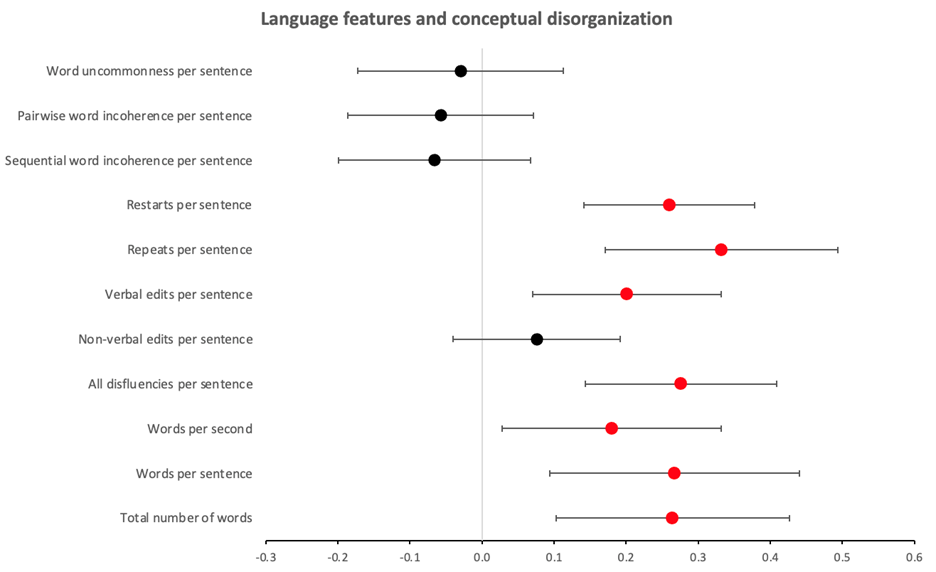}
\caption[Particular linguistic disfluencies predict disorganized thought.]{\textbf{Particular linguistic disfluencies predict disorganized thought.} Forest plot of beta coefficients and 95\% confidence intervals for the general linear regression models representing the relationship between each language feature and the conceptual disorganization clinical score (PANSS P2). Each feature was tested in a separate mixed effects model, including participant as a random variable. Plot points colored in red mark significant models. Figure and caption reproduced from Figure 2 in \citep{disorg22}, on which I am a contributing author.}
\label{fig:disorg-nlp-regression}
\end{figure}

\noindent Although none of the language features were significantly associated with N5 ($p > 0.14$) or G11 ($p > 0.15$) scores, there were a number of significant findings in modeling the P2 rating of conceptual disorganization (Figure \ref{fig:disorg-nlp-regression}).

\noindent The results depicted in Figure \ref{fig:disorg-nlp-regression} were reported as follows by \cite{disorg22}:

\begin{quote}
Six of the features were found to have a significant (Bonferroni corrected) positive association with P2 scores, including ‘ratio of participant words’ (Beta, B = 0.26, 95\% confidence interval, CI = ±0.16, p = 0.002), ‘words per sentence’ (B = 0.27, CI = ±0.17, p = 0.003), ‘total disfluencies per sentence’ (B = 0.28, CI = ±0.13, p = 0.0001), ‘verbal edits per sentence’ (B = 0.20, CI = ±0.13, p = 0.003), ‘repeats per sentence’ (B = 0.26, CI = ±0.16, p = 0.0001), and ‘restarts per sentence’ (B = 0.26, CI = ±0.12, p = 0.00003). The association of ‘words per second’ with P2 did not survive Bonferroni correction (B = 0.18, CI = ±0.15, p = 0.02). The other features were not significantly associated with P2 (p > 0.19). 
\end{quote}

\noindent Furthermore:

\begin{quote}
There were also significant random effects of participant on the associations of all the language features with the P2 scores, suggesting that even though the fixed effects reported above were significant across the group, there was also considerable variation between participants. 
\end{quote}

\noindent Thus, features of verbosity and linguistic disfluencies in patient speech during a clinical interview have predictive power for severity of conceptual disorganization, as measured by the corresponding PANSS item. 

\FloatBarrier

\paragraph{Feature characterizations further inform results.}
The modeling results did not distinguish between explanatory power that was shared by features and explanatory power that was independent, so it is possible that many of the significant results could be purely attributable to differences in overall speech production rather than a specific contribution of e.g. restarts. Because of this, I performed the supplementary characterization of linguistic features (section \ref{subsubsec:disorg-nlp-feats}), to help inform final interpretation of the model results. 

Repeats per sentence had the most predictive power for conceptual disorganization in the general linear model (Figure \ref{fig:disorg-nlp-regression}) and also a minimal linear correlation with the word count related features (Figure \ref{fig:disorg-interview-corr}), increasing confidence in the relevance of repeats to disorganized thought. Restarts were the second most relevant disfluency, and had similar power to the word count features in predicting conceptual disorganization. Restarts were moderately correlated with verbosity and with repeats, so it is unclear if they have explanatory power beyond these two factors. However it is at least likely that they hold clinical relevance unrelated to their relationship with verbosity alone. 

Verbal edits were the final disfluency with a significant connection to conceptual disorganization, though the relationship was weaker than others observed. Given that verbal edits correlated somewhat with the word count features, it is difficult to determine from this information alone whether they hold any independent predictive power. Because of the dataset size, a verbal edit specific factor would probably not survive multiple testing correction as it stands (Figure \ref{fig:disorg-nlp-regression}). However, as verbal edits had minimal correlation with repeats (Figure \ref{fig:disorg-interview-corr}), and also had a distribution with a fairly fat tail (Figure \ref{fig:disorg-disfluency-dists}B), they warrant a closer look in follow up work. 

It is worth noting that mean word uncommonness was not predictive of conceptual disorganization at all (Figure \ref{fig:disorg-nlp-regression}), despite having a moderate negative correlation with words per sentence (Figure \ref{fig:disorg-interview-corr}). More broadly, it's of course not possible to definitively dissect the contributing factors of the different features with the current approach. Some of the predictive power of the verbosity metrics could even be contributed by the increased word count generated directly by disfluencies, though it is unlikely to be a large effect here.  

In future work, it will be important to consider models that fit all four disfluency categories as well as the verbosity measures in a single function. One could then both evaluate the relative weights and test comparison models with specific features in question removed. Alternatively, the linguistic features could be more carefully defined to remove some of the sources of covariance as was discussed above, though that is more difficult to incorporate into the framework of a general use code release. 

Overall, the correlational investigation done here was able to increase confidence in the strength of the repeat result. If the relevance of repeats to conceptual disorganization could be replicated using a count of only word repeats rather than any utterance repeats (e.g. stutters), this would be an especially powerful result, as word repeats are easier to automatically detect without using expensive verbatim manual transcription. Distinguishing verbal edits from true uses of "like" or detecting sentence restarts are more complex algorithmic problems without TranscribeMe notation by contrast. 

The distributional investigation suggested interesting takeaways about properties of disfluency use as well, and opens up future questions about the fatter tails of repeats and verbal edits in this dataset. It is possible that a nonlinear model could improve the strength of the observed relationships with conceptual disorganization, in particular for the verbal edits per sentence feature. A better understanding of subject-specific factors would also be critical to such an investigation. \\ 

\paragraph{Main takeaways.}
The clinical relevance of free speech disfluencies specifically for assessing conceptual disorganization is a novel result. Moreover, this result is driven by a subset of disfluency categories. The breakdown by type of disfluency showed nonverbal edits to be insignificant and verbal edits to be only marginally significant (though promising), while the moderate association of total disfluencies with conceptual disorganization was largely driven by patient use of restarts and especially of repeats. Because filler words have been previously associated with severity of Schizophrenia symptoms more broadly \citep{Tang2021}, our results indicate that observed sum score correlations with different disfluencies may be attributable in part to different symptom dimensions. This demonstrates the utility of item-based approaches and opens up a number of lines of further questioning.

With a larger and more representative set of PANSS scores and language samples, it would be possible to not only verify this result, but also build up a robust model of the relationships between different categories of disfluencies and the many symptom domains covered by PANSS items. It is highly plausible that a shorter snippet of speech would be sufficient to obtain clinically relevant measurements of the linguistic disfluencies (see chapter \ref{ch:1} for results in audio journals), so the cost for same day linguistic analysis need not be excessive to supplement a study that is capable of collecting a large PANSS dataset. 

Another major style of question suggested by this investigation is an even greater focus on longitudinal data collection, so that modeling can include an understanding of individual differences rather than treating them like noise. As one of the first studies of interview linguistics in psychotic disorders to consider a long term longitudinal dataset, we found significant differences between participants in the reported results. 

Although overall associations were found, it is highly plausible that there is a truly stronger association between disfluencies and disorganization in some patients and a truly weaker association in others; this would not only be important to understand for the use of models in clinical practice, but moreover the underlying reason for that difference may in fact be clinically relevant itself. Using a larger longitudinal dataset, these individual differences could be properly scientifically explored -- thereby suggesting yet another great use case for the audio journal format, as participant linguistics could be much more deeply characterized over a long period via diaries than is feasible with a clinical interview. \\

\paragraph{On negative results.}
The characterized associations of verbosity and disfluency features with conceptual disorganization (P2) make for interesting takeaways, and although these results ought to be replicated, they are fairly rigorous for a standalone study of clinical interviews. However, other limitations remain which may be particularly impactful to the interpretation of some of our negative results. 

One limitation is the lack of severe symptom examples. P2 scores were much more balanced within the range of 2-5, while G11 and especially N5 were imbalanced, with N5 lacking examples beyond mild severity (Figure \ref{fig:disorg-scales-dist}). The distribution of G11 was likely not too limiting of the expressive capability of this analysis, though it remains possible that highly severe cases not covered here would display linguistic signs that moderate cases do not. Regardless, the larger concern is in evaluating any associations between the linguistic features and difficulty with abstract thought (N5). Because so few N5 scores $\geq 4$ were available, the negative results obtained in the N5 analysis should be taken with a grain of salt. \\

\noindent The other limitation is in the calculation of the semantic incoherence features. These features do not have a definition that is both universal and concrete, and therefore calculation method between different studies can vary widely \citep{Hitczenko2021}, though results may be qualitatively similar. Thus incoherence features have come under increased scrutiny relative to other linguistic features of clinical note. At the same time, there is sufficient prior evidence of broad importance of this category of feature such that researchers should work to refine the definitions rather than move on entirely. 

The embedding model we used here is relatively interpretable and has seen success in a number of related validation tasks, as was already discussed for the audio journal pipeline in chapter \ref{ch:1} (\ref{subsubsec:diary-val-trans}). It is possible that a different embedding model would change our negative results, however I believe that is highly unlikely. It is unclear to what extent automated local semantic incoherence metrics would be expected to associate with conceptual disorganization, as prior machine learning work has primarily focused on clinical ratings of total, or at best positive subscale, symptom severity. Further, the dataset studied here contained an overwhelming majority of timepoints with mild to nonexistent symptoms, and it is plausible that changes in semantic incoherence generally require more severe symptoms to become apparent than changes in linguistic disfluencies do -- some level of disfluency presence is expected even in healthy controls, whereas perceived incoherence is an immediate warning sign.

That said, I would not interpret the negative results seen here to indicate that the word2vec incoherence method is not related to conceptual disorganization or even that it is not associated with mild symptoms of disorganized thought in these participants. The primary takeaway is that mean incoherence of patient speech across an entire clinical interview is not associated with the clinical scale items considered, which is especially unsurprising given the overall mild severity of the patient population, as well as the stunted nature of large portions of the clinical interview. Adjustments to better normalize for sentence length or ground sentences in the previous interview question did not make a difference; one might consider better data cleaning, such as removing repeats (which introduce 0 angles) before calculation, but it is highly unlikely this would change any results either, as the returned features would remain means over a long semi-structured conversation. We have certainly learned something procedural from this peripheral part of our work, but it is very difficult to say much else concrete about the incoherence-related results of \cite{disorg22}.

Ultimately, the most important decision for incoherence research to make is not necessarily how embeddings will be calculated, but how they will be compared to generate metrics \emph{on a per interview basis}. It is possible that if our investigation had used maximum sentence incoherence rather than mean, a relationship might have turned up. Unfortunately, low sample size is a natural consequence of this dataset type that pairs interviews with clinical scales, and by extension feature counts need to be kept limited to retain any statistical power in predicting the scales. Therefore, there has been minimal systematic work on the various ways an entire interview can be turned into a single measure of incoherence. This is an important area for future research that may not directly yield clinical insights, but would greatly facilitate later clinical investigations. It is also another area where audio diaries could provide helpful guidance, though assessment of incoherence in an interview setting has some nuances that are not possible to study with diaries. \\

Overall, our work represents an important step towards addressing the concerns of \cite{Hitczenko2021}, by modeling specific symptom dimensions with patient language in a longitudinal dataset of recorded interviews from psychotic disorders patients. We both present inherently interesting pilot results, and discuss future directions enabled by these results. 

\section{Discussion}
\label{sec:discussion1}
Structured clinical interviews are an important tool in psychiatry research due to the gold standard clinical scale ratings they help to supply. Interviews more generally also provide opportunities to observe how a patient interacts in dialogue and to ask specific questions of prior interest or targeted follow-up questions based on how the conversation has developed. Recordings of interviews therefore enable a variety of research aims related to analysis of patient acoustics, language, facial expressivity, and posture in different contexts, and are well-grounded in existing clinical evaluations. However, the utility of different styles of interview recording (and subsequently the specific methods best suited for analysis) can also vary widely based on these aims. Furthermore, building even a modestly sized interview recording dataset can be time consuming and costly, and temporal resolution in longitudinal interview work is fundamentally capped. These factors, discussed at greater length within the background section \ref{sec:background1}, greatly limit the feasibility and general applicability of exploratory interview recording research.

Two possible avenues for addressing some such limitations were covered in this chapter. One is to collaboratively compile a much larger interview recording dataset than would be feasible for a standard study. The NIMH's recent AMPSCZ initiative, which aims to study young people at clinical high risk (CHR) for developing psychosis across nearly 40 different global locations, involves collection of both clinical and free-form interview recordings as part of their longitudinal and multimodal project design. Over the lifetime of AMPSCZ, more than 20,000 interview recordings are planned to be collected from a CHR population of nearly 2,000, with an additional 1,500+ interview recordings to be collected from 600+ control subjects \citep{Brady2023}. The other possible avenue is to use interview recordings in a highly hypothesis-driven way, focusing moreso on specific symptom domains and a specific class of interview features that are thought to be of psychiatric interest. While this approach is unlikely to yield brand new measures, it can serve to more deeply characterize and solidify the relevance of features already presumed to be clinically informative, thereby facilitating actionable improvements in symptom evaluation practice and providing additional well-validated features for exploratory work to reference against. 

\subsection{Lessons from AMPSCZ speech sampling}
In section \ref{sec:tool1}, I reviewed the functionalities of my open source code base \citep{interviewgit} for data management and quality control of interview recordings collected from across the large collaborative AMPSCZ initiative \citep{AMPSCZ}. As of late January 2023 - early in the overall project's data collection timeline - 398 interview recordings from 149 subjects across 17 global locations had been successfully processed by the pipeline, which already far exceeds the number of interviews included in prior works studying automatic processing of speech sampled from interviews with psychotic disorders patients (e.g. \citep{Bedi2015,Tang2021,disorg22}). Because of the high volume of interviews expected to be received, the sensitive personal nature of the datatype, and the heterogeneity between sites and scientific aims, it was critical that centralized infrastructure to automatically handle file organization, transcription processing, and availability/quality monitoring was well-built and available at the outset for this project. Furthermore, because of the multimodal component of the project, many of the sites submitting recordings do not have prior experience with this type of research protocol. With a broad and expensive data collection initiative, high quality data is an utmost priority that is important to be proactive about.

During the early phases of the project, I oversaw the data flow and quality monitoring results produced by my pipeline, identifying a range of recording issues, from mistakes in interview folder metadata to concerns with usability of some submitted data in downstream analyses. This included flagging specific interviews requiring further review, flagging sites that were repeatedly making the same mistake so they could receive further training, and compiling a list of observed problems for future monitors to look out for (see section \ref{subsubsec:u24-issues}). Moreover, I extensively documented the code in supplement \ref{subsec:interview-code}, to ensure it will be straightforward for someone else from AMPSCZ to take over the necessary software maintenance and feature updates for the remainder of the project. This documentation, along with the characterization of QC feature outputs from the early stages of the project (section \ref{subsubsec:u24-monitor}), can ideally also serve as a reference point to assist in planning of additional projects.

More generally, a secondary aim of reporting on the AMPSCZ interview collection process is to provide a blueprint for future similar initiatives on how they might structure and eventually monitor an interview recording protocol, whether utilizing my code or not. For example, the extent of basic upload issues has been a surprise for me, as many sites have repeatedly done things against procedure that either make little sense because they introduced additional work in order to mess up the data flow (e.g. taking Zoom files out of the autogenerated folder to upload them individually), or are major protocol violations that should have been extremely obvious to avoid (e.g. inexplicably using Microsoft Teams to record the interviews). Additionally, it has been a struggle to get some of the sites to improve upon some of these behaviors, even after repeated trainings. A future project might consider either putting more resources into software development to make the infrastructure much more robust to mistakes, or changing the incentives/funding structure to be able to better enforce correct following of procedures. The code was intentionally written to look for default Zoom naming conventions to minimize the opportunities for user error, but this has not been especially successful given the high number of cases where folders were renamed anyway, at times removing crucial identifying information.  

Perhaps most important to these results was the detailed early look at how the interview conduct and recording protocols might be fundamentally improved in future iterations, based on properties observed in the pipeline outputs. Indeed, through the use of my code I not only identified major operating and quality problems in early AMPSCZ interview recording collection - enabling those problems to be addressed as described - but I also flagged a number of potential concerns about more subtle issues with content quality. The latter type of problem represents potential challenges for downstream scientific analyses rather than data point breaking flaws, and discussions on how to further investigate such potential issues and address as needed remain ongoing (\ref{subsec:interview-outputs}). 

Still, there have already been impactful changes made to the scientific plan of AMPSCZ speech sampling based on this work. I found open interviews thus far to contain substantially more patient words per minute, substantially more words per turn (indicating speech with richer content), and in fact more overall words per minute than psychs interviews. Meanwhile, psychs interviews have often been so long that the expensive transcriptions obtained were not even of complete length; further, recording quality enforcement is intentionally less strict on this interview type, because there is less of a clear scientific plan for speech sampling analysis in the psychs clinical interview recordings. Despite all of this, the initial proposal was to pay for gold standard professional transcriptions of the (first 30 minutes of) structured interviews across all of the (many more) time points they are recorded. Coupled with the comprehensive overview of and arguments for the daily audio journal datatype in chapter \ref{ch:1}, and the fact these journals have been ignored to date, a large (and expensive) philosophical difference between the original and optimal directions of the project emerges in my opinion. As a result of my arguments, presented in full in Appendix \ref{cha:append-ampscz-rant}, the protocol for obtaining psychs clinical interview transcriptions in AMPSCZ will soon be pared back to save money so that diaries can instead be transcribed, and so that the psychs transcripts that will be obtained going forward are of better quality (fuller length). 

\subsection{Focused analysis of a clinical interview dataset}
In section \ref{sec:disorg}, I reported and expanded upon previously published results from linguistics analyses of interview transcripts in the Bipolar Longitudinal Study (BLS) dataset \citep{disorg22}. The described language features were extracted from the interviews using code adapted from the audio diary feature extraction pipeline of chapter \ref{ch:1}. The original curation of the transcript dataset, including quality assurance and management of the TranscribeMe data flow process, was handled by an early version of the code described in section \ref{sec:tool1}.

In the presented BLS project, which focused on identifying correlates of disorganized thought, we found a significant association between conceptual disorganization (PANSS P2) and both verbosity and disfluency related features of same day clinical interview transcripts. Moreover, we found "repeats per patient sentence" to have the strongest relationship with P2, and one that was independent of the verbosity metrics. Restarts and verbal edits per sentence also had statistically significant associations with P2, while nonverbal edits had no relationship. We also observed a significant effect from subject ID in the mixed effects model, suggesting an important role for individual heterogeneity in understanding clinical/linguistic relationships -- a theme in the results of chapter \ref{ch:1} as well. 

Ultimately, the results of \cite{disorg22} underscore the importance of the item focused approach and the longitudinal approach that we take. They also suggest that future work on categorization of specific linguistic disfluencies could be quite interesting, and should not be ignored despite the greater requirement for human transcription currently imposed by them (there are of course a number of other advantages to professional transcription for psychiatry research at present). As part of reporting these results in section \ref{sec:disorg}, I included a more lengthy discussion of interpretation and limitations along similar lines. I will next summarize a handful of promising future directions for interview recording research more broadly, which will involve an overview of further steps that could improve upon our analysis of disfluencies and conceptual disorganization.  

\subsection{Future directions}
\label{subsec:interview-future}
One large category of future work enabled by the contributions of this chapter are future interview data collection studies focused on other topics of psychiatric interest. The documentation of my code in supplemental section \ref{subsec:interview-code} includes information on updates one might make to the pipeline to accommodate different study scenarios. It also includes discussion of the advantages and shortcomings of the software development process that was used for AMPSCZ speech sampling, such as the mock interview certification process and the use of an initial development server for Pronet; this sort of report could assist future psychiatry data collection projects regardless of target datatype.

Another major category is the continuation of AMPSCZ interview recording collection and the eventual scientific analyses that can result -- including those done by researchers outside the AMPSCZ project through the NIH's data sharing platform. To facilitate this, it is first important to continue maintaining data quality throughout the duration of the project, and to expand on the interview software infrastructure to cover additional functionalities that will be necessary for the later stages of the project. For basic code maintenance and onboarding of more of the planned sites, I have included a number of use instructions and troubleshooting tips in the documentation (\ref{subsec:interview-code}). Regarding future updates to the pipeline and expansion of interview infrastructure, there are a few primary issues to be aware of:
\begin{itemize}
    \item For better tracking of site mistakes and verifying other site-reported interview properties, the pipeline should introduce processing of relevant records logged by sites in REDCap/RPMS and integrate it with existing QC outputs. This would also help identify errors that might be introduced by sites on the end of REDCap/RPMS before it is too late to fix.
    \item Certain European sites are unable to use Zoom due to privacy restrictions, so the pipeline preprocess module likely will need to be updated to support Cisco WebEx as well. 
    \item Most of the included foreign languages have not yet been tested with the pipeline, and in addition to likely not yet anticipated bugs, there are  updates that are known to be necessary for the code. At the very least updates related to detection of inaudibles and other English TranscribeMe notation will need to be made to cover these other languages.
    \item Given the plan to change the parameters for psychs transcriptions, corresponding updates will need to be made and monitored within the pipeline's transcription logic, which at present does not have a mechanism to skip sending interviews to TranscribeMe unless they fail QC or have critical problems with the available metadata. 
    \item More broadly, I have identified in the documentation a number of smaller quality of life updates and bug fixes that will become increasingly important as the pipeline scales to greater and greater volume of uploaded interviews, particularly if the rate of upload mistakes does not substantially improve.
    \item Because raw audio and video cannot be shared with the NIH repository, it is of high priority to implement automatic extraction of some initial audio and video features that can be shared with the broader research community for analysis alongside other datatypes (including redacted transcripts). This will be a separate pipeline on a server with much better compute resources available, but it will need to directly interface with and use information from the data flow/QC pipeline.
\end{itemize}
\noindent More detailed discussion on major need points for hiring a replacement for me in AMPSCZ can be found in Appendix \ref{cha:append-ampscz-rant}. Note also that a great deal of work remains for the audio diaries portion of AMPSCZ, as overviewed in chapter \ref{ch:1}. 

The uniquely large size of the dataset being collected by AMPSCZ will introduce opportunities to fine-tune existing machine learning and language processing techniques in the psychotic disorders setting, thus establishing tools and best practices for the future of digital psychiatry research. This could range from more generally useful technical contributions such as domain-specific improvements to automated transcription techniques (grounded in the gold standard professional transcripts being obtained), to scientific results elucidating properties of speech in CHR patients and perhaps even identifying features with predictive power for the eventual development of psychosis. Because of the sample size, many ideas can be explored without sacrificing the ability of the study to draw reasonably strong conclusions. 

In the same vein as the study design contributions of this chapter, as AMPSCZ speech sampling continues and additional insights are gained in that process, it will also be a valuable opportunity to better establish interview protocols for the digital age. It is not yet clear how interviewer conduct might affect the behavior of and particularly the language used by the participant, but from early QC results presented here it does appear there may be systematic differences between sites in the way interviews are performed -- for example in the typical length of participant conversational turns versus interviewer conversational turns. This is especially relevant for open interviews, as the protocol is highly generalizable to other disorders and easy to compare with controls, but it is also a less well established research datatype. At the same time, there are questions one could ask about how features in the open interviews compare to features from the same participant in the psychs interviews. 

Because of the relative paucity of audio journal research, we know very little about the intersection between acoustic and linguistic features extracted from interview recordings versus from diaries for a given participant. Characterizing features from across these two speech sources would be extremely helpful in guiding future study designs to focus on datatypes best suited to their aims. Additionally, relationships between these features might themselves be salient metrics for CHR, as they would provide some notion of changes to a subject's speech in a social/conversational setting. More broadly, the multimodal design of AMPSCZ lends itself to asking a number of questions about the interaction between interview-derived features and other digital psychiatry datatypes. \\

It is worth noting that the lab's BLS dataset permits a similar analysis style, albeit on a smaller scale. Due to our results linking linguistic disfluency use during interviews with same-day clinical ratings of conceptual disorganization in this population, it would be of great interest to compare disfluency use patterns between the audio journals and the interview recordings of individual BLS participants. In particular because the BLS interviews are semi-structured, it is certainly plausible that disfluency use would be inherently different than in a free-form monologue, but this is an open question and any effects may be subject-dependent (or even somewhat interviewer-dependent). Questions like these will be important to address about a range of speech sampling features in order to really establish the subfield of digital psychiatry. If we do not know the extent to which different datatypes elicit independent information, it will be challenging to move past the exploratory data collection stage of research. Similarly, it will be critical to more carefully dissect between participant heterogeneity in future larger studies. 

One advantage of the longer interview recording format when it involves a targeted question/answer period is that patient speech features can be analyzed in context of the preceding question. While this is currently difficult because of the long duration of interviews and the small number of labels, similar techniques will likely prove useful in some hypothesis-driven research aims once better groundwork for the field is established. An early version of such work that might be done to better understand the various linguistic disfluencies in our BLS dataset would be to consider additional within-interview distributional properties. For example, how evenly spread out are different disfluencies across a transcript? Qualitatively, if we pulled up the portion of each transcript with the greatest number of patient disfluencies, might there be any relevance to the questions being asked at that time? 

Questions like those could be asked in a variety of datasets, but for our study in particular it would make sense to focus primarily on usage of repeats. The identity of repeated utterances and words could be of note, in addition to looking at regions of a transcript with a high number of repeats. Further, we counted repeated utterances and words towards the same repeats metric, but the effect may be more specific -- stuttering over a word versus repeating an entire word are not necessarily the same mechanistically. Along the same lines, we kept the scope simple with a linear model only, but feature relationships may be more complicated; disfluencies (besides nonverbal edits) can be relatively rare and thus would likely be better modeled non-linearly. Additionally, a more precise model would ideally directly account for the relationship between overall verbosity and disfluency use. \\

Ultimately, there remains great potential for analysis of interview recordings for specific psychiatric research questions; critically, these recordings have a direct link to existing gold standard symptom evaluation methods, methods which are still underused in real world practice and could possibly be better spread through technological advances that would make them easier for clinicians to employ \citep{Lewis2019}. As automated interview recording metrics improve, useful summaries could be generated for mental health care practitioners to consult when evaluating patient status and planning treatments. An example report for checking recording quality and screening features generated from a lab interview can be found in Figure \ref{fig:onsite-report}. While the measures included on a clinical report in the future would likely be quite different, the sample PDF provides an idea of the benefits automated reporting could bring as interview research improves. 

\pagebreak

\begin{FPfigure}
\centering
\includegraphics[width=\textwidth,keepaspectratio]{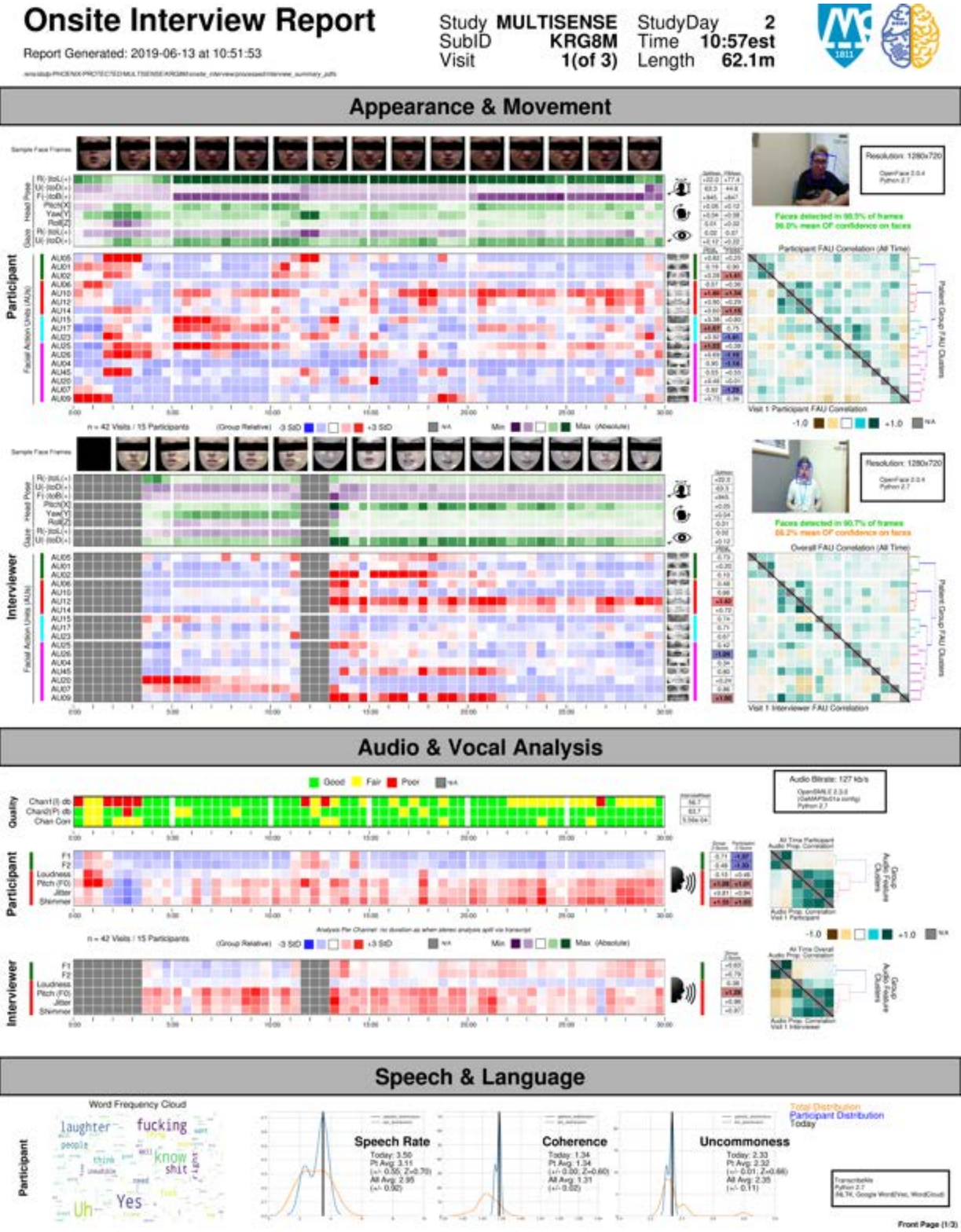}
\caption[Example auto-generated report from a clinical interview recording.]{\textbf{Example auto-generated report from a clinical interview recording.} This report was automatically created for a sample MULTISENSE study interview \citep{Vijay2016} using extracted features from OpenFace \citep{OpenFace}, OpenSMILE \citep{OpenSMILE}, and linguistic features like described in chapter \ref{ch:1}. Individual visualization components were also generated similarly to those described in chapter \ref{ch:1}, and they were constructed into a report automatically using the reportlab python package. I wrote early code to create these reports for all interviews across MULTISENSE, with each PDF containing 1 page per 30 minutes of interview. MULTISENSE featured recording of both interviewer and participant, and the protocol was structured so that the participant would be recorded alone for $\sim 5$ minutes, then undergo a brief interview with a psychiatrist, and finally sit for a longer interview with a trained RA. MULTISENSE aimed to predict discharge-readiness in those admitted to the psychotic disorders unit at McLean. The depicted interview is the first 30 minutes of the first interview with one of the participants who enrolled in the study. The report itself includes information on both quality control and possible features of interest. \newline OpenFace confidence stats and sample extracted faces (with automatically placed black bars for deidentification) are presented along with the video features for participant and then interviewer at the top of the report. Facial features were summarized by taking the mean over 30 second bins, and included pose and gaze estimates presented in a PrGn heatmap with absolute bounds, as well as FAU activations presented in a bwr heatmap that was colored based on the distribution of each FAU (in 30 second means) across the entire study dataset, taking $+/- 3$ standard deviations from the mean as the bounds. When a face was not detected the columns were greyed out instead. The AUs are ordered based on their frame-wise Pearson clustering across the dataset, and a correlation matrix is also included showing on one half the frame-wise correlations for the present interview and on the other half the frame-wise correlations for the current participant across all of their interviews. Summary stat values from the entire interview appear between each heatmap and correlation matrix. OpenSMILE features are summarized in the same way in the acoustics section, following the progression of basic audio QC features over the course of the interview. Finally, the language features summary at the bottom focuses only on participant features, showing a word cloud of their speech from that interview along with the mean speech rate, incoherence, and uncommonness metrics -- which are each marked on a distribution plot containing the study-wide and participant-specific distributions of these interview-level features.}
\label{fig:onsite-report}
\end{FPfigure}

\FloatBarrier

\subsection{Contributions}
\label{subsec:u24-contrib}
In sum, I have gained a great deal of experience with processing and analysis of the interview recording datatype, and have leveraged that experience to not only create useful research tools and enable lab studies, but also to impact important project design decisions for a major research initiative. My specific contributions include:
\begin{itemize}
    \item Built software infrastructure for data management and quality monitoring of interview recordings (both clinical and open-ended) for the NIMH's AMPSCZ project. AMPSCZ is a large, international, and collaborative scientific initiative to study youth at high risk for developing psychosis, and a major component of this project is to compile a quality multimodal digital psychiatry dataset at scale, for eventual use by the wider research community. Given the scope and potential impact of the undertaking, it is imperative (and non-trivial) to have strong foundational software supporting data collection and processing.
    \item Utilized my software to monitor the interview recording collection process through the early stages of AMPSCZ, identifying a host of issues to be addressed by study staff and writing comprehensive documentation of both the code and the monitoring process for whoever will take over my role. Problems uncovered ranged from interview protocol violations to record-keeping mistakes to recording quality concerns, and regular monitoring also involved tracking the flow of good quality transcripts contributed by each site over time, as well as evaluating for qualitative differences between sites in early collection (e.g. site-specific trends in typical open interview length).
    \begin{itemize}
        \item Within this role, I also lead a number of discussions amongst different groups of relevance in the overarching AMPSCZ project, including advising on pipelines for downstream feature extraction on the speech sampling data, highlighting action items for site staff to improve their interview recording process, coordinating with IT at various institutions, mediating the project's relationship with TranscribeMe, creating monitoring dashboards and periodic email updates on collection status for project leaders, and tracking many small details across larger group meetings to ensure that as many questions and action items that could impact eventual dataset quality are addressed as soon as possible. 
        \item Suffice to say that in aggregate these contributions have a large shaping influence on the speech sampling dataset of AMPSCZ and thus the scientific output to eventually result.  
    \end{itemize}
    \item Most importantly along these lines, I spearheaded a successful effort to redirect part of the AMPSCZ project's professional transcription budget from highly structured clinical interview recordings to instead obtain quality verbatim transcriptions of all participant audio journals; this represents an adjustment of well over \textbf{\$100,000} through the next few years. My arguments for the change are presented in full in Appendix \ref{cha:append-ampscz-rant}, and include a detailed comparison of the different speech sampling datatypes theoretically, in addition to supporting evidence from the early stages of AMPSCZ data collection.
    \begin{itemize}
        \item Appendix \ref{cha:append-ampscz-rant} also reviews a number of ideas for the future use of the AMPSCZ speech sampling dataset to be collected, along with broader commentary about the vision of the project.
    \end{itemize}
    \item Internal to the lab, I built and tested a feature extraction pipeline for computing NLP features from clinical interview transcripts, and characterized those features in our Bipolar Longitudinal Study (BLS) dataset.
    \item Performed a deeper dive assessing verbosity and the use of different categories of linguistic disfluency in BLS interviews, contributing to the manuscript by \cite{disorg22}. 
    \begin{itemize}
        \item We found a significant association between conceptual disorganization (PANSS P2) and both verbosity- and disfluency- related features of same day clinical interview transcripts. Repeats (including repeated utterances, words, and phrases) per patient sentence in particular had the strongest correlation with P2, and this relationship was independent of any verbosity-related effects. 
        \item The study direction itself also represents an important direction for future research involving clinical interview language, as it focuses on a specific symptom of interest and breaking down potential correlation with specific linguistic features hypothesized to be of relevance -- this is in contrast to many other recent works analyzing patient interview speech, where more features are often tested than the number of available labels, and the labels themselves are often static and non-specific diagnostic categories \citep{Hitczenko2021}. 
    \end{itemize}
\end{itemize}

\noindent Ultimately, while I strongly prefer the audio diary format of chapter \ref{ch:1} for most scientific questions, interview recordings will remain an important component of many digital psychiatry studies, for good reasons in part (see also section \ref{subsec:interview-history}). Because interview datasets are so much more challenging to construct than audio journals, quality software for managing interview recording data collection is especially critical, as is the large collaborative study format of AMPSCZ. This chapter has in many ways provided a blueprint for future such studies. 

In the next chapter (\ref{ch:3}), I will focus on related problems on an entirely different scale -- a case study involving a single individual, but utilizing highly multimodal longitudinal data collection methods, including neural recording data from a deep brain stimulation device. The data reviewed for the case study will touch on themes from the first two chapters, as both audio journals and interviews were collected. This will provide a proof of concept opportunity to directly link these data with passive digital phenotyping signals, which are being similarly collected across AMPSCZ.

\chapter{Dense phenotypic characterization of a single human with deep intracranial recordings over two years\footnote{This chapter is an extended version of a case report manuscript in preparation that I am leading. That project was a supplement to the trial published by \cite{Olsen2020}, on which I was a contributing author and which I summarize here in the context of this chapter's themes. See Appendix \ref{cha:append-clarity} for detailed attributions.}}\label{ch:3}
\renewcommand\thefigure{3.\arabic{figure}}    
\setcounter{figure}{0}  
\renewcommand\thetable{3.\arabic{table}}    
\setcounter{table}{0}  
\renewcommand\thesection{3.\arabic{section}} 
\setcounter{section}{0}

Treatment resistant mental health conditions are currently addressed in the U.S. using deep brain stimulation (DBS), with recent technologies affording the possibility of simultaneous recording from deep brain sites to enable closed-loop neurostimulation systems. However, little is understood yet about the specific mechanisms by which DBS can alter behavior, a major roadblock for the development of such closed-loop treatments. Towards this end, we followed a single individual with severe affective disturbance who was implanted with two deep brain electrodes, one placed in ventral striatum (VS) and a second in supplementary motor area (SMA). 

For well over a year, we tracked measures of the individual's affective state using a multimodal array of sensors, including intrinsic phone sensors to capture variation in geolocation features, accelerometry measurements to capture changes in psychomotor behavior, daily surveys to probe for subjective productivity levels and related psychological constructs, daily voice diaries to capture changes in linguistic indicators, and a series of videos recorded during repeated study visits with the research team. We combined these metrics both with traditional psychiatric measures like clinical scales, and with local field potential recordings from the implant sites -- to characterize their relevance in treatment evaluation as well as their relationship with neural activity. 

While the intended purpose of the novel DBS paradigm was to treat obsessive-compulsive disorder (OCD) in this patient, which did not succeed \citep{Olsen2020}, our primary goal was to provide a proof-of-concept for the use of dense behavioral data collection in the development of DBS treatments. Indeed, we demonstrate here that the participant experienced numerous changes in affect over the course of the study, from profound depression and immobilization to active engagement in life activities and back. Further, we describe correlation structures amongst both behavioral and neural features, ultimately providing a blueprint for how our data collection scheme can be utilized in future studies. \\

\noindent \textbf{An outline: } In this chapter, I will first give deeper motivation for our work (section \ref{sec:background3}), in addition to reviewing the case report by \cite{Olsen2020} that performed a more traditional evaluation of the stimulation paradigm pilot in question (section \ref{subsec:stim-params}). I will then introduce the numerous new modalities our data collection methods added to the DBS evaluation (section \ref{subsec:ocd-methods}). These preliminaries will lay the groundwork needed for presenting the key results of our analyses in section \ref{sec:results3}.

Our results can be broadly categorized into four major contributions. The first is a demonstration of the power of these behavioral datatypes in longitudinal tracking of a single patient; intricate visualizations present views of key moments for this patient at different timescales and in different modalities, highlighting how our methodology can be used to detect behavioral anomalies that might otherwise go unnoticed, potentially even informing downstream study design decisions. A birds-eye view of the collected data is reported on in section \ref{subsec:ocd-birds-eye}, as is a closer look at the patient's life during the period of major DBS setting transitions and during the onset of the COVID19 pandemic. The longitudinal tracking framework is then revisited in section \ref{subsec:ocd-self-report}, with a deeper look at subjective patient accounts of daily life and some of the interesting subtleties that can be detected on the level of the individual day using our data. 

The second major contribution of our work is the use of a clustering approach for understanding the clinical scale data from this trial. As I will report in section \ref{subsec:clinical-clusters}, we were able to uncover variation in symptom severity that was not seen when looking at the standard total and subscale scores, as was done by \cite{Olsen2020}. We did not find a symptom cluster that clearly improved due to the cortical stimulation, so our results do not change their conclusions. However, isolating periods of higher or lower severity of a particular class of symptoms is a critical component to contextualizing our objective signals. The results of our analysis of clinical scale data are thus an important part of the proof-of-concept impact of this case report.

The third major contribution of our work is the characterization of feature correlation structures in this patient, which I report on in section \ref{subsec:ocd-corr}. Relationships amongst different digital psychiatry metrics are important to understand in downstream feature selection - not only to identify potentially redundant features for future study design, but also because correlation strength between two base features may be a clinically predictive feature in itself. Here, we provide Pearson and Spearman correlation matrices to serve as a pilot study of these relationships. Additionally, we report correlations between behavioral features and clinical outcomes (section \ref{subsubsec:ocd-behav-corr}) and between behavioral features and neural recording features (section \ref{subsubsec:ocd-neuro-corr}); this enables us to identify particular features of note, which we can emphasize in forming predictive models for clinical outcomes of future patients.

The final major contribution in this case report is the proof-of-concept experiment done to causally probe the effects of cortical stimulation status on shorter-term patient behavior. In section \ref{subsec:ocd-causal}, I report on the use of facial and linguistic analysis from video recording to supplement a traditional interview experiment. Again, I outline not only interesting takeaways from this patient's results, but also a broader picture of the promise for our data collection methodologies, as well as considerations to implement them in a quality fashion.

Ultimately, this chapter is a discussion of the applications of digital psychiatry to evaluation of OCD and to development of novel DBS treatments, presented in the form of a case report. Therefore I conclude with a summary of the lessons learned for future data collection and analysis in section \ref{sec:discussion3}, along with a more general overview of the difficult open questions that these tools could help to address.

\section{Background}
\label{sec:background3}
To set the stage for our case report results, I will begin the chapter with background on OCD diagnosis and treatment (section \ref{subsec:ocd-treat}) and on the use of DBS in psychiatry (section \ref{subsec:dbs-rev}). I will also review recent literature on digital phenotyping, with a focus on the phone sensor modalities we most heavily relied on when analyzing this individual's dataset (section \ref{subsec:deep-pheno}). The deep phenotyping review will cover a broad spectrum of psychiatric disorders, as its use in OCD is thus far sparse. Existing results on broader behavior quantification technology for OCD will then be discussed in section \ref{subsec:ocd-dp}, to give context before describing the key open questions for both OCD and DBS applications of digital psychiatry (section \ref{subsec:ocd-motivation}). For additional background on daily patient audio diaries and recorded clinical interviews in psychiatry, see chapters \ref{ch:1} and \ref{ch:2} respectively. While such datatypes have great potential advantages, as described in those chapters, they are fundamentally limited in temporal resolution compared to the digital phenotyping data that will be a major focus of this chapter.

\subsection{Current and emerging OCD treatment options}
\label{subsec:ocd-treat}
Obsessive-Compulsive Disorder (OCD) is a chronic condition characterized by uncontrolled, repetitive thoughts and/or behaviors. Both obsessions and compulsions can take a variety of forms, such as excessive fear of germs or intrusive aggressive thoughts. OCD has a lifetime prevalence of $\sim 2.3\%$ \citep{Kessler2009}, and severe cases that are identified are often unable to be treated: over a quarter of patients do not respond well to cognitive behavioral therapy or serotonin reuptake inhibitors, the first line treatments for OCD. Further, determining treatment strategy is generally difficult, as there is only a weak link between chronicity, severity, and treatment resistance \citep{Fontenelle2019}. 

Many alternative or adjunctive treatments for non-responders have shown promise in trials, including antipsychotics, ketamine, neurosurgical ablation, transcranial magnetic stimulation, and deep brain stimulation \citep{Hirschtritt2017}. However, none is a silver bullet, and it is still unclear how these different options could be used in a personalized approach. Especially for strategies that are more invasive or have stronger side effects, the trial and error involved in modern individual treatment planning carries high risk to patient health, in addition to high costs for the healthcare system.

Thus to improve outcomes for current patients and develop future treatments, procedures for OCD diagnosis and monitoring must improve. The state of the art in symptom evaluation, the Yale-Brown OCD Scale (YBOCS), is resource intensive and reliant on self-report: leaving clinicians and researchers with infrequent, subjective data. This makes it difficult to predict or even evaluate treatment response, as well as decreases the accessibility of care. The use of biomarkers such as EEG has improved prediction of treatment outcome to a point in pilot studies \citep{Krause2016}, but these techniques fail to capture longitudinal trends and are unrealistic for widespread adoption. In practice, clinicians still primarily rely on the YBOCS and other similar clinical scales for symptom evaluation, and treatment decisions often follow a standardized workflow \citep{Fontenelle2019}. 

More concretely, while the YBOCS performs very well on measures of inter-rater and 7 day test-retest reliability (Pearson's $r > 0.9$ across multiple studies), its validity is somewhat questionable. Correlation with other measures of OCD such as the MOCI is moderate ($r \sim 0.5-0.6$ across studies), and extremely similar to the correlation seen with depression ratings such as the Hamilton Depression Scale \citep{Woody1995}. Undoubtedly the YBOCS does contain signal, as many years of treatment research have reliably shown responses in the YBOCS that align with patient self-reported improvement, in addition to the generally acceptable level of convergent validity mentioned. However, the concerns with divergent validity, coupled with the necessarily infrequent nature of the YBOCS data, strongly suggest a need for alternative metrics that can supplement the YBOCS -- to fill gaps in both symptom detail and temporal resolution. 

Fortunately, OCD is especially well-suited for the development of digital phenotyping technology. Patients that have a particular activity or set of activities they feel they need to repeat, like washing hands or turning on/off lights, are said to exhibit ritualizing behavior. This is one of the most common and debilitating symptoms of OCD, and often appears as an overt physical motion \citep{Ruscio2010}. Therefore, a large subset of the OCD population has a major symptom that is observable and possible to systematically define. Rituals are generally not a rare event, and can have long duration. They avoid many of the computational challenges present in quantifying severity of other psychiatric symptoms. Quantification of physical ritualizing behaviors could enable new perspectives for neurobiological data, which could in turn yield insights into mental rituals or even other symptoms of OCD. Given the potential for the use of digital phenotyping in evaluation of OCD, I will review prior literature on the use of these technologies in psychiatry more broadly in an upcoming section.

Ultimately, improvements to symptom evaluation methods for OCD would not only help with assigning patients to treatments well-suited for them, but also with development of novel treatments. A subset of responders within a clinical trial analysis might be drowned out by more frequent non-response, but through using more precise measurements a more precise study design would become possible. Moreover, treatments that require intervention with higher temporal resolution would become feasible too; one area of particular interest is the development of closed-loop deep brain stimulation (DBS) paradigms. Next, I will overview the current state of DBS research, including its promising use cases in OCD to date, and discuss the interplay between improvements to DBS for OCD and improvements to our understanding of the neurobiology of OCD. Without a deeper understanding of behavior, limitations observed in biomarker identification work will remain extremely difficult to overcome, thereby limiting what can be achieved with DBS. 

\subsection{Deep brain stimulation}
\label{subsec:dbs-rev}
Deep brain stimulation (DBS) is a technique where electrodes are inserted into a target brain region, to provide electrical current either constantly at a particular frequency, or as intermittent bursts. DBS has been a great advance so far not only clinically, but also in research, as it allows for direct manipulation in the study of human pathologies, and has demonstrated that changes in local activity can have downstream effects in many brain regions. DBS is standard of care for those with Parkinson’s Disease (PD) and other movement disorders, and is also an experimental treatment for a handful of severe psychiatric disorders. While it has shown much promise, it is not fully effective, even in PD -- where it only treats a subset of the symptoms \citep{Lozano2019}. 

Neuroscientifically, target brain regions and stimulation parameters are still poorly understood. Indeed, advances in DBS have frequently resulted from medical observation rather than neurobiological first principles. DBS was originally intended as a treatment for chronic pain, an application that has yielded comparatively much less success. Research is thus often oriented around scientifically explaining the efficacy of existing DBS treatments \citep{Gardner2013}. It is known that high frequency pulse trains and low frequency stimulation produce fundamentally different effects, but a detailed comparison of these techniques remains to be done. High frequency DBS is hypothesized to be a useful clinical tool because it overrides pathological low frequency signals and has little effect on plasticity \citep{Lozano2019}. However, significantly more research is necessary, especially to understand the role of DBS in psychiatric treatment, where symptom profiles are also poorly understood. \\

Among psychiatric disorders, OCD has been arguably the most effective target for DBS. It was the first psychiatric disorder to receive FDA approval for DBS (under a humanitarian device exemption) back in 2009. While it remains a last resort intended only for OCD cases that are well-documented to be both severe and treatment refractory, there is clear evidence it outperforms the prior last resort of ablative surgery. It is still considered an emerging treatment by consensus, but in its current state it is the gold standard for such severe treatment-resistant patients \citep{Hemmings2021}. Note that most of the approved applications for DBS are for disorders with clear motor symptoms, something that is highly relevant for the subset of OCD patients with physical rituals.   
    
In current DBS paradigms for OCD, implants typically go in the ventral striatum/ventral capsule (VS/VC), and stimulation parameters are personally tuned for each patient via clinical scales and self-reported efficacy. The final stimulation protocol decided on will involve a constant stimulus at the chosen frequency and voltage. $\sim 60\%$ of OCD patients that undergo DBS show improvement, with an average YBOCS reduction of $\sim 45\%$. Moreover, when VC/VS stimulation is turned off unbeknownst to the patient, the positive effects of the treatment subside \citep{Luyten2016}. 

Unfortunately there is currently not much data available to determine whether a patient will respond in advance, although older age at OCD onset was associated with better outcomes \citep{Alonso2015}. A few alternative DBS sites have been tried, but comparison of VS versus subthalamic nucleus found that both sites improved overall symptoms a similar amount, and that dual stimulation did not give further improvement. Subsequent behavioral therapy also was unable to further improve symptoms \citep{Tyagi2019}. \\

Overall, while DBS is able to improve symptoms for a subset of severe OCD patients, there is much to improve as far as both treatment protocol and prediction of response. Because $\sim 60\%$ of OCD patients that undergo DBS improve \citep{Alonso2015}, $\sim 5-10\%$ of all diagnosed patients do not respond to any available FDA-approved treatment. As target brain regions and stimulation parameters are still poorly understood, recordings from neural implants are a key resource. To fully utilize these data though, it will be a necessary first step to characterize behavior in OCD with high temporal resolution. This would enable study of direct relationships between real-time behavior and both stimulation parameters and local field potentials. Without such an understanding, it will be difficult to move away from the current practice of stimulation tuning via trial and error. Furthermore, it will be difficult to develop and evaluate novel stimulation paradigms without behavioral grounding. Conversely, uncovered relationships between well-defined behavioral metrics and simultaneous neural activity could inform basic research into OCD, creating a strong positive feedback loop for disease research.

Of course, it is non-obvious how dense, multimodal behavioral signals ought to be analyzed. Study of different data sources and development of specific computational techniques for them will be necessary to properly leverage the potential of recent sensor technologies. Best approaches may also vary depending on the disorder in question, as well as the end goal. For example, time-alignment with relevant neural features may require completely different tools than those that best characterize OCD symptoms for first line treatment evaluation. Even within the subfield of DBS research, the most appropriate behavioral tools may vary depending on recording site. Psychomotor signs are particularly salient for the current practice of DBS, but emerging therapies targeting other brain regions may require a focus shift to considering digital signs for e.g. executive functioning. \\

Despite the great potential and the many complications that need to be navigated, there is a real paucity of research on the use of passive digital phenotyping in DBS. Even in PD, which has a larger body of literature on automated motor symptom classification with more objective endpoints than OCD, as well as a more established basis for the use of DBS, very little work has linked accelerometry data with DBS \citep{Bhidayasiri2020}. Most recent publications discuss a theoretical roadmap for the usage of digital phenotyping, rather than reporting on actual results in a DBS trial. Indeed, a conference abstract of this case report from 2020 remains on the first page of Google Scholar results for "digital phenotyping DBS" in late 2022. Thus there is a great need for proof-of-concept results in the literature, to stimulate and guide future research. 

Still, digital phenotyping more broadly has shown great promise in recent literature. Next (section \ref{subsec:deep-pheno}), I will provide relevant background on these applications, to ground our case report on the pilot use of passive sensing in DBS. Furthermore, other digital psychiatry techniques -- like the interview recordings described in chapter \ref{ch:2} or the app-based self-report surveys and behavioral tasks touched on in chapter \ref{ch:1} -- have been successfully employed in exploratory works to better understand the mechanisms of DBS for OCD. A review of that research will be provided in the subsequent section \ref{subsec:ocd-dp}.

\subsection{A primer on digital phenotyping}
\label{subsec:deep-pheno}
Working to improve the state of psychiatry, NIMH launched Research Domain Criteria (RDoC) to encourage a precision medicine approach to mental illness. Different dimensions and subdimensions of behavior have been outlined to be characterized through not only observation and self-report, but also biomarkers at multiple levels of abstraction. However, linking biological factors with specific symptoms has been difficult thus far. The recent ability to more objectively quantify behaviors, often passively, will likely be key in bridging the gap between observed symptoms and biological systems. Early work in using data such as speech properties has shown promise in improving evaluation of disorders, but substantial research remains to be done to realize the potential of these technologies \citep{Torous2017}.

Given the direct coupling with gold standard clinical scales, audio/video analysis of recorded clinical interviews is a tractable problem set for early digital psychiatry work. Similarly, acoustic and linguistic analysis of daily patient journal recordings or other sources of more naturalistic speech is a well-grounded direction for future research. Additional background and motivation for these modalities was provided in chapters \ref{ch:1} and \ref{ch:2}, and we fruitfully employ them in this case report. However, they require active data collection work for both patients and study staff, and are fundamentally limited in the objectivity level and temporal resolution they can obtain. 

"Digital phenotyping" usually refers to the use of passively and continuously collected sensor data, a toolset that will certainly be necessary to fully bridge the gap between real-time neural activity and behaviors. In addition to the ability to time-align signals with biomarkers of higher temporal density, digital phenotyping datasets have the advantage of being relatively easy and cheap to collect. This enables tracking more people over longer time periods, and it produces more data that can eventually be used in powerful machine learning models that require a large amount of training material. Moreover, passive data collection in a typical longitudinal study is unlikely to impact participant behavior long term to the same degree that knowledge of active data collection can alter behavior - though this may vary based on the visibility of the device or app. 

Such continuous collection also enables capturing of parts of a patient's day that one may not have originally thought to record, but may turn out to have deep relevance. A notable example of this in the existing literature is the discovery that wrist-measured electrodermal activity (EDA) produces a strong signal during generalized epileptic seizures, with a correlation between amplitude and sudden unexpected death (SUDEP) risk \citep{Johnson2020}. EDA was being continuously measured as part of a study on non-verbal Autistic children when the seizure signal was first detected, which subsequently inspired follow-up studies and eventually the release of an FDA-approved device for Epilepsy monitoring. \\

\subsubsection{Available tools}
In psychiatry studies, digital phenotyping most commonly involves collection of 3-axis accelerometry data from study-provided wrist-worn actigraphy devices and/or GPS signal and phone usage patterns from an app installed on participant phones. Open source packages for extracting features from actigraphy data have already been released by the community, such as activity level estimates generated by GGIR \citep{GGIR} or sleep measures generated by DPSleep \citep{HabibSleep}. Pilot analysis of phone data can similarly be handled by pipelines meant to streamline collection from a specific app, such as Forest for Beiwe \citep{Beiwe}, or from independent open source tools for specific modalities such as DPLocate for GPS feature extraction \citep{HabibGPS}. 

The general features produced by software packages like these form a good base for any study, and as will be described have shown promise in monitoring of certain psychiatric symptoms. Keep in mind though that as the field progresses, so should the set of readily available features, and selection of specific metrics can become more hypothesis-driven. In the interim, we must work to better understand existing features and design new ones to capture a wider range of symptom profiles; for example, developing algorithms to measure repetitiveness and rigidity of movement patterns via accelerometry. Besides expansion of computational techniques, deeper consideration of alternative sensor types is another active area of research development, although it is less of a focus for our work.

Currently, wearable devices for digital phenotyping studies might also collect gyroscope, heart rate, temperature, or EDA data, and often include a button for timestamping events. As Fitbit does not officially support access to raw data, many studies use actigraphy "watches" designed for research collection of accelerometry, from companies such as ActiGraph, Axivity, and GENEActiv (the latter is used in this case report). These devices tend to optimize for robust data collection at a reasonable cost, meaning they are often clunky and ugly with minimal features. If patient participation is imperative or a wider range of sensors is necessary, devices like the Empatica Embrace Plus or Apple Watch should be budgeted for. However for a large study with risk of devices being lost, broken, or stolen, or with a patient population that is minimally technical or unable to regularly charge the device, the straightforward actigraphy "watches" are well-suited. They are also the easiest choices for integration with existing research processing pipelines, as major datasets such as the UKBiobank (Axivity) have already used such devices \citep{UKBWatch}.

Similarly, apps for phone data collection might access additional sensor streams. Phone accelerometer and gyroscope signals can be useful particularly when wearable device data is not possible to collect. Recording of audio from participant calls or even intermittent passive environmental audio recordings have been explored, but can be legally complicated due to potential involvement of individuals besides the consenting participants. Content of participant text messages and screen recordings of phone behaviors were also explored in early pilots but largely abandoned for related reasons. Content from public social media accounts can still be considered though, along with the more generic phone use information collected by modern monitoring apps -- which generally includes unlock/lock times, deidentified call and text metadata, battery state, and/or app usage time. Common apps used for digital phenotyping studies are MetricWire, MindLAMP, and Beiwe (the latter is used in this case report). All three of these options allow collection of various passive datatypes described, as well as administration of active self report survey (EMA) and audio journal prompts like discussed in chapter \ref{ch:1}. 

Note it is standard practice to collect daily self report data alongside continuous digital phenotyping, to use as one source of "ground truth" label that can ideally be predicted by the passive sensor features. The open source nature of Beiwe \citep{Beiwe} and MindLAMP \citep{MindLAMP} can be advantageous for both cost and building out custom pipelines, although an industry app like MetricWire is more likely to have a polished user interface with reliably stable performance. Choice of app should depend not only on available features, but also on the hardware quality and technical abilities expected of the patient population and of the research staff.  \\

\subsubsection{Early results}
While passive data streams collected directly from the participant phone are perhaps the easiest digital phenotyping modalities to set up and represent an exciting field of emerging research, they have greater concerns with patient privacy than wearables do and are largely less grounded in prior psychiatry literature than measurements of motor behavior or autonomic activity from a wearable would be. Thus the majority of early results from applications of deep phenotyping to psychiatry and mental health center around wearable devices, especially data from wrist-worn accelerometers (actigraphy). For example, self-reported stress levels of healthy individuals have been predicted above chance by actigraphy analysis, both alone and in conjunction with phone usage information. In the combined case, phone usage added predictive power above actigraphy alone, but was less predictive in a vacuum than actigraphy was. It is worth noting that fairly simple accelerometer features were identified as most salient to perceived stress, such as the variance of the signal magnitude in each time bin -- a proxy for overall activity levels \citep{Schmidt2018, Smets2018}. 

Additionally, actigraphy can be used to improve measures of sleep in research studies without significantly complicating procedures or increasing costs. While self-reports of sleep time and efficiency were generally well correlated with watch-derived sleep metrics in healthy subjects, the watch metrics were able to uncover other relationships that were not apparent with the self-report data. Most notably, even when controlling for lifestyle and demographic factors, cardiovascular risk was predicted by actigraphy sleep scores but not by self-reported sleep estimates \citep{Teo2019}. 

Furthermore, actigraphy has been successfully applied in specific psychiatric disorders to estimate various facets of mental health and well being, and subsequently better understand how the correlations seen in healthy participants compare and contrast with observed predictive power of actigraphy in patients. For example, wrist accelerometry was used to evaluate Major Depressive Disorder (MDD) and Bipolar Disorder (BD) patients, by characterizing the relationship between detected activity levels and same-day self-reported mood, energy, and sleep duration. All three of the survey measurements were significantly correlated with actigraphy features in both patients and controls, with the strongest correlations detected in BD \citep{Merikangas2018}. This suggests that wrist accelerometry may actually have more expressive power in quantifying severity of certain psychiatric symptoms than it does in measuring healthy habits for the general population, and the latter is already commercially successful via brands such as Fitbit. 

Beyond relating actigraphy features to key psychological constructs for mental health, whether estimated by self-report or clinician rating, accelerometry can also be used to detect specific actions of interest. These actions could be previously known to be of direct clinical relevance, for example hand washing in OCD with contamination fears or cigarette smoking in an addict \citep{Senyurek2019}, or they could simply form a broader picture of patient behavior. Such a picture could uncover behaviors that are important for a particular patient, like participation in a choice hobby, and it could also be utilized by a therapist in personalizing cognitive-behavioral treatment. In the research context, behavior detection tools would enable new study designs for enhancing our understanding of motor behavior, especially when paired with access to real-time neurobiological data. Eventually, behavior detection could be utilized in just-in-time behavioral intervention strategies. On a daily timescale, a similar style of intervention has already shown promise: a pilot study was able to identify whether elderly patients had brushed their teeth that day or not, in addition to detecting hand washing, hair brushing, and medication taking, in order to issue reminders as needed \citep{Cherian2017}.

Overall, early research gives an exciting glimpse of the potential for digital phenotyping. Yet there is a distinct lack of studies that consider multimodal data analysis techniques or truly longitudinal ($> 3$ month) data streams. Moreover, most current studies have small sample sizes, which not only impacts generalizability to different patients, but also the ability to use certain machine learning techniques. In addition to the gaps in existing digital phenotyping research, there are also entire disease categories that have been largely neglected, as has the collection of most biomarkers. Some of these gaps are being worked on in upcoming and in-progress studies, as discussed in chapter \ref{ch:2}. It is not clear though how well these study designs will address all of the critical lines of questioning in making digital phenotyping a viable subfield akin to functional neuroimaging. For more on the development of this subfield at large, see the concluding chapter. The remainder of this chapter will focus on ideas and issues only as they pertain to OCD and/or DBS. \\

\subsubsection{Relevance to neurobiology}
While the above-mentioned research directions linking digital phenotyping with clinical outcomes will indeed require substantial further investigation, there should still be simultaneous consideration for how digital phenotyping features may be linkable to neurobiological properties in parallel. The use of biomarkers to predict clinical metrics has seen success in some areas, which it may be possible to expand on using digital phenotyping. One such study demonstrated that social functioning outcomes over the course of a year were predicted more accurately by analysis of brain imaging alone than by expert opinion alone. This held true in both recent onset depression and psychosis, and while the psychosis predictions were notably more accurate than the depression ones, the use of neuroimaging was a larger improvement over the use of clinical scores for depression outcome prediction than it was for psychosis \citep{Koutsouleris2018}. 

Recent work has indeed shown that ecological momentary assessment surveys self-reporting sociality in psychosis patients significantly correlated with simultaneous phone GPS features like distance from home, and that these features in turn related to severity of negative symptoms measured by clinical scales \citep{Raugh2020}. An obvious next investigation would then be to characterize the relationship between the most salient digital phenotyping features for estimating social functioning from \cite{Raugh2020} and the most salient neuroimaging features for predicting social functioning from \cite{Koutsouleris2018} across time in psychosis patients.
 
Ultimately, the ideal goal for any digital psychiatry question will be a framework that can relate traditional clinical ideas, digital behavior measurements, and neurobiomarkers in one overarching model, which could require steps adjacent to what might constitute successful progress on any single pairwise relation. Thus far there has been very little work including all three perspectives; understandable given the recent emergence of digital phenotyping, but certainly not something that should be completely ignored going forward. In fact, it might be preferable to tackle a hyperspecific behavior or biological anomaly, but to do so from the unified perspective, than to attempt a project on a broader topic that leaves out one of the three perspectives. Nevertheless, the power of the broader research community is the ability for different groups to take different approaches. Here, we report a pilot result that restricts analysis to a single patient, but includes neural recordings and clinical outcomes alongside multiple modalities from both passive and continuous digital phenotyping, as well as app-based self-reporting. To our knowledge, this is the first of its kind in the context of deep brain stimulation. Therefore, utilizing deep phenotyping in the context of DBS is one of our major goals. 

\subsection{Digital psychiatry tools in OCD}
\label{subsec:ocd-dp}
Next, I will discuss the small amount of preliminary work relating deep phenotyping and other digital psychiatry techniques to OCD -- as another major goal of ours is to further this promising application, independent of DBS. Although the more established nature of DBS in OCD does provide additional motivation for furthering development of passive behavioral measurement techniques for OCD.

\subsubsection{Emerging methods for DBS evaluation}
 Some existing work has employed digital psychiatry techniques to evaluate DBS in the context of providing more detailed analysis of an experimental task or interview. While we focus more on longitudinal analysis in our case report, we also perform an experiment inspired by the most notable prior work along these lines in OCD. In that paper \citep{Goodman}, video recordings were taken of 2 OCD patients that had recently received the standard ventral striatal DBS implants, with footage recorded for both stimulation on and stimulation off conditions. Video was automatically processed for facial action units (FAUs), and for each frame an affect score was returned representing the sum of activations for all affect-related FAUs in the frame. 
 
 Significant affect differences were found between the conditions; the 'DBS on' condition increased positive affect while the off condition flattened affect \citep{Goodman}. This was a strong proof of concept for the use of facial analysis software to evaluate DBS in treatment-resistant OCD. However, the study did not blind the participants to stimulation status, meaning the results may not reflect a true effect of stimulation. Additionally, both participants in the study were responders, so it is not clear whether DBS non-responders may have any affect changes from stimulation. Furthermore, their experiments were run using a traditional DBS paradigm, not applied towards evaluating a novel DBS paradigm with additional brain region targets.

 Later work involving the same group did venture into the longitudinal sensing space with OCD patients undergoing DBS: \cite{Provenza2021} collected over 1000 hours of naturalistic LFP recordings from across 5 OCD patients with bilateral VC/VS implants, and analyzed this neural data in conjunction with app-based self-report data on symptom severity and sleep patterns, as well as teletherapy video recordings and data from at home self-administered tasks. Additionally, these analyses were grounded in more traditional onsite assessment and clinical scale rating outcome measures. Because the neural implant device changed part way through the study, the majority of the results were reported on 3 of the subjects, with participation varying widely between them. One participant contributed nearly $\frac{3}{4}$ of the behavioral task data with time-synchronized LFP, and was also the only participant to wear an Apple Watch for part of the study, though that lasted less than 48 hours and was thus very limited relative to the passive sensing done in our case report. 
 
 All 3 participants did report lessened OCD symptoms over the course of the DBS treatment, with 2 of them demonstrating a stable decrease. The most active participant had a much greater level of variation in self-reported symptom severity, though still a decrease overall after beginning DBS. This high variability enabled some interesting analyses on the relationship between LFP features and OCD. Average normalized spectral band powers (Delta: 0-4 Hz, Theta: 4-8 Hz, Alpha: 8-15 Hz, Beta: 15-30 Hz, and Gamma 30-55 Hz) from both the left and right implants were correlated with same day overall severity report. On both sides, delta power had a statistically significant and fairly strong ($r$ between $-0.55$ and $-0.6$) negative linear correlation with reported symptoms \citep{Provenza2021}. These results provide an exciting proof of concept for the use of digital psychiatry tools in OCD, especially when combined with data from DBS implants, and they also provide some pilot observations of interest for future such studies. However they do not include passive digital phenotyping sensing like we propose here. Ideally, relevant lessons from both works can be combined for an improved future framework for digital psychiatry studies incorporating dense neural data. 
 
\subsubsection{The promise of passive sensing for naturalistic symptom measurement}
Overall, passive and continuous digital phenotyping techniques are of greater interest here than techniques for supplementing active experiments -- not only for alignment with DBS data, but also for increased understanding of OCD behaviors. Because rituals present a unique opportunity to detect a relatively frequent symptom as it occurs, OCD is especially well suited for the development of actigraphy-based ‘vital signs’. Furthermore, OCD patients still have fluctuations in variables like mood at a longer timescale, and such properties may influence or be influenced by severity of OCD-specific symptomatology. Changes in movement have been previously linked to mood and stress levels in OCD \citep{Walther2015}, and as discussed actigraphy devices have successfully detected variations in these measures in a variety of other conditions \citep{Tazawa2019}. Thus there are multiple timescales of simultaneous and concrete relevance in movement signals from OCD patients with physical rituals, which makes actigraphy data analysis in OCD a more tractable problem with more clear opportunities to apply existing computational techniques than can be found in most digital phenotyping studies to date. It is in some ways self-grounded, whereas raw actigraphy data from e.g. depression patients with no other context available would be much more difficult to ask interesting questions about.  

 Behavioral interventions for OCD also may be improved via digital technology. Showing patients with contamination-related OCD videos of themselves compulsive hand-washing or touching a dirty object improved symptom scores, while a similar video of themselves doing generic hand motions had no effect \citep{Jalal2018}. Interventions of this style could be directly included in app-based therapies, which have already shown promise for increasing accessibility of cognitive behavioral therapy without significantly reducing efficacy in less severe cases of OCD \citep{Hwang2021}. Further, the results of \cite{Jalal2018} suggest that detection of ritualizing behavior may be usable as part of a just-in-time behavioral treatment, in addition to the clear implications that such tech would have for accuracy and scalability of patient progress monitoring. 

Despite the potential for actigraphy in OCD, however, there is little work towards collecting carefully-designed datasets in this population nor towards developing computational tools for custom analysis of large scale datasets that happen to contain some OCD participants (like the UKBiobank). The prior research that does exist has focused on using actigraphy to quantify sleep in OCD \citep{Donse2017, Jaspers-Fayer2018}, which showed promise in identifying sleep disturbances and even in predicting response to a novel treatment, but did not consider day time movement patterns at all, let alone leverage the unique properties of OCD outlined above. Additionally, these previous studies collected data for only one week per participant, which would prevent modeling of longer-term symptom fluctuations altogether. The UKBiobank dataset similarly includes just one week of actigraphy (if any) for each participant \citep{UKBWatch}.

Other modalities of digital phenotyping data, such as systematic measurement of phone usage, have hardly been studied in OCD either. This is again counterintuitive, as rituals can involve phone behaviors just as they can involve physical ones, introducing an entire line of study for usage data in OCD in addition to more general questions relating phone use and e.g. stress. Custom solutions utilizing new technologies have been developed for specific patients, for example a case report on GPS-based feedback for a patient with checking compulsions that frequently involved going outside \citep{Olbrich2016}. The use of available technology to tackle low hanging fruit for mental wellness of individual patients is a good direction for practitioners to take digital phenotyping in its current state. But to advance the scientific understanding of the field, we will require larger studies testing more general computational approaches, which have not yet been done at all for OCD. 

 It is worth noting that OCD increases risk of other disorders, including a 10x increased chance of developing Major Depressive Disorder compared to the general population \citep{Pallanti2011}. Beyond psychiatric comorbidities, OCD patients have increased risk of cardiovascular and metabolic disease, and treatment of OCD symptoms alone may not address these secondary concerns \citep{Mataix-Cols2020}. Thus the use of wearable and phone sensors to address other medical issues may have increased relevance for OCD patients. More generally, certain features extractable from digital phenotyping data - e.g. step counts or time spent at home - can be interpreted as inherently relevant to physical health and mental well being. In particular when these features become strongly anomalous in a specific patient or have changes coinciding with changes in other modalities like EMA responses from that patient. Therefore, despite the disappointing lack of OCD-specific digital phenotyping research to date, there are potential immediate applications for digital phenotyping results from other populations. 

 In the realm of digital psychiatry as a whole, there are also already non-DBS applications of the recent app-based self-report surveys to OCD. Digital phenotyping study designs should keep in mind the latest EMA literature anyway, as it is a powerful tool for providing more frequent, even if weaker, labels than gold standard clinical outcome measures. One of the more notable reports was the application of EMA in a clinical trial of detached mindfulness therapy for OCD, as compared to the standard cognitive behavioral therapy approach and a waitlist control condition. In both treatment groups, the YBOCS showed similar levels of improvement after therapy, with clinical significance in $\sim 40\%$ of patients. The EMA questions designed to assess severity of obsessions and compulsions also improved, by similar amounts between the two treatment groups, and statistically significantly more than changes in the baseline EMA that was collected from the control group. Meanwhile, EMA questions about general emotional state did not change due to either treatment \citep{Rupp2019}. This study demonstrated treatment sensitivity of EMA designed for OCD, and showed how different categories of EMA question can be used to probe fundamentally different things in OCD patients. Further work on EMA could take greater advantage of customization options to ask patients about personal obsessions and compulsions, and/or target administration of EMA to moments of interest (like exposure events) within a course of therapy. \\  
 
 \noindent We hope that this case report will soon encourage an increase of digital phenotyping work in OCD. Unfortunately the patient we were able to follow exhibited only mental rituals, and refused to where a wrist accelerometer for most of the study, so it was not possible to address many of the questions posed here in our hands-on proof-of-concept work. However, we do provide discussion on how our analyses could be expanded to better capture unique features of OCD in section \ref{sec:discussion3}, alongside our discussion of how future DBS studies can best utilize our methodologies. 

\subsection{Moving forward}
\label{subsec:ocd-motivation}
The above literature review demonstrates that new behavioral metrics are a key path towards improving deep brain stimulation (DBS) results in OCD. Such metrics would have implications in treatment planning and patient monitoring, clinical trial evaluation, development of new treatment technologies like closed-loop DBS, and expansion of basic scientific knowledge about neurobiological correlates of OCD behaviors. In other words, increasing the descriptiveness, robustness, and temporal resolution of behavior quantification would have implications at all levels of the biomedical research abstraction, creating a strong positive feedback loop in research progress. Conversely, continuing with only the current standard for clinical outcome measures would make it impossible to richly contextualize real-time neural recording data or uncover new variables that might explain treatment response differences across seemingly similar individuals.  

\noindent As stated in the recent "BRAIN Initiative: Brain Behavior Quantification and Synchronization" NIH Funding Opportunity [RFA-MH-22-240] --

\begin{quote}
Matching the scientific rigor and precision of measurements of brain activity with equally precise, temporally dense measurements of the functional output of the brain, as expressed in a broad range of behaviors, will accelerate the discovery of brain-behavior relationships in both health and disease. Achieving a comprehensive understanding across these levels of analysis demands the same level of rigor, precision of measurement, and temporal resolution across all levels. At present, tools for measuring behavior in humans and other species lack the necessary precision and resolution to fully capture behavioral dynamics synchronously with data from the environment with which the organism is interacting and which shapes the behavior under study. [...] Achieving higher-resolution quantification of complex behavior in more naturalistic environments in humans is especially crucial for understanding higher-order cognitive functions, whose associated brain activity may be readily available to train and control closed-loop neuromodulatory devices designed to treat complex behavioral disorders and interrupt maladaptive behaviors before they occur (such as self-harm, substance misuse, etc.).
\end{quote}

\subsubsection{Chapter aims}
OCD has both behavioral profiles especially relevant to digital phenotyping technology and the most well-established DBS paradigm to date of all psychiatric disorders, making it a uniquely good fit for the line of questioning posed by the BRAIN Initiative. Yet there is an extreme paucity of research on digital phenotyping in OCD and on passive digital phenotyping in DBS, let alone their intersection. Our case report is a first of its kind step towards addressing some of these gaps. It lays out a blueprint for success to be used by future larger studies on the topic. 

\section{Testing a novel DBS paradigm for OCD}
\label{sec:mgh-intro}
Our case report on relationships between novel behavioral features, neural data, and clinical outcomes was conducted as part of a trial for a novel DBS paradigm for OCD, rather than in a traditional DBS scheme like described above. The clinical results of the pilot study were previously reported by \cite{Olsen2020}; I will summarize their aims and takeaways in this section. 

Current models for many psychiatric diseases involve hyper- or hypo- connectivity between brain regions. Cortico-striato-thalamo-cortical (CSTC) connections in particular have been shown to play a role in numerous behaviors and disorders. Evidence suggests that these regulatory loops integrate sensory input to guide attention and modulate behavior. They are observed to be disrupted in depression, eating disorders, anxiety, substance use, and schizophrenia, among other disorders \citep{Peters2016}. In OCD they have been of especial interest, as the disorder can be modeled as a loss of control in habitual behavior, like an unstable positive feedback loop. Experimentally, decreased activity on functional neuroimaging was observed in the striatum after successful treatment of OCD with SSRIs and CBT. Genetic mouse models of OCD exhibit corticostriatal dysfunction, and optogenetic stimulation of one such pathway was able to ameliorate symptoms \citep{Burguiere2015}. 

Thus there is an abundance of evidence suggesting that CSTC hyper-connectivity is closely linked with OCD. Established surgical lesion treatment for severe OCD as well as emerging DBS approaches that have shown some clinical success indeed already target CSTC structures, and may already function in part because of unintended disruptions to CSTC synchrony \citep{Widge2019}. However, these treatments do not involve the cortical regions that are a critical part of CSTC loops. The proposed DBS paradigm aims to address this gap; \cite{Olsen2020} hypothesized that CSTC hyper-connectivity would result in hyper-synchrony of the relevant brain regions, and moreover that disrupting this hyper-synchrony via external input from neural implants would in turn disrupt the over-expression of habitual behaviors seen in OCD.

\subsection{Disrupting corticostriatal hypersynchrony}
\label{subsec:stim-params}
The target brain regions for this experimental deep brain stimulation paradigm were the ventral capsule/ventral striatum (VC/VS), which is the most common implant site for traditional DBS in OCD, and the supplementary motor area (SMA), a cortical region that is in a CSTC loop with VC/VS. The VS is a subregion of the striatum within the basal ganglia, thought to play a primary role in reinforcement learning and motivated behavior, and implicated in Pavlovian conditioning and addiction. There is increased resting state functional connectivity of the ventral corticostriatal axis in OCD \citep{Macpherson2019}. The SMA is located in the dorso-medial prefrontal cortex, and receives input from the basal ganglia through the globus pallidus. It projects to the striatum, which projects to the globus pallidus, completing a cortico-striatal loop. SMA also has reciprocal connections with the primary motor cortex, and is thought to play multiple roles in movement and potentially movement planning. Lesions to SMA can sometimes cause involuntary actions, and sometimes prevent movement, as does stimulation of the area. Human SMA is active even when just observing a graspable object, and is very involved in sequential action \citep{Nachev2008}. 

\begin{figure}[h]
\centering
\includegraphics[width=\textwidth,keepaspectratio]{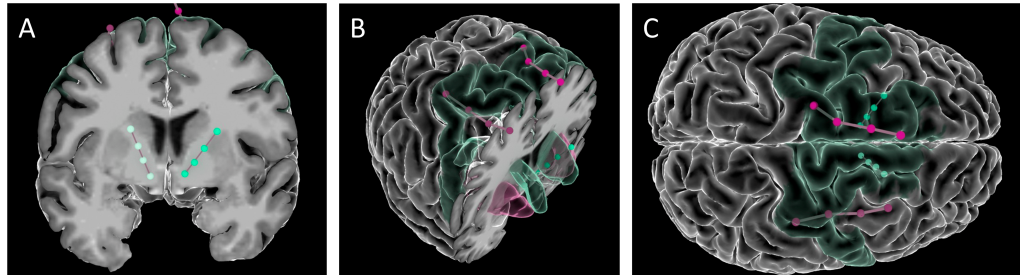}
\caption[Images of the DBS and paddle leads in patient's brain.]{\textbf{Images of the DBS and paddle leads in patient's brain.} Reproduced from \cite{Olsen2020}, on which I am a contributing author. Rendered with the Multi-Modality Visualization Tool \citep{Felsenstein2019}. Position of left (red) and right (blue) SMA leads, and left (green) and right (yellow) VC/VS leads. Brodmann Area 6 is colored in turquoise. \emph{A)} Coronal slice showing position of VC/VS leads (subcortical regions not colored). \emph{B)} Angled view with cortical coronal slice. Caudate nucleus colored in green, nucleus accumbens (NAcc) in blue, and putamen in pink. \emph{C)} Superior view (left on top) showing cortical lead positions. Note that the right VC/VS lead is in the caudate nucleus and NAcc whereas the left VC/VS lead is more laterally placed in the putamen. No adverse effects of lead placement were observed with the patient.}
\label{fig:ocd-mgh-leads}
\end{figure}

To test the hyper-synchrony hypothesis, the Medtronic Activa PC+S device was implanted in the pilot patient (Figure \ref{fig:ocd-mgh-leads}) for simultaneous stimulation and recording from the regions of interest, bilaterally. One major part of evaluating the hypothesis was to track the levels of cortico-striatal hyper-synchrony in the patient over time. Local field potential (LFP) was automatically recorded at each implant site four times each day, at evenly spaced intervals and for a duration of 60 seconds at a time. Additional recordings were also taken via external trigger during clinical site visits. To measure hyper-synchrony, coherence between oscillations recorded from the striatal lead and from the cortical lead was assessed on each side. LFP data were collected for 606 days following the implant operation.

The other major component was to determine the clinical outcome of chronic stimulation that decreases synchrony between VC/VS and SMA in this patient, and to compare that to the efficacy of a standard DBS paradigm for OCD. Therefore, the striatal stimulation was turned on first, and stimulation settings were tuned for that implant as they would be typically. After 6-9 months, cortical stimulation would be turned on, using a frequency setting marginally different from that used in the striatum. By driving the regions at a frequency mismatch, cortico-striatal coherence should decrease, and it was hoped that compulsive symptoms would likewise be reduced. Both patient and experimenter were initially blinded to cortical stimulation status, and upon unblinding the cortical stimulation settings were more precisely tuned. In this patient, it turned out that cortical stimulation was turned on 235 days after the operation. Unblinding was done at the 9 month mark on day 270. 

It turned out that the initial frequency setting for the cortical leads -- 130 Hz, chosen due to striatal frequency of 135 Hz -- actually strongly increased the cortico-striatal coherence metrics. Based on both this and the patient's subjective report of feeling "overstimulated", the cortical stimulation frequency was changed to 100 Hz shortly after unblinding. At this frequency, coherence was sharply reduced, although still above baseline levels. The resulting stimulation paradigm being tested was thus not reflective of the original hypothesis, though it was still a worthwhile investigation into manipulating cortico-striatal synchrony in an OCD patient. 

The patient did not demonstrate any clinically significant improvement in primary rating scales, or even daily self-report survey responses, due to cortical stimulation. The main assessments used were the Yale-Brown Obsessive-Compulsive Scale (YBOCS) and the Montgomery-Asberg Depression Rating Scale (MADRS), rated at bi-weekly site visits. Only the patient global impression scale (PGI) scores, which measures subjective patient opinion of the treatment impact, substantially and robustly improved after cortical stimulation. This effect started just after patient unblinding, suggesting placebo rather than efficacy of the stimulation in this patient. 

In our case report, we used these clinical scale and neural recording data that were collected by the neurotherapeutics team. For details on their methodologies, see \cite{Olsen2020}. We input the scale values directly into our relevant analyses; for the LFP data we used the \cite{Olsen2020} artifact correction and feature extraction code to obtain power band features for each lead as well as frequency-averaged cortico-striatal coherence for each side, computed for each recording. We then took the mean between left and right sides for each output, to reduce our feature set. Where ever we report daily values instead of hourly, we also took the mean over the four recordings in that day. 

\subsection{Quantifying behavior during a DBS trial}
\label{subsec:ocd-methods}
To extend the typical DBS evaluation dataset beyond clinical scales and neural features, we employed a number of digital psychiatry tools. This included audio/video recording and transcribing of clinical interview sessions, daily prompting for app-based self-report survey and free-form audio journal recording, and passive sensing of phone location, movement, and screen-time. A set of core features was extracted from the collected data and visualized longitudinally at different timescales. Correlation structure was also analyzed, both within the extracted feature set as well as between the digital features and the clinical/neural features reported by \cite{Olsen2020}. 

Because the longitudinal data collection scheme here was best suited for only correlational analyses, we additionally performed a much shorter-term experiment involving direct changes to cortical stimulation status, hoping to gain some insight into potential causal effects of the new stimulation paradigm. The patient participated in a series of structured conversations with cortical stimulation on or off, blinded to this status. Facial expression and language were then analyzed over the course of this experimental interview recording. \\

\noindent All described analyses were done using python. Methodology details for data collection and analysis, as well as the interview experiment protocol, can be found in supplemental section \ref{sec:ocd-methods-append}.

\section{Results}
\label{sec:results3}
Clinical evaluation and research into OCD is oft-limited by sparse yet time-consuming cross-sectional evaluations. Here we provide an example of a rich data set collected from a single patient, both actively and passively. The goal of this idiographic study was two-fold: to evaluate the efficacy of a novel dual site deep brain stimulation (DBS) paradigm for treatment of OCD, and to serve as a proof of concept for using digital phenotyping tools to evaluate DBS outcomes more generally. In addition to stimulation settings, local field potential recordings, and clinical outcomes (previously reported on by \cite{Olsen2020}), we collected a variety of temporally denser behavioral data types, including GPS, phone usage, and accelerometry. These data can provide useful information between site visits, as well as additional insight to behaviors not typically observed such as sleep patterns. This is particularly important for evaluating DBS trials, as stimulation parameters can vary over the course of the day, and the eventual development of closed loop systems will rely on understanding stimulation outcomes with greater temporal resolution and finer dimensional detail than infrequent, coarse clinical scales can provide. \\

\noindent The following sections will take a closer look at patterns we found in these data within individual modalities, as well as relationships we found between features across modalities. See supplemental section \ref{sec:ocd-methods-append} for methodological details.

\subsection{Supplementing clinical measures with dense, objective behavioral data in a novel brain stimulation trial}
\label{subsec:ocd-birds-eye}
Over the course of approximately two years (872 days), the patient received 44 total YBOCS (12 questions each) and MADRS (10 questions each) scores, collected approximately every 14 days (Figure \ref{fig:ocd-report}a). To enhance our temporal resolution of the patient’s clinical status between visits, we collected a total of 350 daily self-report surveys about their productivity levels (8 questions each), also known as Ecological Momentary Assessments (EMA), as well as 334 audio diaries describing the events of a given day and their corresponding emotional state. Across all audio diaries, there were 3,587 sentences containing 44,790 words. At an even finer temporal resolution, we were able to passively collect 107,519 minutes of phone usage data, 146,476 minutes of location (GPS) data, and 623,144 minutes of phone accelerometer data over the course of the study, collectively referred to as digital phenotyping data (Figure \ref{fig:ocd-report}a). In total there were 10,822 hours with some phone usage and accelerometer information available, and 10,119 hours with some GPS information available.

\pagebreak

\begin{FPfigure}
\centering
\includegraphics[width=\textwidth,keepaspectratio]{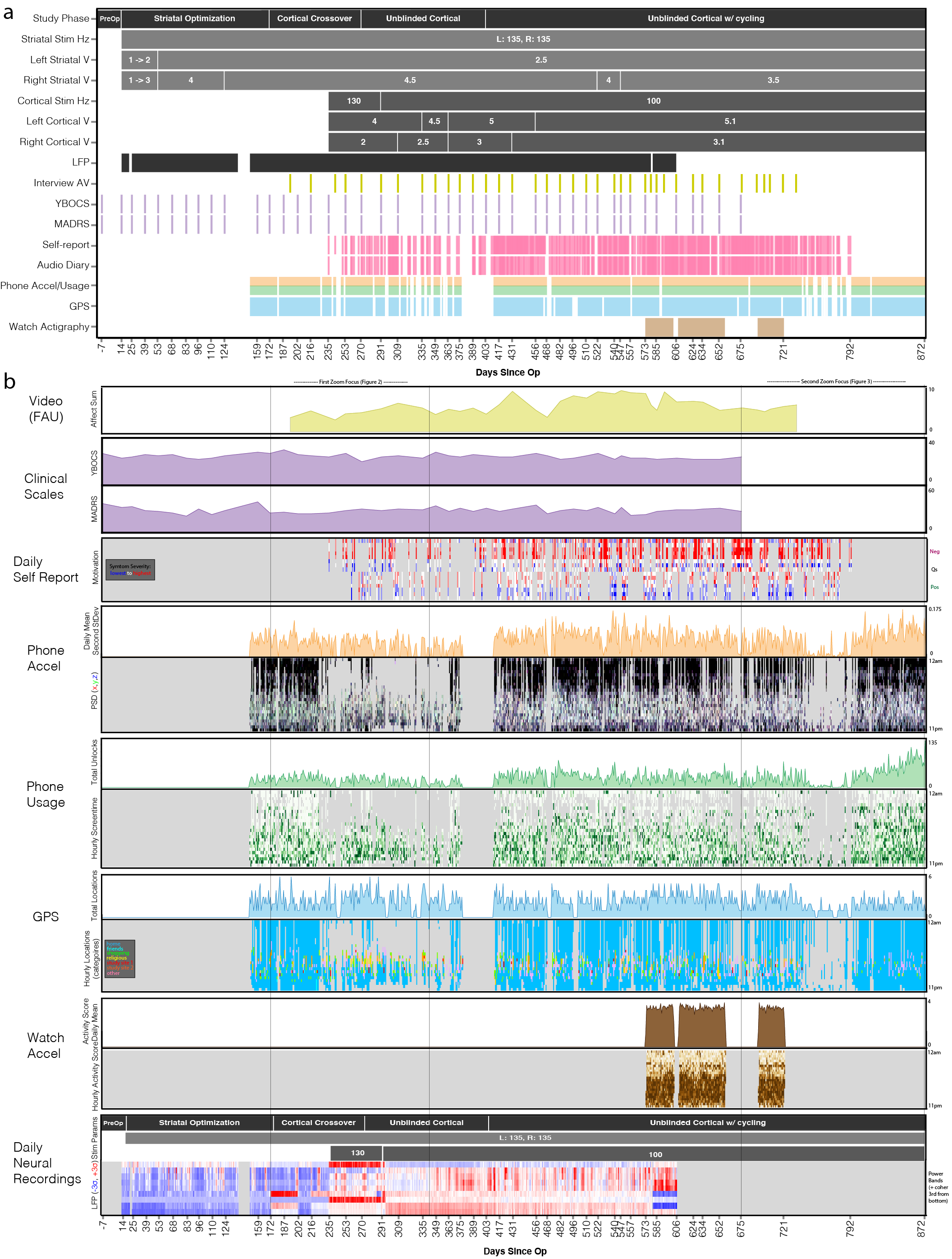}
\caption[Bird’s eye view of behavioral data collected from a DBS participant.]{\textbf{Bird’s eye view of behavioral data collected from a DBS participant.} In addition to stimulation settings, local field potential recordings, and clinical outcomes, we collected a variety of denser behavioral data types, including GPS, phone usage, and accelerometry. We thus adapted the study timeline created by \cite{Olsen2020} to display both the traditional and novel data types collected throughout the study (a). The grey bars denote periods of change in stimulation parameters, while the colored bars indicate each day for which the corresponding modality has data available (for digital phenotyping, counted whenever at least 1 hour was available). We then generated a high level summary of how features extracted from across these modalities varied over the course of the study (b). We created time-aligned area plots and heatmaps covering the following daily values, from top to bottom: total facial affect score from video-recorded clinical interviews (yellow), sum clinical scale scores (purple) for both YBOCS and MADRS, self-report survey responses (bwr heatmap with each question as a row), total phone movement score (orange), phone accelerometer power spectral density (hour by day heatmap with \emph{x/y/z}-axis power contributing red/green/blue respectively), total phone unlocks (green), phone screentime (hour by day heatmap colored $0\%$ of time white to $100\%$ of time dark green), total number of unique phone locations (blue), phone GPS location categories (hour by day heatmap with different colors corresponding to different location types), total wristwatch movement score (brown), watch accelerometer activity score (hour by day heatmap with minimum value white and maximum brown), and neural recording features (bwr heatmap containing different power bands as rows). In all heatmaps data unavailability is denoted by grey cells. In the sparser area plots (video, clinical scales) values are connected across missing days, while in the rest of the area plots entirely missing days are blanked out. Note linguistic features from daily phone diaries are not included in this bird’s eye view, but can be found within section \ref{subsubsec:ocd-diaries} below.}
\label{fig:ocd-report}
\end{FPfigure}

\FloatBarrier

Across the data types we collected, there were substantial differences in both data availability and capability to observe fine temporal patterns (Figure \ref{fig:ocd-report}b). Even during most of the periods with sparse digital phenotyping data, there were still many more data points than could be realistically gathered with existing clinical measures over the same timeframe – for example, when passive data was available only during the daytime, we were still able to generate inferences about patient activity every day. Additionally, by collecting a diverse range of data types, our methodology is more robust against data missingness issues. For example, between days 721 and 792 there was substantial missingness in the passive data due to technical issues, but the patient continued to submit self-report surveys (Figure \ref{fig:ocd-report}b). Conversely, passive data continued to be collected during other periods when patient participation lapsed, such as at the very end of the study (Figure \ref{fig:ocd-report}b). More generally, passive data enabled collection to continue for patient follow-up beyond the main study, and during unforeseen circumstances such as the pandemic.

While the experimental cortical stimulation largely did not result in clinical improvements as measured by clinical scales \citep{Olsen2020}, and our results are generally consistent with that, we were able to notice behavioral changes at certain points in the study period that would have otherwise been missed, demonstrating the potential power of these techniques. Further, some of the passively collected measures could be considered wellness markers in themselves, regardless of association with psychiatric outcomes - such as time spent at home or physical activity level. The bird’s eye view presented here, for example, makes it clear that this patient rarely left their home for more than a few hours a day when not visiting a study site (Figure \ref{fig:ocd-report}b). Indeed, they reported enjoying study visits as one of their main opportunities for socializing.

Two times of particular note were patient behavior right after being unblinded to cortical stimulation status (day 270), where we can see substantially increased attendance at their religious institution for over a week (section \ref{subsubsec:ocd-stim-zoom}), and the onset of the COVID-19 pandemic (end of study), where we can see substantially increased phone usage (section \ref{subsubsec:ocd-covid-zoom}). These trends can already be picked out of the bird’s eye view, but a crucial advantage of our data collection strategy is that we can also zoom in on time periods, or even particular days, of interest. We will next delve deeper into the behavioral, mood, and anxiety symptoms in this patient during these time periods of interest.

\FloatBarrier

\subsubsection{Behavioral data provide new insight to cortical stimulation transition period}
\label{subsubsec:ocd-stim-zoom}
\begin{figure}[h]
\centering
\includegraphics[width=\textwidth,keepaspectratio]{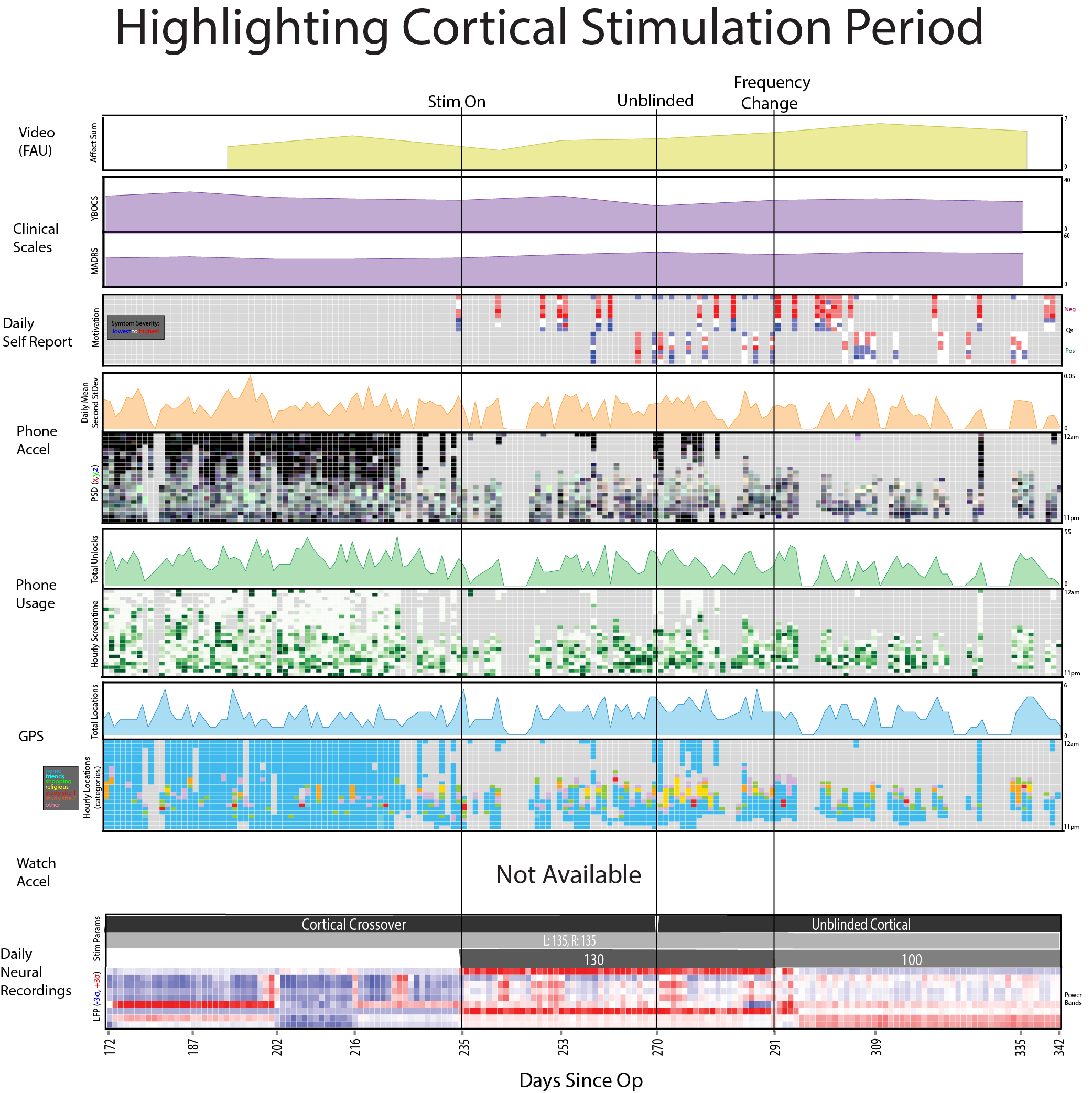}
\caption[Behavioral data provide new insight to cortical stimulation transition period.]{\textbf{Behavioral data provide new insight to cortical stimulation transition period.} Zoomed in version of Figure \ref{fig:ocd-report}, focusing only on data from the first day cortical stimulation could have been turned on through to $\sim8$ weeks after cortical stimulation was put at its final frequency setting. The day that cortical stimulation was actually turned on, the day that stimulation status was unblinded to both researcher and patient, and the day that the frequency for the stimulation was changed are all marked with vertical black bars.}
\label{fig:ocd-report-stim}
\end{figure}

To identify any behavioral changes in the time surrounding initiation of the novel cortical stimulation technique, we focused in on features extracted during the relevant study period, from day 172 through day 342 post-op (Figure \ref{fig:ocd-report-stim}). Day 172 was the first day that cortical stimulation theoretically could have been turned on; this marked the beginning of the “Cortical Crossover” period, during which stimulation status was double blinded. It was later revealed that cortical stimulation was turned on 235 days post-op. Unsurprisingly, there is a stark change in LFP properties (Figure \ref{fig:ocd-report-stim}, bottom panel) at the time stimulation status changed, most notably a large increase in corticostriatal coherence. Because the intended purpose of the stimulation was to disrupt hypersynchrony by driving the two regions at slightly different frequencies, the rise in coherence prompted a change in cortical stimulation frequency to 100 Hz at day 291. Shortly after the switch to 100 Hz, another sharp change can be seen in the LFP features, with coherence dropping substantially.

Despite the major changes in the neural data, we did not find any significant behavioral changes that coincided with change in cortical stimulation status. However, notable aberrations occurred directly after day 270 post-op, when stimulation status was unblinded (Figure \ref{fig:ocd-report-stim}). Abnormalities in GPS signal (ninth and tenth panels) stand out most clearly, as the patient spent multiple hours each day at a religious institution for the entire week following unblinding – note the long yellow bars that appear abruptly and then disappear again. Checking against both the patient audio diaries and the dates of major religious holidays in the patient’s faith, no external explanation for this anomaly could be found. Similarly, the patient’s response pattern to EMA questions (fourth panel) changed substantially (albeit temporarily) after day 270. Not only did they follow the positively worded branch (bottom bars) for more days in the period after unblinding, but they also answered the questions regardless of which branch they took in a more positive fashion (bluer color).

One potentially noteworthy trend in behavior that may relate to cortical stimulation frequency was an increase in facial affect \citep{Goodman} seen during clinical interviews after the stimulation setting was put to 100 Hz. The two interviews occurring after the frequency change were also the two interviews with the highest affect during this 170-day period \ref{fig:ocd-report-stim}, third panel). In the scope of the entire study, we regularly observed affect scores above baseline after day 290 post-op, through the final interview (Figure \ref{fig:ocd-report}b). 

\FloatBarrier

\subsubsection{Observing behavior changes during the COVID19 pandemic}
\label{subsubsec:ocd-covid-zoom}

\begin{figure}[h]
\centering
\includegraphics[width=\textwidth,keepaspectratio]{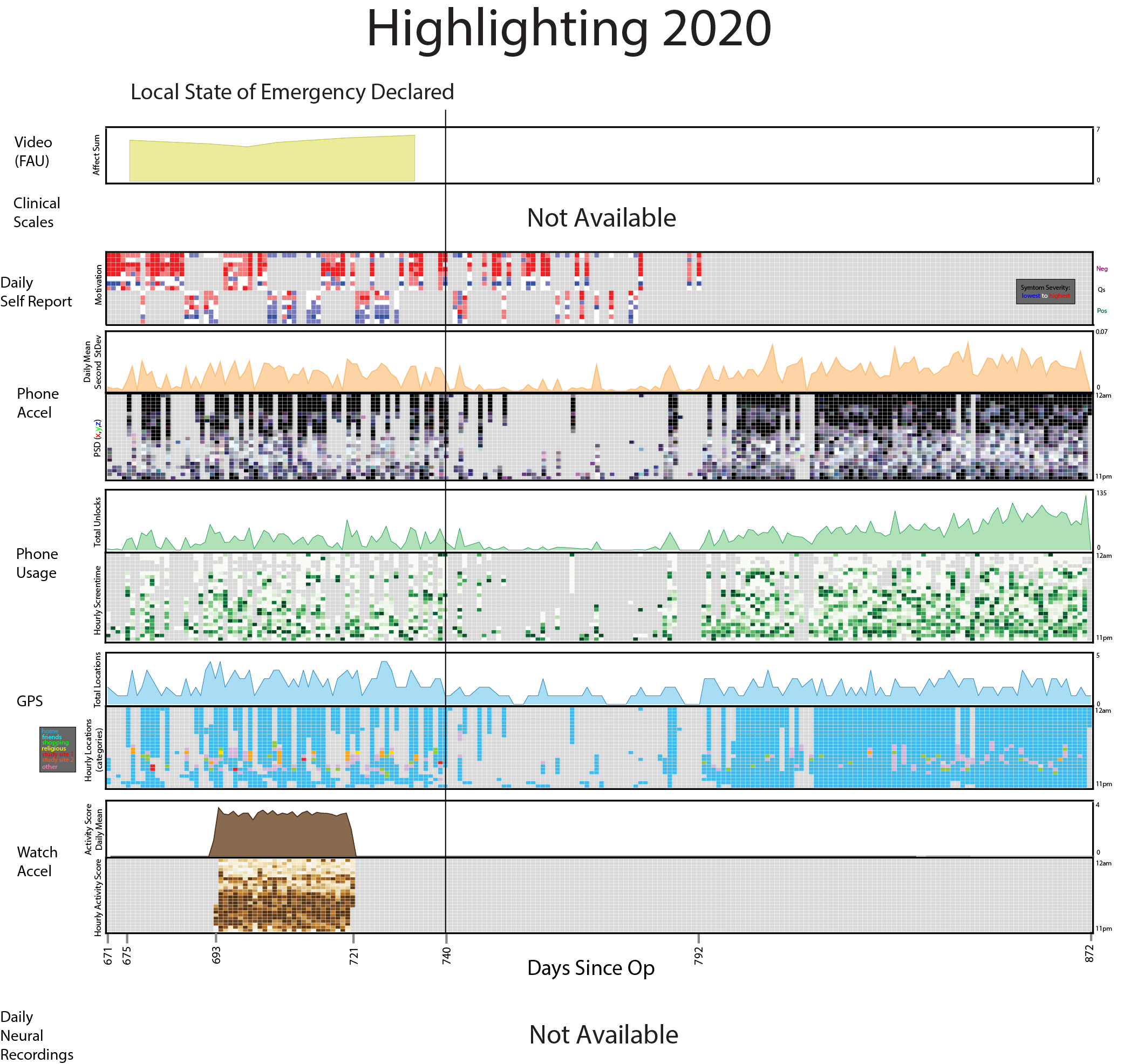}
\caption[Observing behavior changes during the COVID19 pandemic.]{\textbf{Observing behavior changes during the COVID19 pandemic.} Zoomed in version of Figure \ref{fig:ocd-report}, focusing only on data from the beginning of 2020 through the conclusion of the study in July. The day that the patient's city declared a state of emergency due to the pandemic is marked by a vertical black bar.}
\label{fig:ocd-report-covid}
\end{figure}

The COVID19 pandemic, which happened to coincide with the end of our study, caused widespread behavioral changes in the general population and had an outsized impact on those with severe mental illness \citep{Simon2020,Nemani2021}. Therefore, to identify potential behavioral changes caused by the pandemic in this patient, we also focused in on features extracted between day 671 post-op and the conclusion of the study at day 872 (Figure \ref{fig:ocd-report-covid}). Day 740 serves as a marker for the local onset of the pandemic, as a state of emergency was declared that day in the patient’s home city. 

To get a sense of the patient’s own perception of the pandemic, we analyzed changes in their self-report data surrounding this time (Figure \ref{fig:ocd-report-covid}, second panel). At the start of 2020, they went through periods of both high and low EMA lasting for weeks at a time, showing much more consistency in subsequent responses than was seen earlier in the study (Figure \ref{fig:ocd-report}b). In the week leading up to the declaration of a state of emergency, they were in a negative EMA period – although this could have been due to a variety of factors unrelated to COVID19. Interestingly, they submitted relatively positive EMA surveys right when the state of emergency was declared, after which EMA responses begin to fluctuate much more rapidly again. We also examined the audio diaries submitted during this time to get a better understanding of the trends seen in the EMA. However, the patient only mentioned the pandemic twice: once to complain about their sibling having a friend over when they wanted to follow social distancing guidelines, and the other time stating “I played with my dog a little bit, obsessed about the coronavirus”.

Next, we examined the passive behavioral data we collected during 2020 to provide additional insights into the patient’s behavior during the early pandemic, as well as extend our analysis into the summer months (Figure \ref{fig:ocd-report-covid}). GPS data (seventh and eighth panels) showed an increase in time spent at home through much of this period, even when compared to the limited mobility seen earlier in the study (Figure \ref{fig:ocd-report}b). During the pandemic, multiple days would pass between brief trips to a shopping center or doctor. However, by the late summer of 2020, the patient was once again leaving the house for a few hours most days. At the same time, we observed a sharp increase in phone usage over the course of the pandemic, as demonstrated by the patient’s daily number of phone unlocks (Figure \ref{fig:ocd-report-covid}, fifth panel), which remained high through the end of the study. Hourly phone accelerometer (fourth panel) and screen time (sixth panel) data then enabled us to further investigate this phenomenon. Phone usage in the middle of the night was higher during the pandemic period, driving much of the overall increase. Although it was not uncommon throughout the study to see disrupted sleep via phone movement and unlock data, screen time during standard sleeping hours was substantially lower pre-pandemic.

\FloatBarrier

\subsection{Revealing structure in clinical outcomes through clustering}
\label{subsec:clinical-clusters}
As described previously by \cite{Olsen2020}, there was little clinically significant variation in the patient’s total YBOCS score over the course of the study, and no clinically significant changes corresponding to the cortical stimulation treatment in their YBOCS or MADRS totals. While this highlights the potential utility of collecting supplementary behavioral data, in order to best utilize our dataset we still must analyze it in conjunction with the existing gold standard clinical measures. Therefore, we first aimed to maximize the information we can get out of these measures. To do so, we applied hierarchical clustering techniques to the YBOCS and MADRS scores, grouping questions based on how they co-varied within this individual (Figure \ref{fig:ocd-clusters}). Additionally, we compared the data-driven clustering results to more traditionally-defined subscales. 

\pagebreak

\begin{FPfigure}
\centering
\includegraphics[width=\textwidth,keepaspectratio]{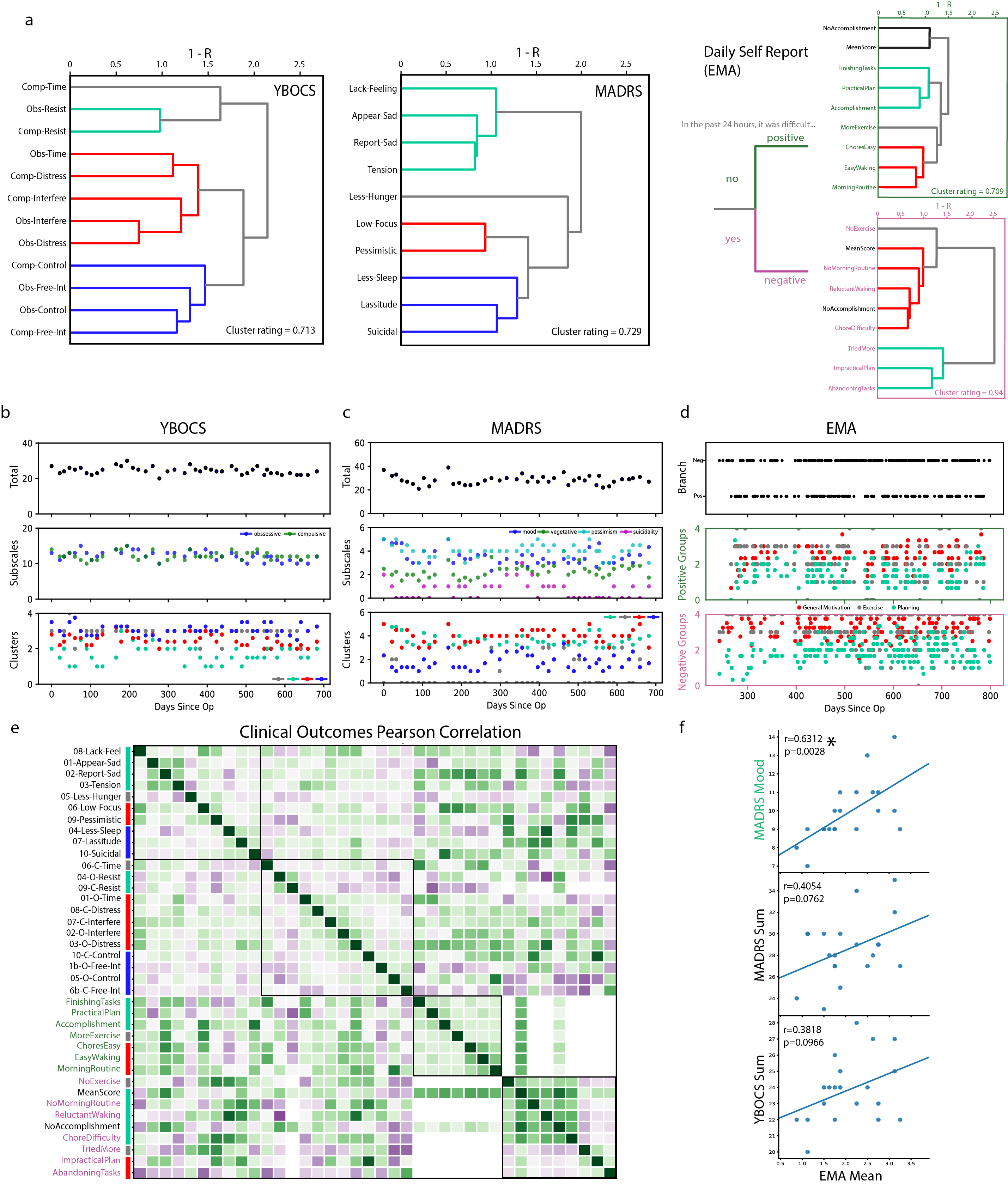}
\caption[Clinical outcomes of treatment and relationships between clinical measures.]{\textbf{Clinical outcomes of treatment and relationships between clinical measures.} We performed hierarchical clustering on the patient's clinical scores (for both YBOCS and MADRS, left/middle of a) and self report EMA surveys (for both positive and negative branches, right side of a), and used the resulting dendrograms to assign each question to a color-coded cluster. The time series of total YBOCS (top of b) and total MADRS (top of c) over the course of the study (top) were then contrasted with the progression of average question score from both clinically defined subscales (middle of b and c) and our clusters (bottom of b and c). Particularly for YBOCS, the clusters highlight symptom variation that is otherwise difficult to see. We similarly created a set of dot plots for EMA (d), including which branch the patient followed (top), and average question score for each positive branch cluster (middle) and each negative branch cluster (bottom). While we obtained the clusters using correlation amongst questions only within each scale, we also considered Pearson correlation of questions between the scales. The correlation matrix (e) has individual questions ordered by cluster, and is colored using the matplotlib PrGn colormap with bounds -1 to 1. Each scale is set apart with a black box, while the outside correlations on the off-diagonal represent cross-scale correlations. Select relationships between EMA and clinical scales were additionally visualized via scatter plots with linear fit line (f). Mean EMA question score is represented on the \emph{x}-axis in all three plots, with \emph{y}-axes corresponding to the same-day YBOCS total (bottom), MADRS total (middle), and average MADRS \emph{mood} cluster question score (top). By considering just the \emph{mood} cluster, we find stronger evidence of a potential relationship between clinical scores and EMA.}
\label{fig:ocd-clusters}
\end{FPfigure}

\FloatBarrier

We refer to the four resulting YBOCS clusters as \emph{time} (1 item), \emph{resistance} (2 items), \emph{interference} (5 items), and \emph{control} (4 items). These groupings were determined by the dendrogram structure (cophenetic $r=0.713$) of the YBOCS questions (Figure \ref{fig:ocd-clusters}a, left; clusters color-coded from top to bottom). Notably, parallel questions between the obsessive and compulsive subscales of the YBOCS tended to cluster together here, despite the fact they are traditionally split up by these subscales. There is both little variation in total YBOCS score (Figure \ref{fig:ocd-clusters}b, top) and little difference between obsessive and compulsive subscale scores (Figure \ref{fig:ocd-clusters}b, middle) over the course of the study. However, by analyzing average questions scores within our clusters, we were able to uncover clinically relevant variation in the \emph{resistance} related OCD symptoms (Figure \ref{fig:ocd-clusters}b, bottom, teal). We observed a coefficient of variation of 0.2989 for \emph{resistance} questions, as compared to coefficients of variation of 0.1114 and 0.0948 for the obsessive and compulsive subscales, respectively.

Analogously, we refer to the four MADRS clusters as \emph{mood} (4 items), \emph{hunger} (1 item), \emph{productivity} (2 items), and \emph{energy} (3 items). These groupings were determined by the dendrogram structure (cophenetic $r=0.729$) of the YBOCS questions (Figure \ref{fig:ocd-clusters}a, middle; clusters color-coded from top to bottom). The patient’s MADRS total score did significantly vary over the course of this study (Figure \ref{fig:ocd-clusters}c, top), although it did not respond specifically to the novel cortical stimulation treatment. By analyzing the time course of the MADRS clusters' average item scores (Figure \ref{fig:ocd-clusters}c, bottom), we were able to isolate the main symptom groups driving disease severity, as well as the symptom groups that varied the most week to week. While the \emph{mood (teal)} and \emph{productivity} (red) clusters were highest throughout the study, changes from week to week were primarily explained by the \emph{energy} (blue) cluster, along with occasional large changes in \emph{hunger} (grey). We observed coefficients of variation of 1.7113 and 0.3569 for \emph{hunger} and \emph{energy} questions respectively, as compared to a coefficient of variation of 0.1361 for the patient’s total MADRS scores. The mean question score in the \emph{mood} cluster was 3.642 and the mean question score in the \emph{productivity} cluster was 3.955, while the mean MADRS question score overall was 2.841. All MADRS questions are on a scale from no symptoms at 0 to most severe symptoms at 6.

We also clustered the questions from this patient’s daily EMA responses, as few significant trends were found in the total EMA score data over the course of the study \citep{Olsen2020}. EMA was by design split into positively worded and negatively worded questions, so that only one set would be administered on a given day, based on the valence of the first submitted answer. Thus we clustered the EMA questions separately for each of the two branches (Figure \ref{fig:ocd-clusters}a, right). While the clustering produced some minor differences between the positive and negative branches, both cases resulted in three clusters relating to \emph{motivation} (red), \emph{planning} (teal), and \emph{exercise} (grey). Overall, the patient was more likely to follow the negative branch (Figure \ref{fig:ocd-clusters}d, top), but the valence of their answers frequently changed between subsequent days. Focusing in on the time course for each cluster within the positive (Figure \ref{fig:ocd-clusters}d, middle) and negative (Figure \ref{fig:ocd-clusters}d, bottom) branches did not reveal any hidden trend. Although EMA did not significantly correlate with the total MADRS or YBOCS, as was previously reported \citep{Olsen2020}, correlating our identified MADRS/YBOCS clusters with total EMA yielded more insight. For example, the MADRS \emph{mood} questions correlate significantly with total EMA score (Pearson's $r=0.6312$, $p=0.0028$), despite the weak relationship observed between total MADRS and EMA scores (Figure \ref{fig:ocd-clusters}f). Note that all EMA questions were designed to assess facets of the patient’s productivity. \\

To further investigate the clinical scale clusters we discovered in this patient, we analyzed a smoothed version of the scores within each cluster by taking the mean and variance within a four-week window, sliding with an overlap of two weeks (Figure \ref{fig:ocd-smooth}b). By increasing the signal to noise ratio in this way, the trends found in the YBOCS \emph{resistance} (teal) cluster are emphasized, and new attention is drawn to different periods of variation in the MADRS total score. For example, \emph{hunger} (grey) largely spikes at different times than any of the other symptoms, while \emph{mood} (teal) tracks well with \emph{productivity} (red) for most of the study. Indeed \emph{productivity} is the only MADRS cluster to have a positive correlation with \emph{mood} (Pearson’s $r=0.2897$, $p=0.1274$), and this is the highest correlation found between any two MADRS clusters. By contrast, the summarized EMA cluster data reveal high variability over short time periods, making it more difficult to draw any meaningful conclusions from changes in EMA scores, even when grouped into clusters. However, the smoothed view does clearly demonstrate that the patient reported extremely low productivity on negative days (mean question score $2.808$), but only neutral productivity on positive days (mean score $1.886$). All EMA questions were analyzed on a scale from least symptoms at 0 to most symptoms at 4. 

\pagebreak

\begin{FPfigure}
\centering
\includegraphics[width=0.9\textwidth,keepaspectratio]{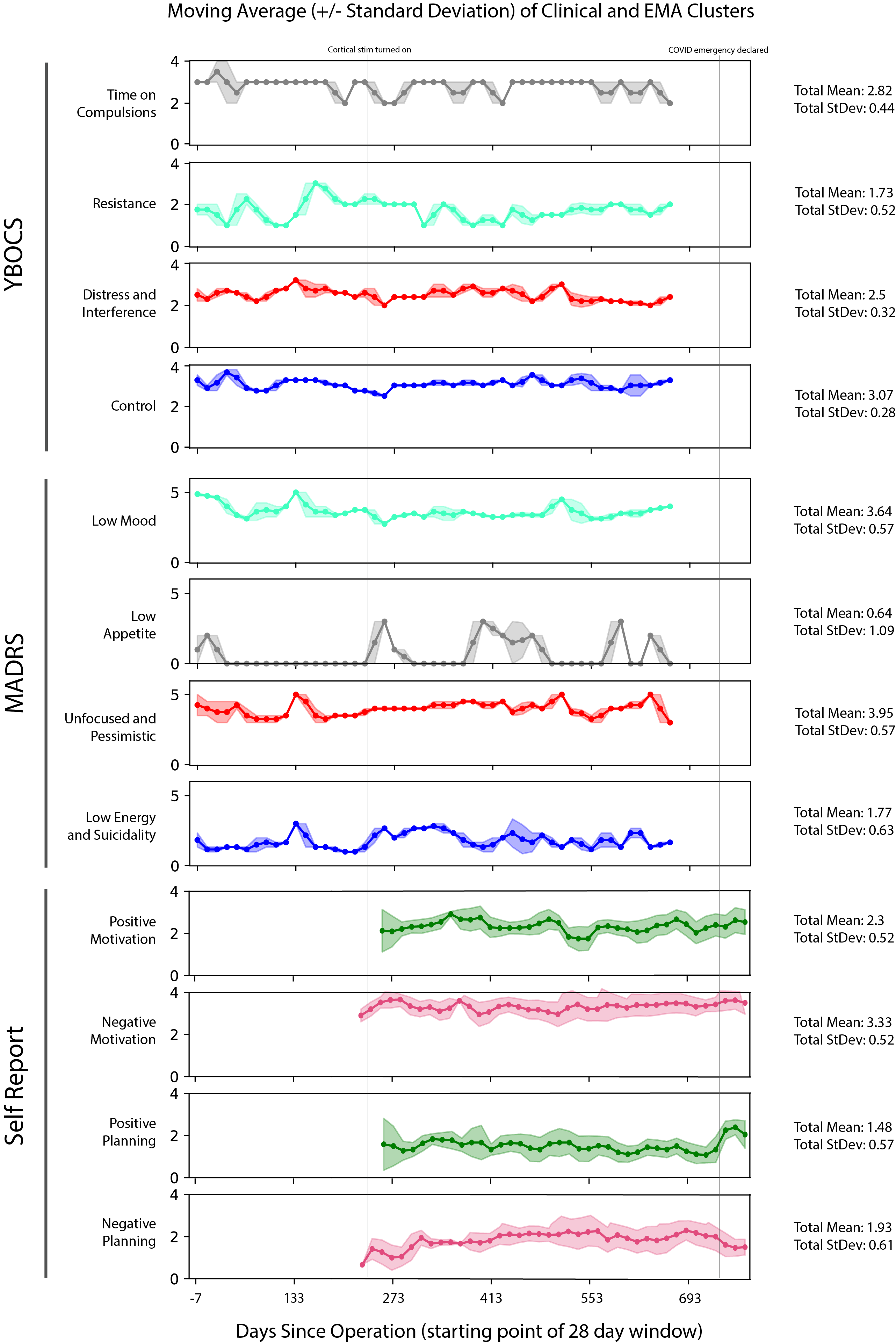}
\caption[Clustering clinical questions reveals structure missed by typical subscale organization.]{\textbf{Clustering clinical questions reveals structure missed by typical subscale organization.} Continuing with the color-coded clusters from Figure \ref{fig:ocd-clusters}, we did a deeper dive into the dynamics of clinical scores over the course of the study, with the goal of characterizing variance and contribution to overall symptom levels for each cluster, as well as comparing clusters to each other at different key points in the study. In order to see such study wide trends, we used smoothed line plots here of moving average with standard deviation error shading. These were computed using a 28 day sliding window with 14 days of overlap; note each window covers a period of real-time days, not the number of data points available. Each data point included before smoothing represented the mean question score for that cluster on that day. The windowing for each cluster covered the exact same days, and thus the plots are temporally aligned. For YBOCS (top), we plot \emph{time} (grey), \emph{resistance} (teal), \emph{interference} (red), and \emph{control} (blue) from top to bottom. For MADRS (middle), we plot \emph{mood} (teal), \emph{hunger} (grey), \emph{productivity} (red), and \emph{energy} (blue) from top to bottom. For self report (bottom), we plot \emph{motivation+exercise} for positive (green) and negative (magenta) branches first, and then \emph{planning} for both branches. The total mean and standard deviation for each category across the study are included to the right of each plot. Vertical lines mark the times that cortical stimulation was turned on and the COVID19 state of emergency was declared.}
\label{fig:ocd-smooth}
\end{FPfigure}

\FloatBarrier

\subsection{Understanding patient experiences through novel data sources}
\label{subsec:ocd-self-report}
Beyond identifying broader trends in symptom severity and effects of stimulation, our dataset paints a richer portrait of the patient’s experience in this study and provides context to certain anomalies seen in the bird’s eye view. This includes content analysis of daily audio diaries, as well as consideration of features from passively collected data at a much finer timescale. In fact, different timescales can be analyzed in a hierarchical way when such dense data is available, enabling a huge variety of questions to be explored. 

\subsubsection{Daily audio diaries provide context for clinical and behavioral measures}
\label{subsubsec:ocd-diaries}
Because our EMA survey questions focused specifically on productivity, and more generally because surveys have limited ability to capture the wide variety of ways in which behavior can change from day to day, we collected audio diaries to provide an entirely new angle to our self-report data. We used an extremely open-ended prompt, with the goal of assessing what the patient was feeling in the moment and what events they considered of particular salience that day. We collected diaries from day 235 post-op through day 792 post-op, allowing for characterization of well over a year of the patient’s life. We utilized these diaries to gain additional insight on abnormalities seen in our other longitudinal data streams, as well as to pinpoint those days where a topic of suspected clinical relevance was discussed. Further, we applied natural language processing (NLP) techniques to extract features from the diaries, allowing deeper analysis of trends seen over the course of the study. 

We first considered patterns of word use from the entire set of diaries submitted, to identify topics that are important to this patient and gain insight into the type of language they generally use to describe their feelings. This patient’s diaries generally follow a similar template each day, as they regularly describe their morning routine and eating habits, and when applicable will discuss any recreational activities from the day. Common themes are playing video games, watching YouTube videos, playing with the dog, religious activities, and going for a walk or bike ride. They also regularly use specific language to describe their symptoms, most notably describing themself as having “low mental energy” on many of the days where they fail to get much done. We later used their own terminology in constructing questions for clinical interviews.

As the participant often spent most of their day at home, the activities described in the diaries can give us information about variance in mobility levels we might otherwise miss. Further, they often recounted going upstairs and downstairs within the house, but some days would mention not leaving the upstairs at all, which serves as a marker of especially high symptom severity days. In addition to common activities of daily life, we also analyzed the diaries for direct mention of doctor visits and various treatments, which is of particular relevance for the evaluation of DBS. Interestingly, the patient frequently described taking Adderall, and this is the only treatment that they repeatedly identified as alleviating their symptoms in the diaries. Discussion of the stimulation paradigm (36 times) was much less frequent than discussion of Adderall (73 times) over the course of the study (Figure \ref{fig:ocd-lang}a, blue/bottom), and it may be associated with worse depressive symptoms (Pearson's $r=0.4502$, $p=0.0464$) (Figure \ref{fig:ocd-lang}d, left), while discussion of Adderall was associated with increased expressiveness in the diaries (Pearson's $r=0.431$, $p<0.0001$) (Figure \ref{fig:ocd-lang}d, right).

\pagebreak

\begin{FPfigure}
\centering
\includegraphics[width=\textwidth,keepaspectratio]{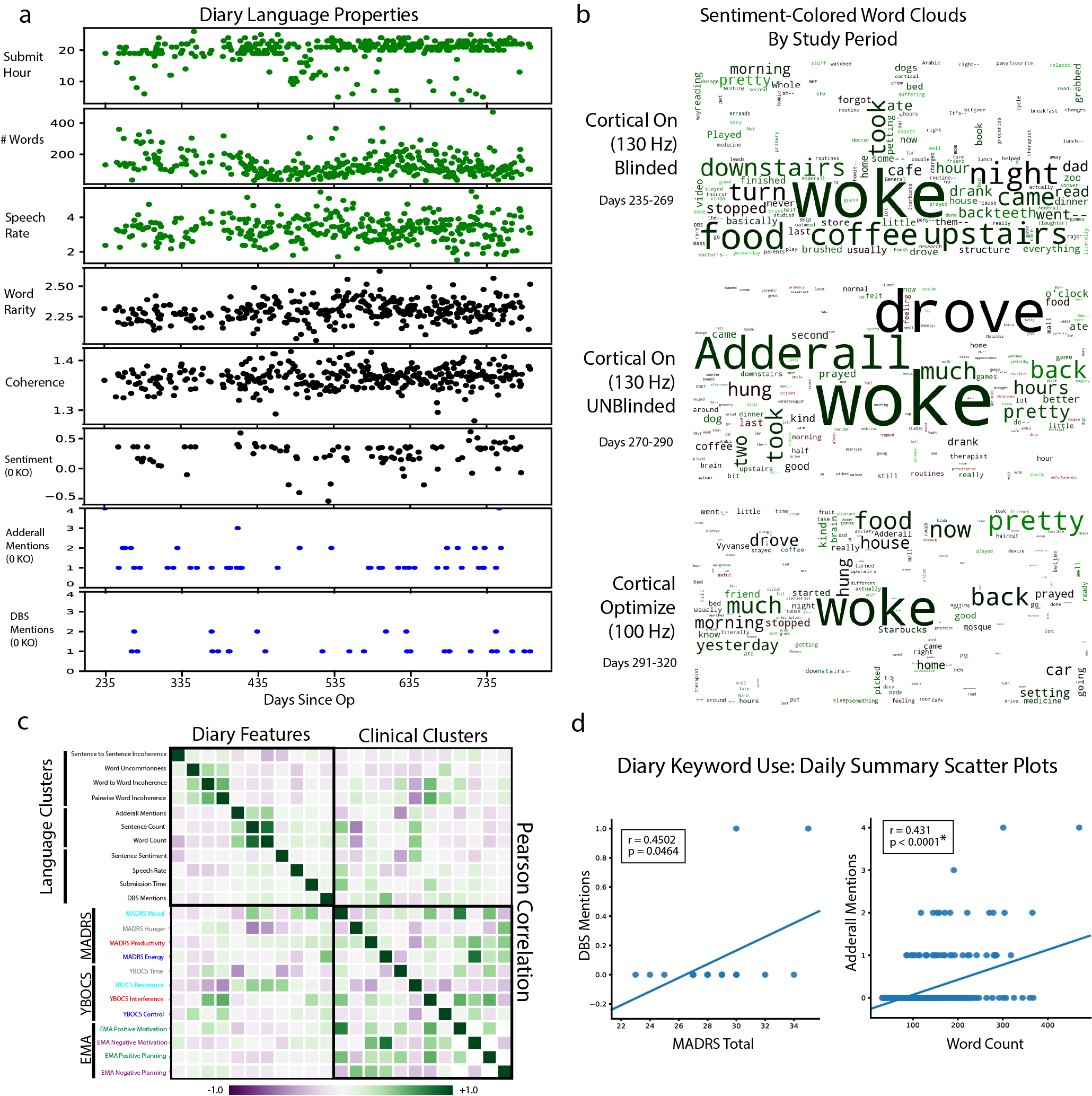}
\caption[Daily audio diaries complement clinical measures.]{\textbf{Daily audio diaries complement clinical measures.} We used daily audio diaries recorded by the patient to both extract quantitative features for correlation analysis and to qualitatively identify recurring topics. Dot plots (a) depict select diary feature values over time (beginning with the first diary at day 235), including basic metadata/accounting features (submission hour, word count, syllables per second; green), natural language processing (word uncommonness, semantic incoherence, sentence sentiment; black), and patient-specific keyword counts (Adderall, DBS; blue). All features besides submission time were computed on the sentence level, and then either summed (counting features) or averaged (others) over the file to get a daily data point. For features with a high concentration of 0 values, they were removed from the depiction here for clarity (denoted by 0 KO on the left). To visualize topics of interest to the patient at different key points in the study, we also generated word clouds (b) from the language used during three timespans - day 235 through 269 (top) when cortical stimulation was on at 130 Hz but the patient was blinded, day 270 through 290 (middle) when stimulation was still at 130 Hz but the patient was unblinded, and day 291 through 320 (bottom) when cortical stimulation frequency was switched to 100 Hz (first month). The size of each word corresponds to how often it was used during the period, while the color corresponds to the average sentiment score of sentences in which the word appeared (with bright red being most negative and bright green most positive). We also looked more directly at the relationship between clinical symptomatology and diary properties, by computing the Pearson correlation between the full set of extracted features and the clinical clusters identified in Figure \ref{fig:ocd-clusters}. The top left block of the matrix (c) depicts the internal correlation structure of diary features, while the off-diagonal shows correlation strength between these features and the average question score of the corresponding clinical cluster. Select relationships involving keyword counts were additionally visualized via scatter plots with linear fit line (d). The total MADRS score was plotted against DBS mentions in same-day diary recordings (left), and the word count from a given diary was plotted against Adderall mentions in that diary (right).}
\label{fig:ocd-lang}
\end{FPfigure}

\FloatBarrier

We next quantified the diaries using computational techniques and visualized the resulting feature progressions over the course of the study (Figure \ref{fig:ocd-lang}a). This included structural properties of the diary (green/top) such as the number of words used and NLP features (black/middle) such as sentence sentiment and coherence. While we were able to identify some trends, no changes mapped to a time of note in the overarching study. Number of words has the most obvious pattern, dropping over time in the first half of the collection period, but with an uptick around day 535 that is sustained for most of the rest of the study. We also noted a few distinct periods where mean diary sentiment was consistently positive or consistently negative. For example, four of the most negatively-worded diaries from across the entire study all appeared within an $\sim 3$ month period between days 435 and 535. Interestingly, the period between days 435 and 535 also stands out as a time of particularly short diary submissions with barely any mentions of Adderall (Figure \ref{fig:ocd-lang}a, blue/bottom).

Due to the variance found in word use and sentiment across the study, we took a deeper look at these properties during three key study periods - diaries submitted during the cortical blinded period (Figure \ref{fig:ocd-lang}b, top), diaries submitted in the period between unblinding and frequency change (Figure \ref{fig:ocd-lang}b, middle), and diaries submitted within the first month after the switch to 100 Hz cortical stimulation (Figure \ref{fig:ocd-lang}b, bottom). It is immediately clear within the word clouds how often the patient mentions Adderall in the diaries directly after unblinding, whereas mentions of Adderall drop considerably when the frequency setting is changed. Meanwhile Vyvanse becomes a more frequent mention during this latter period. Surprisingly, there is little mention of the cortical stimulation in the period directly following unblinding, despite the patient expressing concerns during study visits that they were obsessing over the stimulation parameters. 

Additionally, we compared diary properties across the study periods in the context of behavioral trends previously identified via other modalities. For example, words such as “drove”, and to some extent “prayed”, increased in usage directly after unblinding (Figure \ref{fig:ocd-lang}b, middle). This is consistent with the increased time spent away from home and at a religious institution during the same time frame (Figure \ref{fig:ocd-report-stim}). There was also a decrease during this period in use of “upstairs” and “downstairs”, further supporting the conclusion that the patient was getting out of the house more often after unblinding. In contrast, they appeared more talkative in the diaries prior to unblinding, with higher sentiment scores as indicated by the greener coloring in that word cloud (Figure \ref{fig:ocd-lang}b, top).

Finally, to quantify any relationships between clinical outcomes and daily audio diaries, we correlated the diary language features (Figure \ref{fig:ocd-lang}a) with clusters derived from MADRS/YBOCS and EMA (Figure \ref{fig:ocd-clusters}a) to generate a Pearson correlation matrix (Figure \ref{fig:ocd-lang}c). While we found no statistically significant correlations after multiple testing correction, there were some promising relationships between audio diaries and clinical clusters that could warrant further investigation. For example, the median sentence incoherence had a promising potential relationship with same day YBOCS \emph{interference} score ($r=0.598$, $p=0.005$), suggesting that a less coherent diary could serve as a marker for the current level of impact OCD symptoms are having on the patient’s day to day life - moreso than features like diary length or estimated sentiment.  

\FloatBarrier

\subsubsection{Using passive behavioral data at different timescales}
\label{subsubsec:ocd-timescales}
Passively collected phone usage, phone accelerometer, GPS, and wrist actigraphy data can all provide information about behavior on the minute-by-minute level, in stark contrast to the timescales possible to capture with actively collected data types. In addition to serving as rich data sources that can be utilized to uncover trends in patient behavior at multiple timescales, these modalities also provide context about typical day to day life for the patient. We therefore analyzed our dataset to uncover details of our patient’s routines, including characterizing sleep behavior and identifying activities like bike rides that were important to this patient. We used this approach to better understand their behavior on days of particular interest to the study timeline, as well as to dig into anomalies seen at a broader timescale.  

Here we focus on minute-level digital phenotyping features. We calculated the standard deviation of the phone accelerometer signal within each minute, as this is one of the most commonly used accelerometry feature to estimate overall movement intensity \citep{Picard2017,Smets2018,Teo2019}. We also determined whether the majority of GPS datapoints in a given minute were at the same coordinates as the patient’s home, and calculated the mean phone battery level in a given minute. Note that battery data is only available during minutes where the phone was in use, increasing the utility of the datatype. 

\begin{figure}[h]
\centering
\includegraphics[width=0.33\textwidth,keepaspectratio]{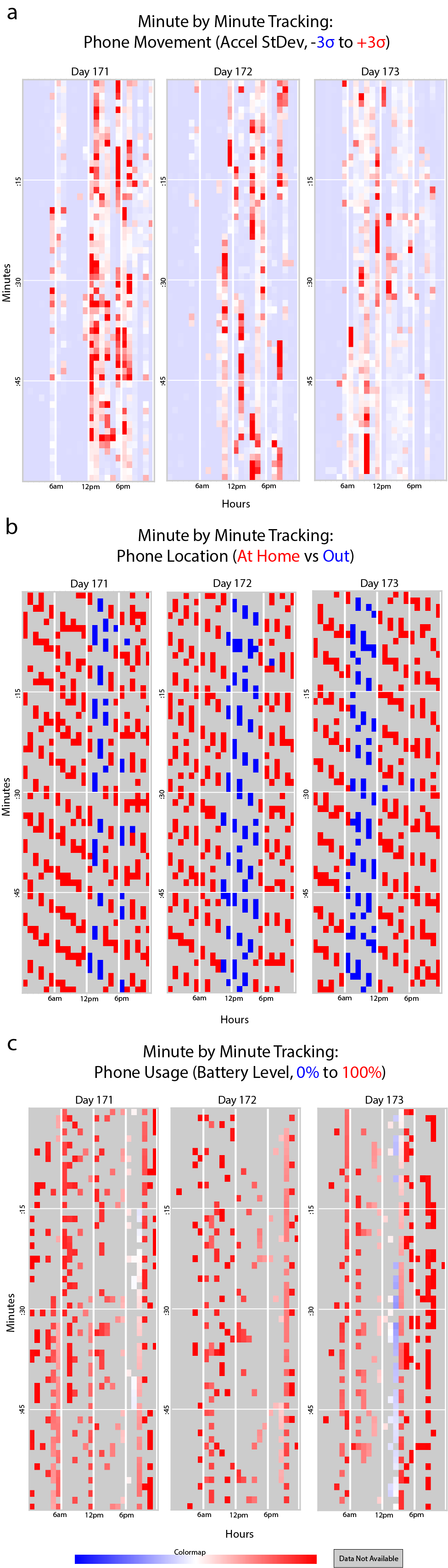}
\caption[Detecting late night waking via brief phone use periods.]{\textbf{Detecting late night waking via brief phone use periods.} As in Figure \ref{fig:ocd-zoom}, we also created minute by hour heatmaps for days 171 (left) through 173 (right) post-op, with 172 (middle) being the first day the cortical stimulation theoretically could have been turned on. This period was chosen over other DBS study transition periods due to good passive data availability. Here we organize the figure panels by phone movement (a), location (b), and usage (c), with each panel containing that datatype for the entire period. Late night phone usage (c; non-grey squares in the early morning hours) is particularly apparent in this view.}
\label{fig:ocd-zoom-sup1}
\end{figure}

\begin{figure}[h]
\centering
\includegraphics[width=0.5\textwidth,keepaspectratio]{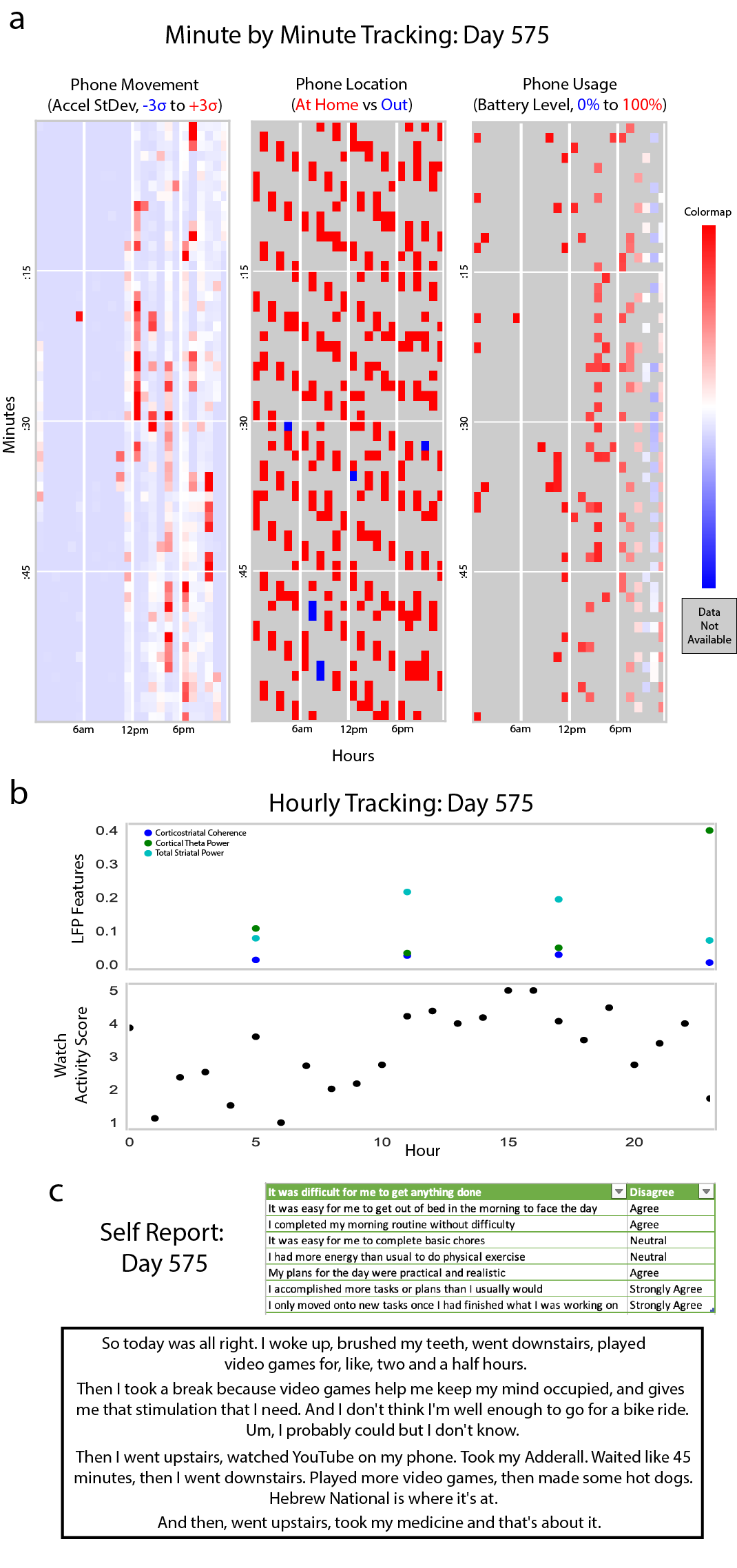}
\caption[The benefits of multimodal digital phenotyping.]{\textbf{Looking for behaviors across different digital phenotyping modalities.} As in Figure \ref{fig:ocd-zoom}, we also created minute by hour heatmaps (a) for day 575 post-op, as this was one of the only study days with full data availability across most modalities. We compare the minute-by-minute phone features (a) with metrics from the same day capturing implant LFP and wrist actigraphy properties (b), as well as self-reported feelings (c). For LFP (top), corticostriatal coherence (blue), cortical theta power (green), and total striatal power (teal) features are plotted at the four time points that recordings were taken that day. For wrist actigraphy (bottom), the hourly activity score is plotted over time, aligned with the LFP features. For self report, we reproduce the patient's response to each EMA question (top) and the transcript of their recorded audio diary (bottom) from that day.}
\label{fig:ocd-zoom-sup2}
\end{figure}

These three minute-wise features were computed throughout the day on a handful of days of particular interest, chosen due to a combination of data availability and relevance within the overall study timeline. We analyzed days 171 through 173 post-op (Figure \ref{fig:ocd-zoom-sup1}), with 172 being the first day of the cortical crossover period; day 575 post-op (Figure \ref{fig:ocd-zoom-sup2}), which offered full availability of every modality considered during this study; and days 869 and 870 post-op (Figure \ref{fig:ocd-zoom}), covering the end of the study including the COVID-19 pandemic. 

\pagebreak

\begin{FPfigure}
\centering
\includegraphics[width=0.6\textwidth,keepaspectratio]{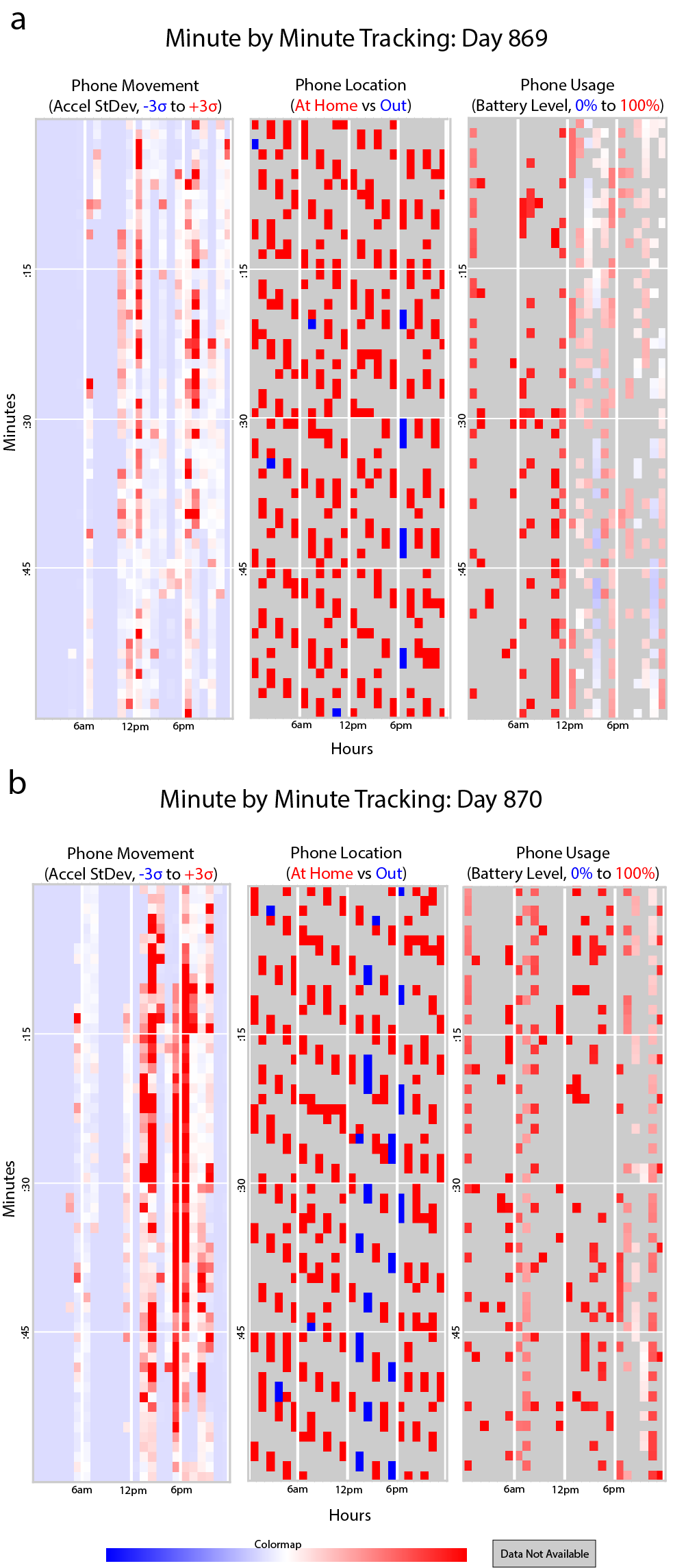}
\caption[Passive digital phenotyping measures densely fill temporal gaps of actively collected data.]{\textbf{Zooming in further - passive digital phenotyping measures fill in temporal gaps of actively collected data.} Here we demonstrate the ability of passively collected phone data to focus more closely on details of a single patient day or even hour. We chose to zoom in on days 869 (a) and 870 (b) post-op, a period at the end of the study in summer 2020, in part because it coincided with the COVID19 pandemic. However, such a detailed view could theoretically be taken of any day in the study. In each panel, we show minute (row) by hour (column) heatmaps of phone movement (left), location (middle), and usage (right) for that day. All heatmaps use grey to indicate when data is unavailable. The total movement score for each minute is colored relative to the mean ($+/- 3$ standard deviations) for this same feature from this patient over the course of the study, using the bwr matplotlib colormap. The location measurement is binary, indicating time at home (red) versus time out of the house (blue) as the patient’s primary location in any given minute. The usage metric colors each minute based on the battery status of the patient’s phone at that time, with $0\%$ to $100\%$ bounds on a bwr colormap. Note battery level is only reported when a phone usage event occurs, so data missingness is itself a signal within the battery data.}
\label{fig:ocd-zoom}
\end{FPfigure}

\FloatBarrier

We not only illustrate the sheer level of detail that can be accomplished with our techniques, but also demonstrate identification of notable behaviors that would likely have been missed if looking at only a broader timescale. In the GPS signal (Figure \ref{fig:ocd-zoom}, middle panels) we identified times when the patient left the house for 30 minutes or less. While the patient discussed walking their dog and going for bike rides around their neighborhood as important activities that they made an effort to engage in on good days, these activities would involve small changes to GPS signal with relatively short duration - making them difficult to detect in general measurements of daily activity level. Yet by considering “zoomed in” behavioral data, the same activities became much more apparent. Further, we leveraged the multimodal nature of our dataset to improve the accuracy of our inferences. For example, intensity of phone movement can distinguish between the aforementioned neighborhood walks and bike rides identified via GPS data. The patient’s phone accelerometer signal on day 869 post-op (Figure \ref{fig:ocd-zoom}a, left) was significantly less intense during the times the patient was out of the house than their phone accelerometer signal while out of the house was on day 870 post-op (Figure \ref{fig:ocd-zoom}b, left), suggesting they took a walk on the first day and went for a bike ride on the second one. 

Multimodal inference can extend to the use of active datatypes such as audio diaries and EMA, to corroborate the presence of activities predicted by digital phenotyping. As described above, we detected an $\sim30$ minute walk beginning shortly after 6:15 pm on day 869 post-op (Figure \ref{fig:ocd-zoom}a), and an $\sim1$ hour bike ride beginning shortly before 5:30 pm on day 870 post-op (Figure \ref{fig:ocd-zoom}b), as well as a shorter bike ride in the mid-afternoon that same day. These predicted labels were based not only on intensity of accelerometer movement and simultaneous location tracking, but also on activities known to be a part of the patient’s habits. Unfortunately, we do not have self-report data from the end of the study, but bike rides were repeatedly mentioned in earlier journals. For example, in the audio diary submitted on day 575 (Figure \ref{fig:ocd-zoom-sup2}c), the patient said that they did not feel well enough to go for a bike ride that day. Interestingly, the EMA survey the patient submitted on day 575 reported abnormally good productivity levels despite skipping their exercise routine, with an average question score of 1 compared to the average over all EMA submissions of 2.42 (standard deviation 0.59). 

An even more striking example of a short duration but clinically important behavior captured by our multiple timescales approach is our patient’s habit of unlocking their phone when they awake in the middle of the night (Figure \ref{fig:ocd-zoom-sup1}c). We have captured that throughout the study the patient would check their phone multiple times during periods of supposed sleep. While the phone use was brief more often than not, it demonstrated a recurring problem of sleep disruption. On some days this behavior was observed multiple times an hour for nearly the entire night, including day 171 post-op (Figure \ref{fig:ocd-zoom-sup1}c, left). The behavior was also consistent with increased movement levels in our watch actigraphy data. Just a single minute of phone checking around 5:15 am on day 575 post-op (Figure \ref{fig:ocd-zoom-sup2}a, right) aligned with a watch activity score near day time levels for the 5 o’clock hour that day (Figure \ref{fig:ocd-zoom-sup2}b, bottom), even while phone movement was low (Figure \ref{fig:ocd-zoom-sup2}a, left).  

\FloatBarrier

\subsection{Identifying correlation structures across datatypes}
\label{subsec:ocd-corr}

Because digital phenotyping provides a copious amount of behavioral data, it can be a challenge to analyze across timescales and measurement modalities in an interpretable and digestible way. To make best use of these new datatypes, we must understand not only what information is provided by individual features of the data, but also how different features interact with each other. 

Correlation structure comprises pairwise correlation of a set of features. From the correlation structure, we can generate a smaller feature set that still captures much of the signal. We previously employed this approach to cluster YBOCS, MADRS, and EMA questions (Figure \ref{fig:ocd-clusters}a). With a larger dataset, this methodology increases in power. Further, correlation structures can also directly contain information of interest, as the cooccurrence of two features can be a more meaningful signal than the value of either feature alone. To identify salient digital phenotyping features and better characterize relationships between these features and neural activity, we examined the correlation structures both within and between our multiple datasets – including digital phenotyping, clinical scales, EMA, and neural recordings. Thus we provide a proof of concept for the use of correlation structure in evaluating patients on an individual level, an approach that will likely be critical due to the heterogeneity in relationships between behavior and clinical outcomes.  

\subsubsection{Relationships between behavioral data and existing clinical measures}
\label{subsubsec:ocd-behav-corr}

\begin{figure}[h]
\centering
\includegraphics[width=\textwidth,keepaspectratio]{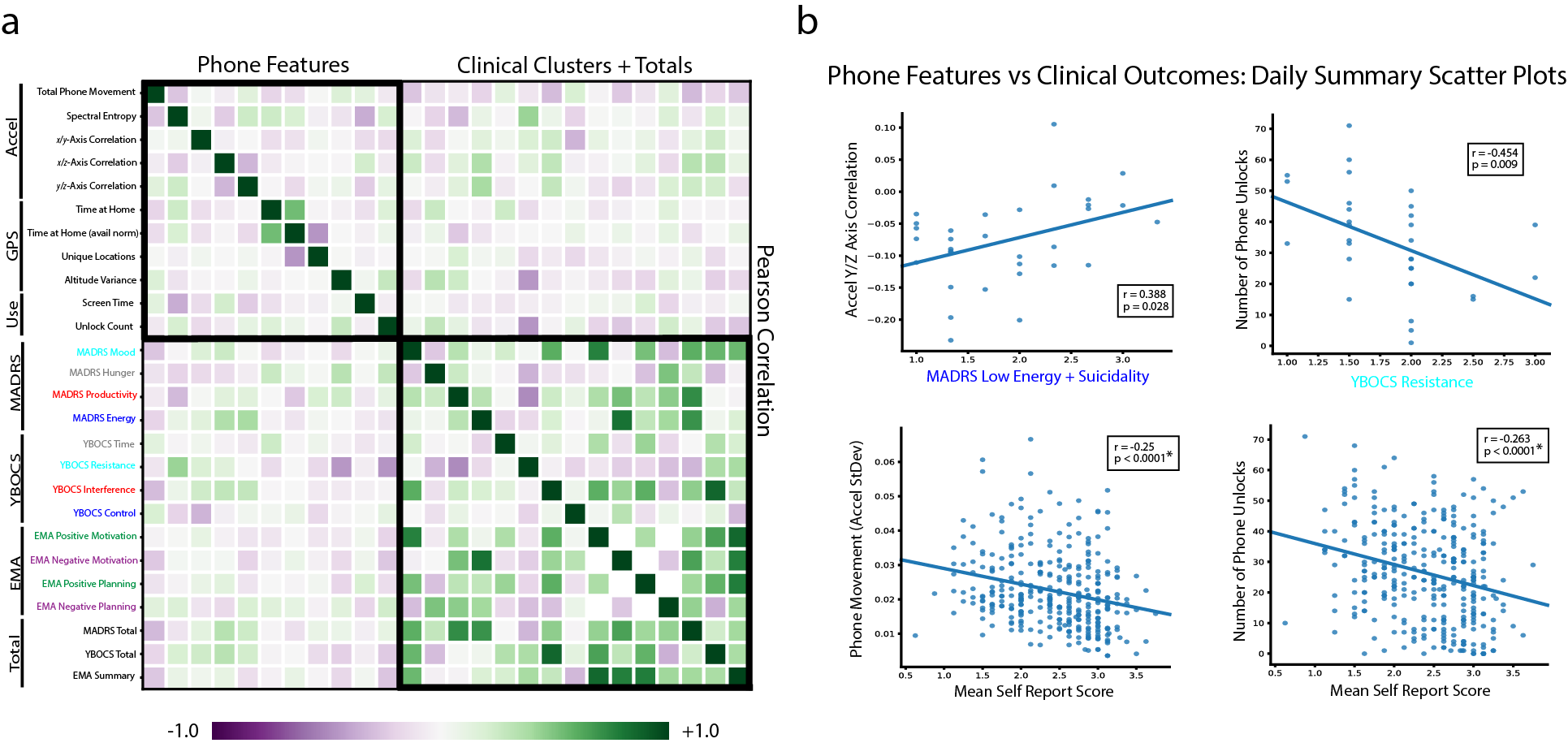}
\caption[Relationship between digital phenotyping results and clinical outcomes.]{\textbf{Relationship between digital phenotyping results and clinical outcomes.} To better understand the passively collected phone data, we analyzed the internal correlation structure of a handful of daily features for each phone modality, as well as their relation with the clinical clusters found in Figure \ref{fig:ocd-clusters}. The resulting Pearson correlation matrix (a) is colored using the matplotlib PrGn colormap with bounds -1 to 1. The phone features (top left) and clinical clusters (bottom right) are set apart with black boxes, while the outside correlations on the off-diagonal represent relationships between the digital phenotyping and the symptomatology. Select pairings of phone features with clinical clusters were additionally visualized via scatter plots with linear fit line (b); $r$ and $p$ values for the Pearson correlation are given within each plot. We considered MADRS \emph{energy} cluster score versus same-day accelerometer \emph{y/z}-axis correlation (top left), YBOCS \emph{resistance} cluster score versus same-day phone unlocks count (top right), and EMA summary score (bottom) versus both same-day total phone movement (left) and same-day phone unlocks count (right).}
\label{fig:ocd-dp-corrs}
\end{figure}

To determine correlation structure for the digital phenotyping features, we selected a handful of daily features from each of the phone modalities (usage, location, and accelerometry). We selected features based on prior use in the device literature \citep{Picard2017,Smets2018,Teo2019}. We then calculated the Pearson correlation between each of the selected digital phenotyping features, as well as between these features and the previously identified clusters of YBOCS, MADRS, and EMA questions (Figure \ref{fig:ocd-clusters}a). To assess whether digital phenotyping features should be treated independently or as clusters, we evaluated correlation structure amongst these features first. We found no significant correlation between a plurality of the selected phone features in this subject, particularly when comparing features from different phone sensors (Figure \ref{fig:ocd-dp-corrs}a, diagonal blocks). Across modalities, the few relationships that did survive Bonferroni correction included a slight negative correlation between time spent at home and overall phone movement (Pearson's $r=-0.1802$, $p<0.0001$), and a moderate negative correlation between phone use time and accelerometer spectral entropy (Pearson's $r=-0.3633$, $p<10^{-20}$). To capture potential non-linear effects, we also computed the analogous Spearman rank correlation matrix. However, the Spearman correlations largely followed the Pearson.

We then evaluated the correlation structure between individual digital phenotyping features and clusters of YBOCS, MADRS, and EMA questions (Figure \ref{fig:ocd-dp-corrs}a, off-diagonal blocks). A handful of these feature pairings exhibited strong relationships of interest. In particular, the MADRS \emph{energy} cluster was positively correlated with phone \emph{y}-axis/\emph{z}-axis correlation (Pearson's $r=0.3883$, $p=0.0281$), and the YBOCS \emph{resistance} cluster was negatively correlated with number of phone unlocks (Pearson's $r=-0.4545$, $p=0.009$). We also saw a moderate negative correlation between total EMA summary score and both overall phone movement (Pearson's $r=-0.2506$, $p<0.0001$) and number of phone unlocks (Pearson's $r=-0.2632$, $p<0.0001$). Thus we more closely evaluated the data points forming each relationship via scatter plots with linear fit lines (Figure \ref{fig:ocd-dp-corrs}b). While only the correlations between digital phenotyping features and EMA were statistically significant after Bonferroni multiple testing correction (Figure \ref{fig:ocd-dp-corrs}b, bottom), the correlations of phone features with YBOCS and MADRS clusters were promising given the many fewer data points collected from those scales (Figure \ref{fig:ocd-dp-corrs}b, top). Once again, repeating the analysis with Spearman rank correlation yielded similar results.

\FloatBarrier

\subsubsection{Neural correlates of clinical outcomes and digital biomarkers}
\label{subsubsec:ocd-neuro-corr}
We next investigated the relationship between neural activity in the ventral striatum and supplementary motor area, assessed via local field potential (LFP) recordings, and digital phenotyping features. We additionally compared neural activity with the identified YBOCS, MADRS, and EMA clusters (Figure \ref{fig:ocd-clusters}a), to reassess the low correlation between LFP features and clinical outcomes in this patient that was previously reported \citep{Olsen2020}. To obtain the LFP features, we computed (using the same methodology as \cite{Olsen2020}) power in the gamma, beta, alpha, and theta bands from both striatal and cortical leads, as well as corticostriatal coherence because of the focus of this DBS trial on hypersynchrony. Computed features represent the arithmetic mean between the right and left hemispheres.

\begin{figure}[h]
\centering
\includegraphics[width=\textwidth,keepaspectratio]{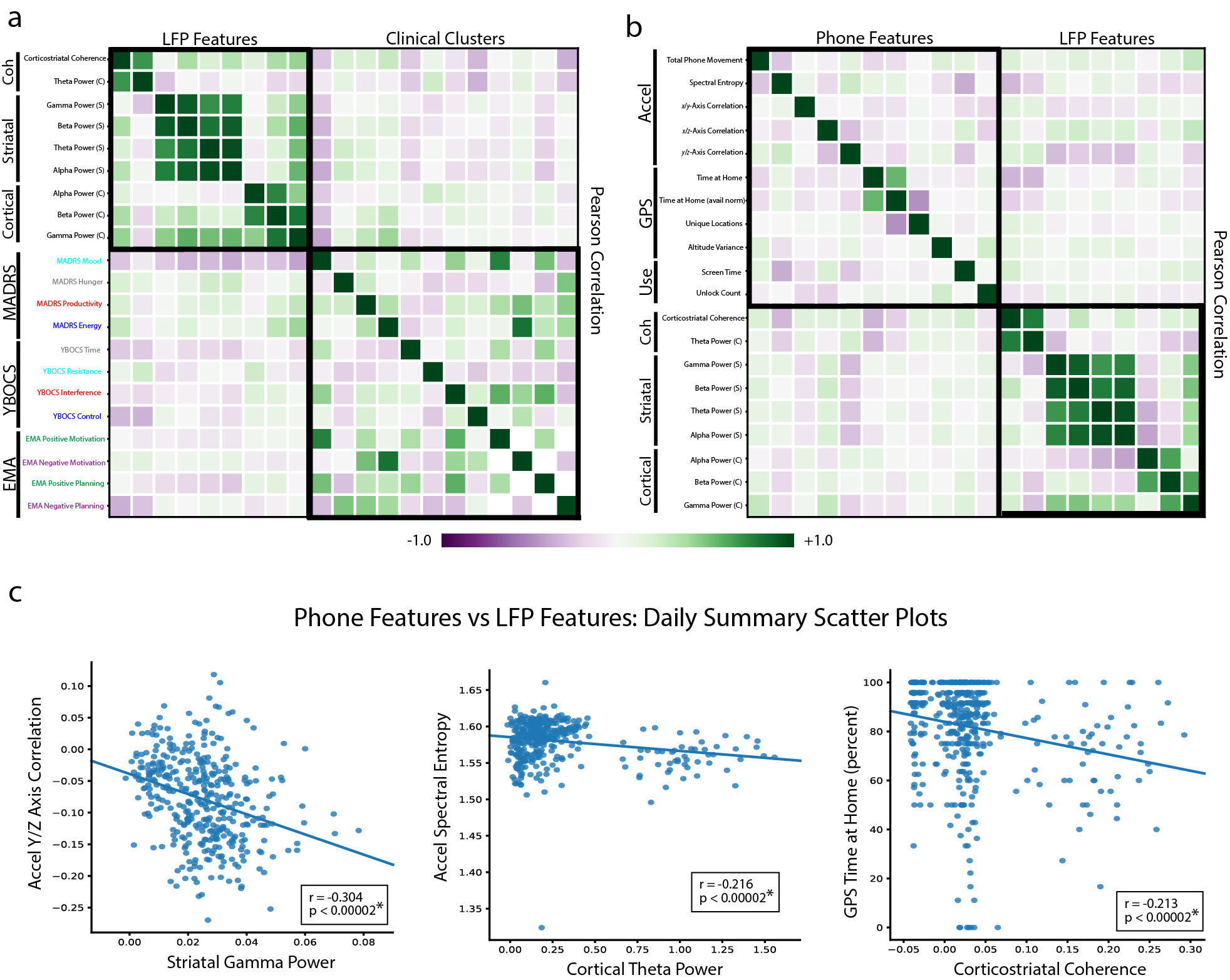}
\caption[Neural correlates of digital biomarkers.]{\textbf{Neural correlates of digital biomarkers.} Analogous to the correlation analysis depicted in Figure \ref{fig:ocd-dp-corrs}a, we investigated the relationship between daily neural recording features and our clinical clusters. The resulting Pearson correlation matrix (a) has the LFP features (top left) and the clinical clusters (bottom right) set apart with black boxes, with the LFP features ordered according to their internal clustering. The outside correlations on the off-diagonal then represent relationships between the neural recordings and the symptomatology. At the same time, we also looked for potential correlations between the digital phenotyping features described in Figure \ref{fig:ocd-dp-corrs} and the LFP features. That Pearson correlation matrix (b) is laid out similarly, with phone features on the top left block diagonal and LFP features on the bottom right. Select pairings of phone features with LFP features were then visualized via scatter plots (c). Here we considered striatal gamma power against accelerometer \emph{y/z}-axis correlation (left), cortical theta power against accelerometer spectral entropy (middle), and corticostriatal coherence against time spent at home (bottom).}
\label{fig:ocd-neuro-corrs}
\end{figure}

Correlation structure amongst the LFP features was very strong (Figure \ref{fig:ocd-neuro-corrs}a, diagonal blocks), in sharp contrast to what we observed with the digital phenotyping features (Figure \ref{fig:ocd-dp-corrs}a, diagonal blocks). All four striatal power bands were highly correlated with each other ($0.688 \leq r \leq 0.947$, all $p < 10^{-60}$), as was the cortical beta power band with both cortical alpha and gamma (both $r \sim 0.64$, $p < 10^{-50}$). Cortical theta and corticostriatal coherence were also highly correlated in the Pearson matrix ($r=0.7746$, $p<10^{-90}$), however these two features exhibited low correlation when computing the Spearman rank correlation instead. All other relationships remained similar between the Pearson and Spearman matrices, indicating a largely linear correspondence between LFP features.

We moved on to evaluate the correlation structure between LFP features and the YBOCS, MADRS, and EMA clusters (Figure \ref{fig:ocd-neuro-corrs}a, off-diagonal blocks). However, we observed minimal correlation here, supporting the conclusions already drawn from the analysis using total clinical scale scores \citep{Olsen2020}. The only relationship that survived Bonferroni correction was that between the \emph{planning} cluster of the negative EMA branch and corticostriatal coherence (Pearson's $r=-0.358$, $p<0.0001$). A negative correlation was also found between corticostriatal coherence and the YBOCS \emph{control} cluster (Pearson's $r=-0.32$, $p=0.0474$), but the smaller sample size of clinical scale scores prevents drawing any real conclusions. It is worth noting that the MADRS \emph{mood} cluster had uncorrected significant negative correlation with a few LFP features as well - striatal beta, striatal alpha, and cortical gamma ($-0.326 \geq r \geq -0.4$, all $p < 0.05$).

To capture potential relationships between behavior and neural activity that were not captured by the clinical scales nor EMA, we evaluated the correlation structure between LFP features and digital phenotyping features (Figure \ref{fig:ocd-neuro-corrs}b, off-diagonal blocks). A few phone location and accelerometer features, most notably the accel spectral entropy and the percent of time spent at home, demonstrated a strong negative correlation with both corticostriatal coherence and cortical theta power ($-0.216 \geq r \geq -0.335$, all $p < 0.00002$). The statistically significant negative relationship between time spent at home and coherence is particularly interesting, as it contradicts the expectation that high corticostriatal coherence is associated with worse symptom severity. However, upon closer inspection, these digital phenotyping/LFP relationships appear to be driven purely by a bimodal distribution (Figure \ref{fig:ocd-neuro-corrs}c, middle/right). As it turns out, the data points with highest theta power and corticostriatal coherence arose when the cortical stimulation was set at 130 Hz, a relatively short period of the study that coincided with unblinding. Thus it is difficult to draw any conclusions about corticostriatal coherence correlations.

In contrast, a few phone accelerometer features exhibited a strong negative correlation with the striatal power bands, which were not as affected by the frequency change as the cortical features. We therefore selected the strongest of these feature pairings – those that were statistically significant (after Bonferroni correction) with the highest magnitude $r$ values – to look more closely at the data points forming each relationship. Striatal gamma power versus the correlation of \emph{y} and \emph{z} axes from phone accelerometry demonstrated the most convincing relationship ($r=-0.304$, $p<10^{-8}$), with an evenly distributed scatter plot (Figure \ref{fig:ocd-neuro-corrs}c, left). Recall that this same \emph{y}/\emph{z}-axis correlation feature displayed a positive relationship with the MADRS \emph{energy} cluster (Figure \ref{fig:ocd-dp-corrs}b).

\FloatBarrier

\subsection{Effects of stimulation status on patient behavior}
\label{subsec:ocd-causal}
Because correlation structure cannot capture causal relationships between cortical stimulation and patient behavior, we conducted eight recorded short interviews – four with stimulation off and four with stimulation on. The interview questions assessed the patient’s perception of current stimulation status, as well as their current anxiety and energy levels. We also used patient-initiated discussion during the periods between structured interview bouts to further evaluate trial-by-trial behavior. Notably, the patient was incorrect about current stimulation status on 7 of the 8 trials (Figure \ref{fig:ocd-video}a, top), suggesting that they might have been able to detect something different between the conditions. 

Human observation of the interviews during this experiment identified a few potential behavioral changes between trials, including increased talking when cortical stimulation was off (\emph{off-trials}), and acting tired much of the time it was on (\emph{on-trials}). To objectively validate these observations and look for more subtle behavioral changes, we extracted facial action units (FAUs) from each frame of the video of the patient. Total FAU activity, a measure of facial expressivity, decreased over the course of the experiment; trials also became shorter (Figure \ref{fig:ocd-video}a). These changes over time are consistent with the patient’s subjective expressions of impatience during later trials. 

\pagebreak

\begin{FPfigure}
\centering
\includegraphics[width=\textwidth,keepaspectratio]{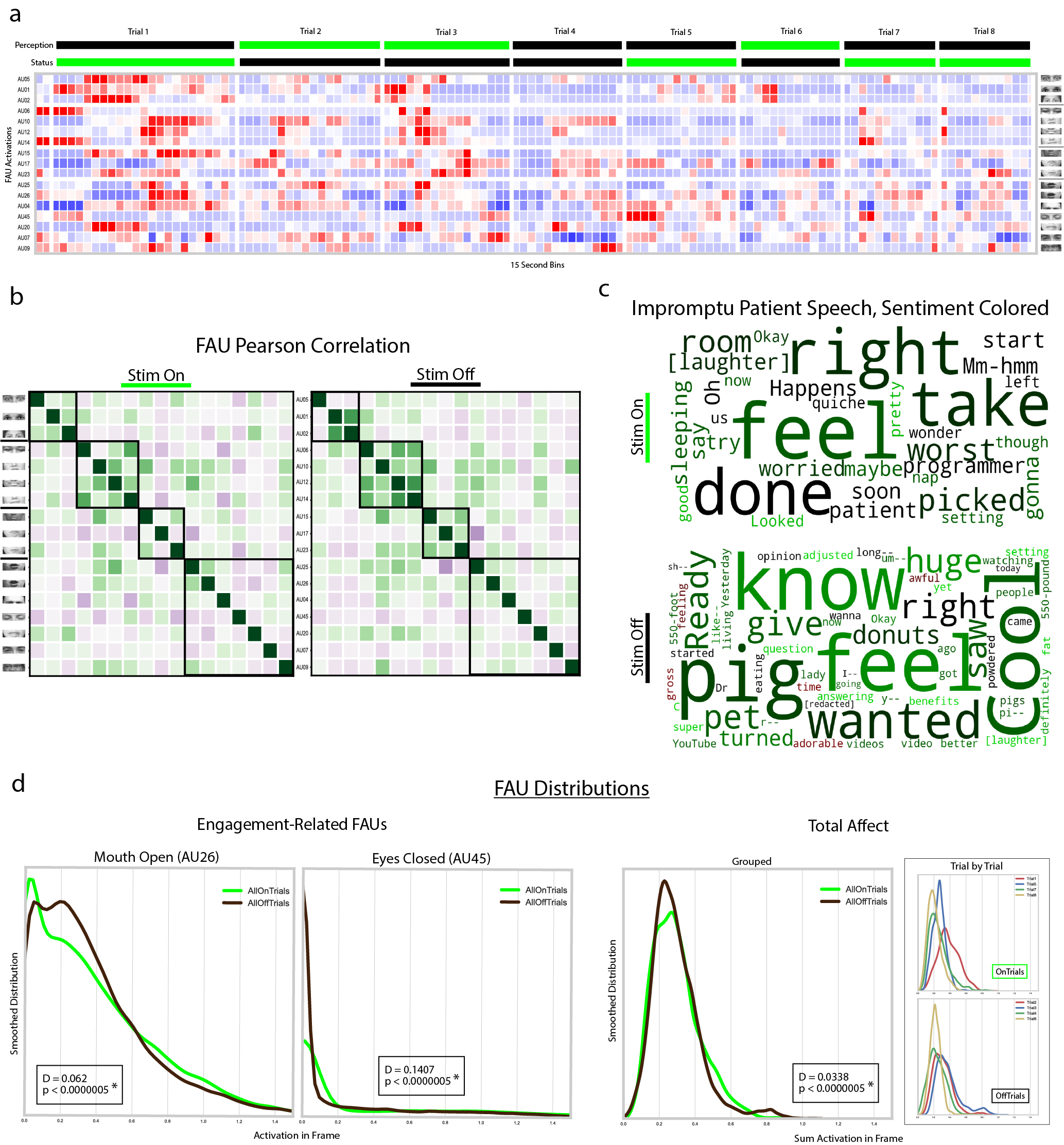}
\caption[Short-term changes in cortical stimulation status suggest improved energy when stimulation is turned off.]{\textbf{Short-term changes in cortical stimulation status suggest improved energy when stimulation is turned off.} To causally probe the effects of cortical stimulation, we conducted a recorded experimental interview consisting of 8 trials with stimulation status blinded to both interviewer and patient. As part of each trial, the patient was asked to predict whether stimulation was currently on or off (a, top; green = on, black = off). Using the video recording, we extracted facial action units, and visualized their progression over the course of the experiment (a, bottom). In this heatmap, each row is an FAU and each column is a 15 second segment of the recording. The color of each cell represents the mean activation of that AU during that time period, colored using the bwr matplotlib colormap bounded by the distribution of the AU from this interview. Thick vertical white lines set apart the trials, and the rows are ordered according to FAU clustering results from a larger video dataset \citep{Vijay2016}. To compare the existing AU clusters with this patient’s correlation structure during this interview, we also generated Pearson correlation matrices (b) independently from the frame-by-frame activations of \emph{on-trials} (left) and \emph{off-trials} (right). The AUs in these matrices are ordered as in the heatmap (a), with black boxes denoting the broader clusters. $r$ values from this interview are colored using the matplotlib PrGn colormap with bounds -1 to 1. We then focused in on a select few FAUs of interest, creating kernel density estimate plots (d) to contrast frame-wise distributions between \emph{on-trials} (green) and \emph{off-trials} (black). This included AUs for mouth opening (left), eye closing (middle), and a total affect score summing emotion-related AUs (right). KS-test results are reported for each plot. For total affect, trial by trial distributions (right inset) were also plotted separately for the on (top) and off (bottom) conditions. In addition to video, we utilized the transcription of the recorded experiment to assess potential differences in language between the conditions. Of particular interest is patient speech outside the scope of the coded interview questions; here we show sentiment-colored word clouds depicting impromptu speech (c) during the \emph{on-trials} (top) and \emph{off-trials} (bottom) respectively.}
\label{fig:ocd-video}
\end{FPfigure}

\FloatBarrier

Prior studies have demonstrated an increase of overall facial affect in response to striatal deep brain stimulation for OCD \citep{Goodman}. We therefore also compared the sum of activations from all FAUs corresponding to facial affect (as identified by \cite{Goodman}) between \emph{on-trials} and \emph{off-trials} for the cortical DBS paradigm. We found minimal relationship between facial affect and cortical stimulation status in this patient though, with only small magnitude differences seen when comparing the AU activation distributions from the two conditions (Figure \ref{fig:ocd-video}d, right). Furthermore, we did not observe anything of note that differed between Trial 4 and the other \emph{off-trials} (inset, bottom; dark green curve versus others), despite the fact that the patient correctly identified stimulation as off during Trial 4 alone. 

However, by focusing on the specific FAUs corresponding to mouth opening (AU26) and eye closing (AU45), we indeed could corroborate the human observation of behavioral differences between \emph{on-trials} and \emph{off-trials} (Figure \ref{fig:ocd-video}d). The patient activated the mouth open AU substantially more often during the \emph{off-trials} than during the \emph{on-trials}, with distributional peaks at approximately $0.2$ activation and less than $0.05$ activation respectively (Figure \ref{fig:ocd-video}d, left). Conversely, the patient activated the eyes closed AU substantially more often during the \emph{on-trials} than during the \emph{off-trials}, with respective activation distribution RMS widths of approximately $0.1$ and less than $0.05$ when modeled as half-normal (Figure \ref{fig:ocd-video}d, middle). For both of these comparisons, a two-sample Kolmogorov-Smirnov (KS) test indicated statistically significant distributional differences (both $D>0.06$, $p<0.0000005$).

It is worth noting that magnitude of AU activation can have relevance in feature interpretation. In the case of the mouth open action unit (AU26), a moderate activation level is consistent with speaking, while a high activation level is actually more consistent with yelling or yawning. We saw a robust effect of stimulation status on AU26 around activation levels of $\sim 0.2$, as described above (Figure \ref{fig:ocd-video}d, left). In the same distribution though, we also saw a slight effect in the opposite direction – that the patient had activation levels greater than $0.75$ more often during \emph{on-trials}, rather than during \emph{off-trials}. This might suggest that yawning was detectable more often when cortical stimulation was on, which would be consistent with observer report. \\

In addition to looking at patterns in individual action units over the course of the trials, we also evaluated the Pearson correlation structure between different FAUs across frames, in \emph{on-trials} versus \emph{off-trials} (Figure \ref{fig:ocd-video}b). This approach was inspired by the use of functional connectivity networks in fMRI literature \citep{Rogers2007}, as well as by evidence that specific combinations of AUs are useful for characterizing meaningfully different expressions - most famously the Duchenne smile as an indicator for genuine enjoyment \citep{Frank1993}. Here, we grouped the action units in the correlation matrix based on clustering results from a larger and more diverse set of 42 clinical interview videos, containing different subjects and both interviewer and interviewee faces \citep{Vijay2016}. We then used our groupings to provide a frame of reference for any differences seen between stimulation conditions. 

When cortical stimulation was off, the 3 main action unit clusters (Figure \ref{fig:ocd-video}b, right, top three diagonal blocks) largely were correlated within this patient. The major exception was AU5 (upper lid raiser), which had no relationship with AU1 or AU2 here (both brow raising units) as it had in the other dataset. When cortical stimulation was on, the 3 main action unit clusters (Figure \ref{fig:ocd-video}b, left, top three diagonal blocks) all had weaker internal structure than they did during \emph{off-trials}, while several correlations from across clusters (off-diagonals) became more strongly negative (Figure \ref{fig:ocd-video}b). 

By visualizing the time course of each FAU’s activation over the course of the trials, grouped into the same clusters, we can pinpoint when an FAU “network” was active (Figure \ref{fig:ocd-video}a, bottom). For example, there was clear simultaneous activation of FAUs in the third cluster (containing features thought to relate to sadness, anger, disgust) during Trial 3. Meanwhile, the FAUs in the third cluster activated independently during Trial 1. There is great potential in being able to identify these moments for further evaluation – is the context of the interview different in these two cases? How much of the difference in facial expression structure might we directly ascribe to stimulation status? To address such questions, one can utilize both expert opinion, as well as behavioral data collected in other modalities. \\

Thus for our proof of concept work here, we went beyond video analysis to also consider language. Specifically, we transcribed the interview audio and then applied natural language processing tools, as was done with the patient audio diaries (Figure \ref{fig:ocd-lang}). As it turns out, even simple features derived from interview transcripts can be of use in quantifying behavioral differences. For example, the total number of words used in \emph{off-trials} was substantially larger than the total number of words used in \emph{on-trials}. While the sum of \emph{on-trial} lengths and the sum of \emph{off-trial} lengths both amounted to 15 minutes, the patient spoke 423 words across 102 sentences when cortical stimulation was off, and only 234 words across 85 sentences when cortical stimulation was on. Furthermore, the larger number of words spoken per sentence in \emph{off-trials} was a consistent effect across the individual trials, with the notable exception of Trial 4 – an \emph{off-trial}, yet the patient spoke the fewest words per sentence of any trial ($\sim 1.6$). It is of particular interest then that Trial 4 was the only trial during which the patient correctly identified the hidden stimulation status. Any paucity of speech in \emph{on-trials} is therefore more likely to relate to their perception that stimulation is off rather than the underlying condition. 

Because human observation of the experiment recording suggested much of the difference between \emph{on-trials} and \emph{off-trials} occurred during spontaneous speech, we focused our next analysis of the audio transcripts on spontaneous speech during each of the trials, rather than on structured interview portions. Indeed, there was a large difference in the amount of unprompted speech, although outside of the questioning period it primarily manifested as fewer sentences rather than fewer words per sentence. In this context, the patient spoke 202 words across 33 sentences when they perceived cortical stimulation to be on (3 trials), and 91 words across 17 sentences when they perceived it to be off (5 trials). 

To dig deeper, we then generated sentiment-colored word clouds representing all patient-initiated speech during \emph{on-trials} versus during \emph{off-trials} (Figure \ref{fig:ocd-lang}c). The \emph{off-trial} word cloud (Figure \ref{fig:ocd-lang}c, bottom) is notably denser with stronger and more varied sentiment than the \emph{on-trial} word cloud (Figure \ref{fig:ocd-lang}c, top). Additionally, the \emph{off-trial} word cloud largely contains discussion of general interest topics (e.g. YouTube videos), while the \emph{on-trial} word cloud primarily focuses on discussion of treatment-related topics (e.g. stimulation parameters). As there was no patient speech outside of the questioning period during Trial 4, the word clouds for perceived stimulation status were identical (with flipped labels).

\FloatBarrier

\section{Discussion}
\label{sec:discussion3}
We've presented a comprehensive case report on a novel deep brain stimulation trial for obsessive-compulsive disorder, by utilizing over 2 years of temporally dense multimodal behavioral data in conjunction with neural implant recordings and standard clinical scale evaluations. Thus, we've provided a powerful proof of concept for analysis of these datatypes in psychiatric clinical trials. To further this contribution, we will now discuss in more detail both the conclusions drawn and the limitations encountered during our pilot study. Our primary aim is to serve as a reference for similar future studies -- to both replicate our successes and improve upon our pitfalls.

The discussion will begin with reflection upon the results from the more traditional evaluation of this DBS trial, as reported previously by \cite{Olsen2020}, with added context provided by our novel data collection and analysis methods (section \ref{subsec:ocd-dbs-disc}). We will then discuss data collection, feature extraction, and visualization design for the many modalities we contributed to this case report, including daily patient self reports and passively collected location, movement, and phone usage signals (section \ref{subsec:ocd-dp-disc}). Further characterization of these features will subsequently be provided in reviewing our uncovered correlation structures (section \ref{subsec:ocd-corr-disc}), and finally a blueprint for causal study of the effects of DBS using some of these new technologies will be discussed (section \ref{subsec:ocd-exp-disc}). The chapter will wrap with broader future directions enabled by our work (section \ref{subsec:ocd-future}) and a summary of key contributions from our reported results (section \ref{subsec:ocd-contrib}).
 
\subsection{The traditional DBS trial}
\label{subsec:ocd-dbs-disc}
The work presented by \cite{Olsen2020} was the first known case of combining cortical stimulation with DBS targeting the ventral striatum. It is thus novel in its own right, and an important step towards understanding how cortico-striatal synchrony can be modulated by DBS systems. However, the clinical outcomes and impact of DBS on behavior more broadly were measured using traditional tools in their report. Here we present some of their important takeaways and discuss how our methodology for evaluating behavior in a DBS trial adds additional context to those conclusions. This portion of the discussion serves as an example of how digital psychiatry could influence discussion of DBS paradigms in the future. In subsequent sections, takeaways and next steps for the methodology itself will be discussed.

Providing evidence for safety of the stimulation paradigm, the patient reported no significant side effects from the cortical stimulation. They did report that it may have impacted their sleep though, and was therefore turned off at night by the patient later in the study. We note that periods of severely disrupted sleep were observed in the digital phenotyping data of this patient before cortical stimulation was even turned on (Figure \ref{fig:ocd-zoom-sup1}). Regardless, the study serves as a pilot demonstration of safety of this potential treatment. Questions on efficacy remain more unclear. 

With striatal stimulation, the patient’s YBOCS and MADRS scores decreased $13\%$ and $26\%$ from baseline, respectively; there was no meaningful further change in these scores with the addition of cortical stimulation. Thus VC/VS stimulation likely had some clinical effect on depression symptom severity but little on OCD, and the novel cortical stimulation paradigm had no clinically relevant impact. Note that striatal stimulation was turned on almost immediately after the surgery, so the baseline clinical data collection contained fewer timepoints than originally intended. This was necessary due to concerns about patient suicidality.    

Nevertheless, the patient reported much stronger symptom improvements from the cortical stimulation than from striatal, with a large effect on PGI and a self-described increased ability to control focus away from obsessive thoughts, seen only after cortical stimulation was on and unblinded. This result suggests a placebo effect, although improved performance on a cognitive control task (the Multi-Source Interference Task) was observed following cortical stimulation. While placebo effect remains the most likely explanation for these results, it is also possible that the clinical scales failed to capture meaningful quality of life improvements that were present. This presents a good test case opportunity for employing digital psychiatry techniques, as many of our features can be used to estimate facets of overall level of functioning. 

Ultimately, we observed few significant and sustained behavioral changes in the patient after cortical stimulation was turned on. For example, the patient remained at home the vast majority of the time and regularly reported low levels of productivity throughout the study. Immediately after unblinding the patient was observed to leave the house for extended periods and self-report high productivity levels, but this effect quickly dissipated. Some other features such as physical movement levels (measured by phone sensor) and facial affect during site visits did appear to increase from baseline for a longer time period after unblinding, but it is less clear how directly these changes map to meaningful changes in functioning, and moreover these changes too did not appear during the blinded cortical stimulation period. Taken together with the original results of \cite{Olsen2020}, this further supports the notion that life improvements seen following the addition of cortical stimulation were minor in magnitude and attributable to the placebo effect rather than the stimulation. Because the period of blinded cortical stimulation was relatively short ($\sim 1$ month) and at a different frequency than the majority of the study, it is impossible to make a conclusive statement however. Additionally, availability of the passive digital phenotyping data during the blinded cortical stimulation phase was non-ideal, as will be discussed further in subsequent sections. \\

The other key component of the report by \cite{Olsen2020} was the study of neural dynamics in this patient. As mentioned, cortico-striatal coherence strongly increased when the driving frequencies were only 5 Hz apart, though this change appeared only chronically rather than acutely. The observation of heightened synchrony may be explained by neural systems having broad frequency response curves, in particular when input amplitudes are high as in DBS, thereby functionally serving as driving input of the same frequency \citep{Frohlich2015}. \cite{Olsen2020} also suggest that separation between driving frequencies could entrain synchronous activity at the difference between them \citep{Grossman2017}, however the driving frequencies used to demonstrate that effect were an order of magnitude higher than the ones used here. Moreover, the large coherence increase during this study period was found across many frequencies much higher than 5 Hz.

Surprisingly, coherence remained above baseline levels even once cortical stimulation was switched to a frequency 35 Hz mismatched from striatal stimulation, albeit to a much smaller extent (Figure \ref{fig:ocd-report}b, bottom heatmap). This suggests that designing stimulation paradigms to disrupt cortico-striatal synchrony may be more challenging than expected. On the other hand, it also suggests driving synchrony as a possible therapeutic approach. Although cortico-striatal hyper-connectivity is well documented in OCD as discussed, hypo-connectivity has also been observed \citep{Gursel2018}. \cite{Olsen2020} point out that this could be explained by differences across different CSTC loops, highlighting the importance of analyzing neural recording data during DBS trials. 

Another potential contributing factor to the existence of evidence for both hypo- and hyper-connected CSTC loops in OCD is individual heterogeneity. As mentioned in section \ref{sec:background3}, there is currently limited information able to predict OCD treatment response, and the gold standard YBOCS has weak discriminant validity. There is a great deal of behavioral variation across people that could contribute to OCD symptom presentation, but is not captured by traditional methods. Development of more precise behavioral quantification techniques could lead to metrics that are more closely linked with differences in neurobiology, thereby providing a non-invasive way to predict response to various DBS paradigms and more broadly a stronger model for the relationship between biomarkers and clinical outcomes. Therefore this result also highlights the importance of further developing methodologies like the one we introduce here.

Indeed, if any improvements were caused by the cortical stimulation in this patient, they would be attributable to increased synchrony between VC/VS and SMA. Notably, the strongest lifestyle changes we observed in our dataset occurred during the 130 Hz cortical stimulation period when coherence was highest (Figure \ref{fig:ocd-report-stim}). Of course, these changes coincided with unblinding, so they are more likely explained by the placebo effect. Still, given the observed improvements in daily life functioning during this period, one might have hesitated to change stimulation parameters to such a large degree at that time. The digital phenotyping data were not being considered as part of stimulation setting decision making during this study, and the increase in synchrony was a direct issue for the core study's hypothesis. However this directly illustrates another possible use case for the digital phenotyping data. Moreover, the patient's EMA responses and geolocation variety both demonstrated a regression to lower levels of functioning immediately after the switch of cortical stimulation to 100 Hz, although the patient was aware of the settings change so this could be another placebo-like effect. \\

Overall, there are interesting takeaways from the results of this pilot patient, but many questions remain about the DBS paradigm. A number of the limitations that caused this are related to suitability of the particular patient for the treatment. As mentioned by \cite{Olsen2020}, the patient did not respond to prior DBS treatment with VC/VS-only stimulation, which might indicate a pathology that is different than those with observed CSTC hyper-connectivity, as VC/VS stimulation alone may already have some effects on synchrony \citep{Widge2019}. Further, the patient does not perform physical rituals, making SMA a potentially less relevant target and ultimately symptom assessment more reliant on subjective report. Especially because the patient also exhibited severe depression symptoms, it is difficult to disentangle any OCD pathology from forms of intrusive thoughts that occur in e.g. MDD.

An additional confounding factor not covered by \cite{Olsen2020} is the patient's ADHD diagnosis and in particular their resultant medication use. A preoccupation with Adderall relative to other topics was uncovered by our audio journal dataset (Figure \ref{fig:ocd-lang}). Furthermore, we found a few time periods during the study where the patient mentioned Adderall less frequently but instead sometimes mentioned Vyvanse. One such period coincided with early cortical optimization, directly after the switch to 100 Hz stimulation frequency. This raises concerns about medication variance as another contributing factor to any behavioral fluctuations observed. The medication itself is of additional interest, as amphetamines have been found to increase prevalence of OCD behaviors in some individuals, including cases where OCD symptoms were not observed prior to chronic amphetamine use and cases where the amphetamines were used at medically prescribed doses \citep{Shakeri2016}. Unfortunately we do not have access to the records necessary to more deeply investigate this patient's medication use, but our observations do underscore the relevance of digital psychiatry methodologies in interpreting patient outcomes. In future workflows, incoming patient journals could be considered as a factor in making potential mid-study adjustments to other aspects of data collection. Here we could have added direct tracking of medication use, for example.    

\subsubsection{More on clinical scales}
\label{subsubsec:ocd-scale-disc}
As mentioned, the patient did not have any clinically meaningful improvement due to the novel treatment. The YBOCS in particular did not show much clinically significant variation in total score over the course of the study, while the MADRS total score fluctuated at times during the cortical stimulation period despite the lack of sustained improvement. Even with the lack of overall improvement from cortical stimulation, utilizing a data-driven approach for analysis of clinical scales could still yield interesting results, including the potential to isolate substructures of the YBOCS that do vary and to identify any patterns in the variations that do occur in the MADRS data. 

Indeed, question clusters identified by our computational approach (Figure \ref{fig:ocd-clusters}a) revealed symptom structure specific to this patient and enabled identification of symptom changes over time that were missed by the standard trial evaluation. Within-scale correlations generally exceeded between-scale correlations, providing reassuring evidence for the hierarchical clustering methodology used. The YBOCS \emph{resistance} cluster showed particularly strong internal correlation among the individual YBOCS questions, further supporting the use of clustering to guide the study’s focus. Especially because \emph{resistance} to obsessions and compulsions actually did vary across the study to a much greater degree than other YBOCS questions, something missed by the standard split into obsessive and compulsive subscales (figure \ref{fig:ocd-clusters}b). Trends that were previously being “drowned out” by the largely static YBOCS totals emerged when clusters were plotted individually over time.

The MADRS on the other hand had a more variable total score that showed a moderate response to the striatal stimulation, and we can see this reflected in a decrease from baseline for all MADRS clusters for most of the striatal optimization period (Figure \ref{fig:ocd-smooth}). Still, our clusters uncovered new relationships in the MADRS data not previously seen. Interestingly, appetite and energy levels were fairly stable at low severity with striatal-only stimulation, but appeared to vary to a much greater degree once the cortical stimulation was added. This is in line with patient self-report of greater sleep difficulties with cortical stimulation on, and \emph{energy} in fact showed a small improvement once day/night cycling started. It is possible based on these results that cortical stimulation additionally affected the patient's appetite. Decreased appetite and sleep disruptions are also side effects of many ADHD medications however, including Adderall. Characterizing any interplay between the medication and the stimulation paradigm would be important to fully disentangle the physiological and behavioral effects of the DBS. 

Although there was no sustained improvement in any of our clinical clusters from the cortical stimulation, the MADRS results suggest that there indeed may have been some non-placebo effects of the treatment observable outside of neural activity patterns, as further supported by the causal stimulation experiment to be discussed in section \ref{subsec:ocd-exp-disc}. Moreover, our primary motivation is detection of behavioral changes that relate to clinical outcomes -- so our identification of YBOCS features that meaningfully varied during different periods of this study was an important find for contextualizing some of our digital psychiatry work, regardless of any relationship with DBS parameters. \\

In addition to allowing us to extract more meaningful information from the time course of clinical measures, the clustering of questions within a particular patient can also have inherent value for understanding the pathology of that patient. Prior factor analysis has found that the YBOCS contains factor structure when split into the canonical obsessive and compulsive categories, but also contains factor structure when separated into disturbance and severity categories \citep{McKay1998}. Despite these results, most clinical applications still utilize only the traditional subscales when grouping questions. While many OCD patients do show independent variation between obsessive and compulsive symptoms, it is clear that for this patient obsession and compulsion severity go hand in hand; thereby highlighting the importance of a personalizable approach like the one reported here.

Unfortunately, most previous research into the structure of the YBOCS has worked with few scores each from many patients, rather than taking a longitudinal approach that enables evaluation on the level of the individual patient. Given that the patient's OCD symptoms largely presented mentally, it may not be surprising that obsessions and compulsions would be deeply intertwined within them, or at least difficult for an external observer to disentangle. Further study will be needed to characterize personal YBOCS clusters across patients, as well as other properties that might predict how symptoms would cluster for a particular patient. Moreover, previous factor analyses generally began with an assumption on the number of factors, most commonly two. Here we suggest that the hierarchical clustering approach can better identify question groupings of interest, allowing more flexibility in designation of clusters -- something that could prove even more fruitful for a project with longitudinal data from many subjects. 

Ultimately, focusing on cluster scores provides an important middle ground between analysis of individual questions and analysis of total scores. We not only used our hierarchical clustering results to contribute the conclusions just discussed, but also as the gold standard correlates for most of our downstream analyses of self report, digital phenotyping, and neural data. Thus the clinical clusters are a critical contribution of our work, and this methodology can serve to guide future studies involving the pairing of clinical scales like the YBOCS and MADRS with denser digital psychiatry datatypes.  

\subsection{Broader implications of our behavioral data}
\label{subsec:ocd-dp-disc}
Of particular note is how flat and sparse the MADRS and YBOCS scores appear throughout the key time period of cortical stimulation starting, unblinding, and tuning (Figure \ref{fig:ocd-report-stim}). The lack of variance is consistent with the previously reported lack of clinical response to the cortical treatment \citep{Olsen2020}, and as discussed above a greater level of variability can be found in the scales when focusing on certain clinical question clusters. Still, it indicates great gaps in the descriptiveness and temporal resolution of symptom tracking available when using only clinical scales, especially during a critical portion of the study where major transition points repeatedly occur separated by only a couple of weeks or months at a time. Not only are these scales not varying at all during a time of substantial change in the patient’s life, they also are only able to vary in a strongly constrained way - in sharp contrast with rich features like continuous geolocation. Thus to truly advance the field, there is a necessity for developing new tools to detect behavioral changes that are subtle, time sensitive, or simply not in line with the preexisting models of psychiatry. Excitingly, recent tech has provided new opportunities for creation of such tools. Identifying major aberrations or shifts in digital phenotyping features may in fact be a powerful clinical tool in some applications even without assigning valence to a detected change. 

In this case report, we added a number of datatypes to supplement the biweekly clinical scales, providing increasingly finer temporal resolution: from the day level surveys and journals all the way to sub-minute information on phone movement and usage. This results in orders of magnitude more data points available over the course of the study, and fills in gaps where key events could have otherwise been missed. Further, such high temporal detail was distinctly relevant for this study, because of the local field potentials being recorded for 60 seconds at a time, 4 times a day. Linking dense, passive behavioral readouts to clinical outcomes is of inherent interest, but another crucial advantage of these readouts is that they can be time-aligned with dense neural data. It is thus imperative that research on longitudinal, passive digital phenotyping in the context of DBS is carried out. While our study is a pilot proof-of-concept report, it is the first of its kind to our knowledge. 

In addition to the planned novelty, our study also had unplanned novelty due to the onset of the COVID-19 pandemic during data collection. We were able to collect over 4 months of data during spring and summer 2020 (Figure \ref{fig:ocd-report-covid}), which in itself highlights one of the unique benefits of digital psychiatry; as most biomedical research was being paused or otherwise disrupted, large portions of this project were able to continue with few issues. Moreover, the close tracking of an individual with OCD during this time period is clinically salient. The very nature of OCD could lead to new symptom presentations that directly revolve around the pandemic, and to symptom severity worsening as a result. Additionally, severe mental illness has been associated with an increased risk of COVID-19 mortality \citep{COVID}, so a detailed account of pandemic behavior in a patient with severe mental illness is noteworthy. \\

\noindent Our primary goal was to lay a foundation for future studies of this type. While our study is a purely ideographic one, the computational techniques enabled by collecting dense data longitudinally lend themselves to future nomothetic approaches that simultaneously contain detailed ideographic perspectives. In this way, our work can inform many potential future directions for the study of digital phenotyping in treatment evaluation, particularly for neurotherapeutics. Therefore, I will discuss in the upcoming sections the lessons we learned about our methodologies and how they can be improved and extended upon going forward.  

\subsubsection{Self reporting}
\label{subsubsec:ocd-self-disc}
A patient's own perception of well-being is an important facet of treatment outcome to consider. With the relatively recent proliferation of smartphones, it is now much more feasible to regularly collect patient survey responses, at targeted times and using personalized questions where relevant. While these prompts require more patient cooperation than continuous passive data collection approaches, they are easier to analyze and interpret, making them a more approachable entry point into digital psychiatry for research groups that do not have members trained in computer science or related fields. 

Beyond inherent value in assessing a patient's self perception, the resulting survey dataset can also in some ways provide a more appropriate prediction target for digital phenotyping signals than clinical scales do. The more frequent availability of app-based survey responses naturally maps better to digital psychiatry timescales, and includes the possibility of asking questions that concretely map to certain feature sets - for example prompting on hours of sleep each night. In this way daily self-report surveys form a critical bridge between passive behavioral readouts and traditional clinical outcomes. \\

\paragraph{EMA.}
Here we focused our ecological momentary assessment (EMA) questions heavily on motivation and productivity, as this was the area the patient struggled most with. Therefore our EMA data captured a large amount of detail about specific symptoms of that class, but did not measure any OCD-specific symptoms. There is always a trade-off between including more questions and keeping the patient engaged with the daily prompts, which is why we kept our survey lightweight. However, we likely would have extracted more value from the EMA if our original question set were trimmed, and other factors were also asked about in replacement. Response rate was largely good (Figure \ref{fig:ocd-report}), so we perhaps could've somewhat increased the number of questions too without much participation drop-off in this patient. 

Because the questions were not designed to overlap with dimensions assessed by the YBOCS or MADRS, the lack of correlation between total scale scores and same-day EMA is not necessarily unexpected. Of course it is more difficult to validate clinical relevance of the EMA without such overlap, and as pointed out by \cite{Olsen2020} we would recommend that future studies do include some questions that are directly in line with scale outcomes of interest in order to perform more robust clinical validation. Still, our EMA was not entirely unrelated to the traditional clinical outcomes, as our scale clustering approach revealed statistically significant correlation between the EMA summary score and MADRS \emph{mood} questions (Figure \ref{fig:ocd-clusters}f): yet another result supporting our methodology for analysis of the YBOCS and MADRS, in addition to providing evidence for meaningfulness found within the survey dataset.

It is worth noting that there is unique value to questions that diverge from the topics covered by the YBOCS and MADRS, as they may be able to detect symptom changes that would otherwise go unnoticed. Productivity levels are of immediate relevance to patient functioning, and potentially could have revealed some response to cortical stimulation. Unfortunately in this study, as will be discussed, we did not begin EMA collection until shortly before the onset of cortical stimulation, so the treatment sensitivity question is impossible to properly assess. It is clear that the patient's reported productivity often oscillates, but we were able to observe a stretch of positive EMA directly following unblinding, as well as a trend of more likely extended periods with negative response towards the end of the study (Figure \ref{fig:ocd-report}b). The latter result could indicate that the steady state cortical stimulation paradigm was not improving functioning in the longer term, though it is difficult to determine if there was a dissipation of placebo effect or if the changes in cortical stimulation settings contributed to this pattern. 

A reasonable way to sanity check EMA questions that do not cleanly correspond to clinical scales is to include some questions that should strongly correlate with specific passive sensor measurements. When the EMA answers reliably align with the expected digital phenotyping features, confidence in both modalities can be greatly reinforced. In our case, we did ask a survey question about exercise that could have theoretically been used to assist with validation. However the question didn't clearly distinguish between exercise routines with varying levels of vigorousness or duration, and for most of the study we only had access to phone accelerometer data, which is a subpar way to detect exercise compared to wrist accelerometry. Many forms of exercise do not involve even carrying a phone at all. Thus we did not attempt this validation method in the present report. We would recommend that future research projects include a few EMA questions carefully designed to align with both the clinical goals and the data availability of the specific study. Here we had more information about sleep because the patient reliably brought their phone to bed, so questions about sleep duration and interruption frequency might have been better suited in place of some of the productivity questions. 

As noted, the full set of productivity questions likely did not contribute much information beyond what a subset would have. This is particularly stark in the negatively-worded branch of the EMA, where most of the individual questions were highly correlated with each other (Figure \ref{fig:ocd-clusters}e). The branching structure itself turned out to be a limitation of the EMA as designed, because the matched positively-worded and negatively-worded questions were not always proper reflections of each other. Rather than simple negation, many of the questions were rewritten using different language between the two branches. This prevented us from performing analyses that coupled positive and negative branch questions to consider each productivity topic across all days available. Instead, we largely treated the positively- and negatively- worded questions as separate datasets. Visually, the switch between branches over days was itself interesting (Figure \ref{fig:ocd-report}b), but for clustering it was a downside to have to treat the two branches as separate "scales" (Figure \ref{fig:ocd-clusters}a). Future studies should not discount the benefits of a branching structure for valence of posed survey statements, but should take more care to write question pairs that are closely matched if they do choose such a design.

While the EMA clustering was indeed not as cleanly separated as that for the clinical scales (Figure \ref{fig:ocd-clusters}a), likely also in part because all of the EMA questions surrounded a single topic, we could still draw additional conclusions from the methodology. For example, the lack of mirrored structure between the branches was especially apparent in the differing relationship between \emph{motivation} and \emph{planning} related questions. The negative branch surveyed for planning qualities like "trying to do too much", which was not a problem frequently encountered by this patient in general, let alone on low productivity days. As a result, there was a negative correlation between the two clusters when the patient took the negative branch, but a positive correlation between the two clusters when the patient took the positive branch, which had framed planning questions to be about having realistic goals for the day (Figure \ref{fig:ocd-clusters}e). Overall we can confidently conclude that the patient felt their motivation level impacted their productivity much more than their ability to plan did. Knowing clinical background on the patient, this outcome is unsurprising. Still, it is reassuring that we found patient perception on a day to day basis was in line with expectations based on severity of MADRS items, and was also in line with information we gained about their habits from other modalities of our methodology to be discussed. Furthermore, there may have been an interesting aberration in this pattern coinciding with the onset of the COVID-19 pandemic, when \emph{planning} question responses shift much more than \emph{motivation} ones, and positive \emph{planning} specifically undergoes a sharp increase in severity above what was seen throughout the rest of the study (Figure \ref{fig:ocd-smooth}). Based on the simultaneous dip in severity of the negatively-worded \emph{planning} items, we presumably captured a period where the patient felt they were planning even less than usual, a belief that was no doubt common for many people at that time.   

In hindsight, the lack of external responsibilities in the patient's life likely contributed to the minimal nuance that could be uncovered by the various productivity related questions. If the participant had substantial work, school, or care-taking/home-making responsibilities, a redesign of the EMA might take a different form. Regardless, future studies should aim to capture a broader set of properties that in sum would contribute opportunity for clinical validation, passive sensing validation, direct treatment assessment, and personalized tracking of the most bothersome symptoms for that patient. In exchange, fewer details would be captured about the primary symptom type, but likely not to an extent that would have measurable impact on modeling outcomes. Further, as mentioned above, we probably could have added a few questions without much impact on participation rate. Ideally future studies would begin with slightly longer EMA surveys, and can always pare back the length if participation is found to be inadequate. Branching functionality could also be used to ask different follow-up questions in different contexts, so that EMA data could contain both a core set of daily questions and a larger set of questions from which a subset is available each day. Branching logic allows that subset to be chosen strategically based on patient state, but one could also imagine study designs where a simpler approach to swapping through secondary questions would be more suitable - including random selection, fixed daily rotation order, or fixed study period differences.

Ultimately, beginning EMA collection well before treatment start and actively monitoring the results should be a key part of future approaches, in addition to the basic lessons discussed above. Not only can survey size be adjusted to obtain maximal information while maintaining engagement, but question content and phrasing can also be actively tuned before the primary experimentation begins. Particularly for multimodal digital psychiatry approaches like the one we describe here, preliminary results from across data streams can be leveraged to improve the personalized EMA. For this patient, there are some additional questions that would have been ideal to ask in retrospect, and if we had been collecting for some time before treatment it is plausible we could have made adjustments to capture those salient aspects of patient life. A few such properties are already mentioned above; additionally, it would have been valuable to know the patient's perception of stimulation status during the cortical crossover period (motivated by Figure \ref{fig:ocd-video}) and to have patient-reported records of daily medication intake, particularly for ADHD medications (motivated by Figure \ref{fig:ocd-lang}). Furthermore, survey prompts might have utility as a personalized intervention type of their own. For example, we could have asked our patient about visiting unique locations outside of their home, to potentially encourage getting out around the time of the EMA prompt; it was clear from the study-wide view of geolocation (Figure \ref{fig:ocd-report}b) that the patient spent far too much of their time at home as it stood. 

Moreover, while the exact topics of the theoretical survey questions proposed here were inspired by results of the present personalized data collection work, they do still provide insight into the types of questions that future DBS trials ought to consider before even designing preliminary EMA. We did find our initial discussions with the patient about their most impactful symptoms and frequent struggles valuable, and this should certainly still be a component of EMA design -- but the discussion should include content about coincident treatments and about what the patient hopes and expects to get out of the DBS treatment, and the final design should take closer note of the gold standard clinical outcome measurements and the nuances of any available digital phenotyping or neurobiological data. \\

\paragraph{Audio journals.}
Besides EMA, we also collected self-report data in the form of short daily audio journals recorded by the patient via the same app. These journals provide the opportunity for less subjective analyses of acoustic and linguistic properties, in addition to the ability to collect more information about patient perceptions using the diary contents. The recording prompt for daily diaries could be designed to ask for more detail on a specific question, although for this study we used an open-ended prompt that asked the patient broadly about their day and current feelings. As will be discussed, we found this option for the patient to provide self-reported data in a free-form manner to be extremely valuable, and strongly recommend the use of open audio journals alongside daily EMA prompts in any digital psychiatry study that aims for detailed characterization of individual participants or is exploratory in nature. Given the high cost and recruitment difficulty of specialized DBS studies, it would be foolish not to supplement EMA with diaries in any such trial if the patient is at all amenable. For much of the rest of this section, I will discuss the specific takeaways from this case report that support our arguments for the utility of audio journal collection. For discussion of diary use cases in other studies, see chapter \ref{ch:1}.

The first property that daily journals are helpful in assessing is patient engagement, which is both important for building a quality dataset and of potential clinical relevance. EMA participation can only be directly assessed as a binary daily value, and even with careful design to validate a participant is answering seriously, there is only so much that can be said about study engagement from EMA responses. Free-form diaries on the other hand allow for submissions of variable length, with content that may vary in properties such as detail or intimacy levels. It is thus crucial to not just encourage regular recordings, but also to encourage genuine participation in the recordings. Richer diaries enable greater insights into topics of particular salience to the patient at a given time, and provide better grounding for extraction of automated features. Diary richness also may be related to engagement in other parts of the study that are harder to measure in this way, for example care taken in responding to the EMA surveys. As mentioned, engagement drop-off might indicate a need for adjustment of study requirements or associated incentive structures, but it also might indicate a corresponding decrease in more general levels of functioning. Therefore analysis of fairly simplistic diary features can be of inherent utility; indeed consideration of word count and submission time alone uncovered interesting trends in this case report. 

It is expected that participant engagement will decrease to some extent as the novelty of a study task wears off. A limitation here is that diary collection began only shortly before the onset of cortical stimulation, which left no time to establish a proper baseline, let alone habituate the patient to submitting recordings. Any trends in diary participation during early tuning of cortical stimulation are thus more difficult to interpret in this patient, as they may be confounded by natural loss of interest effects, in addition to the lack of a true baseline for comparison. Nevertheless, we were able to identify a period of abnormal behavior later in the study via diary submission statistics that aligned with and may have even been predictive of a transient spike in depression symptoms. Specifically, there was an approximately 3-4 month period in the middle of the study during which diary word counts were at their lowest levels and variance in diary submission time was at its height, which can be clearly seen between the day 435 and day 535 markers in Figure \ref{fig:ocd-lang}a. Meanwhile, severely depressed \emph{mood} was found in the MADRS ratings at day 522, an absolute peak for the post-cortical stimulation period of the study (Figure \ref{fig:ocd-smooth}), and the return to a more normal \emph{mood} level for this patient around day 550 was closely aligned with the recovery of diary content level. Notably, the same $\sim 3-4$ month period coincided with submissions of most of the highly negative sentiment diaries, with the record low sentiment submission occurring around the same time of worst \emph{mood}, but additional negative diaries coming earlier in the period (and none after the period, until many months later). This helped us to validate the sentiment analysis as well as to underscore its potential for prediction of upcoming symptom severity, especially when taken together with other diary features. 

Although word count is a simple measure of engagement that does not capture all the available nuance, it is both plainly linked to engagement and demonstrably useful. The decline in diary word count from this patient actually somewhat preceded the other trends observed before the depressed mood occurrence. Still, diary collection had been running for over 150 days before the engagement decrease occurred, a red flag that something has changed besides normal study fatigue. That something might have been related to study protocol, or it might have been a life change that was not inherently negative, but it also might have been an early direct sign of decreasing mood. Regardless, it is impressive how sustained the shorter diary submissions were until shortly after the lowest \emph{mood} point, and how sustained the improved diary engagement was after it did recover. In future studies, a major factor for better evaluating occurrences such as the word count drop described in real-time will be the use of other journal features, and perhaps features from other modalities too. In this case, we observed that about one month into the period of note, submission time changed substantially and language with negative sentiment started popping up, while word count remained depressed. It is highly plausible that these dynamics were intertwined with whatever caused the depressed \emph{mood} incident, although we do not have enough information to estimate if this would be repeatable or was the result of a specific life event.

Regardless of the explanation behind the period of abnormal diary submission times, patterns in submission metadata are also of great interest. Targeting push notifications to the time of day a participant is most comfortable recording the journals can improve participation rate; this issue is more salient for audio diaries than for surveys because a private setting is often a necessity for patients to be willing to submit, and can help to ensure recording quality for our analyses. While location data from this patient does not suggest they were regularly out in the evenings more often during the period of differing submission times, it is possible a change in activities at their home had an effect. More likely in this case is that some other change in or disruption to patient routine precipitated the increased variability of recording time. Forming routines around necessary or otherwise useful activities can be highly beneficial to productivity, which as mentioned was a recurring struggle for this patient. On the other hand, routines can also involve harmful activities, and specifically in OCD can become pathological. Therefore estimation of routine adherence is of clear interest in this study, though its implications could cut either way. Future studies should continue to use metrics like study participation metadata, as well as other modalities to be discussed, to characterize patient habits as is appropriate for the context of their participants and research aims. 

A second major contribution of free-form audio journals, and perhaps their most unique attribute amongst the typical digital psychiatry modalities, is the ability to contextualize different study periods based on what the patient considers important or what comes immediately to their mind, rather than what the clinician considers important. Interview recordings and especially resulting scale values from site visits are largely focused on the latter perspective, and additionally cannot feasibly give context on the daily level. EMA surveys are ultimately designed by clinicians as well, while passive digital phenotyping measures are unable to capture subjective perceptions of the patient. Therefore diaries fill an important gap left not only by traditional clinical measures, but also by many of the emerging methods for behavioral quantification. It is not necessary for all research questions, but it is hard to imagine a scenario where it is not at all useful with a willing participant. Moreover, although this perspective is indeed largely subjective and impossible to completely quantify, there are many computational techniques that can assist in easily extracting portions of relevant context. 

Indeed, in this case report we never read through all diaries, or even a majority of them. We instead took a few "shortcuts". We looked for words that occurred at a high frequency to identify previously unknown topics of interest to the patient, which identified habitual activities like watching YouTube and playing video games, and most critically identified a fixation on ADHD medications that patient was taking -- and as mentioned previously, that result also raised new concerns about experimental confounds. Additionally, we utilized our sentiment-colored word clouds to quickly scan all diaries for notable changes in language sentiment, descriptiveness, and commonly used words. That approach allowed for identification of other topics of more transient interest, such as the patient's used of a sun lamp in the winter time. It also allowed for identification of days that might be of particular interest, isolating a small number of diaries to read more closely. Extracted language features and information gleaned from other modalities similarly highlighted days where the diaries might contain especially salient context, for example when the patient attended their religious institution for many hours throughout an entire week. While we were unable to find an explanation for the abrupt change in behavior within the diaries, the absence of context was itself informative in this instance, because it coincided directly with cortical stimulation unblinding. Diaries recorded during important study transition periods are obvious targets for closer study regardless (Figure \ref{fig:ocd-lang}b). Moreover, there are established NLP techniques such as topic modeling that can be used to automatically summarize the broader context a patient provides to a study through audio journals, with a greater emphasis on relationships between different words and word groupings than we had in our pilot report. As emerging computational tools like large language models are further developed and adapted to a psychiatry context, there could be the ability to probe for specific patient opinions and inquire about their temporal dynamics over a dataset via a chatbot interface. It is in fact plausible that a usable prototype for this would be available within the lifetime of a longitudinal study beginning now, so there is a large amount of upside to diary collection. 

Even with our relatively simple analysis approach, we reported on a number of interesting observations. Patient discussion of Adderall and at times Vyvanse clearly had qualitative importance in this study, which then informed additional longitudinal features that turned out to have quantitative importance. The months leading up to the MADRS \emph{mood} peak discussed above also coincided with a notable gap in mentions of Adderall. Although one would expect shorter diaries to contain fewer key words, the extent to which Adderall ceased to be mentioned at all is unexpected relative to the magnitude of the word count decrease and the number of words still being spoken during that period. Further, in the $\sim 150$ days that make up the nearly blank space in the middle fo the Adderall count timecourse, stimulation related keywords occurred in 5 diaries. This is a typical frequency for most of the dataset, yet mentioning Adderall in just 3 diaries is highly atypical (Figure \ref{fig:ocd-lang}a). Across the entire dataset Adderall mentions significantly correlated with word count, to a degree not explainable by simply the extra words that would be contributed by the sentences about Adderall (Figure \ref{fig:ocd-lang}d). On the other hand, there was hardly any correlation at all between count of stimulation keywords and total diary word count (Figure \ref{fig:ocd-lang}c). Of course, the positive correlation between mentioning Adderall and recording longer diaries across the dataset is largely driven by the pattern seen in this $\sim 150$ day stretch. 

Taken together, there is a compelling hypothesis found in the diary data: that patient medication usage not only changed for a substantial period of time in the middle of the study, but that this change resulted in decreased diary engagement, disrupted routine, and more negative outlook, ultimately culminating with a severely depressed mood event detected by the MADRS, which was then shortly followed by a return to the original medication plan. Obviously there are a number of other plausible explanations for the trends we observed, and as mentioned we do not have access to the medical records from the physician treating the patient for ADHD, so it is impossible to properly test. Either way, this example demonstrates the power of audio journal analysis and the unique angle that it can bring to a study, through generation of a coherent hypothesis to explain a previously mysterious spike in clinical scale score along with identification of new features that may have been a warning sign well in advance. 

Despite the strength of the journal dataset in this case report, we did not really discuss the value of a more systematic extraction of acoustic and linguistic features. As mentioned, language sentiment estimates demonstrated relevant signal here. The other major NLP features we extracted were centered around word embeddings, in particular measurements of semantic incoherence and word uncommonness. We did not observe anything notable in those features, although that is not entirely unexpected for the current use case. In studies of psychotic illnesses, such features likely contain much more signal, so their inclusion in future DBS research should depend on properties of the recruited patient population. There are additionally features not included here, for example categorization of linguistic disfluencies over journals, that might be fruitful to consider in future work. Similarly, we did not pursue any acoustic analyses, which have shown prior relevance in estimating depressive symptoms. The primary reason we did not take an exhaustive feature extraction approach in this case study was the absence of both baseline journal data and a clear clinical outcome to model. That is not a commentary on their potential in a different study, but for the highly exploratory present work it would not have been appropriate to look at a large set of harder to interpret features across only a few hundred diary data points. \\

As the use cases for and design and analysis of both self-report surveys and open-ended audio journals have been extensively discussed, I will now close the section with takeaways about missingness of these active datasets and additional considerations for encouragement of patient participation in similar methods for future work.  

The largest limitation of the self-report data from this patient was the substantially delayed start of collection. The passive phone data to be discussed also did not start collection until months after surgery, which is non-ideal, but crucially it did start about a month before the cortical crossover period and therefore well before the cortical stimulation was actually turned on. Active data unfortunately lagged this by nearly three months, meaning we did not have any baseline data from before the cortical stimulation was added. Because active data collection requires active patient participation, it may require more technical assistance or troubleshooting with a patient before issues with submission protocol are worked out. The daily prompt notifications to remind the patient may also cause more headaches than the passive data portion of the app, particularly for the current implementation of Beiwe. Furthermore, habituation time may be required for certain self-report modalities before novelty effects subside, and given the substantial difference in temporal density between active and passive data, it is advisable to collect as many records at baseline as possible. Daily self-report submissions often vary from day-to-day by design, and do not have the luxury of a large backing in the literature like scales do. Therefore it is most important to get self-report data collection running with a very healthy lead time for a treatment evaluation study. We failed to properly plan for these various issues in advance, exacerbated by our peripheral role in the main clinical trial. It is imperative that future research err on the side of caution by ensuring there is working active data collection as early as possible. 

A positive of our self-report dataset on the other hand is the good participation rate of the patient, which continued even amidst the onset of the COVID-19 pandemic when other parts of the study had already concluded. This is to a large extent just luck in the participation aspect of patient selection, but there are still lessons that can be learned from our experiences with this patient. One key lesson is that it is important to understand patient motivation for participation in a research study. Obviously our patient had strong motivations for enrolling in the DBS trial, but they could have refused to partake in this digital psychiatry supplement. Some patients will enroll in additional data collection for monetary gain, however this patient was from a financially stable background and was not really concerned with the direct compensation. Other motivations we've observed across prior studies have included a general desire to contribute to scientific research, a more specific desire to understand more about themselves through the study, or a perceived connection with study staff. In this case, we suspect that a desire for approval from the researchers was a major driving factor in patient participation levels. The patient specifically reported enjoying the social interaction that resulted from site visits, and at times appeared to be attempting to answer interview questions in a way that they thought would please study staff. 

Understanding this motivation is helpful not only for determining ways to best encourage participation, but also for identifying biases that could result in self-report data collected from a particular patient. In this study, thanking or otherwise praising the patient at regular intervals for their participation in self-reporting could have improved engagement at times, particularly if specific goals like longer audio journal length were encouraged. Even the basic incentive structure outlined in our compensation guidelines could have altered engagement levels in different ways by emphasizing different end points to the patient; while they may not have cared about a few extra dollars a day, they very well might have recorded longer diaries if payment were based on content, due to the message conveyed by that alternative protocol. 

Regardless, varying methods for improving participation must be conducted carefully to avoid potential biases. In a larger cohort, patients motivated by different factors might provide more or less information because of early study design decisions, which could impact data richness in a demographic-specific way. For the longitudinal study here, timing of any intervention meant to improve participation levels would have been critical, and would likely be best suited at a pre-planned regular interval. We did check in with the patient at site visits to ask if they had encountered any technical issues with the Beiwe app, and would recommend future studies do this regularly too. Technical problems were not rare, and the question also served as an indirect reminder. Potential for more direct approaches should be thoughtfully tailored to the specific study.

As discussed extensively above, missing or low quality data can sometimes itself be a relevant signal, so the protocol for checking in with a patient due to observed data collection issues should be concretely defined prior to the study, and should likely require a period of extended missingness or severely bad quality. On the other hand, clearly identifiable technical issues should be noted in advance so that they can be flagged before the end of a longitudinal study. Examples include static noise disrupting journal recordings, which can be monitored by a pipeline like described in chapter \ref{ch:1}, or Beiwe notifications not being received by a participant, which can be a regular check-in question regardless of observed submission frequency. 

It is worth noting that in this study the patient's reported desire to please the experimenters might've directly contributed to the strong and sustained cortical treatment effect seen in the PGI, as well as their tendency to praise the stimulation paradigm when specifically asked. The free-form daily diary prompts and to some extent the EMA surveys were a medium that allowed collection of patient self-report without being quite so susceptible to that bias. Indeed, while their daily perception of productivity improved immediately after unblinding and stayed positive for a time, EMA over the course of the full study was most often negative. It is possible some placebo effects decayed, or even that further cortical settings tuning worsened patient state. However it is also possible that much of the behavior directly after unblinding was the patient doing what they felt they were supposed to, which is not very sustainable and would explain why some of the potential behavioral effects we described in this report wore off within a month -- yet the PGI response to cortical stimulation remained robust.

Ultimately, active patient participation is critical for studies involving daily self-report data collection. Still, that participation need not be sustained for each patient throughout the study, particularly for longer-term longitudinal research. Steps such as those suggested here should be taken to encourage participation in a patient-dependent but systematic way. Crucially, workflows should also be in place to mitigate any impact that technical problems or misunderstandings may have on engagement levels. At the same time, data availability, quality, and richness are themselves interesting signals, so a time-dependent lack thereof should not be assumed a negative. This is especially true when study design includes simultaneous collection of passive phone sensor data, as the features collected there can fill in context during any gaps of self-reporting. 

Indeed, digital phenotyping and daily app-based self-reporting enhance the benefits of each other in numerous ways. In future work we would prioritize beginning our methodologies as much in advance of treatment application as possible. Digital phenotyping results could then also be used to inform EMA questions and audio diary prompts, and vice versa for generating novel analysis perspectives for digital phenotyping data -- thereby encouraging an active study design process. I will next discuss the additional takeaways supported by our passive data collection results, and the lessons learned for future iterations on our methodologies. 

\subsubsection{Passive data collection}
\label{subsubsec:ocd-phone-disc}
Passively collected digital phenotyping features are of great interest due to their relative ease of collection in many people over long time periods, their much higher temporal resolution, their ability to be recorded continuously, their much lesser susceptibility to patient reporting biases, and their generally higher levels of objectivity. In sum, these tools can provide an extremely rich, dense input dataset for characterizing behavior. 

Early applications have shown promise in relating some digital phenotyping features to clinically relevant measures, as was discussed within section \ref{sec:background3}. In addition to the value some phone features may have in predicting gold standard clinical outcomes, certain features are themselves interpretable as health outcomes. OCD is linked with a variety of other illnesses such as cardiovascular disease \citep{Mataix-Cols2020}, so measurement of physical activity and time spent outside can be especially useful in this population. Thus the progression of these features over time is inherently of interest, as is their relationship with other markers such as neural and biological data.

The work presented in this chapter is a first of its kind look at objective and continuously collected digital phenotyping features coinciding with deep brain stimulation treatment and accompanying neural recordings. It is also fairly unique in its highly multimodal and longitudinal nature, relative to much of the other pilot work on digital phenotyping. Because of these distinctive factors, a number of more specific observations in the study could also have broader relevance to future work in the same vein. 

Here, I will discuss observations on missingness from the data collection and analysis processes, and then I will discuss qualitative interpretations of the digital phenotyping data that was available. Both points should serve to outline many relevant considerations for working with passive digital phenotyping datatypes. \\

\paragraph{Lessons on data missingness.}
Perhaps the largest limitation of this work, particularly as it relates to the supplementary data collection that was our focus, was the large amount of missingness in particular modalities or at important points in the study. To some extent, these issues come with the territory of early applications. Over the course of the study we identified numerous ways in which the data collection process could be improved.

One downside of this pilot study was the fact that our passive data came almost exclusively from phone sensors. The patient wore a GENEActiv wrist accelerometer for only a short period, but had offered to wear an Apple Watch months prior. Because we did not have a workflow nor regulatory approval established yet for obtaining raw data from the Apple Watch, it was not feasible to integrate Apple Watch use into the study by the time the patient made the request. But particularly for unique datasets like the DBS one collected here, participant enthusiasm for chosen wearable device should not be underrated.

Another downside was the failure to consent the patient to our portion of the passive phone data collection until 150 days after the implant surgery was completed. By this time the patient was already nearly completed with striatal optimization, which means our dataset essentially began at the same time that cortical stimulation could have theoretically first been turned on. Not only did this miss out on any behavior changes related to striatal stimulation, it also left little time to troubleshoot or fine-tune data collection before the core intervention, so that data availability for most of the cortical baseline is subpar relative to what it could have been. 

However, once passive data collection was fully underway it was rare to encounter a day with no data available -- and even partial passive data can provide rich information. The one extended period of complete passive data missingness in the middle of the study was due to the participant switching to a new phone, something we will be better prepared for in future trials. One change that could help would be to investigate alternative passive data collection apps to Beiwe, as it is frequently necessary to troubleshoot issues with participants in the current setup. As an interesting aside, there was a shift in the coloring of the phone accelerometer heatmap at this time (Figure \ref{fig:ocd-report}b), corresponding to a change in the accelerometer axis with the greatest amount of movement. This suggests that the patient kept the new phone in a different orientation than their previous one, and hints at the types of subtle changes this methodology can detect. 

The late consent date and missing wrist accelerometry modality made it more difficult to fill in missingness gaps due to app or other issues than it otherwise would have been. In general, collecting data from a variety of modalities simultaneously can not only strengthen interpretations (by enabling one modality to provide context for another), but also allow for interpolation of missing data in one modality using another. A simple example suggested in this report was the use phone accelerometer availability information to determine whether phone usage missingness is due to lack of phone usage, or is actual missingness caused by technical or other issues -- but statistical interpolation techniques could be considered in certain applications.    

Data missingness can also be at times a relevant feature itself. Though this is more obviously true for active datatypes like the EMA and audio journals discussed above, it is true for certain passive datatypes as well. Most notably, geolocation tracking requires GPS permissions to remain turned on on the participant's phone. There are a number of reasons one might turn location tracking off on their phone for a time, which may have nothing to do with the study. This can vary widely from person to person and situation to situation, but it is important to keep in mind that there are many considerations in determining how to appropriately handle missingness in a model; considerations which will certainly differ from study to study. In our case, the participant rarely had GPS data missing but other passive datatypes available, and the participant more generally did not leave the house much, so it was not a necessary issue to tackle. \\

\paragraph{Qualitative interpretations from passive data.}
There were many interesting observations resulting from qualitative analyses of the passive data, observations which suggest potential quantitative methods of interest for future works with more robust datasets. 

One observation is that we occasionally saw high phone usage in early morning hours corresponding with patient waking during this period, but more frequently screen time remained low in the early morning despite waking -- which was detected via recorded manual screen unlock events (Figure \ref{fig:ocd-zoom}). In future analyses using engineered features, it will be important to keep in mind that a single late night phone unlock with a short duration of associated screen time could still carry very relevant behavioral information, while during the day time such occurrences would be much less salient.

Notably, there there was a change in the participant's night time phone usage pattern during the COVID19 pandemic (Figure \ref{fig:ocd-report-covid}), where the usage signal indicated the patient was spending hours looking at the phone when they awoke, instead of attempting to fall back asleep. Thus we have a clear demonstration of increasingly disrupted sleep during the pandemic, which could have a number of clinically relevant downstream effects. It is not clear from our data whether the phone use prevented the patient from falling back asleep, or if the patient first experienced significant difficulty in falling back asleep and then spent time on their phone. However based on the context of the pandemic, it seems highly plausible that phone checking started contributing directly to less sleep. If this were the case, storing their phone outside of the bedroom overnight is a simple intervention that might be suggested, and wrist accelerometry or other wearable signals could then be used to measure subsequent sleep interruptions.

We were unfortunately only able to collect wrist-worn accelerometry data for a short portion of the study. When we did have these data available, it could be seen that the patient often woke up in early morning hours, then tried to go back to sleep with mixed success -- something also observed at times in the phone usage data. Interestingly, the period with less early morning movement largely aligned with a period of positive EMA, suggesting quality of sleep indeed related to self-reported patient productivity. This leads to another major potential analysis consideration, the effective use of multimodal data streams. 

In cases where wrist actigraphy is unavailable, phone accelerometer data could also be a useful tool to cross-reference when events of note are identified via location or usage data. For example, it could help determine if the patient checks their phone while lying in bed, as opposed to sitting up or even getting out of bed with their phone - a metric which could serve as one estimate of severity of sleep disruption.

Of course, in the ideal case both phone and wrist accelerometry data would be available. With a larger such dataset, we could have characterized relationships between the two features, to understand how much signal they share and how that correlation might be modulated by another feature like phone location. This kind of characterization very likely needs to occur on a personalized level, as many subject-specific nuances could arise based on when they keep their phone with them and when they do not (e.g. running outside with phone versus putting phone in a gym locker). Nevertheless, understanding those relationships would facilitate the most efficient use of the two modalities, and allow for one modality to be used to infer the other in times of missingness, or one modality to be used to quality check another. 

Another potential application for personalized analysis of multimodal data would be to summarize phone usage and accelerometer data at different locations the patient frequents, to determine how behavior may vary between different settings. While this particular patient spent the majority of their time at home throughout the study, some comparisons between time in and out of the house could still be made. For example, when the patient was out of the house, they tended to check their phone less frequently. 

Because the patient did not leave the house often, there was minimal variation that could be picked up in their geolocation signal as compared to a more typical study (for example the BLS participants discussed in chapter \ref{ch:1}). Further, there was not much variance in most of the patient's clinical evaluations of symptom severity. Given the lack of variance in both areas, it is not particularly surprising that we found few significant results across the study, and it is in fact somewhat impressive that behavioral observations like the aberrations in location after unblinding or the changes in night time phone use during the pandemic were so clearly detectable. In a case with more variance in patient behavior, the power of our approach would only increase as well.

Another downside of this case that limited our analyses was the fact that this patient primarily reported mental rituals and did not describe any specific recurring physical rituals. As such, stereotypy in signals like accelerometery was not a major part of the analysis plan as we had originally hoped. Another datatype that could have been analyzed for stereotyped or otherwise repetitive behaviors would have been phone usage. Though the information we had available was limited to high level usage rather than specific app use times, it did not appear that there was any stereotyped pattern in the way this participant used their phone. Future work in digital phenotyping for OCD should strongly consider both actigraphy and phone use measures for detecting different instances of ritualizing. \\

Ultimately, passive data collection has a number of unique advantages to contribute to understanding behavior - advantages which could be relevant in the context of almost any psychiatry study, if tailored carefully to the design considerations of that study. Furthermore, the temporal resolution of passive sensing data is critical for the ability of future research to link neural signal directly to behavior. 

One way that digital phenotyping tools can be used with back-and-forth iteration between work at different timescales, to personalize assessment and assist clinicians in identifying a variety of potential behavioral patterns in a given patient. This case study demonstrates the potential of such an approach. More broadly, analysis of passive behavioral data streams can be extremely challenging, given their dense multimodal nature and general lack of grounding in concrete labels. Development of techniques for most effectively utilizing digital phenotyping data remains an ongoing process, and on the community level it should certainly involve collaboration between experts across psychiatry, machine learning, and modality-specific processing. 

For this case report, we took a qualitative approach to the use of passive data in many ways. However, we performed a number of quantitative analyses to better understand the correlation structure amongst digital phenotyping features and between digital phenotyping features and other domains like neural recording features in this patient, which I will discuss next. Through multiple approach styles, we took small steps in multiple parallel directions towards better understanding digital phenotyping data for behavioral characterization.  

\subsection{Correlation structures}
\label{subsec:ocd-corr-disc}
Feature correlations could provide value both in improving understanding of the individual features and in perhaps generating relevant features themselves. In the context of this case report we focus on the former. 

While there was little relationship between neural data and clinical outcomes in the study, the availability of digital phenotyping data enabled additional behavioral changes over the course of the study to be uncovered. Some of these behavioral features showed a promising relationship with clinical outcomes, and others were significantly correlated with neural data. Further, many of these features are straightforward to interpret, with their own inherent clinical interest. Collecting such behavioral data allowed a richer look at the outcomes of a novel DBS trial, providing proof of concept for the use of these tools in treatment evaluation, as well as in research on the links between neural activity and behavior. Refinement of the methodologies described here would have the potential to revolutionize clinical trials in psychiatry, and in particular the development of novel DBS treatments, which could leverage behavioral information for development of closed loop systems.

As such, I will now discuss our correlational findings in more depth -- first within behavioral measures as well as between these behavioral measures and clinical outcomes (\ref{subsubsec:ocd-behav-corr-disc}) and then within neural features as well as between the neural features and the behavioral measures (\ref{subsubsec:ocd-lfp-corr-disc}). This discussion will include not only scientific interpretation of the discovered correlations in the dataset, but also a more general discussion of benefits and limitations of such an approach.

\subsubsection{Behavioral measures}
\label{subsubsec:ocd-behav-corr-disc}
The lack of significant correlation between daily features extracted from different phone sensors (Figure \ref{fig:ocd-dp-corrs}) suggests that different passive data modalities are providing distinct information, which supports the comprehensive collection approach we took and emphasizes the value of multimodal approaches. It is intuitive that in general such features would provide independent information, for example daily physical activity levels can only reveal so much about location variety, and vice versa. It's also worth noting that correlation structures looked very similar amongst passive digital phenotyping features (and between these features and clinical outcomes) whether linear Pearson or rank Spearman correlation was used. Because linear models are relatively simple and interpretable, it is scientifically convenient when interesting dynamics can be captured linearly. These results hint that one should not necessarily immediately jump to nonlinear models when dealing with passive sensor data.

On the other hand, it is important to note that such relationships can be highly dependent on the individual and their habits, and the subject of this report is a particularly unique case. The intention of the study was not to draw any specific conclusions but rather to document the capabilities of our methods. Despite our extensive characterization of the dataset, we in many ways did not scratch the surface of possible analyses; in terms of correlation structures, all results presented utilized day-level features, but relationships could look very different on different timescales (e.g. hourly feature correlations) or when restricted to different contexts (e.g. correlation of features at night only). \\

In considering correlations between passive sensor features and clinical features, it is important to recall that the clinical outcomes (and the self-report surveys) did not meaningfully correlate with stimulation parameters nor with the results of neural recording, and for the YBOCS there was minimal variation over the course of the study at all \citep{Olsen2020}. As such, strong correlations between daily digital phenotyping features and clinical outcome were not really expected. Furthermore, because the sample size for clinical scale ratings was relatively small ($n=44$) compared to the number of days with digital phenotyping data ($n=627$), we had much less statistical power with the chosen set of passive features here than we did for other comparisons made. 

I will also note that the correlations focused on same-day digital phenotyping features to limit the number of total comparisons made while keeping the methodology straightforward as well as analagous between clinical and self-report scores (and other considered features). However, a study focusing only on correlation between passive sensing features and clinical scales might try a wider variety of possible summary methods for digital phenotyping, in order to capture possible impact of behavior from preceding days on score prediction. The timescales to consider may additionally depend on the particulars of the scale and the project collection protocol, as different clinical scales may ask the participant to discuss how they've felt over the last 48 hours versus over the last week, etc. It could be possible as well to increase the total number of datapoints by assigning the clinical rating from a particular day independently to all the earlier days it was intended to cover by protocol, or by using extra context like self-report surveys to attempt to impute clinical ratings for every day. Replicating the labels however might cause misleading estimates of the role of random noise, and imputation would itself need enough available data to be verified for accuracy.

Given how difficult it can be obtain a large dataset of clinical ratings and how temporally dense digital phenotyping data can be, how to best handle multiple timescales whilst balancing the need for statistical power and the need to retain relevant signal is a highly complex technical problem. In this pilot work we focused primarily on intuitive "low hanging fruit" phone features/timescales, which can serve as a solid baseline for many studies. For near term practical clinical impact, there remains quite a bit of research to be done establishing details on such features, but for longer term scientific discovery, it will be important to establish good practices on dealing with data complexity, in order to draw rigorous and understandable conclusions as a field. Broadly, I feel the two major avenues for doing so will be studies that focus on testing hypotheses about a very specific behavior, and studies that explore many facets of digital phenotyping data in a bottom-up fashion, only invoking clinical labels to compare against features that e.g. showed an interesting pattern of natural variation over time. 

Additional hypotheses could be generated from these exploratory works or they could be sourced from clinical observations or neurobiological priors. Exploration itself could utilize a variety of modern machine learning tools from unsupervised approaches to supervised learning within the passive sensing data, and it could also leverage a number of qualitative techniques for gaining intution on large multi-variable datasets (like some of the visualizations displayed in this report). One important approach towards understanding a dataset that is in between these two extremes is of course analysis of correlation structure! \\

Despite the challenges, we still found some interesting potential trends in correlations between digital phenotyping features and our clinical scale clusters (Figure \ref{fig:ocd-dp-corrs}). Additionally, we found a few significant correlations between daily EMA survey score and same-day passive sensing features -- which was particularly relevant in cases where the same phone feature demonstrated a possibly intruiging relationship with a much less frequently collected clinical outcome, thereby underscoring the utility of obtaining more frequent labels to fill in gaps, even if imperfect.

For example, the EMA summary's significant negative correlation with phone use appeared surprising at first look, but it was also consistent with a promising negative correlation observed between number of phone unlocks and the YBOCS \emph{resistance} cluster. Considering the scatter plots of these relationships and the days involved, there was no obvious confound like was found with phone location. It thus may be the case that the patient checked their phone more frequently when they were better able to resist their obsessions and compulsions, which is a reasonable causal explanation for someone who did not report phone-related rituals or fixations, and it is reasonable that this could interact with self-reported productivity. Further, we found a signficiant negative correlation between the EMA summary score and the daily phone movement too.

We of course cannot say for sure what was behind these observed relationships, but regardless it is a powerful example of how the same behavioral feature can have vastly different interpretation in different people, with oppositely signed clinical correlations. Along these lines, it is worth noting that phone unlocks overnight allowed us to infer levels of sleep disruption and at times indicated objectively unhealthy sleep behaviors. But because the number of phone unlocks is much higher during the day than at night, any adverse signs from nighttime phone behavior would have minimal impact on these tested daily feature correlations.

Perhaps the most surprising non-result was that daily location features displayed the weakest correlations with clinical outcomes including self-report, with no real meaningful relationships found. This is despite the fact that the most obvious behavioral abberation when visualizing the data was in the GPS signal directly after stimulation unblinding, as well as that location variety has an intuitively obvious relationship with depressive symptoms, something the patient not only struggled with but also displayed meaningful levels of variation in over the course of the study. It is possible that suspected "common sense" relationship is not actually real, though there is a good body of preliminary scientific work suggesting it indeed is. In particular, the recent emergence of digital phenotyping work has enabled a number of new studies closely quantifying this, and associations between depression severity and simple GPS features like increased time spent at home and decreased location variety have so far shown to be robust results \citep{Saeb2015,Masud2020,Laiou2022}. 

While it is most likely that location variety just does not correlation with symptoms in \emph{this particular patient}, which may in part relate to their limited geolocation variety overall, there are other limitations in our pilot correlational analyses that might have missed a relevant relationship. There was more hourly missingness in GPS than the other phone modalities, and for location variety it was difficult to calculate a summary score that did not either underrate nor overrate days with a moderate amount of missingness. For correlation with clinical scales especially, matching data points by day as was done here will skew GPS results more than other phone modalities, because the patient spent a couple hours at the hospital in any day where clinical scales are collected. Ultimately, while a relationship between location variety (and time spent at home) and clincial outcome was probably absent in this participant, an analysis considering more nuance in the types of locations visited and the time of day might have uncovered more signal. \\

On the topic of correlation between self-report behaviors, we found high overlap in survey submission and journal submissions in this participant, like was the case for the participants in chapter \ref{ch:1}. However, despite these submissions often occurring around the same time and the fact we had a reasonably powered sample size for the features tested, we found no significant relationships between EMA scores and diary features in the current patient. This is in sharp contrast to the results of chapter \ref{ch:1}; it is of course once again the case that individuals can display huge variety in their profiles, and there will be some participants who submit largely uninformative diaries and/or largely uninformative EMA surveys. In this instance though, it is difficult to say whether the difference is due to subject-dependent factors or to the large difference in EMA design between the two studies. It is possible that the diaries would be better for assessing symptoms like mood and level of obsessive thoughts in this patient, but did not relate specifically to their self-reported productivity (which had minimal relation with clinical scores anyway). 

\subsubsection{Local field potentials (LFP)}
\label{subsubsec:ocd-lfp-corr-disc}
One initial observation from our LFP correlation work was that there was a strong internal correlation structure between most of the daily power band features considered (Figure \ref{fig:ocd-neuro-corrs}), suggesting that a clustering approach to cut back on the number used in downstream analyses might have been worthwhile. At the same time, we could have utilized LFP features from different times of day separately to perhaps enhance the amount of independent information we were considering. However, part way through the present study the stimulation paradigm was changed to allow the patient to turn it off before going to sleep and then turn it back on upon waking, which introduced additional confounds at different times of day over the course of the study, beyond the scope of the current work. 

We unsurprisingly did not find many meaningful correlations between daily LFP features and clinical scale ratings or self-report surveys, in line with the results reported by \cite{Olsen2020}. However, this highlights the potential relevance of digital phenotyping features to find relationships between behavior and neural activity that might have otherwise been missed, something that has the potential to identify new digital psychiatry features of clinical interest for further study. The temporally denser nature of the phone data gives it more power to detect subtle interactions with LFP data than the clinical scales possibly could, and as mentioned above some of the behaviors identified via phone data are inherently interesting. In the future, one could even consider a high resolution alignment between e.g. a recorded LFP segment and simultaneous wrist actigraphy signals. That was limited here not only by the particulars of the patient, but also by the nature of the recording scheme, which captured 60 seconds of LFP at a time only 4 times a day -- small windows to attempt clinically meaningful real time analyses within, though of course we could have tractably looked at e.g. hourly digital phenotyping features with our present methodology. 

While we found a handful of significant relationships between daily phone features and striatal power bands, there were minimal relationships between the phone features and cortical LFP or the corticostriatal coherence summary feature, which would have been of particular interest given the hypothesis behind the DBS paradigm trial. Significant correlation with some of the GPS features and corticostriatal coherence was observed, but it is highly likely to be a coincidence: the scatter plot of Figure \ref{fig:ocd-neuro-corrs} shows the relationship being driven by differences in GPS feature between the pieces of the bimodal distribution of corticostriatal coherence, a distribution split that relates to the change in stimulation frequency part way through the study. It is obvious from Figure \ref{fig:ocd-report-stim} that the smaller amount of time being spent at home follows unblinding much more closely than it does the start of cortical stimulation, such that it is most likely the placebo effect caused a transient change in geolocation behaviors, and the placebo just so happened to overlap with a relatively short time period where coherence was extra high. Ultimately, this shows the importance of visualizing data carefully to interpret any potential trends, even in the case where enough data is theoretically available to do well-powered statistics. It also reiterates some of the weaknesses of the current case report, as the participant not only did not respond to the major treatment nor display much variation in clinical state (and in many ways behavior more generally) over the two+ years, but additionally there were major confounding factors within parts of the study timeline. 

\subsection{Causal experiment}
\label{subsec:ocd-exp-disc}
Because of such unavoidable limitations of this case study - which included that the participant would not consent to having cortical stimulation off for more than a short period and would not consent to having striatal stimulation turned off at all - it was challenging to ask many casual questions about the effects of the stimulation paradigm on behavior or even to test dynamical models that could at least hypothesize on causal structure within the longer term data. To address this in part, we conducted the cortical stimulation interview experiment of Figure \ref{fig:ocd-video}, aiming to determine how cortical stimulation being on versus off (as well as the patient's belief about it) impacted speech and facial affect. While this particular stimulation paradigm is novel, there is prior literature utilizing tech to demonstrate that DBS treatment for OCD can increase patient affect in a causal manner, though the separation in cause between stimulation itself and belief of stimulation has not been well-studied yet \citep{Goodman,Provenza2021}. Nevertheless, we took a similar approach here to study the cortical stimulation effects (rather than striatal stimulation effects) in this patient, whilst ensuring participant and interviewer were blinded to stimulation status in each trial.

In performing this causal experiment, we encountered a few additional limitations. One is that the patient quickly became bored with the protocol, and had declining engagement over the course of the 8 trials at a rate much more rapid than anticipated. This can be seen both in the length of the trials and in the overall levels of facial expressiveness. We had hoped to run a few shortened versions of the experiment trying additional variations in the ordering of trials, but unfortunately the original experiment was not long before the onset of the COVID19 pandemic, so it was never repeated.

One of the most interesting results of the report was that the participant was quite good at guessing the incorrect stimulation status, and expressed little uncertainty in these guesses. Further, a similar effect was also observed by the MGH group during an EEG experiment with this participant, where they were blinded to cortical stimulation status and adamantly stated the stimulation was on during multiple trials where the stimulation was actually off. While it would require more follow-up, these results certainly suggest that the participant might be capable of detecting cortical stimulation status (or at least changes in cortical stimulation status) even while blinded, but assigns the opposite status because of a strong belief that the stimulation helps them.

At the same time, the consistent guessing of the patient presented an analysis limitation, as it was not feasible to separate behavioral differences coinciding with true stimulation status from behavioral differences coinciding with believed stimulation status. Overall, the patient was more talkative and alert when cortical stimulation was off, as supported by both human observation and computational analysis of video and speech from the recorded interview. It appears that in the short term the cortical stimulation could actually have made them more subdued and perhaps even sleepy, with increased yawning and periods of eye closing coinciding with the stimulation on state as well. Still, it is entirely possible that these effects were primarily caused by belief the stimulation was off, and the fact that belief overlapped with stimulation being on could have been caused by something else entirely (maybe even random chance, though that is the less likely explanation at this point). 

Regardless, it is important to underscore that the timescale of these experiments could have had a large effect on the results, and in the context of standard DBS treatments the timescale of stimulation status switching was entirely contrived here. As stated we were unable to do a truly longer timescale causal study with this participant, and it is unclear whether waiting 2 or 5 minutes between status changes before conducting the interview would have made a big difference versus the 1 minute waiting period. Additionally, a longer wait period would need to be balanced with the observed participation fatigue, which contributed to the selection of this time in consult with the MGH neurotherapeutics team, and turned out to be worse than expected. Still, there remains a major potential concern that would have warranted repeated experiments with somewhat different wait periods and number of trials: lagged effects of stimulation status changes impacting downstream trials. Through random permutation, we happened to have a period of three straight stimulation off trials towards the beginning of the experiment, and it is certainly possible that having stimulation off for $\sim 10$ straight minutes could have continuing effects on the subsequent trials. 

It is worth noting that the patient did have cortical stimulation on all day preceding the experiment, and entered the hospital with a subdued demeanor, yet an uptick in observed energy levels was quickly detectable at the start of trial 2 when stimulation was first turned off. The results of the experiment were thus quite salient despite being very preliminary, and they suggest that future such studies might consider including trials like this one earlier on in their timeline, so they can be properly repeated with different parameters to robustly identify the short term causal effects of stimulation. Moreover, the use of facial expressiveness features from recorded video and language features from recorded audio demonstrate the multitude of ways that data from this style of experiment could be utilized objectively and quantitatively. For a longitudinal series of experiments or in a larger trial, there are many more detailed analyses that could be employed beyond what was even done here - including methods discussed in chapters \ref{ch:1} and \ref{ch:2}. 

\subsection{Future directions}
\label{subsec:ocd-future}
The major goal of this case report was to demonstrate the possibilities of our style of methodology, and to work out a number of limitations that can be better addressed by similar studies in the future, a theme throughout the rest of this discussion section. Therefore, a primary future direction is simply to iterate on this work by applying an improved version of the methodology to more patients. One focus area that would make it relatively easier to recruit a participant population would be to report only on the DBS response characterization perspective, tracking behaviors and stimulation paradigms over time in OCD (or depression) patients receiving standard DBS, but not requiring implants with neural recording capabilities nor looking to assess novel DBS paradigms. Such an approach could elucidate clinical mechanisms of DBS efficacy, improve the stimulation parameter tuning process in practice, and in a psychiatry context could pair especially well with study of larger cohorts receiving other more common treatments for the same disease. 

Separately, there remains the goal of linking behavioral features and neural activity, which is of course more challenging to recruit for at present and also more challenging scientifically, due to the early stages that longitudinal naturalistic behavior quantification tech is itself in. Still, another "n of 1" study in this context has the potential to be a powerful starting point, if an OCD participant with plainly observable rituals can be recruited. For a single participant with high value neural data available, it is realistic to tailor behavior classification tech around their clinical profile. There are already algorithms for e.g. accurately detecting hand washing with an Apple Watch, so though they would require verification in a ritualizing context, it is quite tractable to propose quantifying time of all hand washing occurrences in a patient with hand washing rituals. It would even be feasible with the current technology to sometimes trigger an LFP recording in the DBS implant device in response to hand washing detection. Plausibly there could be a device that employs a similar recording practice as modern game consoles, such that a recording from the previous few minutes can always be saved if immediately prompted, but otherwise storage would be kept free by regular deletion of non-prompted recordings. This of course would depend on the extent to which battery life impacts limit the number of recordings versus the storage available within the implant, but if possible it would extend the analysis to be able to include neural activity directly preceding ritual onset. 

Regardless, even with only the four 60 second timed LFP recordings available with the current protocol, having specific times of ritualized behaviors would carry great opportunity. The presence or absence of ritual (as well as how long it had been going on and how long it would continue after) during a particular LFP recording would already be of interest, and during times identified to contain rituals the raw watch accelerometer data might also be aligned with the raw neural recording data for deeper mechanistic analyses. In the current protocol neural recordings were additionally triggered at moments during some site visits, which could be leveraged for related task experiments to be performed. If not detrimental to the patient's health - particularly if the DBS treatment has significantly reduced symptoms - recording of neural activity during prompted normal hand washing and during prompted simulated ritualized hand washing could be compared to neural activity during a true hand washing ritual. Further, the wrist accelerometer data from these trials could be compared to evaluate objective motor differences between the distinct hand washing contexts, in order to contextualize any observed neural differences. A major downstream goal of such work would be not only to improve neurobiological models of OCD, but also to provide a framework for future evaluations of and improvements to DBS. Particularly for OCD, one could image a scenario where real-time behavioral feedback based on ritual detection might be used to trigger stimulation changes for improved outcomes. 

More generally, ritualized behavior detection in OCD is one of the more tractable and hypothesis-driven goals that digital psychiatry could attempt to tackle at this time. Physical rituals are often frequent and/or prolonged, and they are often externally observable behaviors with a clear beginning and end. At the same time, they are themselves a core disease symptom that can inherently cause distress. There are not many psychiatric disease symptoms that directly (and definitionally) map to observable behaviors and occur frequently enough to be a tractable detection problem. Ritualizing behaviors are thus well-suited for digital phenotyping methods, and moreover OCD patients will experience fluctuations in other symptoms like general anxiety levels too -- which means that targeted ritual direction could be paired with more generalist passive sensing techniques, thereby enabling detection of symptom changes at multiple timescales in a strategic way. As mentioned, for a patient of particular note or for rituals that are especially common (most notably hand washing), supervised methods for detecting the behavior in question are very realistic. Certainly context-specific questions would need to be worked out, but that is a major purpose of research in the first place. One could then explore other digital psychiatry metrics alongside the characterization of rituals, though ritual labeling in itself could be highly useful in a clinical trial setting or in a study using neurobiological markers. Beyond this, there are many interesting questions that could be asked about unsupervised approaches to detecting ritualized behaviors. Motif finding algorithms within raw (or transformed e.g. cleaned spectrograms) actigraphy data might allow for identification of specific rituals based on the movement data alone, eventually facilitating detection of a variety of ritual types within and between subjects. This is of course an exploratory aim, but more tractable variations could also be explored. Existing methods for detecting repetition in signals can be utilized to test engineered features for scoring abnormally repetitive movements within watch data. Such metrics if successful could not only be used as a heuristic for estimation of ritualizing behavior times, but it might also be adaptable to characterize specific types of symptoms in other disorders of interest, for example stereotypy in ASDs. \\

Not all OCD patients demonstrate physical rituals, and not all physical rituals would be feasibly detectable with typical passive sensing technology, but it remains a significant symptom for many. The fact of the matter is that all psychiatric diseases as they are presently defined will not be wholly capturable by any study with specific aims that meaningfully narrow the analysis goals or the collection modalities, but that does not mean this type of study should not be conducted -- it will in fact be an important piece in pushing the field forward. It should be kept in mind though that longer term success with digital phenotyping is more likely to occur when starting with a problem space it already maps well to, for both political and scientific reasons. By demonstrating concrete results that utilize tech to accomplish something not previously possible and something with near term clinical relevance for at least some patients, this will buy exploratory works more time before they fall out of favor for being associated with a field that is not going anywhere. More importantly, it can inform those exploratory works on a variety of best practices in data collection logistics, patient interaction, generalization across institutions, and so forth, and it can also open the door for digital psychiatry datatypes to be used more frequently in exploratory neurobiology contexts, especially through the connection of OCD with DBS. As an interesting but relatively tractable and practical problem, ritual detection additionally has the potential to draw some additional machine learning collaborators or software engineers to the field, which would pay long term dividends in shaping the scientific culture and ultimately how the less well-defined problems might be addressed. 

I suspect another issue directing most digital phenotyping work away from OCD is that it is not the most popular funding topic. While there are other clinical behavior quantification problems that could fill a similar role as the proposed OCD work, I can't think of any that are less niche within psychiatry. The most "popular" diseases are some of the least clearly defined and/or least presently well understood, and because digital psychiatry itself remains a small portion of funding, it is problematic but unsurprising to find little at the intersection of niches. It is imperative in my opinion that some space be carved out for research that focuses on the most suitable starting points for digital psychiatry rather than the most ambitious ones, and for there to be a reasonable amount of diversity between these extremes. Work focused on ritual detection in OCD would be a promising step towards this direction.

\subsection{Contributions}
\label{subsec:ocd-contrib}
In sum, this case report is a first of its kind look at passive digital phenotyping data collected in conjunction with naturalistic neural recording data from a deep brain stimulation trial over 2+ years. It is also one of the first studies to employ digital psychiatry measures in evaluating a novel DBS paradigm for OCD -- in the first patient to pilot this paradigm. Despite the challenges encountered with the patient's clinical profile and the efficacy of the new treatment itself, we were able to demonstrate a number of advantages of this experimental approach, as well as establish a number of lessons for related studies to reference in the future. \\

\noindent Within the project, I cleaned, analyzed, visualized, and interpreted all of the digital psychiatry data, and designed and oversaw the causal stimulation experiment. My contributions through this chapter include:
\begin{itemize}
    \item An overview of pros and cons of different digital psychiatry datatypes within the background of section \ref{sec:background3}, paired with a variety of pitfalls to avoid and considerations to make for future studies preparing a similar data collection process within the discussion of section \ref{sec:discussion3}. 
    \item A demonstration of the potential power of digital phenotyping techniques to capture behaviors of interest that require a naturalistic setting or better temporal resolution than typical clinical scale ratings can provide. Through the example methodology for qualitative review of multimodal data at multiple timescales, I observed:
    \begin{itemize}
        \item In the bird's eye view, a week-long stretch of highly abnormal and otherwise unexplained geolocation data directly after the patient was unblinded to stimulation status (Figure \ref{fig:ocd-report}). This participant spent the vast majority of their time at home overall, but spent multiple straight long days at religious institutions immediately following the cortical stimulation status reveal.
        \item Zooming in on passive sensing data from over the course of individual days, a pattern where on certain nights the patient would briefly unlock their phone nearly every hour and sometimes multiple times per hour, indicating periods of seriously disrupted sleep (Figure \ref{fig:ocd-zoom-sup1}). This amount of phone usage would register as a small amount during the day (for most people), so uncontextualized daily summary features would not pick up on it.
        \item Looking at topics of particular interest to the patient based on their choice of focus in audio diary submissions, a potential confound in the stimulation trial became apparent. The patient was prescribed ADHD medication by an unaffiliated clinician, and they often made a point of talking about their Adderall and how much they felt it helped them. For a period of about 4 months in the middle of the study, the patient stopped mentioning Adderall and started more often bringing up Vyvanse in their recordings (Figure \ref{fig:ocd-lang}). A severe mood episode occurred towards the end of this period (Figure \ref{fig:ocd-smooth}), and then Adderall mentions picked back up again.
        \item Interestingly, even basic journal metadata displayed meaningful changes during this period, preceding the episode of low mood and shifting back to normal shortly following it -- submission hour became much more variable and diary word counts dropped off. Most of the lowest sentiment diaries also occurred in this period. This underscores the relevance of audio diaries, discussed at much greater length in chapter \ref{ch:1}. Further, it highlights that seeming drops in engagement partway through a study may not be pure participation fatigue, but rather could be a clinically relevant behavior (and regardless could rebound without interceding). 
        \item Note as well that these datatypes are much easier to collect, and it is often easier to add additional modalities once already using this study format due to multiple functionalities of e.g. Beiwe. In addition to the much greater temporal resolution that digital psychiatry formats can provide, we can often use one modality to fill in certain gaps in another when missingness arises. We can even continue to collect such data in the face of unprecedented extenuating circumstances: in this case, the onset of the COVID19 pandemic (Figure \ref{fig:ocd-report-covid}).
    \end{itemize}
    \item Suggestions for and examples of core analysis components that could improve tractability of quantitative approaches for such longitudinal multimodal studies, including:
    \begin{itemize}
        \item A patient-specific hierarchical clustering approach to clinical scale items, to capture the nuance of variability in different symptom domains while maintaining a reasonable scope for the number of labels considered. This is particularly important due to much lower sample sizes of gold standard clinical ratings. As demonstrated here, there should be a patient-dependent component: because the YBOCS is typically grouped into separate obsessive and compulsive subscales, but for this patient similar obsessive and compulsive items instead grouped together, thereby revealing variation over time in \emph{resistance} to OCD symptoms that would have otherwise been missed (Figure \ref{fig:ocd-smooth}).
        \item Utilizing self-report survey scores to bolster interpretations of clinical outcome relationships, which are difficult to otherwise explore with digital phenotyping features due to the sample size mismatches. For examples, total number of phone unlocks over the course of the day showed a promising potential relationship with same-day YBOCS \emph{resistance} score, but it was a relatively small number of points and not significant after multiple testing correction. The fact that the same phone usage feature demonstrated a significant negative correlation with same-day EMA summary scores over many more days (and spanning somewhat different parts of the study) was thus comforting.
        \item Considering correlation structure within the passive sensing modalities to not only better understand the dataset, but also as a potentially interesting feature in itself, akin to functional connectivity profiles. More broadly, context from across modalities can be highly informative in identifying behaviors, e.g. using geolocation to inform classification of physical activities. 
        \item Taking greater care to visualize relationships of particular note in scatter plot form, and to cross-check those relationships with other features. There are many reasons a relationship could be "real" in a statistical sense but not "real" in implication. Here, the highly relevant neural feature of corticostriatal coherence was significantly correlated with same-day GPS features like time spent outside of the home (Figure \ref{fig:ocd-neuro-corrs}), but this relationship was actually the result of a bimodal coherence distribution caused by stimulation frequency changes which was inadvertently confounded by the unblinding. The aforementioned placebo effect period occurred during a number of the highest coherence days, driving the quantitative relationship.  
        \item Performing causal experiments to supplement the observational work wherever possible, which here corresponded to brief periods of blinded cortical stimulation status with accompanying recorded interviews (Figure \ref{fig:ocd-video}). The recordings allowed for objective analyses of facial expressions and language to more thoroughly support what was reported by the patient and observed by the interviewer. Interestingly, the patient seemed to be capable of telling the different between cortical stimulation statuses, but consistently reported the opposite to true status. Given that they were also more talkative and appeared less tired during the stimulation \emph{off} trials, this might suggest that they felt better in the short term without cortical stimulation.
    \end{itemize}
\end{itemize}
\noindent It is important to note that this work encountered a number of challenges that limited the potential scope of analysis for the collected data. There was difficulty with recruiting eligible subjects, the patient that was recruited displayed minimal physical rituals which eliminated much of the original analysis plan, and then there were delays in setting up collection of many of the digital psychiatry modalities (further exacerbated when striatal stimulation was turned on immediately for medical reasons). Once data collection was underway, it turned out that for a number of features (including some clinical ones) the patient did not display much meaningful variation, and ultimately did not respond to the novel cortical stimulation paradigm. The only documented clinical effects were a response in depression symptoms to the striatal stimulation that began immediately after implant, and an apparent placebo effect where the patient self-reported feeling much better only after unblinded to cortical stimulation status \citep{Olsen2020}. 

Of course it is also the case that the stimulation did not have the intended effect on corticostriatal coherence, and in fact inadvertently increased it; this increase was initially very high, which caused a major settings change to occur shortly after unblinding, and even after the frequency adjustment coherence remained significantly above baseline. The unexpected impact of stimulation on coherence highlights the need for continued work on understanding synchronization in the brain, through both biological study and model formulation. The work to be presented in the next chapter (\ref{ch:4}) uses control theory tools to study robustness and interpretability properties of artificial neural networks, tools which can also be used in mathematical models of systems synchrony. Control theory on the whole is of strong potential utility for understanding and improving on DBS systems, and a long term vision of the style of work here is to develop closed loop DBS paradigms that integrate behavioral measures in determining stimulation inputs. \\

In the end, this study was always intended to be a highly preliminary work as an 'n of 1' case report. It has therefore accomplished a number of its goals in characterizing the opportunities provided by longitudinal and multimodal digital psychiatry data collection, in addition to identifying successful early techniques and a variety of practical roadblocks to consider for future works. It lays important groundwork for the extensive use of passive sensing and self-report technology to improve clinical trial evaluations and to contribute to new neuroscientific insights. Digital phenotyping is especially important for the latter, because without developing such tools we will not have relevant and validated behavioral measures at an appropriate timescale for analysis with human neurobiological readouts. This is already true, and as emerging neurotechnologies steadily improve we will encounter a point where the biggest bottleneck is behavior -- unless research building on this case report and similar lines of questioning becomes more commonplace. 

Indeed, the utility of methodologies like the one outlined here is abundantly apparent with just a bird's eye look at the data (Figure \ref{fig:ocd-report}). If this were a typical DBS trial, only the 44 bi-weekly scale time points containing 22 discrete ratings each would have been available to evaluate and attempt to understand any potential clinical effects of stimulation. Instead we were able to characterize 627 unique days with multiple different features usable down to the minute level. We covered more than 10,000 hours of the patient's life over the course of the trial, with the flexibility to capture detail of aberrations in e.g. nighttime behavior that would often be overlooked or poorly remembered in a traditional trial setting. It is an important message to spread that these techniques hold both great long term promise as well as potential for straightforward short term improvements, and that it is feasible to start using and iterating on them now, just as we have here.  

\subsubsection{An aside on the McLean OCD Institute}
\label{subsubsec:ocdi}
Prior to COVID, a goal of this project was to launch a study of ritualizing behaviors in a larger patient population at McLean's OCD Institute (OCDI), in order to develop relevant actigraphy metrics for estimating ritual severity, in at least some subset of patient profiles. Such metrics then could have informed digital phenotyping in future DBS trials, like the one forming the basis of this chapter. 

Ideally, we would have recruited a full dataset of $\sim 20$ participants with physical rituals in residential rehab at the OCDI. Those participants would have undergone a period where they pressed a button on the watch directly following all ritual occurrences as best they could, and then would have worn the watch passively for a longer period, through the study filling out short daily EMAs about ritual frequency and duration in the last 24 hours. We also would have conducted a brief recorded interview with each participant to ask them for more details about their ritualizing behaviors, perhaps with a visual demo. Unfortunately, even pilot data collection was hardly able to get off of the ground before I had to formalize the contents of my thesis (and focus on those analyses).  

One hope that I still maintain for that project is that metrics of repetitive movement derived from wrist accelerometry - perhaps somewhat feature engineered or perhaps learned in an unsupervised manner via e.g. motif finding algorithms - could provide usable estimates of onset and offset times for ritualizing behaviors across a decent variety of physical ritual characteristics. This would be similar in spirit to the themes of personalized modeling components in chapter \ref{ch:1}, via a procedure that could fit individual ritual nuances while simultaneously covering a meaningful subset of OCD ritual manifestations. Such a tool would enable a variety of downstream scientific and clinical studies like described in the future directions above, as well as holding some inherent technical interest.

To facilitate investigation into generalizable ritual detection algorithms with personalizable components, it would be useful to supplement the more deeply labeled proposed OCDI dataset with a larger dataset of OCD patients (and controls) contributing mostly unlabeled wrist accelerometry signal. Towards that end, the UKBiobank contains data from over 1000 OCD patients, and also has a large set of week-long actigraphy signal from a wide variety of their participants. Nothing else specific is known about those days, but a number of other datatypes are available describing these patients more broadly, including demographic info and diagnostic notes. The sheer size of the accelerometer dataset available there could go a long way towards enabling exploratory unsupervised learning and feature search approaches whilst maintaining rigor in the more exhaustive characterization of the obtained OCDI dataset.

At the onset of the COVID19 pandemic, I did not anticipate just how long clinical research would be disrupted for. As mentioned, digital psychiatry can enable a number of fully remote study options, and of course the UKBiobank dataset could be analyzed entirely remotely. On the other hand, the intended OCDI study was an important piece of the proposal that would not have been feasible to begin full data collection for up until very recently. Despite the setbacks, I did continue to perform a number of duties through spring 2022 to further set up this OCD project to occur down the road, and in fact actigraphy data has now been collected for a new pilot participant from the OCDI at the time of writing, with recruitment now ongoing. 

\noindent My contributions towards enabling this research to happen include:
\begin{itemize}
    \item Drafting a full application for the desired UKBiobank data including specific analysis proposals and needed logistical information, and seeing it through to eventual data access being granted. At that time I also compiled resources on how to download various UKBiobank datatypes and query the dataset for particular participant subsets of interest, as well as documented expected needs for compute power and storage space in order to carry out intended analyses, thus setting up the lab to utilize the obtained data more easily.
    \item Collecting many weeks of continuous bilateral wrist actigraphy (GeneActiv) data from myself, with labeled times of potentially notable naturalistic activities including various grooming behaviors, as well as a snippet of intentionally repetitive simulated ritualizing behaviors. I passed these data, along with some earlier ritual simulation accelerometry data from OCDI staff, on to a high school student I was supervising -- who collected some of her own naturalistic watch actigraphy data too. The personal datasets provided a chance to perform some initial experimentation on possible supervised detection of behaviors commonly included in rituals, and to mentor a student who contributed some actigraphy analysis code back to the lab. A summary of that work can be found in supplemental section \ref{sec:lacey}.
    \item Writing code to pull OCDI patient information from REDCap and ultimately automate a number of the OCDI's weekly patient tracking and reporting tasks, as well as generate an ongoing broader census of OCDI enrollment. This has saved many hours of RA time from performing boring, repetitive administrative tasks, and it improved the outlook of our own research plans by not only bolstering the Baker Lab's collaborative relationship with the OCDI, but also providing regular, easily accessible updates on the patient population for potential study recruitment. More generally, there are likely a large number of menial record keeping tasks that could free up quite a lot of hospital research time if automated properly. With software engineers on staff, it would be feasible to identify those tasks that are safely and tractably automatable. \newline Note that my code in particular serves a variety of functions using the pulled REDCap data, and thus assists a variety of people affiliated with the OCDI. These primary functions are:
    \begin{itemize}
        \item Generating PDF reports with information about weekly assessment values (e.g. YBOCS, HAMD-6, PSQI, etc. collected via patient self-report) over time for each patient (using matplotlib and reportlab python packages), along with other relevant metadata about the patient and their reported medication use while at the OCDI. These reports are used as part of clinical evaluation over the course of rehab program enrollment. The reporting code runs on a weekly basis on the latest REDCap information, every Tuesday morning.
        \item Ensuring those PDF reports get shared with OCDI-affiliated clinicians, through secure upload of the latest reports (clearing any old ones) to Partners Healthcare Dropbox, generation of temporary shareable Dropbox links, composition of all active patient report links mapped to corresponding first name in an easy to read HTML body, and finally automatically sending that HTML body out as an email to an OCDI mailing list using Outlook's SEND SECURE functionality.
        \item As part of the reporting code, updated summary stats are similarly generated to be emailed to those managing OCDI clinical enrollment or recruiting from the OCDI for their research study. This census update email includes distributional information about program type (residential, virtual, or day), research consent status (for use of the weekly assessment responses in active research projects), and assessment response rate (e.g. compliance in filling out the aforementioned weekly assessment values). The stats are provided across all active OCDI patients, broken down separately for newly admitted patients within the last week, currently enrolled patients, and newly discharged patients within the last week. A subset of these stats are also provided from across the history of the OCDI's current research program protocol. 
        \item When constructing the census during the weekly reporting run, the code additionally compiles a final email that lists the individual ID of the specific active patients who did not respond to any of their most recent weekly assessment prompts, also noting which of those patients were admitted or discharged within the last week. This email is sent to the RA that manages the patient REDCap records, to assist in ensuring assessment compliance is as good as possible. 
        \item Besides the weekly reporting on Tuesday, I have also written code to manage prompting for the latest weekly assessments each Thursday morning. Time is set aside on Friday mornings for patients to fill out these assessments, which they have always done via REDCap forms issued by a unique link accessed from their email inboxes. With the new code, RAs no longer need to manually put together each individual patient's prompt email, as it is automatically composed with the correct REDCap link for the current patient and week and then automatically sent to the corresponding email address. To ensure any bugs (whether in the code or in the initial REDCap data entry done manually at enrollment time) don't cause weekly assessments to be missed needlessly, an easily readable log of the code's main outputs, with warnings at the top if there were any, is emailed to the relevant RAs each week at the conclusion of the script. That log can be briefly reviewed each week so manual intervention can still occur when necessary. 
    \end{itemize}
    \item Managing bi-weekly meetings between OCDI research staff, Baker Lab members, and new collaborating groups, including regularly preparing a meeting agenda and taking detailed meeting notes on topics ranging from regulatory and recruitment concerns to troubleshooting and expanding on software tools being used by the OCDI (most commonly those written by me). 
    \item Maintaining the IRB protocol approved for the proposed collection of wrist actigraphy from $\sim 30$ subjects to be recruited at the OCDI, which was originally set to kickoff in early spring 2020. As part of maintaining this protocol, I also updated it to accommodate possible pandemic-era research changes, the needs of new collaborative funding sources, and general improvements to the proposed procedures based on scientific aims written in the process of drafting the UKBiobank application. When it came time to begin preparing for new study recruitment, I assisted with other administrative work too, including updates of Rally advertisements and communicating with onsite OCDI staff about the recruitment/consenting workflow. 
\end{itemize}

\chapter{Recursive construction of stable assemblies of recurrent neural networks\footnote{Portions of this chapter were co-authored with \cite{NIPS22} for our jointly lead manuscript published at NeurIPS. A substantial amount of extra background and discussion was written by me to add to the chapter for my thesis. See Appendix \ref{cha:append-clarity} for detailed attributions.}}\label{ch:4}
\renewcommand\thefigure{4.\arabic{figure}}    
\setcounter{figure}{0}  
\renewcommand\thetable{4.\arabic{table}}    
\setcounter{table}{0}  
\renewcommand\thesection{4.\arabic{section}} 
\setcounter{section}{0}

The preceding chapters all focused on laying groundwork for future digital psychiatry studies, providing discussion of and methods for data collection and monitoring across a number of modalities, including considerations for large collaborative dataset building initiatives such as the AMPSCZ project described in chapter \ref{ch:2}. They also characterized a variety of processing tools for automatic quantification of many different features of prior interest in the psychiatry literature, providing a comprehensive picture of how these features can be leveraged in the digital age. Towards this end, such features were analyzed in a few different novel contexts, including daily audio journals from Bipolar disorder patients in chapter \ref{ch:1} and passive sensing signals from an OCD patient undergoing a deep brain stimulation (DBS) clinical trial in chapter \ref{ch:3}. 

With just a psychiatric science perspective, much can be accomplished via digital tools, and many interesting future directions were identified along those lines in the associated chapters. However, emerging computational methods also open up entirely new scientific approaches to psychiatry; approaches that not only have the potential to revolutionize clinical practice, but even to potentially identify features of psychiatric relevance that remain wholly unsuspected at present. At the same time, there are many difficult challenges in applying machine learning (ML) to digital psychiatry data. While the data processing techniques mentioned can facilitate certain ML approaches, and some are themselves utilizing ML, applications of ML to psychiatry largely either remain far from the cutting edge or attempt to use "state of the art" (SOTA) models inappropriately. As such, it is crucial to make clear the unique challenges facing ML for healthcare and especially ML for psychiatry. 

The work in this chapter develops novel deep learning theory and demonstrates strong performance of a novel recurrent neural network (RNN) architecture derived from the theory on benchmark sequence learning tasks. Within the background of section \ref{sec:background4}, I will characterize many of the specific roadblocks that ML for psychiatry projects face, highlighting the points that our RNNs could help to one day address. In particular, they are provably stable, a property that is not often guaranteed of RNNs, but is highly important in safety-critical applications. Furthermore, they perform very well on benchmarks relative to their size, which along with the multi-area structure we impose lends itself to network interpretability -- something that is generally quite difficult for SOTA models. 

More broadly, the work in this chapter represents a strategic combination of concepts from control theory, machine learning, and neuroscience. We utilized control theory to prove novel stability conditions for continuous-time RNN models of interest for theoretical neuroscience, and then adapted those results to a deep learning optimization context for proof of concept ML application. Throughout that process, prior results in neurobiology were utilized to inspire architecture design decisions and eventually interpretation of our results. Different expansions on our work will focus both on improvements to machine learning applications and on addressing open questions in neuroscience theory (as will be covered in the discussion section \ref{sec:discussion4}).

Interestingly, the control theory tools described in this chapter could be very well suited towards improving models of human behavior, even outside of the scope of directly modeling the brain. Thus the theoretical background provided has what is in my opinion an underrated relevance for the field of computational psychiatry. Additionally, there are opportunities for close connection between theoretical neuroscience and neurological practice in the near future if control theory is well employed, as the described contraction analysis methodology has immediate applicability to predicting synchronization of systems. Recall that one major hypothesis for the efficacy of DBS in OCD is the disruption of corticostriatal hypersynchrony, yet the novel dual site stimulation paradigm evaluated in chapter \ref{ch:3} accidentally increased synchrony instead of decreasing it. With a deeper understanding of control systems, design and understanding of DBS paradigms would become a more tractable problem space. \\

\noindent \textbf{An outline: } After establishing the necessary machine learning background and connecting it to the problems addressed throughout the rest of the thesis, I will provide a more comprehensive overview of the close bidirectional ties that exist between all of psychiatry, neurobiology, control theory, and machine learning. This serves to further justify the strong potential in exploring ideas that can be unified across the four domains, setting up the rest of the chapter. To close the background section (\ref{sec:background4}), I will then review in greater detail a few specific neuroscience topics of interest and the open questions that remain, alongside introducing those aims that we specifically addressed in \citep{NIPS22,sparsegit}. The most critical themes covered include the role of stability in neurobiology and the emerging study of multi-area RNNs to facilitate theoretical understanding of interacting brain regions.

To begin the core content of this chapter, I will focus on contraction analysis, the key nonlinear control tool used to obtain our results. Section \ref{sec:math-intro} provides additional conceptual motivation for and a technical overview of contraction analysis, as well as more concretely defining the neuroscience theory models we work with and the prior mathematical results existing for the nonlinear case that is of particular salience. Our novel stability conditions for the nonlinear network are then presented in section \ref{sec:math-results}, in addition to counterexamples for some previously published "conditions". 

Once the stability results for the single nonlinear RNN model are covered, I will next present stability results for the provably stable "RNN of RNNs" model, which we constructed out of individual networks meeting our conditions, using principles of stable combination derived in contraction analysis. This is the other major theme covered in the mathematical results of section \ref{sec:math-results}, and will also include some early theoretical work on the inverse consideration of neuron pruning. 

Finally, the last major result of this work - the empirical success of stable "RNNs of RNNs" on sequential processing benchmarks - will be presented in section \ref{sec:dl-results}. That section will first cover the extension of the theoretical results, which are conditions on static network weights in continuous time, to the practice of deep learning, which requires a discrete optimization technique that will maintain the stability conditions throughout training. I will then report on the experimental results that we obtained, not only demonstrating a high level of performance for the architecture, but also analyzing the effects of different design choices on performance. Most notably, sparsity in component RNNs was associated with a strong boost in test accuracy in our stable multi-area architecture.

At the end of the chapter (section \ref{sec:discussion4}), I will discuss the broader implications of our work, as well as many related future directions of promise. The initial discussion will be more technically focused, but I will end with high level conceptual takeaways from our contributions. Within the concluding chapter, I will present a broader outlook on future intersections between neuroscience and deep learning, both scientifically and meta-scientifically. \\

\noindent Ultimately, we contributed mathematical results in the field of neuroscience theory, and then showed how those results could be leveraged in a deep learning context to produce relatively small vanilla RNNs that are both provably stable and capable of competing with larger, more engineered, and not necessarily stable architectures on benchmark sequence learning tasks. It is highly informative when theoretical results align well with subsequent empirical performance, and taken together with the neurobiological priors that drew us to multi-area networks in the first place, I believe we have made a strong case for a wide range of future research on architectures like the SparseComboNet presented. That work has the potential to not only inform computational neuroscience as well as deep learning practice, but also to address some current concerns in the space of applied machine learning for psychiatry -- and do so much better than most lines of research on artificial neural networks could.

\section{Background}
\label{sec:background4}
As mentioned, I will begin the background by motivating my recurrent neural network research within the themes of the rest of the thesis, reviewing the challenges of using deep learning tools for digital psychiatry data at present (section \ref{subsec:technical-difficulties}). I will then focus on the technical background needed to understand the work and to ultimately connect it back to broader questions in neuroscience. Thus in section \ref{subsec:neuro-to-ml}, I overview the historical importance brain and cognitive sciences have had in the formation of AI/ML breakthroughs, and provide a primer on the technical details of deep learning most relevant here. Conversely, I give examples of the various ways deep learning has been used to further neuroscience research in section \ref{subsec:ml-to-neuro}, and then in section \ref{subsec:control-intro} I tie both these fields together with control theory, highlighting concepts most salient for understanding our RNN architecture. Finally, I will close the background with a discussion of open questions at these interdisciplinary intersections, with a focus on those questions that \cite{NIPS22,sparsegit} aimed to address (section \ref{subsec:rnn-motivation}).

\subsection{Challenges in applying machine learning to digital psychiatry data}
\label{subsec:technical-difficulties}
Deep neural networks (DNN) are a machine learning method that iteratively fits a composition of non-linear functions to a labeled training dataset. They can be represented as layers of “neurons”, where each layer takes input from the previous, and each unit generates output by applying a non-linearity to the weighted sum of its inputs. The use of multiple hidden layers makes learning a complex function much more plausible, as shallow networks would need an unreasonable number of units in a given layer to achieve the same fit. One reason for the recent successes of DNNs is that they are fairly resistant to overfitting despite having very high model complexity. Given the nature of much data in biological sciences -- high dimensional, heterogeneous, temporally dependent, sparse, and irregular -- DNNs provide a potentially revolutionizing method for the field. Deep learning has already been used successfully in pilot healthcare trials, particularly for certain medical imaging applications; it has also facilitated a number of new basic research directions in the biological sciences, for example in the field of computational genomics \citep{Miotto2018}. More recently, the AlphaFold neural network architecture introduced by \cite{jumper2021alphafold} soundly beat prior academic work on protein folding prediction, entirely shifting the direction of that subfield away from the previous contest dataset that used to drive much research.

It is clear that deep learning has massive potential to eventually revolutionize medicine, and for some digital psychiatry signals like wrist actigraphy there are certainly industry groups already applying neural networks to parse huge amounts of input data (though not in a psychiatric context). However, we have seen quite underwhelming results to date when these tools have been applied to many biomedical datasets, for example electronic health records or fMRI scans, and many of the limitations are further amplified when dealing with psychiatric disease. Challenges arise both in how interdisciplinary healthcare researchers apply modern machine learning tools and in the properties of the tools themselves. Thus in this section, I will overview many of these challenges on both sides, and for those challenges directly related to recent deep learning architectures, I will discuss how our "RNNs of RNNs" work could help to attenuate concerns. \\

\noindent Problems in applied ML for psychiatry at present include:
\begin{itemize}
    \item Psychiatry datasets invariably include personal medical information, with content that is at times highly sensitive. Due to these very real privacy concerns, it is difficult to build large datasets for shared research community use or even to share data for reproducibility purposes. Depending on the consent forms used for the study, there might be limitations on involving collaborators with varying expertises from other institutions as well, and when it is permitted the onboarding process can be slow with a number of logistical hurdles. Similarly, it may not be possible to utilize certain tools to analyze the data, for example external APIs that involve sending any data to a different server (e.g. using Whisper through OpenAI's API) will each require their own approval process that very well may be denied. For hypothetical models that might make it into the clinic in the future, it will also be critical to ensure that sensitive personal information used to learn cannot be extracted from within the trained model by a determined adversary. Even now, it is unclear whether it is appropriate to share certain trained models with the research community, and these are not particularly large nor well performing networks. 
    \item When it comes to any ML project that aims to directly progress clinical practice, there are also major potential safety concerns - concerns that are challenging to assess for models that largely amount to a black box. Incorrect predictions could have meaningful medical implications, and if the rates of different types of incorrect predictions are different for a model than for a human clinician, there will need to be careful ethical discussions about e.g. systemic biases. With complex models it may not even be feasible to identify in advance the sorts of mistakes the system will be predisposed to though, especially as generalization from individual medical studies to the broader population is itself a difficult topic. An additional major shortcoming of most current deep learning models closely related to this problem is their inability to accurately assess their own certainty levels, as well as their susceptibility to adversarial attacks.  
    \item Along the same lines as safety concerns, but of importance for present research too, is interpretability concerns. With the substantial number of open questions remaining in psychiatric science and the problematic nature of some of the currently available "ground truth" labels, it is crucial to the advancement of the field that computational modeling studies at large do not eschew explainability (see the introductory chapter for more on diagnostic categories in psychiatry). While understanding neural network behavior holds inherent interest, it is not critical for an image recognition or language translation application to be interpretable. However, those are not only generally lower stakes problems, they are also problems where the primary aim is task performance. In digital psychiatry there could be room for some studies with such an aim, but broadly speaking it would be hard to call accurate psychiatric disease diagnosis prediction a foundational scientific success, if it does not involve any explainability and quite possibly comes at the expense of resources for more fundamental research. In theory, neural networks have great potential to teach us a lot about a dataset, but interpretability tools are mostly lacking at this time, and many of the best performing cutting edge models are very large and employ a network structure that does not help much in understanding their behavior. 
    \item As mentioned, there is a core problem in framing digital psychiatry ML research that centers around the lack of clear ground truth. This has far-reaching impacts beyond interpretability concerns, as it makes it difficult to construct a good supervised learning dataset in the first place. Some of the potential non-ideal options for labels include:
    \begin{itemize}
        \item Diagnosis, which is a generally accessible and understood label, but is a static label with a host of problems in meaningfully distinguishing people -- including both groups that are very different under the same label and groups that are quite similar under different labels (exacerbated by high comorbidity rates, rates that are likely partially explainable by the poor diagnostic definitions). 
        \item Clinical scale symptom ratings, which are not exactly a perfect ground truth but are a reasonable gold standard today. However, these ratings are time-consuming and expensive to obtain, and as such it is difficult to build large datasets across patients, as well as practically impossible to obtain more than one data point per patient per week. Realistically, this will be a monthly label in most studies, and it will involve many fewer patients than could have been recruited for a phone app only study. Furthermore, much of the advantage in the clinical scales is the symptom-specific breakdown by item score; yet with a quite small set of labeled timepoints, it is not tractable to leverage the full range of item scores in modeling.
        \item For certain disorders, there are potential objective and externally observable markers of extreme disease severity: for example, a suicide attempt in a depressed patient or a Bipolar patient hospitalized for manic behaviors. Such markers are rare occurrences over a longitudinal study even for patients with severe disease however, and many patients with less severe disease will never experience them. This limits the scope of the labels, and perhaps more importantly makes for a very difficult problem even if the study population is restricted to those at most risk. Rare event prediction within a dense, longitudinal signal is an active challenge for machine learning when dealing with clean benchmark datasets, let alone to tackle such a messy real world problem.  
        \item As healthy habits like exercise and socialization are related to disease severity and probably have a bidirectional causal relationship with it, there is an argument for instead tracking health-related behaviors using digital psychiatry tools and taking that as a sort of ground truth for functioning level. Although this may be appropriate for some studies (it certainly has the potential to help people), it is not at all capable of disentangling psychiatric mechanisms, and would thus struggle to further scientific models within psychiatry past a point. 
        \item Patient subjective experiences with psychiatric symptoms (and daily life) can be rated quantitatively via survey or can be provided qualitatively via journaling on a regular basis using current digital psychiatry apps. These can be a reasonable label for individualized models over long time periods, but they are difficult to compare across different people due to their heavily subjective nature and other possible reporting biases. They also may be difficult to utilize even within the individual for some people, as reliability in seriously answering these prompts as well as ability to accurately introspect and remember recent events can be quite different across people. 
        \item Related to a few of the above points, it is also plausible that "common sense" metrics like the sentiment of language used in daily diary recordings could serve as a label for better understanding less interpretable digital psychiatry features. If well-validated technically (e.g. ensuring sentiment scores align with sentiment ratings from clinicians in a random sample), this could be a useful path for some research. However, this framework suffers from a chicken/egg problem if it is not coupled with other label sources, and "common sense" is not exactly a concrete definition for selecting such metrics anyway.
    \end{itemize}
    \item On the input side, there are also technical and logistical challenges towards defining a tractable prediction problem. When taken together with the often small sample size of labeled timepoints, these issues include:
    \begin{itemize}
        \item In practice we often end up with temporally dense data streams that contain a variety of different missingness patterns, and those patterns may or may not be disease relevant. How to best deal with data missingness remains an open topic in applied machine learning, and because the data are also much denser temporally than the available labels, deciding how to handle the multiple possible timescales for input analysis is a complex question. Frequently it results in an unreasonably large number of summary input variables, far too many to retain any meaningful statistical power for the size of the label set. 
        \item It is not yet clear which types of digital psychiatry data will be most relevant for which questions, so studies often involve highly multimodal data collection. This further increases input complexity without improving the labels situation, and when it comes to efficiently leveraging multimodal data sources deep learning is still in early stages of research. On the other hand, naturalistic data can be very messy scientifically, something that multimodal context can help to combat. 
        \item Digital psychiatry datatypes are also often multimodal in a truly interdisciplinary way. While e.g. image recognition and language processing are such major problem domains in modern deep learning, datatypes like wrist accelerometry, heart rate, electrodermal activity, phone usage patterns, and so on (not to mention neurobiological markers such as EEG or fMRI) are much more specialized. These are all either niche interests within machine learning, or interests where progress has been very restricted to company internals thus far (e.g. Fitbit). Although there are certainly fundamental principles that can assist with creating a good e.g. sequence classification model in general, it is often the case that domain specific knowledge about the nuances of the datatype are necessary to optimize performance. It is difficult to find legitimate machine learning experts within digital psychiatry as it stands, but to try to corral advice about processing multiple different relatively uncommon datatypes is an entire additional challenge.
        \item Psychiatric disease symptoms could hypothetically have a direct impact on the accuracy of a number of established ML methodologies with good standard benchmark performance, but this remains largely uncharacterized despite feature extraction using these methods being pervasive in e.g. psychiatric speech sampling "explorations". Furthermore, if this is a broader problem, it is unclear if a typical study would even have enough data to attempt to tune models to address such issues, and it is unclear how these issues might vary across different study contexts. One simple example is that some of the audio diaries submitted by mood disorder patients in the study of chapter \ref{ch:1} contained a whispering voice with extremely garbled enunciation at times, possibly associated with fluctuating symptom severity. Human transcribers were not always capable of accurately transcribing all of the words in these diaries, but they clearly marked their uncertainty. A model like Whisper that was certainly not benchmarked on this kind of abnormal speech could very well insert many incorrect words into the transcript, without denoting the uncertainty and in a way that could bias some of the most clinically relevant transcripts to contain the most egregious errors. 
    \end{itemize}
    \item Trends within and between groups can vary in numerous ways. It is possible for an association to be clinically relevant across people but not within people over time, or vice versa. It is possible for a real association to go one direction in one person, the opposite direction in another person, and not exist at all in a third. It is even possible for an association to exist over time in a person one year and not the next, perhaps due to sustained changes in the environment not captured in the dataset or perhaps due to some sustained effect of a transient intervention. To hope to measure any of this it will require a strong longitudinal data collection protocol -- but even then, how to effectively manage personalized versus overall (and possibly groupings in between) models remains a complex topic, as does continual learning. Furthermore, for some research questions and study populations there will be highly relevant clinical factors that will not be feasible to reliably capture. Substance abuse is one example of such a factor, and more generally even basic medication adherence can be challenging to track, especially when patients are seeing additional clinicians outside of the study's institution. 
    \begin{itemize}
        \item There is some evidence to suggest that the patient in the deep brain stimulation clinical trial of chapter \ref{ch:3} changed to a different ADHD medication at one point in the middle of the trial, which could confound interpretation of the stimulation settings, but this information is not reflected either way in the patient's medical records at our hospital, because these medications were prescribed by an outside clinician.       
    \end{itemize}
    \item Coherence between medical records across institutions is a broader problem in clinical practice, and it is just one issue that could complicate the ability of models trained for a specific study (or even the overarching training methodology) to generalize to other groups. Psychiatry data in particular can be highly susceptible to confounding factors related to culture, socioeconomic status, etc. which will make it even less likely that a complicated ML model will generalize in the ways necessary for wide-spread clinical practice. Large collaborative endeavours such as the AMPSCZ project described in chapter \ref{ch:2} could help to address these concerns, but this is itself an open question that will partially depend on how computational psychiatry researchers choose to proceed with the collected data. 
    \item The field itself lacks a centralized means for communicating results. While certainly not perfect, sites like PapersWithCode help ensure ML researchers are aware of the work most relevant to theirs and the corresponding results that have come before them. Between the poor cohesion and incentives to superficially keep up with cutting edge machine learning, subfields of digital psychiatry like speech sampling are littered with exploratory papers whose "promising features" were never followed up on and do not have enough statistical power to be taken at face value. In addition to the paucity of ML experts meaningfully collaborating within digital psychiatry, it is much harder for psychiatry groups to hire software engineers at the salaries they can offer, so the academic code situation is worse than it is within deep learning -- and psychiatry does not have the same industry research resources helping to push forward the field either.
\end{itemize}
\noindent Ultimately there are some major tractability problems here, with difficulty both in identifying predictors of interest (especially that are grounded in psychiatry, collectible at scale, and rigorous time-varying measures) and in handling the many complex (and dirty) naturalistic input data sources that one might consider. This is compounded by (and perhaps a cause for) the lack of proper ML experts in the space; when coupled with the numerous ethical and logistical challenges that arise for anything that might make it to clinical practice, it is clear the road ahead will be difficult. Deep learning is not currently capable of addressing many of these problems, and in fact can exacerbate some of them. By utilizing deep learning appropriately, there remains great potential for its use in psychiatry research in the near future, but it must be applied intelligently \emph{when the problem calls for it} and not akin to a hammer looking for a nail. 

At the same time, progress in machine learning research that focuses on topics such as neural network interpretability and building efficient network architectures can improve the capabilities of future deep learning methods to tackle these problems head on. It may not appear as flashy to the NIH as e.g. large language models, but this is an underrepresented angle for application of the latest deep learning research to psychiatry at present. The work in this chapter is just one contribution within that subfield of deep learning, but it is an example of relevant research -- stability guarantees are important for robust and predictable RNN behavior, the "network of networks" modular structure we employ lends itself to interpretability, and our architecture was able to perform competitively on sequence learning tasks against a number of architectures $10x$ (or more) the size.  

\subsection{Applying neuroscience to machine learning}
\label{subsec:neuro-to-ml}
The history and development of artificial neural networks is deeply intertwined with brain and cognitive sciences. The basic idea came from neuroscience theory, several major developments in the relatively recent success of deep learning were directly inspired by neuroscience (e.g. the receptive field-esque structure in convolutional neural networks), and several of the major players in establishing deep learning had prior training in psychology or neuroscience. In fact one of the premier conferences for machine learning today, NeurIPS, began as a small gathering for computational neuroscientists. With the practical explosion of machine learning, largely driven by the capabilities of deep learning, most research on neural networks today has its roots squarely in computer science and engineering. However, artificial intelligence had been a largely interdisciplinary field even well before neural nets caught on, with research in the days of expert systems heavily influenced by ideas from domains such as cognitive science and linguistics \citep{sejnowski}.

Although biological plausibility will likely remain a niche subfield of deep learning going forward (insofar as it is considered an inherently desirable property), there remains a rich opportunity space for loosely neuropsychology-inspired ideas to make a strong impact on modern machine learning. Most of the recent paradigm shifts in deep learning have resulted from novel architectural designs, designs that are often motivated by similarities to hypotheses about the brain \citep{sejnowski}. Several of the most popular papers in the last few years have also contained neuroscientific analogies, for example the lottery ticket hypothesis of \cite{frankle2019lth} and its parallel to synaptic pruning. Further, despite the fact that the weight learning process is the most biologically problematic part of modern neural network models, networks have demonstrated emergent properties in their learning process that resemble known properties of the brain (e.g. critical periods \citep{critperiod}), and thus might be leveraged to improve network performance through additional neuro-inspired ideas. 

Conversely, neural networks have demonstrated some behaviors antithetical to human expectations, such as their susceptibility to nonsense adversarial perturbations \citep{Shamir2022}. Inspiration from brain and cognitive sciences has the potential to address these concerns, whether directly through curriculum learning methods \citep{sinha} or architectural designs \citep{GWT} that might improve adversarial robustness via brain-like methods, or indirectly through novel learning algorithms that are more biologically plausible than gradient descent based backpropagation -- something being worked on at present by \cite{hinton}, a "founding father" of deep learning.

Computational and theoretical neuroscience research is thus well-positioned to remain a highly influential factor in the future of machine learning, which is in turn important to the futures of both scientific inquiry and medical practice. In this chapter, I present mathematical results that would be better characterized as neuroscience theory contributions than as deep learning contributions. At the same time, I demonstrate proof of concept results that utilize the theory to improve performance of a deep learning system in practice. The work is ultimately interdisciplinary and cannot be entirely disentangled; as such, themes from both neuroscience and machine learning will recur throughout. To ensure that necessary background to understand the work is established for a diversity of possible audiences, the next subsection will provide an overview of deep learning that includes relevant technical details.

\subsubsection{A primer on deep learning}
\label{subsubsec:neural-nets}
To begin, we will introduce notation used throughout the chapter in terms of a very simple two "neuron" system, which has output equivalent to $\mathbf{W}\mathbf{v}$, where $\mathbf{v}$ is a vector of the unit states and $\mathbf{W}$ is a $2 \times 2$ matrix specifying the values of the weights connecting the units. In particular:

\begin{align*}
\mathbf{v} = \begin{bmatrix}
v_{1} \\
v_{2}
\end{bmatrix} 
\end{align*}

\noindent and

\begin{align*}
\mathbf{W} = \begin{bmatrix}
w_{1\rightarrow1} & w_{2\rightarrow1} \\
w_{1\rightarrow2} & w_{2\rightarrow2}
\end{bmatrix} 
\end{align*}

\noindent The output associated with the first neuron (having state $v_{1}$) would then be, by matrix multiplication: 

\begin{align*}
= w_{1\rightarrow1}*v_{1} + w_{2\rightarrow1}*v_{2}
\end{align*}

\noindent Of course there will be an analogous output for the second neuron, and the overall output will therefore be a vector of size 2 as well. \\

Note that I have not yet included any notion of dynamics in these equations. That is because we will start with the simplest example of a feedforward neural network, a class of architectures that has seen and continues to see great success in a variety of applications. Typical feedforward networks are comprised of layers, where neurons in layer $i$ send outgoing weights only to neurons in layer $i+1$. There is usually an input layer that has weights to "translate" from the format of the raw input to the state of the units in the first hidden layer, and then the values will cascade through the layers until the output of the final hidden layer is passed through a similar "translating" output layer to get to a valid prediction format. When an input is evaluated, it is presented to the network all at once, the math is worked out, and the output is returned; there is no concept at all of a persistent state in a traditional feedforward network, and thus no need for time.

Returning to the example $\mathbf{W}$, we can think of this as a network with 2 hidden layers, each containing 1 unit. As such, the feedforward network must have $w_{1\rightarrow1} = w_{2\rightarrow1} = w_{2\rightarrow2} = 0$. These values will always be frozen at 0, and the trainable value here would be the weight going from unit 1 to unit 2, $w_{1\rightarrow2}$. Similarly, feedforward networks typically do not allow external inputs to the network outside of the initial input to hidden layer 1 connection, so $v_{2}$ would not have any value associated. Of course this does not matter, because the weights that might act on $v_{2}$ in the equation are all fixed at 0 anyway. Likewise, there is no externally utilized output of unit 1 in layer 1 as it is not the final hidden layer, and indeed the first output equation (produced above) contains only 0 weights. 

For simplicity we can consider both the input and output function here to be the identity function, so that a single variable input is given to unit 1 through the state $v_{1}$, and the goal is to predict some other single variable as the output of unit 2. This works out to be $= w_{1\rightarrow2} * v_{1}$, a simple linear regression with no intercept. One could add an additional unit to the first layer with a constant input of 1 in order to get a bias term, as per matrix multiplication the weight from that unit to the layer 2 unit would end up added to the final equation. Regardless, this might seem anticlimactic -- and it is not only because it is such a small toy example. Rather, it is because I made the initial example a linear network. \\ 

Linear feedforward networks are indeed just matrix multiplication. Stringing together layers linearly provides minimal benefit, because for matrices $\mathbf{A}$,$\mathbf{B}$ and vector $\mathbf{x}$, $\mathbf{A}(\mathbf{B})\mathbf{x} = (\mathbf{A}\mathbf{B})\mathbf{x}$. This holds for sequences of many layers too, because matrix multiplication in general is associative. Nonlinearities have been critical to the success of artificial neural networks, and although linearity assumptions can make theory much more tractable (and in certain specific cases are warranted) it is critical that we continue to develop theory that can deal with nonlinear networks. For control theory in particular, industry applications have largely remained linear to date, but the rise of neural networks has the potential to change that. 

There is also neurobiological reason to introduce nonlinearities into the neural network model, as many neuroscientific interpretations for unit output would not allow for negative signal. While of course outgoing \emph{weights} can be negative, the output signal itself ought not to be for most representations, as neurons cannot negatively fire. Very early artificial neural network models used a step function nonlinearity to represent the "all or nothing" principle in neurobiology, but this posed challenges for optimization methods that aim to learn useful weights in order to complete a particular supervised learning task, because such methods typically rely on calculus-based gradient descent. The sigmoid function was thus the natural successor, as a smooth approximation to the step function that could be differentiated for weight update calculations. Now the output of a given unit would essentially be either 0 or 1, thresholded based on the sum of all the inputs passed to it, with those inputs expressible in terms of a linear combination of outputs from neurons in the preceding layer. With all hidden layers originally using sigmoid activation, said outputs would themselves all be either 0 or 1, but then each would be multiplied by the corresponding incoming weight within that input sum.

Accordingly, early theoretical work on artificial neural networks within machine learning focused on sigmoid activation, and on demonstrating the expressive power that networks with this activation could hypothetically have. Relatively simple feedforward nonlinear networks, when their size is not constrained, can actually cover remarkable ground. \cite{universal} proved that a neural network with a single arbitrarily wide hidden layer using sigmoid activation can serve as a universal function approximator: i.e. for any possible continuous function $\mathbb{R} \Rightarrow \mathbb{R}$ there exists some set of weight assignments to use the hypothetical network to approximate that function. Nonlinearities have thus been critical to the utility of deep learning, and through the more efficient expressive power of compositionality have benefited greatly from multi-layer architectures, earning neural networks the "deep" moniker. Of course there are tradeoffs between layer width and depth in theory, and deep networks are actually much wider than they are deep in practice. But the ability to compose complex functions through a multi-layer nonlinear structure has still been critical to the practical success of artificial neural networks, and empirically it is common to balance depth and width during model tuning to achieve the best results \citep{Nguyen2021}. 

 More recently, sigmoid-based activation has largely fallen out of favor as other nonlinear activation functions demonstrated better performance empirically. However, neuroscience has still remained a key source of inspiration, with the popular rectified linear unit (ReLU) activation of special interest here. The use of ReLU as an activation function in deep learning was the result of framing unit output as a firing rate instead of as an "all or nothing" single spike. ReLU therefore outputs a constant 0 for values $\leq 0$ and is otherwise linear. The shift from sigmoid to ReLU was likely of particular importance for the feedforward architectures that do not have a concept of time and receive a given input for a single pass through. Empirically though, ReLU has improved prediction performance in a variety of architectures (even those that do contain dynamics), and ReLU can also be further motivated by its improved computational efficiency during the weight optimization process \citep{Talathi2016}. \\

 The typical learning process for a neural network involves randomly shuffling a large training dataset, dividing it into batches to perform a weight update step iteratively after each batch, and then repeating this process (referred to as an epoch) tens or even hundreds of times. Batching is largely done instead of using the entire dataset for all weight updates due to efficiency reasons, and there are of course tradeoffs in deciding batch size, just as there are also tradeoffs in a number of other parameters like the size of the network or how aggressively to update the weights based on the feedback from each batch (referred to as the learning rate). For this reason, it is common to begin a project by performing hyperparameter tuning, where a grid search is run to test such settings starting from priors and eventually settling on the combination of options most suited to the current problem. When performing hyperparameter tuning one will often hold out validation and test sets separately, so that generalization ability can be tested for each combination of settings on the validation set, but a final test set can be still be run to check for meta overfitting. 

 In performing the actual weight updates, it is critical that an appropriate loss function is defined in order to specify how incorrect the network was on a given output. For a problem like vector regression, the loss may be something simple like squared distance, but there are many real world situations where a more complicated loss function is required. For example, if the labels in a challenging classification dataset are highly imbalanced, the network would not "worry" about getting rare events wrong unless the loss function penalized that accordingly (or alternatively the dataset itself could be augmented to make it more even). Similarly, the network obviously does not have any concept for how "bad" it is for downstream outcomes to make one type of mistake over another, unless that is encoded within the loss function. Loss functions can also be used to impose secondary penalties that push the network towards e.g. smaller weights, which often helps with generalization ability. The optimization process uses the loss function directly in determining how to change the weights for the next time step. 
 
 Thinking in terms of a single input at a time for simplicity, we can easily express the output of the network given that input as a function of weight multiplications and layer nonlinearities, which I will now walk through. For a feedforward-only architecture, the expression above can be simplified by making each weight matrix represent only the weights from layer $n$ to layer $n+1$, so that the top left entry at index $1,1$ represents the weight from unit 1 of layer $n$ to unit 1 of layer $n+1$. Looping connections are therefore not possible to express in this representation. We then have that the output of a 2 hidden layer nonlinear feedforward network in response to input $\mathbf{x}$ is equal to $out(f(\mathbf{W}_{l1 \rightarrow l2}f(in(\mathbf{x}))))$, where $f$ is an activation function like ReLU and $in$ and $out$ are the input and output layers respectively. If another hidden layer were added we would have the output equal to $out(f(\mathbf{W}_{l2 \rightarrow l3}f(\mathbf{W}_{l1 \rightarrow l2}f(in(\mathbf{x})))))$ instead, and so on. 
 
 The loss function by definition takes a prediction output and the true known label and returns a single number indicating the penalty associated. We therefore hope to minimize the loss over the course of training. As such, given the input $\mathbf{x}$ and the known label $\mathbf{y}$, we take the partial derivative of the loss evaluated at that particular data point (e.g. $L(\mathbf{y},out(f(\mathbf{W}_{l1 \rightarrow l2}f(in(\mathbf{x})))))$) with respect to each weight and use this to determine how much to step in each "direction" (i.e. by adding something to each weight for the next iteration) -- in order to ultimately minimize the loss by moving some amount in the most downward sloping direction of the loss function for this sample within the high dimensional weight space. This is like an iterative approximation of the principle behind finding local minima in calculus, though it has not turned out to be a practical concern in deep learning applications that the network might get stuck at a local minimum \citep{goodfellow2016deep}.

 In practice, the derivatives computed for the gradient descent optimization process are done using the aforementioned backpropagation algorithm. This is highly important for the efficiency of the training process, as it avoids a great deal of recomputation by calculating derivatives in a layer-wise fashion, focusing on the derivatives with respect to the last layer weights first and then using that information in computing the derivatives with respect to the second to last layer of weights, and so on. Note also that input and output layers typically do contain trainable weights. Backpropagation is thus essentially an application of dynamic programming to the chain rule in calculus. Variants on this core principle remain the major method for training neural networks to this day, though it is also a major sticking point for computational neuroscience, where similar model definitions to modern machine learning are often used but weight update algorithms may substantially differ. Unfortunately, alternatives like Hebbian-inspired learning rules have generally struggled to achieve performance at scale anywhere near competitive levels \citep{Amato2019}. \\

 While biologically plausible network learning algorithms remain an active area of research at the intersection of computational neuroscience and deep learning, it is worth emphasizing that they are largely irrelevant to the work presented in this chapter. All of our theoretical results (\ref{sec:math-results}) - like much of the early theory on the expressive capabilities of neural networks - focus on properties of neural networks with static weights. They apply equally well regardless of how the network weights were obtained, whether that was through gradient descent, Hebbian learning, evolutionary algorithms, or divined from some oracle. Our proof of concept experimental results (\ref{sec:dl-results}) do utilize modern deep learning optimization methods, particularly in training the linear negative feedback connections between the subnetworks to be described. Still, this proof of concept does not conceptually rely on gradient descent. \\

\noindent Feedforward networks generally constrain their structure in the layer to layer manner described, but do not constrain which units in one layer send weights to which units in the next -- it is typical to assume that such a network will be fully connected, though some weights may learn to be quite weak. Many architectural advances in deep learning have broken some of these assumptions to better fit a particular style of data. I will now close this primer by briefly reviewing 3 architectures of salience to the themes in my thesis: recurrent neural networks, convolutional neural networks, and transformers, with recurrent architectures being the most critical as the topic of the results in this chapter. For more on the role of neuroscience in the development and early success of artificial neural networks, see the recent historical account by \cite{sejnowski}. For a full mathematical and practical primer on deep learning, see the canonical introductory textbook by \cite{goodfellow2016deep}. \\

\paragraph{Recurrent neural networks (RNNs).}
As discussed, the feedforward model has no concept of dynamics, and is thus not naturally suited for sequential processing tasks. While single variable sequences could be input to a feedforward network as a vector of values with indices corresponding to timestamp, this has a number of drawbacks. It requires simultaneous input of the entire sequence, so that it is not natural to obtain an updating prediction in an online manner. It also requires consistency in the length of the input sequences, though some case-dependent engineering hacks may work around this to a degree. Additionally, when dealing with multi-variable sequences there is not a natural way to make clear the structural assumptions inherent in the way the input is formatted (though a convolutional approach could help to some extent in that problem without introducing recurrence). Similarly, very long sequences will introduce a large amount of required complexity for a feedforward network to process, because this will correspond to a very large number of input variables. 

A solution to these drawbacks is to introduce recurrent weights, meaning that within a set of neurons, connections can form between any pairing and in either (or both) directions. For recurrent weights to be meaningful, we also now need a concept of neuron state -- thus after each timestep, the new state of a given unit will be a function of its prior state as well as the weighted inputs it received from the outputs of the other units at that timestep. The state update may include additional terms as well, for example both a mechanism for external input and (in computational neuroscience) a leak term are common. In a sense, the network is now sharing weights across time when processing a sequence, and the sequential updates to the state of the network serve as a sort of integrative memory of the earlier timesteps. Using the language of the initial 2 unit example that opened up this primer: in an RNN all entries of the $\mathbf{W}$ matrix can be non-zero trainable parameters, and for it to be recurrent functionally there must be some loop in the connections, which for a $2 \time 2$ matrix would correspond to either both off-diagonal entries or at least one of the diagonal entries being non-zero (or both). The product $\mathbf{Wv}$ would be a contributor to the state update equation at each timestep, with the state vector $\mathbf{v}$ now a dynamically updating variable. Predictions can be made from this network through evaluation of the state of all units at a time of interest. 

\noindent For a concrete example of a recurrent weight connection matrix, consider the following $\mathbf{W}$ for a 3 unit RNN, which represents a network with circular shape (i.e. neuron 1 outputs to neuron 2 which outputs to neuron 3 which outputs to neuron 1):
\begin{align*}
\begin{bmatrix}
   0 & 1 & 0 \\
   0 & 0 & 1 \\
   1 & 0 & 0
\end{bmatrix}
\end{align*}
\noindent It is common for RNN weight matrices to be trained with a fully connected assumption, similar to feedforward networks; of course this is not the case when working with sparse RNNs, a topic of importance to our work. \\

Regardless of the practical advantages of RNNs, they are of great relevance to theoretical neuroscience, because the brain contains recurrent connections. Indeed, computational neuroscientists have been studying recurrent models for many decades in attempting to understand how e.g. biological memory is implemented \citep{Amit1989}. RNNs for neuroscience theory are often defined quite similarly to RNNs in deep learning, though generally in continuous-time dynamical systems form. The specific RNN model we focus on is introduced in detail below, in section \ref{sec:math-intro}. The most important conceptual note is that when utilizing the RNN weights to update neuron state, a nonlinearity is applied to the mentioned $\mathbf{Wv}$ term. Like with feedforward networks, nonlinearities can greatly improve the expressive power of RNNs \citep{goodfellow2016deep}, and - most importantly for our purposes - are critical to biological interpretations of network dynamics, for the same reasons as described above.

A complication of recurrence is that it enables instability -- while feedforward networks are trivially stable because they do not have dynamics (and using a feedforward structure in an RNN model definition will cause the effects of any input to "bleed out" over time as it has no way to reinforce its state), even very simple recurrent network topologies can "blow themselves up" in response to the wrong inputs. The 3 unit example weight matrix provided is one such topology: with highly positive weights in this configuration and non-zero (transiently positive) input to the network, the general recurrent model will get caught in a positive feedback loop that causes the neuron states to grow unbounded at an exponential pace. This is obviously undesirable behavior for an engineered system, and in a computational neural network model can prevent learning of longer sequences that may trigger such instability. The biological brain has fundamental physical limitations on activity levels, but even within the bounds of biological limits an analogous phenomenon would negatively impact learning ability. Furthermore, extremely high activity levels can be detrimental to the health of a neuron and has the potential to cause cell death; excitotoxicity has indeed been observed in a number of disease states, including prolonged epileptic seizures \citep{Fujikawa2005}.

Although stability conditions are very well established for linear RNN models (as is linear control theory more generally), nonlinearities make it much harder to assess stability. Nonlinearities in fact make it much harder to even define stability, as many different notions of stability that are entirely overlapping for linear time-invariant (LTI) systems can apply to some nonlinear systems and not others. This includes properties that have the potential to make stable nonlinear systems substantially more expressive in their convergent behavior than LTI systems could ever be, but it of course also greatly complicates proofs of stability (or of instability for that matter). A major aim of this work is to introduce new stability conditions for common classes of nonlinear RNNs that increase the flexibility in networks that we can guarantee to be stable. Though it is worth noting that these conditions remain quite strict relative to what we might expect to hold true for the biological brain, they still take an important step towards expanding our theoretical understanding, and they also provide a way to obtain a stability certificate for safety-critical ML applications seemingly without much detriment to the network's ability to learn. \\

Besides working on stability conditions for the nonlinear RNN dynamical system having all connections described by the weight matrix $\mathbf{W}$, another major contribution of our work is in extending the stability analysis to a multi-area RNN framework. In doing so, each individual RNN is described by a nonlinear RNN model equation like described, and then those units are connected between subnetwork "areas" via a different component of the overall model, in a way that keeps the overall network stable so long as the subnetworks are stable (here we specifically use linear negative feedback connections). This allows for stable "RNNs of RNNs" to be constructed where the individual subnetworks can be quite different in a number of fundamental properties, including activation functions, the timescale on which they are stable, the gain on the neuron leak term, etc. It also provides a potential mechanism by which pretrained stable components could be combined and trained on a larger task whilst preserving stability of the overall RNN. Further, such a procedure might better fit with learning algorithms that take an evolutionary approach to building up network structures, as component modules could be found and reused in a recursive way.

In some sense, one might parallel the "RNNs of RNNs" to the idea of the layer structure in deep feedforward networks. For machine learning applications it is typical to have feedforward input and output layers going into and out of the RNN (processing inputs and producing outputs for each timestep), and it is also fairly common to see additional hidden feedforward layers surrounding the RNN, especially after it. However it is not common at all to find "RNNs of RNNs", perhaps because by the most general definition "RNNs of RNNs" are functionally no different than a single RNN. Yet when imposing a differentiating structure, "RNNs of RNNs" have great expressive potential as well as strong ties to neuroscientific (and computer engineering) principles. From this perspective, it warrants reiterating that in theory shallow but arbitrarily wide networks are universal function approximators for a number of possible nonlinear activation choices -- that does not mean it is at all tractable to find good weights for many problems with such an architecture. The move to multi-layer structures and, as will be discussed shortly, the leap to create convolutional layers that do not use a fully connected weight matrix were both examples of new architecture styles that did not change much in terms of what they could hypothetical compute. It is possible to express a convolutional layer in a fully connected feedforward network equation, it is just highly, highly unlikely that training a feedforward network would result in that structure. \\

Besides the potential downside of instability in RNNs (which is something our architecture avoids though plausibly may over-enforce to an extent), there are other practical reasons to favor feedforward architectures in most machine learning tasks. As you might imagine, optimization of RNN weights is significantly more complicated than optimization of feedforward weights, as the loss function needs to be "rolled out" over time. RNNs took longer to make a mark in deep learning than feedforward architectures like the convolutional neural network, as optimization algorithms struggled with the "vanishing gradient" problem, which prevented RNNs from successfully processing long sequences in practice \citep{goodfellow2016deep}. This was in large part solved by altering model definitions to add components like a gated memory system, though more recent research has developed solutions that do allow certain "vanilla RNNs" to learn challenging sequence tasks. As will be discussed in section \ref{sec:dl-results}, such research includes other architectures utilizing different stability guarantees (like the work of \citep{erichson2021lipschitz}).

Though the ability for "vanilla RNNs" to learn has vastly improved within the last 5 years, it is still most common in practice to use engineered RNN models such as the long short-term memory (LSTM) network with a gated memory cell system. It is also the case that with the introduction of the feedforward Transformer architecture for sequence processing by \cite{transformers}, many of the problem domains that LSTMs and related architectures showed early promise in were soundly overtaken by the empirical performance of Transformers, most notably in the language processing space. Because for the majority of large scale sequence learning problems it is currently most sensible to instead use Transformers, RNNs have slowly become somewhat of a niche topic in the latest deep learning research. 

This does not change the high relevance that RNN models hold for computational neuroscientists, and I suspect some flavor of multi-area RNN architecture will eventually cause a strong shift in practice for one or more problem areas of ML that remain relatively unaddressed (and especially challenging) today -- perhaps for a dense multimodal and sequential datatype like video. It is also highly plausible that a recurrent structure connecting feedforward module components might be an important component towards effectively utilizing existing tools such as convolutional layers and transformer attention mechanisms within the broader context of a multitask agent. If this is the case, it is yet another important reason to continue making machine learning advances in the RNN domain. 

Additionally, there remain a number of specific applied use cases that could benefit from the advantages of an RNN, and many of those use cases would also call for an RNN architecture that has the properties we study. The best performing Transformer networks have leveraged the general architecture's strong ability to scale, and therefore involve huge networks and huge training datasets. This makes it near impossible to interpret exactly how a Transformer is learning from a particular dataset. RNNs on the other hand are capable of succeeding with sequence learning at much smaller sizes, as exemplified by our pilot results (\ref{sec:dl-results}). Small network size of course itself carries inherent value in a number of applications, because it enables model deployment and especially model training using only modest compute resources.  

While the dynamics introduced by recurrence can make interpreting RNN behavior more difficult in some ways, for inputs that take a natural sequential form (unlike the challenging but contrived benchmark tasks we tested on), the temporal "weight sharing" of RNNs can actually lend itself to understanding the patterns found by the network in the sequence data. It would likely remain difficult to interpret weights in a moderately sized fully connected RNN, but as our architecture enforces sparse connections and constrains the overall RNN to take a modular structure with negative feedback between regions, it has the potential to greatly facilitate RNN interpretability in sequence learning. Beyond the supervised learning sequence classification framework, reinforcement learning contexts still have clear uses for RNNs, and for such applications stability guarantees can often be particularly relevant in order to reasonably constrain agent behavior. \\

\paragraph{Convolutional neural networks (CNNs).}
Like RNNs, CNNs have a history in computational neuroscience well preceding their widespread application to practical machine learning problems. \cite{OGCNN} first reported on a CNN-like artificial network architecture that was explicitly modeled after the structure of the visual cortex. It was composed of multiple layers of weights inspired by the receptive fields found in neurobiology, where a weight sharing scheme enabled the feedforward network matrix to act as a series of filters being "dragged" over the input image, detecting e.g. edges within the pixels. Between each of these layers was a downsampling operation that smoothed and shrunk the filtered image outputs, which from there could then have another set of filters applied to them. Variations on these two components would later become the convolutional and pooling layers of the modern CNN.

While the earliest CNNs did not involve the most effective algorithms for learning convolutional filters, and at times even involved hand design of these filter layers, the core ideas that were modeled from neuroscience remained central to the CNN architecture when it eventually achieved its remarkable practical success, thereby demonstrating the importance of fundamental computational neuroscience research in downstream deep learning advances. Through use of the backpropagation optimization procedure, combined with the availability of much more powerful hardware, CNNs became a state of the art technique within computer vision \citep{goodfellow2016deep}. 

Perhaps the most important landmark moment in the foundation of deep learning as a major paradigm shift for ML was the 2012 ImageNet competition, a premiere computer vision benchmark at the time, where \cite{ImageNet} significantly beat all the best existing techniques using a CNN. Their architecture contained five convolutional layers interleaved with several pooling layers, followed by 3 fully connected feedforward layers. By switching to the not yet popularized ReLU nonlinearity, employing bespoke regularization techniques to improve CNN generalization, and creating an implementation that was possible to train in a parallelized manner across multiple GPUs, they were able to leverage the advantages of CNNs to robustly surpass prior image recognition methods \citep{ImageNet}. \\

Computationally, it is not particularly surprising that CNNs are capable of performing as well as they do. Through the constraints placed on the feedforward weight structure of the convolutional layers, information about the 2D structure of image inputs has been "baked into" the network. More specifically, because natural images tend to have localized pixel dependencies, there is great benefit to processing input in a way that explicitly treats pixels within small contiguous 2D patches of the image as highly related to each other. Through compositionality, successive summaries of summaries produced by convolutional layers can efficiently represent patterns within the 2D image at multiple levels of spatial resolution. Further, although it is highly unlikely that a feedforward network would naturally learn such a well-suited topology, it is also the case that convolutional layers require many fewer weights to be updated than a corresponding fully connected feedforward layer would, which made it significantly more practical to train large scale CNNs on last decade's hardware -- thus compounding the benefits, as CNNs could not only perform better than typical feedforward networks at similar sizes, but the largest CNN tractable to train would additionally contain many more units than the largest such fully connected network \citep{ImageNet}. 

CNNs remain extremely possible for computer vision applications today, and they remain quite relevant through a neuroscience lens as well. Given the inspiration of CNNs, computational neuroscience research has since investigated overlap between recorded primate neural activity in the visual cortex and CNN "activity" in response to the same image \citep{yamins2014performance}. Results have been promising, and indeed CNNs are still utilized as models for the visual cortex in neuroscience studies. CNNs even demonstrate other neurobiological-like phenomena such as a critical period \citep{critperiod}, which warrants further research at this interdisciplinary intersection. Of course there are ways in which CNNs are not good biological models, and certainly not all CNNs should be taken as having neuroscientific relevance. At the same time, they are probably better models than many others currently studied within computational neuroscience, and they are perhaps the best model that is extremely relevant to modern machine learning practice at this time. Although CNNs are in large part restricted to modeling visual processing, it is possible to design successful architectures that incorporate both convolutional layers and recurrent elements. In the future directions of section \ref{subsec:rnn-future}, I thus review some large next steps that our architecture has the potential to take within the image recognition landscape. \\

\paragraph{Transformers.}
Though not as relevant to neurobiology as CNNs and not used in the deep learning work I have done, Transformers have recently enabled large strides in natural language processing, a field critical to the work of chapters \ref{ch:1} and \ref{ch:2} -- so I provide a brief introduction to them here as well. As mentioned above, the Transformer is a feedforward architecture designed to efficiently learn long sequences at scale, and as such has replaced RNNs in a number of practical applications including language processing. 

The key insight of Transformers is their attention mechanism, as conveyed by \cite{transformers} in the title of the paper introducing them: "Attention is All you Need". Like a standard feedforward network, Transformers receive an entire sequence as input simultaneously, but through the attention mechanism the network is able to much more efficiently utilize the large amount of information within the sequence, selectively drawing from context far earlier in the sequence as needed when making later processing decisions. The core attention unit building block that is used within Transformers involves training 3 different weight matrices, query weights, key weights and value weights. The result of applying the query and key weights to a particular point in a high dimensional input sequence (like words) is two distinct vectors. To perform an evaluation at a particular point in the sequence, the attention unit will obtain the query vector for the current word and the key vectors for all of the preceding words, and will use the dot product between the query and each of the keys to determine how much the information from each of those preceding points should be weighted for consideration by the value computation. Note that Transformers contain a number of such trainable attention units. \\

Because a Transformer not only performs a fixed number of "steps" in evaluating an input sequence at different "time"points, but also does so in a way that is parallelizable across the different points, it is significantly more efficient to train Transformers on long sequences at scale than it would be to train an RNN \citep{transformers}. This was likely one of the most important factors in the success Transformers have achieved for language modeling, though Transformers additionally have the advantage of a functionally perfect "working memory" to draw from, something that must be achieved through system design in an RNN. Of course on the other hand, one could argue that imperfect memory serves as a weaker form of "attention" mechanism within the RNN. This argument is especially relevant for non-vanilla RNNs that are not accomplishing sequence learning purely through network state. Ultimately though, future artificial intelligence will likely leverage both some form of attention mechanism and some form of memory, as well as both feedforward and recurrent elements. However it is unlikely that these mechanisms will look all that similar to what we have today, just as artificial neural networks have already evolved substantially from what they were 15 years ago.  

From a neurobiology perspective, it is worth noting that while attention mechanisms in deep learning were originally inspired by cognitive science and implemented in many forms years before the breakout success of the Transformer architecture \citep{Niu2021}, this current implementation makes no attempt to be even loosely neuro-inspired \citep{GWT}. Indeed, the paper by \cite{transformers} itself does not make even a passing reference to neuroscience nor to humans, and their innovations were fully based in mathematics and engineering -- as they utilized existing attention mechanism ideas to develop a new sequence processing architecture capable of overcoming a number of prior challenges. Moreover, even from a cognitive science perspective the "attention" driving Transformers is only a narrow form of top-down attention; from a bottom-up perspective, there is no real mechanism for Transformers to selectively attend to surprising or otherwise inherently salient stimuli \citep{GWT}. 

Nevertheless, there are many strong opportunities going forward to integrate work in the computational cognitive and neuro- sciences with study of attention mechanisms in modern deep learning systems, and of course many opportunities to apply Transformer architectures fruitfully (though with much care) in the study of psychiatric patient speech.

\subsection{Applying machine learning to neuroscience}
\label{subsec:ml-to-neuro}
Broadly, advances in machine learning have a number of practical implications for neuroscience. Open questions become increasingly complex over time in science, and because it is (as far as we know) more difficult to neatly abstract principles in study of the brain than it is in e.g. physics, it is uniquely important for future neuroscientific advancement to develop tools that can deal with large, dense datasets both spatially and temporally. Physical technology for e.g. recording neural activity simultaneously across multiple different brain regions has continued to make great strides, and it will be important that techniques for analyzing these data keep pace. As discussed, advances in machine learning could also greatly contribute to advances in psychiatry, which in turn inform neuroscientific research directions. Such useful advances could include improved performance on tasks that remain very challenging (e.g. problems in multimodal ML), or they could involve improved interpretability features that would enable additional scientific conclusions to be drawn from \emph{how} the algorithm made predictions. To both motivate machine learning research in directions of particular relevance to neuroscience and to ensure that relevant machine learning advances are used (and used properly) within neuroscience, it will be important that interdisciplinary collaboration between these fields remains strong. In the concluding chapter, I will discuss factors that uniquely impact interdisciplinary work, especially at this intersection, to hopefully encourage steps that will foster such collaboration even as the fields expand in size and complexity.

In the case of neural networks, there is the additional component that these machine learning models began as early theoretical models of (parts of) the brain, and in some forms still do serve as explicit neuroscientific models -- like much of the work of the present chapter. Though deep learning has now branched off enough that many advances will not have a direct effect on neuroscience, there remains many modern advances that can. Work on efficient and interpretable architectures - along the lines of the (relatively) small and sparse, modular vanilla RNNs reported on here - has the potential to lead to new insights on the dynamics of connectionist systems along with the pros and cons of various network topologies, which in turn can serve neuroscientific hypothesis generation. Furthermore, there is still an active subfield within deep learning that focuses directly on modifying existing neural network architectures and training paradigms to be more biologically plausible, both in the hopes of practical ML improvements and for the sake of better neuroscience modeling. To this day there is an explicit neuroscience track at the popular ML conference NeurIPS, and in fact at the most recent meeting \cite{hinton} presented a keynote on a possible future replacement for backprop-based optimization that is much more biologically inspired. 

Under this same umbrella there is new consideration for designing tasks that are themselves neuroscience-inspired as well; for example, the cognitive task set produced by \cite{yang2019task} and recently expanded on by \cite{Khona2022}. Although these tasks are not as difficult as most current ML benchmarks, they provide many opportunities to experiment with properties of smaller RNN architectures in a setting more directly comparable to neurobiological research. Prior (and continually developing) intuition from both brain and cognitive sciences and machine learning can guide experimental questions, and the artificial setting enables much more extensive exploration to in turn help guide questions that might warrant consideration in the biology lab. In addition to evaluating task performance in terms of the architectural properties, this framework also opens further possibilities for comparison of real neural activity and network "activity", akin to (relatively) early work demonstrating that the visual cortex-inspired convolutional neural network architecture and the real macaque visual cortex indeed shared similar activity patterns \citep{yamins2014performance}. Related techniques have been applied more recently to human fMRI data too, suggesting improvements to existing scientific models of the human auditory cortex and demonstrating a functional learning advantage \emph{in silico} for a topology with hierarchical pathways separating processing of music and speech inputs \citep{Kell2018}. \\

\noindent Within the future directions section \ref{subsec:rnn-future}, I include a subsection on specific ways our "RNNs of RNNs" architecture could be studied in such a framework, including discussion of lines of questioning that might improve our ability to model psychiatric disease mechanisms. 

\subsection{A control theory perspective}
\label{subsec:control-intro}
The roots of control theory are in designing control systems for physical machines, with wide reach in industries such as manufacturing and aerospace engineering. However, control theory provides many mathematical tools to better understand any dynamical systems that contain feedback, and thus can be highly useful across the biological sciences. On the other side, it is unsurprising that control theory - especially adaptive control - was once tightly linked with machine learning work, given its history in machine control. Furthermore, while linear control is a well established engineering practice, many open questions on nonlinear control remain to be researched; though strong foundations exist to enable immediate application of nonlinear results, there is also great opportunity for work in the biological sciences and in machine learning to contribute more generally useful nonlinear theory \citep{slotine1991applied}.

There has been a relative paucity of recent literature exploring uses of control theory with deep learning, despite the fact that neuroscience theory has often focused on dynamical systems modeling and was involved in the development of artificial neural networks. Meanwhile, the pace at which deep learning has improved is far more rapid than the pace at which deeper understanding of these networks has been gained. This is of course not inherently bad, but there is an increasing need for more research focused on understanding neural nets. The interdisciplinary intersection of control theory, systems neuroscience, and deep learning is one avenue well-poised to address that need; at the same time, it is the perspective at the core of the research reported on in this chapter. As such, I will introduce here more background on the current relationships between control theory and ML (subsection \ref{subsubsec:control-theory-ml}) and between control theory and theoretical neuroscience (subsection \ref{subsubsec:neuro-theory}), and discuss how future work might strengthen these relationships. Because of both the themes of the overarching thesis as well as the broader phenomenon that behavior is too often treated as a neuroscientific afterthought, I will additionally introduce some ways that modeling of human behavior - in health and in disease - could contribute to research at this intersection (subsection \ref{subsubsec:behav-control}). 

\subsubsection{Control theory in machine learning}
\label{subsubsec:control-theory-ml}
As will be discussed, control theory has many applications within theoretical neuroscience, and control theorists' involvement with artificial neural networks dates back almost as far as neuroscientists' does. Many early proofs about the capabilities of the Perceptron, a precursor to modern artificial neural networks, utilized control theory tools and published in control theory venues \citep{perceptron}. Of course, the use of control theory in development and characterization of machine learning methodologies extended well beyond neural networks. 

Supervised machine learning has long focused on minimizing a loss function in terms of some adjustable parameters, and historically it was considered important for that function to be convex within the problem space, meaning that any local minima must also be a global minima. The study of convex optimization - methods to find the minima, ideally efficiently and provably - is closely linked with nonlinear stability analysis, because it deals with convergence to an equilibrium point. For example, the Bregman divergence is a distance function specialized for convex functions that serves as a Lyapunov function (an important tool in nonlinear stability) for proving convergence within convex optimization \citep{slotine1991applied}. Such work was critical to the success of some of the more powerful early machine learning tools, like support vector machines (SVMs), and it remained foundational within machine learning through the turn of the century \citep{Young1997,Fradkov2020}. Modern neural networks do not typically have convex loss functions, a subject of early debate that has obviously not caused many problems in practice \citep{goodfellow2016deep}; though it is worth noting that nonlinear stability tools like contraction analysis can still be fruitfully applied to studying theoretical properties of current neural network optimization techniques \citep{Singh2017}, it is no longer central to practical ML advances.

For some specific applications that might fall under "artificial intelligence", such as robotic motion planning and a number of other use cases that are intertwined with physical hardware, control theory does remain central to research. However for much of modern machine learning, ironically because of the rise of deep learning, control theory has largely taken a back seat. As pointed out by \cite{Recht2019}, reinforcement learning and continuous control researchers are in many ways studying the same problem, yet due to differences in technical language as well as the silos that form when fields scale, there is hardly any collaboration or even inspiration-drawing that occurs between the two fields today. This is one of the most notable gaps, but control theory contributions are also much less prevalent in modern machine learning broadly; \cite{Fradkov2020} hypothesizes that because deep learning is the first branch of machine learning to really take off empirically well past the limits of our current theoretical understanding, it has relegated many control theorists to working on models far removed from modern practice, which is unsurprisingly a more niche corner of science. However progress in that space remains important to the overall outlook of machine learning, and as artificial neural networks become more integrated into our lives, the demand for explainability research is likely to significantly rise. Further, intuition from control theory has the potential to improve architecture designs without invoking formal proofs, if the fields were to collaborate more again.

\subsubsection{Theoretical neuroscience}
\label{subsubsec:neuro-theory}
Because much of neuroscience theory revolves around describing neural systems in terms of continuous-time dynamical systems models and then subsequently working to better understand those models, theoretical neuroscience is inextricably linked with control theory in many ways \citep{AbbottBook}. Advances in understanding recurrent neural network (RNN) models are critical to both forming \emph{in vivo} predictions that can be tested to iteratively improve our model definitions, as well as generating brand new hypotheses that can influence the direction of future wet lab research. Moreover, improved understanding of RNNs now also has strong potential utility in further advancing modern machine learning. In many ways, control theory could be at the heart of deeper and more rigorous connections between neuroscience and deep learning, a relationship that both of these fields should hope to foster. Eventually, an understanding of control theory will be requisite for development of many neuroscience-based psychiatric treatments as well; the currently poorly understood yet seemingly effective deep brain stimulation paradigms emerging for OCD and depression are a prime example of a control system framework within neurology, and in turn within psychiatry \citep{Lozano2019}.

As part of the discussion section \ref{sec:discussion4}, I will overview a number of specific ways in which the work of this chapter could be utilized in addressing open computational neuroscience questions (\ref{subsubsec:neuro-rnn-next}). Here, I focus more broadly on why nonlinear control theory tools, and especially the tool that we primarily used in our work (contraction analysis), are so relevant for neuroscience theory. This will include discussion of neural stability along with other critical neuroscientific concepts that are connected to stability -- such as synaptic sparsity and neural synchronization. Indeed, recent analysis of dynamical RNN models showed that anti-Hebbian learning, sparse connectivity, and excitatory/inhibitory balance all play a role in ensuring contraction, a form of stability that is particularly biologically relevant \citep{kozachkov2020achieving}. \\

\paragraph{Stability analysis.}
Stability can provide a system with many advantages in theory, and in practice ensuring stability is a critical part of control theory applications across many industries (e.g. aerospace). Stability can have multiple different meanings for a nonlinear model, but possible benefits of stability (and guarantees of contractive stability) include:
\begin{enumerate}
    \item Bounded input always leading to bounded output in a well-defined manner, which is important for constraining neural activity in a reasonable way. This has clear parallels to epileptic seizures, the prototypical example of unstable activity in the human brain. Photosensitive Epilepsy in particular might be directly modeled through this framework.
    \item The effects of transient perturbations on network state will always exponentially decay, resulting in activity that is more robust to noise -- of obvious importance for agents needing to navigate a real world environment in a multi-faceted way. 
\end{enumerate} 
\noindent Stability can also make system behavior more predictable and improve generalization ability, in addition (and for related reasons) to these major two properties \citep{slotine1991applied,lohmiller1998contraction}.

In a neurobiology context, stability can be motivated by the fact that neural firing rates are reproducible with low variability across trials, despite different initial conditions, noise, and input \citep{Churchland2010}. Indeed, stability is a central component in several influential neuroscience theories \citep{hopfield1982neural,seung1996brain,murphy2009balanced}, perhaps the most well-known being that memories are stored as stable point attractors \citep{hopfield1982neural}. Furthermore, recent technological advances that have enabled neural activity to be tracked in freely moving animals have associated stability in neural activity with the ability to perform complex real world behaviors. \cite{Liberti2022} found that the hippocampal code in free-flying bats was highly stable over an extended period of days to weeks, and that this code remained stable even in response to environmental perturbations like changing the lighting level of the room. As computational and systems neuroscience progress to elucidate more and more about observed stability properties in biological brains, it is more important than ever that stability theory for RNN models progresses along with it. 

As will be clear throughout the results of this chapter (and has already been well established in theory, see e.g. the discussion by \cite{kozachkov2020achieving}), both excitatory/inhibitory (E/I) balance and synaptic sparsity are closely related to stability. Empirically, excitatory/inhibitory imbalance has been observed in human disease states, most saliently network hyperexcitability in Epilepsy \citep{Ziburkus2013}. E/I imbalance in different brain regions has also been linked to psychiatric illnesses, including both Schizophrenia and Autism Spectrum Disorders (ASD); because E/I perturbations can be localized to particular brain circuits and can thus manifest themselves differently in observed behaviors \citep{Lam2021}, a multi-area approach to RNN stability theory is an important future direction which we lay groundwork for through our results. \\

\paragraph{The role of sparsity in theory and practice.}
Sparsity is especially relevant to our work, as we found achieving stability in our architecture through sparsifying the component RNNs produced the best empirical results (section \ref{sec:dl-results}). Sparsity is a recurring theme in neuroscience, both enabling stability in theory \citep{kozachkov2020achieving}, and occurring in practice, with only $\sim 0.00001\%$ of potential connections forming synapses in the human brain \citep{slotine2012links}. Sparsity may be of additional interest in our "RNNs of RNNs" framework because it encourages individual subnetworks to take on truly unique topologies that can be much more efficiently distinguished from each other. Recent neurobiological evidence suggests that great diversity in the number and spatial arrangement of thalamocortical synapses across neurons is what enables strong information processing capabilities despite the extreme level of sparsity, and that this confers benefits in robustness and flexibility \citep{Balcioglu2023} -- properties closely linked to stability analysis.

Like E/I balance, synaptic sparsity has also been linked to disease states, and synaptic pruning has even been hypothesized as a major underlying mechanism in Schizophrenia and ASD, with excessive pruning believed to occur in Schizophrenia and - inversely - insufficient pruning in ASD \citep{Stevens,Sakai}. While there is likely a great deal of nuance in how this manifests due to the massive amount of heterogeneity between brain regions and even between different neuron types, that is all the more reason for closer study of sparsity in theory and particularly in a multi-area RNN context, which will of course be intertwined with study of stability properties. Pruning-related hypotheses are so compelling because they can connect observations from across genomics, immunology, developmental neuroscience, connectomics, and cellular neurobiology with ideas from cognitive science, machine learning, and theoretical neuroscience, thereby creating an opportunity for rich interdisciplinary crosstalk and model iteration. 

The relevance of sparsity has indeed been a discussion topic in neuroscience theory for many decades, including results on the energy efficiency of sparse coding, which reduces the total amount of neural activity needed to represent information \citep{Levy1996}, as well as results on learning rules that can lead to sparse representations \citep{Foldiak1990}. Of course when taken to an extreme sparse coding is not without its drawbacks, and it is not the case that concepts are each represented by activity of a single neuron in the brain. However, neuroscientific evidence does suggest that the brain leans more towards a sparse code than a dense one, and a number of computational neuroscience works have demonstrated that observed neural responses can be emergent properties of a nonnegative sparse coding scheme \citep{Beyeler2019}. \cite{Beyeler2019} additionally point out that the degree to which neural activity matches a sparse representation model can vary by brain region, yet another argument for greater study of "networks of networks".

Despite the fact that sparse coding has been a popular topic in the computational neuroscience literature, there has only recently been a reemergence of works focused on sparsity in the weights of trainable RNN models \citep{Khona2022}. This is likely a result of the practical success of artificial neural networks over the last decade, which have in large part focused on fully connected models. While it is known that sparsity can improve generalization ability (for reasons akin to Occam's razor), and regularization techniques like $L1$ loss are directly related to sparsity, most deep learning architectures remain ultimately very densely connected \citep{goodfellow2016deep}. A subfield of deep learning research that has recently increased in popularity focuses on achieving strong (albeit often slightly worse) performance on benchmark tasks using networks that are much smaller than the typical state of the art architectures employed. This might involve building up networks with fewer units and fewer weights between them via an evolutionary inspired algorithm \citep{gaier2019wann}, or it might involve "pruning" the majority of weights based on early training results \citep{frankle2019lth}. It also can involve architecture design that draws from theoretical and/or neuroscience inspired principles, such as the work presented in this chapter. 

Ultimately, there are many practical reasons for a deep learning researcher to care about sparsity of network weights, including efficiency of training, ability to easily deploy a trained network on weak/small hardware, and the interpretability of network behavior. It is plausible that artificial neural networks could one day be a common tool not only for predictions, but also to assist data scientists in forming mechanistic insights. However it will first be necessary to substantially further the mathematics and the \emph{science} of deep learning; sparse network topologies will be an important component to understand for this pursuit, and given the clear connections to neurobiology this is an obvious area for computational neuroscientists and machine learning experts to join forces. \\

\paragraph{Understanding neural synchronization.}
Besides stability analysis, control theory also has deep ties to study of system synchrony -- a recurring topic in neurobiology and likely a critical topic to elucidate if we hope to understand sleep, itself a critical component of healthy (and diseased) human functioning. Because contraction analysis defines stability in terms of convergence of different system trajectories to each other as opposed to convergence of system trajectories to equilibrium states, it is possible to pose questions about whether connected systems will synchronize instead as questions about whether some "auxiliary system" representing a difference between their states will contract (see section \ref{subsec:contraction-intro} for more on contraction). 

Sleep broadly synchronizes neural activity, with certain stages of sleep driving slow oscillations at frequencies that are otherwise not usually observed in a healthy, alert brain \citep{Timofeev2013}. It is well known that sleep plays a key role in learning and memory, with experience replay observed via neural activity in sleep, and clear cognitive deficits measured with even a moderate lack of sleep \citep{Alhola2007}. Further, it is likely that sleep plays some role in maintaining neural stability, possibly by "cleaning up" neural state while closing off inputs/outputs; interestingly, sleep deprivation has been a well-documented seizure trigger for Epilepsy patients and animal studies have demonstrated that REM sleep has a protective effect against seizures \citep{Malow2004}. It can be difficult to probe the causal direction of sleep perturbations, so it would be immensely helpful to gain a better understanding of how synchronization can theoretically affect both network state and weight updates (and vice versa) through nonlinear dynamical systems modeling. 

Unsurprisingly, an array of different sleep abnormalities have been found across psychiatric disorders. While it is of course possible that altered sleep dynamics are an outcome of psychiatric disorders rather than a cause, it is more plausible that causation runs in both directions, generating an unwanted positive feedback loop in patients. For example, poor sleep can increase the severity of hallucinations in Schizophrenia, and psychotic episodes can even occur in otherwise healthy adults due to severe sleep deprivation \citep{Waters2018}. Interestingly, sleep abnormalities seen prior to a seizure (e.g. abnormally long sleep spindles in slow wave sleep \citep{Picard2007}) are the opposite of those typically seen in Schizophrenia patients (i.e. shortened sleep spindles \citep{Ferrarelli2007}). This is consistent with the more general "mirror image" phenomena of Autism and related disorders having many inverse properties of Schizophrenia and related disorders, most notably results related to hyper- versus hypo- connectivity and excitatory versus inhibitory imbalance. It perhaps hints that architectural differences may lead to the differing sleep dynamics, which may in turn be a causal factor for associated symptoms. 

Besides sleep, there are additional emerging reasons to be interested in periodic, synchronizing inputs to neural systems. The use of flashing lights or tones of a particular frequency can entrain brain activity, and may have effects such as clearing protein aggregates associated with Alzheimer’s Disease \citep{Iaccarino2016}. A theoretical understanding of how such inputs may alter brain activity and even plasticity would be beneficial in guiding future research on behavioral and medical applications of this type of protocol. Along the same lines, invasive procedures to alter brain activity - most notably deep brain stimulation (DBS) - often involve periodic inputs in practice as well. More broadly, DBS can be thought of as a control system, and thus control theory tools have obvious potential relevance. Recall that the DBS system trialed in chapter \ref{ch:3} aimed to disrupt hypothesized corticostriatal hypersynchrony in OCD by driving ventral striatal and motor cortical regions at slightly different frequencies, but inadvertently increased coherence empirically observed between the regions in the pilot subject. 

Design of DBS systems is of course highly complex in practice, and even the settings of established paradigms require experimental fine tuning for each patient. At the same time, DBS often involves chronic stimulation at a particular frequency, including the psychiatric treatment applications with the most evidence to date in OCD and depression. As such, theoretical study of model system synchronization behavior in response to a recurring input could indeed prove useful for future understanding of why these DBS paradigms work, possibly even assisting directly in design of new paradigms. By contrast, while stability properties are highly relevant for the biological brain and studying stability of RNNs provides valuable insights into network topologies of interest, provably stable systems can still demonstrate a wide variety of initial responses to transient inputs, and thus convergent behavior is not always of the utmost practical relevance. One could also imagine leveraging both of these perspectives by asking questions on which network configurations are more or less likely to encourage "resting state" hyper- or hypo- synchrony between regions in a multi-area RNN model, to further our theoretical understanding of connection patterns that might lead to psychiatric symptoms. This could then be coupled with analysis of convergent behavior of such a system in response to extended DBS-like input, in order to form new hypotheses for specific DBS system designs. By working closely with neurologists, network models could be built iteratively based on their ability to predict properties of neural activity like corticostriatal coherence.

\subsubsection{Dynamical models of behavior}
\label{subsubsec:behav-control}
It is highly plausible that network configurations resulting in abnormal connectivity levels between brain regions may also require different inputs to modulate synchrony than would be effective in a more typical topology. Additionally, it is highly plausible that starkly different topology aberrations could produce observable animal behavior that is only subtly different -- especially in terms of our current classification systems. There is a great deal of nuance in the extent to which a disease state is actually pathological in a particular environment as well, and although DBS is presently reserved for incredibly severe treatment resistant conditions, as technology advances and less invasive alternatives become available there will be many advantages to a closed loop system that can utilize information about system outputs and their environmental context, not just the system state information available through neural recording. A better understanding of behavior is therefore a crucial (and often underappreciated) component to the future of not only DBS (as discussed in chapter \ref{ch:3}), but also neuroscientific modeling more broadly. 

In the introductory chapter, I presented arguments that behavior in psychiatry disease is particularly underrated amongst computational neuroscientists. Study of aberrant system output can be a highly effective reverse engineering tool across technical disciplines, yet careful characterization of human "output" in disease states is frequently labeled as not neuroscientific, and the vast majority of computational modeling work focuses on normal learning (which incidentally, normal is not the best either). Whether one cares about disease for medical purposes or not, naturally occurring variation is a prime opportunity to drive hypothesis generation in entirely different directions than typical incremental question answering would. In my experience with non-psychiatrist neurobiology researchers, it is an open secret that "disease analogies" are mostly used disingenuously and somewhat begrudgingly for the purposes of grantsmanship. Both psychiatry and neuroscience could substantially benefit if models of disease were more often seriously engaged with for the inherent opportunities that they hold.  

On the other side, psychiatry has largely failed to formalize behavioral observations in terms of dynamical systems. As pointed out by \cite{Salvi2021}, while psychiatrists often utilize information about behavioral fluctuations qualitatively when evaluating a patient, the vast majority of psychiatry research focuses on cross-sectional studies that treat disease properties as static traits. In order to build dynamical models of behavior that can be analyzed using control theory and that can be integrated with established work in computational neuroscience, it will first be necessary to collect richer and - most critically - more longitudinal behavioral data, in both humans and animal models and in both health and disease. With more precise and much more temporally dense ways of measuring system output, we can begin the iterative process of forming dynamical systems models of psychiatric symptomatology and updating them based on their ability to predict real world readouts. Collection of longitudinal naturalistic behavioral signals from psychiatric patients was indeed a theme throughout the rest of the thesis.

Note that although this framework could directly involve neural network models (and perhaps should in the case of animal models), it is also possible to define models of psychiatric symptom dynamics and employ control theory to better understand them whilst entirely abstracting away any neurological components. Such work could have immediate relevance to clinical aims, for example evaluation of treatment planning as a control system -- what sort of external inputs might be able to inhibit a "bad" positive feedback loop, and what sort might trigger one? Studies often discuss correlation versus causation, but complicated causal webs are not as often considered, at least not beyond the mere mention they might exist. Dynamical models of behavior could go a long way in forming intuition about how such systems operate, bringing much needed nuance to study interpretation. Further, dynamical models of behavior - even when not explicitly neuroscientific - would provide fundamental groundwork for eventually connecting complicated neural network models to the behavior that they produce. 

It is often the case in theory that a proof involves "translating" a system to some other equivalent representation in which it is much easier to prove the desired claim. Mathematics can serve as a rigorous and interdisciplinary language for describing ideas and forming downstream predictions based on them. The rest of this chapter will focus on neural network theory and deep learning science, but it is worth emphasizing that for phenomena as complex and dynamic as human (or any animal) behavior, we should really be putting a lot more focus on establishing such a language. This applies both to the neuroscientists that think of behavior as an afterthought and to the psychiatrists that focus so heavily on static diagnostic labels. 

\subsection{Moving forward}
\label{subsec:rnn-motivation}
Neuro-inspired machine learning has profoundly altered many fields such as computer vision, natural language processing, and computational neuroscience \citep{bengio2017deep,hassabis2017neuroscience}. While deep learning models can be remarkably powerful, they are for the most part `black boxes'. This opaqueness may be dangerous in safety-critical applications like autonomous driving or human-centered robotics, and it limits conceptual progress, progress that would have the potential to in turn inform neuroscience. Such drawbacks are of particular relevance for machine learning applications in psychiatry, which encounter roadblocks with both practical clinical concerns and limitations in the foundational knowledge collectively available at present. Therefore, not only do many open questions remain to be addressed in understanding artificial neural networks, but neuroscientists ought to be triply motivated to address them. 

Stability carries a number of theoretical benefits, especially for the robustness and predictability of system behavior, and is thus a crucial component in the future of deep learning science. As mentioned, stability also has implications for the biological brain, and important neuroscience concepts like sparsity go hand-in-hand with stability; it is likely that some notion of stability will be necessary for fully understanding brain function. There are many reasons to study stability in RNN models, and many open questions that remain.

While there is a rich history of stability-related results in neuroscience theory \citep{hopfield1982neural,seung1996brain,murphy2009balanced,wiener2019cybernetics}, proven stability conditions for continuous-time nonlinear RNNs remain quite restrictive -- whether in the weights they allow or in other prior assumptions such as restricting the allowed nonlinearities to exclude some popular activation functions like ReLU. At the same time, incorrect RNN stability theorems have been found in the machine learning literature, and the field as a whole has struggled to prove certain seemingly simple conjectures. 

After introducing the mathematical background on the RNN dynamics we study in section \ref{sec:math-intro}, I will discuss existing stability conditions for this RNN model (\ref{subsec:rnn-lit-rev}). Within our results in section \ref{sec:math-results}, I will present counterexamples to a handful of problematic previously published conditions, in addition to dissecting the technical challenges that have made proving conjectured stability conditions in this model so difficult (\ref{subsec:counters}). Suffice to say there are still a large number of open theoretical questions about continuous-time nonlinear RNN stability conditions, which is one area our work makes progress in. \\

\noindent Of course, perhaps the largest contribution of our work is not in the study of individual nonlinear RNN stability properties, but rather in the study of stability conditions for combinations of RNNs, as well as the application of such conditions in machine learning practice. There is a shockingly small amount of prior literature on "RNNs of RNNs" across the board, whether considering neuroscience theory or deep learning science. In the next subsection I will thus overview why modular systems are so salient and some of the major research questions about networks of networks the community should hope to address. 

\subsubsection{Study of network combinations: a forefront for future interdisciplinary breakthroughs}
\label{subsubsec:combos-intro}
The combination and reuse of primitive "modules" has enabled a great deal of progress in computer science, engineering, and biology. Modularity is particularly apparent in the structure of the brain, as different parts are specialized for different functions \citep{kandel2000principles}, and these areas must perform in concert with each other to achieve the goals of the organism. Furthermore, connectivity patterns between such regions likely play an important role in human intelligence, as models accounting for global representations have predicted cognitive ability beyond what could be explained through local representations alone \citep{Anderson2022}. Recent work in modeling macaque neural recording data has also shown that modular neural networks were much more capable of predicting  parietofrontal cortex activity during a grasping task than traditional neural networks were \citep{Michaels2020}.

Most experimental studies throughout the history of neuroscience have focused on a single brain area in association with a single behavior \citep{Abbott_Svoboda_2020}. Similarly, RNN models of brain function have mostly been limited to a single RNN modeling a single area. However, neuroscience is entering an age where recording from many different brain areas simultaneously during complex behaviors is possible. As experimental neuroscience has shifted towards multi-area recordings, computational techniques for analyzing, modeling, and interpreting these multi-area recordings have begun to emerge \citep{mashour2020conscious,Abbott_Svoboda_2020,Perich_2021,semedo2019cortical,yang2021towards}. Despite this, RNN theory has lagged behind.

In the field of deep learning, there has been a similar paucity of study on so called "networks of networks", despite the clear neuroscience analogies and the small conceptual distance from the idea of "deep" multi-layer architectures that also take advantage of compositionality. There have been successful architectures with hints of modularity -- such as the Generative Adversarial Network (GAN), which typically utilizes a two network structure where one network is rewarded for "tricking" the other \citep{goodfellow2016deep}. The groundbreaking success of AlphaGo was similarly enabled by a multi-network structure \citep{AlphaGo}. However, there has been minimal work to date on leveraging modularity in highly recurrent systems, as well as minimal mathematical or scientific work on how modularity impacts network learning. 

At the same time, there is increasing demand for machine learning systems that can accomplish multimodal tasks, as well as for deep learning agents that can multi-task, continually learn, accurately assess uncertainty, behave in an interpretable way, etc. Artificial neural networks have lead to paradigm shifting advances in a number of domain-specific areas like image recognition and language processing, but applications that require "putting it all together" have not been quite so earth shattering to date. In the light of the many open questions that remain about modular deep learning systems, this is not surprising. For example, while one might intuitively expect that training a particular module to perform better (e.g increasing visual acuity) would improve the performance of the overall system, this is not necessarily the case. In the context of self-driving cars, it has been shown recently that additional training for one module while holding the other modules' parameters fixed can actually make overall performance worse \citep{wu2021fixes}. 

Establishing sufficient conditions for preservation of functionality in the construction and training of modular systems is therefore an interesting and open question. Our research makes several advances in this domain by providing conditions under which training one (or several) modules while holding the others fixed provably preserves the \textit{stability} of the overall system -- a minimal requirement for robust and predictable behavior. To do so, we utilize a special notion of nonlinear stability called contractive stability. \\

\paragraph{A conceptual introduction to contractive stability.}
In section \ref{subsec:contraction-intro} below, I will provide a mathematical primer on contraction analysis, along with additional context from nonlinear stability theory. Here I focus on contraction analysis from the perspective of how it enabled our work, as well as its ties to biological sciences. Most notably contractive stability has potential applications in evolutionary biology and - as we show throughout this work - neurobiology. However, contraction analysis can also be used in many other domains, such as in modeling the dynamics of bacterial quorum sensing or the spread of infectious disease. 

A major benefit of contractive stability is that there are a number of ways in which contracting systems can be combined whilst preserving stability of the overall system. Indeed, themes of modularity are apparent in biological evolution and in the structure of the brain. It is thought that the majority of traits that have developed over the last 400+ million years are the result of evolutionary forces acting on \emph{regulatory elements that combine core components}, rather than mutations to the core components themselves. This mechanism of action makes meaningful variation in population phenotypes much more feasible to achieve, and is appropriately titled "facilitated variation" \cite{gerhart2007fv}. In addition to the biological evidence for facilitated variation, computational models have demonstrated that this approach produces populations that are better able to generalize to new environments \cite{parter2008fv}, an ability that will be critical to further develop in deep learning systems. 

\begin{figure}[h]
\centering
\includegraphics[width=\textwidth,keepaspectratio]{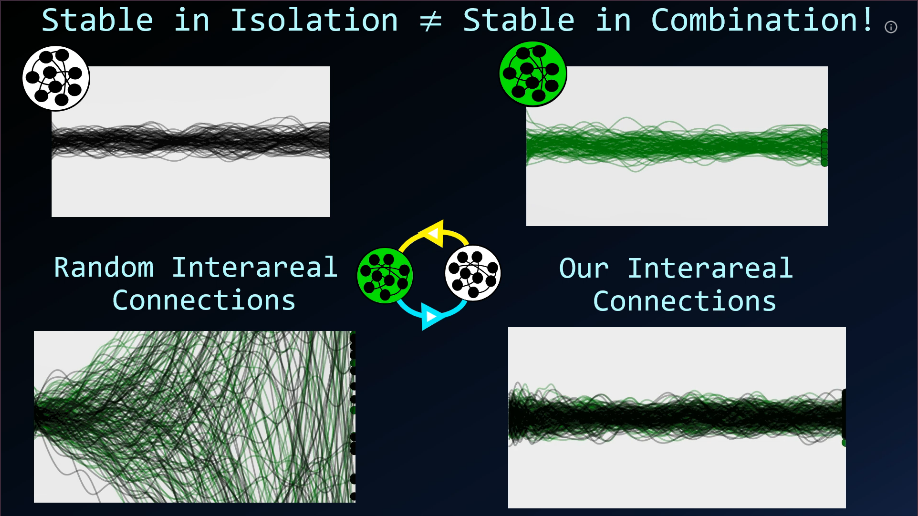}
\caption[Combinations of stable systems are not necessarily stable.]{\textbf{Combinations of stable systems are not necessarily stable.} When simulating neural dynamics in MATLAB, it can be easily seen that maintaining stability in the final network when combining two stable networks requires great care. Here we see how each of the neuron voltages varies over an input to two different stable RNNs (top), with values staying in check. When the two (black and green) RNNs are connected together, we can see that random conjoining connections will cause the voltages in the new combined RNN to rapidly rise in response to the same input and ultimately "explode" (bottom left), whereas using negative feedback conjoining connections derived from the contraction combination properties will keep the new combined RNN's behavior stable (bottom right).}
\label{fig:stab-demo}
\end{figure}

Although the importance of building modular systems is clear, in general there is no guarantee that a combination of stable systems will itself be stable (Figure \ref{fig:stab-demo}). Thus the tractability of these proposed evolutionary processes hinges on some mechanism for ensuring stability of combinations. As contraction analysis tools allow complicated contracting systems to be built up recursively from simpler elements, this form of stability is well suited for neural and other biological systems \cite{slotine2012links}. Moreover, contracting combinations can be made between systems with very different dynamics, as long as those dynamics are each contracting -- thereby facilitating a great deal of flexibility in constructing contracting modular systems \citep{slotine2001combos}. In our particular case, this also means that the contracting RNNs discussed herein could be combined with physical systems with quantified contraction properties.

\subsubsection{Chapter aims}
\label{subsubsec:ch4-aims}
As discussed, cognitive models are moving increasingly towards the study of multi-area dynamics, but many questions remain on how to best train and evaluate such networks \citep{yang2018dataset,yang_molano-mazon_2021}. Understanding how different brain regions interact harmoniously to produce a unified percept/action will require new ideas and analysis tools. 

In the presented work, we leveraged the hierarchical and especially negative feedback combination properties of contracting systems in order to develop a framework for constructing complex nonlinear "RNNs of RNNs" that preserve stability features of their component subnetworks, something not generally guaranteed with system combinations (Figure \ref{fig:stab-demo}). In this process, we also pushed forward the literature on stability conditions for the individual nonlinear RNN dynamics in continuous time, and subsequently provided empirical proof of concept results for a deep learning implementation of our architecture design on challenging sequence learning tasks. \\

\noindent Ultimately, our primary contributions are four-fold: 
\begin{itemize}
    \item Novel contraction conditions for \emph{continuous-time \textbf{nonlinear} RNNs}, to use in conjunction with the combination condition (section \ref{subsec:novel-conditions}).  
    \item A novel parameterization for \emph{linear negative feedback combination} of contracting nonlinear RNNs that enables direct optimization using standard deep learning libraries (section \ref{subsec:combo-conditions}).
    \item Identification of flaws in stability proofs from prior literature on nonlinear continuous-time RNNs, along with technical discussion of some of the challenges inherent in proving popular conjectured conditions for RNN stability (section \ref{subsec:counters}). 
    \item Experiments demonstrating that our 'network of networks', implemented based on our theoretical results, sets a new \emph{state of the art} (SOTA) for stability-constrained RNNs on benchmark sequential processing tasks -- also performing well against overall SOTA benchmarks with a much smaller number of trainable parameters (section \ref{sec:dl-results}). 
\end{itemize}
\noindent As part of my thesis chapter, I additionally provide extended commentary on the numerous future directions that our work can enable (section \ref{subsec:rnn-future}). 

\FloatBarrier

\section{Mathematical preliminaries}
\label{sec:math-intro}
Before presenting our mathematical results (section \ref{sec:math-results}), I will first establish the necessary mathematical background to understand them. Recall that we focus on proving the stability of a given fixed RNN (or RNN of RNNs), not on the stability of the weight update process. Though at times we will discuss how to guarantee that a network remains provably stable throughout the training process per some of our conditions, the primary perspective remains on stability certificates for networks with static connections. The individual nonlinear RNNs we study, characterized mainly by their weight matrix $\mathbf{W}$, will be defined along with further model explanation in section \ref{subsec:rnn-defs}. I will then introduce details of the specific nonlinear control tool we use - contraction analysis - and present background on the combination properties enjoyed by contracting systems (\ref{subsec:contraction-intro}), which we leverage in conjunction with the individual RNN model to create our modular "RNNs of RNNs". Finally, section \ref{subsec:rnn-lit-rev} overviews prior literature relevant to the chosen RNN model and to applications of contraction analysis in neuroscience, in greater technical specifics enabled by the rest of the mathematical preliminaries.

\subsection{Nonlinear model definitions}
\label{subsec:rnn-defs}
This chapter focuses on analysis of rate-based neural networks. Unlike spiking neural networks, these models are continuous and smooth. We consider the following RNN with update dynamics introduced by \cite{Wilson_Cowan_1972}, which may be viewed as an approximation to a more biophysically-detailed spiking network: 
\begin{equation}\label{eq:RNN}\tag{nonlinear RNN dynamics}
\begin{split}
\tau \dot{\mathbf{x}} = f(\mathbf{x}) = -\mathbf{x} + \mathbf{W}\phi(\mathbf{x}) + \mathbf{u}(t)
\end{split}
\end{equation}
\noindent Here $\tau > 0$ is the time constant of the network \citep{dayan2005theoretical}, and the vector $\mathbf{x} \in \mathbb{R}^n$ contains the "voltages" of all $n$ neurons in the network. The voltages are converted into firing-rates through a static nonlinearity $\phi$. We only consider monotonic activation functions with bounded slope: in other words, $0 \leq \phi' \leq g$. We do not restrain the sign of the nonlinearity. The matrix $\mathbf{W} \in \mathbb{R}^{n \times n}$ contains the synaptic weights of the RNN. It is this matrix that ultimately determines the stability of (\ref{eq:RNN}), and will be a main target for our analysis. Finally, $\mathbf{u}(t)$ is the potentially time-varying external input into the network, capturing both explicit input into the RNN from the external world, as well as unmodeled dynamics from other brain areas. 

Common example nonlinearities $\phi(x)$ that satisfy the given constraints include hyperbolic tangent (tanh) and rectified linear unit (ReLU), both of which have $g=1$. Though most of our theorems can handle more general $g$ (as will be included within the proof details), I will assume $g=1$ in much of the discussion in this chapter, for more simple interpretability. Note that some prior literature on the (\ref{eq:RNN}) network instead restricts $\phi' > 0$, for example the stability analyses by \cite{matsuoka1992stability}. Such results will not necessarily apply to networks using activation functions like ReLU, which has 0 slope for negative input values.

\subsubsection{Network Jacobian}
Contraction analysis, to be reviewed next (section \ref{subsec:contraction-intro}), requires the Jacobian $\mathbf{J}$ of the system in question. The Jacobian is the matrix of partial derivatives of the dynamics with respect to the system state -- in a 2 unit system, we could express (\ref{eq:RNN}) in terms of $f(x_{1})$ and $f(x_{2})$ separately, and then see the Jacobian would be $\mathbf{J} =$ 
\begin{align*}
    \begin{bmatrix} 
    \frac{\partial f(x_{1})}{\partial x_{1}} & \frac{\partial f(x_{1})}{\partial x_{2}} \\
    \frac{\partial f(x_{2})}{\partial x_{1}} & \frac{\partial f(x_{2})}{\partial x_{2}} 
  \end{bmatrix}
\end{align*}
\noindent We can see that for our system (\ref{eq:RNN}), the leak term $-\mathbf{x}$ will contribute only diagonal entries to the Jacobian, as each unit's leak only directly affects itself. Similarly, the input $\mathbf{u}(t)$ is not state-dependent and thus will have 0 impact on the Jacobian. However the connection weights $\mathbf{W}$ allow different units to influence each other, so that this term accounts for most of $\mathbf{J}$ here. Note that when we take $\phi(x) = x$ i.e. a linear system, the partial derivative matrix of the term would be $\mathbf{W}$. The nonlinearity of course complicates things. \\

\noindent Ultimately, we find through the chain rule that the system (\ref{eq:RNN}) has Jacobian: 
\begin{equation*}
\begin{split}
\mathbf{J} = \mathbf{WD} - \mathbf{I}
\end{split}
\end{equation*}
\noindent where $\mathbf{D}$ is a diagonal matrix with entry $D_{ii} = \phi'(x_{i})$. $\mathbf{D}$ thus represents the derivative of the activation function evaluated at the current state. Note that we can allow different activation functions for different units without loss of generality, as long as each function satisfies our constraint $0 \leq \phi' \leq g$. $\mathbf{D}$ is therefore a matrix that can vary over time due to inputs or internal dynamics, but it will always be nonnegative and each entry will always be upper bounded by $g$. \\

\paragraph{Equivalent representations.}
(\ref{eq:RNN}) is equivalent to another commonly used class of RNNs where the term $\mathbf{W}\mathbf{x} + \mathbf{u}$ appears inside the nonlinearity $\phi$. See the proof by \cite{miller2012mathematical} for details. Our mathematical results therefore apply equally well to both types of RNNs. Because this other class of RNNs has Jacobian $\mathbf{DW} - \mathbf{I}$, we can use either ordering when plugging $\mathbf{J}$ into contraction analysis, and any resulting conditions will apply to both dynamics formats. 

\subsection{Contraction analysis}
\label{subsec:contraction-intro}
Unlike linear stability analysis, there are many different notions of stability for a nonlinear system, and it can sometimes be difficult to say anything at all about the stability of a particular nonlinear dynamics. Lyapunov analysis for example relies on finding an appropriate Lyapunov function to prove that a system is Lyapunov stable, but failing to find a Lyapunov function does not prove that a general system is \emph{not} Lyapunov stable \citep{slotine1991applied}. As discussed within the background section \ref{sec:background4}, another complication of nonlinear stability analysis is that combinations of stable systems need not be stable in general, and even when combined in a seemingly "stable" way, when nonlinearities are involved it can be extremely difficult to verify that stability was actually maintained. 

\begin{figure}[h]
\centering
\includegraphics[width=\textwidth,keepaspectratio]{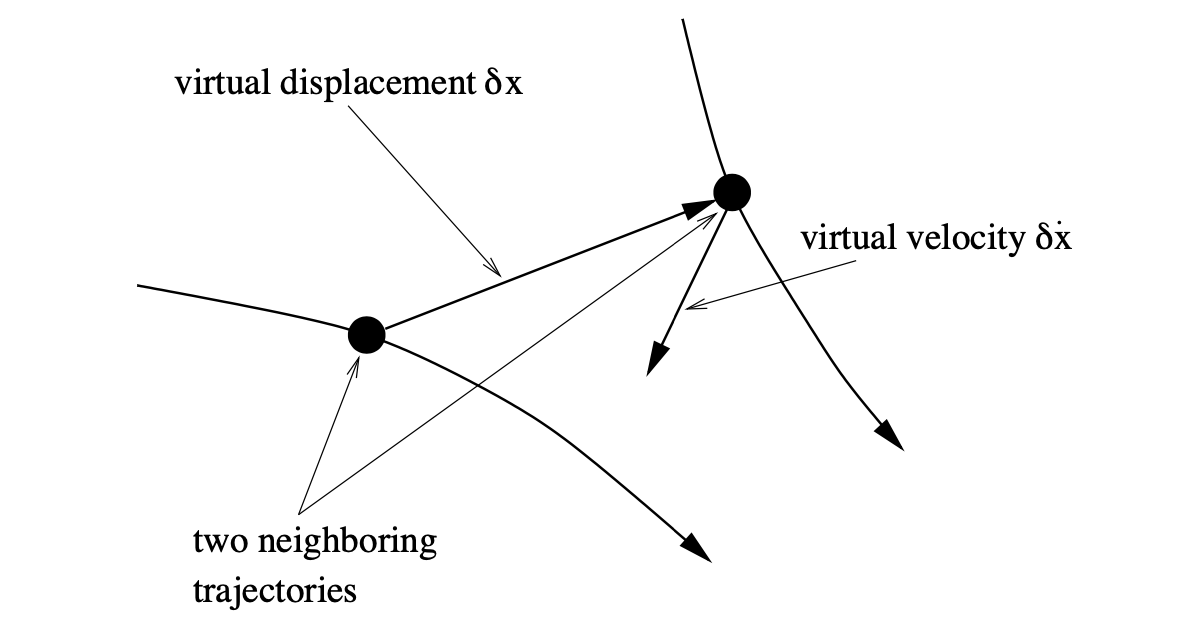}
\caption[Contraction theory is based on a differential analysis of convergence.]{\textbf{Contraction theory is based on a differential analysis of convergence.} By focusing on exponential convergence of any two neighboring trajectories, contraction analysis allows convergence and limit behavior to be treated separately. While the traditional Lyapunov analysis for stability of nonlinear systems can be seen as "virtual classical mechanics", contraction analysis can be seen as "virtual fluid mechanics". This explanatory cartoon has been reproduced from \cite{lohmiller1998contraction}, the paper introducing contraction. It is fair use, and the senior author is of course also the senior author on our paper \citep{NIPS22}}
\label{fig:contracting-trajectories}
\end{figure}

Here we focus on a special type of nonlinear stability, known as contractive stability, as introduced by \cite{lohmiller1998contraction}. Contractive stability is a strong form of dynamical stability which implies many other forms of stability, such as input-to-state stability \citep{sontag2010contractive}. Loosely speaking, a dynamical system is said to be globally contracting if any two of its trajectories converge to each other exponentially, regardless of initial conditions (Figure \ref{fig:contracting-trajectories}). Contraction analysis is convenient because it allows us to analyze convergence behavior without worrying about the identity of equilibrium points like in e.g. Lyapunov analysis. This also enables the combination properties of contractive stability, which we make heavy use of throughout the present work.

A primary advantage of contraction analysis is that it is directly applicable to non-autonomous systems, which the vast majority of recurrent models are \citep{slotine2001combos,modular}. Contraction analysis has found wide application in nonlinear control theory \citep{manchester2017control}, synchronization \citep{pham2007stable}, and robotics \citep{chung2009cooperative}, but has only recently begun to find application in neuroscience and machine learning \citep{boffi2020learning,wensing2020beyond,kozachkov2020achieving,revay2020contracting,jafarpour2021robust,kozachkov2022robust, centorrino2022contraction,burghi2022distributed,kozachkov2022generalization}. Contraction analysis is particularly useful for neuroscience because it is directly applicable to systems with external inputs. It also has the potential to be closely linked with evolutionary biology through its combination properties \citep{gerhart2007fv,slotine2012links}. \\

\noindent I will next provide a summary of the math behind contraction analysis as originally reported by \cite{lohmiller1998contraction}, along with the specific contraction condition that will be explored in this chapter for our (\ref{eq:RNN}) network (subsection \ref{subsubsec:contraction-primer}). I will then review two of the core contraction combination properties - hierarchical combos and negative feedback combos - in terms of technical details, to set up how we will later make our 'networks of networks' (subsection \ref{subsubsec:contraction-primer-combos}). 

\subsubsection{Contraction math}
\label{subsubsec:contraction-primer}
We will begin by considering contractive stability in Euclidean space. The core idea behind contraction analysis is that stability can be analyzed differentially by evaluating if nearby trajectories will converge to each other -- so we consider an infinitesimal displacement at a fixed time and then "observe" the subsequent behavior. Specifically, we will measure the squared distance between the displaced and original trajectories over time, and aim to demonstrate exponential convergence (Figure \ref{fig:contracting-trajectories}). If they do exponentially converge, and we can say this for similar displacements of any trajectory of the system, then the system is globally contracting in Euclidean space. 

\cite{lohmiller1998contraction} use this to derive the condition that if $\mathbf{J} + \mathbf{J^{T}} \preceq -\beta$ uniformly, then the system with Jacobian $\mathbf{J}$ is globally contracting at rate $\beta$ ($> 0$). Note that throughout we take $\prec$ to correspond to an inequality on the maximum eigenvalue of the matrix on the left hand side. Recall also that the symmetric part of a matrix $\mathbf{A} = \frac{\mathbf{A} + \mathbf{A^T}}{2}$, and all matrices can be broken down into the sum of their symmetric and skew-symmetric parts. Symmetric matrices have all real eigenvalues, and a matrix is referred to as 'positive definite' if it is symmetric with only positive eigenvalues. \\

The core idea of contraction analysis does not require such a specific distance metric as Euclidean space however. Contractive stability can involve a wide variety of distance functions, and these distance functions can even vary over the course of the "observation". It is of course necessary that the distance metric always evaluates to 0 when given identical points, and evaluates to a positive number when not given identical points. Basically, contraction is guaranteeing that even in Euclidean space the trajectories will converge, but it is possible the path will look squiggly, while in some other valid space the trajectories are in fact exponentially converging to each other with no "backtrack" possible -- like what we saw in Euclidean space for the original simple case. 

\noindent Specifically, \cite{lohmiller1998contraction} show that the non-autonomous system

\[\dot{\mathbf{x}} = \mathbf{f}(\mathbf{x},t) \]
\noindent is contracting if there exists a metric $\mathbf{M}(\mathbf{x},t)  = \mathbf{\Theta}(\mathbf{x},t)^T\mathbf{\Theta}(\mathbf{x},t) \succ 0$ such that uniformly

\[\dot{\mathbf{M}} + \mathbf{M}\mathbf{J} + \mathbf{J}^T\mathbf{M} \preceq -\beta \mathbf{M} \]
\noindent where $\mathbf{J} = \frac{\partial \mathbf{f}}{\partial \mathbf{x}}$ and $\beta > 0$. \\

Essentially, we want to find some uniformly positive definite $\mathbf{M}$ that will make our system Jacobian $\mathbf{J}$ uniformly satisfy $\dot{\mathbf{M}} + \mathbf{M}\mathbf{J} + \mathbf{J}^T\mathbf{M} \prec 0$ in order to prove global contraction. This chapter focuses entirely on using a constant metric to show global contraction of our network (for reasons that will be discussed in section \ref{subsec:counters}), so we deal exclusively here with a static $\mathbf{M}$ and thus $\dot{\mathbf{M}} = 0$.

\noindent This leads us to the following contraction condition for the (\ref{eq:RNN}) system, which we will work to prove is true for particular restrictions on $\mathbf{W}$ and with particular valid metrics $\mathbf{M}$:
\begin{equation}\label{eq:JacobCond}
\begin{split}
\frac{\mathbf{MWD} + \mathbf{DW^{T}M}}{2} - \mathbf{M} \prec 0
\end{split}
\end{equation}
\noindent In other words, we want to show that there exists a positive definite $\mathbf{M}$ such that $\mathbf{MWD}_{sym} - \mathbf{M}$ has all negative eigenvalues for a given network, for all possible values that $\mathbf{D}$ can take. As $\mathbf{MWD}_{sym} - \mathbf{M}$ is a symmetric matrix, an equivalent statement is that if we can show this is negative definite uniformly, we have proven that (\ref{eq:RNN}) with weights $\mathbf{W}$ is globally contracting in metric $\mathbf{M}$. 

\FloatBarrier

\subsubsection{Modularity principles}
\label{subsubsec:contraction-primer-combos}
\begin{figure}[h]
\centering
\includegraphics[width=\textwidth,keepaspectratio]{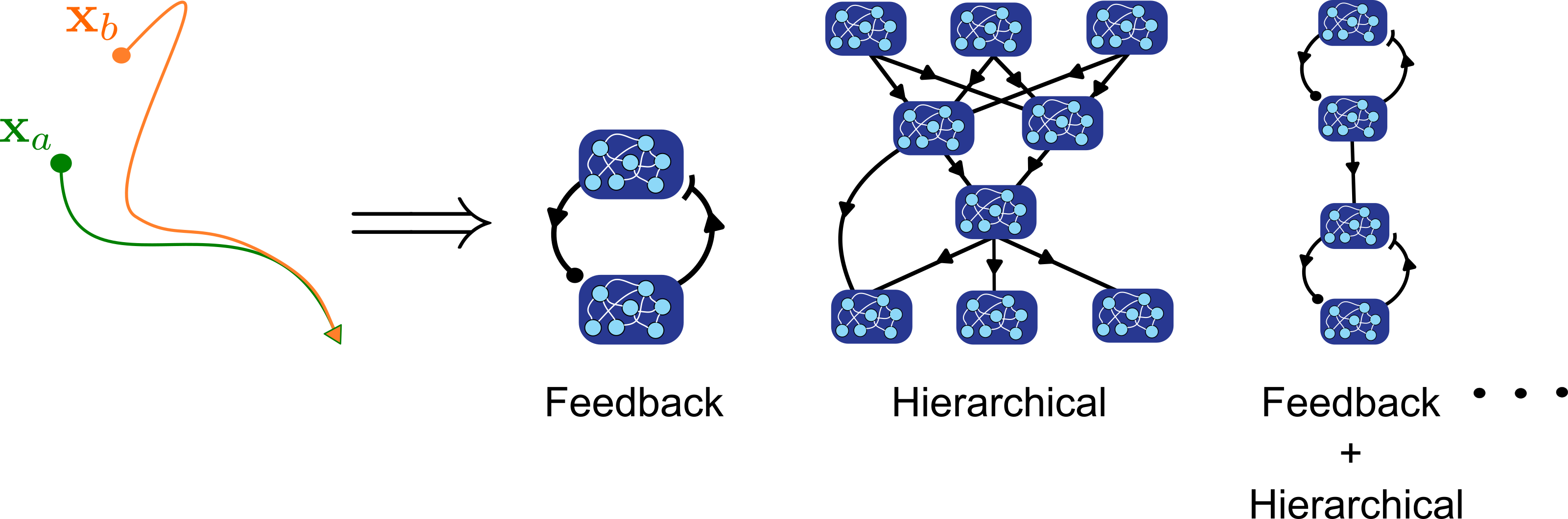}
\caption[Contractive stability implies a modularity principle.]{\textbf{Contractive stability implies a modularity principle.} Because contraction analysis tools allow complicated contracting systems to be built up recursively from simpler elements, this form of stability is well suited for understanding biological systems. Contracting combinations can be made between systems with very different dynamics, as long as those dynamics are contracting.}
\label{fig: Cartoon}
\end{figure}

As mentioned, a number of combination properties for contracting systems have been previously proven \citep{slotine2001combos,modular}. These properties allow for different types of connections between different stable systems to be formed while preserving contractivity in the combined system. It is possible to combine systems with different underlying dynamics, systems that are contracting in different metrics, etc. through these properties. Here we focus on negative feedback and hierarchical system combinations, which have straightforward neuroscientific interpretations and which can allow for highly complex combined networks to be built up through recursive construction (Figure \ref{fig: Cartoon}). 

A major goal of our results (section \ref{sec:math-results}) was to take individual RNNs described by (\ref{eq:RNN}) that are provably contracting, and then apply these combination properties to build up "RNNs of RNNs" that are themselves contracting. I will therefore introduce the prior more general results on contracting combinations of relevance here. \\

\paragraph{Feedback combination.}
Consider two systems, independently contracting in constant metrics $\mathbf{M}_1$ and $\mathbf{M}_2$, which are combined in feedback:

\[\dot{\mathbf{x}} = \mathbf{f}(\mathbf{x},t) + \mathbf{B}\mathbf{y} \]

\[\dot{\mathbf{y}} = \mathbf{g}(\mathbf{y},t) + \mathbf{G}\mathbf{x} \]

\noindent If:

\begin{equation*}
\mathbf{B} = -\mathbf{M}^{-1}_1\mathbf{G}^T\mathbf{M}_2 
\end{equation*}

\noindent then the combined system is contracting as well. This may be seen as a special case of the feedback combination derived by \cite{tabareau2006notes}. In particular, this describes a linear \emph{negative} feedback connection between the systems. \\

\paragraph{Hierarchical combination.}
Consider again two systems, independently contracting in some metrics, which are combined in hierarchy:

\[\dot{\mathbf{x}} = \mathbf{f}(\mathbf{x},t)\]

\[\dot{\mathbf{y}} = \mathbf{g}(\mathbf{y},t) + \mathbf{h}(\mathbf{x},t) \]
\noindent where $\mathbf{h}(\mathbf{x},t)$ is a function with \textit{bounded} Jacobian. Then this combined system is contracting in a diagonal metric, as shown by \cite{lohmiller1998contraction}. By recursion, this extends to hierarchies of arbitrary depth. Note also this permits for nonlinear connection between the systems.

\FloatBarrier

\subsection{Prior research on RNN model stability}
\label{subsec:rnn-lit-rev}
While there has been minimal theory to date on stability conditions for multi-area nonlinear RNNs, there of course has been a good deal of prior literature on the individual RNN model described by (\ref{eq:RNN}), as well as prior application of contraction analysis to other recurrent models. This section serves to establish more technical background on such prior results, to better contextualize our work on (\ref{eq:RNN}). \\

We first note that the 'Echo-State Condition' introduced by \cite{jaeger2001echo} is equivalent to discrete-time contraction in the identity metric. A later generalization of this condition included a diagonal metric \citep{buehner2006tighter}. In the context of neuroscience, contraction analysis has been applied to analyzing the dynamics of winner-take-all networks \citep{rutishauser2011collective,rutishauser2015computation} as well as networks with synaptic plasticity \citep{kozachkov2020achieving}. In the machine learning context, Miller and Hardt recently rederived an `echo-state property' for discrete recurrent models, and went on to prove that these contracting recurrent models could be well-approximated by feedforward networks in certain cases \citep{miller2018stable}.
More recently still, in a series of papers Revay, Wang, and Manchester applied contraction analysis to discrete-time recurrent networks \citep{revay2020contracting, revay2021recurrent,revay2020convex}, expanding the class of models considered by \cite{miller2018stable}.

In addition to contraction (which amounts to a strong form of non-autonomous exponential stability) there has been a considerable amount of work attempting to enforce weaker forms of stability in RNNs, such as \textit{autonomous} stability \citep{erichson2021lipschitz, chang2019antisymmetricrnn}. Unfortunately, as we discuss in section \ref{subsec:counters}, autonomous stability does not in general imply non-autonomous stability, so it is not clear what stability properties these RNNs possess when driven with external input. 

There is also a line of work which uses orthogonal weight matrices to avoid the vanishing/exploding gradient problem during training. Orthogonality is typically ensured during training via a parameterization \citep{arjovsky2016unitary,lezcano2019cheap} -- for example, by exploiting the fact that the matrix exponential of a skew-symmetric matrix is orthogonal, as is done in \cite{lezcano2019cheap}. These works focus on parameterizations of \textit{individual} RNNs which ensure stability is preserved throughout training, and our use of Theorem \ref{theorem: singularvaluetheorem} does draw from these works. By contrast, while we also introduce novel stability conditions for individual RNNs, our work focuses on 'combined' RNNs (i.e network of networks) and provides a parameterization on the \textit{connections between stable subnetworks} such that training preserves the overall network stability (Figure \ref{fig: Cartoon}). 

\section{Mathematical results}
\label{sec:math-results}
The mathematical results of this chapter fall under 4 major categories, to be covered in the following subsections:
\begin{enumerate}
    \item New stability conditions on the weights of the general network equation (\ref{eq:RNN}), each guaranteeing global contraction in a constant metric when satisfied (subsection \ref{subsec:novel-conditions}).
    \item Application of prior contraction analysis results to develop a deep learning applicable methodology for combining contracting networks described by (\ref{eq:RNN}) into a larger recurrent "RNN of RNNs" that is itself provably stable (subsection \ref{subsec:combo-conditions}).
    \item A discussion of conditions under which pruning might disrupt stability, to facilitate future work on directly sparsifying multi-area RNNs, a topic of potential relevance both for more organic extensions on our pilot experimental results to be discussed, as well as for better understanding neuroscientific principles (subsection \ref{section:subnets}).
    \item Counterexamples to a few previously published stability "conditions" for (\ref{eq:RNN}), alongside a deeper explanation for why certain conjectured stability conditions on these dynamics have been so hard to prove thus far (subsection \ref{subsec:counters}).
\end{enumerate}

\subsection{Novel contraction conditions for individual, continuous-time RNNs}
\label{subsec:novel-conditions}
This subsection focuses on presenting, explaining, and proving our novel stability conditions for the nonlinear RNN described by (\ref{eq:RNN}). As previously proven conditions have been quite restrictive, these theoretical results are valuable in their own right. However, they are also important to our broader aims, because the new conditions can guarantee that (\ref{eq:RNN}) is globally contracting in a constant metric, and moreover can provide such a metric -- the properties that allow us to apply prior results in contraction analysis to construct stable "networks of networks" out of the component RNNs satisfying the conditions here. The two major contraction conditions on (\ref{eq:RNN}) that we derive have quite different pros and cons, and were employed in different ways experimentally. 

As will be covered in section \ref{sec:dl-results}, Theorem \ref{theorem: absolutevaluetheorem} was used for the "Sparse Combo Net" stable multi-area architecture, which demonstrated the best performance in our hands. The condition is very simple to check because of its deep link to linear stability, and it is satisfied by a number of matrices with interesting sparse structures. Theorem \ref{theorem: absolutevaluetheorem} will thus be the major focus of exposition in this subsection, covered in depth in \ref{subsubsec:theorem-1-exp} next. The other condition I will review here (in \ref{subsubsec:theorem-svd-exp}) was used for our "SVD Combo Net" architecture. Theorem \ref{theorem: singularvaluetheorem} has the advantage that matrices satisfying it can be written in terms of straightforward operations between other matrices, the latter of which are all either expressions that do not require special constraints or contraction properties that can be preset. The condition can therefore be enforced throughout training of a network by performing optimization on components of said parameterization. By contrast, we could find no clear way to directly parameterize Theorem \ref{theorem: absolutevaluetheorem}.

\subsubsection{Linear stability of weight magnitudes as a nonlinear contraction condition}
\label{subsubsec:theorem-1-exp}
As will be discussed in more depth in section \ref{subsec:counters}, interpretation of negative feedback is the major complication in proving stability of a nonlinear system described by (\ref{eq:RNN}). We can think of this intuitively using the rectified linear unit (ReLU) activation function as an example. ReLU is a valid $\phi$ for this system and it is a very common activation function to use in modern deep learning systems. From a neuroscience perspective, ReLU was originally designed to move on from the classic sigmoid function interpretation of unit outputs as an "all or nothing" signal, instead framing them to be interpreted as neuron firing rates. As such, ReLU outputs 0 for inputs $\leq 0$, but otherwise just outputs the inputs, making the derivative instead the binary function. 

When we compare ReLU activation to an otherwise identical linear network, it becomes obvious that if everything were constrained to be positive, the networks would be functionally equivalent. In reality we do not assume control of initial conditions or of inputs, but if the network can "survive" the most adversarially large all-positive inputs, it makes intuitive sense that negative inputs would not "break" global contraction. Linear stability is already well-understood and easy to check, so taken together this is what inspired the question of whether we could find stability conditions on $\mathbf{W}$ in terms of the linearized version of the system: first beginning with the narrow class of excitatory-only RNNs, but then extending the idea to a more general condition on magnitudes of weights in the network. \\

\noindent We ultimately ended up with the following theorem, which shows that a linearly stable $\mathbf{|W|} - \mathbf{I}$ implies the original nonlinear system with weight matrix $\mathbf{W}$ is also stable, and in fact will be stable in the same constant metric, a property we will leverage later.

\begin{theorem}[Absolute Value Restricted Weights]
\label{theorem: absolutevaluetheorem}
Let $\mathbf{|W|}$ denote the matrix formed by taking the element-wise absolute value of $\mathbf{|W|}$.  If there exists a positive, diagonal $\mathbf{M}$ such that:
\[\mathbf{M}(\mathbf{|W|}-\mathbf{I}) + (\mathbf{|W|}-\mathbf{I})^T\mathbf{M} \prec 0 \]
then (\ref{eq:RNN}) is contracting in metric $\mathbf{M}$. If $W_{ii} \leq 0$, then $|W|_{ii}$ may be set to zero to reduce conservatism.
\end{theorem}

\noindent This condition is particularly straightforward in the common special case where the network does not have any self weights, with the leak term driving stability. While it can be applied to a more general $\mathbf{W}$, the condition will of course not be met if the network was relying on highly negative values on the diagonal of $\mathbf{W}$ for linear stability. As demonstrated by counterexample in the proof of Theorem \ref{theorem: absolutevaluetheorem}, it can be impossible to use the same metric $\mathbf{M}$ for the nonlinear RNN in such cases, which breaks the proof method. However as the issue arises when only some of the self weights are highly negative, one could consider instead allowing a gain on the leak term (which would scale the identity matrix $\mathbf{I}$) if designing a system. \\

Theorem \ref{theorem: absolutevaluetheorem} allows many weight matrices with low magnitudes or a generally sparse structure to be verified as contracting in the nonlinear system (\ref{eq:RNN}), by simply checking a linear stability condition.  This is particularly useful because sparsity is a recurring theme in neuroscience, both enabling stability in theory \citep{kozachkov2020achieving}, and occurring in practice, with only $\sim 0.00001\%$ of potential connections forming synapses in the human brain \citep{slotine2012links}. This could explain why there has been little issue approximating nonlinear systems with linear ones in many computational neuroscience papers.

Beyond verifying contraction, Theorem \ref{theorem: absolutevaluetheorem} actually provides a metric, with little need for additional computation. Not only is it of inherent interest that the same metric can be shared across systems in this case, it is also of use in machine learning, where stability certificates are becoming increasingly necessary for some applications. More generally, there are a variety of systems of interest that meet this condition but do not meet the well-known maximum singular value condition, including those with a hierarchical structure. 

One downside of the condition is that it is not straightforward to enforce throughout training in a deep learning framework. However, there are a number of ways that one could encourage the condition to be maintained throughout training, and it is easy to confirm the final network is stable under these criteria before using it in applications. A number of regularization techniques that work to keep network weights small and/or sparse would likely help with maintaining the condition, as would strategic initialization methods or training paradigms that restrict which weights can be updated. This is even more relevant in light of the recent impacts of the Lottery Ticket Hypothesis, which states that given a neural network with a particular initialization, aggressive pruning after training can identify much smaller networks that are similarly capable of learning the task from scratch \citep{frankle2019lth}. One could imagine using pruning to enforce Theorem \ref{theorem: absolutevaluetheorem} post-hoc, or in an inverse scenario, aiming to build up small stable RNNs capable of strong task performance via mechanisms like the Weight Agnostic Neural Networks introduced by \cite{gaier2019wann}.

Of course in our multi-area framework (to be detailed in section \ref{subsec:combo-conditions}), it is possible to train the linear negative feedback connections between nonlinear RNNs meeting the Theorem \ref{theorem: absolutevaluetheorem} condition without updating the nonlinear RNNs themselves, like a more complex version of the historical echo state network idea \citep{jaeger2001echo} and reminiscent of the evolutionary concept of facilitated variation \citep{gerhart2007fv}. This is the approach we took with our pilot experiments, which demonstrated great early success particularly in the context of the simplicity of this system (\ref{sec:dl-results}). \\

I will now restate the Theorem \ref{theorem: absolutevaluetheorem} in different but equivalent terms, before proceeding to provide a proof of the claims. The primary principle underlying the proof is the fact that linear stability is equivalent to diagonal stability for Metzler (i.e. nonnegative off-diagonal) matrices \citep{narendra2010metzler}. This principle is used in conjunction with well-established linear algebra inequalities. To skip these details, please see instead the next Theorem's introduction below, in subsection \ref{subsubsec:theorem-svd-exp}. \\

\paragraph{Proof of Theorem \ref{theorem: absolutevaluetheorem}.}
Stated another way:

 \begin{quote}
 Given an anti-diagonal (i.e. no self weights) $\mathbf{W}$ in the nonlinear RNN, if its elementwise absolute value $\mathbf{|W|}$ would produce a linearly stable system (i.e. $\mathbf{|W|} - \mathbf{I}$ is Hurwitz), then the nonlinear system with $\mathbf{W}$ is guaranteed to be globally contracting. Furthermore, it is globally contracting in the same diagonal metric(s) in which the $\mathbf{|W|}$ system is linearly stable. 
 \end{quote}

 \begin{proof}
Recall that if a positive definite $\mathbf{M}$ exists such that $\frac{\mathbf{MWD} + \mathbf{DW^{T}M}}{2} - \mathbf{M} \prec 0$ uniformly for a given $\mathbf{W}$, then the RNN system is contracting in that metric $\mathbf{M}$. As noted, '$\prec$' refers throughout this chapter to a bound on the maximum eigenvalue of the matrix on the left hand side. Recall also that $\mathbf{D}$ is a diagonal matrix containing the activation function derivatives evaluated at the current state for each unit (which must be nonnegative and at most 1 for the functions considered here). See section \ref{sec:math-intro} for further introduction on mathematical notation, contraction analysis, and the Jacobian of the system in question. 
 
For proof of the main claim, we will first prove that given a nonnegative (i.e. all elements nonnegative) Hurwitz $\mathbf{W}$, the system is contracting. Note that if a nonnegative $\mathbf{W}$ is linearly stable at all, it must be diagonally stable \citep{narendra2010metzler} -- it then follows that the system with linear activation (meaning $\mathbf{D}=\mathbf{I}$ for all time) must be contracting in some diagonal metric. So, there exists a positive definite diagonal $\mathbf{M}$ such that $\frac{\mathbf{MW} + \mathbf{W^{T}M}}{2} - \mathbf{M} \prec 0$. To extend to the nonlinear activation case, We will show that given any diagonal $\mathbf{D}$ with all elements between 0 and 1 (inclusive), we can guarantee that $\frac{\mathbf{MWD} + \mathbf{DW^{T}M}}{2} - \mathbf{M} \prec 0$, meaning the nonlinear system is contracting in the same diagonal metric $\mathbf{M}$. Note that the proof applies equally well regardless of the $\mathbf{M}$ used, as long as it is a diagonal contraction metric for the linear system; therefore the nonlinear system will share all diagonal contraction metrics with the linear counterpart. 

To prove the claim for the nonnegative Hurwitz $\mathbf{W}$, we will make use of the spectral radius function, $\rho$. Per Perron-Frobenius, for a real nonnegative symmetric matrix $\mathbf{A}$, $\rho(\mathbf{A})$ will always be equivalent to $\lambda_{max}(\mathbf{A})$ (i.e. the maximum eigenvalue of $\mathbf{A}$), which will be real. Furthermore, the spectral radius is monotone for nonnegative matrices $\mathbf{A}$ and $\mathbf{B}$, meaning $\rho(\mathbf{A}) \leq \rho(\mathbf{B})$ if $\mathbf{A} \leq \mathbf{B}$ elementwise. More broadly, we will use the fact that $\rho(\mathbf{A}) \leq \rho(\mathbf{|A|}) \leq \rho(\mathbf{B})$ if $\mathbf{|A|} \leq \mathbf{B}$ elementwise at multiple times throughout this proof. A proof of that inequality can be found in the Linear Algebra text by \cite{Meyer}, in Example 7.10.2.

Now observe that $\frac{\mathbf{MWD} + \mathbf{DW^{T}M}}{2} - \mathbf{M}$ is the sum of two symmetric matrices, and therefore symmetric. It is Metzler, but the diagonals can (and will) contain negative values due to the $-\mathbf{M}$ contributed by the leak term (otherwise it would not be linearly stable in the first place). To work with a fully nonnegative matrix instead, we will choose some positive integer $c$ such that $\frac{\mathbf{MWD} + \mathbf{DW^{T}M}}{2} - \mathbf{M} + c\mathbf{I}$ has all nonnegative diagonals. Given the system assumptions, it will always be possible to use $c = max(\mathbf{M})$ i.e. the maximum element in the diagonal contraction metric used for the linear system. Adding $c\mathbf{I}$ will increase the eigenvalue of the resulting matrix by exactly $c$ \citep{Meyer}, so that without loss of generality we can instead prove $\frac{\mathbf{MWD} + \mathbf{DW^{T}M}}{2} - \mathbf{M} + c\mathbf{I} \prec c$ uniformly to show that the nonlinear RNN system is contracting in metric $\mathbf{M}$. 

$\frac{\mathbf{MWD} + \mathbf{DW^{T}M}}{2} - \mathbf{M} + c\mathbf{I}$ meets all our desired conditions for utilizing the spectral radius. Furthermore, we know that $\frac{\mathbf{MW} + \mathbf{W^{T}M}}{2} - \mathbf{M} + c\mathbf{I} \prec c$ by the same logic, and that $\frac{\mathbf{MW} + \mathbf{W^{T}M}}{2} - \mathbf{M} + c\mathbf{I}$ is also a real symmetric nonnegative matrix. We will therefore aim to show that $\frac{\mathbf{MWD} + \mathbf{DW^{T}M}}{2} - \mathbf{M} + c\mathbf{I} \preceq \frac{\mathbf{MW} + \mathbf{W^{T}M}}{2} - \mathbf{M} + c\mathbf{I}$ uniformly via the spectral radius inequality, thereby demonstrating that $\frac{\mathbf{MWD} + \mathbf{DW^{T}M}}{2} - \mathbf{M} + c\mathbf{I} \prec c$, as desired.

To do so, note that for any nonnegative $\mathbf{W}$, the diagonal $\mathbf{D}$ with values between 0 and 1 can only possibly decrease the elements of $\frac{\mathbf{MWD} + \mathbf{DW^{T}M}}{2}$ (recall $\mathbf{M}$ in this case is a positive diagonal matrix). Because the other terms on the left hand side of the above inequality are constant regardless of $\mathbf{D}$, $\mathbf{D}=\mathbf{I}$ is guaranteed to produce the largest possible matrix elementwise. Due to the monotonicity of spectral radius in nonnegative matrices, the max eigenvalue of $\frac{\mathbf{MWD} + \mathbf{DW^{T}M}}{2} - \mathbf{M} + c\mathbf{I}$ must then always be $\leq$ to that when $\mathbf{D}=\mathbf{I}$ -- which is exactly the linear RNN condition given. Thus the maximum eigenvalue of $\frac{\mathbf{MWD} + \mathbf{DW^{T}M}}{2} - \mathbf{M} + c\mathbf{I}$ can never go above $c$ regardless of how $\mathbf{D}$ varies, completing our lemma for Hurwitz nonnegative networks. \\

To return to the main claim, suppose instead the given was that $\mathbf{|W|}$ is Hurwitz, where $\mathbf{|W|}$ represents the elementwise absolute value of the weight matrix $\mathbf{W}$. Analogous to the simpler case, we know based on this information that there is a diagonal $\mathbf{M}$ such that $\frac{\mathbf{M|W|} + \mathbf{|W|^{T}M}}{2} - \mathbf{M} \prec 0$. If we plug in the preceding spectral radius inequality here, we have that $\rho(\frac{\mathbf{MWD} + \mathbf{DW^{T}M}}{2} - \mathbf{M} + c\mathbf{I}) \leq \rho(|\frac{\mathbf{MWD} + \mathbf{DW^{T}M}}{2} - \mathbf{M} + c\mathbf{I}|)$. Our aim is to use these known properties to show that the original $\mathbf{W}$ is contracting in the nonlinear system with metric $\mathbf{M}$, i.e. $\frac{\mathbf{MWD} + \mathbf{DW^{T}M}}{2} - \mathbf{M} \prec 0$ uniformly. 

We will begin with an assumption that $\mathbf{W}$ has nonnegative diagonals. In that case, we can pull $c\mathbf{I} - \mathbf{M}$ (which will be a positive diagonal matrix) outside of the absolute value sign. Then observe that taking the absolute value after taking the symmetric part of a matrix will always produce a result that is elementwise less than or equal to the matrix that results from taking the absolute value beforehand (because anti-symmetric elements cannot cancel out when the absolute value is taken first). $\mathbf{M}$ and $\mathbf{D}$ are nonnegative diagonal matrices, so that $\mathbf{M|W|D} + \mathbf{D|W|^{T}M} = \mathbf{|MWD|} + \mathbf{|DW^{T}M|}$. This gives us the final relationship $|\frac{\mathbf{MWD} + \mathbf{DW^{T}M}}{2} - \mathbf{M} + c\mathbf{I}| \leq \frac{\mathbf{M|W|D} + \mathbf{D|W|^{T}M}}{2} - \mathbf{M} + c\mathbf{I}$ elementwise. 

These matrices can now be plugged into the above spectral radius inequality to get $\rho(\frac{\mathbf{MWD} + \mathbf{DW^{T}M}}{2} - \mathbf{M} + c\mathbf{I}) \leq \rho(|\frac{\mathbf{MWD} + \mathbf{DW^{T}M}}{2} - \mathbf{M} + c\mathbf{I}|) \leq \rho(\frac{\mathbf{M|W|D} + \mathbf{D|W|^{T}M}}{2} - \mathbf{M} + c\mathbf{I})$. Thus the maximum magnitude eigenvalue of the original system will always be less than or equal to that of the system with $\mathbf{|W|}$, where we can already show stability across possible $\mathbf{D}$ per our first lemma. \\

The above assumed that $\mathbf{W}$ had no negative diagonal entries. To still apply the technique even with a $\mathbf{W}$ that has negative self weights, we can simply 0 those out -- as this only makes the system less likely to be stable. More formally, we will express $\mathbf{W}$ as $\mathbf{C} + \mathbf{S}$, where $\mathbf{S}$ is a diagonal matrix containing all negative self weights (0 on the corresponding diagonals that do not have a negative value), and $\mathbf{C} = \mathbf{W} - \mathbf{S}$ is all the connections and any remaining positive self weights (note these must be $< 1$ for hope at stability anyway). 

Plugging the new expression for $\mathbf{W}$ into the symmetric part of the Jacobian and rearranging terms, we have $\mathbf{J} = (\mathbf{M}(\mathbf{C} + \mathbf{S})\mathbf{D})_{sym} - \mathbf{M} = (\mathbf{MCD})_{sym} + (\mathbf{MSD})_{sym} - \mathbf{M}$. Applying Weyl's inequality, $\mu_{2}(\mathbf{MCD} + \mathbf{MSD} - \mathbf{M}) \leq \mu_{2}(\mathbf{MCD} - \mathbf{M}) + \mu_{2}(\mathbf{MSD})$. With $\mathbf{M}$ and $\mathbf{D}$ as nonnegative diagonal matrices and $\mathbf{S}$ a nonpositive one, $\mu_{2}(\mathbf{MSD}) \leq 0$, so that we have the final inequality $\lambda_{max}(\mathbf{J}_{sym}) \leq \lambda_{max}((\mathbf{MCD})_{sym} - \mathbf{M})$. Therefore any stability condition that holds for the connection matrix $\mathbf{C}$ must also hold for the larger $\mathbf{W}$ matrix. \\

It is worth noting that highly negative diagonal values in $\mathbf{W}$ will prevent the same metric $\mathbf{M}$ from being used for the nonlinear system. Therefore the method used in this proof cannot feasibly be adapted to further relax the treatment of the diagonal part of $\mathbf{W}$. The intuitive reason behind this is that in the symmetric part of the Jacobian, $\frac{\mathbf{M}\mathbf{W}\mathbf{D} + \mathbf{D}\mathbf{W}^{T}\mathbf{M}}{2} - \mathbf{M}$, the diagonal self weights will also be scaled down by small $\mathbf{D}$, while the leak portion $-\mathbf{M}$ remains untouched by $\mathbf{D}$.

\noindent Now we actually demonstrate a counterexample, presenting a $2 \times 2$ symmetric Metzler matrix $\mathbf{W}$ that is contracting in the identity in the linear system, but cannot be contracting \textit{in the identity} in the nonlinear system (\ref{eq:RNN}): 

\begin{align*}
\mathbf{W} = \begin{bmatrix} 
    -9 & 2.5 \\
    2.5 & 0 
  \end{bmatrix}
\end{align*}
  
\noindent To see that it is not possible for the more general nonlinear system with these weights to be contracting in the identity, take 
  $\mathbf{D} = \begin{bmatrix} 
    0 & 0 \\
    0 & 1 
  \end{bmatrix}$. Now
  
\begin{align*}
    (\mathbf{W}\mathbf{D})_{sym} - \mathbf{I} = \begin{bmatrix} 
    -1 & 1.25 \\
    1.25 & -1 
  \end{bmatrix}
\end{align*}

\noindent which has a positive eigenvalue of $\frac{1}{4}$.
\end{proof}

\noindent Note that this proof has focused on the case $g=1$ (i.e. activation function derivative upper bounded at 1), which most of the commonly used activation functions at present satisfy. However, a version of the condition can be derived for more general $g$ as well (supplemental section \ref{sec:thm1-alt-pf}).

\subsubsection{A contraction condition for guaranteed stability throughout training}
\label{subsubsec:theorem-svd-exp}
While there are many interesting network structures that satisfy Theorem \ref{theorem: absolutevaluetheorem}, and a number of potential methods to push networks towards satisfying Theorem \ref{theorem: absolutevaluetheorem} during training, strictly enforcing that condition throughout optimization is not straightforward. This motivates our next theorem, which derives a different continuous-time contraction condition that does have a direct parameterization that can easily be plugged into modern optimization libraries. \\

\noindent The theorem states:

\begin{theorem}\label{theorem: singularvaluetheorem}
If there exists a positive diagonal matrix $\mathbf{P}$ such that:

\[g^2\mathbf{W}^T\mathbf{P}\mathbf{W} - \mathbf{P} \prec 0 \]

\noindent then (\ref{eq:RNN}) is contracting in metric $\mathbf{P}$.  
\end{theorem}

\begin{figure}[h]
\centering
\includegraphics[width=0.5\textwidth]{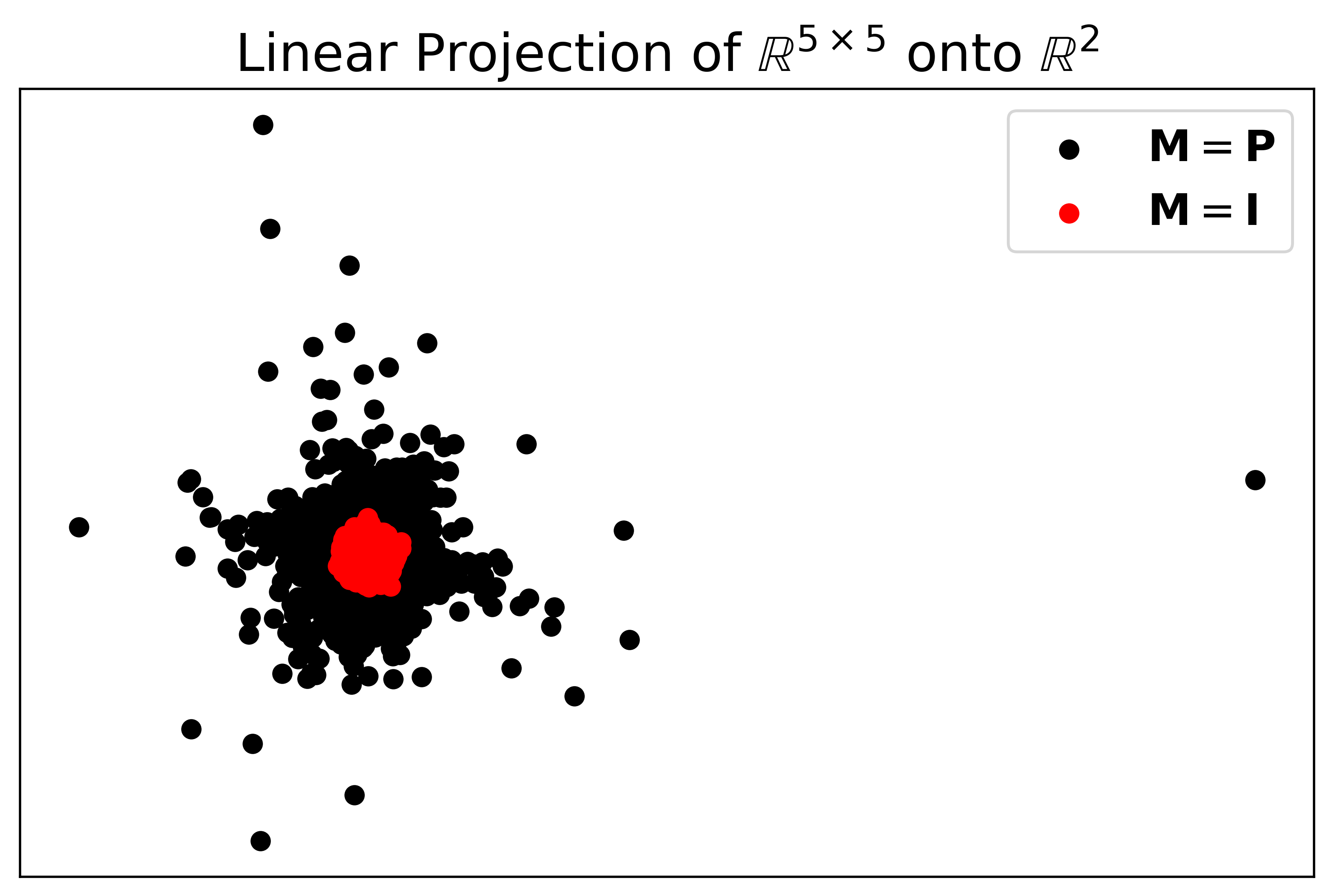}
\caption[Our novel single-RNN conditions capture a larger set of stable nonlinear network weight matrices.]{\textbf{Considering different contraction metrics can capture a larger set of stable network weight matrices.} Linear projection (using principal components analysis) of randomly generated matrices satisfying the stability condition in Theorem \ref{theorem: singularvaluetheorem} with $\mathbf{M} = \mathbf{I}$ and $\mathbf{M} = \mathbf{P}$.} 
\label{fig:projection-map}
\end{figure}

\noindent This condition is an extension of checking the maximum singular value of $\mathbf{W}$, but allowing a constant metric that can expand the set of matrices it can detect as stable -- demonstrating another advantage of contraction analysis (Figure \ref{fig:projection-map}). It reduces to the former condition when $\mathbf{M}=\mathbf{I}$. A discrete time version of this contraction condition had been previously proven by \cite{revay2020contracting}, but this is the first proof in continuous time to our knowledge. \\

\noindent For a direct parameterization of the theorem, note the above condition on $\mathbf{W}$ is satisfied if $\mathbf{W}$ can be written as:

\[\mathbf{W} = g^{-2}\mathbf{P}^{-1/2}\mathbf{U}\mathbf{S}\mathbf{V}^T\mathbf{P}^{1/2}\]
where $\mathbf{U},\mathbf{V}$ are orthogonal matrices and $\mathbf{S}$ is a non-negative diagonal matrix with $ 0 \leq S_{ii} < 1$. The matrix $e^{\mathbf{N}}$ with $\mathbf{N} = -\mathbf{N}^T$ is an orthogonal matrix, so this can be used to parameterize $\mathbf{U}$ and $\mathbf{V}$.

The proof of Theorem \ref{theorem: singularvaluetheorem} is brief and presented next. Our work on combination properties will be reported on following the proof, in section \ref{subsec:combo-conditions}. \\

 \paragraph{Proof of Theorem \ref{theorem: singularvaluetheorem}.}
 \begin{proof}
Consider the generalized Jacobian:

\[\mathbf{F} = \mathbf{P}^{1/2}\mathbf{J} \mathbf{P}^{-1/2} = -\mathbf{I} + \mathbf{P}^{1/2}\mathbf{W}\mathbf{P}^{-1/2}\mathbf{D}  \]

where $\mathbf{D}$ is a diagonal matrix with $\mathbf{D}_{ii} = \frac{d\phi_i}{dx_i} \geq 0$. Using the subadditivity of the matrix measure $\mu_2$ of the generalized Jacobian we get:
\[\mu_2(\mathbf{F}) \leq -1 + \mu_2(\mathbf{P}^{1/2}\mathbf{W}\mathbf{P}^{-1/2}\mathbf{D}) \]

Now using the fact that $\mu_2(\cdot) \leq ||\cdot||_2$ we have:

\[\mu_2(\mathbf{F}) \leq -1 + ||\mathbf{P}^{1/2}\mathbf{W}\mathbf{P}^{-1/2}\mathbf{D})||_2 \leq -1 + g||\mathbf{P}^{1/2}\mathbf{W}\mathbf{P}^{-1/2}||_2 \]

Using the definition of the 2-norm, imposing the condition $\mu_2(\mathbf{F}) \leq 0$ may be written:
\[g^2\mathbf{W}^T\mathbf{P}\mathbf{W} - \mathbf{P} \prec 0 \]
which completes the proof.
\end{proof}

\FloatBarrier

\subsection{Generalizing to the network of networks: "RNNs of RNNs"}
\label{subsec:combo-conditions}
The RNN in (\ref{eq:RNN}) is a \textit{subnetwork} of the overall model we introduce here. Our goal is to combine these subnetworks into a 'network of networks' in a manner that preserves stability of the overall network as long as the underlying modules are stable. We will make use of both of the theorems from section \ref{subsec:novel-conditions} to enforce contraction in subnetworks for our experiments, but other contraction conditions could also be used. Through the rest of this section on network combinations, we will thus assume we are given a set of subnetworks that are provably contracting in a given set of corresponding metrics, and we will focus only on combining them in a provably stable way.

Unsurprisingly, we will utilize the contraction combination properties discussed within section \ref{sec:math-intro} to do so. We focus on negative feedback combinations in particular due to their inherently recurrent structure. For reasons hinted at in discussing Theorem \ref{theorem: absolutevaluetheorem}, which will be covered in substantially more detail with Theorem \ref{theorem: Wdiagstabcounterexampletheorem} below, it is difficult to utilize negative feedback connections in a provably stable way when those connections are nonlinear. However, stability of linear negative feedback connections between the subnetworks follows directly from the existing contraction combination properties. 

Note that the connections within each subnetwork will remain nonlinear as described by (\ref{eq:RNN}), it is only the connections \emph{between} subnetworks that take the form of linear negative feedback. Therefore the system on the whole remains nonlinear. The linearity assumption can also be motivated by the fact that RNNs have been found to be well-approximated by linear systems in many neuroscience contexts \citep{sussillo2013opening,langdon2022latent}. \\

\noindent Here, we formally apply the negative feedback contraction combination property to our model, and prove that it can be parameterized in a way that allows the strengths of different negative feedback connections within the network to be learned using modern optimization tools. We thus make extensive use of the following theorem:

\begin{theorem}[Network of Networks]
\label{theorem: network_of_networks}
Consider a collection of $p$ subnetwork RNNs governed by (\ref{eq:RNN}). Assume that these RNNs each have hidden-to-hidden weight matrices $\{ \mathbf{W}_1,\dots,\mathbf{W}_p \}$ and are independently contracting in metrics $\{ \mathbf{M}_1,\dots,\mathbf{M}_p \}$. Define the block matrices $\Tilde{\mathbf{W}} \equiv \text{BlockDiag}(\mathbf{W}_1,\dots,\mathbf{W}_p )$ and $\Tilde{\mathbf{M}} \equiv \text{BlockDiag}(\mathbf{M}_1,\dots,\mathbf{M}_p )$, as well as the overall state vector $\Tilde{\mathbf{x}}^T \equiv (\mathbf{x}^T_1 \cdots \mathbf{x}^T_2)$. Then the following 'network of networks' is globally contracting in metric $\Tilde{\mathbf{M}} $:

\begin{equation}\label{eq:comboRNN}
\begin{split}
\tau \dot{\Tilde{\mathbf{x}}} = -\Tilde{\mathbf{x}} + \Tilde{\mathbf{W}}\phi(\Tilde{\mathbf{x}}) + \mathbf{u}(t) + \mathbf{L}\Tilde{\mathbf{x}} \\
\mathbf{L} \equiv \mathbf{B} - \Tilde{\mathbf{M}}^{-1}\mathbf{B}^T\Tilde{\mathbf{M}}
\end{split}
\end{equation}

Where $\mathbf{B}$ is an arbitrary square matrix, and can be treated as a trainable tensor in a deep learning framework. 
\end{theorem}

\noindent In the experiments of section \ref{sec:dl-results}, we freeze the block diagonal elements of $\mathbf{B}$ at 0, to prevent linear negative feedback connections from forming within the subnetworks, thereby preserving the interpretation of our model as presented conceptually. \\

In theory, the feedback combination property, taken together with hierarchical combinations, may be used as combination primitives for recursively constructing complicated networks of networks while automatically maintaining stability. The recursion comes from the fact that once a modular system is shown to be contracting it may be treated as a single contracting system, which may in turn be combined with other contracting systems, ad infinitum. Note that while the negative feedback currently requires linear interareal connections, hierarchical interareal connections can be nonlinear \citep{lohmiller1998contraction}. Although we do not formally explore this in the paper, it is another potentially fruitful direction in which our architecture could be taken. \\

To prove Theorem \ref{theorem: network_of_networks}, we show that \emph{in the appropriate metric} the connections in $\mathbf{L}$ boil down to pure antisymmetric negative feedback, i.e. $\mathbf{L}_{ij} = -\mathbf{L}_{ji}^T$. Recall that the metric refers to the "space" in which we know the system has all trajectories exponentially converge. When $\mathbf{M} = \mathbf{I}$, this is simply Euclidean space and we would indeed look for skew-symmetry directly. It is of course possible for a system to be contracting in multiple different metrics, though we only need one to use the Theorem and it is considered a given input there. The intuition behind the negative feedback combination property itself is that because contraction analysis relies on analyzing the symmetric part of Jacobian matrices, the skew-symmetry of $\mathbf{L}_{ij}$ 'cancels out' when  computing the symmetric part, and leaves the stability of the subnetwork RNNs untouched. 

\subsubsection{Proof of Theorem \ref{theorem: network_of_networks}}
\begin{proof}
Consider the differential Lyapunov function:

\[V = \frac{1}{2}\delta\mathbf{x}^T\Tilde{\mathbf{M}}\delta\mathbf{x} \]

The time-derivative of this function is:

\[\dot{V} = \delta\mathbf{x}^T\Tilde{\mathbf{M}}\delta\dot{\mathbf{x}} = \delta\mathbf{x}^T (\underbrace{\Tilde{\mathbf{M}}\Tilde{\mathbf{J}} + \Tilde{\mathbf{J}}^T\Tilde{\mathbf{M}}}_{\text{Jacobian of RNNs before interconnection}} + \underbrace{\Tilde{\mathbf{M}}\mathbf{L} + \mathbf{L}^T\Tilde{\mathbf{M}}}_{\text{Interconnection Jacobian}} )\delta\mathbf{x}\]

Since we assume that the RNNs are contracting in isolation, the first term in this sum is less that the slowest contracting rate of the individual RNNs, which we call $\lambda > 0$. The second term in the sum is the zero matrix, by construction. Thus the time-derivative of $V$ is upper-bounded by:

\[\dot{V} \leq -2\lambda V\]
This implies that the network of networks is contracting with rate $\lambda$. 
\end{proof}

\subsection{Relation of stability conditions with neuron pruning}\label{section:subnets}
Throughout this chapter, the word "subnetwork" is generally used to refer to an individual nonlinear RNN as defined in (\ref{eq:RNN}), a set of which will then be combined using linear connections that meet the combination property constraints detailed above in \ref{subsec:combo-conditions}. However, for this subsection only, "subnetwork" will refer to a network defined by taking only a subset of the units in a given RNN with (\ref{eq:RNN}), and considering an RNN defined by the same dynamics but with the unchosen neurons entirely excluded (i.e. considering $\mathbf{W}$ to be the principal submatrix corresponding to only the selected units, and keeping the states and inputs of only the selected units). 

While we have mainly focused on sufficient conditions for the system (\ref{eq:RNN}) to be contracting regardless of (eligible) activation function, identifying \emph{necessary} conditions is also of great potential use. Interestingly, one such necessary condition, presented in Theorem \ref{theorem: subnetworkstabilitytheorem}, has clear interpretations in neuroscience and machine learning, mirroring some of the implications of the combination properties described above, but now in the inverse case of pruning a system rather than building one up. Furthermore, the necessary condition again stems from the presence of both positive and negative connections between neurons. \\

\begin{theorem}\label{theorem: subnetworkstabilitytheorem}
In order for a linearly stable RNN to be contracting in all of the nonlinear activation functions permitted by (\ref{eq:RNN}), every subnetwork must also be linearly stable. Put another way, it is necessary for all principal submatrices of the matrix $\mathbf{W}$ to have only eigenvalues with negative real part. \\

\noindent Additionally, we show that if the nonlinear system (\ref{eq:RNN}) is contracting in a diagonal metric, all of its subnetworks are also contracting in a diagonal metric.
\end{theorem}

\noindent Note that as linear activation is a possible function for $\phi$ in (\ref{eq:RNN}), it is definitionally true that any $\mathbf{W}$ must be Hurwitz to possibly be stable in the more general case we consider. Thus the necessary condition here could also be formulated as "$\mathbf{W}$ and all of its principal submatrices must be linearly stable". Obviously it may be possible for the network to still be stable in a specific activation function of interest, but that is not the goal of our investigation, and furthermore the popular ReLU function (which we also use in our experiments) is strictly more complicated than the linear $\phi$, and for nonlinear stability with ReLU we do require this necessary condition, as can be seen in its proof. 

Theorem \ref{theorem: subnetworkstabilitytheorem} also connects back to Theorem \ref{theorem: absolutevaluetheorem}. All diagonally stable linear networks will meet the necessary condition for potential nonlinear stability, because diagonal stability guarantees that all principal submatrices are also diagonally stable \citep{narendra2010metzler}. This will of course include all Hurwitz Metzler matrices, as we know it must due to the sufficient condition of Theorem \ref{theorem: absolutevaluetheorem}. Furthermore, Theorem \ref{theorem: subnetworkstabilitytheorem} shows that the same property is true in the (\ref{eq:RNN}) system -- if it is contracting in a diagonal metric, all of its subnetworks are also contracting in a diagonal metric. Because Theorem \ref{theorem: absolutevaluetheorem} guarantees a diagonal metric for the nonlinear system, we know that removing units from an RNN that satisfies Theorem \ref{theorem: absolutevaluetheorem} will not impact stability. This is not generally true, as one can imagine a system that is relying on a highly inhibitory unit to enforce stability. \\

Enforcing stability of subnetworks has a number of functional benefits. Robustness to loss of units has obvious safety implications, and furthermore the ability to remove units may also play a role in learning and development. Dropout, a powerful regularization technique for deep neural networks, involves randomly removing a fraction of units for each training round \cite{bengio2017deep}. Many Neural Architecture Search algorithms include removing units as a potential step. Even in biology, a large number of neurons undergo programmed cell death during human development, and this pattern is conserved across many animals \cite{hutchins1998death}. It is likely important not only to be able to combine stable blocks without concern for stability of the overall system, but also to be able to take blocks away without concern. Additionally, work on conditions related to the pruning of neurons could pave the way for more complex conditions dealing with pruned weights, a topic of great recent interest in neurobiology and machine learning \citep{frankle2019lth}. \\

\noindent Recall that our system (\ref{eq:RNN}) allows for different activation functions to be used with different units. The core idea of the Theorem \ref{theorem: network_of_networks} proof exploits that fact -- although a similar effect could be replicated in a network with all ReLU activation, through targeting highly negative inputs that will put the unit below the activation threshold only at those units we want to "deactivate". Further intuition and then the mathematical details of the proof follow. To skip to the next topic on incorrect prior "stability" theorems in machine learning, see section \ref{subsec:counters}.

\subsubsection{Proof of Theorem \ref{theorem: subnetworkstabilitytheorem}}
From an adversarial perspective, unstable subnetworks could be exploited by the activation functions purely via linear gains, so this necessary condition does not even require a dynamic $\mathbf{D}$. If there is an unstable subnetwork within our network, we can simply assign linear activation to those units in that subnetwork, and constant activation to all other units. This will produce a $\mathbf{D}$ in the Jacobian that zeros outputs from any units not in the unstable subnetwork, essentially generating a network that is the hierarchical combination of the unstable subnetwork with each other individual unit. Therefore the unstable portion of the network will have its behavior unchecked, and as it provides inputs to the other units can drive the entire network to "blow up".

\begin{proof}
Suppose that some principal submatrix $\mathbf{W}_{S}$ of an $n \times n$ weight matrix $\mathbf{W}$ has an eigenvalue with real part in the right half plane. Without loss of generality (as units can be renumbered), assume that indices 0 through $m-1$ of $\mathbf{W}$ are also in $\mathbf{W}_{S}$, and indices $m$ through $n-1$ are not in $\mathbf{W}_{S}$. Then we can set $\phi_{i}$ to be the identity function for $i < m$, and set $\phi_{i}$ to be a constant function $f(x) = 1$ for $i \geq m$. Now $\mathbf{D}$ is a static diagonal matrix with 1 on diagonals up to index $m-1$, and 0 on diagonals after. $\mathbf{W}\mathbf{D}$, which can also be viewed as a linear system and therefore evaluated by linear stability criteria, is equal to:
 \begin{align*}
 \begin{bmatrix} 
    \mathbf{W}_{S} & \mathbf{0} \\
    \mathbf{W}_{S_{out}} & \mathbf{0} 
  \end{bmatrix}
  \end{align*}
  \noindent where $\mathbf{W}_{S_{out}}$ is a matrix containing the connections from the first $m$ neurons to the last $n-m$ neurons. 
  
  Observe that this is a hierarchical combination, with $\mathbf{W}_{S}$ receiving no inputs. Therefore it is impossible for the overall system $\mathbf{W}\mathbf{D}$ to be stable, as it contains an unstable system at the top of its hierarchy. This completes the proof of the necessary condition. \\
  
\noindent As mentioned, we will additionally prove that diagonal stability of the nonlinear system (\ref{eq:RNN}) guarantees all subnetworks of that system will also be stable in a diagonal metric. This extends a known result for linear diagonal stability \cite{narendra2010metzler} to the nonlinear RNN, and is of particular relevance here due to both results such as Theorem \ref{theorem: absolutevaluetheorem} that guarantee diagonal stability of the nonlinear system, and also the theoretical benefits of subnetwork stability described in section \ref{section:subnets}.

To begin, suppose we are given an RNN with weights $\mathbf{W}$, contracting in known diagonal metric $\mathbf{M}$. Then we have:
\begin{align*}
(\mathbf{M}\mathbf{W}\mathbf{D})_{sym} - \mathbf{M} \prec 0
\end{align*}
\noindent uniformly. 

Now suppose we want to keep an arbitrary subset, $S$, of the neurons in $\mathbf{W}$. We will call the principal submatrix of $\mathbf{W}$ containing only the indices in $S$ $\mathbf{W}_{S}$, and we will prove that this network is contracting in the analogously defined $\mathbf{M}_{S}$. As of course only the activation functions that act on the kept neurons will remain relevant, we will need to show:
\begin{align*}
(\mathbf{M}_{S}\mathbf{W}_{S}\mathbf{D}_{S})_{sym} - \mathbf{M}_{S} \prec 0
\end{align*}
\noindent uniformly.

Observe that with diagonal $\mathbf{M}$, taking the principal submatrix of the overall Jacobian is equivalent to taking the principal submatrix of each component independently, i.e.
\begin{align*}
    (\mathbf{M}_{S}\mathbf{W}_{S}\mathbf{D}_{S})_{sym} - \mathbf{M}_{S} = ((\mathbf{M}\mathbf{W}\mathbf{D})_{sym} - \mathbf{M})_{S}
\end{align*}
\noindent This can be seen in steps, first focusing on the $\mathbf{M}_{S}\mathbf{W}_{S}\mathbf{D}_{S}$ term, as this equals $(\mathbf{M}\mathbf{W}\mathbf{D})_{S}$ due to $\mathbf{M}$ and $\mathbf{D}$ both being diagonal. $\mathbf{M}$ simply weights each row by the corresponding diagonal element, so it is irrelevant whether a row is discarded before or after multiplication - and analogously for $\mathbf{D}$ on the columns.

Next, note that the symmetric part can also be taken before or after subsetting, as it is simply the sum of a matrix and its transpose multiplied by a scalar. The sum operates elementwise, so again there is no reason for discard time to matter. This can also be applied to subtraction of the $\mathbf{M}_{S}$ term, completing the transformation.

To complete the proof, we use Cauchy's Interlace Theorem as described by \cite{hwang2004interlace} to show that:
\begin{align*}
    ((\mathbf{M}\mathbf{W}\mathbf{D})_{sym} - \mathbf{M})_{S} \prec (\mathbf{M}\mathbf{W}\mathbf{D})_{sym} - \mathbf{M}
\end{align*}
\noindent with the latter term already known to be uniformly negative definite. 

Because we are concerned with the maximum eigenvalues of symmetric matrices, we can guarantee we will only be working with real values. Thus Cauchy's Interlace Theorem can be applied directly, as it states that given a square Hermitian matrix $\mathbf{A}$ and a principal submatrix of $\mathbf{A}$, $\mathbf{B}$, then $\lambda_{max}(B) < \lambda_{max}(A)$.

\end{proof}

\subsection{Identifying counterexamples to previously published RNN stability theorems}
\label{subsec:counters}
Several recent papers in ML, e.g  \citep{haber2017stable,chang2019antisymmetricrnn}, claim that a sufficient condition for stability of the nonlinear system:
\[\dot{\mathbf{x}} = \mathbf{f}(\mathbf{x},t)\]
is that the associated Jacobian matrix $\mathbf{J}(\mathbf{x},t) = \frac{\partial \mathbf{f}}{\partial \mathbf{x}}$ has eigenvalues whose real parts are strictly negative, i.e:
\[\max_i \text{Re}(\lambda_i(\mathbf{J}(\mathbf{x},t)) \leq -\alpha\]
with $\alpha>0$. However, this claim is generally false - see Section 4.4.2 in \citep{slotine1991applied}. Recall that contraction analysis evaluates only the symmetric part of the Jacobian. For any matrix $\mathbf{A}$, the largest eigenvalue of the symmetric part will always be $\geq Re(\lambda_{max}(\mathbf{A}))$. 

It is of course not necessarily the case that certain claimed conditions are inherently false, as we do not have counterexamples yet that directly demonstrate that e.g. a diagonally stable $\mathbf{W}$ was actively unstable in the (\ref{eq:RNN}) system. We simply point out that some existing proofs make demonstrably false claims, and thus we have no guarantee that (\ref{eq:RNN}) will be globally contracting just because $\mathbf{W}$ is diagonally stable. In the rest of this section, we provide more detail on why certain conditions have been difficult to prove, and what additional constraints might be imposed in order for such properties to guarantee stability. \\

\noindent In the \textit{specific} case of the RNN (\ref{eq:RNN}), it appears that the eigenvalues of the symmetric part of $\mathbf{W}$ do provide information on global stability in a number of applications. For example, \cite{matsuoka1992stability} showed that if $\mathbf{W}_{sym} = \frac{\mathbf{W}+\mathbf{W^T}}{2}$ has all its eigenvalues less than unity, and the input $\mathbf{u}$ is constant, then (\ref{eq:RNN}) has a unique, globally asymptotically stable fixed point. This condition also implies that the real parts of the eigenvalues of the Jacobian are uniformly negative. Moreover, it has been shown that setting the symmetric part of $\mathbf{W}$ almost equal to zero (yet slightly negative) led to rotational, yet stable dynamics in practice \citep{chang2019antisymmetricrnn}. This leads us to the following theorem, which shows that if the slopes of the activation functions change \emph{sufficiently slowly} as a function of time, then the condition in \citep{matsuoka1992stability} in fact implies global contraction of (\ref{eq:RNN}). 

\begin{theorem}\label{theorem: Wdiagstabtheorem}
Let $\mathbf{D}$ be a positive, diagonal matrix with $D_{ii} = \frac{d\phi_i}{dx_i}$, and let $\mathbf{P}$ be an arbitrary, positive diagonal matrix. If:

\[ (g\mathbf{W}-\mathbf{I})\mathbf{P} + \mathbf{P}(g\mathbf{W}^T-\mathbf{I}) \preceq -c\mathbf{P} \hspace{.5cm} \text{and} \hspace{.5cm} \dot{\mathbf{D}} - cg^{-1}\mathbf{D} \preceq -\beta\mathbf{D}\]
for $c,\beta > 0$, then (\ref{eq:RNN}) is contracting in metric $\mathbf{D}$ with rate $\beta$. 
\end{theorem}

\noindent We stress however, that it is an open question whether or not diagonal stability of $\mathbf{W}$ implies that (\ref{eq:RNN}) is contracting. Note that the ReLU activation function has slope only $=$ 0 or 1, such that targeted inputs can theoretically cause rapidly changing $\mathbf{D}$, thus making it much less likely to satisfy Theorem \ref{theorem: Wdiagstabtheorem}. \\

It has in fact repeatedly been conjectured that diagonal stability of $g\mathbf{W}-\mathbf{I}$ is a sufficient condition for global contraction of the (\ref{eq:RNN}) system, for example by \cite{revay2020lipschitz}. However this has been difficult to prove, with multiple mistaken proof attempts appearing and later disappearing in various preprints, and many months ourselves struggling with what seems like a relatively simple claim. To better characterize this conjecture, we present Theorem \ref{theorem: Wdiagstabcounterexampletheorem}, which shows by way of counterexample that diagonal stability of $g\mathbf{W}-\mathbf{I}$ can \emph{not} imply global contraction in a \textbf{constant} metric for the general (\ref{eq:RNN}) system. 

\begin{theorem}
\label{theorem: Wdiagstabcounterexampletheorem}
Satisfaction of the condition
\begin{align*}
g\mathbf{W}_{sym} - \mathbf{I} \prec 0
\end{align*}
is \textbf{not} sufficient to show global contraction of the general nonlinear RNN (\ref{eq:RNN}) in any \textbf{constant} metric. High levels of antisymmetry in $\mathbf{W}$ can make it impossible to find such a metric, which we demonstrate via a $2 \times 2$ counterexample of the form
\begin{align*}
\mathbf{W} = \begin{bmatrix} 
    0 & -c \\
    c & 0 
  \end{bmatrix}
 \end{align*}
 with $c \geq 2$ when $g=1$.
\end{theorem}

\noindent Note that $g\mathbf{W}_{sym} - \mathbf{I} = g\frac{\mathbf{W} + \mathbf{W}^{T}}{2} - \mathbf{I} \prec 0$ is equivalent to the condition for contraction of the system (\ref{eq:RNN}) with \textit{linear} activation in the identity metric. It is also possible to show that if $g\mathbf{W}_{sym} - \mathbf{I} \prec 0$ \emph{were} a sufficient condition, it would follow that contraction of the linear system in any diagonal metric is a sufficient condition -- hence the discussion of diagonal stability of $\mathbf{W}$ as a conjectured condition. Recall that linear diagonal stability of $\mathbf{W}$ does at least meet the necessary condition of Theorem \ref{theorem: subnetworkstabilitytheorem}, whereas for Hurwitz $\mathbf{W}$ without diagonal stability, it may still be impossible to achieve the more general nonlinear stability. It would be necessary to check for linear stability of all the principal submatrices independently in that case, as described in section \ref{section:subnets} (and that would only be a necessary condition, not a sufficient one). \\

Theorem \ref{theorem: Wdiagstabcounterexampletheorem} means that attempts to prove the conjectured condition with a constant contraction metric are in actuality futile. It is possible for a system to be contracting in a varying metric, and it is certainly possible that the conjecture is true because of this. However, it will be extremely difficult to prove in the general case, because with a metric that varies, the derivative of that metric also appears in the contraction condition inequality. For the present system, that will involve a $\dot{\mathbf{D}}$ term; but to bound $\dot{\mathbf{D}}$ is complicated, because it is implicitly a function of the input (as $D_{ii} = \phi_i''\dot{x}_i$). 

Theorem \ref{theorem: Wdiagstabcounterexampletheorem} thus suggests that future work might want to either focus more on local notions of contraction or introduce additional constraints on the (\ref{eq:RNN}) system to make proving '$\mathbf{W}_{sym} - \mathbf{I} \prec 0 \implies$ global contraction' more approachable. One might also consider searching for other forms of conditions that can serve new purposes despite being less permissive than $\mathbf{W}_{sym} - \mathbf{I} \prec 0$ -- akin to those we presented in section \ref{subsec:novel-conditions}. 

It is worth highlighting that Theorem \ref{theorem: Wdiagstabcounterexampletheorem} has deep ties to our methods for combining RNNs as well. If we allowed the negative feedback connections to be nonlinear in the same way that we allow nonlinearities in (\ref{eq:RNN}), the exact same problem would be encountered. Because the combined system would be fully expressible in the terms of (\ref{eq:RNN}) in that case, and Theorem \ref{theorem: Wdiagstabcounterexampletheorem} has just shown we cannot easily prove global contraction with high levels of skew-symmetry, we would therefore lose our stability guarantee. This also underscores the relevance of continued theoretical research along these lines. \\

\noindent The proof for Theorem \ref{theorem: Wdiagstabtheorem} follows in subsection \ref{subsubsec:conjecture}. I will then present the proof for Theorem \ref{theorem: Wdiagstabcounterexampletheorem}, along with intuition on the dynamics that make this Theorem true, in subsection \ref{subsubsec:my-theorem}. That will conclude the mathematical results; to proceed to the experimental work, where the theory was applied for proof of concept success in sequence learning benchmarks, see section \ref{sec:dl-results}.

\subsubsection{Proof of Theorem \ref{theorem: Wdiagstabtheorem}}\label{subsubsec:conjecture}
\begin{proof}
Consider the differential, quadratic Lyapunov function:
\[V = \delta\mathbf{x}^T\mathbf{P}\mathbf{D} \delta\mathbf{x}\]

where $\mathbf{D} \succ 0 $ is as defined above. The time derivative of $V$ is:
\[ \begin{split} \dot{V} = \delta\mathbf{x}^T \mathbf{P}\dot{\mathbf{D}}\delta\mathbf{x} +\delta\mathbf{x}^T (-2\mathbf{P}\mathbf{D} + \mathbf{P}\mathbf{D}\mathbf{W}\mathbf{D} + \mathbf{D}\mathbf{W}^T\mathbf{D}\mathbf{P} )\delta\mathbf{x}  \end{split} \]

The second term on the right can be factored as:

\[\begin{split}\delta\mathbf{x}^T (-2\mathbf{P}\mathbf{D} + \mathbf{P}\mathbf{D}\mathbf{W}\mathbf{D} + \mathbf{D}\mathbf{W}^T\mathbf{D}\mathbf{P} )\delta\mathbf{x} = \\
\delta\mathbf{x}^T\mathbf{D} (-2\mathbf{P}\mathbf{D}^{-1} + \mathbf{P}\mathbf{W} + \mathbf{W}^T\mathbf{P} )\mathbf{D} \delta\mathbf{x} \leq \\
\delta\mathbf{x}^T\mathbf{D} (-2\mathbf{P}g^{-1} + \mathbf{P}\mathbf{W} + \mathbf{W}^T\mathbf{P} )\mathbf{D} \delta\mathbf{x} = \\
\delta\mathbf{x}^T\mathbf{D} [\mathbf{P}(\mathbf{W} - g^{-1}\mathbf{I}) + (\mathbf{W}^T - g^{-1}\mathbf{I})\mathbf{P} ]\mathbf{D} \delta\mathbf{x} \leq \\
-cg^{-1}\delta\mathbf{x}^T\mathbf{P}\mathbf{D}^2 \delta\mathbf{x}
\end{split} \]

where the last inequality was obtained by substituting in the first assumption above. Combining this with the expression for $\dot{V}$, we have:
\[ \begin{split} \dot{V} \leq \delta\mathbf{x}^T \mathbf{P}\dot{\mathbf{D}}\delta\mathbf{x} -cg^{-1}\delta\mathbf{x}^T \mathbf{P}\mathbf{D}^2\delta\mathbf{x}  \end{split} \]

Substituting in the second assumption, we have:
\[ \begin{split} \dot{V} \leq \delta\mathbf{x}^T \mathbf{P}(\dot{\mathbf{D}} -cg^{-1} \mathbf{D}^2)\delta\mathbf{x}  \leq -\beta \delta\mathbf{x}^T\mathbf{P}\mathbf{D}\delta\mathbf{x} = -\beta V \end{split} \]
and thus $V$ converges exponentially to $0$ with rate $\beta$.
\end{proof}

\subsubsection{Proof of Theorem \ref{theorem: Wdiagstabcounterexampletheorem}}\label{subsubsec:my-theorem}
The main intuition behind the counterexample of this theorem is that high levels of antisymmetry can prevent a constant metric from being found in the nonlinear system. This is because (in the simplest and most ubiquitous $g=1$ case) $\mathbf{D}$ is a diagonal matrix with values between 0 and 1, so the primary functionality it can have in the symmetric part of the Jacobian is to downweight the outputs of certain neurons selectively. In the extreme case of all 0 or 1 values (which is in fact the possible values that can be found in $\mathbf{D}$ when $\phi = $ ReLU), we can think of this as selecting a subnetwork of the original network, and taking each of the remaining neurons to be single unit systems receiving input from that subnetwork. For a given static configuration of $\mathbf{D}$ (think linear gains), this is a hierarchical system that will be stable if the subnetwork is stable (which it is for diagonally stable $\mathbf{W}$). But as $\mathbf{D}$ can evolve over time with many of the considered nonlinearities -- including ReLU, where targeted inputs could theoretically switch an entry of $\mathbf{D}$ to 0 or 1 at will -- we would need to find a constant metric that can serve completely distinct hierarchical structures simultaneously in order to accommodate those cases where negative feedback enabled high weight magnitudes. As it turns out, such a constant metric does not exist even in a simple 2 neuron negative feedback example.

To sketch the intuition of the proof in terms of matrix algebra, observe that in the constant $\mathbf{M}$ contraction condition for the system (i.e. $\frac{\mathbf{MWD} + \mathbf{DW^{T}M}}{2} - \mathbf{M} \prec 0$), the matrix $\mathbf{D}$ can zero out columns of $\mathbf{W}$ but not their corresponding rows. So for a given weight pair $w_{ij}, w_{ji}$, which has entry in $\mathbf{W}_{sym} = \frac{w_{ij} + w_{ji}}{2}$, if $D_{i} = 0$ and $D_{j} = 1$, the $i,j$ entry in $(\mathbf{W}\mathbf{D})_{sym}$ will be guaranteed to have lower magnitude if the signs of $w_{ij}$ and $w_{ji}$ are the same, but guaranteed to have higher magnitude if the signs are different (because we are removing one and not the other). Thus if the linear system would be stable based on magnitudes alone $\mathbf{D}$ poses no real threat (as we already saw with Theorem \ref{theorem: absolutevaluetheorem}), but if the linear system requires antisymmetry to be stable, $\mathbf{D}$ can make proving contraction quite complicated (if possible at all). 

\begin{proof}
The nonlinear system is globally contracting in a \textit{constant} metric if there exists a symmetric, positive definite $\mathbf{M}$ such that the symmetric part of the Jacobian for the system, $(\mathbf{M}\mathbf{W}\mathbf{D})_{sym} - \mathbf{M}$ is negative definite uniformly. Therefore $(\mathbf{M}\mathbf{W}\mathbf{D})_{sym} - \mathbf{M} \prec 0$ must hold for all possible $\mathbf{D}$ if $\mathbf{M}$ is a constant metric the system \textit{globally} contracts in with any allowed activation function, as some combination of settings and inputs to obtain a particular $\mathbf{D}$ can always be found, with ReLU in particular allowing for $\mathbf{D}$ to take the form of various permutations of 0s and 1s over time as desired by using targeted inputs. We will move forward assuming that $\phi = ReLU$ then, and demonstrate by counterexample that there exists a diagonally stable $\mathbf{W}$ that cannot meet the contraction condition using the same $\mathbf{M}$ when $\mathbf{D}$ takes different configurations of 0s and 1s, values that could naturally occur for the system as defined. \\

\noindent Thus to prove the main claim, we present here a simple 2-neuron system that is contracting in the identity metric with linear activation function, but which can be shown to have no $\mathbf{M}$ that simultaneously satisfies the $(\mathbf{M}\mathbf{W}\mathbf{D})_{sym} - \mathbf{M} \prec 0$ condition for two different possible $\mathbf{D}$ matrices. 

\noindent To begin, take 
\begin{align*}
    \mathbf{W} = \begin{bmatrix} 
    0 & -2 \\
    2 & 0 
  \end{bmatrix}
\end{align*}

  \noindent Note that any off-diagonal magnitude $\geq 2$ would work, as this is the point at which $\frac{1}{2}$ of one of the weights (found in $\mathbf{W}_{sym}$ when the other is zeroed) will have magnitude too large for $(\mathbf{W}\mathbf{D})_{sym} - \mathbf{I}$ to be stable. \\
  
  \noindent Looking at the linear system, we can see it is contracting in the identity because
  \begin{align*}
     \mathbf{W}_{sym} - \mathbf{I} = \begin{bmatrix} 
    -1 & 0 \\
    0 & -1 
  \end{bmatrix} \prec 0 
  \end{align*}

  \noindent Now consider $(\mathbf{M}\mathbf{W}\mathbf{D})_{sym} - \mathbf{M}$ with $\mathbf{D}$ taking two possible values of 
  \begin{align*}
      \mathbf{D}_{1} = \begin{bmatrix} 
    1 & 0 \\
    0 & 0 
  \end{bmatrix}
  \hspace{5mm} and \hspace{5mm}
  \mathbf{D}_{2} = \begin{bmatrix} 
    0 & 0 \\
    0 & 1 
  \end{bmatrix}
  \end{align*}
   
  \noindent We want to find some symmetric, positive definite 
  $\mathbf{M} = \begin{bmatrix} 
    a & m \\
    m & b 
  \end{bmatrix}$
  such that $(\mathbf{M}\mathbf{W}\mathbf{D}_{1})_{sym} - \mathbf{M}$ and $(\mathbf{M}\mathbf{W}\mathbf{D}_{2})_{sym} - \mathbf{M}$ are both negative definite. \\
  
  \noindent Working out the matrix multiplication, we get 
  \begin{align*}
  (\mathbf{M}\mathbf{W}\mathbf{D}_{1})_{sym} - \mathbf{M} = \begin{bmatrix} 
    2m - a & b - m \\
    b - m & -b 
  \end{bmatrix} 
  \end{align*}
  and 
  \begin{align*}
    (\mathbf{M}\mathbf{W}\mathbf{D}_{2})_{sym} - \mathbf{M} = \begin{bmatrix} 
    -a & -(a + m) \\
    -(a + m) & -2m - b 
  \end{bmatrix}
  \end{align*} \\
  
  \noindent We can now check necessary conditions for negative definiteness on these two matrices, as well as for positive definiteness on $\mathbf{M}$, to try to find an $\mathbf{M}$ that will satisfy all these conditions simultaneously. In this process we will reach a contradiction, showing that no such $\mathbf{M}$ can exist.
  
  A necessary condition for positive definiteness in a real, symmetric $n \times n$ matrix $\mathbf{X}$ is $x_{ii} > 0$, and for negative definiteness $x_{ii} < 0$. Another well known necessary condition for definiteness of a real symmetric matrix is $|x_{ii} + x_{jj}| > |x_{ij} + x_{ji}| = 2|x_{ij}| \hspace{.2cm} \forall i \neq j$. See \citep{wolframPD} for more info on these conditions. \\
  
  \noindent Thus we will require $a$ and $b$ to be positive, and can identify the following conditions as necessary for our 3 matrices to all meet the requisite definiteness conditions:
  \begin{align}
  2m < a \label{D1diag}
  \end{align}
  \begin{align}
  -2m < b \label{D2diag}
  \end{align}
  \begin{align}
  |2m - (a + b)| > 2|b - m| \label{D1offdiag}
  \end{align}
  \begin{align}
  |-2m - (a + b)| > 2|a + m| \label{D2offdiag}
  \end{align}
 
  Note that the necessary condition for $\mathbf{M}$ to be PD, $a + b > 2|m|$, is not listed, as it is automatically satisfied if \eqref{D1diag} and \eqref{D2diag} are. \\
  
  \noindent It is easy to see that if $m = 0$, conditions \eqref{D1offdiag} and \eqref{D2offdiag} will result in the contradictory conditions $a > b$ and $b > a$ respectively, so we will require a metric with off-diagonal elements. To make the absolute values easier to deal with, we will check $m > 0$ and $m < 0$ cases independently.
  
  First we take $m > 0$. By condition \eqref{D1diag} we must have $a > 2m$, so between that and knowing the signs of all unknowns are positive, we can reduce many of the absolute values. Condition \eqref{D1offdiag} becomes $a + b - 2m > |2b - 2m|$, and condition \eqref{D2offdiag} becomes $a + b + 2m > 2a + 2m$, which is equivalent to $b > a$. If $b > a$ we must also have $b > m$, so condition \eqref{D1offdiag} further reduces to $a + b - 2m > 2b - 2m$, which is equivalent to $a > b$. Therefore we have again reached contradictory conditions.
  
  A very similar approach can be applied when $m < 0$. Using condition \eqref{D2diag} and the known signs we reduce condition \eqref{D1offdiag} to $2|m| + a + b > 2b + 2|m|$, i.e. $a > b$. Meanwhile condition \eqref{D2offdiag} works out to $a + b - 2|m| > 2a - 2|m|$, i.e. $b > a$. \\
  
  \noindent Therefore it is impossible for a single constant $\mathbf{M}$ to accommodate both $\mathbf{D}_{1}$ and $\mathbf{D}_{2}$, so that no constant metric can exist for $\mathbf{W}$ to be contracting in when a nonlinearity is introduced that can possibly have derivative reaching both of these configurations. One real world example of such a nonlinearity is ReLU. Given a sufficiently high negative input to one of the units and a sufficiently high positive input to the other, $\mathbf{D}$ can reach one of these configurations. The targeted inputs could then flip at any time to reach the other configuration. \\
  
  \noindent An additional condition we could impose on the activation function is to require it to be a strictly increasing function, so that the activation function derivative can never actually reach 0. We will now show that a very similar counterexample applies in this case, by taking 
    \begin{align*}
      \mathbf{D}_{1*} = \begin{bmatrix} 
    1 & 0 \\
    0 & \epsilon 
  \end{bmatrix}
  \hspace{5mm} and \hspace{5mm}
  \mathbf{D}_{2*} = \begin{bmatrix} 
    \epsilon & 0 \\
    0 & 1 
  \end{bmatrix}
  \end{align*}
  
  \noindent Note here that the $\mathbf{W}$ used above
 produced a $(\mathbf{W}\mathbf{D})_{sym} - \mathbf{I}$ that just barely avoided being negative definite with the original $\mathbf{D}_{1}$ and $\mathbf{D}_{2}$, so we will have to increase the values on the off-diagonals a bit for this next example. In fact anything with magnitude larger than 2 will have some $\epsilon > 0$ that will cause a constant metric to be impossible, but for simplicity we will now take 
 \begin{align*}
 \mathbf{W}_{*} = \begin{bmatrix} 
    0 & -4 \\
    4 & 0 
  \end{bmatrix}
  \end{align*}
  
  \noindent Note that with $\mathbf{W}_{*}$, even just halving one of the off-diagonals while keeping the other intact will produce a $(\mathbf{W}\mathbf{D})_{sym} - \mathbf{I}$ that is not negative definite. Anything less than halving however will keep the identity metric valid. Therefore, we expect that taking $\epsilon$ in $\mathbf{D}_{1*}$ and $\mathbf{D}_{2*}$ to be in the range $0.5 \geq \epsilon > 0$ will also cause issues when trying to obtain a constant metric.
  
  We will now actually show via a similar proof to the above that $\mathbf{M}$ is impossible to find for $\mathbf{W}_{*}$ when $\epsilon \leq 0.5$. This result is compelling because it not only shows that $\epsilon$ does not need to be a particularly small value, but it also drives home the point about antisymmetry - the larger in magnitude the antisymmetric weights are, the larger the $\epsilon$ where we will begin to encounter problems. \\
  
  \noindent Working out the matrix multiplication again, we now get 
  \begin{align*}
  (\mathbf{M}\mathbf{W}_{*}\mathbf{D}_{1*})_{sym} - \mathbf{M} = \begin{bmatrix} 
    4m - a & 2b - m - 2a\epsilon \\
    b - m - 2a\epsilon & -4m\epsilon - b 
  \end{bmatrix}
  \end{align*}
  and 
  \begin{align*}
    (\mathbf{M}\mathbf{W}_{*}\mathbf{D}_{2*})_{sym} - \mathbf{M} = \begin{bmatrix} 
    4m\epsilon - a & -(2a + m - 2b\epsilon) \\
    -(2a + m - 2b\epsilon) & -4m - b 
  \end{bmatrix}
  \end{align*}
  
  \noindent Resulting in two new main necessary conditions:
  \begin{align}
  |4m - a - b - 4m\epsilon| > 2|2b - m - 2a\epsilon| \label{epsD1offdiag}
  \end{align}
  \begin{align}
  |4m\epsilon - a - b - 4m| > 2|2a + m - 2b\epsilon| \label{epsD2offdiag}
  \end{align}
  \noindent As well as new conditions on the diagonal elements:
  \begin{align}
  4m - a < 0 \label{epsD1diag}
  \end{align}
  \begin{align}
  -4m - b < 0 \label{epsD2diag}
 \end{align}
 
 \noindent We will now proceed with trying to find $a,b,m$ that can simultaneously meet all conditions, setting $\epsilon = 0.5$ for simplicity. 
 
 Looking at $m=0$, we can see again that $\mathbf{M}$ will require off-diagonal elements, as condition \eqref{epsD1offdiag} is now equivalent to the condition $a + b > |4b - 2a|$ and condition \eqref{epsD2offdiag} is similarly now equivalent to $a + b > |4a - 2b|$. 
 
 Evaluating these conditions in more detail, if we assume $4b > 2a$ and $4a > 2b$, we can remove the absolute value and the conditions work out to the contradicting $3a > 3b$ and $3b > 3a$ respectively. As an aside, if $\epsilon > 0.5$, this would no longer be the case, whereas with $\epsilon < 0.5$, the conditions would be pushed even further in opposite directions. 
 
 If we instead assume $2a > 4b$, this means $4a > 2b$, so the latter condition would still lead to $b > a$, contradicting the original assumption of $2a > 4b$. $2b > 4a$ causes a contradiction analogously. Trying $4b = 2a$ will lead to the other condition becoming $b > 2a$, once again a contradiction. Thus a diagonal $\mathbf{M}$ is impossible \\
 
So now we again break down the conditions into $m > 0$ and $m < 0$ cases, first looking at $m > 0$. Using condition \eqref{epsD1diag} and knowing all unknowns have positive sign, condition \eqref{epsD1offdiag} reduces to $a + b - 2m > |4b - 2(a + m)|$ and condition \eqref{epsD2offdiag} reduces to $a + b + 2m > |4a - 2(b - m)|$. This looks remarkably similar to the $m = 0$ case, except now condition \eqref{epsD1offdiag} has $-2m$ added to both sides (inside the absolute value), and condition \eqref{epsD2offdiag} has $2m$ added to both sides in the same manner. If $4b > 2(a + m)$ the $-2m$ term on each side will simply cancel, and similarly if $4a > 2(b - m)$ the $+2m$ terms will cancel, leaving us with the same contradictory conditions as before. 
 
 Therefore we check $2(a + m) > 4b$. This rearranges to $2a > 2(2b - m) > 2(b - m)$, so that from condition \eqref{epsD2offdiag} we get $b > a$. Subbing condition \eqref{epsD1diag} in to $2(a + m) > 4b$ gives $8b < 4a + 4m < 5a$ i.e. $b < \frac{5}{8}a$, a contradiction. The analogous issue arises if trying $2(b - m) > 4a$. Trying $2(a + m) = 4b$ gives $m = 2b - a$, which in condition \eqref{epsD2offdiag} results in $5b - a > |6a - 6b|$, while in condition \eqref{epsD1diag} leads to $5a > 8b$, so \eqref{epsD2offdiag} can further reduce to $5b - a > 6a - 6b$ i.e. $11b > 7a$. But $b > \frac{7}{11}a$ and $b < \frac{5}{8}a$ is a contradiction. Thus there is no way for $m > 0$ to work. \\
 
Finally, trying $m < 0$, we now use condition \eqref{epsD2diag} and the signs of the unknowns to reduce condition \eqref{epsD1offdiag} to $a + b + 2|m| > |4b - 2(a - |m|)|$ and condition \eqref{epsD2offdiag} to $a + b - 2|m| > |4a - 2(b + |m|)|$. These two conditions are clearly directly analogous to in the $m > 0$ case, where $b$ now acts as $a$ with condition \eqref{epsD2diag} being $b > 4|m|$. Therefore the proof is complete.
\end{proof}

\noindent It is worth mentioning that an additional condition we could impose on the activation function is to require it to be a strictly increasing function, so that the activation function derivative can never actually reach 0. This is sometimes an assumption made on the (\ref{eq:RNN}) network in the literature for simplicity. However, even with such an assumption, the counterexample of Theorem \ref{theorem: Wdiagstabcounterexampletheorem} still holds, as I prove in supplemental section \ref{sec:my-theorem-continued}. An example activation function that would fit this definition in real life is softplus, a smooth approximation to ReLU. 

\section{Sequence learning experiments}
\label{sec:dl-results}
As an empirical proof of concept, we used our mathematical results to train provably stable "network of networks" architectures on sequence learning benchmark tasks. We focused on three sequential image classification tasks often used to evaluate RNN performance: sequential MNIST, permuted sequential MNIST, and sequential CIFAR10. These tasks measure information processing ability over long sequences \citep{le2015simple}, and are particularly common to use in benchmarking provably stable RNNs, as stability in theory has a tradeoff with the ability to integrate information over long timescales -- though in practice stable RNNs have largely performed well with sequence processing \citep{miller2018stable}. 

In these tasks, images are presented pixel-by-pixel and the network makes a prediction at the end of the sequence. For permuted seqMNIST, pixels are input in a fixed but random order, to increase the distance between relevant pixels and thus make the sequence learning harder. Note that while MNIST (black and white handwritten single digit recognition) and CIFAR10 (RGB photos belonging to one of 10 classes - 6 different animals and 4 different vehicle types) are extremely easy problems for modern computer vision, we are turning them into quite difficult sequence learning tasks. Even the simpler MNIST involves 784 pixels when flattened, and the network receives just 1 pixel per timestep, only making a prediction at the very end of the sequence. Because our network is a vanilla RNN, meaning it does not include any engineered mechanism for memory and can only process sequential information based on the state evolution described by (\ref{eq:RNN}), a high level of performance on sequence classification would be especially impressive. 

In fact, the long short-term memory (LSTM) recurrent architecture with engineered gating mechanisms for retaining information was designed because vanilla RNNs were not capable of learning even moderately difficult sequence tasks in practice, including the original sequential MNIST \citep{le2015simple}. The work by \cite{le2015simple} developed a weight initialization method that allowed vanilla RNNs to compete with LSTMs on sequential MNIST; in that paper, they also introduced the harder permuted version of the task, as standard seqMNIST performance had exceeded $90\%$. Their performance on permuted seqMNIST saturated at $\sim 80\%$, while our architecture with a similar number of units (but many fewer trainable weights) achieved $\sim 90\%$ on the same task -- supplemental section \ref{Appendix:all-tables} contains an exhaustive account of our experimental results across all trials. 

Of course, models in the years since the work of \cite{le2015simple} have gotten increasingly large and in many cases have added further engineering solutions such as imposing attention mechanisms. Though permuted seqMNIST remains a good task for experimentation on smaller vanilla RNNs, it is not a difficult problem for modern machine learning at large. However, seqCIFAR10 remains quite challenging, with exceedingly few methods (none of which are vanilla RNNs) exceeding $75\%$ best test accuracy performance to date, despite often using network sizes $10x$ (or more) larger than the largest network we include here. As such, our experiments show not only state of the art (SOTA) performance amongst provably stable RNNs on these tasks, but also demonstrate highly competitive sequence learning performance against a wider range of methods. 

Thus beyond the benefits of having a stability certificate, our architecture stands on its own due to the sequence learning performance it achieves for its size and for its straightforward vanilla RNN formulation, not to mention the potential for interpretability afforded by the multi-area structure. While the tasks are ultimately toy tasks in the sense that their difficulty is entirely contrived, it is important to highlight that they are indeed challenging benchmarks, and our pilot experimental results bode well for future applications of the network to real world sequence learning problems. More broadly, the experiments highlight the relevance of our major theoretical contributions, and motivate a number of other ways they could be applied (to be covered in detail within the discussion section \ref{sec:discussion4}). \\

\noindent In this section, I will first detail how we formalized our theoretical results within a modern deep learning framework, to train stable "RNNs of RNNs" in practice (\ref{subsec:methods4}). I will then report on our best performing architecture, including comparison to prior benchmark scores and results on the consistency of its performance over repeated trials (\ref{subsec:results4}). Finally, I will present results on how architecture settings such as network sparsity levels (\ref{subsec:sparsity-tests}) and network size (\ref{subsec:size-tests}) modulated performance, alongside discussion of how these results might inform future applications of "RNNs of RNNs".

\subsection{Network architecture and methods summary}
\label{subsec:methods4}
Recall that a major theme of our work is establishing foundational results for RNNs with multi-area structure. In the mathematical results of section \ref{subsec:combo-conditions}, we focused on linear negative feedback connections between (given) stable nonlinear subnetworks, because such connections were shown to guarantee stability of the larger network, and can be easily trained using the derived parameterization with modern deep learning libraries. Here, we realized that work empirically. In particular, the matrix $\mathbf{B}$ in Theorem \ref{theorem: network_of_networks} was made a trainable tensor in PyTorch, used in conjunction with contraction metrics of the individual subnetworks to calculate the network's linear all-to-all negative feedback weight matrix $\mathbf{L}$ (as defined in the Theorem). 

In order to enforce the linear negative feedback weights $\mathbf{L}$ to form connections only \emph{between} different subnetworks (but not within a subnetwork), we masked the diagonal blocks of $\mathbf{B}$ to freeze those self-connections at 0; thereby preserving an interpretation of network structure that separates the longer range feedback that different "regions" provide to each other from the weights internal to each "region". Note that because each weight has dependency on its opposite direction counterpart, we can train only half of the matrix $\mathbf{B}$ without losing degrees of freedom in the network weights $\mathbf{L}$, so to improve the efficiency of the training process we ultimately only trained the lower-triangular blocks of $\mathbf{B}$ while masking the other entries. 

That was one major piece of our "RNN of RNNs" architecture design, and was used consistently throughout these pilot experiments. The other major piece would of course be the subnetworks making up the combined network, taken as a given when defining the feedback connection scheme. A benefit of our overarching framework is the level of flexibility allowed in the makeup of these subnetworks, flexibility which we leveraged to only a small extent here. It is required that every subnetwork included be definable in terms of the nonlinear RNN dynamics (\ref{eq:RNN}), and that every subnetwork included be provably globally contracting in an obtainable metric $\mathbf{M}$. But subnetworks could span a range of sizes, activation functions, time constants, and even contraction metrics. Furthermore, for any subnetwork defined in a way that allows contraction to remain guaranteed throughout training, it is possible for those internal nonlinear subnetwork weights to be updated jointly with the negative feedback training.

In these experiments, we used our novel contraction conditions for individual nonlinear RNNs - proved in section \ref{subsec:novel-conditions} - to enforce the needed constraints on the subnetworks. For all models, the subnetworks' dynamics were each defined according to (a discrete-time version of) equation (\ref{eq:RNN}), using ReLU activation function across the board. Because the two considered stability conditions have quite different pros and cons however, they led to two distinct "network of networks" approaches that we explored empirically. 

\begin{figure}[h]
\centering
\includegraphics[width=\textwidth,keepaspectratio]{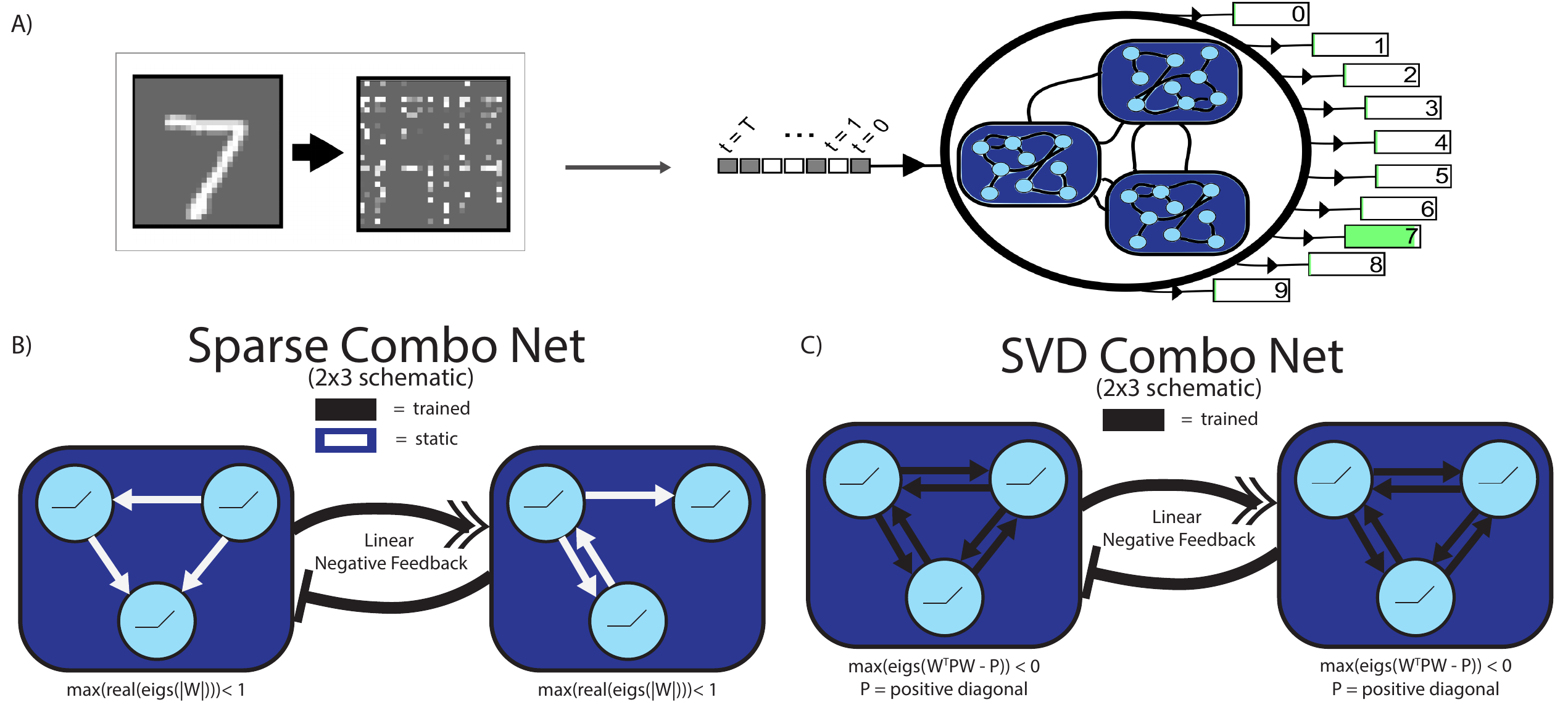}
\caption[Summary of proof-of-concept task structures and implemented "network of networks" architectures.]{\textbf{Summary of task structure and network architectures.} Images from MNIST (or CIFAR10) were flattened into an array of pixels and fed sequentially one by one into the modular 'network of networks', with classification based on the output at the last timestep. For MNIST, each image was also permuted in a fixed manner (A). The subnetwork 'modules' of our architecture were constrained to meet either Theorem \ref{theorem: absolutevaluetheorem} via sparse initialization (B) or Theorem \ref{theorem: singularvaluetheorem} via direct parameterization (C). Linear negative feedback connections were trained between the subnetworks according to Theorem \ref{theorem: network_of_networks}.}
\label{fig:network-cartoon}
\end{figure}

For Theorem \ref{theorem: absolutevaluetheorem}, we used a bespoke initialization scheme for $\mathbf{W}$ that would guarantee contraction of each subnetwork (and provide a corresponding contraction metric), but we did \emph{not} train the internal subnetwork weights at all (Figure \ref{fig:network-cartoon}B). For Theorem \ref{theorem: singularvaluetheorem}, we used a more standard initialization and jointly trained all parameters of the model (Figure \ref{fig:network-cartoon}C). 

We refer to networks with subnetworks constrained by Theorem \ref{theorem: absolutevaluetheorem} as 'Sparse Combo Nets' (because sparse hidden-to-hidden weights more readily satisfy the condition), and to networks with subnetworks constrained by Theorem \ref{theorem: singularvaluetheorem} as 'SVD Combo Nets'. Throughout the experimental results we use the notation '$p \times n$ network' - such a network consists of $p$ distinct subnetwork RNNs, with each such subnetwork RNN containing $n$ units.

Note that task input at each timestep was first passed through a (trainable) linear input layer, represented by a weight matrix of size 1 (or 3 for RGB CIFAR pixels) by $N$, where $N = p*n$ is the total number of units in the Combo RNN. On the flip side, the outputs of the Combo RNN neurons at the final timestep in the sequence were used for prediction via passing them through a (trainable) linear output layer. The output layer was represented by a matrix of size $N$ by 10, with 10 being the number of different classes contained in both the MNIST and CIFAR10 datasets. This allowed for computation of cross entropy loss during training based on the true class of each input. At test time, the output index with the maximum value was selected as the predicted label (Figure \ref{fig:network-cartoon}A). 

Unless specified otherwise, all networks were trained for 150 epochs, using an Adam optimizer with initial learning rate 1e-3 and weight decay 1e-5. The learning rate was cut to 1e-4 after 90 epochs and to 1e-5 after 140 epochs. Once we identified the most promising settings through our experiments, we ran final repetition trials training the best architecture for 200 epochs with learning rate cuts after epochs 140 and 190. \\

\noindent Because Sparse Combo Net reliably demonstrated better performance on the sequence learning tasks, and this was also the architecture that I primarily worked with, I will only briefly report on the SVD Combo Net results in this chapter. For details on the direct parameterization that allowed those subnetwork weights to be trained while maintaining contraction guarantees, see \cite{NIPS22}. On the other hand, I will next elaborate on the implementation of Sparse Combo Net in greater detail, before moving on to results (\ref{subsec:results4}).

\subsubsection{More on sparsity}
\begin{figure}[h]
\centering
\includegraphics[width=\textwidth,keepaspectratio]{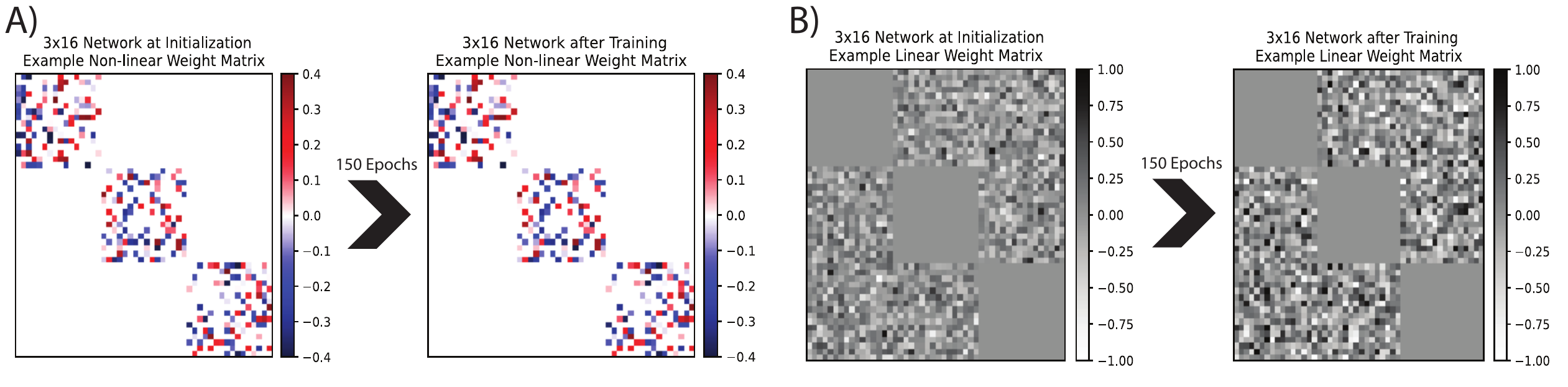}
\caption[Example Sparse Combo Net weights before and after training.]{\textbf{Example $3$x$16$ Sparse Combo Net weight matrices, before and after training.} Nonlinear intra-subnetwork weights are initialized using a set sparsity, and do not change in training (A). Linear inter-subnetwork connections are constrained to be antisymmetric with respect to the overall network metric, and are updated in training (B).}
\label{figure:example-training}
\end{figure}

As mentioned, we were not able to find a parameterization to continuously update the internal nonlinear subnetwork weights during training in a way that preserved a contraction guarantee through Theorem \ref{theorem: absolutevaluetheorem}, and thus the Sparse Combo Net did not have the weights within its individual modules trained -- relying only on training the linear connections between modules (Figure \ref{figure:example-training}), as well as the linear input and output layers. This is reminiscent of the facilitated variation theory of evolution discussed within the background section \ref{sec:background4}, in that we train this combination network architecture \textit{only} through training connections between contracting subnetworks, without changing the component modules at all. \\

\noindent Fortunately, it is easy to randomly generate matrices with a particular likelihood of meeting the Theorem \ref{theorem: absolutevaluetheorem} condition by selecting an appropriate sparsity level and limit on entry magnitude. Sparsity in particular is of interest due to its relevance in neurobiology and machine learning, so it is convenient that the condition makes it easy to verify stability of many different sparse RNNs. As $g=1$ for ReLU activation, we check potential subnetwork matrices $\mathbf{W}$ by simply verifying linear stability of $|\mathbf{W}| - \mathbf{I}$, keeping only those subnetworks that satisfy. 

Because every RNN meeting the condition has a corresponding well-defined stable linear time-invariant (LTI) system contracting in the same metric, it is also easy to find a metric to use in our training algorithm, another advantage of Theorem \ref{theorem: absolutevaluetheorem}. Solving for $\mathbf{M}$ in $-\mathbf{I} = \mathbf{M}\mathbf{A} + \mathbf{A}^{T}\mathbf{M}$ will produce a valid metric for any stable LTI system $\mathbf{A}$ \citep{slotine1991applied}. Thus the $\mathbf{M}$ calculated when $\mathbf{A} = |\mathbf{W}| - \mathbf{I}$ will be a contraction metric for our nonlinear subnetwork, and can be used as needed in training the negative feedback weights per Theorem \ref{theorem: network_of_networks}. 

To solve for this $\mathbf{M}$ in practice, I integrated $e^{\mathbf{A}^{T} t} \mathbf{Q} e^{\mathbf{A} t} dt$ from 0 to $\infty$ (with scipy.integrate.quad). For efficiency reasons, and due to the guaranteed existence of a diagonal metric in the case of Theorem \ref{theorem: absolutevaluetheorem} (as Hurwitz Metzler matrices are diagonally stable), integration was only performed to solve for the diagonal elements of $\mathbf{M}$. For safety, a check was added prior to training to confirm that the initialized network indeed satisfied Theorem \ref{theorem: absolutevaluetheorem} with metric $\mathbf{M}$. However, it was never triggered by our initialization method.

To randomly generate the fixed nonlinear subnetworks to use for each Sparse Combo Net instantiation, I required each individual subnet weight matrix $\mathbf{W}$ to have all diagonal entries $= 0$ and only $s\%$ of off-diagonal entries non-zero. Each non-zero entry was then sampled uniformly between $-z$ and $z$. Throughout this chapter, I will therefore refer to Sparse Combo Nets containing subnetworks with just $s\%$ of their weights non-zero as having an "$s\%$ sparsity level". Both $s$ and $z$ were settings of the architecture explored during our experiments, though they were always the same across subnetworks in a given Sparse Combo Net. 

Selection of these values should depend to some degree on the size of the subnetworks being used, but there are a wide variety of options that can be considered. While $s$ and $z$ need to be reasonable enough values for a subnetwork of a given size to possibly meet the Theorem \ref{theorem: absolutevaluetheorem} condition, the random sample and the verification of the Theorem are both computationally cheap, so that the proportion of random $\mathbf{W}$ satisfying Theorem \ref{theorem: absolutevaluetheorem} can be small if desired, as the code will continue to resample until the intended number of contracting subnetworks have been found. Note that with a smaller $s$ a larger $z$ (i.e. entry magnitude) can still be used while maintaining stability -- and vice versa, though there are other theoretical reasons to prefer sparser weights with higher magnitudes to a significant extent. \\

\noindent The code to reproduce experiments with Sparse Combo Net can be found at \citep{sparsegit}. Additional experimental details for Sparse Combo Net can also be found in appendix \ref{cha:append-chapt-refch:4}. 

\FloatBarrier

\subsection{Setting a new state of the art for provably stable RNN architectures}
\label{subsec:results4}
The Sparse Combo Net architecture ultimately achieved the highest overall performance on both permuted seqMNIST and seqCIFAR10, with $96.94\%$ and $65.72\%$ best test accuracies respectively -- thereby setting a new state of the art (SOTA) for provably stable recurrent neural networks on these challenging sequential image classification benchmarks (Table \ref{table:sota}). This was accomplished with a very similar number of trainable parameters as the previous stable RNN SOTA, and performance was also competitive with a number of other methods utilizing much larger networks.  

\begin{table}[h!]
\centering
\caption[Sparse Combo Net performance versus published benchmarks.]{\textbf{Sparse Combo Net performance versus published benchmarks for sequential MNIST, permuted sequential MNIST, and sequential CIFAR10 test accuracy.} Architectures are grouped into three categories: modern baselines, best performing RNNs with claimed(*) stability guarantee, and methods achieving overall SOTA. Because very recently published work soundly beat the overall SOTA for all tasks considered, the methods that previously held SOTA for permuted seqMNIST and for seqCIFAR10 are included rows as well. Within each grouping, networks are ordered by number of trainable parameters (for CIFAR10 if it differed across tasks). Where possible, we include information on performance repeatability in addition to the best achieved test accuracy columns for each task, though this is unfortunately uncommon to report in the prior literature. Our network is highlighted, and SOTA best test accuracy scores are bolded, for both overall SOTA and SOTA in the class of provably stable recurrent neural networks -- which we presently hold for permuted seqMNIST and seqCIFAR10.  \newline
(*)For more on stability guarantees in machine learning, see section \ref{subsec:counters}.}
\label{table:sota}

\small
\begin{tabular}{ | m{2.25cm} || m{0.85cm} | m{1cm} || m{1.3cm} | m{1.3cm} | m{1.35cm} || m{0.95cm} | m{0.95cm} | m{0.95cm} | }
\hline
 Name & Stable RNN? & Params & sMNIST \newline Repeats \newline Mean (n) \newline [Min] & psMNIST \newline Repeats \newline Mean (n) \newline [Min] & sCIFAR10 \newline Repeats \newline Mean (n) \newline [Min] & Seq \newline MNIST \newline Best & PerSeq \newline MNIST \newline Best & Seq \newline CIFAR \newline Best \\
\hline\hline
LSTM \citep{chang2019antisymmetricrnn} & & 68K & \centering --- & \centering --- & \centering --- & 97.3\% & 92.7\% & 59.7\% \\
\hline
Transformer \newline \citep{trinh2018cifar} & & 0.5M & \centering --- & \centering --- & \centering --- & 98.9\% & 97.9\% & 62.2\% \\
\hline\hline
Antisymmetric \newline \citep{chang2019antisymmetricrnn} & ? & 36K & \centering --- & \centering --- & \centering --- & 98\% & 95.8\% & 58.7\% \\  
\hline
\rowcolor{Gray}
Sparse Combo Net & \checkmark & 130K & \centering --- & \textbf{96.85\%} (4) \newline [\textbf{96.65\%}] & 64.72\% (10) \newline [63.73\%] & 99.04\% & \textbf{96.94\%} & \textbf{65.72\%} \\  
\hline
Lipschitz \newline \citep{erichson2021lipschitz} & \checkmark & 134K & 99.2\% (10) \newline [99.0\%] & 95.9\% (10) \newline [95.6\%] & \centering --- & \textbf{99.4\%} & 96.3\% & 64.2\% \\  
\hline\hline
CKConv \newline \citep{romero2021ckconv} & & 1M & \centering --- & \centering --- & \centering --- & 99.32\% & 98.54\% & 63.74\% \\
\hline
S4 \citep{gu2022s4} & & 7.9M & \centering --- & \centering --- & \centering --- & \textbf{99.63\%} & \textbf{98.7\%} & \textbf{91.13\%} \\
\hline
Trellis \citep{bai2019trellis} & & 8M & \centering --- & \centering --- & \centering --- & 99.2\% & 98.13\% & 73.42\% \\
\hline
\end{tabular}
\end{table}

The strong performance result of Sparse Combo Net is in itself a powerful proof of concept for the theoretical results of the chapter, and coupled with other beneficial properties of the "network of networks" architecture structure suggests there are a number of domains where extensions of Sparse Combo Net could be fruitfully applied. I will keep this as a recurring theme throughout presentation of the full experimental results.

Furthermore, we were able to reproduce SOTA scores over several training repetitions, including 10 trials of seqCIFAR10. Along with repeatability of results, we also showed that the contraction constraint on the connections between subnetworks ($\mathbf{L}$ in Theorem \ref{theorem: network_of_networks}) was important for performance, particularly in the Sparse Combo Net -- results to be detailed in the upcoming subsection \ref{subsubsec:rnn-rigor}. \\ 

\noindent \textbf{Summary of additional results.} As mentioned, during our experimentation we profiled how various architecture settings impacted performance of our networks. For both Sparse Combo Net and SVD Combo Net, we found that increasing the total number of neurons improved task performance, but with diminishing returns in the current structure. We also found that the sparsity of the hidden-to-hidden weights in Sparse Combo Net had a large impact on the final network performance, in line with intuitions from neurobiology.

\subsubsection{Repeatability and controls}
\label{subsubsec:rnn-rigor}
Because Sparse Combo Net does not have the connections within its subnetworks trained, network performance could be particularly susceptible to random initialization. Thus we ran repeatability studies on permuted sequential MNIST and sequential CIFAR10 using our best network settings ($16 \times 32$ with subnetwork sparsity level of $3.3\%$) and an extended training period. Fortunately, the strong performance of the network turned out to be quite reliable. Mean test accuracy over the 4 trials of permuted seqMNIST was $96.85\%$ with variance $= 0.019$, while mean test accuracy over the 10 trials of seqCIFAR10 was $64.72\%$ with variance $= 0.406$. 

\begin{figure}[h]
\centering
\includegraphics[width=0.8\textwidth,keepaspectratio]{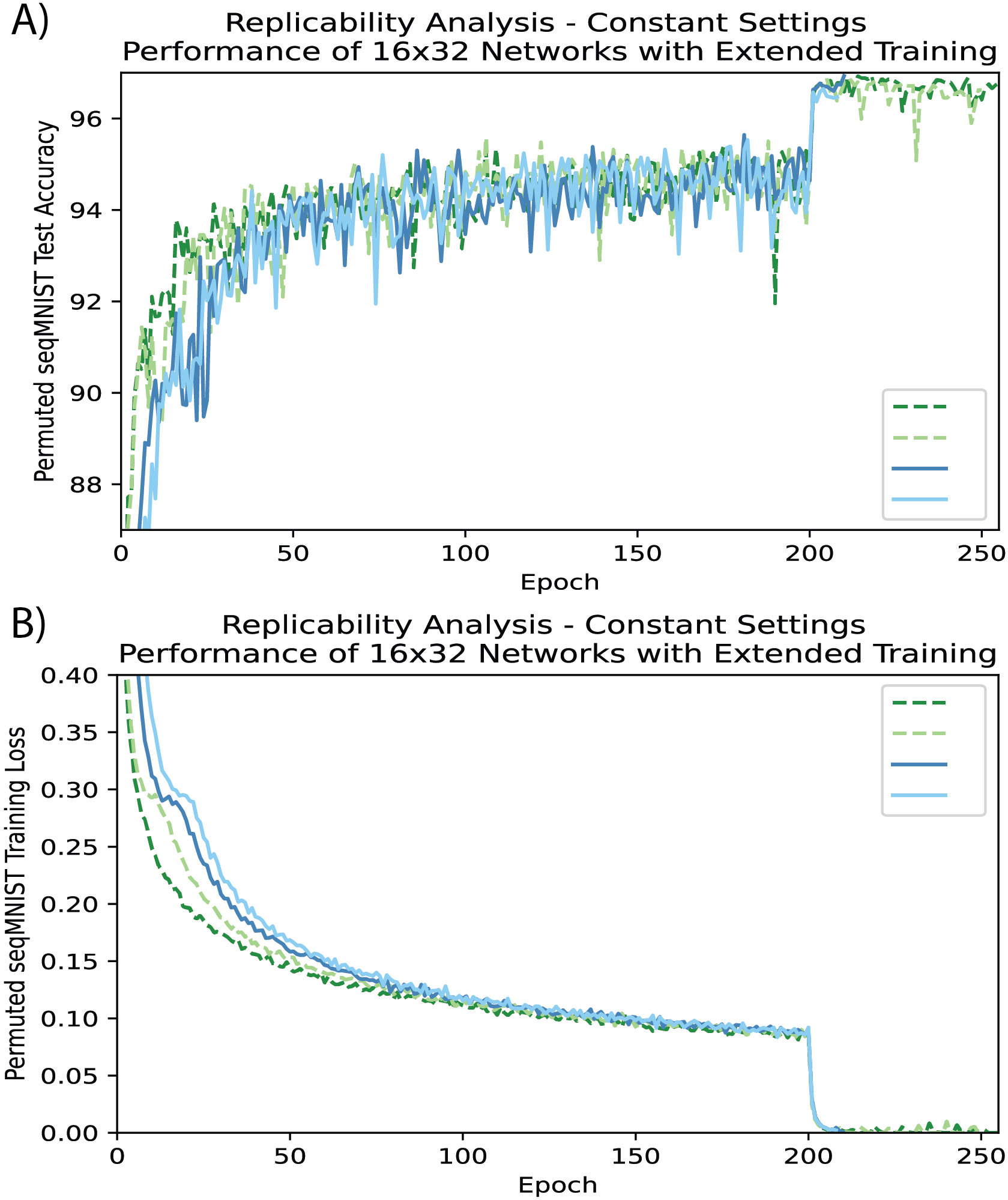}
\caption[Permuted seqMNIST performance on repeated Sparse Combo Net trials.]{\textbf{Permuted seqMNIST performance on repeated Sparse Combo Net trials.} Four different $16 \times 32$ networks with 3.3\% density and entries between -6 and 6 were trained for 24 hours, with a single learning rate cut after epoch 200. (A) depicts test accuracy for each of the networks over the course of training. (B) depicts the training loss for the same networks.}
\label{figure:test-reproduce}
\end{figure}

\begin{figure}[h]
\centering
\includegraphics[width=\textwidth,keepaspectratio]{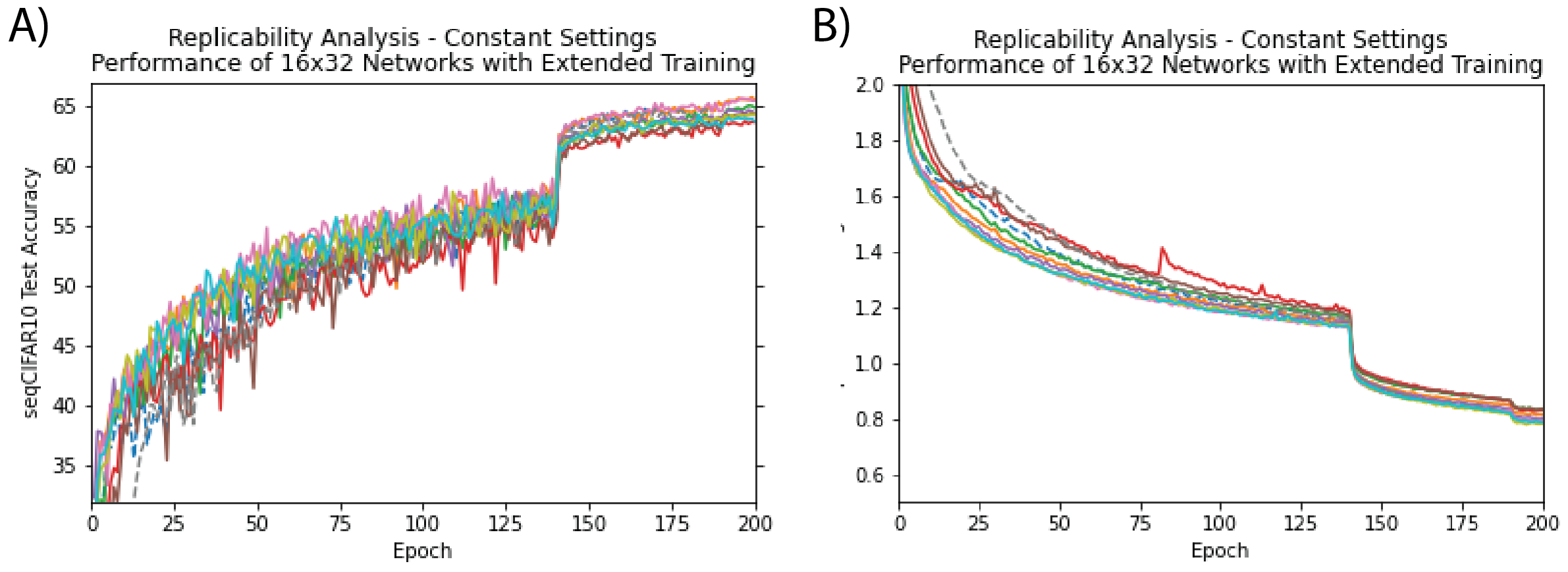}
\caption[seqCIFAR10 performance on repeated Sparse Combo Net trials.]{\textbf{seqCIFAR10 performance on repeated Sparse Combo Net trials.} Ten different $16 \times 32$ networks with 3.3\% density and entries between -6 and 6 were trained for 200 epochs, with learning rate divided by 10 after epochs 140 and 190. (A) depicts test accuracy for each of the networks over the course of training. (B) depicts the training loss for the same networks.}
\label{figure:cifar-reps}
\end{figure}

Across these permuted seqMNIST trials, best test accuracy always fell between $96.65\%$ and $96.94\%$, a range much smaller than the differences seen with changing sparsity settings and network size. Three of the four trials showed best test accuracy $\geq 96.88\%$, despite some variability in early training performance (Figure \ref{figure:test-reproduce}). Similarly, eight of the ten seqCIFAR10 trials had test accuracy $>64.3\%$, with all results falling between $63.73\%$ and $65.72\%$ (Figure \ref{figure:cifar-reps}). This robustly establishes a new SOTA for provably stable RNNs, comfortably beating the previously reported (single run) $64.2\%$ test accuracy achieved by Lipschitz RNN \citep{erichson2021lipschitz}. \\

\noindent As a control study, we also tested how sensitive the Sparse Combo Net was to the stabilization condition on the interconnection matrix ($\mathbf{L}$ in Theorem \ref{theorem: network_of_networks}). To do so, we initialized the individual RNN modules in a $24 \times 32$ Sparse Combo Net as before, but set $\mathbf{L}=\mathbf{B}$ and did not constrain $\mathbf{B}$ at all during training, thus no longer ensuring contraction of the overall system. This resulted in $47.0\%$ test accuracy on the permuted seqMNIST task, a stark decrease from the original $96.7\%$ test accuracy - thereby demonstrating the utility of the contraction condition.

\FloatBarrier

\subsection{Additional results on network sparsity}
\label{subsec:sparsity-tests}
Because of the link between sparsity and stability as well as the biological relevance of sparsity, we explored in detail how subnetwork sparsity affected the performance of Sparse Combo Net. We ran a number of experiments on the permuted seqMNIST task, varying sparsity level while holding network size and other hyperparameters constant. We began our original testing with 32 unit subnetworks using a $26.5\%$ sparsity level (i.e. 26.5 percent of inter-subnet weights non-zero), and thus did not see much improvement when switching from 16 unit subnetworks to 32 unit subnetworks (appendix \ref{cha:append-chapt-refch:4}). The benefit of larger subnetworks though is that they can have many fewer non-zero weights proportionally, while maintaining expressivity -- making it easier to meaningfully test sparsity in this setting. 

\begin{figure}[h]
\centering
\includegraphics[width=\textwidth,keepaspectratio]{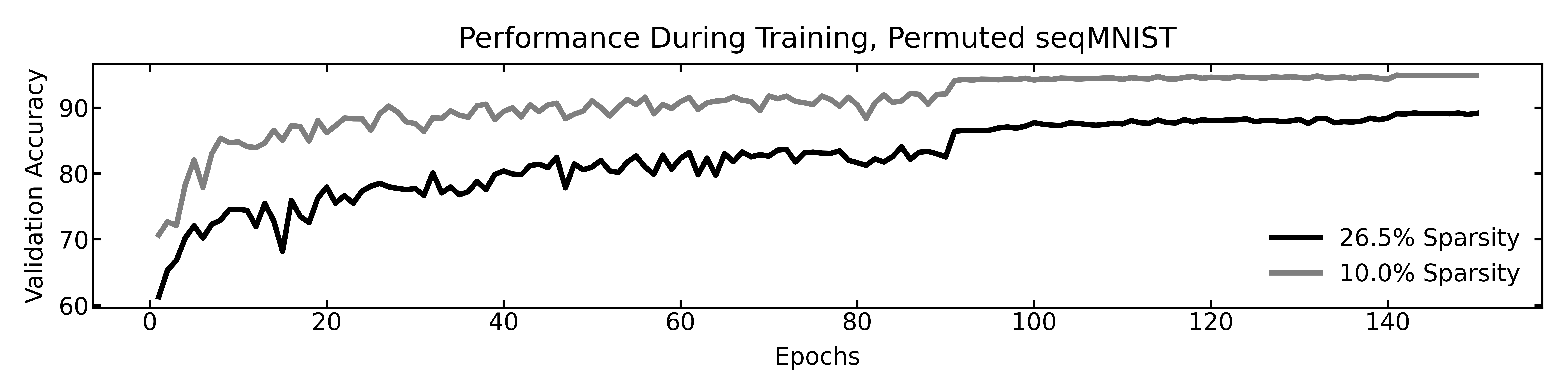}
\caption[Sparser component networks improve Sparse Combo Net performance.]{\textbf{Sparser component networks improve Sparse Combo Net performance.} Permuted seqMNIST test accuracy over the course of training for two $11 \times 32$ Sparse Combo Nets with different sparsity levels is plotted here. The network performed notably better when only $10\%$ of the within-subnetwork weights were non-zero, as opposed to the originally tested setting ($26.5\%$) for subnetworks with 32 units each. Where comparing sparsity, networks with fewer weights are mapped to a lighter grey.}
\label{figure:test-sparsity}
\end{figure}

Interestingly, our very first test of decreasing sparsity had a surprisingly large impact on performance. In particular, we observed a large ($>5$ percentage point) performance boost when switching from a $26.5\%$ sparsity level to a $10\%$ sparsity level in the $11 \times 32$ Sparse Combo Net (Figure \ref{figure:test-sparsity}). We therefore proceeded to test significantly sparser subnetworks in a $16 \times 32$ Sparse Combo Net, experimenting not only with sparsity level but also with entry magnitude ranges, as sparser subnets allow larger weights within the bounds of Theorem \ref{theorem: absolutevaluetheorem}, and it was not clear to what extent this was driving the performance increase versus the differing subnetwork topologies between different sparsity levels. 

\begin{figure}[h]
\centering
\includegraphics[width=\textwidth,keepaspectratio]{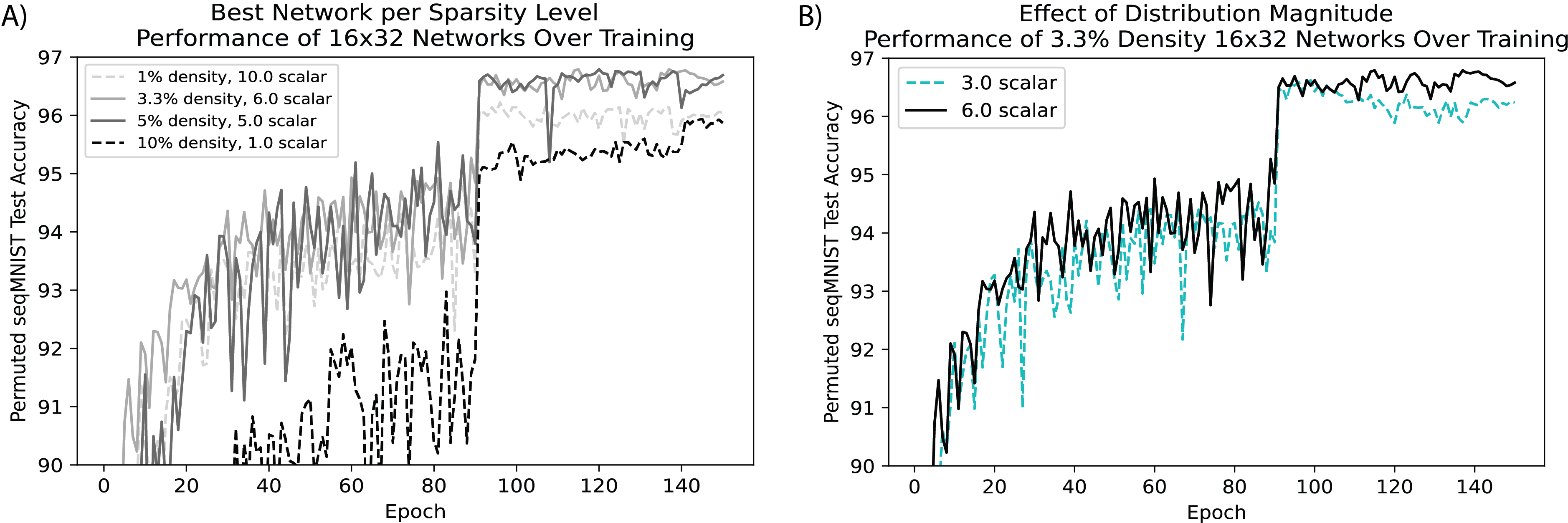}
\caption[Characterizing the effects of component network sparsity level on training.]{\textbf{Characterizing the effects of component network sparsity level on training - permuted seqMNIST performance by component RNN initialization settings.} Test accuracy was plotted over the course of training for four $16 \times 32$ networks with different sparsity levels and entry magnitudes (A), highlighting the role of sparsity in network performance. Performance continued to increase as sparser subnetworks were tested, until a small performance drop-off at the $1\%$ level. Recall that each subnetwork has just 32 units, so with a larger network the ideal proportion of non-zero weights could likely be much lower, something we hope to explore in the future. Because subnets with fewer weights would permit higher entry magnitudes while still meeting the stability condition, we also did a preliminary exploration of tbe role of weight size (B). Test accuracy was plotted over the course of training for two different $16 \times 32$ networks, both with $3.3\%$ sparsity level, but using different entry magnitudes (-3 to 3 versus -6 to 6). There was indeed a small benefit to the higher magnitude, and this likely partially explains the benefits of sparsity observed here. However, it does not seem to fully explain the increase, and when the magnitude became too high, performance cratered out of view of the current axis limits due to training difficulties.}
\label{figure:test-sparsity-sup}
\end{figure}

Ultimately, both sparsity and magnitude had a clear effect on performance (Figure \ref{figure:test-sparsity-sup}). Increases in both had a positive correlation with test accuracy and training loss through most parameters tested, though eventually caused declines at more extreme values. Best test accuracy overall was quite similar for both a $16 \times 32$ network with $5\%$ sparsity level and final entries between -5 and 5, and for a $16 \times 32$ network with $3.3\%$ sparsity level and final entries between -6 and 6. The latter also achieved the best epoch 1 test accuracy observed in these trials at $86.79\%$, which is an impressive result for a single round of training. As such, we chose to go with these settings for our extended training repetitions on permuted seqMNIST, and later seqCIFAR10. Note that with a component RNN size of just 32 units, a $3.3\%$ sparsity level is quite small, containing only one or two directional connections per neuron on average.

Not only did the $16 \times 32$ Sparse Combo Net with $3.3\%$ sparsity level (and an entry magnitude cap of $6$) achieve test accuracy on sequential CIFAR10 that is the highest to date for a provably stable RNN, it was higher than the 1 million parameter CKConv network, which held the overall SOTA for permuted seqMNIST accuracy until very recently (Table \ref{table:sota}). Our network had $\sim 130,000$ trainable parameters by comparison, and is defined in terms of purely vanilla RNNs. Thus we achieve very impressive results given the characteristics of our architecture, and both the sparsity and multi-area structure are well-suited for applications requiring a greater level of interpretability. On top of that, it is perhaps even more surprising that Sparse Combo Net performed so well without training any of the nonlinear subnetwork weights at all. This motivates additional future research directions related to sparsity in RNNs, not just for deep learning practice but also for advancing neuroscience theory, where empirically we know connections in the brain are highly sparse. \\

\noindent After training was completed, we inspected the weights of networks used in the sparsity trials, pulling the nonlinear ($\mathbf{W}$) and linear ($\mathbf{L}$) matrices from both initialization time and the final model. For $\mathbf{L}$, we checked the maximum element and the maximum singular value before and after training, to try to gain any insights on how the negative feedback connections were learning to classify the sequences. In general, both went up over the course of training, but by a modest amount. This can be visualized in the toy example of Figure \ref{figure:example-training} as well, where $\mathbf{W}$ remains unchanged but $\mathbf{L}$ contains a number of larger weights after training. Note that when all nonlinear subnetworks contract in the identity metric (i.e. normal exponential convergence in Euclidean space), the negative feedback matrix $\mathbf{L}$ will be literally antisymmetric per our definition. The sparser subnetworks at the core of the paper were all indeed contracting in the identity, such that the main results only functionally involve the most literal definition of negative feedback, where each between-subnetwork linear connection was $x$ in one direction and $-x$ in the other direction. Thus training these $x$ values (and the linear input/output layers) was the only way in which Sparse Combo Net learned.

Besides checking the corresponding $\mathbf{M}$ for each $\mathbf{W}$, we also inspected the properties of internal subnetwork weights more directly. We first confirmed once again that $\mathbf{W}$ did not change over training, to verify there were not mistakes in the implementation that could break the stability guarantee. We then inspected the maximum real part of the eigenvalues of $\mathbf{|W|}$, in accordance with Theorem 1, to determine if this might correlate with final performance. The least sparse matrices - despite mostly using much lower entry magnitudes - tended to have $\lambda_{max}(\mathbf{|W|}) > 0.9$, while the sparsest ones tended to have $\lambda_{max}(\mathbf{|W|}) < 0.1$. As this relates to the rate at which the network state will contract, it was the case that stronger levels of stability largely led to better performance in Sparse Combo Net. This is of course contrary to the theoretical concern that stability could hinder sequence learning, as by definition the network is exponentially "forgetting" its initial conditions and any perturbations. 

Although notions of nonlinear stability can include many equilibrium states, including time-varying stable states (i.e. limit cycles), and thus a system could theoretically use stability to its advantage in sequence learning via an attractor mechanism such as that described by \cite{uhler2020assoc}, we use a much stricter stability definition. By enforcing global contraction, we are indeed guaranteeing the network will always eventually converge to the same point after a transient perturbation. Contraction combination properties could be leveraged in the future with local contraction conditions instead, allowing for a multi-area architecture that could take advantage of diverse equilibria. While this would indeed be an interesting direction, I think our results suggest that some of the concerns about expressivity under strict stability conditions are overblown, especially when taken together with the strong results of other practical experiments with stable RNNs \citep{miller2018stable}. There are certainly tradeoffs that come with greater stability, and it is very unlikely the human brain is as strictly stable as our networks, but at the same time it is possible to obtain quite interesting behavior in response to time-varying inputs even with global contraction, and even if the contraction rate is not especially slow. Questions remain on the extent to which stability itself enabled our results, as opposed to other benefits of the mechanisms we used to enforce it; regardless, we have clearly demonstrated that the strict stability does not preclude sequence learning, and the factors underlying the performance we obtained warrant additional investigation.

\subsubsection{Negative feedback sparsity}
The use of sparsity in subnetworks to improve performance suggests another interesting direction that could enable better scalability of total network size (to be discussed subsequently): enforcing sparsity in the linear feedback weight matrix $\mathbf{L}$. This could be done by controlling which units of two given subnetworks are connected in negative feedback, instead of allowing all-to-all training there, and for engineering purposes that is worth exploring in the future. More interesting from an interpretability perspective though, this could also be done by limiting which entire subnetworks are connected in negative feedback, potentially allowing discrete sub-roles for individual module RNNs to be identified. I performed preliminary testing of the latter idea, by sparsifying the adjacency matrix in the code that specifies the subnetwork pairings that are to be trained in negative feedback with each other. In particular, different feedback sparsity levels were tested in $24 \times 32$ Sparse Combo Nets on the sequential CIFAR10 task, with other settings held static (including the same best performing 32 unit subnetwork initialization method as before).

Sparsifying the negative feedback connections indeed showed promise in mitigating the saturation effect seen with added subnetworks in Figure \ref{figure:test-sizes} (presented below). Performance results by feedback adjacency sparsity level took an inverse U shape here, where the network with only $50\%$ of possible feedback connections non-zero had the best test accuracy of $65.14\%$ on seqCIFAR10, achieved in just 124 epochs of training (Table \ref{table:scalability}). As an added benefit, training was (unsurprisingly) more efficient on average with fewer non-zero negative feedback weights. On the other hand, given the limited resource setting our experiments were run in, it was unfortunately impossible to make an entirely fair accuracy comparison as the densest negative feedback settings could not get close to completion within the allotted training time.

\begin{table}[!htbp]
\centering
\caption[Considering sparsity in the between-network feedback connections.]{\textbf{Considering sparsity in the between-network feedback connections.} Results from pilot testing on the sparsity of negative feedback connections in a $24 \times 32$ Sparse Combo Net and a $16 \times 32$ Sparse Combo Net. Feedback Density refers to the percentage of possible subnetwork pairings that were trained in negative feedback, while the remaining inter-network connections were held at 0. All networks were trained with the same 150 epoch training paradigm as mentioned in the main text, but were stopped after hitting a 24 hour runtime limit. Decreasing Feedback Density is a promising path towards further improving performance as the size of Sparse Combo Nets is scaled. The ideal amount of feedback density will likely vary with the size of the combination network.}
\label{table:scalability}

\begin{tabular}{ | m{2.25cm} | m{1.5cm} || m{1.25cm} | m{2.5cm} | m{2.5cm} | }
\hline
Size & Feedback Density & Epochs & Best Overall Test Acc. & Best Test Acc. Through 85 Epochs \\
\hline\hline
$24 \times 32$ & 100\% & 86 & 52.7\% & 52.7\%  \\  
\hline
$24 \times 32$ & 75\% & 88 & 56.49\% & 56.48\%  \\  
\hline
$24 \times 32$ & 66.6\% & 89 & 58.84\% & 58.84\%  \\  
\hline
\rowcolor{LightCyan}
$24 \times 32$ & 50\% & 124 & 65.14\% & 58.01\%  \\   
\hline
$24 \times 32$ & 33.3\% & 129 & 61.86\% & 56.05\%  \\ 
\hline
$24 \times 32$ & 25\% & 92 & 54.26\% & 50.54\%  \\ 
\hline
$24 \times 32$ & 0\% & 130 & 39.8\% & 38.38\%  \\ 
\hline\hline
\rowcolor{Gray}
$16 \times 32$ & 100\% & 150 & 64.63\% & 55.82\%  \\ 
\hline
$16 \times 32$ & 75\% & 150 & 64.12\% & 57.23\%  \\ 
\hline
$16 \times 32$ & 50\% & 127 & 59.87\% & 54.26\%  \\ 
\hline
\end{tabular}
\end{table}

As an aside, Table \ref{table:scalability} also sheds some light on the role of the linear input and output weights in the performance of Sparse Combo Net. With no feedback connections between modules at all, the $24 \times 32$ network reached a peak test accuracy of $39.8\%$ over 130 epochs. All other feedback sparsity levels tested ($25\%$ or more) exceeded $50\%$ test accuracy within 85 epochs, with the majority exceeding $55\%$ by that point.

\FloatBarrier

\subsection{Additional results on network size and modularity}
\label{subsec:size-tests}
\begin{figure}[h]
\centering
\includegraphics[width=\textwidth,keepaspectratio]{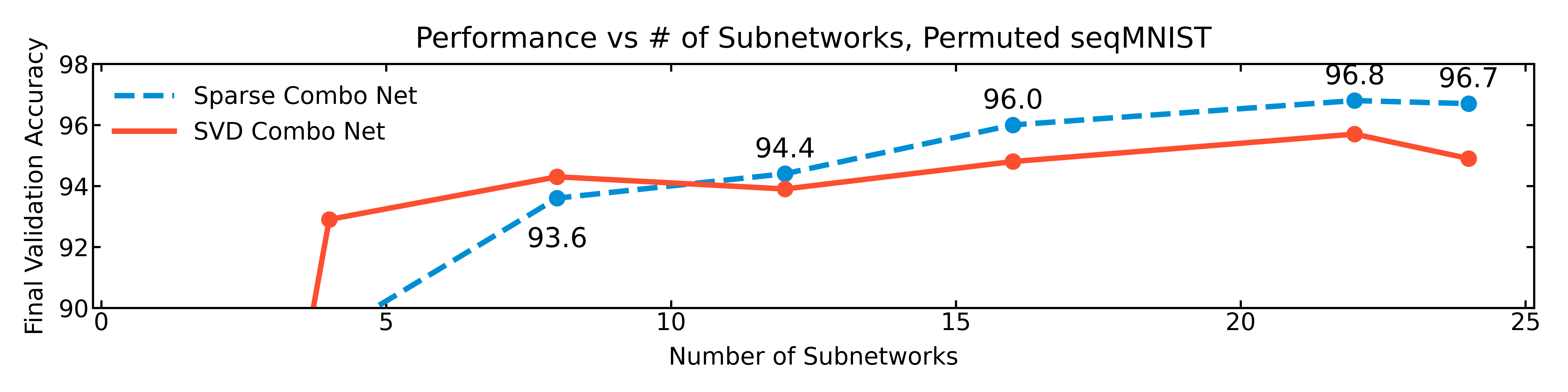}
\caption[Sparse Combo Net beat SVD Combo Net at most network sizes.]{\textbf{Sparse Combo Net beat SVD Combo Net at most network sizes.} Final permuted seqMNIST test accuracy plotted against the number of subnetworks used in training. Each subnetwork had 32 neurons. Results are shown for both Sparse Combo Net and SVD Combo Net. For Sparse Combo Net, the subnetworks used the best performing $3.3\%$ sparsity level initialization settings described.}
\label{figure:test-sizes}
\end{figure}

Understanding the effect of size on network performance is important to practical application of these architectures. For both Sparse Combo Net and SVD Combo Net, increasing the number of subnetworks while holding other settings constant (including fixing the size of each subnetwork at 32 units) was able to increase network test accuracy on permuted seqMNIST to a point (Figure \ref{figure:test-sizes}). The greatest performance jump occurred when increasing from one module ($37.1\%$ Sparse Combo Net, $61.8\%$ SVD Combo Net) to two modules ($89.1\%$ Sparse Combo Net, $92.9\%$ SVD Combo Net). After that the performance increased steadily with number of modules until saturating at $\sim 97\%$ for Sparse Combo Net and $\sim 95\%$ for SVD Combo Net. 

As the internal subnetwork weights are not trained in Sparse Combo Net, it is unsurprising that its performance was substantially worse at the smallest sizes. However Sparse Combo Net surpassed SVD Combo Net by the $12 \times 32$ network size, which contained a modest 384 total units. Note also that the SVD Combo Net never reached $55\%$ test accuracy for seqCIFAR10 in our early experiments. Due to the better performance of the Sparse Combo Net, we focused much of the analysis in \citep{NIPS22} there. \\

\begin{figure}[h]
\centering
\includegraphics[width=0.7\textwidth,keepaspectratio]{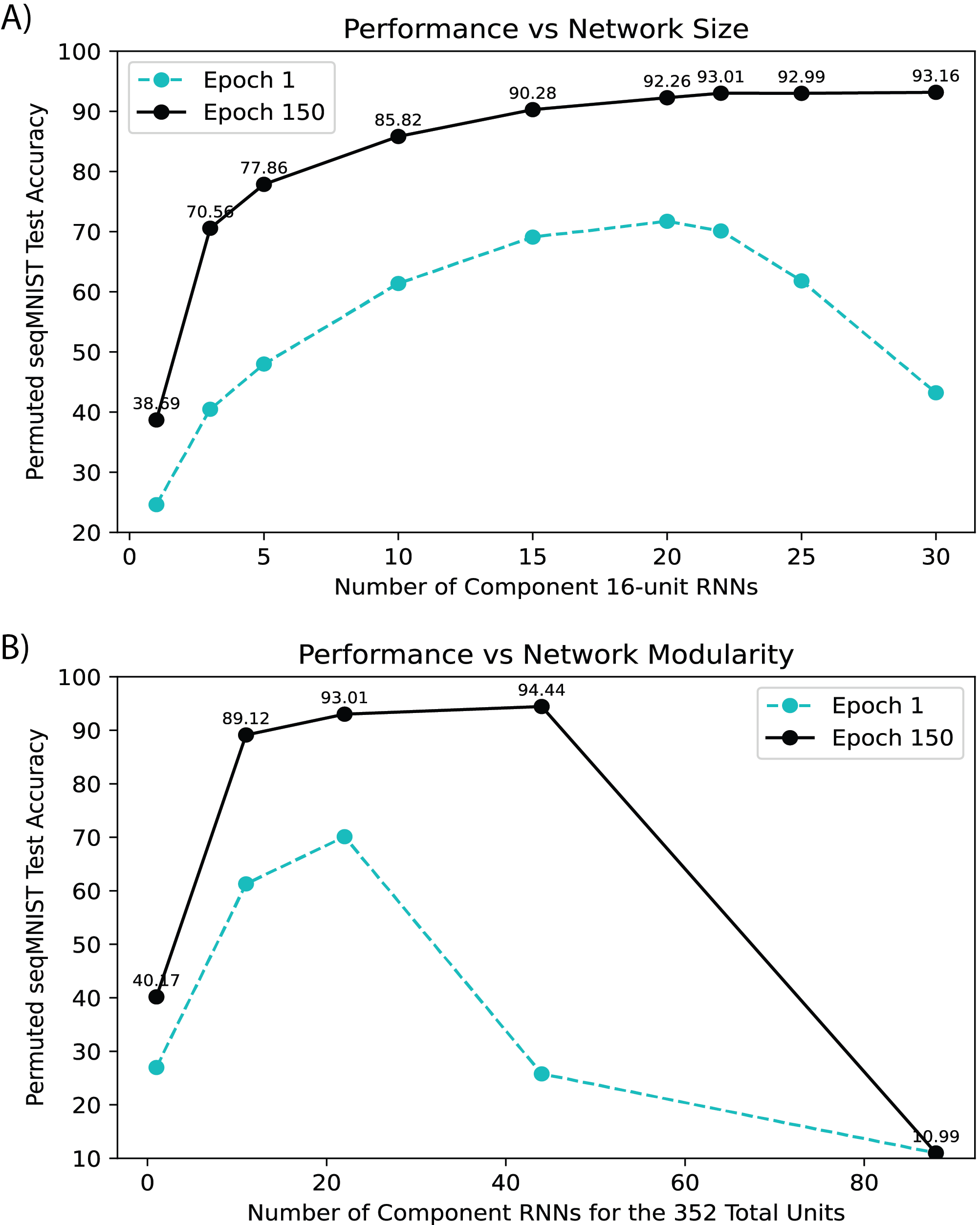}
\caption[Performance of Sparse Combo Net varies with both network size and modularity.]{\textbf{Performance of Sparse Combo Net varies with both network size and modularity.} We tested the effects on final and first epoch permuted seqMNIST test accuracy of both total network size and network modularity. The former is assessed by varying the number of subnetworks while each subnetwork is fixed at 16 units (A), and the latter by varying the distribution of units across different numbers of subnetworks with the total sum of units in the network fixed at 352 (B). Note that these experiments were run prior to optimizing the sparsity initialization settings. Experiments on total network size were later repeated with the final sparsity settings, as reported on above (Figure \ref{figure:test-sizes}). The results of both the size experiments are consistent.}
\label{figure:test-sizes-sup}
\end{figure}

Another natural question related to network size in the "RNN of RNNs" context is the effect of different subnetwork sizes, particularly when the total number of units is otherwise held constant -- thereby testing the impact of the modularity we are enforcing with our multi-area structure. Chronologically, the first thing I tested with Sparse Combo Net after hyperparameter tuning was in fact the role of size and of modularity, using the original sparsity level initialization settings that produced denser subnetworks with individual weights of lesser magnitude (Figure \ref{figure:test-sizes-sup}).

Consistent with the results of the optimized experiment repeated in Figure \ref{figure:test-sizes}, increasing the number of subnetworks in the original tests (when the individual RNNs each contained 16 units at $40\%$ sparsity level) led to improvements in test accuracy with diminishing returns, eventually entirely saturating. The accuracy consistently hit $\sim 93\%$ with a large enough number of subnetworks, but neither training loss nor test accuracy showed meaningful improvement past the $22 \times 16$ network (Figure \ref{figure:test-sizes-sup}A). Interestingly, early training loss and test accuracy became substantially worse once the number of components increased past a certain point, falling from $70\%$ to $43\%$ epoch 1 accuracy between the $22 \times 16$ and $30 \times 16$ networks. This raises the concern that the network may actually have declining performance with scale, despite our tests dealing with networks of at most modest size across the board. However, as discussed within section \ref{subsec:sparsity-tests} above, sparsity in the negative feedback structure has the potential to substantially mitigate the issue, and more generally there are many architecture setting adjustments that could be considered to combat potential issues as larger sizes are attempted in the future. So while scale may indeed be a problem for the very specific details of the present network, I do not believe it will hinder further work with the broader Sparse Combo Net architecture (see discussion section \ref{sec:discussion4} for more on this topic). 

Based on the results of the original network size experiment, I performed the network modularity experiments to center around the $22 \times 16$ Sparse Combo Net. Thus the modularity of the networks were varied while the total number of units was fixed at 352. In this case, we observed an inverse U shape (Figure \ref{figure:test-sizes-sup}B), with poor performance of a $1 \times 352$ net and an $88 \times 4$ net, and best overall performance from a $44 \times 8$ net. Given the two different extremes in testing modularity, an inverse U shape is entirely unsurprising and entirely unconcerning here. The permuted seqMNIST test accuracy after the first epoch of training was quite bad for the $44 \times 8$ network though, suggesting the $22 \times 16$ and $11 \times 32$ networks that had close final accuracy scores and much better early accuracy scores would likely be better candidates for further exploration. 

Note that because larger networks require different sparsity settings to meet the Theorem \ref{theorem: absolutevaluetheorem} condition, the initialization settings were not held entirely constant between trials in the modularity comparison experiment, but rather selected in the same way between trials. I looked for settings that would keep sparsity level ($s$) and entry magnitude ($z$) balanced and result in $\sim 1\%$ of the randomly generated matrices meeting the condition. This resulted in $s = 7.5\%$ and $z = 0.077$ for the 352 unit subnetwork, $s = 26.5\%$ and $z = 0.27$ for the 32 unit subnetworks, $s = 60\%$ and $z = 0.7$ for the 8 unit subnetworks, and $s = 100\%$ and $z = 1.0$ for the 4 unit subnetworks. However, with larger subnetworks it is possible to have many fewer weights than were tested here and still maintain potential for expressivity, so there is no way to do an objectively "fair" comparison without running an order of magnitude more experiments covering a large grid of parameter combinations. I therefore chose to test a few other sparsity levels using the 32 unit subnetworks, and because of the strong results stuck with this subnetwork size for the rest of the experimental results reported. A future more empirically focused work might consider revisiting the size, modularity, and sparsity questions more exhaustively, in conjunction with other questions raised in this section.

\FloatBarrier

\section{Discussion}
\label{sec:discussion4}
Biologists have long noted that modularity can provide organisms with stability and robustness \citep{Kitano_2004}. The other direction -- that stability and robustness can provide modularity -- is well known to engineers \citep{khalil2002nonlinear,slotine1991applied,modular}, but has been less appreciated in biology. We use this principle to build and train provably stable assemblies of recurrent neural networks. Like real brains, the components of our "RNN of RNNs" can communicate with one another through a mix of hierarchical and feedback connections. In particular, we theoretically characterized conditions under which an RNN of RNNs will be stable, given that each individual RNN is stable. We also provided several novel stability conditions for single nonlinear RNNs that are compatible with these stability-preserving interareal connections. Our results contribute towards understanding how the brain maintains stable and accurate function in the presence of massive interareal feedback, as well as external inputs. 

While primarily motivated by neuroscience, our approach is relevant for machine learning too. Deep learning models can be as inscrutable as they are powerful -- opaqueness that limits conceptual progress and may be dangerous in safety critical applications like autonomous driving or human-centered robotics. In the case of RNNs, one difficulty is that providing a certificate of \textit{stability} is often impossible or computationally impractical. However, the stability conditions we derive here allow for recursive construction of complicated RNNs while automatically preserving stability. By parameterizing our conditions for easy optimization using gradient-based techniques, we successfully trained our architecture on challenging sequential processing benchmarks. The high test accuracy achieved with a small number of trainable parameters demonstrates that stability does not necessarily come at the cost of expressivity. Given that stability is a fundamental property of dynamical systems -- and is intimately linked to concepts of control, generalization, efficiency, and robustness -- the ability to guarantee stability of a recurrent model will be important for ensuring deep networks behave as we expect them to \citep{richards2018lyapunov,choromanski2020ode,revay2021recurrent,rodriguez2022lyanet}. Our results likewise contribute towards understanding stability certification of RNNs.  \\

In future work, we will explore how our contraction-constrained RNNs of RNNs perform on a variety of neuroscience tasks, in particular tasks with multimodal structure \citep{yang2019task}. One desiderata for those future models is that they learn representations which are formally similar to those observed in the brain \citep{yamins2014performance, schrimpf2020brain, williams2021generalized}, in complement with the structural similarities already shared. Moreover, a "network of networks" approach will be especially relevant to challenging multimodal machine learning problems, such as the simultaneous processing of audio and video. Thus the advancement of neuroscience theory and machine learning remain hand-in-hand for our next lines of questioning. Indeed, combinations of trained networks have already seen groundbreaking success in DeepMind's AlphaGo \citep{silver2016mastering}.

As well as the many potential experimental applications, there are numerous theoretical future directions suggested by our work. Networks with more biologically-plausible weight update rules, such as models discussed in \citep{kozachkov2020achieving}, would be a fruitful neuroscience context in which to explore our conditions. One promising avenue of study there is to examine input-dependent stability of the learning process. In the context of machine learning, our stability conditions could be applied to the end-to-end training of multidimensional recurrent neural networks \citep{graves2007multi}, which have clear structural parallels to our RNNs of RNNs but lack known stability guarantees. \\

In sum, recursively building network combinations in an effective and stable fashion while also allowing for continual refinement of the individual networks, as nature does for biological networks, will require new analysis tools. Here we have taken a concrete step towards the development of such tools, not only through our theoretical results, but also through their application to create stable combination network architectures that perform well in practice on benchmark tasks. In the rest of this section, I will further build on our contributions by providing an extended discussion of both the impact (\ref{subsec:impact4}) and limitations (\ref{subsec:limitations4}) of the present work, as well as a multi-disciplinary characterization of specific future directions of interest (\ref{subsec:rnn-future}). Finally, I will close the chapter by reiterating its major accomplishments (\ref{subsec:rnn-contrib}).

\subsection{Impact}
\label{subsec:impact4}
Most work on stability of task-trained RNNs has focused on \textit{single} RNNs. Here we leveraged tools from nonlinear control theory to derive novel single-RNN stability conditions which enable the recursive construction of stable \textit{assemblies} of RNNs. In particular we showed that certain stability conditions for individual nonlinear RNNs, including a few of our own novel conditions, allow for a simple parameterization of linear connections between these RNNs that automatically preserves stability during training. We then showed that these modular 'networks of networks' repeatedly perform better than existing stable RNNs on key sequence classification tasks such as permuted sequential MNIST and sequential CIFAR10. We also provided control studies that show the stabilizing parameterization of connections between subnetwork RNNs is important for the observed high performance. Additionally, we found that sparsity in the component networks was a key feature in attaining our highest accuracy levels -- levels that set a new state of the art for provably stable recurrent neural networks. 

Given the paucity of existing theory on modular networks, our novel stability conditions and proof of concept combination architectures are a significant step in an important new direction. The "network of networks" approach is evident in the biological brain, and has seen early practical success in applications such as AlphaGo. There is much evidence this line of questioning will be critical in the future, and our work is the first on stable \emph{modular} networks. Furthermore, we developed an architecture based on such combinations of "vanilla" RNNs that is both stable and achieves high performance on benchmark sequence classification tasks using \emph{few trainable parameters} (small particularly for sequential CIFAR10). When considering just the facts about the Sparse Combo Net, it really has no business performing anywhere near as well as it did. In this way, our work represents a departure from certain prior expectations about what factors might maximize network performance, thereby opening a number of future directions in the empirical domain. Moreover, the multi-area structure is inherently ripe for applications research that focuses on interpretability, a major need area for deep learning at this time.

Theoretically, it is important to highlight that our contributions extend beyond the explicit stability conditions we derived. Not only did we identify flaws in previously published stability "proofs" and clarify the underlying confusion that might have lead to them, but we also performed a more careful dissection of the challenges inherent in proving certain stability conditions for the (\ref{eq:RNN}) condition. It is quite uncommon to find results like these published at all, let alone in the main text of a paper at a major venue. However, such reporting is critical to ensure that future works do not accidentally build on false results, as well as to save countless hours of researcher time chasing futile leads. In my opinion, results along the lines of Theorem \ref{theorem: Wdiagstabcounterexampletheorem} - which shows that a popular conjectured stability condition cannot be proved for the general system in \emph{any constant metric} - are also underrated in the level of understanding that they can directly supply. This style of work is a crucial piece towards tackling the high dimensional search problem we face, as it suggests new search vectors that would not be implied by typical literature containing only proofs (or counterexamples) related to the main stability claim. The implication of time-varying metrics is in fact inherently interesting, and points towards an important shift for the field at large to give input-dependent factors much greater attention in theoretical study of RNN stability. 

\subsection{Limitations}
\label{subsec:limitations4}
Of course our work is not without its drawbacks, limitations which should be the subject of near future research to iron out concerns and lingering questions, thereby enabling the longer term vision to be more fully realized. Most of these immediate drawbacks are related to the present scope of our empirical claims, though there are also theoretical limitations that ought to be kept in mind in the medium term. It is also the case that some limitations are context dependent -- in addressing machine learning shortcomings we may move further from biological plausibility, whereas to make our findings more directly applicable to systems neuroscience work we may move further from practical data science applicability. As such, I divide the limitations into 3 categories for consideration dependent on the aims of the particular follow up study. \\

\noindent Machine learning experimentation:
\begin{itemize}
    \item Our tests were restricted to sequential image classification benchmarks, which are challenging but ultimately toy sequence learning tasks. It will be important to demonstrate that the architecture can perform well on real world sequence datasets like wrist accelerometry signals, and also to show that the architecture (or variations on it) can perform well in other ML contexts like reinforcement learning. 
    \item Our preliminary results suggested a performance saturation effect at a scale much smaller than most modern deep learning applications. As noted within section \ref{sec:dl-results}, there are good reasons for optimism that the architecture can scale with appropriate adaptations, but that remains to be tested.
    \item The best performing Sparse Combo Net architecture does not currently allow for its nonlinear subnetwork weights to be updated while maintaining the stability guarantee. This could limit its applicability in a number of contexts, and requires further investigation -- both on the exact downsides of this restriction and on how the Sparse Combo Net can be adapted to ease this restriction.
    \item The SVD Combo Net that does allow joint training of within and between subnetwork connections was very slow to train in addition to its worse observed performance. The way that the SVD Combo Net $\mathbf{W}$ matrices are parameterized as the product of several other matrices requires training many more parameters to obtain ultimately the same number of weights. While this probably cannot be avoided in order to maintain the stability of the SVD Combo Net, there may be implementation changes that could improve training efficiency.
    \item Although stability has a number of theoretical benefits, and we did show that removing the stability condition on our inter-network connections decreased overall performance, it is unclear the extent to which stability itself was beneficial versus a by-product of the benefits of having structured negative feedback connections. Future work should include a more thorough set of control cases, and should also seek to explore performance properties with clearer theoretical ties to stability than overall test accuracy has. Perhaps most notably, adversarial robustness could be investigated.
    \item Despite the small size and structural interpretability advantages of our networks, we did not yet attempt to demonstrate that e.g. the decisions of Sparse Combo Net can be understood. For the purposes of not only enabling data science applications that require interpretability, but also better understanding the architectural properties that lead to our good performance, it will be important to take more time to dissect the network in future experiments.
\end{itemize}

\noindent Neuroscientific parallels:
\begin{itemize}
    \item The current approach focuses on describing complicated networks that can be built from simpler ones while ensuring stability of the larger system, such that the priority is in efficiently utilizing nonlinear control theory tools. To instead construct multi-area networks that incorporate known anatomical detail (i.e. projections between V4 and PFC) while still preserving stability will require further research. Further, our experimental approach enforced all-to-all negative feedback between subnetworks, even more removed from possible anatomical detail than what might be possible with our early theoretical results.
    \item In our theoretical work, we focus on networks with fixed weights, and at times parameterizing those weights in terms of matrices that could be trained with modern deep learning libraries. This entirely neglects biological plausibility in the learning process -- something that could be of particular interest in the case of multi-area networks due to decisions that might be made about when weights in different parts of the network are updated or not during learning. 
    \item Although tasks inspired by cognitive and systems neuroscience are available for training RNNs \citep{Khona2022}, we did not make any attempt to test our architectures on such tasks yet. While these tasks are not very useful for benchmarking performance of a new architecture in the context of artificial neural networks, they allow for much more detailed comparisons to be made with neurobiological data, which will be necessary to empirically ground any neuroscientific conclusions we may be tempted to draw.
\end{itemize}

\noindent Theoretical takeaways:
\begin{itemize}
    \item We focused here only on global contraction, and moreover global contraction in a constant (and for the most part diagonal) metric. While it is possible for such a globally contracting system to display a diversity of behaviors in response to persistent and heterogeneous input signals, they are fundamentally limited in their expressivity in response to transients. It is extremely unlikely that the human brain is globally contracting in this way, and it is also probable that as we consider much larger networks performing much more complex and multimodal tasks we will not want such a strict notion of stability. There may still be a desire for certain individual modules to be globally contracting, but ultimately theoretical results will remain limited until we can better deal with less restrictive and more nuanced notions of contraction within the RNN context.
    \item Despite discussing the fact that hierarchical connections can apply to our framework, and that our framework can theoretically be applied recursively, we did not formalize either of these claims mathematically nor empirically. Including nonlinear hierarchical connections mixed in with linear negative feedback ones requires some care in how the contraction metrics are used, but this could then enable a wide range of unique structures to be considered in a way that would remain interesting after repeated recursion. 
    \item There are also other types of connections that can be used to combine contracting systems in a stability-preserving way, for example some known small gain theorems in the contraction analysis literature \citep{modular}.  However, parameterizing these conditions is less straightforward than parameterizing the negative feedback condition. Additionally, negative feedback connections between brain regions are an important concept in neuroscience (as are hierarchical connections), and empirically it is unclear how well small gain connections would perform against negative feedback connections that can assume much larger values -- along the lines of our observation that increased weight magnitude in Sparse Combo net subnetworks correlated with better performance (to a point).
\end{itemize}
\noindent Of course for all of these list categories, our aims were quite foundational. I will therefore next provide an extended characterization of future directions of possible interest across the 3 domains, including details on some of the specific questions we might ask to address the limitations identified here. 

\subsection{Future directions}
\label{subsec:rnn-future}
To begin this section, I will review the immediate follow-ups that ought to be done to fill gaps in the pilot experimental work of \cite{NIPS22,sparsegit}, as our early contributions were primarily theoretical (subsection \ref{subsubsec:rnn-paper2}). I will then discuss a number of interesting broader questions we might ask about extensions to our architectures in the machine learning context (\ref{subsubsec:rnn-ml-vision}). Finally, I will preview some of the neuroscientific (\ref{subsubsec:neuro-rnn-next}) and mathematical (\ref{subsubsec:theory-rnn-next}) directions that our work could enable in the future. There are a great deal of opportunities to continue the research of this chapter in all 3 directions and various intersections of them, and I hope to lay additional groundwork along these lines to facilitate collaborators in continuing this work in the long run. 

\subsubsection{Immediate empirical questions}
\label{subsubsec:rnn-paper2}
An initial future direction for the architecture described in this chapter is to clear up some of the empirical questions about how the network functions raised by our pilot experiments, as well as to make the code more realistic for use by others on practical sequence learning problems. There are a number of functionalities we could add by first testing on sequential CIFAR10, both to further improve performance and to better enable network interpretability. Because our networks perform very well for their size and have a modular structure that makes breaking down behavior more tractable, they could be especially well suited to applications that require interpretability - but we must explicitly test this first. Once we have a better empirical understanding via toy task, extension to other sequence learning problems would be straightforward, allowing us to confirm the broader methodology can generalize in this way, thereby opening the door for applied data science groups to utilize our code. Furthermore, a deeper understanding will also assist in adapting the architecture to other problem types down the road, and could even generate new hypotheses for theoretical or neuroscientific directions to explore. \\

\noindent The major points I would want to address in the sequence learning context include:
\begin{itemize}
    \item A more thorough characterization of different architecture settings would be helpful, and in particular how they relate to each other. Due to resource limitations, the original paper largely ran each experiment with settings otherwise fixed, but a grid search approach would have been preferable, especially as sparsity level has different interpretations at different network sizes. There were also some settings hardly explored at all, such as the role of the time constant $\tau$, and we did not take advantage of the flexibility contraction could offer in mixing and matching settings within one combo network yet either.
    \item One of the bigger empirical concerns in the original paper was the indication that performance started to saturate and possibly even curve back down as we added more subnetworks (and we stayed relatively small compared to a number of other architectures). I suspect a way to combat this is to introduce sparsity in the negative feedback adjacency structure, but it could take a number of different forms that require investigation.
    \begin{itemize}
        \item One could test varying levels of sparsity in the adjacency structure, i.e. the number of subnetwork pairings that are connected at all, but one could also test varying levels of sparsity in the number of units that are connected in negative feedback given two subnetworks that will be joined with negative feedback.
        \item Sparsity in feedback connections could be randomly decided upon, but it could be enforced based on known properties of the subnetworks and of the task as well, or even based on neuroanatomical priors. Using an approach similar to the Lottery Ticket Hypothesis (LTH), it could also be possible to allow the network to learn a useful sparse structure itself, by performing some rounds of all-to-all negative feedback training and then freezing some desired percentage of the smallest magnitude weights at 0, to subsequently restart training from the beginning with only the narrowed set of weights available \citep{frankle2019lth}. Inversely, one might attempt to build up a negative feedback structure of relevance from nothing using an efficient evolutionary approach, such as that described by \cite{gaier2019wann}.
    \end{itemize}
    \item We performed only minimal interpretation research on the trained networks in the first paper, despite a number of architecture-specific questions of interest. It was largely beyond the scope of that paper, but future works could benefit greatly from a deeper dive into how the network was successfully learning sequences and how similar this was between repeated training runs. Simple tests could include identifying any patterns in the inputs that the network was getting wrong and determining how much this would change when evaluations were performed with one or more specific subnetworks ablated. One might also look at network activation over the course of a particular input, to determine if particular "regions" were more active for certain types of inputs or at different times within the input sequence.
    \begin{itemize}
        \item If specific subtask roles can be identified for particular subnetworks, it could then be asked how consistent that role is when the same subnetworks have their negative feedback connections retrained (all repetitions in the original paper are entirely new instantiations). Similarly, it could be asked what happens if only a specific subnetwork is removed or rerandomized before training, and to what extent the overall network can learn to make up for a missing subnetwork if training is resumed without it (or with it rerandomized), rather than restarted. This could speak to early questions about the "critical period" of the architecture.
    \end{itemize}
    \item Another present limitation for Sparse Combo Net is that the nonlinear within-subnetwork weights are not able to be trained while maintaining the stability guarantee. This could limit its performance ceiling on various tasks, and regardless a number of the more interesting questions related to the multi-area structure would require the subnetworks to be trained in some capacity. There are many possible directions we could take to address this concern, so it will require a careful characterization of pros and cons before moving forward.  
    \begin{itemize}
        \item A major area of opportunity for the multi-area architecture is in testing pretraining paradigms. For the Sparse Combo Net, one could imaging obtaining subnetworks through a pretraining process that involves heavy pruning, again similar to LTH. 
        \item An LTH-style approach could also be employed in the joint training context, allowing subnetworks to learn unconstrained along with the negative feedback connections, and then performing subnetwork pruning to obtain modules that satisfy Theorem \ref{theorem: absolutevaluetheorem} for another round of negative feedback training. Such an approach could potentially be integrated with an LTH strategy for sparsifying negative feedback as discussed above.
        \item It is an open question whether the performance of SVD Combo Net could be improved by strongly enforcing sparsity there, in which case training could proceed without worry for the stability condition. Training might still be slow if this is done through only sparse initialization and $L1$ regularization, but it is possible to enforce sparsity by strategically masking component weights too. One could also imagine a paradigm where some subnetworks are allowed to train through an SVD Combo Net mechanism while other subnetworks follow the original Sparse Combo Net protocol.
        \item It is likely that many of the \emph{sparser} Sparse Combo Nets, which displayed the best performance in our hands, contained a number of directed acyclic graphs (DAGs) as subnetworks. As DAGs contain no actual recurrence, they will be contracting by the hierarchical principle, no matter their weight magnitudes. Guaranteeing they remain contracting in the same metric for efficient training will require some care, but realistically weight magnitudes only go so high during deep learning training, so it should not be difficult to find sparse subnetwork initializations which we can use in practice and allow their non-zero weights to learn without constraint. Because the linear negative feedback connections remain a strong source of recurrence, the overall network will still be an RNN with strong potential to learn long sequences, though it may be better described as an "RNN of DAGs" in this case. 
        \begin{itemize}
            \item Prior to pursuing this angle, it might be warranted to test training the original Sparse Combo Net across different sparsity levels and magnitude limits but with a modified instantiation procedure that either guarantees the subnetworks will be DAGs or guarantees the subnetworks won't be DAGs. With higher magnitudes and sparser structure it may become increasingly difficult to find subnetworks that aren't DAGs, but in the ranges we tested in the original paper this should not be a problem. 
            \item Note that while DAGs are in a sense a feedforward network, they do not match the typical structure implied by feedforward in a deep learning context. Furthermore, the concept of state is not often reflected in feedforward systems, where a unit's output is usually based only on the inputs it receives at the present time, not influenced by some preexisting "voltage" value. \newline For reference, a traditional feedforward network of 3 layers with 2 units each would have an adjacency structure like this, as the units in the first layer go only to the second and the second only to the third:
            \begin{align*}
            \begin{bmatrix}
                0 & 0 & 0 & 0 & 0 & 0 \\
                0 & 0 & 0 & 0 & 0 & 0 \\
                1 & 1 & 0 & 0 & 0 & 0 \\
                1 & 1 & 0 & 0 & 0 & 0 \\
                0 & 0 & 1 & 1 & 0 & 0 \\
                0 & 0 & 1 & 1 & 0 & 0
            \end{bmatrix}
            \end{align*} 
            \newline While a DAG of 3 layers with 2 units each could have an adjacency structure including skips between layers or acyclic connection within layers, like this:
            \begin{align*}
            \begin{bmatrix}
                0 & 0 & 0 & 0 & 0 & 0 \\
                1 & 0 & 0 & 0 & 0 & 0 \\
                1 & 1 & 0 & 0 & 0 & 0 \\
                1 & 1 & 1 & 0 & 0 & 0 \\
                1 & 1 & 1 & 1 & 0 & 0 \\
                1 & 1 & 1 & 1 & 1 & 0
            \end{bmatrix}
            \end{align*} 
        \end{itemize}
    \end{itemize}
    \item With pretraining implemented, some of the above questions about subnetwork functionalities could be addressed in a much more targeted way, and many possibilities for task-specific pretraining paradigms as well as subnetwork connection topologies would arise. In the case of Sparse Combo Net subnetwork pretraining for sequential CIFAR10 (and then subsequent overall training with subnetworks frozen), there are already numerous questions we could ask about what methodologies might obtain the subnetworks topologies most well-suited for overall network learning.
    \begin{itemize}
        \item Individual subnetworks are unlikely to be able to learn sequential CIFAR10 much above chance on their own. However it isn't clear how good a subnetwork would need to be at a task in order to carry some useful information for the larger network. Still, we could imagine breaking down the task input in a number of ways.
        \begin{itemize}
            \item Pretraining on shorter sequences that contain more information per time step, for example row or column CIFAR10 instead of pixel by pixel.
            \item Pretraining on shorter sequences by presenting a particular subnetwork with only part of the original sequence, which could involve a contiguous cropped region or skipping of timesteps (e.g. only show every third pixel).
            \item Pretraining on a smoother version of the sequence such as a moving average, which could involve reducing the sequence length (e.g. no window overlap) but doesn't need to.
            \item For sequences where individual timesteps have multiple values (like RGB signal for CIFAR10), one could also consider pretraining using methods to simplify that to a single value per timestep.
        \end{itemize}
        \item Similarly, one could simplify the labels that the network is asked to learn, whether in combination with the input simplification above, or standalone. One possible example of a subtask in the sequential CIFAR10 context is to ask a subnetwork to classify each input as an animal or a vehicle using a smoothed version of the black and white sequence.
        \begin{itemize}
            \item Besides making the labels simpler via abstraction, labels could also be simpler by focusing on a specific distinction, for example binary classifying the CIFAR inputs as either containing a horse or not containing a horse.
            \item By pretraining a specific subset of the training dataset, it would be possible to make the distinction even more concrete, for example classifying only between horses and ships. 
            \item In addition to utilizing the provided labels, it is also possible to generate alternative subtasks, for example asking a subnetwork to predict the next pixel over the course of an entire sequence. Relatedly, one might allow certain subnetworks during pretraining to make a final prediction based on some summary of the network output over the entire sequence course, rather than only using the state at the last timestep. 
        \end{itemize}
        \item It is also not clear how hyperparameter tuning should work during pretraining, as the best learning subnetwork on the subtask may not be the most informative when it comes to negative feedback training. Most critically, it is an open question whether networks with substantially different learning rates would be capturing different information in how they go about a subtask, such that including them both in overall training would be worthwhile regardless of typical hyperparameter tuning results. Along the same lines, it is not clear how long we should want to pretrain for and to what extent this should be based on a pretraining performance analysis versus a hard-coded number of epochs. Early stopping has been a longstanding method for combating overfitting \citep{goodfellow2016deep}, so especially as our goal here is to find sparse subnetwork topologies that would be useful in the larger negative feedback RNN, we might want to err on the side of a shorter training run -- but that needs to be empirically verified. It will be important to keep in mind as well that the process being employed to ensure meeting of Theorem 1 (likely a LTH-inspired approach) will influence the answers to these questions.
        \item So far I have focused on pretraining using a training set that overlaps with the real training set. However one might want to consider for real world applications a situation where pretraining is done on a smaller holdout set that is used \emph{only} during pretraining. Beyond this concern, it is an interesting question the extent to which pretraining on different sequence classification tasks (e.g. permuted sequential MNIST) would be useful for the Sparse Combo Net, both in a vacuum when compared to random initialization as well as whether added modules from different tasks might provide benefits not obtainable from pretraining with the same task alone (e.g. better adversarial robustness). 
        \item In addition to questions about setting up the negative feedback training between pretrained modules to best take advantage of task structure knowledge and priors about the roles of subtasks, it is also an open question how to decide which subnetworks to include amongst a large set of pretrained ones. The ideas listed above enable a huge number of different pretraining paradigms to be tested, which can be multiplicatively expanded by trying different architecture settings for different subnetworks (e.g. sparsity, size, etc.). Further, we do not know whether inclusion of redundant subnetwork module designs (i.e. different pretraining runs with the exact same protocol) might be of benefit, nor the extent to which we might conditionally eliminate modules because existing ones can already serve a similar purpose. Ideally task knowledge can help to narrow this search space greatly, but for the initial experimentation a more exhaustive characterization is likely warranted. Methods similar to the LTH and evolutionary algorithms mentioned above could perhaps be useful in the context of subnetwork exclusion/inclusion decisions, to allow a good set to be decided in part by the network itself.
    \end{itemize}
    \item Finally, another potential advantage of the modular structure is in the ability to add new subnetworks as circumstances change. A "critical period" effect has been well-documented to occur in many deep learning systems \citep{critperiod}, which could make it challenging to adapt to new tasks without entirely retraining. Transfer learning has shown some success in enabling e.g. niche image datasets to be learned by transplanting some layers from large general purpose image classification networks that have already been trained \citep{goodfellow2016deep}. While not exactly biologically inspired, targeted insertion or excision of subnetworks in a trained "RNN of RNNs" could serve as a conceptual extension to transfer learning, where the Combo Net agent becomes a multi-purposed learner. The subnetwork structure is of course also well-suited to multimodal problems, which could allow for deep learning systems to become more human-like in their ability to be a "jack of all trades". That said, many questions on how this modification process might be best implemented, let alone how well it will work, remain to be addressed.
    \begin{itemize}
        \item Along these same lines, one might wonder whether adaptive add-ons could be leveraged to improve performance at the current task, based on known deficiencies of the model. For example, if most of the mistakes made by a particular Combo Net on permuted seqMNIST are mix-ups of 1s and 7s, might we be able to integrate a subnetwork pretrained on this distinction subtask after the fact somehow in order to improve the overall performance further than additional training on the existing network possibly could?
        \item Speaking of this, permuted MNIST has also been used in the traditional image classification domain to measure a network's ability to simultaneously perform multiple tasks, by training on multiple datasets of permuted MNIST with different fixed permutations. When presenting different permutations interleaved, the network may struggle to learn all of them to its best ability, but when presenting different permutations sequentially, some networks tend to lose their ability to perform the earliest learned permutations well - deemed "catastrophic forgetting" \citep{Goodfellow2013}. Considerations in architecture design, training parameters, and curriculum learning can all play a role in catastrophic forgetting, as can the nature of the tasks attempting to be transferred. As critical period effects in convolutional neural networks tend to relate to the "receptive fields" they form \citep{sinha}, it is not surprising that a very similar style of image task could still be learned by an already trained network, though possibly at the expense of its original skill. Ultimately continual learning requires a great deal of further research overall, but given the existing link to the permuted sequential MNIST task, we might be able to answer some early questions about continual learning with sequences in the stable "RNN of RNNs". Questions related to the roles of both modularity and stability could be of much interest in this context.
        \item On a meta level, and more meaningfully utilizing the idea of "recursive construction", might we be able to emulate a "self-reflection" sort of system by training connections between multiple Combo Nets? One could imagine training a second Combo Net on a dataset that is generated from the first, where inputs are the same but labels are now a binary classification task to determine whether the original network was correct or not on that image during its last training epoch. A final training round connecting these two larger networks in negative feedback could then be performed on the original task. This might be an interesting alternative way to scale the architecture, though it would also be worth testing whether two smaller Combo Nets could achieve better than the sum of their parts through a mechanism like this.  
    \end{itemize}
\end{itemize}
\noindent After more thoroughly exploring such questions on the benchmark sequence task, a small subset of baseline and best performing comparison points for the model settings could be chosen to be tested on additional common sequence classification tasks, with a focus just on test accuracy and consistency in repeated trials. A final chosen architecture could then be fine tuned for some real world sequence task, and in that context the interpretability tools could also be reapplied to show practical use cases. 

\subsubsection{Machine learning extensions}
\label{subsubsec:rnn-ml-vision}
In the long term, it is an aim to try our architecture on more powerful machines with extended training runs, to address complicated problems with large datasets and compete directly with overall SOTA scores on such tasks. The modular structure of the network makes multimodal ML problems involving datatypes like video particularly salient as a future direction. The benefits of the stability guarantee - as well as the broader parallels with neuroscience and with control theory - also suggest that reinforcement learning problems might be a fruitful avenue to pursue. Furthermore, there are a number of applied problem spaces where interpretability is critical, including digital psychiatry, which could warrant their own extended domain-specific follow ups. Along these lines, one might consider pursuing difficult computational problems like instrument source separation (and the link with the human cocktail party effect) that have an intuitive parallel with multi-area structuring. 

There will also very likely be more questions that result from the research plan described in section \ref{subsubsec:rnn-paper2}, in addition to the potential for a similar framework to be applied to investigate different task styles. Explorations of continual learning and out of distribution adaptation in particular would likely warrant its own paper extending on the initial pretraining evaluations. Such questions could connect to concepts in control theory as well, for example last layer adaptive control; we might consider a framework that allows for some weights to be perpetually updatable in real time and even re-initializable, while other weights are locked in after early training. Overall, there are a vast many possible extensions of the work in this chapter, far more than we can personally achieve. We should therefore work to maintain a well-documented and flexible central code base throughout, to facilitate applications of our architectures to new contexts by others, encouraging new functionalities to be contributed back to said central repository. \\

One medium term direction I have a high level of personal interest in pursuing is the use of these architectures on difficult image recognition tasks. In our sequence learning work, we utilized image tasks that are very easy in modern computer vision, and made them difficult by presenting them to the RNN one pixel at a time. Each pixel was processed by the linear input layer at each timestep to send inputs to each of the units in the RNN. While it would be beyond the scope of current sequence learning capabilities (even the overall SOTA) to perform image classification on more difficult image datasets sequentially, we can instead reframe the task to input an entire image to the network at once, thereby allowing tests to occur on modern machine vision benchmarks. It could be difficult for the network to learn with every unit receiving input information from all pixels, but convolutional preprocessing layers with pooling downsampling components could precede recurrent layers, or recurrent layers could be interleaved with convolution like was done in the ConvRNN architecture designed by \cite{YaminsColab}. One might also consider inputting specific pixel(s) to different units in the RNN architecture and then solving the dynamics in the manner of a neural ODE \citep{choromanski2020ode}. Efficiency wise, it might be worth testing a similar approach but where the image is instead held constant for some number of steps and the network prediction is taken based on the state after the acclimation time. Approaches involving differing amounts of convolutional preprocessing, inputs explicitly targeted at different units, and different ways of handling the dynamics might be mixed and matched in early testing as well. 

A large number of questions will need to be addressed in all these domains -- for example, should any convolutional layers used here be trained from scratch along with the network, or perhaps sourced from a pretrained feedforward network? Moreover, many of these questions interact with questions about how to best leverage the stable multi-area RNN architecture. Thus this is a complicated future direction with a broad number of potential paths. Nevertheless, extension to challenging image recognition problems is a crucial direction for our architecture to take in my opinion, because this domain contains many of the most straightforward links between both our work and its neuroscience inspiration, and our work and its theoretical underpinnings in stability analysis. Convolutional neural networks were motivated by knowledge of visual processing systems in the brain, and their function has subsequently been itself correlated with real neural activity \citep{yamins2014performance}, as well as scientifically studied for neurobiological-like properties such as the critical period \citep{critperiod,sinha}. With the many pretraining questions enabled by our architectures, as well as the inverse question of studying learned properties of different subnetworks within a joint training scheme, we would have many opportunities to further curriculum learning research in the image classification domain, and to link such results back to neurobiological foundations.

Convolutional networks have also already found their way into daily life, for example search features in Google Photo, yet they demonstrate startling weaknesses to adversarial attacks, which make them a concern in safety-critical applications and weaken some of the human brain analogies. Not only will networks classify completely random white noise as specific objects like a penguin with $99\%$ confidence \citep{fooling}, but small pixel perturbations, sometimes at levels not even detectable by a human, can be added over an image and cause the network to drastically flip labels \citep{Brendel2018}. This remains a large open issue, and even as defenses are built against simpler methods for generating these perturbations, qualitatively similar attacks continue to challenge convolutional neural networks \citep{Shamir2022}. While there may be no free lunch, this style of attack is so problematic because it can cause the network to make an incorrect prediction with extremely high confidence, at the same time evading easy human detection. It would be conceptually satisfying to engineer networks that perform competitively with existing methods in most cases without falling prey to white noise perturbations, even if that does introduce new failure modes -- particularly if that introduces more human-like failure modes. Because a theoretical benefit of stability is resistance against perturbations, and there are also theoretical reasons to suspect a modular structure could improve adversarial robustness \citep{GWT}, this could very well be the machine learning topic on which we are best equipped to advance the overall state of the art with our current tools; an added benefit of course being the opportunity to strengthen connections between the computational neuroscience and deep learning communities. \\

\paragraph{Image recognition questions.}
There are a number of specific questions we might ask about how a stable multi-area architecture learns an image classification task, including:
\begin{itemize}
    \item Full joint training of Combo Nets on the task, perhaps just using SVD Combo Net, should be performed to use as a comparison point for different pretraining tests. Application of previously discussed interpretability tools to the resulting jointly trained networks might be of great interest as well. 
    \item We will want to perform some characterization of basic network settings as well as traditional hyperparameter tuning on this new task type. It will also be important to run careful control cases studying the role of the stability constraints throughout different training paradigms of interest. Stability constraints can be removed within subnetwork, between subnetworks, and both.
    \item In pretraining on specific image subtasks, how does breaking up the image classification task by input versus by label versus by both impact final performance?
    \begin{itemize}
        \item For this question I am assuming all subnetworks use the same high level pretraining convention, but one could also consider a mix of subnetworks pretrained with different conventions. 
        \item Beyond that, there are many ways to test breaking up an image task through input modifications, and independently many ways to test breaking up an image task through label modifications (including potential data subsetting), so there is probably no single answer; but I think this split is important conceptually in considering how the architecture can be leveraged.
        \item Questions to consider first independently in the context of subtasks defined as input modifications include the following (as well as various combinations of them) --
        \begin{itemize}
            \item Different subnetworks could receive input derived from only the same cropped region of every image.
            \item Different subnetworks could receive input from only a certain subset of the convolutional filters used in overall preprocessing.
            \item Different subnetworks could receive input images that have been augmented in some specific way (Gaussian blurring, pixelated downsampling, occlusion, adversarial-style perturbations, rotations, etc.).
        \end{itemize}
        \item On the label side alone, one could imagine a "20 Questions" or "Guess Who" sort of game. Subtasks might include the following (as well as various combinations of them) --
        \begin{itemize}
            \item Different subnetworks could be pretrained on different subsets of the data including only a subset of the possible labels, potentially strategically grouped to make more difficult distinctions.
            \item Different subnetworks could be pretrained to do binary classification on the presence of a particular label or set of labels, possibly covering different levels of abstraction; for example detecting if something is an animal or not, a mammal or not, etc.
            \item With a richer set of labels available, different subnetworks could be pretrained based on a number of other subquestions, such as the size of the main object, the color of the main object, where the main object is in the frame, what the background landscape looks like, etc. Exactly what to ask here could depend a lot on the types of images being considered in the first place though. A purely face dataset might ask things like eye color, is the person smiling, nose shape, and so on. 
        \end{itemize}
        \item A combined framework also enables new subtasks to be defined that wouldn't fit in either of the more focused scopes; a highly salient example here would be a subnetwork pretrained to determine whether an input image had been perturbed (perhaps as part of an adversarial attack on a different network). 
    \end{itemize}
    \item How should pretraining fit in the overall Combo RNN training process?
    \begin{itemize}
        \item Given a large set of pretrained subnetworks, is there a disadvantage to including all of them in the combined task training? If so how can we filter?
        \begin{itemize}
            \item Can pretraining accuracy, correlation between pretraining accuracy on corresponding examples across different subnetworks, intuition from prior task knowledge, etc. help to inform this decision making?
            \item How much could an additive evolutionary approach or subtractive pruning approach to subnetwork inclusion help improve the performance of final networks? How would these impact overall efficiency of the training process? Is there a tradeoff where this can be done quickly and still help at least somewhat?
        \end{itemize}
        \item Subnetwork weights could be held static after pretraining or they could be allowed to continually update alongside the connections between subnetworks, so long as we are using a stability preserving learning method. 
        \item One could also imagine a situation where some pretrained subnetworks are held static and others are not, or perhaps entirely yet untrained subnetworks could be added that would allow to train alongside the negative feedback training while the pretrained ones are frozen (this could include exploration of combining SparseComboNet and SVDComboNet style subnetworks in one MetaComboNet). Such an architecture might be paralleled to low level sensory regions being more "locked in" earlier in development, while higher level reasoning areas might continue to "update".
        \item In the case of pretraining subtasks that use modifications to and/or parts of images, there is also a question of whether the subnetwork during full task training should continue to be given the exact same type of input through artificially orchestrating the input layer, or if the full network training should be able to figure out how to use the pretrained "knowledge" more organically. 
        \item Are there priors about the image dataset or about how image recognition works in general that could inform which pretrained subnetworks we want to explore connecting in negative feedback? Can results from pretraining or perhaps a more nuanced breakdown of mistakes in pretraining (or even early full training) be used to inform this? Conversely, what can the results of a more organic pruning approach to decide negative feedback pairings tell us about how the network is making predictions, what mistakes it might tend towards, or possibly properties of the image dataset overall?
    \end{itemize}
    \item For all questions, performance can be taken to mean a multidimensional property including test accuracy and robustness to the selected adversarial attacks. Performance should be compared across cases mentioned, as well as against prior adversarial defense literature (overall and those using architecture with recurrent elements).
    \item For select cases of interest based on the preliminary results, a more careful characterization of the network could be done in both accuracy and robustness contexts, hopefully enabling some further iterations to improve robustness without too much test accuracy tradeoff (or vice versa).
    \begin{itemize}
        \item Considering the types of mistakes made not only in standard image classification (e.g. confusion matrix, maybe hierarchical) but also repeated in the context of successful versus unsuccessful adversarial attacks (e.g. see if perturbations disrupted certain labels more often) could enable development of new pretrained subnetworks to be integrated with the current network, or perhaps pretrained subnetworks that would be a benefit to include in a from-scratch full task retraining. 
        \item Comparisons of typical adversarially-produced perturbations that worked against a CNN versus against the Combo RNN could be done as part of this too. An example of something that might be worth testing would be the distribution of distances between perturbed pixels. In many classic adversarial examples the perturbations tend to be more diffuse, which is part of what makes these perturbations look so stupid to humans. Relatedly, label switches from these small perturbations are often comically shown to span wide semantic distances (e.g. koala to firetruck, not koala to sloth). 
        \item Ablation studies and other interpretation tools like mentioned in the prior section could similarly be extended to understanding in what way the network is most susceptible to adversarial attacks, as well as how consistent that is when the same training paradigm is repeated. Though we (for the most part) have prior knowledge this time of which subnetworks were pretrained how, it is unclear that they would serve exactly that role in the final network nor how that might relate to adversarial weakness. 
        \item An additional full training phase that includes some adversarial examples generated against the original and also allows the network to say "I don't know" at a reduced loss penalty (moreso when example was perturbed) could be effective, particularly in conjunction with a small number of added subnetworks targeting weaknesses. This could be a way to integrate those new subnetworks while still allowing for other negative feedback connections to be updated in a way that isn't as redundant for those older connections. 
        \item It is unclear to what extent making changes like this after the fact might introduce new adversarial weaknesses, perhaps weaknesses outside the scope of the adversarial test of focus of the updates -- so it will be important to have some variety represented in the adversarial attacks that are used for evaluation. Such variety could be purely implementation-related, but it could also reflect different resources or objectives: including black box versus transparent and typical "make incorrect" attacks versus more targeted "make this classify as a car, no matter what it is" attacks. 
    \end{itemize}
    \item For a handful of chosen training paradigms, I also think it could be very interesting to run some tests inspired by the CNN critical period study of \cite{critperiod}.
    \begin{itemize}
        \item Can the critical period of "RNNs of RNNs" be broken into phases more like humans, with potential separate critical periods for the pretrained regions and then for the training of connections between the subnetworks? In general how does the critical period relate to what we would get from a typical CNN if trained on the same dataset?
        \item Does it matter the extent to which we might be relying on an initial convolutional layer that is sourced from an existing conv network? If we retrained that CNN with the initial layer static what sort of critical period might it demonstrate?
        \item Within our own model set, how do the stability conditions impact the nature of the critical period? And what happens to the critical period when we do full training simultaneously versus in some of the more constructed pretraining paradigms? Might there still be some sort of phase structure arising organically?
        \item Does allowing the subnetworks to train further during the full task training process change properties of the overall network's critical period? How might this compare to critical period in a transfer learning setting?
        \item Do different image subtasks themselves have fundamentally different critical periods? Can this information be used to further improve the pretraining process? Where applicable, what if those same subtasks were presented to a standard CNN? 
        \item How consistent are these critical period properties across repeated training procedures? 
        \item How (if at all) do clear differences in critical period properties correlate with overall task or adversarial defense performance characteristics of the architecture? Can this be connected with ideas from theoretical explanations of adversarial examples such as the Dimpled Manifold model of \cite{Shamir2022}? Is it possible that any potential improvements in a multi-area RNN context could be related to why humans have such fundamentally different adversarial weaknesses, described within such present theoretical frameworks?
    \end{itemize}
    \item Because this process involves a number of dataset-specific considerations, it will be important to show that a version of the paradigm that worked best in the main experiments can be replicated to at least moderate success when (more briefly) tuned on a different image dataset.  
\end{itemize}
\noindent Design decisions in pursuing some of this work could potentially be informed by the results of \ref{subsubsec:rnn-paper2}, but given the context of an entirely new task structure it will be important to repeat a number of those experiments. This might also give us an initial hint of how well certain principles about the Combo Nets will broadly generalize.

\subsubsection{Neuroscience connections}
\label{subsubsec:neuro-rnn-next}
The question of neural stability is one of the oldest questions in computational neuroscience. Indeed, cyberneticists were concerned with this question before the term 'computational neuroscience' existed \citep{wiener2019cybernetics,ashby2013design}. Stability is a central component in several influential neuroscience theories \citep{hopfield1982neural,seung1996brain,murphy2009balanced}, perhaps the most well-known being that memories are stored as stable point attractors \citep{hopfield1982neural}. Our work shows that stability continues to be a useful concept for computational neuroscience as the field transitions from focusing on single brain areas to many interacting brain areas. 

To fully leverage computational modeling in neurobiology, it will be necessary for such models to represent the modular structure of the biological brain. Despite this being widely acknowledged in computational neuroscience, there remains a paucity of prior results on the topic \citep{yang_molano-mazon_2021}. Not only has the work in this chapter taken a step towards addressing the gap, it has included an open source multi-area RNN architecture implementation \citep{sparsegit}, and shown that this architecture is provably stable \citep{NIPS22}. Thus a natural extension of the work would be to apply our networks to learn tasks that mirror those used with mouse models in neurobiology labs. 

\begin{figure}[h]
\centering
\includegraphics[width=0.75\textwidth,keepaspectratio]{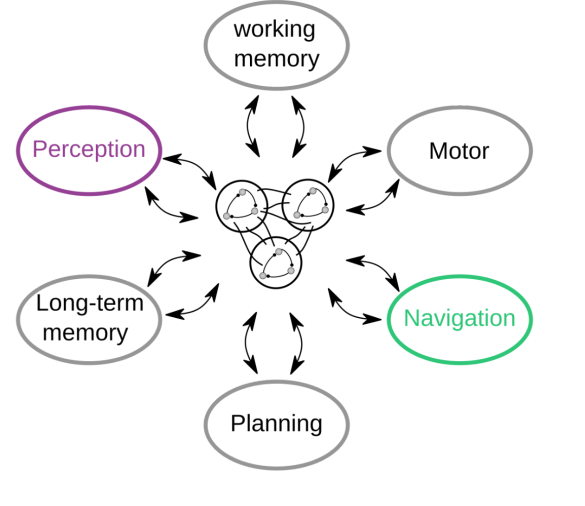}
\caption[Multi-area RNNs in the future of neuroscience.]{\textbf{Multi-area RNNs in the future of neuroscience.} As experimental neuroscience and machine learning move increasingly towards studying connectionist systems in a modular fashion, computational neuroscience will soon need to include more models with a structure akin to our "RNNs of RNNs". Indeed, recent proposals have called for neuroscientific modeling to focus on training multi-area networks on multiple different multimodal tasks, instead of simulations based around a single region and single task \citep{yang_molano-mazon_2021}. Though progress in the multi-task space has already begun to emerge \citep{Khona2022}, there remains a paucity of results on complex multi-area systems. However, our networks are well suited to addressing questions of neuroscience modeling like those posed by \cite{yang_molano-mazon_2021}. This diagram, reproduced from their proposal, depicts a global workspace feedback topology between subnetworks specialized for different roles. While our pilot tests largely involved all-to-all negative feedback, the code allows for setting an adjacency matrix to describe which subnetworks will be joined in negative feedback. One could therefore have a subnetwork (or set of interconnected subnetworks) act as a central hub like depicted, paralleling executive function control from e.g. the prefrontal cortex, and then have a different set of subnetworks act like the spokes here, possibly pretrained on specific subtasks to emulate some of the specializations depicted. \newline [reproduced from Figure 3b of the perspective piece by \cite{yang_molano-mazon_2021}, covered by fair use].}
\label{figure:multi-area-nets}
\end{figure}

\cite{Khona2022} recently built on the "NeuroGym" reinforcement learning tasks released by \cite{yang2019task}, to expand the test battery to include more difficult learning problems whilst still using the same input/output structure, which includes representation of multimodal inputs. There are now multiple distinct directions that could be explored using these tasks and our networks. One such direction involves the intentional partitioning of subnetwork modules into distinct roles analogous to localization in the brain, with a negative feedback adjacency structure reflecting neuroanatomical priors (Figure \ref{figure:multi-area-nets}). 

This would likely involve jointly training the combined network on all tasks, but restricting which subnetworks receive external input (partitioning modalities), which subnetworks can produce external output, which subnetworks are able to update their weights during different parts of training, and the timescale of the processing dynamics within each subnetwork (Figure \ref{figure:multi-area-nets}). It could also be informative to try pretraining different modules on subsets of tasks from different categories, perhaps focusing pretraining on only the easier original NeuroGym tasks, and then subsequently training negative feedback between modules across all tasks. Additionally, given the focus of our initial best performing architecture on sparsity, it is likely that computational neuroscience directions focused on highly sparse networks and/or network pruning would intermix well with some of these pretraining-related questions.

Although it might be inherently interesting to evaluate how well a network can perform across these tasks as settings are changed and different training paradigms are tested, their advantage of course is in their neuroscientific grounding. Thus another broad direction relates to training a multi-area architecture jointly without enforcing any specific structural assumptions, and studying the structure that arises. By utilizing methodologies like ablation studies, the role of different subnetworks in performing different tasks can be dissected. Additionally, analysis tools analogous to those used with neural recording data could be implemented to study how "neuron" activity in different subnetworks varies over time and in response to different inputs. One might be able to obtain a simulated parallel to a functional connectivity map for the RNN of RNNs, for example. A major advantage of the synthetic data methodology is that this can be repeated numerous times and over numerous different settings at little cost. 

Eventually we would like to connect such representations back to real mouse brain recordings collected during these same tasks, to determine what architecture settings produce activity most similar to biology, as well as identify major remaining differences. This could then allow for hypothesis formulation via further exploration with the selected model, and subsequent testing in real animals to determine the extent to which the model could predict real behavior and/or neural activity that had not been part of the original comparison process. However, such work would require closer collaboration with a systems neurobiology group, and is likely a target for the longer term. In the meantime, the network analyses described could still be compared between networks that were trained with fewer restrictions and those that were trained with subnetwork specialization or neuroanatomically-inspired structure explicitly enforced.

Furthermore, even in a vacuum it would be possible to link the synthetic "neural activity" data to task performance across networks in careful detail. Between looking for patterns in the mistakes that different networks make and determining how much performance suffers when different areas are ablated, it would be possible to characterize final trained networks in a data-driven way behaviorally and neuro-behaviorally. This could be done similarly for clustering of networks based only on neural activity patterns, and one could then investigate the degree to which these groupings overlap with each other and with known underlying settings differences. Once again, such an exploration could involve many repetitions and great heterogeneity in settings considered, from small hyperparameter differences to major conceptual distinctions in the training paradigm. Because we do have priors about task behavior from mouse studies, it also would be possible to contextualize some of these results in terms of real animal models, by identifying how "life-like" the network mistake patterns appear and potentially whether errors typical in mouse models of certain diseases might be inducible via specific mechanistic adjustments hypothesized to be disease-relevant. 

Similarly, another possible direction is to study variation in these networks in phenomena that we know occur in artificial neural networks and that are fundamentally analogous in some way to phenomena we observe in biological brains -- most notably the critical period. The human critical period has a phased structure, and is also intertwined with sensory development; for example the "hardware"-based limitation on visual acuity in early life, which may or may not be a benefit to the system's development \citep{sinha}. Early work on critical periods in deep learning demonstrated some parallels to the human critical period, but some differences as well \citep{critperiod}. It could therefore be salient to characterize critical periods in multi-area RNNs and particularly in multi-area RNNs with different topologies and different training methods, as this could serve to directly inform future neurobiology hypotheses to be tested in mouse development and task learning. The large number of distinct tasks in the NeuroGym expansion curated by \cite{Khona2022} would be well-suited to this line of questioning, because continual multi-task learning has many links with the critical period. \\

\paragraph{Global workspace models.}
In the realm of (loosely) replicating modular neuroanatomy, the global workspace (GW) theory originally proposed by \cite{GWOG} has had a recent resurgence in computational neuroscience and neuro-inspired deep learning. Figure \ref{figure:multi-area-nets} depicts a GW-style structure where individual modules are all recurrently connected to a central network, but not to each other \citep{yang_molano-mazon_2021}. The core idea is that the modules communicate with each other instead through a shared latent space in the GW, a prefrontal cortex-like structure that in a human cognitive interpretation represents conscious experience and sensory imagination \citep{GWOG}. As such, the GW should have higher dimensionality than any one subnetwork, but much lower dimensionality than the sum of the subnetworks, to serve as a translation bottleneck of sorts \citep{GWT}.

In addition to the GW themes arising with discussion of emerging multi-area RNN research in computational neuroscience, \cite{BengioGW} recently utilized the concept of a shared global workspace in the context of Transformer attention mechanisms. While their work demonstrated great success and brought a spotlight to GW ideas in machine learning, their model does not have any dynamics and is ultimately a vaguely neuro-inspired engineering advance rather than an architecture directly applicable to computational neuroscience questions. At the same time, there have been additional calls for modular and especially GW architectures from those at the intersection of computational neuroscience and deep learning, such as a recent proposal by \cite{GWT}. 

It is therefore quite convenient that our general network can be easily applied to a global workspace framework, and this direction is a possible priority for near future research exploring stable "RNNs of RNNs" through a computational neuroscience lens. Note that in our combo RNN methodology one could represent a GW with an adjacency matrix structure like the following (which specifies those subnetworks to be connected in negative feedback), ensuring that the subnetwork corresponding to the GW has the appropriate size and other desired settings when compared to the other subnetworks (which of course may also vary with respect to each other). This example represents an architecture with a global workspace and 3 specialized subnetworks:
\begin{align*}
    \begin{bmatrix}
        0 & 0 & 0 & 1 \\
        0 & 0 & 0 & 1 \\
        0 & 0 & 0 & 1 \\
        1 & 1 & 1 & 0 
    \end{bmatrix}
\end{align*}

Dissecting the latent space of the GW could be highly salient in an interdisciplinary manner, whether investigating neuroscience tasks or more typical machine learning benchmarks. There are also questions on which engineering and neuroscience perspectives might substantially diverge -- particularly those related to how a GW model implements perceptual attention and how (as well as the extent to which) it enforces cycle consistency of its representations between sensory modalities. The perspective piece by \cite{GWT} presents a detailed proposal on implementing a modern version of a GW model using cutting edge deep learning tools whilst attempting to balance with cognitive science principles. Though their roadmap might have strong merit for future machine learning advances, I felt it was a little over-engineered for the purposes of using an "RNN of RNNs" architecture within computational neuroscience. As such, I will close the discussion of neuroscientific future directions with some additional commentary on their proposal and how it fits (or doesn't) with our work. \\ 

I would personally be most interested in seeing what might arise naturally given different architecture and curriculum learning choices in RNNs of RNNs, and especially to what extent the stability conditions might facilitate the network to meet other criteria without ever directly imposing them. For example, artificially imposing a transformer-style attention mechanism might perform better in an initial GW application, but what does that really tell us? It would be much more interesting if well-designed task training with a Combo RNN GW structure could generate some attention-like behavior. One could obviously argue the entire situation is artificial, but I think there is fundamental scientific value in trying to understand connectionist systems and how other (perhaps imperfect) properties might emerge from them, even if the training process is not at all biological. Whereas the attention mechanism described by \cite{GWT} is an engineering solution built on top of the rest of the system, which would make sense if the hypothesis were "global workspace theory will lead to the next major deep learning architecture", but I don't think it fits as well if the goal is to train models on neuroscience tasks and interpret results through a neuroscience lens.

Similarly, I find it very hard to believe that humans optimize primarily for cycle consistency in the way proposed by \cite{GWT}. It is much more plausible to me that cycle consistency is a secondary goal -- like an added regularization term from an ML perspective, or maybe something that sleep attempts to do from a biological one. It's also possible that any cycle consistency in humans arises entirely inadvertently as the larger system is tasked with coordinating on complicated external problems. But to suggest that the core training goal should be maintaining cycle consistency seems causally backwards to me, as without rich external feedback it would be possible to obtain cycle consistency in useless ways, and conversely in the real world cycle consistency could be a useful tool for accomplishing actual end goals but is not a goal in itself. So I think the overall network absolutely should be required to perform more complex tasks as its primary objective. Then task performance could be studied as a secondary cycle consistency term is varied, or perhaps more interestingly, changes in cycle consistency levels of trained networks without any explicit term could be studied as other experimental variables are tested.

I will also note that in addition to underrating external tasks, \cite{GWT} mention that even from a cycle consistency perspective one should not need many real matched examples to learn. The major argument they provide for this is that different language latent spaces can be mapped between with very few supervised matching examples. However, human language did not arise in a vacuum, it is the result of a collective "training" process that involved countless multimodal real world observations. The fact that language latent spaces have similar topologies is a result of shared human experiences, and I don't see how that reflects on what is most important for an entirely naive agent to develop coherent associations in their latent space. If you train multiple GW-shaped networks on the same training dataset, you may or may not be able to map between their latent spaces in an unsupervised manner, and perhaps how well you can do this is an interesting architecture property. But regardless, this test would not tell you anything about whether the network could be successfully trained from scratch using few supervised modality-matched examples.

Furthermore, it is not clear to me how they would define cycle consistency across all of the different modalities mentioned. Most people do not smell in high resolution, and animals with a good sense of smell are usually lacking in other senses. There is not going to be a bijective map between smells and visualizations, and this is even obvious from the example they give: someone might imagine a smell when playing a video game if the situation is similar to something they've seen in real life, but if a month later you present that smell I doubt most people would think of the video game at all, let alone that being the first thing that comes to mind. Yet if the game scene were presented again in that same moment instead, certainly the smell would come back. Of course the larger phenomenon there is (to at least some extent) a feature, not a bug -- if you did have a perfect high resolution bijective mapping, you'd probably be very bad and/or very slow at many forms of out of distribution generalization. On the other extreme, if you gouge out your eyes and chop off your nose you could also have a technically perfect bijective mapping. I don't see how it could be thought of as an inherently sufficient property. 

In a deep learning context, it's important as well to watch out that the network doesn't "cheat" in its methodologies when using cycle consistency loss. \cite{GANattack} found forcing pixel-level cycle consistency too harshly in contexts where a perfect one-to-one map can't actually exist will lead the CycleGAN to "hide" small perturbations in the lower information image that allow it to still reconstruct the original detailed image, through a similar mechanism as adversarial attacks. This is the exact kind of thing our work should hope to attenuate, not worsen.

Maybe humans do have hierarchies of bijections at different levels of abstraction, and this could be replicated with cycle consistency learning in a more complex module structure containing workspace tiers. But I suspect the situation biologically is a lot messier than that, and regardless it would seriously complicate any proposed computational architecture to attempt to implement this. Sticking with the existing GW structure though, I can't think of a great way to navigate the cycle consistency tradeoffs while using a learning paradigm that prioritizes it. Pushing too hard for cycle consistency in high detail I don't think would make for an interesting latent space, especially not with a neurobiological motivation. Pushing too hard for cycle consistency in low detail can lead to entirely spurious associations. I find it hard to imagine a perfect level of detail existing, certainly not perfect across situations. It is reminiscent to me of many conversations I've heard over the years about "enhancing human intelligence" that sure sound a lot like potential mechanisms for ASD or psychotic disorders. So I really do not think cycle consistency should be \emph{the} goal, for many reasons. Obviously that doesn't mean it can't be a component, but \cite{GWT} seemed to want it to be the crux of the GW loss.

\FloatBarrier

\subsubsection{Expansions to theoretical results}
\label{subsubsec:theory-rnn-next}
Beyond the numerous future directions enabled by this work in machine learning research, scientific data analysis applications, and computational neuroscience modeling, there are also a number of new theoretical questions of inherent interest opened up by our mathematical results, as well as additional theoretical questions identified to be of salience due to the role they could serve in further advancing the noted machine learning or neuroscience applications. 

Most notably, Theorem \ref{theorem: Wdiagstabtheorem} requires further investigation. As argued above, we suspect that diagonal stability of $g\mathbf{W}-\mathbf{I}$ is a sufficient condition for global contraction. However, as we showed in Theorem \ref{theorem: Wdiagstabcounterexampletheorem}, proving this will require a time-varying metric, thereby laying important groundwork towards finding such a condition. The counterexample we found is mainly exploited by antisymmetric connections and changing inputs, so research into input-dependent metrics may prove particularly fruitful. Developing a less restrictive contraction condition for $\mathbf{W}$ via input-dependent contraction analysis would be a major contribution in the stability theory literature and could have far-reaching implications in downstream applications. In the context of our experimental architecture, a proof that overcomes the challenges presented by Theorem \ref{theorem: Wdiagstabcounterexampletheorem} would have great value in facilitating the formation of provably stable nonlinear negative feedback connections between subnetworks, which would also bring much deeper potential meaning to the recursive construction of these networks.   

While our conjecture focuses on further improving the conditions for global contraction of the general RNN model (\ref{eq:RNN}), another way to uncover new conditions would be to update assumptions to align with a more specific question of interest. One such path forward would be to focus on the dynamics of a particular activation function, like the common ReLU or tanh functions, which have 0/1 derivative and bounded value respectively; tanh in particular may be an easier case in which to prove conditions like the conjectured diagonal stability claim. Another path would be to add biologically-inspired (or application-specific) assumptions about the network topology, for example studying Daleian networks that constrain each neuron to have either all excitatory or all inhibitory outgoing weights. Yet another would be to focus on local regions of contraction, which could make the Jacobian in the time-varying metric case more tractable, and would also enable us to build networks with weaker stability criteria, something of interest regardless of mathematical differently.

There are also a number of theoretical questions that could be pursued by directly extending our current conditions in conjunction with other prior contraction analysis results. For example, exploring more combination properties to potentially apply in the RNN context, like the small gain theorem mentioned above, would enable study of a much greater variety of modular network topologies. One might consider extending the present combination properties to combining an even more diverse set of subnetwork RNNs as well, such as exploring feedback combination of continuous and discrete systems, or application of singular perturbation isolation to better characterize combinations of systems that contract on very different timescales (as demonstrated by \cite{nguyen2018primaldual}). 

Our conditions could additionally be used in other applications distinct from those already outlined. For instance, questions about synchronization of networks might be addressed via a virtual systems perspective to contraction, as demonstrated by \cite{wang2005oscillators}. Related methods could even be applied to models that include weight updates, such as the RNN with Hebbian and anti-Hebbian learning rules discussed by \cite{kozachkov2020achieving}. Input-dependent stability criteria for \emph{updating networks} are of particular interest, to provide theoretical groundwork on topics such as curriculum learning and transfer learning, as well as better understand development in the biological brain.

Further, while our work focuses primarily on stability of combinations, there are a variety of other theoretical questions that might be asked about combinations of neural systems. In particular, preservation of function is an important property to address in a model of evolution that acts on edges rather than nodes \cite{slotine2012links}. Therefore investigating whether our conditions can produce any guarantees on functionality of combinations is another future aim. Combination properties resulting from this work could be employed in research of e.g. Neural Architecture Search algorithms, and the converse property of stable subnetworks could be investigated in the context of mass pruning techniques. \\

\noindent Breakthroughs in deep learning have often coincided with major developments in architecture design, so such avenues of questioning are critical to future research; inspiration for deep learning practice should indeed continue to draw from the intersection of neuroscience and mathematics.

\subsection{Contributions}
\label{subsec:rnn-contrib}
Throughout this chapter, I've made sure to underscore the importance of the reported research as well as its high multi-disciplinary relevance. In this section, I will close by reiterating many of these points with a slightly different perspective -- a focus on the scientific contributions I have personally made via the work presented here.

\noindent In summary, my major contributions were:
\begin{itemize}
    \item Writing the chapter, I laid out extensive background on links between nonlinear control theory, machine learning, neurobiology, and digital psychiatry, in order to contextualize the results of the chapter in terms of potential impact for both neuroscientific contributions as well as applied ML on the datatypes described through the rest of the thesis. 
    \begin{itemize}
        \item As part of this, I also ensured that technical concepts outside of the scope of the assumed thesis audience were explained in greater detail and with more intuitive remarks than would be typical for a manuscript on RNN stability, essentially writing an entirely new "paper" around the results of \cite{NIPS22}. 
    \end{itemize}
    \item I developed and proved a novel sufficient condition for global contraction of the individual nonlinear RNN in terms of the linear stability of its weight magnitudes. The condition is conceptually easy to understand, computationally easy to check, and allows for an associated contraction metric to be obtained via straightforward calculations common in linear control.
    \item I then leveraged this sufficient condition in conjunction with a neuroscience-inspired focus on sparsity to design our best performing "RNN of RNNs" architecture, which set a new state of the art for provably stable RNNs on multiple sequential processing benchmarks.
    \item I planned and performed pilot experiments characterizing the effects that network size, network modularity, subnetwork sparsity, subnetwork weight magnitude distribution, and negative feedback sparsity had on the aforementioned Sparse Combo Net performance, as well as running relatively extensive repeatability experiments for the best Sparse Combo Net settings and ensuring that this final architecture code was well-commented and publicly released \citep{sparsegit}. 
    \item These early empirical results have provided important groundwork for determining next steps in both understanding how Sparse Combo Net is working and in best applying the architecture to more practical machine learning tasks in the future. As such, I characterized a number of these future experimental plans in detail in this chapter, as well as discussing a variety of longer term future directions that could be of particular interest to computational neuroscientists and to machine learning specialists.
    \item Theoretically, I also set up future work via finding and proving a counterexample to the original proof method we had spent many months attempting to apply to a conjectured less restrictive nonlinear stability condition (related to linear stability of the RNN weight matrix). While this of course does not show that the conjectured stability condition is not true, and it also does not show that the method is useless (we in fact used it in proving the new conditions we did provide, though they are more restrictive than what we originally set out for), it does demonstrate that more complex methods - e.g. contraction in a distance metric that varies based on input and/or neuron state - will be needed for any future research that attempts to tackle said conjecture.  
    \begin{itemize}
        \item Similarly, I proved a necessary condition on the network weight matrix for the individual nonlinear RNN to be contracting for all possible activation functions we permit. This condition connects deeply with the other two mathematical results I proved, and could serve to narrow the search space for future such research. I also contextualized this condition in terms of neuron pruning, and discussed how contractive combination properties can be mirrored in the use case of taking apart systems, which while less suited to evolutionary parallels has possible implications for other important neurobiological mechanisms.
        \item Additionally, note that the counterexample I provided is not only of relevance to future study of stability in the individual nonlinear RNN, but also to future study of multi-area RNN stability: the fact that we had to use linear connections when forming the negative feedback between different nonlinear subnetworks is directly related to this counterexample, and a new method that hypothetically proved the underlying conjecture would thus expand the capabilities of the stable "RNN of RNNs" architecture as well. 
    \end{itemize}
\end{itemize}
\noindent Ultimately, there has been a surprising lack of prior work on modular networks both in neuroscience theory and in deep learning practice thus far. In my opinion this line of research will be important to both fields going forward, and is a great opportunity for increasing bidirectional interdisciplinary contributions in the modern deep learning era. My research presented in this chapter has contributed to groundwork on both fronts, and I expect to one day see that extensions along these lines have enabled progress on interpretable, safety-critical machine learning: a necessary component for the vision underlying my other thesis chapters to come to fruition.

\addtocontents{lof}{\protect\addvspace{0.3cm}}
\chapter*{Conclusions and observations}\label{ch:conclusion}
\addcontentsline{toc}{chapter}{Conclusions and observations}
\renewcommand\thefigure{C.\arabic{figure}}    
\setcounter{figure}{0}  
\renewcommand\thetable{C.\arabic{table}}    
\setcounter{table}{0}  
\renewcommand\thesection{C.\arabic{section}} 
\setcounter{section}{0}

Current criteria for psychiatric disease diagnosis produce labels that are inconsistent between clinicians, unable to predict treatment outcomes, and difficult to track over time. Heterogeneity between and longitudinal fluctuations within humans make this task especially difficult, but also provide great opportunity for new insights about the brain. Consequently, the development of robust behavioral analyses is critical to medicine and research \citep{Fisher2018}. Until recently, it was not feasible to address these concerns, but with ongoing advances in sensor and computing technology, behavior can increasingly be studied in-depth in a natural setting. For these reasons, it is crucial to develop metrics for quantification of psychiatric symptoms that leverage continuous, passive data. 

At the same time, it will be necessary to formulate research problems - especially in the earliest stages of the digital psychiatry subfield - in a tractable way, whether focusing on data science and machine learning methods or on utilizing emerging tech in the context of hypothesis-driven psychiatry research. It is of the utmost importance that future works \emph{in aggregate} consider both dimensions in a diversity of ways. There is a general lack of diversity at certain levels of the study design process across science, and the negative impact on emerging and interdisciplinary fields can be exponential. \\

\noindent Through this thesis, I have attempted to address a number of broad concerns in early digital psychiatry and especially speech sampling research, including:
\begin{itemize}
    \item A tendency for some recent speech modeling papers in psychiatry to be so exploratory that they lose truly all meaning when they are not followed up on - which they functionally won't be because CS model development (entirely insensitive to domain-specific nuances incidentally) will continue to easily outpace real data collection for some time. However, there are a variety of ways to address this (in part) that also happen to be of direct relevance to the thesis:
    \begin{itemize}
        \item Use a better datatype, which as I've hopefully made clear in chapter \ref{ch:1}, is probably audio journals for many research applications that are presently neglecting them. At the very least, a committee of experts should be capable of justifying why they think interviews are the best source for patient speech for their intended aims. Audio journals are much easier to obtain a large scale dataset for and enable truly longitudinal modeling.
        \item Develop a set of core benchmark features so that the research community can better share a common technical language and build shared distributions for reference in more thoroughly dissecting and interpreting works. If centered around a quality software release, this could also facilitate better data collection practices and ultimately enable researchers to contribute their exploratory works back to a more centralized structure. These are some of the themes I tried to focus on when reviewing my audio journal processing work internal to the lab in chapter \ref{ch:1}, through code publication/documentation \citep{diarygit}, scientific explanations for feature choices, and extensive figures and metrics to characterize the low level results on our dataset.
        \item Collect a much larger and diverse dataset by instead participating in a major collaborative data collection initiative, like the one described for the AMPSCZ project in chapter \ref{ch:2}. It is important to both funding efficiency and the integrity of downstream analyses to be especially careful with quality control and processing of the interview recording datatypes here, as privacy concerns limit their sharing much moreso than the other multimodal datatypes involved in the project. My code \citep{interviewgit} for facilitating AMPSCZ data collection across nearly 40 global sites has therefore played an important role in the future of this sort of direction, with a large number of problems already caught in early submitted interviews. 
        \begin{itemize}
            \item Though perhaps the largest contribution of the entire thesis has been the somewhat indirect result of re-prioritizing the speech sampling research plan for AMPSCZ more towards the audio journals and in particular away from the structured clinical interview recordings with their myriad of technical and scientific limitations (Appendix \ref{cha:append-ampscz-rant}).
        \end{itemize} 
        \item Select and adapt modern machine learning techniques for use in psychiatry in a principled way that focuses on e.g. working with limited feature sets as efficiently as possible. While I did not directly do this in my thesis, I did contribute results (along with pilot model code) related to achieving good performance while maintaining higher levels of robustness and interpretability in recurrent neural networks, as part of my work in chapter \ref{ch:4} \citep{NIPS22,sparsegit}. 
    \end{itemize}
    \item A tendency for a great deal of psychiatry research to focus on diagnosis classification and an even greater proportion to focus on other less bad but still highly fuzzy and inherently temporally limited readouts. This indeed makes it difficult for hypothesis-driven research to appear anything but highly incremental. However, there are also many ways that existing labels could be reworked in a more principled way to improve on various short comings, again of relevance to the work in this thesis:
    \begin{itemize}
        \item Longitudinal datasets and patient-specific models were a large theme in the audio journal dataset characterization of chapter \ref{ch:1}, with demonstration of large individual differences across most considered diary feature distributions, from looking at only a handful of participants to begin with. Additionally, even the simplest of diary verbosity features had opposite direction significant relationships for modeling self-reported symptom severity scores in two different subjects. 
        \item Longitudinal data was of course also a big theme of the DBS case report in chapter \ref{ch:3}, where the relevance of multimodal data was highlighted as well. To have appropriate timescale behavioral readouts necessary for furthering neurobiology (arguably regardless of psychiatry), it will be critical for a major general research direction to be the characterization of passive data sources for behavior sensing. Regardless, multimodal data collection can provide important sources for implicit labels that could enable exploratory methods without sacrificing statistical rigor on final analysis of clinical relationships. At present multimodal data already helps greatly with addressing data missingness and interpreting surprising aberrations. 
        \item The scientific interview results \citep{disorg22} reported in chapter \ref{ch:2} focused on modeling that used a relatively small dataset to ask questions about a specific class of linguistic feature (disfluencies) and a specific domain of psychotic disorders symptoms (conceptual disorganization). While in some sense this is definitionally incremental, there is a paucity of research that focuses on specific questions related to specific symptoms, which is critically needed to work towards improving disease classification. More broadly, there seems to be a huge lack of functionally incremental work in the brain and cognitive sciences, ironically despite how common intellectually incremental work is. 
    \end{itemize}
    \item A major underlying problem for both cases is that we have encountered a search space in neuroscience(/psychiatry) where it is really just not feasible to do anything but average out most interesting stuff or find spurious correlations in a data-driven approach, and where a fully formalized modern hypothesis-driven approach wouldn't be capable of scratching the surface in our lifetimes. This is further compounded by the field's inherently interdisciplinary nature at present. However, there are a large number of ways that research directions could be diversified and hypotheses could be generated in less traditional ways. 
    \begin{itemize}
        \item One obvious source for inspiration is the patient's own account, which is a large part of what the audio diary is. Throughout chapters \ref{ch:1} and \ref{ch:3} self-reported content was highly relevant to interpretation in many forms, for example describing helpful details about medication usage or making  directly relevant statements that were missed by the resolution of the clinical scales - like the observed report of a suspected manic episode that coincided with large otherwise unexplained abnormalities in the diary features. It could also be an entirely unprompted argument (insofar as diaries are free-form) about why the person is not currently manic. The specific content in the diaries is relevant in a myriad of personal but potentially meta-generalizable ways.
        \begin{itemize}
            \item Interesting subconscious language use patterns can be found in a similar way via open ended journals, for example the participant who said "kinda down" in multiple diaries in a row preceding both of their two major mood episodes. Each episode spanned multiple months, yet they did not use these words otherwise, and they were not self-reporting mood out of the ordinary at those times via the checkbox surveys.
        \end{itemize}
        \item Another potential source is through further development of and mechanistic study of interpretable neural network models, like the work of chapter \ref{ch:4}. Surely this should at least be an easier problem than understanding the human brain, and many models are capturing patterns in data in a way that we don't yet fully understand - but perhaps could. Small scale examples of this have already occurred both in the sense of a modern adaptation to looking at linear regression parameters, and also in the sense of e.g. pro Go players learning new moves by playing against AlphaGo.
        \item More systemically, there are likely plenty of good old ideas that got thrown out with the bathwater over the years, a cycle I would imagine was particularly likely to occur in psychiatry. The most promising neuro-related health tech results I have personally seen to date have been with electrodermal activity, which of course used to be called galvanic skin response. Yet in certain circumstances (e.g. seizure-related spikes) EDA has such a ridiculously strong signal there is nothing really to even statistically analyze \citep{Johnson2020}. Regardless, bad practices shouldn't be overly associated with inadvertently affected lines of questioning where it can be avoided. Many bubbles started for a real reason (Figure \ref{fig:fads}).
        \begin{itemize}
            \item Along these lines, "exploratory" shouldn't translate to "aimless model fitting", though I just cast it in that light above. In theory, exploratory could involve a lot of things, including old fashioned iterative curiosity and meaningful attempts to challenge results from both oneself and others. 
        \end{itemize}
    \end{itemize}
\end{itemize}
\noindent Overall, I have tried to outline a number of scientific challenges throughout this thesis as they pertain to the future of digital psychiatry. While I made arguments and presented resources and results related to a few specific directions for future work that I believe have promise, my broader hope is simply that nuances across the stages of study design - like those I detailed in this document - will be more openly discussed and ultimately better diversified upon. \\

\noindent I will therefore focus the remainder of the concluding chapter on a more general scientific scaling problem that I believe has made a number of the roadblocks I described substantially more challenging to address. For more exhaustive summaries of thesis results and my specific contributions from each chapter, see the following sections (respectively): 
\begin{enumerate}
    \item Audio journals - \ref{subsec:diary-contrib}
    \item Interview recordings - \ref{subsec:u24-contrib}
    \item OCD DBS case report - \ref{subsec:ocd-contrib}
    \item Stable RNNs of RNNs - \ref{subsec:rnn-contrib}
\end{enumerate}

\pagebreak

\noindent \textbf{On Figure C.1:}

\noindent Anecdotally, I thought galvanic skin response (GSR) was total pseudoscience when first learning about research in high school, and now I'm a real believer in electrodermal activity (EDA) for a number of applications; yet GSR = EDA. What I suspect happened is that GSR was a popular research topic a long time ago, to the point it became over-hyped, and then people (whether scientists or not) misused it entirely in real world applications, most notably the lie detector test. This then lead to a "pseudoscience" misconception about the entire concept of GSR, and ultimately discouraged/defunded scientific research on it for a period of decades, until it enjoyed some interesting results under the EDA name. But it felt weird for me to speculate on what scientist attitudes towards GSR were before I was born without \emph{any} evidence at all, so I did a cursory look at publication trends related to these terms, presented in Figure \ref{fig:fads} (A/B). 

I also provide in Figure \ref{fig:fads} (C/D) an example of a bubble in the financial markets that was full of misrepresentation at its peak, yet at the same time a truly important breakthrough was underlying it -- the dotcom era at the turn of the century, which included the founding of giants like Amazon. Obviously there are many defunct internet companies that were hyped in the late 90s as well, and more broadly there have been bubbles surrounding topics that never reemerged as internet tech did. The point here is not some sort of generalizable insight in itself, it is just to recognize that when a bubble bursts, it doesn't necessarily mean the underlying idea was bad. It just means somewhere along the way, something was systematically misrepresented or otherwise misunderstood about the value (or its timescale). Once the value of a thing becomes primarily driven by the group's perception of that value, there will predictably be a period of unstable dynamics, hardly connected to the true long term value at all. Depending on how such a state arose though, the fact it was possible might be useful signal on the topic.

While it is difficult for human-involved systems to avoid hype cycles entirely (and arguably they aren't entirely a bad thing either), it is probably advisable to avoid designing your system to literally encourage second order evaluations over first order ones. Unfortunately, the modern academic funding structures are extremely conducive to formation of scientific bubbles, because grantsmanship has become so uniform and so critical. In practice, biomedical project planning seems to start much too often with the question "what is most likely to be funded by the NIH?", which means that scientific decision making becomes based primarily on perception rather than fundamentals (or curiosity). Obviously communication is a critical component of science, and it's also true that having some synergy in the questions different groups work on can be highly valuable. But communication here should most often be framed as "how can I best communicate what I already have?", not "what should I work on so that I can communicate it best?". Furthermore, there are many ways that cohesion in research topics across certain groups could arise, and I'm hard-pressed to think of a mechanism that would be more phony than the current one.

\pagebreak

\begin{FPfigure}
\centering
\includegraphics[width=0.9\textwidth,keepaspectratio]{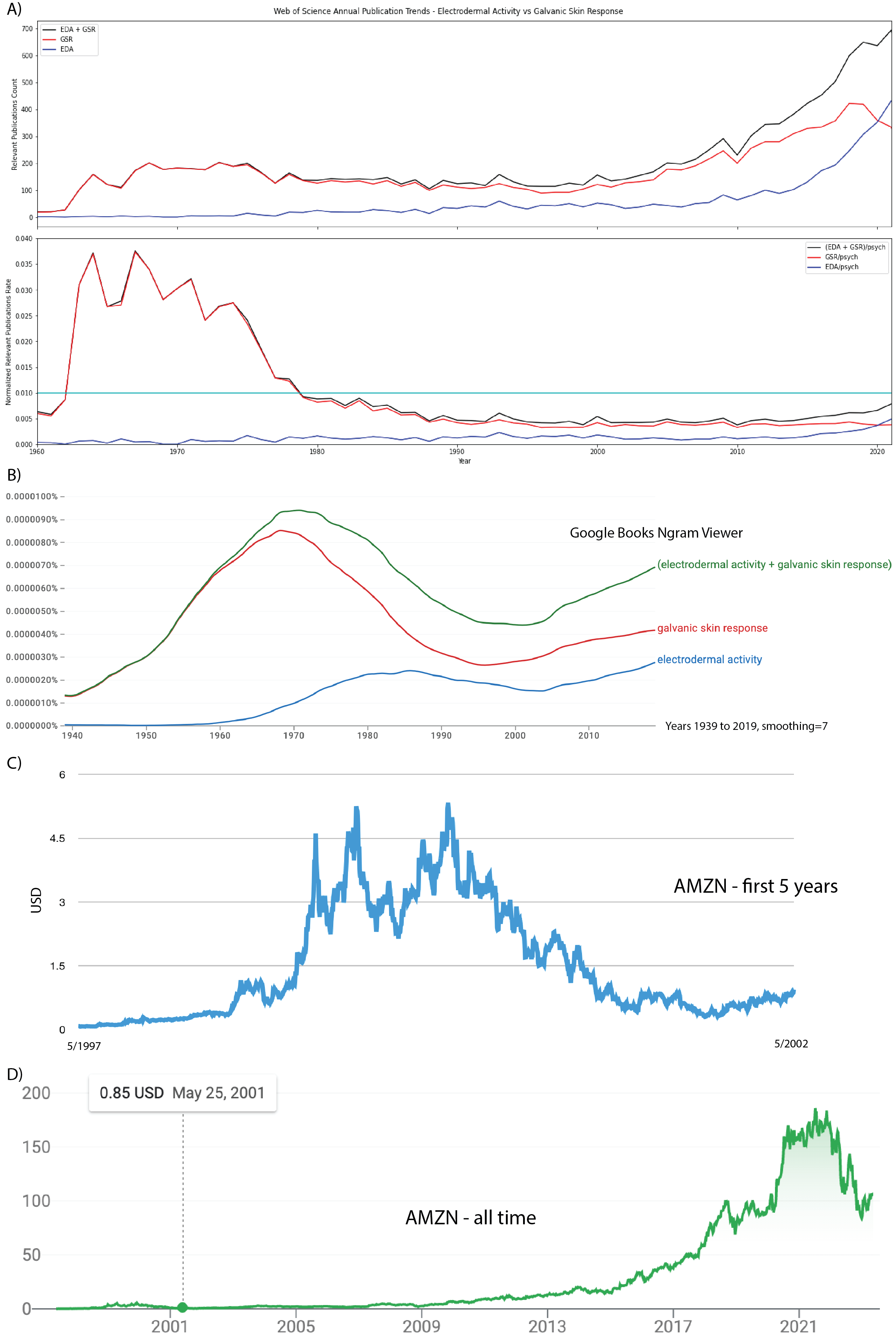}
\caption[Bubbles in academic science.]{\textbf{Bubbles in academic science.} \newline A) Web of Science (WoS) publication counts by year (1960-2021, queried in 10/2022) for articles relating to keywords "galvanic skin response" (red), "electrodermal activity" (blue), and "galvanic skin response" OR "electrodermal activity" (black). Both raw publication counts (top) and publication counts divided by each year's count for "psychology" (bottom) are depicted. The cyan line on the bottom plot marks the point where the number of records on WoS relating to the keywords is 1\% of the "psychology" records count. From this view, GSR research output per year stagnated sometime in the mid-1970s, and its relative popularity sharply decreased over the back half of the 70s, continuing a slow decline throughout the 80s. In the mid-2010s a notable uptick in EDA popularity began, with a further drop in GSR popularity approaching 2020. Interestingly, scientific articles using GSR vs EDA terminology are largely disjoint - In 2021 there were 334 records on WoS for "galvanic skin response", 432 records on WoS for "electrodermal activity", and 694 records covering either. Thus only 72 of these 694 mentioned both terms. It is not surprising that those shifting to EDA would largely not use the GSR term in publication, but I am surprised that there remain hundreds of papers annually mentioning GSR that do not use the term EDA at all. It does appear that researchers are beginning to switch terms, but it is concerning that research output might be needlessly siloed because of terminology changes that are reminiscent of the euphemism treadmill (in fact my search is entirely missing at least one other name, "skin conductance"). Regardless, it is encouraging that a new uptick in this line of work is emerging, and many promising open questions remain about phenomena observed in EDA signals, for example "sleep storms" \citep{Johnson2020}. \newline \newline B) A screenshot of the results for a similar keyword search from the public Google Books Ngram Viewer tool, taken from 1939 to 2019 (latest available date) with smoothing set to 7. Given that topics covered in a book may lag those in scientific publication outputs, and books are more likely to discuss a particular term in a historical or cultural context regardless of perceived scientific merit, this plot could provide a quite different perspective on the relative popularity of GSR and EDA terminology over time. However, similar overall trends emerge. There was a large peak in GSR interest in the late 1960s/early 1970s, yet by the 80s its popularity was quite diminished. More recently we see the beginnings of another uptick on the broader topic, with the EDA term in particular more popular now than it ever has been. \newline \newline C) A plot of the close price of Amazon's stock (ticker AMZN) per day over the first 5 years it was publicly traded (May 1997 - May 2002). Data exported from Yahoo Finance and plotted in Numbers. The dotcom bubble and its subsequent burst are apparent here. Buying AMZN relatively early in the surge and then selling at the peak would have $\sim 5x$ returns, while buying during the bubble might've left you with 20 cents on the dollar when marked to the end of this period. In hindsight of course, the most regrettable thing you could have possibly done during this time would be to sell AMZN because of the early 2000s dotcom crash. \newline \newline D) A screenshot of the Google Search Market Summary for AMZN over all time. The dynamics of (C) are an incredibly small blip here, and if sold at present AMZN bought in May 2001 would have more than $100x$ returns.}
\label{fig:fads}
\end{FPfigure}

\FloatBarrier

\section*{What is my actual thesis?}
\addcontentsline{toc}{section}{C.1 \hspace{0.2cm} What is my actual thesis?}
\label{sec:real-thesis}
There are many well-intentioned policies that have had counterintuitive results. Unfortunately, it is difficult to argue against these policies until far too late, because it is easy to conflate the intention and the policy. In the face of growing concerns about reproducibility, a call for fewer expectations of rigor sounds nonsensical. In the face of growing concerns about novelty, a call for fewer expectations of creativity sounds nonsensical. As more and more publications get churned out every year, it's tempting to argue for fewer papers with higher standards in both novelty and rigor -- the end goal of science after all. However, that perspective remains rooted in the assumptions of current scientific practice, a system that is unsustainable when open problems inevitably grow in complexity. \\

\noindent \textbf{The central hypothesis.} Scientific processes have become increasingly formalized. Scientific processes also remain extremely individualistic. Both of these factors could potentially explain many of the cited issues with modern academia, which I will discuss in the following sections; at the same time, neither is an inherently bad quality. What is truly pernicious though is the combination of these two properties in a system that needs to scale. Because much of academic research has formalized around the individual, or at least around the individual lab group, there is an expectation that every lab and even most single publications will present outputs of completed "science" per the formal guidance. Combination of the output scientific units across groups is thus expected to naturally fall in line towards a greater total scientific output. But this betrays either a naive belief that important open questions will forever remain addressable on a per lab basis, or a deep misunderstanding of how subsystems can affect the final outputs in a complicated \emph{system of systems (of systems ...)}. \\

\noindent Ultimately, formalizing an entire process on the level of the small subsystem is going to generate homogeneity, which makes a system susceptible to changing environment. Further, homogenization breeds more homogenization. As new generations telephone the best practices to each other, they stop being understood as valuable tools to be applied in the right situations, and they instead become definitional. As nearly every paper now follows the introduction/results/discussion format, and grant applications have come to expect quite specific proposal structures, the very nature of scientific thought has been increasingly constrained within those bounds. 

While these arguments may seem entirely tangential to the rest of the thesis, they're deeply relevant to my PhD experience. Not only are they drawn from observations I've had while trying to do science, they in fact draw in many ways directly from the technical ideas I've explored. A major theme of chapter \ref{ch:4} is how the properties of individual systems inform the behavior of their combination. A major theme of the rest of the thesis is the unprecedented types of work a psychiatry study could theoretically do with modern technology, and the necessity of not only interdisciplinary collaboration, but also a diversity of meta-scientific perspectives, to make that work succeed. Furthermore, prior results on human behavior are of course intertwined with what behavior one might expect to come from the incentives and pressures of the current scientific system. In that respect, improved behavioral characterization would be all the more relevant to improved modeling of \emph{hierarchies} of structures that will produce the best science. 

In parts of this concluding chapter, I will write a little about meta-science, about science history, about human operations research, about economic incentives, and so on. I am not officially trained in any of these topics; my beliefs started with personal observations and friends' anecdotes, they grew through introspection and analogies, and they only became moderately grounded as I searched for existing evidence after the fact. I'm biased on these topics, but I write this anyway because it would be wholly against the spirit of the outlined beliefs if I didn't. An individual bias is not inherently bad. Publishing a preliminary idea that turns out to be wrong is also not inherently bad. By trying to eliminate such instincts, the scientific literature has incentivized a duality of conservativeness and equivocation, thus creating the "paradox" that modern science moves too carefully yet simultaneously produces too many careless false results. 

It is of the utmost importance that multiple perspectives exist in science on a meta level, while instead we hardly learn or discuss anything to do with meta-science when "training" to be a "scientist". The exact same form of overarching problem exists here too, in that PhD and subsequent stages of academic training are oddly homogenized, and labor sources outside the typical career lotto are oddly rare. It isn't unpopular to complain about that topic, but almost any concrete suggestion seems to get some subset riled up. So all I'd really like to argue here is that there is absolutely 0 fundamental reason for nearly every department across every institution to implement a similar $X$. Whatever $X$ is, stop suggesting everybody do it, because that is inherently suboptimal for the long term future of science and education. I call this my actual thesis not because it is my main scientific output, or even a scientific output at all. I call it so because it is the closest thing to what I would like my thesis to be - a clean overarching hypothesis that I argue can cohesively explain many downstream phenomena of interest. \\

\noindent To be clear, there are still a number of existing scientific paths that I'm optimistic about, and a number of established scientists I greatly respect, some of whom do closely follow the playbook I'm criticizing. They run the playbook really well though, and they focus on problems that are well suited for it. I don't want this model to be discarded, rather I want other models to coexist. Towards that end, I will make the following arguments throughout the upcoming sections: 

\begin{itemize}
\item "\nameref{sec:the-problems}" presents prior research on the worsening novelty and reproducibility issues across many scientific fields, and their relationship with a possible relative increase in open problem complexity. Furthermore, I discuss how even the mere belief that these problems exist could manifest them in the next generation of scientists. I then argue that the explanations the cited authors propose are themselves symptoms of my hypothesis, and include related evidence about trends in publication language, the job market, and so on. From this, I also build an argument that simply slowing publication speed and increasing per publication demands within an otherwise stagnant framework will largely not solve the identified problems, nor would throwing money at an otherwise stagnant NIH.
\item "\nameref{sec:solutions}" presents arguments that methodological diversity is an increasingly critical need area in academic science, from labor sources to funding mechanisms to publication credit systems to institutional philosophies. This includes additional discussion of the following themes, all of which relate to natural consequences of the current system as I've described it:
\begin{itemize}
    \item The effects of evaluation metric diversity on long term outcomes, both direct and indirect.
    \item Job scarcity doesn't only add noise to the selection process, it fundamentally changes behavior.
    \item How should a scientific unit be defined? With better (or really any) categorization schemes, more things could be one.
    \item Formal criticism - and I don't just mean debunking of blatantly bad work - as negative feedback connections between scientific "modules". 
    \item The shortcut of evaluating work based on its fit to acceptable research templates, and how this exacerbates the harm that fear of judgement can have on scientific creativity.   
    \item Even most proponents of publishing negative results undervalue negative results, with connections to the paucity of high dimensional language in scientific planning.
   \item The NIH's misunderstanding of existential risk, including the correlated tail risk inherent in the homogeneity of popular "low risk" study methods like the highly controlled mouse model.
\end{itemize}
\item "\nameref{sec:conclusions}" summarizes takeaways and some broad remaining questions on these topics, and closes with some perspectives on writing the PhD thesis.
\end{itemize}

\section*{Recent accounts of problems in science}
\addcontentsline{toc}{section}{C.2 \hspace{0.2cm} Recent accounts of problems in science}
\label{sec:the-problems}
Over the last couple of decades, a number of alarming results have come out about the reproducibility of and creativity in science. These concerns appear to be gaining steam within community perception at a remarkable rate lately, and have now become intertwined with growing concerns about a depleted academic science labor force. I was genuinely surprised by some of the articles and quotes I came across while looking into this topic more formally, and so here I will highlight a few issues of note and then discuss why many of the proposed solutions for such problems cannot fundamentally fix the larger underlying flaw.  

\subsection*{Reproducibility crisis, and the NIH}
\addcontentsline{toc}{section}{\hspace{0.8cm} C.2.1 \hspace{0.2cm} Reproducibility crisis, and the NIH}
Many papers have reproducibility issues for non-malicious reasons. Some have reproducibility issues for absolutely wild reasons, like the work by \cite{choi} that found maternal immune activation models of ASD did not replicate between genetically identical mouse lines from two different suppliers, because of differences in the maternal gut microbiota. That style of reproducibility issue is likely far more widespread than anyone would like to admit, and the discouragement of curiosity by the formal academic system has played a part. Not only is there less ability to test random things for the sake of it these days (especially to probe one's own results, which no one wants to do \emph{too} much), but also the studies themselves are so tightly controlled that many real but highly narrow results can be found; it is unsurprising then that those results could be perturbed by unaccounted for explanations just as well as they could fail to translate to a more diverse mouse population. When these observations are followed up on, like by \cite{choi}, it is a strong scientific contribution. But the real question is: how often are these discrepancies even noticed? Let alone looked into? 

On the complete opposite end of the spectrum, we have the recent controversy surrounding fabricated Alzheimer's research results. Those results undoubtedly lead the field astray, and while I recall skepticism about the role of amyloid beta going back to high school, it is clear that funding for its research had continued to be plentiful over the last decade. The problematic paper from 2006 was cited over 2300 times, and a detailed investigation by Science has supported the evidence that images were likely fabricated \citep{ab-contro}. This is of course bad, and I was originally going to cite it in a passing sentence about reproducibility. But I was blown away by this quote from the \cite{ab-contro} article:

\begin{quote}
    [Lesne] became a leader of UMN’s neuroscience graduate program in 2020, and in May 2022, 4 months after Schrag delivered his concerns to NIH, Lesne received a coveted R01 grant from the agency, with up to 5 years of support. The NIH program officer for the grant, Austin Yang - a co-author on the 2006 Nature paper - declined to comment.
\end{quote}

\noindent Lesne is of course the lead author of the problematic paper in question, and Schrag is the whistleblower. 

R01 success rates are around 20\%, and the NIH will find many dumb ways to reject people for these grants. Yet they can't be bothered to double check legitimate fraud claims, and apparently didn't find it problematic at all that a co-author of the paper in question was the program officer. I can't say I'm too surprised that this was possible, but I was stunned by the fact that it has been a relatively neglected part of the story -- at least I've been around dozens of conversations about the controversy and seen a number of Tweet threads and article headlines, and I had no idea. Of course, it is much more convenient to get angry at a single bad actor than to call attention to the massive problems with the NIH. I found this to be a trend in my reading, that whatever issue was front and center felt less glaring than the implications the article had about the negative impacts of the NIH. For example, the following quote from an interview conducted by \cite{china}:

\begin{quote}
    “If NIH says there’s a conflict, then there’s a conflict, because NIH is always right,” says David Brenner, who was vice chancellor for health sciences at the University of California, San Diego (UCSD), in November 2018 when the institution received a letter from Lauer asking it to investigate five medical school faculty members, all born in China. 
\end{quote} 

\noindent What happened to the researchers in question was very unfortunate, but the underlying statement is seriously alarming to the integrity of much of biomedical research. That NIH funding has created a situation with such unilateral power over university departments is deeply concerning. I very much knew that there were perverse incentives at play in the grant review process, but I primarily imagined them as diffuse unintended consequences. It seems increasingly worth questioning that assumption, and by consequence, further increasing skepticism towards the common arguments suggesting we need more money to go to the NIH to resolve these scientific crises. 

In any event, reproducibility has been a problem to at least some extent in many sciences, not just biology. Nature surveyed 1576 researchers from a diversity of fields about replications \citep{repro-survey}, and more than 70\% of them reported failing to reproduce another scientist's results. More than half reported failing to reproduce their own results, yet less than 15\% had published a negative results replication and less than 20\% had ever been contacted by another researcher with questions about reproducing any of their results. So not only is it common to have replication issues, but we don't talk about them, leaving others to keep building on potentially weak work. It's a bit reminiscent of grade inflation -- instead of being accepting of mediocre grades, we've decided to pretend most people deserve A's most of the time, which has turned C's into marks of shame, which now makes it hard to go back to the old ways because you're kind of an ass if you give the 50th percentile a C in 2023. 

And yes, official replication experiments have repeatedly dug up disappointing numbers, even when focused on highly cited works or higher impact journals. Of 100 psychology papers tested by \cite{psych-osc}, 97 reported significant results yet only 36 had significant results in the repeated study. An effort by \cite{repro} to replicate 193 experiments from 53 impactful cancer biology papers (from 2010-2012) resulted in only 50 experiments from 23 papers actually occurring. No, that wasn't the reproducibility rate, that was the fraction of papers they were even able to \emph{attempt} to replicate. All papers had missing protocol details and the majority required reagent requests from the original authors. I can't imagine there was a bias in who responded... Of the experiments they did test, replication effect sizes were an incredible 85\% smaller on average, and only 40\% of positive results replicated with significance. In the papers with cooperative authors!

As mentioned, there are many real reasons that a result would not reproduce when run by a different group in a different location. For example, the sheer amount of noise in psychiatric disease diagnosis could easily introduce a bias whereby different demographic groups and recruitment locations with technically the same diagnosis are meaningfully different clinically. I would guess that the majority of the basically useless papers are not basically useless because there was anything incorrect about what the authors did. They're basically useless because of the system, with its ridiculously misaligned expectations and its ultimately homogeneous publication structure. Incidentally, a great transition to the creativity crisis.

\subsection*{Creativity crisis, and modern scientific writing}
\addcontentsline{toc}{section}{\hspace{0.8cm} C.2.2 \hspace{0.2cm} Creativity crisis, and modern scientific writing}
Accurately quantifying this is arguably impossible, but by a broad swath of metrics, it is a problem. I was also happy (in a bittersweet way) to find many notable old scientists complaining about the way the system has changed to strip away opportunities for creative contribution. I particularly appreciated this work by \cite{Hubel}, in part because of how happy HMS is to claim Hubel and Wiesel. Rightfully so, but I'd love to see what they would say about the current Program in Neuroscience. I can already get a pretty good sense:

\begin{quote}
    In thinking about all this I have been struck by the radical nature of the changes in science styles — at least biological science — over the last few decades. My own career provides an example of the old style, and I suspect that my experience is far from unique.

    ...

    I have no illusions that biological science is likely to return overnight to the system that prevailed a generation ago, but I believe a start could be made in that direction. If I were 40 years younger and a group leader and found myself imprisoned in an office most of the time, I would adopt a 5-year plan to change my scientific style. I would choose a project to share with one partner, put aside a lab bench that I could call my own, and submit a research proposal to fund that project. I would encourage any postdocs in my lab to do the same, with their own funding and independent projects. I would give advice gently and sparingly, realizing that strongly worded advice from a senior person can be hard to ignore and that in science making one's own mistakes can be an important part of learning.
\end{quote}

\noindent Unfortunately, I think \cite{Hubel} would be disappointed to see how things have further progressed in the last $\sim 15$ years. I know students who quite literally received daily experiment instructions. Fortunately my experience was not anything like that, but I hardly stepped foot on the HMS campus after year 1, intentionally so.

There is no shortage of opinion articles and books expressing concern about an increasingly incremental nature of science, often reflecting on ways academia has changed for the worse in the authors' experiences \citep{selfies,book1,book2,Alberts2012}. There has also been a rapid increase in (still largely anecdotal but very convincing) reports about an exodus of labor and talent from academic science \citep{postdoc1,postdoc2,postdoc3,postdoc4}. Given the NIH recently held a request for information from postdocs about this exact issue, there is clearly some truth to the fears. Not only are people leaving, but I have to imagine creative people are leaving \emph{and} rigorous people are leaving, because if you're not willing to compromise on either of those principles and you aren't Einstein, you are basically playing the lottery with your career as a present-day trainee. 

I say that not because I know with 100\% certainty that it is true, but because I see on the ground that this is the perspective many trainees have. There is time pressure to rush out incremental-but-spinnable and sloppy-but-by-the-book results. That part does appear widely acknowledged and theoretically addressable, though I have my doubts. What I suspect gets left behind regardless though, without more radical paradigm shifts that seem much less common to be proposed, are those with unevenly distributed abilities. If humanity could only have a single scientist you may not want these people, but in an effective large system you very much want some of these people. In present day science, where exactly does someone who is amazing at carefully addressing a specific question but who has literally no scientific vision fit? Where does someone who has great gut instincts or truly creative ideas but minimal ability to carefully carry out a study fit? How do we manage interdisciplinary skillsets? I've lost count of the peer reviews I've seen devolve into an argument about what novelty really is or an argument about why something wasn't shown 3 other ways with an order of magnitude larger sample size. Which is all quite obviously a facade, based on the practical results that have been obtained.

Speaking of which, a relevant quote from Nobel Prize winner Sydney Brenner \citep{peer}:
\begin{quote}
    I don’t believe in peer review because I think it’s very distorted and as I’ve said, it’s simply a regression to the mean. I think peer review is hindering science. In fact, I think it has become a completely corrupt system. 
\end{quote}
\noindent It is nice to see people that have "made it" actually make bold claims like this, which was supposed to be the point of tenure I thought. Nevertheless, peer review is not an inherently bad thing, but it is implemented so incredibly poorly it's comical. A cumulative 100 million hours of global scientist labor going for free to peer review, every year \citep{Aczel2021}. And not just free in the financial sense, but free in the academic credit sense too, which is what makes this such a clear oversight in the non-scalable design of present science. Imagine having no prior knowledge of the system and hearing that this is how it worked! 

Double blind peer review simply shouldn't be the universal way to publish in my opinion, which would combat some of its downsides not only through use of the alternatives, but also through the lightened system burden that would result if people took alternative forms of publication seriously. It's interesting that there seems to be a fear about further exacerbating bad quality science without peer review, yet fake papers make it through peer review all the time, and even the most prestigious journals have let through fundamentally flawed or partially fabricated works. Formal experiments have been run to show such issues too: submitting intentionally erroneous papers to BMJ reviewers resulted in less than 30\% of the problems actually being caught, and this included some glaring mistakes central to the supposed paper's claims \citep{Schroter2008}.

More importantly to me, peer review has unfortunately homogenized all of academic writing, which in turn I believe has contributed to the homogenization of scientific thinking. There is no inherent reason for peer review to require a specific article format, but the rise of the Introduction/Method/Results/Discussion (IMRaD) template coincides pretty well with the rise of peer review. \cite{IMRAD} studied papers across BMJ, Lancet, NEJM, and JAMA over a 50 year period and found that there were 0 papers following this format in 1935, and by 1985 100\% of the papers across these journals were IMRaD. Incidentally, IMRaD numbers were under 50\% across all of these journals at the time Hubel and Wiesel were doing their major work. Again, the problem is not with IMRaD itself but with its ridiculous ubiquity. Not all of science best fits into that framework. It encourages separation of arguments across sections, which disconnects both prior literature and post-hoc scientific reflection from the paper's contributions. It also encourages the use of brief jargon-filled descriptions, because in the part where you could carefully explain your results, you are instead supposed to state them as dryly and to the point as possible, avoiding invoking others' works. There is a time and a place, but emerging interdisciplinary fields are not that time/place -- it is far too easy to obfuscate flaws in one discipline when propagating works around the other discipline with typical IMRaD papers. 

The language actually used by papers to describe their contributions has even changed over the years, with increasing focus on improvement/optimization-related terms and less on discovery/creation-related terms \citep{Editor}. Some suggest that overpublishing is inherently to blame for the perceived drop in creativity. Indeed, publishing practices have gotten out of hand in many ways -- \cite{crazy} found more than 9000 different authors who published $>72$ full papers in a single calendar year (since 2000). Still, this assumes that papers should be complete scientific units, which is not necessarily how a creative process works. The judgement inherent in this assumption that publications need to be reduced to increase ultimate scientific quality could very easily be one of the factors that contributes to suppression of deviating ideas. We should care less about the quality of individual papers and more about the quality of scientific outcomes. 

So why not publish slowly then? Because feedback can be valuable, letting others build on your ideas can be valuable, and practically speaking, we do still need some input on which to evaluate people. We can apply a wide variety of evaluation functions to that input, but without any input at all we'd have a big problem with the tenure track scarcity as it is. A purely holistic system will be challenging to scale and open to all kinds of bias/nepotism issues. A subset of the system being purely holistic, sure - but the majority is still going to require some output(s) to be judged. It simply need not be a packaged "conclusion" as we try to enforce now. Ideally, "publication" wouldn't have a single clear meaning anymore.

\subsection*{Is productivity unnecessarily dropping, or is science just getting harder?}
\addcontentsline{toc}{section}{\hspace{0.8cm} C.2.3 \hspace{0.2cm} Is productivity unnecessarily dropping, or is science just getting harder?}
 A recent piece of evidence about the creativity crisis that made the rounds was the analysis by \cite{Park2023} of citation trends across 45 million papers. They found an enormous decrease in their disruption index since 1945, regardless of scientific field. They defined breakthrough papers as those that were frequently cited by future works, and where said future works did not cite the predecessors of (i.e. references in) the breakthrough paper. Such disruptive publications have become far less frequent, which the authors hypothesized was largely due to narrowing of research fields. However, as publication practices in general have changed substantially over the years, it is difficult to read into this too much. On the other hand, when coupled with the many other hints of declining creativity, eyebrows remain raised.
 
 While it may not be a direct indicator of research creativity (and in fact may suffer in part from rigor issues as well), there have been very clear signs of declining research productivity. Pharmaceutical output has been declining since the mid-1990s in spite of increasing research expenditures \citep{pharma}. Notably, neuroscience has fared particularly poorly in pharma trials, resulting in many companies cutting neuro branches -- including AstraZeneca, Bristol-Myers, GSK, Pfizer, and Amgen \citep{pharma-n}. Many early psychiatric drugs were discovered sort of by accident, but somehow things are less productive now than they were then.
 
 Of course, as more discoveries are made it can become increasingly difficult to address the remaining open questions. I find it hard to believe we are entirely there yet given the low hanging fruit that are still largely untouched in digital medicine, but it is certainly something that should be thought about when future funding and research group structures are designed, so that they may scale as it becomes increasingly necessary. Additionally, this phenomenon probably does already have some real effect: on the cost and the labor required for the average research advance, which has gone up across disciplines. \cite{stanford} found that "the number of researchers required today to achieve the famous doubling of computer chip density is more than 18 times larger than the number required in the early 1970s". \cite{Jones2009} found through patent data that the age of first time inventors has gone up, the scope of tech covered from an individual inventor has gone down, and teamwork between inventors has gone up over time.
 
 These trends would be consistent with the behavior needed to combat increasing complexity of open questions and an increasing body of prior literature to be learned. The problem in academic science is how poorly these exact changes go in the system as designed, coupled with the capacity limits of the system's intended (who knows why) labor structure. Taking a longer amount of time before your first patent is the equivalent of a never-ending postdoc, except the postdoc is a much bigger ask. Narrowing individual scope, longer project timelines, and increased project costs have resulted in decreased intellectual freedom for PhD students and postdocs, and added an unfortunate non-linearity into the career pipeline where laborer trainees are now supposed to magically transform into visionary mentors. Teamwork on patents is fine because you can come up with an agreeable profit split, but there is no ideal way to split academic credit in the current publication model, so teamwork occurs in a highly suboptimal way in science; which in turn further exacerbates the lack of intellectual freedom, because one is better off plugging away at a single incremental solo project than trying to be creative on a smaller portion of a bigger project. Regardless, many PIs don't give trainees the choice.  

Many have argued that problems should not actually be harder relative to where we are because of the advances we have to build on \citep{selfies,Park2023} or that problems are primarily harder because they are more expensive and funding increases haven't kept pace \citep{Alberts2012,cap}. Quite frankly though, it is functionally not that important whether science has gotten worse or problems have gotten harder or some mix of both. It's a silly chicken/egg argument that distracts from the real issue, which is that science is currently very poorly set up to handle increasing relative problem complexity, to whatever extent it is already here. Those systemic issues would continue to exist irrespective of increased funding as well, and it is even arguable that the massive rate of increase in NIH funding between 1998 and 2003 exacerbated some of these issues by bringing more trainees into an unsustainable labor structure without any plan for long-term change. I sincerely hope that the NIH does not receive large additional funding increases until those increases are going to meaningfully different funding methodologies.

\section*{There is no solution, but there may be solutions}
\label{sec:solutions}
\addcontentsline{toc}{section}{C.3 \hspace{0.2cm} There is no solution, but there may be solutions}
I've noticed most perspective pieces tend to suggest a particular solution, or to argue for something vague and generally agreeable like treating postdocs as human beings. I've found very little that sounded radical to me though, and I've also found very little that acknowledged the importance of intellectual diversity. If the latter is supposed to be implicit, it is extremely unclear. 

I was required to take a science ethics course this year, which involved a number of peer group discussions, and I was surprised at how many topics had fundamental opposition from one "side" or the other. I would love nothing more than to see science involve more genuinely curious screwing around, but I recognize that practically there is great value in formats like the pre-registered study. Suggesting that basic biology ought to have more pre-registered studies as part of the funding pool received immediate counterarguments about how bad creativity already is as it stands. But pre-registered studies could in fact take some pressure off of creative studies, because with a proper mechanism for verifying interesting observations downstream, there would be less need to put on a rigorous mask. But for some reason there is a pervasive underlying assumption of system-wide homogeneity being indefinitely preserved. Ironically, the exact issue of incrementalism is being meta-applied to how we think about adjustments to science.

The same goes for the debate about whether PhD students and postdocs are trainees or laborers. How the hell does anyone expect science to function if the answer to that question has to be only one or the other? If it is one or the other, then either we have near-zero labor or we have no one being trained for the next generation of science. I recognize that the awkward mix of both that everyone is apparently expected to be has not turned out well, but that is exactly why there must be more diversity in scientific positions. There is no reason that all PhD students need to be expected to go into academia. Even of those expected to go into academia, there could be massive differences in training depending on goals to be hyper-focused versus to be interdisciplinary, to be a manager PI versus to be a bench PI like Hubel wished for. Why not diversify the degree programs and the funding structures offered for them? Ideally in tandem with diversifying what end academic goals look like, which is the part that probably requires changes originating from higher up the power hierarchy. Though individual institutions, when taken in aggregate, could certainly have a hand if they stopped trying to copy each other so much. 

More importantly, there is no reason that all labor has to come from the same career ladder. Professional baseball is the only other legitimate career I can think of that functions this way. I don't understand why introducing outside labor sounds radical (besides the high upstart cost I suppose), because without prior knowledge of the academic system, the current state of things is what would be thought of as radical. Large scale construction projects would never get safely finished if everyone on site was trying to compete to show that they should be an architect some day. The next time the government feels the need to do a jobs program, they should seriously consider funding labor for labs. There is absolutely no reason why a healthy person couldn't be trained to move mice around and pipette some things. 

Ultimately, diversifying labor will be critical to scaling up for larger biology research problems, but it won't solve the philosophical academic crisis in itself. That still requires quite a bit of rethinking how grants are awarded, how publications are done, and how intellectual ego stroking currency gets allocated.

\subsection*{Goodhart's Law}
\addcontentsline{toc}{section}{\hspace{0.8cm} C.3.1 \hspace{0.2cm} Goodhart's Law}
"When a measure becomes a target, it ceases to be a good measure" is a popular saying, and it obviously rings true in science. \cite{measurements} describes the metric-gaming treadmill from overvaluing of top journal publications alone to the introduction of individual citation tracking metrics that have simply created new system gaming behaviors. Furthermore, these behaviors have escalated to the point that it is increasingly difficult to avoid enabling them, even if one does not directly partake -- pilot data from studies of peer review suggest that outright refusing to add questionably relevant citations during a paper submission process can \emph{halve} your probability of acceptance \citep{phony-citations}.

Realistically, we need measurements though. A very obvious way to attenuate the impact of Goodhart's law is in fact through diversification. If science were more meaningfully different across groups within a field, there would be more meaningfully different metrics and less sense of direct competition. It would not be feasible to try to simultaneously game metrics that value e.g. major single paper robust contributions and e.g. spreading news about surprising observations. Maybe groups would start gaming a specific submetric, like getting superficially great at pre-registered studies, but the good thing about having a number of fairly independent measures is that they can be weighted in a variety of ways and combined linearly or non-linearly. From year to year and department to department, different things may be valued, and that could include qualitative components as well, or dark horse metrics like teaching reviews. 

At some point, one becomes better off just being genuine about trying to contribute their best to the scientific community. Might some get unfairly missed by chance in that system? Sure, but that pretty obviously happens already. With a high enough degree of selectivity, it becomes literally impossible to avoid false negatives. It remains critical though to prevent the scarcity from impacting people's behaviors in bad ways, as it seems to now. Similarly, it is important that false negatives are distributed in as minimally a biased way as possible, otherwise the system on the whole suffers. A homogeneous selection scheme on a handful of gameable metrics will trivially produce entire classes of false negatives. 

\subsection*{Alternative funding mechanisms and higher-dimensional language}
\addcontentsline{toc}{section}{\hspace{0.8cm} C.3.2 \hspace{0.2cm} Alternative funding mechanisms and higher-dimensional language}
As mentioned, pre-registered studies are one good alternative subformat that could help to solidify suspected hypotheses in a much more rigorous (and exhaustive for practical applications) way. Replication studies, especially focused on characterizing broader generalization, are a closely linked rigor-related subformat. A subformat that connects back to concerns about a lack of curiosity would be the question-driven approach described by \cite{Glass}. Instead of posing a hypothesis to test, grants could pose questions that do not have a specific desirable answer, but instead are to be carefully examined from all angles. In my opinion, this would ideally involve some degree of entirely discretionary funding that researchers could use to pursue curious follow-ups to whatever comes up during their questioning. 

Within existing hypothesis- and data-driven formats, there is room to diversify as well. Where can hypotheses be sourced from besides a narrow body of subfield prior literature? How can data-driven approaches find smaller sub-patterns instead of focusing on whole group effects only? On a meta level, how are these proposals evaluated and by whom? For funding that is allocated for "high risk/high reward" research, how is that actually being defined? 

To answer these questions, science needs to adopt more multidimensional language in its proposals/evaluations. It is well known that "risk" is not a single thing -- for example, the low default risk bonds bought by Silicon Valley Bank were very obviously risky on other dimensions. A project that answers a question where the answer is more likely to be no is not some massively risky undertaking; if the results are published, that can be a useful contribution to science given the question was considered so plausible in the first place, as most funded projects seem to be. There needs to be a better way to evaluate the cost/benefit tradeoffs of different project outcomes, so that they can be discussed along with estimated outcome probabilities. But regardless, that isn't the real risk the NIH should be worried about.

A frequent source of recent financial downfalls has been correlated "low risk" events. One might assume in a vacuum that certain risks are uncorrelated (because in normal times the values aren't much correlated), and make many large inadvertently intertwined "safe" bets. If there is a 1\% chance that a given thing will fail, and you need 100 of them to fail at once to go bankrupt, you'd think it would take a ridiculous black swan event to take you out. But if the tail risks are correlated, now there is a very real chance of catastrophic failure. Myron Scholes, a Nobel Prize winner and a namesake of the Black-Scholes equation for options pricing, was involved in such a catastrophic failure, with the firm Long Term Capital Management. So yes, this can happen to extremely smart people. 

While the NIH may not be able to implode like a hedge fund can, it certainly has been making a lot of intellectually homogeneous "low risk" bets for many years now. When the system stops working, it could really stop working. Of course, a way to counteract this would be to meaningfully diversify. To invest in projects whose primary upside is exactly what you would need in the case that those typical moves don't work out. This appears to me to be sorely lacking. 

A specific example is the serious rarity of mouse studies with a diverse population. What if a big part of the generalization to pharma problem is because we actually can't even generalize to other mice? We won't know, because it is hardly done. It would be too risky to eschew tight controls, and there isn't such an extensive existing biotechnology base to draw tools from. But picture doing all human subjects research by cloning a handful of people and raising them in cages. Would that generalize? Probably not. This is a massively correlated risk across so many of the "low risk" neurobiology projects.

That example connects back to the lack of multidimensional language as well. How are we searching a high dimensional space here? It seems by taking little steps along some existing vectors, or trying to average over the whole damn thing. More diversity in project directions could be framed in terms of how different research directions advance the search in such a multidimensional space. This could also facilitate more tool-building projects insofar as they propose improvements to the search process itself. Software to help synthesize prior works from across groups will be very important as problems do get more complicated and more interdisciplinary. Cross-checking of results against other expectations could additionally help in preventing overfitting with lower sample sizes. Neural networks have managed to defy expectations with their ability to generalize despite such a large number of parameters, so it is perhaps not entirely necessary to be "by the book" in the typical way all of the time. Again, it is really most critical that methods be diversified and hopefully eventually synergized.

Cookbook following becomes especially bad when training paradigms start simply passing on the cookbook instead of the fundamentals that lead to it. What is even worse though is the backlash (or at least perception of backlash) in questioning it. What I would define as risky research from the perspective of a lab group, is a proposal that makes most people think \emph{what the fuck}. Does that happen anymore? Looking for digital biomarkers of psychosis might be kind of a moonshot at present, but it is a highly plausible idea, not one I think anyone would be embarrassed about trying decades from now. What you don't see anymore is stuff like that time Alan Turing included a section on extra sensory perception in "Computing Machinery and Intelligence". I suppose if you believed in ESP it would be important to clarify how it could affect the Turing test. But consider in recent times trying to publish an academic article with a section about how the presence of the Holy Spirit might've affected your results (the scent of experimenters can). No way that would get past peer review, but also no way anyone would have the guts to do that, even if they did believe it. Obviously that in itself would not further science, but the ability to plausibly do it seems quite important to breakthrough developments, as well as to combating the irrational hype/stigma cycle that seems especially prevalent in psychiatry over the decades \citep{dsm3}. 

Relatedly, a natural counter to bad arguments that doesn't require overzealous peer review nor conservative funding agencies would be a legal checks and balances style of approach. As mentioned, it is rare for people to publish negative replications they accidentally happen across, so clearly it is extremely rare for people to go out of their way to try to invalidate someone else. Individuals are too attached to their individual results, but with a more collaborative credit assignment system it would perhaps seem less brutal to do this. It would make it less bad when individual groups are biased towards a hypothesis, because there would no longer need to be a facade of objectivity. As long as some ground rules for serious misconduct were established so that outputs of the "nodes" were known to be honest in what they did, an effective negative feedback system could form, as yet another subcomponent of science. 

\section*{Final conclusions}
\label{sec:conclusions}
\addcontentsline{toc}{section}{C.4 \hspace{0.2cm} Final conclusions}
In sum, I'm pretty convinced that much of academic science is in the process of a slow motion train wreck, but not because any individual is doing anything incorrect (though a few are it's hardly the effect of interest). Instead, the problem is that the collective works simply don't fit together well enough anymore. Good faith efforts to formalize science have homogenized it far too much, with too much focus on individual lab groups and not enough consideration for the system-wide long term risks involved. These problems are far from irreversible, but I worry that most discourse I hear is not getting at what I feel to be the core issue: we need to reframe scientific optimization around outputs of the collective system, and accept the fact that in complicated connected systems there should be diverse - and maybe even counterintuitive - roles for individual nodes. 

Specifically, I would be very pleased to see any updates to diversify the ways that NIH distributes funding, whether that budget goes to reproducibility studies of the highest rigor or completely off the walls radical proposals or some other currently uncovered space. Ideally this diversification will happen more thoroughly, but you can only call for so much at once. Anything legitimately different would be a great first step. This could occur on the publication level as well, enabling legitimately different types of contributions to receive credit in an honest fashion. 

Additionally, there should be an adjustment to the language used when writing and evaluating scientific proposals, in order to acknowledge the high dimensional search space that we are ultimately trying to characterize. Different lines of questioning will cover that search space in different ways, which will hopefully prompt deeper consideration of data collection and analysis design decisions. These are intellectual tools that ought to be considered pieces of a puzzle that can be strategically chosen for a given goal -- just as technical tools can (and should) be. These are also far from easy considerations, and they are considerations that are entirely glossed over by most academic training. 

As such, it will be important for the research community to work together, in order to best collectively leverage the different methods being employed to further our understanding. This is a relevant role for the NIH too, and it is critical that deeper consideration of different types of risk occurs so that the larger research portfolio of a subfield can avoid massively correlated tail risks. If mouse models generalize poorly to the natural mouse population across a large set of neuroscientific questions, that would be a very bad NIH bust. Obviously there are great arguments for using these models for certain questions, and it makes perfect sense for individual researchers to want to do so. The problem, if one does turn out to exist, falls squarely on the funding agency shoulders, for allowing such a large amount of research to rely on a single assumption. Not only that, but they've simultaneously failed to fund many "high risk" projects that would likely be huge successes in the case that this is a failure mode. 

More broadly, I would hope that an emphasis on diversification and on more exact language for describing project costs/benefits could attenuate some of the extreme bubble-like behavior seen in academic funding and publishing trends. In my opinion, the scientific collective is remarkably good at throwing the baby out with the bathwater (see also Figure \ref{fig:fads} above); doubly so for a community that tries so hard to legislate away many other human cognitive biases. Improving the tendency for visceral reaction to anything so much as vaguely associated with prior pseudoscience could revive a number of older hypotheses of value, and it could prove important for the future of digital psychiatry, as the field certainly has the potential for hype and the potential to bust in the short term. It would be very disappointing if that resulted in a downturn in any such research, because there are a wide variety of directions that could be taken and that should not be overly associated with each other.

Concretely, I haven't answered at all how to best distribute methodological diversity over levels of such a complicated system. It's a question that would clearly require some trial and error, but perhaps results from evolutionary biology and engineering systems design could be informative for foundational ideas. More knowledge of economics can't hurt either. It seems that meta-scientist should be a more respected role, that even if somewhat ignored on the university level, is a key part of overseeing funding agencies like the NIH. It is sensible that biologists would be involved too of course, but realistically the training to be a biologist just doesn't address these topics, at least not nowadays. Yet such topics are an important component of effective field-wide decision making.   
 
\subsection*{On writing the thesis}
\addcontentsline{toc}{section}{\hspace{0.8cm} C.4.1 \hspace{0.2cm} On writing the thesis}
 Despite my tone here, I've had a good experience overall doing a PhD. Obviously, many of these issues did not apply to me - the sheer fact I was able to spend time writing on such a broad range of topics and that nobody was looking over my shoulder as I repeatedly criticized e.g. the dysfunctional structure of AMPSCZ \emph{is} a testament to that. Still, I've observed enough troubling trends across multiple fields from afar that it would be very difficult to not be exceedingly jaded. 
 
 Even MIT, which I have always thought of as a little bit different, has unfortunately accelerated its move towards being more like "peer institutions" in the last couple of years. Historically beloved classes with high enrollment and good course evaluations are now being permanently cancelled because they don't fit the academic agenda of a few individuals with concentrated power. Any intellectual uniqueness has been literally dying off. I never liked the implication that PhD students are not students, and should just be working hard on some single-minded research problem from day 1. But I didn't imagine that attitude accelerating to the point that now we can't have undergraduate classes that cover old school AI topics, because apparently anything that can't immediately be regurgitated into a modern research project is a waste of time, including for 19 year olds. I know that sounds ridiculous enough that you'd think I'm exaggerating, but I'm not.
 
 In any event, I wanted to put in a good word for writing a thesis like this one. I realize it is ridiculously long, and in large part not going to be read. But that isn't the point of doing it, and I wish I'd started writing some of this sooner. I'm definitely glad I did not just staple papers together (not that I totally could have). Some benefits that came out of sitting down and writing this thing out:
 \begin{itemize}
     \item Without a constraint on the chapter background section lengths, I read a lot more interesting slightly tangential stuff than I ever would have otherwise, and I got a chance to really think through my arguments from multiple angles (in a way that is not at all encouraged by the IMRaD format). I always sort of preferred the audio journals just because the software part was more tractable, but I feel if I realized sooner how underresearched they are and how many problems with interviews they could address, I would have pivoted my focus there way sooner and perhaps accomplished much more.
     \item Having an active thesis document and knowing I could always put additional arguments in my thesis made me way more inclined to actively look into and document concerns I had with some of the pilot AMPSCZ data collection. If I hadn't been writing so much to begin with, I doubt I would've made such a point to argue about the audio journals, and we'd probably be wasting money on literal half-finished stunted clinical interview transcripts for the duration of the project.
     \item Taking the time to write additional introductory material for different audiences for chapter \ref{ch:4} was actually quite helpful in finding new analogies to motivate future work on related projects. In particular, convolutional neural networks are just feedforward neural networks with some neuro-inspired constraints enforced on the weight structure. A fully connected network could theoretically learn to be a convolutional network but in practice that is extremely unlikely (and less efficient). There has not really been a parallel development for recurrent neural networks, but I feel unique multi-area constraints could one day revive the practical popularity of RNNs through a similar mechanism.
     \item Finally, formalizing some of my existential concerns about academic science was helpful in removing any doubt about my next steps. If I'd stapled together hypothetical papers, maybe in 2 years I'd think about how I left things unfinished or how perhaps an unstable postdoc wouldn't be so bad. Instead, I'm more concerned about scientific outlook, but very confident in myself. 
 \end{itemize}
 \noindent So, I apologize for the long document, but if any student asks I will definitely recommend doing something similar. The process of writing it was highly relevant to my learning, and it's kind of a shame that apparently this is rare to do.




\begin{singlespacing}
  \renewcommand{\bibname}{References}

  \bibliographystyle{plainnat}
  \bibliography{references}
\end{singlespacing}

\begin{appendices}

\chapter{Supplement to Chapter \ref{ch:1}}\label{cha:append-chapt-refch:1}
\renewcommand\thefigure{S1.\arabic{figure}}    
\setcounter{figure}{0}  
\renewcommand\thetable{S1.\arabic{table}}    
\setcounter{table}{0}  
\renewcommand\thesection{S1.\arabic{section}} 
\setcounter{section}{0}

\section{Additional results sections}

\subsection{Diary pipeline QC metrics validations}
\label{sec:more-qc}
To further characterize the pipeline's QC features, we both used manual quality review methods and looked at relationships between audio and transcript quality metrics. This supplementary section will detail the results of that work.

\subsubsection{Cross-checking with manual review}
A major part of our workflow in characterizing said dataset was careful manual review of a subset of audio diaries and, where available, matching transcripts. Both student volunteers and a lab RA conducted this process, cross-referencing with outputs from my pipeline where applicable. The manual review of quality of TranscribeMe's outputs that was described above is one such example. Now I will report the results of a few different approaches for manual assessment of audio quality done by the lab.

Because manual review looked at a few different subsets of diaries from BLS, I will first review the overall file counts. Of the 11324 BLS audio files, 11101 were deemed non-empty (i.e. volume not NaN). Of those 11101, intentionally forgiving pipeline cutoffs filtered out another 648 of extreme low quality. Thus 10453 files were sent to TranscribeMe over the duration of BLS. 10271 were able to be transcribed, while 182 were returned to us by the service as non-transcribable. 

Prior to most files being sent for transcription, a student reviewed a sample of audio recordings to determine what QC cutoffs we might use. The primary goal was to determine what we could filter out while restricting the false positive rate to near-zero (i.e. send liberally). Original cutoff proposals were volume of at least 30 db and length of at least 5 seconds. Thus from the batch of audio recordings available at the time, the student listened to 14 files with db $\leq 30$ and 53 files with duration $\leq 5$ seconds. Of the 14 with low volume, all were found to be nonsense recordings. However 31 of the 53 short files were found to contain meaningful speech, and of the 22 that did not all had volume below 40 db. 

The review was therefore expanded to include 108 files with db $\leq 40$. Of these 108, only 4 were found to contain any transcribable speech, and only 1 was deemed by the student to have no quality issue whatsoever. Based on these results, the QC cutoff used for the majority of transcription decisions became volume of at least 40 db but no duration minimum. \\

\noindent The next major round of manual QC review was to assess the 182 files that were rejected by TranscribeMe, to better characterize false negatives produced by the implemented $40$ db threshold. An RA listened to each audio that was returned to us without a transcript and took notes on the contents. They found that 177 of the files were indeed entirely untranscribable, while 1 was an acceptable recording and 4 were difficult but perhaps possible to at least partially transcribe. They also found that the majority (119) of the files were extremely short in duration ($<1$ second) and contained an unrecognizable sound, while most of the remaining files contained excessive background noise of some form (Table \ref{table:bls-qc-returned-transcripts}).

\begin{table}[!htbp]
\centering
\caption[Manual review of recordings with acceptable volume that were unable to be transcribed.]{\textbf{Manual review of recordings with acceptable volume that were unable to be transcribed.} The 182 files that passed a volume threshold of 40 db but were subsequently returned by TranscribeMe were reviewed and categorized by an RA. Counts for each quality problem category are provided here. 177 of these 182 were verified to be completely unsalvageable by the RA. The 119 of short length (row 1) could be easily filtered in the future.}
\label{table:bls-qc-returned-transcripts}

\begin{tabular}{ | m{8cm} | m{3cm} | }
\hline
\textbf{Category} & \textbf{File Count} \\
\hline\hline
Unrecognizable sound (duration $<1$ second) & 119 \\
\hline
Static noise & 2 \\ 
\hline
Ambient non-speech background noise (e.g. traffic) & 30 \\ 
\hline
Ambient speech background noise (e.g. TV/radio) & 11 \\
\hline
Poor enunciation or otherwise inaudible speech from participant & 15 \\
\hline
Presence of another human speaker & 5 \\ 
\hline
\end{tabular}
\end{table}

Therefore, we can generally trust TranscribeMe to return bad files that do slip through. Even if they did choose to partially transcribe a largely inaudible file and charge us for that, the total fee here would have been less than 40 dollars for the 182 bad uploads, and the transcript QC would have easily detected such problematic transcripts. Still, the automated QC cutoffs can be improved from this experience. We chose to not filter length because we decided even a single sentence could contain interesting information, and the cost is of course dependent on file length anyway. However, it is clear from this that sometimes when a participant hits stop immediately after hitting start there may be noise picked up that will cause the recording to pass the volume threshold. For the purposes of file organization, on both our end and TranscribeMe's, it makes sense to instead filter out any records of less than 1 second.

With an extremely short length cutoff added, just 62 files of the 11324 originally screened files would have been sent to TranscribeMe and subsequently deemed non-transcribable, while 990 low quality files would have been filtered out automatically before the TranscribeMe transfer -- indicating that our basic QC screening has been largely successful in identifying the problematic files. Furthermore, 10271 files were successfully transcribed of 11324 submissions, indicating that quality of recordings is much more often good than not. 

It is unclear if light weight QC metrics would be able to accurately detect the issues in the untranscribable files of longer duration (broadly identified in Table \ref{table:bls-qc-returned-transcripts}). With the metrics in the present pipeline, there are some correlations, but nothing strong enough to justify filtering entirely. For example, of the 62 files in the untranscribable set that had duration of at least 1 second, the mean volume was 57.44 db, which is significantly smaller than the mean of 66.82 db from across all files with defined volume (Table \ref{table:bls-qc-means}). There were a number of files with volume between 40 and 60 that were transcribable, so it is not necessarily recommended to filter these out; at the same time, it could be recommended to monitor for subjects that are regularly submitting journals in this range via DPDash, as well as for RAs to be on the lookout for and advise participants about the quality concerns listed in Table \ref{table:bls-qc-returned-transcripts}.

It is worth noting also that 110 of the 182 problematic recordings that were sent to TranscribeMe, including 22 of the 62 with druation greater than one second, came from the same subject ID (M8MXM). Another subject (GFNVM) submitted 10 of the 62 remaining low quality recordings, despite only accounting for 11 of the larger set of 182 (with a mean duration over the 11 of $\sim 30$ seconds). This suggests that participant GF was unlikely to be submitting bad audio intentionally, but still managed to account for a disproportionate chunk of the low quality files. Thus paying careful attention to subject ID is of clear relevance in managing data collection in an audio diary study. \\

\FloatBarrier

\subsubsection{Automatic methods for validating QC measures}
The other major part of our workflow to evaluate potential QC features was statistical analysis of the complete dataset. While manual review can produce a more nuanced characterization of particular diaries, it may be subject to human biases, and it would be a massive undertaking to review the entire corpus. Ultimately, one of the core goals of building this pipeline was to reduce manual label required from data collectors. To fully utilize the tool then, it is necessary to understand how generated features are distributed and what correlation structure exists amongst them. Furthermore, because we obtained a large number of high quality transcriptions via TranscribeMe's premium services, I was able to treat certain transcript quality metrics as a sort of "ground truth" about the audio's suitability for transcribing.

To provide initial context, I calculated summary stats for the distributions of the main QC measures in consideration: duration, volume, flatness, word count, and inaudible, questionable, and redacted counts (Table \ref{table:bls-qc-means}). Mean recording duration was $\sim 1 \frac{1}{2}$ minutes, but with high variance across BLS journals. Unsurprisingly, the related transcript word count feature also showed high variance, with a mean $\sim 180$ words. This is consistent with the variation in engagement seen across patients, as well as possible diminished engagement within some patients over time (discussed further in Chapter \ref{ch:3}). 

Many of the other QC features, such as spectral flatness and inaudible count, also had high variance, but additionally had low means - suggesting a majority of diaries that would safely pass a quality threshold, and a smaller set that produces a long feature tail. Indeed, a histogram of inaudible count distribution (Figure \ref{fig:diary-qc-dist-init}D) shows exactly this pattern. Because inaudibles occurred much more frequently than questionables in the TranscribeMe results and are more concretely defined, I utilized the inaudible count metric as the primary endpoint for measuring transcript quality in subsequent analyses. As short free-form patient journals are not likely to contain much PII, the reported redaction counts are in line with expectations, and thus not of interest for further study. See chapter \ref{ch:2} for discussion of PII in the context of long clinical interviews.  \\

I will now discuss the distributions of volume (Figure \ref{fig:diary-qc-dist-init}C) and duration (Figure \ref{fig:diary-qc-dist-init}B) in more detail. About 250 recordings had volume near 0 and could be directly discarded. Recordings with volume $< 30$ dB were otherwise very rare, as were recordings $> 80$ dB. Fewer than 1000 diaries were submitted with volume $< 55$ dB; while some of these may have been transcribable, they in general produced lower quality transcripts, as reported below. Thus the relatively small number of recordings $< 55$ dB in our dataset suggests that 55 could be a suitable dB threshold for future Beiwe studies.  

\begin{figure}[h]
\centering
\includegraphics[width=0.8\textwidth,keepaspectratio]{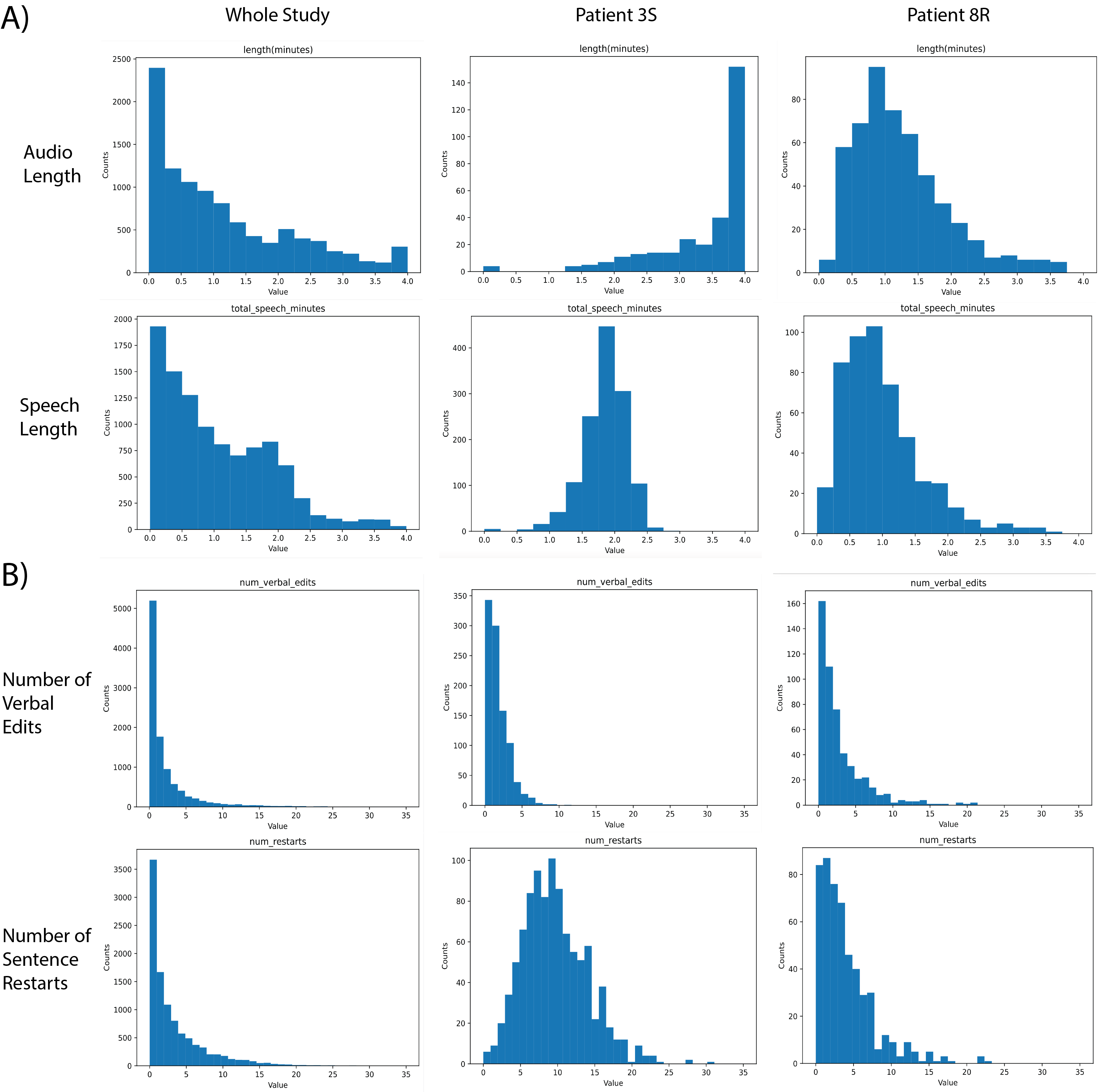}
\caption[\hspace{0.075cm} Distributions of a few diary features of clinical interest for the BLS study as a whole, compared to two patients of note.]{\textbf{Distributions of a few diary features of clinical interest for the BLS study as a whole, compared to two patients of note (3S and 8R).} Audio length and speech length distributions are compared to illustrate differences in patient diary recording times and pause times (A). The distribution of recording length across the study (left) is skewed towards shorter submissions, but both 3S (middle) and 8R (right) make longer submissions in general (top) - particularly 3S. However it is also clear from these distributions that 3S takes up much more time with pauses in speech than 8R does, as can be seen in the speech length histograms (bottom). Similarly, transcript-detected disfluency occurrences also vary by patient and by disfluency type (B). Patient 3S has a relatively low number of verbal edits (top) compared to diaries of a similar length in the larger study, yet they frequently engage in restarts (bottom).}
\label{fig:diary-dist-comp-init}
\end{figure}

Recording length is a quality metric where the distribution of specific patient's diaries can be plainly found making up part of the larger study distribution. Nearly 2500 submitted diaries had length $< 15$ seconds, and there is a clear drop-off in the histogram from that point up to 2 minutes. However a second hump can be seen for recordings between 2 minutes and 2 minutes 15 seconds, as well as a final bump at the maximum recording length of 4 minutes. The histograms of diary duration in particular patients of note (3SS93, 8RC89) make this even clearer, as the different participants have distributions that peak at clearly different lengths, from both each other and the larger study distribution (Figure \ref{fig:diary-dist-comp-init}A, top row).

It is worth noting that recording length need not correspond to the actual amount of content in the journal submission. One big confound, to be discussed in section \ref{subsubsec:diary-val-aud}, is the presence of pauses and the tendency of particular participants to pause longer than others. Indeed, the estimated distributions of speech duration for 3SS93 and 8RC89 (Figure \ref{fig:diary-dist-comp-init}A, bottom row) are vastly different when taken in the context of their respective submission duration distributions. In sum, 3SS93 submitted 3768 minutes of successfully transcribed audio, but registered only 1837 speech minutes when run through voice activity detection. 8RC89 submitted 662 total minutes, with 513 minutes registered as speech time. 

A different simple metric that does not run into this problem is the transcript word count. 3SS93 had a total of 328,040 words across their transcripts, while 8RC89 had a total of 110,679 words. Thus their ratios of word count, speech duration, and recording duration were $\sim 0.34$, $0.28$, and $0.18$, respectively. This lends credence to the pipeline pause detection algorithm, which will be further validated in subsequent sections. Of course word count is not a perfect metric, as simpler language can generate higher word counts for similar content, and pause time information may itself be clinically meaningful. Moreover, recording duration is much easier to screen than word count as it does not rely on transcript availability. \\

\begin{figure}[h]
\centering
\includegraphics[width=0.8\textwidth,keepaspectratio]{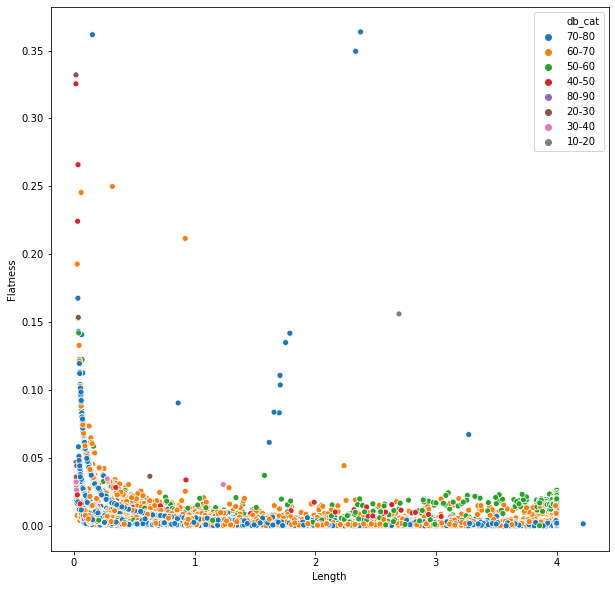}
\caption[Relationship between audio diary length, spectral flatness, and volume in the BLS dataset.]{\textbf{Relationship between audio diary length, spectral flatness, and volume in the BLS dataset.} All audio recordings of at least 1 second submitted by BLS participants, whether transcribed or not, were assessed for potential relationships between audio quality control features. Each point on this scatter plot represents a single diary submission, with x location corresponding to the recording's length in minutes, y location corresponding to its mean spectral flatness, and color corresponding to its volume (binned - see legend, ranges in dB).}
\label{fig:diary-audio-qc-scatter}
\end{figure}

\noindent As mentioned, in addition to inspecting the individual quality control features, I also looked at relationships between audio QC features within the BLS dataset. Both shorter diaries and quieter diaries tended to have higher mean spectral flatness across the entire audio set (Figure \ref{fig:diary-audio-qc-scatter}), suggesting a correlation between red flags in different measures of submission quality. However, given the primary goal was to identify features which improve accuracy of the pipeline's quality screening, I was more interested in whether diaries of standard length and volume, but with higher spectral flatness, tended to be of lower quality. Therefore I focused most of my investigation into the connections between audio quality control metrics and the corresponding transcript quality control features. These final analyses were thereby restricted to diaries that were successfully transcribed.

For each diary, the number of inaudibles marked by TranscribeMe was divided by the total length of the recording in minutes to mitigate bias against longer diaries. I then used inaudibles/minute as my primary outcome feature. Pearson correlation between inaudibles/minute and volume was very minimally negative ($r = -0.0413$, $p = 0.00003$), and Pearson correlation between inaudibles/minute and duration was only slightly stronger in effect ($r = -0.0878$, $p < 10^{-18}$). Because the distributional analysis suggested potential nonlinear relationships between features, I also computed the Spearman rank correlation, which resulted in stronger correlations and in the case of duration flipped directionality ($r = -0.0747$, $p < 10^{-13}$ for volume; $r = 0.1627$, $p < 10^{-61}$ for length). These correlations were calculated using the entire set of transcribed non-empty diaries (minimum 1 second duration).

\begin{figure}[h]
\centering
\includegraphics[width=\textwidth,keepaspectratio]{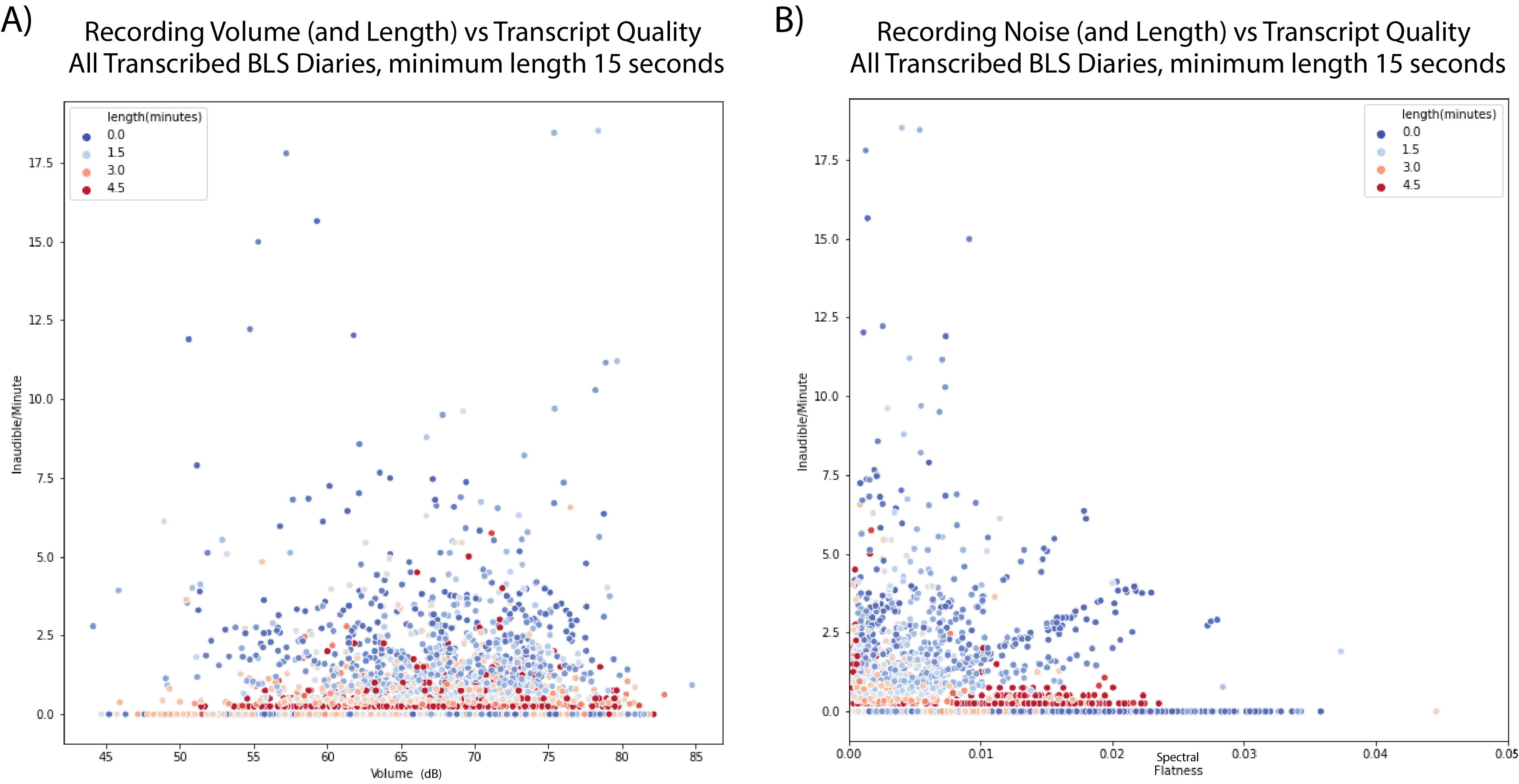}
\caption[Audio QC metrics versus resulting transcript QC from across BLS diaries.]{\textbf{Audio QC metrics versus resulting transcript QC from across BLS diaries.} All transcribed BLS diaries of at least 15 seconds were analyzed for correlation between audio QC features and transcription quality. Here, various audio metrics are compared against the number of inaudibles marked in the transcript per minute. Overall volume (in dB) serves as the x-axis for one such scatter plot (A), while mean spectral flatness serves as the x-axis for the other (B). For clarity, points with flatness > 0.05 were excluded. Both plots have their points colored by the diary length, with the shortest recordings dark blue and the longest dark red.}
\label{fig:diary-aud-trans-qc-scatter}
\end{figure}

Given the limited utility of very short journals for automated analyses, as well as the large outliers such journals were able to produce in the inaudibles/minute metric, I did a deeper dive into these feature correlations using only submissions of duration $\geq 15$ seconds. A scatter plot of volume versus inaudibles/minute with points colored by duration (Figure \ref{fig:diary-aud-trans-qc-scatter}A) shows many diaries of near-maximum length with a non-zero number of inaudibles, regardless of volume. This explains the statistically significant positive Spearman correlation between length and transcript quality, and it is likely a result of poor recording environment and/or enunciation in the submissions from 3SS93 -- who accounted for over $\frac{1}{3}$ of all max length diaries (Figure \ref{fig:diary-dist-comp-init}A) and also submitted a large number of untranscribable diaries (Figure \ref{fig:diary-pt-submit-chart}C). Additionally, many moderate length diaries had an even higher rate of inaudibles, particularly when dB was outside the 65-75 peak distributional range.    

Because of the clear relationship between mean spectral flatness and recording duration (Figure \ref{fig:diary-audio-qc-scatter}), I also scattered flatness against inaudibles/minute with the same duration hue (Figure \ref{fig:diary-aud-trans-qc-scatter}B). The lowest quality transcriptions of long recordings actually tended towards having particularly low mean flatness. \\

Although scatter plots can be informative in understanding the interplay between different values, they are not able to cleanly represent large datasets that have many overlapping values. Here, the majority of returned transcripts had no inaudible words, but that entire part of the distribution is minimized in Figure \ref{fig:diary-aud-trans-qc-scatter}. As potential rank relationships were noticed at both extremes of the various quality metrics, I next investigated the distribution of inaudibles/minute in sets of diaries falling into different QC categories.

\begin{figure}[h]
\centering
\includegraphics[width=\textwidth,keepaspectratio]{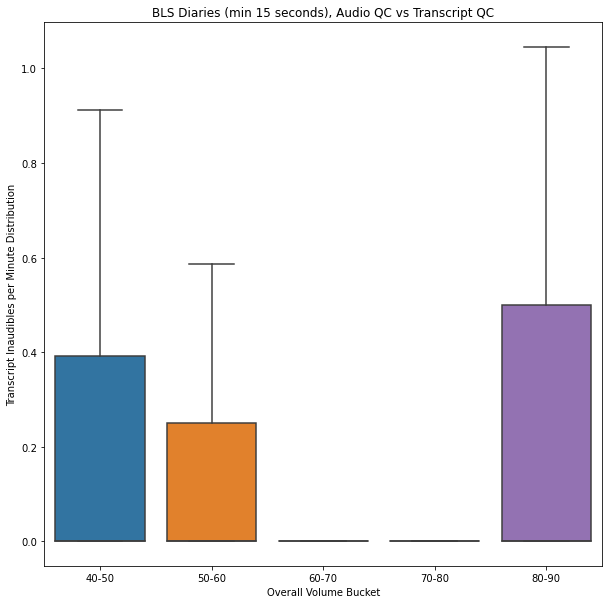}
\caption[Distribution of words marked inaudible by TranscribeMe in BLS audio diaries of different volume levels.]{\textbf{Distribution of words marked inaudible by TranscribeMe in BLS audio diaries of different volume levels.} Transcribed BLS audio journals with duration at least 15 seconds were binned based on their overall volume metric. This Seaborn box and whisker plot shows for each category (40-50 dB, 50-60 dB, 60-70 dB, 70-80 dB, and 80-90 dB) the distribution of the inaudibles/minute diary metric. For clarity, outliers (defined as outside 1.5 times the interquartile range) were excluded from this plot. The median in all cases is 0, but we can see variation in the 75th percentiles (tops of colored boxes) as well as the maximum non-outlier values (tops of whiskers).}
\label{fig:diary-qc-box-plot}
\end{figure}

Of the 8557 transcribed recordings containing 15 or more seconds of audio, 28 had volume between 40 and 50 dB, 886 had volume between 50 and 60 dB, 4147 had volume between 60 and 70 dB, 3472 had volume between 70 and 80 dB, and 24 had volume between 80 and 90 dB. The distribution of inaudibles per minute across these categories shows that volume between 60 and 80 dB is the most likely to result in high quality transcriptions (Figure \ref{fig:diary-qc-box-plot}). While volume $> 80$ dB or $< 60$ dB should not necessarily be automatically skipped, it may be worth flagging these recordings for RA review or holding them back subject to transcription budget constraints. 

\begin{figure}[h]
\centering
\includegraphics[width=\textwidth,keepaspectratio]{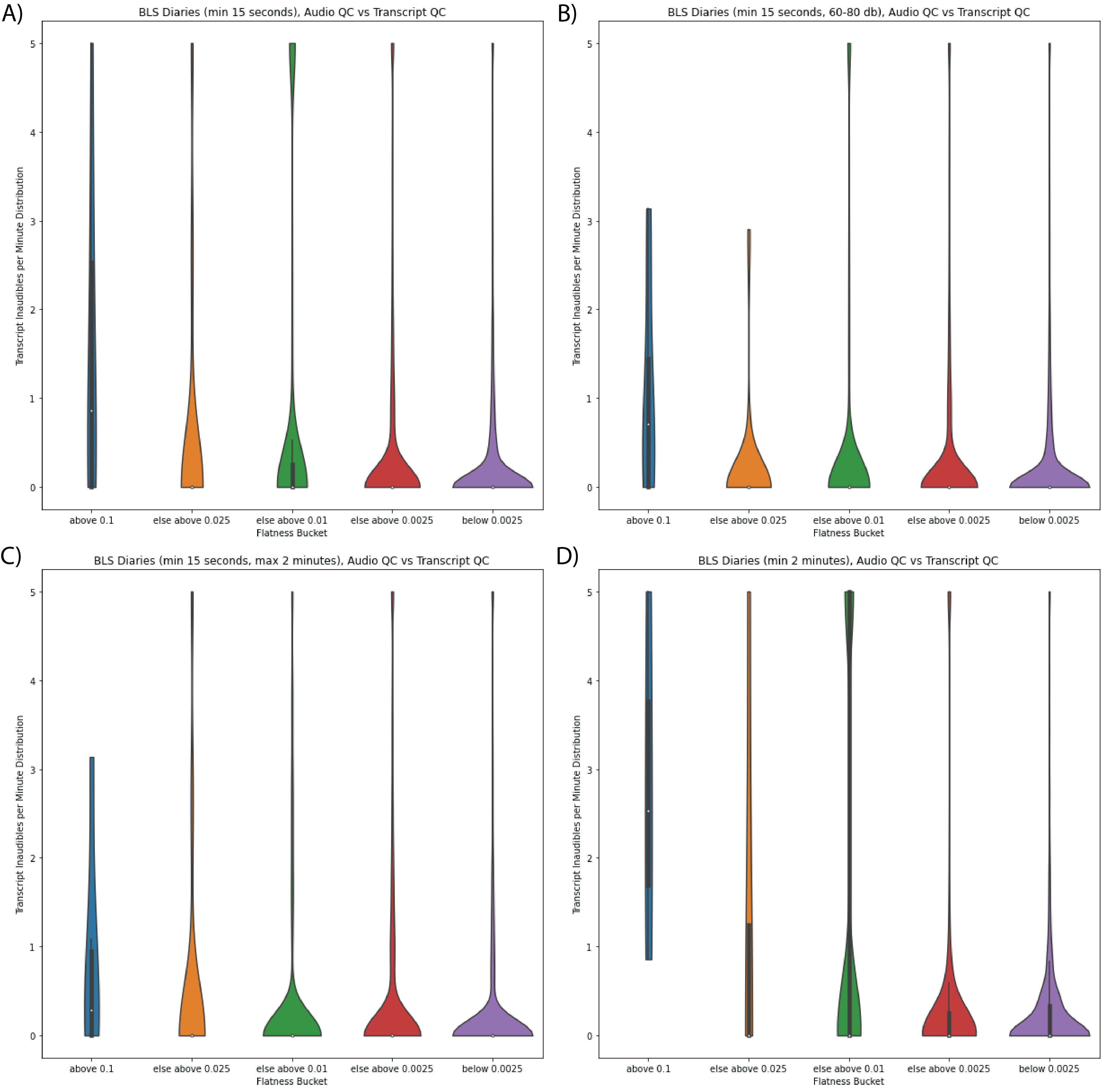}
\caption[Distribution of inaudible frequency in BLS diary transcripts varies with recording spectral flatness.]{\textbf{Distribution of inaudible frequency in BLS diary transcripts varies with recording spectral flatness.} Transcribed BLS audio journals with duration at least 15 seconds were also binned based on their mean spectral flatness metric. The distribution of the inaudibles/minute diary metric was visualized for each flatness category via violin plot (A). For clarity in the violin plots, inaudibles/minute was capped at 5 - with any exceeding that number rounded down here. An additional violin plot restricting the dataset to only those diaries with volume between 60-80 dB was also produced (B), to determine if any relationship remains when only the samples with volume corresponding to the highest quality are considered. Because of the demonstrated relationship between mean flatness and length, the distributions were examined with further restrictions on length as well, with violin plots considering only diaries between 15 second and 2 minutes (C) and only diaries longer than 2 minutes (D).}
\label{fig:diary-flatness-violin}
\end{figure}

In the same set of transcribed audio recordings, 8 had mean spectral flatness above 0.1, 61 had mean spectral flatness between 0.025 and 0.1, 1756 had mean spectral flatness between 0.01 and 0.025, 2926 had mean spectral flatness between 0.0025 and 0.01, and 3806 had mean spectral flatness below 0.0025. In general, the lower the flatness bin, the higher proportion of quality transcripts produced (Figure \ref{fig:diary-flatness-violin}A), but only $> 0.1$ had systemic quality issues worthy of a hard cutoff. When focusing in only on diaries within the best dB range (60-80), a similar relationship between flatness and inaudible words remained, indicating that flatness may indeed be a feature worth independently considering (Figure \ref{fig:diary-flatness-violin}B). Flatness between 0.025 and 0.1 underwent the largest change in quality distribution between the two datasets, so that using a harsher flatness cutoff when dB is outside the 60-80 range may be advisable. Similarly, for diaries of duration at least 2 minutes, mean flatness below 0.025 (and arguably even 0.01) becomes a recommended hard requirement for quality (Figure \ref{fig:diary-flatness-violin}D). In diaries between 15 seconds and 2 minutes however (Figure \ref{fig:diary-flatness-violin}C), the relationship is more similar to that seen in the full transcript set (Figure \ref{fig:diary-flatness-violin}A). For future study, it may be worth considering an additional flatness bin to capture values nearer 0. There is still some weak evidence, as well as theoretical justification, for concern over extremely low spectral flatness in longer files. 

\begin{figure}[h]
\centering
\includegraphics[width=\textwidth,keepaspectratio]{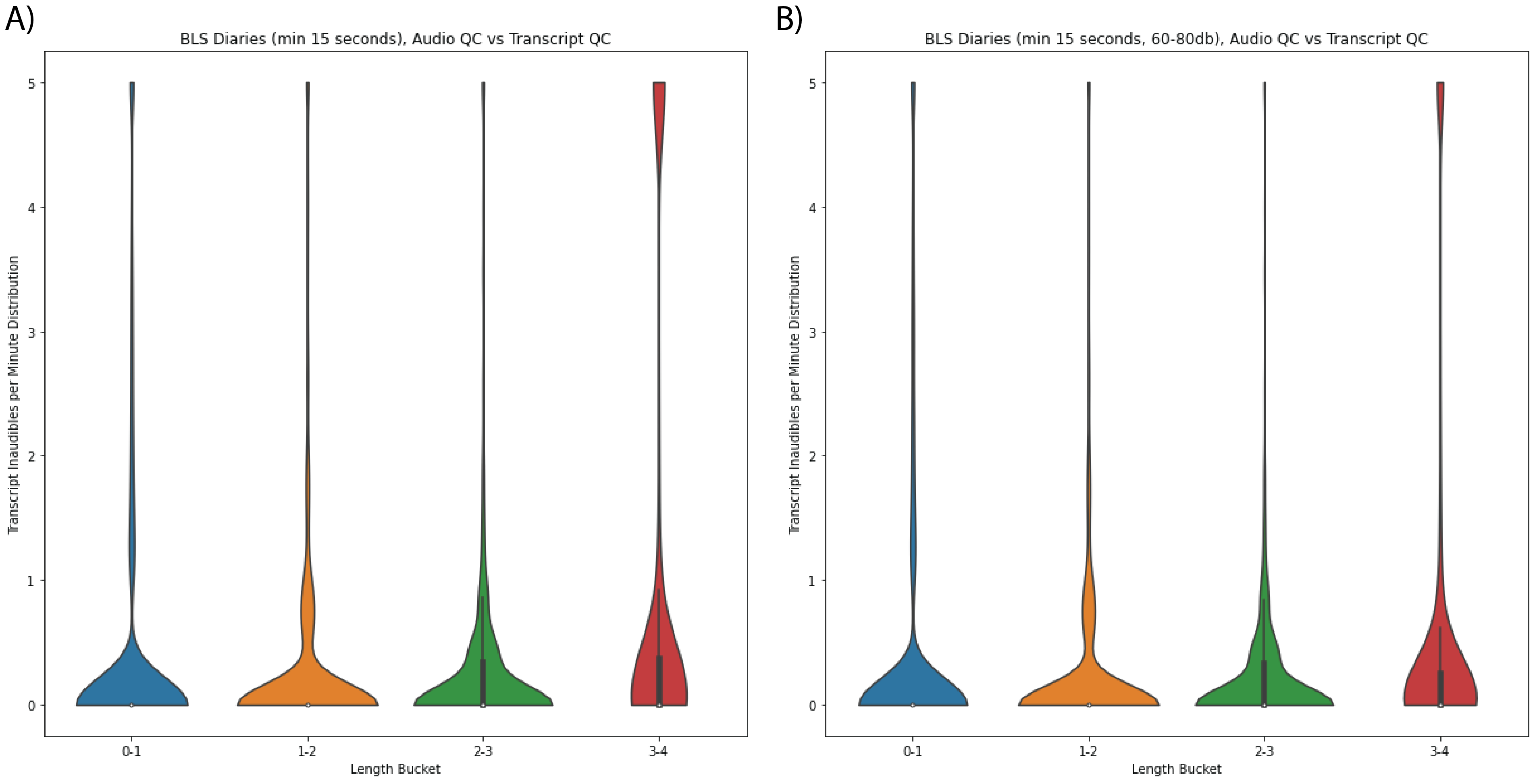}
\caption[Distribution of inaudible frequency in BLS diary transcripts of different lengths.]{\textbf{Distribution of inaudible frequency in BLS diary transcripts of different lengths.} In the same fashion as Figure \ref{fig:diary-flatness-violin}A, a violin plot of the inaudibles/minute distribution for the different diary length bins (in minutes) was produced (A). As before, a second violin plot further restricting the dataset to only those diaries with volume between 60-80 dB was also produced (B).}
\label{fig:diary-length-violin}
\end{figure}

As a final characterization of this quality metrics dataset, I looked more closely at the suggested bimodal relationship between diary length and transcript inaudible frequency. Indeed, diaries $> 3$ minutes in length fared notably worse with transcript quality (Figure \ref{fig:diary-length-violin}A), even when considering only diaries with volume in the best-identified 60 to 80 dB range (Figure \ref{fig:diary-length-violin}B). This highlights the value of considering features in addition to volume when filtering for QC. However, the specific relationship here is very unlikely to be generalizable to other datasets; as discussed, it is very likely driven by submission patterns of specific BLS patients. Besides that, small correlations could also be an artifact of the normalization method, as short diaries have fewer opportunities for an inaudible to present, but when one does will be more heavily penalized by the length denominator. This latter effect is clearly visible in the shape of the violin plots for the first three length buckets in Figure \ref{fig:diary-length-violin}.

Still, it is important to be aware of how recording length can confound various quality features, as there are flaws both in normalizing by it and in not doing so. It is also important to more generally know your own dataset. Hence any quality cutoffs or screening techniques suggested here are meant only as a rough starting point, to be tailored to specific studies based on patient population, primary research questions, and budget (among other things). The above results may be utilized as a blueprint for doing so. The key contribution is not only in providing foundational code and corresponding expected outputs for sanity checks, but also in the detailed documentation of issues to watch out for and potential additional use cases of produced features. This will remain a theme throughout the chapter. \\

\FloatBarrier

\subsection{Detailed distributions of select diary features across the BLS dataset}
\label{subsec:plain-hists}
\begin{figure}[h]
\centering
\includegraphics[width=\textwidth,keepaspectratio]{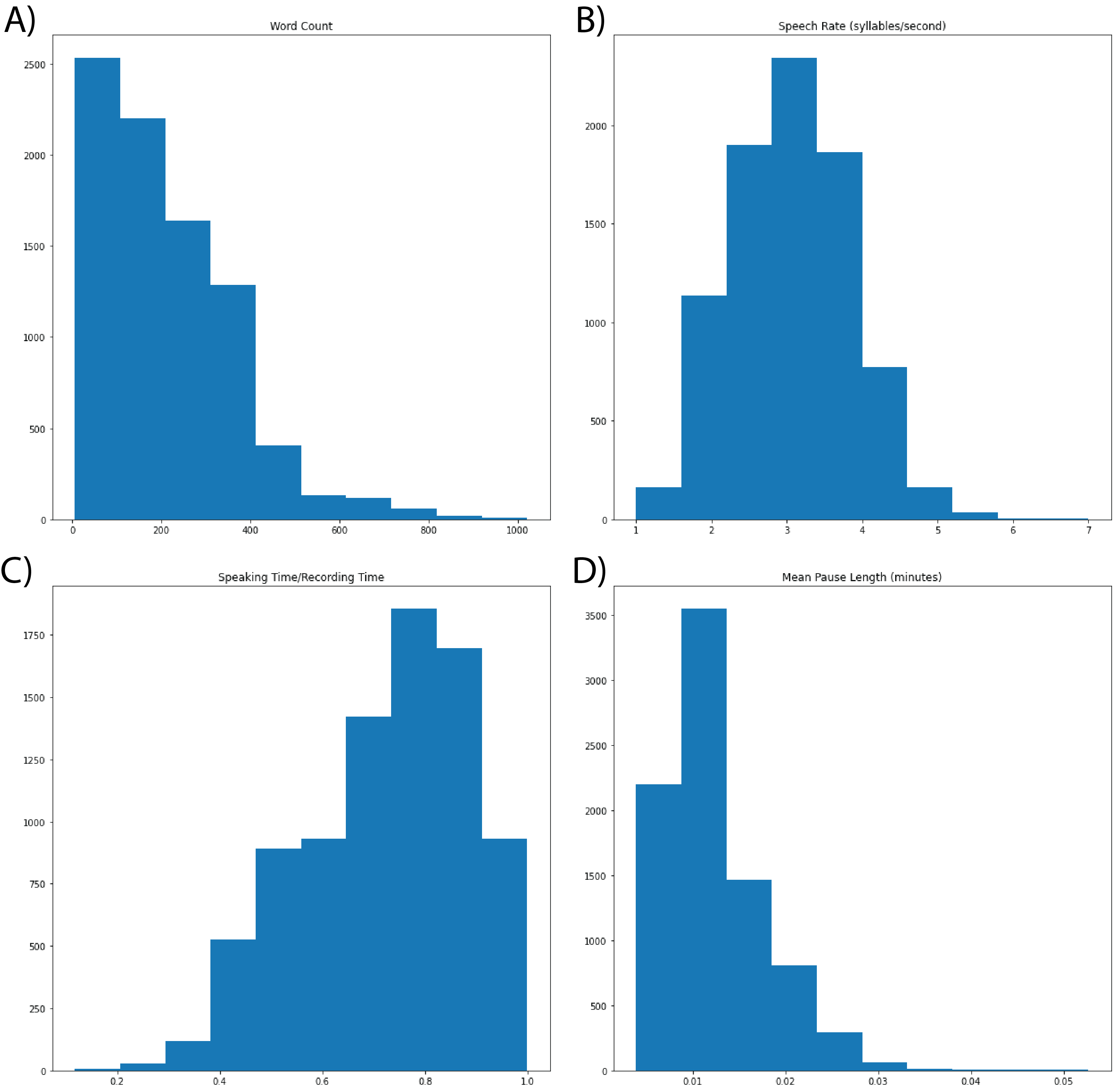}
\caption[Distribution of speech production related features in the final BLS audio journal dataset.]{\textbf{Distribution of speech production related features in the final BLS audio journal dataset.} Histograms of each per diary feature were generated over the set of quality-filtered BLS submissions ($n=8398$). Features relevant to speech production are shown here: the total number of words (A), the fraction of recording duration that was spent speaking (B), the mean pause duration in minutes (C), and the mean sentence speech rate in syllables per second (D). The plot in (D) was axis limited for clarity, so that 5 diaries with estimated syl/sec $<1$ and 15 with syl/sec $>7$ were excluded. Note that this speech rate metric was derived from transcript timestamps, so it incorporates longer pauses; for a pause-normalized version of the feature, see Figure \ref{fig:new-rate-per-pt}A.}
\label{fig:diary-verbosity-dists}
\end{figure}

Using the dataset and features described at the beginning of section \ref{sec:science2} in the main chapter, I first evaluated overall distributions by visualizing histograms with matplotlib. To contextualize the data, it is particularly important to understand how speech production related features are distributed (Figure \ref{fig:diary-verbosity-dists}). As discussed, information like verbosity and pause duration have been well-established to hold clinical relevance, and these properties also affect other features in a way that is difficult to perfectly normalize for. 

Total word count across journals (Figure \ref{fig:diary-verbosity-dists}A) varied widely, with $\sim 2500$ of the transcripts containing fewer than 100 words and $\sim 2000$ of the transcripts containing at least 300 words. A question for upcoming sections is thus how much of the word count variance occurs within versus between subjects. The percent of time spent speaking feature (Figure \ref{fig:diary-verbosity-dists}C) also demonstrated a high level of meaningful variance across journals, indicating a strong potential for downstream analyses. About half of the diaries contained participant speech during at least $75\%$ of the recording, while well over 1000 diaries contained participant speech for just $60\%$ or less of the recording duration.

Estimated speech rate (Figure \ref{fig:diary-verbosity-dists}B) was distributed like a Gaussian centered at $\sim 3$ with standard deviation $\sim 1$. With 817 recordings having speech rate below 2 syllables/second, this distribution is quite different than what one would expect from prior knowledge on human speech rate. However, it is not inconsistent with expectations here, because the methodology focused on transcript syllable count and timestamps, such that long pause times were incorporated into the present speaking rate metric. Because this feature distribution looks promising for further analysis, another question for upcoming sections is the separation of long pauses from the speech rate feature. 

\begin{figure}[h]
\centering
\includegraphics[width=\textwidth,keepaspectratio]{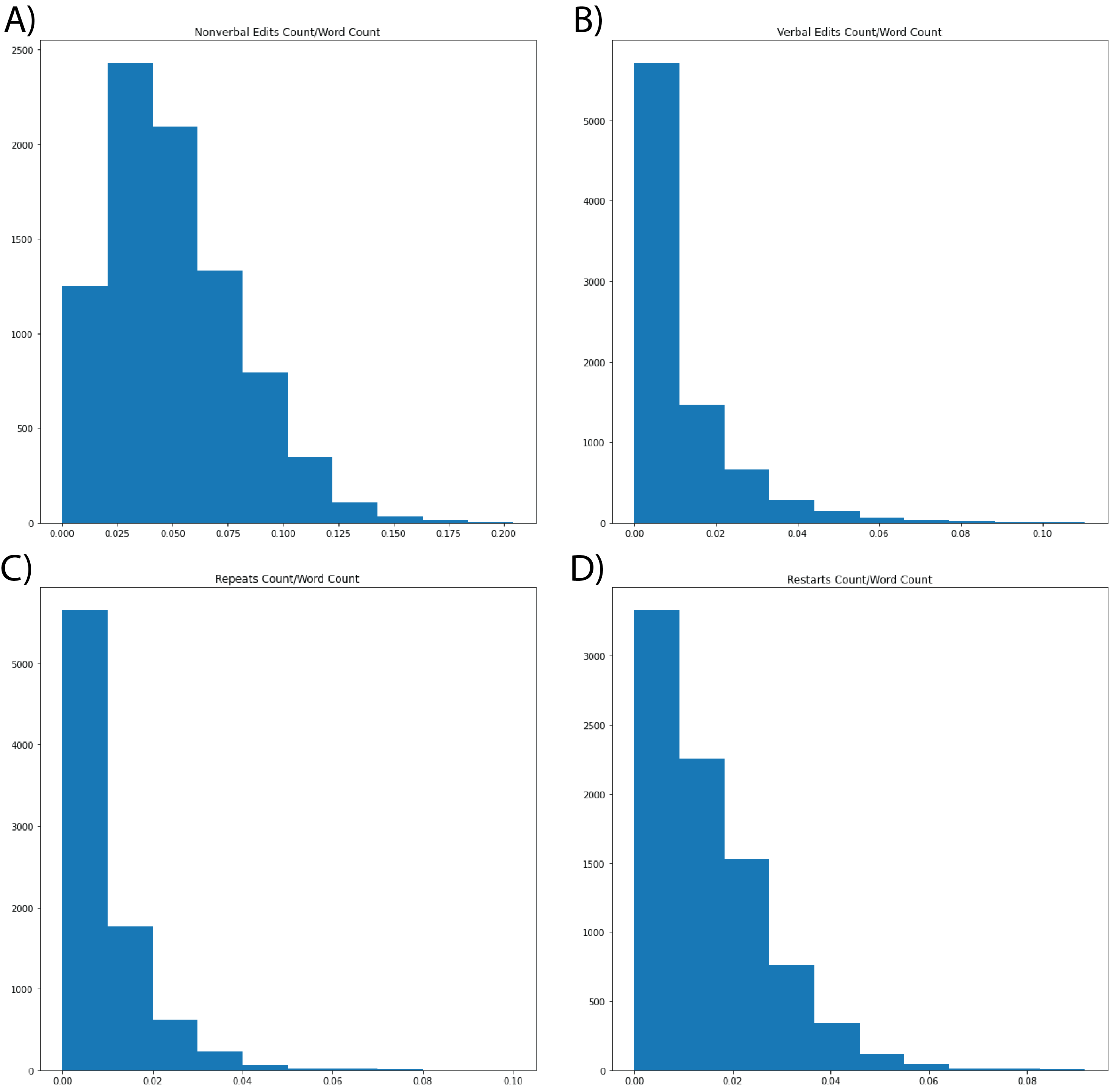}
\caption[Distribution of linguistic disfluency features in the final BLS audio journal dataset.]{\textbf{Distribution of linguistic disfluency features in the final BLS audio journal dataset.} Histograms of each per diary feature were generated over the set of quality-filtered BLS submissions ($n=8398$). Counts for the four major categories of linguistic disfluency are depicted here: nonverbal edits (A), verbal edits (B), repeats (C), and restarts (D). Recall that each of these features was normalized by transcript word count for analysis. The plot in (C) was axis limited for clarity, so that 3 diaries with repeats/word $>0.1$ were excluded.}
\label{fig:diary-disfluency-dists}
\end{figure}

Linguistic disfluencies (Figure \ref{fig:diary-disfluency-dists}) are of particular interest due to their relevance for the onsite interview language analysis presented in chapter \ref{ch:2}. Characterizing the differences between disfluency frequency in the two formats, especially in a patient-specific way, would greatly facilitate future work on the questions posed in section \ref{sec:disorg}.

Nonverbal fillers (Figure \ref{fig:diary-disfluency-dists}A) were substantially more common than verbal fillers (Figure \ref{fig:diary-disfluency-dists}B) in this dataset, which is similar to our observations in Figure \ref{fig:disorg-disfluency-dists}. Nonverbal fillers had no relationship with conceptual disorganization in the analysis of chapter \ref{ch:2}, and the relationship of verbal filler word use with conceptual disorganization was questionable at best. Because prior work suggests fillers are relevant to other symptom domains in psychotic disorders, it will be interesting to further evaluate them in this broader audio journal dataset. The lack of correlation between nonverbal and verbal edits in the interview dataset (Figure \ref{fig:disorg-interview-corr}) further motivates their dissection here.

Restarts (Figure \ref{fig:diary-disfluency-dists}D) were also noticeably more common than repeats (Figure \ref{fig:diary-disfluency-dists}C) in the diary dataset. This difference existed in the interview distributions of Figure \ref{fig:disorg-disfluency-dists}, but was much less pronounced. As repeats and restarts were the most impactful linguistic features for modeling conceptual disorganization in chapter \ref{ch:2}, the healthy distribution of restarts across these audio journals inspires confidence in follow up work using the diary format to assess this kind of disfluency.

It is important to note that the transcribed interview set used in chapter \ref{ch:2} (\ref{sec:disorg}) contained many fewer timepoints from fewer patients than the audio journals do, so distributional differences could be attributable to these factors rather than anything different about the format of a semi-structured interview -- though that does have the potential to cause change in disfluency usage. Regardless, the broad consistencies observed between the two formats here is good justification for more research on disfluencies in audio journals, as the journals are much easier to collect and cheaper to carefully verbatim transcribe. 

As was discussed in section \ref{subsubsec:diary-val-trans}, sentence-level sentiment outputs were often 0, and in our hands skewed positive when non-zero. Of course, both the semi-structured interview protocol and the daily audio journal prompt we used for BLS would lend themselves to many neutral sentences, so future research using a prompt of stronger valence ought to evaluate sentiment distribution with that in mind. 

\begin{figure}[h]
\centering
\includegraphics[width=\textwidth,keepaspectratio]{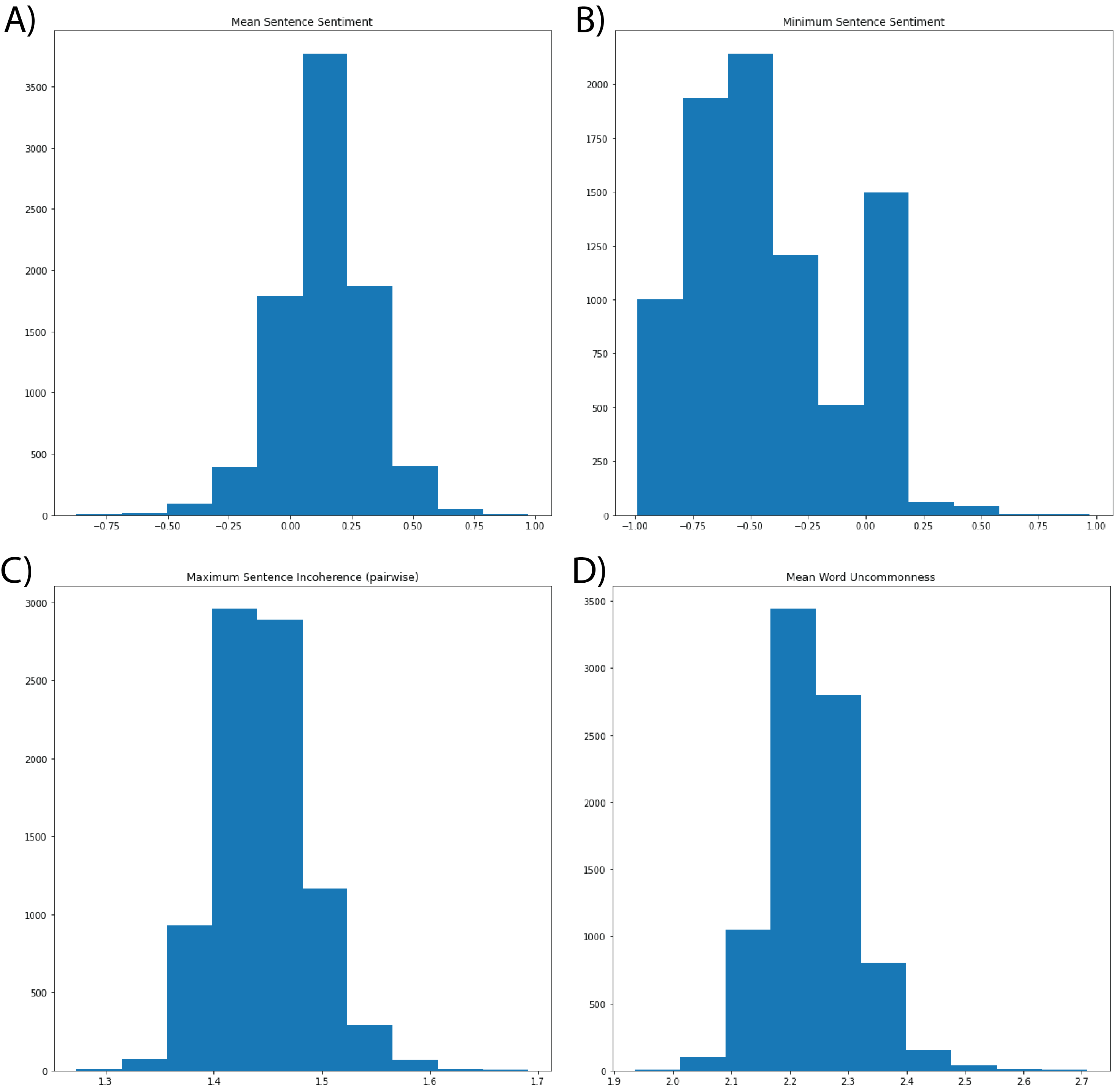}
\caption[Distribution of sentiment and incoherence features in the final BLS audio journal dataset.]{\textbf{Distribution of sentiment and incoherence features in the final BLS audio journal dataset.} Histograms of each per diary feature were generated over the set of quality-filtered BLS submissions ($n=8398$). Outputs from the VADER sentiment (top) and Google News word2vec (bottom) NLP tools are represented here. For sentiment, the diary-level features of focus are the mean (A) and minimum (B) sentence sentiment scores. For the word embeddings, the diary-level features of focus are the maximum sentence incoherence (C) and the mean word uncommonness (D). Incoherence for a sentence was defined pairwise, as the mean angle between all words in the sentence, while word uncommonness was the mean vector magnitude.}
\label{fig:diary-nlp-dists}
\end{figure}

Because of these observations in our dataset, I included both mean (Figure \ref{fig:diary-nlp-dists}A) and minimum (Figure \ref{fig:diary-nlp-dists}B) sentence sentiment as diary-level features under final consideration. The mean was quite evenly centered around $\sim 0.2$, and the vast majority of journals had mean sentiment fall between $-0.25$ and $+0.5$. More than 6500 journals had a negative minimum sentence sentiment, while 1471 had minimum sentiment exactly 0. The length of the journal will have a clear impact on these features, so while the minimum sentiment appears to have differential utility from the mean sentiment based on their distributions, this will be a question for additional study in the upcoming sections. 

The word2vec-derived features (Figure \ref{fig:diary-nlp-dists}C/D) under consideration have units that are much less immediately interpretable than all the previously discussed features. We are very likely only interested in abnormally high levels of incoherence rather than abnormally low (one of the reasons for taking the maximum instead of the mean), so the 500+ submissions with incoherence $>1.5$ in Figure \ref{fig:diary-nlp-dists}C will be a topic of interest in upcoming sections. 

Word uncommonness (Figure \ref{fig:diary-nlp-dists}D) has been less of a focus of prior literature in psychotic disorders, though high levels of uncommonness have been related to positive symptoms in a task context. It is plausible that low levels of word uncommonness would associate with negative symptoms, but it's unclear how strong the effect would be when verbosity is controlled for. As such, word uncommonness is included here as a more experimental feature. 

\FloatBarrier

\subsection{Highlighted relationships between diary features in BLS}
\label{sec:scatters}
\begin{figure}[h]
\centering
\includegraphics[width=0.8\textwidth,keepaspectratio]{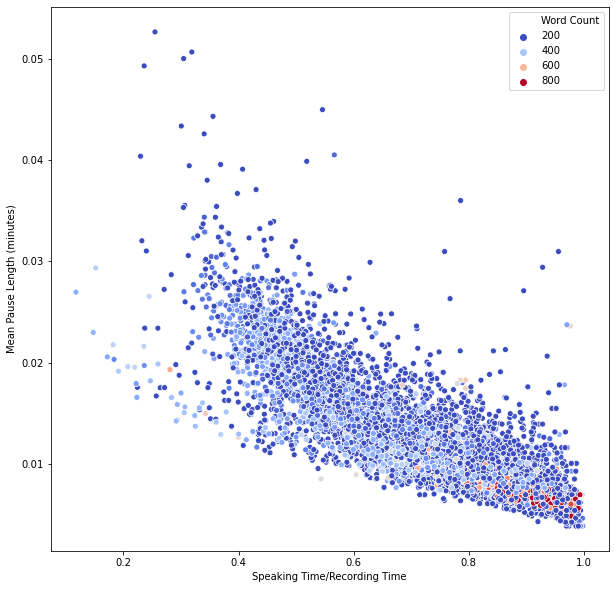}
\caption[Fraction of recording duration spent speaking versus mean pause length, word count hue.]{\textbf{Fraction of recording duration spent speaking versus mean pause length, word count hue.} Each diary in the final BLS dataset is plotted as a point here, with x-coordinate corresponding to the Speaking Time/Recording Time value for that diary and y-coordinate corresponding to the Mean Pause Length in minutes for that diary. Each diary point is colored according to its word count, from dark blue at $\leq 200$ crossing over to dark red at $\geq 800$.}
\label{fig:pause-feat-scatter}
\end{figure}

\begin{figure}[h]
\centering
\includegraphics[width=0.8\textwidth,keepaspectratio]{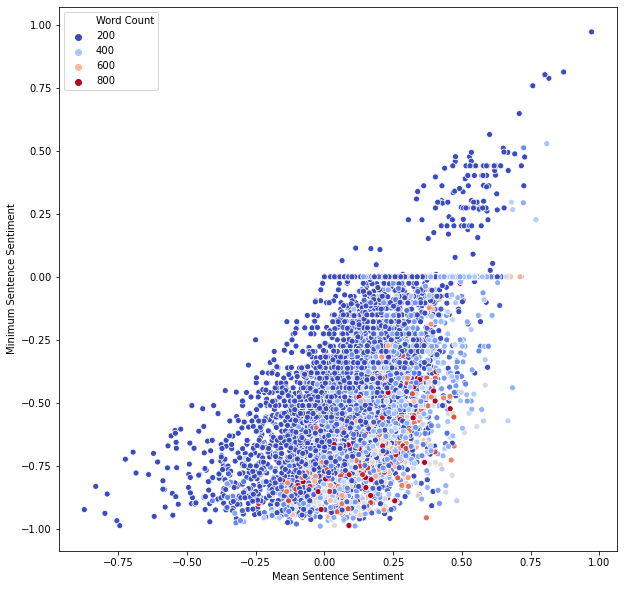}
\caption[\hspace{0.075cm} Mean sentence sentiment versus minimum sentence sentiment, word count hue.]{\textbf{Mean sentence sentiment versus minimum sentence sentiment, word count hue.} Each diary in the final BLS dataset is plotted as a point here, with x-coordinate corresponding to the Mean Sentence Sentiment for that diary and y-coordinate corresponding to the Minimum Sentence Sentiment for that diary. Each diary point is colored according to its word count, from dark blue at $\leq 200$ crossing over to dark red at $\geq 800$. Note there is no indicator for perfectly overlapping dots, so it may be difficult to interpret the role of 0 sentiment in this relationship.}
\label{fig:sentiment-scatter}
\end{figure}

\begin{figure}[h]
\centering
\includegraphics[width=0.8\textwidth,keepaspectratio]{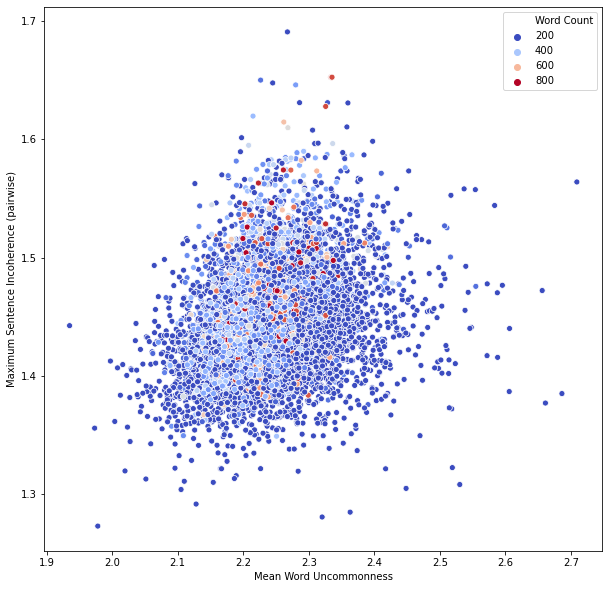}
\caption[\hspace{0.075cm} Mean word uncommonness versus maximum sentence incoherence, word count hue.]{\textbf{Mean word uncommonness versus maximum sentence incoherence, word count hue.} Each diary in the final BLS dataset is plotted as a point here, with x-coordinate corresponding to the Mean Word Uncommonness value for that diary and y-coordinate corresponding to the Maximum Sentence Incoherence value for that diary. Each diary point is colored according to its word count, from dark blue at $\leq 200$ crossing over to dark red at $\geq 800$.}
\label{fig:word2vec-scatter}
\end{figure}

\begin{figure}[h]
\centering
\includegraphics[width=0.8\textwidth,keepaspectratio]{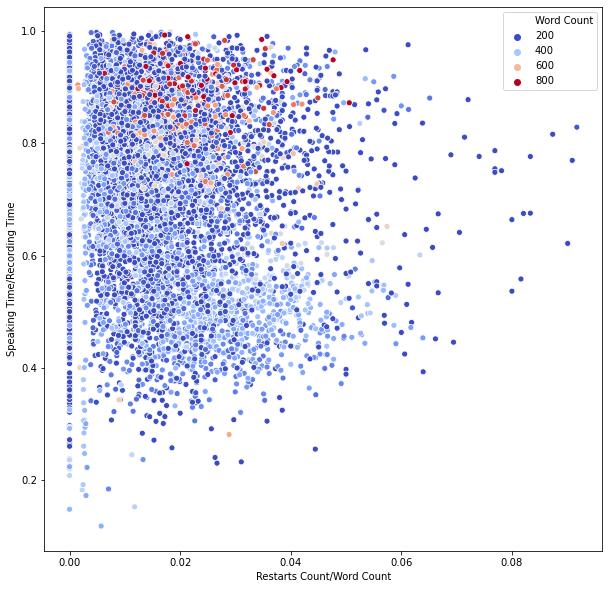}
\caption[\hspace{0.075cm} Frequency of restarts versus fraction of recording duration spent speaking, word count hue.]{\textbf{Frequency of restarts versus fraction of recording duration spent speaking, word count hue.} Each diary in the final BLS dataset is plotted as a point here, with x-coordinate corresponding to the Restarts Count/Word Count value for that diary and y-coordinate corresponding to the Speaking Time/Recording Time value for that diary. Each diary point is colored according to its word count, from dark blue at $\leq 200$ crossing over to dark red at $\geq 800$. Note there is no indicator for perfectly overlapping dots, so it may be difficult to interpret the role of 0 restarts in this relationship.}
\label{fig:restart-pause-scatter}
\end{figure}

For the following select feature pairings with interesting Pearson correlation results (Figure \ref{fig:pearson-diary-final}), I created scatter plots to better understand each relationship:
\begin{itemize}
    \item Fraction of recording duration spent speaking versus mean pause length (Figure \ref{fig:pause-feat-scatter})
    \item Mean sentence sentiment versus minimum sentence sentiment (Figure \ref{fig:sentiment-scatter})
    \item Mean word uncommonness versus maximum sentence incoherence (Figure \ref{fig:word2vec-scatter})
    \item Frequency of restarts versus fraction of recording duration spent speaking (Figure \ref{fig:restart-pause-scatter})
\end{itemize}
\noindent All scatters also had their data points colored based on word count, to further elucidate any potential dependency each relationship may have on overall verbosity.

The strongest observed correlations (Figure \ref{fig:pearson-diary-final}A) of speaking fraction with mean pause length (Figure \ref{fig:pause-feat-scatter}) and of mean versus minimum sentence sentiment (Figure \ref{fig:sentiment-scatter}) had the expected clear trends on display in scatter plot form. In future work, a closer investigation into outliers on these plots could prove quite valuable; outliers may be used to uncover interesting time points in a case study, detect bugs or unwanted noise in feature extraction methods, further dissect relationships between features into interpretable independent components, and serve as important inputs for rare event detection algorithms -- among other purposes. However for the aims of the EMA modeling to be performed here in section \ref{subsec:diary-ema}, there is lesser value in including both features when they demonstrate a strong relationship, in particular when that relationship itself has some correlation with verbosity (as in Figures \ref{fig:pause-feat-scatter} and \ref{fig:sentiment-scatter}). 

For mean word uncommonness versus maximum sentence incoherence (Figure \ref{fig:word2vec-scatter}), it is unsurprising that shorter diaries populated both extremes in uncommonness, as well as dominated the low end of max incoherence. However even when considering only short or only moderate length diaries, there was a clear relationship between uncommonness and incoherence, in particular with the lower mean word uncommonness records typically also having lower maximum sentence incoherence. As moderate length diaries (word count $\sim 400$) reflect the broadest swath of participants, this correlation captured a fairly general relationship in the dataset. Given the nature of the trend, it is most likely a property of the maximum incoherence summary stat in part: if max incoherence is high, it means at least one sentence must contain at least somewhat uncommon words (the most common words will have low angle between each other), and because diaries contain a relatively small total number of sentences (even at max 4 minute duration) the presence of that sentence will likely prevent the overall mean uncommonness score from being especially low. Of course, understanding other factors that very well may be at play -- such as a link between certain properties across sentences in the same diary -- would require further investigation. \\

\FloatBarrier

\subsection{Pilot sentiment analyses in the journals of participant 3SS93}
\label{sec:3s-sentiment}
As mentioned in the main chapter, early investigations of VADER sentence sentiment and derived diary-level sentiment scores helped to shape the pipeline, and in part involved looking at item-level EMA Spearman correlations with sentiment in a small subset of 3S's data. Because this provides a small preview into some extensions to the EMA summary score linear modeling performed on a much larger dataset within chapter \ref{ch:1}, the results of the early sentiment validation work are reported in a bit more detail in this section.

\begin{figure}[h]
\centering
\includegraphics[width=\textwidth,keepaspectratio]{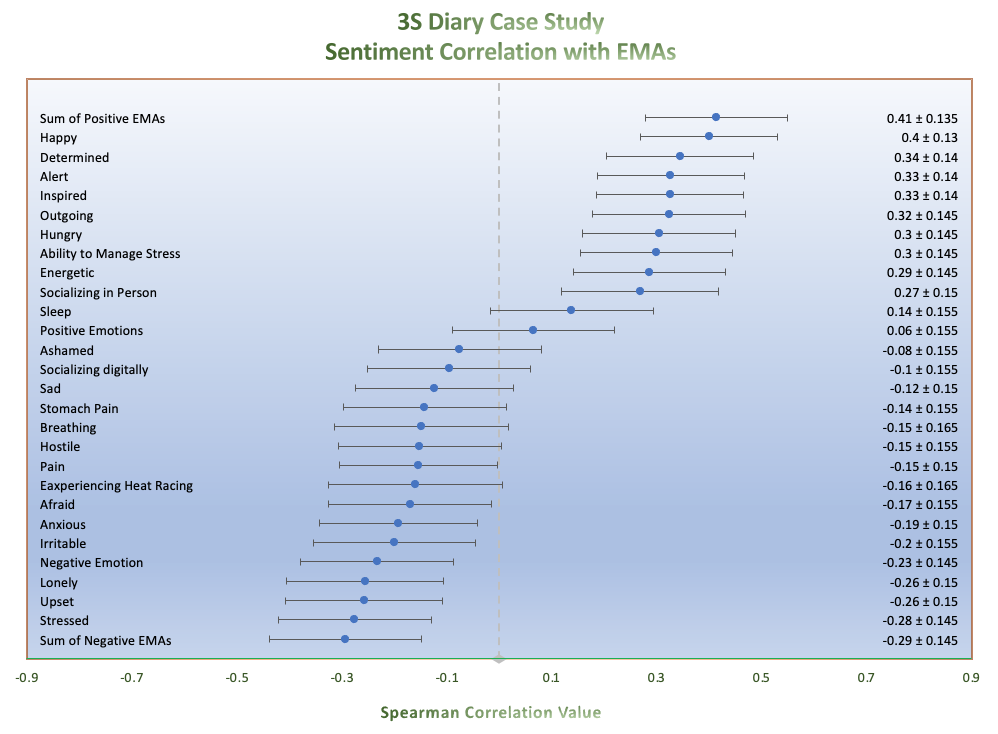}
\caption[\hspace{0.075cm} A closer look at BLS patient 3SS93: diary sentiment scores correlate with same-day mood reported on daily surveys.]{\textbf{A closer look at BLS patient 3SS93: diary sentiment scores correlate with same-day mood reported on daily surveys.} The same $\sim 6$ month journal dataset from 3SS93 as shown in Figure \ref{fig:vader-manual-3s} was also used in a pilot correlational analysis with same-day self reported EMA survey responses. For each EMA question, as well as the sum of all positively-worded questions and the sum of all negatively-worded questions, the Spearman correlation between survey score and mean diary sentiment is reported here, with $95\%$ confidence intervals. Questions are ordered based on $r$ value. Note that the EMA scores by default are non-negative integers, with stronger agreement corresponding to a higher score.}
\label{fig:vader-ema-3s}
\end{figure}

The pilot correlational analysis between language sentiment (mean sentence VADER score) and EMA responses in the first $\sim 6$ months of participant 3S's data uncovered some interesting trends, particularly in consideration of item-level EMA correlations (Figure \ref{fig:vader-ema-3s}). Note that in this analysis, EMA responses were not adjusted to align positive and negative scoring -- so a high score on a negative property was bad while a high score on a positive property was instead (generally) good. Overall, rating of happiness was the highest positively correlated question with diary language sentiment, and rating of stress was the most negatively correlated question.

Of course there were some EMA items with little immediate relevance to diary sentiment, such as stomach pain on the negative side or sleep quality on the positive side. Presence or absence of these item-based relationships with diary properties in an individual patient may itself be informative, as there can be a modulating effect from both the topics a participant chose to talk about in the diaries as well as the level of interference that a specific symptom assessed by EMA (e.g. sleep disruption) had on their daily life. 

\FloatBarrier

\subsection{Characterization of the pipeline's sentence to sentence incoherence feature}
\label{sec:sen-to-sen}
\begin{figure}[h]
\centering
\includegraphics[width=\textwidth,keepaspectratio]{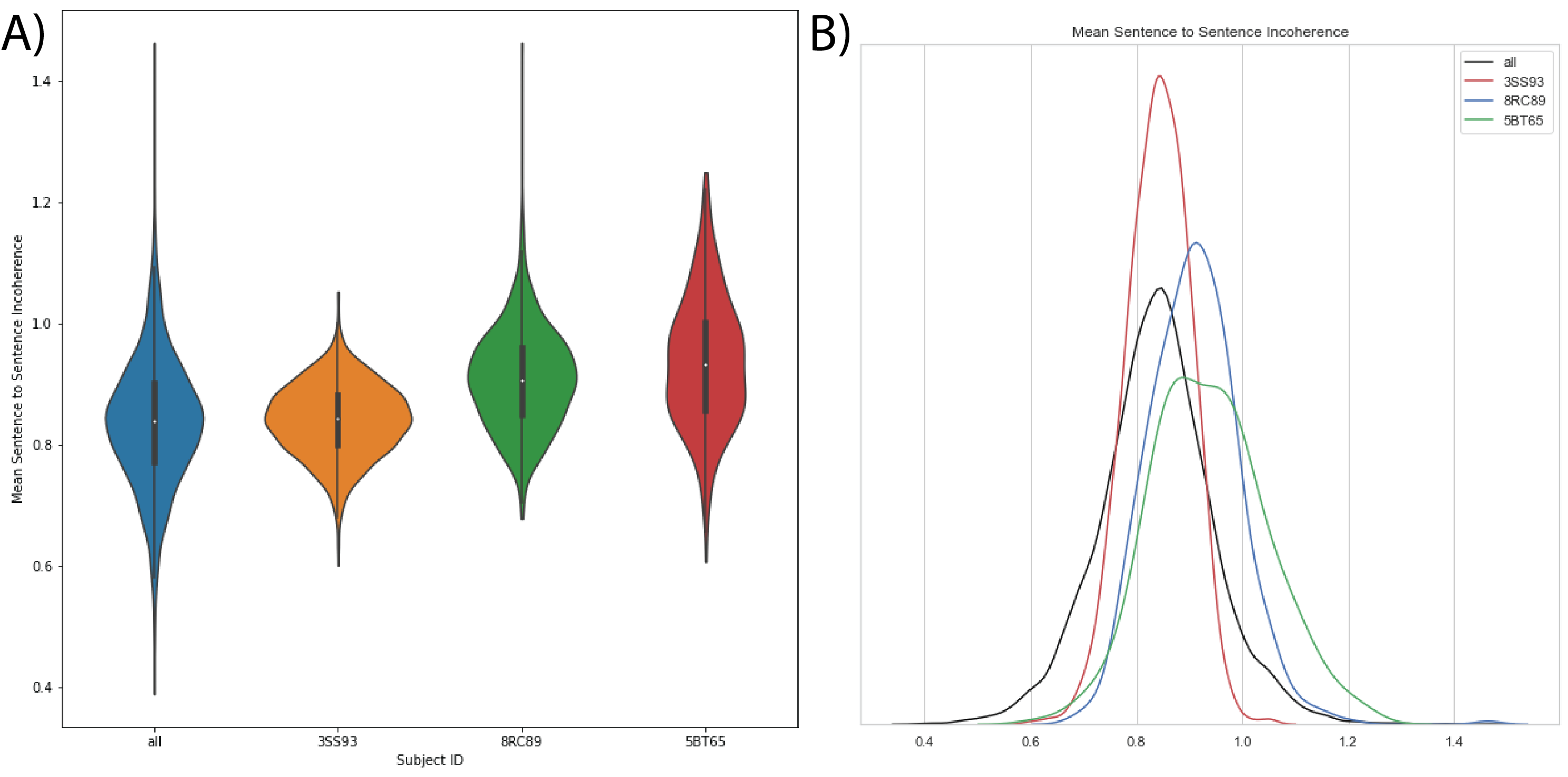}
\caption[\hspace{0.075cm} Comparison of sentence-to-sentence incoherence distributions between different patients of interest.]{\textbf{Comparison of sentence-to-sentence incoherence distributions between different patients of interest.} Using the same methodology as in Figure \ref{fig:word-count-per-pt}, additional subject-specific feature distributions are compared against each other and against the distribution across the final BLS dataset, with both violin plots (A) and smoothed distribution curves (B). Here we consider the mean sentence-to-sentence incoherence score of each transcript, as a potential supplement to the previously discussed mean word uncommonness and maximum (within-)sentence incoherence metrics.}
\label{fig:between-sentence-pt-dists}
\end{figure}

Of the $n=8398$ journals in the final BLS analysis dataset of section \ref{subsec:diary-dists} in the main chapter, $15$ contained only one sentence as marked by TranscribeMe. Therefore we had $n=8383$ final records with a defined mean sentence to sentence (sequential) incoherence score. As described, this between sentence incoherence metric was calculated by obtaining a vector for a given sentence via the mean of all word2vec-derived word vectors in the sentence. The angles between the vectors for subsequent sentences were then averaged across the transcript. Significant subject-specific distributional differences were again seen in the between sentence incoherence feature (Figure \ref{fig:between-sentence-pt-dists}). The KS test results for all 3 subject comparisons were significant by the previous described criteria, with $D=0.13$ ($p < 10^{-10}$) for 3S versus all of BLS, $D=0.32$ ($p < 10^{-35}$) for 8R versus all of BLS, and $D=0.36$ ($p < 10^{-35}$) for 5B versus all of BLS. 

\begin{figure}[h]
\centering
\includegraphics[width=\textwidth,keepaspectratio]{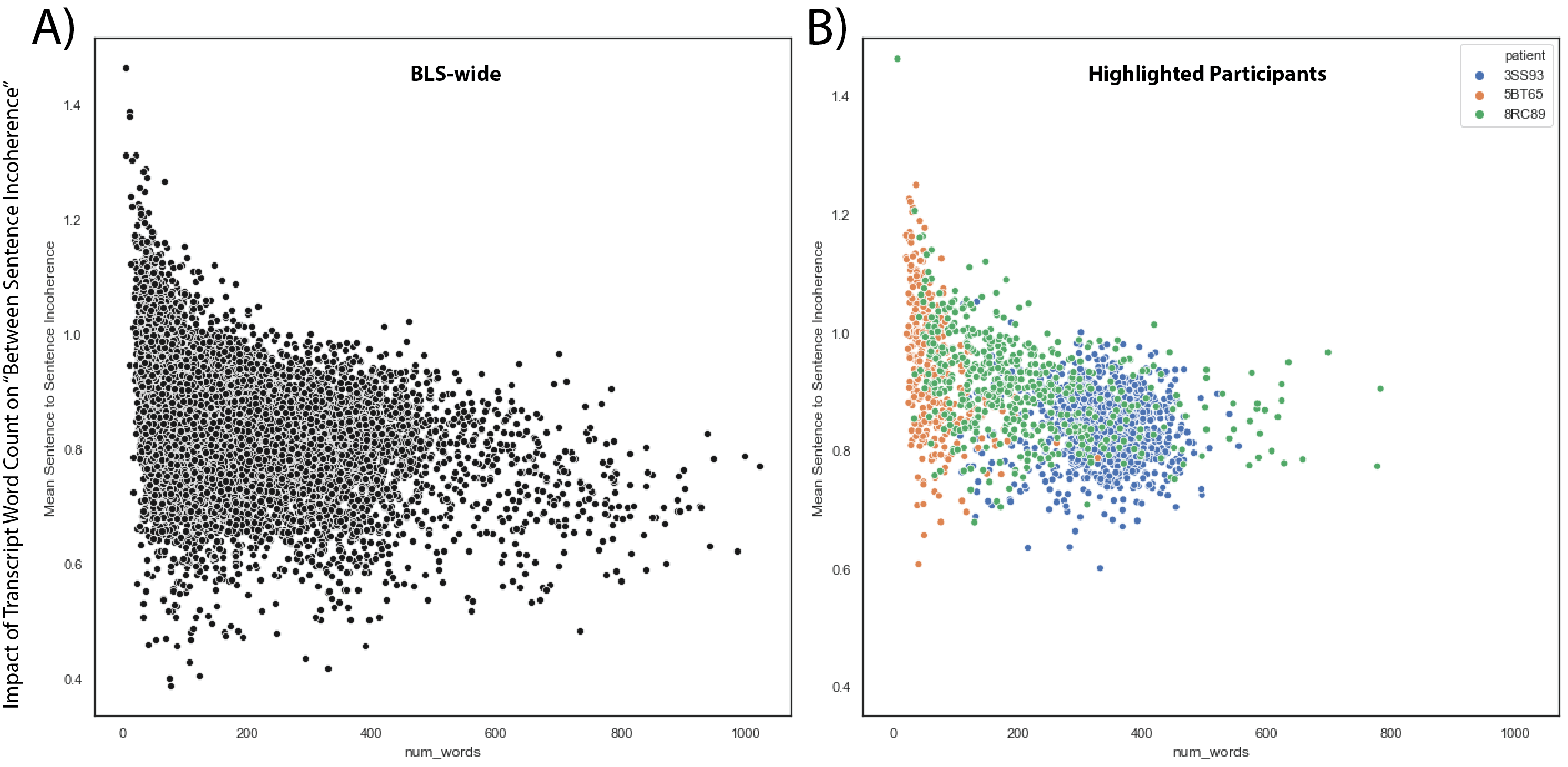}
\caption[\hspace{0.075cm} Participant-dependent distributional variation in sentence-to-sentence incoherence is not entirely explainable by verbosity differences.]{\textbf{Participant-dependent distributional variation in sentence-to-sentence incoherence is not entirely explainable by verbosity differences.} Mean between sentence incoherence values were negatively correlated with transcript word count across the final BLS dataset (Pearson's $r=-0.338$), as can be seen in the scatter of all transcripts presented here; each point has x-coordinate corresponding to word count and y-coordinate corresponding to mean sentence-to-sentence incoherence score (A). However, the subject-specific distributional differences seen in Figure \ref{fig:between-sentence-pt-dists} were not purely an artifact of the verbosity differences amongst the highlighted subjects. I also scattered word count against the between sentence incoherence measure in the participant subsets only (B). 8RC89 (green) in particular displayed an upward shift in mean sentence-to-sentence incoherence across the range of observed transcript lengths.}
\label{fig:between-sentence-scatters}
\end{figure}

Of course, to some extent significant subject-specific distributional differences were to be expected based on relative diary lengths. For a mean over sentences we would expect lower variance in longer diaries, which did appear to be the primary driver of the distributional difference between 3S and BLS at large. As the angle for each sentence was determined by the mean of the contained words, we would also expect sentence length to play a role in this feature distribution. However the fact that 8R and 5B both had sentence-to-sentence incoherence distributions shifted upwards towards higher incoherence (Figure \ref{fig:between-sentence-pt-dists}) does suggest a potential participant-dependent role for this feature independent of verbosity. This would be consistent with prior clinical expectations for 8R, and also would align with preliminary conclusions from the in depth review of the maximum within sentence incoherence metric for 5B. Indeed, differences in the sentence-to-sentence incoherence between 3S, 8R, 5B, and the rest of BLS can be clearly seen to transcend the total transcript word count when visualized as a scatter plot (Figure \ref{fig:between-sentence-scatters}).

 In order to further understand the role of verbosity in the reported results, as well as the relationship between the sentence-to-sentence incoherence and the other word2vec-derived features, I next looked at the correlations of these features of interest with the sentence-to-sentence incoherence metric across BLS. There were moderate magnitude and highly significant correlations (all $p < 10^{-200}$) found with mean uncommonness (Pearson's $r=0.383$), max sentence incoherence (Pearson's $r=0.358$), and word count (Pearson's $r=-0.338$). On the other hand, the correlation with sentence count was negligible (magnitude of $r < 0.04$) and insignificant after multiple testing correction -- though we of course would still expect higher variance in the feature in journals with smaller sentence counts. Note that the results of the nonlinear Spearman rank correlation between all of these feature pairings were very similar to the linear results. 

\begin{figure}[h]
\centering
\includegraphics[width=0.8\textwidth,keepaspectratio]{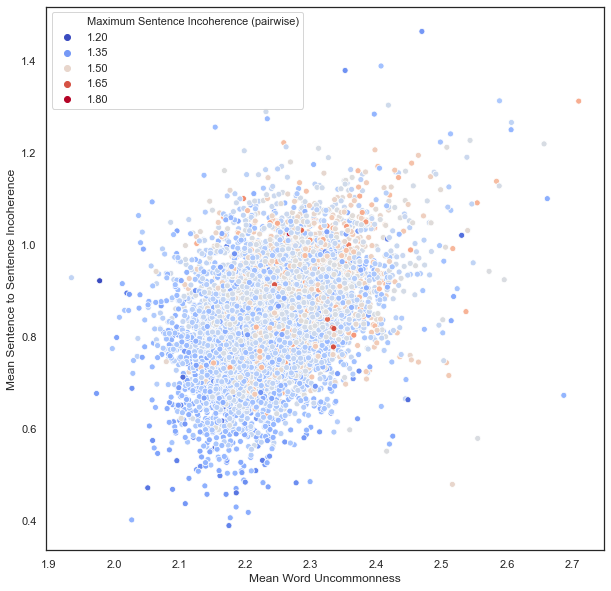}
\caption[\hspace{0.075cm} Mean word uncommonness versus mean sentence-to-sentence incoherence, maximum sentence incoherence hue.]{\textbf{Mean word uncommonness versus mean sentence-to-sentence incoherence, maximum sentence incoherence hue.} Each diary in the final BLS dataset is plotted as a point here, with x-coordinate corresponding to the Mean Word Uncommonness value for that diary and y-coordinate corresponding to the Mean Sentence to Sentence Incoherence value for that diary. Each diary point is colored according to its Maximum Sentence Incoherence (pairwise) metric, from dark blue at $\leq 1.2$ crossing over to dark red at $\geq 1.8$. The between sentence incoherence feature turned out to more strongly correlate with both mean uncommonness and max (within-)sentence incoherence than those two features did with each other (Figure \ref{fig:word2vec-scatter}).}
\label{fig:word2vec-tri-scatter}
\end{figure}

 As was presented in section \ref{subsubsec:diary-corrs} of the main chapter, correlation magnitude $> 0.3$ was relatively high amongst the considered journal features. Furthermore, note that the correlation between maximum within sentence incoherence and word count was positive, as would be mathematically expected, so it is interesting that the between sentence coherence had a fairly strong positive correlation with max sentence incoherence alongside a fairly strong negative correlation with word count. The correlation between mean uncommonness and sentence-to-sentence incoherence was also notably stronger than that between mean uncommonness and within sentence incoherence. Taken together, it is unclear how much information the between sentence incoherence would actually add that is independent of the originally considered feature set (Figure \ref{fig:word2vec-tri-scatter}).

Taken together with the lack of correlation with transcript sentence count, there was likely an especially strong relationship between mean sentence-to-sentence incoherence and words per sentence. Such a relationship would also be consistent with expectations based on the definition of the sentence-to-sentence incoherence feature. It is therefore important to consider how the mean transcript words per sentence varied over the final dataset and how it related to our other features of interest before moving on to further analyses. This is especially true because the sentence splitting was based on human transcriber judgement rather than automatic tokenization, which may capture interesting trends in speaking patterns but may also somewhat confound a subset of the downstream features extracted. 

\begin{figure}[h]
\centering
\includegraphics[width=0.8\textwidth,keepaspectratio]{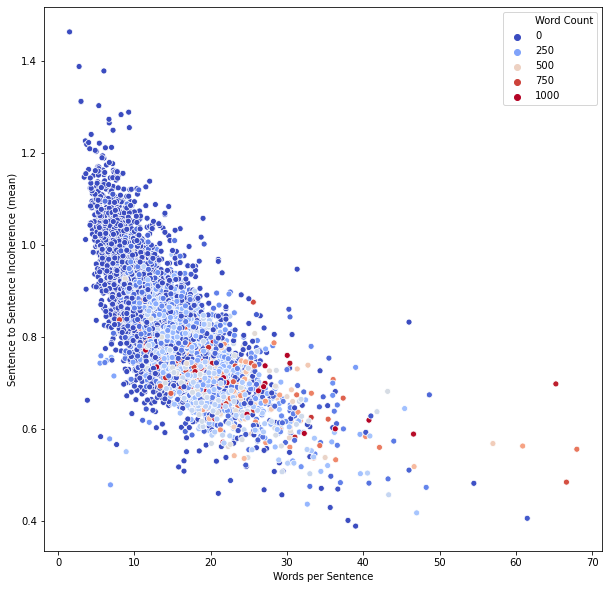}
\caption[\hspace{0.075cm} Words per sentence versus sentence-to-sentence incoherence, word count hue.]{\textbf{Words per sentence versus sentence-to-sentence incoherence, word count hue.} Each diary in the final BLS dataset is plotted as a point here, with x-coordinate corresponding to the mean words per sentence in that diary and y-coordinate corresponding to the mean between sentence (sequential) incoherence value for that diary. Each diary point is colored according to its word count, from dark blue at $\leq 100$ crossing over to dark red at $\geq 800$. Of the NLP journal features considered in this section, the sentence to sentence incoherence feature was thus by far the most correlated with basic verbosity and transcript structure measures.}
\label{fig:splits-vs-incoherence}
\end{figure}

To finish the evaluation of the between sentence incoherence metric, I reviewed the relationship of sentence length with sentence-to-sentence incoherence (Figure \ref{fig:splits-vs-incoherence}). The negative correlation was exceptionally strong for this dataset (Pearson's $r = -0.733$); combined with the relationship of sentence-to-sentence incoherence with the other word2vec features reported above (Figure \ref{fig:word2vec-tri-scatter}), the marginal value of including sentence-to-sentence incoherence in the intentionally limited modeling feature set is highly questionable. 

\FloatBarrier

\subsection{Model results for psychosis-related EMA responses from GFNVM}
\label{sec:gf-models}
Hallucinations EMA summary scores were binned into binary labels indicating some self-reported hallucination experience or no self-reported hallucination experience. For the delusions EMA score, binning was also performed, into baseline ($= 2$), lower symptoms ($< 2$), and higher symptoms ($> 2$) categories. This was done despite the nicer distribution because there was only a single delusions-related EMA item in the primary BLS EMA, so linear regression would have high error due to the large gaps between possible scores. 0/1 and 3/4 responses were combined due to the relative infrequency of 0 and 4. I then performed linear regression with discrete labels using the logits (and mnlogits) function of statsmodels. I tested verbosity-only models first, but prediction was hardly above chance and model fit was insignificant for both EMA categories. 

\begin{figure}[h]
\centering
\includegraphics[width=0.75\textwidth,keepaspectratio]{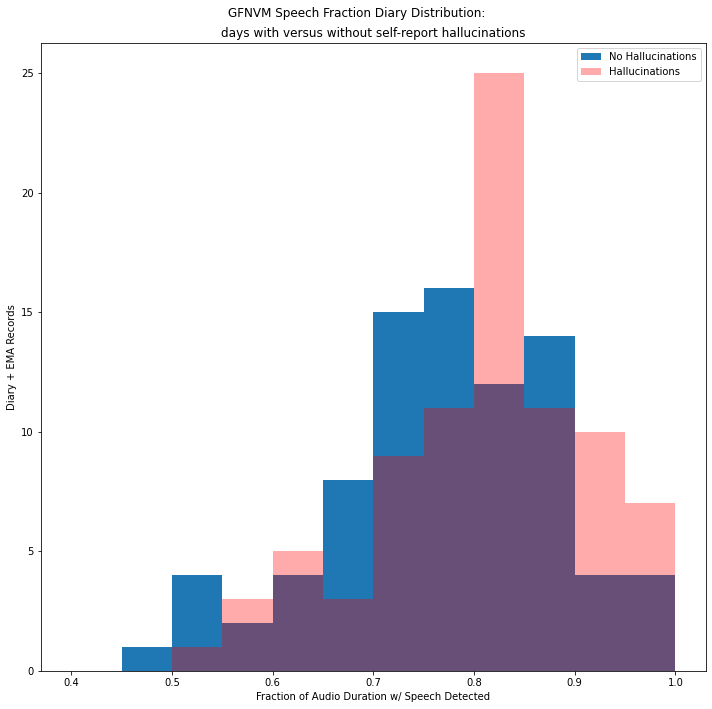}
\caption[\hspace{0.075cm} Fraction of audio diary containing detected speech versus self-reported hallucinations from GFNVM.]{\textbf{Fraction of audio diary containing detected speech versus self-reported hallucinations from GFNVM.} In the psychosis-related EMA training set, the speech fraction feature derived from the pipeline's pause detection was the most relevant journal feature for linear prediction of participant GF's survey responses, though this relationship was weak (if real). Here the distribution of speech fraction over diaries where the corresponding EMA self-reported no hallucinations (blue) is compared against the distribution of speech fraction over diaries where the corresponding EMA self-reported some level of hallucinations (transparent red). Journals with $> 80\%$ of their duration filled with detected vocals were more likely to match to a non-zero hallucinations EMA, while those with $< 80\%$ were more likely to match to a zero hallucinations EMA.}
\label{fig:gf-speech} 
\end{figure}

I then independently tested pause-related, disfluency-related, incoherence-related, and sentiment features in separate models for each of the feature categories and each of the 2 EMA types. However training classification performance remained generally poor at near chance for most of these models. Only the model of pause-related features for hallucinations EMA seemed potentially promising, with the speech fraction feature the main contributor. A logistic regression model using speech fraction alone to predict hallucinations EMA response produced the best performance of this set, though still not at all strong. It could classify $\sim 60\%$ of the training points correctly without class bias, and had model fit $p = 0.013$. Considering the distribution of speech fraction in the two different hallucinations cases (Figure \ref{fig:gf-speech}), the difference being fit by the model was clear, with more of the overall records that had speech fraction $> 0.8$ corresponding to days where non-zero hallucinations were self-reported. 

In the test set there were 9 of 15 total non-zero hallucination records with speech fraction $> 0.8$ (7 of those exceeding $0.85$), but there were only 9 of 18 zero hallucinations records with speech fraction $< 0.8$ (with 13 below $0.85$). It is therefore not possible to draw any clear conclusion about the relationship between speech fraction and self-reported psychosis symptoms in GF. Recall also that GF tended to speak with particularly poor enunciation at times, and some of their most mumbling files caused pauses to be missed by the current pause detection algorithm. It is therefore plausible that there is a real relationship here that is concentrated in the extremes, though it may not relate directly to pausing.  

Ultimately, there were no meaningful relationships between diary features and same day psychosis-related EMA summary scores in participants GF that could be expressed via basic logistic regression. It is unclear to what extent this is a lack of information in the diary features for this type of prediction versus a timescale mismatch for such predictions versus a weakness of linear coefficients versus a random noise problem in GF's hallucinations and delusions EMA responses. 

\FloatBarrier

\subsection{Extra diary examples from manual review}

\subsubsection{Sentiment}
\label{subsubsec:sentiment-diary-trans}

\begin{figure}[h]
\centering
\begin{tcolorbox}[top=-0.25cm,bottom=0.4cm,left=-0.25cm,right=-0.25cm]\begin{quote}\small
00:01.32    \hspace{2mm}     Um, yesterday I ended up feeling kind of sick in the morning, but then I wasn't a-- um, getting manic like I thought I would when I saw my boyfriend and felt, um, [inaudible] hypermania, um, and it's a good feeling, I guess.

00:21.07    \hspace{2mm}    I mean, again, I know it's not natural, but, um, it's been a while since I've been hypermanic so it's kind of nice to have a-- I don't feel terrible right now, um, but I know I'm kind of sleep deprived because I couldn't fall asleep very easily and get much sleep the past few nights, and I'm, I'm hoping I can just get some sleep.
\end{quote}\end{tcolorbox}

\begin{tcolorbox}[top=-0.25cm,bottom=0.4cm,left=-0.25cm,right=-0.25cm]\begin{quote}\small
00:03.28    \hspace{2mm}     I understand it all-- it started off-- I have to record it, my day, which started off draining a little bit and dragging, but I drank a lot of caffeine, and I felt better seeing my boyfriend and should be able to sleep well tonight.
\end{quote}\end{tcolorbox}
\caption[\hspace{0.075cm} The two most positive GFNVM diaries by mean sentence sentiment were not in the manual review top 5.]{\textbf{The two most positive GFNVM diaries by mean sentence sentiment were not in the manual review top 5.} The 2 GF transcripts with the highest mean VADER sentence sentiment scores are reproduced here, in rank order. These journals were not selected as especially positive by our manual reviewer, and were likely assigned an inflated sentiment by the pipeline due to the way TranscribeMe split these sentences -- though that likely itself contains information of interest on GF's pattern of speech on these days, and it is particularly interesting in light of the multiple mentions of hypomania in the $\#1$ diary.}
\label{fig:positive-diary-gf-disagree}
\end{figure}

\FloatBarrier

\subsubsection{Incoherence (within sentence)}
\label{subsubsec:incoh-diary-trans}

The top 4 journals by \emph{within sentence} incoherence are reproduced for GFNVM in Figures \ref{fig:incoherent-diaries-gf-disagree}-\ref{fig:incoherent-diaries-gf-disagree2}, as 1 overlapped with the between sentence list. Finally, the top 5 journals by \emph{within sentence} incoherence are reproduced for 8RC89 in Figures \ref{fig:incoherent-diaries-8r-disagree}-\ref{fig:incoherent-diaries-8r-disagree2}.

\pagebreak

\begin{FPfigure}
\centering
\begin{tcolorbox}[top=-0.5cm,bottom=0.1cm,left=-0.5cm,right=-0.5cm]\begin{quote}\tiny
00:02.78    \hspace{2mm}     Um, my boyfriend had, um, made a comment, um, that he was only attracted to skinny girls and that I could lose some weight.

00:29.07    \hspace{2mm}     And I know I'm fat and all and I, I have body issues and I have eating disorder issues from my past.

00:38.18    \hspace{2mm}     And I don't know.

00:42.28    \hspace{2mm}     It really hurt me because I thought-- I thought he thought I was pretty.

00:54.45    \hspace{2mm}     And then I was like, "Why are you with me?"

00:57.07    \hspace{2mm}     And I mentally broke.

01:01.68    \hspace{2mm}     And I still-- thoughts won't leave my head.

01:05.67    \hspace{2mm}     They keep attacking me and telling me just to run and just to kill-- like, I don't wanna die.

01:17.49    \hspace{2mm}     But this is like, why live?

01:19.29    \hspace{2mm}     It's just like all these hopeless despairing thoughts, you know?

01:24.72    \hspace{2mm}     They just won't shut up, no matter what.

01:30.11    \hspace{2mm}     And it's driving me mad ever since.

01:32.84    \hspace{2mm}     And I also found out the, the month, um, he had broken up with me, he really had like a serious relationship with a girl who was like, [inaudible] away and that was sexual.

01:51.85    \hspace{2mm}     And, like, I don't know.

01:57.0   \hspace{2mm}     I just-- and then he just comes crawling back to me and it's like, "How dare you?"

02:02.16    \hspace{2mm}     Apparently just didn't go well with her.

02:08.49    \hspace{2mm}     And it's like, wha-what am I then?

02:11.25    \hspace{2mm}     A-and I freaked out on him and, um, [inaudible] and just, was, like, all was just crying.

02:19.26    \hspace{2mm}     And it was-- I couldn't mentally handle it.

02:22.45    \hspace{2mm}     I couldn't mentally keep it inside anymore.

02:24.72    \hspace{2mm}     And then he had a panic attack while driving, thinking I was gonna leave him and he didn't want me to leave him.

02:32.34    \hspace{2mm}     And I just don't know-- I don't know what to do anymore.

02:39.83    \hspace{2mm}     I'm just, like-- my perfect world is destroyed.

02:43.95    \hspace{2mm}     And I'm trying not to let my eating disorders come back.

02:51.99    \hspace{2mm}     But I'm trying-- I, I am really not trying.

02:57.68    \hspace{2mm}     I am reducing my diet and restricting and monitoring things closely again.

03:03.96    \hspace{2mm}     And I just need to make sure it doesn't get out of control because, like, I would want to think I'm pretty.

03:12.28    \hspace{2mm}     Other than that, at the same time, I'm like-- it's like I want to hurt him.

03:19.52    \hspace{2mm}     And then I-- and the-- but I want to, like, just cut myself and make him happy with me.

03:26.11    \hspace{2mm}     Like, it's just, like-- It's a-- I'm just all-- I don't know what to do.

03:31.52    \hspace{2mm}     And I'm so lost and sad and confused.

03:36.45    \hspace{2mm}     And I hate myself again, more than ever.

03:38.95    \hspace{2mm}     And I never wanted to be back in this position.

03:42.65    \hspace{2mm}     And I didn't think-- I didn't think he was like that.

03:47.1      \hspace{2mm}    I would s-- I don't know.

03:51.58    \hspace{2mm}     Guess I just wish it was said earlier.

03:54.03    \hspace{2mm}     So just don't get back with me if you're not physically attracted to me.

03:58.61    \hspace{2mm}     I don't know.
\end{quote}\end{tcolorbox}
\vspace{-0.5cm}
\begin{tcolorbox}[top=-0.5cm,bottom=0.1cm,left=-0.5cm,right=-0.5cm]\begin{quote}\tiny
00:02.22    \hspace{2mm}     Well, I ended up coming off my Abilify, so, um, I started off just kind of oversleeping and feeling like crap and then started having really vivid dreams and, um, after that, I was able to finally kinda get up, get showered, try to get myself moving.

00:21.86    \hspace{2mm}     I missed an appointment.

00:22.88    \hspace{2mm}     I didn't really care.

00:23.73    \hspace{2mm}     I was just so out of it and slept through my alarm and was just too late by the time I realized, and, um, now I'm just, like, really antsy, and, um, I feel like crap.

00:39.47    \hspace{2mm}     I mean, it's just, I guess from stopping the medicine, I'm just, like, weird, like, at night now.

00:44.9     \hspace{2mm}     I'm feeling antsy and, like, jittery and can't move or stop moving, and, like, all day I've been just, like, oversleeping and, like, acting like my depressed state where I just sleep too much when I get in that kinda zone.

00:58.64    \hspace{2mm}     And now I'm, like, antsy but tired and-- but I can't sleep 'cause I'm antsy 'cause I'm moving my legs and stuff, but I'm just, like, so unsettled and, like, can't feel restful, can't feel, like, relaxed.

01:10.15    \hspace{2mm}    It's just, like, constant just, like, ugh, this weird energy, and I just, I don't know, I'm just-- I'm not looking forward to going back to how I was, but at the same time I am in so many ways 'cause I can't stand the way that I feel right now, and that's how I've been feeling.

01:24.15    \hspace{2mm}    And, and so, like, I've experienced both today and it's like well, they both suck, but, you know, what am I gonna do?

01:30.2     \hspace{2mm}     And I'm just in a weird [inaudible].
\end{quote}\end{tcolorbox}
\caption[\hspace{0.075cm} Within sentence incoherence metrics identify a different kind of anomalous journal for GFNVM.]{\textbf{Within sentence incoherence metrics identify a different kind of anomalous journal for GFNVM.} The $\# 1$ most incoherent GFNVM journal as ranked by our maximum within sentence incoherence (pairwise) summary feature was the same as the $\# 5$ most incoherent journal chosen by the sentence to sentence incoherence feature in Figure \ref{fig:incoherent-diaries-gf}, discussing orientation. The remaining 4 transcripts in the top 5 of the within sentence incoherence ranking were not particularly high on sentence to sentence incoherence and spanned a fairly wide range. However they appeared to identify a different type of speaking pattern sometimes used by GF that may also be of interest. Those 4 journals are thus reproduced in rank order, with the first two in this Figure and the following two in Figure \ref{fig:incoherent-diaries-gf-disagree2}. The connection between sentences is generally more clear than what was seen in the transcripts selected by the sentence to sentence ranking (Figure \ref{fig:incoherent-diaries-gf}), with more total sentences here as well. On the other hand, the transcripts selected by this within sentence ranking tended to contain more "run on" sentences, and more extreme thoughts can be found within some of them. For example, within the first journal depicted here, GF says: \newline "And I still-- thoughts won't leave my head. \newline They keep attacking me and telling me just to run and just to kill-- like, I don't wanna die."}
\label{fig:incoherent-diaries-gf-disagree}
\end{FPfigure}

\begin{figure}[h]
\centering
\begin{tcolorbox}[top=-0.25cm,bottom=0.3cm,left=-0.25cm,right=-0.25cm]\begin{quote}\scriptsize
00:00.41    \hspace{2mm}    Uh, a totally good day overall with my boyfriend.

00:02.81    \hspace{2mm}     I got my homework done even though it was a little late.

00:05.94    \hspace{2mm}     Um, I did have one mental breakdown just because I was triggered by, um, my new habits and [inaudible] and feeling guilty.

00:16.94    \hspace{2mm}     I wasn't respecting my diet as I should and just being triggered and crying and just-- this was just over the top emotional for really no reason.

00:30.95    \hspace{2mm}     He apologized and said it was [inaudible].

00:34.43    \hspace{2mm}     Like [inaudible] is fine now and [inaudible] kind of passed on to, like, a [inaudible] just to kind of review my system [inaudible].

00:47.43    \hspace{2mm}     And we watched a movie and, um, just hang around [inaudible] time together [inaudible] again tomorrow.

01:00.35    \hspace{2mm}     I'm glad to have had the day off [inaudible] even though I did get a little kind of irritable and probably frustrated with him earlier where I didn't [inaudible] come across mean, but I was kinda mean.

01:14.52    \hspace{2mm}    [inaudible] best interest [inaudible].

01:19.88    \hspace{2mm}     I apologized.
\end{quote}\end{tcolorbox}

\begin{tcolorbox}[top=-0.25cm,bottom=0.3cm,left=-0.25cm,right=-0.25cm]\begin{quote}\scriptsize
00:01.25    \hspace{2mm}     Um, I didn't have a good day today.

00:03.29    \hspace{2mm}     I, uh-- I'm not--

00:08.98    \hspace{2mm}     I'm, I'm fat, you know?

00:12.06    \hspace{2mm}     My boyfriend just casually said that he is only attracted to pretty girls.

00:18.16    \hspace{2mm}     And then he realized what he said was offensive.

00:20.62    \hspace{2mm}     And he was like, "But you could just lose a little weight, you know?"

00:26.61    \hspace{2mm}     And [I'm like?], "[inaudible] spot."

00:31.72    \hspace{2mm}     I don't--

00:39.77    \hspace{2mm}     [Making?] sense.
\end{quote}\end{tcolorbox}
\caption[\hspace{0.075cm} More journals from GFNVM displaying high within sentence incoherence.]{\textbf{More journals from GFNVM displaying high within sentence incoherence.} Follow-up to Figure \ref{fig:incoherent-diaries-gf-disagree}, now presenting the number 4 and 5 ranked transcripts by within sentence incoherence for subject GF.}
\label{fig:incoherent-diaries-gf-disagree2}
\end{figure}

\FloatBarrier

\pagebreak

\begin{FPfigure}
\centering
\vspace{-0.25cm}
\begin{tcolorbox}[top=-0.5cm,bottom=0.1cm,left=-0.5cm,right=-0.5cm]\begin{quote}\tiny
00:00.41    \hspace{2mm}     Uh, Monday, the 25th.

00:03.01    \hspace{2mm}     Um, we wen-- we got coffee with my mom and her brother, Uncle [redacted].

00:08.84     \hspace{2mm}     He's visiting.

00:09.54    \hspace{2mm}     He's going back tomorrow.

00:11.62    \hspace{2mm}     Uh, it's good.

00:13.17    \hspace{2mm}     Things went well.

00:13.85    \hspace{2mm}     [redacted] got along with everybody, got along me, [redacted], them.

00:18.68    \hspace{2mm}     Then I worked from 5 to 11, pretty easy shift, um, just reading through people's files, was learning stuff about them.

00:27.21    \hspace{2mm}     Um, that's about all.

00:28.87    \hspace{2mm}     Hanging in there.

00:30.23    \hspace{2mm}     Not sure if I'm gonna stop at the bar after work or not.

00:32.38    \hspace{2mm}     I have the car this time.
\end{quote}\end{tcolorbox}
\vspace{-0.6cm}
\begin{tcolorbox}[top=-0.5cm,bottom=0.1cm,left=-0.5cm,right=-0.5cm]\begin{quote}\tiny
00:00.68    \hspace{2mm}    Uh, Wednesday, I, uh, got up at the last possible moment, but I managed to get on the bus, get to a psychiatrist's appointment at [redacted], which is like two miles away, um, on time.

00:10.74    \hspace{2mm}     Saw her, um, told her about my vitamin regimen, said my sleep is getting better but still not totally normal.

00:17.39    \hspace{2mm}     Sometimes I sleep way more than I should, but it's getting more regular.

00:20.78    \hspace{2mm}     I can track it with my Fitbit now.

00:22.59    \hspace{2mm}     Um, also saw my, uh, s-- therapist.

00:27.45    \hspace{2mm}     Um, then I went to group therapy.

00:29.61    \hspace{2mm}     All good times.

00:31.2     \hspace{2mm}     I didn't tell anyone about hooking up with my ex.

00:33.39    \hspace{2mm}     Um, keeping that to myself.

00:36.68    \hspace{2mm}     Um, I mean, I could, but I chose not to.

00:41.66    \hspace{2mm}     Um, you know, no one's perfect.

00:43.64    \hspace{2mm}    Uh, yeah, so then I'm going to [redacted] at, um, 7:00.

00:49.09    \hspace{2mm}     There's a speaker.

00:51.08    \hspace{2mm}     Um, so yeah, I had an okay day, got what I needed to get done, done.

00:55.93    \hspace{2mm}     I was supposed to hang out with [redacted], uh, tonight, but, um, she said she wasn't feeling well and postponed it to Friday.

01:03.35    \hspace{2mm}     I said, "I gotta be up early. I'm doing the [inaudible] walk on Saturday, so can we do Saturday?" and she said, "Yeah. That sounds great."

01:09.53    \hspace{2mm}     So, yeah, looking forward to that.

01:11.37    \hspace{2mm}     I'll be good to see her, um, you know, dinner and a movie, and hopefully some time alone together will be nice.

01:19.64    \hspace{2mm}     So yeah, that's about all.

01:21.66    \hspace{2mm}    I'm doing well.

01:24.05    \hspace{2mm}     Um, gonna work tomorrow and the next day.

01:26.6     \hspace{2mm}     Um, so, yeah, doing all right.
\end{quote}\end{tcolorbox}
\vspace{-0.6cm}
\begin{tcolorbox}[top=-0.5cm,bottom=0.1cm,left=-0.5cm,right=-0.5cm]\begin{quote}\tiny
00:00.81    \hspace{2mm}     Uh, Wednesday I didn't go to DBSA.

00:04.67    \hspace{2mm}     Um, I, uh-- was that the day I was so tired I went home at 2:00?

00:10.67    \hspace{2mm}     I don't know.

00:11.01    \hspace{2mm}     I didn't ask my boss, but somehow word got around that I was just dead

00:14.7     \hspace{2mm}     She let me go early.

00:16.94    \hspace{2mm}     So I guess I'm getting some, uh, you know, benefits to having a-- being a new parent.

00:26.47    \hspace{2mm}     So that's all right.

00:28.01    \hspace{2mm}     I just wish they wouldn't have scheduled me six days in a row.

00:30.59    \hspace{2mm}     It just it gets very tiring.

00:34.01    \hspace{2mm}     And then having, like, more than two or three days off, there's just kind of a long [inaudible] return.

00:40.32     \hspace{2mm}     Not, not that beneficial, but that's my whole-- that's what I am.

00:45.23    \hspace{2mm}     I'm a relief employee.

00:46.33    \hspace{2mm}     I'm a relief staff, so my whole purpose is to fill in the gaps.

00:49.95    \hspace{2mm}     So I don't know.

00:51.75    \hspace{2mm}     Hopefully, my schedule will be more regular as time goes on, but I'm not married to this job.

00:57.37    \hspace{2mm}     It'd be nice if I get a higher-paying job that's not horrible.

01:02.92    \hspace{2mm}     We'll see.
\end{quote}\end{tcolorbox}
\caption[\hspace{0.075cm} Within sentence incoherence metrics identify a different kind of anomalous journal for 8RC89.]{\textbf{Within sentence incoherence metrics identify a different kind of anomalous journal for 8RC89.} The top 5 most incoherent 8RC89 journals as ranked by our maximum within sentence incoherence (pairwise) summary feature are reproduced in rank order, with the first three in this Figure and the following two in Figure \ref{fig:incoherent-diaries-8r-disagree2}. They were quite different than those identified by the mean sentence to sentence incoherence score for 8R (Figure \ref{fig:incoherent-diaries-8r}). However the within sentence metric appeared to uncover interesting thoughts buried in longer diaries that were not identified in the sentence to sentence case -- like the same metric ranking did for GFNVM (Figure \ref{fig:incoherent-diaries-gf-disagree}). From this review we can also identify recurring topics of importance to the patient, as they often pop up in journals even when disconnected from other points. For 8R, a closer look into mentions of romantic relationships and of alcohol consumption over time could be of particular salience.}
\label{fig:incoherent-diaries-8r-disagree}
\end{FPfigure}

\begin{figure}[h]
\centering
\vspace{-0.3cm}
\begin{tcolorbox}[top=-0.55cm,bottom=0.05cm,left=-0.5cm,right=-0.5cm]\begin{quote}\tiny
00:02.13    \hspace{2mm}     Um, been doing all right.

00:05.46    \hspace{2mm}     Uh, working.

00:06.68    \hspace{2mm}     Worked today.

00:07.04    \hspace{2mm}     It was pretty calm.

00:08.8     \hspace{2mm}    People acted up only a little bit, um.

00:13.28    \hspace{2mm}     I, uh, tidied up my room because I wanna have this girl [redacted] over that I'm dating.

00:16.78     \hspace{2mm}     We're supposed to hang out Saturday, tomorrow, but I haven't heard from her about time and place to meet up.

00:21.87    \hspace{2mm}     I hope I hear from her.

00:22.7     \hspace{2mm}     I hope she didn't decide, for some reason, that she doesn't like me or doesn't want to see me or talk to me.

00:29.3     \hspace{2mm}     Um, you know, I'm probably just being paranoid, though.

00:32.3     \hspace{2mm}     She's probably just busy with her own stuff.

00:36.83    \hspace{2mm}     You know, she's running around trying to get some benefits from social security, um, so it's a lot of paperwork and lawyers and stuff.

00:46.1     \hspace{2mm}     Every night this week, I've slept over at my ex [redacted]'s, and, um, she's, like, more willing to cuddle, which is nice, uh, s-- you know, sort of giving me some of what I need, I guess.

00:57.65    \hspace{2mm}     But, you know, I fee-- I kind of feel like saying to her sometimes, "Regardless of the reason you, uh-- you dumped me, you really hurt me."

01:06.14    \hspace{2mm}     "I'm not over it."

01:09.34    \hspace{2mm}     "On some level, I still wanna be with you, but I know you'll probably just break my heart a third time."

01:14.64    \hspace{2mm}     But almost like what's the point?

01:18.71    \hspace{2mm}     Gonna go see, um-- see her later.

01:20.63    \hspace{2mm}     Her friend [redacted]'s in town.

01:23.54    \hspace{2mm}     So I'm gonna see her and [redacted] and her friend [inaudible] and this random guy that likes [redacted].

01:28.68    \hspace{2mm}     And, uh, you know, last time [redacted] was here, we kind of hooked up.

01:31.8     \hspace{2mm}    Um, probably nothing will happen between us, but, um, you know, be tempting if she makes any overtures, you know, like.

01:42.49    \hspace{2mm}     She checks a lot of boxes for me, the red hair, the pale skin, um, yeah, so.

01:51.22    \hspace{2mm}     We'll have to see what happens.

01:52.4     \hspace{2mm}     Oh, I've been drinking off and on.

01:55.68    \hspace{2mm}     Uh, my mom found out.

01:56.98    \hspace{2mm}     She was not happy, but at least I don't have to go to a program so far.

02:04.04    \hspace{2mm}    Um, I might have to go back [inaudible] tomorrow, so I might have some drinks tonight, stop and buy something on the way.

02:11.35    \hspace{2mm}     I don't know, we'll see how it goes.

02:14.41    \hspace{2mm}     Doing all right.
\end{quote}\end{tcolorbox}
\vspace{-0.35cm}
\begin{tcolorbox}[top=-0.55cm,bottom=0.05cm,left=-0.5cm,right=-0.5cm]\begin{quote}\tiny
00:00.97    \hspace{2mm}     Thursday I worked.

00:03.03    \hspace{2mm}     Um, I think I spent the night at my mom's house.

00:07.0     \hspace{2mm}     Um, it was a smooth day at work.

00:09.76    \hspace{2mm}     [redacted] didn't get out, unfortunately.

00:11.32    \hspace{2mm}     Unfortunately, I'm still banned from seeing her.

00:13.74    \hspace{2mm}     Got a real Nurse Ratched type up there at the psych ward.

00:16.03    \hspace{2mm}     They're very strict.

00:17.27    \hspace{2mm}     Um, yeah, hanging in there.

00:21.41    \hspace{2mm}     Hanging out with the cats when I'm there doing laundry, washing their dresses, hanging them up to dry.

00:26.74    \hspace{2mm}     The ones that, you know, have special care instructions.

00:30.17    \hspace{2mm}     It's the least I can do.

00:31.86    \hspace{2mm}     Um, yeah, that's about it.

00:35.08     \hspace{2mm}     I'm not getting that much rest.

00:36.4      \hspace{2mm}     I'm not sleeping that well.

00:38.01    \hspace{2mm}     Fall asleep watching Netflix.

00:40.53    \hspace{2mm}     Then the cats wake me up digging their razor sharp little claws into me.

00:44.87    \hspace{2mm}     Yeah, hard times.

00:47.16    \hspace{2mm}     We'll maintain, we'll survive.

00:49.89    \hspace{2mm}     Hopefully, stay together.

00:51.12    \hspace{2mm}     I love her so much.
\end{quote}\end{tcolorbox}
\vspace{-0.6cm}
\caption[\hspace{0.075cm} More journals from 8RC89 displaying high within sentence incoherence.]{\textbf{More journals from 8RC89 displaying high within sentence incoherence.} Follow-up to Figure \ref{fig:incoherent-diaries-8r-disagree}, now presenting the number 4 and 5 ranked 8R transcripts by within sentence incoherence.}
\label{fig:incoherent-diaries-8r-disagree2}
\end{figure}

\FloatBarrier

\subsection{Other supplemental figures}

\subsubsection{More example pipeline outputs}
\label{subsec:sup-code-outs}

\begin{figure}[h]
\centering
\includegraphics[width=0.9\textwidth,keepaspectratio]{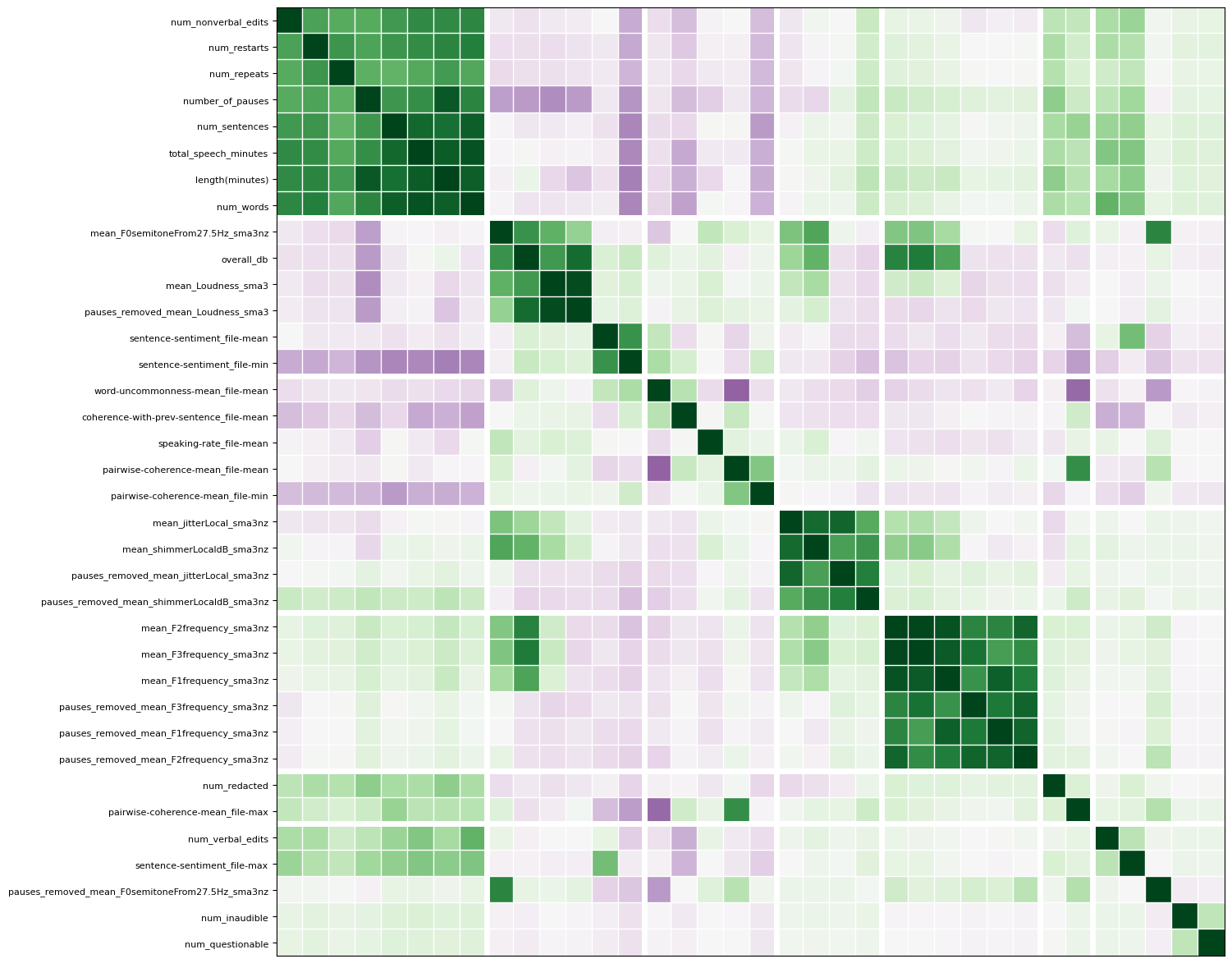}
\caption[\hspace{0.075cm} Pearson correlation matrix of key diary features from across the BLS study.]{\textbf{Pearson correlation matrix of key diary features (both acoustic and linguistic) from across the BLS study.} Features are ordered using hierarchical clustering results from the same dataset. Matrix is colored using the PrGn matplotlib colormap, with limits of -1 to 1. Note these are the core day-level features output by the standard pipeline, described in section \ref{subsec:diary-code}.}
\label{fig:diary-corr-init}
\end{figure}

\FloatBarrier

\pagebreak
\clearpage

\begin{FPfigure}
\centering
\includegraphics[width=0.9\textwidth,keepaspectratio]{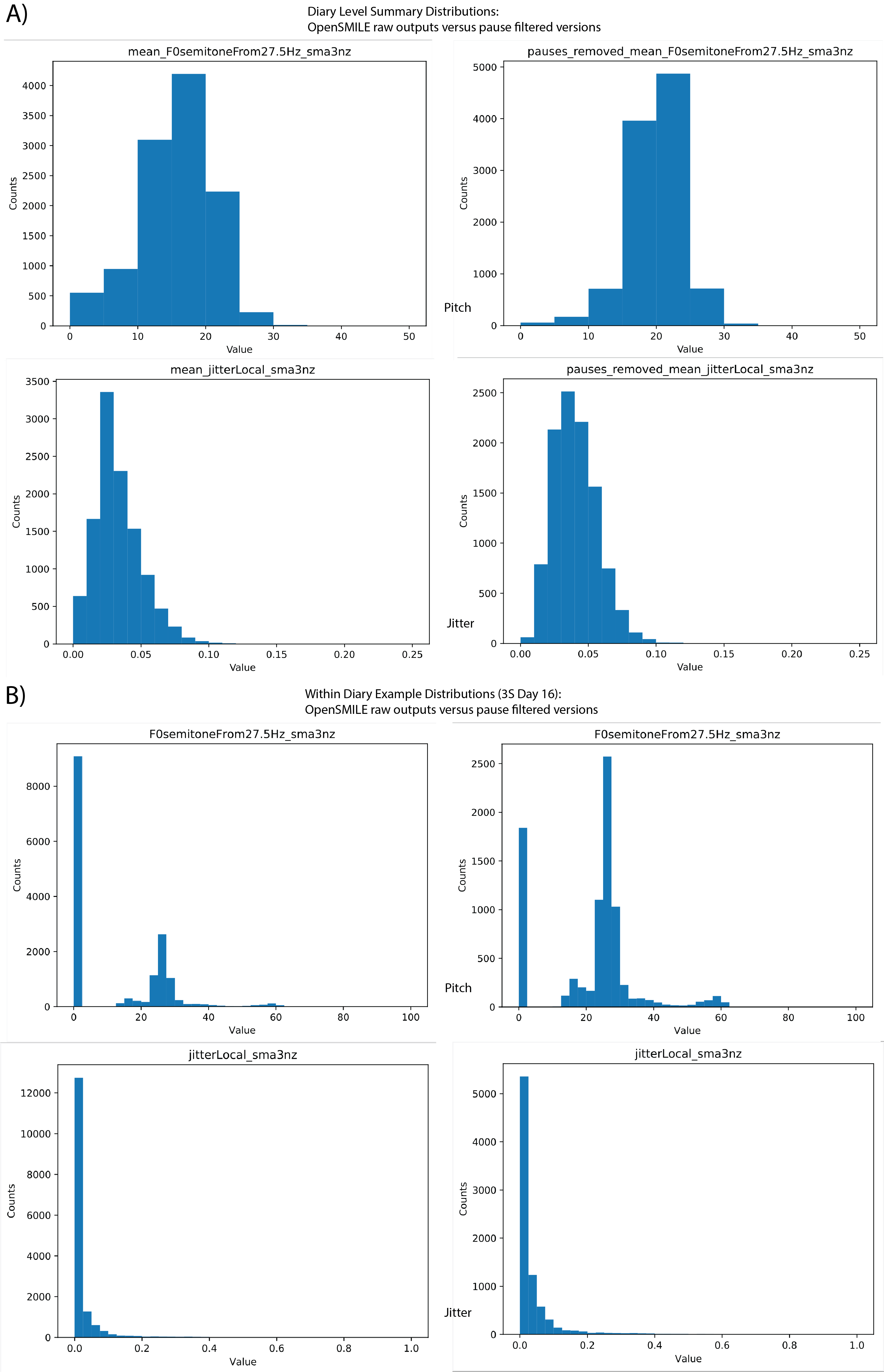}
\caption[\hspace{0.075cm} Effects of pause filtering on BLS diary OpenSMILE results.]{\textbf{Effects of pause filtering on BLS diary OpenSMILE results.} Distributions of select features of particular relevance (pitch and jitter) from Table \ref{table:bls-aud-means} (A, right) are contrasted with the same features computed without labeled pauses removed from the OpenSMILE results i.e. taking the mean over the raw output (A, left). To illustrate what is occurring on a diary level, distributions over the 10ms OpenSMILE output bins are plotted for the underlying pitch (top) and jitter (bottom) features in an example diary from patient 3SS93 (B). This patient was chosen due to their frequent use of pausing. The particular diary shown is from day 16, the same used to demonstrate VAD spectrograms in Figure \ref{fig:diary-vad-spec}.}
\label{fig:os-filter-comp}
\end{FPfigure}

\FloatBarrier

\subsubsection{More on participant-specific diary submission dynamics}
\label{subsec:participation-fig-sup}

\begin{figure}[h]
\centering
\includegraphics[width=0.8\textwidth,keepaspectratio]{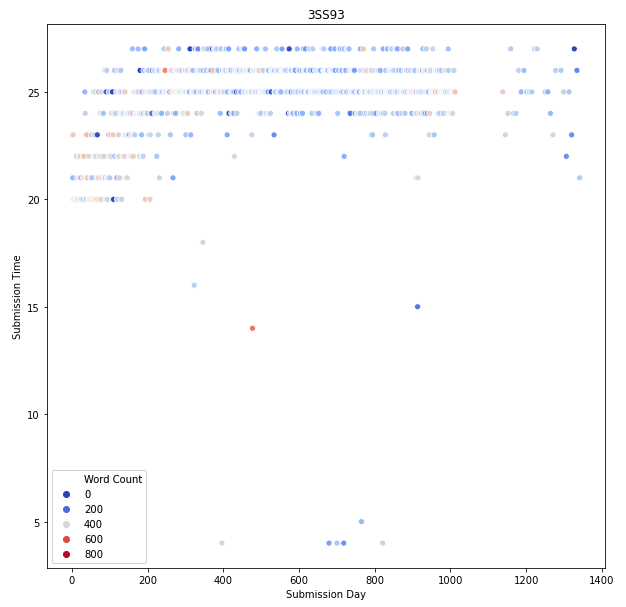}
\caption[\hspace{0.075cm} Submission day versus submission time with word count hue for BLS subject 3SS93.]{\textbf{Submission day versus submission time with word count hue for BLS subject 3SS93.} The methodology of Figure \ref{fig:diary-day-vs-time-2yr} is repeated here for only the diaries submitted by participant 3SS93. However, the x-axis is no longer artificially limited, so that the full time course of data from 3S is depicted here.}
\label{fig:submit-time-3s}
\end{figure}

\begin{figure}[h]
\centering
\includegraphics[width=0.8\textwidth,keepaspectratio]{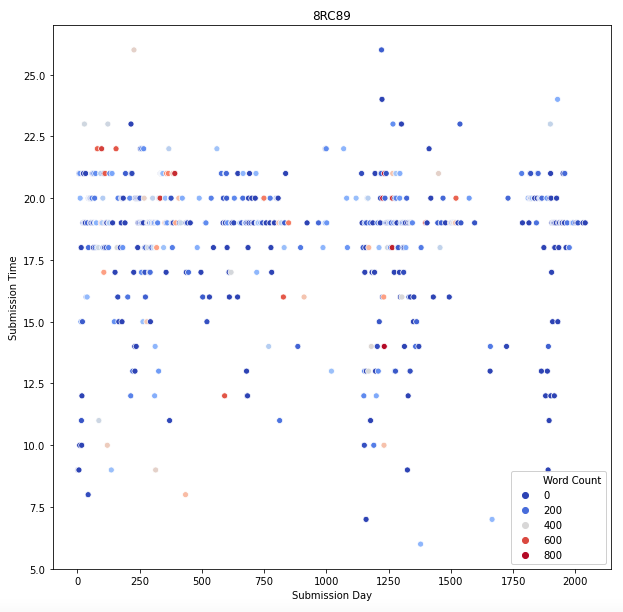}
\caption[\hspace{0.075cm} Submission day versus submission time with word count hue for BLS subject 8RC89.]{\textbf{Submission day versus submission time with word count hue for BLS subject 8RC89.} An analogous plot to the one created for participant 3SS93 in Figure \ref{fig:submit-time-3s} was also created for 8RC89 here.}
\label{fig:submit-time-8r}
\end{figure}

\begin{figure}[h]
\centering
\includegraphics[width=0.8\textwidth,keepaspectratio]{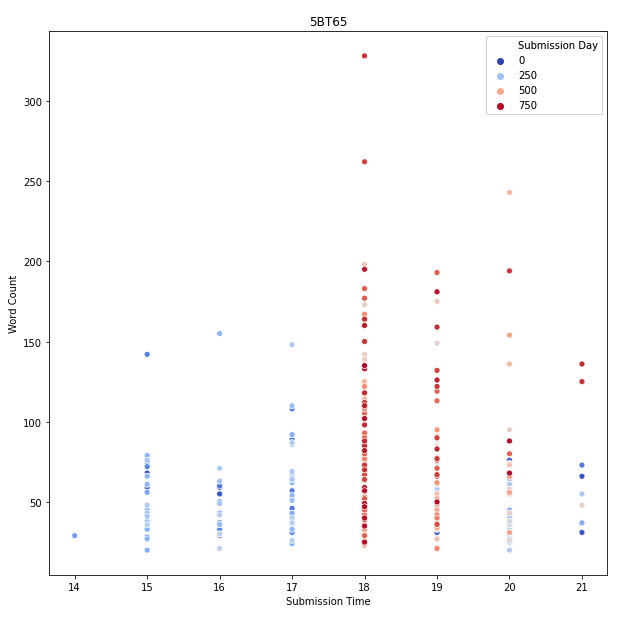}
\caption[\hspace{0.075cm} Submission time versus word count with study day hue for BLS subject 5BT65.]{\textbf{Submission time versus word count with study day hue for BLS subject 5BT65.} The methodology of Figure \ref{fig:diary-et-vs-words} is repeated here for only the diaries submitted by participant 5BT65. Thus this scatter represents a small subset of the points on the original Figure \ref{fig:diary-et-vs-words}, demonstrating an interesting subject-specific dynamic in the relationships between these features.}
\label{fig:submit-time-5b}
\end{figure}

\FloatBarrier

\subsubsection{Further characterization of modeling results: participant-dependent effects}
\label{sec:sup-stupid}

\begin{figure}[h]
\centering
\includegraphics[width=\textwidth,keepaspectratio]{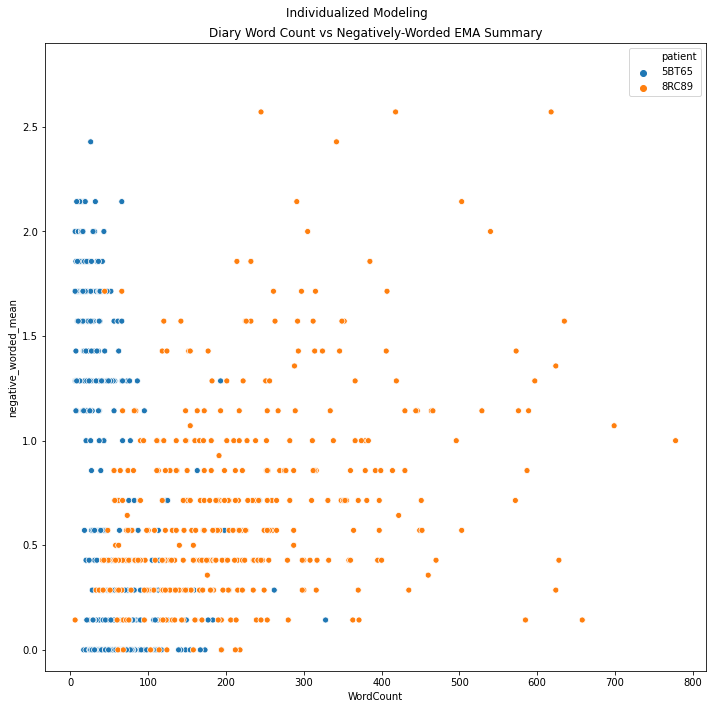}
\caption[\hspace{0.075cm} Subject-specific effects in word count versus negative EMA.]{\textbf{Subject-specific effects in word count versus negative EMA.} Figure \ref{fig:8r-5b-comp} from within chapter \ref{ch:1} is reproduced here without a limit on the x-axis, to show the full range of 8R's data.}
\label{fig:8r-5b-sup} 
\end{figure}

\begin{figure}[h]
\centering
\includegraphics[width=\textwidth,keepaspectratio]{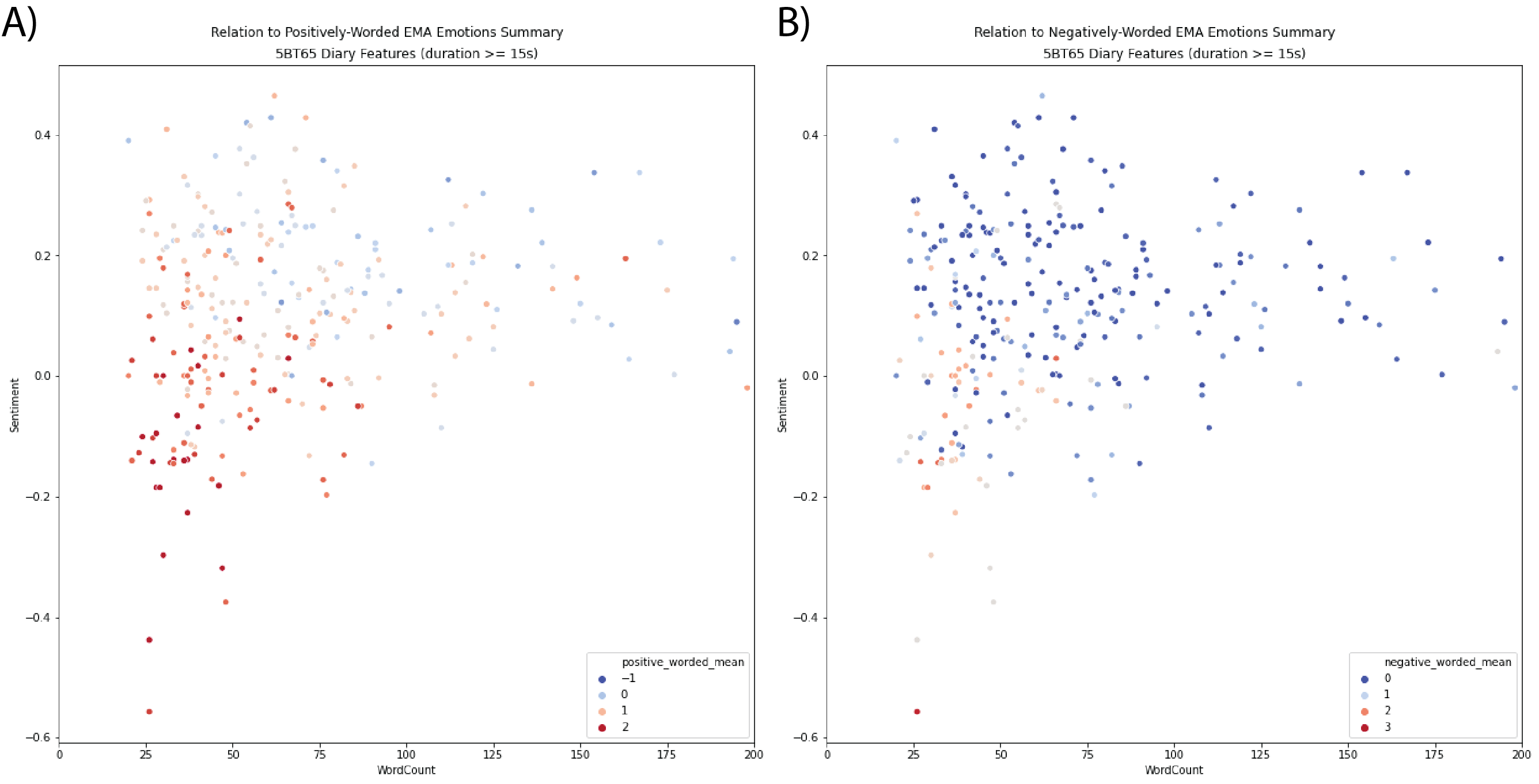}
\caption[\hspace{0.075cm} Diary word count and mean sentence sentiment had a strong nonlinear relationship with self-reported mood scores in longer 5BT65 recordings.]{\textbf{Diary word count and mean sentence sentiment had a strong nonlinear relationship with self-reported mood scores in longer 5BT65 recordings.} For the primary model training dataset, data points with journal duration $< 15$ seconds were filtered out, which removed a large chunk of records for subject 5B in particular. Despite this, word count remained relevant to EMA scores in participant-specific model fitting. Of the new feature set, sentiment was also found to be highly relevant. Although linear model fits were already quite strong, scattering word count and sentiment against each other with EMA hue revealed additional potential nonlinear structure. Here, diary word count is plotted on the x-axis and mean diary sentiment is plotted on the y-axis in each panel, across the 5B diaries in the duration-filtered modeling training set. Positively-worded (A) and negatively-worded (B) EMA item means color the two scatters using the coolwarm colormap with seaborn. The positive map (A) runs from -1 to 2 here and the negative map (B) from 0 to 3, as 5B rarely submitted extremely high levels of positive emotions or negative emotions.}
\label{fig:5b-ema-fin}
\end{figure}

\begin{figure}[h]
\centering
\includegraphics[width=0.75\textwidth,keepaspectratio]{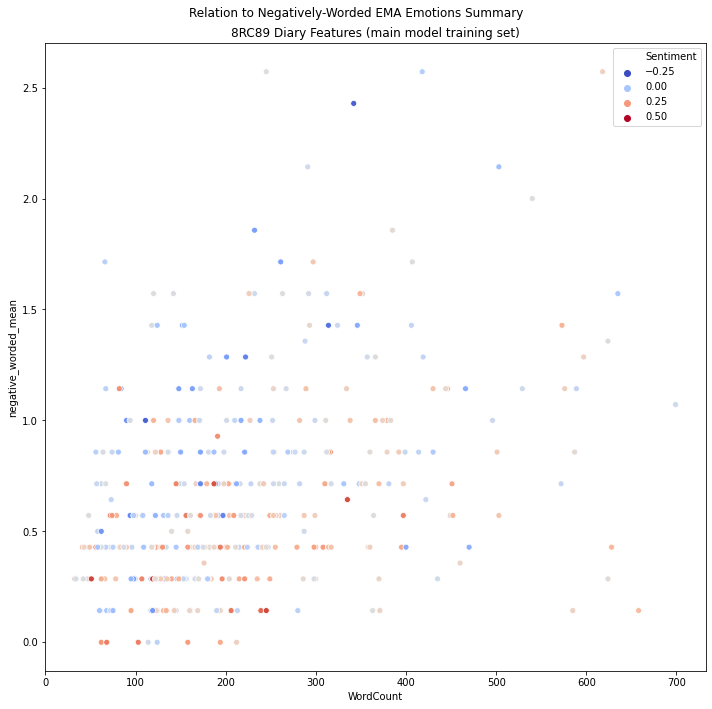}
\caption[\hspace{0.075cm} Adding sentiment context to the positive correlation between diary word count and 8RC89 negative emotions self-report.]{\textbf{Adding sentiment context to the positive correlation between diary word count and 8RC89 negative emotions self-report.} Word count (x-axis) of diary submissions by 8R in the main model training set was plotted versus mean negatively-worded EMA score (y-axis) from the corresponding day. This is very similar to the 8R relationship depicted in Figure \ref{fig:8r-5b-comp}, but points are now colored according to the diary's mean sentiment score, with range -0.25 to 0.5. It is clear here that sentiment can help to distinguish between journals with a similar word count in predicting negative EMA rating from 8R, though the word count relationship with EMA is stronger in this subject.}
\label{fig:8r-ema-fin}
\end{figure}

\FloatBarrier

\subsection{Other supplemental tables}

\subsubsection{Pipeline summary stats for BLS}
\label{subsec:bls-diary-summary-sup}

\begin{table}[!htbp]
\centering
\caption[Summary of diary QC metric distributions in BLS.]{\textbf{Summary of diary QC metric distributions in BLS.} The first three rows are the key audio quality metrics considered, with mean and standard deviation taken over 11,101 submitted recordings with defined volume (a small subset of the submitted diaries were completely empty and thus returned a dB of -inf). The final four rows are key transcript quality metrics, with mean and standard deviation taken over all 10,271 successfully transcribed diaries. For exact feature calculation details, see section \ref{subsec:diary-code}.}
\label{table:bls-qc-means}

\begin{tabular}{ | m{6cm} || m{4cm} | m{4cm} | }
\hline
\textbf{Feature} & \textbf{Mean} & \textbf{Standard Deviation} \\
\hline\hline
Duration (minutes) & 1.49 & 1.325 \\
\hline
Volume (dB) & 66.82 & 9.51 \\ 
\hline
Spectral Flatness & 0.0209 & 0.0804 \\ 
\hline\hline
Total Word Count & 180.3 & 156.2 \\
\hline
Inaudible Word Count & 0.34 & 1.11 \\ 
\hline
Questionable Word Count & 0.15 & 0.52 \\
\hline
Redacted Word Count & 0.78 & 1.7 \\
\hline
\end{tabular}
\end{table}

\begin{table}[!htbp]
\centering
\caption[Summary of highlighted audio feature distributions in BLS.]{\textbf{Summary of highlighted audio feature distributions in BLS.} The first three rows are diary-level features derived directly from the pause detection algorithm. The final five rows are diary-level features derived from the OpenSMILE results, with pause times filtered out. The raw OpenSMILE features are returned in 10ms bins, and pause times are used to label those bins that correspond to speech. Each of the features in the table represents the mean of these speech bins over the recording. For both the pause-derived features and the OpenSMILE features, the table presents the mean and standard deviation taken over all 10,271 successfully transcribed diaries (diaries unable to be transcribed were considered also too low quality for acoustic analyses). For exact feature calculation details, see section \ref{subsec:diary-code}.}
\label{table:bls-aud-means}

\begin{tabular}{ | m{6cm} || m{4cm} | m{4cm} | }
\hline
\textbf{Feature} & \textbf{Mean} & \textbf{Standard Deviation} \\
\hline\hline
\emph{Speech} Duration (minutes) & 1.040 & 0.836  \\
\hline
\emph{Pause} Duration (minutes) & 0.481 & 0.609 \\ 
\hline
Pause Count & 33.3 & 32.2 \\ 
\hline\hline
Mean Pitch (F0) & 20.03 & 3.97 \\
\hline
Mean F1 & 565.28 & 55.10 \\ 
\hline
Mean F2 & 1560.61 & 77.22 \\
\hline
Mean Jitter & 0.0407 & 0.0163 \\
\hline
Mean Shimmer & 1.025 & 0.242 \\
\hline
\end{tabular}
\end{table}

\begin{table}[!htbp]
\centering
\caption[Summary of highlighted transcript feature distributions in BLS.]{\textbf{Summary of highlighted transcript feature distributions in BLS.} The first row is an estimation of sentiment. The following four rows measure different types of linguistic disfluencies. The next two rows are features derived from semantic coherence metrics. The final row is an estimate of speaking rate based on computed syllables and the TranscribeMe timestamps. For all of these diary-level features, the table presents the mean and standard deviation taken over all 10,271 successfully transcribed diaries. For exact feature calculation details, see section \ref{subsec:diary-code}.}
\label{table:bls-trans-means}

\begin{tabular}{ | m{6cm} || m{4cm} | m{4cm} | }
\hline
\textbf{Feature} & \textbf{Mean} & \textbf{Standard Deviation} \\
\hline\hline
Mean Sentence Sentiment & 0.164 & 0.219  \\
\hline\hline
Nonverbal Edit Count & 8.52 & 8.60 \\ 
\hline
Verbal Edit Count & 2.09 & 4.25 \\ 
\hline
Restart Count & 3.03 & 4.26 \\
\hline
Repeat Count & 1.66 & 2.71 \\ 
\hline\hline
Mean Word Uncommonness & 2.27 & 0.20 \\
\hline
Mean Sentence Incoherence (pairwise) & 1.35 & 0.05 \\
\hline\hline
Mean Sentence Speaking Rate (syllables/second) & 3.14 & 1.48 \\
\hline
\end{tabular}
\end{table}

\FloatBarrier

\subsubsection{Participant-specific distributional comparison (KS test results)}
\label{subsec:ks-sup}

\begin{table}[!htbp]
\centering
\caption[Subject-specific differences observed in BLS audio journal feature distributions.]{\textbf{Subject-specific differences observed in BLS audio journal feature distributions.} For each of the highlighted BLS participants, the two sided Kolmogorov-Smirnov (KS) test, performed with the scipy.stats python package, was used to quantify how different the empirical distribution of each feature in the patient's dataset was from the empirical distribution of the same feature in the overall BLS dataset. The obtained KS D statistics are reported in this table, with one results column per considered subject ID and one results row per considered journal feature. For all D stats that met the conservative Bonferroni-corrected cutoff of $p \leq 0.0005$, p-value information is also given in parentheses. The vast majority of these comparisons were highly significant with large effect size, demonstrating that subject-specific factors are highly relevant to analysis of audio journal features.}
\label{table:bls-ks-test}

\begin{tabular}{ | m{8cm} || m{1.75cm} | m{1.75cm} | m{1.75cm} | }
\hline
\textbf{Feature Distribution Compared \newline (subject vs all BLS)} & \textbf{3SS93 \newline D stat} & \textbf{8RC89 \newline D stat} & \textbf{5BT65 \newline D stat} \\
\hline\hline
Word Count & 0.553 ($p<10^{-14}$) & 0.104 ($p<10^{-4}$) & 0.57 ($p<10^{-101}$) \\
\hline
Speaking Time/Recording Time & 0.724 ($p<10^{-5}$) & 0.198 ($p<10^{-16}$) & 0.282 ($p<10^{-22}$) \\
\hline
Mean Pause Length (minutes) & 0.71 ($p<10^{-6}$) & 0.27 ($p<10^{-30}$) & 0.219 ($p<10^{-13}$) \\
\hline
Speech Rate (syllables/second) & 0.636 ($p<10^{-8}$) & 0.488 ($p<10^{-105}$) & 0.291 ($p<10^{-24}$) \\
\hline
Restarts Count/Word Count & 0.476 ($p<10^{-15}$) & 0.105 ($p<10^{-4}$) & 0.352 ($p<10^{-36}$) \\
\hline
Repeats Count/Word Count & 0.509 ($p<10^{-15}$) & 0.096  ($p<0.0003$) & 0.348 ($p<10^{-35}$) \\
\hline
Nonverbal Edits Count/Word Count & 0.3 ($p<10^{-15}$) & 0.102 ($p<10^{-4}$) & 0.07 NS \\
\hline
Verbal Edits Count/Word Count & 0.276 ($p<10^{-15}$) & 0.148 ($p<10^{-8}$) & 0.501 ($p<10^{-76}$) \\
\hline
Mean Sentence Sentiment & 0.222 ($p<10^{-15}$) & 0.187 ($p<10^{-14}$) & 0.108 NS \\
\hline
Minimum Sentence Sentiment & 0.294 ($p<10^{-15}$) & 0.075 NS & 0.294 ($p<10^{-25}$) \\
\hline
Maximum Sentence Incoherence (pairwise) & 0.157 ($p<10^{-15}$) & 0.206 ($p<10^{-17}$) & 0.071 NS \\
\hline
Mean Word Uncommonness & 0.194 ($p<10^{-15}$) & 0.166 ($p<10^{-11}$) & 0.12 ($p<0.0002$) \\
\hline
\end{tabular}
\end{table}

\begin{table}[!htbp]
\centering
\caption[Using random simulation to identify subject-specific distributional differences attributable to verbosity (KS test results).]{\textbf{Using random simulation to identify subject-specific distributional differences attributable to verbosity (KS test results).} The same protocol used to compare the patient-specific feature distributions to the study-wide ones in Table \ref{table:bls-ks-test}, including the same significance cutoff, was repeated here on simulated datasets. Using the true distribution of diary sentence counts for each subject and for all of BLS, sentence-level feature values were randomly and independently sampled from underlying distributions designed to simulate the sentiment and incoherence features (exact methods detailed above). Distributions of analogous diary features were then obtained by taking the appropriate summary stats over the simulated data. This table thus presents the D stats obtained by running KS tests on the simulated subject-specific distributions versus the simulated study-wide ones. The stats here are from the final run of simulations, but for all iterations on the underlying sentence distribution parameters, every D stat was within $+/- 0.05$ of the numbers reported here, and the non-significant (NS) comparisons remained consistent. These D stats serve as an estimate of the contribution of verbosity to the observed D stats for the corresponding sentiment and coherence features in Table \ref{table:bls-ks-test}.}
\label{table:random-ks-test}

\begin{tabular}{ | m{8cm} || m{1.75cm} | m{1.75cm} | m{1.75cm} | }
\hline
\textbf{Feature Distribution Compared \newline (subject vs all BLS)} & \textbf{3SS93 \newline D stat} & \textbf{8RC89 \newline D stat} & \textbf{5BT65 \newline D stat} \\
\hline\hline
Randomized Mean Sentiment & 0.073 NS & 0.058 NS & 0.104 NS \\
\hline
Randomized Minimum Sentiment & 0.214 ($p<10^{-15}$) & 0.117 ($p=0.000003$) & 0.238 ($p<10^{-16}$) \\
\hline
Randomized Maximum Incoherence & 0.205 ($p<10^{-15}$) & 0.085 NS & 0.272 ($p<10^{-21}$) \\
\hline
Randomized Mean Uncommonness & 0.09 ($p<10^{-5}$) & 0.072 NS & 0.084 NS \\
\hline
\end{tabular}
\end{table}

\FloatBarrier

\subsubsection{Participant-specific submission stats}
\label{subsec:participation-table-sup}

\begin{table}[!htbp]
\centering
\caption[Diary participation stats for top BLS subjects - frequency of back to back submissions.]{\textbf{Diary participation stats for top BLS subjects - frequency of back to back submissions.} For each of the $25$ patients with at least 100 journals submitted of sufficient length, summary stats were computed about the temporal dynamics of the submissions. The first two counts provided are the total number of recordings $\geq15$ seconds submitted by that subject, and then the number of those days where the subject also submitted a $\geq15$ second recording the following day. The next two counts are the same statistics but restricted to only the first year of the study period for each participant. Then the final four columns are fractions derived from these counts: the observed probability of diary submission given a diary was submitted the prior day, the same probability given it is also year one, the fraction of days in year one where a diary was submitted, and the fraction of the total dataset for that subject that is covered by year one. Heterogeneity in journal participation behavior is observed between participants.}
\label{table:bls-submit-probs}

\begin{tabular}{ | m{1.5cm} || m{1cm} | m{1cm} | m{1.35cm} | m{1.35cm} || m{1cm} | m{1.35cm} || m{1.25cm} | m{1.25cm} | }
\hline
\textbf{Patient ID} & \textbf{Diary Count} & \textbf{Next Day Count} & \textbf{Diary Count \newline (Year 1)} & \textbf{Next Day Count \newline (Year 1)} & \textbf{Next Day / Total} & \textbf{Next Day / Total \newline (Year 1)} & \textbf{Year 1 / $365$} & \textbf{Year 1 / Total} \\
\hline\hline
3SS93 & 1287 & 1233 & 345 & 325 & 0.958 & 0.942 & 0.945 & 0.268 \\
\hline
7NE49 & 744 & 436 & 164 & 89 & 0.586 & 0.543 & 0.449 & 0.22 \\
\hline
SBPQM & 560 & 351 & 268 & 181 & 0.627 & 0.675 & 0.734 & 0.479 \\
\hline
8RC89 & 515 & 172 & 182 & 79 & 0.334 & 0.434 & 0.499 & 0.353 \\
\hline
GMQBM & 451 & 265 & 211 & 134 & 0.588 & 0.635 & 0.578 & 0.468 \\
\hline
2XC43 & 369 & 160 & 129 & 59 & 0.434 & 0.457 & 0.353 & 0.35 \\
\hline
5BT65 & 345 & 216 & 141 & 72 & 0.626 & 0.511 & 0.386 & 0.409 \\
\hline
5KXYM & 341 & 291 & 314 & 275 & 0.853 & 0.876 & 0.860 & 0.921 \\
\hline
CYG4M & 333 & 222 & 231 & 163 & 0.667 & 0.706 & 0.633 & 0.694 \\
\hline
GFNVM & 329 & 200 & 231 & 137 & 0.608 & 0.593 & 0.633 & 0.702 \\
\hline
DQDCM & 311 & 165 & 156 & 94 & 0.531 & 0.603 & 0.427 & 0.502 \\
\hline
8ADCM & 266 & 153 & 134 & 67 & 0.575 & 0.5 & 0.367 & 0.504 \\
\hline
Z8YRM & 262 & 198 & 262 & 198 & 0.756 & 0.756 & 0.718 & 1.0 \\
\hline
M8MXM & 256 & 176 & 247 & 174 & 0.688 & 0.705 & 0.677 & 0.965 \\
\hline
V8GMM & 203 & 154 & 203 & 154 & 0.759 & 0.759 & 0.556 & 1.0 \\
\hline
8KX53 & 199 & 93 & 178 & 90 & 0.467 & 0.506 & 0.488 & 0.895 \\
\hline
ACMCM & 170 & 84 & 163 & 79 & 0.494 & 0.485 & 0.447 & 0.959 \\
\hline
B3CRM & 147 & 97 & 147 & 97 & 0.66 & 0.66 & 0.403 & 1.0 \\
\hline
5QXNM & 130 & 27 & 65 & 14 & 0.208 & 0.215 & 0.178 & 0.5 \\
\hline
5CR39 & 126 & 98 & 126 & 98 & 0.778 & 0.778 & 0.345 & 1.0 \\
\hline
TQJKM & 117 & 63 & 117 & 63 & 0.539 & 0.539 & 0.321 & 1.0 \\
\hline
9SU83 & 111 & 41 & 88 & 33 & 0.369 & 0.375 & 0.241 & 0.793 \\
\hline
V65HM & 109 & 68 & 95 & 61 & 0.624 & 0.642 & 0.26 & 0.872 \\
\hline
JVBKM & 108 & 69 & 108 & 69 & 0.639 & 0.639 & 0.296 & 1.0 \\
\hline
F6VVM & 103 & 63 & 103 & 63 & 0.612 & 0.612 & 0.282 & 1.0 \\
\hline
\end{tabular}
\end{table}

In addition to daily submission availability, I also looked at patterns in submission timing and potential relation with resulting diary content richness, as mentioned in the main chapter section \ref{subsubsec:diary-time}. For each possible Eastern Time (ET) submission hour, Table \ref{table:bls-submit-times} presents the total number of diaries in the final BLS dataset assigned to that hour bin, as well as the mean and standard deviation of the total word counts of all such diaries. Recall that recordings submitted prior to 4am were counted to the previous day, so that assigned hours ranged from 4 to 27 - thereby ensuring that late night submissions would not be counted as early morning in the correlational analyses mentioned.

\begin{table}[!htbp]
\centering
\caption[Diary metadata stats by submission time in the final BLS dataset.]{\textbf{Diary metadata stats by submission time in the final BLS dataset.} The number of diaries assigned to each possible submission time integer is reported here; the diary counts match the number of points that fall onto that submission time in Figure \ref{fig:diary-et-vs-words}. The mean and standard deviation of diary word counts across the subset of each submission hour is also reported, to provide additional context for Figure \ref{fig:diary-et-vs-words} - as overlapping points are not clearly visible on the scatter.}
\label{table:bls-submit-times}

\begin{tabular}{ | m{2cm} | m{2cm} | m{2cm} | m{2cm} | m{2cm} | }
\hline
\textbf{Submit Time \newline (integer)} & \textbf{Submit Time (ET)} & \textbf{Diary Count} & \textbf{Mean Word Count} & \textbf{StDev Word Count} \\
\hline\hline
4 & 4am & 98 & 210.2 & 164.1 \\
\hline
5 & 5am & 86 & 210.8 & 186.4 \\
\hline
6 & 6am & 73 & 218.5 & 214.5 \\
\hline
7 & 7am & 94 & 190.1 & 163.6 \\
\hline
8 & 8am & 144 & 224.1 & 160.1 \\
\hline
9 & 9am & 282 & 240.6 & 140.6 \\
\hline
10 & 10am & 240 & 233.4 & 176.2 \\
\hline
11 & 11am & 295 & 193.1 & 174.4 \\
\hline
12 & 12pm & 474 & 166.5 & 159 \\
\hline
13 & 1pm & 357 & 225.3 & 182.6 \\
\hline
14 & 2pm & 270 & 224 & 190.3 \\
\hline
15 & 3pm & 239 & 181.5 & 155.6 \\
\hline
16 & 4pm & 381 & 212.3 & 136.3 \\
\hline
17 & 5pm & 349 & 186.4 & 140.4 \\
\hline
18 & 6pm & 477 & 160.3 & 147.7 \\
\hline
19 & 7pm & 521 & 180.5 & 146.3 \\
\hline
20 & 8pm & 804 & 184.8 & 142.4 \\
\hline
21 & 9pm & 579 & 203.5 & 142.4 \\
\hline
22 & 10pm & 675 & 205.8 & 121.9 \\
\hline
23 & 11pm & 520 & 235.4 & 137.7 \\
\hline
24 & 12am (next day) & 407 & 250.1 & 138.1 \\
\hline
25 & 1am (next day) & 475 & 293 & 107.9 \\
\hline
26 & 2am (next day) & 393 & 286.3 & 109.4 \\
\hline
27 & 3am (next day) & 165 & 243.2 & 128.4 \\
\hline
\end{tabular}
\end{table} 

\FloatBarrier

\section{Additional information on data collection methods}
The BLS protocol was sanctioned by the Mass General Brigham Institutional Review Board. Protocol Number: 2015P002189 Circuit dynamics underlying longitudinal fluctuations in mood and cognition in bipolar patients, Principal Investigator Name: Baker, Justin T.

\subsection{BLS recruitment and demographics}
Patients interested in BLS were screened for eligibility over the phone by a trained research assistant. The only inclusion criterion was current diagnosis of a severe affective or psychotic disorder, established based on the Structured Clinical Interview for Diagnosis of DSM-V Disorders. Schizophrenia, Schizoaffective Disorder, Bipolar Disorder (with or without psychosis), and Major Depressive Disorder were among the eligible diagnoses. Individuals under legal or mental incompetence or legal guardianship were excluded. In order to partake in the audio diary collection portion of the study, a prospective patient must have also owned their own smartphone capable of downloading the Beiwe app. The resulting study-wide demographics breakdown can be found in Table \ref{table:bls-demographics}.

\begin{table}[!htbp]
\centering
\caption[BLS demographics.]{\textbf{BLS demographics.} From the time of enrollment, for n = 74 enrolled study participants. Reproduced from \citep{disorg22}.}
\label{table:bls-demographics}

\begin{tabular}{ | m{5cm} || m{2cm} | m{2cm} | }
\hline
& \textbf{Count} & \textbf{Percent} \\
\hline\hline
\rowcolor{Gray}
\textbf{Sex} & & \\
\hline
Female & 44 & 59.5\% \\
\hline
Male & 30 & 40.5\% \\ 
\hline\hline
\rowcolor{Gray}
\textbf{Race} & & \\
\hline
African American & 8 & 10.8\% \\
\hline
American Indian & 1 & 1.4\% \\ 
\hline
Asian & 9 & 12.2\% \\
\hline
White & 52 & 70.3\% \\
\hline
Not Reported & 4 & 5.4\% \\
\hline\hline
\rowcolor{Gray}
\textbf{Education} & & \\
\hline
4 Years College & 25 & 33.8\% \\
\hline
Part College & 31 & 41.9\% \\ 
\hline
Professional School & 11 & 14.9\% \\
\hline
High School & 5 & 6.8\% \\
\hline
Not Reported & 2 & 2.7\% \\
\hline\hline
\rowcolor{Gray}
\textbf{Age} & & \\
\hline
48-51 & 3 & 4.1\% \\
\hline
40-47 & 3 & 4.1\% \\ 
\hline
30-39 & 11 & 14.9\% \\
\hline
24-29 & 24 & 32.4\% \\
\hline
18-23 & 23 & 31.1\% \\
\hline
Not Reported & 10 & 13.5\% \\
\hline
\end{tabular}
\end{table}

\FloatBarrier

\subsection{Patient diary collection tips}
\label{sec:diary-tips-supp}
When recording an audio diary, remember to:

\begin{enumerate}
    \item Keep your device level with your voice throughout the entire recording 
    \item Limit the movement of your device throughout the entire recording (keep your device steady)
    \item Maintain a small distance between your mouth and the microphone on your device throughout the entire recording 
    \item Attempt to limit background noise as much as possible throughout the entire recording 
\end{enumerate}

\noindent Ideally, when recording an audio diary, the participant will be seated in an area with limited background noise. Being seated allows for the participant to limit movement of the device and subsequent feedback from the device. 

\subsection{EMA survey design for BLS}
\label{sec:bls-ema-qs}
Through May 2022, ecological momentary assessment (EMA) self-report surveys were collected from BLS participants using the described Beiwe app \citep{Beiwe}. The primary survey was a daily questionnaire with 30 items, where each item was a radio button (i.e. select a single option) prompt with 5 response choices ranging from strong disagreement to strong agreement, though individual questions could also be skipped. The aim was for participants to submit a survey response daily, and as such payment was based on daily completion. However, participants were free to submit multiple responses to the same survey in a day if desired. Additionally, supplemental surveys were administered in select situations that could be completed for extra compensation. 

Here, I provide detailed information on the EMA items used across different surveys in BLS, broken into subsections. Where applicable, I note areas where EMA design or Beiwe encoding methodology varied over time, and discuss related caveats to be aware of in processing these data. While the dataset used in the EMA modeling of chapter \ref{ch:1} was carefully curated to include only some of the most consistent and clean EMA items with good availability, understanding the properties of the broader dataset is still important for future study and analysis design.

\subsubsection{30 question EMA}
As mentioned, the 30 question was the original and primary EMA survey used in BLS; though there were a few changes in the design over time, the bulk of the questions remained very similar through the duration of the Beiwe study, and many BLS participants submitted answers to these items regularly for multiple years (which does bode well for EMA usability given the length of the survey). The 30 question survey always began with the following info box, before launching into the items.

\begin{quote}
    Over the past 24 hours how much were you:
\end{quote}

\noindent Each item was then a radio button question, many of which had the following standardized options.
\begin{itemize}
    \item Very slightly or not at all
    \item A little
    \item Moderately
    \item Quite a bit
    \item Extremely
\end{itemize}

\noindent The following 15 items used the above mentioned standardized radio button options -
\begin{itemize}
    \item Happy
    \item Stressed
    \item Hostile
    \item Irritable
    \item Alert
    \item Ashamed
    \item Inspired
    \item Determined
    \item Upset
    \item Afraid
    \item Lonely
    \item Outgoing
    \item Bothered by hearing voices
    \item Bothered by seeing things others could not
    \item Bothered by feeling like other people are out to get you or cause you trouble
\end{itemize}

\noindent There were also 17 items that had their own unique radio button options (usually with 5 options but sometimes more), as follows - 
\begin{itemize}
    \item Energetic
    \begin{itemize}
        \item Little energy or motivation to do much of anything
        \item Enough energy to get by but not enough to be very productive
        \item Typical energy level with usual productivity
        \item Plenty of energy to be even more productive than usual
        \item Unusually high energy feeling hyper or even agitated at times
    \end{itemize}
    \item Anxious
    \begin{itemize}
        \item Completely relaxed
        \item Relaxed
        \item Typical some moments of anxiety 
        \item High anxiety
        \item Extreme anxiety that dominated my thoughts
    \end{itemize}
    \item Able to manage stress
    \begin{itemize}
        \item Very unsuccessful
        \item A little bit successful
        \item Moderately successful
        \item Very successful 
        \item Extremely successful
    \end{itemize}
    \item Social in Person
    \begin{itemize}
        \item I spent almost all of my time alone
        \item I interacted with others but little more than superficial interactions
        \item I interacted with others in a meaningful way
        \item I extensively interacted with close friends or family or a significant other
        \item I experienced an unusually deep connection with another person
    \end{itemize}
    \item Social Digitally
    \begin{itemize}
        \item I spent almost no time interacting with others on my phone or computer
        \item I interacted with others but little more than superficial messages
        \item I interacted with others on my phone or computer in a meaningful way
        \item I extensively interacted with close friends or family or significant other
        \item I experienced an unusually deep connection through a digital interaction
    \end{itemize}
    \item Physically Active
    \begin{itemize}
        \item Minimal movement (didn't get off the couch)
        \item Little more than getting around my living space 
        \item Only did what was necessary (walking to do errands)
        \item Exercised 30 min or less 
        \item Exercised 60 min or less
        \item Exercised more than 60 min
    \end{itemize}
    \item Hungry
    \begin{itemize}
        \item Not hungry at all and ate much less than I normally would
        \item Not as hungry as usual and ate slightly less than normal
        \item Typical level of hunger and ate all my normal meals and foods
        \item More hungry than usual and ate a little more than usual (extra portion or snack)
        \item Surprisingly hungry and ate a lot more than usual (extra meal or binge)
    \end{itemize}
    \item Bothered by your stomach
    \begin{itemize}
        \item Not at all
        \item It was a little upset
        \item It was mildly upset
        \item It was very upset
        \item Severe illness
        \item I did not notice one way or the other
    \end{itemize}
    \item Experiencing your heart racing
    \begin{itemize}
        \item I am not generally aware of my heart rate
        \item I am aware of my heart rate and did not notice any changes
        \item I noticed my heart racing after physical exertion (stairs or exercise)
        \item My heart was racing but I was able to control it and it went away
        \item My heart was racing and I felt unable to control it
    \end{itemize}
    \item Experiencing difficulty breathing
    \begin{itemize}
        \item I am not generally aware of my breathing
        \item I am aware of my breathing and did not notice any changes
        \item I noticed I was out of breath after physical exertion (stairs or exercise)
        \item I noticed I had difficulty breathing periodically throughout the day
        \item I felt unable to take deep breaths or catch my breath for most of the day
    \end{itemize}
    \item Having a headache or other physical pain
    \begin{itemize}
        \item Not at all
        \item A little but it did not interfere with my day
        \item I had some mild pain that stopped me from engaging in certain activities
        \item The pain interfered with my ability to work or do other activities
        \item The pain was severe enough to not leave my house or for me to seek medical attention
    \end{itemize}
    \item Consuming Caffeine
    \begin{itemize}
        \item None
        \item One cup of coffee or tea or soft drink
        \item Two cups of coffee or tea or soft drink
        \item Three cups of coffee or tea or or soft drink
        \item More than three cups of coffee or tea or soft drink
    \end{itemize}
    \item Consuming Alcohol
    \begin{itemize}
        \item None
        \item One drink
        \item Two drinks
        \item Three to five drinks
        \item Five or more drinks
    \end{itemize}
    \item Consuming cannabis or CBD
    \begin{itemize}
        \item Zero times
        \item Once
        \item Twice
        \item Three times
        \item Four or more times
    \end{itemize}
    \item Sleeping
    \begin{itemize}
        \item Terribly: little or no sleep
        \item Not so well: got some sleep but not enough
        \item Sufficient: got enough sleep to function
        \item Good: got a solid night's sleep and felt well rested
        \item Overslept: slept more than usual and/or feel groggy
    \end{itemize}
    \item Taking medications
    \begin{itemize}
        \item None/I don’t usually take prescribed medication
        \item Less than usual (forgot/skipped or reduced a dose)
        \item As prescribed (same as usual)
        \item More than usual (increased dose or took an extra dose)
        \item Substantially more than usual (took 2 or more extra compared to the usual amount)
    \end{itemize}
    \item Menstruating
    \begin{itemize}
        \item No 
        \item Uncertain (may be just beginning)
        \item Yes with light flow or minimal cramping
        \item Yes with medium flow or moderate cramping
        \item Yes with heavy flow or very painful cramping
        \item Not Applicable
    \end{itemize}
\end{itemize}

\noindent Note that 32 questions are listed here -- the survey began with 30 questions, but a couple more were added over time. A handful of questions also had slight wording changes over time, usually to fix typos or grammatical errors. The EMA questions focused on in the modeling of chapter \ref{ch:1} (drawn from within these 30) were not altered, but any future analyses that consider more of the EMA items will need to take care to document all of the wording changes that did occur and account for any that might have scientific impact. Changes occurred at times in both question prompt and answer option wording. \\

\paragraph{More on analysis caveats.}
A number of challenges in analyzing the BLS EMA responses could be prevented through more careful study setup and improvement (or alternatives) to Beiwe conventions. Question IDs issued by Beiwe do not necessarily align across subjects even when the prompted questions have identical text and options, and they also did not necessarily align across time in a subject if the question was edited. It is not clear to me whether this is a Beiwe issue or an issue in the way the lab utilized it, but nevertheless it means that those IDs are functionally useless in aligning questions. It is thus important to minimize unnecessary changes mid-study, which would be easier if initial survey design was not only more carefully reviewed for typos, but also tested on both the app interface and the resulting exported data prior to rollout. 

Some of the inconsistencies in survey design lead to small discrepancies that could have had a large impact on results if not for careful data cleaning review. For example, some questions had an explicit "N/A" option at one point, independent of Beiwe's function for marking empty items (which directly specifies not presented or no response as appropriate). The problem is that this option was encoded by default in different ways at different times, and either of these ways was a potential problem -- marking it as a 0 would shift the actual strongly disagree response up to a 1 encoding, and marking it as a 5 would imply it is a high agreement response by convention, if one focused on only the numbers and did not cross check the text. 

More generally, it is important to think carefully about question wording and how the provided answer options will scale. Though asking about abnormally high energy is an interesting topic for a Bipolar disorder cohort, it also created an odd dynamic in the analysis as the answer options followed the trend of most other questions in being monotonic with respect to symptom severity (as it pertains to low energy in depression here), but then abruptly switched to being a different type of symptom marker with the final option. While it is important to keep the number of questions asked reasonably low, it is also important to obtain a clean dataset, and for such a relevant symptom in both directions it would perhaps have been better to split out the item. 

Furthermore, wording is of course highly important to the actual responses elicited and thus the overall dataset quality. Some of the items included in the 30 question BLS EMA that I did not use had in my opinion quite shaky wording. For example, there is a big difference between using cannabis (which implies THC content) and using pure CBD, so responses to a question asking about cannabis/CBD use are not especially meaningful. The answer options to that question were similarly oddly defined, because no units that would be typically associated with a daily amount of marijuana use were supplied. 

Similarly, items that centered their options around "typical" amounts are more difficult to interpret; although we are more concerned with longitudinal fluctuations in an individual rather than comparing across individuals, "typical" can change drastically over the course of a few years, and so questions focused on this may be less comparable across the entire study timeline. It is akin to the "exceeds expectations" model of performance review at many tech companies, where for some managers it becomes difficult to keep exceeding expectations, as they literally adjust their expectations each quarter. EMA questions benchmarked in this way can still have appropriate use cases, but it is an important discussion to have, particularly when other questions asking about different symptoms in the same EMA set will not be using that wording. 

Ideally issues like the above could have come up in a careful initial design phase for the EMA surveys, rather than after the dataset was already collected and data cleaning/filtering needed to occur -- so this is an important lesson for future studies, both in the EMA design process and in how critical it is to still perform careful data review as automated techniques remove some of the labor burden for research data collection. Automated techniques can at times introduce their own data cleanliness issues, e.g. the requirements for writing Beiwe questions that do not interfere with the automatic CSV export process for the submitted responses. The questions themselves should not contain commas, so the lab has mostly used semi-colons. However for the answer option text, it is important to not use semi-colons either, because Beiwe export denotes the split between different answer options in the CSV column using semi-colons. 

Broadly, it is worth paying careful attention to details like punctuation early on in the study process, as well as performing testing from start to finish of EMA prompting and data compilation. Considering potential nuances of the intended statistical analyses and the impact of wording on patient response in parallel would likely prove fruitful too. Such planning might also involve testing of multiple app platforms for administering the EMA surveys (and other digital psychiatry data collection).

\subsubsection{7 question EMA}
More recently, another daily EMA survey was designed for BLS, this one containing only 7 items -- all radio button questions with the same 5 answer options. Note that 4 of the 7 items have analogous questions on the 30 count EMA. 

\noindent The 7 question EMA began with the info box prompt "Over the past day; how \_\_\_\_\_\_ did you feel?", and then presented the following items:
\begin{itemize}
    \item Happy
    \item Tense
    \item Relaxed
    \item Energetic
    \item Stressed
    \item Sad
    \item Fatigued
\end{itemize}

\noindent The response options for the 7 item survey were consistent across questions, and only slightly different from the generic 5 options provided for half of the 30 question survey EMA items: 
\begin{itemize}
    \item Not at all
    \item A little
    \item Moderately
    \item Quite a bit
    \item Extremely 
\end{itemize}

\noindent As this briefer EMA was more recently designed, availability skews towards newer subjects, and thus has minimal overlap with the Beiwe audio journal dataset that we transcribed for BLS. For that reason, the 7 question EMA was not used in any of the analyses of chapter \ref{ch:1}, despite being a cleaner dataset than the older 30 question EMA described above. \\

\paragraph{Shortening EMA.}
Note that Beiwe also provides a checkbox option for question answers instead of a radio button, which means that each option can be independently ticked off or not, and not checking a given box is an implicit no. This does confound lack of response with an actual no response, and it lacks some of the resolution that the radio button questions can have. At the same time, it can compact a number of items into one question, making it easier for participants to respond to more different prompts. An example that was used briefly in the BLS dataset was the following prompt --
\begin{quote}
    Consider how you have felt over the past 24 hours and check all that apply:
\end{quote}
\noindent with the following options --
\begin{quote}
    Irritable; Energetic; Hostile; Inspired; Ashamed; Afraid; Upset; Happy; Lonely; Outgoing; Alert
\end{quote}
\noindent such that Beiwe would return a subset of this list that indicates which answers were selected. That could then be automatically converted into the same format as the related items in the previously described EMA, but instead encoding each as a binary variable. 

\subsubsection{Patient-specific EMA supplements}
Another option enabled by Beiwe that was used at times in BLS is the administration of surveys about specific topics to target patients. One type of patient-specific supplement administered in BLS was an EMA about drug usage, which would be set up to ask about a specific substance and prompts directed to only relevant participants. This survey used a radio button format but with only yes/no answers for each question, along with a final free form response box - another EMA feature that could be better leveraged in some future studies.   

\noindent More specifically, the drug survey would ask for some particular substance X:
\begin{quote}
    In the past 24 hours have you used X in any form?
\end{quote}
 
\noindent If yes, the following additional yes/no radio button questions would be presented:
\begin{itemize}
    \item Did you use X for the sole purpose of avoiding withdrawal symptoms?
    \item Would the amount of X you used be considered excessive?
    \item Did you experience any side effects?
    \item Did you feel guilty about using X?
\end{itemize}
\noindent If no, instead an open ended question "What happened to make you choose not to use X?" would be presented with a text box, and then a few other yes/no items:
\begin{itemize}
     \item Did you experience any cravings or felt like you needed to use X?
    \item Did you experience any withdrawal symptoms?
\end{itemize}
\noindent In BLS, this survey was most commonly administered to ask about tobacco use. \\

Note other types of participant-specific surveys with a similar format included items centered around about appetite/eating behavior, and items centered around experienced hallucinations. While this sort of EMA has the potential to be highly relevant in practice, it was not collected consistently or robustly enough during BLS to be used in larger scale analyses. It is good for future study design to be aware of the many possible EMA options however. 

\subsubsection{COVID EMA supplement}
An advantage of Beiwe EMA administration is that it was feasible to release a new type of survey mid-study, in order to capture feelings and behaviors related to the COVID-19 pandemic. While analysis of these EMAs was beyond the scope of the present thesis, they could be particularly interesting when considered in conjunction with passive sensing data collected around the same time. The COVID surveys were weekly EMAs typically prompted on Sundays, and involved both radio button can checkbox items.

\noindent The radio button questions included the following, rated in response to "Over the past week..."
\begin{itemize}
    \item To what extent has COVID-19 affected your work or school situation? 
    \item To what extent has COVID-19 affected your living situation?
    \item To what extent has COVID-19 affected your physical health?
    \item To what extent have you been following recommendations to prevent spread of COVID-19?
\end{itemize}
\noindent with the following answer options available:
\begin{itemize}
    \item Not at all
    \item A little
    \item Moderately
    \item Quite a bit
    \item Extremely
    \item N/A - Prefer not to answer
\end{itemize}
\noindent For each of the above questions there were also corresponding checkbox question(s) for further elaboration. Those questions were as follows, with the answer options for each contained in the associated nest listed (the participant being prompted was asked to "check all that apply"):
\begin{itemize}
    \item Which of the following are true about your work or school situation?
    \begin{itemize}
        \item Working or taking classes from home with full load as before COVID19
        \item Working or taking classes from home with reduced load from before COVID19
        \item Working away from home
        \item Lost job or was furloughed
        \item Started a new job or educational program
        \item Not working - same as before COVID19
    \end{itemize}
    \item Which of the following are true about your living situation?
    \begin{itemize}
        \item Living alone with no pets or roommates
        \item Living mainly on my own but with a pet or a roommate
        \item Living with a significant other or family member(s)
        \item Living with someone (e.g. child or elder) for whom I am the primary or sole caregiver
        \item Living with someone (e.g. child or elder) where I am responsible for some caregiving
        \item Living with someone who is my primary caregiver
    \end{itemize}
    \item Which of the following are true about your relationships?
    \begin{itemize}
        \item Increased interpersonal stress with someone you live with
        \item Improved interpersonal relationships with someone you live with
        \item Increased interpersonal stress with someone you are not living with
        \item Improved interpersonal relationships with someone you are not living with
        \item Had a breakup or ended a longterm relationship
        \item Started a new relationship
    \end{itemize}
    \item Which of the following are true about your physical health? Viral symptoms typically include fever; body aches; cough; fatigue; chills; digestive issues.
    \begin{itemize}
        \item I am not experiencing viral symptoms and as far as I know have not acquired COVID19
        \item I am not currently experiencing viral symptoms but I know or suspect I have acquired COVID19
        \item I am currently experiencing mild-moderate viral symptoms which I know or suspect are due to COVID19
        \item I am currently experiencing severe viral symptoms which I know or suspect are due to COVID19
        \item Someone in my home or close physical contacts has acquired COVID19
        \item I previously had COVID19 but am no longer experiencing symptoms
    \end{itemize}
    \item Which of the following behaviors have you been performing to prevent the spread of COVID-19?
    \begin{itemize}
        \item Staying home as much as possible
        \item Avoiding indoor social gatherings with anyone who does not live with me
        \item Avoiding outdoor social gatherings with anyone who does not live with me
        \item Avoiding any places with 10 or more people
        \item Staying at least 6 feet away from other people who do not live with me
        \item Avoiding physical contact with other people who do not live with me
    \end{itemize}
    \item Over the same period, which of the following behaviors have you been performing?
    \begin{itemize}
        \item Preventing contact with others who live with me
        \item Using protective gear (e.g. masks or face covering and/or gloves) when around those who live with me
        \item Using protective gear (e.g. masks and/or gloves) when around those who do not live with me
        \item Isolation in a medical facility
        \item Self-monitoring symptoms (e.g. regularly taking temperature)
        \item Washed hands with soap and water whenever I return home
    \end{itemize}
\end{itemize}

\section{Analysis method details}

\subsection{Distributional comparisons}
\label{subsec:dist-comp-methods}
 I compared the distributions of the three subsets with the reported distributions of the full BLS dataset. For each of the 12 key features, I generated violin plots and kernel density estimate plots using the seaborn python package to visually compare the distribution of that feature across these 4 datasets. I also performed a two sided Kolmogorov-Smirnov (KS) test using the scipy python package, to measure the difference between the distribution of each feature in each highlighted participant with the distribution of that feature across the entire final dataset (Table \ref{table:bls-ks-test}). To correct the KS test results for multiple testing, I used a conservative significance cutoff of $p=0.0005$. This was based on the Bonferroni-based corrected cutoff for 3 comparisons on each of 12 features: $\frac{0.05}{12*3} \approx 0.00139$, which I then divided by 2 to account for the two sidedness of the test and finally rounded down. Table \ref{table:bls-ks-test} uses the $p=0.0005$ cutoff to mark the many subject-specific distributions that are significantly different than the study-wide one.

 The KS test can be used to compare empirical distributions of different sizes and can capture distributional differences related to both shape and location. The returned $D$ statistic represents the largest observed distance magnitude between the cumulative distribution functions of the two empirical datasets, and therefore ranges between 0 and 1. The corresponding $p$ value represents the probability of observing a $D$ statistic at least as large as actually was, if two samples of the same sizes were drawn from the same underlying population.

 \subsubsection{Simulated distributions based on diary lengths}

 I defined the sentence-level feature distributions to be used with random sampling based on the corresponding ones reported for interviews in chapter \ref{ch:2}, along with the properties of the sentiment and incoherence features observed in the diaries. I then used the real distribution of sentence counts over diaries in the present datasets (i.e. the 8398 BLS-wide diary time points, and the 3SS93, 8RC89, and 5BT65 subsets of size 981, 510, and 344 respectively) to create a dataset of fake "diaries" as described. In the simulated dataset, each record thus had a sentence count corresponding to an actual diary in the real dataset, and for each sentence in that record there was an associated sentiment score, word uncommonness score, and pairwise incoherence score, which had been obtained from random sampling of the defined distributions. 

Simulated distributions of diary-level features were then obtained directly from the "diary" records by taking the mean over the randomly sampled sentiment and uncommonness scores, as well as the minimum of the sentiment scores and the maximum of the incoherence scores of each "diary". I ran KS tests (Table \ref{table:random-ks-test}) and generated violin plots and KDE plots (Figure \ref{fig:random-dists-per-pt}) comparing the subject-specific simulated distributions to the BLS-wide simulated distributions, just as was done for the real distributional comparisons above. Those results were subsequently utilized to inform interpretation of the corresponding results from the real diary feature distributions. 

The simulation protocol was run 5 separate times, each with slightly different settings in the definitions of the underlying sentence distributions. However, the results were very similar each time, so only the final round of simulated distributions (which most closely matched the real BLS-wide distributions) are reported in depth here. The sentence pairwise incoherence and word uncommonness scores had real distribution neatly resembling a Gaussian both in the journal and interview transcript datasets (see Figure \ref{fig:disorg-sentence-dists}C/D), so all tested sentence-level distributions for those simulated features were randomly sampled from a Gaussian defined using numpy's random.normal function. The final settings used in the depicted simulation round were mean 1.35 with standard deviation 0.075 to represent incoherence, and mean 2.25 with standard deviation 0.325 to represent uncommonness.

To define the sentence-level distribution used to represent sentiment score in simulation, a more complicated scheme was employed. As can be seen in the journal distributions above and in the interview sentence distributions of chapter \ref{ch:2} (see Figure \ref{fig:sentence-sentiment-pt-dist}), the sentence sentiment distribution does not closely match a convenient theoretical shape like the Gaussian. Zero sentiment sentences are very common, and of sentences with a non-zero sentiment, we found positive scores to be notably more common, and also found the magnitude of non-zero sentiment scores to be biased away from both extremes (0 and 1). Based on these observations, I used the following methodology to randomly sample sentence sentiment in the final round of simulated distributions. 

Numpy's random.rand function was used to uniformly sample between 0 and 1, and this output was turned into a coin flip result by checking if it was $< 0.5$ or not. Thus half of the time the protocol proceeded on the positive branch and half of the time on the negative branch. On the positive branch, a random sample was taken from a Gaussian centered around 0 with standard deviation 0.3, then 0.25 was added to the absolute value of this sample to get a final simulated sentiment score. If the score was $> 1$, it was rounded down to exactly 1 to keep in line with the range of possible sentence sentiment values. On the negative branch, another uniform sample was taken, this time to perform an 80/20 split by checking the new sample for value $< 0.8$. Through this process, $80\%$ of the negative branch sentiment scores were assigned to be exactly 0. The other $20\%$ were assigned a negative value by again sampling from a Gaussian with mean 0 and standard deviation 0.3 and then adding 0.25 to the absolute value of the result (capping at 1), but then multiplying the final score by $-1$.

\subsection{Correlations}
\label{sec:sup-corr-meth}
Correlation matrices were computed for the 12 key features using both linear (Pearson) and rank (Spearman) correlation methods, across the final quality-filtered BLS journal dataset ($n=8398$). Because each matrix is symmetric and correlation ($r$) with self $=1$, the number of comparisons per matrix is $\frac{12^{2}-12}{2} = 66$. Thus the Bonferroni-corrected significance threshold for the correlation of each feature pairing was $0.000\overline{75}$. The obtained Pearson and Spearman distances ($1-r$) were input to cluster the diary features, using the same methodology as described for clinical scale clustering in the supplement to chapter \ref{ch:3}, section \ref{sec:ocd-methods-append}. The ordering output by the clustering algorithm for Pearson inputs was used to order the features in the Pearson correlation matrix, and analogously for Spearman. When displaying the matrices, any relationships that were not significant after multiple testing correction (i.e. with $p \geq 0.000\overline{75}$) or that had $\lvert r \rvert < 0.1$ (i.e. less than $1\%$ of variance explained) were masked to $r=0$, in order to highlight only potentially meaningful relationships.

\subsubsection{Pearson/Spearman comparison}
\begin{figure}[h]
\centering
\includegraphics[width=0.8\textwidth,keepaspectratio]{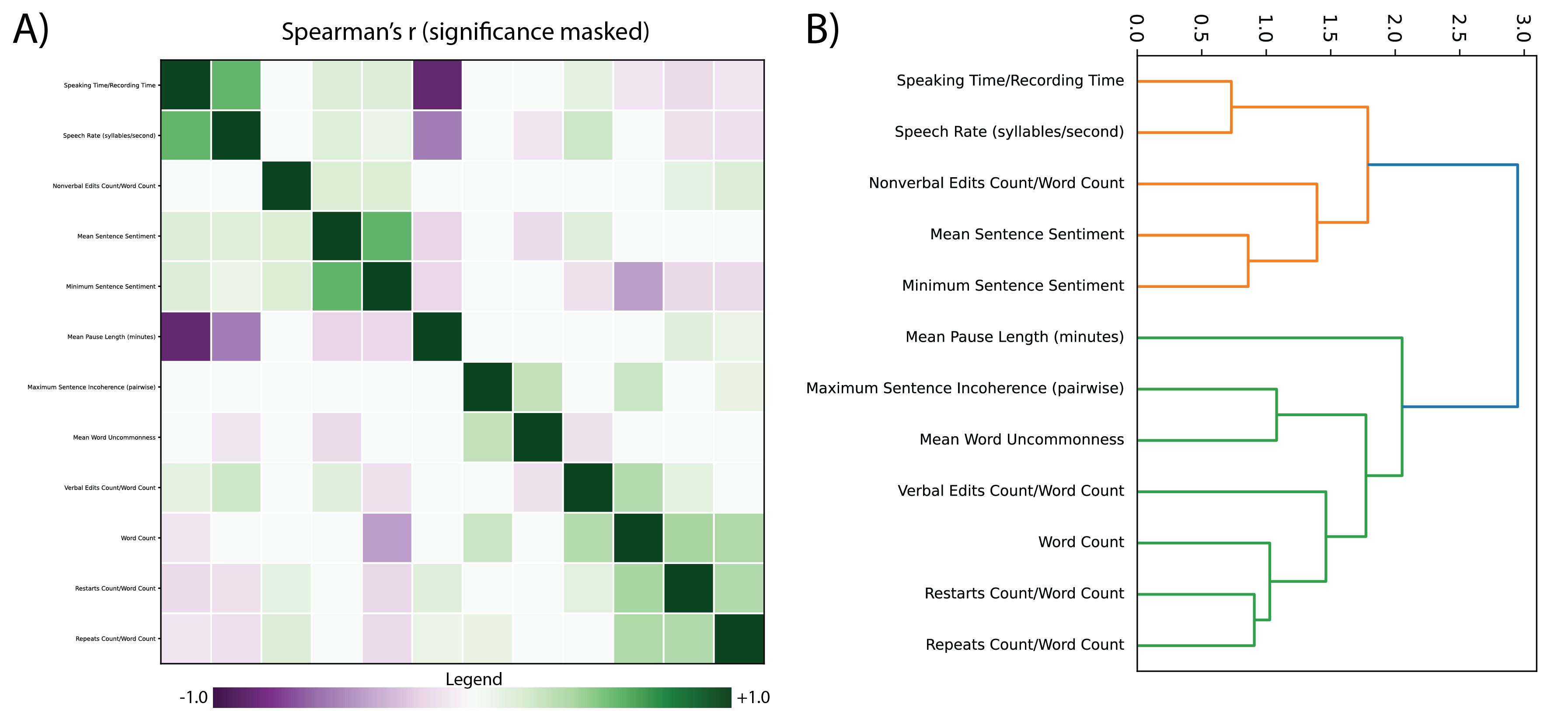}
\caption[\hspace{0.075cm} Spearman rank correlation structure of curated diary features in final BLS dataset.]{\textbf{Spearman rank correlation structure of curated diary features in final BLS dataset.} To identify potential nonlinear feature relationships, the same methodology as in Figure \ref{fig:pearson-diary-final} was repeated using the Spearman rank correlation instead (again implemented via the scipy.stats package). The clusters derived from Spearman distance (B) were analogously used to order the features in the correlation matrix visual (A).}
\label{fig:spearman-diary-final}
\end{figure}

The cophenetic correlation, a measure of the strength of clustering, was $0.628$ for the Pearson dendrogram (Figure \ref{fig:pearson-diary-final}B) and $0.714$ for the Spearman dendrogram (Figure \ref{fig:spearman-diary-final}B). The clustering of features was largely similar between the two algorithms, with the first major branch ("Speaking Time/Recording Time", "Speech Rate", "Nonverbal Edits Count/Word Count", "Mean Sentence Sentiment", and "Minimum Sentence Sentiment") essentially identical between the two. The other major branch was also similar in both methodologies, with "Restarts Count/Word Count" and "Repeats Count/Word Count" grouping relatively closely, "Mean Word Uncommonness" and "Maximum Sentence Incoherence" grouping relatively closely, and "Mean Pause Length" somewhat distant from the rest of the group. However, the main difference between the Pearson and Spearman clusterings was the location of "Word Count" and "Verbal Edits Count/Word Count" -- in both cases they were near each other, but in the Pearson results they are a standalone cluster while in the Spearman results they are grouped near the restarts and repeats disfluency subtypes. 

The correlation strengths were also largely similar between the Pearson (Figure \ref{fig:pearson-diary-final}A) and Spearman (Figure \ref{fig:spearman-diary-final}A) matrices, with differences in line with those described in the clustering results. Word count more strongly rank correlated with a number of the disfluencies, particularly repeats. The positive correlation between speech fraction and speech rate was also notably higher in the rank correlation results, although that is not entirely surprising and was later corrected for in the modified speech rate feature reported on within the prior section. Overall, there was little reason to believe there are nonlinear rank relationships between any of these features that would substantially change interpretations from the linear correlations observed. Though there may still be nonlinear relationships that are not capturable by Spearman correlation (i.e. not mediated by rank order), we do not have any prior reason to suspect this here and consider it out of scope. 

\FloatBarrier

\subsection{EMA modeling}
\label{sec:ema-meth}
I preprocessed all EMA responses from all 24 subjects of initial interest identified in Table \ref{table:bls-ema-avail}. Each EMA submission was assigned a day number using the same methodology as is used for audio journals in my pipeline, because the Beiwe convention for denoting submission date/time in filenames is the same for both audio recordings and survey responses. 

\subsubsection{Early preprocessing}
I then filtered the answers in the selected submissions down to the 15 questions identified for further study, and transformed the response encodings to range from -2 to 2 for least severe to most severe symptom experience (from "very slightly or not at all" to "extremely" for most items, and inverted for the positively worded ones). Questions that a participant chose not to answer in a survey submission were marked with NaN. If a participant submitted multiple EMA responses in a single day, the recorded response to each item was set to the mean (ignoring NaNs) of the values for that item from each submission that day. Because all of the emotion-related questions selected were always asked together, as were all of the psychosis-related questions selected, I proceeded to analyze the availability and distributional properties of these categories separately. Within these two broad categories I dropped all submissions where the participant did not answer all included questions, as this represented only a small portion of the dataset and was impacting data cleanliness.

\begin{figure}[h]
\centering
\includegraphics[width=0.85\textwidth,keepaspectratio]{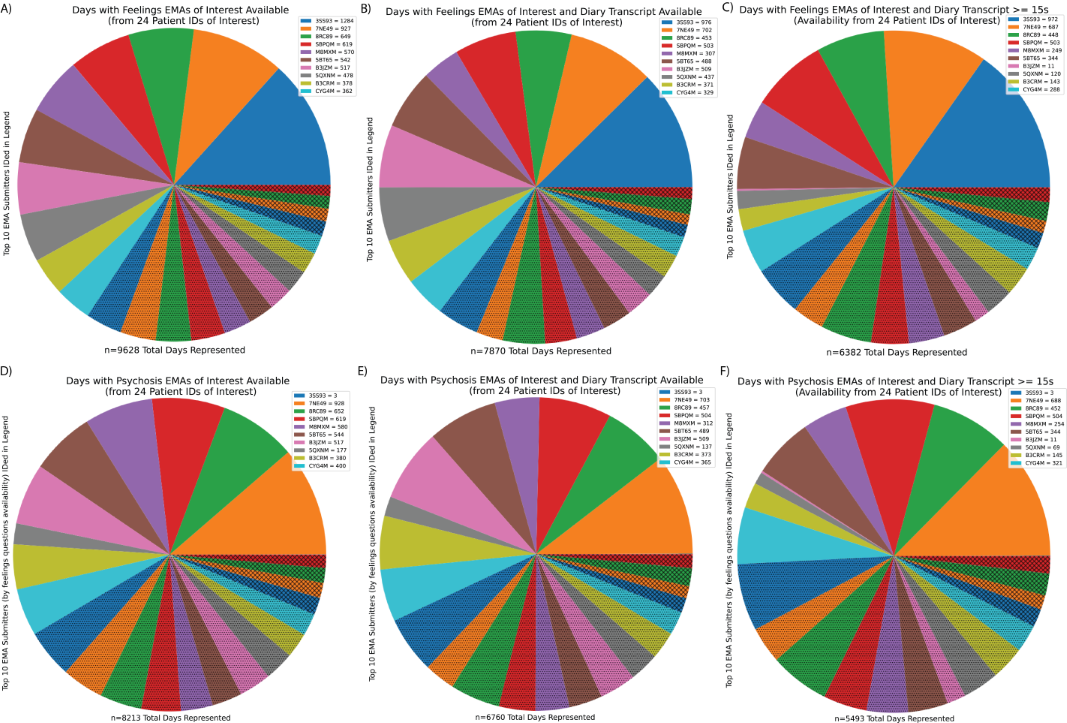}
\caption[\hspace{0.075cm} Breakdown of daily survey (EMA) availability by specific question categories of interest.]{\textbf{Breakdown of daily survey (EMA) availability by specific question categories of interest.} To better characterize the availability numbers reported in Table \ref{table:bls-ema-avail}, I considered the available dataset sizes for specific categories of EMA question and for different restrictions on diary transcript, with breakdown by participant ID. Because only a small subset of participants will be included in final modeling, I did this pilot review of data availability focused only on the 24 subjects listed in Table \ref{table:bls-ema-avail}. Each of the 24 subjects was assigned a particular color and pattern in this figure to mark their share of each dataset across the pie charts. The total number of time points making up the entirety of a given chart is denoted below it. Coloring was assigned based on the rank order of (A), and the top 10 participants in (A) have numbers for each dataset category reported in the respective legends across the figure. As described above, two main types of EMA item were investigated independently: emotion-related questions (A/B/C) and psychosis-related questions (D/E/F). For each of these categories, I checked the overall size of the daily EMA datasets (A/D), the size of the datasets when including only days with overlapping journal transcript availability (B/E), and the size of the datasets when further filtering to exclude any journals of recording duration $< 15$ seconds (C/F). In general, availability of EMA and availability of EMA with a diary transcript were in line with each other in these participants, though a few of them - especially B3JZM (solid pink) - had a tendency to submit extremely short diaries. Note also that while 3SS93 (solid blue) was a top contributor of both audio journals and emotion-related EMA responses, they hardly ever answered the psychosis-related EMA items.}
\label{fig:ema-pie-24}
\end{figure}

The number of days with complete EMA responses available for each of the 24 selected subjects can be found in Figure \ref{fig:ema-pie-24}A/D for emotion-related/psychosis-related questions respectively. With the exception of 3SS93, who answered psychosis-related questions on only a very small number of occasions, the number of psychosis-related responses available was similar to the number of emotion-related responses. Merging these EMA availability tallies with same day audio journal availability unsurprisingly showed just a small drop-off in the number of days included for each subject (Figure \ref{fig:ema-pie-24}B/E), which is consistent with the expectations from the overlap statistics sourced before data cleaning (Table \ref{table:bls-ema-avail}). Recall though that the subjects considered here were chosen due to good audio diary participation rates. Fortunately, outside of a couple of subjects with a tendency to submit very short diaries, further limiting the dataset to only days where the recorded journal had duration at least 15 seconds did not remove many data points (Figure \ref{fig:ema-pie-24}C/F), again boding well for modeling plans. 

\FloatBarrier

\subsubsection{Reviewing distributions}
For the described 2 emotion-related EMA summary score labels (positively worded and negatively worded), I generated histograms of the distribution over the considered dataset of $9628$ patient days (Figure \ref{fig:ema-wider-dists}A/B). Similarly, for the described 2 psychosis-related EMA summary score labels (hallucinations and delusions), I also generated histograms over the considered dataset, of $8213$ patient days (Figure \ref{fig:ema-wider-dists}C/D). Based on the distributional characterization of Figure \ref{fig:ema-wider-dists}, I chose to shift all but the positively worded summary score to be encoded on a range between 0 and 4, instead of -2 and 2. For the negatively worded emotion-related EMA items as well as the psychosis-related EMA items that were shifted before proceeding, this new range is consistent with the original Beiwe encoding convention (though that implementation is based on technical considerations only for the general use of Beiwe, not a reflection on specific study design). 

To further characterize the dataset, I also reviewed the same summary score distributions amongst the diary-aligned and length-filtered datasets of Figure \ref{fig:ema-pie-24} (B/E and C/F). The overall shape of these distributions was similar, with a few subtle shifts upon length filtering. I will report on label distributions when compiling the final modeling dataset in the next subsection (\ref{subsubsec:bls-ema-methods}), after selecting the smaller subset of participants to include in this pilot work.

\subsubsection{Final subject selection}
I aimed to compile a final dataset including $\sim 3000$ participant days of audio journal features and EMAs, representing $\sim \frac{1}{3}$ of the records accounted for in Table \ref{table:bls-ema-avail}. One easy way to narrow the considered subjects was to focus exclusively on those with a primary diagnosis of Bipolar disorder. Because 3SS93, 8RC89, and 5BT65 were closely studied in section \ref{subsec:diary-dists}, and will be the subjects of pilot case reports in section \ref{subsec:diary-case-study}, they were obvious inclusions in the EMA modeling dataset. GFNVM was also a fairly obvious inclusion in the final modeling dataset, given their relevance in the manual validation of section \ref{subsec:diary-val} as well as the fact their diagnosis is Bipolar type II, less common amongst the 24 participants initially reviewed. 

From there, I selected 3 more subjects with varying diary/EMA dataset sizes, to meet the target dataset size. SBPQM was selected due to a large number of samples in combination with their Bipolar type I diagnosis, GMQBM was selected due to the note of clear psychotic features as part of their Bipolar I symptomatology, and ACMCM was selected as a participant with similar dataset size to GFNVM but with Bipolar type unspecified. The 7 total subjects included in the modeling subset are highlighted within Table \ref{table:bls-ema-avail}. Although this dataset narrowing was largely done for the sake of other's future projects, it does indeed have the convenient side effects of bringing the focus onto Bipolar disorder to a greater extent, as well as enabling more interpretable analyses on the participant-dependent results of mixed effects modeling.

At the same time, the fact that even after this fairly aggressive dataset narrowing a large number of data points were still available is yet another strong demonstration of the research value of the audio journal format. As will be detailed below, the primary modeling will involve 11 diary features and will occur for each of the 4 EMA summary score labels, thus amounting to 44 total comparisons if considered in isolation (a simplification of the actual modeling technique). With 50 comparisons the Bonferroni-corrected significance threshold for model fit would be $p < 0.001$, a threshold that any meaningful relationships in a dataset of size $\sim 3000$ should have no problem meeting. While it may be difficult to make strong claims about generalization to other patients with a population of just 7 subjects, a major portion of this EMA modeling section will involve comprehensive breakdown of participant-dependent effects, providing a proof of concept for a general approach to include personalized components in audio diary feature modeling. 

\subsubsection{Modeling details}
The final modeling dataset sizes were as follows --
\vspace{-0.25cm}
\begin{center}\begin{tabular}{c}\begin{lstlisting}
Emotion-related EMA with no diary duration filter
(Verbosity only modeling): 
n=2660 / n=480 
train / test split
\end{lstlisting}\end{tabular}\end{center}

\begin{center}\begin{tabular}{c}\begin{lstlisting}
Emotion-related EMA with diaries of at least 15 seconds
(Full feature modeling): 
n=2502 / n=456 
train / test split
\end{lstlisting}\end{tabular}\end{center}

\begin{center}\begin{tabular}{c}\begin{lstlisting}
Psychosis-related EMA with diaries of at least 15 seconds
(Full feature modeling): 
n=1724 / n=318 
train / test split
\end{lstlisting}\end{tabular}\end{center}
\vspace{0.25cm}
\noindent Recall that the psychosis-related EMA was usually not submitted by subject 3SS93, so that dataset spans 6 participants while the others span 7. \\

\paragraph{Diary features chosen for full modeling.}
The following 11 journal properties will be explored for their predictive value in estimating same-day EMA summary scores in the described final full modeling datasets. All of the listed measures were either features output by my pipeline, or easy transformations of those features:
\begin{itemize}
    \item Total word count of diary transcript
    \item Mean words per sentence of diary transcript
    \item Fraction of recording duration spent speaking (as detected by the pipeline pause labeling algorithm)
    \item Estimated speaking rate from transcript-derived syllables/second (adjusted to remove long pauses via normalization by speech fraction)
    \item Rate of nonverbal edit usage per word (all disfluencies counted using TranscribeMe verbatim notation)
    \item Rate of verbal edit usage per word
    \item Rate of repeats per word
    \item Rate of restarts per word
    \item Mean transcript sentence sentiment (VADER calculated)
    \item Mean transcript word uncommonness (word2vec-derived)
    \item Maximum within sentence incoherence (word2vec-derived, word vectors compared pairwise)
\end{itemize}
\noindent Features were chosen due to good evidence validating them in our hands (section \ref{subsec:diary-val}), as well as good reason to believe they have clinical relevance (section \ref{sec:background2}). Their broad distributional properties were then characterized across BLS (section \ref{subsec:diary-dists}), resulting in small adjustments to create this final list of 11 inputs for consideration, removing a handful of metrics that were more strongly correlated with others already included. The total number of features was intentionally kept small so that relationships could be realistically dissected and statistical power on the modeling subset would be sufficient to explore various model choices with these features without losing the ability to make confident results claims. \\

\paragraph{Modeling tools.}
For the scope of the present thesis, I will primarily be utilizing straightforward linear regression tools to perform the modeling of these EMA summary scores from select diary features, via the "statsmodels" Python package \citep{statsmodels}. Goodness of fit will primarily be measured by the $R^{2}$ and F-stat derived p-values for each linear model. Individual feature coefficients and their corresponding t-stats will also be reviewed. I will begin with general modeling that does not take into account subject identity. 

As mentioned, it is difficult to interpret most audio journal features on the very short diaries of duration $< 15$ seconds. Thus for the dataset with no diary length requirement, I will perform an in depth characterization of the relationships of total word count and words per sentence with positively and negatively worded emotion-related survey summary scores on the training data. 

I will then proceed to the more complete feature modeling datasets, where I will model each EMA label independently, but fit the input journal features simultaneously, to estimate individual contributions of different features. On these training sets, I may also fit alternative models with a few choice feature subsets to further elucidate independent contributions, in conjunction with referring back to the correlation structure results of section \ref{subsec:diary-dists}. To assist in interpretation of any model fitting results, I will create visualizations over the training data as needed. The full list of diary features to be included in this analysis is provided next. Recall that for the psychosis-related EMA summary scores I will be using a logistic regression model on simplified symptom classification labels due to the distribution of that EMA.

For estimating subject-dependent effects, I will first run another general model fit (informed by the prior results) where the feature values are normalized to a baseline relative to the submitting participant, utilizing a z-score. For example, if the word count feature for subject $x$ at time point $d$ was $f(d)$, the total words feature for the new model fit became $\frac{f(d) - avg(f(D_{x}))}{std(f(D_{x}))}$, where $D_{x}$ is the full set of points in the training set for ID $x$. This has the potential to identify features with possible broad clinical relevance, as long as they are compared to the patient's own baseline as opposed to an absolute scale. Of course this cannot capture participant-specific differences in the underlying relationships, for that will require separate modeling terms -- something else to be explored. 

After analyzing model fits, I will choose just a handful of final model methods (and at most 1 fit per method) to use for prediction on the held out test set, balancing between simplicity in the number of features included and training accuracy. I will report on the overall test accuracy as well as breaking down the generalization ability based both on hold out time period and on subject ID. I will also investigate how test accuracy might relate to the severity level of the true label, and recreate any interesting feature relationship plots from the training process on the held out test data. 

\subsection{Case study methodologies}
\label{subsubsec:case-study-methods}
As a number of interesting distributional differences were found in diary features between these three patients (\ref{subsubsec:diary-pt-dists-comps}), there are particular features of heightened interest for investigation in each. 

For each case report, I will qualitatively evaluate the time courses of select audio journal features of interest, and compare them with the progression of EMA and clinical scales over the course of the study. Such temporal information was beyond the scope of the pilot quantitative EMA modeling, but even a simple heatmap view of features over time provides the opportunity to identify periods that are of note for individual subjects; time periods that might not be identified otherwise, as major insights of interest may be found in how a particular feature changes from day to day rather than in the set of values themselves. Furthermore, the case reports present an opportunity to dig deeper into the gold standard clinical scale ratings for these subjects, and ideally discover evidence of the alignment of these scales with all of the other evidence found for salient diary features above.

Additionally, I will do a deeper dive on diary content, identifying topics of recurring interest for each patient. This will include additional visualization tools like the sentiment-colored wordclouds mentioned in section \ref{subsec:diary-outputs}. For the case reports in this section, I will not only consider individual diary wordclouds like are created by the pipeline by default, but also wordclouds that represent combined diary content from more extended time periods of interest. Where relevant, I will use observations from a bird's eye review of wordclouds to identify keywords for further quantitative analysis of usage patterns over time. Such work more generally highlights the unique value in digital psychiatry tools that can be utilized at multiple distinct timescales.

Although I will connect the observations of each pilot case report back to broader facts known about that patient's story, I will not include many other datatypes that could be used to turn these reports into exhaustive characterizations. For example, passive digital phenotyping signals, longitudinal brain imaging data, recorded clinical interviews, and more detailed medical record information were all components of the larger BLS study. Connecting these datatypes could further elucidate anomalous time periods or relationships found between features that are discussed in the upcoming subsections. 

\subsubsection{Recapping subject-specific hypotheses}
To set up each individual case report, I will now recap relevant observations about the highlighted subjects from the dataset review throughout the rest of this chapter. These observations have uncovered the following interesting research directions for each particular participant, to be addressed as part of the corresponding pilot case studies: 
\begin{itemize}
    \item 3SS93 
    \begin{itemize}
        \item Subject 3S expressed a belief that study staff they had interacted with at site visits were actively listening to their submitted journals, which may have contributed to their exceptional participation rates, but may also be related to psychiatric symptomatology, something that warrants further investigation within the diary content. 
        \item During manual transcript review, a few recurring topics of potential salience for 3S were identified. Mentions of these topics over time could be evaluated -
        \begin{itemize}
            \item Migraine headaches.
            \item Severe symptoms of anxiety.
            \item Family members.
        \end{itemize}
        \item In considering quality control metrics, it was found that 3S submitted journals with a higher rate of inaudible words than most other BLS subjects. Recording quality variance has the potential to relate to pathology, whether because of changes in enunciation or speaking rate, or because of changes in the recording environment brought on by clinically relevant changes in lifestyle or other habits. As such, inaudible occurrences over time also warrant further investigation for this participant. 
        \item Key pipeline features identified with quantitative differences for 3S versus the rest of BLS -
        \begin{itemize}
            \item Abnormally low speech fraction.
            \item Frequent use of most disfluency types, but not verbal edits.
            \item Lower mean sentence sentiment.
        \end{itemize}
    \end{itemize}
    \item 8RC89
    \begin{itemize}
        \item During manual transcript review, a few recurring topics of potential salience for 8R were identified. Mentions of these topics over time could be evaluated -
        \begin{itemize}
            \item Their mother.
            \item Drinking alcohol.
            \item Various girlfriends, which also lead to somewhat frequent redactions in their diaries, such that redaction count alone might be a feature of interest for 8R.
            \item Hypomania, which was explicitly mentioned less frequently than some of these other topics, but obviously carries inherent interest for the study.
        \end{itemize}
        \item It was also noticed during manual review that 8R tended to end diary recordings with a vague concluding statement that was often positive regardless of the sentiment in the rest of the diary; it therefore may be interesting to look at trends in how 8R ended their recordings over the course of the study. 
        \item Key pipeline features identified with quantitative differences for 8R versus the rest of BLS -
        \begin{itemize}
            \item Abnormally high speaking rate.
            \item Frequent use of verbal edits.
            \item Higher mean word uncommonness and maximum sentence incoherence.
        \end{itemize}
    \end{itemize}
    \item 5BT65
    \begin{itemize}
        \item Many diaries of highly short duration, repeated evidence of ties between transcript length and clinical state.
        \begin{itemize}
            \item Based on this, an investigation into patterns of diary missingness may prove fruitful for subject 5B in particular.
            \item Regardless, the deeper dive into 5B should not involve a duration filter wherever possible, and should focus quite a bit on further characterization of the relationship between journal verbosity and depressive symptoms.
        \end{itemize}
        \item Preliminary evidence of a relationship between submission time and verbosity in this patient, so that diary metadata should also be a bigger consideration in this particular case report.
        \item Key pipeline features identified with quantitative differences for 5B versus the rest of BLS -
        \begin{itemize}
            \item Abnormally low speaking rate.
            \item Lack of positive sentiment diaries expected in distribution.
            \item Possibly anomalous distribution of maximum sentence incoherence.
        \end{itemize}
    \end{itemize}
\end{itemize}

\FloatBarrier

 \section{Code architecture (Baker Lab audio diaries pipeline documentation)}
\label{subsec:diary-code}
The two scripts "phone\_audio\_preprocess.sh" and "phone\_transcript\_preprocess.sh" in the top level of the GitHub repository \citep{diarygit} are for running the full phone audio preprocessing and full phone transcript preprocessing workflows. There is now also a third top-level script, "phone\_diary\_viz.sh", which is intended to be run periodically to generate visualizations of the outputs created by the audio and transcript processing scripts. Currently the weekly pipeline runs need to be queued manually in the Baker Lab, as they require entering sensitive password information.

Under the \hl{individual\_modules} subfolder of the \hl{process\_audio\_diary} repo are bash scripts called by these three primary scripts, which can also be used directly to perform individual steps of the pipeline on an input study as needed. For example, if the pipeline gets interrupted in the middle of a run, the modules can be used to complete the preprocessing of that batch without needing to restart the entire pipeline (everything besides email generation is currently robust to interruptions). The modules can also be used to rerun only a subset of the preprocessing steps if a major change is made and need to be updated in legacy data - such as addition of new QC features. The python scripts called by the modules are in the \hl{individual\_modules/functions\_called} subfolder, although there should be no need for the user to run these directly. If the code needs to be run only on particular patients, the wrapping bash script modules can be edited to use a whitelist or blacklist of patient IDs when looping over a study. \\

The next two subsections will provide more detailed instructions for installing (\ref{subsubsec:diary-install}) and running (\ref{subsubsec:diary-run}) the code as is. Then I will proceed to explain implementation details for each module of the audio (\ref{subsubsec:diary-aud-deets}), transcript (\ref{subsubsec:diary-trans-deets}), and summary/visualization (\ref{subsubsec:diary-viz-deets}) pipeline paths, and report on estimated runtimes from our internal use of the code (\ref{subsubsec:diary-comp-time}). That will complete the documentation for the pipeline as is and its future use in the Baker Lab or with replica systems. For justification on the features that were included, see section \ref{subsec:diary-val} above.

Drawing on the context provided by these implementation details, I will close the code documentation with section \ref{subsubsec:diary-adapt}, which reviews a number of potential differences between our studies and those of other groups, and outlines what changes might need to be made to the code to adapt the pipeline accordingly. The list of possible updates also more generally covers potential issues to look out for and broad code improvements that might be a next step for future researchers working directly on software development.  

\subsection{Installation}
\label{subsubsec:diary-install}
Initial installation of the code should be straightforward and is described here. Note we run this code on a standard compute node of the Partners Healthcare ERIS cluster, so it does not have any special hardware requirements - one of the intentions of the design. It does not use a GPU and easily runs on a machine with just a single CPU and 4 GB of RAM. \\

\noindent Most necessary Python (3.9) dependencies can be found in the setup/audio\_process.yml file in the GitHub repository. As long as you have Anaconda installed, you can easily create a Python environment for running this code by moving to the setup folder and running:

\begin{center}\begin{tabular}{c}\begin{lstlisting}
conda env create -f audio_process.yml
\end{lstlisting}\end{tabular}\end{center}

\noindent To activate the environment, which should be done each time before running the code, use:

\begin{center}\begin{tabular}{c}\begin{lstlisting}
conda activate audio_process
\end{lstlisting}\end{tabular}\end{center}

\noindent Note that before the first time running on a new machine, it may be necessary to start Python (in the activated conda environment) and enter the following two commands, in order for the NLP features script to work:

\begin{center}\begin{tabular}{c}\begin{lstlisting}
import nltk
nltk.download('cmudict')
\end{lstlisting}\end{tabular}\end{center}

\noindent Similarly, a handful of the required packages are not conda installable, so that before the first the run the following should be entered on the command line (in the activated conda environment):

\begin{center}\begin{tabular}{c}\begin{lstlisting}
pip install soundfile
pip install librosa
pip install vaderSentiment
pip install wordcloud
\end{lstlisting}\end{tabular}\end{center}

\noindent It will also be necessary to install the lab encryption package before first running the pipeline, as raw audio files are expected to be encrypted (something done by default if using Lochness to organize the raw data as described below). To do so, enter the following command after activating the environment:

\begin{center}\begin{tabular}{c}\begin{lstlisting}
pip install cryptease
\end{lstlisting}\end{tabular}\end{center}

\noindent Other dependencies are OpenSMILE (2.3.0) and ffmpeg (4.3.2), as well as the ability to run bash scripts and a working mail command - the latter two should be available on any standard Linux server. 

For both OpenSMILE and ffmpeg it will be important to ensure the path to their executables are part of your PATH environment variable before running the code. To setup this sort of initialization to occur automatically upon server login for regular pipeline runs, look into editing your \emph{.bashrc} file in your home directory. Regardless, for OpenSMILE the path to the default config files downloaded into your installation location will need to be updated directly in this code base - to the GeMAPS config for full compatibility with the current pipeline. 

\noindent Look for \hl{/data/sbdp/opensmile/opensmile-2.3.0/config/gemaps/GeMAPSv01a.conf} in this repo to find path locations to update.

Finally, it is necessary for the NLP features as implemented to download the publicly available Google News 300 dimensional word2vec model, and ensure the path to that model is updated in the code to reflect the path on your machine.

\noindent Look for \hl{/data/sbdp/NLP\_models/GoogleNews-vectors-negative300.bin} in the current code. \\

\noindent Note that the use of TranscribeMe is also functionally a dependency for the entire transcript side of the pipeline at this time, though the audio features can still be extracted regardless. To set up the pipeline to work with TranscribeMe, you must first contact them to ask for an account on their SFTP server like is used for the Baker Lab and the AMPSCZ project. For full functionality, you should request the same transcription quality and notation settings as described in section \ref{subsec:diary-methods} for all uploads. From there no further "installation" work is required, and the same TranscribeMe account can be used for multiple lab studies as desired. \\

\noindent Optionally, the Lochness \citep{lochness} and DPDash \citep{dpdash} open source tools used by the Baker lab can be integrated with the pipeline to further facilitate operations. Lochness will pull audio diaries from Beiwe to the compute server and enforce the expected input folder structure and naming conventions automatically once set up. On the other end, the QC summary CSV output by the pipeline is formatted for compatibility with DPDash, so if that is set up the outputs of the pipeline can be easily monitored via the online GUI with automatic updates.

\subsection{Running the code}
\label{subsubsec:diary-run}
Instructions on how to easily run the entire pipeline are provided here, along with details on a few additional setup requirements that need to be ensured for each individual study to be processed. For more on the role of specific modules and details of their implementations, see the associated subsections below (\ref{subsubsec:diary-aud-deets}-\ref{subsubsec:diary-viz-deets}). \\

\subsubsection{Expected file structure} 
The code as a whole expects a particular set of data organization and naming conventions to be met for raw data -- conventions that will be satisfied by default if using Lochness \citep{lochness} to pull the journal recordings from Beiwe. If running the code in a more bespoke manner, it will be necessary to either replicate the expected input structure or edit the code. For convenience, input expectations are specified here. 

The root path for the main Baker Lab storage is currently hard coded, but could be very easily changed across the scripts, from \hl{/data/sbdp/PHOENIX} to any other desired root folder. However, the file organization from there (especially the nested PHOENIX folder structure) will be more difficult to change in the code, so that should be avoided if possible. The expectations for the nested folder structure are as follows:
\begin{itemize}
    \item Under \emph{root} there must be two folders, \hl{GENERAL} and \hl{PROTECTED}. It is expected that GENERAL will contain deidentified data while PROTECTED will contain sensitive identifiable information. 
    \item Under both GENERAL and PROTECTED will be the top level folders for each study. For example, \hl{BLS} is one of our study folders. When running the pipeline, the \emph{study} to process audio journal data for will be a prompted argument, but obviously if the study folder is not set up properly it will not be able to run.
    \item Under the study folder will be folders for each subject ID enrolled in that study e.g. \hl{3SS93}. 
    \begin{itemize}
        \item For the GENERAL side study folder, there also must be a csv file named \newline \hl{[study]\_metadata.csv} (i.e. BLS\_metadata.csv in our case). That csv needs to map subject IDs to their consent dates. If using Lochness \citep{lochness} to pull the raw data from e.g. Beiwe, this will already be set up (as should the entire folder structure). 
    \end{itemize}
    \item Subject ID folders must then contain folders for each major data source that is being collected from that participant. For data being pulled from Beiwe, this should be called \hl{phone} (and is what the pipeline here expects).
    \item Under the data source folders come the \hl{raw} and \hl{processed} folders. The input data should of course be found under raw, and outputs of the pipeline will go under processed. As audio journal recordings are a sensitive datatype, this all occurs under PROTECTED for the current code. 
    \begin{itemize}
        \item For the code to be run by a particular account, either a folder named \hl{audio} will need to already exist under all of the relevant PROTECTED side processed folders (and be accessible/writable by the current account), or the processed folders themselves would need to be writable by the current account to create \emph{audio}. 
        \item The phone folders under each raw folder will need to be accessible for the code to run as well, but the code does not need write permissions to any part of raw. It is recommended that the code not be able to make changes in raw -- while there is no reason it should modify or delete anything there, it is best for data safety reasons to minimize these permissions for the purest form of the data. 
    \end{itemize}
    \item For processing the raw audio diaries, the code assumes there will be mp4 files pulled from Beiwe by Lochness and encrypted with cryptease. 
    \begin{itemize}
        \item The convention for Beiwe folder structure is that under a given raw folder there will be one subfolder (with unique ID) for each device the participant has (or had) Beiwe registered on. The pipeline will automatically go through and check for/process new diaries from any registered device.
        \item Under each device folder, there will be subfolders for different datatypes Beiwe provides. In the pipeline's case, the relevant folder name is \hl{audio\_recordings}.
        \item For audio diaries, Beiwe allows multiple different prompts to be administered to a participant in parallel. As such, there are folders under audio\_recordings that each correspond to a unique prompt ID. Under those folders are where the (encrypted) .mp4.lock files are found. 
        \begin{itemize}
            \item Per the current Beiwe $\Rightarrow$ Lochness protocol, these are not necessarily shared across subjects or even devices, regardless of if a prompt is the same on the study design end. Because of that, the pipeline presently treats these folders just as it treats the phone ID folders -- tracked in the highest level metadata, but other than that entirely ignored. The pipeline will search through all of the available prompt ID folders for possible new journal submissions. 
            \item Note that the code right now enforces a 1 diary per study per subject ID convention, so if multiple submissions are made in one day (whether under the same prompt or not), only the first chronologically is counted towards that study day.
        \end{itemize} 
        \item The raw audio file names themselves are expected to follow Beiwe convention for providing metadata, i.e. \hl{YYYY-MM-DD hh\_mm\_ss} followed either directly by the file extension (including the .lock), or sometimes a "+00" or other two digit code after a plus sign and then the file extension.
        \begin{itemize}
            \item Note that this time provided by Beiwe is always in the UTC timezone per our protocol. The pipeline thus assumes this is UTC, and for our purposes converts to ET. 
        \end{itemize}
    \end{itemize}
\end{itemize}
\noindent So to recap, the code will look for new input diaries at the following file paths:
\begin{center}\begin{tabular}{c}\begin{lstlisting}[breaklines=true]
[root]/PROTECTED/[study]/[subjectID]/phone/raw/*/audio_recordings/*/YYYY-MM-DD hh_mm_ss.mp4.lock
\end{lstlisting}\end{tabular}\end{center}
\noindent Where [root] is an easy to change path throughout the code that currently corresponds to /data/sbdp/PHOENIX, [study] is the name of the study prompted by the main pipeline upon launch, [subjectID] is all IDs with audio journal data and a properly registered consent date under the input study (will of course be carefully tracked throughout), * are wildcards that will match any folder name as they correspond to random Beiwe-generated IDs, and finally the filename contains date and time information as presented and is an mp4 encrypted by cryptease. 

\noindent Similarly, all outputs created by the pipeline can be found under the following folders:
\begin{center}\begin{tabular}{c}\begin{lstlisting}
[root]/PROTECTED/[study]/[subjectID]/phone/processed/audio
\end{lstlisting}\end{tabular}\end{center}
\noindent Recall that if you are pulling data from Beiwe with Lochness set up per \cite{lochness}, much of this structure will be automatically setup. 

\subsubsection{Storage considerations.} 
As individual audio journals are quite short, they do not generally require much careful consideration about storage relative to related datatypes like interview recordings (chapter \ref{ch:2}). However, file sizes can add up, and pipeline intermediates may require a small (but $> 1$) multiplier on raw data storage, so it is worth a quick thought whether extra disk space might need to be looked into before getting too deep in a study.

A rough estimate for raw audio-only MP4 storage is $\sim 1 MB$ per minute of audio, so with a mean diary length of $\sim 1.5$ minutes, we would expect on average $1.5 MB$ per diary recorded, which for a multi-year study the size of BLS could reach $\sim 15 GB$. Of course this is extremely small for an entire study relative to many other datatypes like video recorded interviews or MRIs. Even wrist accelerometry data when continuously collected will likely take up substantially more space than the raw diaries. Long term pipeline outputs are also largely not very big at all, and shouldn't take more total space than the raw recordings themselves. For the entire project then one could account a permanent $2 MB$ per minute of audio, which would be $\sim 3 MB$ for the average diary submission in our experience with BLS. 

Transiently, the pipeline will require an order of magnitude more space per individual audio file however, so this could add up if processing a massive batch all at once. This is because the pipeline converts the MP4 files to WAV format after decrypting them, as required for use with OpenSMILE. It also temporarily saves foreground and background audio versions of the WAV produced by the voice activity detection part of the code. While these latter audios will be deleted once the script successfully finishes processing all newly recognized audio for the current run, the main WAV will not be immediately deleted for some time. That is because the file is retained on the server until the corresponding transcript processing has been successfully completed, which requires waiting for TranscribeMe to return that transcription. It is important that the WAV files left pending transcription are not deleted manually, as it will interfere with the pipeline's operations. 

WAV files take about 10 times more space than audio MP4s do, so while awaiting transcription each diary will account for an expected $15 MB$ of additional storage space, not counting the $3 MB$ above. In the middle of running, the entirely temporary VAD audios written to disk could contribute another $30 MB$, but this is unlikely to cause a problem in normal use. Note that raw audios could be exported from Beiwe as WAVs instead if desired, but that would obviously substantially increase the permanent storage required on the server for these journals. \\

\subsubsection{Launch commands}
\paragraph{Audio side pipeline wrapper.} 
To initiate the audio side of the pipeline, navigate to the repository folder and run:

\begin{center}\begin{tabular}{c}\begin{lstlisting}
bash phone_audio_preprocess.sh
\end{lstlisting}\end{tabular}\end{center}

\noindent The script will then prompt for study name and decryption password. It will also ask if the user wants audio to be automatically uploaded to TranscribeMe, and if so will further prompt for the TranscribeMe account password, and minimum length/volume requirements for audio to send. For our lab studies, I recommended a 40 db minimum for total audio volume, and a length requirement of only 1 second. However, this may change with a different data collection procedure or study aim. See section \ref{subsubsec:diary-val-qc} for more information on the review process for deciding our cutoffs.

Additionally, the user can set a limit on the total number of minutes of audio that will be uploaded to TranscribeMe. If the processed amount exceeds this "budget", no files will be uploaded and the decrypted files will instead be left as they would in the case of the no transcription setting. Note we pay \$1.75 for the specific transcription service described in section \ref{subsec:diary-methods}.

Logging from the script is both printed to the console and (for now) saved in the file "audio.log" newly created for each run under the folder from which the pipeline was called. A summary of the processed files is also emailed when transcript upload is enabled. To get an email alert without transcription, enable automatic upload with the maximum number of minutes set to 0. \\

\paragraph{Transcript side pipeline wrapper.} 
To initiate the transcript side of the pipeline, navigate to the repository folder and run:

\begin{center}\begin{tabular}{c}\begin{lstlisting}
bash phone_transcript_preprocess.sh
\end{lstlisting}\end{tabular}\end{center}

\noindent The script will then prompt for study name and TranscribeMe account password. It expects that the audio side pipeline has been run with auto transcription on recently, and new transcripts are now returned by TranscribeMe - if there are no new transcripts available to pull it will recognize this and exit. If the later steps of the pipeline need to be run on older transcripts or other transcripts sent outside the scope of the automatic push code, individual modules used in the pipeline can be called separately as needed. Logging from the main script is both printed to the console and (for now) saved in the file "transcript.log" newly created for each run under the folder from which the pipeline was called. A summary of the new transcripts is also emailed. \\

\paragraph{Visualization/summary pipeline wrapper.}
To create all pipeline-related visualizations, navigate to the repository folder and run:

\begin{center}\begin{tabular}{c}\begin{lstlisting}
bash phone_diary_viz.sh
\end{lstlisting}\end{tabular}\end{center}

\noindent The script will then prompt for study name. Details on each type of visualization generated follow. All visualization code draws from functions defined in viz\_helper\_functions.py under the functions\_called subfolder. Logging from the main script is both printed to the console and (for now) saved in the file "viz.log" newly created for each run under the folder from which the pipeline was called.

\subsection{More on audio processing components}
\label{subsubsec:diary-aud-deets}
In this section, I describe each individual module on the audio side of the pipeline (first set of boxes in Figure \ref{fig:diary-arch}), in the order in which they are run. Each step is linked to a wrapping bash script in the individual\_modules subfolder of the GitHub repository, which would allow that step to be run independently if needed. Most steps also rely on helper functions under individual\_modules/functions\_called. \\ 

\noindent \textbf{Step 1: run\_metadata\_generation.sh}

\noindent The first step creates a metadata CSV that maps to each study day a single recording file (if any), and the submission time info in Eastern Time (ET). Look for the "ETFileMap" output in each patient's \hl{phone/processed/audio} folder to see this metadata, which is of course also used as input to subsequent pipeline steps. Thus for any work involving the organization of raw audio journal files, running this step first is a must. To work, this wrapping bash script calls the phone\_audio\_metadata\_format.py helper. It utilizes known Beiwe naming conventions to do the metadata compilation, including the UTC timestamp found in the raw diary filenames. It also utilizes consent information for each patient ID found in the study metadata.

Note that submission time variable produced here is coded as an integer between 4 and 27, because any submission prior to 4 am ET will be considered a night time submission counted towards the previous day. As of now, only the first recording submitted in a day is considered for downstream processing. \\

\noindent \textbf{Step 2: run\_new\_audio\_decryption.sh}

\noindent The next step decrypts all audio diaries under \hl{phone/raw/$*$/audio\_recordings} that have not yet been analyzed (determined by checking for an OpenSMILE output for that file, see Step 4 below). It uses the crypt\_exp helper so that the study password entered at the beginning can be used for every audio file that needs to be decrypted. Decrypted files are then converted to WAV format (as our Beiwe default is MP4), and kept in a temporary folder for each patient located at \hl{phone/processed/audio/decrypted\_files}. WAV conversion is done for compatibility with all processing tools used by the pipeline, in particular OpenSMILE. 

At the end of this step, files are still named according to the original raw Beiwe convention, so decryption can technically be run without running Step 1 first. The WAV files generated here are used by subsequent steps to extract features, and then the metadata CSV is used to organize these features for study analysis. \\

\noindent \textbf{Step 3: run\_audio\_qc.sh}

\noindent This step computes audio quality control stats for all of the currently decrypted audio files, as found in each patient's described "decrypted\_files" folder. If run directly after Step 2, it should therefore add audio QC stats for any newly uploaded diaries to the existing record for the corresponding patient, or create a new record with current diaries for any newly recognized patient ID. QC computation occurs via calling the phone\_audio\_qc.py helper, and the resulting (or updated) audioQC output can be found in the patient's \hl{phone/processed/audio} folder, as is for other outputs. The features at the end of this step are still mapped to raw Beiwe file names, but the audioQC CSV created is used downstream (in Steps 5 and 6) to generate sets of features merged with relevant metadata.

\noindent The primary features computed are:
\begin{itemize}
    \item Length of audio (in minutes)
    \item Overall volume (in dB)
    \item Standard deviation of amplitude (as mean amplitude of audio files is generally 0, this is often a direct transformation from the log scaled dB)
    \item Mean of spectral flatness computed by librosa  (ranges between 0 and 1, with higher values indicating more white noise-like sound)
    \item Max of spectral flatness computed by librosa
    \item There is also a check that the file is mono (i.e. a single channel, which is expected for the diaries)
\end{itemize}
\noindent Note that while in the BLS dataset the standard deviation of amplitude feature was perfectly related to the overall volume, it still had utility in better distinguishing files of high volume, as the log scale of dB saturates more at higher values. Additionally, as the maximum spectral flatness was largely the same value (near 1) across audios, we really only proceeded to use mean spectral flatness, and often refer to this as just "spectral flatness". \\

\noindent \textbf{Step 4: run\_opensmile.sh}

\noindent This step computes OpenSMILE features for all currently decrypted audio files, again as found in each patient's described "decrypted\_files" folder. It is thus expected to be run after Step 2, to process any newly uploaded diaries; but it may be run in parallel with Step 3. As mentioned, the outputs from this step are used by future runs of the decryption script (Step 2) to identify when a file has already been processed, so that only new files are decrypted. It is therefore quite critical to overall operation of the pipeline at this time and cannot be skipped if the code is to be run on an ongoing basis.

One feature CSV for each audio file will be saved, in the \hl{opensmile\_feature\_extraction} subfolder of \hl{phone/processed/audio}. Note the naming of these initial OpenSMILE outputs will reflect the raw Beiwe audio name. They are considered intermediate outputs, as they will be filtered to remove non-speech times (and given final names) in Step 5. 

For specifics of feature extraction, the script uses the command line interface of OpenSMILE to extract low level descriptor features (i.e. 10 millisecond bins) using the OpenSMILE-provided GeMAPS configuration. This configuration was chosen due to its prior success in emotion recognition competitions \citep{GeMAPS}. The low level descriptor setting is used to enable future analyses where careful alignment of acoustics and language will be necessary. In this pipeline, summary stats using the low level descriptor features will also be computed. \\

\noindent \textbf{Step 5: run\_vad.sh}

\noindent The next step runs voice activity detection (VAD) on all the currently decrypted files, again as found in each patient's described "decrypted\_files" folder. In addition to computing outputs directly from the available WAV files, this step also creates outputs based on filtering the available OpenSMILE feature CSVs. It is therefore expected to be run after Step 4. 

This module first generates a temporary foreground audio file by using a nearest neighbors filtering technique on the audio spectrogram, as described in the Librosa \citep{librosa} tutorial on vocal separation, and then running an inverse Fourier Transform again using Librosa. The resulting foreground audio file is subsequently used for pause detection, by identifying times of silence within the file. A list of all pause times across a given patient's processed diaries can thus be found in the: \newline \hl{[study]\_[subjectID]\_phone\_audioVAD\_pauseTimesOutput.csv} file on the top level of the \hl{phone/processed/audio} folder for the corresponding patient ID. 

Specifically, the pause time detection uses a sliding window of width 250 milliseconds, moving 50 milliseconds at each timestep, to detect conversational pauses as defined by $\geq 250$ consecutive milliseconds of vocal silence. It takes the spectrogram of the VAD-derived foreground audio using the same Librosa spectrogram function as was used for the original input as the first step of VAD. It then runs the sliding window over the spectrogram, taking the root mean square (RMS) overall all indices of that part of the spectrogram (flattened). The RMS is then thresholded to determine if the window contains any speech or not. The current value for that threshold is $0.03$, tuned on a small subset of $\sim 50$ journals from 1 week of BLS submissions, and then verified more broadly across the study. 

The pause detection function then uses various numpy array manipulation options to join together overlapping bins detected to be silent, find the start and stop indices for each contiguous silence bin, and convert detected start/stop indices to indices that can actually be used into the original raw audio WAV. It also estimates a corresponding pause length in milliseconds based on each start/stop bin index. Note that while the pipeline itself does not have settings to change the sliding pause duration window's width, step size, or threshold, these are function arguments within the underlying python script, so they are very easy settings to change. However, we did find the current settings to work well across a range of times and subjects. 

There are also three spectrogram images generated by the VAD scripts, to assist in quick manual validation of collected data - a figure comparing the original, foreground, and background audios, as well as a figure showing only the portions selected as speech and a figure showing only the portions marked as pauses. All 3 images for each diary can be found under the \hl{vad\_spectrogram\_comparisons} subfolder of \hl{phone/processed/audio}. Notes on manual review of VAD outputs can be found in section \ref{subsubsec:diary-val-aud}. 

The calculated pause times are next used to generate additional QC metrics for each diary, saved to \hl{[study]\_[subjectID]\_phone\_audioVAD\_pauseDerivedQC\_output.csv} in the same folder, to be merged with the traditional audio QC measures in Step 6. The pause-derived QC measures may be useful for both validating clear presence of patient speech in the diaries, and for clinical evaluation of changes in speech production. 
 
\noindent These features include:
\begin{itemize}
    \item Total minutes marked as speech
    \item Total number of pauses
    \item The average pause length in seconds
    \item The maximum pause length in seconds
    \item The decibel level during pause times
    \item The mean spectral flatness during pause times
\end{itemize}
 
Finally, the calculated pause times are used to filter the raw OpenSMILE results produced in Step 4, by NaNing out all rows of the OpenSMILE output that map to a bin wholly contained in one of the identified pause periods. The resulting filtered OpenSMILE outputs are saved in the \hl{opensmile\_features\_filtered} subfolder of the corresponding \hl{phone/processed/audio} folder, with each individual file now named using study day number convention: \newline \hl{[study]\_[subjectID]\_phone\_audioSpeechOnly\_OpenSMILE\_day[cur\_day].csv}, where cur\_day is formatted to 4 digits. Day number is defined as days since the patient consented to the study $+ 1$, and is found by looking up the file in the metadata CSV produced by Step 1. If the file cannot be found, the filtered result is discarded. The set of OpenSMILE outputs in this filtered output folder are intended to be one of the end use case outputs of the pipeline, while so far everything else that has been described is an intermediate. 

The phone\_audio\_vad.py script, called by this module, executes all of the above processing through the three functions diary\_vad, diary\_pause\_detect, and diary\_pause\_qc. \\

\noindent \textbf{Step 6: run\_dpdash\_format.sh}

\noindent In this step, the current "ETFileMap" CSV for each patient is left-merged with the current "audioQC" and "pauseDerivedQC" output CSVs. Some additional metadata columns (like weekday) are also added. The phone\_diary\_dpdash\_compile.py helper is utilized to run the step.

The file accounting in this step is done in order to prepare and save a new up to date DPDash-formatted CSV for each patient, so that the diary QC features can be easily visualized by lab staff using another one of our internal tools (DPDash). If you would like to use DPDash, see that documentation \citep{dpdash} for setup instructions. However, the produced CSV can still be used regardless, as a final clean output version of the quality control related features. DPDash is simply a dashboard for easily visualizing clinical-study related CSVs, so the outputs can of course be checked in a similar way without it. 

The formatted QC CSV is additionally used by some of the code on the visualization side of this pipeline, to create additional resources for screening diaries - such as QC feature distribution histograms. \\

\noindent \textbf{Step 7: run\_audio\_selection.sh}

\noindent The next part of the pipeline identifies those newly decrypted audio files (output by Step 2) that are acceptable to be sent for transcription, moving them to the \hl{phone/processed/audio/to\_send} folder for the corresponding patient. This script utilizes the phone\_audio\_send\_prep.py helper. 

Files are considered acceptable if they are the first audio diary submitted for a particular day (i.e. able to be looked up in the formatted QC CSV produced by Step 6), and they are above the requisite length and volume thresholds that were specified by the user via prompts when the code was queued. If called from the larger pipeline with auto-transcription off, the thresholds will be assumed 0, and files will need to be manually inspected within to\_send before an upload decision is made. 

In all cases, files moved to the to\_send folder are renamed here to match expected naming conventions for processed lab files. WAV files kept in the temporary decrypted\_files folder will be deleted at the end of the run if called from the larger pipeline. \\

\noindent \textbf{Step 8: run\_transcription\_push.sh}

\noindent The final primary module of the audio side, Step 8, is run by the larger pipeline whenever auto transcription is on. It can also be run independently to automatically upload to TranscribeMe all files already curated in patient "to\_send" processed phone audio folders for a given study. Before beginning the upload process, the code ensures that the sum of audio lengths found in the to\_send folders does not exceed a total limit, if one was specified. When that occurs, the code simply takes a hard stop, and it is up to the user to determine how to proceed given their budgeting constraints. This is one use case where a later rerun of a pipeline step outside the main script might make sense - to do manual curation of files contained in to\_send, and then rerun automatic upload for whatever is left by queuing Step 8 directly. 

Once upload safety is verified, the phone\_transcribeme\_sftp\_push.py helper manages the upload using the pysftp package (built on Paramiko). As mentioned, TranscribeMe SFTP account details are required for this step. When a file is successfully uploaded, it is moved to the corresponding patient \hl{phone/processed/audio/pending\_audio} folder, which is used for tracking files on the transcript side of the pipeline. To maintain the code's data flow expectations, it is important that files in pending\_audio folders are \emph{not} modified or deleted outside of this pipeline - manually or by other software.

In the rare case where a file upload fails, it should be detected by the pipeline and an according error message logged. When this occurs, the script keeps the files in the to\_send folder, so the upload can be reattempted as needed (by direct running of the SFTP push module). When called from the larger pipeline, this script as well as the audio identification script (Step 7) utilize targeted renaming of the audio files with coded prefixes, to help in constructing the email alert described in the wrap up steps. \\

\noindent \textbf{Pipeline wrap-up:}

\noindent When the entire pipeline is called (as described in section \ref{subsubsec:diary-run}), there are a few additional wrap up steps that occur but are not implemented as standalone modules. For example, when automatic transcription is set to on, the pipeline script will generate the bodies of the data monitoring email alerts via the run\_email\_writer.sh module; however this module is not suitable to be run outside of the entire pipeline at all like the rest can be.

The email writing steps works through the phone\_audio\_email\_write.py helper. It creates one email to be sent to lab members and one email to be sent to TranscribeMe sales and tech support addresses (with all addresses specified near the top of the audio side pipeline script). The email to the lab includes information on each file processed and the outcome (sent for transcription vs upload failed vs rejected), while the email to TranscribeMe simply summarizes how many total files were successfully uploaded, and how many minutes they sum to. Before this script finishes it will revert any tracking-related name changes made to the audio files.

Further, at the end of the audio side pipeline wrapper script, a few additional commands are included for data cleaning purposes. The "decrypted\_files" folder for each subject ID is deleted, removing all the rejected decrypted audios along with the generated foreground audios. Any empty "to\_send" folders are also deleted. It is important to be mindful of these data cleanliness steps if running individual modules, as decrypted versions of patient audio should not be accidentally left on the server long-term. The pipeline also has protections in place to prevent running a brand new iteration if it appears a prior run was not fully completed, so this has the (somewhat intended) consequence that careless data management could stall future processing. 

Note again though that the pending\_audio subfolder needs to remain intact for the transcript side of the pipeline to work correctly, so this should certainly be left as is after running any audio side code. Pending audio WAVs will be cleaned up as needed after corresponding transcripts are pulled back by the transcript side.

\subsection{More on transcript processing components}
\label{subsubsec:diary-trans-deets}
In this section, I describe each individual module on the transcript side of the pipeline (second set of boxes in Figure \ref{fig:diary-arch}), in the order in which they are run. Each step is linked to a wrapping bash script in the individual\_modules subfolder of the GitHub repository, which would allow that step to be run independently if needed.  \\

\noindent \textbf{Step 1: run\_transcription\_pull.sh}

\noindent The first step of the transcript side checks the TranscribeMe server for text (.txt) files available that match the expected names of new transcripts,  based on currently available diary audio in the \hl{pending\_audio} subfolder for each subject ID. Thus this step expects previous use of the audio side of the pipeline, with SFTP upload to TranscribeMe completed. 

Per TranscribeMe operating procedure, any new transcripts generated from an audio file (which the audio side of the pipeline would have deposited as a WAV with appropriate study day naming convention in the audio folder of the SFTP server) will be deposited by the transcriber in the output folder of the SFTP server. This transcript will have the same filename as the audio, but with appropriate filetype extension. For use of the pipeline, it is important to request that TranscribeMe uses plain text. 

The SFTP pull step of the pipeline operates analogously to the push step (now via the phone\_transcribeme\_sftp\_pull.py helper), pulling any available transcripts that it can based on the above expectations. Any newly pulled transcripts are placed in the \hl{phone/processed/audio/transcripts} subfolder for the corresponding patient. 

Upon successful pull, the script will delete the raw decrypted audio WAV from both the TranscribeMe server and the matching "pending\_audio" folder on PHOENIX, as well as moving the transcript txt file to the appropriate study archive subfolder on the TranscribeMe server. This ensures that dataflow remains organized and no personal patient data is left unencrypted longer than necessary. 

If a transcript cannot be found for a particular pending audio, it is left as is here. Downstream steps of the pipeline will compile this information for logging purposes, so that if a file is kept pending for too long manual intervention can be initiated. This will sometimes occur if TranscribeMe makes a mistake in the transcript file format or file upload procedure, or simply misses a file from a large batch. It is not a common issue, and the resolution is generally quick upon contacting our TranscribeMe rep. \\

\noindent \textbf{Step 2: run\_transcript\_csv\_conversion.sh}

\noindent The next step simply converts any new transcripts found under \hl{phone/processed/audio/transcripts} to a CSV, placing it in the \hl{csv} subfolder of that "transcripts" folder. New transcripts are detected in this case by only processing one if there is not a matching name already under the csv subfolder. This step expects .txt text files returned by TranscribeMe, which is the only file type that would be pulled by the previous SFTP step anyway. Note these text files should not include any header information, as all needed metadata to proceed is encoded in the filename. 

The CSV conversion step is implemented purely as a bash script, which uses known TranscribeMe conventions to separately parse the speaker IDs, timestamps, and actual text from each sentence in the transcript, and removes unnecessary white space and other unusable characters. It will also detect any transcript text files that are not ASCII-encoded and skip them to prevent later errors -- but will ensure a detailed warning identifying the problematic characters within the file is logged, and if email alerting is turned on, the ASCII issue will be flagged there as well. 

\noindent Columns obtained by the script include the following separated information:
\begin{itemize}
    \item Speaker ID to mark who is currently speaking. This is labeled S1 to denote the first unique speaker in the audio, S2 to denote the second unique speaker in the audio, and so on. Of course for audio journals specifically, it is rare to have more than one speaker, so this will largely just say "S1" the whole way through. It is still worthwhile to double check for QC however. 
    \item Timestamp to denote the time within the audio file that the current sentence began. This will have whatever resolution TranscribeMe provides, which may be giving digits down to the millisecond (not really varying beyond the 10 ms digit though) or it may be stopping at seconds. The pipeline can handle either, but second resolution timestamps with short sentences can cause issues in speech rate estimation. We typically receive millisecond formatted timestamps.
    \item The actual transcribed text of the current sentence. This is reproduced as is outside of cleaning up excess whitespace characters and a few other text artifacts sometimes produced by TranscribeMe's internal annotation software. The text column is of course the primary "feature" used in all later steps. 
\end{itemize}
\noindent Note obtained rows are assumed by the pipeline to be sentences produced by the sentence-level timestamps TranscribeMe splitting setting. If TranscribeMe is instead splitting into larger chunks for a cheaper fee, there will be many fewer rows in the CSV, and some downstream features may not make sense without other pipeline adjustments. \\

\noindent \textbf{Step 3: run\_transcript\_qc.sh}

\noindent This step computes the transcript quality control metrics for all existing phone diary transcripts found under patient \hl{phone/processed/audio/transcripts/csv} folders, as created by Step 2 of the transcript side pipeline. Therefore the transcript QC module is dependent upon completion of transcript CSV conversion for TranscribeMe transcripts, or CSV formatting of some other transcript source to match the expected output of Step 2. Note this code runs quickly and thus to facilitate future feature additions is rerun on all available transcript CSVs for a given study each time the module is called. This occurs via calling the phone\_transcript\_qc.py helper, which saves the computed metrics as an intermediate CSV in the patient's \hl{phone/processed/audio} folder. This CSV is later merged with the quality control metrics from the audio side, in Step 4 of the transcript pipeline (DPDash formatting). 

\noindent The primary transcript QC features computed are:
\begin{itemize}
    \item The number of subjects (different speakers identified by TranscribeMe, for this should usually be 1).
    \item The number of sentences and the number of words, as well as the minimum and maximum number of words found in a sentence.
    \item The number of occurrences of [inaudible] and [redacted] markers, as well as questionable transcriptions where the word is surrounded by a question mark and brackets to denote the transcriber's uncertainty.
    \item Counts of both non-verbal (uh/um) and verbal (like/you know/I mean) edits aka filler words.
    \item Counts of repeated utterances i.e. when words or word fragments occur multiple times in a row (with words defined based on space or comma splitting here, and stutter fragments denoted by single dash directly between letters).
    \item Counts of sentence restarts (via use of double dashes by TranscribeMe).
    \item The number of commas and the number of dash characters appearing in the transcript (factors related to disfluencies per our manual transcript review).
    \item The final timestamp that occurs in the transcript (will be at the start of the last sentence).
    \item The smallest and largest number of seconds that occurs between two sentence timestamps in this transcript.
    \item The smallest and largest number of seconds that occurs between two sentence timestamps in this transcript, when weighted by the number of words in the intervening sentence. The smallest absolute value is also included as occasionally a negative time is found between timestamps.
    \item The number of sentences spoken by speaker ID 1 (which should be the vast majority of the sentences even if there were additional subjects identified).
\end{itemize}
\noindent Note that for finding nonverbal edits, the code uses the following regular expression pattern:
\begin{center}\begin{tabular}{c}\begin{lstlisting}
"[^a-z]u+[hm]+[^a-z]"
\end{lstlisting}\end{tabular}\end{center}
\noindent in the re python functionality run on each sentence, with extra spaces first added to pad between words for full accuracy of this regex. For finding verbal edits on the other hand, it is a simple count function for python strings, except a comma after each of the 3 considered verbal edit phrases is required. Basic count and split functions are used in general for all of these features. 

Recall that many of these features, such as words per sentence and occurrence rates of various disfluencies, could have implications in both transcription quality and clinical practice. To use these measures in a scientific context, we are relying on the accuracy of the particular manual transcript, as all of these features have some dependence on decisions made by TranscribeMe. However in combination with other quality control metrics, we can be proactive about identifying transcription issues, and be much more confident in the analysis of these features for our studies. In section \ref{subsubsec:diary-val-trans}, we showed that with the correct transcription settings, the transcripts produced by TranscribeMe have a high level of accuracy in encoding of verbatim speech. It is therefore important that future applications of this code take into account any differences with transcription service for evaluating these particular features.  \\

\noindent \textbf{Step 4: run\_dpdash\_format.sh}

\noindent This step left-merges the current DPDash-formatted audio QC CSV (produced by the audio side pipeline) with the current transcript QC output CSV from Step 3 of this workflow, producing an up to date diary quality control CSV with language features, viewable in DPDash if set up. The same script as on the audio side is called, because it performs both the audio QC formatting and then the transcript QC formatting where applicable. \\

\noindent \textbf{Step 5: run\_transcript\_nlp.sh}

\noindent The final major module of the transcript side of the pipeline computes natural language processing (NLP) features on the level of the sentence for each transcript CSV (again supplied via Step 2), and then summarizes the computed features on the transcript level for each patient. The script uses phone\_transcript\_nlp.py to perform these operations, drawing from specific helper functions in language\_feature\_functions.py for the different features extracted. The enhanced transcript CSVs with sentence-level feature columns are saved under a new subfolder of the patient's "transcripts" folder, called "csv\_with\_features". 

\noindent The aforementioned sentence features include:
\begin{itemize}
    \item Incoherence and word uncommonness scores computed using the Google News 300 dimensional word2vec model, applied via the gensim Python package. The path to the saved model needs to be uploaded into the code to where you have downloaded this Google News (or if desired, other) word2vec model file. The model is used to embed individual words, and will skip any words not recognized. 
    \begin{itemize}
        \item Uncommonness for a sentence is defined as the mean vector magnitude of the embedding of each word in the sentence.
        \item Incoherence for a sentence is defined in two different ways - sequential and pairwise. Sequential takes the mean of angles computed between only between consecutive words (where both have an embedding available). Pairwise takes the mean of angles computed from all possible pairings of the word vectors within the sentence. 
        \item The elementwise mean over all vectors in a sentence is also taken, to get a representative vector for the sentence. Then for each sentence, the angle between its mean vector and the mean vector of the previous sentence (where available) is computed. This results in a between sentence incoherence estimate. 
    \end{itemize}
    \item Compound sentiment scores computed by the VADER Python package.
    \item Number of syllables and associated speech rate (syllables/second) estimated using the NLTK Python package and TranscribeMe's provided timestamps.
    \item Counts of any specified keywords - which words to include are edited by the lab depending on the current study. This needs to be done within the NLP module's code, not currently available as an option on the broader pipeline interface. 
\end{itemize}
\noindent Note the keyword functionality is a simple count as implemented, so it will capture parts of words where the same letters appear consecutively even if it is not the whole word. This can be used advantageously, but at the same time requires care to not count something else accidentally. \\

\noindent For each transcript and each sentence-level feature, the following summary statistics are then computed over sentences, to contribute to a dataset of summary diary-level features:
\begin{itemize}
    \item Mean
    \item Standard deviation
    \item Minimum
    \item Maximum
\end{itemize}
\noindent The file \hl{[study]\_[subjectID]\_phone\_transcript\_NLPFeaturesSummary.csv} saved in the top level of the corresponding patient's \hl{phone/processed/audio} folder contains all of these transcript-level summary features. 

Both the feature summaries and the transcripts with sentence-level features could be used directly in analysis and are therefore key outputs of the pipeline. They are also utilized by the pipeline's visualization component to better characterize language use; from within patients over time, to across studies. A key subset of summary features (as well as of transcript QC features from Step 3) is identified there for further consideration in study of psychiatric patient audio journals.

The diary-level NLP summary features, particularly the mean, were the primary NLP module output investigated in greater depth (both for validation and scientific inquiry) within the scope of the main chapter \ref{ch:1}, but I also discuss a number of possible future directions for use of the sentence-level features in other projects there. \\

\noindent \textbf{Pipeline wrap-up:}

\noindent When the entire pipeline is called (as described in section \ref{subsubsec:diary-run}), there are a few additional wrap up steps that occur but are not implemented as standalone modules. 

On the transcript side, the wrapping pipeline script compiles and sends an email to the lab with information on which pending transcripts were successfully pulled and processed, and which were not yet available from TranscribeMe. The email will also include information on any errors encountered during processing that could require some manual intervention. As the email for this side of the pipeline is simpler logic than the audio side, it is just built into the main pipeline bash script instead of warranting its own module for composition. Like on the audio side, the lab email address list to send the status update to is specified near the top of the wrapping top level bash script.

\noindent Note no additional file management or intermediate clearing is required at the end of the transcript side of the pipeline, unless explicit code errors were encountered. 

\subsection{More on summary and visualization components}
\label{subsubsec:diary-viz-deets}
In this section, I describe each individual module in the visualization portion of the pipeline (final set of boxes in Figure \ref{fig:diary-arch}), in the order in which they are run. Each step is linked to a wrapping bash script in the individual\_modules subfolder of the GitHub repository, which would allow that step to be run independently if needed. \\ 

\noindent \textbf{Step 1: run\_distribution\_plots.sh} 

\noindent The first step of the visualization workflow is to compile study-wide distributions for the audio and transcript QC features of interest. These outputs are used not only to plot histograms of said distributions, but also in later steps of the visualization pipeline - such as for creating a correlation matrix of features. The study-wide distribution CSVs and histogram PDFs can be found under a different root file structure, in our case titled \newline  \hl{/data/sbdp/Distributions/phone/voiceRecording} (with this distributions root presently hard-coded). Simultaneously, histogram PDFs are also saved for each patient's own distributions using their DPDash-formatted QC CSV as input, and output saved under the top level of their \hl{phone/processed/audio} folder.

Under the hood, this step calls phone\_audio\_per\_patient\_distributions.py and phone\_transcript\_per\_patient\_distributions.py to take a particular patient's journal QC features and update the study-wide distribution with any newly added values, for audio and transcripts respectively. These two scripts also generate the patient-specific histogram PDFs, drawing from the distribution plotting function of viz\_helper\_functions.py; that function utilizes matplotlib and saves one feature histogram per page of output PDF. Histogram bin settings are hard-coded for each feature of interest, to keep distribution \emph{x}-axes the the same across patients. The hard-coded values were decided upon based on both review of the study-wide distribution from our primary study of interest (BLS), and any theoretical limits that might exist for a given feature. In addition to setting up the respective QC distributions, phone\_transcript\_per\_patient\_distributions.py puts together another set of distributions from the \emph{NLP} feature outputs, while phone\_audio\_per\_patient\_distributions.py does the same for OpenSMILE results. 

For OpenSMILE specifically, the script will first generate a summary CSV for the patient with mean and standard deviation of all low level GeMAPS features per diary saved to the file \hl{[study]\_[subjectID]\_phone\_audio\_OpenSMILEFeaturesSummary.csv} in the top level of the patient's \hl{phone/processed/audio} folder, and then use this to proceed as with the other datatypes. Also specific to OpenSMILE, the code will create a PDF of 10 ms bin based feature distributions for each individual audio diary, found under a new subfolder in the folder containing the OpenSMILE results CSVs, called \hl{per\_diary\_distribution\_plots}. Note that all OpenSMILE-related distribution and visualization generation is done on both the raw OpenSMILE results and the results filtered to contain only speech times. The later uses the same naming conventions but with "FilteredOpenSMILE" in the place of "OpenSMILE".

Finally, this module calls phone\_diary\_total\_distributions.py to create histogram PDFs for both audio and transcript features (QC metrics and all other extracted features) from the study-wide distributions as mentioned (replacing any old versions), and to generate a pared down distribution CSV containing only select key features combined across modalities (and save an updated histogram PDF for that as well). 

\noindent The features focused on in the pared down CSV at this point, contained across a given study in the \hl{[study]-phoneDiaryKeyFeatures-distribution.csv} under the described distributions summary folder root, are:
\begin{itemize}
     \item day
 \item patient
 \item ET\_hour\_int\_formatted
 \item overall\_db
 \item length(minutes)
 \item total\_speech\_minutes
 \item number\_of\_pauses
 \item num\_sentences
 \item num\_words
 \item num\_inaudible
 \item num\_questionable
 \item num\_redacted
 \item num\_nonverbal\_edits
 \item num\_verbal\_edits
 \item num\_restarts
 \item num\_repeats
 \item mean\_Loudness\_sma3
 \item mean\_F0semitoneFrom27.5Hz\_sma3nz
 \item mean\_jitterLocal\_sma3nz
 \item mean\_shimmerLocaldB\_sma3nz
 \item mean\_F1frequency\_sma3nz
 \item mean\_F2frequency\_sma3nz
 \item mean\_F3frequency\_sma3nz
 \item pauses\_removed\_mean\_Loudness\_sma3
 \item pauses\_removed\_mean\_F0semitoneFrom27.5Hz\_sma3nz
 \item pauses\_removed\_mean\_jitterLocal\_sma3nz
 \item pauses\_removed\_mean\_shimmerLocaldB\_sma3nz
 \item pauses\_removed\_mean\_F1frequency\_sma3nz
 \item pauses\_removed\_mean\_F2frequency\_sma3nz
 \item pauses\_removed\_mean\_F3frequency\_sma3nz
 \item speaking-rate\_file-mean
 \item word-uncommonness-mean\_file-mean
 \item pairwise-coherence-mean\_file-mean
 \item pairwise-coherence-mean\_file-max
 \item pairwise-coherence-mean\_file-min
 \item coherence-with-prev-sentence\_file-mean
 \item sentence-sentiment\_file-mean
 \item sentence-sentiment\_file-max
 \item sentence-sentiment\_file-min
\end{itemize}
\noindent Note that these are the exact column names corresponding to the various features discussed leading up to this point. The BLS version of the key features CSV was the source for the work described in the main text of chapter \ref{ch:1} (filtered further down to some extent and normalized where stated). 

Using the key features of interest, another easier to review histogram PDF is generated as mentioned, along with a basic summary CSV containing a few high level stats per participant. The latter is saved under \hl{[study]-phoneDiarySummary-perOLID.csv} in the \hl{/data/sbdp/Distributions/phone/voiceRecording} folder, and can be used to quickly assess which patients are submitting diaries of good quality at a regular rate. \\

\noindent \textbf{Step 2: run\_heatmap\_plots.sh} 

\noindent While Step 1 in the visualization scheme ought to be run first, the remaining steps can be executed in parallel. 

The wrapping pipeline next creates heatmaps showing progression of select audio and transcript QC features over time for each patient, also representing diary missingness on days where applicable. This module loops through patients in the input study, calling phone\_diary\_qc\_heatmaps.py to generate a heatmap-formatted pandas DataFrame for each patient that has existing audio QC output, where rows are features and columns are days. This DataFrame will also include transcript QC features where available. The code then uses the functionality in viz\_helper\_functions.py to create an actual heatmap image, colored using the bwr map provided in matplotlib. The minimum and maximum values for each feature in this map are currently hard-coded, in the same manner as the distribution histogram bins described above. To improve image readability, a new heatmap image is saved for every 13 weeks of the study.

The audio features found in the heatmap are diary duration (minutes), overall decibel level (db), mean spectral flatness, number of pauses, and total speaking time, while the transcript features are number of sentences and words, number of inaudibles, questionables, and redacteds, and minimum timestamp distance between sentences (weighted by the number of words in the sentence). Output heatmaps are saved for a given patient in the heatmaps subfolder of their corresponding \hl{phone/processed/audio} folder.

These heatmaps are primarily meant for data submission and quality monitoring, as an alternative to checking DPDash. However the same code could be easily adapted to create additional heatmaps of other pipeline features, which may have more clinical interest. \\

\noindent \textbf{Step 3: run\_wordclouds.sh} 

\noindent The wrapping pipeline next creates frequency-based wordclouds for each available transcript. The module loops through patients in the input study, calling phone\_transcript\_wordclouds.py for each patient. That script will use VADER sentiment along with the Python wordcloud package (drawing from viz\_helper\_functions.py) to generate a sentiment-colored wordcloud for each transcript CSV available for that patient, if such a wordcloud does not already exist. Word size in the wordcloud is determined by the number of occurrences of the word in that transcript, while word coloring is based on the average sentiment of sentences in which the word appears. The color runs from bright green for most positive to bright red for most negative, with black in the middle. The wordcloud images for a given patient's diaries can be found in a subfolder of that patient's \hl{phone/processed/audio} folder called \hl{wordclouds}. \\

\noindent \textbf{Step 4: phone\_diary\_correlations.py} 

\noindent Finally, the wrapping pipeline utilizes the phone\_diary\_correlations.py Python script to generate study-wide Pearson correlation matrices, drawing from helper functions defined in correlation\_functions.py. As this operation only occurs on the study-level and not on the patient-level, there is no bash script needed to facilitate calling of this module. 

More specifically, for each of the study-wide distributions generated by the distribution module in Step 1 (those saved under \hl{/data/sbdp/Distributions/phone/voiceRecording}), a Pearson correlation matrix is computed by calculating Pearson's $r$ pairwise for each feature in the given input CSV. Then for each such CSV, the Pearson matrix is visualized using the purple-white-green matplotlib colormap, where the darkest green represents a correlation of 1.0, the darkest purple represents a correlation of -1.0, and white represents a correlation of 0.0. These matrix images are saved under \hl{/data/sbdp/Distributions/phone/voiceRecording} with prefix naming convention matching that of the source study distribution. 

For the selected key features distribution described at the end of Step 1, the features are clustered using the Pearson distance ($1-r$) as well, and the resulting dendrogram image is saved in that same folder. A second correlation matrix image with the key summary features ordered based on prior cluster results (currently hard-coded to represent results from our largest study, BLS) is saved alongside this.

\subsection{Runtime expectations}
\label{subsubsec:diary-comp-time}
Based on our internal use of the code on a default node of the Partners Healthcare ERIS compute cluster, expected runtime for the audio side of the full pipeline is about 2 minutes per 1 minute of audio diary newly submitted. This estimate is rounded up from repeated lab observations, to be conservative for planning a workflow on a basic machine. In a typical week of diary data flow for the lab, audio processing took $\sim 1$ hour of compute time. The bulk of the runtime on this side comes from the VAD module - the rest of the audio pipeline without this step can be run in about 30 seconds per 1 minute of audio.

The transcript side of the pipeline is expected to run much faster overall than the audio side, although it is more impacted by the number of diaries than by their length. Most steps are negligible in runtime compared to the time that would be spent on the same diary on the audio side ($< 10\%$), however there is some constant overhead in obtaining the word embeddings from the word2vec model for a given diary. Therefore the incoherence measures take the bulk of the transcript-side time, and the overall runtime on a basic machine should be estimated at about 1 minute per transcript to be conservative. In a typical week of diary data flow for the lab, the transcript processing took $\sim 30$ minutes.

Note we typically ran audio processing on Mondays only and transcript processing on Fridays only. In this framework, TranscribeMe turnaround time was very good, with the entire batch of transcriptions always finishing in time for the transcript side to pull everything back the same week. Even with longer term bulk uploads we were generally impressed with transcription return times for audio diaries, which is likely attributable to the short length per file, allowing more parallelization of the human transcription process.

The visualization portion of the pipeline was not built out until after the audio and transcript sides were already regularly running. Visualization could be run every week after transcript processing finishes, but it is also suitable to run less frequently, only as needed for analysis. In our case, we measured visualization runtime on a large set of diaries, including over 10,000 audio files and over 1,000 transcripts. Distribution generation of per-diary summary stats was able to complete in under 5 minutes, and heatmap generation in under 10. Study-wide correlation matrices for the summary stats could also be completed in under 5 minutes. 

The wordclouds on the other hand were much slower, only getting through $\sim 7$ transcripts each minute - however, wordclouds are only generated once for a given transcript, so if the plan is to run regularly there will not be additional update costs as the dataset grows, like there could be for distributions, correlations, and to some extent heatmaps. The per-diary OpenSMILE histograms were somewhat similar to wordclouds, with the full dataset taking a little over an hour to process. But this is still an order of magnitude faster runtime per example, and yet it has the same advantage of not requiring any recomputation. Therefore the main bottleneck for visualization are the sentiment-colored wordclouds. 

\subsection{Adapting the code to run in new contexts}
\label{subsubsec:diary-adapt}
As mentioned, the current pipeline has flexible settings for use across different Baker Lab studies, but it does not account for possible changes across labs, and was really just written with internal lab use in mind initially. I therefore provide here a high level summary of the main changes that might be needed for another group to utilize this software. The closer the data collection process mimics that of our lab, the simpler these needed changes are. Indeed, there are a number of possibly necessary cleanup steps described as part of the expected file structure information near the start of this section. Current assumptions and associated updates that would be required to change those assumptions include:
\begin{itemize}
    \item Ensuring the root path is updated to match your root (look for \hl{/data/sbdp/PHOENIX} in the current code). Similarly, make sure there is some root folder for storing the study-wide distributions to, currently look for \hl{/data/sbdp/Distributions}.
    \item Following the described nested folder structure, or changing the code to have different folder expectations, which could require care depending on the needed updates. Note also that I enforce in the folder structure here an expectation that subject IDs for a given study will always have exactly 5 characters, as that is a lab convention. This is something else about the input/output folder structure assumptions that may need to be updated for use by other groups.
    \begin{itemize}
        \item As far as outputs of the pipeline go, currently everything is stored on the PROTECTED side for ease of use, which doesn't materially change anything for our internal lab server. However for a project like AMPSCZ (described in chapter \ref{ch:2}), the GENERAL versus PROTECTED distinction is important for downstream dataflow. It therefore may be necessary to redirect non-PII outputs to the corresponding GENERAL side \hl{processed} folders instead, via updates to this codebase and parts of its assumed folder structure. 
    \end{itemize}
    \item Encrypting the raw audio diary files with cryptease (easy via Lochness \citep{lochness}), or updating the pipeline to remove the decryption steps and no longer look for .lock files. 
    \begin{itemize}
        \item Manual entry of the decryption password is also the main reason the code is not intended for use with a regular cron job. Setting up the code to use a config file instead of answering command line prompts to determine the options for each run would be a good improvement to the code, but it would require care in how the decryption password is stored, if there will be one.
        \item The TranscribeMe SFTP account password is manually entered for similar reasons, but the TranscribeMe SFTP account username is currently hard coded, something that will obviously need to be updated for other groups. Look for the username "partners\_itp" in the code here to find where this should be replaced (or ideally abstracted out to improve the general use of the code going forward - and similarly for changes to e.g. the root path). 
    \end{itemize}
    \item Using MP4 format, or updating the ffmpeg file conversion steps as needed (unnecessary if starting with WAV, but these are much bigger in storage). 
    \item Using Beiwe, or emulating their described folder structure and file naming conventions (or adapting the code to accommodate new file naming and related conventions for raw audio journals). 
    \item Accounting for different timezones as needed (code assumes Eastern Time only). 
    \item If multiple different prompt types will be submitted by the same subjects on possibly the same days (or need to be carefully distinguished in tracking across different days), the pipeline will need to be reworked to track prompt IDs. Regardless, one may want to rework the pipeline to handle multiple submissions in a day differently -- perhaps retaining the longest recording instead of the first, or at least confirming the first does not fail QC before rejecting subsequent submissions from that day. 
    \item If receiving email alerts is desired, the email lists at the top of the audio and transcript branch wrapping bash scripts used to run the pipeline should be edited, as they are currently hard coded there. Ideally these email lists would eventually also be abstracted out to a more general settings file for other groups to immediately be able to use this code. \begin{itemize}
        \item For the email to TranscribeMe sent by the audio side of the pipeline, the address it will appear to be sent \emph{from} (for reply purposes) additionally will need to be updated, as it is currently hard coded closer to the bottom of the audio wrapper bash script (search for mennis@g.harvard.edu). 
    \end{itemize} 
\end{itemize}
\noindent Another major assumption is the use of TranscribeMe for transcription, with SFTP account set up and all the same settings as us. If switching from TranscribeMe entirely this could introduce many new requirements past the early audio processing steps, and might warrant rewriting parts of the pipeline entirely. On a smaller scale, there are also some specific settings differences that may need to be adapted for depending on project budget: 
\begin{itemize}
    \item If transcription notation will not be verbatim then the disfluency features don't make sense, and there might be other notation things to tweak within the QC functions. TranscribeMe markings and notation should be confirmed before proceeding with this code regardless.
    \item If TranscribeMe will not do sentence by sentence splitting, then the code needs to be adjusted to either split into sentences automatically and keep all the same features, or handle the individual transcript CSVs as containing chunks of text in the rows instead, changing the feature calculations themselves. For different features different processes may be needed. 
    \item The timestamp-based QC and speaking rate features may need to be revisited if the provided timestamps will have lower resolution than 100 ms. 
    \item The code will not work without tweaking if TranscribeMe does not provide speaker IDs as described. 
    \item Currently the pipeline expects ASCII encoding, but this should likely be relaxed to UTF-8, especially for any projects that will involve non-English languages.
    \item More broadly, foreign languages are entirely untested and would be very unlikely to work smoothly with this pipeline out of the box.
\end{itemize}
\noindent Note also the pipeline does not support return of non-txts, but TranscribeMe is fine to do this as long as they are clearly told to. \\

Throughout the above implementation descriptions, I've mentioned places where there are "settings" currently only internal to the code. Ideally more settings will be built into the actual wrapping script, but in the meantime if one wants to make updates to - for example -  visualization settings (e.g. bins used for the different feature histograms, changing included features or normalizing features before generating the visualizations, etc.), that will require going into the individual modules and making updates. This is also true of core process settings like the chosen OpenSMILE config, the word2vec model used for incoherence, the minimum vocal silence duration needed to label something as a conversational pause, and so on. Keep in mind too that only the provided settings have been validated against a pilot diary dataset, or even tested at all. It will be important to read the above implementation information to get a handle on what aspects might be desirable and/or easy to change, as well as where they can be found and which other component python functions they might impact. 

It is worth emphasizing that while adding entirely separate new features can be relatively straightforward for an active study, if intermediates need to be updated for existing diaries it may be more difficult to add those new features part way through a study with data being processed by this pipeline. That applies too to mistakes that could interfere with the accuracy of downstream data, like incorrectly entered and later updated consent dates in the metadata CSV (sourced in our workflow from REDCap by Lochness). Improving on the ability to automatically update intermediates with metadata corrections and/or new features could be a worthwhile improvement to the code more generally.

Finally, note that there are places where the efficiency of the current code might be improved. Only some of the individual module wrappers actually use bash in a meaningful way, and converting the ones that don't to python wrappers instead could prevent some unnecessary re-importing of python packages for every subject with new diary data; this is especially relevant for the loading of the word2vec model for computing incoherence metrics. There are also places where computation may be unnecessarily repeating for simplicity, like the OpenSMILE summary code. There are of course potential wish list expansions to the code as well, but these are not really software engineering considerations but rather scientific project design questions, which are covered at much greater length for audio journals in the main text of chapter \ref{ch:1}.

\section{Considerations for implementing an audio diary pipeline for AMPSCZ}
\label{sec:ampscz-diary-todo}
It is important that data flow and quality monitoring for diaries is launched soon, in order to identify potential QC issues before it becomes too late to address them, as well as to better monitor participation rates and intervene where suitable to improve long term subject submission habits. \\

\noindent The list of differences to keep in mind between my pipeline described in chapter \ref{ch:1} and what needs to be done for AMPSCZ -- which is intended for use by someone familiar with the project's interview architecture details provided in chapter \ref{ch:2} -- is as follows:
\begin{itemize}
    \item Update the preprocess steps to handle the MindLAMP expectations instead of Beiwe.
    \begin{itemize}
        \item The files for both Pronet and Prescient are located on PROTECTED side of PHOENIX under raw in the phone data type for each subject. For a given participant, each submission is found as a loose MP3 file there.
        \item Each filename begins with an ID that is unique to the subject based on their MindLAMP account, which doesn't have any clear use for the present pipeline, though it could be another field to sanity check.
        \item The only other real information provided in the diary filename is the submission date and at the end of the name the index of the recording - in case the participant submitted multiple in a given day.
        \item An example path from within PHOENIX is \newline \hl{PROTECTED/PronetYA/raw/YA01508/phone/} \newline \hl{U3627455308\_PronetYA\_activity\_2022\_07\_04\_sound\_0.mp3}
        \item Unfortunately it is not clear from what timezone the date is assigned without further follow-up with the MindLAMP team. Regardless, it will not be possible to replicate anything the pipeline does with regards to submission times without further investigation.
        \begin{itemize}
            \item It is unclear if the timestamp in file metadata on the server is the recording time or if it reflects a timestamp from later steps of the upload process.
            \item If it is not accurate to recording time, then it is unclear if recording time can be obtained through some other presently available data, or even whether it is something that can be incorporated for future submissions by updating certain settings.
            \item If there is not a good way to find true recording time, then the date in the filename needs to just be taken at face value to find the study day number -- which has clear downsides relative to the current diary pipeline's approach to adjust for late night submissions. It is also unfortunate because recording time can be an inherently interesting variable. 
        \end{itemize}
    \end{itemize}
    \item The decryption part of the diary pipeline needs to be removed as it is not relevant for AMPSCZ. 
    \begin{itemize}
        \item The pipeline then needs to be adapted to use a config file instead of manual settings entry, so it can be on a cron job analogous to the present AMPSCZ interview recording pipeline.
    \end{itemize}
    \item Someone needs to decide how to handle multiple submissions in one day. If the plan is not just to ignore all submissions after the first, then that design assumption needs to be changed in the diary code and the naming conventions used for outputs need to be updated to reflect that change.
    \item Update the folder structure to match the AMPSCZ PHOENIX conventions instead of the Baker Lab ones, as the ordering of folder levels is quite different for AMPSCZ. 
    \begin{itemize}
        \item Also update diary pipeline outputs to go to GENERAL where appropriate (i.e. PII is safely removed), so they can make their way towards the data sharing repository per the overarching AMPSCZ data flow structure.
    \end{itemize}
    \item Within the wrapping commands that run the diary pipeline, the more compute intensive modules need to be broken out entirely, as only lightweight QC is possible to run on the data aggregation server where Lochness runs.
    \begin{itemize}
        \item The intensive feature extraction can instead be implemented on the AV processing server where interview feature extraction will be occurring. The dependencies for this should not be installed on the data aggregation server, so the repository organization needs to also be updated to reflect the split structure. 
        \item As technically the AV processing server is meant only for low level feature extraction on those data types that cannot go directly to the \emph{predict} analysis server for privacy reasons, only the lowest level acoustics feature extraction functions should really be implemented as a package there. Summary stats and visualizations computed by this pipeline should separately be reviewed to decide what is worth including, and then implemented downstream on \emph{predict}. This also includes all feature extraction running on redacted transcripts.
        \item The data flow infrastructure and server set up steps required to adapt those pieces of the diary pipeline will need to be considered similarly to the ongoing work to generate a full feature set from interview recordings (see chapter \ref{ch:2} future directions discussion), though the diaries should turn out to be much less complicated.
        \item In the meantime, the process of working with the diary datatype for AMPSCZ should begin with just getting the initial data management and quality monitoring steps to be run on the data aggregation server fully working, which is by and large the focus of this list.
    \end{itemize}
    \item Storage requirements need to be reviewed for the proposed AMPSCZ diary pipeline, and in that process it should be established with DPACC which files will be protected from clearing by Lochness, as well as what subfolder structure and permissions are expected in processed for phone-related datatypes. 
    \item Transcription settings for the journals need to be verified with TranscribeMe, and the code accordingly updated to handle any differences.
    \begin{itemize}
        \item Implement process for curly brace marked redactions as described for interviews in chapter \ref{ch:2}.
        \item Push for the diary transcripts to have sentence splitting in the same way that we do internally (not the case for AMPSCZ interviews, which are turn-based due to cost). If the splitting process will remain coarse however, then certain current diary pipeline features need to be redefined, or an automated sentence splitting method could be used to preprocess transcripts into sentences for those features. Though the latter would still mean missing timestamps on some of those sentences compared to our original method, which means a few features that rely on those timestamps would need to use the original transcripts and be updated still to assume not sentence-level. 
        \item Review the different TranscribeMe transcript formatting conventions covered in the AMPSCZ interview CSV conversion script, and ensure they are all included in the version of that function for this diary pipeline, as exact TranscribeMe notation can sometimes vary between projects.
        \item Support for the additional non-English languages included in the AMPSCZ project needs to be built.
        \begin{itemize}
            \item Handle the labeling of language by site in the transcript filename for TranscribeMe accounting. 
            \item Change the code so it does not reject non-ASCII transcripts by default, instead rejecting non-UTF8 only.
            \item Will also need to handle language-specific differences that may arise in the actual QC (e.g. "inaudible" markings in other languages), which is not yet done for interviews either and requires further observation/testing to be exhaustive. 
            \item Note for any eventual transcript feature extraction, the use of other languages will be an entirely different issue that cannot be accommodated by my pipeline at present.
        \end{itemize}
    \end{itemize}
    \item The process for site manual redaction review needs to be decided on, analogous to AMPSCZ interviews. What needs to be reviewed initially (if anything) and with what frequency should diaries be sent for review? Then that needs to be implemented within the pipeline as is done in the interview code, including creating the site-specific reminder emails about manual review. 
    \begin{itemize}
        \item Support for this will additionally need to be built into the existing push tool on Pronet/Prescient for sending interviews back to sites for manual review. It will need to ensure the diaries are kept separated from the interview transcripts that are to be manually reviewed. Then Lochness must be set up to pull the successfully reviewed transcripts separately. 
    \end{itemize}
    \item The automated emails that are currently implemented by the diary pipeline for single lab use need to be adjusted to include more relevant updates per site on the audio and transcript ends, and to not send these updates when nothing new has occurred.
    \begin{itemize}
        \item The design of the existing diary emails should be tweaked like the interview emails were, to include the info we decide is most salient for daily tracking (see chapter \ref{ch:2}), and then the code can be updated accordingly.
        \item An additional email that sends when there are major protocol or QC warnings also needs to be implemented, along with improved logging for the needs of AMPSCZ. Though the number of SOP problems should be less than for interviews because more of the recording/upload process is automated by MindLAMP/Lochness for diaries. 
        \item In order to implement improved logging in the diary pipeline, more accounting on when files are first found, what the consent was at that time, etc. needs to be added to the code first -- again referring to the interview code and problems encountered with interviews thus far to assist in design.
    \end{itemize}
    \item The site-wide combined QC CSV and combined accounting and warning CSVs need to be implemented into a weekly email for diary progress updates, as is being done currently for interviews. Alongside this, HTML tables with relevant counting stats and histograms with QC distributional comparisons should be replicated from the interview pipeline where appropriate for the diaries. 
    \item Installation and initial testing will need to occur on both Pronet and Prescient data aggregation servers (though dependencies should all already exist per the interview pipeline).
    \begin{itemize}
        \item Also determine if TranscribeMe needs to be validated indpendently on a few example diaries like was done for interviews before embarking on that production pipeline setup. This could occur as part of the experimental comparison of manual with automated methods for diary transcription that is planned (described in subsequent discussion). 
    \end{itemize}
    \item The QC outputs need to be integrated with the AMPSCZ DPDash as was done for interviews. This requires both interfacing with DPACC for the standard DPDash view of QC features over time in each subject, as well as writing additional computation functions to create the stats needed for the new DPChart views.
    \item The key QC features and acceptable ranges might be different for AMPSCZ diaries than they were for BLS diaries or for AMPSCZ interviews, so this needs to be kept in mind as the code is written.
    \begin{itemize}
        \item Note that the maximum duration for AMPSCZ diaries is 2 minutes instead of the 4 minutes used to cap the BLS Beiwe diaries. Although that shouldn't change anything about the core code it will change to some extent how results are visualized and interpreted.
        \item Once initially written, decisions related to QC feature interpretation in the diary pipeline can of course be iterated on, and periodic updates to DPDash configs and email summary methods are expected.
    \end{itemize}
\end{itemize}
\noindent It is likely that this adaptation process will be easier if beginning with the relevant (non-video) pieces of the interview pipeline and referring to the parallel portions of the present journal pipeline, rather than vice versa. That will more directly tackle the pressing need for adaptation of the data flow and quality monitoring code to diaries for use on the data aggregation server. Other parts of the diary pipeline from this chapter can then later be used somewhat more directly for feature extraction on the AV server. The renamed and converted WAV files created as intermediates under the PROTECTED side of processed by the data flow pipeline can largely be input to feature extraction functions provided in the existing diary pipeline pipeline without major changes necessary.

\subsection{On estimating audio journal participation rates}
One relatively minor scientific limitation of the audio journal that can cause much greater logistical challenges for a project with the reach of AMPSCZ is the difficulty in estimating expected participation rates and thus data collection costs early on in a study. This in part comes with the territory of a psychiatry humans research project, and doubly so a datatype that is fairly uncharted to date. Still, as longitudinal journal participation rates are mostly subject-driven, and compensation per submission plus added professional transcription costs can accumulate to meaningful amounts, it is harder to have a narrow margin of error in direct pricing estimates of diary collection than most other digital psychiatry datatypes. This is even more true for AMPSCZ, where consenting participants to record audio journals is a bit of an afterthought in the project workflow. 

Ultimately, final journal dataset costs could vary highly depending not only on how many participants end up consenting, but how regularly they end up recording journals, what durations their submissions typically are, and to what extent such factors change over the course of a year enrolled in the study. While a number of these questions were characterized in the pilot BLS dataset in the main chapter \ref{ch:1} (\ref{subsubsec:diary-time}), and we were generally very satisfied with that overall pilot data collection process, there are many distinctions between BLS and AMPSCZ that make it especially difficult to predict how well any participation observations will translate. New considerations for AMPSCZ include:

\begin{itemize}
    \item Demographics -- while BLS had diversity amongst some demographic categories (Table \ref{table:bls-demographics}), it was less so in others. This was limited by available recruitment demographics at McLean, and ultimately single site studies conducted here are also going to reflect some cultural factors related to its location in the Boston area. AMPSCZ covers such a wide variety of sites globally that we would expect a quite different overall demographic profile, and likely more heterogeneous cultural properties, some of which may relate to diary participation rates. This is especially striking when considering the many international sites collecting data for AMPSCZ, where major considerations can pop up such as differences in recording language or in local laws. Some factors that could affect participation rates in this domain that could not only differ between AMPSCZ and BLS but also between distinct AMPSCZ sites are as follows:
    \begin{itemize}
        \item Attitudes surrounding mobile phone app usage and especially digital privacy.
        \item Attitudes surrounding openness about mental health issues (though this could moreso affect content).
        \item Availability of private spaces for recordings/attitudes surrounding physical privacy.
        \item Availability of free time for reflection.
        \item Perception of the value of psychiatry research/sense of duty to research contributions (or to carry through on a commitment). 
        \item How much a study's monetary compensation actually means to a given person.
        \item Likelihood of discussing participation with friends or family, and type of response that might be received.
        \item Possible belief in journaling as a good personal habit, or overall level of self-interest.
    \end{itemize}
    \item Youth -- this is of course just another demographic category, but it is one that is a direct target of AMPSCZ, while BLS did not recruit anyone under the age of 18. Any AMPSCZ participants that are minors living at home with guardians could have a completely different engagement profile with audio journals than an adult with otherwise similar demographics might, and it could also depend on the relationship they have with their parents and their parents' attitudes towards the study. 
    \item Psychiatric diagnosis -- while BLS also had a fairly diverse population of mood and psychotic disorders, AMPSCZ is targeting a diagnostic category not covered by BLS at all, i.e. clinical high risk for eventual development of psychosis. Additionally, AMPSCZ will recruit some controls alongside clinical participants. Symptom severity levels could differ between the study populations as well, and BLS mostly covered moderate disease, so it is unclear how those with severe symptoms or healthy controls might differ in participation trends. The specific set of symptoms that are server may be yet another factor affecting engagement too.
    \item The app being used -- all of the BLS data analyzed in the main chapter \ref{ch:1} came from the Beiwe platform, but AMPSCZ will be collecting diaries through MindLAMP instead (in addition to EMA and passive phone sensing datatypes). There might be differences in app interface or notification style that make it more or less cumbersome to submit a journal or affect how rewarding the recording/submission process feels. There also might be differences in what sort of glitches arise that cause data loss on the back end or require participant troubleshooting on the front end, and this could vary by operating system as well, which presents another opportunity for demographic effects. 
    \begin{itemize}
        \item Recall that there is no software in place yet for tracking the diary recording submissions across AMPSCZ, so we have little insight yet on any of these possible issues or how they might be counteracted during the study. 
    \end{itemize}
    \item The protocol being used -- the prompt for AMPSCZ diary recordings ("how are you doing lately") is slightly different than the prompt used for BLS, and using the word "lately" instead of "in the last 24 hours" might plausibly encourage people to submit less often because they are less aware of the daily expectation. Further, some BLS patients continued to submit journals for many years, but AMPSCZ subject enrollment is supposed to last for only one year per person. There are also differences in both the protocol for compensating participation and for approaching/consenting participants in the first place, and ultimately audio journals are even less of a priority for AMPSCZ than they were for BLS. It is not clear in fact what sort of instructions are being given to subjects about recording the diaries in practice, which might interact with some of the other above concerns like patient privacy beliefs or possible recording software bugs, and might impact final recording quality too.
    \item The researchers running the study -- just as there are many possible cultural, legal, and other differences between site participant population pools, there are also differences in attitudes between the researchers running the study across sites, which is quite distinct from a single group project like BLS where we are very aware of all data collection and patient interaction details. As can be seen by the wide variety of site mistakes and other site-specific patterns found in early interview data collection for AMPSCZ (chapter \ref{ch:2}), there are differences in training quality, prior experience with (or care for) a particular datatype, recruitment rate and rapport with participants, etc. across the sites. Different sites' handling of the consent process might change the rate of diary enrollment irrespective of other between site differences in e.g. study population. Different sites' handling of the app setup process might similarly change how many audios get correctly uploaded or how frequently (and with what typical duration) people who do participate tend to record. More general rapport between site staff and participants can impact how engaged certain subjects are with the diary recording process too.  
\end{itemize}
\noindent Future single lab studies, even if shifting to an entirely different disease population or working more to target different demographics or e.g. changing a prompt would still be able to much more tightly control many of these things, and can track such differences in much more detail/with more personal knowledge. So although any future works will need to take care with participation estimates, it is especially tricky for AMPSCZ. 

Even with the pilot data we do have from AMPSCZ to date, a number of features would need to be considered to make any attempt at a decent long term participation estimate. Knowing how many diaries their are and their typical lengths says only so much without a proper pipeline in place. We would want to know how many subjects are the diaries so far coming from and what frequency have they been submitting at/when were they enrolled and when did they last submit. We would also want to look for site-specific differences in submission trends so far, and ensure that large chunks of data are not missing because of site errors with the MindLAMP app/data import setup (overall or specifically for journals). Additionally, many foreign sites have not started the study yet or started the study only recently, so estimates of language distribution across recordings may not be accurate. But the cost difference between languages can be upwards of a dollar per minute, and so uncertainty in site/cultural participation variance might introduce extra noise in cost estimates because of the language pricing discrepancies. 

Nevertheless, I do provide some very rough early estimates of possible AMPSCZ dataset size and associated transcription costs in Appendix \ref{cha:append-ampscz-rant}. On the bright side, if costs exceed this estimate it means diary participation was great, which would be a nice surprise in some respects. Conversely, if diary participation is disappointing, that would at least mean a cost savings from the expected budget. Overall I am obviously hoping for very good diary participation in AMPSCZ, especially as it seems the project coordinators might slowly be taking the datatype more seriously. It is a nice opportunity for AMPSCZ to lead the way in characterizing many properties of this under-studied format. 

It is true also that biases in participation rates/data availability can have scientific impact and seriously complicate computational analyses, but that problem is near universal in humans subject research. Fortunately for diaries, they are bite-sized enough that they are easy (and sometimes even enjoyable) for many participants to do, and some amount of missingness in the overall dataset is not necessarily problematic. Their suitability for longitudinal work is very relevant here, because for any features being grounded in a participant's own baseline, we can control for baseline properties of participation likelihood too. In fact missingness in diary recordings may itself be a clinically relevant signal for some. There will of course be people that just don't really participate at all, and not much can be done about that -- but this is a limitation of most datatypes, and the power of diaries for certain analyses doesn't mean they should be some sort of silver bullet used to study all people. Multi-modal digital psychiatry work - like that of chapter \ref{ch:3} - can protect against the between subjects bias issue somewhat by diversifying data sources, and in fact within the discussion of chapter \ref{ch:3} I consider more ways to potentially further improve diary participation rates as well as the sorts of bias different participation schemes could introduce.

\chapter{Supplement to Chapter \ref{ch:2}}\label{cha:append-chapt-refch:2}
\renewcommand\thefigure{S2.\arabic{figure}}    
\setcounter{figure}{0}  
\renewcommand\thetable{S2.\arabic{table}}    
\setcounter{table}{0}  
\renewcommand\thesection{S2.\arabic{section}} 
\setcounter{section}{0} 

\section{Details of AMPSCZ interview conduct and recording procedures}
\label{sec:ampscz-pro}

\subsection{Onsite EVISTR recording (psychs only)}
\label{sec:evistr}
From the AMPSCZ standard operating procedure (SOP):

\begin{quote}
Turn on the recorder using the ON/OFF slider on the left-hand side. Press the MENU button on the front and use the PREVIOUS/NEXT buttons on the right-hand side to scroll between menu options. Select the highlighted menu option using the MENU button; you can backtrack with the STOP button. To set the date, go to: System settings > Date and time > Set date. The VOLUME UP/DOWN buttons on the right side (below PREVIOUS/NEXT) can then be used to move back and forth between the year, month, and day categories. Within each category, PREVIOUS/NEXT can be used to change the value. Once the date is set correctly, pressing the MENU button will save the new settings and exit. The time can be set analogously, using the "System settings > Date and time > Set time" option.
\end{quote}

\begin{quote}
To ensure recording quality, next navigate to "Recording settings > Recording quality" under the menu, and confirm that "1536.WAV" is selected, using the same menu navigation system as described above. Then navigate to "Recording settings > Voice activation" and ensure that "AVR mode" is selected.
\end{quote}

To collect recordings with the device at each interview, place the recorder between you and the subject (but closer to the subject) and press the red REC button to start. Only do so once the interview is actually ready to begin. Once the interview is completed, pressing STOP will terminate the recording, which will be automatically saved on the device. The resulting WAV file should then be transferred to a computer as soon as possible using the USB 2.0 to Micro B cable provided by EVISTR. A folder will pop up once the device is connected to the computer, and the recordings can be found in the RECORD folder labeled with the corresponding date/time. Files should be transferred as is to the lab computer and ultimately to whatever upload mechanism is used for interview data management by the group. In the AMPSCZ project, this will be either Box or Mediaflux, as will be described. Sites should upload files as they become available, and follow the protocol outlined in the project standard operating procedure (SOP). 

\subsection{Zoom interviews protocol}
\label{sec:zoom-settings}

To ensure quality and organization of the recordings obtained from Zoom, the project SOP provides instructions on settings to use within the Zoom app. For the interviewer, many of these suggested settings can be set up once on your work computer with the Zoom client logged into your work account. If the account is signed out or the Zoom client updates, the settings should be reconfirmed, but they generally should remain set correctly once initially done. 

\noindent The first category of account settings, which are only needed on the interviewer's end and which should not be changed for any interview, are the settings for saving of recordings. These can be found on the "recordings" tab of the settings menu on the Zoom computer app:
\begin{itemize}
    \item Preset the folder path where recorded interviews will be locally stored on the interviewer's computer. It is important that this folder is located on an encrypted hard drive (see instructions specific to your machine to confirm this). Do not select "Choose a location to save the recording after the meeting ends".
    \begin{itemize}
        \item Note that this preset path should not be within a cloud storage folder. Upload of completed interviews to the server via Box/Mediaflux (or Dropbox if internal to Baker lab) is a separate process to do once the interview recording folder is successfully saved by Zoom.
        \item Note also that the automatically generated Zoom file and folder names should not be changed. It is important to be recording research metadata notes (for AMPSCZ via REDCap/RPMS runsheets) about each interview as soon as they happen, including the date/time the interview occurred along with the subject ID involved and the type of interview conducted. 
        \item Interview folders should be moved or copied to the appropriate upload location as soon as possible, which as detailed in the project SOP will involve putting them under the appropriate subfolder in cloud storage for the subject ID and interview type. This needs to be done carefully, and it should absolutely not be done until Zoom indicates the entire interview recording has successfully completed saving.
    \end{itemize}
    \item Ensure "Record a separate audio file for each participant" is checked.
    \item Ensure "Add a timestamp to the recording" is checked.
    \item Do not check "Keep temporary recording files".
    \item There is also a screen share setting, which should be irrelevant as screen sharing should not occur. But in case it does, it is advisable to check "Record video during screen sharing" to ensure the interview does not get interrupted. 
\end{itemize}
\noindent Additionally, you should ensure your display name remains consistent so that the diarized audios can be easily distinguished by automatically generated filename, which will include this display name. You can change your display name before beginning recording if desired by clicking the dots next to your current name in the participant list and selecting "rename". Zoom should then continue to use this same display name for you across meetings hosted from your account. Set this before the first interview you conduct, and then keep it consistent after that. 

\emph{Important note on display names: } If conducting an onsite open interview, please ensure that the participant's display name has been changed to their subject ID, or at the very least to not match a known lab member's name. As onsite interviews use lab equipment for both ends of the recording, if care is not taken both diarized audio files might match expected interviewer display names, making it impossible to determine interviewer and participant identities based on metadata. Similarly, when conducting an offsite psychs interview that includes additional non-interviewer attendees (e.g. parents), please ensure the display name of the participant is changed to subject ID so they can be distinguished from the other non-interviewers.

Note that if display name conventions are not followed, the backup method for automatic detection of subject and interviewer requires that a different SOP request is followed -- that the primary interviewer is the first speaker in the recording and the participant is the second speaker. Please make a concerted effort to meet both of these SOP requirements. \\

The other major category of settings is related to recording quality. These settings need to be initially confirmed and may need to be occasionally adjusted on the interviewer account, but they also need to be walked through with the participant before beginning recording any interview. Even for repeat subjects, these quality checks should be performed every time, as there is no way of knowing if they have changed their account settings or logged in from a different device since the last interview. A few of these settings additionally may require tweaking from default based on the specifics of the current interview. Remember to revert them to default on your own account after the interview recording is complete if they were altered. For these reasons, it is also worth confirming your own settings before each interview as part of the quality assurance workflow.

\noindent The settings that should be adjusted under the "audio" tab of the Zoom client on both computers are:
\begin{itemize}
    \item If using an external microphone rather than the built-in one, ensure that it is selected.
    \item Uncheck "automatically adjust microphone volume", then set input volume using Zoom's "test mic" function to ensure that the input audio is not too quiet nor too loud.
    \item Put "suppress background noise" on the low setting. The interviewer should of course always be in a quiet setting and hopefully not need to tweak this. If the participant is also in a quiet space, they should do the same, but a higher background suppression setting may be needed in some circumstances if you notice a high level of noise coming from their environment.
    \item Check the box that says “show in-meeting option to ‘turn on original sound’ from microphone.” Then uncheck the boxes underneath that say “high-fidelity music mode,” “echo cancellation”, and “stereo audio”. 
    \begin{itemize}
        \item This will cause a "turn original sound on” button to appear in the top left of every video call. It is important to make sure this is clicked before starting recording, for every interview.
        \item Once you have turned original sound on, please confirm that the participant's audio quality sounds acceptable, and ask the participant to confirm that your audio quality sounds acceptable, before beginning the recording.
        \item One thing in particular that may need to be adjusted is that depending on participant setup, it may be necessary to turn back on the "echo cancellation" option mentioned. 
    \end{itemize}
\end{itemize}
\noindent It is important that settings are confirmed and that audio quality is sanity checked on both ends before beginning recording. Note that not all of these options are available for audio input on the phone Zoom app, which some participants may be using for some interviews, though this should be advised against. If a participant needs to use their phone, please encourage them to use a headset with the phone if possible.

For video recording, there are not many account or app settings to adjust, however there are quality checks that need to be made on the level of the individual interview. On the participant's end, it is important to ensure that their entire face is within the frame, and that the camera is stationary. The latter problem is mostly an issue with phones; the participant can be advised to lean their phone against something to stabilize the camera, but in this case it is important to ensure that the audio input on the phone is not obstructed, unless they are using a headset. Also, for those using a headset, it should be positioned in a way that prioritizes audio quality first, but ideally in a way that does not obstruct the view of the face. Similarly, participants should be asked not to wear a mask if possible, nor any hat that might obstruct view of their eyes.

On the interviewer's end, it is important that the Zoom client is in gallery view. Unfortunately there is not an account wide setting to force all meetings to default to gallery view, so this has to be confirmed manually before beginning recording of any interview. Gallery view can be selected via the "view" panel in the top right of the meeting window. For open interviews in particular, if there are people sitting in on the call besides the subject and the interviewer, such as a subject's parents, please ask them to keep their cameras off. You should then be sure to select "hide non-video participants" in the same "view" panel. Before beginning an open recording, you should confirm that there are exactly two video feeds showing on the call with faces visible: yourself and the participant. Note that "hide non-video participants" is available as an account-wide setting under the "video" panel of the settings menu in the Zoom client, so this could be set as the default for your account to save time during interview set up. 

For psychs interviews there are no specific requirements for video, but if anyone will have the camera on, please ensure you have the meeting in gallery view. Regardless, please ensure that no one with their camera off (but a box visible in the call) has a profile picture icon that features a face, as this could cause our initial QC metrics to overestimate the number of psychs interviews that we have available with face data.

\subsubsection{On pauses in Zoom recordings}
Note that if the recording needs to be paused for a time in the middle of the interview, we recommend using the pause functionality rather than stopping and starting a new recording, as will be discussed in the AMPSCZ project report to come. Stopping recording and then starting again within the same Zoom meeting will create split files within one Zoom interview folder, which can be difficult to systematically handle appropriately with software (and is not supported by the code described here). Pausing on the other hand will cause produced content to have a gap with silence/a black screen during the pause times, so that the final recording accurately reflects the entire duration of the recorded call. If it is desired to have separate sessions within one interview, it is much preferable from a code perspective to end the first Zoom meeting entirely, save that recording, and then start a new Zoom meeting where recording can begin again when ready. This will produce two separate interview folders with different start timestamps, which can be processed normally by my pipeline.

\subsection{REDCap/RPMS forms}
\label{subsec:redcap}
For clinical interviews in particular, scale ratings for each time point will be recorded via the REDCap/RPMS platform. Regardless of interview format, information on the recording should be carefully entered into the interview runsheet for each session, similar to how a lab notebook would be used for wet lab data collection. This includes standard metadata as well as any notes of potential interest on recording quality, abnormal settings used for interview collection, etc. Details of the REDCap/RPMS instruments within these interview recording protocols are beyond the scope of the current chapter, as handling of REDCap/RPMS data is not performed by my software. However, info found (or unfortunately not found) on REDCap/RPMS has been utilized in conjunction with the pipeline outputs to interpret certain data monitoring anomalies, in addition to their obvious importance in eventual downstream research questions.

\subsection{Conducting open interviews}
\label{subsec:open-sup-ra}
Open interviews are intended to occur at baseline and at a 2 month follow-up point, for each participant enrolled in the study. The goal is to elicit interesting free speech from the participant on any topic possible. Some participants may not want to talk much, and this can of course be clinically relevant, but it is important to make a good attempt to get them to open up during the allotted time. 

\noindent The AMPSCZ SOP recommends beginning the open interview recording with the following script, or something similar to it:
\begin{quote}
First, I’d like to thank you for taking the time to talk with me. Like I mentioned, our conversation will be recorded for analysis. This interview is different from the other interviews we do. I would really like to get to know you and learn what your life is like. So, how have things been going for you lately?
\end{quote}

\noindent The SOP goes on to provide the following tips on conducting a quality open interview:
\begin{itemize}
    \item You should continue with elaboration questions: 
    \begin{itemize}
        \item Restate or rephrase part of what they say as a question to prompt elaboration.
        \item Prompt them with a follow up question such as, “What is/was it like to [...]?"
    \end{itemize}
    \item You are trying to elicit stories that are meaningful in the subject’s life. The depths of these stories are more important, rather than the number of stories you can elicit. 
    \begin{itemize}
        \item Ideally, the interviewer disappears into the story; it may help to avoid phrases like “tell me”, which draw attention to the interviewer.
        \item However, sometimes it can help to put yourself into the conversation to express understanding and empathy.
    \end{itemize} 
    \item Make an effort not to derail or redirect the flow of the conversation: 
    \begin{itemize}
        \item If the subject is bringing a theme to the conversation, it is best to follow it.
        \item Leading questions are best and yes/no questions should be avoided.
        \item If you don't know what to say you can always reiterate what the subject says.
    \end{itemize} 
    \item Try to make the subject know they are being heard. You can do this by acknowledging the emotion behind the story, mirroring what they have said, and expressing gratitude for their openness.
    \item Sometimes subjects may speak in a way that the interviewer finds confusing. If you lose track of their train of thought, you can request help by saying, “I want to make sure I follow you, can you help me understand...”
    \item If someone stops speaking altogether, you can repeat what they last said as a question, or even look at them expectantly, but being comfortable with silence is important too. Often people are formulating their thoughts and will continue after a pause.
\end{itemize}

\noindent The SOP also provides the following suggested ending to the open interview:
\begin{quote}
    Thank you so much for taking the time to talk with me [or thank you for sharing with me]. I’m going to stop the recording now.
\end{quote}

\noindent Remember throughout this process that the main goal is to have subjects speak freely without interruption and to show that you are listening and are interested in what they are saying. This involves not only the prompting used while conducting the interview, but also body language and tone throughout. It is important to let the participant guide the conversation, but at the same time not to let the participant lose interest in the conversation.

By contrast, psychs interviews are much more structured in the prompts required and the responses expected of participants. The goal of the psychs interview is to be providing specific clinical ratings, and these interviews may run for a much longer time but include much briefer (and less rich) individual answers as a result. Many prompts are in fact questions directly about experienced symptom severity and ask for a yes/no style response, which is discouraged in the open context. I will now review the plan for conducting of psychs interviews in the AMPSCZ project. \\

\subsection{Conducting psychs interviews}
\label{subsec:psychs-sup-ra}
The psychs interview is a clinical interview based around a new rating scale for clinical high risk (CHR) of Schizophrenia. Per \cite{Brady2023}, AMPSCZ aims to recruit 1,977 individual between 12 and 30 years old who meet the CHR criteria defined by the new "Positive SYmptoms for CAARMS Harmonized with SIPS" (PSYCHS) psychometric instrument. The enrolled CHR subjects will complete screening and baselines psychs interviews, and then will participate in 7 follow-up psychs interviews over the course of 2 years. In parallel, the project aims to recruit 640 healthy controls to complete screening and baseline psychs interviews. $\sim 200$ of the control subjects will also participate in follow-up psychs interviews at 2 month, 12 month, and 24 month study time points \citep{Brady2023}. Psychs interviews entail both the interview recording protocol detailed, as well as PSYCHS scale ratings based on this interview. All enrolled subjects will also participate in the described open interviews, as well as other project data collection modalities (though subjects may participate in a subset of modalities without consenting to all of them). 

Note the PSYCHS scale is based on two existing scales for rating CHR: the Comprehensive Assessment of At Risk Mental State (CAARMS) and the Structured Interview of Prodromal Symptoms (SIPS). The CAARMS and SIPS are the two primary scales used in current practice for clinical high risk of psychosis. These scales are reliable and well-validated, but have only moderate success in actually predicting future development of psychosis, particularly as a large number of people that meet CHR criteria - even with quite high scores - do not end up developing a psychotic disorder \citep{Malda2019}. Thus one goal of AMPSCZ is to improve on the false positive rate of CHR assessment.

The SIPS identifies three CHR syndromes: Attenuated Psychotic Symptoms Syndrome (APSS), Brief Intermittent Psychosis Syndrome (BIPS), and Genetic Risk and Deterioration (GRD) \citep{Woods2014}. Per the Orygen Youth Health Research Center's manual for conducting the CAARMS, the full scale involves assessing the following domains.
\begin{enumerate}
    \item Positive Symptoms. Requires rating of:
    \begin{itemize}
        \item Unusual Thought Content
        \item Non-bizarre Ideas
        \item Perceptual Abnormalities
        \item Disorganised Speech
    \end{itemize}
	\item Cognitive Change – Attention/Concentration. Requires rating of:
    \begin{itemize}
        \item The subjective experience 
        \item Any observed cognitive change
    \end{itemize}
    \item Emotional Disturbance. Requires rating of:
    \begin{itemize}
        \item Subjective emotional experience
        \item Observed blunted affect
        \item Observed inappropriate affect
    \end{itemize}
    \item Negative Symptoms. Requires rating of:
    \begin{itemize}
        \item Alogia
        \item Avolition/apathy
        \item Anhedonia
    \end{itemize}
    \item Behavioural Change. Requires rating of:
    \begin{itemize}
        \item Social isolation 
        \item Impaired role function
        \item Disorganising/odd/stigmatising behaviours
        \item Aggression/dangerous behaviour
    \end{itemize}
    \item Motor/Physical Changes. Requires rating of:
    \begin{itemize}
        \item Subjective complaints of impaired motor functioning
        \item Informant reported or observed changes in motor functioning
        \item Subjective complaints of impaired bodily sensation
        \item Subjective complaints of impaired autonomic functioning
    \end{itemize}
    \item General Psychopathology. Requires rating of:
    \begin{itemize}
        \item Mania
        \item Depression
        \item Suicidality and self-harm
        \item Mood swings/lability
        \item Anxiety
        \item Obsessive-compulsive disorder symptoms
        \item Dissociative symptoms
        \item Impaired tolerance to normal stress
    \end{itemize}
\end{enumerate}

\noindent The PSYCHS, which is a new scale building upon these two rating systems, is therefore a quite structured interview where the goal is to assess symptom domains and possible syndromes such as those topics covered in the core SIPS and CAARMS scales. This means that many of the interview questions are probing about specific experiences and asking for subjective symptom ratings that do not necessarily elicit more elaborate patient responses like the open interviews do. On the other hand, the content of the recorded interview can be directly mapped to the resulting PSYCHS scale scores that are reported, in that they are rated based on this clinical interview. 

\subsection{Details of TranscribeMe settings used for AMPSCZ}
\label{subsec:u24-tm-ap}
As mentioned, the AMPSCZ project uses similar transcription settings with TranscribeMe as our lab does (described in the methods of chapter \ref{ch:1}). However, there are a few key differences that affect pipeline operations and/or the immediate next steps required for extracting features from the returned transcripts. In this supplemental section, I go through these small but important distinctions. 

The requested transcription quality level and verbatim notation - including the conventions for marking of disfluencies and of inaudible words/other noises - are identical, and speaker identification is still performed, with IDs and timestamps formatted in the same way. A key difference for the AMPSCZ interviews is a reduction in the number of timestamps provided, to cut back on cost. These transcripts have lines split by conversational turn rather than sentence, so that a long interviewee response will be contained all on the same line with a single timestamp provided at the start of the response. Additionally, the timestamps report only second-level resolution, rather than providing millisecond digits with accuracy intended up to 100 ms like for BLS. For most analyses, the changes will not be a problem; however, it does alter how one might interpret some of the secondary QC features in the AMPSCZ interview pipeline, which were ported from the work with BLS reported in chapter \ref{ch:1}.

Another difference between the AMPSCZ transcription protocol and the BLS one is the notation for marking redactions. This has been handled by my interview pipeline and is not something necessary to consider for those monitoring outputs or planning to do downstream analyses on the deidentified transcript data. However, it is worth noting because it was an intentional study design decision made to hopefully generate a good training dataset for future development of automated redaction systems by those with access to the sensitive raw data. In BLS, any redactions of personally identifiable information (PII) are made directly in the returned transcript, so that we only receive a copy with "[redacted]" in place of whatever words were PII. For AMPSCZ, the same guidelines are used by transcribers to identify PII -- but instead of removing it, they are instructed to include the verbatim words in the transcription, simply enclosing it {in curly braces}, which are not otherwise used by TranscribeMe. Thus we can automatically create a version of each transcript that looks just like the BLS redaction style, but we also now have a version that reveals what the PII was for research applications where that is appropriate.  

Finally, there is a change in the AMPSCZ transcription protocol that does not affect the form of the outputs in any way, but can have a major impact on what is available for analysis, particularly if sites do not remain aware of it as they are conducting interviews. To prevent running up a large bill from extremely long interviews (as will be discussed it is not uncommon to see psychs recordings exceeding 90 minutes, despite the recommendation being about half that length), TranscribeMe has been instructed not to exceed 35 minutes of transcribed content for a given interview audio. The transcriber will thus intelligently cut off each transcription of a long interview between $30$ and $\sim 32$ minutes based on the content, to prevent a transcript from ending in the middle of a participant response. This cutoff point will begin at the 30 minute mark, but such transcripts will always be charged as a 32 minute submission, so that the vast majority of psychs transcripts will cost $\$58.56$ in English and Spanish, and over $\$100$ in other languages. The resulting transcript can be analyzed like any other, but it is important to keep in mind when comparing across transcripts that the interviewer content might vary more than expected because of the duration cutoff. Note that this issue is functionally exclusive to the psychs interview type. \\

Once transcripts are returned by TranscribeMe, there are manual review processes in place to ensure PII is correctly redacted. This is a particularly important thing to check, through both human and automated methods, because the redacted transcripts will eventually be shared as part of a broader database that any researchers can apply to access. The pipeline includes some redaction-related sanity checks that will be described; however, the major protocol component of redaction review is an expectation of the sites. 

Specifically, sites are expected to carefully review a subset of the transcripts returned from the interviews they conducted, and mark any PII for redaction that was missed by TranscribeMe. They should use the same curly brace convention as TranscribeMe if a redaction needs to be added, and whenever that does occur they should also log the event so there is a clear awareness of the frequency of missed redaction issues. Sites are only expected to review for redactions and not other potential quality issues, and they in fact should not be making any changes to files outside of the scope of redactions. If an egregious quality issue unrelated to redactions is found by a site during the review process, they should contact their point person at the central server to communicate the issue. Central monitors can handle the situation from there, and will likely request a re-transcription of that audio file by TranscribeMe. 

Each site will perform a manual redaction review on the first 5 interview transcripts that are produced from interviews submitted by that site, and from that point forward they will review a randomly selected $10\%$ of TranscribeMe transcripts from their interviews. Transcripts for review are returned to sites using the same cloud-based file system as they use for uploading interviews to the central server, and once a manually reviewed transcription is finalized, it will be returned to the central server through a similar mechanism, as sites move the file from the "for review" side of the folder structure to the appropriate spot on the "approved" side. Note that the pipeline (section \ref{subsec:interview-code}) will not process a transcript that was sent for review at all until the final version is returned by the site -- though in practice the final version is the same file, as redaction changes are rarely needed. 

\section{AMPSCZ interview monitoring supplement}
\subsection{Early results from accounting of open interview availability by subject ID and timepoint across sites}
\label{sec:open-timepoint-counts}
As described in the main text of chapter \ref{ch:2}, the interview pipeline has enabled detailed accounting of site progress, to identify and address a number of issues -- including discrepancies in open interview availability and expected timepoints, documented in detail here for all 14 active Pronet sites as of 1/31/2023. 

\noindent These stats focus in particular on available open interview recordings and the associated study days tallied for each participant ID with some documented interview involvement (related to the DPChart monitoring view presented in the main text of chapter \ref{ch:2}, in Figure \ref{fig:dpchart}):
\begin{itemize}
    \item There are 102 subjects with at least one recognized interview (i.e. an interview of either type with at least one modality successfully processed by QC).
    \item Of those, 79 subjects have at least one open interview processed (i.e. at least one modality successfully processed by QC). 
    \begin{itemize}
        \item The 23 subject IDs with some interview but no open interview is likely explained in part by site delays in conducting and uploading open interviews versus psychs interviews, though it also may reflect on issues with certain sites incorrectly submitting interviews at a relatively high rate. It is possible as well that a small portion of subjects will be screened out of study participation after a single psychs interview, but the site will still upload the psychs screening, resulting in an ID that will therefore never partake in an open interview.
    \end{itemize}
    \item Of those, 61 subjects have at least one open interview with a redacted transcript available. 
    \begin{itemize}
        \item The 18 subjects with a correctly submitted open interview but no transcription yet are primarily accounted for by the sites that have not been returning the subset of transcripts they were assigned for manual redaction review. There are also a handful of new subjects that are awaiting an output for their first open interview from TranscribeMe, but this process should generally not introduce more than a 1 week delay, especially for English language sites, which represent the vast majority of active Pronet sites to date.
    \end{itemize}
    \item From those 61 subjects, 14 have transcripts at both expected open interview time points (baseline and 2 month follow-up), and another 27 have a single open transcript from the baseline time point. 
    \begin{itemize}
        \item Of course, subjects with only baseline transcripts are a natural occurrence as the study progresses, and we expect to see follow-up transcripts eventually populate for each of these 27.
    \end{itemize}
    \item There are thus 20 subjects with open transcripts available but that are labeled with potentially questionable time points. 
    \begin{itemize}
        \item 4 of these subjects were identified above as clearly problematic submissions from site LA.
        \item 7 of these subjects have two open interview transcripts with adequate separation (at least 48 days apart), so they are likely intended to represent the two open time points, and are largely acceptable transcripts. However it remains to be confirmed how long after consent it is acceptable for the baseline open interview to occur. All 7 of the mentioned subjects had their first interview after day 30 of study enrollment, including as late as day 65. If this is a concern for the project, sites should be warned even though the data from both open sessions will of course be included in downstream analyses.
        \item The final 9 subjects of these 20 have a single open interview transcript from an interview conducted past day 30 of study enrollment. It is unclear how many of these subjects will eventually have another interview conducted to join the aforementioned group of 7 subjects (i.e. the current transcript is intended to be a baseline interview even if late), versus how many represent only a follow-up open transcript with unexpectedly missing baseline transcript. The latter could occur due to various issues during the interview upload process or could simply be explained by a baseline transcript that has been produced by TranscribeMe but happened to be assigned for manual review which was subsequently ignored by the site. 
    \end{itemize}
\end{itemize}
\noindent As such, a number of potential problems for further follow-up have been identified in this way (to join the many other issues we have identified with various monitoring tools, reported on throughout the main chapter \ref{ch:2}).

\subsection{Constructing the weekly server-wide interview quality accounting tables}
\label{sec:dumb-table}
The values reported in the weekly server summary table described in the main text of chapter \ref{ch:2} (with methods details here) are as follows, in order of columns presented (Tables \ref{table:pronet-counts} and \ref{table:prescient-counts}). They are reported separately for each site and interview type: 
\begin{itemize}
    \item The total number of interviews that have QC metrics available for any modality. Note that many of the file management mistakes to be discussed in more depth later (section \ref{subsubsec:u24-issues}) will prevent an interview or a particular interview modality from being processed, so that this number is not necessarily equivalent to the number of interviews recorded.
    \item The number of interviews with video QC metrics available. This might not match the preceding count due to a file upload mistake that impacts video but not audio, though recall that psychs interviews need not include video and will not include video by design if the hardware recorder is used.
    \item The number of processed videos with high quality QC results, as defined by an average of between $1.8$ and $2.3$ faces detected per QC-extracted frame. 
    \item The number of processed videos with acceptable QC results, but outside the high quality range. This includes an average of $1.1-1.8$ faces per frame or an average of $> 2.3$ faces per frame. Videos with $1$ or fewer faces per frame have a high chance of being recorded in speaker view rather than gallery view, and regardless are an SOP violation for the open interview format, as the interviewer should have their camera on in gallery view.
    \item The number of interviews with audio QC metrics available. If there are no errors, this should always match the total number of processed interviews.
    \item The number of interviews with transcript QC metrics available. This may lag the processed audio count due to the time required for TranscribeMe transcription and in some cases manual site redaction review. It is also possible for this number to be less than the audio count if a recording was rejected because of QC (i.e. volume $< 40$ db), however that has not occurred for any files processed in the AMPSCZ project to date. Otherwise, if there are no mistakes in the TranscribeMe and site review transcript upload processes, the numbers should eventually match.
    \item The number of returned transcripts with high quality QC results, as defined by fewer than $0.01$ inaudible occurrences per word (i.e. $< 1$ inaudible label per 100 words).
    \item The number of returned transcripts with acceptable QC results, but outside the high quality range. Here, this is defined as between $0.01$ and $0.05$ inaudible occurrences per word. So far, all finalized transcriptions have had QC of at least acceptable quality.
    \item Finally, the mean duration (in minutes) is reported from across all of the processed audio files. Recall that open interviews should be \emph{at least} 10 minutes, and ideally closer to 20. Psychs interviews on the other hand are expected to be $\sim 40$ minutes, and this is also the approximate limit of what will be manually transcribed per the study budget. However, sites are free to upload much longer psychs recordings, with the understanding only the first $\sim 30$ minutes (cropped intelligently by TranscribeMe to avoid cutting off a turn) will be transcribed.
\end{itemize}
\noindent This is of course not an exhaustive consideration of the quality issues that could occur, nor is it even an exhaustive consideration of the quality categories we could define using the pipeline's QC features. The table is meant to be a bird's eye account of dataset collection status, and other tools like the site-specific emails or the DPDash heatmaps described above are necessary for a more nuanced understanding of quality control outputs. 

\section{Code architecture (AMPSCZ interviews dataflow/QC pipeline documentation)}
\label{subsec:interview-code}
In this section, I provide implementation details and other technical documentation, targeted at both those in the AMPSCZ project that may need to troubleshoot the code in the future and those who may be considering using a version of this code to manage data collection for a different interview recording study. As such, references to inputs and outputs herein will focus on the inputs and outputs of this specific pipeline on the data aggregation servers, with discussion of the larger AMPSCZ infrastructure setup kept as minimal as possible. 

 The documentation will begin with information on installing (subsection \ref{subsubsec:interview-code-install}) and running (\ref{subsubsec:interview-code-run}) the code as is. It will then dive into implementation details for each module within the audio (\ref{subsubsec:interview-code-aud}), transcript (\ref{subsubsec:interview-code-trans}), and video (\ref{subsubsec:interview-code-vid}) paths of the pipeline. Unique to this section, I will report the security information presented for the necessary security reviews that were done before installation on the Pronet and Prescient central servers (\ref{subsubsec:interview-code-sec}). Finally, the section will close with an account of the most likely updates that might be needed to adapt the code to run in somewhat different study contexts, as well as quality of life improvements suggested for the future of AMPSCZ (\ref{subsubsec:interview-adapt}). 

 \subsection{Installation}
\label{subsubsec:interview-code-install}
It is highly recommended that this code is run on a Linux machine, but there are minimal hardware requirements as the pipeline was designed with low compute resources in mind. Given the naturally large file sizes of recorded video interviews however, a decent amount of system storage space is expected (more detail on required space can be found in the main chapter section \ref{subsubsec:u24-storage}). 

The code requires FFmpeg (tested version 4.0.2) and Python3 (tested version 3.9.6) to be installed, as well as use of standard bash commands, including realpath and dirname. For the email alerting to work, the mailx command must be configured. The python package dependencies can be found in the \hl{setup/audio\_process.yml} file, with the exception of SoundFile and librosa on the audio side and PyFeat on the video side. If Anaconda3 is installed this YML file can be used directly to generate a usable python environment. \\

\noindent For the AMPSCZ project, the above dependencies were already installed and the sendmail command configured on the relevant servers by IT. Thus I installed my code with the following steps (once in the directory location intended for software installs):
\begin{itemize}
    \item git clone https://github.com/dptools/process\_offsite\_audio.git
    \item cd process\_offsite\_audio/setup
    \item conda env create -f audio\_process.yml
    \item conda activate audio\_process
    \item pip install soundfile
    \item pip install librosa
    \item pip install opencv-python
    \item pip install torch {-}{-}no-cache-dir
    \item pip install py-feat==0.3.7
    \item sudo yum -y install libsndfile
    \begin{itemize}
        \item This last step was done on Prescient data aggregation server only, to get SoundFile and librosa to work there. It may or may not be necessary to do this then, depending on what libraries are already installed by IT on your machine. 
        \item If it is missing, it might require contacting IT to do the libsndfile installation, if you do not have sudo access (which ideally the account officially running daily AV processing should not, to protect raw file integrity).
    \end{itemize}
\end{itemize}
\noindent Note I installed a few PyFeat dependencies (OpenCV and PyTorch) separately before the PyFeat installation, because we are using low resource servers for data aggregation and they were unable to install PyTorch without the {-}{-}no-cache-dir flag. On a different machine those two steps before the PyFeat install may not be necessary. 

Similarly, version 0.3.7 of PyFeat is the one used across AMPSCZ, and it is known that the latest version is not compatible with the pipeline, because it is no longer possible to specify None for the pose, FAU, and emotion models when initializing a PyFeat detector. On the Pronet data aggregation server there is not enough RAM however to even load those models, and so it is important for the pipeline to be able to load \emph{only} the lightest weight face detection model when running video QC. For other projects this may not be a concern, in which case it would be fine to install the latest version of PyFeat (as of spring 2023), but would require changing the line in the video QC python function that specifies the models to be loaded. However, for an isolated conda environment to be used for the sole purpose of running this pipeline, there would be minimal benefit to upgrading the PyFeat version at present.

As time goes on it will of course be important to be aware of updates to any of the pipeline dependencies, and maintain the repository accordingly. In addition to the direct dependencies, there are also indirect dependencies on assumptions about Zoom recording settings, and for AMPSCZ there are interactions with other tools written for the purposes of digital psychiatry study management as well -- mainly the Lochness \citep{lochness} and DPDash \citep{dpdash} codebases described in section \ref{subsec:diary-code}, which were further adapted specifically for the use of the AMPSCZ project. Lochness is used for pulling the data submitted by sites and enforcing the expected folder structure, but it would be possible to recreate the folder structure manually or with different software if desired. DPDash is used for visually monitoring updates to the dataset via a regularly refreshing web portal, but is not strictly necessary. These are all helpful tools for managing AMPSCZ and would likely be helpful for other future projects too; for the most part, I will assume throughout this section that this pipeline will be integrating with Lochness and DPDash equivalents wherever it is used, and will thus make clear what design decisions are influenced by those tools throughout.  

Finally, note that for the transcript portion of the pipeline it will be necessary to have a TranscribeMe SFTP account setup with expected transcription settings. This is largely similar to what was already described for the phone diary code in section \ref{subsec:diary-code}, but a few small (yet important) distinctions were made for AMPSCZ - these are detailed in section \ref{subsec:u24-tm-ap} above. Relatedly, the folder structure enforced by the AMPSCZ implementation of Lochness and therefore expected by this pipeline is somewhat different than that expected for the audio journal code. The input expectations for the interview code are provided below (\ref{subsubsec:u24-folder-paths}). 

\subsection{Running the code}
\label{subsubsec:interview-code-run}
The pipeline will generally be run using the four wrapping bash scripts provided in the top level of the repository \citep{interviewgit}, with the path to a settings file as the single argument (shared across branches of the pipeline). The four scripts cover audio preprocessing, transcript preprocessing, video preprocessing, and finally subject-level summarization for monitoring. Each such overarching pipeline branch script will save detailed logging information from each run to a logs subfolder, which is made inside the installation folder of the repository. The \hl{audio\_process} conda environment should be activated prior to launching these scripts. In AMPSCZ, each site is treated by the code as an individual study, and so has its own settings file and will have each part of the pipeline run independently on it. For a template that can be edited into a settings file for a particular site in a project like AMPSCZ (or a particular study on a lab server), see \hl{setup/example\_config.sh}. The available settings in this file will be outlined shortly. 

Each major step in a given branch of the pipeline is initiated using a script in the individual\_modules folder, see this folder if specific steps need to be run independently. More details on the separate steps will be provided in subsequent sections. First, basic instructions on running the full pipeline to completion will be provided though, for each branch separately (after more details on the config to be used across branches). Note that in addition to the main pipeline, a few utility scripts are provided for combining info across studies. To see how these fit together, refer to the wrapping scripts found within the repo, currently running on a daily cronjob for each of the two AMPSCZ central servers: \newline  \hl{amp\_scz\_launch/production\_pronet\_all\_sites\_cron\_script.sh} and \newline \hl{amp\_scz\_launch/production\_prescient\_all\_sites\_cron\_script.sh} respectively. \\

\noindent Specifically for AMPSCZ, the repository folder itself can be found at the path: \newline \hl{/opt/software/process\_offsite\_audio} on the Pronet data aggregation server, and at the path: \newline \hl{/home/cho/soft/process\_offsite\_audio} on the Prescient data aggregation server. 

\subsubsection{Available settings}
The settings that can be specified for the pipeline (shown in the example\_config.sh file) currently are as follows:
\begin{itemize}
    \item \hl{data\_root}, which is the root path to the main data structure folder. For a given AMPSCZ server this is the same across sites, and within the Baker Lab this is the same across most studies (but not all, as data from some different collaborations are kept on different briefcases). In our terminology, this would be the path on the server to the \textbf{PHOENIX} folder.
    \item \hl{study}, which is the name of the study (e.g. BLS) or site (e.g. PronetLA) on the server. This must match the folder name corresponding to the study/site.
    \item \hl{auto\_send\_on}, which for AMPSCZ is expected to always be "Y". If "N", then QC will be computed for the audio without sending anything to TranscribeMe.
    \item \hl{transcribeme\_username}, which should be the username used to login to the TranscribeMe SFTP server where data transfer for the project is intended to take place. For AMPSCZ this is one email for Pronet and one email for Prescient. For a lab group this may always be the same username across studies, or it may vary depending on payment allocation details.
    \item \hl{length\_cutoff}, which is the minimum duration for an individual interview recording to be sent to TranscribeMe, in minutes. For example, one might consider an interview of less than 10 minutes to likely be a fragment or incorrectly conducted and not worth sending for transcription. However for AMPSCZ this is always expected to be 0. 
    \begin{itemize}
        \item Note that a maximum length per transcript of $\sim 30$ minutes is enforced on TranscribeMe's end for the project, as the science committee wants these audios to be transcribed but does not want to spend too much on them. For a similar setup please discuss with TranscribeMe (though I do not necessarily endorse this approach personally, see Appendix \ref{cha:append-ampscz-rant}). 
    \end{itemize}
    \item \hl{db\_cutoff}, which is the minimum volume for a mono interview audio recording to be considered acceptable for transcription. In our hands on prior projects we decided on a 40 db cutoff, so this is also what is used for AMPSCZ. To date it has only been triggered once while $\sim 500$ interviews have now been successfully transcribed, and inaudible rates in the transcribed interviews have largely been very good. Future projects might consider further tweaking this number, though it may be largely a non-issue for both Zoom audios and handheld EVISTR recordings in the context of transcribability. For lab onsite interview recordings using other audio collection methodologies, we have seen more variance in quality and more frequently unusable audio though.
    \item \hl{auto\_send\_limit\_bool}, and if this is "Y", \hl{auto\_send\_limit}. For AMPSCZ and for internal projects we have kept this setting to be "N", but it is an option provided for safety with research expenditures, especially if the code is run less frequently or left to run only at the end of a study. When the setting is turned on, auto\_send\_limit should be an integer specifying the maximum number of total minutes that it is acceptable to upload to TranscribeMe in a given run of the code with these settings. If the sum of audio lengths selected for upload exceeds this value, the files will be left set aside in the respective to\_send folders (described more in implementation details below), but will not yet be uploaded to TranscribeMe - so the decision can be sorted out manually instead.
    \item \hl{lab\_email\_list}, a comma-separated string of email addresses for the detailed monitoring updates to go to. This was originally intended to contain site-specific addresses for AMPSCZ, but to date monitoring has been largely centralized, and so the addresses here are similar across configs. All of the daily (as needed) site-specific emails will go to the addresses listed under the current site config. For central monitoring across many sites/studies, I would highly recommend setting up targeted email filters, but if monitoring is well-distributed than the rate of messages should be extremely manageable.
    \item \hl{site\_email\_list}, another comma-separated string of email addresses specifying those that should receive emails when there are outstanding transcripts awaiting manual redaction review for that site (first 5 transcripts from each site and a random 10\% thereafter). These are the only emails that actually go to site contacts for AMPSCZ at this time. Obviously this is only relevant if manual redaction review will be a part of your project.
    \item \hl{server\_version}, which is a label that will be applied to email headers to help email filters differentiate between different projects. For AMPSCZ, we primarily used this to easily distinguish between development testing alerts and alerts from real production data when those servers were running simultaneously for Pronet (and later for Prescient) near the start of data collection ramp up, as central server name (e.g. Pronet) is implicit in site name (e.g. PronetLA). In other large scale projects, this string could be used to make a variety of other designations as needed.
    \item \hl{transcription\_language}, which is a string (currently in all caps) that specifies the language the site will be using to conduct most of their interviews. The pipeline uses this string in the filename it uploads to TranscribeMe, to assist their process in assigning the correct transcription team to a given upload. When transcripts are pulled back this part of the name is removed, so it is used only in the TranscribeMe upload process and not on actual pipeline outputs (or intermediates). This could theoretically be used to convey other messages to TranscribeMe such as transcription settings, in future adaptation of the code. The information here will also be compiled as part of metadata that goes to the NIH data repository for AMPSCZ.
\end{itemize}
\noindent The configs being actively used (or for some sites prepared to be used) for AMPSCZ can be found in the \hl{amp\_scz\_launch/production\_pronet\_site\_configs} and \hl{amp\_scz\_launch/production\_prescient\_site\_configs} folders for Pronet and Prescient respectively. 

To add new sites to the existing processing for this project, one can simply add another config in the appropriate folder for that central server, working from the existing configs for that server as a template (and referring to the above where additional reference is needed). One would then just need to make sure that there are not issues with folder permissions when Lochness initializes the structure for that site. Obviously for setting up a brand new process, one would aim to replicate the current setup for Pronet and Prescient as far as settings files and cron job go, after following the installation instructions in the preceding section to get the code and its dependencies onto the new central server. In the case of a new central server, it would obviously be necessary to change some further settings within each new config -- such as the TranscribeMe username, which would require creating a separate SFTP account with TranscribeMe reps first. \\

\noindent On that note, an additional component built into the config .sh file is the management of the password for the specified TranscribeMe SFTP account, which I will next describe how to set up for new servers/projects.

\subsubsection{Providing TranscribeMe password to the pipeline}
In order for the code to run automatically on a cronjob, the password for the TranscribeMe SFTP server needs to be stored somewhere on the source server. By default, this is assumed to be in a file named \hl{.passwords.sh} directly under the repo folder where the code is being run. The .gitignore included with the code will ensure that the .passwords.sh file is never pushed back to the central repository (as well as any logs generated by the code more broadly). On both Pronet and Prescient data aggregation servers, this passwords file has 700 permissions and is owned by the account that is set to run the pipeline code. For more on privacy/security, see section \ref{subsubsec:interview-code-sec} below. \\

\noindent If one will use an identical setup to the AMPSCZ central Lochness servers, then it is straightforward to just create the \hl{process\_offsite\_audio/.passwords.sh} file directly and then paste in the following two lines:

\begin{center}\begin{tabular}{c}\begin{lstlisting}
transcribeme_password="[password]"
export transcribeme_password
\end{lstlisting}\end{tabular}\end{center}

\noindent Where of course [password] is replaced with the TranscribeMe SFTP account password corresponding to that server's account. Note that TranscribeMe will also periodically require this password to be updated, so for all projects (including both accounts for AMPSCZ) it will be necessary to occasionally update the password contained in the .passwords.sh file directly on the applicable server. 

If one wants to store the password file elsewhere or needs to use different password files for different studies/sites operating out of the same server (because different TranscribeMe SFTP accounts are being used), the path to the passwords file can be specified using the \hl{passwords\_path} variable found at the bottom of each config file (like example\_config.sh). The passwords file itself needs to be an \emph{sh} file type with content matching the above to work out of the box, but there is no other restriction on the naming/folder placement (as long as the running account can access) when its path is appropriately specified within the corresponding config.sh settings file.

\subsubsection{Using the audio portion of the pipeline}
As the particular study (or in the case of the AMPSCZ project, site) of interest is specified as part of the settings files along with all other relevant parameters, the audio branch of the pipeline in Figure \ref{fig:zoom-arch} can be easily run with the following command:

\begin{center}\begin{tabular}{c}\begin{lstlisting}
bash interview_audio_process.sh example_config.sh
\end{lstlisting}\end{tabular}\end{center}

\noindent Of course the user should substitute the path to the config file with the settings they actually want to use in the place of "example\_config.sh" here. 

\subsubsection{Using the transcript portion of the pipeline}
Similarly, to run the transcript branch of the pipeline in Figure \ref{fig:zoom-arch}, the following analogous command can be used:

\begin{center}\begin{tabular}{c}\begin{lstlisting}
bash interview_transcript_process.sh example_config.sh
\end{lstlisting}\end{tabular}\end{center}

\noindent Note that the config file here is identical to the config file used for the audio portion of the pipeline, as the settings were designed to work across all phases of the code. 

\subsubsection{Using the video portion of the pipeline}
Finally, to run the video branch of the pipeline in Figure \ref{fig:zoom-arch}, the following analogous command can be used:

\begin{center}\begin{tabular}{c}\begin{lstlisting}
bash interview_video_process.sh example_config.sh
\end{lstlisting}\end{tabular}\end{center}

\noindent Note that the video processing can be run in parallel with the audio in theory, as they do not rely on each other - but the transcript code of course will only work after the audio part of the pipeline has been run and TranscribeMe has returned some transcripts. 

\subsubsection{Generating additional summary stats for a single site}
Recently, a smaller fourth branch was added to the pipeline to facilitate generation of the site-level summary view in DPDash (Figure \ref{fig:dpdash}) and ultimately to facilitate the site-wide summary supplemental utility scripts mentioned in the main chapter \ref{ch:2}. However, even when applying this code to a single study scenario, the summary stat portion of the pipeline provides information helpful for monitoring data collection more regularly via quick scans of the summary outputs, so this branch of the code is likely worth running regardless.

\noindent To run these summaries on a study of interest, the following command analogous to the above can be used:

\begin{center}\begin{tabular}{c}\begin{lstlisting}
bash interview_summary_checks.sh example_config.sh
\end{lstlisting}\end{tabular}\end{center}

\noindent This script enables the monitoring that is done for each individual site creating subject-wide summaries. The other pipeline components focus on participant-level processing, looping across all participants for a given study/site. 

\subsubsection{Input expectations/folder structure}
\label{subsubsec:u24-folder-paths}
For the AMPSCZ project, we expect two different types of interviews (open and psychs), which are both handled by the full pipeline and will both be included in the same email alerts, but are processed independently by each module with outputs saved separately. These expectations are built into the wrapping pipeline scripts, and so at present the names of these two interview types are somewhat hard-coded. 

The code expects to work within a PHOENIX data structure matching the conventions described in the AMPSCZ standard operating procedure (SOP). The pulling of data submitted by sites into this folder structure is handled by Lochness in our case, though it is still possible for sites to violate some interview naming conventions within this framework, as Lochness does not enforce e.g. the presence of a date in the folder name as a requirement for upload to the central server. Regardless, I will detail exactly what inputs are expected by the code from the perspective of already being on the central server in this section. 

Note in addition to the upcoming expectations for interview naming conventions, the code also requires a metadata file directly within the site/study level folder contained on the GENERAL side, and will only process a particular participant ID if that ID appears as a row in said metadata file with a corresponding valid consent date (per Lochness guidelines, though again this could be replicated independently). If there are issues encountered with the basic study setup appropriate error messages will be logged.

Note the pipeline only ever reads from raw datatypes. All pipeline outputs are saved/modified as processed datatypes. The deidentified data that make it under the GENERAL side of processed will subsequently be pushed by Lochness to the \emph{predict} server for downstream analyses and eventual upload to the NIH data repository. 

More specifically, the raw audio and video data available for an interview are expected to meet the following criteria on the central server, first described for Zoom interviews and then for EVISTR (psychs-only) recordings. For more on the actual conduct of these interviews, see section \ref{subsec:interview-methods}. \\

\noindent For Zoom interviews, each session should have all files under a single folder, [Zoom\_folder\_name]. It should contain the interview components like so, matching automatic naming conventions currently used by Zoom:
\begin{itemize}
    \item Single video should then be found directly under [Zoom\_folder\_name]/video*.mp4
    \item Single collapsed all-speakers (mono) audio file should be directly under [Zoom\_folder\_name]/audio*.m4a
    \item Diarized speaker-specific audios should be under a subfolder, [Zoom\_folder\_name]/Audio Record/*.m4a
    \begin{itemize}
        \item Be aware there is a chance these filenames contain participant names
    \end{itemize}
\end{itemize}
\noindent * represents a wildcard that could match any additional text in the filename string.

We require [Zoom\_folder\_name] to begin with \hl{YYYY-MM-DD hh.mm.ss }. Again, this is default Zoom formatting, and the code utilizes the date and time info provided, so it is important to not change it. Anything can come after the date/time metadata (but note the second space at the end to split out the time). 

For psychs interviews only, the SOP also allows for interview sessions to be recorded by a single device, so single audio files (each representing one session) of the following form are also allowed directly as a raw interview submission. Like the Zoom interviews, we also require timestamp metadata in the top level name. The SOP requirement here is \hl{YYYYMMDDhhmmss.WAV}.

In this case, the code is currently strict in requiring exactly this naming format with nothing appended. Note the capitalization of WAV, which is also required at this time (per the default of the approved recording device). \\

\noindent The transcript side of the pipeline primarily relies on outputs from the audio side. It also expects a box\_transfer/transcripts folder on the top level of the PHOENIX structure's PROTECTED side, to facilitate transferring completed transcripts to the corresponding sites for correctness review. In finalizing the transcripts that are subsequently returned by sites, it looks for the interviews/transcripts datatype on the PROTECTED side of raw, and it expects these transcripts to have the same name as those originally returned by the pipeline via TranscribeMe. 

On that note, the code also expects TranscribeMe to return transcripts with a name identical to the uploaded audio filename, just with the extension switched to .txt. So it is possible, though not an issue we've encountered much, that TranscribeMe could cause a stall in processing by violating naming conventions.

\subsubsection{Using cross-site utility functions}
\label{subsubsec:cross-site-utilities}
In cases where multiple different studies - or for projects like AMPSCZ, sites - are being tracked simultaneously, it will often be desirable to have some monitoring at the server-wide level, summarizing progress per study/site in one place. For the present codebase, I have included supplemental utility scripts for AMPSCZ (but easily adaptable to other contexts) to generate such Pronet- and Prescient- wide summaries. They continue to run automatically for this project and have had minimal issues since implementation, in addition to helping us catch a number of quality issues as a complement to the views in DPDash.

These utility scripts create the weekly histograms, tables, and combined QC + accounting CSVs that are sent out by the code, as described in the primary monitoring report of the main chapter \ref{ch:2}. They also integrate with DPDash tools to facilitate central monitoring. Note while these functionalities are particularly helpful for monitoring quality of successfully processed interviews (and noticing site-specific trends), data upload issues and other more fundamental SOP violations may not be reflected in these summaries, which focus on those interviews with QC information available. Please continue to refer to core pipeline features on the site-specific level for tracking of logistical problems with interviews.

\noindent Recall that the primary resources attached to the weekly monitoring emails are:
\begin{itemize}
    \item A PDF containing histograms of select QC feature distributions across the server, with open and psychs interview types represented separately via a stacked histogram approach.
    \item A PDF containing histograms of select QC features distributions across the server for open interview only, but with different sites now represented separately.
    \item An analogous PDF for psychs interview distribution visualization across sites.
    \item An HTML table containing counts of good, acceptable, and bad quality interviews by site and type. 
    \item The abridged version of that table focuses only on transcript quality, but there is another table attached with similar counts giving more detailed interview counts across different modalities. 
    \item A CSV containing information on how many subject IDs for each site have transcripts available for both open interview timepoints (baseline and 2 month follow-up) versus just one or the other (or neither, for any IDs that are known to have some interview upload processed across types).
    \item The combined QC CSV with key QC metrics from across modalities provided for all interviews processed on the server to date -- this is similar to what is imported in the combined DPDash view.
\end{itemize}
\noindent A concatenated SOP violations list CSV and overall processed accounting CSV (more detailed contents of which will be described subsequently) are then attached to a second weekly monitoring email for those looking into possible upload/file organization mistakes or who are otherwise doing more detailed data evaluation.

The script coordinating the cross-site monitoring utilities for AMPSCZ can be found at \hl{amp\_scz\_launch/final\_all\_sites\_utility.sh}. The arguments it takes are specified within the \hl{amp\_scz\_launch/production\_pronet\_all\_sites\_cron\_script.sh} and \hl{amp\_scz\_launch/production\_prescient\_all\_sites\_cron\_script.sh} scripts that control the running of the entire pipeline for each respective AMPSCZ central server. The final\_all\_site\_utility.sh script organizes the summary CSVs on a daily basis - optionally emailed to a specified list, which is currently just "mennis2@partners.org,philip.wolff@yale.edu". Those CSVs are used on a weekly basis by another script that is called by the top level cron scripts, again with settings specified therein. Histograms and formatted tables are restricted to this weekly update. That script is \hl{amp\_scz\_launch/weekly\_logging\_utility.sh}, which manages the generation and emailing to a broader list of project staff of the weekly figures and tables every Monday. Both the described utility scripts draw on python functions found under the \hl{amp\_scz\_launch/helper\_functions} repo folder. 

In the main chapter \ref{ch:2}, I also investigated relationships between some of the QC features in the early AMPSCZ interview dataset. Although scatter plots are not part of automated monitoring updates, the notebooks used to generate them can be found in the \hl{amp\_scz\_launch/preliminary\_stats} repo folder. By downloading the \hl{combined-QC.csv} sent by the weekly logging utility, it should be possible to rerun the notebook at any time to get the latest scatter plots, though one may want to tweak some of the visual settings as the number of sites and the number of data points increase over time.

As far as the automatic daily and weekly summarization scripts though, little should need to be updated by AMPSCZ, and the main update for future projects looking to use the code would be the broader issue of handling cases where the interview types are not specifically "open" and "psychs". The primary change that may need to be made periodically for AMPSCZ will simply be to change the list of addresses receiving emails for each central server. The settings for that, particularly relevant for the more widely dispersed weekly figures, can be found towards the bottom of each of production\_pronet\_all\_sites\_cron\_script.sh and production\_prescient\_all\_sites\_cron\_script.sh central bash scripts. There the pii\_email\_list specifies the addresses to receive combined CSVs with SOP violation and processed accounting logs to date -- CSVs that contain dates and could potentially contain real names in certain violation scenarios, but are important for logistics tracking. Then the deid\_email\_list specifies the much longer list of addresses to receive the QC figures primarily described here.

For AMPSCZ members making changes to these sorts of settings, please be sure to update the code on the repository for any longer term updates and also ensure that the new code is correctly pulled to the relevant data aggregation servers. A small additional note to be aware of is that the Pronet data aggregation server uses -A flag for mail command attachments currently whereas Prescient uses the more standard -a flag. 

For changes that might need to be made on the DPDash displays of combined QC, note that the utility scripts directly in this repo run daily on the data aggregation servers and only function to inform the email alerting system that is the major component of monitoring implemented directly by this pipeline. Saved combined CSVs and records of sent emails can be found within the pipeline's logging outputs. But because Lochness pushes only outputs stored directly within the GENERAL side of the PHOENIX data aggregation folder structure to \emph{predict}, which is defined very clearly on the level of the individual subject ID (and interview type), combined CSVs are not pushed to \emph{predict}. The combined CSVs generated for DPDash then are actually created directly on \emph{predict}. This is done using very similar code to portions of the cross-site utility scripts that were described here, adapted directly from those scripts. However they are run independently by DPACC, and stored in a different repository related to AMPSCZ's DPDash \citep{dpdash-amp} utilities. That repository additionally includes some other utility functions originally written by me to format other summary information for display on DPDash, in particular the code defining CSVs in the needed form for our DPChart views. In sum, for changes that may need to be made to DPDash steps of the interview monitoring workflow, please contact DPACC instead of working directly in my repository. \\

\noindent Now that the pieces of the pipeline have been overviewed with the most key instructions provided, I will provide more detailed information on the steps contained within each branch and their respective implementations. After that I will discuss security considerations, and finally wrap with a detailed account of various future updates suggested for this codebase.

\subsection{Audio processing components}
\label{subsubsec:interview-code-aud}
The major steps of the audio side of the pipeline are as follows, presented alongside their corresponding wrapping bash modules (as subbullets), modules which should be reviewed in greater detail if one plans to run only pieces of the pipeline (Figure \ref{fig:zoom-arch}):
\begin{enumerate}
    \item Identify new interview audio files in raw by checking against existing QC outputs and interview protocol upload expectations, first for open interviews and then for psychs interviews. This involves converting any newly found top level interview audio files to WAV format (if not already), thus saving WAV versions (a copy if EVISTR WAV format on raw) of all new mono audio files within the corresponding folder on the processed side of PROTECTED.
    \begin{itemize}
        \item These tasks are handled by the run\_new\_audio\_conversion.sh module, which relies on ffmpeg but does not utilize python.
    \end{itemize}
    \item Run quality control functions on the new audio files, which at the start of this step are found with default date/time naming under a temporary folder on processed. A sliding window audio QC is run to create a single output for each file that will also allow for easy checking of whether a raw interview has already been processed. Then the new audio files are renamed to match SOP convention for processed datatypes, including removing date/time info and instead naming in terms of the study day that the interview occurred (where the corresponding participant's consent date is defined as day 1). Finally, interview-level QC is calculated on the renamed files, to produce final shareable QC outputs that can assist with downstream steps.
    \begin{itemize}
        \item These tasks are handled by the run\_audio\_qc.sh module, which calls the following python functions to initiate each of the 3 described steps for a given participant ID and interview type:
        \begin{enumerate}
            \item sliding\_audio\_qc\_func.py (run per file by the module instead of per participant/type)
            \item interview\_audio\_rename.py
            \item interview\_audio\_qc.py
        \end{enumerate}
        \item Details on the exact outputs produced by these steps will be provided subsequently.
    \end{itemize}
    \item Check for sufficient volume levels to determine which mono interview audio files should be uploaded to TranscribeMe (40 db for AMPSCZ). This will also enforce other per interview upload requirements if they are specified in the current settings file (e.g. minimum individual interview duration), and it will ensure that uploaded files indeed met the basic SOP metadata requirements as well, denoting reason for rejection for any file that is not approved to be sent. For psychs interview type specifically, it will prevent the sending of any interview that is not the first from its day number for that subject ID (so that sites cannot get around the hard cap on transcription duration for individual interview sessions that TranscribeMe implements for this project).
    \begin{itemize}
        \item These tasks are handled by the run\_audio\_selection.sh module, which calls the interview\_audio\_send\_prep.py python script for each subject ID/interview type to actually identify and set aside approved files for the next part of the audio pipeline branch.
    \end{itemize}
    \item Upload all approved new audio files to TranscribeMe SFTP server (with temporarily appended site-based language marker to the end of the processed file name convention, to assist TranscribeMe in assigning transcribers). Any audio that fails to upload will be kept in the folder with approved audios so that a future run of the code can hopefully successfully upload it. All properly uploaded audio (which in practice is basically all selected audio) are moved to a different subfolder to be used by other parts of the pipeline in tracking pending transcriptions.
    \begin{itemize}
        \item These tasks are handled by the run\_transcription\_push.sh module, which primarily uses the interview\_transcribeme\_sftp\_push.py script (called for each subject ID and interview type with files awaiting upload) to handle the SFTP push. More details on the upload implementation can be found in security/privacy section \ref{subsubsec:interview-code-sec} below.
        \item This step can be turned off via settings file, where it is also possible to add a "stop loss" of a certain max duration sum for uploads from the current run (the latter done where applicable by the same module calling the overall interview\_audio\_length\_check.py python script). However AMPSCZ does not restrict interview uploads to TranscribeMe in any such way.
    \end{itemize}
    \item Compile and then send an email listing all files that were newly processed, indicating which were successfully uploaded to TranscribeMe and otherwise documenting possible quality issues. Note that this step requires "build up" of content from all previous steps, and should really only be run in the context of full usage of the audio branch of the pipeline at present. Additionally, it only runs in cases where upload to TranscribeMe is turned on (though one could hack around this by setting the mentioned "stop loss" maximum total sending limit to 0 sum minutes).
    \begin{itemize}
        \item These tasks are completed by the run\_email\_writer.sh module, which calls the study-level interview\_audio\_email\_write.py python helper to add further information to the drafted email body.
    \end{itemize}
    \item Update the file accounting log to specify all of the raw audio files that were newly recognized and (hopefully) processed on the current run, taking note not only of the present date but also metadata that could be (incorrectly) changed in the future, such as the consent date that is listed for each relevant subject ID at the time of file processing. Further, this step maps the identified day and session numbers to the raw real date/time, it denotes whether a file was rejected for upload to TranscribeMe or not, and it collects extended information on the contents of the corresponding raw folder -- such as the number of speaker-specific audio files present in any Zoom interview folder.
    \begin{itemize}
        \item These tasks are completed by the run\_final\_audio\_accounting.sh module, which calls the python script interview\_audio\_process\_account.py on each subject ID and interview type. Note this step, unlike the daily email alerting, occurs even if automatic TranscribeMe upload is turned off. 
        \item The primary use case for the outputs of this step within the broader pipeline will be downstream site-wide monitoring/alerting on possible QC and SOP concerns, processes to be described soon as part of the details on the final branch of the pipeline. 
    \end{itemize}
\end{enumerate}
\noindent For implementation details, one should refer to the python functions (found under \hl{individual\_modules/functions\_called} within the repo folder) that are used by the wrapping bash modules (found directly under individual\_modules), as described in this list. 

Additionally, there are a number of similarities between the present pipeline and the preprocess portions of the audio diary pipeline detailed in section \ref{subsec:diary-code}, at least with regards to the main audio QC functionality and the interfacing with TranscribeMe. One might refer to that section (as well as other components of chapter \ref{ch:1}) to complement the knowledge found here and in the main text of chapter \ref{ch:2}. However for now, I will focus on describing the important outputs produced by the audio branch of the AMPSCZ interview pipeline. 

\subsubsection{Key direct outputs of audio branch of pipeline}
Aside from the renamed and converted audio files that can be used directly for future processing by AMPSCZ code, the primary immediate outputs of this part of the pipeline are the audio QC features. Of course the audio branch also uploads audio to TranscribeMe and does other file accounting to facilitate downstream processes, but those steps (and their subsequent outputs) will be focused on in the upcoming sections on other branches of the pipeline. As such, I will close this section with more detailed information on the audio QC outputs produced by my code.

The primary audio QC output is the interview-level quality summary stats for each mono interview audio, mapped to the final date-removed filenames produced by the renaming step, with one row per interview session. The interview\_audio\_qc function creates (and subsequently updates) one such QC CSV per subject ID and interview type, saved on the \textbf{GENERAL} side of processed. I will first explain the filenaming conventions and use cases for these CSVs, and then will go into the specific QC metrics contained within.

Under GENERAL, the main interview-level CSV for a given participant and interview type can be found at: \newline \hl{[siteFull]/processed/[subject]/interviews/[type]/[site2Digit]-[subject]-interviewMonoAudioQC\_[type]-day[start]to[end].csv}. To illustrate, an example such path would be: \newline \hl{PronetOR/processed/OR03988/interviews/open/OR-OR03988-interviewMonoAudioQC\_open-day78to78.csv}. Of course as new interviews are processed, the day numbers will expand to reflect the latest range: so the example file mentioned might one day have suffix "day78to138.csv" for instance, if OR03988 has a second open interview conducted 60 days after their initial open interview. 

The pipeline will automatically only retain the latest audio QC summary CSV for a given ID and interview type, ensuring that prior records are properly concatenated to the current CSV before deleting CSVs with outdated day numbers in the filename. It is important that outside processes never delete these QC CSVs however, as the persistence of prior QC records is vital to the operation of the pipeline over the course of a project. For example, these CSVs are loaded by the audio selection step of the code, to determine which audios successfully passed the metadata needed for renaming as well as the simple quality control threshold(s).

Because the mentioned CSVs are stored on the GENERAL side of processed, they will additionally be pulled by Lochness to the fully centralized \emph{predict} server, where DPDash can be used to visually monitor progress updates and where data will get staged for later sharing with the NIH data repository. These QC results that are shared there are indeed used as part of the high level interview metadata mapped to redacted interview transcripts that are planned to be released (prepped by the transcript branch of the pipeline to be described below). 

It is also on \emph{predict} where further merge operations and QC summary stat filtering/binning occur, in order to construct the CSVs that are imported into DPDash to create the central server-wide views shown in the main text of chapter \ref{ch:2} -- including both the combined longitudinal heatmap of all sessions and the per subject counting stats used in the open interview bar charts. Because Lochness does not handle renaming of files (for safety against data deletion), \emph{predict} will by default store all of the outdated DPDash CSVs with different day numbers in filenames. Future projects using DPDash might consider reevaluating the CSV naming convention that it requires, to avoid this problem with constantly changing filenames; first and last available day numbers are directly contained in the CSV anyway. 

Recall that my code does not handle any processing or maintenance on \emph{predict}, and is not influenced by any operations on \emph{predict} (though these could hinder the usage of integrated tools like DPDash with my outputs). \\

\noindent Critically, the columns found in each interviewMonoAudioQC file (with values mapped per interview) are as follows:

\begin{itemize}
\item Metadata required by DPDash \citep{dpdash} or to otherwise assist in downstream merges. 
\begin{itemize}
    \item Note that only a necessary subset of DPDash columns are actually filled with values however.
\end{itemize}
\item Length of audio (in minutes).
\item Overall volume (in dB).
\item Standard deviation of amplitude (though as mean amplitude of audio files is generally 0, this is often a direct transformation from the log scaled dB).
\item Mean of spectral flatness (computed by librosa).
\end{itemize}

\noindent The specific column names used for these features are:

\begin{center}\begin{tabular}{c}\begin{lstlisting}[breaklines=true]
["reftime","day","timeofday","weekday","study","patient","interview_number","length_minutes","overall_db","amplitude_stdev","mean_flatness"]
\end{lstlisting}\end{tabular}\end{center}

\noindent With reftime always containing np.nan as the value but present as the first column for DPDash compatibility. The rest of the columns are appropriately filled out based on available interview information. Recall that this QC output focuses on mono interview audio, which will be the top level audio file containing all speakers in the case of Zoom.

It is worth highlighting that an interview might be rejected by the audio selection based on the content of the CSV (too low volume for the corresponding row, multiple psychs interview sessions uploaded on the same day from one ID, etc.), reasons that will be logged in the email alerts. However a file also may be rejected because the queried filename could not be found within this CSV, or because the corresponding CSV was not found in the first place. The former will log as a metadata issue and the latter as a permissions issue, as these are possible reasons for a detected interview audio that passed initial SOP screening to later fail. In practice, especially for long interviews on Pronet, the same error messages might also trigger though because the audio QC function crashed before the information about that interview could be saved. More on troubleshooting this issue (among others) can be found within the future directions of the upcoming supplement section \ref{subsubsec:interview-adapt}. \\

\noindent As mentioned, the audio QC processing includes a sliding window QC operation as well, which is performed on the identified new (WAV-converted) mono audio files as a first step under the PROTECTED side of processed before they are renamed according to typical output conventions. One such sliding window QC CSV is saved for each detected mono interview audio file, and the CSV filename matches the date and time given for the corresponding interview in raw -- which is then used for filtering out previously processed interview audios on subsequent pipeline runs. 

These sliding window QC outputs are not currently used for any other purposes besides processing documentation (and thus should not be deleted - they take up minimal storage space anyway). The filename map txt files giving a very direct command line accessible link between raw interview name and processed interview renaming are similarly important to retain through the duration of a project. Because they stay on the PROTECTED side of processed, these outputs within AMPSCZ are never shared with \emph{predict} either, and in some sense are thus only intermediates. Still, they could assist in the future for quickly identifying certain finer timescale issues, like psychs interview audios that contain a notable break for a time in the middle of the recording. It is particularly plausible that such outputs might be used to facilitate certain steps of the upcoming feature extraction pipeline, which is to be written for obtaining shareable acoustic measures of possible clinical relevance from interview audio files.

Indeed, a given sliding window output CSV will contain decibel level and mean flatness values as the primary metric columns, with each row corresponding to a particular slice of the interview. By default (and therefore for all AMPSCZ interviews), the sliding window QC computes metrics in 3 second increments with no overlap between these bins. However the default could easily be changed for a future project by supplying additional arguments when the sliding QC python helper is called by the wrapping audio QC module.

\subsection{Transcript processing components}
\label{subsubsec:interview-code-trans}
The transcript processing branch of the pipeline of course generally relies on successful completion of the audio processing branch of the pipeline, and subsequent return of TranscribeMe transcriptions as specified. Most of the steps thus rely on assumptions about TranscribeMe, though one might still utilize the latter modules in a case where obtained transcripts are analogous but the project uses a different upload/download process, or conversely the earliest modules in a case where TranscribeMe is being used but the specified settings for the project are very different. 

For AMPSCZ, all branches of the pipeline are obviously being run daily for each site, and if no new transcripts have been returned by TranscribeMe for a particular site, the transcript branch of the pipeline will simply flag any pending transcriptions in the monitoring steps (if relevant) and gracefully exit. Nevertheless, I will now provide a parallel list of major code steps and their corresponding wrapping bash modules for the transcript branch of the pipeline, as I did for audio above:

\begin{enumerate}
    \item Check the TranscribeMe server for any pending transcriptions, and pull back all those that have been newly completed. Of the newly pulled transcripts, direct them either towards the site manual review process or towards the next steps of the processing pipeline (bypassing manual review), as per the procedure defined in the SOP. This step additionally takes note of the current status of each such transcript as it works through them, to compile an up to date monitoring email (if there is any actively relevant information to share for the site's transcriptions status). 
    \begin{itemize}
        \item These tasks are handled by the run\_transcription\_pull.sh module, which primarily uses the interview\_transcribeme\_sftp\_pull.py script (called for each subject ID and interview type with files awaiting transcripts) to handle the SFTP pull. More details on the download implementation can be found in security/privacy section \ref{subsubsec:interview-code-sec} below.
        \item The run\_transcription\_pull wrapper itself handles manual review assignment, first determining whether the site has had at least 5 interview transcriptions sent for manual review yet, and then if not assigning all new transcripts to manual review, otherwise assigning each new transcript to redaction review with 10\% probability while allowing the rest to skip manual site checking. 
        \item To keep track of transcripts that are still awaiting TranscribeMe transcription, the corresponding audio WAV files are kept in a pending\_audio subfolder within each participant and interview type's PROTECTED side processed folder by the overall pipeline. These files should obviously not be deleted then, but as transcripts are successfully pulled back to the data aggregation server from TranscribeMe's SFTP server, this first step of the transcript branch of the pipeline will move them to a different (completed\_audio) subfolder. The contents of those completed\_audio subfolders can safely be deleted at any time if storage concerns arise.  
    \end{itemize}
    \item Identify any transcripts that were sent for manual site redaction review and have been newly returned, to copy them to the appropriate location on the data aggregation server for file organization and downstream pipeline steps. Also identify those transcripts (if any) that are still awaiting manual review for a given site, to put together a list for a daily email alert to the site contacts (as needed).
    \begin{itemize}
        \item The former task is handled by the run\_transcription\_review\_update.sh module, and the latter by the run\_transcription\_review\_alerts.sh module -- both of which are simple bash scripts written for these AMPSCZ-specific accounting issues.
    \end{itemize}
    \item Create redacted versions of any newly finalized transcripts, whether just returned by TranscribeMe and bypassing manual redaction review or just returned from the site review.
    \begin{itemize}
        \item This task is handled by the run\_transcript\_redaction.sh module, primarily through calling the redact\_transcripts\_func.py python helper on each new transcript identified to have a redacted copy generated.
        \item Recall that for the AMPSCZ project, TranscribeMe returns transcripts with redacted markings surrounding all PII/PHI words, but not obfuscating the words themselves (a good idea to consider for future projects). To mark words here we use {curly braces} as our convention, because they are not otherwise used in TranscribeMe's notation. Therefore TranscribeMe is the one performing the redacting, but this step is needed in the code to create shareable fully redacted versions by replacing all marked words with REDACTED. 
    \end{itemize}
    \item Convert any newly redacted transcript text files to CSV format, with each row a line in the transcript (here mapping to individual turns in the interview) and columns corresponding to TranscribeMe speaker ID, TranscribeMe turn start timestamp, and the verbatim text itself. This step will also confirm that the returned txt is correctly UTF-8 encoded and clean up excess white-space characters.
    \begin{itemize}
        \item These tasks are handled by the run\_transcript\_csv\_conversion.sh module, a pure bash script written to parse transcript text files formatted according to TranscribeMe's convention(s).
    \end{itemize}
    \item Compute interview-level transcript QC stats across the redacted transcript CSVs, to produce shareable QC outputs for transcription monitoring on the level of the subject ID and interview type -- analogous to the final audio QC CSVs described in the preceding section.
    \begin{itemize}
        \item These tasks are handled by the run\_transcript\_qc.sh module, through calling the interview\_transcript\_qc.py python function for each participant ID and interview type with transcripts available. Details on the exact outputs produced by this QC function will be provided subsequently.
    \end{itemize}
    \item Send the daily monitoring/alerting email(s) when applicable. The main monitoring email, to go to central monitors at present, will list all the transcripts that were successfully pulled from TranscribeMe for the current site, and those still awaiting transcription, as well as the transcripts newly returned from manual redaction review. The alerting email listing any transcripts actively awaiting manual redaction review goes directly to site contacts by contrast. 
    \begin{itemize}
        \item Both of these emails were compiled as part of earlier pipeline steps described, and are only sent near the end of the transcript pipeline process because major warnings/errors that arise in later conversion and QC steps may be appended, to assure they are not missed.
        \item The sending of these emails (when created due to relevant updates being available) to the email addresses specified in the corresponding config occurs directly in the top level transcript pipeline branch wrapper (interview\_transcript\_process.sh), rather than occurring in a separated module. 
    \end{itemize}
    \item Update the file accounting log with information on current status of all of the newly recognized/moved/processed transcripts, including additional metadata like whether the transcript was marked for manual redaction review, the date that each major step in the transcript flow was completed (thus far), and specific transcript txt encoding info. This is an analogous step to the final accounting step on the audio branch of the pipeline, with similar downstream uses. 
    \begin{itemize}
        \item These tasks are completed by the run\_final\_transcript\_accounting.sh module, which calls the python script interview\_transcript\_process\_account.py on each subject ID and interview type. 
    \end{itemize}
\end{enumerate}

\noindent Note that to transfer files back to sites for the described manual review, separate code was written - by the Pronet IT team for push from their data aggregation server to Box and by DPACC for push from the Prescient data aggregation server to Mediaflux. 

When my code designates a transcript for manual redaction review, it will put a copy of it into the \hl{PROTECTED/box\_transfer/transcripts} folder on the respective data aggregation server's main PHOENIX data structure. This folder is called box\_transfer on both servers for simplicity. The supplemental transfer scripts (both found outside my repository) will then use information in the transcript file names along with Box/Mediaflux API tools to upload the transcripts for review to the corresponding site's Box (or Mediaflux) transcript review folder, moving them on the server from box\_transfer/transcripts to box\_transfer/completed to denote the transfer was successful. As my code checks this \hl{completed} subfolder to see how many have been transferred so far for a given site, it is important that the transcripts stay there. 

When my code designates a transcript for manual redaction review, it will also move the version of the transcript returned directly by TranscribeMe into a \hl{prescreening} subfolder that is contained within the folder where both the raw transcripts that bypassed site review and the returned versions of manually reviewed transcripts are kept on the top level. This way we can always compare versions of transcripts before and after manual site review, in which sites should add curly braces around any words they feel were missed PII by TranscribeMe before returning them back to the data aggregation server.

To return successfully reviewed transcripts, the sites should simply move finished transcript text files from the "For review" subfolder they have in their transcript review Box/Mediaflux folder, to the corresponding spot under the "Approved" subfolder within the transcript review Box/Mediaflux -- without changing the file path name under these subfolders when doing the move between the two. For more specific instructions on performing this transfer (and the manual review itself), please see the AMPSCZ project SOP. There is also some information on this process as part of the common site mistakes list provided in the main text section \ref{subsubsec:u24-issues}. 

Lochness pulls any transcripts that have been correctly moved by sites to within "Approved" in this way, uploading them to the PROTECTED side of raw on the data aggregation server. Then my code copies newly returned transcripts into the corresponding places on the PROTECTED side of processed to maintain a complete set of final transcript txts with PII visible, handling the rest of the transcript processing steps from there (as were described above). Note that the only transcripts under raw will be those that underwent manual review and were subsequently returned, those that bypassed manual review will stay only under processed folders. 

\subsubsection{Key direct outputs of transcript branch of pipeline}
The primary outputs of this branch of the pipeline include the transcripts at various stages of processing, as well as the final transcript-derived QC measures. In this section, I will provide more information about the contents of these outputs and their locations on the data aggregation servers.

We receive full verbatim transcriptions from TranscribeMe, with turn-level timestamps of second resolution and speaker identification included. As mentioned above, any PII in the transcription is marked by curly braces. Cost per minute varies by language. Saved transcripts can be found both with PII visible on the PROTECTED side of processed, and with marked PII fully redacted on the GENERAL side of processed. The former transcripts cannot be shared with the wider group but when needed can be further processed by those with access to the data aggregation server data structure. The latter transcripts are automatically pushed to \emph{predict} by Lochness, and will eventually be staged for regular transfers to the NIH data repository. The redacted GENERAL side transcripts are also converted to CSV format for easier processing, as described above, which facilitates the transcript QC process as well.

The final PII-visible transcript txts can be found on the main PHOENIX data structure of each central aggregation server at paths of the form: \newline \hl{PROTECTED/[siteID]/processed/[subject]/interviews/[type]/transcripts/}\newline\hl{[siteID]\_[subject]\_interviewAudioTranscript\_[type]\_day[day\#]\_session[session\#].txt}, where day number and session number are formatted as 4 and 3 digit integers respectively, defined based on the study day and interview number determined by the renaming function in the audio branch of the pipeline. Recall that there are many reasons (like split interview folders or backdated uploads) that could cause the automatically-determined session numbers to diverge from the true session number per the protocol for each interview type. This is discussed at greater length within main text section \ref{subsubsec:u24-issues}, but ultimately it does not affect the workings of the interview dataflow/QC pipeline, and can easily be taken into account in downstream interpretation. 

The final redacted transcript txts can be found analogously at paths of the form: \newline \hl{GENERAL/[siteID]/processed/[subject]/interviews/[type]/transcripts/}\newline\hl{[siteID]\_[subject]\_interviewAudioTranscript\_[type]\_day[d\#]\_session[s\#]\_REDACTED.txt}. Under each GENERAL side processed transcripts folder, there is also a \hl{csv} subfolder that holds the CSV-converted versions of each redacted transcript text file. In the future, AMPSCZ analysis work taking place on the data aggregation briefcase might include comparing the fully redacted transcripts to their PII-visible counterparts.

Of course, analogous to the audio branch of the pipeline, transcript QC features are stored on the GENERAL side of processed using DPDash naming conventions. Thus for all participants with interview transcript(s) of a particular type available, there will be a CSV containing interview-level transcript QC metrics for those sessions, found on the data aggregation servers at paths of the form: \newline \hl{GENERAL/[siteFull]/processed/[subject]/interviews/[type]/}\newline\hl{[site2Digit]-[subject]-interviewRedactedTranscriptQC\_[type]-day[start]to[end].csv}. The overall construction/structure of these QC CSVs is similar to those for audio QC, and they are obviously also pushed by Lochness to \emph{predict} and subsequently imported for viewing in DPDash. Additionally, they play a critical role in the quality monitoring processes that occur outside of DPDash, including some that utilize the data aggregation server more directly (such as email alerting). \\

\noindent The specific metrics calculated in the transcript quality control step are as follows: 

\begin{itemize}
\item Metadata for DPDash formatting and related tracking (analogous to the audio QC step above).
\item Number of unique speakers identified by TranscribeMe (labeled in the txt using speaker IDs "S1" to "SN" for N speakers, numbered in the order they first spoke).
\begin{itemize}
    \item Note that per the AMPSCZ SOP, S1 should be the primary interviewer and S2 the participant, and open interviews in particular should only have 2 speakers total - though these instructions have likely not been perfectly followed by sites to date.
\end{itemize}
\item For each of the first three speaker IDs (S1-S3), the following counting stats:
\begin{itemize}
    \item Total number of conversational turns.
    \item Total number of words.
    \item Smallest number of words in a turn.
    \item Largest number of words in a turn.
\end{itemize}
\item Number of times throughout the transcript a segment of the audio was marked [inaudible].
\item Number of times throughout the transcript a segment was marked as uncertain (where TrancribeMe places the word/phrase followed by a ? in brackets).
\item Number of times throughout the transcript [crosstalk] between two speakers was marked.
\item Number of words marked by TranscribeMe as {PII} throughout the transcript.
\item Total number of commas appearing in the transcript, as well as the total number of dashes appearing in the transcript (both related to TranscribeMe's verbatim notation, so this is a quick check on that).
\begin{itemize}
    \item For more on TranscribeMe's verbatim convention and leveraging this in feature extraction steps in order to detect different categories of linguistic disfluency, see the audio journal pipeline documentation of supplemental section \ref{subsec:diary-code} above.
\end{itemize}
\item Timestamp for the start of the final turn, in minutes.
\item Shortest gap between the starts of two subsequent turns, in seconds
\item Longest gap between the starts of two subsequent turns, in seconds
\item The shortest and longest such gaps weighted by the number of words in the intervening sentence.
\end{itemize}

\noindent Thus each column in an interviewRedactedTranscriptQC DPDash CSV is one of these individual metrics, and each row is a particular transcript (corresponding to one interview session). The specific column names used for these features are:

\begin{center}\begin{tabular}{c}\begin{lstlisting}[breaklines=true]
["reftime","day","timeofday","weekday","study","patient","interview_number","transcript_name","num_subjects","num_turns_S1","num_words_S1","min_words_in_turn_S1","max_words_in_turn_S1","num_turns_S2","num_words_S2","min_words_in_turn_S2","max_words_in_turn_S2","num_turns_S3","num_words_S3","min_words_in_turn_S3","max_words_in_turn_S3","num_inaudible","num_questionable","num_crosstalk","num_redacted","num_commas","num_dashes","final_timestamp_minutes","min_timestamp_space","max_timestamp_space","min_timestamp_space_per_word","max_timestamp_space_per_word"]
\end{lstlisting}\end{tabular}\end{center}

\noindent All of these features are imported into DPDash to assist in interview progress monitoring as needed. With results reported in the main chapter \ref{ch:2}, a subset of the transcript QC features (in addition to normalized versions e.g. redactions per word) have been studied in more depth in the early AMPSCZ dataset, and where useful have subsequently been more closely monitored as part of regular data quality assurance. \\ 

As an aside, if you are in charge of the TranscribeMe SFTP account you will by default receive a separate email alert from them for every single completed transcript they upload to the SFTP server. You can turn these off (or redirect them to other addresses) by contacting your TranscribeMe rep. Regardless, please note that these emails have no direct relation to the interview pipeline described here, which has its own email alerts. Not every transcript returned by TranscribeMe will end up in manual site review, and there can be up to a one day delay between TranscribeMe returning a transcript and it hitting the data aggregation server, as my code runs on a once daily cron job to check for new transcriptions to pull back from the TranscribeMe SFTP server. From there, there could be an additional delay of up to a day for the external utility script to transfer any transcripts that were selected for manual review to Box/Mediaflux, and for transcripts not selected for manual review there could be an additional delay of up to a day for produced redacted transcripts on the GENERAL side of processed to be copied over to \emph{predict}. 

Usually gaps between these separated steps will be shorter than a full day, but please wait a reasonable amount of time before raising a potential missingness concern based on TranscribeMe's email alerts. More importantly, please do not raise a concern about missing transcripts based only on what gets transferred to the sites' manual review Box/Mediaflux folders, as the clear majority of transcripts do not get sent for manual review at all and thus never get synced back to Box/Mediaflux. Finally, if you are performing data monitoring for AMPSCZ, please do not contact TranscribeMe directly looking for transcript outputs unless you have first confirmed the missingness is a TranscribeMe issue. Transcripts for this project need to flow through the designed pipeline, for privacy and data integrity reasons, and as far as actual unintended missingness is concerned it is far more likely to be a site mistake or even a code bug -- not things TranscribeMe has any control over. There have of course been a few times where we had to contact TranscribeMe over real issues with the way they returned a file, but these instances have been very rare. If you are not the person responsible for monitoring the operation of this software, you should probably not be the person initiating contact with TranscribeMe about a possible missing transcript.  

\subsection{Video processing components}
\label{subsubsec:interview-code-vid}
The final relevant modality for interview recordings is video -- assumed to be obtained from Zoom for the present code, and a required component of the open interview protocol (optional for psychs) in the AMPSCZ project. 

The video parts of the pipeline do not rely on outputs from the audio/transcript branches to be available, so for basic data monitoring and quality assurance purposes it is possible to run only the video branch, if desired by a future project. Obviously for each AMPSCZ site all branches are run daily, and in the cron job we use the video modality comes third after the audio and transcript code has been run for that site. 

Given the format of Zoom meetings, the main goal of video QC here is to determine how many faces are detectable throughout the interview, along with some other basic facts about detected faces like their estimated size in the frame. Of course the pipeline also handles basic metadata compilation for file renaming and monitoring purposes on the video branch too. For more clarity, I will now provide a parallel list of major code steps and their corresponding wrapping bash modules in the video branch of the pipeline, as I did for audio and transcripts above:
\begin{enumerate}
    \item Identify new video files that have been correctly uploaded to raw within a Zoom interview folder, additionally initializing an email body for the video update daily site message if there are indeed any new videos available for the current site. From each new video, extract 1 frame for every 4 minutes of recording. The extracted frames of course go into the PROTECTED side of processed, contained within a subfolder that is named using the date and time of the interview extracted from the Zoom title. The presence of such extracted frames is used in subsequent runs to determine whether a video has been analyzed yet or not.
    \begin{itemize}
        \item These tasks are handled by the run\_new\_video\_extract.sh module, which relies on ffmpeg but does not utilize python.
        \item Note frames are extracted periodically so that face detection can be run whilst keeping the pipeline feasible to use on the lightweight data aggregation servers and quick to complete processing of even longer recordings.
    \end{itemize}
    \item Run PyFeat's basic face detection model on each newly extracted frame, saving details about any detected faces on the frame level, as well as using those results to compile per interview summary stats about face detection. This step also utilizes site and interview metadata to map each raw video name to an appropriate processed file name based on our conventions, ultimately creating/updating a DPDash-formatted video QC CSV for each subject ID and interview type with new recordings. That info is used in adding details about successfully processed (or not) video recordings to the daily video site update email as well.
    \begin{itemize}
        \item These tasks are handled by the run\_video\_qc.sh module, through calling the interview\_video\_qc.py python function for each participant ID and interview type with newly extracted video frames. Details on the exact outputs produced by this QC function will be provided subsequently.
    \end{itemize}
    \item If a daily video monitoring email has been constructed due to new video uploads being detected, send out that finalized email. This email will provide a list of all new video recordings processed for QC, in terms of their final renaming for processed outputs. It will also include any error messages that may have arisen that prevented QC from running on other newly detected video files. 
    \begin{itemize}
        \item The email body was compiled (when relevant) as part of earlier pipeline steps described, and is then sent near the end of the top level pipeline branch wrapper (interview\_video\_process.sh), analogous to the transcript email alerts step. 
    \end{itemize}
    \item Update the file accounting log with metadata information related to the newly detected videos, very similarly to the final accounting steps at the ends of the audio and transcript branches of the pipeline. 
    \begin{itemize}
        \item This task is completed by the run\_final\_video\_accounting.sh module, which calls the python script interview\_video\_process\_account.py on each subject ID and interview type. 
    \end{itemize}
\end{enumerate}

\noindent Note that if the daily video update message sends for a site but includes no video names and no specific error messages, it means that a new video was detected but the code failed prior to reaching the video QC step - thus indicating a likely error with ffmpeg's frame extraction command on one of the site's uploaded interview recordings. For more interpreting possible edge case errors within the existing monitoring infrastructure (for all parts of the pipeline), please see the troubleshooting information in section \ref{subsubsec:u24-next-steps-interview} below.

\subsubsection{Key direct outputs of video branch of pipeline}
The primary output of the video branch of the pipeline is again the DPDash CSVs for quality assurance monitoring, although the interview video renaming maps, as well as the extracted frame images and their corresponding PyFeat outputs, have possible downstream uses too. The details of these outputs follow, first file naming conventions and then additional content information where applicable. 

When a new, correctly formatted, video has been uploaded by Lochness to raw within a (likely new) Zoom interview folder, the first step of the pipeline will create a corresponding subfolder for storing extracted frames from that interview on the PROTECTED side of processed. These folders can be found on the main PHOENIX data structure of each central aggregation server at paths of the form: \newline \hl{PROTECTED/[siteID]/processed/[subject]/interviews/[type]/video\_frames/[date]+[time]}, where date and time are extracted from the Zoom folder name based on expected conventions. Each extracted frame from the interview is saved as a JPG under such a folder and named based on its timestamp within the interview; the sampled frames are taken from pre-determined recording timestamps that occur literally every 4 minutes, starting from the 1st second of the video (to avoid the very first frame), and then proceeding with timestamp 00:04:00, 00:08:00, and so on. These extracted frames are expected to be kept on the PROTECTED side of processed for the duration of the study, for both accounting purposes and for quick spot checking videos as needed.

As part of video QC, more detailed PyFeat information is stored for each frame, to reference in possible expanded QC if needed in the future. The individual frame CSVs are named according to their corresponding frame JPGs, under a subfolder with the extracted frames i.e. \hl{video\_frames/[date]+[time]/PyFeatOutputs}. Also as part of the video QC there is a raw to processed file name link stored, analogous to the audio file name maps previously described. This mapping can be found at \hl{video\_frames/[date]+[time]/[date]+[time].txt}, and it is especially important for ongoing accounting that it is not deleted for any previously processed interview videos. The file name of the txt itself, when taken along with folder names contained in the rest of the file path structure, is directly linkable to a specific raw interview folder upload. The content of the txt is what that raw video should be renamed to in a processed setting, based on SOP conventions.

Of course all of the video outputs contained within a video\_frames subfolder are on the PROTECTED side of the data aggregation server storage structure, and are thus only accessible to a limited number of individuals working on AMPSCZ interview recording collection. The primary output type that is pushed along to \emph{predict} is the DPDash CSV with per interview video QC stats. Like audio and transcript QC, those metrics can now be visualized using the DPDash web interface, and they are additionally included in compiled form as part of the weekly summary updates with broader reach (reported on extensively in the main chapter \ref{ch:2}). Again, there is one such DPDash CSV created by the video branch of the pipeline for each subject ID and interview type with at least one video recording available on PHOENIX. Each of these CSVs contains one row per detected interview session, and can be found on the data aggregation servers at paths of the form: \newline \hl{GENERAL/[siteFull]/processed/[subject]/interviews/[type]/}\newline\hl{[site2Digit]-[subject]-interviewVideoQC\_[type]-day[start]to[end].csv}. \\

\noindent The interview QC metric columns for the video datatype include:

\begin{itemize}
    \item Metadata for DPDash formatting and related tracking (analogous to the audio and transcript QC steps above).
    \item Number of frames extracted, and thus used for computing the QC summary stats (should equal the floor of interview minutes divided by 4).
    \item Minimum number of faces detected across the extracted frames from the interview.
    \item Maximum number of faces detected across the extracted frames from the interview.
    \item Mean number of faces detected across the extracted frames from the interview.
    \item Minimum confidence score across all faces that were detected from across all frames.
    \item Maximum confidence score across all faces that were detected from across all frames.
    \item Mean confidence score across all faces that were detected from across all frames.
    \item Minimum area across all faces that were detected from across all frames.
    \begin{itemize}
        \item Area for a particular face was computed by multiplying the height and width variables returned by the PyFeat face detection model, describing the face box associated with the detected face.
    \end{itemize}
    \item Maximum area across all faces that were detected from across all frames.
    \item Mean area across all faces that were detected from across all frames.
\end{itemize}

\noindent The specific column names used for these features are:

\begin{center}\begin{tabular}{c}\begin{lstlisting}[breaklines=true]
["reftime","day","timeofday","weekday","study","patient","interview_number","number_extracted_frames","minimum_faces_detected_in_frame","maximum_faces_detected_in_frame","mean_faces_detected_in_frame","minimum_face_confidence_score","maximum_face_confidence_score","mean_face_confidence_score","minimum_face_area","maximum_face_area","mean_face_area"]
\end{lstlisting}\end{tabular}\end{center}

\noindent We also have the CSVs (one per extracted frame) saved in associated PyFeatOutputs subfolders on the PROTECTED side. Those CSVs contain one row per detected face, with the following additional metrics provided as columns:

\begin{itemize}
    \item Face ID (assigned by PyFeat).
    \item Face location X coordinate in frame.
    \item Face location Y coordinate in frame.
    \item Face box's rectangle width.
    \item Face box's rectangle height.
    \item Face's detection confidence score.
\end{itemize}

\noindent Recall that it is not possible to load the standard face pose or action unit detection models on the data aggregation servers (particularly for Pronet) at this time, and it would generally be incompatible with the lightweight philosophy of this pipeline. Therefore any QC code that might use information about e.g. FAUs would need to be separately written, and we focused only on basic detection here. 

\subsection{Site-specific daily monitoring additions}
The final major branch of the pipeline, which also runs on a per-site (or per-study) basis using the config file system described, runs a set of tools added onto the codebase for more detailed file accounting and monitoring.  While parts of the individual modules within this file accounting branch may be useful to run independently on a different project, much of this work relies on existing pipeline outputs from the other three branches -- primarily the outputs of the final accounting step within each of those branches, along with the DPDash QC CSVs for each modality.

The main goal in adding this branch to the pipeline was to call additional attention to SOP violations and clear cut quality red flags, as well as to create some more streamlined progress updates for each site that could assist in higher level monitoring than the detailed emails sent by the other three branches. As part of that, this step also improves the DPDash views available by creating QC CSVs that are combined across modalities for each subject/interview type, with a filtered down set of key columns.

\noindent Like with the other branches, I will now provide a list of major code steps and their corresponding wrapping bash modules for the additional monitoring components of the pipeline:
\begin{enumerate}
    \item Identify file formatting and organizational issues that would prevent anything newly uploaded to raw from being processed by the pipeline at all, so that monitoring can easily notice such violations.
    \begin{itemize}
        \item This task is handled by the run\_raw\_folder\_check.sh module, through calling the raw\_interview\_sop\_check.py python function on each relevant subject ID and interview type. Obviously it relies on assumptions about folder structure and file naming being used for the AMPSCZ interview data collection protocol, which has already been described at length.
        \item Many of the types of site mistakes that have prevented processing and were caught by this step in early AMPSCZ monitoring have been identified within the main chapter section \ref{subsubsec:u24-issues}. The exact info tracked by this module will be detailed subsequently. 
    \end{itemize}
    \item Using the SOP CSV from step 1 and the respective processed accounting CSVs created in the final step of each of the audio, transcript, and video branches of the pipeline, isolate all interview session records that have a newly detected upload issue or have some new processing update. If there are any processing updates, initialize (or add to) a summary stats update email with a header containing a list of subject IDs and interview types that have some processing updates from the current code run. Then for each specific interview record with updates, look up its QC metrics in the corresponding DPDash CSVs for each modality, checking for a few specified major QC red flags as well as some possible metadata inconsistencies (from both QC and accounting CSVs). If any warnings were triggered in that process, or if new basic upload SOP violations were detected, set up (or add to) a warning email alert to be sent today for this site. Ultimately, this step identifies a number of important warnings to be notified about, saving/updating a CSV for each subject ID and interview type where relevant with an account of all warnings detected to date. Also saves a combined version of the processed accounting CSV containing merged info from all 3 modalities (or whichever modalities are available) for each session. 
    \begin{itemize}
        \item These tasks are handled by the run\_processed\_accounting\_check.sh module, through calling the interview\_warnings\_detection.py python function for each subject and interview type with any available interview outputs under processed (including just a raw SOP violation log). A more specific list of the warnings that are compiled by this step can be found subsequently.
    \end{itemize}
    \item If a summary stats email was initialized for this site as part of step 2, it means that there are new QC metrics for at least one modality of one interview for the site. In that case, the pipeline will compute new site-wide and per-subject summary stats for the current site, with select metrics from the former also being added into the summary stats update email to be sent. Regardless (to account for backlogged updates in a timely manner because it is an important output being pushed to \emph{predict}), all subject IDs and interview types with DPDash QC CSVs available will then have a new modality-merged QC CSV saved containing the latest information. 
    \begin{itemize}
        \item The wrapping script for this pipeline branch (interview\_summary\_checks.sh) checks for a new summary stats email body, and if one has been created it will use the run\_qc\_site\_stats.sh module to handle all of these tasks. Otherwise, it will use the run\_qc\_combine\_only.sh module to handle only the task of modality merging QC CSVs.
        \item Both run\_qc\_site\_stats and run\_qc\_combine\_only call the interview\_qc\_combine.py python function on all subjects/interview types with QC available. This is the only major functionality in the run\_qc\_combine\_only module.
        \item The run\_qc\_site\_stats module then also calls the interview\_qc\_statistics.py python script, first for all open interviews across the present site and then for all psychs interviews across the present site. The details of the outputs produced by this module will also be further detailed below.
    \end{itemize}
    \item If a summary stats update and/or new issues warning email alert was generated, send them to the site-specific email update addresses specified in the corresponding config. Email sending is again performed directly in the wrapping script for this pipeline branch. Here it is the last step of the pipeline for a given site (when applicable).
\end{enumerate}

\noindent Note that while the code is in progress, the described email bodies (when relevant) are drafted directly within the top level of the repository installation folder. After any finalized emails are successfully sent (or if the sending fails due to any issue with the mail command or server connectivity), the corresponding txt files will have a unique Unix timestamp appended to their name and will be moved into the logs folder of the repository for record-keeping purposes. These are of course also part of what will be excluded from any git pushes via the .gitignore. A similar process occurs for email alerts in all branches of the main pipeline.  

\subsubsection{Resulting accounting CSVs and monitoring alert details}
A major result of this branch of the pipeline is the more salient daily site email updates produced, particularly the alerts that will highlight new warnings that arise (to be elaborated on next), which has been a key part of my monitoring workflow to date - especially for catching violations that prevent an interview upload from appearing in downstream QC records at all. Another major result is the creation of additional CSVs with new curated file accounting and quality control metrics, many of which are later used in server-wide (i.e. cross-study/site) summaries created by the utility scripts written for further AMPSCZ monitoring. Those utilities were introduced in more detail above, in section \ref{subsubsec:cross-site-utilities}. Here I will elaborate on the specifics of both the emails and CSVs created by the site-specific monitoring branch of the main pipeline in the above mentioned steps.

The first step, utilizing raw\_interview\_sop\_check, creates (or updates) a CSV ending with "RawInterviewSOPAccountingTable.csv" on the PROTECTED side of processed for each subject ID and interview type, i.e. directly within the folders of the form: \newline \hl{PROTECTED/[siteID]/processed/[subject]/interviews/[type]/} \newline The CSV contains one row for each problematic file or folder found under the raw interview uploads for that subject ID and interview. Note that only raw paths corresponding to an SOP violation are included in the list, and it focuses only on those SOP violations that completely stall processing of one or more modalities focused on by this interview pipeline (i.e. mono audio and/or video). 

\noindent Thus the major issues caught here are:

\begin{itemize}
    \item An interview upload is a standalone file but it is not named according to EVISTR convention.
    \item An interview upload is a folder but it is not named according to Zoom recording convention. 
    \item An interview upload is a correctly named folder but it does not contain a top level M4A file with the expected mono audio naming convention for Zoom, or alternatively it contains more than one such file. 
    \item An interview upload is a correctly named folder but it does not contain a top level MP4 with the expected video file naming convention for Zoom, or alternatively it contains more than one such file. 
\end{itemize}

\noindent It is possible for a given upload to commit more than one of these violations, but because the various columns contained in the RawInterviewSOPAccountingTable will describe all such issues, there will still only be one row for that upload in the CSV. The columns of the CSV are as follows:

\begin{center}\begin{tabular}{c}\begin{lstlisting}[breaklines=true]
["raw_name", "folder_bool", "valid_folder_name_bool", "num_audio_m4a", "num_video_mp4", "num_top_level_files", "date_detected"]
\end{lstlisting}\end{tabular}\end{center}

\noindent Conversely, note that if the files from a Zoom folder are uploaded loosely (outside of their folder), there will be one row in this CSV for each such loose file. When the warnings email is constructed by the next step, it will list the raw name of each record newly added (i.e. detected today) to any RawInterviewSOPAccountingTable from across the given site, where applicable. All such issues are labeled as "SOP Violation" in that warnings email. \\

\noindent The other types of issues that will be included in the warnings email put together by the interview\_warnings\_detection function of the second step are:

\begin{itemize}
    \item "Audio Rejected by QC"
    \item "Audio Failed SFTP Upload"
    \item "Missing Expected Speaker Specific Audios"
    \item "Transcript Encoding Not UTF-8"
    \item "English Transcript Encoding Not ASCII"
    \item "Video Day Inconsistent with Audio Day"
    \item "Video Number Inconsistent with Audio Number"
    \item "Consent Date Changed with New Files"
    \item "Session and Day Numbers Inconsistent"
    \item "Session Number Repeated"
    \item "Interview Under 4 Minutes"
    \item "No Faces Detected"
    \item "No Redactions Detected"
\end{itemize}

\noindent The first 10 problem types listed are derived from the processed file accounting CSV outputs, while the last 3 come from checking QC outputs. This list is of course not exhaustive, particularly when it comes to monitoring of quality metrics; other tools like DPDash and weekly histograms should also be used as part of the quality review workflow.  

Recall that the above issues are only reported here when detected in interviews that were found to have had some update on the current pipeline run. If multiple problems are detected in such an interview though, each will be independently logged -- in the email alert by listing the study day and code-labeled session number corresponding to each problem on a new line. The warnings generation step additionally creates (or updates) a CSV ending with "InterviewProcessWarningsTable.csv" on the PROTECTED side of processed for each subject ID and interview type i.e. under the same location as the RawInterviewSOPAccountingTable CSVs.

\noindent The InterviewProcessWarningsTable contains one row per detected processed accounting/QC warning from a run of this pipeline branch, with the following columns:

\begin{center}\begin{tabular}{c}\begin{lstlisting}[breaklines=true]
["day", "interview_number", "interview_date", "interview_time", "warning_text", "warning_date"]
\end{lstlisting}\end{tabular}\end{center}

\noindent As such, the CSV can be revisited at any time for a complete record of previously issued warnings for a given subject ID and interview type. 

In the same processed folder, the code also saves a CSV merging the full set of file accounting information that has been compiled to date by other pipeline branches, under a file name ending with "InterviewAllModalityProcessAccountingTable.csv". Additionally, the interview\_qc\_combine function saves an analogous CSV that represents the merging of all DPDash QC CSVs produced from across modalities for each subject ID and interview type, under a file name ending with "combinedQCRecords.csv". However the combined QC metrics can be saved on the GENERAL side instead, so it may be pushed by Lochness to \emph{predict}. That CSV is therefore found under \hl{GENERAL/[siteID]/processed/[subject]/interviews/[type]/} instead. \\

\noindent The final major step, utilizing interview\_qc\_statistics when new processing updates have occurred, creates (or updates) a CSV ending with "InterviewSummaryStatsLog.csv" on the GENERAL side of processed for each subject ID and interview type. This CSV has 2 new rows added each time it is updated, one for subject-specific summary stats and one for site-wide summary stats, both specific to the interview type and each with a column denoting the current computation date (and obviously a column indicating whether it is a subject summary or a study i.e. site-wide summary). Besides other basic metadata columns, the main columns of this CSV are summary stats over each QC metric available from across the interview modalities. The four summary stats calculated for every QC feature are the mean, standard deviation, minimum, and maximum of the feature to date. 

While each InterviewSummaryStatsLog CSV covers summary stats for all QC features included by the pipeline, a select few on the site-wide level for each interview type (open and psychs here) are included in the daily summary stats update email, when there is new information to be conveyed. In that summary update email, the mean (+/- the standard deviation) are reported for open and then for psychs for the following features:

\begin{itemize}
    \item "Interview Length in Minutes"
    \item "Mono Audio Decibels"
    \item "Audio Spectral Flatness"
    \item "Video Faces Detected per Frame" (mean per video)
    \item "Face Detection Confidence" (mean per video)
    \item "Transcript Number of Speakers"
    \item "Speaker 1 Word Count"
    \item "Speaker 2 Word Count"
    \item "Number of Inaudible Words"
    \item "Number of Redacted Words"
\end{itemize}

\noindent Note that each InterviewSummaryStatsLog CSV will get pushed to \emph{predict} by Lochness as a GENERAL side output. Concatenated versions of these CSVs (filtered to be on subject and site levels separately) are also attached to a daily server-wide email sent by the AMPSCZ utility scripts described above (section \ref{subsubsec:cross-site-utilities}).

The reason that site-wide stats are stored in every single one of these individual CSVs on the data aggregation server is because there is no good way to store data that is summarized on anything above the subject level with the PHOENIX data structure. Combined versions of CSVs on the site or server level that are computed for email alerts or other logging thus get stored within the repo installation folder on each server, in the logs subfolder. These outputs are of course protected from being pushed back to the central repository via the .gitignore defined in the repo. But still do take care of permissions on the installation folder, to ensure that summary info cannot be seen by anyone who should not have access to the server's interview data on even a summary level. In particular, real dates can be found via these files. \\

\noindent Ultimately, the participant-wide accounting branch of the pipeline has been important for the monitoring workflow of AMPSCZ. While the specific warning alerts are not exhaustive and sometimes can be triggered by different mistakes than they intend to encode, they have succeeded in capturing a great number of issues across sites, and continue to be checked regularly to find new SOP violations. For more on the specific warnings messages of greatest relevance and the likely root causes behind different types of warnings, please see the troubleshooting information within the future directions section below (\ref{subsubsec:u24-next-steps-interview}). 

The summary stat updates on the site-level sent by this branch of the pipeline can be of use as well, though for reviewing other possible QC issues I largely use the stats from cross-site server-wide information emailed out by the mentioned utility scripts (\ref{subsubsec:cross-site-utilities}). Primarily I review the weekly histograms and tables described at length in main chapter \ref{ch:2} to detect possible new problems, but I also often refer to the combined QC CSVs from daily email updates if more specific stats need to be investigated for a given interview. 

\subsection{Security review}
\label{subsubsec:interview-code-sec}
Audio files are pushed to TranscribeMe using the following process. We connect to the TranscribeMe server via SFTP. The pySFTP python package, a wrapper around the Paramiko package specific for SFTP is used to automate this process. The host is sftp.transcribeme.com. The standard port 22 is used. The username is provided as a setting to the code, but will generally be one account for all sites on the Pronet side and one account for all sites on the Prescient side, with set up of this facilitated by TranscribeMe. These accounts are created only to be used with this code and will serve no other purpose.

The account password is input using a temporary environment variable. It was previously prompted for using read -s upon each run of the code, but for the purposes of automatic scheduled jobs, the password may now be stored in a hidden file with restricted permissions on a secure portion of our machine instead with permissions fully limited to only the account meant to run the interview pipeline. Presently this has been approved to be stored directly in the folder where the code is installed, but if needed it could instead be stored directly on the PROTECTED side of PHOENIX. In that case, there would be a further layer of guarantee, as anyone who could access the password file necessarily must also have access to the raw files themselves, so use of the TranscribeMe password would provide no additional benefit.

Interfacing with TranscribeMe is done using the same SFTP protocol as described in the audio section. Note that once the transcript side of the pipeline has detected a returned transcript on this SFTP server, the corresponding uploaded audio is deleted by our code on TranscribeMe's end using the same SFTP package. TranscribeMe also automatically deletes files on their server (including the transcripts this code moves to "archive" subfolders) after a certain amount of time has passed. TranscribeMe's server maintenance practices have also been reviewed and approved by Partners Healthcare.

The code itself (in the state presented here) was approved by both Pronet and Prescient central servers for safe installation and use on AMPSCZ data, and it is of course fully available for open source review as the need arises \citep{interviewgit}. The other privacy/security considerations relevant to AMPSCZ interview processing relate to other steps of the data collection and analysis process, which I will now touch on briefly as they pertain to possible future changes for this data management/QC pipeline. 

\subsubsection{Commentary on the use of Zoom}
The use of HIPPA-compliant institutional Zoom accounts for recording AMPSCZ interviews (as detailed in the SOP instructions and reviewed in supplemental section \ref{sec:ampscz-pro} above) has been IRB approved for all US-based sites and most foreign sites in this project. 

Zoom was originally chosen as the remote interview recording platform because of its ability to save speaker-specific audio files, something absent from other major options (e.g. Microsoft Teams, Cisco WebEx). Of course, it is also desirable for sites across a collaborative study to agree on the recording platform. Fortunately, an acceptable secure protocol for using Zoom has largely been established across sites, and thus the pipeline was designed around Zoom for offsite (and onsite open) interviews. Practically speaking, the current implementation does not carry major risk of privacy violations and is considered suitably secure by institutions across the country (and world).

There are still a handful of sites that have not been permitted to use Zoom for recording participant interviews though, some of whom in fact may not be allowed to contribute video of participants at all. These restrictions are primarily a result of legal differences in some European countries, and for this project it is mostly German sites that are impacted. As an edge case, early work on the AMPSCZ interview pipeline did not prioritize accommodating these sites, but it will be necessary in the future to do so. 

Microsoft Teams was one proposed alternative, but not all affected sites could agree on this. Cisco WebEx was the only platform that all such sites were able to get approved by their IRBs. It is therefore recommended that preprocess/quality control code add in handling of Cisco WebEx as another supported platform, for both these AMPSCZ sites and possible future projects that may encounter similar issues. It is worth reiterating though that Zoom should be consistently used for all interviews wherever possible, and WebEx should only be a fallback. 

When adding support for WebEx to the pipeline, it will be important not only to include the expected WebEx file naming conventions and formats in the new audio identification code, but also to update all relevant downstream renaming/accounting steps to ensure that Cisco naming is now considered an acceptable session type rather than an SOP violation. Whoever works on this should walk through all steps of the data flow and monitoring (along with a sample Cisco interview input) to double check that any part of the code potentially impacted by the addition of a new interview platform (or specific differences in output between WebEx and Zoom e.g. lack of diarized audios) are properly updated. Once implemented, the initial stages of monitoring should be extra careful for these interviews as well. \\

\paragraph{Other TODOs for accommodating site differences.}
As mentioned, some of the sites not allowed to use Zoom are also potentially not allowed to submit video at all to the central server (a subject of ongoing IRB debate). In the case that video upload is not allowed, one option is simply for the site to exclude video entirely in their recording settings, which would only be a violation under the typical protocol requirements for the open interview type. 

Even for open interviews, video analysis has been somewhat of a backburner topic thus far, and by excluding the video entirely for only the impacted site(s) it would ease software engineering burden, in addition to seriously minimizing the likelihood that a video accidentally gets uploaded by the site to the central server. However if there is an insistence on video processing for any such sites, then it would be possible to set up a system whereby video feature extraction occurs on a local machine for the site, with the produced feature CSVs uploaded to Box/Mediaflux under the corresponding interview folder. On this topic, I have reproduced here a plan I sent to the Jena site for implementing such a system if desired in the future. \\

\noindent My proposed plan is to have a local folder structure that matches the Mediaflux folder structure. After an interview, the RA would put the automatically saved interview folder \emph{as is} into whatever local backup structure Jena wants to use to preserve the raw interviews, and then \emph{copy} the same interview folder into the "fake" local version of Mediaflux described, otherwise following the same SOP instructions (i.e. no modifications, and make sure it is copied under the folder that corresponds to correctly matching subject ID and interview type). I could then help them install a script that would work from this "fake" Mediaflux to integrate properly with the central server, by performing the following steps for each subject ID and interview type available:
\begin{enumerate}
    \item Identify interview folders that still have any MP4 (or other video formats) within. If the chosen interview recording platform saves the recording as a standalone MP4 instead of within a folder, move this file into an automatically created folder that is named after the MP4 file. 
    \item Run PyFeat feature extraction locally on these identified MP4s, with whatever desired settings. This would save a CSV directly to the interview folder, containing selected face features (e.g. AUs, landmarks, pose/gaze for every face detected) for each included frame.
    \begin{itemize}
        \item One might try to replicate the exact feature extraction paradigm (once it is planned) that will be used for the rest of the files that do go directly to the Prescient data aggregation server. However depending on the local machine specs at Jena it might be better instead to decide on a modified feature extraction protocol -- perhaps by focusing on a subset of the most critical features or by decreasing the number of frames covered to get a temporally courser but overall very similar feature set.
    \end{itemize}
    \item For each newly processed interview folder, confirm that the CSV was successfully saved with all expected features, and if so delete the MP4 from the folder. If the interview recording platform does not produce a separated mono audio file automatically, then prior to deleting the MP4 the script would also need to use FFmpeg to extract an audio-only WAV track from the provided single file. 
    \item Finally, for all interview folders, identify those that do not have any MP4 (or other video format) present, and then check for the same folder path (and all specific audio/CSV contents) in the real Mediaflux folder structure for that subject ID and interview type. For any folder/files missing from the real Mediaflux, copy them over accordingly. 
    \item For safety, one might also consider a part of the script where MP4s (or other applicable video formats) within the real Mediaflux are searched for and moved out if found, along with triggering an alert. The script should make it impossible for video to accidentally be uploaded if the work goes through the local structure intended, but of course it is plausible that an RA might accidentally upload an interview directly to Mediaflux.
\end{enumerate}
\noindent Once in place, I would just need to confirm that Lochness will pull CSVs within interview folders, and then integrate a step into the video pipeline that uses the uploaded CSVs directly to get QC stats when a video file isn't available. Of course whatever interview platform is used for this local process should be consistent with interview platforms otherwise supported by the pipeline (though the local WAV extraction from MP4, if necessary, would require a few additional tweaks to the audio side of the code too). \\

\noindent A downstream consideration along the same lines as site-specific central server upload restrictions would be site-specific restrictions on uploads to the more public NIH data repository. Of course raw audio and video do not get uploaded regardless, but redacted transcripts and QC/metadata do, and eventually extracted features from audio and video should also be part of the data repository contribution. 

These issues will impact data repository staging done on \emph{predict}, and could potentially impact decisions for the feature extraction code that is to be implemented on the AV processing server. Neither of these steps are directly involved with my code, but they still will integrate with the pipeline and ultimately they relate to project design decisions of interest. 

There has not yet been a larger group decision about which extracted features can go to the NIH data repository. As far as professional transcriptions with redacted PII, it has already been agreed upon that these may be uploaded, although it is worth noting that most of the international sites have not begun data collection yet and concerns from these sites/their review boards have regularly flip flopped when it comes to other topics. 

Still, unless brought up otherwise, it is fair to assume redacted transcriptions are shareable with the research community, and thus aside from linguistic features that require access to the (infrequently) redacted words or multimodal analyses that require access to the raw audio or video, the vast majority of work on interview language can be done by anyone with access to the NIH data repository. This is especially true for the open interviews that will not encounter the same sort of duration cutoff issues that psychs interviews have; besides, open interviews appear to have much greater value for speech sampling science. 

The more serious discussion will be surrounding the shareability of various extracted features from raw AV, something that requires more careful thought from the planning committee as a whole, and something that has a decent chance of unearthing more site-specific regulatory differences. Certainly there are broad summary-level stats that no one will have an issue with sharing to the data repository, but as most people will not be able to analyze the raw signals directly, it is highly desirable to provide as temporally fine and spatially descriptive features as possible. Summary stats should in fact be left in large part to individual groups that will be analyzing the shared data, to compute themselves as they see fit, per the AMPSCZ project guidelines.

On the other hand, acoustic features such as the lowest level descriptors output by e.g. OpenSMILE and video features such as the exhaustive facial landmark coordinates available from e.g. MediaPipe are both potentially capable of revealing subject identity if carefully analyzed -- especially if done so in conjunction with the time-aligned redacted transcripts or with broader information from day-aligned passive phone signals. These considerations involve open questions in digital healthcare forensics and biomedical research ethics. It is relieving that the NIH data repository has strict access controls for more sensitive datatypes that do become available to researchers (by application), though given the younger study population here it perhaps remains important to err on the side of caution for AMPSCZ data. 

Regardless of the general decisions made for data repository sharing, it will be critical to ensure that across the many international institutions there are not unique regulatory conflicts for any chosen feature extraction scheme. If there are, then it will require further discussion on how infrastructure should be written to ensure staging for the NIH repository covers any edge cases robustly (i.e. not easily susceptible to human mistakes). If there are a large number of institution-specific considerations, then one might even consider adding a secondary feature set to extract from all interviews: a feature set that would be as exhaustive as possible whilst maintaining shareability across all sites. Then for US-based (and other compatible) sites both primary and secondary core features would be extracted for sharing. 

\subsubsection{Permissions on the data aggregation servers}
For both Pronet and Prescient production data aggregation servers, my account has sudo access. As far as privacy of interview data is concerned this is not an issue, because I was part of a very small group that was granted access to all recordings across sites, in order to implement and test my pipeline. It has also turned out to be extremely helpful in troubleshooting various issues without relying on a slow back and forth with server IT groups. 

However, for data safety reasons (e.g. avoiding accidental deletion), this setup is not ideal for the longer term project, and is something that could maybe be changed when the pipeline is transferred to a different account upon my graduation. Though the code as implemented is careful to not delete anything from raw and it is unlikely a bug could cause the modification of anything under raw, there is no reason why it should be even capable of making any changes to raw side folders. On the Prescient aggregation server, the code is run on cron from my account without the use of sudo currently, so it runs less risk in that respect. 

For Pronet data flow though, the code is now run on cron from a root account. This has the benefit of avoiding some permissions issues that have primarily impacted Prescient project status, but the long term project goal should involve use of specialized processing accounts with tailored permissions. A major roadblock there is gaps in Lochness functionality, something that will come up again in the next section on suggested future code updates. Unfortunately, there has been an under-supply of software engineering time for AMPSCZ infrastructure in general, and this extends to a long wish list for Lochness updates too.

\noindent Some specific examples of permissions issues that have arisen when using my account to run the pipeline without sudo access, and that therefore required manual intervention to solve, include:
\begin{itemize}
    \item Under the GENERAL side of processed for a given subject ID, that ID did not have an "interviews" subfolder, which my account could not create due to lacking write permissions for \hl{GENERAL/[siteID]/processed/[subjectID]} -- which is in itself reasonable, but Lochness then needs to setup all of the expected datatype folders.
    \begin{itemize}
        \item Note that this is particularly likely to occur for the interview datatype, because there will never be raw interview-related files under GENERAL. But that does not mean that it does not need a processed folder under GENERAL for the datatype! In fact it definitely does here, as this is how we separate out the shareable products of the interview pipeline.  
        \item The same issue could theoretically apply to the PROTECTED side of processed as well, though it appears less common in practice.
    \end{itemize}
    \item Along the same lines, the [subjectID] folder for a particular subject with interviews under the PROTECTED side of raw does not always exist on the GENERAL side of processed, and sometimes there will not even be a processed subfolder for that ID on the PROTECTED side. My code will try to create such [subjectID] folders, but again my account without sudo often does not (and really should not) have the needed permissions to create folders directly under "processed".
    \item It has sometimes also happened at the beginning of a site's enrollment that a "processed" folder itself did not even exist yet for that site, seen at times under PROTECTED and at times under GENERAL sides. This is despite a raw interview folder being available under PROTECTED. Of course, the solution to this should be improved folder creation processes within Lochness -- and in this case my pipeline will not try to create the missing folders but will instead just exit the process for that site, logging an appropriate error message. 
    \item On the other end of the folder hierarchy, it is necessary that on both GENERAL and PROTECTED sides there are \hl{open} and \hl{psychs} (matching the interview types) subfolders under \hl{[siteID]/processed/[subjectID]/interviews}. It should be possible for the interview code running account to write outputs under these type-specific folders. My pipeline will create them if they do not already exist, and has not encountered many permissions issues at this level; but one might also consider building the interview type folders into the Lochness-generated structure instead.
    \item While my account being able to read raw interview folder contents has not really been a problem, there has been one recurring permissions issue with access to the raw data pulled onto the aggregation server by Lochness, particularly on the Prescient one. This has been the ability to change directories into a particular existing \hl{PROTECTED/[siteID]/raw/[subjectID]/interviews/[type]} folder. It appears to happen when Lochness assigns S permissions to the group's executable bit on one or more of the folders making up that folder path. It has occurred on a number of newer subjects' folders created by Lochness, and so this is an ongoing issue to watch out for (and in the meantime manually fix with sudo-enabled account as needed).
\end{itemize}
\noindent Ideally the outlined issues would one day be addressed by more robust folder structure setup functionality built into Lochness, so that processing code could focus on generating processed outputs and not core folder structure. This would in turn enable tighter permissions control on different AMPSCZ infrastructure accounts (though the unique lock on raw audio/video to not leave the aggregation server makes it largely irrelevant for other modalities in this project).  

As far as outputs generated by the account running my code, it is important that any on the GENERAL side of processed are accessible by the main account that runs the Lochness pushes to \emph{predict}. The final output updates to group assignment (pronet/prescient respectively for AMPSCZ) and file permissions (770 used here) are lines in the cron job script that runs daily (one for Pronet and one for Prescient). There is no clear reason that this would need to be changed any time soon, though it may still be worth hashing out the exact permissions that are needed for each input/intermediate/output once a final account for running the data flow and QC code has been decided on. Such decisions may additionally be dependent on the plan for central monitoring responsibilities going forward, particularly with regards to permissions on the pipeline products kept on the PROTECTED side of processed (which of course do not go to \emph{predict}). 

\subsection{A roadmap for next steps}
\label{subsubsec:interview-adapt}
In the future directions of chapter \ref{ch:2}, I outlined some broader next directions that included additional software that will eventually be needed for processing AMPSCZ interviews. In this section, I provide a much more focused list of specific updates that should be made to the interview data flow and quality monitoring pipeline over the course of the next few years, going into much greater detail for the reference of whoever may take over these responsibilities -- including information that is particularly likely to be relevant for troubleshooting ongoing and emerging bugs (subsection \ref{subsubsec:u24-next-steps-interview}). I then discuss changes that another future project may want to make to the present pipeline to suit different needs (subsection \ref{subsubsec:u24-dif-project}), an already relevant topic due to consideration of my tool for use in a Wellcome Leap project currently being planned.

\subsubsection{Building on the pipeline for ongoing AMPSCZ interview collection}
\label{subsubsec:u24-next-steps-interview}
Potential issues and future code improvements to be aware of for management of the AMPSCZ project (along with tips on monitoring existing email alerts and other troubleshooting tools) include:
\begin{itemize}
    \item May need to add sites to the configs available on the existing servers (e.g. Santiago at Prescient) or possibly eventually set up a third central server. The process for setting up the current sites on the current servers is documented in this section above and should be referenced in this process. 
    \begin{itemize}
        \item Note that when sites that are already set up with config files newly begin their interview collection process, it is still necessary to confirm that the correct top level folders exist and that permissions are okay, as the code may not start working for that site automatically otherwise.
        \item Even for existing sites on Prescient, there may be a recurring issue with S permissions on folders created by Lochness preventing the pipeline from running on some newly enrolled subjects.
    \end{itemize} 
    \item It is necessary to periodically clean up incorrect site uploads on the raw side of PHOENIX manually, once the site has removed or fixed the offending files in Box/Mediaflux. For site mistakes that silently propagate through the pipeline's outputs (e.g. an interview uploaded under the wrong type), fixes need to occur manually on the processed end at this time, ensuring that all relevant intermediates are correctly updated on both the GENERAL and PROTECTED sides of PHOENIX. This includes not just moving or renaming outputs that map to the specific interview, but also editing all affected CSVs and other accounting materials. 
    \begin{itemize}
        \item A number of early issues that have been documented have not yet been manually fixed in this way, and new ones continue to be periodically detected by the existing monitoring tools. Ideally updates to the code would introduce quality of life tools for semi-automation of such fixes to previous pipeline outputs, though it is of course important that running this sort of tool is done with caution so as to not introduce new mistakes. On the raw side, manual intervention will remain necessary, as automated deletion of raw interviews is highly discouraged. 
        \item A number of site mistakes that can cause issues with processed intermediates (or can cause processing to be skipped over entirely) are listed as part of the common problems provided in the main chapter (section \ref{subsubsec:u24-issues}). 
        \item Note it is also not currently possible to explicitly and automatically detect when a prior interview upload issue was fixed. In the general case, this is a challenging problem that is unlikely to be fully automated. But there are certainly quality of life improvements that could be made to the code and the way it interfaces with the monitoring process, to work towards addressing this. 
    \end{itemize}
    \item A major problem with the code currently that should be dealt with in the near term (particularly after psychs transcriptions settings updates are done to limit the number transcribed) is to deal with periodic crashing issues that cause some interviews to not receive audio QC and thus not be sent to TranscribeMe. The problem seems to be generally caused by interviews that are extremely long, and so it is in part a problem with server reasons, and also has only really impacted psychs interviews. Still, it is important that the code be updated to directly handle such long interviews differently. Furthermore, it is possible for crashing to occur for other reasons, though much less frequent, and there is currently not a good way to make the pipeline go back and redo these rejections without having to manually delete the intermediate files that indicated audio rejection. Additional notes to potentially help troubleshoot the issue follow:
    \begin{itemize}
        \item Note that for this random crashing, audio and video seems to fail somewhat independently, and audio crashing is notably more common. This also primarily occurs on Pronet, which makes sense because the data aggregation server for Pronet is weaker and Pronet ha more sites submitting problematic and/or absurdly long recordings.
        \item For audio, if kernel dies during QC attempt (which seems directly related to very long duration) then the WAV file will be left under rejected and the monitoring email will say there was a file naming issue because it couldn't locate the QC record in the DPDash CSV.
        \item For video, the pipeline seems to be failing at ffmpeg frame extraction stage occasionally, which means it keeps getting reattempted every time, but apparently will systematically fail on the affected interviews for months on end. This is why the monitorig emails will say a new video was detected but list none for a handful of sites. I am still not sure what exactly causes this.
        \item It is also possible for audio to fail at the ffmpeg conversion stage repeatedly, but since it is never successfully converted this will just be an ongoing silent failure.
        \item When audio QC crashes for a non-kernel reason it will catch it and delete the evidence that the file was attempted so that it will try again next time, and it will email about this every day until fixed. This never occurred on Pronet to date, but it had occurred for two Prescient interviews for some reason. It happened for a few straight days before the code crashed in a way that instead caused those two files to be rejected.
        \item If the audio QC says the file was rejected for unknown reason (check permissions), then that means that there was some other failure point after the file was converted to WAV besides its ability to be looked up in the DPDash CSV. In the past it often came up when folder permissions were incorrect, hence the added parenthetical, but this is rare to encounter now.
        \item For now, the way to trigger a redo of audio processing for audios rejected due to kernel death is to go in and delete a handful of files on the PROTECTED side of processed corresponding to that particular interview: the file name map text file, the sliding window QC CSV, and the audio WAV under the rejected subfolder. Once a solution to the longer interview issue has been implemented, this will need to be done for all the older rejected files. A number of these can be identified by simply looking in the current DPDash CSV for those records with video QC available but not audio QC. 
    \end{itemize}
    \item As far as storage space is concerned, it is important that processed WAV files are not deleted while they are still in the process of awaiting transcripts. However, they can be deleted if needed once they make it all the way through the pipeline and get placed into the completed subfolder. They are kept as of now to facilitate easier implementation of a feature extraction pipeline, as they are already renamed and converted to the necessary format for OpenSMILE. But because the primary focus of most acoustic feature extraction will be the diarized audios not included at present, it may be worth just clearing the completed WAVs out from time to time. The file name map text files can be used instead to handle mapping new feature extraction outputs to the naming conventions of the pipeline. 
    \begin{itemize}
        \item On a related note, implementing some QC to run on each of the diarized audios when available would be another valuable addition to the current pipeline. 
    \end{itemize}
    \item The email alerts sent by the pipeline now could use some cleaning up to better serve the monitoring workflow as it has developed over the course of the project. Particularly once a final manual monitoring procedure is decided upon and responsibilities are more specifically allocated, the emails could be very helpful if directed to the right individuals at the right frequency, but this will require some adjustment. In the interim, some points to be aware of when reviewing the daily monitoring alerts are as follows:  
    \begin{itemize}
        \item A warning will say that video and audio days don't align when one of the file types is missing or failed processing, so this is rarely an actual metadata issue.
        \item More often than not the warning about consent date being changed is not because the consent date was actually changed, but because some bug in the code occurred that disrupted the file accounting. Those records do not have a consent data nor a raw audio file path listed in the standard accounting table, but I am not yet sure what causes this, so it is a pipeline bug that needs to be followed up on. In the meantime, it has also happened that a consent date was changed, so it is necessary to look at the dates in the accounting table when this warning occurs and determine whether it is a real problem or not. When consent dates do change files will become misaligned, and it is not clear how this should be handled systematically, so it is important sites minimize this sort of mistake. 
        \item When an SOP violation is listed in the warning email and it is not obvious from the folder/file names supplied, it is currently necessary to look either at the more detailed SOP violation CSV or the raw folder itself to identify the exact problem. It could be that the single folder contained multiple mono audio or video files (split interview) or it could be that the folder was missing one or both of those files (or had them renamed or changed the file format in a way that prevented the pipeline from recognizing it). 
        \item In the context of this project, it is generally not important if there is a warning about non-ASCII encoding. However if a warning about non-UTF8 encoding comes up (which it hasn't yet), that will entirely prevent transcript processing and will require manual follow-up. 
        \item When the warning email notes that an audio file was rejected, it is necessary to check that day's audio process email update to determine the reason for rejection. It may be a true QC rejection or it could relate to code crashing issues for long recordings mentioned above. 
        \item Note that SOP violations in how speaker specific files are named or even in the number of speaker specific files present is not a part of the current email alert system. The total number of speaker specific files is logged in the file accounting CSV if it needs to be checked on, but as long as at least 1 m4a is found under 'Audio Record' there will be no SOP violation logged (and subsequently no warning).
        \begin{itemize}
            \item The naming of the speaker specific m4a files is not at all taken into account by the present pipeline. As the feature extraction pipeline becomes increasingly necessary to implement, problems with naming of these diarized audios looms large, as discussed in the main text of chapter \ref{ch:2}. Thus aside from the infrastructure work required to deal with speaker assignment in the extraction pipeline, there is also room for improvement in what the QC pipeline's monitoring can detect. Review of the speaker specific m4a names could help to catch issues more quickly and ensure longer term data quality by intervening quickly where needed. 
        \end{itemize}
    \end{itemize}
    \item DPDash does not accommodate sorting of columns within the interface, and the current combined upload is alphabetized by subject ID, only providing chronological order within a subject. This makes DPDash only really helpful for checking on specific files rather than for quickly looking for new issues. It would therefore be a big quality of life monitoring improvement to implement overall chronological ordering for the DPDash views. However this will require pulling in additional information to what is currently output for the GENERAL side, as that removes all external data references. It is also not yet entirely clear to what extent this is allowed, but there are other DPDash views containing real dates at present. 
    \item For legal reasons, it will be necessary to update the initial stages of the pipeline to accommodate Cisco WebEx for Germany. To do so, we will need to first obtain example inputs for reference, as well as confirm what exactly is expected of the corresponding outputs. There will not be diarized audios, but aside from that and the obvious differences in file name and formatting conventions, it is not yet clear what other differences might need to be dealt with. Along these lines, it may be important to start tracking the source of individual interviews once other services will be supported. 
    \begin{itemize}
        \item Relatedly, there may be German sites that cannot have video uploaded to Lochness server at all, so any feature extraction on video if it were to occur would need to be set up to run locally. That is a whole separate problem, but in the interim the audio could be uploaded alone. Unclear if there needs to be a careful check for this first though as part of restrictions or if they would just trust their RAs to not put the video part in Mediaflux. 
        \item It is also worth noting along these lines that if Zoom has any software updates in the next few years that will change its file recording conventions, the protocol and/or preprocessing code may need to be revisited and adapted accordingly. It will be important to update the code to handle the new case without disrupting the old format, something that already happened to Zoom naming conventions at one time early in the project's development. 
    \end{itemize}
    \item As more international sites onboard, it will be important as well to be more on the lookout for problems in the code caused by foreign languages that have not yet been tested properly. There will likely need to be updates to the QC code to handle a greater variety in TranscribeMe notation, even if the code doesn't fail in a highly visible way on these other languages. 
    \begin{itemize}
        \item Still waiting on TranscribeMe to answer about how their notation might change. We know redactions will be the same across languages, but for everything else including marking inaudible words and uncertainty, it is not clear. It is also not clear (especially for languages that do not use an English-like alphabet) whether verbatim notation related to disfluencies (e.g. stutters) would be the same. This is less immediately relevant to the pipeline, but it is the type of consideration that will need to be made as the project expands. 
        \item As far as potential Zoom naming convention issues in other countries, the date/time metadata part should probably be fine, but do need to check about 'Audio Record' subfolder, word audio in front of the main file name, and other related raw file convention assumptions. 
        \item Note that some sites (thus far Montreal) might use English in some interviews and a foreign language in others. In this case the pipeline labels the upload with the foreign language, and then TranscribeMe redirects to the right transcriber as needed. For downstream file organization and eventual analyses this will remain important to keep in mind.  
    \end{itemize}
    \item For improving the pipeline's tracking of issues, integration with related information pulled from REDCap/RPMS would be extremely helpful. We do not currently have a good estimate for how many interviews we should actually be expecting for a particular subject at any given time, and we also cannot make entirely accurate assignments to study timepoints (e.g. baseline interview) based on day number alone. This means the monitoring process is likely missing some issues at present, both in how sites are filling out interview metadata information in the research record keeping software and in how certain interviews might be missing for various reasons without our knowledge.
    \begin{itemize}
        \item At this time, information is available in raw form on Lochness, but it is not at all cleaned. Parts of the process for distilling these records might be a responsibility for other groups within AMPSCZ, but identifying various site issues and ultimately integrating with the interview QC pipeline will require someone that is working directly on interviews to be involved. 
    \end{itemize}
\end{itemize}
\noindent There are also of course a number of action items for the mentioned downstream feature extraction code, some of which integrate directly with this QC pipeline. Similarly, potential future changes to e.g. Lochness will impact what role the current interview pipeline plays. As such, I will now comment in more detail on a handful of topics relating to the future use of my pipeline as it fits within other AMPSCZ processes (whether active or planned). Note that further discussion of work required for Zoom alternatives can be found instead in the security/privacy-focused section \ref{subsubsec:interview-code-sec} above. \\

\paragraph{Improving DPDash integration.}
As mentioned, there are some quality of life improvements that could be made to the current DPDash views, primarily by way of integrating existing AVL DPDash CSVs with CSVs reflecting key interview info pulled from REDCap/RPMS (to be discussed more next). When new features are added to a DPDash CSV, it is important that the configuration files on the web platform are appropriately updated to display those new metrics. Iterative DPDash configuration updates are also recommended for fine tuning which metrics are worth displaying most prominently (or at all), as well as the color scheme/bounds for each. Updating configurations on the web platform does not require direct software knowledge, just reference to tool-specific documentation and background on the columns found in the DPDash CSVs produced by my pipeline. Those column names are made clear in the above audio, transcript, and video QC implementation sections. Distributions of many of the feature values can be found in the weekly histogram PDFs, and otherwise can be checked more directly in the weekly combined-QC.csv spreadsheet email attachment. One might also get a feel for the values to date directly via AMPSCZ's DPDash web portal.

Recall that DPDash \citep{dpdash} has been an open source visualization tool (used by the Baker Lab) for monitoring digital psychiatry data for quite some time, but has been more recently adapted to the needs of AMPSCZ by \cite{dpdash-amp}. It is therefore recommended to refer to the documentation at \citep{dpdash-amp} and contact those individuals if needed when working with DPDash for the AMPSCZ project. While the current view should continue to automatically update without much issue, there are potential additions to be made for entirely new DPDash and/or DPChart views related to interviews (or the upcoming audio diaries). When entirely new views are to be added, it is necessary to follow the instructions for importing new CSVs (once generated) to DPDash via the command line and setting up brand new configurations to define the corresponding webpage. When these details are ironed out, the regular import process should be transferred to a centralized account, so this will require interfacing directly with the DPACC team regardless. \\

\paragraph{Processing REDCap/RPMS interview notes.}
Sites log a variety of information about subjects via REDCap (for Pronet) and RPMS (for Prescient) interfaces. Forms relating to the conduct of each interview session are included as part of this battery. However, although some forms are already being used in project monitoring, none of the interview-related information is yet being processed or even systematically referenced by central monitors -- so obviously the information is not yet being integrated with outputs from the interview dataflow/QC pipeline. 

Initial processing of these data likely falls under responsibilities of a different role than mine. Still, once done the cleaned form outputs will need to be integrated with the existing interview AV code, which opens up a number of possible action items for my replacement. Additionally, the lack of this integration is hindering monitoring of interview processing, so if someone with time to properly dedicate to the project were to actually replace me, they could pick up this task and ultimately greatly improve our current monitoring workflow (both of incoming data and of possible software bugs). 

The raw JSON files exported from REDCap and the raw CSV files exported from RPMS are currently pulled to the Pronet and Prescient data aggregation servers respectively, where they are then passed along to the PROTECTED side of raw on \emph{predict} (under the surveys datatype). Processing of these raw form exports should thus likely occur on \emph{predict}, not the data aggregation servers - in contrast with everything I've done to date. Therefore the merge with available AVL QC data and any subsequent monitoring warnings would also occur on \emph{predict}, and so this should probably be a separate downstream code repository entirely. Unsurprisingly, the exported data at present are unprocessed with ugly default formatting; the initial undertaking here would be to clean up the AV-related form exports from both services and get this running regularly on \emph{predict} along with basic integration with the interview pipeline outputs that are available on \emph{predict}. 

From there, it would be necessary to better understand the information that is found in the interview forms. Not just the questions asked and the possible answer options, but also the trends found in site responses. Given the many mistakes in the interview recording and upload process encountered to date, it is highly likely that there will be mistakes in site responses. The types of mistakes made, how frequently they are made, and by which sites will all need to be identified, as well as how any mistakes relate back to raw interview data (i.e. which form discrepancies are a form-filling mistake versus an interview recording mistake?). If it is anything like the raw interview data, a very wide variety of failure points will be found, and some may need to be protected against in software, while others will require interfacing with site contacts to hopefully improve behavior. As more is learned about the datatype, it may require further changes to how AVL QC is integrated, what sort of monitoring alerts are sent out, and so forth. 

Because of all these challenges, it is important that some legitimate software development time is dedicated to the issue; this is not something that can be thrown together if it is to be used effectively. Of course, the reason it is important that REDCap/RPMS integration for interviews gets implemented is precisely because of the tracking difficulties caused by the many heterogeneous site protocol violations observed with the raw interview data. At present we do not know which interview timepoints sites claim to have conducted, so we can not systematically identify missing sessions -- which can fall through the cracks at many different steps when SOP is not followed, as detailed in main chapter section \ref{subsubsec:u24-issues}. 

In part as a result of this, I still do not even have an answer about what acceptable timeframes for conducting different interview sessions are. Obviously the 2 month follow-up open interview need not happen 2 months later on the dot, but is 4 months later too long? Is one month later too short? Beyond that, currently I don't know for sure whether an interview submitted 60 days after consent is intended to be the baseline open protocol or is a 2 month follow-up with a missing baseline (digging into a few specific cases, both have been observed). So simply having site-logged timepoint information would be greatly helpful to determining what truly is missing and isolating any presently ignored software bugs as well as protocol (following) issues.

Intended session number and corresponding date (incidentally even a brief manual spot check on Pronet identified some date recording discrepancies that are likely errors on the REDCap end) are not the only useful features found within an interview reporting form though. Another major piece of information is the identity of the interviewer. This could help us identify specific individuals that may be most responsible for their site's common errors, which would allow for more targeted training intervention that would ultimately be more likely to actually change behavior. More importantly for the integrity of the present data, we ought to be verifying interviewer identity is properly recorded so that subject and interviewer can be easily identified from the diarized audios produced by Zoom. \\

\paragraph{Continual tracking of SOP violations.}
One wish list item that would be much better facilitated if reliable REDCap/RPMS information were available is the continual tracking of SOP violations. At present, the pipeline will log newly detected SOP issues, and it will separately log newly uploaded interviews that were successfully processed, but it does not track updates on previously problematic timepoints. It would be infeasible to do so in general with the information currently available, so when a fixed version of an interview is uploaded by a site, it will not be linked to the previous failed version in any way automatically. 

Tracking of SOP violations is handled semi-manually with a spreadsheet monitored by multiple groups, but ideally some of this burden could be lifted if interview form information were available, as there would be a clearer idea of which interviews are actively missing at a given time. This would also facilitate further communication with sites, because it is already abundantly clear when a site keeps repeating the same mistake, but we have less insight into which sites have addressed their old mistakes and which have not. The more time passes after an interview was conducted the more likely other mistakes could lead to losing that recording entirely if it has not yet been successfully uploaded to data aggregation. I've included next some notes I sent about monitoring to a primary central site contact about how we might track protocol issues over time without access to the interview REDCap/RPMS forms. However it should be clear from this that integration with study notes would be much preferable as a sustainable strategy. \\

\noindent As far as long term tracking, there are SOP violation and processed warning CSVs that do keep track of all issues that have been notified, and a processed accounting CSV keeping track of things like the date when a file was first registered and the consent date at the time. These CSVs are included in the automated weekly accounting email, which you should get alongside the one with the QC stats. Though the CSVs are kind of confusing the way they are currently formatted, and they include a number of warnings that were sent to me but I filtered from the weekly update because upon further inspection they seemed unimportant (or things like 0 redactions which we've decided to stop checking). 

So technically a history of noticed violations can be reviewed and filtered by e.g. site or date of detection. However this might not be what you are asking for? It is not currently possible to check whether an interview that had a previous upload issue was later fixed, only whether the site continued to have issues in other uploads. I don't think this would be systematically possible, because there is no systematic way to know what interview a site even intended to upload when they remove date metadata or put things in the wrong folder or whatever. Lochness does not modify files that were already pulled, if you delete something in Box/Mediaflux it stays on the server, and if you change a name in Box/Mediaflux it will upload the file as a completely new upload so that both the wrong name and the right name will exist in data aggregation.

A specific problem like missing diarized audios would be feasible to periodically recheck in isolation, and then update if marked interviews have been fixed, but the way it is implemented now it only checks for those audios when a valid upload is first recognized. If there are specific types of problems that it would be good to have updates on maybe we could discuss further how to do that?

For interviews that weren't processed at all or had a QC modality missing, if you know what subject ID and interview type (+ ballpark day number) is expected from previously incorrect files, they could be looked for in DPDash. If the interview is now correct, QC values would be available for that time point. Another thing I can do is send you info from the processing update emails, which for each site will update on new files that successfully had QC complete that day, if there were any. Alternatively, the processed accounting CSV I mentioned could be sorted by the columns that record when audio QC, video QC, and transcript QC were completed, and you could see what subject IDs/interview types/day numbers come up at the top to determine if that matches the expectations based on issues sites were supposed to fix. \\

\paragraph{Updating transcription decision points.}
It will hopefully be the case that in the near future some scheme will be adopted to limit the number of psychs interviews sent for transcription (see Appendix \ref{cha:append-ampscz-rant} for more on the salient funding-related considerations). Likely AMPSCZ still wants baseline psychs interviews transcribed, and given that we supposedly want this, it seems silly to only get the first 30 minutes of those - they're also particularly common to be very long duration. 

So there are multiple good reasons to limit the number of other psychs interviews that get uploaded for the highest quality professional TranscribeMe service. To do this, one could consider a protocol where a subset of the 8 other clinical timepoints are left out consistently for all subjects, where a consistent fraction of the other timepoints are randomly selected for each subject separately, or where a fraction of subjects are randomly selected to get all of their timepoints transcribed with the rest receiving only baseline. Whatever the organizing committee decides on though, this will need to be implemented within the current code. As an aside, note it would be possible for the pipeline to set aside such "rejected" psychs interviews in a different folder, which would make it very easy for additional uploads to occur in the future if new evidence of scientific uses for these particular transcripts comes to light. To implement a basic version of psychs timepoint filtering in the current framework, please see the interview\_audio\_send\_prep.py file for a note on where such a statement can be added and what other parts of the code would require subsequent update.

To actually implement such a protocol with maximal correctness, it would be necessary to know what timepoint each site is marking a given interview upload as. Thus we are back to the issue of REDCap/RPMS integration. Furthermore, even if such a tool were available as described in an above paragraph, the information compiled on \emph{predict} would not be available by default to the data aggregation server (where the TranscribeMe upload needs to occur from), as information generally flows off of those servers to \emph{predict}, not the other way around. One could instead try to implement the form-processing code on the data aggregation server, but given the typical protocol for that data flow, it is likely that not all forms are maintained in perpetuity on Pronet/Prescient aggregation servers, as raw data that can make it to \emph{predict} are intended to have that be their final storage place. Because the forms are not especially large files, this could feasibly be changed, but it would require additional coordination with those at DPACC working on Lochness. Alternatively, they may recommend keeping the form-processing code on \emph{predict} as I originally proposed, and then adding a utility that gives the data aggregation server access to the combined summary of interview form responses to be regularly generated on \emph{predict}.

Once these issues with assessing timepoint have been sorted out, other pipeline updates will still need to be made to account for the new case, across steps. Logging of e.g. whether an interview was rejected or not will need to add an entirely separate case (that probably should not be coded as a simple rejection sub-type) to denote psychs interviews that were held back for efficiency/budgetary reasons. That will affect monitoring infrastructure all the way from daily site-directed email alerts up to weekly server-wide summary tables, and thus will require careful checking/testing of many steps of the pipeline (really all of them outside of the earliest audio processing steps and the video branch). Additionally, it will remain pertinent to keep an eye out for a myriad of site mistakes that could affect the transcription decision making process described, especially early on. \\

\paragraph{Better synergizing with Lochness.}
In addition to the email alerts created by my pipeline, the Lochness infrastructure for pulling the data to the aggregation servers will send out email alerts for certain SOP violations/upload mistakes, in particular those mistakes that prevent upload to the aggregation server at all. Such mistakes would of course be impossible for my pipeline to catch, so at present we require someone who is monitoring interview progress to also be interfacing with Lochness monitoring. 

Even if a REDCap/RPMS integration tool were available, this would still likely be required (albeit as a lesser time commitment), because some of the caught site mistakes involve incorrectly registered subject IDs -- meaning those IDs will not have any forms pulled to central servers either, until the registration mistakes are fixed. However other mistakes, for example incorrectly naming the subject ID folder in Box/Mediaflux only, would be quickly noticeable as missing interviews when cross checking with form data. In the interim, the Lochness email alerts are a good resource for catching all such problems. 

As mentioned, the core issue it will check is that each subject ID folder in Box/Mediaflux meets all expectations. This folder should be in the form \hl{XX\#\#\#\#\#}, where XX is the two digit site ID and \#\#\#\#\# is the five digit within-site subject ID. Besides meeting basic formatting though, Lochness will also verify that the particular subject ID is actually registered for the project, and that it has a realistic consent data correctly entered into the study metadata (pulled from REDCap/RPMS). 

Additionally, Lochness will prevent upload and flag in email alert any file formats that are not expected to appear within the folder structure of the interview datatype. This was not always the case: hence the "double click to convert" .zoom files mentioned in main chapter section \ref{subsubsec:u24-storage}, which are created when a site does not wait for a Zoom recording to properly save before trying to upload it. But it has now been fixed to prevent upload of some of the most egregiously useless files that were accidentally being included in interview uploads. That is largely a nice positive, as it prevents clutter and storage waste on the server and reduces possible issues for the AV pipeline to have to consider. However it does make it more important to be monitoring Lochness alerts now, as there may be some files that sites tried to upload under a valid subject ID and interview type, but still got excluded from the data aggregation server entirely. 

In the future, it would be ideal for Lochness to do further filtering of what gets uploaded, for example preventing upload of loose M4A and MP4 files so that only compiled Zoom interview folders (and standalone recorder WAVs) actually get uploaded. It would be especially useful if Lochness prevented the upload of any interview folders that did not begin with the Zoom naming convention \hl{YYYY-MM-DD HH.MM.SS } (and similarly for EVISTR WAV file convention), because this would help filter out a large number of incorrectly named folders that end up rotting on the server long term, due to how Lochness handles source name changes. Having more careful filters would facilitate more efficient data flow for the eventual AV feature extraction operations as well. 

This is something that a replacement for AV software development could perhaps help with, as development resources are stretched thin across the board right now -- a problem that would only worsen if there is no replacement hired for me! The long term efficiency of the project would be much improved if more of the very basic upload issues were filtered by Lochness, and the interview dataflow/QC infrastructure could focus on tracking issues that it is better suited to with the info it has available (e.g. do we have all the expected diarized audio files, what is the recording quality like, etc.). Lochness on the other hand has direct visibility into when, for example, a Box/Mediaflux folder has been renamed with contents otherwise the same.

Recall as well that Lochness handles the pull of transcripts returned by sites after manual redaction review back to the data aggregation server for further processing by my pipeline. In this case, Lochness will look for .txt files that are found directly under a subject ID subfolder in the site's "Approved" Box/Mediaflux folder. That does filter out the most common site mistakes in trying to return the transcripts, but it does remain possible for a transcript text file to be pulled back by Lochness and not recognized by the pipeline, as my code expects the filename to be the same as when it was issued to the site, but Lochness will pull text files with any names. Fortunately, it is easy to track potential issues on this front, because we have an ongoing curated list of transcripts that are awaiting manual redaction review sent out every day by the pipeline. \\

\paragraph{Careful data cleaning on the aggregation servers.}
Because Lochness does not delete files deleted from source nor rename files renamed on source (instead uploading a copy under the new name), the myriad of site interview SOP violations have lead to a number of incorrect interview recording uploads indefinitely taking up space on the data aggregation server -- this was discussed at length in main chapter section \ref{subsubsec:u24-storage}. Improvements to Lochness's upload filtration process discussed in the preceding paragraph would help to mitigate the issue going forward, but all existing mistake uploads (and any additional ones in the interim) ought to be periodically cleaned out from raw interview folders on the PROTECTED side of PHOENIX manually. 

Interviews are a special case precisely because they cannot go directly to \emph{predict}, and so it is very important that any manual deletion process takes great care, and automated methods for this task should probably be avoided. Similarly, correct raw interview uploads should likely be kept on the respective Pronet and Prescient data aggregation servers for the lifetime of the AMPSCZ project. Perhaps eventually old uploads could be backed up on cheaper (slower to access) storage, but for the foreseeable future it is necessary that access to all raw interview recordings remains easy for the planned feature extraction processes -- the specifics of which continue to be actively changed/updated.

Related to the raw interview data is cleaning of processed interview outputs on the data aggregation servers. There is less of a data cleanliness issue there, but still possible storage optimization considerations. On the flip side, there are some files under processed folders that are very important \textbf{not} to delete for the duration of the AMPSCZ project. Thus the main purpose of this section is to emphasize files that should certainly be kept, and reiterate the largest files that could safely be deleted if freeing up space becomes necessary.

On the GENERAL side processed folders, stored interview pipeline outputs are very small and so deletion would provide minimal storage benefit. For most datatypes, everything on the GENERAL side can be passed to \emph{predict} and then safely deleted from the data aggregation server, because most processing will occur on \emph{predict}. This is obviously not the case with interview data. The interview-level summary stat tracking CSVs are especially important not to be deleted, as the code expects them to always contain a fully up to date account of the processed interviews thus far. Deleting them at any time could cause unexpected pipeline behaviors, and would almost certainly erase data from DPDash web views downstream. Regenerating them in some cases would be a bit of a pain in the ass too, so this is seriously \textbf{not} recommended. While redacted transcripts - the other major interview code output on the GENERAL side - could always be regenerated from their unredacted counterparts, this is not something the pipeline currently expects, and so it could confuse the accounting logs. 

Simply put, there is no reason for any interview outputs on the GENERAL side to ever be deleted, with only possible downsides - so it should not happen. An exception (relative to other modalities) has already been built into Lochness to avoid deleting any processed outputs under the interview folders on the GENERAL side, and this should be kept in place permanently. Future projects that decide to use Lochness and may have similar privacy restrictions specific to AV flow (or decide to use the current interview QC pipeline for any other reason) should be mindful that the default behavior for AMPSCZ Lochness is to delete GENERAL side processed outputs as soon as they are successfully moved to \emph{predict}, and an exception for the interview data flow to work correctly was later added.

On the PROTECTED side processed folders, Lochness does not delete data automatically, as of course these outputs will never be moved off of the data aggregation server onto \emph{predict}. As such, any deletion from these processed folders should be carried out with clear intentions and great caution. For interview outputs specifically, it is important that unredacted transcripts do not get deleted, as TranscribeMe will remove them from their server after a time, and so they may be unrecoverable if removed on the respective data aggregation servers. Note that these transcripts are indeed considered a processed output under each interview datatype. 

Transcripts of course do not take up much space anyway, and along the same lines there are some file accounting CSVs and text files (including the per interview audio/video file name map text files and the sliding audio QC CSVs) that would cause serious issues in pipeline functionality if deleted partway through processing of an ongoing study. This set of files additionally includes some subject ID/interview type level CSVs containing row-wise per interview summary info on processing logistics and early pipeline stage SOP violations. None of these processed files should therefore ever be deleted from the data aggregation servers during the course of the AMPSCZ project. The reason they stay on the PROTECTED side to begin with (and so are not even backed up to \emph{predict}) is that they have the potential to contain PII, such as real dates or file names that may contain real participant names. However the per interview summary info is emailed to a small group of individuals with the appropriate data access level, to assist in monitoring the low level organization-related issues. 

The other major type of intermediate interview recording output found on the PROTECTED side of processed on the AMPSCZ data aggregation servers are converted versions of the raw data. For video this is simply a single frame sampled once from every 4 minutes of the interview recording, so even for very long interviews it corresponds to a small number of photos. As the extracted frame images make it much easier to quickly spot check anything that might seem weird about video QC output, I would not recommend deleting these through the duration of the project, though removing the image files alone should not have any pipeline-interfering impact. 

More important are the WAV files, most often converted from the Zoom top-level mono audio M4As, though also sometimes copied directly from uploaded EVISTR output. They are renamed to reflect the intended processed naming convention, and they are directly ready for analysis by tools that are incompatible with many other audio formats, such as OpenSMILE. That was my original intention in retaining them, but ultimately for any Zoom audio we will want to focus most acoustic processing on the diarized audios instead. Because WAV files can take 10x more storage space than their M4A counterparts, this adds a nonnegligible additional storage burden that could mostly be avoided (though raw videos are even larger, they can't exactly be deleted safely). As such, one might consider deleting all WAV files that have made their way to the complete\_audio subfolder on the PROTECTED side of processed for interview datatypes; there is currently no real plan for further using those specific files.

It is critical to underscore though that \textbf{not} all WAV files under processed can be safely deleted at any time. Files that were put under the rejected\_audio folder should probably at least be manually inspected before deleting (and are a small fraction of the total processed interview mono audios). Files under to\_send either are actively in the process of being uploaded to TranscribeMe, or had SFTP fail for some reason and should be reattempted shortly (and perhaps code logs should be checked on). Files under pending\_audio are especially important not to delete, because this is at present how the pipeline tracks which audios are awaiting a TranscribeMe transcription, and accordingly pulls them back when completed. If pending WAVs are deleted the matching transcript will never be pulled onto the data aggregation server and there will also never be an explicit log indicating that a transcript is missing. These are of course temporary storage only, and TranscribeMe typically returns transcriptions quickly at that. 

Given all of that, safely updating many of the processed (whether final or intermediate, PROTECTED or GENERAL) interview pipeline outputs when a site or code mistake has propagated through the whole pipeline for a particular record can unsurprisingly be a can of worms, and will depend on the specific issue. While a nice code expansion would be to facilitate safe semi-automatic updating of existing records for various common errors, it is not the highest priority at this time. Likely for the duration of AMPSCZ such problems will need to be handled fully manually on a case by case basis and in a careful manner, with reference to the importance of maintaining many of these records that was outlined here. \\

\paragraph{Facilitating data management for future feature extraction pipelines.}
Beyond the currently active pipeline interactions (or desired interactions), there is also the major feature extraction pipeline project that needs to be developed and will likely need to utilize some of this pipeline's outputs for managing its data flow efficiently. As detailed, it will be important to extract more acoustic and facial feature information to share with the NIH data repository and even facilitate internal AMPSCZ analyses, because statistical status-aware analysis is not to occur on the data aggregation briefcase, and raw audio and video cannot leave that briefcase. 

Of course, the data aggregation servers do not have the compute resources needed to run more intensive feature extraction processes, even the relatively modest functions implemented for the Baker Lab audio diary feature extraction pipeline of chapter \ref{ch:1}. Additionally, it is important that feature extraction would not hog resources from the daily data flow and quality control operations that need to occur on those servers. For these reasons, an AV processing server has been set up for Pronet, with the data aggregation server's main briefcase mounted on it. An analogous system will need to be replicated for Prescient, and though they are reasonably waiting for more concrete functionality on Pronet before investing into this, there could be a number of additional troubleshooting issues or further logistical delays involved in fully rolling out any such feature extraction pipeline because of that. 

Still, here I will focus on the many hurdles piloting interview AV feature extraction systems specifically on Pronet, as this is the major task at hand. Further, I will focus primarily on concrete software-related hurdles, especially those that connect to the use of my pipeline. There are many scientific and bureaucratic hurdles in generating the feature extraction pipeline code too, but despite how shocking the behavior of the supposedly expert speech sampling advisory committee has been in this regard, it is quite frankly not the problem of my direct replacement. 

On the Pronet AV processing server, it is much slower to load the raw interview recording data directly from the mounted briefcase than it would be directly on the data aggregation server; because more people are involved with designing feature extraction code there are greater privacy and data safety concerns when it comes to account permissions as well. For these reasons, the plan for a feature extraction pipeline is to have infrastructure to manage the identification of interviews that still need to be processed (and that are processable e.g. not mislabeled), to maintain a reasonable rate of copying these interviews to the native storage on the processing server where feature extraction accounts can work with them, to ensure that generated outputs are correctly named and transferred back to the appropriate locations on the data aggregation briefcase structure, and to delete the raw files that have completed the intended feature extraction. 

It is therefore obviously useful to draw from existing interview data flow/QC outputs generated by my pipeline. Without doing so, it would be necessary for the feature extraction pipeline to reinvent the wheel with all of the SOP violation tracking and output renaming. It could also lead to some misalignment between outputs generated by the different pipelines, because distinct bugs might be present in each, and more pressingly the various site violations we have observed might be handled differently by each. Thus the best course of action will be to refer directly to the list of interviews that have successfully produced interview QC outputs (per DPDash CSVs) as those that should be copied to the local AV server storage for feature extraction, cross-referencing with the file renaming accounting produced by my pipeline (per file name mappings) to find the correctly corresponding raw paths and subsequently ensure that outputs are named accordingly before their copy back to the data aggregation structure.

On the topic of keeping a reasonable rate of data flow, particularly as the large backlog of interviews to be processed for features gets addressed initially, there will likely be some amount of manual monitoring required, and exact implementation details for marking completed interviews and how to prioritize the not yet transferred ones will require more discussion with the other involved parties. Pronet IT can be very helpful in assisting with some of this data flow code and ensuring that metrics like server resource usage are properly tracked, and will be an important initial resource for this process. The other main reference point will be any of the researchers that are implementing feature extraction functions for AMPSCZ raw audio or video interview recordings. Building the feature extraction pipeline will obviously require composing their various functions, or if they will be the accounts to run their own functions (not the most sustainable long term arrangement) then the data management infrastructure will regardless need to know exactly what outputs to expect for transfer back and for marking the raw file copies for deletion. This remains an ongoing discussion and will probably require some amount of iteration over time. 

Unfortunately, besides interfacing with various different groups and carefully implementing (and then testing) the functionalities needed for utilizing the interview QC/accounting outputs and ultimately handling the downstream data transfers, there are additional complications that will need to be addressed for managing outputs of feature extraction functions for AMPSCZ interviews. Many of the described site protocol violations (and maybe many not yet identified) could pop up in different forms when it comes to impact on feature extraction, and for audio in particular there are new considerations related to the diarized Zoom audio files, files that are largely ignored by the data flow/QC pipeline. 

The largest concern is identification of interviewer versus participant. For transcripts this would be easy when sites follow the protocol, and fortunately because the redacted transcripts can be shared directly with the data repository, there is no issue in later revisiting the automatic assignments, whether adding additional automated considerations to enhance these assignments or doing manual sanity checking where it seems necessary. It is also the case that most language features can be computed in terms of a TranscribeMe speaker ID, and whether a given ID was participant or interviewer (or in the case of psychs interviews some other attendee) can be one of the final steps of a processing pipeline. 

For audio and video however, there is not the same convenience. Data sent to \emph{predict} with a given speaker identification cannot be validated post-hoc in the same way, because no one from that point forward will have access to the corresponding raw audio. The transcription timestamps could perhaps assist (and similarly for determining which face ID is who), but it is a much more difficult problem than it needs to be at that stage. The ideal situation would be for feature extraction code to treat all speaker-specific audios and all face IDs as similarly abstract entities, but then as part of the feature extraction pipeline outputs, a map to the suspected identities would be provided too. That way updated identifications, whether manually adjusted by central monitors or determined by improved automated techniques, could be propagated to old outputs as well as new. This is important for maintaining the highest level of robustness, because it would not be particularly hard to get a mostly accurate identification algorithm working today -- but a huge number of edge cases do complicate complete correctness, and giving purportedly iron clad IDs should require as close to complete correctness as possible, something I suspect the initial version of a feature extraction pipeline will not have. Exactly how it is implemented will require collaboration between the aforementioned groups though. 

Aside from enforcing principled infrastructure decisions, the key role for my replacement in this process would be handling the speaker identification labels directly related to data flow, particularly if they will also end up working closely with the REDCap/RPMS interview forms. This is because the file names of the speaker-specific M4A files found in the 'Audio Record' subfolder of a given Zoom interview recording folder should contain the display name of the corresponding meeting attendee, which if sites follow the protocol should provide a clean indication of who is who. In practice adherence to attendee naming suggestions has been subpar, though as long as interviewers keep consistent display names it remains feasible for code to be written to handle many of the less than ideal cases. 

Even still, there are certain cases that cannot be identified from saved display names alone, and will require reference to forms (if properly filled out) or potentially to other signals based on the actual interview content. The main such case is the onsite open interview, which uses Zoom on two different lab machines. Some interviewers have not bothered to update the display name of the participant, so two known site interviewer names will come up as the two diarized audio file names. On top of that, there are psychs interviews (or sometimes SOP-violating open interviews) where additional meeting attendees will be present, in which case it may not be immediately obvious how to label the primary interviewer versus other interviewers or the participant versus e.g. the participant's parents. Further, and again much more likely to occur for much longer duration psychs protocols, there are instances where an attendee may leave an interview and rejoin or where an entire interview will be split into multiple recordings (which can occur in a variety of ways as previously outlined) -- additional cases that will create more diarized files to handle in the data flow. Fortunately split audios should not generally cause any problem with identification, but they do introduce a number of time alignment issues, and regardless require additional consideration when it comes to data management code. 

For face identification from videos, estimated labels will be more relevant for the feature extraction team to handle, as they will be based more directly on outputs, likely as a first line technique based on face location within the interview frame. Still, those doing quality assurance/data monitoring should very much be aware of and regularly checking on such outputs. Because Zoom will put the gallery view faces in a consistent order from the host's view who is recording the meeting, it should be easy to identify the primary interview from coordinates alone. The problem is pretty well solved then when the only other attendee is the participant, as should be the case for the open interviews (though this is not always followed and interviewers also don't always use gallery view as they are supposed to). However for interviews with more than 2 participants - primarily the psychs interviews, which also have less specific video recording rules overall - it is less clear how to be fully confident in labels assigned automatically based on simple location. Location may even change over the course of an interview as attendees may leave and may or may not come back in the larger interview setting. 

As such, pipeline outputs should be used to monitor for interviews that may cause difficulties to a straightforward face identification algorithm, and invoke manual review of those interviews and corresponding feature extraction outputs where relevant, especially early in the feature extraction launch process. More broadly, this brings up concerns about the lesser quality standards and fewer protocol consistencies across the board with psychs interview recordings. Just one (large) example is that onsite psychs recordings are only mono by default and would require testing automated diarization methods to attempt to extract comparable features to the Zoom interviews' speaker-specific audios. 

While there is no real downside to simply collecting the psychs recordings, and if a plan develops for how to use extracted features then the steps needed to deal with further processing them could certainly prove fruitful. At this moment though, it seems clear to me that psychs feature extraction should be of the lowest priority. I do not think it is a good idea to release poorly validated features, and validating in the case of psychs introduces many challenges given the software engineering resources currently available to the project. The clinical interviews are so much longer than the open ended ones too, and occur at many more timepoints, so they will require much more compute time on top of all of that. Personally I would suggest focusing feature extraction efforts purely on open interviews, first for Pronet and then for Prescient, and return to psychs interviews when there is more bandwidth to handle many of the described complications.  

In any event, the key point is that the existing dataflow/QC pipeline will need to be closely linked to future endeavours to implement feature extraction pipeline(s), something critical to the long term future plan for AMPSCZ speech sampling. Moreover, an eventual feature extraction pipeline could in turn supplement our existing QC monitoring workflows and perhaps suggest expansions to QC metrics that could allow future large scale interview data collection projects to build on the groundwork of AMPSCZ. \\

\paragraph{Longer term goals for AMPSCZ interview analysis.}
It is important to remember that only those steps that need raw interview recording (or unredacted transcript) input should be run on the data aggregation and by extension AV processing servers. Most processing steps should in fact occur downstream, and are not really in scope for the AMPSCZ infrastructure team - certainly not at this time, nor without getting the requisite approvals for any scientific analyses. 

\noindent For reference, here is a list of some of the possible future (and ongoing) directions that would relate to AV server processing work:

\begin{itemize}
    \item As mentioned, the acoustic and facial feature extraction flows to be used remain themselves in development. For acoustics, we have generally agreed on using GeMAPS config with OpenSMILE (lld), but other details remain to be sorted out. Additional OpenSMILE configs may be considered, as well as various PRAAT scrips and other tools like Montreal Forced Aligner. Further, how to utilize info from across the different speaker-specific data streams simultaneously for feature computation related to conversational flow remains of interest. Regardless, the feature extraction scripts should not actually need to know which diarized audio file corresponds to which identity, as all features should be extractable for all present speakers in an interview. However this remains to be formalized via actual feature extraction function(s).
    \item For facial feature extraction, next steps remain even less clear. IT has vetoed the use of OpenFace \citep{OpenFace} due to its Python 2.7 implementation and lack of recent updates, and as mentioned it is also just quite difficult to install on a Linux machine. PyFeat \citep{pyfeat} has since been tested, but runs much to slow to feasibly be used across frames for all videos. While PyFeat theoretically has GPU capabilities, it has reportedly not helped with evaluation speed in the AMPSCZ AV server testing. It is important to remember though that PyFeat has multiple different models that could be loaded in (and continues to be updated) for each feature set it extracts (including landmarks, pose, AUs, and emotion detection). Group discussion to date has largely neglected to specify which models were being tested, and whether we could leave some feature (most notably the emotions) out entirely. Nevertheless, PyFeat is likely infeasibly slow for the project at this time, which leaves open many questions.
    \begin{itemize}
        \item We may run PyFeat with frame skipping on an initial implementation. Given the lack of clarity on scientific aims, it is unclear how much of meaning is lost by skipping every $n$ frames in the temporal resolution of PyFeat outputs.
        \item As GPUs do not seem helpful here, the AV processing server will not include them as a cost-saving measure. However if model building or even tuning is to happen later, or PyFeat releases updates (or incorporates new models) that are a significant improvement, it is not entirely clear how easy it will be to add back a GPU to the workflow. More generally, monitoring of updates to PyFeat and any models that may be loaded into PyFeat should be ongoing.
        \item The face detection package MediaPipe can run more quickly and obtain facial landmarks with higher resolution, so there is a plan (though not yet a clear function to use in the pipeline) for incorporating MediaPipe. This could potentially run on a frame-by-frame basis to complement PyFeat, and extracted landmarks from MediaPipe could be used to estimate AU activations, something unfortunately currently missing from its functionalities. However it will be important before proceeding with this that the accuracy of the MediaPipe landmark-derived AUs is sufficiently good, as it will be strung together in a bespoke way per the current plan, and is thus not an independently-validated tool.   
    \end{itemize}
    \item Running automatic transcription tools, to assist in evaluating and later fine-tuning those tools against the gold standard unredacted professional transcriptions we are obtaining. With some tools (though not the state of the art Whisper at this time \citep{whisper}), we may also be able to improve the timestamp resolution of our transcripts and thus their time alignment with raw signals and ultimately any higher temporal resolution extracted features.
    \begin{itemize}
        \item Much about the present and future of automated transcription for these applications is discussed in Appendix \ref{cha:append-ampscz-rant}.
        \item For the interview format, this might also involve development of a tool to interleave transcripts automatically generated from the speaker-specific audios saved by Zoom (and later perhaps from outputs of diarization methods for non-Zoom mono recordings).
        \item In the case of psychs interviews in particular, we will want to explore auto transcription further due to the current 30 minute duration cutoff being imposed on the professional transcriptions. The utility of this setup should also be more generally discussed with greater care.
    \end{itemize}
    \item Working on tools for automatic redaction of transcriptions using the unredacted versus redacted versions of the TranscribeMe transcripts. We might also consider more sophisticated methods for redacting the words TranscribeMe has marked PII rather than blanket replacing them with REDACTED. Words might be hashed so that repeats of the same word across transcripts for a particular individual can be detected. Alternatively, PII words might be replaced with categorical info like PLACE or PERSON.
    \item Figuring out how to handle automated speaker separation from recorded interviews where only mono audio is available - including both the EVISTR recorder and the possible use of other remote interviewing platforms (latter discussed above in section \ref{subsubsec:interview-code-sec}). In the interim, it is unclear to what extent it is even worth extracting acoustic features from those mono files when we have many other TODOs to prioritize. On the flip side, if mono audio is to have acoustics directly extracted, we ought to extract features from all mono files regardless of diarized audio availability, for comparison purposes.
    \begin{itemize}
        \item In the past, the lab has experimented with summarizing mono acoustic audio for participant versus interviewer based on the timestamps and speaker IDs provided by TranscribeMe. This is likely of sufficient accuracy for many summary analyses, especially for groups with small datasets that need to limit the scope of features they consider. For the much larger interview recording dataset to be collected by AMPSCZ, we would be much more inclined to want methods that can be more accurate at higher temporal resolution - though much of this hinges on the scientific aims the project wants to prioritize.
    \end{itemize}
\end{itemize}

\noindent Again, this list focuses on those operations that need to occur on the AV processing server. Higher level summary features to be computed from lower level features, as well as any downstream visualizations, should occur at later stages of data flow where information is more available to a wider group of scientists. It is also not clear to what extent any such work should be done by an AMPSCZ team on behalf of the project, as the data shared with the NIH repository should largely be as raw as possible. Research groups accessing the data are free to summarize as they see fit, and may be able to contribute back summaries that they complete as part of their work with the AMPSCZ dataset. 

Towards that end, note that any researchers who will be working with AMPSCZ transcripts sent to the NIH data repository will want to familiarize themselves with the TranscribeMe professional verbatim transcription conventions, so that they can process these raw data using all relevant information. This includes accurate extraction of usage patterns for different types of linguistic disfluencies, something discussed in greater detail for the audio diary feature extraction pipeline documented in section \ref{subsec:diary-code}.

\subsubsection{Adapting the code to run in new contexts}
\label{subsubsec:u24-dif-project}
In addition to the many future directions for processing and eventual analysis of AMPSCZ interview recording data, there are also a number of potential future directions for using this pipeline in other similar data collection projects. As such, in this section I will discuss the major improvements that could be made to the pipeline to facilitate more different interview projects. At the same time, the code could in many ways be immediately applied to other future projects, and so the following list will emphasize the changes to the recording format that would require specific updates for such projects. Finally, in the process of setting up the code there have been a number of operational lessons learned that might be of relevance to any future interview recording projects, and so I will close the section with some discussion of the launch of my data flow and quality assurance code in AMPSCZ.

\noindent The primary pipeline implementation details that may need to be updated for future projects include:

\begin{itemize}
    \item Generalizing to handle up to $n$ interview types with names specified in settings/matching subfolder names of "interviews" folders. Currently the software is hard-coded to process open and psychs interview types, and at times uses specific type-specific expectations about input format. How to flexibly enforce interview type-specific requirements via a settings file will be a software engineering question to address if needed by a future project, but even just generalizing the interview types to be treated the same in basic data flow/QC and have specifiable name would be a good improvement to the broader usability of my code.
    \item Supporting other interview recording sources and possibly other input/output folder structures. Both of these adaptations would be done on a case by case basis, though we recommend avoiding major changes to the overall PHOENIX folder structure that is set up by AMPSCZ's Lochness (which we also recommend using, in synergy with this code).
    \begin{itemize}
        \item For more on adding support for other interview recording services, see the discussion of Cisco WebEx (which may need to be handled by AMPSCZ in the future) in section \ref{subsubsec:interview-code-sec} above.
        \item Along these lines, interview recording files on raw are not independently encrypted like is done for our internal lab storage (described for audio journals in section \ref{subsec:diary-code}), but this might be something a future project wants to add back.
    \end{itemize}
    \item Supporting other transcription methods, including other TranscribeMe settings as well as different services entirely (which may or may not support the SFTP server transfer mechanism for audio uploads). This again will need to be done on a case by case basis, and could involve changes both to the audio branch of the pipeline and the transcript branch of the pipeline. How audio gets uploaded for transcription and how transcripts get pulled back is one obvious portion to potentially change here, but also the details of transcript file post-processing and transcript QC metric computation may need to change in various ways. Furthermore, one might want to make changes to what audio properties/interview metadata are checked before deciding if a file should be sent for transcription or not.
    \begin{itemize}
        \item Along these lines, the transcript manual redaction review process may or may not be relevant for a given project, and even if it is relevant the exact implementation here may need to be changed.
    \end{itemize}
    \item Another functionality that might be more broadly generalized is the server-wide summary utilities described in section \ref{subsubsec:cross-site-utilities} above, with outputs reported on at length in the main chapter \ref{ch:2}. While these utilities written for AMPSCZ could theoretically be used for other projects without too much adaptation, they were some of the last modules written and are not as fully integrated into the pipeline structure. There is also a lot of room for flexibility in settings as far as which features get included in summaries, how they are binned in histograms, etc. Because these summaries have been so useful in our final monitoring workflow, further improvements to the robustness and optionality of their implementation could pay dividends.
    \begin{itemize}
        \item More broadly, the implementation of this pipeline might benefit from better distinguishing between an internal lab server that is running the pipeline for multiple different studies at a single location versus a collaborative project central server that is running the pipeline for multiple different sites collecting data for a common project. Theoretically it can handle both at this time, but is has been primarily built up for the latter, and thus may not handle nuances of the former as best it could.
    \end{itemize}
    \item One might consider expanding the QC metrics that are included in the pipeline, which could benefit many projects but might be particularly salient to a given one. A more general pipeline improvement that would facilitate the addition of QC features would be a module that allows for back-application of new QC metric computation on previously processed interviews when new metrics are added, allowing for old outputs to be updated for seamless integration with ongoing pipeline runs. Some possibilities for expansions to QC at present include:
    \begin{itemize}
        \item For videos, especially if some video collection will return to an in-person camera recording format, it may be important to expand QC to cover other features besides just the face. Basic features like video duration, resolution, and fps could be added easily using ffprobe directly on videos, but one might consider other methods too. For lighter weight QC like intended for AMPSCZ, other packages such as OpenPose could perhaps be run on the extracted frames to obtain summary info like confidence scores. 
        \begin{itemize}
            \item Note as well that the PyFeat QC could be expanded on in various ways. One uncommon but recurring issue is PyFeat picking up phantom faces, which can often be suspected based on existing QC features, but could be more rigorously detected (and perhaps even root causes identified) with an expanded suite of QC metrics related to properties of the faces detected in extracted frames.
        \end{itemize}
        \item For the mono audio, QC would benefit from detection of echos and loud plosives. The former is especially relevant in the remote offsite interview context, because poor internet connections can cause odd echoing noises in the audio. The already computed sliding window QC could be helpful for improving what we can currently detect, for example by looking at maximum or 75th percentile volume over the analyzed bins. However for echo detection it may require returning to the raw audio computation, potentially considering autocorrelation-derived features from the mono audio waveform.  
        \item As an additional QC module, it would certainly be beneficial to have a way to assess available speaker-specific audios. This could be implemented as an entirely independent module, though to integrate fully with the pipeline it would require handling a number of new logistical edge cases related to diarized audio naming, presence of more than 2 speakers, time alignment and split files with "meeting" attendees coming and going, etc. Many of those problems should be addressed as part of the feature extraction pipeline implementation though, and fortunately most of them are much less often relevant in the valuable open interview format. However for future projects this might take many forms, and might be entirely different if onsite lapel or headset microphones are being used for the in-person recording format instead. Regardless, a first pass that just replicates existing mono QC features on each available diarized audio file independently could certainly be of use. For our internal onsite audios large difference in volume between different speaker's microphones were often indicative of lower quality transcriptions. For a more elaborate implementation, looking for crosstalk between channels by correlating their waveforms is also of potential utility. 
    \end{itemize}
    \item Another longer term future direction might involve actual filtering or other preprocessing of audio files to improve their quality where needed.
    \item Other new functionalities might grow out of results from AMPSCZ, and could in turn expand what this dataflow/QC pipeline can do. Such new functionalities might relate directly to the feature extraction steps to be implemented on the AV processing server, which might inform the outputs of this pipeline and the AMPSCZ monitoring process more generally on an ongoing basis. They could relate as well to models that are produced at the conclusion of AMPSCZ, along the lines of some of the long term ideas presented in Appendix \ref{cha:append-ampscz-rant}.
\end{itemize}

\noindent Of course it will also be important to handle many smaller details, like any changes to code dependencies that may arise over time. This, as well as some of the above items, may end up being addressed as part of ongoing AMPSCZ work -- so before proceeding with work to adapt the pipeline of \citep{interviewgit}, it is recommended to contact current members of the AMPSCZ groups responsible for software infrastructure (DPACC) and speech sampling data collection (Team H).  \\

\paragraph{Launching a large scale collaborative data collection project.}
In the initial phases of AMPSCZ data collection, in order for a site to begin real participant enrollment they had to complete a mock data certification, including conducting a mock open ended interview. In parallel, those developing software infrastructure were testing code on the mock data via development versions of the servers to eventually be used for the project, set up by the corresponding central site IT teams. This process was overall a good way to address early bugs in the code, iron out details of the TranscribeMe transcription settings, identify some early issues with site recording procedures and data organization tendencies, and add new features to code where relevant. It of course also provided initial training for participating sites on how to collect various datatypes, in particular here how to appropriately conduct an open-ended style of interview. As different AMPSCZ sites have different areas of prior expertise, that component of the project setup was indeed quite important. 

For the purposes of informing similar future initiatives, I will now provide some additional information on the specific workflow used for AMPSCZ mock testing, highlighting benefits of the approach along with limitations that could be addressed in other projects down the road. Where possible, I will also try to distinguish between additional development testing that could have been particularly helpful from a software perspective (i.e. not necessarily something to be required of all sites and perhaps even something that could be run through a fake site for development only, alongside real site mock certification) versus additional mock data collection testing requirements that might have been helpful towards reducing some of the site protocol errors we've observed in production. In the latter case, there of course remains a balance between requiring useful training and inadvertently producing training that is treated like bureaucratic nonsense and thus functionally ignored. There is only so much that can be done to strike the right balance when people fundamentally do not care however, so a major overarching issue is carefully defining incentive structures (and enforcement mechanisms) across such a multi-faceted project.

Nevertheless, a major shortcoming of the dev testing and mock certification process was how quickly it got shut down. The development servers are no longer being used at all as far as I can tell, and to get mock certified is now a much easier process than it originally was (and it never was asking especially much). Clearly, someone felt pressure to rush to production data collection for as many sites as possible. It is hard for me to imagine where that pressure came from, because there has been quite obviously no pressure to successfully submit interviews. Months have passed at times before anyone noticed entire sites were missing from interview collection lists because of widespread SOP violations or server permissions bugs. Half a year has passed with dozens of transcripts held up by the manual redaction review at a few sites, despite repeated contacts. I don't think I can reiterate the absurdity of these issues too much. Anyway, maybe it was the teams responsible for other modalities that resulted in rushing the process for more sites to get to production data collection before they were prepared, though from the teams I am aware of that does not appear to be the case. Maybe it was the sites themselves, which if so says a lot -- how about you take 2 seconds to learn to put a file into a folder correctly first? 

But most likely, it was a misguided attempt to pretend to the NIMH that great progress was being made, something they are seemingly obsessed with on the surface but really goddamn bad at legitimately checking on. Ironically, progress would have been much better by now if more \emph{quality} enforcement happened back then. Both more time spent on training the sites (delaying them as needed) and on building out further software to more gracefully handle a lot of the unanticipated edge cases (and better testing to identify more of them more quickly). Maybe the interview feature extraction plan would be a bit more concrete now too if organizational factors made more sense back then. As it stands goals are not exactly clear. Anyway, the development/mock collection testing process was pretty good while it lasted.

For interviews, dev testing with Pronet sites began in early February 2022. In late May 2022, the production version of the Pronet data aggregation server was launched, and an additional single test interview was run through for one site (YA) to confirm no new issues arose in migrating servers. Then the mock production interview was cleared out and real prod data collection began. Additional Pronet sites could begin immediately with submitting real data to the production Pronet aggregation server, as soon as they'd been mock certified. However (at least on the interview front) this amounted to finalizing the certification for those sites that had already made it decently far into the development process, but not requiring new sites to complete the whole workflow and in fact not even requiring the sites in progress to properly complete everything they were supposed to. One example of a skipped step by a number of sites was successfully performing a manual transcript redaction review and then correctly returning it to the data aggregation server for final processing. Interestingly, that step remains an issue for some sites, despite the addition of new code functionalities like targeted email warnings to try to combat it.

In the end, 4 Pronet sites completed the full interview mock workflow on the dev server - one with 3 interviews, one with 2 interviews, and two with 1 interview. 10 Pronet sites made it mostly through the mock interview process, but got stalled on the aforementioned manual site redaction review. 5 of those sites had attempted at one point to return a transcript after reviewing it, but did this incorrectly and failed to fix it. I can't imagine it would have taken a big time commitment for them to move a file to a different folder using clarified, detailed instructions, but instead the issue was just dropped. Fortunately, the lack of timely manual redaction review problem has not been quite so bad on a Pronet-wide basis in production, though a handful of sites have been very bad about it. Moving on, 4 Pronet sites failed to have their uploaded mock interview processed by the interview QC pipeline, because the folder name was missing required metadata and the issue was simply never resolved on the site's end. Another 3 Pronet sites failed to have their recorded mock interview pulled by Lochness to the dev server at all, because they messed up the registration of the fake subject ID in REDCap or misnamed the subject ID's folder in Box. 

Note as well that many of the sites that had been creating subject IDs correctly in development REDCap for the mock interview submissions had not necessarily been providing a valid, realistic mock consent date (if entering any consent info at all). This occurred even for some of those sites that otherwise successfully passed mock certification. It is important in real data collection practice that sites enter correct consent dates for each subject ID from the outset, which was not at all enforced by the mock process. Additionally, because the mock process did not enforce entering of a reasonable consent date for the mock subjects, bizarre software issues cropped up that are not really relevant to the production code. Lochness for production data will refuse to pull a record if the consent date for the subject ID in REDCap/RPMS is not valid, but for development data it was inserting a dummy consent date instead that was many, many years in the past. This produced ridiculously large day numbers that interfered with expected naming schemes and stretched the view in DPDash to a point that made it essentially unusable. Granted, catching that sort of problem is a good outcome for the dev testing process in some sense, but the main downside again comes back to the fact it wasn't addressed for follow-up testing purposes in the dev environment and was also never required for sites to do properly for obtaining mock certification. Perhaps 2 years ago I would have said such a simple field in a form is straightforward enough to gloss over, but knowing the many errors that sites have managed to commit on a recurring basis, my opinion has been updated.

Beyond those many unresolved interview-related issues at the closure time of the Pronet development infrastructure, there were also 5 Pronet sites that as far as I can tell from Lochness/REDCap records never tried to conduct a mock interview at all. Some of those sites may have been further from being ready for data collection, and indeed there are still a number of Pronet sites that have not yet begun recording interviews. But then the development processes should not have been halted the way they were and as early as they were. Additionally, many of those missing sites were foreign language sites, which means that mock software testing was never even performed for clear cut cases of relevance to the long term health of the interview collection project, nor was spot checking of some of the foreign languages we will require from TranscribeMe. That problem persists today, but at least some manual mock interview certification and transcription quality check will occur on the side for each site that joins from a yet to be represented language. Hopefully there will be a replacement hired for my software-related roles, so that more careful code monitoring (and potential tweaks) for new foreign languages will not be neglected going forward, doubly so for those with a character set very distinct from English's. 

The mock testing infrastructure shut down much too soon for a number of reasons, in addition to the already identified premature halting of the protocol for adequately clearing sites for real collection responsibilities. Another logistical consideration is that AMPSCZ is intended to span $\sim 5$ years, and real data collection will probably span at least 3, such that the RAs handling certain interview recording duties at certification time (not all of whom even needed to be tested) are unlikely to be the same as the RAs that will be handling similar duties later in the project. This is yet another reason that there is no guarantee site compliance will be smooth sailing a year or two from now, and it additionally calls into question whether interview conduct or recording quality patterns, particularly open ones, may change as the project continues. In my opinion, anyone intending to handle collection of many AMPSCZ interviews, or at the very least anyone who repeatedly makes similar mistakes directly advised against in the written protocol, should be required to undergo a basic mock interview certification process like what was enforced(-ish) on only the site level during early phases of Pronet data aggregation. 

Importantly, there were many ways in which the mock testing did not sufficiently cover relevant cases of interest for software verification either. These issues need not be handled on the level of the individual site, but the process on the whole would have benefited from better coverage of available test data (and an extended period for testing and for corresponding software updates):
\begin{itemize}
    \item During dev testing, there was never a mock subject with more than 1 interview of the same type submitted. This caused a glaring bug to be missed until months into production deployment, whereby all interviews of a particular type for a given subject after the first baseline upload were being rejected by the pipeline, due to an oversight when saving updated versions of existing DPDash QC CSVs.    
    \item There was also never a mock site with more than 5 interviews submitted aross mock subject IDs and interview types. This meant that the full protocol for manual redaction review assignment was never tested on the dev server, as the first 5 are all automatically sent back.
    \item In fact there were no psychs mock interview recordings at all tested on the Pronet dev server.
    \begin{itemize}
        \item It is the case that psychs have fewer expectations of consistency and overall lesser quality restrictions per the SOP, but it still seems odd to not check the interview conduct and recording procedure at all from a site performance perspective during development testing. 
        \item The lack of psychs uploads in testing caused another bug to be missed where the pipeline would simply skip over any subjects that did not have a single open interview available, regardless of how any psychs interviews they had uploaded.
        \item Psychs onsites have an entirely different recording procedure than open onsites, at least for those sites that would plan to use the EVISTR recording device for the clinical onsites. That is a different case for the pipeline to handle (and a different protocol for sites to make mistakes on), yet it was not tested in any way during the development server phase. 
        \item The disagreement surrounding psychs transcription protocol and the duration cutoff associated might also have been lessened if mock clinical interviews were included in early testing in some capacity, or if more abstractly a better monitoring process and convention for raising issues had been established during dev testing. It is funny that mock transcriptions did not include any of the semi-structured format, despite that accounting for the vast majority of the interview transcript budget at the time (and to some extent still now). 
    \end{itemize}
    \item Finally, many properties of the interviews themselves were non-exhaustive because of missing sites or lack of mock subject variety (e.g. cultural/language differences) and otherwise non-reflective of real data due to being needlessly contrived mock scenarios. Really any healthy adult should be able to participate in an engaging open interview of a reasonable duration with a modicum of effort. However, I wrote the following near the end of Pronet development testing about one year ago: 
    \begin{itemize}
        \item "All sites have uploaded mock interviews that are notably shorter than what we will expect during real data collection. We should test at least one longer interview to ensure there are no compute issues and to get a better sense for expected runtimes and storage requirements."
        \item As it turned out, a variety of issues related to interview durations have appeared over the course of production data processing thus far, so it is unfortunate that my request for more realistic interview durations during mock testing went entirely unaddressed by sites back then. 
    \end{itemize}
\end{itemize}
\noindent These concerns also bring up the point that the human operations factor in this project was not really stress tested whatsoever. No one ever tested whether sites were actually performing manual redaction review when they returned their transcripts, for one example. But other issues like lapses in monitoring and unclear responsibilities were recurring in the Pronet dev server phase as well. 

Throughout much of this material, I have focused on Pronet systems because that is the central aggregation server that has had much quicker progress with data collection (and on the other hand a ridiculous variety and frequency of site mistakes). I will now focus for a bit on the Prescient side, under which many of the foreign sites are organized.

Unfortunately, Prescient development processes moved even slower and ended even more abruptly. There appeared to be a greater lack of IT personnel and a greater disconnect between the Prescient server IT staffs' overall responsibilities and the specific targets for AMPSCZ. DPACC team members ended up taking over some of the server administrative responsibilities that were run by dedicated Yale IT staff on the Pronet side. Furthermore, Prescient did not have a clear plan for how audio files would be transferred to TranscribeMe until past the point that development testing was already well underway on Pronet's data aggregation dev server. For a long time the plan was apparently for TranscribeMe to be granted full access to all (ultimately disorganized) site Mediaflux folders and to go in an find the audios to be transcribed themselves. Obviously that plan would carry a number of extra organizational and data security concerns, but more important was the lack of coordination between relevant parties in developing and (not really) updating this plan. 

Prescient was eventually dragged onto the same page as Pronet, at least to the extent that I could largely just migrate my code to their data aggregation server and tweak a few things. Still, discrepancies remained. Some of their earliest sites were officially mock certified using a very outdated SOP, and there was a long delay before new mock sessions were recorded by each such site to go through the dev code testing and TranscribeMe verification process. By the time things got underway for Prescient dev testing, they were essentially already ending. There was never a separate dev server machine, which did make transition to production data structure more efficient, but made it more difficult to keep running mock/development processes directly in parallel with real/production processes, and also introduced more opportunities for questionable account permissions to spread. 

Suffice to say, the clear majority of Prescient sites never even attempted to submit an interview recording for dev testing, and the vast majority of Prescient sites have still not begun any interview data collection at this time. Only 3 Prescient sites had submitted any production interview recordings at all by mid-March of 2023, and only 1 of those sites had submitted a notable (and regularly progressing) number of them. Again, there remains a serious lack of non-English sites for testing (Melbourne, Birmingham, and Singapore all record in English). 

One contributing factor is the privacy issues and constantly flip flopping information about resulting restrictions on acceptable recording software from certain European countries (especially Germany), as was described within section \ref{subsubsec:interview-code-sec} above. However that does not explain the wider phenomenon that Prescient has mostly lagged Pronet. In all likelihood, Prescient would have no interview QC or data flow software running now if it were not for them being pestered by Pronet to use the software that was being developed for their aggregation server.

To their credit, the Prescient sites that have gotten ramped up have so far done a much better job of following the intended protocol than Pronet sites have. Given the small sample size, one can only hope that this persists when later sites finally join in for Prescient interview recording, but at the very least Melbourne has been one of the best sites across the project at consistently submitting interviews that are not often severely messed up. Prescient could certainly use an expansion to their IT services along the lines of Pronet's IT resources, but perhaps Pronet sites could benefit from the interview training or incentives or who knows what is going on at the early Prescient sites. 

And across the board, there is a dire need for more people that have the time and the inclination to perform some kind of regular basic check on the continually incoming pipeline results, results that require follow-up unfortunately a bit too often at present. While there is some reason to believe protocol violations will get better over time, it has been slow going so far for a number of established sites, many other sites have still barely started yet, and there is simply not enough monitoring bandwidth nor enough teeth to enforcement to truly stamp out most recurring problems. Seriously, I was away for an internship in summer 2022, and when I left production interview code was only turned on for the first approved Pronet site, Yale. It was not until the fall, more than a month after the end of my internship and multiple months after a few other sites had started collection, that someone else noticed other sites were missing expected data. I didn't realize other sites had started yet, and since I've resumed deeper involvement I've taken over more of the awareness responsibility. 

It is unclear who exactly will replace me in that capacity though. I don't have a ton of faith in the situation precisely because huge chunks of missing data have gone unnoticed for long periods of time when I was not the one to bring them up. How does this connect back to mock testing? Well for one thing expectations on the sites could have been higher from the outset -- and maybe resources for the sites too, but I can't speak to how much they are understaffed for this project or not from the information I've had access to. Regardless, many fundamental issues with interview recording procedure were noticed during the mock process, and were just totally ignored. Moreover, large gaps in the monitoring workflow were abundantly clear at that time as well, and yet I foolishly continued to believe that once real data collection started sites would begin using the monitoring tools to keep track of their own data quality and collection progress. 

If limitations were instead addressed head on, the development stage of the project could have been more fruitfully used to inform later monitoring decisions: from software design decisions to allocation of personnel to codified manual review "algorithms" to agreed upon responsibility assignments. Unfortunately, what actually happened is that the confusion over individual roles festered, many site mistakes went unchecked (such that downstream code then went literally unchecked and bad habits formed), and absolutely zero plan was drafted for what should be prioritized in the manual monitoring components of the project. The mock certification/dev server system was a great idea that very plausibly could have helped AMPSCZ to avoid a lot of the interview processing headaches that continue to arise, but it unfortunately was barely leveraged as it should've been. 

To be clear, the people that put a lot of effort into the project, particularly from the perspective I have been able to observe with AMPSCZ speech sampling, are \emph{very} minimally at fault here, despite the massive number of technical (many of which are technically on me) and logistical shortcomings. There just were too few such people, and at the same time others nominally involved that have done hardly anything they were supposed to nor even genuinely attempted as far as I can tell. Then there are the people who - because they haven't been paying attention to relevant details, whether that's in the project at large or only in speech sampling because it's a side note for them - spread misinformation to other hardly involved people at the subcommittee meetings, thereby creating a completely dysfunctional decision making environment and unfortunately tying the hands of some of the members that have made good faith and sometimes above and beyond efforts towards this part of the project. This also muddies the waters to such an extent that it would be quite easy to act in pure career-climbing (and/or hypothesis-married) self-interest with AMPSCZ speech sampling project recommendations, and yet obfuscate that intention amongst the overarching chaos.

What is the take-home then? Well, proper testing and software practices are uniquely important to this style of data collection project, and you should try to improve on as many of the little details I've mentioned throughout this section where you can when launching a new project. But really at the end of the day, my takeaway is you can't trust the NI(M)H to competently spearhead this sort of undertaking in 2023. They created ridiculous incentive structures, they've failed to enforce basic procedural standards on sites or basic contribution expectations on funded subcommittee members, they've even failed to do anything despite knowing some speech sampling members have been breaking the stated rule that no clinical outcome, case/control, or related analyses should occur on the access-restricted infrastructure servers. Maybe bleaker than that, they severely undervalue software contributions across the board, both in the respect they show them and the absurd funding caps they place on hiring software engineers, which is trivially a very bad combination for productive infrastructure building. I would say the project doesn't have enough resources, but what I think is much closer to the truth is that the resources that have been given to AMPSCZ (at least for speech sampling) have been very poorly allocated, something I would assume the government funding agency is supposed to oversee.  

To close this appendix with slightly lesser manifesto undertones, I will circle back to the fact that the mock certification/dev software testing framework used in the pilot phases of AMPSCZ infrastructure setup was a good idea that at the time surprised me with how official it appeared. I thought it boded very well for the project, and was perhaps less alarmed than I should have been when 3 months later many site mistakes and implementation question marks identified during the initial testing process were never addressed. Instead the project plowed forward into production, abandoning the (good) idea that later onboarding sites should still need to complete the certification process or that new major code updates should still be tested in a development environment. 

In conclusion, many of the early challenges encountered in site mock interviewing and development software testing turned out to be challenges encountered regularly with real data in production. Some have slowly been improved upon, but it is a testament to both the success and failure of the AMPSCZ launch that major problems were well-foreshadowed and simultaneously utterly ignored for months on end. 

\section{Extra materials for BLS interview analysis results}
While the bulk of this supplemental chapter (and its corresponding main chapter \ref{ch:2}) have focused on the AMPSCZ project and the code I wrote for dataflow/QC of interview recordings collected for that project, there are some additional materials to provide related to BLS interview analysis as well -- which was the focus of section \ref{sec:disorg} in particular \citep{disorg22}, and also holds relevance to potential future directions for the audio journal work on the BLS dataset from chapter \ref{ch:1}. 

\subsection{Conducting BLS clinical interviews}
\label{sec:sup-bls-prot}
Information on recording hardware used for BLS interview collection was provided in the main text of chapter \ref{ch:2}. Here I provide methods details for the actual conduct of the interviews used in BLS interview recording analyses. \\

\noindent The recorded BLS interviews followed the format of a structured clinical interview, and each interview time point included rating of the following clinical scales:
\begin{itemize}
    \item Young Mania Rating Scale (YMRS)
    \item Montgomery-Asberg Depression Rating Scale (MADRS)
    \item Scale for Assessment of Positive Symptoms (SAPS)
    \item Scale for Assessment of Negative Symptoms (SANS)
    \item Positive and Negative Syndrome Scale (PANSS)
    \item Multnomah Community Ability Scale (MCAS)
\end{itemize}
\noindent The MADRS and PANSS were discussed at length as part of the scientific results in chapter \ref{ch:1}, and have been a common focus of our analyses to date due to their prevalence in assessing depressive and psychotic symptoms, respectively. 

To capture potential Bipolar disorder symptoms not well represented by the PANSS, the YMRS is an 11 item scale intended for assessing manic symptoms over the previous 48 hours. Similarly, the MCAS captures how psychiatric symptomatology is affecting a patient's real world functioning in more detail than is possible with the other scales. It is a 17 item scale intended to reflect on functioning over the past month, including questions on topics such as alcohol/drug abuse and the ability to manage money. 

The SAPS and SANS on the other hand are alternative scales for assessing positive and negative symptoms of psychotic disorders. Overall SAPS and SANS scores have been shown to correlate highly with overall PANSS positive and PANSS negative scores, respectively \citep{Erp2014}. However the SAPS and SANS cover many more specific symptom details than the PANSS does, with 34 and 24 items each. For example, the SAPS breaks down hallucinations into different modalities and delusions into a number of different types like "religious delusions" and "delusions of being controlled". \\

Because the BLS interviews are all clinical interviews, there are always scale values directly associated with each recording. This also means that each recording contains a number of questions that expect a somewhat specific answer format, resulting in periods of briefer conversational turns that may be less rich for linguistic analyses than a more open ended dialogue could be. Still, the guidelines for the trained RAs conducting and scoring these interviews include a number of follow-up questions to consider asking for each symptom dimension, many of which can elicit longer patient responses and potentially interesting introspective descriptions -- well suited for many natural language processing (NLP) techniques. Furthermore, note that the high quality audio and video collected from our interview recordings enable a breadth of acoustic and facial feature analyses throughout the duration of each interview. 

To provide context for the BLS interview dataset reported on in this chapter (section \ref{sec:disorg}), a summary of the requirements for conducting a BLS interview is reproduced here from the study guidelines written for lab RAs. \\ 

\noindent You should be able to rate the scales based on the interview questions from the interview checklist (derived from the relevant scales' items). Remember when taking notes on the interview checklist that you are assessing endorsement of the following within the last month:

\begin{itemize}
    \item Mania
    \item Depression
    \item Positive and Negative psychosis symptoms
    \item Patient support and community functioning (MCAS questions)
\end{itemize}

\noindent During the interview be sure to ask follow-up questions regarding:

\begin{itemize}
    \item Details of the symptoms - what it was like for the patient.
    \begin{itemize}
        \item For example, a patient saying they felt “depressed” is insufficient. Ask how they felt! What thoughts accompanied that feeling? If they felt it localized in part of their body? How did it effect their day to day lives?
    \end{itemize}
    \item Severity - subjective intensity and effect on patient functioning.
    \item Duration - how long was each episode, how much of each day?
    \item Frequency - how often per week?
\end{itemize}

\subsection{Additional methods details for BLS interview transcript processing}
\label{sec:disorg-sup}
To compute the features from the TranscribeMe transcripts used in the modeling of the main text section \ref{sec:disorg}, I first calculated the following base features for every sentence:
\begin{itemize}
    \item Number of words, determined by splitting the sentence text on spaces and taking the length of the resulting list.
    \item Sentence length in seconds, determined by subtracting the start timestamp of the sentence from the start timestamp of the following sentence.
    \item Number of nonverbal edits, determined by counting the number of matches to a regular expression capturing individual words with any number of ‘u’ characters followed by any number of ‘h’ and/or ‘m’ characters.
    \item Number of verbal edits, determined by counting the number of times “like”, “I mean”, and/or “you know” appeared enclosed by commas in the sentence (for example “I like this” would not be counted, while “I, like, think that” would).
    \item Number of repeats, determined by counting the number of times a word (split on spaces with punctuation removed) was identical to the directly subsequent word, plus the number of times a stutter (denoted by single dash within a word) had all characters from before the dash also appear directly after the dash.
    \item Number of restarts, determined by counting the number of times “{-}{-} ” appeared in the sentence.
    \item Number of disfluencies, which is the sum of nonverbal edits, verbal edits, repeats, and restarts. Note each individual disfluency detection method was developed after consulting with TranscribeMe’s verbatim transcription style guide. 
    \item Mean pairwise word incoherence, calculated using the Google News 300 dimension word2vec model. A vector (word embedding) was generated for each word in the sentence (again split on spaces with punctuation removed), and then the angles between each word embedding and the embeddings of every other word in the sentence were computed in radians. The mean of all these angles is the resulting pairwise incoherence measure for the sentence.
    \item Mean word uncommonness, also calculated using the Google News word2vec model. For this feature, the vector magnitude (2-norm) of each word embedding in the sentence was computed. The mean of these magnitudes is the resulting mean word uncommonness value for the sentence. 
\end{itemize}
\noindent I also saved the vector embedding for each full sentence, by taking the vector mean of all the word embeddings found in the sentence. 

I used the above sentence-level features to obtain the final interview-level features for each transcript, in the following manner. To get \emph{total words in the interview}, I summed up the number of words across sentences. I then filtered the sentence-level features to keep only the rows corresponding to patient speech. To determine for a given interview which TranscribeMe speaker ID (of the primary two) is the patient, I counted the number of question marks that occurred for each ID in the interview, assigning the patient to the ID with fewer questions. A subset of interviews was manually reviewed by a student volunteer to confirm speaker assignment was correct across the board. 

The filtered sentence features were then summed to get \emph{total patient words} and \emph{total seconds of patient speech}, which together with \emph{total words in the interview} provided the "patient words/total words" and "patient words/second" features. As the \emph{number of patient sentences} is simply the number of rows in the filtered interview, the other nine features were computed by taking the mean of the corresponding sentence-level feature. 

For the incoherence and uncommonness features, I additionally computed a weighted mean for each transcript using the number of words per sentence as the weight. Further, I used the sentence-level vector embedding to add two additional coherence features -- one representing the mean angle between patient sentences that directly follow an interviewer sentence and said interviewer sentence, and one representing the mean angle between patient sentences that occur sequentially. These features were generated to follow up on the null result from the original incoherence metrics, but were not reported in the results as they did not meaningfully change any of the results. 

All computation was done in Python 3.9, utilizing the pandas, numpy, and nltk packages. The NLP code described in chapter \ref{ch:1} was adapted to the interview dataset for this study. The features chosen for focus here were decided upon after careful consideration of prior scientific results and existing technical evidence for feature accuracy, both of which are discussed at great length in chapter \ref{ch:1}.

\subsection{Sentence-level linguistic feature correlation results in BLS interviews}
\label{sec:sentence-disorg-feats}

\begin{figure}[h]
\centering
\includegraphics[width=\textwidth,keepaspectratio]{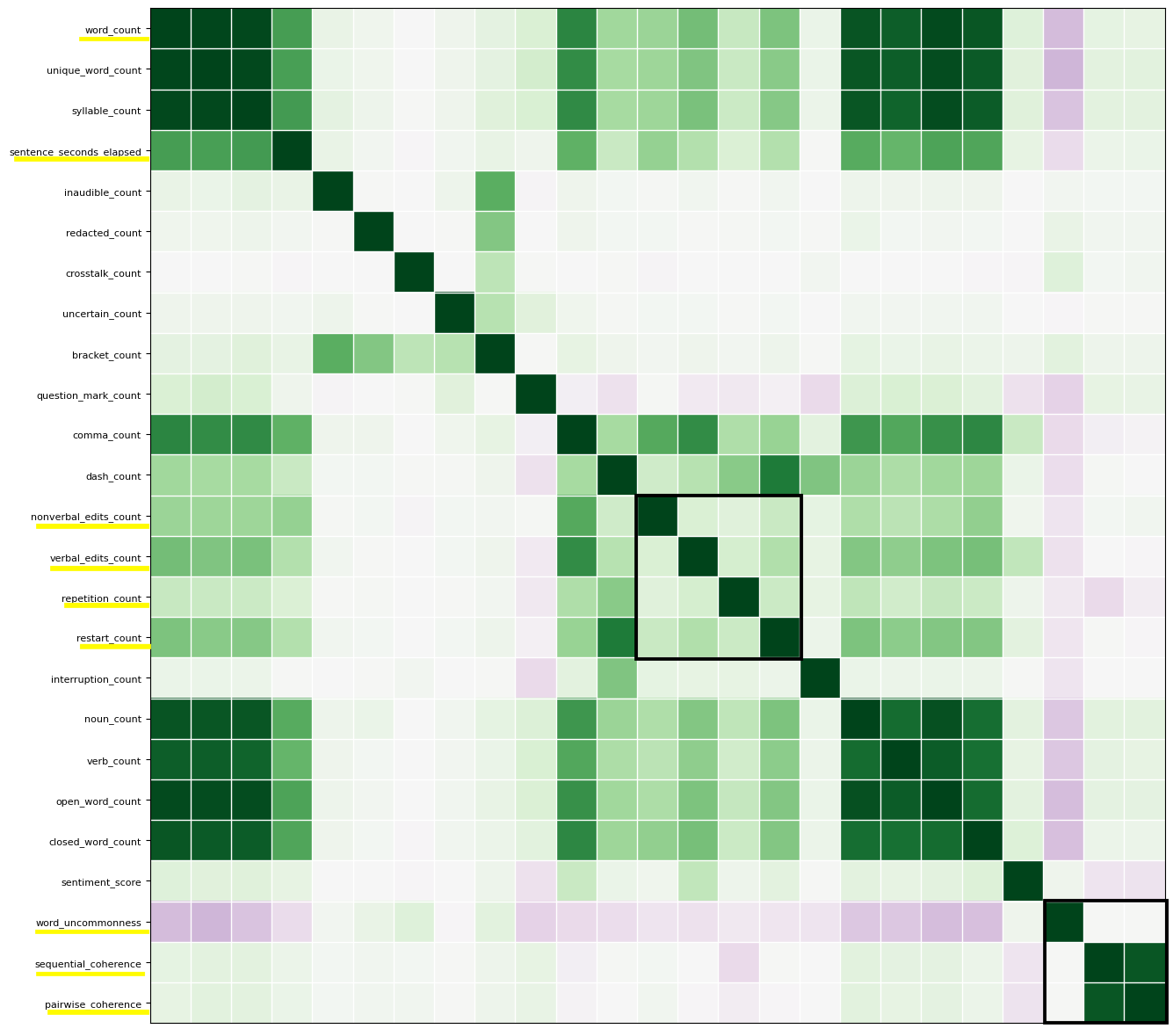}
\caption[Correlation structure of base sentence-level features across interview transcripts.]{\textbf{Correlation structure of base sentence-level features across interview transcripts.} Using all sentences in the BLS interview transcript set, whether from interviewer or patient, a Pearson correlation matrix for the sentence-level feature set was generated, via the same general method as in Figure \ref{fig:disorg-interview-corr}. As was explained in Figure \ref{fig:sentence-sentiment-pt-dist}, many per sentence features were computed that were never summarized on the interview level nor included in downstream analyses, but they are included here. The key sentence features that underlie the interview features used by \cite{disorg22} are thus highlighted with a yellow underline. From top to bottom: sentence word count, sentence duration in seconds, number of nonverbal edits, number of verbal edits, number of repeats, number of restarts, mean word uncommonness, sentence incoherence (sequential method), and sentence incoherence (pairwise method). The block diagonal submatrices containing the disfluency and semantic incoherence related features' internal correlation structures are also denoted with a black box.}
\label{fig:disorg-sentence-corr}
\end{figure}

To expand a bit more on the feature relationships we did find in the pilot interview-level analysis in the main text of chapter \ref{ch:2} (for the disorganization results within section \ref{sec:disorg}), a Pearson correlation matrix was also constructed for the sentence-level features described there (Figure \ref{fig:disorg-sentence-corr}). Recall that a number of features extracted on the sentence level were not included in the interview summaries or subsequent analysis of this research -- because they may still have broader interest, I've included them in the correlation matrix here. Still, the focus of this investigation is on those sentence features that make up the modeled interview features.

Many of the correlation patterns observed in interview features (Figure \ref{fig:disorg-interview-corr}) replicated with the sentence features (Figure \ref{fig:disorg-sentence-corr}). That included a stronger positive correlation between word count and both verbal edits and restarts than the other two disfluency types, a very strong positive correlation between pairwise and sequential incoherence, no relationship between incoherence and uncommonness, and a negative correlation between word count and uncommonness. 

Interestingly, the correlations between disfluency categories were even weaker on the sentence level, although of course there could be a real correlation on a per interview basis that doesn't manifest on the individual sentence level. A future analysis with a higher powered dataset might consider timing of disfluencies within an interview, for example to see if there is any increased likelihood of using a particular disfluency type in a sentence given the particular type that was used in the preceding sentence. Nevertheless, there was a potential small positive correlation found between the sentence counts of verbal edits and restarts, which was one of the interview correlations of moderate strength. 

\FloatBarrier

\chapter{Supplement to Chapter \ref{ch:3}}\label{cha:append-chapt-refch:3}
\renewcommand\thefigure{S3.\arabic{figure}}    
\setcounter{figure}{0}  
\renewcommand\thetable{S3.\arabic{table}}    
\setcounter{table}{0}  
\renewcommand\thesection{S3.\arabic{section}} 
\setcounter{section}{0}

\section{Data collection and analysis}
\label{sec:ocd-methods-append}
The data collected in chapter \ref{ch:3} was part of an early feasibility study of the described deep brain stimulation paradigm, clinical trial ID NCT03184454. The supplemental protocol for our digital psychiatry data collection and analysis was sanctioned by the Mass General Brigham Institutional Review Board under protocol number 2018P000717.

In this section, I will describe the methods we employed to collect and analyze our dataset. In subsection \ref{subsubsec:ocd-data} I will overview the modalities we added and how they were collected. Availability of each of the datatypes over the course of the study is depicted in Figure \ref{fig:ocd-report}a. In subsection \ref{subsubsec:ocd-analysis} I will then provide methods details for analysis of these longitudinal data. Finally, in subsection \ref{subsubsec:stim-exp-methods} I will explain the experimental interview trials we ran to causally probe the effects of cortical stimulation, as well as the techniques applied to analyze the resulting data.

\subsection{Longitudinal data collection methodology}
\label{subsubsec:ocd-data}
All procedures described below were approved by the Institutional Review Board at McLean Hospital. Partners Healthcare guidelines for preserving patient confidentiality and securely storing the data were followed. 

\subsubsection{Clinical interview recordings} 
As part of the DBS study \citep{Olsen2020}, the patient was interviewed bi-weekly for a total of 44 interviews. We recorded audio and video from 36 of these interviews, using a ZOOM Q8 handy video recorder and Sennheiser HSP4 headworn microphones. Interviews were later transcribed using the TranscribeMe full verbatim transcription service with speaker identification, sentence-level timestamps, and PII redacted. 

\subsubsection{Passive sensing (smartphone)} 
The patient downloaded the Beiwe mobile application at 151 days post surgery. This platform passively collects continuous accelerometer data at 10 Hz (independently for \emph{x/y/z}-axes) for $\sim 30$ seconds out of every minute, pings GPS at regular intervals (latitude, longitude, altitude), and logs basic phone usage events (current battery state and timestamp for each instance of screen on/off, charger plugged in/unplugged, battery filled, phone rebooted) as well as periodically sampling battery percentage when in the middle of active use. For more information on Beiwe, see \cite{Beiwe}. 

\subsubsection{Daily self-report questions} 
Smartphone-delivered surveys inquiring about the patient’s momentary mental state were sent once daily through the Beiwe mobile platform as well. One survey contained a series of text prompts (ecological momentary assessment i.e. EMA) that probed for subjective motivation, productivity, and related psychological constructs. As described by \cite{Olsen2020}, the survey contained eight questions regarding the patient’s energy and ability to perform tasks that day, derived from their own description of their primary symptoms. Each question was presented as a statement that the patient would then rate agreement level with, from "strongly disagree" to "strongly agree". The item responses were then converted to scores on a scale from 0 to 4, with 0 indicating the highest level of functioning and 4 indicating the lowest. The scores for the eight questions were averaged to create a summary EMA score. Note that the answer submitted to the first question determined whether the rest of the 7 statements presented to the patient would be positively worded or negatively worded. Regardless, the scores we used in analysis always mapped a higher number to more severe symptoms, whether that corresponded to "agree" or "disagree" for a given statement. 

\noindent The item that started every EMA and determined the branch that the rest of the items would be pulled from was as follows:
\begin{quote}
    In the past 24 hours; it was difficult for me to get anything done
\end{quote}
\noindent Then if the patient's response indicated difficulty with productivity, the following 7 items would be presented, referred to throughout as the "negative branch":
\begin{itemize}
    \item I dreaded getting out of bed in the morning to face the day
    \item It was difficult for me to complete my morning routine (i.e. brushing teeth; washing face; showering)
    \item It was difficult for me to complete basic chores (i.e. laundry)
    \item It was difficult for me to do physical exercise
    \item I had a plan that was impractical or unrealistic
    \item I changed my plans or moved onto new tasks before finishing what I was working on
    \item I tried to accomplish more tasks or plans than I usually would
\end{itemize}
\noindent Otherwise, the following 7 items would be presented, referred to as the "positive branch":
\begin{itemize}
    \item It was easy for me to get out of bed in the morning to face the day
    \item I completed my morning routine without difficulty (i.e. brushing teeth; washing face; showering)
    \item It was easy for me to complete basic chores (i.e. laundry)
    \item I had more energy than usual to do physical exercise
    \item My plans for the day were practical and realistic
    \item I only moved onto new tasks once I had finished what I was working on
    \item I accomplished more tasks or plans than I usually would
\end{itemize}
\noindent When initially prompting the survey, Beiwe would ask the participant to rate their agreement with each statement over the past 24 hours, using the app's radio button interface. 

\subsubsection{Daily diary recordings} 
The other survey given by Beiwe asked the patient to record an audio diary for up to four minutes, using the prompt “Please describe how you've felt over the last 24 hours in relation to any uncontrollable reoccurring thoughts and behaviours or significant events that may have occurred. Remember to focus on how the events made you feel, and to avoid names of specific people or places.”. Submitted daily audio diaries were later transcribed using the same service as the clinical interviews. 

\subsubsection{Wrist accelerometry} 
For a shorter portion of the study, the patient wore the GeneActiv actigraphy watch (ActivInsights, UK) for continuous wrist accelerometer data collection. This device collected x, y, and z axis accelerometry data at 40 Hz, along with temperature and light sensors to verify when the watch was being worn.

\subsection{Data analysis details}
\label{subsubsec:ocd-analysis}
Analyses were performed on the ERIS High Performance Computing (HPC) Linux cluster that is maintained by Partners Healthcare for use by researchers at affiliated hospitals. The cluster includes 240 compute nodes, most having 12 CPU cores with memory ranges from 48GB to 128GB RAM. This work was completed using Python 3 on a single compute node. Intermediate outputs were saved as CSVs, and all data management within the Python scripts was done using pandas. Other packages used for data processing were numpy (np) and scipy. Packages used for visualization were matplotlib and seaborn (sns). 

Metadata compilation and alignment of all datatypes in terms of days post-op was done similarly to the early stages of the audio diary pipeline described in chapter \ref{ch:1}. Day assignment for all passive data utilized the current date in Eastern Time (the patient's timezone), while EMA and audio journal submissions were assigned to days using the same 4 am cutoff used by the lab diary pipeline.

For each passive sensing datatype (phone accelerometer, GPS, phone usage, wrist accelerometer), I will first describe how low level features were extracted and why those features were chosen. I will then detail how the extracted features were summarized at different timescales -- primarily the day level for downstream analyses, but also the minute and hour levels for various visualizations. For the actively collected features (audio diaries, bi-weekly site visit recordings), I will similarly describe the low level feature extraction process and the resulting output values that were generated per recording.

Once the methods behind the extracted features are made clear, I will detail the final analysis techniques that used both these features and the ones previously described by \cite{Olsen2020} (LFP, clinical scales, EMA) as input. At that time, I will also explain exactly how all feature visualizations presented in this report were created. 

\subsubsection{Base feature extraction}
Initial phone accelerometer processing extracted low level features using a sliding window of 2 seconds in width that slid with 1 second of overlap. For each second then there was a bin that represented the feature values over the course of that second and the next second. At a 10 Hz sampling rate, this worked out to 20 samples per axis per bin to compute each feature. 

The following features were first computed independently for each axis on the second level, as described above: standard deviation of the accelerometer signal (np.nanstd), mean power spectral density (np.nanmean of powers returned by scipy.signal.periodogram), and spectral entropy (scipy.stats.entropy of powers returned by scipy.signal.periodogram, after norming them by dividing by the sum of powers). These features were selected based on prior use in the literature on accelerometry analysis for mental health, and (except where specified otherwise) they were then combined across axes using the root mean square, due to the common finding that axis-specific information has minimal relevance in the context of these simple activity level estimates \citep{Schmidt2018,Smets2018,Ghandeharioun2017}. Interpreting individual accelerometer axis data for a phone is also more difficult than for a watch, because the phone may be found in many different orientations relative to the individual, while a watch can be consistently worn in the same way. 

The other main features computed within these same second bins of the accelerometer data were the three pairwise Pearson correlations between the signals from each of the different axes (scipy.pearsonr). Axis correlation was included here because much of the prior literature mentioned was focused on wrist accelerometry data, and was not interested in identifying specific behaviors. We did not collect the appropriate data to do a deep dive on relationship between wrist and phone accelerometry or on identifying specific behaviors from phone data, but we did want to ensure that we computed some summary features that might be able to capture potential differences in types of motion. \\

Moving on to GPS, low level feature extraction took place on the minute timescale. A sliding window with length of 1 minute and no sliding overlap was used. Beiwe does GPS sampling only during some minutes ($\sim \frac{1}{3}$ of them), in order to preserve battery life. Note that if the participant ever turned off the location setting on their phone this would also prevent Beiwe from sampling GPS, even while other phone sensor modalities were still being collected. Any minute without GPS data available was marked with np.nan in the data frame of processing results. For minutes with data, the sample rate was consistent at 10 Hz. 

Within each active minute, all of the available longitude and latitude measurements were rounded to the nearest thousandth (i.e. 3 significant figures after the decimal), and then the mode of compiled coordinate pairs was taken to get a single location for the minute (scipy.stats.mode). The resolution of 3 sigfigs was chosen to maximize signal to noise -- consumer GPS is typically only accurate to $\sim 4$ sigfigs, and the fourth decimal place is not very robust to noise, which can lead to jumpiness in the recorded signal. As the fourth significant figure can separate out standard buildings but not location within a building, while the third can separate out areas of a couple blocks ($\sim 360$ foot radius), the importance of accurately estimating out to four digits will vary based on living environment \citep{Deakin2004}. 

This patient lived in a suburb and rarely visited the city besides sessions required for the study, so we preferred to confidently identify locations such as a neighborhood, shopping mall, or campus, without trying to assign exact addresses. In addition to enabling high accuracy with simple techniques, this approach also makes patient geolocation data less highly sensitive at an early stage in processing. Moreover, we wanted to isolate the most frequent discrete patient locations, which is why we used the mode of coordinates. However the mode could be much more skewed by noise in the GPS signal than a summary statistic like mean would be, so it was especially important to smooth the coordinates here.  

Altitude was also measured by the GPS, with reported accuracy to the 10s digit. On the minute level, altitude was treated similarly to the position coordinates, as it was rounded to the nearest 10 and then the mode over the minute was taken. To get an hourly altitude estimate, the mode over available minute altitude estimates was taken.

Note that we observed a large number of bizarre outliers here that were sustained over multiple minutes: for example altitude levels much higher than possible at the patient's home when the latitude and longitude indicated they were indeed still at home, as well as altitude levels varying large amounts at night when the phone accelerometer and usage measures suggested the patient was not touching the phone at all and likely asleep. Because of this and the substantially less prior information available about phone altitude sensing, we have little faith in the altitude features and only included one related summary stat in a handful of our final analyses. \\

The final phone modality, usage logging, had features directly extracted at both the minute and hour timescales. On the minute level, the same window bins were used as for GPS, and the features computed were mean battery level (np.nanmean of power column) and number of unlocks (pandas count of rows with the event type "unlock"). As mentioned, the phone use data contains timestamps, event type label, and current battery level, and only records the info when specific events occur. This means if the phone remains locked and untouched for a long period of time, there may be no phone use data for those many minutes - even while Beiwe is still actively recording other data. Like with GPS, minutes without data were marked with np.nan.

On the hour level, screentime was estimated via lock and unlock events. "Unlock" and "lock" events were isolated within that hour's CSV, and then the number of seconds elapsed between each unlock and subsequent lock event was calculated. During this process we also ensured that all unlock events were directly followed by a lock, and vice versa. However if the hour started with a lock event we assumed an unlock event at time 0, and if the hour ended with an unlock event we assumed a lock event at time 3600 (seconds). The seconds totals from across the screen use blocks were then summed, and finally that sum was divided by 3600 to report a fraction between 0 and 1 that reflected the amount of screentime during that hour. \\

The other source of passive data, wrist actigraphy, was only available for a short period of time towards the end of this study. Thus we simply applied existing lab code published by \cite{HabibSleep} to get an activity score for each hour with available data. The score ranges from 1 to 5 for least to most activity, and will instead return a 0 if the algorithm detects that the watch is not being worn. These data were used in some visualizations, but were not included in correlational analyses due to the window of the study they were extracted from. \\

For active data, natural language processing (NLP) features were extracted from the transcribed audio diaries using the linguistics portion of the pipeline already described in chapter \ref{ch:1}. The features considered here were the core diary-level outputs of submission time, word count, speech rate, word2vec uncommonness score, word2vec sequential incoherence score, and VADER sentiment. The keyword count tool was used here as well, with custom settings to track the number of mentions of "Adderall", and also to estimate the number of times that the DBS treatment was brought up by tracking counts for a set of words that were used by study staff when discussing stimulation parameters with the patient ("DBS", "cortical", "striatal", and any word beginning with the letters "stimula"). 

The other main active datatype was the video recordings from study site visits. For low level feature extraction, each video was run through OpenFace \citep{OpenFace} to obtain frame by frame facial action units (FAUs) and head pose. This is a well-validated tool for automatically producing facial action coding system labels, a framework for analyzing facial expressions that has been used extensively across psychology literature - in the past done by a systematic manual labeling process \citep{Ekman1978}. 

\subsubsection{Summary features}
All of the above features were further summarized at various timescales appropriate for use in visualizations and statistical analyses of interest.

For the phone accelerometer data, per minute features were obtained by taking the mean over available second features within that minute. Hourly features were similarly computed using the mean over minute features, and finally daily feature were obtained from the mean over hourly ones. The numpy nanmean function was used to compute the mean, so that missing data points were excluded from each mean. \\

The base geolocation features were more complicated, as the per minute coordinate estimates calculated are sensitive personal information that can't be directly used. For visualizations that directly required a minutely feature, we identified the coordinates of the patient's home property and used that to convert the longitude/latitude tuples to a binary feature indicating whether they were home or out at a given time. 

To create more descriptive hourly features, we returned to the extracted minute outputs from above, and tallied the number of minutes that each estimated coordinate set appeared in the data. There were only 20 unique locations that could be attributed to at least two hours of patient time. These 20 coordinates were looked up on Google Maps to determine the types of locations the patient frequented and assign them a category. The primary categories were the patient's "home", other residences (labeled "friends"), "shopping" centers, "religious" institutions, and the two hospitals that served as "study sites" for this trial. Any location coordinates besides these 20 that occurred were classified as "other". Note that the initial review of patient location was done at the end of the main study in late 2019, so if new locations became recurring for the participant during the COVID19 pandemic they would be marked as "other". 

Hourly features were then obtained from the category labels by taking the mode over the minutes in the hour of the location coordinates, and subsequently mapping each hour's coordinates to the map of location categories. The per hour category labels were used both directly in visualization, and finally to generate daily location summary features.

Four daily features were considered from GPS data. Two related directly to time spent at home, computed using different normalization techniques: one was the number of hours in the day where the hourly location category was "home" divided by 24 and another was the number of hours in the day where the hourly location category was "home" divided by the number of hours that had hourly location data available. The other daily location feature, number of unique locations visited, was derived from the personally identifiable raw GPS hourly features (mode of rounded coordinates) by counting the number of hours with a different approximate location assignment. 

The final daily GPS feature was extracted from the altitude information, which as mentioned above may not have been reliable. But as there was clinical reason to consider altitude (the patient sometimes did not leave the upstairs on bad days), we still attempted further investigation. Based on this prior information, the daily altitude feature was the variance in altitude observed, computed by using pandas to take the standard deviation of the hourly altitude estimates as defined above. \\

The final phone features - related to phone usage - had 3 total hourly and 2 total daily summary features computed. Percent screentime was extracted on the hour level to begin with, and battery level and unlock count features from the minute level were summarized per hour via the mean and sum respectively. On the day level, number of unlocks was computed by summing the hourly counts, and phone usage was estimated by the mean (np.nanmean) of the hourly screentime fractions available for that day.

Note that if a participant's phone was completely unused (including no charging or other events that would register in the log) over the course of an hour, there would be no CSV available in the phone usage category for that hour. In order to distinguish between true unavailability and simple lack of phone use, one could align the usage availability map with the accelerometer availability map, and mark any missing times that had accelerometer available as 0 phone use. However, marking these hours as missing in the hourly heatmap actually made it much easier to see when the participant very briefly checked their phone in the middle of the night versus not at all in the heatmaps of Figures \ref{fig:ocd-report}-\ref{fig:ocd-report-covid}. \\

To get daily watch activity estimates for days the participant wore the wrist accelerometer, the mean over the above described hourly activity scores was taken. For the audio diary features, the results were of course already on a daily timescale. 

On days with an interview recording available, the OpenFace features could also be considered. While there were relatively few days with interviews, the interviews themselves contain a rich amount of information that would need to be summarized. To obtain a single summary feature per interview, the activations of the action units comprising affect-related facial expressions as defined by \cite{Goodman} were summed in each frame of the video, and the mean affect activation over frames with a detected face was taken. 

\subsubsection{Creating visualizations}
All visualization elements were created using the matplotlib and seaborn packages in python 3.7. Figure components were exported as SVGs and then put together in Adobe Illustrator, with additional clarification on axis labels, legends, and other text details and highlighted markings done there, as well as fine tuning a consistent colorscheme. 

\begin{figure}[h]
\centering
\includegraphics[width=0.5\textwidth,keepaspectratio]{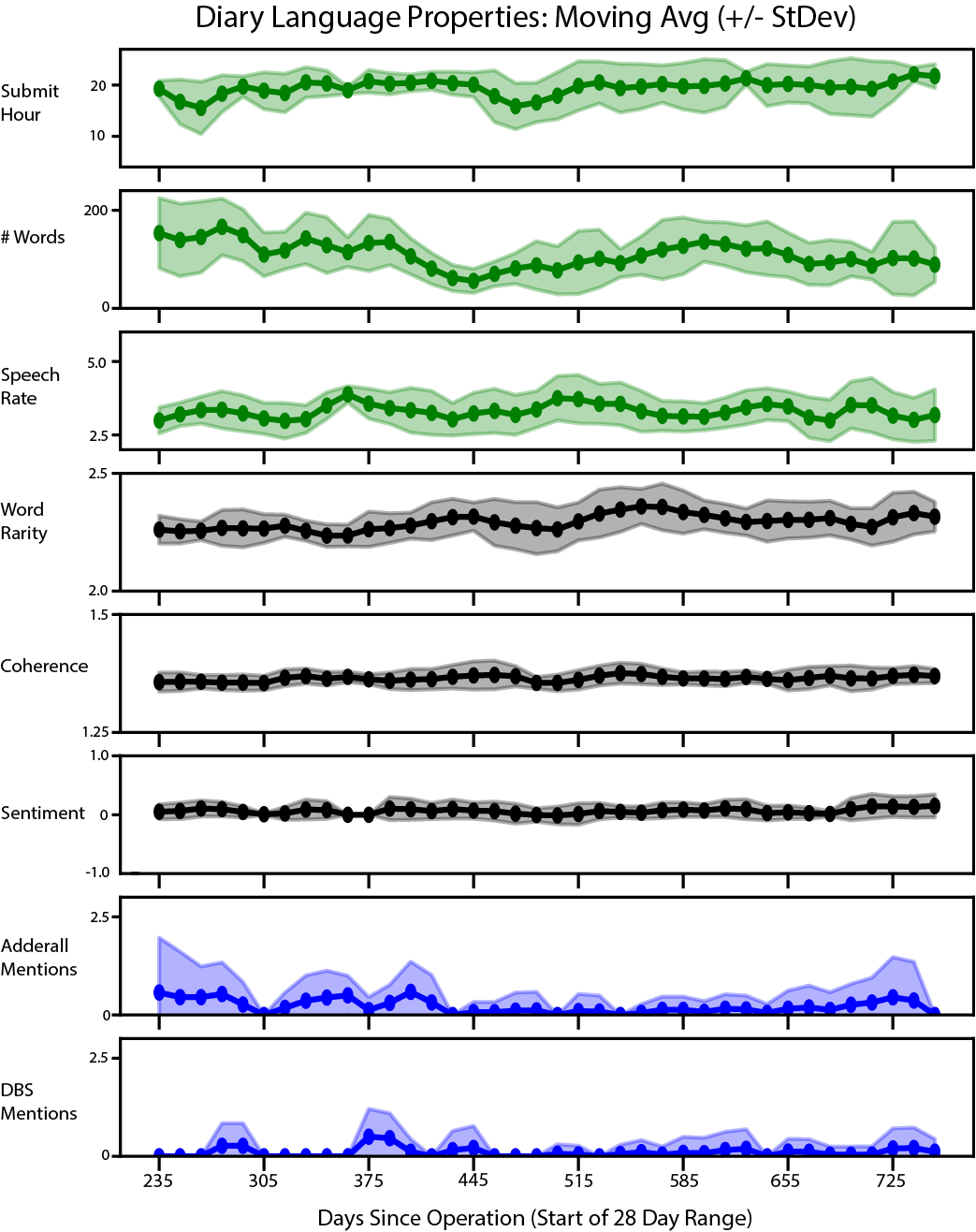}
\caption[Capturing trends in audio diary features over time.]{\textbf{Capturing trends in audio diary features over time.} The data presented in Figure \ref{fig:ocd-lang}a were smoothed over time to more easily spot changes across the study timeline. Here we show line plots of the moving average, with error shading representing the standard deviation, for our key diary features. Each point indicates the start of a 28 day window, which was slid over the data 14 days at a time.}
\label{fig:ocd-lang-line-plot}
\end{figure}

Heatmaps were created using the matplotlib imshow function with custom colormaps, to be described next. Area plots (Figure \ref{fig:ocd-report}-\ref{fig:ocd-report-covid}) were created using both the fill\_between and plot functions of matplotlib on the same day-level features; for the temporally dense passive digital phenotyping features, gaps were left in the plot for days with entirely missing data, while for rarer features available days were connected. Time course dot plots throughout were also created using the plot function, with marker argument instead of line. They were stacked for a given figure element using matplotlib's subplots functionality with linked x-axes. For the dot plots of Figure \ref{fig:ocd-lang}, daily features with a large number of zero-valued days (sentiment score and keyword counts) had those points left off for clarity. Similarly, the smoothed time course plots in Figures \ref{fig:ocd-smooth} and \ref{fig:ocd-lang-line-plot} used the plot function with x-linked subplots, to plot the described moving average values as dots at each time point and additionally connect them with lines. To include the variation in the same window as a shaded region, $+/-$ 1 standard deviation from the mean was then input to the fill\_between matplotlib function on the same plot. 

That covers all included time course visualizations. Other visualizations throughout included the scatter plots of select features against one another, which were done with seaborn's lmplot function using the linear fit line option, and correlation matrices, which were direct visualizations of correlation matrices computed in numpy, using matplotlib's imshow with a purple/green diverging colormap and absolute bounds of -1 to 1. There were then two additional visualizations that used other python packages - dendrograms (Figure \ref{fig:ocd-clusters}) and word clouds (Figure \ref{fig:ocd-lang}). The dendrograms were a direct output of the cluster.hierarchy.dendrogram function in the scipy package. The word clouds utilized the wordcloud python package, with word size corresponding directly to relative frequency of word use, and word color assigned on a bright red (-1) to black (0) to bright green (1) scale based on the mean sentence sentiment (computed by VADER) of sentences the word was included in. Word clouds for each period were generated using concatenated text from all patient diary transcripts from within that time range, after removing the default set of stop words as well as custom stop words corresponding to filler disfluencies and TranscribeMe markings (e.g. inaudible words), and a few other extremely common words used by this patient (today, day, yeah, went, got). \\

 For all heatmaps, grey was not included in the colormap and instead used to indicate missingness. The heatmaps of Figure \ref{fig:ocd-report} each have columns corresponding to days; they were generated over the same period and then aligned in Illustrator. EMA survey submissions (rows = questions, more severe responses correspond to higher numbers), phone usage (rows = hours, percent of hour with screen unlocked), and wrist accelerometry (rows = hours, integer activity level score) were colored using absolute bounds based on the minimum and maximum values of the feature in each cell. EMA used a diverging colormap where the lowest score corresponded to dark blue, the middle score to white, and the highest score to dark red; the other two used a sequential colormap from white to the respective color, with darker color corresponding to higher score. The LFP heatmap (rows = different power band and coherence features) also used a blue/white/red colormap, but with upper and lower bound corresponding to $+/- 3$ standard deviations from the mean of each feature, based on the patient's distribution. When zooming in to a shorter time period in Figure \ref{fig:ocd-report}, the patient's overall distribution was still used to set the bounds. The GPS and phone accelerometry heatmaps both had rows corresponding to hours like phone usage and wrist actigraphy, but each used a unique colorscheme. For GPS, hours were assigned a location category, and the discrete location category labels were then mapped to specific colors: blue for home, cyan for other residences (likely friends/family), green for malls/shopping centers, yellow for religious institutions, red for study site 1, orange for study site 2, and pink for other locations. For phone accelerometry, the mean power spectral density was taken separately for each axis (the only place where summary features were treated separately across axes), normalized to be between 0 and 1, and logarithmically scaled; the final values between 0 and 1 were then used like RGB pixel values for displaying an image in matplotlib, with \emph{x}/\emph{y}/\emph{z} axes corresponding to red/green/blue, respectively. For the scaling itself, I multiplied each cell by $133.3$ to get between 0 and 1, leading to the following final norm operation on each mean PSD cell $p$: $\frac{ln(1+255*(p))}{ln(256)}$, where ln is the natural log computed using math.log. Note Figures \ref{fig:ocd-report-stim} and \ref{fig:ocd-report-covid} are just versions of Figure \ref{fig:ocd-report} zoomed in on particular portions of the study timeline. 

The heatmaps of Figures \ref{fig:ocd-zoom}-\ref{fig:ocd-zoom-sup2} display minute-level feature values over an individual day, with columns corresponding to hours and rows corresponding to minutes within the hour. They were created using the same helper function as the previously mentioned heatmaps, with missing data marked in grey, as well as blue/white/red heatmaps. The phone movement heatmap used the standard deviation of the accelerometer signal over the minute (per axis and then RMS), with upper and lower bound determined by $+/- 3$ standard deviations from the mean of the minute-level distribution over the patient's entire dataset. The phone usage heatmap instead had absolute bounds between 0 and 100, representing the minimum and maximum possible phone battery levels (with missing data whenever the phone is going unused/untouched). The minute-level GPS feature was simply a binary classification of "at home" (red) or "out" (blue). \\

\noindent For more details on specific contents of each figure, see the corresponding legends. 

\subsubsection{Evaluating correlation structures}
For the clinical scales (MADRS and YBOCS independently) and the EMA (positive and negative branches independently), items were clustered based on their Pearson correlation with each other. The stats.pearsonr function in the scipy python package was used to obtain $r$ (and associated $p$) values for the patient's scores from each relevant question pair, to construct a correlation matrix. From the correlation matrix, a Pearson distance matrix was generated (where distance between the questions in each pairing is $1 - r$). The distance matrix was then input to the cluster.hierarchy.linkage function from scipy, with method setting 'complete' to obtain a hierarchical clustering result. From there a dendrogram image was generated, and the default cutoff point was used to define the clusters used in practice. To rate the clustering, scipy's spatial.distance.pdist was used together with cluster.hierarchy.cophenet to calculate the cophenetic correlation of the clustering result. The ordering of the dendrogram was then used to permute the correlation matrix for figure generation, and for each clinical outcome time point the mean item score for each cluster was calculated, to be used in downstream comparisons. These cluster scores were also evaluated over time and compared to more traditional clinical subscale groupings. Results can be found in Figures \ref{fig:ocd-clusters} and \ref{fig:ocd-smooth}.  

Note that the MADRS "subscale" question groupings used in Figure \ref{fig:ocd-clusters}c to compare against our data-driven patient-specific clusters were decided by a psychiatrist working on this project, based on a more general view of how clinicians might typically group MADRS items. This was done prior to any of the clustering analysis and was not based on this patient's data, though the results largely turned out similarly (with the primary difference being some of the vegetative-related questions clustering instead with pessimism/suicidality-related ones). The 4 "subscales" were labeled as:
\begin{itemize}
    \item Mood, corresponding to MADRS questions 1-3
    \item Vegetative, corresponding to MADRS questions 4-7
    \item Pessimism, corresponding to MADRS questions 8-9
    \item Suicidality, corresponding to MADRS question 10
\end{itemize}
\noindent Because the number of questions in these groupings is uneven, the score for each "subscale" was the mean of the scores of the contained items - just as was done for our derived cluster scores. By contrast, YBOCS subscales each contain the same number of items and are typically reported as a sum score, so we kept with that convention when presenting the YBOCS subscale comparison.

To revisit the correlational analysis of clinical scales and EMA self-report done by \cite{Olsen2020}, we performed Pearson correlation of same-day mean EMA summary score (as all questions were ultimately dealing with a similar topic of productivity/motivation) with each of the clinical cluster scores, along with repeating correlation of the YBOCS and MADRS sums with EMA summary. The comparisons of Figure \ref{fig:ocd-clusters}f thus use a significance cutoff of 0.005, Bonferroni corrected based on a total of 10 comparisons. 

Due to the highly focused nature of the EMA items, we primarily dealt with the EMA summary score rather than the clusters, whereas we used our clinical scale clusters extensively in downstream analyses. When EMA clusters were visualized or included in cross-modality analyses, we combined the \emph{exercise} and \emph{motivation} clusters on each branch that were obtained from our original splitting method, so that 4 total EMA clusters would be highlighted across the positive and negative branches. \\

To evaluate correlation structure of daily digital phenotyping and their relationship with clinical/self-report cluster scores, and similarly for daily LFP features and digital phenotyping features, the same scipy correlation method was used. In this case clustering was not performed, with features instead ordered based on the modality assessed, and thicker lines boxing in the diagonal blocks containing the within-category (i.e. digital phenotyping) relationships. To check for potential nonlinear rank relationships in these data, corresponding Spearman correlation matrices were also constructed, using scipy's stats.spearmanr function.

Significance was evaluated for each matrix using a $p$-value cutoff determined by Bonferroni correction on $0.05$. We took $\frac{n^{2} - n}{2}$ as the number of hypotheses tested, where $n$ was the number of features included in the matrix. This resulted in a significance threshold of $\frac{0.1}{n^{2} - n}$ for each such analysis. While the correction for multiple comparisons was not explicitly applied across all correlations tested in the paper, many of the matrices contained highly similar components (e.g. the upper left block diagonal of digital phenotyping internal correlations in Figure \ref{fig:ocd-dp-corrs}a and the upper left block diagonal of digital phenotyping internal correlations in Figure \ref{fig:ocd-neuro-corrs}b), and Bonferroni correction is a particularly conservative method to begin with. \\

One might notice that the bottom right diagonal block matrix of both Figure \ref{fig:ocd-dp-corrs}a and Figure \ref{fig:ocd-neuro-corrs}a should both show the internal correlation structure of the clinical cluster scores, yet are not exactly identical. This is because Figure \ref{fig:ocd-dp-corrs}a restricts the clinical scales to only dates where we also had passive data, while Figure \ref{fig:ocd-neuro-corrs}a restricts the scales to only the days we had neural data. Due to the differing start and stop times for these two collection protocols, \ref{fig:ocd-dp-corrs}a and \ref{fig:ocd-neuro-corrs}a contain somewhat different clinical data points. It is therefore promising that the general structure seen is extremely similar between the two. 

\subsection{Stimulation on/off experimental interview protocol}
\label{subsubsec:stim-exp-methods}
All of the above described data collection and processing was done longitudinally over the course of the study, and all analyses were correlational. To get some insight into causal effects of the stimulation paradigm, the participant additionally completed a series of structured interviews with cortical stimulation on or off, blinded to this status. Analysis of facial expression and language use over the course of the interviews was then used to identify potential behavioral changes. In this section, I will describe both the experimental protocol and the subsequent analyses. The audio-video recording and transcription procedure was identical to that used for the bi-weekly site visits. 

Note the experiment was conducted focusing only on cortical leads, as the participant did not consent to a parallel experiment with all stimulation turned off. We therefore ran this experiment only later in the study so that all settings tuning would be finished and cortical parameters would be stable. We were not able to do a repetition experiment as originally planned because of the timing of the COVID19 pandemic.

The experiment was designed to take $\sim 30$ minutes, and consisted of 8 double-blind trials. Cortical stimulation was on for half of the trials and off for half of the trials, in random order determined by running np.random.permutation on an array containing four 0s and four 1s. Within each trial, the same structured interview was given, after a 1 minute delay for stimulation change to "sink in". The stimulation settings were managed by an experimenter observing out of view of the patient and interviewer, who signaled the end of each waiting period using an alarmed timer. After each interview, the interviewer confirmed with the participant that they were ready to move on to the next trial, and then communicated to the experimenter to access the stimulation settings and set the cortical implants to the appropriate status (whether changing or repeating the previous state). Once the settings interface was closed, the experimenter started the waiting period timer as described. The auditory cues in the recording were used to mark timestamps for the start of each trial (one minute before the waiting period end), the end of the waiting period/start of the structured interview, the end of the structured interview/start of downtime, and the start of the next trial/end of the current one. After the last trial, stimulation settings were accessed again to set cortical stimulation on (whether it was already on or not), which served as a marker for the end of the experiment. 

At the start of the visit, the interviewer explained the structure of the experiment to the participant, and gave instructions for the interview portion. The participant was to rate a series of questions on a 1 to 5 scale, where 1 corresponded to strongly disagree and 5 to strongly agree, and would also be asked a handful of open-ended questions. All questions centered around their impression of the current stimulation status or their current feelings, as described using a few phrases they 
had commonly used in past open-ended interviews to explain their symptom severity. Because the patient was having trouble quickly mapping their agreement to the 1 to 5 scale, they were instructed to switch to explicitly stating their level of agreement with each question midway through the trials. Engagement with the rated responses was generally okay and the patient was talkative at times, but their responses to the specific open ended questions were very limited across trials.  

The exact questions asked in the structured interview were as follows, with each listed statement posed once in every trial. Each interview was intended to take $\sim 3$ minutes, and assess a few different feelings using positive, negative, and neutral framings.

\begin{itemize}
    \item Please rate the following statements about what happened in the last minute:
    \begin{itemize}
        \item The stimulation was just turned off
        \item The stimulation was just turned on
        \item There was no change in the stimulation
        \item There was no change in my mental energy level
        \item I just started feeling a bit less mental energy
        \item I just started feeling a bit more mental energy
        \item I just started feeling a bit less anxious
        \item I just started feeling a bit more anxious
        \item There was no change in my anxiety level
    \end{itemize}
    \item Can you describe any other thoughts or feelings that came to mind in the last minute?
    \item Please rate the following statements:
    \begin{itemize}
        \item Cortical stimulation is currently on
        \item Cortical stimulation is currently off
    \end{itemize}
    \item Have your feelings changed over the course of this discussion? If so, how?
\end{itemize} 

\noindent Answers to the stimulation-related interview questions were used to assign perceived stimulation status to each trial, to compare with true stimulation status and assist in downstream analyses. The ratings given for the introspective assessment questions were evaluated to see if there was a persistent difference in any feelings between the \emph{on} and \emph{off} trials. Opinions from the experiment observer were also collected to add a professional's subjective interpretations to the psychological reporting. 

\subsubsection{Analysis of interview recording}
A major purpose of this experiment was to run computational analyses on the audio and video recording, to uncover any ways in which the digital psychiatry behavioral features may be impacted by cortical stimulation status. For video, base feature extraction was performed in the same way it was with the bi-weekly clinical interviews, by obtaining frame-wise OpenFace \citep{OpenFace} features. Using the trial timestamps discussed above, each frame was labeled with a trial number and whether it took place during the 1 minute waiting period, the structured interview, or the end of trial downtime, based on the second in which it occurred.

To visualize facial expressivity over the course of the experiment, the mean of frame-wise action unit activations was taken in 30 second increments. With each 30 second bin as a column and each AU as a row, the above described heatmap function was utilized, with the same blue to red divergent colormap and bounds again based on $3$ standard deviations from the mean, for that action unit in that interview. Thicker bars were used to divide the different trials periods, and additional labeling about stimulation status was later added in Illustrator (Figure \ref{fig:ocd-video}).

Frame by frame activations for each AU pairing were also correlated across the set of interview frames taking place during \emph{on} trials, and separately across the set of interview frames taking place during \emph{off} trials. The Pearson correlations were calculated and visualized in the same way throughout the study. Here the ordering of the AUs (within the correlation matrix and the heatmap) was based on prior clustering results from OpenFace frame-wise AU activations across a much larger set of videos from a prior study including healthy interviewer and psychotic disorders interviewees \citep{Vijay2016}.

For the summed affect score computed for each frame described (based on the affect feature definition by \cite{Goodman}), as well as for the frame-wise activations for select AUs of particular interest (e.g. open mouth, eyes closed), distributions over the course of the interview were compared overall, in \emph{on} trials only, and in \emph{off} trials only. The smoothed distribution plots were created using seaborn's kdeplot (kde = kernel density estimate) function, and the \emph{on} versus \emph{off} distributions were statistically compared with a two sample Kolmogrov-Smirnov test (using scipy's stats.kstest function). Kernel density estimate plots were additionally made to compare the frame-wise affect score distribution separately across each individual trial. \\

Further, the transcription produced from the audio was used for a few analyses, besides just easily identifying the trial period delimiters. Transcript sentences were labeled with trial number and segment type using the same methodology as the video frames, now considering the sentence start timestamp provided by TranscribeMe. Because TranscribeMe was told for this interview to explicitly transcribe the specified auditory experiment cues as separate sentences, there was no issue with a sentence spanning multiple segments. These "sentences" that marked e.g. the alarm going off were filtered out before analysis. 

Patient verbosity was measured by automatically counting words (defined by splitting on spaces) from transcript sentences within each trial, both overall and broken down by trial portion (initial waiting period, interview, cool down period). Patient sentence sentiment scores were computed using the VADER python package and subsequently input to the same word cloud generation function as detailed above, to create separate word clouds of impromptu (i.e. not during the structure interview portion) speech when stimulation was \emph{on} versus \emph{off} (Figure \ref{fig:ocd-video}). \\

\noindent Tables with additional information about the interview responses by the participant in each trial can be found in section \ref{subsec:interview-answers}.

\subsection{Data security}
\label{subsec:ocd-irb}
In accordance with McLean guidelines, participant confidentiality was ensured throughout the study. All data was stored on a secure Partners Healthcare server, and all personally identifiable data was additionally encrypted using the Harvard Research Computing cryptease package. Data obtained was accessible only to the research staff listed on the approved protocol. All devices used to access any study data were encrypted and connected to Partners network (either directly or through VPN), by hospital policy. All server transactions were transferred over a public-private key encrypted channel, based on the HTTPS/TLS web protocol. Coded data stored on the secure server was only accessible to trained researchers, and secured with a firewall and other security measures. All software written by the lab was backed up and documented via the internal Partners Gitlab, to maintain good organizational practices while maintaining security.

\section{Smoothed diary feature timecourses (supplemental figure)}

\FloatBarrier

\section{Additional stimulation experiment results (supplemental table)}
\label{subsec:interview-answers}

\begin{table}[!htbp]
\centering
\caption[Participant responses across trials in the causal cortical stimulation interviews experiment.]{\textbf{Participant responses across trials in the causal cortical stimulation interviews experiment.} For each structured question posed in each of the 8 interview trials, it was recorded whether the participant agreed with the statement (Agr), disagreed with the statement (Dis), or expressed uncertainty (?). This table presents that response for each statement (rows) in each trial (columns). The last row notes whether the true cortical stimulation status was on or off. Note the participant believed the opposite of the true stimulation status in 7 of the 8 trials.}
\label{table:ocd-responses}

\begin{tabular}{ | m{5cm} || m{0.75cm} | m{0.75cm} | m{0.75cm} | m{0.75cm} | m{0.75cm} | m{0.75cm} | m{0.75cm} | m{0.75cm} |}
\hline
\textbf{Statement posed} & \textbf{Trial 1} & \textbf{Trial 2} & \textbf{Trial 3} & \textbf{Trial 4} & \textbf{Trial 5} & \textbf{Trial 6} & \textbf{Trial 7} & \textbf{Trial 8} \\
\hline\hline
Stimulation just turned off & Agr & ? & Dis & Agr & Agr & Dis & Dis & Dis \\
\hline
Stimulation just turned on & Dis & ? & Agr & Dis & Dis & Agr & ? & Dis \\
\hline
No stimulation change & Dis & Dis & Dis & Dis & Dis & Dis & Agr & ? \\
\hline
No mental energy change & Dis & Dis & Dis & Dis & Dis & Dis & Dis & Dis \\
\hline
Less mental energy & Agr & Dis & Dis & Agr & Agr & Dis & Dis & Agr \\
\hline
More mental energy & Dis & Agr & Agr & Dis & Dis & Agr & ? & Dis \\
\hline
Less anxious & ? & Agr & ? & Dis & Dis & ? & Agr & Dis \\
\hline
More anxious & Dis & Dis & ? & ? & Agr & ? & Dis & Agr \\
\hline
No anxiety change & ? & Dis & ? & Dis & Dis & Dis & ? & Dis \\
\hline
Cortical stimulation is on & Dis & Agr & Agr & Dis & Dis & Agr & Dis & Dis \\
\hline
Cortical stimulation is off & Agr & Dis & Dis & Agr & Agr & Dis & Agr & Agr \\
\hline\hline
True stimulation status & On & Off & Off & Off & On & Off & On & On \\
\hline
\end{tabular}
\end{table}

\FloatBarrier

\section{Supervised detection of specific behaviors of interest in wrist-worn accelerometry data from a mock participant}
\label{sec:lacey}

\begin{figure}[h]
\centering
\includegraphics[width=\textwidth,keepaspectratio]{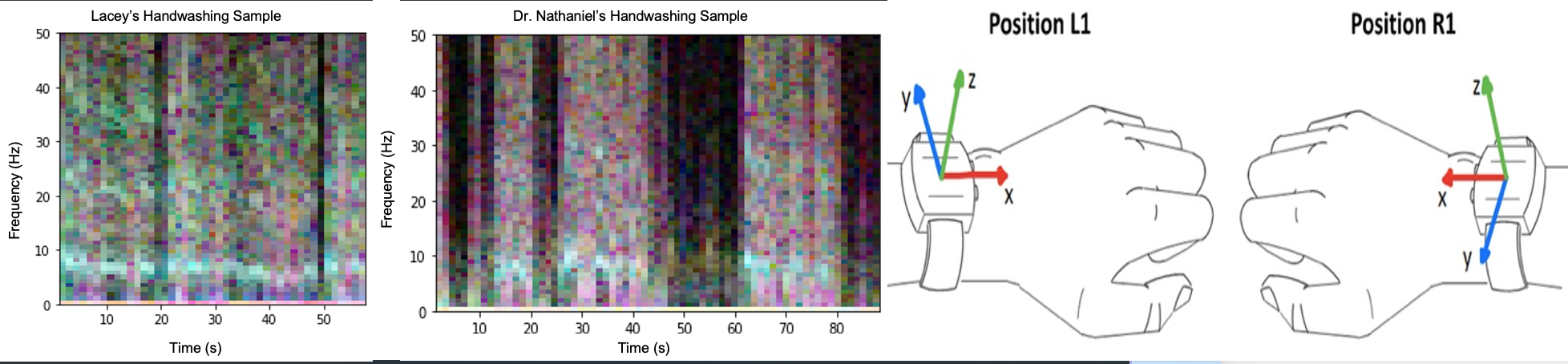}
\caption[Rhythmic wrist movements during hand washing are clearly visible in actigraphy spectrograms.]{\textbf{Rhythmic wrist movements during hand washing are clearly visible in actigraphy spectrograms.} Example spectrograms generated from wrist-worn accelerometry data that was recorded using the GeneActiv watch while two different individuals were washing their hands (left). Individual spectrogram matrices were created for each axis of the wrist accelerometer, and then composed as an RGB image with the \emph{x}-axis mapping to red, the \emph{y}-axis to green, and the \emph{z}-axis to blue (right). For both individuals, there was a clear rhythmic movement at frequency ranging between 5 and 10 Hz during active hand washing periods, and this was by far the most powerful signal in the frequency space for the behavior. It also involved movement with the same rhythm across multiple axes, with some component from all 3 axes and especially strong signal for \emph{y} and \emph{z}. Hand washing is therefore plausibly detectable with a very simple algorithm using actigraphy spectrogram inputs, a hypothesis that was explored further using bilateral continuous actigraphy recordings with labeled grooming behaviors from a third individual (Figure \ref{fig:handwash-pred}).}
\label{fig:handwash-spec}
\end{figure}

\begin{figure}[h]
\centering
\includegraphics[width=\textwidth,keepaspectratio]{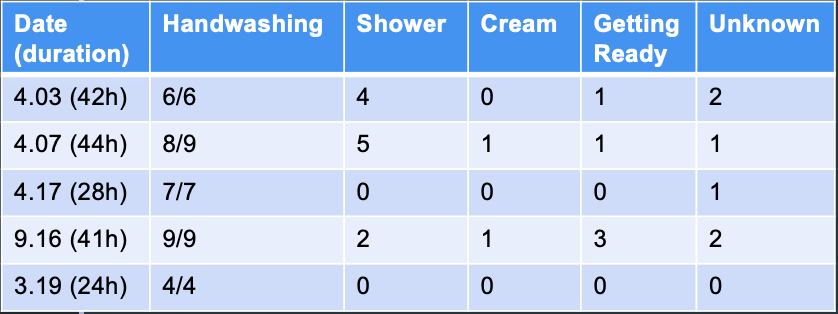}
\caption[Hand washing periods are accurately detectable within continuous naturalistic actigraphy recordings via simple algorithm (n of 1 pilot).]{\textbf{Hand washing periods are accurately detectable within continuous naturalistic actigraphy recordings via simple algorithm (n of 1 pilot).} Wrist accelerometry data were continuously collected from both wrists in a sample control participant for 1 to 2 days at a time, using the basic GeneActiv watch with 40 Hz sampling rate. This was repeated for 5 distinct time periods, and during each an activity log was kept, taking particular care to record the start and stop times of all grooming-related behaviors with minute resolution. A couple example hand washing clips from a different day were referenced in conjunction with the examples from the other two individuals depicted in Figure \ref{fig:handwash-spec} to come up with a hand washing algorithm for broader testing. A threshold on spectral power in the $\sim 5-10$ Hz band (shown in Figure \ref{fig:handwash-spec}) to enforce bilaterally was determined based on the personal training examples, and then a rolling window algorithm was applied to identify contiguous minutes that contained suspected hand washing. Of 35 documented hand washing occurrences in the test set, 34 were detected by the algorithm, an impressive false negative rate. There were also 24 other times flagged by the algorithm that were not hand washing, but given the full dataset spanned 8 entire days of recording, this is a minuscule false positive rate. The positive predictive power (PPV) was just okay in official terms at around 60\%, but looking at the behaviors that were documented to be happening during false positive "hand washing" it is plausible that the true PPV is even better -- as nearly half occurred while showering, an activity that probably involves hand washing. Moreover, this algorithm worked quite well to detect normal daily hand washing, and ritualized hand washing has some properties that could make it easier to detect automatically in actigraphy data.}
\label{fig:handwash-pred}
\end{figure}

The detection of grooming behaviors relevant for OCD rituals in actigraphy signal introduced at the end of the main chapter \ref{ch:3} (\ref{subsubsec:ocdi}) is elaborated on here, with promising results from a pilot work to isolate hand washing times automatically. The methods used and outcomes observed are presented via the Figures \ref{fig:handwash-spec} and \ref{fig:handwash-pred}.

\FloatBarrier

\subsection{Potential for exploratory tinkering with personal (and pet?) actigraphy signals}
In addition to holding much promise for quantification of symptomatology in serious psychiatric illness, wrist-worn accelerometry has many potential use cases of interest. Obviously exercise tracking and basic physical activity monitoring is a large mainstream one with relevance to public health, but other uses have emerged in the realm of fitness and sports. This includes tracking of activity during practices to inform strength and conditioning programs at the professional level, labeling of game outcome related information for hobbyists (e.g. collecting stats about golf strokes), and even applications in youth athletics. The latter is particularly interesting -- not because people need to take childhood sports team performance so seriously, but because it does present a real educational opportunity if the kids were able to use their own data directly. With modern education typically so formalized and ultimately pre-packaged, and modern childhood schedules so ridiculously packed, we are in great need of situations where scientific and computational skills can be learned organically and actually for fun. \\

\noindent An entirely separate market for fun accelerometer analysis is pet tracking. I think there are plenty of people out there who would legitimately buy a "Fitbit for dogs" if one existed. But I mostly bring this up so I can include a picture of my dog (Figure \ref{fig:juke}).

\pagebreak

\begin{FPfigure}
\centering
\includegraphics[width=0.5\textwidth,keepaspectratio]{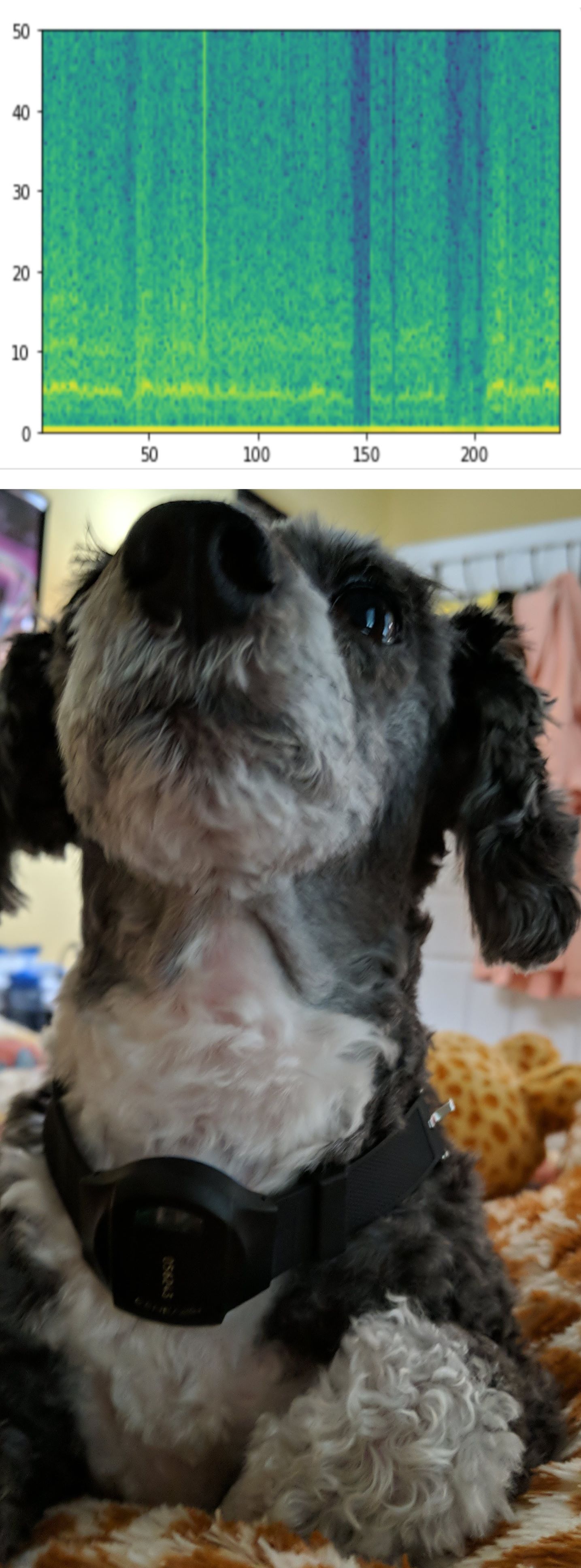}
\caption[My dog.]{\textbf{My dog.} One day during COVID, Juke wore the GeneActiv watch attached to his collar for a little while (bottom). He went on a walk around the block, and you can see in a spectrogram snippet from this time (top) the frequency the watch was bouncing up and down as he walked, and the time periods where he stopped to sniff something. An enlarged version of this spectrogram is also shown in Figure \ref{fig:juke-spec}. Juke is turning 16 this August, and still sprints around the house like a puppy once or twice a day, so that's pretty impressive.}
\label{fig:juke}
\end{FPfigure}

\begin{figure}[h]
\centering
\includegraphics[width=\textwidth,keepaspectratio]{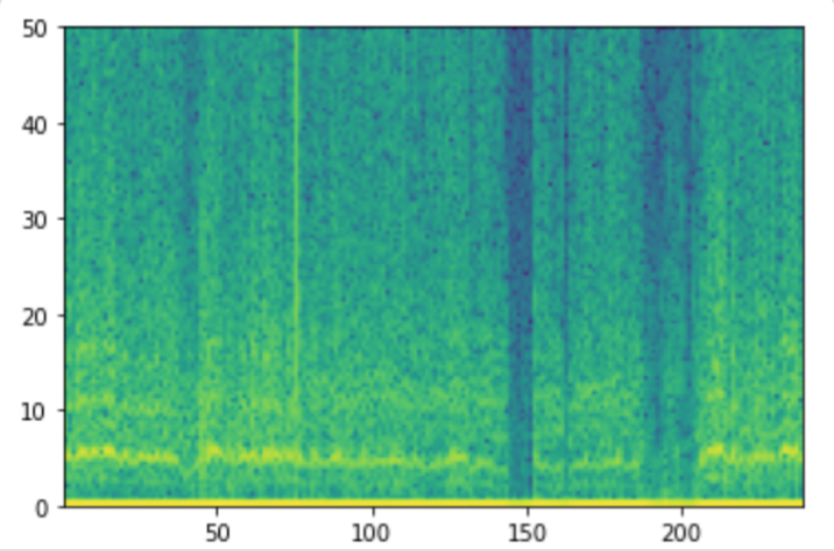}
\caption[Characterizing Juke's walk around the block.]{\textbf{Characterizing Juke's walk around the block.} A closer look at the spectrogram shown in Figure \ref{fig:juke}}
\label{fig:juke-spec}
\end{figure}

\FloatBarrier

\chapter{Supplement to Chapter \ref{ch:4}}\label{cha:append-chapt-refch:4}
\renewcommand\thefigure{S4.\arabic{figure}}    
\setcounter{figure}{0}  
\renewcommand\thetable{S4.\arabic{table}}    
\setcounter{table}{0}  
\renewcommand\thesection{S4.\arabic{section}} 
\setcounter{section}{0}

Here I provide further details on the experimental methodology and results from chapter \ref{ch:4}, as well as some additional mathematical details. \\

\noindent Figure \ref{figure:architecture} below gives a 
more in depth diagram of what both of our architectures actually look like for a small example network. \\

\begin{figure}[h]
\centering
\includegraphics[width=\textwidth,keepaspectratio]{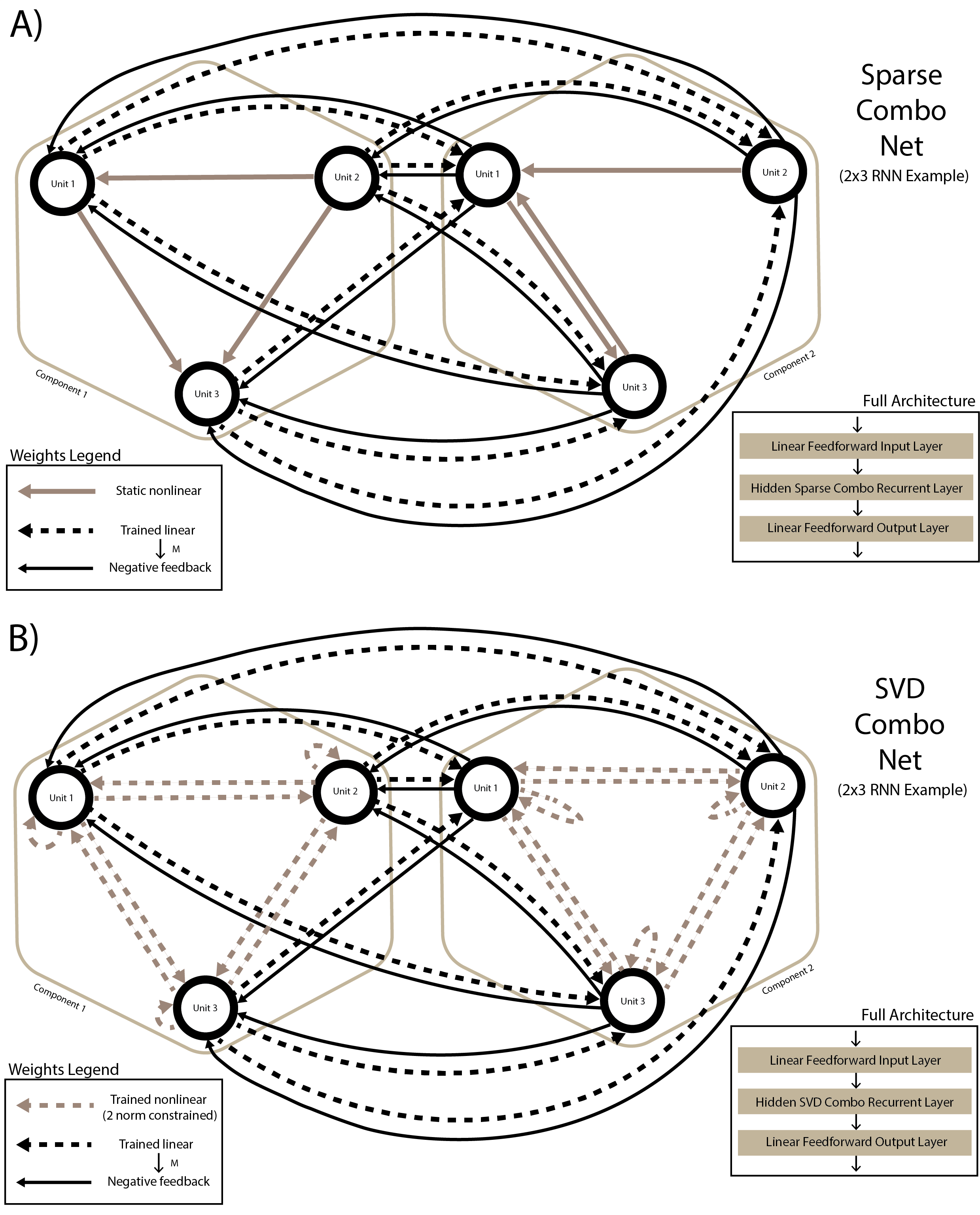}
\caption[Detailed network architecture diagrams.]{\textbf{Detailed network architecture diagrams.} Depicted are a $2 \times 3$ Sparse Combo Net (A) and a $2 \times 3$ SVD Combo Net (B). The main diagrams show the detailed "RNN of RNNs" weight connections, while the small inset boxes on the right show how the networks are fit for the benchmark tasks}
\label{figure:architecture}
\end{figure}

\noindent The subsequent experimental sections will focus solely on the Sparse Combo Net, with full information needed to replicate the methods and a complete log of results from all networks trained in my hands. \\

\noindent At the end of this chapter's appendix, I also provide an extended proof to bolster the claims of Theorem \ref{theorem: Wdiagstabcounterexampletheorem}. I then reproduce the proof of Theorem \ref{theorem: absolutevaluetheorem} that is published in our NeurIPS paper, as an alternative proof of that theorem is given in the main body of my thesis. See supplemental section \ref{sec:ch4-explain} for an explanation on the two different proofs of our theorem. 

\FloatBarrier

\section{Experimental procedures supplement}
\label{sec:experiments-ch4-supp}
\subsection{Code and data availability}
All Sparse Combo Nets reported on in the main text were trained using a single GPU on Google Colab. Code to replicate all experiments can be found via the GitHub repository at \citep{sparsegit}. Runtime for the best performing architecture settings on the sequential CIFAR10 task was $\sim 24$ hours. Colab Pro+ was used to limit timeouts and increase access to quality GPUs. Multiple accounts were created to run repeat experiments in parallel. Note that without a Pro+ account, the shared notebook would be very difficult to run to completion directly in Colab. However the shared notebook can be easily set up to run on a personal computer with GPU or a different cloud compute platform of choice if desired.

The base datasets we used for our tasks were MNIST and CIFAR10, downloaded via PyTorch. MNIST is a handwritten digits classification task, consisting of 60,000 training images and 10,000 testing images (each 28x28 and grayscale). It is made available under the terms of the Creative Commons Attribution-Share Alike 3.0 license. CIFAR10 is a dataset of 32x32 color images, split evenly among the following 10 classes: airplane, automobile, bird, cat, deer, dog, frog, horse, ship, and truck. It also contains 60,000/10,000 training/test images, and is distributed under the MIT License. The use of these datasets is included in the above link to our code, where they are processed to be presented to the network in pixel-by-pixel form (and first permuted where applicable) as described in the main text. 

\subsection{Network settings and hyperparameter tuning}
Note that across the whole paper, Sparse Combo Net was run with $\tau = 1$ and $dt = 0.03$, such that after each timestep (i.e. after each pixel was presented), the current state was updated to be the previous state $+ 0.03 * f(x)$, where $f(x) = -x + \mathbf{W}*\phi(x) + \mathbf{L}*x + u(t)$ with $x$ corresponding to the previous state of the neurons and $u(t)$ corresponding to the values returned by the linear input layer when given the pixel that was just input to the larger network. $\mathbf{W}$ and $\mathbf{L}$ are of course the \emph{intra-} and \emph{inter-} subnetwork weights respectively, as described in the main text -- so they are static throughout this process of evaluating an individual image sequence. Recall that before each new sequence, network starting state was reset.

At initialization, the nonlinear RNN weights for Sparse Combo Net were always randomly generated based on given sparsity ($s$) and entry magnitude ($z$) settings, and then confirmed to meet the Theorem \ref{theorem: absolutevaluetheorem} condition (or discarded and sampling repeated if not) -- as described in the main text. Uniform sampling was done for non-zero entries using scipy.sparse.random. For various potential $s$ and $z$ settings, I quantified the likelihood that a generated $\mathbf{W}$ would satisfy Theorem \ref{theorem: absolutevaluetheorem}, to assist in deciding on starting settings. As the random sampling and condition checking process was extremely fast at the sizes we were considering, we preferred to select a density that would result in a relatively small percentage of usable subnetworks for early testing. Of course this is also dependent on subnetwork size, as larger subnetworks are both able to be less dense in a meaningful way and required to be less dense in order to maintain the stability condition. Nevertheless, I used this info to choose $40\%$ density with weights between $-0.4$ and $0.4$ to start performing experiments with. These settings were selected as an early baseline because they resulted in $\sim1\%$ of randomly sampled $16$ by $16$ weight matrices meeting the Theorem \ref{theorem: absolutevaluetheorem} condition. \\

With base settings out of the way, tuning of typical training hyperparameters (initial learning rate, number of epochs, learning rate schedule, Adam weight decay) was done primarily with $10 \times 16$ combination networks on the permuted seqMNIST task, starting with settings based on existing literature on this task, and verifying promising settings using a $15 \times 16$ network (Table \ref{table:hyperparams}). The described initialization settings were held the same throughout tuning. 

Once hyperparameters were decided upon, the trials reported on in the main text began. Most of these experiments were done on permuted seqMNIST, where we characterized performance of networks with different sizes and sparsity levels/entry magnitudes. When we moved to the sequential CIFAR10 task, we began by simply training with the same best settings that were found from these experiments. After achieving test accuracy that was already SOTA for stable RNNs on the first run, we did try a few hyperparameter tweaks, but ultimately did not change the settings for our final runs. The results of all attempted trials are reported in section \ref{Appendix:all-tables}. \\

As part of the experiments on the subnetwork sparse initialization settings - after the early results demonstrating a large performance boost for sparser networks (with higher entry magnitude, to a point) - I revisited the initialization settings to better dissect what properties might be contributing to the performance variance. Specifically, I split the previously described magnitude setting $z$ into two different scalars: one ($z_{1}$) used the same way as $z$, i.e. to initialize the non-zero matrix entries between $-z$ and $z$ before checking the matrix against the Theorem \ref{theorem: absolutevaluetheorem} condition, and one ($z_{2}$) applied to all entries after a matrix was actually selected. Thus each subnetwork now had non-zero entry magnitude functionally sampled between $-(z_{1}*z_{2})$ and $z_{1}*z_{2}$, but would need to be contracting per Theorem \ref{theorem: absolutevaluetheorem} even if it were divided by $z_{2}$, with $z_{2}$ of course being $> 0$ and $\leq 1$.

I chose to separate the settings in this way after noticing that at a $5\%$ sparsity level, random $32 \times 32$ weight matrices met the Theorem \ref{theorem: absolutevaluetheorem} condition roughly $1\%$ of the time, whether the original magnitude setting $z$ was $10$ or $100000$. Conversely, $\sim 85\%$ of sampled matrices using scalar $10$ would continue to meet the condition even if multiplied by a factor of $10000$. These numbers could vary quite a bit depending on the sparsity level, but ultimately when testing moderately sparse matrices I wanted a mechanism that could focus selection on those that are stable due to the structure of their weights and not due to magnitude constraints. At the same time, it is not feasible in practice to train a network with such massive weights, even when sparse. This style of selection process was then used throughout the final repeatability experiments.

\subsection{Trainable parameters accounting}
To report on the number of trainable parameters in Sparse Combo Net, we used the following formula:

$\frac{n^{2} - M*C^{2}}{2} + i * n + n * o + n + o$

\noindent Where $n$ is the total number of units in the $M \times C$ combination network, $o$ is the total number of output nodes for the task, and $i$ is the total number of input nodes for the task. Thus for the $16 \times 32$ networks highlighted, we have 129034 trainable parameters for the MNIST tasks, and 130058 trainable parameters for sequential CIFAR10.

Note that the naive estimate for the number of trainable parameters would be $n^{2} + i * n + n * o + n + o$, corresponding to the number of weights in $\mathbf{L}$, the number of weights in the feedforward linear input layer, the number of weights in the feedforward linear output layer, and the bias terms for the input and output layers, respectively. However, because of the combination property constraints on $\mathbf{L}$, only the lower triangular portion of a block off-diagonal matrix is actually trained, and $\mathbf{L}$ is then defined in terms of this matrix and the metric $\mathbf{M}$. Thus we subtract $M*C^{2}$ to remove the block diagonal portions corresponding to nonlinear RNN components, and then divide by 2 to obtain only the lower half. 

\section{Additional experimental results}
\subsection{Supplemental figures}
\begin{figure}[h]
\centering
\includegraphics[width=0.8\textwidth,keepaspectratio]{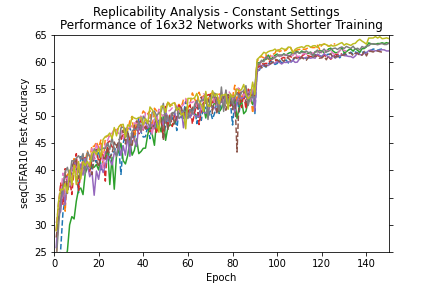}
\caption[Sparse Combo Net performance on additional seqCIFAR10 tasks with shorter training time.]{\textbf{seqCIFAR10 performance on repeated trials with shorter training time (done to complete more trials).} Nine different $16 \times 32$ networks with 3.3\% density and entries between -6 and 6 were set up to train for 150 epochs, with learning rate divided by 10 after epochs 90 and 140. Most of these networks hit runtime limit before completing, however they all got through at least 100 epochs and all had test accuracy exceed 61\%. This figure depicts test accuracy for each of the networks over the course of training. Networks that completed training are plotted as solid lines, while those that were cut short are dashed.}
\label{figure:cifar-reps-shorter}
\end{figure}

As we encountered difficulties with Colab GPU assignment when we started working on these repetitions, we also trained nine networks over a smaller number of epochs (Figure \ref{figure:cifar-reps-shorter}). These networks were trained using the 150 epoch paradigm previously described, although only four of the nine completed training within the 24 hour runtime limit. Complete results for these trials can be found in Table \ref{table:cifar-repeats-short}. Mean performance among the shorter training trials was $62.82\%$ test accuracy with variance $= 0.95$. \\

\begin{figure}[h]
\centering
\includegraphics[width=0.8\textwidth,keepaspectratio]{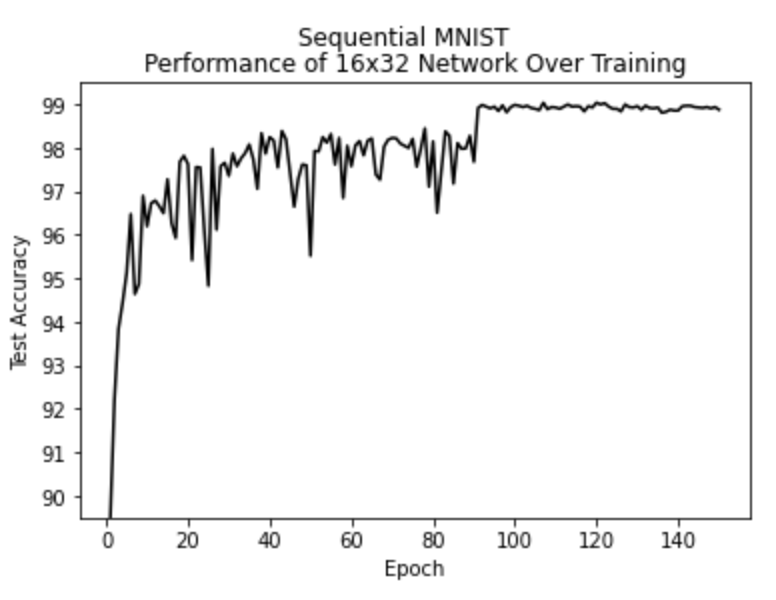}
\caption[Performance over training on the seqMNIST task.]{\textbf{Performance over training on the seqMNIST task.} This shows test accuracy over time of a $16 \times 32$ Sparse Combo Net with best settings (using 150 epoch training protocol). Final test accuracy exceeded 99\%.}
\label{figure:seqmnist}
\end{figure}

\noindent To complete our benchmarking table, we also ran a single 150 epoch trial of our best network settings on the sequential MNIST task. Test accuracy over the course of training for this trial is depicted in Figure \ref{figure:seqmnist}.

\FloatBarrier

\subsection{Tables of results by trial}\label{Appendix:all-tables}
Results from across all Sparse Combo Net tests are documented in the following tables:
\begin{itemize}
    \item Table \ref{table:hyperparams} shows all trials run on permuted sequential MNIST before beginning the more systematic experiments reported on in the main text. Notably, our networks did not require an extensive hyperparameter tuning process. 
    \item Tables \ref{table:sizes-compare} and \ref{table:modules-compare} report additional details on the initial size and modularity experiments (Figure \ref{figure:test-sizes-sup}). Table \ref{table:compare-sizes-again} reports the results of the repeated experiment on Sparse Combo Net size, this time using 32 unit component subnetworks with best sparsity settings (Figure \ref{figure:test-sizes}A).
    \item Tables \ref{table:sparsity-compare} and \ref{table:best-results} report results from all trials related to our sparsity experiments (Figures \ref{figure:test-sparsity} and \ref{figure:test-sparsity-sup}). 
    \item Table \ref{table:psmnist-repeats} provides further information on the four trials in the permuted seqMNIST repeatability experiments (Figure \ref{figure:test-reproduce}). 
    \item Table \ref{table:cifar} reports the results of all trials of different hyperparameters on the sequential CIFAR10 task, in chronological order. Ultimately the same settings as those used for permuted seqMNIST were chosen.
    \item Table \ref{table:cifar-repeats} shows the results of all ten trials in the seqCIFAR10 repeatability experiments (Figure \ref{figure:cifar-reps}). Table \ref{table:cifar-repeats-short} shows results from nine additional seqCIFAR10 trials of shorter training duration (Figure \ref{figure:cifar-reps-shorter}), run to increase sample size while unable to access the higher quality Colab GPUs.  
    \item Finally, Table \ref{table:scalability} reports the results of our pilot trial on introducing sparsity into the linear feedback connection matrix - as this table was presented in the main text of chapter \ref{ch:4} however, we do not reproduce it here.
\end{itemize}

\subsubsection{Permuted seqMNIST trials}

\begin{table}[!htbp]
\centering
\caption[Permuted seqMNIST hyperparameter tuning trials.]{\textbf{Training hyperparameter tuning trials, presented in chronological order.} * indicates that training was cut short by the 24 hour Colab runtime limit. LR Schedule describes the scalar the learning rate was multiplied by, and at what epochs. The best performing network is highlighted, and represents the training settings we used throughout most of the main text.}
\label{table:hyperparams}
\begin{tabular}{ | m{2cm} | m{1.25cm} | m{1cm} | m{1cm} | m{4cm} | m{1cm} | }
\hline
 Size & Epochs & Adam WD & Initial LR & LR Schedule & Final Test Acc. \\
\hline\hline
$10 \times 16$ & 150 & 5e-5 & 5e-3 & 0.1 after 91 & 84\% \\
\hline
$10 \times 16$ & 150 & 1e-5 & 1e-2 & 0.1 after 50,100 & 85\% \\
\hline
$15 \times 16$ & 150 & 2e-4 & 5e-3 & 0.1 after 50,100 & 84\% \\ 
\hline
$10 \times 16$ & 150 & 2e-4 & 1e-2 & 0.5 every 10 & 81\% \\ 
\hline
$10 \times 16$ & 200 & 2e-4 & 1e-2 & 0.5 after 10 then every 30 & 81\% \\ 
\hline
$10 \times 16$ & 171* & 5e-5 & 1e-2 & 0.75 after 10,20,60,100 then every 15 & 84\% \\ 
\hline
\rowcolor{LightCyan}
$15 \times 16$ & 179* & 1e-5 & 1e-3 & 0.1 after 100,150 & 90\% \\ 
\hline
\end{tabular}
\end{table}

\begin{table}[!htbp]
\parbox{.45\linewidth}{
\centering
\caption[Network size experiment trials.]{\textbf{Results for combination networks containing different numbers of component 16-unit RNNs.} Training hyperparameters and network initialization settings were kept the same across all trials, and all trials completed the full 150 epochs.}
\label{table:sizes-compare}

\begin{tabular}{ | m{1cm} || m{1cm} | m{1cm} | m{1cm} | }
\hline
 Size & Final Test Acc. & Epoch 1 Test Acc. & Final Train Loss \\
\hline\hline
$1 \times 16$ & 38.69\% & 24.61\% & 1.7005 \\  
\hline
$3 \times 16$ & 70.56\% & 40.47\% & 0.9033 \\
\hline
$5 \times 16$ & 77.86\% & 47.99\% & 0.7104 \\  
\hline
$10 \times 16$ & 85.82\% & 61.38\% & 0.4736 \\
\hline
$15 \times 16$ & 90.28\% & 69.09\% & 0.3156 \\  
\hline
$20 \times 16$ & 92.26\% & 71.72\% & 0.2392 \\
\hline
\rowcolor{LightCyan}
$22 \times 16$ & 93.01\% & 70.11\% & 0.2073 \\
\hline
$25 \times 16$ & 92.99\% & 61.81\% & 0.2017 \\  
\hline
$30 \times 16$ & 93.16\% & 43.21\% & 0.1991 \\
\hline
\end{tabular}}
\hfill
\parbox{.45\linewidth}{
\centering
\caption[Network modularity experiment trials.]{\textbf{Results for different distributions of 352 total units across a combination network.} This number was chosen based on prior $22 \times 16$ network performance. For each component RNN size tested, the same procedure was used to select appropriate density and scalar settings. All networks otherwise used the same settings, as in the size experiments.}
\label{table:modules-compare}

\begin{tabular}{ | m{1cm} || m{1cm} | m{1cm} | m{1cm} | }
\hline
 Size & Final Test Acc. & Epoch 1 Test Acc. & Final Train Loss \\
\hline\hline
 $1 \times 352$ & 40.17\% & 26.97\% & 1.662 \\  
\hline
 $11 \times 32$ & 89.12\% & 61.29\% & 0.3781 \\
\hline
\rowcolor{Gray}
 $22 \times 16$ & 93.01\% & 70.11\% & 0.2073 \\ 
\hline
\rowcolor{LightCyan}
$44 \times 8$ & 94.44\% & 25.78\% & 0.1500 \\
\hline
$88 \times 4$ & 10.99\% & 10.99\% & 2E+35 \\  
\hline
\end{tabular}}
\end{table}

\begin{table}[!htbp]
\centering
\caption[Repeated network size experiment trials with updated sparsity settings.]{\textbf{Results for Sparse Combo Nets containing different numbers of component 32-unit RNNs, now with best found initialization settings.} Still using the standard 150 epoch training paradigm. This experiment was run to demonstrate repeatability of the size results seen in Table \ref{table:sizes-compare}. All trials were run to completion. A control trial was also run with the largest tested network size - the connections between subnetworks were no longer constrained, and thus this control combination network is not certifiably stable.}
\label{table:compare-sizes-again}

\begin{tabular}{ | m{4cm} | m{2cm} || m{1cm} | }
\hline
Name & Size & Final Test Acc. \\
\hline\hline
Sparse Combo Net & $1 \times 32$ & 37.1\% \\  
\hline
Sparse Combo Net & $4 \times 32$ & 89.1\% \\ 
\hline
Sparse Combo Net & $8 \times 32$ & 93.6\% \\ 
\hline
Sparse Combo Net & $12 \times 32$ & 94.4\% \\ 
\hline
Sparse Combo Net & $16 \times 32$ & 96\% \\ 
\hline
\rowcolor{LightCyan}
Sparse Combo Net & $22 \times 32$ & 96.8\% \\ 
\hline
Sparse Combo Net & $24 \times 32$ & 96.7\% \\ 
\hline\hline
Control & $24 \times 32$ & 47\% \\ 
\hline
\end{tabular}
\end{table}

\begin{table}[!htbp]
\centering
\caption[Testing component network sparsity settings within different network sizes.]{\textbf{Results for different sparsity initialization settings in networks of different sizes.} Varying sparsity and magnitude of the component RNNs for different network sizes. All other settings remained constant across trials, using our selected 150 epoch training paradigm.}
\label{table:sparsity-compare}

\begin{tabular}{ | m{1cm} | m{1.25cm} | m{1cm} || m{1cm} | m{1cm} | m{1cm} | }
\hline
 Size & Density & Scalar & Final Test Acc. & Epoch 1 Test Acc. & Final Train Loss \\
\hline\hline
\rowcolor{Gray}
$11 \times 32$ & 26.5\% & 0.27 & 89.12\% & 61.29\% & 0.3781 \\  
\hline
$11 \times 32$ & 10\% & 1.0 & 94.86\% & 70.67\% & 0.1278 \\
\hline\hline
\rowcolor{Gray}
$22 \times 16$ & 40\% & 0.4 & 93.01\% & 70.11\% & 0.2073 \\  
\hline
\rowcolor{LightCyan}
$22 \times 16$ & 20\% & 1.0 & 95.27\% & 76.58\% & 0.0924 \\  
\hline
$22 \times 16$ & 10\% & 1.0 & 94.26\% & 71.53\% & 0.1425 \\  
\hline\hline
\rowcolor{Gray}
$44 \times 8$ & 60\% & 0.7 & 94.44\% & 25.78\% & 0.1500 \\
\hline
$44 \times 8$ & 50\% & 1.0 & 95.05\% & 30.52\% & 0.1180 \\  
\hline
\end{tabular}

\vspace{0.5cm}

\caption[Extended sparsity experiment trials in network of fixed size.]{\textbf{Further optimizing the sparsity settings for high performance using a $16 \times 32$ network.} The final scalar is the product of the pre-selection and post-selection scalars. Note that the 5\% density and 7.5 scalar network was killed after 18 epochs due to exploding gradient. All other trials ran for a full 150 epochs.}
\label{table:best-results}
\begin{tabular}{ | m{1.25cm} | m{1cm} | m{1cm} || m{1cm} | m{1cm} | m{1.25cm} | }
\hline
 Density & Pre-select Scalar & Post-select Scalar & Final Test Acc. & Epoch 1 Test Acc. & Final Train Loss \\
\hline\hline
10\% & 1.0 & 1.0 & 95.87\% & 73.67\% & 0.074 \\  
\hline\hline
5\% & 10.0 & 0.1 & 95.11\% & 73.10\% & 0.1311 \\
\hline
5\% & 10.0 & 0.2 & 96.15\% & 82.50\% & 0.0051 \\
\hline
\rowcolor{LightCyan}
5\% & 10.0 & 0.5 & 96.69\% & 75.76\% & 0.0001 \\
\hline
5\% &  6.0 & 1.0 & 96.41\% & 21.55\% & 3.3E-5 \\
\hline
5\% & 7.5 & 1.0 & 16.75\% & 11.39\% & 3068967 \\
\hline\hline
3.3\% & 30.0 & 0.1 & 96.24\% & 83.89\% & 0.0005 \\
\hline
\rowcolor{LightCyan}
3.3\% & 30.0 & 0.2 & 96.54\% & 86.79\% & 4E-5 \\
\hline\hline
1\% & 10.0 & 1.0 & 96.04\% & 81.2\% & 0.0001 \\
\hline
\end{tabular}
\end{table}

\begin{table}[!htbp]
\centering
\caption[Permuted seqMNIST repeatability trials.]{\textbf{Repeatability of the best network settings on permuted seqMNIST.} Four trials of $16 \times 32$ networks with 3.3\% density and entries between -6 and 6, trained for a 24 hour period with a single learning rate cut (0.1 scalar) after epoch 200. All other training settings remained the same as our selected hyperparameters. Trials are presented in chronological order. The mean test accuracy achieved was 96.85\% with Variance 0.019.}
\label{table:psmnist-repeats}

\begin{tabular}{ | m{1.25cm} | m{1.5cm} | m{1.5cm} | m{1.5cm} | }
\hline
Epochs & Best Test Acc. & Epoch 1 Test Acc. & Best Train Loss \\
\hline\hline
208 & 96.65\% & 61.15\% & 0.00224 \\  
\hline
255 & 96.94\% & 84.19\% & 5e-5 \\  
\hline
250 & 96.88\% & 81.08\% & 5e-5 \\  
\hline
210 & 96.93\% & 67.19\% & 0.00069 \\  
\hline
\end{tabular}
\end{table}

\FloatBarrier

\subsubsection{seqCIFAR10 trials}

\begin{table}[!htbp]
\centering
\caption[seqCIFAR10 hyperparameter tuning trials (ultimately kept unchanged).]{\textbf{Additional hyperparameter tuning for the seqCIFAR10 task, presented in chronological order.} * indicates that training was cut short by the 24 hour Colab runtime limit, or in the case of high learning rate was killed intentionally due to exploding gradient. LR Schedule describes the scalar the learning rate was multiplied by, and at what epochs. The best performing network is highlighted. Ultimately we decided on the same network settings and training hyperparameters for further testing, just extending the training period to 200 epochs with the learning rate cuts occurring after epochs 90 and 140.}
\label{table:cifar}

\begin{tabular}{ | m{1.25cm} | m{1cm} | m{1cm} | m{1.1cm} | m{1cm} | m{1cm} | m{3cm} | m{1.1cm} | }
\hline
Density & Pre-select Scalar & Post-select Scalar & Epochs & Adam WD & Initial LR & LR Schedule & Best Test Acc. \\
\hline\hline
3.3\% & 30 & 0.2 & 150 & 1e-5 & 1e-3 & 0.1 after 90,140 & 64.63\% \\  
\hline
3.3\% & 30 & 0.2 & 34* & 1e-5 & 5e-3 & 0.1 after 90,140 & 35.42\% \\
\hline
5\% & 6 & 1 & 150 & 1e-5 & 1e-3 & 0.1 after 90,140 & 60.9\% \\ 
\hline
5\% & 10 & 0.5 & 150 & 1e-5 & 1e-4 & 0.1 after 90,140 & 54.86\% \\ 
\hline
3.3\% & 30 & 0.2 & 150 & 1e-5 & 5e-4 & 0.1 after 90,140 & 61.83\% \\ 
\hline
3.3\% & 30 & 0.2 & 200 & 1e-6 & 2e-3 & 0.1 after 140,190 & 62.31\% \\ 
\hline
\rowcolor{LightCyan}
3.3\% & 30 & 0.2 & 186* & 1e-5 & 1e-3 & 0.1 after 140,190 & 64.75\% \\ 
\hline
3.3\% & 30 & 0.2 & 132* & 1e-6 & 1e-3 & 0.1 after 140,190 & 62.31\% \\ 
\hline
5\% & 10 & 0.5 & 195* & 1e-5 & 1e-3 & 0.1 after 140,190 & 64.68\% \\ 
\hline
\end{tabular}
\end{table}

\begin{table}[!htbp]
\parbox{.475\linewidth}{
\centering
\caption[seqCIFAR10 repeatability trials.]{\textbf{Repeatability of the best network settings on seqCIFAR10.} Ten trials of $16 \times 32$ networks with 3.3\% density and entries between -6 and 6, trained for 200 epochs with learning rate scaled by 0.1 after epochs 140 and 190. All other training settings remained the same as before. Trials are presented in chronological order. The mean test accuracy achieved was 64.72\% with Variance 0.406. Most trials completed all 200 epochs, but two were cut short due to runtime limits.}
\label{table:cifar-repeats}

\begin{tabular}{ | m{1.1cm} | m{1cm} | m{1cm} | m{1cm} | m{1cm} | }
\hline
Epochs & Best Test Acc. & Epoch 1 Test Acc. & Best Train Loss \\
\hline\hline
186 & 64.75\% & 10.53\% & 0.83 \\  
\hline
200 & 64.04\% & 26.35\% & 0.787 \\ 
\hline
200 & 64.32\% & 26.76\% & 0.778 \\ 
\hline
170 & 64.88\% & 20.65\% & 0.857 \\ 
\hline
200 & 65.72\% & 27.28\% & 0.813 \\ 
\hline
200 & 63.73\% & 10.47\% & 0.826 \\ 
\hline
200 & 65.03\% & 18.75\% & 0.83 \\ 
\hline
200 & 64.71\% & 32.36\% & 0.799 \\ 
\hline
200 & 64.4\% & 10.09\% & 0.83 \\ 
\hline
200 & 65.63\% & 30.25\% & 0.792 \\ 
\hline
\end{tabular}
}
\hfill
\parbox{.475\linewidth}{
\centering
\caption[\hspace{0.1cm} Additional seqCIFAR10 repeatability trials with shorter training time.]{\textbf{Repepatability trials with shorter training time.} An additional nine trials investigating the repeatability of our results on the seqCIFAR10 task. For these trials we used the same 150 epoch training paradigm as previously, although only four of the networks were able to fully complete training. These trials were done to expand our sample size while access was limited to only slower GPUs. The mean observed test accuracy among the shorter trials was 62.82\%, with variance of 0.95.}
\label{table:cifar-repeats-short}

\begin{tabular}{ | m{1.1cm} | m{1cm} | m{1cm} | m{1cm} | m{1cm} | }
\hline
Epochs & Best Test Acc. & Epoch 1 Test Acc. & Best Train Loss \\
\hline\hline
150 & 64.63\% & 28.91\% & 0.837 \\  
\hline
150 & 63.51\% & 9.7\% & 0.9 \\ 
\hline
148 & 62.35\% & 17.51\% & 0.89 \\ 
\hline
126 & 63.33\% & 23.61\% & 0.903 \\ 
\hline
129 & 61.45\% & 28.93\% & 0.937 \\ 
\hline
150 & 62.21\% & 25.82\% & 0.9 \\ 
\hline
150 & 63.42\% & 23.51\% & 0.86 \\ 
\hline
147 & 62.05\% & 27.67\% & 0.882 \\ 
\hline
131 & 62.41\% & 30.33\% & 0.912 \\ 
\hline
\end{tabular}
}
\end{table}

\FloatBarrier

\section{Proof of extended claims in Theorem \ref{theorem: Wdiagstabcounterexampletheorem}}
\label{sec:my-theorem-continued}
As mentioned in the main text section \ref{subsubsec:my-theorem}, the counterexample of Theorem \ref{theorem: Wdiagstabcounterexampletheorem} can also be extended to the case where the activation function is not allowed to have derivative 0 (as is sometimes used to make proving certain conditions on the \eqref{eq:RNN} network easier). In this case, we think more along the lines of softplus, a fully differentiable analog of ReLU (i.e. a smooth approximation of ReLU), where the derivative of the activation can be very small, but never zero. The extension of the proof to show this is as follows.

\begin{proof}
\noindent An additional condition we could impose on the activation function is to require it to be a strictly increasing function, so that the activation function derivative can never actually reach 0. We will now show that a very similar counterexample applies in this case, by taking 
    \begin{align*}
      \mathbf{D}_{1*} = \begin{bmatrix} 
    1 & 0 \\
    0 & \epsilon 
  \end{bmatrix}
  \hspace{5mm} and \hspace{5mm}
  \mathbf{D}_{2*} = \begin{bmatrix} 
    \epsilon & 0 \\
    0 & 1 
  \end{bmatrix}
  \end{align*}
  
  \noindent Note here that the $\mathbf{W}$ used above
 produced a $(\mathbf{W}\mathbf{D})_{sym} - \mathbf{I}$ that just barely avoided being negative definite with the original $\mathbf{D}_{1}$ and $\mathbf{D}_{2}$, so we will have to increase the values on the off-diagonals a bit for this next example. In fact anything with magnitude larger than 2 will have some $\epsilon > 0$ that will cause a constant metric to be impossible, but for simplicity we will now take 
 \begin{align*}
 \mathbf{W}_{*} = \begin{bmatrix} 
    0 & -4 \\
    4 & 0 
  \end{bmatrix}
  \end{align*}
  
  \noindent Note that with $\mathbf{W}_{*}$, even just halving one of the off-diagonals while keeping the other intact will produce a $(\mathbf{W}\mathbf{D})_{sym} - \mathbf{I}$ that is not negative definite. Anything less than halving however will keep the identity metric valid. Therefore, we expect that taking $\epsilon$ in $\mathbf{D}_{1*}$ and $\mathbf{D}_{2*}$ to be in the range $0.5 \geq \epsilon > 0$ will also cause issues when trying to obtain a constant metric.
  
  We will now actually show via a similar proof to the above that $\mathbf{M}$ is impossible to find for $\mathbf{W}_{*}$ when $\epsilon \leq 0.5$. This result is compelling because it not only shows that $\epsilon$ does not need to be a particularly small value, but it also drives home the point about antisymmetry - the larger in magnitude the antisymmetric weights are, the larger the $\epsilon$ where we will begin to encounter problems. \\
  
  \noindent Working out the matrix multiplication again, we now get 
  \begin{align*}
  (\mathbf{M}\mathbf{W}_{*}\mathbf{D}_{1*})_{sym} - \mathbf{M} = \begin{bmatrix} 
    4m - a & 2b - m - 2a\epsilon \\
    b - m - 2a\epsilon & -4m\epsilon - b 
  \end{bmatrix}
  \end{align*}
  and 
  \begin{align*}
    (\mathbf{M}\mathbf{W}_{*}\mathbf{D}_{2*})_{sym} - \mathbf{M} = \begin{bmatrix} 
    4m\epsilon - a & -(2a + m - 2b\epsilon) \\
    -(2a + m - 2b\epsilon) & -4m - b 
  \end{bmatrix}
  \end{align*}
  
  \noindent Resulting in two new main necessary conditions:
  \begin{align}
  |4m - a - b - 4m\epsilon| > 2|2b - m - 2a\epsilon| \label{epsD1offdiagAppend}
  \end{align}
  \begin{align}
  |4m\epsilon - a - b - 4m| > 2|2a + m - 2b\epsilon| \label{epsD2offdiagAppend}
  \end{align}
  \noindent As well as new conditions on the diagonal elements:
  \begin{align}
  4m - a < 0 \label{epsD1diagAppend}
  \end{align}
  \begin{align}
  -4m - b < 0 \label{epsD2diagAppend}
 \end{align}
 
 \noindent We will now proceed with trying to find $a,b,m$ that can simultaneously meet all conditions, setting $\epsilon = 0.5$ for simplicity. 
 
 Looking at $m=0$, we can see again that $\mathbf{M}$ will require off-diagonal elements, as condition \eqref{epsD1offdiagAppend} is now equivalent to the condition $a + b > |4b - 2a|$ and condition \eqref{epsD2offdiagAppend} is similarly now equivalent to $a + b > |4a - 2b|$. 
 
 Evaluating these conditions in more detail, if we assume $4b > 2a$ and $4a > 2b$, we can remove the absolute value and the conditions work out to the contradicting $3a > 3b$ and $3b > 3a$ respectively. As an aside, if $\epsilon > 0.5$, this would no longer be the case, whereas with $\epsilon < 0.5$, the conditions would be pushed even further in opposite directions. 
 
 If we instead assume $2a > 4b$, this means $4a > 2b$, so the latter condition would still lead to $b > a$, contradicting the original assumption of $2a > 4b$. $2b > 4a$ causes a contradiction analogously. Trying $4b = 2a$ will lead to the other condition becoming $b > 2a$, once again a contradiction. Thus a diagonal $\mathbf{M}$ is impossible \\
 
 \noindent So now we again break down the conditions into $m > 0$ and $m < 0$ cases, first looking at $m > 0$. Using condition \eqref{epsD1diagAppend} and knowing all unknowns have positive sign, condition \eqref{epsD1offdiagAppend} reduces to $a + b - 2m > |4b - 2(a + m)|$ and condition \eqref{epsD2offdiagAppend} reduces to $a + b + 2m > |4a - 2(b - m)|$. This looks remarkably similar to the $m = 0$ case, except now condition \eqref{epsD1offdiagAppend} has $-2m$ added to both sides (inside the absolute value), and condition \eqref{epsD2offdiagAppend} has $2m$ added to both sides in the same manner. If $4b > 2(a + m)$ the $-2m$ term on each side will simply cancel, and similarly if $4a > 2(b - m)$ the $+2m$ terms will cancel, leaving us with the same contradictory conditions as before. 
 
 Therefore we check $2(a + m) > 4b$. This rearranges to $2a > 2(2b - m) > 2(b - m)$, so that from condition \eqref{epsD2offdiagAppend} we get $b > a$. Subbing condition \eqref{epsD1diagAppend} in to $2(a + m) > 4b$ gives $8b < 4a + 4m < 5a$ i.e. $b < \frac{5}{8}a$, a contradiction. The analogous issue arises if trying $2(b - m) > 4a$. Trying $2(a + m) = 4b$ gives $m = 2b - a$, which in condition \eqref{epsD2offdiagAppend} results in $5b - a > |6a - 6b|$, while in condition \eqref{epsD1diagAppend} leads to $5a > 8b$, so \eqref{epsD2offdiagAppend} can further reduce to $5b - a > 6a - 6b$ i.e. $11b > 7a$. But $b > \frac{7}{11}a$ and $b < \frac{5}{8}a$ is a contradiction. Thus there is no way for $m > 0$ to work. \\
 
 \noindent Finally, trying $m < 0$, we now use condition \eqref{epsD2diagAppend} and the signs of the unknowns to reduce condition \eqref{epsD1offdiagAppend} to $a + b + 2|m| > |4b - 2(a - |m|)|$ and condition \eqref{epsD2offdiagAppend} to $a + b - 2|m| > |4a - 2(b + |m|)|$. These two conditions are clearly directly analogous to in the $m > 0$ case, where $b$ now acts as $a$ with condition \eqref{epsD2diagAppend} being $b > 4|m|$. Therefore the proof is complete.
 \end{proof}

\section{Alternative proof for Theorem \ref{theorem: absolutevaluetheorem}}
\label{sec:thm1-alt-pf}
\begin{theoremalt}
Let $|\mathbf{W}|$ denote the matrix formed by taking the element-wise absolute value of $\mathbf{W}$.  If there exists a positive, diagonal $\mathbf{P}$ such that:
\[\mathbf{P}(g|\mathbf{W}|-\mathbf{I}) +(g|\mathbf{W}|-\mathbf{I})^T\mathbf{P} \prec 0 \]
then \eqref{eq:RNN} is contracting in metric $\mathbf{P}$. Moreover, if $W_{ii} \leq 0$, then $|W|_{ii}$ may be set to zero to reduce conservatism.
\end{theoremalt}

\begin{proof}
Consider the differential, quadratic Lyapunov function:

\[V = \delta\mathbf{x}^T\mathbf{P} \delta\mathbf{x}\]
where $\mathbf{P} \succ 0 $ is diagonal. The time derivative of $V$ is:

\[\dot{V} = 2\delta\mathbf{x}^T\mathbf{P} \dot{\delta\mathbf{x}} = 2\delta\mathbf{x}^T\mathbf{P} \mathbf{J}\delta\mathbf{x} = -2\delta\mathbf{x}^T\mathbf{P} \delta\mathbf{x} + 2\delta\mathbf{x}^T\mathbf{P}\mathbf{W} \mathbf{D}\delta\mathbf{x} \]

where $\mathbf{D}$ is a diagonal matrix such that $\mathbf{D}_{ii} = \frac{d\phi_i}{dx} \geq 0$.  We can upper bound the quadratic form on the right as follows:

\[\begin{split} \delta\mathbf{x}^T\mathbf{P}\mathbf{W} \mathbf{D}\delta\mathbf{x} = \sum_{ij} P_i W_{ij} D_j \delta x_i \delta x_j \leq  \\ \sum_{i}P_i W_{ii} D_i |\delta x_i|^2 + \sum_{ij, i \neq j} P_i |W_{ij}| D_j |\delta x_i| |\delta x_j| \leq g|\delta \mathbf{x}|^T\mathbf{P}|\mathbf{W}||\delta \mathbf{x}| \end{split}\]

If $W_{ii} \leq 0 $, the term $P_i W_{ii} D_i |\delta x_i|^2$ contributes non-positively to the overall sum, and can therefore be set to zero without disrupting the inequality. Now using the fact that $\mathbf{P}$ is positive and diagonal, and therefore $\delta\mathbf{x}^T\mathbf{P} \delta\mathbf{x} = |\delta\mathbf{x}|^T\mathbf{P} |\delta\mathbf{x}|$,  we can upper bound $\dot{V}$ as:
\[\dot{V} \leq |\delta\mathbf{x}|^T (-2\mathbf{P} + \mathbf{P}|\mathbf{W}| + |\mathbf{W}|\mathbf{P})|\delta\mathbf{x}| = |\delta\mathbf{x}|^T[ (\mathbf{P}(|\mathbf{W}|-\mathbf{I}) + (|\mathbf{W}|^T -\mathbf{I})\mathbf{P})]|\delta\mathbf{x}|\]

where $|W|_{ij} = |W_{ij}|$, and $|W|_{ii} = 0$ if $W_{ii} \leq 0$ and $|W|_{ii} = |W_{ii}|$ if $W_{ii} > 0$.

  \end{proof}

\chapter{Included publications and collaborator contributions}\label{cha:append-clarity}
\renewcommand\thefigure{E.\arabic{figure}}    
\setcounter{figure}{0}  
\renewcommand\thetable{E.\arabic{table}}    
\setcounter{table}{0}  
\renewcommand\thesection{E.\arabic{section}} 
\setcounter{section}{0}

In this section, I note any existing publications associated with each core thesis chapter, as well as any plans for future submissions based on the work. Further, I provide details on any results that I report which were not primarily contributed by me, and in the case of co-authored manuscripts I also explain how different parts of the text were used within my thesis. \\

\noindent In brief - 
\begin{itemize}
\item Chapter 1 (\ref{sec:ch1-explain} here) was written from scratch for the thesis and may or may not later become a paper in some capacity.
\item Chapter 2 (\ref{sec:ch2-explain} here) was largely written from scratch for the thesis, based on two distinct contributions. The first is a primarily software work being used by a large multi-site NIMH study, which will likely be published as a formally citeable piece of code documentation. The second pulls from results I helped produce for a \emph{Schizophrenia Research} paper on which I am second author.
\item Chapter 3 (\ref{sec:ch3-explain} here) was a case report already in preparation, which I finished in an expanded form for this thesis. It draws in part from a different case report on the same patient, on which I am a contributing author. As that was published in \emph{Frontiers in Human Neuroscience}, I will be submitting an edited down version of my chapter there soon.
\item Chapter 4 (\ref{sec:ch4-explain} here) was a co-lead authored work at the most recent \emph{NeurIPS}. However even this chapter was written specifically for the thesis; I include some additional background and interpretation content, and exclude a few less essential parts of the original publication that were not primarily my work, subsequently reediting the whole text for flow. 
\end{itemize}

\noindent Note that all publications mentioned in this section have author agreements allowing use in materials such as theses, so long as the published version is appropriately cited, which I have done for all content within this thesis that was reproduced or directly adapted from those papers.

\section{Chapter 1}
\label{sec:ch1-explain}
This chapter was written by me, with the primary contributions based on code designed and written by me and the resulting outputs organized by me. As clearly noted when describing each instance, other lab members performed some of the manual review tasks and case report work that led to many of the internal validation results I report. The core contributors here were Claire Foley, Bryce Gillis, and Burhan Khan. 

None of the work described is presently published, and the code itself is primarily for internal lab use at this time. Ideally the completed chapter can be edited down to a publishable manuscript that argues for the use of similar audio diaries in future studies, with an accompanying code release that others can use as a starting point for their work. Of course is this does occur I would be a lead author, and it probably will occur in some form. \\

\noindent The current code repository has not yet been peer reviewed, but is already publicly viewable, and citeable via Zenodo DOI \citep{diarygit}.

\section{Chapter 2}
\label{sec:ch2-explain}
The section \ref{sec:disorg} reports results from a recent publication on which I was a contributing author:

\begin{quote}
E. Liebenthal, \textbf{M. Ennis}, H. Rahimi-Eichi, E. Lin, Y. Chung, and J.T. Baker. Linguistic and non-linguistic markers of disorganization in psychotic illness. \emph{Schizophrenia Research}, 2022. https://doi.org/10.1016/j.schres.2022.12.003
\end{quote}

\noindent For that paper, I ran initial inventorying and quality control on audio from the BLS interview dataset, which played a part in the final dataset selection criteria described. I also was involved with the discussions surrounding how to analyze the chosen interviews, and ultimately selected the final language feature set while balancing input from others on the clinical goals and on the recommended limits to feature count for sufficient statistical power. I then handled obtaining the needed transcripts, extracting these linguistic features of interest, and ensuring the final outputs were sanity checked and correctly aligned with clinical scores. 

The dataset I compiled was then used by Einat (the lead author) to run the reported statistical models. For the final manuscript, I drafted a summary of my methodologies, and assisted with interpretation of some of the language results, including by generating additional visualizations depicting properties of the transcript feature set for internal reference. 

In writing section \ref{sec:disorg}, I pulled from my notes to include some of the above-mentioned supplemental figures, as well as to include a more detailed version of my methodology write-up and to expand on a few extra discussion points. I also reproduced a few relevant figures from the published manuscript, noted in the respective legends, and reported on the clinical and statistical methods and results that I was only peripherally involved with. However, I largely wrote my own version of any text providing context for or interpreting those results. Direct quotes are clearly marked and used sparingly, for relaying concrete results statements. \\

All other portions of the chapter were written by me for use in this thesis. Parts were adapted from the already public GitHub repository for the AMPSCZ interview preprocess code described, which I wrote in full. That code continues to be actively used by the project, and I plan to more formally release the code along with a version of the documentation written for the thesis, to serve as a peer-reviewed citable record (such as JOSS) and a major reference material for whoever takes over my current maintenance role in the project. Meanwhile, the code repository is already publicly viewable, and citeable via Zenodo DOI \citep{interviewgit}.

Within this chapter, I made it clear wherever the code interacts with other pieces of software, focusing on technical details only for my own work. SOP design and the AMPSCZ-wide server structure were obviously decisions made prior to my joining the project, but the software design decisions described for my code architecture were made by me. I also did much of the initial manual monitoring, but was increasingly assisted by others later in the project, so not everything mentioned was caught by me. I wrote all documentation though that is provided in the main body. 

\section{Chapter 3}
\label{sec:ch3-explain}
The figures and results used in this chapter were previously drafted by me, with initial edits made by Josh Salvi. The data collection portion of the methods was previously drafted by the study RAs and then edited by me. The rest of the chapter was put together by me for this thesis. 

A shorter version of it now forms the case report in preparation for \emph{Frontiers in Human Neuroscience}, on which I will be lead author. It is meant to be a companion piece to the associated DBS trial report that was previously published there, on which I was a contributing author: 

\begin{quote}
S.T. Olsen, I. Basu, M.T. Bilge, A. Kanabar, M.J. Boggess, A.P. Rockhill, A.K. Gosai, E. Hahn, N. Peled, \textbf{M. Ennis}, I. Shiff, K. Fairbank-Haynes, J.D. Salvi, C. Cusin, T. Deckersbach, Z. Williams, J.T. Baker, D.D. Dougherty, and A.S. Widge. Case report of dual-site neurostimulation and chronic recording of cortico-striatal circuitry in a patient with treatment refractory obsessive compulsive disorder. \emph{Frontiers in Human Neuroscience}, 2020. https://doi.org/10.3389/fnhum.2020.569973
\end{quote}

\noindent In the chapter, I provide my own summary of the core DBS trial results, as well as further discussion of those results in the context of the insights our additional data collection was able to bring. All the datatypes exclusive to our report were processed, visualized, and analyzed by me. I did all of the reported clustering work for the clinical scales and EMA raw data as well. Neural features were used as provided by the DBS trial team. They also ran the stimulation experiment interview session based on my design, then from there I handled the data. 

\section{Chapter 4}
\label{sec:ch4-explain}
This chapter was adapted from the following NeurIPS paper on which I was co-lead:

\begin{quote}
L. Kozachkov* and \textbf{M. Ennis*}, and J.J. Slotine. RNNs of RNNs: Recursive Construction of Stable Assemblies of Recurrent Neural Networks. \emph{Advances in Neural Information Processing Systems}, 2022. https://openreview.net/forum?id=2dgB38geVEU
\end{quote}

\noindent For the purposes of the thesis chapter, I wrote an extended version of the introduction and discussion. Most of the manuscript introduction and discussion is contained within, using clearly demarcated sectioning; as my co-author and I wrote various parts of the paper's background/interpretation and then edited each other's writing, those portions are intermingled in a way that would be difficult to solely attribute.

As far as the reported results, I removed some parts contributed by my co-author that were less critical to the story of the chapter, and also added to the main chapter some more minor supplemental results and intuition-building remarks that were contributed by me. We worked together to varying degrees on different portions of the project, which were sometimes self-driven and sometimes suggested by our senior author. For all the main results I report on in the thesis, I provide an approximate breakdown here. First for the mathematical results, and then for the experimental results.

When I started collaborating with the group, my co-author (Leo) was already working with our PI to try to find better stability conditions for the single RNN described in equation \eqref{eq:RNN}. I eventually came up with an initial proof for Theorem \ref{theorem: absolutevaluetheorem}, which is the proof I provide in the main body of the thesis. Leo later wrote a different much cleaner proof of the same statement, which is the one used in the paper. That alternative proof is reproduced here in the appendix \ref{cha:append-chapt-refch:4}. The other main condition used in our subsequent multi-area RNN applications was Theorem \ref{theorem: singularvaluetheorem}, which I include in the thesis for that reason, but it was fully Leo's work. 

For the discussion of counterexamples, Leo and I worked together. We went over issues from a few different previous reports, though the most glaring one that we focus on in the published paper was found by Leo in \cite{chang2019antisymmetricrnn}. Leo also provided Theorem \ref{theorem: Wdiagstabtheorem}, which is a more limited variation on the conjectured stability condition that was the motivation behind many of the detected problems. I include this as context for Theorem \ref{theorem: Wdiagstabcounterexampletheorem}, which was my work demonstrating an entire class of proof methods cannot work for the general form of that conjecture - explaining why we were having such a hard time with the problem in the first place, and making the conditions we did already have feel much more publishable. 

Finally, I put Theorem \ref{theorem: subnetworkstabilitytheorem} into the main body of my thesis. It was early work of mine that was included in our first preprint but cut for space as we added more work on the implications that contracting systems can have for multi-area networks. I reintroduced it here because I think it helps supplement the intuition behind Theorem \ref{theorem: Wdiagstabcounterexampletheorem}, and because it provides some ideas for future work that might more closely connect the stability theory used to combine contracting systems with stability-preserving conditions for subsetting contracting systems. Subnetwork stability is a topic with obvious relevance to pruning concepts within both neuroscience and deep learning, which can also be connected back to the themes of sparsity seen in our work. 

Moving on from the single RNN of equation \eqref{eq:RNN}, we used contraction combination properties previously proved by our PI (JJ) to extend our work naturally to the domain of multi-area networks. The initial push to do this was provided by JJ. Leo then applied the existing nonlinear control systems results to concretely parameterize a provably stable way to combine the individually contracting nonlinear RNNs in linear negative feedback. This resulted in Theorem \ref{theorem: network_of_networks}, which was Leo's work that enabled the described experiments.

Leo subsequently wrote the core code implementing the custom trainable "RNN of RNNs" described by that parameterization, when using his Theorem \ref{theorem: singularvaluetheorem} to parameterize the component RNNs. As we tested the networks on existing benchmarks, the task part of the code was pulled from existing sources. Leo produced pilot results using what is described in the manuscript as the "SVD Combo Net". I reproduce a few of those results in the thesis in order to compare and contrast architectures, but it is not a major focus of the chapter's experiments section (nor of the publication's). However, it is also important to note this because I made only simple modifications to the code base Leo created in order to run all of my experiments.

I wanted to test the application of Theorem \ref{theorem: absolutevaluetheorem} within the multi-area framework, and also wanted to experiment more with ideas about network sparsity. Therefore I created the described "Sparse Combo Net" architecture, which was able to surpass the prior SOTA for stable RNNs on our chosen tasks. I also ran a larger set of experiments on those tasks using different model settings, to assist in interpreting the model's high performance despite its relative simplicity. The majority of the experimental methods and results section is thus written by me, both in the publication and in my thesis, and all the extended experimental details provided in the supplement were put together by me. 

Ultimately, much of the thesis chapter is written by me, and I made a number of key contributions to the paper. At the same time, it was impossible to remove most of Leo's contributions - and in the case of his mathematical results it made the most sense to directly reproduce his words. I genuinely think we both made unique contributions such that the paper would not have been accepted to NeurIPS or a similar venue with either one of us working on it alone at $2x$ time. 

\chapter{On transcription costs in the NIMH's AMPSCZ project}\label{cha:append-ampscz-rant}
\renewcommand\thefigure{F.\arabic{figure}}    
\setcounter{figure}{0}  
\renewcommand\thetable{F.\arabic{table}}    
\setcounter{table}{0}  
\renewcommand\thesection{F.\arabic{section}} 
\setcounter{section}{0}

\begin{quote}
    \textbf{Appendix F Preface:}

    As a result of the arguments herein, audio journals have now been prioritized over structured clinical interviews in the professional transcription budget for the project, and discussion on changing the parameters for those clinical interviews that will still be manually transcribed is also ongoing. While questions remain about software resources and the broader administrative structure of AMPSCZ, this budgetary shift is a highly positive development in the future outlook for the project.
\end{quote}

\noindent The goal of this supplement is to redirect the transcription budget planned for speech sampling in the NIMH's Accelerating Medicines Partnership Program - Schizophrenia (AMPSCZ) project, a sum easily exceeding 1 million USD over the planned lifetime of the project. In particular, I feel there are serious shortcomings with the current procedure for transcription of the collected clinical (PSYCHS) interviews, and inversely that audio journals have been severely undervalued to date. 

These observations also connect back to broader concerns about the structure of the project moving forward. Most notably, a presently inadequate plan for hiring a replacement for my role in the coming year -- something else I hope to spark a discussion about. I don't really understand where the money that is funding me is evaporating to, and regardless this is an important cost to cover in my opinion. Theoretically, the transcription budget could cover all open interviews, all audio diaries, and my replacement if the clinical interviews were entirely foregone for manual transcription. However, the budget for software building is either much too low or not being efficiently allocated, and that is a problem that should not need to be covered by the transcription budget. \\

\noindent To make my case, I will cover the following topics in the upcoming sections:
\begin{enumerate}
    \item A list of problems I see with the current psychs interview transcription plan (\ref{sec:psychs-dumb}).
    \item A list of downsides if only automated transcription techniques were to be used on the audio journals, coupled with additional potential opportunities enabled with TranscribeMe transcription available (\ref{sec:journals-again}).
    \item A report on observed properties of open versus psychs interviews in the early AMPSCZ dataset, connecting back to listed psychs problems (\ref{sec:open-psychs-data}).
    \item A concrete estimation of the total budget planned to go towards interview transcriptions and more specific questions on how the transcription plan should be decided on (\ref{sec:blowing-money}). 
    \item An estimation of the cost to get all AMPSCZ audio journals transcribed, including a comparison in light of the previously proposed psychs transcription budget (\ref{sec:diary-cost-estimates}).
    \item A very early overview of the composition of the AMPSCZ journal dataset, along with proof-of-concept demonstration of some of the lowest hanging fruit features that TranscribeMe easily provides but would be extremely difficult, if not impossible, to obtain via automated methods (\ref{sec:diary-dataset-u24}).
    \item An overview of the infrastructure work I believe I will have time to finish versus the many remaining tasks for this part of the project, contextualized by the roadblocks facing hiring a replacement (\ref{sec:nih-sucks}).
    \item Finally, a summary of the key claims made here, coupled with more abstract arguments on why audio journals are currently greatly underrated and what sort of vision large scale data collection initiatives ought to have (\ref{sec:guillermo-is-annoying}).
\end{enumerate}
\noindent I can be reached for the foreseeable future at mennis@mit.edu if there are follow-up questions about my role in this project.

\section{Arguments against transcription of psychs interviews}
\label{sec:psychs-dumb}
The structured clinical interview protocol includes a number of quantitative-style questions that elicit matter-of-fact responses, which would be expected to lead to an overall more stunted conversation that has a variety of limitations from a pure speech sampling perspective. These interviews can also be very long at times, so for AMPSCZ a budgetary "stop loss" was implemented to break all transcriptions at $\sim 30$ minutes. For any submission over 30 minutes, TranscribeMe will end the transcription somewhere between 30 and 32 minutes that represents as natural of a stopping point as possible, and we will always be charged the equivalent of 32 minutes. Taken together with other protocol differences between the recording standards for open ended versus psychs interviews, there are already a large number of theoretical concerns with the psychs budget, many of which have appear to have come to fruition in the early phases of data collection. \\

\noindent In upcoming sections, I will argue for redirection of part of the budget presently intended for psychs to specific other targets (mainly audio journals), as well as present more detailed data on the inefficiency of early psychs transcript costs. Here, I provide an exhaustive list of my concerns about the psychs transcription plan: 
\begin{itemize}
    \item Psychs interview transcripts in practice have had significantly shorter conversational turns, significantly lower participant speech fraction, and even fewer total words per minute of recording (section \ref{sec:open-psychs-data}). Thus the cost per transcribed word and especially per transcribed participant word is much higher for psychs, and the words themselves likely carry less information due to the stunted nature of parts of the psychs interview conversation.
    \item The majority of psychs interviews so far have exceeded 30 minutes, and a sizeable minority have exceeded 1 hour -- so most transcripts are missing some information, and a concerning number of transcripts are missing at least half of the interview content (section \ref{sec:open-psychs-data}). \item Exceptionally long psychs interviews ($> 2$ hours) appear to be a fairly site-specific phenomenon, which raises questions about biases that might be introduced to analyses by current procedural inconsistencies, and seriously compounded by the $\sim 30$ minute transcript limit.
    \item The nature of structured question responses limits the signal that can feasibly be captured by many common linguistic analyses, some of which were factors in wanting to pay for the highest quality professional transcriptions in the first place (i.e. psychs is the wrong place to be spending on the TranscribeMe advantages - to be discussed in section \ref{sec:journals-again}). A few examples of language features dampened by a clinical interview format are:
    \begin{itemize}
        \item Turn to turn structure is less inherently interesting, because it is in large part driven by the needs of the clinical scale protocol.
        \item Certain categories of disfluencies are affected, for example it is difficult to restart a sentence if the sentence is actually just a yes or no response, and there are fewer occasions to stutter over words or phrases as one repeatedly answers such questions.
        \item There are also fewer chances for participants to use uncommon words or speak incoherently, and fewer reasons for certain parts of speech that may carry clinical relevance to be used. 
        \item Sentiment of participant speech is largely uninformative without using the context of the preceding question, as e.g. "I'd strongly disagree" would likely carry a small negative sentiment score on its own, but could go to either extreme depending on the question asked. If the interviewer questions were incorporated into sentiment analysis though, it is unclear how much information that would carry beyond what is already being rated in the corresponding clinical scale scoring.
    \end{itemize}
    \item The last point about sentiment connects to a broader issue in feature engineering for structured clinical interview transcript analysis requiring much greater care with the content of the interviewer questions. Certainly there are some interesting aims that could be investigated within this framework, but it would be a separate software problem entirely, and it is not clear in this project who would pursue that, nor exactly what those scientific aims would be. Recall also that we have no guarantee on which questions are included in which transcripts due to the duration cutoff. 
    \item For AMPSCZ in particular, there are additional data quality concerns independent of the transcripts that make psychs interviews less suited for various multimodal analyses than the open interviews are. By design, fewer quality standards are enforced on psychs, including:
    \begin{itemize}
        \item Onsite psychs interviews will not include video nor diarized audio by design.
        \item Even when recorded over Zoom, the inclusion of video and speaker specific audio files in psychs uploads has not appeared to be much of a point of contention - some sites have in fact been intentionally leaving out video files across all psychs interviews.
        \item The psychs procedure does not have a standardized convention for the number of interview attendees, like there is for open interviews. 
        \item How the attendees participate has also varied widely - the number of faces detected, the number of voices detected, and the overlap between these two numbers covered a wide range in the early psychs dataset.
        \item As detailed, psychs interviews are also much longer on average (with substantially more variance in duration beyond that) and will therefore require a much larger share of compute resources to fully process. For each CHR participant, the anticipated total duration of recorded psychs interviews will be more than 6 times the anticipated total duration of recorded open interviews, by a very conservative estimate. 
    \end{itemize}
    \item Interviews require significantly more manual action on the part of sites than audio diaries do, so there are many more opportunities for mistakes to occur and waste money. The QC/data flow pipeline aims to catch some of these issues, but for interviews they have spanned such a range of different forms it is only possible to cover so much. Especially for psychs interviews which regularly surpass the 30 minute duration cap, some site mistakes that affect an entire session will waste over $\$55$ every time they occur. Fortunately most site mistakes prevent transcription upload by the pipeline entirely, but a couple I have observed repeatedly to date that did waste money and would be difficult to entirely automate away are:
    \begin{itemize}
        \item If an interview is put under the open folder accidentally and then later moved to psychs on Box/Mediaflux, that interview will be processed twice by the pipeline (Lochness does not delete files from the server just because they were deleted on source), resulting in two of the same transcription being purchased. 
        \item If a site has a particularly long psychs interview, they will sometimes intentionally break it up into multiple Zoom folders, which they are not supposed to do. The pipeline treats separate interview sessions as separate interview sessions, so they will both be uploaded to TranscribeMe and will therefore be taking the budget of two interviews for just one. Even more concerning, it is usually the case that such a very long interview was split into two parts because each half approaches an hour+ in duration. In this case, $\$115$ is then spent on a very awkward end result where only the first quarter and third quarter of content are transcribed.
    \end{itemize}
\end{itemize}
\noindent Despite the many reasons for concern on the value of professional transcriptions for the structured clinical interview recordings here, there are still many ways that professional transcription provides great value, for example in the open interviews we are also currently having transcribed. I will next explain the many advantages in using TranscribeMe and in particular why TranscribeMe should be used for the AMPSCZ audio journals.

\section{Arguments for transcription of audio diaries}
\label{sec:journals-again}
Without undergoing an entirely separate security review process - which would definitely introduce complications at the scale of AMPSCZ and with the many countries involved - any automated transcription method we might use for audio journals must be run locally. Realistically, the only such option that could maybe perform well enough currently is Whisper \citep{whisper}. It was argued that diaries are in least need of manual transcription because a major remaining hurdle of automated methods is speaker assignment. I suspect accurate speaker assignment would indeed still be difficult with Whisper, but I think this argument is missing the many other reasons professional transcription is preferable for this project, as well as the many ways audio journal language can be more informative than clinical interview language. \\

\noindent There are logistical issues with relying solely on Whisper for diaries, related to both software development and privacy concerns, and there are also a number of scientific concerns that in my opinion have not been remotely adequately addressed by those arguing for sole use of Whisper (or for waiting for some unspecified future software, which I will cover too). The points that make TranscribeMe preferable include:
\begin{itemize}
    \item The data aggregation servers do not have the necessary resources to run Whisper, so this would become another task for the AV processing server. The diary QC pipeline would then be forced to interface with the AV processing server in a way that was not necessary for the interview code, creating new data flow considerations. Furthermore, the produced transcripts would require entirely different quality monitoring considerations than included in the transcript QC functions previously written, which were designed around the conventions of TranscribeMe. If the project does choose to use Whisper as the primary diary transcription method, then it is advisable that an entirely new pipeline be built for the datatype. This task would then fall fully under the responsibilities of my (yet to be specified) replacement.  
    \item Methods for automated redaction of PII remain subpar, are not exactly a bustling area in the grand scheme of machine learning research, and regardless of hypothetical quality involve thorny (international) legal considerations for the foreseeable future if one wants to entirely replace human oversight. \textbf{For a project with the scope and goals of AMPSCZ, this is an exceptionally important point.} The plan for any redacted transcripts generated by TranscribeMe is that they will eventually be shared with the NIH data repository for broader use by the research community; they are also immediately transferred by Lochness to the downstream \emph{predict} server, which allows a larger set of the researchers involved in AMPSCZ to access them now. Unredacted transcripts on the other hand are accessible by only a handful of people, which has inadvertently created an odd incentive structure within AMPSCZ speech sampling. Having redacted transcripts available would offset some of the roadblocks introduced by the fact that the data repository will only ever receive whatever acoustic features the handful of people involved will choose to extract. 
    \item Whisper cannot accurately transcribe most linguistic disfluencies, which is in fact (not surprisingly) considered a feature of their model \citep{whisper}. TranscribeMe on the other hand will transcribe disfluencies carefully and with a consistent notation for easy automated accounting on our end. A wide body of literature has found disfluency usage to be relevant in psychotic disorders \citep{Andreasen1986,Liddle2002,Kuperberg2010,Hong2015,Tang2021,disorg22}.
    \begin{itemize}
        \item Note that disfluencies are particularly unlikely to be captured by upcoming automated techniques, unless those automated techniques are developed by a yet to be identified interdisciplinary group in the digital psychiatry space. There is little reason for a machine learning group or anyone working on machine learning applications in most other domains to want disfluencies transcribed. 
        \item In fact there are many reasons for them to \emph{not} want disfluencies transcribed. The majority of commercial users would consider it a failure if their returned transcript contained a bunch of stutter markings, so it is not a goal of the model. The cost to obtain a large enough training set that contains accurately labeled disfluencies is higher than one without, so including disfluencies to a strong level of accuracy would also have greater cost for the producer. There is no incentive whatsoever.
        \item On the bright side, a dataset like the one being developed by AMPSCZ could introduce the opportunity to actually develop a model - particularly on top of existing work like Whisper - capable of labeling disfluencies. If successful, this would greatly reduce costs for any future studies interested in these patterns of speech. But I don't see how this tool happens within the next decade if not done by some large scale psychiatric speech sampling project. Many things that are theoretically possible with machine learning do not happen, and medicine in particular is filled with examples of this.
        \item In a similar vein, one might wonder if various disfluency categories would pop up as relevant features in psychiatry data science studies more often if they were feasible to automatically classify correctly and in detail from raw audio. What is researched is inherently impacted by what capabilities (and at what cost) the field presently has, and AMPSCZ seems like a setting that should be paving the way for future standards, not falling back on only what is already normally done.
        \item Incidentally, a similar argument about driving the field forward applies to the larger message here about audio journals, and I think disfluencies have more potential in the daily diary setting than they do within interviews, both because of the recording context for the participant and the data science tractability of the longitudinal signals that can be produced with diary disfluency counts. Breaking down by category like we can easily do with TranscribeMe verbatim outputs introduces a number of opportunities in this space.  
    \end{itemize}
    \item One scientifically relevant question about Whisper that would not be fully addressed by typical machine transcription benchmarks \citep{whisper} is how accurate the outputs are in terms of illness-related jargon. Though journal recordings are sometimes generic accounts of a participant's day, it was not uncommon in our work to find submissions where a specific medication was mentioned in relation to the patient's functioning, at times one of the most informative features. In my opinion, it is especially critical to detect these instances in the diaries, as the participants have full control over what they bring up in a diary entry -- so that the mere mention of medication can provide insight into the state of mind for some individuals. The high quality professional TranscribeMe transcriptions take care to accurately transcribe technical terms, making it easy to search the resulting dataset for keywords of interest and enabling approaches like topic modeling to capture technical jargon without input data quality concerns. To thoroughly investigate whether Whisper can do the same thing, it would require a pretty large dataset of matching human-transcribed patient diaries, which I sense becoming a theme...
    \begin{itemize}
        \item More generally, a careful breakdown of the types of mistakes that we find Whisper to make across this large journal set would have much value for planning budgets of future psychiatry studies. The benchmarking done by machine learning groups will not be sufficient for e.g. the NIMH to determine to what extent manual transcription is needed or not for an individual study tackling a particular linguistic question. But an exhaustive characterization on a large psychiatry dataset would go a long way towards intelligently budgeting for manual transcription when it is needed and not when it is not. Both kinds of errors become costly for research funding over time, and AMPSCZ seems the type of initiative that would ideally be adding value across the field in ways like improved downstream proposals, rather than acting as if it were a massive version of a single group study. 
        \item Note that this sort of breakdown would be very practical to complete given both transcript sets, so it carries less risk than the proposed aim of eventually generating adjusted automatic transcription models that can tackle psychiatry-specific issues. Having such a breakdown would also help with guiding the proposed model engineering in the slightly longer term. 
    \end{itemize}
    \item When obtaining naturalistic diary recordings from a large number of people across a variety of locations, there will certainly be high variance in accents, background noise, enunciation, and other factors that can affect transcription quality. TranscribeMe professional transcribers have clear ways to mark uncertainty on certain words as well as truly inaudible portions of the recording. They will also mark noises that could be clinically relevant but would not be included in (and might even disrupt the accuracy of) automated outputs, like crying.
    \begin{itemize}
        \item Recall that these TranscribeMe annotations have enabled the development of relatively straightforward real-time quality monitoring infrastructure.
        \item Whisper meanwhile has been the subject of many anecdotal community accounts (yes I am referencing Twitter) of entirely hallucinated outputs when lower quality audio or audio with long pause periods is input. 
        \item Especially with a clinical population where symptoms can legitimately lead to less understandable speech, the ability for TranscribeMe to not only work through difficult speech pretty well but also to mark robustly when they can't should not be undersold. Whisper has not been sufficiently evaluated on these issues, doubly so in a patient context.
        \item This effect may go beyond vocal properties like mumbling, extending to linguistic abnormalities too -- a model like Whisper is certainly weighing context provided by surrounding words in how it "chooses" to transcribe. It remains entirely unanswered, because again this is not something a machine learning group has the ability or desire to study, whether current out of the box models might disproportionately introduce errors in less coherent speech. Yet another topic that could be addressable by a project like AMPSCZ, were it to obtain professional transcriptions as the primary language source.
    \end{itemize}
    \item Another question warranting further debate is the role of TranscribeMe's manual sentence splitting in developing certain diary summary features. It is clear in our internal diary dataset that the human-determined sentence breaks are influenced by speaking rate, pause usage, and possibly other speech patterns, in addition to the linguistic content. The words per sentence feature was demonstrated to be of clinical relevance, and many of the more interpretable transcript feature extraction techniques available are in turn influenced by input sentence structure. As it remains to be seen how Whisper-derived transcripts will be broken into sentences, experimenting with this in comparison to the TranscribeMe-obtained transcripts would be another useful angle for a pilot investigation to establish baseline best practices for the field, particularly in projects focusing on interpretable models. This is not something we can do with the AMPSCZ interview transcripts due to their dialogue format. 
    \item Note the dataset size anticipated for AMPSCZ diaries is much more in line with modern machine learning expectations than any individual study could hope to be, and moreover the dataset will cover not only an exceptional number of subjects, but also exceptional range in languages and cultures for a digital psychiatry study. Not obtaining strong labels on such a dataset would be a missed opportunity in my opinion. Conversely, the strength in source diversity also introduces further complications with existing automated methods: Whisper's word error rate is much worse on some languages than others \citep{whisper}, and we do not have a clear plan for how we would validate automated transcriptions from non-English sites.
    \item The phone app collection methodology can result in a wide variety in audio quality and background contexts, and as such a labeled interview dataset is not a replacement for the audio journal one in a machine learning context. As far as future scientific directions for the field, I think I've made it clear at this point that I feel focusing too much on interviews would be a mistake.
    \item There are large and entirely open questions on the relationship between speech from interview recordings and speech from audio journals, and how individual or temporal heterogeneity in speech patterns might connect to the two different speech sampling modalities. AMPSCZ could provide an unprecedented opportunity to do comprehensive linguistic analyses in both domains, but if the transcription methodology for journals is not comparable to that for interviews it would confound such work. It would also restrict that work to a much smaller set of possible researchers due to the aforementioned data access situation with unredacted transcripts.
\end{itemize}
\noindent There are a large number of potential opportunities for the AMPSCZ audio journal dataset that would require transcripts to be obtained from TranscribeMe. Unless a different large scale psychotic disorders speech sampling project steps in, those opportunities will not be obtainable with automated transcription outputs in the lifetime of this project. \\

\noindent AMPSCZ speech sampling has the ability to lead the way in establishing the high relevance of the journal format and in developing an applied machine learning framework that would begin to allow future psychiatry research to systematically use a number of ML tools in an appropriate way. If journals are not professionally transcribed, the repository journal dataset will have a gaping hole where redacted transcripts could have been, the domain-specific validation on the automated methods that ought to occur won't, and the features that are extracted for sharing will be driven by the status quo with minimal community input -- all of which would be squandering the advantages a project like AMPSCZ has. Fortunately, there is an easy way to fund audio journal transcription, and I will next present the more specific evidence from the dataset so far on the serious concerns about the psychs interview transcription plan that were listed above (section \ref{sec:psychs-dumb}).

\section{Comparing open and psychs in the early transcript dataset}
\label{sec:open-psychs-data}
As mentioned, major reasons that psychs interviews have turned out to be a quite inefficient use of transcription budget to date include how long their durations have been in practice, how stark the stunted conversational nature has been in the turn-based structure of psychs transcripts as opposed to open ones, how different the balance of patient/interviewer speech has been between the formats, and how much more expensive the cost per word has been in aggregate due to pauses during psychs interviews. In this section, I report the specific preliminary results on these topics. \\

\begin{figure}[h]
\centering
\includegraphics[width=\textwidth,keepaspectratio]{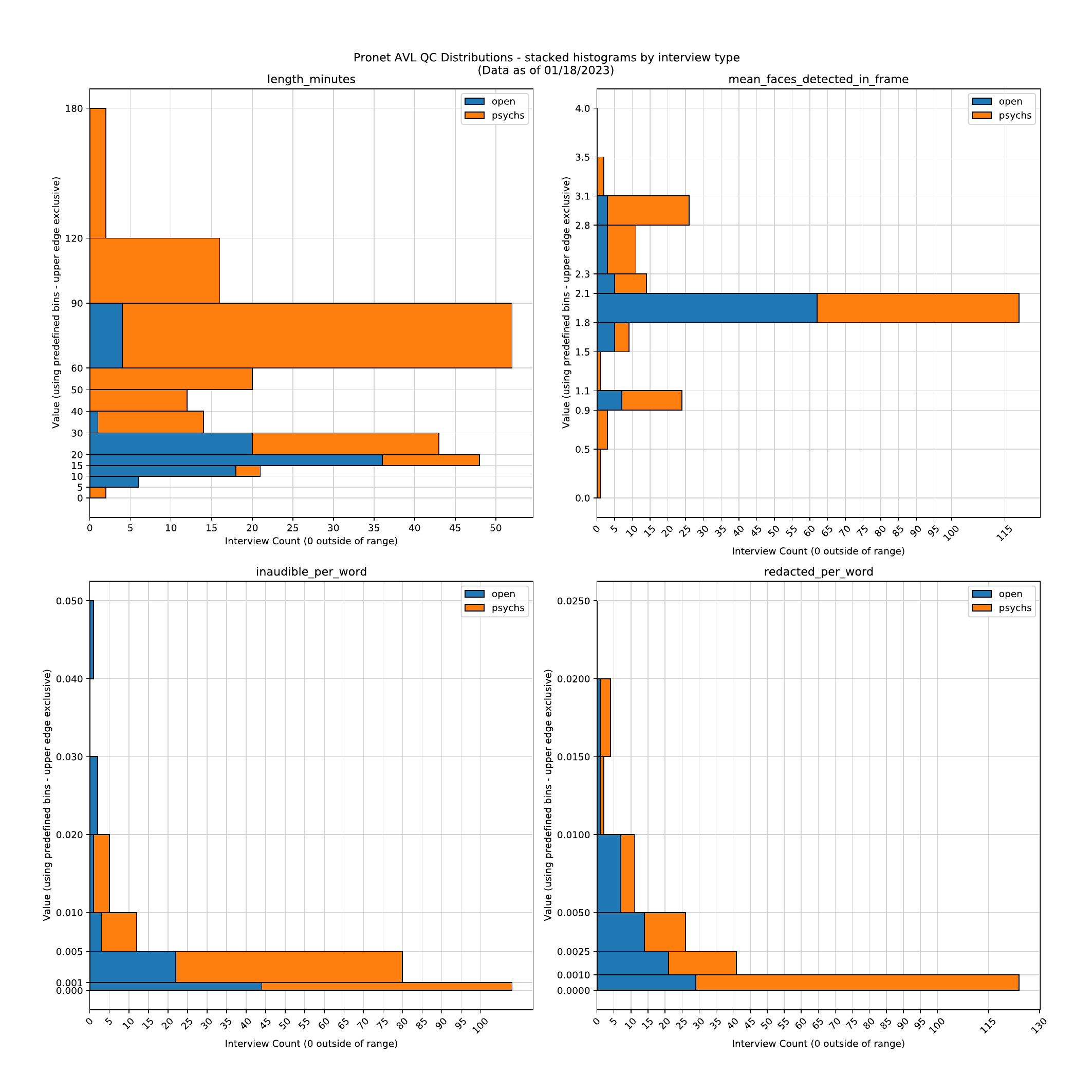}
\caption[Snapshot of key interview QC feature distributions from the Pronet server, colored by interview type (reproduced from chapter \ref{ch:2}).]{\textbf{Snapshot of key interview QC feature distributions from the Pronet server, colored by interview type (reproduced from chapter \ref{ch:2}).} Stacked histograms depicting the distributions of relevant quality control metrics across interviews are generated weekly for each central server. Within each distribution, interviews are marked by their type - for the AMPSCZ project, open (blue) versus psychs (orange). This figure is an example taken from Pronet 1/18/2023. It presents the four most core features for high level monitoring: interview duration in minutes (top left), mean number of faces detected from extracted video frames (top right), number of words marked inaudible per total transcript words (bottom left), and number of words redacted per total transcript words (bottom right).}
\label{fig:pronet-key-dists2}
\end{figure}

\begin{figure}[h]
\centering
\includegraphics[width=\textwidth,keepaspectratio]{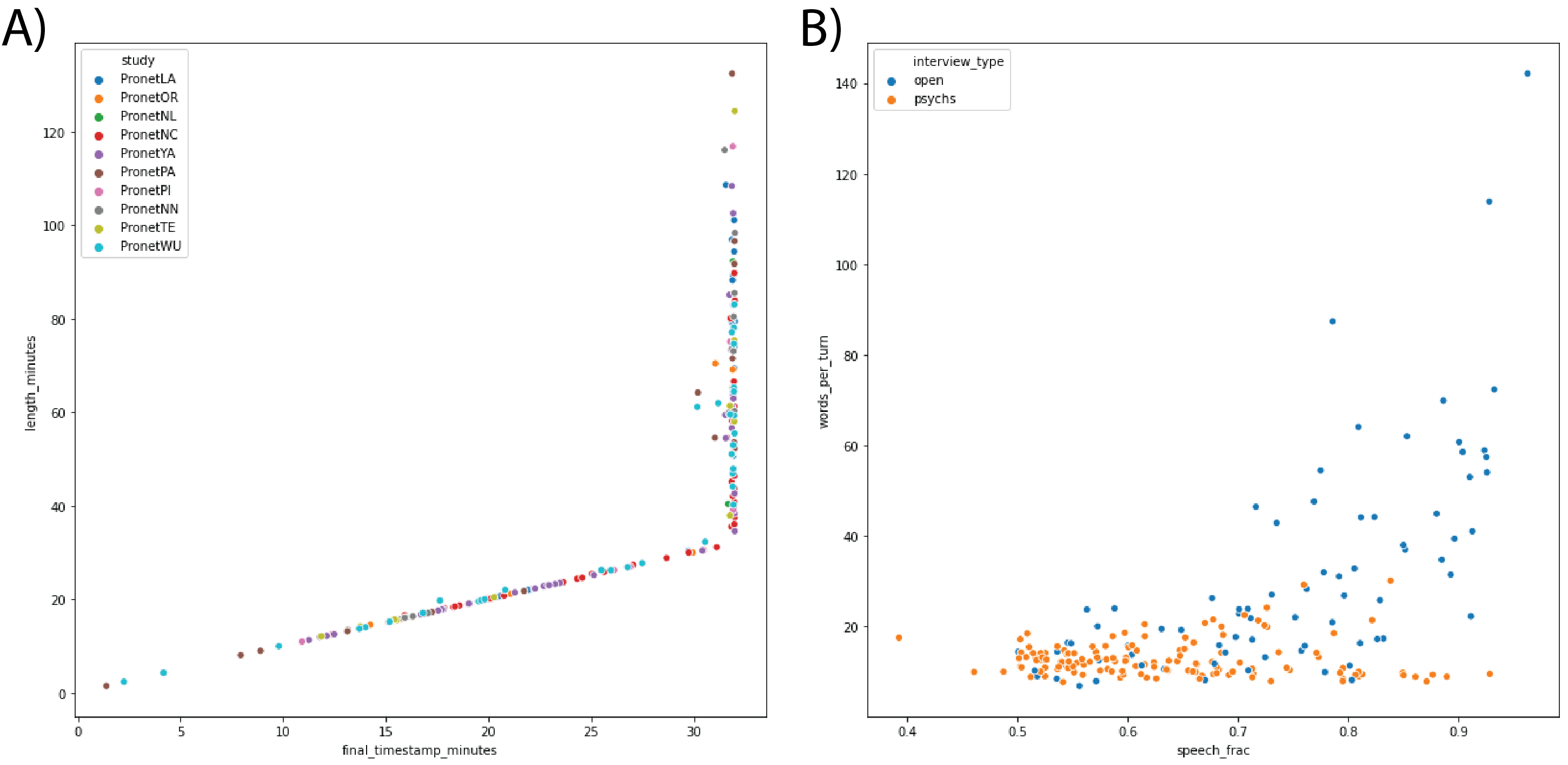}
\caption[Using transcript summary features to better understand interview structure (reproduced from chapter \ref{ch:2}).]{\textbf{Using transcript summary features to better understand interview structure (reproduced from chapter \ref{ch:2}).} Scatter plots were generated to visualize transcript summary feature relationships across all Pronet interviews as of 1/23/2023. First, the start timestamp of the final turn in the transcript (in minutes) was plotted against the corresponding audio length (in minutes) and colored by contributing site (A). This was done to verify TranscribeMe's implementation of the length cutoff described earlier. Then, additional features about the structure of the transcription were derived from the existing QC metrics: the word count from the most verbose speaker ID was divided by the total word count to obtain "speech\_frac", and the total number of words was divided by the total number of turns to obtain "words\_per\_turn". Note that the former feature does not contain any information about participant versus interviewer, it simply relays how balanced the verbosity was amongst speakers. These two features were plotted against each other, with points colored by interview type (B).}
\label{fig:pronet-trans-scatter2}
\end{figure}

\noindent As of late January 2023, the AMPSCZ interview dataset had the following properties related to psychs interview length:
\begin{itemize}
    \item More than $
\frac{2}{3}$ ($n > 110$) of successfully processed psychs interviews on Pronet exceeded 30 minutes in duration, which is the approximate TranscribeMe saturation point (Figure \ref{fig:pronet-key-dists2}, top left).
    \item More than $\frac{1}{3}$ ($n > 65$) of the Pronet psychs transcripts were from recordings exceeding 1 hour in duration (Figure \ref{fig:pronet-key-dists2}, top left) -- meaning they contained $\leq \frac{1}{2}$ of the true interview content!
    \item Every point on the right side of Figure \ref{fig:pronet-trans-scatter2}A (transcript duration versus true duration) cost over $\$55$, for an incomplete transcription with in many cases a great deal of missingness.
    \item Open interviews on the other hand rarely exceeded $30$ minutes thus far, and per the SOP are intended to last $15-20$ minutes in most cases. The four longest "open" interviews have already been identified as mistaken uploads that were meant to go under psychs (Figure \ref{fig:pronet-key-dists2}, top left).
    \item  Therefore the vast majority of open interviews are transcribed in full, while the majority of psychs interviews are \textbf{not} transcribed in full, often not even close (Figure \ref{fig:pronet-trans-scatter2}A).    
\end{itemize}
\noindent Critically, it is not clear to what extent sites plan around or are even aware of the 30 minute transcription limit; given the overall levels of SOP adherence we've observed, which remain low at some sites despite months of repeated warnings, I would not be surprised to find that most sites pay no mind to the duration limit. Regardless, the value of a partially transcribed structured clinical interview is unclear, as is the consistency in how sites conduct such interviews. \\

\begin{figure}[h]
\centering
\includegraphics[width=\textwidth,keepaspectratio]{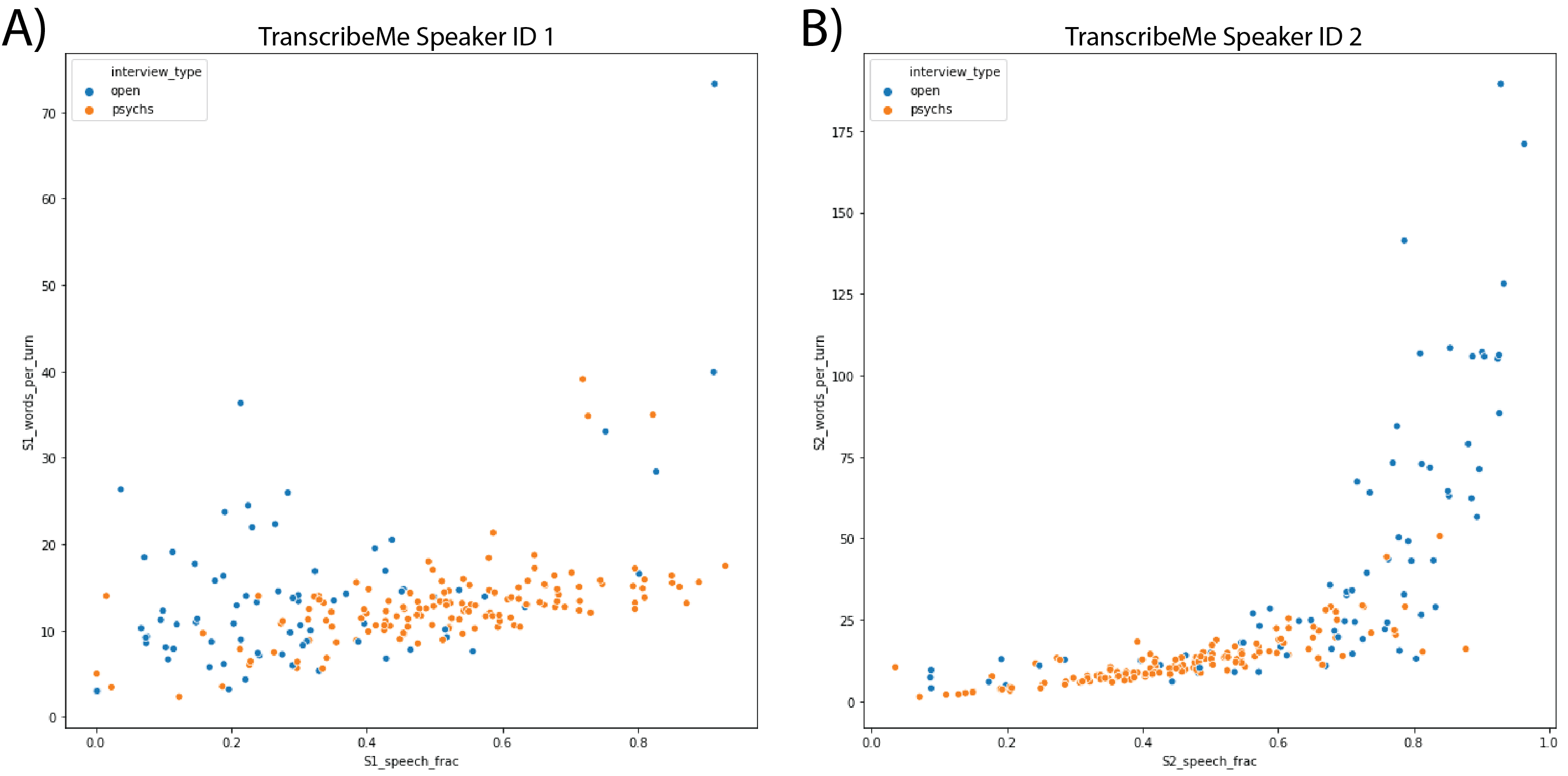}
\caption[Interview transcript structure by TranscribeMe speaker ID across Pronet (reproduced from chapter \ref{ch:2}).]{\textbf{Interview transcript structure by TranscribeMe speaker ID across Pronet (reproduced from chapter \ref{ch:2}).} Using the speaker ID specific versions of the features depicted in Figure \ref{fig:pronet-trans-scatter2}B, I broke down the turn structure trends seen in the open (blue) and psychs (orange) interviews by the first (A) versus second (B) unique speakers chronologically. By protocol, the first speaker should be the primary interviewer and the second the participant. Though this has not been perfectly followed, it has been consistent enough to identify potential high level trends attributable to interviewer versus participant. Note that the y-axis of (A) extends to just under 80 words per turn, while the y-axis of (B) extends as far as nearly 200 words per turn.}
\label{fig:pronet-all-ids-scatter2}
\end{figure}

\noindent Stark differences between open and psychs interviews are already abundantly clear in Figure \ref{fig:pronet-trans-scatter2}B. This disparity can be broken down much further when looking at differences by TranscribeMe speaker ID, differences in extended pause durations marked by TranscribeMe, and so on. Specific observations on the flow of psychs interviews to date are as follows:
\begin{itemize}
    \item Words per turn are significantly higher in most open interviews across speakers, and especially so for the speaker ID that per the SOP should be the participant (Figure \ref{fig:pronet-all-ids-scatter2}).
    \item The fraction of speech contributed by each speaker is also much closer to even in most psychs interviews, while open interviews typically skew towards containing a much greater percentage of (likely) participant words (Figure \ref{fig:pronet-all-ids-scatter2}). 
    \item Indeed, there were large differences in the cost per participant word of transcribing open versus psychs interviews using a number of different methods, to be elaborated below.
    \item Interestingly, there was even a moderate difference observed in the cost per transcribed word overall. I believe this is in part due to psychs sessions sometimes including short interview breaks, which results in some transcripts having 5 straight minutes of silence "transcribed". However, a meaningful difference persists when comparing median cost per word, so there is likely also an effect of greater conversational pausing in psychs interviews.
\end{itemize}  
\noindent So open interviews not only contain many more participant words in a given timeframe, they also contain more total words in a given timeframe. The words in open interviews are furthermore contained in significantly longer segments of speech, meaning that the true value per word (that we should be willing to pay) is likely \emph{higher} for open interviews, despite the true cost we pay TranscribeMe being less. \\

\noindent To translate these differences to real costs, I looked at price per word estimates for open versus psychs transcripts on Pronet as of 1/31/2023. Though transcription cost is $\$1.83$ per minute for English (and Spanish) for AMPSCZ and the early dataset is overwhelmingly English, the cost for other languages will be between $\$3.30$ and $\$3.90$ per minute. To make things easy, I assumed $\$2$ per minute in the following calculations, with the duration of interviews capped at 30 minutes (though 32 would be more precise to the length and cost entailed, it makes only a small difference in the comparison):
\begin{itemize}
    \item Assuming that the participant has the expected speaker ID per the SOP, it cost on average $\sim 2.8$ cents per participant word for open interview transcripts and $\sim 4.7$ cents per participant word for psychs transcripts. 
    \begin{itemize}
        \item This means that 1 dollar spent on open transcriptions produced $\sim 36$ participant words while 1 dollar spent on psychs transcriptions produced $\sim 21$ participant words.
        \item Thus the anticipated budget of 1 million (or maybe more) USD for psychs transcriptions over the lifetime of the project would produce an estimated 21,000,000 participant words. If open interview collection were a larger component of the project (though this is too late to change), the estimated transcription cost for a dataset with the same number of participant words would be under 600K USD.
    \end{itemize}
    \item One concern with the first estimate is that CHR patients will do many more psychs interviews than controls will per the protocol, but all participants will follow the same open interview protocol. Therefore the results could be partially confounded by the cost per word of patient speech versus control speech. However when restricting the estimates to only the first open versus first psychs interview recorded for each subject ID, the results change only minimally.
    \begin{itemize}
        \item In this case open interview participant words had an average estimated cost of $\sim 2.9$ cents, while psychs interview participant words had an average estimated cost of $\sim 4.4$ cents.
        \item The difference in subject words per dollar was therefore $\sim 34$ expected for open transcripts and $\sim 23$ words expected for psychs transcripts.
        \item With a budget of 1 million USD, the final dataset would therefore be expected to contain $> 10,000,000$ additional participant words if the open interview protocol were used - a nearly $1.5x$ multiplier.
    \end{itemize}
    \item A remaining concern with these estimates is that outlier interviews could disproportionately drive the cost differences, which is most important to account for in the case of SOP violations like pausing Zoom recording within the first 30 minutes of a psychs session, as we hope that such violations will decrease in frequency as the study continues. I therefore repeated the estimate of the cost per participant word for the first interview of each type, but using the median instead of the mean. This again produced minimal difference in the comparative results.
    \begin{itemize}
        \item The median estimated transcription cost per participant word in baseline open interviews was $\sim 1.9$ cents, while the median estimated transcription cost per participant word in baseline psychs interviews was $\sim 3.2$ cents.
        \item This represented an even larger difference at scale, producing $\sim 52$ subject words per dollar for the median open transcript and $\sim 31$ subject words per dollar for the median psychs transcript.
        \item It also represented a larger relative difference, with the multiplier from psychs to open $> 1.5x$ in the median estimate.
    \end{itemize}
    \item Finally, the assignment of participant versus interviewer ID based on SOP expectations was not perfect, though it was accurate a majority of the time. Still, it is worth considering the cost per word overall for open versus psychs, so I also estimated that - again using median estimates taken from only baseline interviews. The difference was unsurprisingly more subtle, but still present, and highly relevant at the scale of AMPSCZ.
    \begin{itemize}
        \item The median estimated cost per transcribed word was $\sim 1.3$ cents for open baseline interviews and $\sim 1.6$ cents for psychs baseline interviews.
        \item Thus the total transcribed words per dollar spent on the median open transcript was estimated at $\sim 77$, while the same estimate for psychs came in at $\sim 63$ words.
        \item For an expense of 1 million dollars the difference in expected total words remains $> 10$ million, suggesting the higher number of participant words per minute observed in open interviews was not due only to a counter decrease in the number of interviewer words per minute.
        \item Even in this case it appears that $> 20\%$ more transcribed words are obtained for the same cost if using the open instead of psychs protocol.
    \end{itemize}
\end{itemize}
\noindent Of course, the data are not perfectly clean due to a number of site SOP errors, but very similar trends emerge even when adjustments are made, such as alternative methodologies for identifying the likely participant. Moreover, some of the observed upload mistakes would actually bias the data away from the identified issues with lesser content of psychs interviews, for example the 4 psychs uploads masquerading as open here. \\

On the topic of words per minute in the open versus psychs interviews, I checked on the dataset in mid-March 2023. I aimed to evaluate the extent to which the more expensive cost per word was explainable by long unnecessary recording pauses before starting or in the middle of (the first 30 minutes of) psychs interviews, versus perhaps just slower speech patterns or more incremental pausing -- and it does appear that wasteful pausing is a factor at play. For one thing, TranscribeMe will mark extended periods with no speech using a "[silence]" indicator. TranscribeMe included a silence label in 1080 turns across 222 psychs transcripts, yet only in 133 turns across 120 open transcripts. The average transcribed length of psychs interviews is a little less than $2$ times the average transcribed length of open interviews, but here we see $\sim 4.87$ silence lines per psychs transcript and only $\sim 1.11$ silence lines per open transcript, a difference of more than $4x$. 

Additionally, 11 psychs interviews and \emph{no} open interviews had a turn with timestamps suggesting $> 1$ \emph{minute} per word. At $> 30$ seconds per word, there were 29 psychs transcripts containing at least one such turn, yet just 2 open transcripts meeting the same criteria. Some psychs transcripts contained turns exceeding 5 minutes in length, an effect certainly not explainable by a long rich monologue. For some of the less egregious examples, it is not entirely clear whether the interviewer could have been more careful with how they planned the timing of the recording, or whether it was simply a participant taking a minute or two to think in silence before providing a short answer to a factual question about e.g. symptom frequency. However this latter case is a greater indictment of the psychs transcription plan, because it is not an addressable operational failure but rather another hard to avoid complication of the structure of the clinical interview. 

\section{Interview transcription budget estimates}
\label{sec:blowing-money}
It is important to note that in addition to the much longer duration and likely lesser speech sampling value of psychs interviews, they are also a significantly greater portion of the data collection protocol than open interviews are. So not only are psychs interviews costing notably more per session for notably less upside, there will also be many more psychs sessions. This exacerbates the detailed budgetary concerns. More concretely, AMPSCZ intends to enroll 1977 CHR subjects to complete the full battery, which involves 9 psychs interview timepoints and 2 open interview timepoints \citep{Brady2023}. \\

\noindent Using a lower bound estimate of an average $\$50$ cost per psychs transcription, we arrive at the following expense to obtain psychs transcripts per the current protocol for CHR subjects over the lifetime of the study:
\begin{center}\begin{tabular}{c}\begin{lstlisting}
1977 * 9 * 50 = 889650 USD
\end{lstlisting}\end{tabular}\end{center}
\noindent As control subjects will also complete a baseline psychs interview, and $\$50$ per CHR psychs transcript is almost certainly an underestimate, it is highly likely that if the present plan is continued the money spent on psychs transcriptions will end up \textbf{exceeding 1 million dollars}. \\

\noindent By contrast, using the average duration of open interviews to date and a rough average transcription cost (weighted across languages) of $\$2$ per minute, a very safe expectation for the cost of open transcriptions per interview is $\$35$, thus leading to the following estimated cost of CHR participant open transcripts over the lifetime of AMPSCZ:  
\begin{center}\begin{tabular}{c}\begin{lstlisting}
1977 * 2 * 35 = 138390 USD
\end{lstlisting}\end{tabular}\end{center}
\noindent This means that the money spent on transcripts for the CHR population of AMPSCZ is anticipated to be \emph{at least} $6.5x$ higher for the psychs interview type than for the open interview type. \\

All open interviews are already being transcribed, so there are not any realistic actions that can be pursued at this point in the project based purely on the advantages of open versus psychs. However, these numbers and the broader background on disadvantages of structured clinical interviews for speech sampling science do bring up the question of whether the money currently planned to be used for psychs transcriptions should be spent on interviews at all. It is highly advisable in my opinion that a portion of it be used to obtain high quality transcriptions for the full AMPSCZ audio journal set. 

Regardless, I have yet to see a good justification for the choice to transcribe every single interview at a cap of $\sim 30$ minutes instead of transcribing a subset of interviews at full length or potentially considering a cheaper version of manual transcription (e.g. no verbatim notation, no timestamps, lesser expectations for a very low word error rate) to be used on the full length of all psychs interviews. Budgetary choices could of course mix and match between these various trade-offs, but I think it is important a deeper discussion is had about how to best spend this money while addressing the questions raised here in the context of the project's scientific aims. \\

\FloatBarrier

\noindent More concretely, I would raise all of the following before I would give someone $\$1,000,000$ for this project:
\begin{enumerate}
    \item What is the scientific plan for using the psychs interview recordings? 
    \begin{enumerate}
        \item What value is supposed to be obtained from them beyond the value that can already be obtained from the PSYCHS clinical scale ratings?
        \item What is the concrete and technically sound plan for achieving any such goals?
         \item To what extent are any of these goals actually better suited to psychs recordings, versus being tacked onto psychs from the open speech sampling plan (at a likely much lower efficiency level) only because the recordings already exist?
    \end{enumerate}
    \item How much of this scientific plan relies on linguistic analyses?
    \begin{enumerate}
        \item To what extent do the linguistic analyses proposed require some of the more expensive luxury features of TranscribeMe such as verbatim notation and labeling of events like interviewer/participant crosstalk?
        \item Are psychs interviews actually a good environment for performing any of the said linguistic analyses? Are they a uniquely good environment? 
    \end{enumerate}
    \item How much does this scientific plan benefit from having the whole dataset of psychs recordings transcribed instead of a well-crafted subset of say $50\%$ size?
    \item How much is this scientific plan going to suffer because of the current transcription duration cutoff, a cutoff which has not been sufficient to cover a typical psychs interview in practice?
    \item What needs to be seriously enforced about psychs interview procedures across sites for the dataset to be acceptable for any proposed analyses that do require a large transcription budget?
    \begin{enumerate}
        \item Who will enforce these requirements and what power will they actually have to influence sites? 
        \item When in the project timeline will enforcement actually begin?
        \item Whether durations of $\geq 2$ hours are clinically necessary or not, are they scientifically necessary to record in full for analysis? Surely there must be some point at which the cost to transcribe becomes too ridiculous.
        \item How will procedural variation that is allowed between sites be assessed for potential analysis biases, and how will the problematic ones be handled in concrete terms? 
    \end{enumerate}
    \item If the transcription budget on the whole is capped, why exactly are any psychs linguistic analyses that do justify the present expense estimates worth more than the analyses that would be enabled by redirecting some of this budget towards the audio journals? 
\end{enumerate}
\noindent It is not that psychs interview recordings are not without some unique benefits; in a resource unlimited world I would have them transcribed in full to facilitate the scientific questions that they are most suited for. Because of the redaction issue with data sharing, it is worth considering also that 1 million is a tiny piece of the overall AMPSCZ budget. If the main priority of the project is building a high quality NIH data repository dataset, then I would hope that all audio could be transcribed to the highest quality in full. But the reality of the situation is that quality has not been sufficiently enforced on AMPSCZ interview recordings to date, and there is a very real chance that trade-offs need to be made with regards to psychs transcriptions versus diary transcriptions, or even the nature of what does get transcribed within the psychs recording set. When it comes to diaries especially, I have a very strong opinion that they are being mistakenly overlooked. 

\section{Diary transcription budget estimates}
\label{sec:diary-cost-estimates}
With a current target of $\sim 270$ transcribed minutes of psychs interviews per CHR participant, 6 straight months of quality daily audio journal recordings could be transcribed for each participant instead. Realistically, audio diary participation will not be quite so wonderful that every subject submits a diary of well over a minute in duration every single day for half of a year, so in most cases an entire journal set could be transcribed for only a fraction of the intended psychs interview cost. To add further perspective, site mistakes described above have already wasted enough money to obtain transcriptions of 7 entire months of journals by a conservative estimate. Mistakes are much more likely with interview upload, and thus far some sites have not shown much improvement despite repeated notifications. 

In terms of current data collection pace, as of mid-February 2023, just over 1000 diaries had been successfully submitted by AMPSCZ subjects from across 19 locations. The speech sampling project is in early stages, with only $\sim 2\%$ of planned interviews for the lifetime of AMPSCZ processed at that same time -- suggesting that the journal dataset is on pace to reach an impressive size of 50,000 recordings. Despite the large amount of data collected to date, there remains no pipeline for data flow or quality monitoring of the diaries, let alone a plan for feature extraction for the project. This is exemplary of a more general underrating of the diary format across digital psychiatry, as I will discuss a bit more below. 

Though I hope to see AMPSCZ diary collection exceed this estimate of $n \sim 50000$ once more infrastructure is in place and more sites get up and running, this is a solid estimation for final dataset at present. To be safe, I will assume an average duration of 2 minutes across all journals, which is very likely a moderate overestimate based on the early data as well as prior longitudinal internal studies that ended with mean diary duration a little under $1.5$ minutes. Therefore, a rough estimate erring on the side of over-budgeting would be a final AMPSCZ journal dataset containing $100000$ minutes. Cost per minute varies by language, but with $70\%$ of sites at the lowest $\$1.83$ per minute and non-English sites being disproportionately delayed in starting data collection thus far, an estimate of $\$2$ per minute across the dataset seems reasonable, and $\$2.50$ per minute average would be a generous upper bound. \hl{Thus we can expect the final cost to transcribe all audio diaries to be below or perhaps at worst within the $\$200000 - \$250000$ range.} \\

Meanwhile, the planned cost of transcribing all the psychs interviews over the lifetime of this project will easily exceed $\$800000$ and probably surpass 1 million dollars, as detailed above. By reducing the number of psychs interviews sent for transcription by just $\sim 25\%$, it is likely that the \emph{entire set} of AMPSCZ journals could receive high quality transcriptions. Cutting back on the number of psychs interviews sent would be very easy, because the value obtained from transcribing all 9 from all CHR patients is highly suspect for the many reasons covered. 

I feel continuing to fund psychs transcription at the expense of diaries is a clear misstep, and a major aim of mine here is to rectify that. Next, I will provide a preview of the audio journal dataset collected in AMPSCZ so far. For more general scientific arguments on preferring audio journals as a speech sampling source, see section \ref{sec:guillermo-is-annoying} below.

\section{Diary dataset preview}
\label{sec:diary-dataset-u24}
As mentioned, over 1000 diaries had already been collected and pulled to the AMPSCZ data aggregation servers as of mid-February 2023, but no processing pipeline is yet implemented, and a concrete plan for next steps still needs to be outlined. To give a taste of the data flow thus far along the lines of the interview accounting performed by that pipeline, I provide a breakdown of diary counts by site in Table \ref{table:u24-diaries-submit2}. \\

\begin{table}[!htbp]
\centering
\caption[Early journal data collection numbers show promise in the large collaborative AMPSCZ project.]{\textbf{Early journal data collection numbers show promise in the large collaborative AMPSCZ project.} As of 2/13/2023, there were 1031 successful diary recording submissions from across participants in the AMPSCZ project (described in chapter \ref{ch:2}). 952 total audio journals were successfully pulled by Lochness to the Pronet data aggregation server across Pronet sites, and 79 total audio journals were successfully pulled by Lochness to the Prescient data aggregation server across Prescient sites. In this table, the number of diaries from each contributing site is reported.}
\label{table:u24-diaries-submit2}

\begin{tabular}{ | m{2cm} | m{2cm} || m{3.5cm} | }
\hline
\textbf{Server} & \textbf{Site} & \textbf{Audio Diary \newline Submission Count} \\
\hline\hline
Prescient & ME & 59 \\
\hline
Prescient & SG & 20 \\
\hline
Prescient & All & $= 79$ \\
\hline\hline
Pronet & CA & 8 \\
\hline
Pronet & CM & 1 \\
\hline
Pronet & HA & 3 \\
\hline
Pronet & IR & 77 \\
\hline
Pronet & LA & 123 \\
\hline
Pronet & NC & 166 \\
\hline
Pronet & NL & 4 \\
\hline
Pronet & NN & 91 \\
\hline
Pronet & OR & 42 \\
\hline
Pronet & PA & 126 \\
\hline
Pronet & PI & 69 \\
\hline
Pronet & PV & 3 \\
\hline
Pronet & SF & 18 \\
\hline
Pronet & SI & 60 \\
\hline
Pronet & TE & 21 \\
\hline
Pronet & WU & 81 \\
\hline
Pronet & YA & 59 \\
\hline
Pronet & All & $= 952$ \\
\hline
\end{tabular}
\end{table}

\noindent To obtain a preview of the TranscribeMe audio diaries to further inform decision making, a sample of $\sim 50$ recordings were sampled as follows:
\begin{enumerate}
    \item Obtain a list of all recorded Pronet diaries so far exceeding 1 MB, as a quick proxy for files of at least 1 minute in length. This list contained only a few hundred recordings to begin with.
    \item Filter the list to remove non-English sites (presently only PV from the Pronet sites in Table \ref{table:u24-diaries-submit2}), as I am not able to adequately evaluate such transcripts.
    \item Narrow the list to include only 1 recording per unique subject ID, which lead to a final journal set of $53$ recordings.
    \item Send the recordings to TranscribeMe, ensuring the same high quality professional transcription settings would be followed as are presently for interviews, but also using sentence-level splitting/timestamps.
\end{enumerate}
\noindent When those transcripts were completed, I summarized a number of transcript QC results across them, both to paint a picture of this sample and to demonstrate some of the more straightforward metrics that are uniquely enabled by TranscribeMe. \\

\begin{figure}[h]
\centering
\includegraphics[width=\textwidth,keepaspectratio]{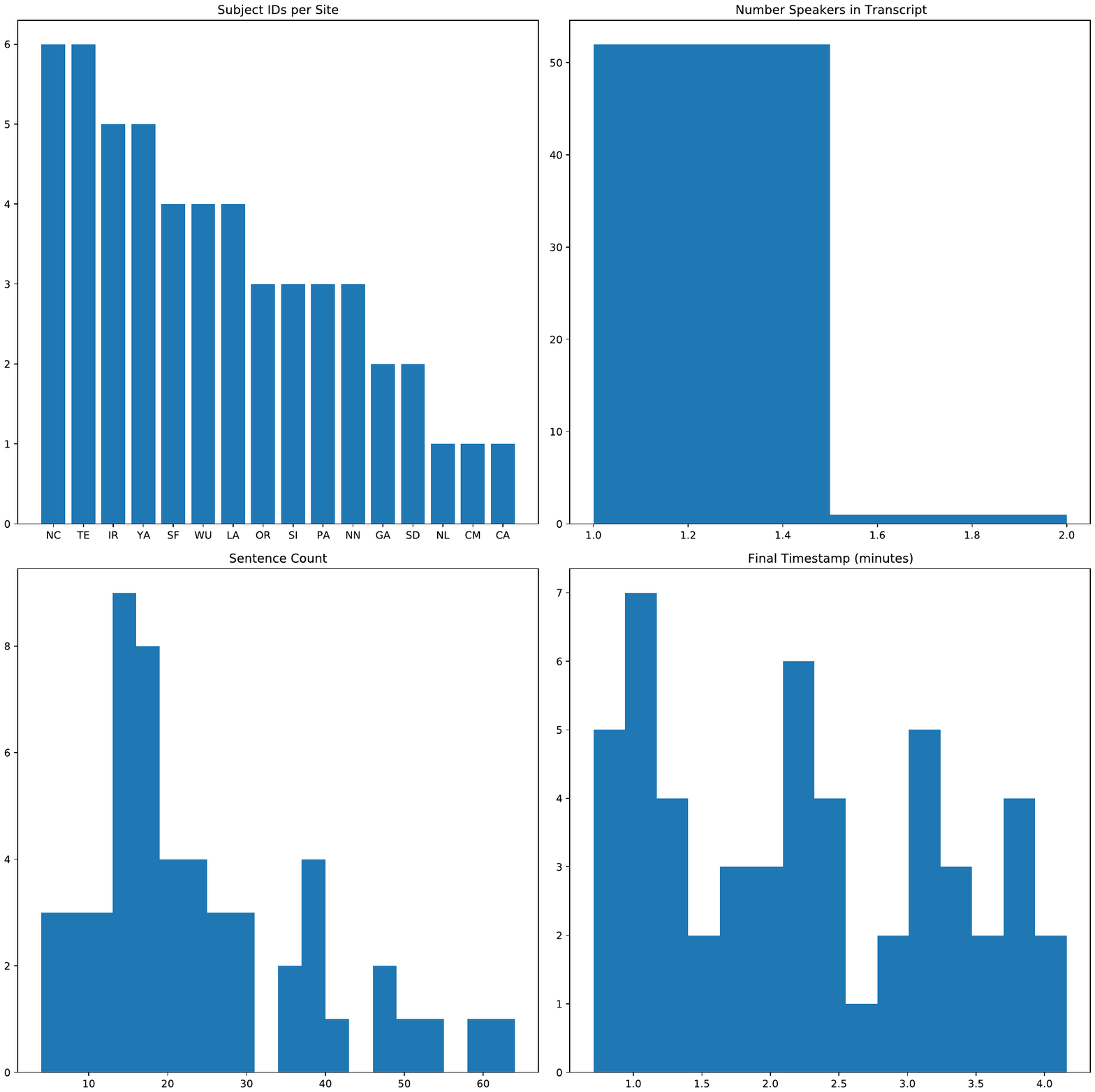}
\caption[Distribution of basic transcript features from a pilot sample of AMPSCZ diaries.]{\textbf{Distribution of basic transcript features from a pilot sample of AMPSCZ diaries.} In mid-March 2023, $53$ English language audio journal submissions on the Pronet server were sampled, each from a unique subject and meeting a minimal duration requirement. The histograms here present distributions of basic transcript QC features across the resulting dataset produced by TranscribeMe: the contributing site (top left), the number of unique speaker IDs (top right), the number of sentences (bottom left), and the final sentence timestamp in minutes (bottom right).}
\label{fig:diary-pdf1}
\end{figure}

\begin{figure}[h]
\centering
\includegraphics[width=\textwidth,keepaspectratio]{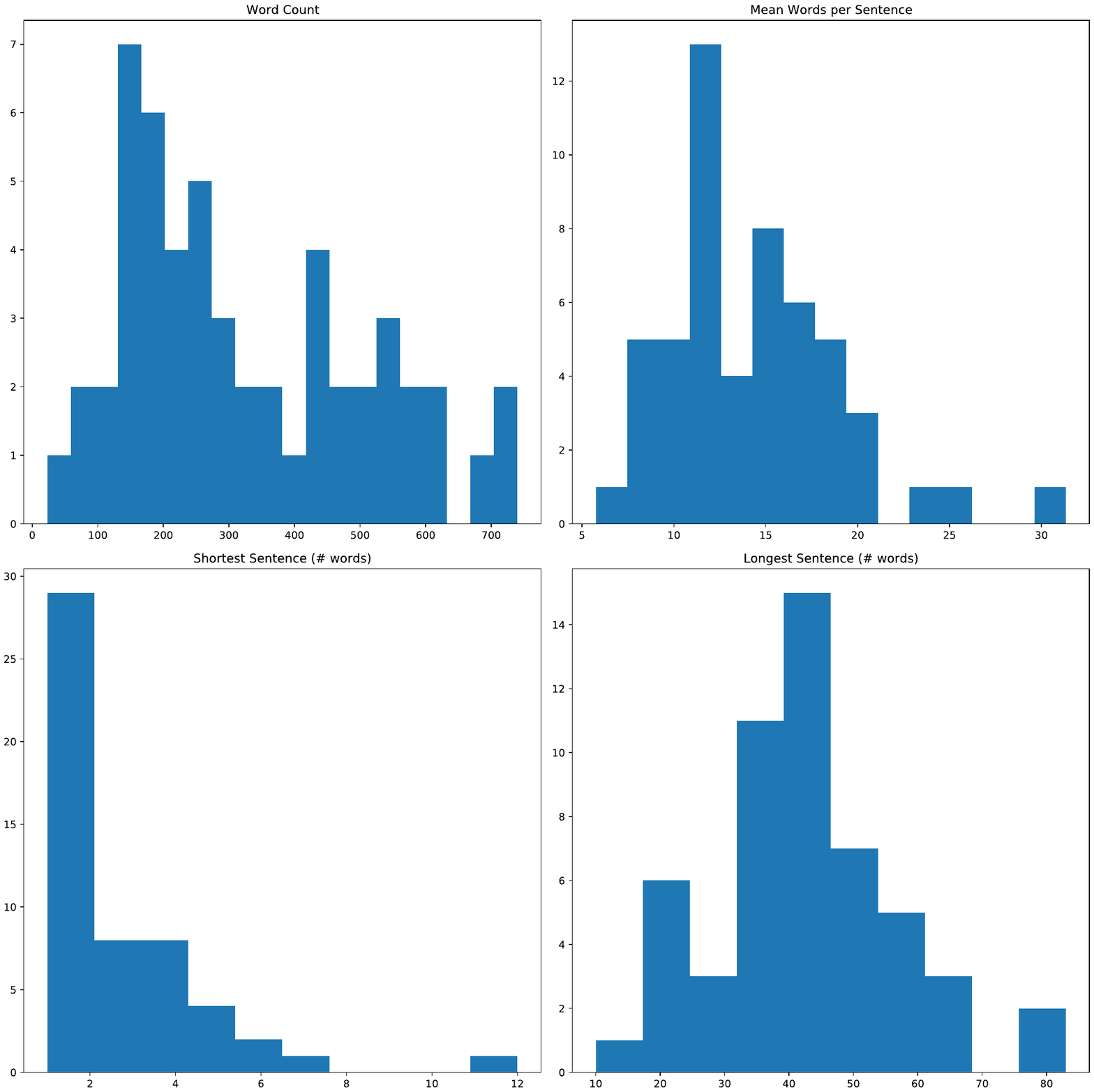}
\caption[Distribution of transcript sentence structure features from a pilot sample of AMPSCZ diaries.]{\textbf{Distribution of transcript sentence structure features from a pilot sample of AMPSCZ diaries.} Using the same transcript set as Figure \ref{fig:diary-pdf1}, I also generated histograms of features related to verbosity and TranscribeMe's sentence splitting. The distributions presented over transcripts ($n=53$) here are: the total word count (top left), the mean words per sentence (top right), the number of words in the shortest sentence (bottom left), and the number of words in the longest sentence (bottom right).}
\label{fig:diary-pdf2}
\end{figure}

\begin{figure}[h]
\centering
\includegraphics[width=\textwidth,keepaspectratio]{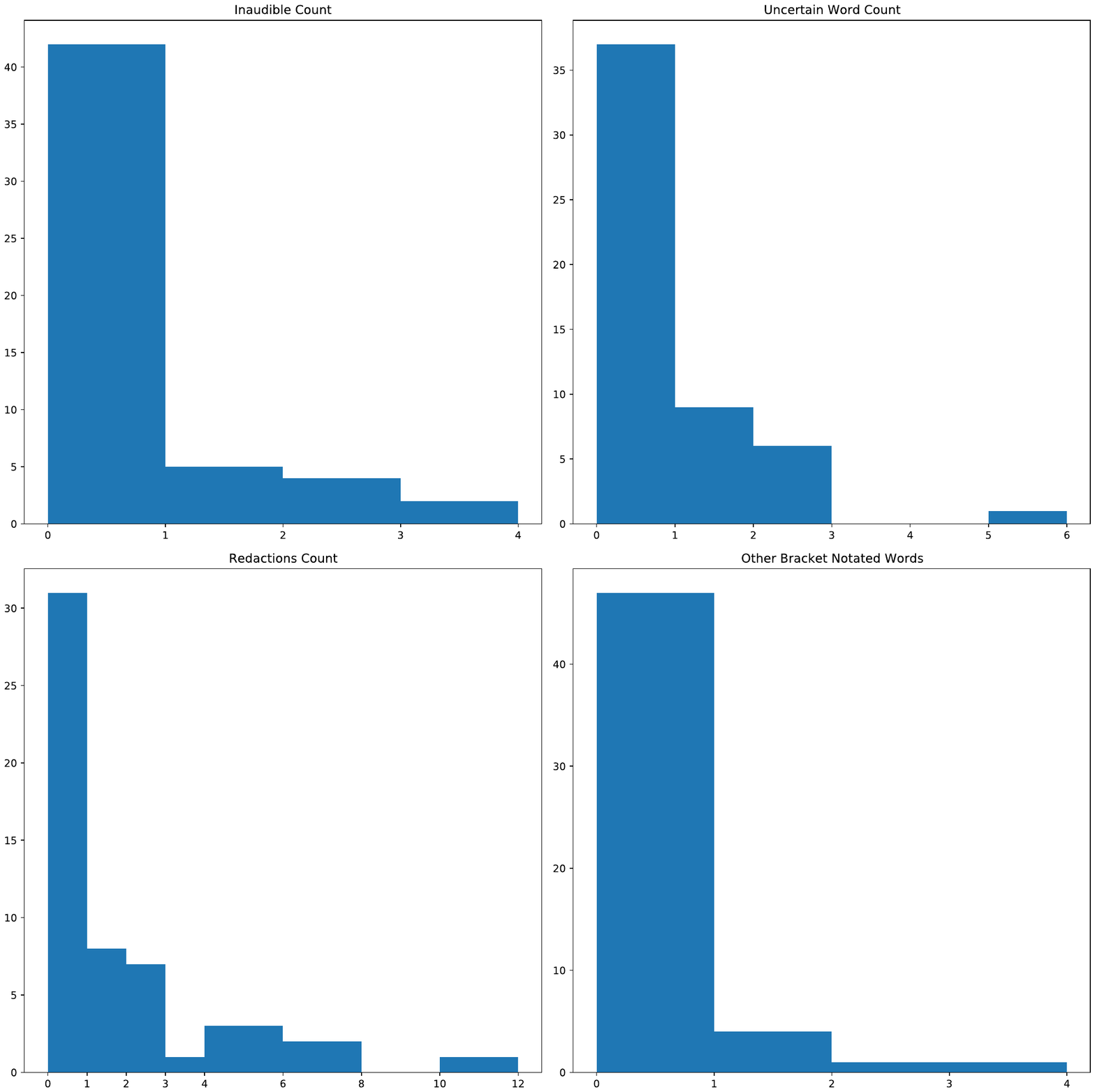}
\caption[Distribution of transcript quality assurance features from a pilot sample of AMPSCZ diaries.]{\textbf{Distribution of transcript quality assurance features from a pilot sample of AMPSCZ diaries.} Using the same transcript set as Figure \ref{fig:diary-pdf1}, I also generated histograms of features related to transcript quality control. The distributions presented over transcripts ($n=53$) here are: the number of times a part of the recording was marked inaudible (top left), the the number of words denoted with uncertainty (top right), the number of words redacted (bottom left), and the number of occurrences of other bracketed markings (bottom right). Such bracketed markings may include labels of laughter, crying, coughing, and so on, as well as labels of extended periods of silence.}
\label{fig:diary-pdf3}
\end{figure}

\begin{figure}[h]
\centering
\includegraphics[width=\textwidth,keepaspectratio]{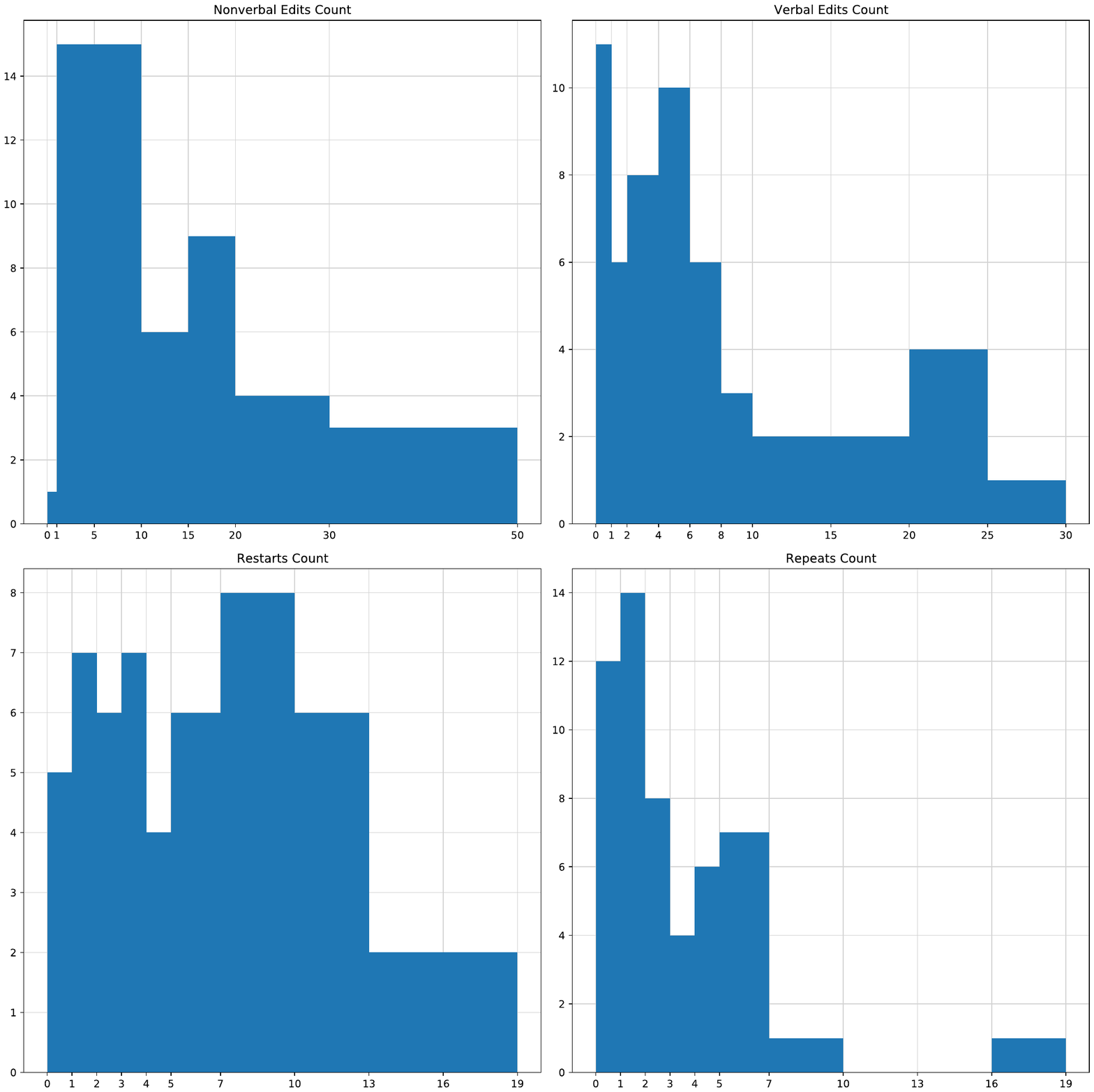}
\caption[Distribution of transcript linguistic disfluency features from a pilot sample of AMPSCZ diaries.]{\textbf{Distribution of transcript linguistic disfluency features from a pilot sample of AMPSCZ diaries.} Using the same transcript set as Figure \ref{fig:diary-pdf1}, I also generated histograms for counts of different categories of linguistic disfluency. The distributions presented over transcripts ($n=53$) here are: the number of nonverbal edits (top left), the number of verbal edits (top right), the number of sentence restarts (bottom left), and the number of word or other utterance repeats (bottom right). Note histogram bin ranges are inclusive on the lower bound and exclusive on the upper bound.}
\label{fig:diary-pdf4}
\end{figure}

\noindent Those features included basic metadata (Figure \ref{fig:diary-pdf1}), a deeper look at TranscribeMe sentence structure (Figure \ref{fig:diary-pdf2}), quality control counts using TranscribeMe notation (Figure \ref{fig:diary-pdf3}), and counts of detected linguistic disfluencies of various categories (Figure \ref{fig:diary-pdf4}). 

Many interesting details can be seen in these data, including observations I was not even expecting to find in this small pilot dataset. For example, the count of speaker IDs present in each transcript identified a handful of journals with multiple speakers (Figure \ref{fig:diary-pdf1}). I manually confirmed that those journals indeed had conversation going on between multiple people, yet another quality issue that automated methods are highly unlikely to pick up on. Background conversation can result in abnormally high incoherence scores in my experience with a prior diary dataset, which with Transcribe speaker IDs is easy to account for both in computational analyses and in interpreting the transcripts in a case report context. With only automated text files available, it would be difficult to determine which extreme examples were real versus an out of context conversation. While this is a minor point in the overall scope of the project, it is important to remember that different types of small quality issues can add up quickly. 

More broadly, counts of markers of uncertainty in transcript quality and of redacted words (Figure \ref{fig:diary-pdf3}) represent information that would be impossible to obtain to a high enough degree of accuracy with automated methods, yet are factors of deep importance in managing a quality data collection process. This sample demonstrates that PII is found in a number of journal submissions, and even though the majority of journals contained no PII, the non-negligible minority that did (at times to a great extent) would "spoil the whole batch" of automated transcriptions so to speak, as far as transfer to the NIH data repository is concerned. 

Similarly, we can see a high level of variance in the transcript sentence structure (Figure \ref{fig:diary-pdf2}) and in the usage rate of all types of linguistic disfluencies (Figure \ref{fig:diary-pdf4}). Of course this does not show that any of these features will turn out to be especially interesting in the proposed research, but it does show that the features vary enough even in a small pilot sample to theoretically contain interesting signals. I find it hard to believe that the entire psychotic disorders speech sampling research community would be unable to find good use for a relatively high temporal resolution signal of different disfluency usage rates over time in the diaries of each of the AMPSCZ subjects.

Note also that Whisper versions of these transcriptions were generated, and a simple check of word error rate apparently yielded very bad results per a recent meeting, though specific numbers and methodological details were not provided to the group, so it is difficult to comment much further. The analysis probably does require a more nuanced review, but it does not bode well and regardless it is fundamentally not capable of addressing many of the questions of section \ref{sec:journals-again}, and cannot address many more at the scale this "experiment" was done at.

One might have expected the advantages of TranscribeMe to be more subtle, and to require more digging to try to demonstrate worth. As it turns out, there is not only an extensive list of nuances that TranscribeMe can capture beyond automated methods, but also a couple hours with a transcript dataset shows numerous advantages for TranscribeMe over Whisper on a surface level too. \\

Much work remains to be done to implement an audio journal processing pipeline however, regardless of transcription decision made. I have concerns about the plan for the future of AMPSCZ in this regard as well, and as such I will next summarize the software engineering resource situation.

\FloatBarrier

\section{Software funding}
\label{sec:nih-sucks}
I will be graduating at the end of May, and at that time my involvement in AMPSCZ will become extremely limited. By the end of summer 2023, any questions I might help with for transitioning the project should be fully resolved. Unfortunately, it is unclear at this time who my role is transitioning to. There seems to be a mistaken assumption that this role only involves maintenance of existing uniformly well written software, which is absolutely not the case. Forgetting the active quality monitoring burden, which really should be reassigned anyway, there is a long list of action items ranging from necessary to highly recommended that remain -- not to mention all the things that would be nice to have, particularly if the NIMH may hope to replicate a similar project structure for another disease area at lesser cost. 

I suspect that many of these issues have arisen out of inadequate high level planning for the unique privacy concerns that impact AV but not many other AMPSCZ datatypes. The fact that raw AV data cannot go to the \emph{predict} analysis server has introduced more software infrastructure challenges and simultaneously made those challenges more important to tackle in a timely fashion than is the case for other modalities. This restriction is unavoidable, but what is not unavoidable is failing to allocate enough software development resources to properly handle necessary infrastructure. The situation was already a stretch, but by not replacing me it will become extremely problematic. 

In the coming weeks, I will focus on finishing my thesis and preparing my defense. I will then have about a month's worth of full time work I can put into wrapping up my involvement in AMPSCZ. In that time I hope to iron out a number of bugs remaining in the present interview data flow and QC code, and finish documenting that code carefully so that others would have the information needed to continue working on it. I also hope to implement a basic version of the audio diary pipeline that can do straightforward accounting and audio quality assurance, and ideally interface with TranscribeMe to get redacted transcripts and transcript QC metrics to the \emph{predict} server as occurs with interviews. \\

\noindent That leaves a large number of major TODOs with no owner, including:
\begin{itemize}
    \item Sites that are using Microsoft Teams or Cisco WebEx for interviews are currently not having their uploads processed at all, support for both these platforms is supposed to be built into the pipeline.
    \item We are not currently utilizing per interview information that is input by sites to REDCap/RPMS, including the AV runsheets. Writing software to clean these data, identify the ways sites have made mistakes in them, determine the parts that would be useful to use in the AV pipeline, and then integrating it all together is an entire project. 
    \item The AV feature extraction plan to occur for interviews on the AV processing server remains unclear, which is another partially code-related problem that could in itself use more resources, but which also requires data flow infrastructure support and the ability to integrate with the QC/accounting pipeline.
    \begin{itemize}
        \item Implementation of the data flow plan drafted with Pronet IT needs to occur (and probably first be refined on) once core feature extraction functions are available, which necessitates an understanding of the data structures being used and which would be greatly facilitated by interfacing with existing pipeline outputs. 
        \item The new SOP violation problems introduced in the feature extraction data flow will need to be better characterized and then intelligently addressed. The naming of diarized Zoom audio files (already a bit of a mess) is the largest such problem, as the identity of the speaker that extracted features came from should be attached to shared outputs. It should not be required by the core feature extraction code, but really should be done as part of the data flow. Similar issues may possibly arise in labeling face identities based on Zoom layout. More complicated computational methods could fill in gaps here, but then that is yet another software task, and one that will require testing with a more in depth manual validation component.
        \item Parts of the feature extraction data flow will be implemented by the Pronet tech team, but there is unlikely a parallel situation available for Prescient. The Prescient AV processing server still needs to be set up in conjunction with Prescient IT, but then it may be a process that falls fully on my theoretical replacement to get the entire data flow set up working there, including adaptation of some code written by Pronet for Pronet.
        \item This entire process will almost certainly involve careful monitoring and adjustments to code in the early stages, which means that it is not just a software time commitment but also a time commitment to organizational tasks and paying close attention to largely boring details. 
    \end{itemize}
    \item There is even less of a plan for any sort of acoustics feature extraction on audio journals to date, and even on the side of data flow/monitoring I suspect my initial diary QC pipeline will not be in as complete of a state as the interview pipeline is when I leave. For example, it will probably require additional work to integrate the outputs with DPDash and to generate all of the expected summary views. It will also probably require additional work to implement monitoring emails with features such as the figures with QC distributional properties per site, which have been a key component of identifying site interview issues in the early stages of the project. While diaries do avoid many of the complications that we encountered with the interview code, I have spent substantially more time on the interview code too, and there are a variety of datatype-specific considerations at play. 
    \item Additionally, there are many TODOs that really should be addressed in the interview code if it is remain robust throughout the duration of AMPSCZ, and doubly so if it is to be successfully used for other AMP-style projects. Some of the primary issues to watch out for include:
    \begin{itemize}
    \item We still lack any way to monitor if sites have fixed most of the SOP issues that prevent interview processing, so we do not have a good way to truly monitor missingness or identify backlogs at particular sites.
    \item As date information is currently being entirely scrubbed from the interview data that goes to \emph{predict}, there is no way for the DPDash combined monitoring dashboard to be viewed in chronological order, so it is very difficult to do recording quality checks using this tool. I am still not entirely clear on the privacy implications, but other modalities are displaying dates in this dashboard, so it seems something changed about what is allowed.
    \item More generally, there is not a functionality currently built to automatically or even semi-automatically update existing intermediates. Therefore when a site inexplicably changes a participant consent date or accidentally uploads a psychs interview under the open type, it requires a tedious manual process to fix the many pipeline outputs affected. 
    \item The Pronet data aggregation server cannot handle processing some of the extremely long psychs uploads, which are currently being rejected by QC due to the memory error -- so some rethinking of how to deal with this scenario is required. 
    \item Most of the non-English languages that are to be covered by AMPSCZ have yet to be tested by site interviews. New bugs are sure to arise, and the transcript QC code is definitely not currently covering markings of e.g. inaudibles in other languages. The only thing that is language-agnostic is the redaction markings, but many other notations will need to be reviewed for each foreign language separately and integrated into the code.
    \item As the manual monitoring workflow is ironed out, there will be a better idea of how the present email alerts could be used most effectively. Refining the content of these and who they go to would pay dividends if clear specifications can be decided on first.
    \end{itemize}
\end{itemize}
\noindent It is important to emphasize that the larger a software undertaking is the more difficult seemingly trivial tasks become. The interview pipeline for example has been built to check for a number of Zoom-dependent SOP violations, so to replicate that for differing conventions in other meeting software is not as simple as converting an MP4. Even if we were to agree that we will treat the edge cases with less care, it still requires updates in multiple spots of the Zoom-related code to ensure that the Teams interviews are not picked up as a Zoom violation without accidentally missing actual Zoom violations. This is just one example to illustrate that throwing together a script for a single lab dataset is not the same thing as actually integrating that process into a large piece of software. \\

While the data processing might not grind to a halt without tackling these problems, it certainly does impact the completeness and quality level of the final AMPSCZ dataset, something I would hope would be a major priority to stay on top of in a large scale data collection project. Furthermore, I would be very hesitant to recommend the use of the pipeline to another project as is without knowing someone is actively working on maintaining it and improving these limitations. With the right investment, good open source infrastructure would pay dividends for digital psychiatry projects for a long time, but at present I do not see this software being especially useful beyond AMPSCZ. 

At a high level, I think future projects might also want to rethink the structure dividing DPACC/Pronet/Prescient, especially in the context of the data transfer limitations of the AV modality. Responsibilities and capabilities are very oddly partitioned, in a way that (even with a group of only the most well-intentioned members) can result in some dropped balls amidst the confusion. DPACC should really be responsible for hiring my replacement, but they are focused primarily on writing more general use infrastructure like Lochness and DPDash, and are quite busy as is. For the AV modalities to fall to further neglect wouldn't be an especially surprising outcome. 

Meanwhile, those who have been working tirelessly to oversee the interview data collection process and even manage the AV relevant software development are not in a position that actually grants them access to the data aggregation servers, nor the ability to hire a replacement for me that they can continue to work with. It makes sense that those most interested in language would have more of a vested interest in the AV software specifically, and would thus put in a large amount of work to keeping things going -- in particular Phil Wolff. The unfortunate side effect of the AMPSCZ structure is that these individuals are needlessly limited because they are not technically part of DPACC. 

It would be one thing to partition roles in a well-established project with a very concrete scientific and engineering plan, ideally based on specific proposals that are actually being carried out, but AMPSCZ is not this. When barriers are so artificially constructed in a collaborative project, it is natural they would lead to awkward and inefficient dynamics. To what extent this pervades the project I can't say, but I do feel it is relevant to the hiring of my replacement.

\section{Closing arguments}
\label{sec:guillermo-is-annoying}
In summary, I've laid out many arguments both against the current AMPSCZ psychs transcription plan (\ref{sec:psychs-dumb}) and for obtaining professional transcriptions of the AMPSCZ audio journal recordings (\ref{sec:journals-again}). This included both a number of theoretical points as well as concrete demonstrations of the value inefficiency of present psychs transcriptions (\ref{sec:open-psychs-data}) and proof of concept demonstration of a few TranscribeMe advantages in a small journal subset (\ref{sec:diary-dataset-u24}). I also outlined estimated budgets for various transcription proposals (\ref{sec:blowing-money}-\ref{sec:diary-cost-estimates}), identifying an easy way that all journals could be transcribed by cutting back only in part on which psychs interview recordings are sent for transcription (though one might certainly consider cutting back more). Given, I will be leaving soon, I also reviewed concerns about the great deal of work that remains for AV modality software infrastructure in AMPSCZ (\ref{sec:nih-sucks}), and why it is necessary that the incentive structures created by AMPSCZ be more carefully revisited in the discussion about whether/how to hire a replacement for my role. 

I will now close this supplement with a brief discussion of the value of audio diaries in a more general scientific sense, and a few thoughts on how this ought to relate to the vision of AMPSCZ. \\

\noindent Audio journals are uniquely well-suited to provide scientific opportunities for a wide range of psychiatry studies. Compared to interview recordings, diaries have better temporal resolution, are easier to collect from a larger number of participants, are easier to collect from a given participant over a longer study period, enable capture of open ended self-reporting in a truly patient-driven manner, and produce individual time points that are more realistic to adequately summarize without requiring a large feature set or complicated context-aware algorithms. Although there are advantages to the interview recording format for some specific types of scientific question, in the general case audio journals are substantially more tractable for exploratory data science work as well as for hypothesis-driven work that addresses longitudinal questions or requires a more heterogeneous study population. Furthermore, they have great promise for novel insights both qualitative and quantitative, as they remain much less studied than clinical interviews; to truly realize the potential of emerging digital psychiatry tools, it will be necessary to obtain data that is more naturalistic and more temporally dense.

By contrast, digital phenotyping datatypes like phone GPS or wrist accelerometry have much greater temporal coverage than audio journals and are even easier to collect from many subjects due to their passive nature. However, it is difficult to interpret these datatypes without clinical or self-report context, and it is difficult for primarily psychiatry-focused groups to carefully analyze such large timeseries with relatively infrequent labels. There are certainly scientific questions of interest to be addressed with passive sensing datatypes alone, but there are also many relevant questions that would benefit from the availability of audio journal data. Analysis of patient speech is an important ongoing direction for psychiatry due to its established clinical evidence base and its relationship with well characterized biological principles. The connection between psychomotor signs and impacted speech patterns alone warrants more work linking audio journals with passive sensing data. Additionally, the open ended patient self-reporting enabled by daily diaries can be used to contextualize passive data in unique ways that would be difficult to achieve with surveys alone, for example detailed accounts of exercise routines. Meanwhile, we have found submission rates of EMA to coincide very well with submission rates of audio journals, such that it is straightforward to obtain time-aligned datasets of these two modalities that can be used powerfully in conjunction with digital phenotyping (or just by themselves!). 

Audio journals are therefore highly versatile in their abilities, capable of simultaneously bringing data science tractability to more traditional psychiatric speech sample studies and bringing qualitative clinically interesting insights to studies centered on processing dense data streams. A speech sampling study could feasibly perform an exploratory analysis with actual statistical power if audio journals were used, while a passive sensing digital psychiatry study would be much more equipped to generate interesting follow-up hypotheses with access to journals. While it may not be appropriate for every group, the digital psychiatry field at large should not underrate the potential value in autobiographical accounts, as it has thus far. Quantitative and qualitative work need not be mutually exclusive, nor do objective and subjective metrics. The daily audio diary format can be well utilized in all 4 ways, and it can be well utilized in conjunction with many different datatypes or even entirely on its own. 

There is a broader philosophical point here as well, in that hypothesis-driven work which sticks closely to formal scientific processes during hypothesis generation (or in order to justify a hypothesis for a grant) is much more likely to be incremental. Incremental work is an important part of science, but it cannot overtake too much, lest it will limit the paths we can possibly take. On the other hand, exploratory works often tend to take a "throw shit at the wall" approach that may be appropriate at times, but is not ideal in a field where great resources are needed to obtain a dataset of even modest size. AMPSCZ is the type of project that could allow for exploratory work that achieves reasonable breadth without being too woefully underpowered, but more broadly, I feel there should be more acceptable "official" avenues for hypothesis generation and even results interpretation. On obvious such avenue in the realm of psychiatry is to consider patient autobiographical accounts in a more formal research context, and audio journals are a great way to do this (among many other things). What participants choose to discuss in these open ended daily recordings is entirely up to them, and I have seen quite a range of interesting comments in the limited amount of time I've been looking at these. Thus it is a way of discovering ideas where the patient narrows the search space, instead of beginning narrow or hardly pruning the search space at all. \\

There are many reasons then that I feel it is critical for high quality journal transcripts from across AMPSCZ to reach the NIH data repository. Audio journals are one of the most fundamentally interdisciplinary datatypes I've come across in digital psychiatry so far, yet there is such a paucity of literature on them to date. A large dataset available to the broader research community along with AMPSCZ's support could go a long way towards driving interdisciplinary collaborations in this space, and pushing forward the psychiatric speech sampling field by encouraging more frequent use of journals. Regardless of your agreement with my views on audio journals, I would hope we can all agree that this is the \emph{type} of contribution AMPSCZ should aim to be making. It would be a big disappointment to spend over 100 million dollars on an initiative that does not adequately leverage its large scale/collaborative/multimodal advantages and instead pushes along the status quo, especially if that project doesn't produce an accessible high quality dataset either. The relative cost of transcribing some diaries and paying for some software maintenance is minuscule, and on top of that there are already obvious inefficiencies that could be budgeted out. 

Finally, it is worth reiterating that machine learning advances are \emph{not} being benchmarked on psychiatry considerations by default, and at times have aims that are in fact the opposite of what a clinician might find worthwhile. Unless there are hidden plans to consult directly with pure machine learning research groups (which would be great!), AMPSCZ cannot assume that those groups are developing technology that will work the way it should for psychiatry research applications out of the box -- no matter how long of a wait. This project indeed has the opportunity to provide a dataset that might be large enough to actually bring some ML advances to psychiatry in an appropriate way. But it needs to take the actions that will enable that, because large initiatives should be leading the way at their cost, not waiting around for an unspecified period of time for someone else to get models to an unspecified performance level. There is plenty of work remaining to be done (as I've detailed), and there are applied ML fields much further along than psychiatry that do continue to do research, because in most cases there are many domain specific issues to be resolved. For example, take a look at the many authors and the heterogeneity in their academic backgrounds on the AlphaFold paper. If you believe that there are big differences in speech in psychotic illness, then you should be fundamentally opposed to throwing models at the problem without proper domain-specific validation. If you don't, then what is the hypothesis of this subgroup? Regardless, what exactly is AMPSCZ speech sampling hoping to accomplish, if it is not leading the way on ML adaptations for psychiatric speech and it is not leading the way on establishing tools and standards for emerging speech sources?

\end{appendices}

\end{document}